 \documentclass[a4paper,twoside,12pt]{mybook}
\pdfoutput=1
\UseRawInputEncoding
\usepackage{amssymb,amsmath}
\usepackage{makeidx}
\usepackage{subeqnarray}
\usepackage{graphicx}
 \usepackage[colorlinks=false,hidelinks]{hyperref}
\usepackage{fancyh1}
\usepackage{color}
\usepackage{natbib}

\newcommand{\beq}{\begin{equation}}  
\newcommand{\eeq}{\end{equation}}    
\newcommand{\beqa}{\begin{eqnarray}} 
\newcommand{\eeqa}{\end{eqnarray}}   
\renewcommand{\chaptermark}[1]{\markboth{#1}{#1}}

\makeindex

\begin{document}

\def\today{\number\day\space\ifcase\month\or
  January\or February\or March\or April\or May\or June\or
  July\or August\or September\or October\or November\or December\fi,
  \number\year}

\def\etal{{\it et al.\/}}
\def\ie{{\it i.e.\/}}
\def\eg{{\it e.g.\/}}
\def\etc{{\it etc.\/}}
\def\1o2{\textstyle {\frac{1}{2}}}
\def\alphab{\mbox{\boldmath $\alpha$}}
\def\bcb{\mbox{\boldmath ${\cal B}$}}
\def\betab{\mbox{\boldmath $\beta$}}
\def\d{{\rm d}}
\def\dcb{\mbox{\boldmath ${\cal D}$}}
\def\deltab{\mbox{\boldmath $\delta$}}
\def\Deltab{\mbox{\boldmath $\Delta$}}
\def\dotprod{\!\cdot\!}
\def\e{{\rm e}}
\def\ecb{\mbox{\boldmath ${\cal E}$}}
\def\epsilonb{\mbox{\boldmath $\epsilon$}}
\def\fcb{\mbox{\boldmath ${\cal F}$}}
\def\gammab{\mbox{\boldmath $\gamma$}}
\def\Gammab{\mbox{\boldmath $\Gamma$}}
\def\hcb{\mbox{\boldmath ${\cal H}$}}
\def\intd{\int\!\!\int}
\def\intt{\int\!\!\int\!\!\int}
\def\kappab{\mbox{\boldmath $\kappa$}}
\def\lambdabar{\lambda \! \! \! \!
^{\rule{0.8mm}{0mm}\rule[0.25mm]{1.6mm}{0.12mm}}\rule{0.15mm}{0mm}}
\def\mcb{\mbox{\boldmath ${\cal M}$}}
\def\me{m_{\rm e}}
\def\nablab{\mbox{\boldmath $\nabla$}}
\def\omegab{\mbox{\boldmath $\omega$}}
\def\Omegab{\mbox{\boldmath $\Omega$}}
\def\pcb{\mbox{\boldmath ${\cal P}$}}
\def\phib{\mbox{\boldmath $\phi$}}
\def\Phib{\mbox{\boldmath $\Phi$}}
\def\pib{\mbox{\boldmath $\pi$}}
\def\Pib{\mbox{\boldmath $\Pi$}}
\def\psib{\mbox{\boldmath $\psi$}}
\def\Psib{\mbox{\boldmath $\Psi$}}
\def\sigmab{\mbox{\boldmath $\sigma$}}
\def\Sigmab{\mbox{\boldmath $\Sigma$}}
\def\taub{\mbox{\boldmath $\tau$}}
\def\thetab{\mbox{\boldmath $\theta$}}
\def\varepsb{\mbox{\boldmath $\varepsilon$}}
\def\varphib{\mbox{\boldmath $\varphi$}}
\def\vecprod{\!\times\!}
\def\xib{\mbox{\boldmath $\xi$}}
\def\ycb{\mbox{\boldmath ${\cal Y}$}}
\def\zetab{\mbox{\boldmath $\zeta$}}
% ...
\newcounter{listr}
\def\blr{\begin{list} {(\roman{listr})} {\usecounter{listr}}}
\def\elr{\end{list} \setcounter{listr}{0}}
\def\bla{\begin{list} {(\arabic{listr})} {\usecounter{listr}}}
\def\ela{\end{list} \setcounter{listr}{0}}
% ...
\def\boxeq#1{$$\mbox{\fbox{$\displaystyle{#1}$}}$$}
\def\boxeqn#1{\begin{equation}\mbox{\fbox{$\displaystyle{#1}$}}
\end{equation}}
% ...
%
\def\req#1{(\ref{#1})}
%


\def\title#1{\vglue0.05truein{\baselineskip=24 truept
    \pretolerance=10000
    \raggedright\noindent \LARGE\bf #1 \par}\vskip1.25cm}
%
\def\author#1{{\pretolerance=10000\raggedright\noindent
               {\large\bf #1}\par}\vskip-0.25cm}
%
\def\address#1{\bigskip\noindent {\rm #1} \par}
%
\def\addinfo#1{\bigskip\noindent {\rm #1} \par \smallskip}
%


\thispagestyle{empty}

\begin{center}

\vspace*{3.5cm}


{\Huge\bf
COLLISIONS AND STOPPING  \\ [2mm]
OF FAST CHARGED PARTICLES  \\ [6mm]
IN MATTER
}


\vspace*{2cm}

{\LARGE\bf Francesc Salvat}

\end{center}

\vfill

\begin{minipage}[b]{10cm}
\noindent
{\bf Universitat de Barcelona. Facultat de F\'{\i}sica \\
Departament de F\'{\i}sica Qu\`{a}ntica i Astrof\'{\i}sica \\
\rule{0mm}{0mm}}
\end{minipage}

\newpage

\thispagestyle{empty}

\vspace*{45mm}

\vspace*{25mm}

\noindent Francesc Salvat \\
Facultat de F\'{\i}sica (FQA) \\
Universitat de Barcelona \\
Diagonal 645 \\
08028 Barcelona

\noindent e-mail: {\tt francesc.salvat@ub.edu} \\
\phantom{e-mail:} {\tt cesc@fqa.ub.edu} \\ [4mm]
Last edit: \today

\vspace{10mm}

  \clearpage

  \pagenumbering{roman}
  \lhead[\fancyplain{}{\sl\thepage}]{\fancyplain{}{\sl Contents}}
  \rhead[\fancyplain{}{\sl Contents}]{\fancyplain{}{\sl\thepage}}
  \newpage
  \pagestyle{fancyplain}
  \addtolength{\baselineskip}{-1.0mm}
  \tableofcontents
  \addtolength{\baselineskip}{+1.0mm}
  \pagestyle{fancyplain}
\newpage\hbox{}\thispagestyle{empty} 


\lhead[\fancyplain{}{\sl\thepage}]{\fancyplain{}{\sl\rightmark}}
\rhead[\fancyplain{}{\sl\leftmark}]{\fancyplain{}{\sl\thepage}}


\chapter*{Foreword}
\addcontentsline{toc}{chapter}{Foreword}
\markboth{\sl Foreword}{\sl Foreword}
\pagestyle{fancyplain}

Basic concepts of the theory of atomic collisions and of the stopping of
charged particles used to be part of undergraduate courses on atomic and
nuclear physics. Unfortunately, they have practically been removed from
most engineering and physics degree studies, with the result that
researchers in applied fields find it difficult to obtain information on
fundamental aspects of the interaction of particle beams with matter.
Although an elementary introduction to the theory of collisions of
charged particles with atoms is given in textbooks of classical and
quantum mechanics, numerical results are seldom available. The theory of
stopping is found only in advanced textbooks and monographs, and too
frequently limited to the classical approach of Bohr. Additionally, a
good deal of information is dispersed in original publications, which
may not be accessible or known to the non-specialist.

At the same time, with the availability of fast computers, Monte Carlo
simulation codes of radiation transport have been developed and made
freely available to users who are frequently unaware of the
approximations involved in these codes. As a consequence, simulations
may be run inadvertently for situations that are at the edge of, or
beyond the range of applicability of a code.

The present text is intended to offer a consistent presentation of the
theory of collisions and stopping of charged particles in matter,
limited to the range of intermediate kinetic energies where atomic
aggregation effects are relatively unimportant and processes such as the
creation of particle-antiparticle pairs are not likely to occur. I have
made an effort to address the main aspects of interest to developers and
users of Monte Carlo codes for the simulation of the transport of
charged particles, under the assumption that the reader has followed
undergraduate courses on classical mechanics, classical electrodynamics,
and quantum mechanics. The main results of the theory are derived in
much detail and presented in a form suited for numerical evaluation.

The first three Chapters contain introductory material on the classical
description of electromagnetic fields in matter, an overview of quantum
wave equations for a particle in a central potential, and an account of
elementary atomic-structure models. The concepts in these preliminary
Chapters are the basis of most of the arguments presented in later parts
of the text. Chapter 4 is devoted to the classical theory of scattering
by a central field and elastic two-body collisions. The quantum
theory of elastic collisions is presented in Chapter 5. The theory of
inelastic collisions and stopping is split into two parts. First, the
case of collisions with atoms is considered within the plane-wave Born
approximation in Chapter 6, which includes a derivation of the Bethe
stopping power formula. The theory of inelastic collisions in dense
materials is based on the dielectric formalism, which is formulated for
the electron gas, and extended to arbitrary materials by
means of optical-data models in Chapter 7. Chapter 8 offers a detailed
review of the theory of stopping, starting with the classical study by
Bohr and ending with derivations of the Bloch and Barkas corrections to
the stopping power.

Chapter 9 deals with general aspects of transport theory. It includes
derivations of energy-straggling distributions and multiple-scattering
distributions, which are the basis for condensed simulation schemes of
charged particle transport. Finally, Chapter 10 describes the Fortran
programs {\sc elastic} and {\sc sbethe} which implement the main
theoretical models presented in Chapters 4 to 9 for computing elastic
collisions of particles with atoms and the stopping of particles in
matter. These programs give results in fair agreement with
experiments in wide energy ranges, with a minimal amount of input
information. They can be used not only to generate reliable interaction
data but also as pedagogical tools for illustrating quantitative aspects
of the theory.

I would like to thank Josep Llosa for his continuous help and advise on
the essentials of classical electrodynamics, and to Antonio M. Lallena
for reading various parts of the manuscript and suggesting
clarifications. I am also indebted to Pedro Andreo, who read the
manuscript in its early stages and suggested various improvements of the
text, and to my son, Francesc Salvat-Pujol, for his critical reading of
the Chapter on transport theory.

\begin{flushright}
F.S. \\ [2mm]
Barcelona, June 2026.
\end{flushright}

\chapter*{List of acronyms}
\addcontentsline{toc}{chapter}{List of acronyms}
\markboth{\sl List of acronyms}{\sl List of acronyms}

The following acronyms and abbreviations occur repeatedly in the text.

\vspace*{5mm}
\noindent
BEA: binary-encounter approximation. \\ [2mm]
CM: center-of-mass reference frame. \\ [2mm]
CSDA: continuous slowing-down approximation. \\ [2mm]
DIMFP: differential inverse mean free path. \\ [2mm]
DCS: differential cross section. \\ [2mm]
DDCS: doubly-differential cross section.\\ [2mm]
DF: dielectric function. \\ [2mm]
DHFS: Dirac--Hartree--Fock--Slater self-consistent method, and screening
function. \\ [2mm]
DWBA: distorted-wave Born approximation. \\ [2mm]
ELF: energy-loss function. \\ [2mm]
GOS: generalized oscillator strength. \\ [2mm]
IA: impulse approximation. \\ [2mm]
L: laboratory reference frame. \\ [2mm]
LPA: local-plasma approximation. \\ [2mm]
ODF: optical dielectric function. \\ [2mm]
OELF: optical energy-loss function. \\ [2mm]
OOS: optical oscillator strength. \\ [2mm]
PWBA: plane-wave Born approximation. \\ [2mm]
TFM: Thomas--Fermi--Moli\`{e}re screening function. \\ [2mm]
TGOS: transverse generalized oscillator strength. \\ [2mm]
WKB: Wentzel--Kramers--Brillouin approximation. \\  [2mm]

\newpage




\chapter{Classical electromagnetic fields in matter \label{chapt1}}

   \pagenumbering{arabic}
   \pagestyle{fancyplain}

In this Chapter we present the concepts and methods of classical
electrodynamics that are needed for describing the slowing down of fast
charged particles in material media. Although a consistent theoretical
description of the stopping process can be obtained only by using the
methods of quantum electrodynamics and many-body theory, classical
electrodynamics provides conceptually simple models that are useful for
understanding the physical principles that govern the interactions of
charged particles with matter.

\index{Gaussian system of units}
We use the Gaussian-cgs system of units (see Appendix \ref{appC})
throughout the text, with all fundamental constants indicated
explicitly. The numerical values of fundamental constants are taken from
the latest CODATA least-squares adjustment
(\url{https://physics.nist.gov/cuu/Constants/index.html}).


\section{Electromagnetic fields in vacuum \label{sec1.1}}

\index{Maxwell equations!in vacuum|(}
The electromagnetic field produced by arbitrary distributions of
electric charge and current, $\rho({\bf r},t)$ and ${\bf j}({\bf r},t)$,
in vacuum satisfies the Maxwell equations, which in the Gaussian system
of units read \citep[see, \eg][]{Jackson1975}
\begin{subequations}
\label{1.1}
\beqa
& \nablab \dotprod \ecb({\bf r},t) = 4 \pi \rho({\bf r},t)
& \rule{5mm}{0mm} \mbox{(Coulomb law),}
\label{1.1a} \\ [2mm]
& \displaystyle{\nablab \vecprod \bcb({\bf r},t)
- \frac{1}{c} \frac{\partial \ecb({\bf r},t)}{\partial t}
= \frac{4\pi}{c} {\bf j}({\bf r},t)}
& \rule{5mm}{0mm} \mbox{(Amp\`{e}re-Maxwell law),}
\label{1.1b}\\ [2mm]
& \displaystyle{\nablab \vecprod \ecb({\bf r},t)
+ \frac{1}{c} \frac{\partial \bcb({\bf r},t)}{\partial t} = 0}
& \rule{5mm}{0mm} \mbox{(Faraday law),}
\label{1.1c} \\ [2mm]
& \nablab \dotprod \bcb({\bf r},t) = 0
& \rule{5mm}{0mm} \mbox{(there are no magnetic monopoles),}
\rule{11mm}{0mm}
\label{1.1d}\eeqa
\end{subequations}
where $\ecb$ is the electric field and $\bcb$ is the magnetic induction;
$c$ is the speed of light in vacuum.
The charge and current distributions satisfy the continuity
equation, \index{continuity equation}
\beq
\frac{\partial \rho}{\partial t} + \nablab \dotprod {\bf j} = 0,
\label{1.2}\eeq
which follows from the first pair of Maxwell equations and expresses the
conservation of charge. The force produced by the electromagnetic field
on a particle of charge\footnote{The symbol $e$ denotes the elementary
charge, \ie, the charge of a proton. See Appendix \ref{appC}.} $Z_0 e$
that moves with velocity  ${\bf v}$ is given by the Lorentz formula,
\index{Lorentz force}
\beq
\textbf{F} = Z_0 e \left( \ecb + \frac{1}{c} {\bf v} \vecprod \bcb \right).
\label{1.3}\eeq

Equations \req{1.1c} and \req{1.1d} imply that we can introduce
scalar and vector potentials $\varphi$ and ${\bf A}$ such that
\index{electromagnetic potentials}
\beq
\ecb = - \nablab \varphi -
\frac{1}{c} \frac{\partial {\bf A}}{\partial t}, \qquad
\bcb = \nablab \vecprod {\bf A}.
\label{1.4}\eeq
These potentials are not uniquely defined; a gauge transformation of the
potentials leaves the fields $\ecb$ and $\bcb$ unchanged \citep[see,
\eg,][]{Jackson1975}. Here we shall adopt the Coulomb gauge, \ie, we shall
require the potential vector to be solenoidal or transverse:
\index{Coulomb gauge}
\beq
\nablab \dotprod {\bf A} = 0.
\label{1.5}\eeq
Then, Eq.\ \req{1.1a} implies that the scalar potential
satisfies the Poisson equation \index{Poisson equation}
\beq
\nabla^2 \varphi = - 4 \pi \rho,
\label{1.6}\eeq
whose solution is the instantaneous (unretarded) Coulomb potential
\beq
\varphi({\bf r},t) = \int \frac{\rho({\bf r}',t)}{|{\bf r} - {\bf r}'|}
\d {\bf r}'.
\label{1.7}\eeq
Inserting the expressions \req{1.4} into the Maxwell
equation \req{1.1b}, with the aid of the relation
\beq
\nablab \vecprod ( \nablab \vecprod {\bf A} ) = \nablab
(\nablab \dotprod {\bf A} ) - \nabla^2 {\bf A},
\label{1.8}\eeq
which is valid for any well-behaved vector field ${\bf A}$,
we obtain the following inhomogeneous wave equation for the vector
potential,
\beq
\nabla^2 {\bf A} - \frac{1}{c^2} \frac{\partial^2 {\bf A}}{\partial t^2}
= - \frac{4\pi}{c} {\bf j} + \frac{1}{c} \nablab \frac{\partial
\varphi}{\partial t}.
\label{1.9}\eeq

It is worth mentioning that the equations of electrodynamics in vacuum
are covariant, that is, they are valid in any inertial reference frame
\citep[see, \eg,][]{Jackson1975}. Under Lorentz transformations, the
densities of charge and current, $(c\rho, {\bf j})$, and the potentials,
$(\varphi,{\bf A})$, transform as four-vectors. The components of the
electric and magnetic fields are the elements of a second-rank
antisymmetric field-strength tensor.

\index{Maxwell equations!in vacuum|)}


\subsection{Electromagnetic field of a moving charged particle
\label{sec1.l.1}}

\index{electromagnetic field!of a moving charge|(}
The electromagnetic field produced by a moving charged particle is of
fundamental interest in stopping theory.
Let us consider a particle of charge $Z_0 e$ that follows a certain trajectory
${\bf r}_0(t)$, with velocity ${\bf v}_0(t) = \d {\bf r}_0/\d t$
and acceleration ${\bf a}_0 (t) = \d {\bf v}_0 / \d t$, in vacuum.
The electric field and the magnetic induction produced by that particle
at the position ${\bf r}$ and at the time $t$ are given
by the Li\'{e}nard-Wiechert formulas \index{Li\'{e}nard--Wiechert field}
(see Section 14.1 of
\citeauthor{Jackson1975}, \citeyear{Jackson1975}; or \textsection 63 of
\citeauthor{LandauLifshitz1971}, \citeyear{LandauLifshitz1971}),
\begin{subequations}
\label{1.10}
\beqa
\ecb({\bf r},t) &=&
A_0 \left[ \rule{0mm}{4mm} \hat{\bf n}_0 - \betab_{0}(t_R) + B_0 \,
\hat{\bf n}_0 \vecprod
\left\{ \rule{0mm}{4mm}
\left[ \hat{\bf n}_0 - \betab_{0}(t_R) \right] \vecprod {\bf a}_{0}(t_R)
\right\} \right],
\label{1.10a} \\ [2mm]
\bcb({\bf r},t) & = & \hat{\bf n}_0 \vecprod \ecb({\bf r},t),
\label{1.10b}\eeqa
\end{subequations}
with
\begin{subequations}
\label{1.11}
\beq
A_0 = \frac{Z_0 e}{\left| {\bf r} - {\bf r}_0(t_R) \right|^2
\gamma_{0}^2(t_R)
\left[1 - \betab_{0}(t_R) \dotprod \hat{\bf n}_0 \right]^3}
\label{1.11a} \eeq
and
\beq
B_0 = \frac{\left| {\bf r} - {\bf r}_0(t_R) \right|
\,\gamma_{0}^2(t_R)}{c^2}\, ,
\label{1.11b}\eeq
\end{subequations}
where
\beq
\hat{\bf n}_0 = \frac{{\bf r} - {\bf r}_0(t_{R})}{\left|{\bf r} - {\bf
r}_0(t_{R}) \right|}
\label{1.12}\eeq
is the unit vector in the direction from the particle at ${\bf
r}_0(t_R)$ to the
field point ${\bf r}$, $\betab_0 (t_R)= {\bf v}_0(t_R)/c$ is the velocity of
the particle\footnote{Throughout the text we use the
relativistic quantities $\beta=v/c$ and $\gamma = (1-\beta^2)^{-1/2}$
(see Appendix \ref{appA}).} in units of $c$, and $\gamma_0^{-2} (t_R) = 1
-\beta_0^2(t_R)$. Notice that the quantities referring to the
moving particle are ``retarded'': they are evaluated at
the instant $t_R$ such that
\beq
c \left( t - t_R \right) = \left| {\bf r} - {\bf r}_0(t_R) \right| \, .
\label{1.13}\eeq
That is, the electromagnetic field at ${\bf r},t$ is determined by the
state of the particle at the earlier time $t_R$. The difference $t-t_R$
is the time it takes a light pulse emitted from the position of the
particle ${\bf r}_0(t_R)$ at time $t_R$ to arrive at the field point
${\bf r}$ at the time $t$.

Because the implicit equation \req{1.13} is generally difficult to
solve, we limit our considerations to particles with small acceleration,
${\bf a}_0 \simeq {\bf 0}$, so that $\betab_{0} (t_R) = \betab_0$ is
constant with time. In
this situation, the electromagnetic field reduces to
\beq
\ecb ({\bf r},t) =
\frac{Z_0 e \, (\hat{\bf n}_0 - \betab_0)}
{\left| {\bf r} - {\bf r}_0(t_R) \right|^2 \gamma_0^2
\left(1 - \betab_0 \dotprod \hat{\bf n}_0 \right)^3},
\qquad
\bcb({\bf r},t) = \hat{\bf n}_0 \vecprod \ecb({\bf r},t),
\label{1.14}\eeq
and a formal solution of Eq.\ \req{1.13} can be readily obtained as
follows. We have
\beq
{\bf r}_0(t_R) =  {\bf r}_0 (t) - (t-t_R) \, c \betab_0.
\label{1.15}\eeq
Therefore,
\beq
{\bf r} - {\bf r}_0(t_R) = {\bf d} + (t-t_R) \, c \betab_0
\label{1.16}\eeq
with
\beq
{\bf d} = {\bf r}-{\bf r}_0(t).
\label{1.17}\eeq

The unit vector $\hat{\bf n}_0$, Eq.\ \req{1.12}, can be expressed
as
\beq
\hat{\bf n}_0 = \frac{{\bf r} - {\bf r}_0(t_{R})}{\left|{\bf r} - {\bf
r}_0(t_{R}) \right|}
= \frac{{\bf d} + (t-t_{R})\, c \betab_0}{ c (t-t_{R})} = \xi \hat{\bf
d} + \betab_0.
\label{1.18}\eeq
The quantity
\beq
\xi \equiv \frac{d}{c(t-t_{R})}
\label{1.19}\eeq
is positive. Its value is determined by the fact that $\hat{\bf n}_0$ is
a unit vector, \ie,
\beq
\xi^2 + 2 (\betab_0 \dotprod \hat{\bf d}) \xi + \beta_0^2 = 1.
\nonumber \eeq
The positive root of this equation is
\beq
\xi = - \betab_0 \dotprod \hat{\bf d} +
\sqrt{ (\betab_0 \dotprod \hat{\bf d})^2 + \gamma_0^{-2}}.
\label{1.20}\eeq
Now the expressions \req{1.14} of the electromagnetic field
can be reduced to a form that does not involve retarded times.
This is accomplished by noting that, from Eq.\ \req{1.19},
\beq
\left| {\bf r} - {\bf r}_0 (t_R) \right| = c(t-t_{R}) = \frac{d}{\xi}
\label{1.21}\eeq
and, from Eq.\ \req{1.18},
\beqa
1 - \betab_0 \dotprod \hat{\bf n}_0 &=&
1 - \beta_0^2 - \xi (\betab_0 \dotprod \hat{\bf d})
= \gamma_0^{-2} +
(\betab_0 \dotprod \hat{\bf d})^2 -
(\betab_0 \dotprod \hat{\bf d})
\sqrt{ (\betab_0 \dotprod \hat{\bf d})^2 + \gamma_0^{-2}}
\nonumber \\ [2mm]
&=& \xi \sqrt{ (\betab_0 \dotprod \hat{\bf d})^2 + \gamma_0^{-2}},
\label{1.22}\eeqa
where we have used the equality \req{1.20}. With all these
relations, the electromagnetic field \req{1.14} is expressed as
\begin{subequations}
\label{1.23}
\beq
\ecb ({\bf r},t)
= \frac{Z_0 e \, \xi \hat{\bf d}}
{\left( d/\xi \right)^2 \gamma_0^2 \, \xi^3
\left[ (\betab_0 \dotprod \hat{\bf d})^2 + \gamma_0^{-2}
\right]^{3/2}}
= \frac{Z_0 e \, \gamma_0 \, {\bf d}}
{\left[ \gamma_0^2 (\betab_0 \dotprod {\bf d})^2 +
d^2 \right]^{3/2}}
\label{1.23a}\eeq
and
\beq
\bcb({\bf r},t) = \hat{\bf n}_0 \vecprod \ecb({\bf r},t)
= \frac{Z_0 e \, \gamma_0 \, \betab_0 \vecprod {\bf d}}
{\left[ \gamma_0^2 (\betab_0 \dotprod {\bf d})^2 +
d^2 \right]^{3/2}} \, .
\label{1.23b}\eeq
\end{subequations}
For a particle at rest, $\beta_0=0$, the electromagnetic field reduces to
the familiar {\it Coulomb field},
\beq
\ecb ({\bf r},t)
= \frac{Z_0 e \, {\bf d}}{d^3}
\quad \mbox{and} \quad
\bcb({\bf r},t)
= {\bf 0}.
\label{1.24}\eeq

Introducing the angle $\psi$ between the vectors $\betab_0$ and ${\bf
d}$, \ie, $\cos\psi = \hat{\betab}_0 \cdot \hat{\bf d}$, the denominator in
the expressions \req{1.23} can be modified as follows,
\beqa
\gamma_0^2 (\betab_0 \dotprod {\bf d})^2 + d^2
&=& d^2 \left[
(\gamma_0^2 \beta_0^2+1) \cos^2 \psi + \sin^2\psi \right]
= d^2 \left[ \gamma_0^2 \cos^2 \psi + \sin^2\psi \right]
\nonumber \\ [2mm]
&=& d^2 \gamma_0^2 \left[ \cos^2 \psi + (1 - \beta_0^2)
\sin^2\psi \right]
= d^2 \gamma_0^2 \left( 1 - \beta_0^2 \sin^2\psi \right).
\label{1.25}\eeqa
Hence, we can write
\begin{subequations}
\label{1.26}
\beq
\ecb ({\bf r},t)
= \frac{Z_0 e {\bf d}}{d^3 \gamma_0^2
\left( 1 - \beta_0^2 \sin^2 \psi \right)^{3/2}} \, ,
\label{1.26a}\eeq
and
\beq
\bcb({\bf r},t) =
\frac{Z_0 e \, \betab_0 \vecprod {\bf d}}
{d^3 \gamma_0^2
\left( 1 - \beta_0^2 \sin^2 \psi \right)^{3/2}} \, .
\label{1.26b}\eeq
\end{subequations}
The electric field is ``radial'' (\ie, parallel to ${\bf d}$) but differs
from the isotropic field of a charge at rest. The field along the
direction of motion ($\psi=0,\pi$) is diminished by a factor
$\gamma_0^{-2}$, while the field in the transverse direction
($\psi=\pi/2$) is larger by a factor of $\gamma_0$. The magnetic
induction is perpendicular to the plane that contains the vectors
$\betab_0$ and ${\bf d}$.

Because the formulas \req{1.26} have been derived without referring to a
particular reference frame, they hold in any inertial frame, provided
only that the acceleration of the particle is small. In particular, for
a particle moving in the direction of the $z$ axis with velocity ${\bf
v}_0 = \beta_0 c \hat{\bf z}$ that passes the origin of coordinates at
$t=0$ we have
\beq
{\bf d} = {\bf r} - \beta_0 ct \hat{\bf z} = ( x, y, z-v_0 t).
\label{1.27}\eeq
Since $\gamma_0^2\beta_0^2+1=\gamma_0^2$, we can write
\begin{subequations}
\label{1.28}
\beqa
\ecb ({\bf r},t)
&=& \frac{Z_0 e \, \gamma_0 \, ( x, y, z-v_0 t)}
{\left[ \gamma_0^2 \beta_0^2 (z-v_0 t)^2 + x^2 + y^2 +
(z-v_0 t)^2 \right]^{3/2}}
\nonumber \\ [2mm]
&=& \frac{Z_0 e \, \gamma_0 \, ( x, y, z-v_0 t)}
{\left[ x^2 + y^2 + \gamma_0^2 (z-v_0 t)^2 \right]^{3/2}},
\label{1.28a}\eeqa
and
\beq
\bcb({\bf r},t)
= \frac{Z_0 e \, \gamma_0 \beta_0 \, ( -y, x, 0)}
{\left[ x^2 + y^2 + \gamma_0^2 (z-v_0 t)^2 \right]^{3/2}}.
\label{1.28b}\eeq
\end{subequations}
The result \req{1.28} is also obtained as the Lorentz transform of the
Coulomb field in the rest frame of the particle \citep[see, \eg][Section
11.10]{Jackson1975}.
\index{electromagnetic field!of a moving charge|)}

It is worth noticing that retardation effects may be removed by
setting $c \rightarrow \infty$. The unretarded field of a charged
particle moving at constant velocity ${\bf v}_0$ is given by Eqs.\
\req{1.14} with $\beta_0 = 0$ and $\hat{\bf n}_0 = \hat{\bf d}$,
\beq
\ecb ({\bf r},t) =
\frac{Z_0 e \, \hat{\bf d}}
{d^2},
\qquad
\bcb({\bf r},t) = \hat{\bf d} \vecprod \ecb({\bf r},t) = {\bf 0}.
\label{1.29}\eeq
That is, when retardation is ignored, the electromagnetic field reduces
to the instantaneous Coulomb field.


\subsection{Fourier analysis \label{sec1.1.2}}

\index{Fourier transform|(}
It is convenient to consider the Fourier transforms of the different
fields and potentials. Thus, for instance, the transform of the
scalar potential is
\beq
\varphi({\bf q},\omega) \equiv (2\pi)^{-2} \int \d {\bf r} \int \d t \,
\exp\left[ - {\rm i} \left( {\bf q} \dotprod {\bf r} - \omega t \right)
\right] \, \varphi({\bf r},t),
\label{1.30}\eeq
where the {\bf r} integral extends over the entire space ${\mathbb R}^3$ and the time
integral is over the interval from $-\infty$ to $\infty$.  The inverse
transform of $\varphi({\bf q},\omega)$ is
\beq
\varphi({\bf r},t) \equiv (2\pi)^{-2} \int \d {\bf q} \int \d \omega \,
\exp\left[ {\rm i} \left( {\bf q} \dotprod {\bf r} - \omega t \right)
\right] \, \varphi({\bf q},\omega),
\label{1.31}\eeq
with the integrals extending, respectively, over ${\mathbb R}^3$ and
${\mathbb R}$.
To simplify notation, we use the same symbol for the original field and
its transform, which are distinguished by their arguments.

\index{plane waves}
The right-hand side of Eq.\ \req{1.31} can be regarded as the expansion
of the potential function in terms of the plane waves $\exp\left[ {\rm
i} \left( {\bf q} \cdot {\bf r} - \omega t \right) \right]$ with wave
vector ${\bf q}$ and angular frequency $\omega$. These plane waves
form a complete basis of complex functions of the variables ${\bf r}$
and $t$. Evidently, we have
\beqa
\varphi^\ast({\bf r},t) &=& (2\pi)^{-2} \int \d {\bf q}' \int \d \omega' \,
\exp\left[ -{\rm i} \left( {\bf q}' \dotprod {\bf r} - \omega' t \right)
\right] \, \varphi^\ast({\bf q}',\omega')
\nonumber \\ [2mm]
&=& (2\pi)^{-2} \int \d {\bf q} \int \d \omega \,
\exp\left[ {\rm i} \left( {\bf q} \dotprod {\bf r} - \omega t \right)
\right] \, \varphi^\ast (-{\bf q},-\omega).
\nonumber\eeqa
Hence, when the field $\varphi({\bf r},t)$ is real [\ie, when,
$\varphi^\ast({\bf r},t)=\varphi({\bf r},t)$], its Fourier transform
satisfies the relation
\beq
\varphi(-{\bf q}, -\omega) =
\varphi^\ast({\bf q}, \omega).
\label{1.32}\eeq

If a field $\varphi(r,t)$ depends on the vector ${\bf r}$ only through
its magnitude $r=|{\bf r}|$, the Fourier integral \req{1.30} can be
partially evaluated by using polar coordinates  [Eqs.\ \req{B.20} to
\req{B.22} in Appendix \ref{appB}] with the polar axis in the direction
of the vector ${\bf q}$. We have
\beqa
\varphi({\bf q},\omega) &=& (2\pi)^{-2} \int_0^\infty r^2 \, \d r \;
2\pi \int_{-1}^{+1} \d (\cos\theta) \int \d t \, \exp\left[ - {\rm i}
\left( q r \cos\theta - \omega t \right) \right] \, \varphi(r,t)
\nonumber \\ [2mm]
&=& \frac{1}{\pi} \int_0^\infty r^2 \, \d r
\, \frac{\sin(qr)}{qr}
\int \d t \, \exp\left( {\rm i} \omega t \right)
\, \varphi(r,t) \equiv \varphi(q,\omega).
\label{1.33}\eeqa
That is, the Fourier transform of $\varphi(r,t)$ depends on ${\bf q}$
only through the wave number, $q=|{\bf q}|$.

\index{Fourier transform!of the screened Coulomb potential}
Thus, for example, the Fourier transform of the ``screened'' Coulomb
potential
\beq
\varphi_{\rm sC}(r) = \frac{Z_0 e}{r}\,  \exp(-a r)
\label{1.34}\eeq
is
\beqa
\varphi_{\rm sC}(q,\omega) &=& \frac{1}{\pi} \int_0^\infty r^2 \, \d r
\, \frac{\sin(qr)}{qr}
\int \d t \, \exp\left( {\rm i} \omega t \right)
\, \frac{Z_0 e}{r} \, \exp(-ar)
\nonumber \\ [2mm]
&=& \frac{2Z_0 e}{q} \, \delta(\omega) \int_0^\infty \d r
\, \sin(qr) \exp(-ar)
\nonumber \\ [2mm]
&=& \frac{2Z_0 e}{q^2+a^2} \, \delta(\omega) \, ,
\label{1.35}\eeqa
where $\delta(\omega)$ is the Dirac delta function [see Appendix
\ref{appB}, Eq.\ \req{B.8c}].
Considering the limit $a\rightarrow 0$, we conclude that the Fourier
transform of the Coulomb potential $\varphi_{\rm C}(r)=Z_0 e /r$ is
\beq
\varphi_{\rm C}(q,\omega) = \frac{2Z_0 e}{q^2} \, \delta(\omega) \, .
\label{1.36}\eeq

\index{longitudinal vector field} \index{solenoidal vector field}
\index{transverse vector field} \index{irrotational vector field}
As the differential operators $\nablab$ and $\partial/\partial t$ acting
on the exponentials $\exp\left[ {\rm i} \left( {\bf q} \cdot {\bf r} -
\omega t \right) \right]$ simply add the respective factors ${\rm i}
{\bf q}$ and $-{\rm i}\omega$, the Fourier transforms of $\nablab
\varphi$ and $\partial \, \varphi/\partial \, t$ are ${\rm i} \, {\bf q}
\, \varphi({\bf q},\omega)$ and $-{\rm i} \, \omega \, \varphi({\bf
q},\omega)$, respectively.  A solenoidal or divergence-free vector field
${\bf F}({\bf r},t)$ is called {\it transverse} because the equation
$\nablab\, \cdot \, {\bf F}({\bf r},t) =0$ is equivalent to ${\bf q}
\cdot {\bf F}({\bf q},\omega)=0$, \ie, the Fourier transform ${\bf F}
({\bf q},\omega)$ is perpendicular to the wave vector ${\bf q}$. An
irrotational vector field ${\bf F}({\bf r},t)$, characterized by the
property $\nablab \vecprod {\bf F}({\bf r},t)=0$, is called a {\it
longitudinal} field since its Fourier transform ${\bf F} ({\bf
q},\omega)$ satisfies ${\bf q} \vecprod {\bf F}({\bf q},\omega)=0$ and,
consequently, ${\bf F}({\bf q},\omega)$ is parallel to ${\bf q}$. An
arbitrary vector field can always be decomposed into its longitudinal
and transverse parts,
\beq
{\bf F}({\bf r},t) =
{\bf F}^{\rm (L)}({\bf r},t) +
{\bf F}^{\rm (T)}({\bf r},t),
\label{1.37}\eeq
and the decomposition is unique. When stated in terms of
Fourier transforms, this theorem becomes a triviality. We only need to
recall the vector identity
\beq
{\bf F} = \hat{\bf q} (\hat{\bf q} \dotprod {\bf F}) -
\hat{\bf q} \vecprod (\hat{\bf q} \vecprod {\bf F}),
\label{1.38}\eeq
and write
\beqa
{\bf F}({\bf r},t) &=& (2\pi)^{-2} \int \d {\bf q} \int \d \omega \,
\exp\left[ {\rm i} \left( {\bf q} \dotprod {\bf r} - \omega t \right)
\right] \, {\bf F}({\bf q},\omega)
\nonumber \\ [2mm]
&=& (2\pi)^{-2} \int \d {\bf q} \int \d \omega \,
\exp\left[ {\rm i} \left( {\bf q} \dotprod {\bf r} - \omega t \right)
\right] \, \hat{\bf q} [\hat{\bf q} \dotprod {\bf F}({\bf q},\omega)]
\nonumber \\ [2mm]
&& \mbox{} - (2\pi)^{-2} \int \d {\bf q} \int \d \omega \,
\exp\left[ {\rm i} \left( {\bf q} \dotprod {\bf r} - \omega t \right)
\right] \,
\hat{\bf q} \vecprod \left[ \hat{\bf q} \vecprod {\bf F}
({\bf q},\omega)\right],
\nonumber\eeqa
to get the decomposition of ${\bf F}$ into its longitudinal and
transverse parts,
\begin{subequations}
\label{1.39}
\beq
{\bf F}^{\rm (L)}({\bf r},t) =
(2\pi)^{-2} \int \d {\bf q} \int \d \omega \,
\exp\left[ {\rm i} \left( {\bf q} \dotprod {\bf r} - \omega t \right)
\right] \, \hat{\bf q} [\hat{\bf q} \dotprod {\bf F}({\bf q},\omega)],
\label{1.39a}\eeq
\beq
{\bf F}^{\rm (T)}({\bf r},t) =
- (2\pi)^{-2} \int \d {\bf q} \int \d \omega \,
\exp\left[ {\rm i} \left( {\bf q} \dotprod {\bf r} - \omega t \right)
\right] \,
\hat{\bf q} \vecprod \left[ \hat{\bf q} \vecprod {\bf F}
({\bf q},\omega)\right].
\label{1.39b}\eeq
\end{subequations}
From here it follows that
\begin{subequations}
\label{1.40}
\beqa
{\bf F}^{\rm (L)}({\bf q}, \omega) &=&
\hat{\bf q} \left[ \hat{\bf q} \dotprod {\bf F}({\bf q},\omega) \right]
\label{1.40a}\\ [2mm]
{\bf F}^{\rm (T)}({\bf q}, \omega) &=& -
\hat{\bf q} \vecprod \left[ \hat{\bf q} \vecprod {\bf F}
({\bf q},\omega)\right].
\label{1.40b}\eeqa
\end{subequations}

\index{longitudinal vector field!integral formula}
\index{transverse vector field!integral formula}

Integral formulas for the transverse and longitudinal components of the
field ${\bf F}$ at (${\bf r},t$) can be obtained by considering the
relations
\beq
\nabla^2\left( \frac{1}{r} \right) = - 4\pi \, \delta({\bf r}) \quad
\mbox{and} \quad
\nablab_{{\bf r}'} |{\bf r} - {\bf r}'|^{-1} = -
\nablab |{\bf r} - {\bf r}'|^{-1},
\label{1.41}\eeq
where $\delta({\bf r})$ is the three-dimensional Dirac delta function
[Appendix \ref{appB}, Eq.\ \req{B.24}]. Assuming that the field
decreases
faster than $r^{-1}$ at large distances ($r \rightarrow \infty$), we have
\beqa
{\bf F}({\bf r},t) &=& \int {\bf F}({\bf r}',t) \delta({\bf r}-{\bf r}')
\, \d {\bf r}' =
- \frac{1}{4\pi} \int {\bf F}({\bf r}',t)\, \nablab_{\bf r'}^2
\left(
\frac{1}{ |{\bf r}-{\bf r}'|} \right) \, \d {\bf r}'
\nonumber \\ [2mm]
&=& - \frac{1}{4\pi}  \int \frac{\nablab_{\bf r'}^2{\bf F}({\bf r}',t)}{
|{\bf r}-{\bf r}'|} \, \d {\bf r}'
\qquad \mbox{[double integration by parts]}
\nonumber \\ [2mm]
&=&
- \frac{1}{4\pi} \int \left[ \nablab_{\bf r'} \left(\nablab_{\bf r'}
\dotprod {\bf F}({\bf r}',t) \right) ) - \nablab_{\bf r'} \vecprod \left(
\nablab_{\rm r'} \vecprod {\bf F} ({\bf r}',t) \right) \right]
\frac{1}{|{\bf r}-{\bf r}'|} \, \d {\bf r}'
\nonumber \\ [2mm]
&& \qquad \mbox{[using the relation \req{1.8}]} \nonumber \\ [2mm]
&=&
- \frac{1}{4\pi} \nablab \int
\frac{\nablab_{\bf r'} \dotprod {\bf F}({\bf r}',t)}
{|{\bf r}-{\bf r}'|} \, \d {\bf r}'
+ \frac{1}{4\pi}
\nablab \vecprod \left( \nablab \vecprod
\int \frac{{\bf F}({\bf r}',t)}
{|{\bf r}-{\bf r}'|} \, \d {\bf r}' \right), \nonumber
\eeqa
where the last equality results from integrations by parts. Evidently,
the first term in this expression is a longitudinal field [$\nablab
\vecprod (\nablab \phi)\equiv 0$], while the second term is a transverse
field [$\nablab \cdot (\nablab \vecprod {\bf G}) \equiv 0$]. We have
thus decomposed the field into its longitudinal and transverse
components. That is,
\begin{subequations}
\label{1.42}
\beqa
{\bf F}^{\rm (L)}({\bf r},t) &=& - \frac{1}{4\pi} \nablab \int
\frac{\nablab_{\bf r'}
\dotprod {\bf F}({\bf r}',t)} {|{\bf r}-{\bf r}'|} \, \d {\bf r}',
\label{1.42a}\\ [2mm]
{\bf F}^{\rm (T)}({\bf r},t) &=& \frac{1}{4\pi} \nablab \vecprod \left(
\nablab \vecprod \int \frac{{\bf F}({\bf r}',t)} {|{\bf r}-{\bf r}'|} \,
\d {\bf r}' \right).
\label{1.42b}\eeqa
\end{subequations}
It can be verified that the Fourier transforms of these
fields coincide with the results \req{1.40}.


\subsection{The wave equation \label{sec1.1.3}}

\index{wave equation!of the electromagnetic field}
The continuity equation \req{1.2} implies that the Fourier
transforms of the charge and current distributions satisfy the
equation
\index{charge density} \index{current density}
\beq
\rho({\bf q}, \omega) = \frac{q}{\omega} \, \hat{\bf q} \dotprod {\bf
j} ({\bf q}, \omega).
\label{1.43}\eeq
Moreover, from the Poisson equation \req{1.6}, we have
\beq
\varphi({\bf q},\omega)
= \frac{4\pi}{q^2} \, \rho({\bf q}, \omega)
= \frac{4\pi}{q\omega}
\, \hat{\bf q} \dotprod {\bf j} ({\bf q}, \omega).
\label{1.44}\eeq
That is, the Fourier transform of the scalar potential is proportional
to the Fourier transform of the longitudinal component of the current density. Conversely,
\index{current density!longitudinal}
\beq
{\bf j}^{\rm (L)} ({\bf q}, \omega) = \hat{\bf q} \left[ \hat{\bf q} \dotprod
{\bf j} ({\bf q}, \omega) \right] =  \hat{\bf q} \, \frac{\omega}{q} \,
\rho({\bf q}, \omega) = \hat{\bf q}\,
\frac{q\omega}{4\pi} \, \varphi({\bf q}, \omega).
\label{1.45}\eeq

The Fourier transform of the wave equation \req{1.9} gives
\beqa
- q^2 {\bf A} ({\bf q},\omega) + \frac{\omega^2 }{c^2} {\bf A} ({\bf q},
\omega) &=& - \frac{4\pi}{c} \, {\bf j} ({\bf q},\omega) +
\frac{\omega}{c}\, {\bf q} \varphi({\bf q},\omega)
\nonumber \\ [2mm]
&=&
- \frac{4\pi}{c} \, {\bf j} ({\bf q},\omega) + \frac{4\pi}{c}
\, {\bf j}^{\rm (L)} ({\bf q}, \omega).
\nonumber \eeqa
That is,
\beq
\left[ q^2 - \frac{\omega^2 }{c^2} \right] {\bf A}
({\bf q},
\omega) = \frac{4\pi}{c} \, {\bf j}^{\rm (T)} ({\bf q},\omega) .
\label{1.46}\eeq
The equivalent equation in coordinate space is
\beq
\nabla^2 {\bf A} ({\bf r},t) - \frac{1}{c^2} \frac{\partial^2 {\bf A}
({\bf r},t)} {\partial
t^2} = - \frac{4\pi}{c} {\bf j}^{\rm (T)}({\bf r},t).
\label{1.47}\eeq
We see that the purely longitudinal term with $\nablab \varphi({\bf
r},t)$ on the right-hand side of Eq.\ \req{1.9} exactly cancels
the longitudinal component of the current density.
The source term in the wave equation for the transverse vector
potential thus reduces to the transverse component of the current
density. This was to be expected, since the left-hand side of Eq.\
\req{1.47} is a transverse field.

Using the second of Eqs.\ \req{1.4} and Eq.\ \req{1.45},
the Fourier transform of the electric field is expressed in the form
\beq
\ecb({\bf q},\omega) = - {\rm i} {\bf q} \varphi({\bf q},\omega) + {\rm
i} \, \frac{\omega}{c} \, {\bf A}({\bf q},\omega)
= - {\rm i} \, \frac{4\pi}{\omega} \, {\bf j}^{\rm (L)}({\bf
q},\omega) + {\rm i} \,
\frac{\omega}{c} \, {\bf A}({\bf q},\omega),
\label{1.48}\eeq
which displays the longitudinal and transverse parts separately,
\beq
\ecb^{\rm (L)}({\bf q},\omega)
= - {\rm i} \, \frac{4\pi}{\omega} \, {\bf j}^{\rm (L)}({\bf
q},\omega) , \qquad
\ecb^{\rm (T)}({\bf q},\omega) =
{\rm i} \, \frac{\omega}{c} \, {\bf A}({\bf q},\omega).
\label{1.49}\eeq
The Fourier transform of the magnetic induction can be obtained from
the first of Eqs.\ \req{1.4}, which gives
\beq
\bcb({\bf q}, \omega) = {\rm i} {\bf q} \vecprod {\bf A}({\bf q},
\omega) = \frac{c}{\omega} \, {\bf q} \vecprod \ecb^{\rm (T)}
({\bf q}, \omega) .
\label{1.50}\eeq

\index{localized charge distribution}
Very frequently, electromagnetic fields are produced by localized
distributions of charges and currents (which vanish beyond a certain
radius, $R$), and they are considered at points ${\bf r}=(r_1,r_2,r_3)$
that are far
from the distributions ($r \gg R$). To simplify the calculation of the
fields at large distances, we can expand the electromagnetic potentials
in powers of the reciprocal distance $r^{-1}$ as follows. At small ${\bf
r}'$ the function
\beq
f({\bf r}') = \frac{1}{|{\bf r} - {\bf r}'|}
= \frac{1}{\left[(r_1-r'_1)^2+(r_2-r'_2)^2+(r_3-r'_3)^2 \right]^{1/2}}
\nonumber \eeq
can be approximated by its Taylor expansion,
\beqa
f({\bf r}') &=& f(0) + \sum_{i=1}^3
\left[ \frac{\partial f}{\partial r'_i} \right]_{{\bf r}'= {\bf 0}} r'_i
+ \frac{1}{2!} \sum_{i,j=1}^3
\left[ \frac{\partial^2 f}{\partial r'_i \, \partial r'_j}
\right]_{{\bf r}'= {\bf 0}} r'_i r'_j + \cdots
\nonumber \\ [2mm]
&=& \frac{1}{r} + \frac{1}{r^3} \sum_{i=1}^3 r_i r'_i
+ \frac{1}{2} \frac{1}{r^5} \sum_{i,j=1}^3  \left( 3 r_i r_j -
\delta_{i,j} r^2 \right) r'_i r'_j + \cdots
\nonumber \eeqa
where $\delta_{a,b}$ denotes the Kronecker delta,
\index{Kronecker delta}
\beq
\delta_{a,b} = \left\{
\begin{array}{ll}
1 & \mbox{if $a=b$}, \\ [1mm]
0 & \mbox{if $a \ne b$}.
\end{array} \right.
\label{1.51}\eeq
Since
\beq
\nabla^2 \left( \frac{1}{r} \right)
= \sum_{i=1}^3  \frac{\partial^2}{\partial^2 r_i} \, \frac{1}{r}
= \frac{1}{r^5} \sum_{i=1}^3 \left( 3 r_i r_i - r^2 \right) = 0,
\nonumber \eeq
we can subtract the null quantity
$$
\frac{1}{6} \frac{1}{r^5}
 \sum_{i,j=1}^3  \left( 3 r_i r_j -
 \delta_{i,j} r^2 \right) \delta_{i,j} r'^2
$$
and write the more symmetric expansion
\beqa
\frac{1}{|{\bf r} - {\bf r}'|}
&=& \frac{1}{r} + \frac{1}{r^3} \sum_{i=1}^3 r_i r'_i
+ \frac{1}{6} \frac{1}{r^5} \sum_{i,j=1}^3  \left( 3 r_i r_j -
\delta_{i,j} r^2 \right) \left( 3 r'_i r'_j - \delta_{i,j} r'^2
\right).
\label{1.52}\eeqa
The scalar potential \req{1.7} can now be expressed in the form of a
multipole expansion
\index{localized charge distribution!scalar potential of a}
\index{localized charge distribution!scalar potential of a!multipole expansion}
\begin{subequations}
\label{1.53}
\beq
\varphi({\bf r},t) = \frac{1}{r} \, q_e
+ \frac{1}{r^3} \, {\bf r} \dotprod {\bf d}
+ \frac{1}{r^5} \frac{1}{6} \sum_{i,j=1}^3 Q_{ij}
\left( 3 r_i r_j - \delta_{i,j} r^2 \right)
+\cdots
\label{1.53a}\eeq
where
\beq
q_e = \int \rho({\bf r}',t) \, \d {\bf r}', \quad
{\bf d} = \int {\bf r}' \, \rho({\bf r}',t) \, \d {\bf r}', \quad
Q_{ij} = \int \left( 3 r'_i r'_j - \delta_{i,j} r'^2 \right) \,
\rho({\bf r}',t) \, \d {\bf r}',
\label{1.53b}\eeq
\end{subequations}
are the total charge, the dipole moment vector, and the quadrupole
moment tensor, respectively. On the other hand,
the transverse current at large $r$ is [see Eq.\ \req{1.42b}]
\index{current density!transverse}
\beq
{\bf j}^{\rm (T)}({\bf r},t)
= \frac{1}{4\pi} \nablab \vecprod \left( \nablab \vecprod \int \frac{{\bf
j}({\bf r}',t)} {|{\bf r}-{\bf r}'|} \, \d {\bf r}' \right)
\simeq \frac{1}{4\pi} \nablab \vecprod \left( \nablab \vecprod
\left[ \frac{{\bf C}(t)}{r} + O(r^{-2})\right] \right) ,
\nonumber\eeq
with
\beq
{\bf C}(t) =  \int
{\bf j}({\bf r}',t) \, \d {\bf r}'.
\label{1.54}\eeq
Using the relation \req{1.8},
and performing the derivatives, we find that
\beqa
{\bf j}^{\rm (T)}({\bf r},t)
&\simeq& \frac{1}{4\pi} \left[ \nablab \left( \nablab \dotprod  \frac{{\bf
C}}{r} \right) - \nabla^2  \frac{{\bf
C}}{r} +O(r^{-4}) \right]
\nonumber \\ [2mm]
&\simeq&  \frac{1}{4\pi} \left[
- \frac{{\bf C}}{r^{3}} + \frac{3({\bf C}\dotprod
{\bf r}) {\bf r} }{r^{5}} + O(r^{-4}) \right].
\label{1.55}\eeqa
It is worth noticing that ${\bf j}^{\rm (T)}({\bf r},t)$ may differ from
zero at large distances from the distribution of charges, where the
current density ${\bf j}({\bf r},t)$ effectively vanishes. Indeed, this
is necessary to compensate the instantaneous (unretarded) nature of the
scalar potential. A short proof that the Coulomb-gauge potentials do
yield the retarded electromagnetic fields has been published by
\citet{Heras2011}.

For neutral distributions (with total charge equal to zero) the scalar
potential and the transverse current density decrease rapidly with the
distance $r$, at least as $r^{-3}$. In
space regions that are far from the charges ($r \gg R$), $\varphi \simeq
0$, ${\bf j}^{\rm (T)} \simeq 0$, and the vector potential satisfies the
Helmholtz vector equation, \index{Helmholtz vector equation}
\beq
\nabla^2 {\bf A} - \frac{1}{c^2} \frac{\partial^2 {\bf A}}{\partial t^2}
= 0.
\label{1.56}\eeq
The electromagnetic fields in such distant regions are called {\it
radiation fields}. \index{radiation fields!electromagnetic}
\index{Fourier transform|)}


\section{Electromagnetic fields in material media \label{sec1.2}}

Let us now consider the electromagnetic fields produced by
arbitrary charge and current distributions $\rho_{\rm ext}
({\bf r},t)$ and ${\bf j}_{\rm ext} ({\bf r},t)$ in an infinite,
homogeneous, and isotropic medium. These fields satisfy the Maxwell
equations \citep[see][]{Jackson1975}
\index{Maxwell equations!in material media|(}
\begin{subequations}
\label{1.57}
\beqa
&& \nablab \dotprod \dcb({\bf r},t) = 4 \pi \rho_{\rm ext}({\bf r},t),
\label{1.57a} \\ [2mm]
&& \nablab \vecprod \hcb({\bf r},t)
- \frac{1}{c} \frac{\partial \dcb({\bf r},t)}{\partial t}
= \frac{4\pi}{c} {\bf j}_{\rm ext}({\bf r},t),
\label{1.57b}\\ [2mm]
&& \nablab \vecprod \ecb({\bf r},t)
+ \frac{1}{c} \frac{\partial \bcb({\bf r},t)}{\partial t}
= 0,
\label{1.57c} \\ [2mm]
&& \nablab \dotprod \bcb({\bf r},t) = 0,
\label{1.57d}\eeqa
\end{subequations}
where
\begin{subequations}
\label{1.58}
\beq
\dcb = \ecb + 4 \pi {\bf P}
\label{1.58a}\eeq
is the {\it electric displacement} and
\beq
\hcb = \bcb - 4 \pi {\bf M}
\label{1.58b}\eeq
\end{subequations}
is the {\it magnetic field}. ${\bf P}$ is the {\it dielectric
polarization}
(electric dipole moment per unit volume) and ${\bf M}$ is the
{\it magnetization} (magnetic moment per unit volume) of the medium.
Note that the equation
\req{1.57a} for the electric displacement $\dcb$, which determines only
the longitudinal component of $\dcb$, is identical to the Eq.\
\req{1.1a} for the electric field produced by the external charge
distribution $\rho_{\rm ext}({\bf r},t)$ in vacuum.

We shall assume that the external charges represent only a small
disturbance of the original charge distribution of the material and,
consequently, that the induced fields vary linearly with the external
charge and current densities. This {\it linear response approximation}
implies that
\beq
\dcb = \hat\epsilon \ecb, \qquad
\hcb = \hat\mu^{-1} \bcb,
\label{1.59}\eeq
where $\hat\epsilon$ and $\hat\mu$ are linear operators that
transform the electric field and the magnetic induction into,
respectively, the electric displacement and the magnetic field.  The
operators $\hat\epsilon$ and $\hat\mu$ are called the {\it dielectric
``constant''} and the {\it magnetic permeability}, respectively. In the
next Section we shall show that the linear response approximation
determines the way in which these operators act on the fields $\ecb$ and
$\bcb$. Evidently, in the vacuum
$\hat\epsilon=\hat\mu=1$. With the equalities \req{1.59}
the first two Maxwell equations, \req{1.57a} and \req{1.57b}, take the
form
\begin{subequations}
\label{1.60}
\beqa
&& \nablab \dotprod \hat\epsilon \ecb({\bf r},t) =
4 \pi \rho_{\rm ext}({\bf r},t),
\label{1.60a} \\ [2mm]
&&
\nablab \vecprod \hat\mu^{-1} \bcb({\bf r},t)
- \frac{1}{c} \frac{\partial \hat\epsilon \ecb({\bf r},t)}{\partial t}
= \frac{4\pi}{c} {\bf j}_{\rm ext}({\bf r},t).
\label{1.60b}\eeqa
\end{subequations}
From these two equations, it follows that the charge density and the current
satisfy the continuity equation \index{continuity equation}
\beq
\frac{\partial \rho_{\rm ext}}{\partial t} + \nablab \dotprod {\bf
j}_{\rm ext} = 0\, .
\label{1.61}\eeq
The force on a particle of charge $Z_0 e$ that moves with
velocity  ${\bf v}$ within the medium is given by the Lorentz formula
\req{1.3}, \index{Lorentz force}
\beq
\textbf{F} = Z_0 e \left( \ecb + \frac{1}{c} {\bf v} \vecprod \bcb \right).
\label{1.62}\eeq

\index{Coulomb gauge}
Notice that the Maxwell Eqs.\ \req{1.57c} and \req{1.57d} are
independent of the properties of the medium. Hence, scalar and vector
potentials can be defined according to Eqs.\ \req{1.4}. As pointed out
by \citet{Fano1956b}, regarding electromagnetic interactions in
homogeneous material media, a covariant formalism, which would require
using the Lorentz gauge, has no practical advantages because matter
constitutes a special frame of reference. It seems most adequate to use
the Coulomb gauge since it provides a natural decomposition of the
fields into their longitudinal and transverse parts, the latter being
observable as electromagnetic radiation under certain circumstances,
\eg, as Cherenkov radiation (Section \ref{sec8.2.2}). Accordingly, we
set $\nablab \cdot {\bf A} = 0$.  Expressed in terms of the
potentials, the Maxwell equations \req{1.60} then take the form
\begin{subequations}
\label{1.63}
\beqa
&& \nabla^2 \hat\epsilon \varphi({\bf r},t) = - 4 \pi \rho_{\rm ext}({\bf
r},t),
\label{1.63a}\\
&&
\nabla^2 \, \hat\mu^{-1} {\bf A}({\bf r},t)
- \frac{1}{c^2} \frac{\partial^2 }{\partial t^2} \, \hat{\epsilon} {\bf
A}({\bf r},t) =
- \frac{4\pi}{c} \, {\bf j}_{\rm ext}({\bf r},t) + \frac{1}{c} \nablab
\, \hat{\epsilon} \varphi({\bf r},t) \, ,
\label{1.63b}\eeqa
\end{subequations}
because, by the assumed homogeneity and isotropy of the medium, and by the
homogeneity of time, $\hat\epsilon$ and $\hat\mu$ are independent of
${\bf r}$ and $t$.

Following \citet{Lindhard1954}, we can adopt an alternative view and express
the Maxwell equations in terms of only the electric field $\ecb$ and the
magnetic induction $\bcb$. This is accomplished by noting that in vacuum
the sources $\rho_{\rm ext}$ and ${\bf j}_{\rm ext}$ would produce
fields $\ecb_{\rm ext}$ and $\bcb_{\rm ext}$ given by the Maxwell
equations
\begin{subequations}
\label{1.64}
\beqa
&& \nablab \dotprod \ecb_{\rm ext} = 4 \pi \rho_{\rm ext},
\label{1.64a}\\ [2mm]
&& \nablab \vecprod \bcb_{\rm ext}
- \frac{1}{c} \frac{\partial \ecb_{\rm ext}}{\partial t}
= \frac{4\pi}{c} {\bf j}_{\rm ext},
\label{1.64b}\\ [2mm]
&& \nablab \vecprod \ecb_{\rm ext}
+ \frac{1}{c} \frac{\partial \bcb_{\rm ext}}{\partial t}
= 0,
\label{1.64c}\\ [2mm]
&& \nablab \dotprod \bcb_{\rm ext} = 0.
\label{1.64d}\eeqa
\end{subequations}
Under the action of the external charges, the medium is polarized and
generates induced fields
\begin{subequations}
\label{1.65}
\beqa
\ecb_{\rm ind} &=& \ecb - \ecb_{\rm ext} = \hat\epsilon^{-1} \dcb -
\ecb_{\rm ext},
\label{1.65a} \\ [2mm]
\bcb_{\rm ind} &=& \bcb - \bcb_{\rm ext} = \hat{\mu} \hcb -
\bcb_{\rm ext} ,
\label{1.65b}\eeqa
\end{subequations}
which satisfy Maxwell equations similar to those in vacuum:
\allowdisplaybreaks{
\begin{subequations}
\label{1.66}
\beqa
&& \nablab \dotprod \ecb_{\rm ind} = 4 \pi \rho_{\rm ind},
\label{1.66a} \\ [2mm]
&& \nablab \vecprod \bcb_{\rm ind}
- \frac{1}{c} \frac{\partial \ecb_{\rm ind}}{\partial t}
= \frac{4\pi}{c} {\bf j}_{\rm ind},
\label{1.66b}\\ [2mm]
&& \nablab \vecprod \ecb_{\rm ind}
+ \frac{1}{c} \frac{\partial \bcb_{\rm ind}}{\partial t}
= 0,
\label{1.66c} \\ [2mm]
&& \nablab \dotprod \bcb_{\rm ind} = 0,
\label{1.66d}\eeqa
\end{subequations}
}

\vspace*{-3mm}
\noindent
where $\rho_{\rm ind}$ and ${\bf j}_{\rm ind}$ are the induced charge
and current densities. Finally, the Maxwell equations for the fields
$\ecb = \ecb_{\rm ext} + \ecb_{\rm ind}$ and $\bcb = \bcb_{\rm ext} +
\bcb_{\rm ind}$ produced by the external distributions in the medium can
be expressed as
\begin{subequations}
\label{1.67}
\beqa
&& \nablab \dotprod \ecb = 4 \pi \rho,
\label{1.67a}\\ [2mm]
&& \nablab \vecprod \bcb
- \frac{1}{c} \frac{\partial \ecb}{\partial t}
= \frac{4\pi}{c} {\bf j},
\label{1.67b}\\ [2mm]
&& \nablab \vecprod \ecb
+ \frac{1}{c} \frac{\partial \bcb}{\partial t}
= 0,
\label{1.67c}\\ [2mm]
&& \nablab \dotprod \bcb = 0,
\label{1.67d}\eeqa
\end{subequations}
where
\beq
\rho \equiv \rho_{\rm ext} + \rho_{\rm ind} \qquad \mbox{and} \qquad
{\bf j} \equiv {\bf j}_{\rm ext} + {\bf j}_{\rm ind}
\label{1.68}\eeq
are the total charge and current distributions, respectively. Of course,
Eqs.\ \req{1.67} and \req{1.68} are completely equivalent to the set of
Eqs.\ \req{1.57} with the definitions \req{1.59}.

\index{external electromagnetic fields} \index{induced electromagnetic
fields} \index{electromagnetic potentials}
Thus, the external (ext), induced (ind) and total fields
satisfy Maxwell equations with the same structure as those in vacuum.
The first pair of each set of Maxwell equations imply that the
corresponding charge and current distributions
satisfy continuity equations of the type
\beq
\frac{\partial \rho_\circ}{\partial t}
+ \nablab \dotprod {\bf j}_\circ = 0,
\label{1.69}\eeq
where the subscript $\circ$ stands for ``ext'', ``ind'' or
``blank'' (total). The second pair of Maxwell equations imply that we
can introduce corresponding potentials $\varphi_\circ$ and ${\bf
A}_\circ$ such that
\beq
\bcb_\circ = \nablab \vecprod {\bf A}_\circ, \qquad \ecb_\circ =
- \nablab \varphi_\circ -
\frac{1}{c} \frac{\partial {\bf A}_\circ}{\partial t}.
\label{1.70}\eeq
The scalar and vector potentials (in the Coulomb gauge) satisfy the
Poisson equation, \index{Poisson equation}
\beq
\nabla^2 \varphi_\circ = - 4 \pi \rho_\circ,
\label{1.71}\eeq
and the wave equation
\beq
\nabla^2 {\bf A}_\circ - \frac{1}{c^2}
\frac{\partial^2 {\bf A}_\circ} {\partial
t^2} = - \frac{4\pi}{c} {\bf j}^{\rm (T)}_\circ\, ,
\label{1.72}\eeq
respectively. Evidently,
\beq
\varphi = \varphi_{\rm ext} + \varphi_{\rm ind}
\qquad \mbox{and} \qquad
{\bf A} =  {\bf A}_{\rm ext} + {\bf A}_{\rm ind}.
\label{1.73}\eeq

\index{Fourier transform}
From the continuity equation \req{1.69} and the Poisson
equation \req{1.71} (which are valid for the external, induced and
total charge distributions and fields), we obtain the following
relationships between Fourier transforms of the
charge densities, the current densities, and the scalar potentials,
\begin{subequations}
\label{1.74}
\beq
\rho_\circ ({\bf q},\omega) =
\frac{q}{\omega}\, \hat{\bf q} \dotprod
{\bf j}_\circ ({\bf q},\omega)
\label{1.74a}\eeq
and
\beq
\varphi_\circ ({\bf q},\omega) =
\frac{4\pi}{q^2} \rho_\circ ({\bf q},\omega) .
\label{1.74b}\eeq
\end{subequations}
The Fourier transforms of the electric fields and the magnetic
inductions are given by [see Eq.\ \req{1.48} and \req{1.50}]
\begin{subequations}
\label{1.75}
\beq
\ecb_\circ({\bf q},\omega) =
- {\rm i} {\bf q} \varphi_\circ ({\bf q},\omega)
+ {\rm i} \, \frac{\omega}{c} \, {\bf A}_\circ ({\bf q},\omega) \, ,
\label{1.75a}\eeq
and
\beq
\bcb_\circ ({\bf q}, \omega) =
{\rm i} {\bf q} \vecprod {\bf A}_\circ ({\bf q}, \omega) \, ,
\label{1.75b}\eeq
\end{subequations}
respectively. Notice that the two terms on the right-hand side of Eq.\
\req{1.75a} are, respectively, the transforms of the longitudinal and
transverse components of the electric field.


\subsection{Dielectric functions \label{sec1.2.1}}

\index{dielectric functions!definitions|(}
Let us now turn to the dependence of the potentials $\varphi$
and ${\bf A}$ of the total field on the external charge and
current. In its most general form, the assumption of a linear response
of the medium can be formulated as
\begin{subequations}
\label{1.76}
\beqa
\varphi({\bf r},t) = \int \d {\bf r}' \int \d t' \, G^{\rm (L)}({\bf r},
{\bf r}';t,t') \, \rho_{\rm ext}({\bf r}',t'),
\label{1.76a} \\ [2mm]
{\bf A}({\bf r},t) = \int \d {\bf r}' \int \d t' \, G^{\rm
(T)}({\bf r},
{\bf r}';t,t') \, {\bf j}_{\rm ext}^{\rm (T)}({\bf r}',t'),
\label{1.76b}\eeqa
\end{subequations}
where the superscripts in the propagators $G^{\rm (L)}$ and $G^{\rm
(T)}$ refer to the longitudinal and transverse character of the fields.
Evidently, owing to the homogeneity of time, the propagators can only
depend on $t$ and $t'$ through their difference, $t-t'$. Moreover, for a
homogeneous medium, they may depend only on the relative position of the
source and observation points, ${\bf r} - {\bf r}'$. Here and in what
follows we assume that the medium is isotropic and, consequently, the
propagators are scalar quantities. \index{Fourier transform}
Introducing the Fourier transforms, we have
\beqa
\varphi({\bf r},t) &=& \int \d {\bf r}' \int \d t' \, G^{\rm (L)}({\bf
r}- {\bf r}';t-t') \,
\nonumber \\ [2mm]
&& \times
(2\pi)^{-2} \int \d {\bf q} \int \d \omega \,
\exp\left[ {\rm i} \left( {\bf q} \dotprod {\bf r}' - \omega t' \right)
\right] \, \rho_{\rm ext}({\bf q},\omega)
\nonumber \\ [2mm]
&=&
\int \d {\bf q} \int \d \omega \,
\exp\left[ {\rm i} \left( {\bf q} \dotprod {\bf r} - \omega t \right)
\right] \, \rho_{\rm ext}({\bf q},\omega) \;
(2\pi)^{-2} \int \d ({\bf r}-{\bf r}') \int \d (t-t') \,
\nonumber \\ [2mm]
&& \times
\exp\left\{ - {\rm i} \left[ {\bf q} \dotprod ({\bf r} - {\bf r}') -
\omega (t-t') \right] \right\} \,
G^{\rm (L)}({\bf r}- {\bf r}';t-t')
\nonumber \\ [2mm]
&=&
\int \d {\bf q} \int \d \omega \,
\exp\left[ {\rm i} \left( {\bf q} \dotprod {\bf r} - \omega t \right)
\right] \,
G^{\rm (L)}({\bf q},\omega)
\rho_{\rm ext}({\bf q},\omega),
\label{1.77}\eeqa
where $G^{\rm (L)}({\bf q},\omega)$ is the Fourier transform of
$G^{\rm (L)}({\bf r}, t)$. Comparing Eqs.\ \req{1.31} and
\req{1.77}, we obtain
\beq
\varphi({\bf q},\omega) = (2\pi)^2
G^{\rm (L)}({\bf q},\omega) \,
\rho_{\rm ext}({\bf q},\omega).
\label{1.78}\eeq
In a similar manner, we derive the following relation between
Fourier transforms of the vector potential and the transverse external
current,
\beq
{\bf A}({\bf q},\omega) = (2\pi)^2
G^{\rm (T)}({\bf q},\omega) \,
{\bf j}_{\rm ext}^{\rm (T)}({\bf q},\omega).
\label{1.79}\eeq
By defining the quantities \index{longitudinal dielectric function}
\index{transverse dielectric function}
\beq
\epsilon^{\rm (L)} ({\bf q},\omega) \equiv  \frac{1}{\pi q^2 \,
G^{\rm (L)}({\bf q},\omega)} \quad \mbox{and} \quad
\epsilon^{\rm (T)} ({\bf q},\omega) \equiv \frac{c^2}{\omega^2}
\left( q^2 - \frac{1}{\pi c \, G^{\rm (T)}({\bf q},\omega)} \right),
\label{1.80}\eeq
we may write Eqs.\ \req{1.78} and \req{1.79} in the respective forms
\begin{subequations}
\label{1.81}
\beqa
q^2 \epsilon^{\rm (L)} ({\bf q},\omega) \varphi({\bf q},\omega) &=&
4 \pi \rho_{\rm ext}({\bf q},\omega)
\label{1.81a}\eeqa
and
\beq
\left[ q^2 -
\frac{\omega^2}{c^2} \,
\epsilon^{\rm (T)}({\bf q},\omega) \right] {\bf A}({\bf q},\omega) =
\frac{4\pi}{c}\,
{\bf j}_{\rm ext}^{\rm (T)}({\bf q},\omega).
\label{1.81b}\eeq
\end{subequations}
These equations relate the Fourier transforms of the total scalar and
vector potentials with those of the external sources. They are
equivalent to the following differential equations in coordinate and
time variables,
\begin{subequations}
\label{1.82}
\beqa
\nabla^2 [\hat\epsilon^{\rm (L)}\,  \varphi({\bf r},t)]
&=& - 4 \pi \rho_{\rm ext}({\bf r},t),
\label{1.82a} \\ [2mm]
\nabla^2 {\bf A}({\bf r},t) - \frac{1}{c^2} \,
\frac{\partial^2 [\hat\epsilon^{\rm (T)}\, {\bf
A}({\bf r},t)]}
{\partial t^2} &=& - \, \frac{4\pi}{c} \,
{\bf j}_{\rm ext}^{\rm (T)}({\bf r},t),
\label{1.82b}\eeqa
\end{subequations}
where the linear operators $\hat\epsilon^{\rm (L)}$ and
$\hat\epsilon^{\rm (T)}$ act in the form made explicit in Eqs.\
\req{1.81}, \ie, the Fourier transforms of $\hat\epsilon^{\rm (L)}
\varphi({\bf r},t)$ and $\hat\epsilon^{\rm (T)} {\bf A}({\bf r},t)$
are $\epsilon^{\rm (L)} ({\bf q},\omega) \varphi({\bf q},\omega)$ and
$\epsilon^{\rm (T)} ({\bf q},\omega) {\bf A} ({\bf q},\omega)$,
respectively. In the following,
the quantities $\epsilon^{\rm (L)}({\bf q},\omega)$ and $\epsilon^{\rm
(T)} ({\bf q},\omega)$ will be referred to as the {\it longitudinal and
transverse dielectric functions} (DFs).

As $\varphi({\bf r},t)$ and $\rho_{\rm ext}({\bf r},t)$ and the
components of ${\bf A}({\bf r},t)$ and
${\bf j}_{\rm ext}^{\rm (T)}({\bf r},t)$ are all real functions, their Fourier
transforms satisfy the property \req{1.32}. According to Eqs.\
\req{1.81}, we have a similar relation for the DFs:
\beq
\epsilon^{\rm (L)}(-{\bf q}, -\omega) = \left[
\epsilon^{\rm (L)}({\bf q}, \omega) \right]^\ast, \qquad
\epsilon^{\rm (T)}(-{\bf q}, -\omega) = \left[
\epsilon^{\rm (T)}({\bf q}, \omega) \right]^\ast.
\label{1.83}\eeq
For the important case of homogeneous isotropic media, the
propagators $G^{\rm (L)}$ and $G^{\rm (T)}$ in Eqs.\ \req{1.76}
depend on the vector ${\bf
r}-{\bf r}'$ only through its magnitude $|{\bf r}-{\bf r}'|$. This
implies that the DFs only depend on $q=|{\bf q}|$ and $\omega$.
Therefore,
\beq
\epsilon^{\rm (L)}(q, -\omega) = \left[
\epsilon^{\rm (L)}(q, \omega) \right]^\ast, \qquad
\epsilon^{\rm (T)}(q, -\omega) = \left[
\epsilon^{\rm (T)}(q, \omega) \right]^\ast.
\label{1.84}\eeq
These relations imply that the real (imaginary) parts of $\epsilon^{\rm
(L,T)}(q, \omega)$ are even (odd) functions of $\omega$.  In particular,
in vacuum we have
\beq
\epsilon^{\rm (L)}_{\rm vacuum}(q,\omega) =
\epsilon^{\rm (T)}_{\rm vacuum}(q,\omega) = 1.
\label{1.85}\eeq

The occurrence of two DFs may seem paradoxical at first glance. These
DFs express the proportionality between the Fourier transforms of the
scalar and vector potentials and those of the associated sources. As
$\varphi$ and ${\bf A}$ are solutions of the decoupled Eqs.\ \req{1.82},
the associated DFs need not be equal. Alternatively, we could have
started from Eqs.\ \req{1.57a} and \req{1.57b}, and define the
dielectric constant
and magnetic permeability operators so that the Fourier transforms of
the magnetic field $\hcb$ and the electric displacement $\dcb$ are given
by \index{magnetic permeability}
\begin{subequations}
\label{1.86}
\beqa
\dcb({\bf q}, \omega) = \epsilon ({\bf q},\omega) \, \ecb({\bf q},\omega),
\label{1.86a}\\ [2mm]
\hcb({\bf q}, \omega) = \mu^{-1}({\bf q},\omega) \, \bcb({\bf q},\omega).
\label{1.86b}\eeqa
From the general relations \req{1.58} it follows that
\beq
{\bf P}({\bf q},\omega) = \frac{1}{4\pi} \left[
\epsilon({\bf q},\omega) -1 \right]
\ecb({\bf q},\omega)
\label{1.86c}\eeq
and
\beq
{\bf M}({\bf q},\omega) = \frac{1}{4\pi} \left[ 1 -
\mu^{-1}({\bf q},\omega) \right]
\bcb({\bf q},\omega).
\label{1.86d}\eeq
\end{subequations}
As we have seen, when expressed in terms of the Fourier transforms, the
continuity equation \req{1.61}, implies that
\beq
\rho_{\rm ext}({\bf q}, \omega) = \frac{q}{\omega} \,
\hat{\bf q} \dotprod {\bf j}_{\rm ext} ({\bf q}, \omega) \, .
\label{1.87}\eeq
Similarly, the Fourier transforms of Eqs.\ \req{1.63} are
\begin{subequations}
\label{1.88}
\beqa
q^2 \epsilon({\bf q},\omega)\,  \varphi({\bf q},\omega)
&=& 4 \pi \, \rho_{\rm ext}({\bf q},\omega)
\label{1.88a}\eeqa
and
\beqa
q^2 \mu^{-1}({\bf q},\omega) {\bf A}({\bf q},\omega)
- \frac{\omega^2}{c^2} \epsilon({\bf q},\omega) \, {\bf
A}({\bf q},\omega) &=&
\frac{4\pi}{c} \, {\bf j}_{\rm ext}({\bf q},\omega) - \frac{\omega}{c} {\bf
q}
\, \epsilon ({\bf q},\omega) \,  \varphi({\bf q},\omega)
\nonumber \\ [2mm]
&=& \frac{4\pi}{c} \, {\bf j}_{\rm ext}^{\rm (T)}({\bf q},\omega)
\, .
\label{1.88b}\eeqa
\end{subequations}
We see that Eqs.\ \req{1.81} and \req{1.88} are equivalent if we made
the identifications
\begin{subequations}
\label{1.89}
\beq
\epsilon({\bf q},\omega) = \epsilon^{\rm (L)}({\bf q},\omega)
\label{1.89a}\eeq
and
\beq
q^2 \mu^{-1}({\bf q},\omega)
- \frac{\omega^2}{c^2} \epsilon({\bf q},\omega)
= q^2 - \frac{\omega^2}{c^2} \epsilon^{\rm (T)}({\bf q},\omega)\, .
\label{1.89b}\eeq
\end{subequations}
Thus, the properties of the medium can be accounted for by using either
the longitudinal and transverse DFs or the dielectric ``constant''
$\epsilon= \epsilon^{\rm (L)}$ and the magnetic permeability
\beq
\mu({\bf q},\omega) = \left\{ 1 + \left( \frac{\omega}{c q} \right)^2
\left[\rule{0mm}{4mm}\epsilon^{\rm (L)}({\bf q},\omega)
- \epsilon^{\rm (T)}({\bf q},\omega) \right] \right\}^{-1} .
\label{1.90}\eeq

\noindent {\bf Note:} The relation \req{1.86a} can be expressed in tensor
form
$$
{\cal D}_i({\bf q}, \omega) = \epsilon_{ij} ({\bf q},\omega) \, {\cal
E}_j({\bf q},\omega),
$$
with the dielectric-function tensor of an isotropic medium defined as
$$
\epsilon_{ij} ({\bf q},\omega) = \left( \delta_{ij} -
\frac{q_i\, q_j}{q^2} \right) \epsilon^{\rm (T)}(q,\omega)
+ \frac{q_i\, q_j}{q^2} \, \epsilon^{\rm (L)}(q,\omega).
$$
Evidently,
\beq
\dcb ({\bf q},\omega) = \left\{ \ecb({\bf q},\omega)
-\left[ \ecb({\bf q},\omega) \dotprod \hat{\bf q} \right] \hat{\bf q}
\right\} \epsilon^{\rm (T)}(q,\omega)
+ \left[\ecb({\bf q},\omega) \dotprod \hat{\bf q} \right] \hat{\bf q}
\, \epsilon^{\rm (L)}(q,\omega),
\tag{{\rm 1.86a'}}\eeq
which explicitly shows that the transverse and longitudinal dielectric
functions act on the corresponding components of the electric field.


\subsection{Electromagnetic fields \label{sec1.2.2}}
\index{electromagnetic field!in material media}
\index{electromagnetic potentials!in material media}
\index{Fourier transform|(}
We can now determine the electromagnetic fields and potentials
originated by external distributions of charge and current. The Fourier
transforms of the scalar and vector potentials are given by Eqs.\
\req{1.81},
\beq
\varphi ({\bf q},\omega) =
\frac{4\pi}{q^2} \,
\frac{1}{\epsilon^{\rm (L)} ({\bf q},\omega)} \,
\rho_{\rm ext} ({\bf q},\omega)
= \frac{4\pi}{q \omega} \,
\frac{1}{\epsilon^{\rm (L)} ({\bf q},\omega)} \, \hat{\bf q} \dotprod
{\bf j}_{\rm ext} ({\bf q},\omega)
\label{1.91}\eeq
and
\beqa
{\bf A} ({\bf q},\omega) &=& \frac{4\pi}{cq^2}\,
\frac{q^2}{q^2 - (\omega/c)^2 \,
\epsilon^{\rm (T)}({\bf q}, \omega)} \,
{\bf j}_{\rm ext}^{\rm (T)}({\bf q},\omega)\, .
\label{1.92}\eeqa
The Fourier transforms of the potentials of the external charges in
vacuum are given by similar expressions with $\epsilon^{\rm (L,T)}=1$,
that is,
\beq
\varphi_{\rm ext} ({\bf q},\omega)
= \frac{4\pi}{q^2} \, \rho_{\rm ext} ({\bf q},\omega)
= \frac{4\pi}{q \omega} \, \hat{\bf q} \dotprod
{\bf j}_{\rm ext}^{\rm (L)} ({\bf q},\omega)
\label{1.93}\eeq
and
\beqa
{\bf A}_{\rm ext} ({\bf q},\omega) &=& \frac{4\pi}{cq^2}\,
\frac{q^2}{q^2 - (\omega/c)^2} \,
{\bf j}_{\rm ext}^{\rm (T)}({\bf q},\omega)\, .
\label{1.94}\eeqa
The Fourier transforms of the potentials of the induced electromagnetic
field are then
\beq
\varphi_{\rm ind} ({\bf q},\omega) =
\varphi ({\bf q},\omega) - \varphi_{\rm ext} ({\bf q},\omega) =
\frac{4\pi}{q \omega} \,
\left( \frac{1}{\epsilon^{\rm (L)} ({\bf q},\omega)} - 1 \right)
\hat{\bf q} \dotprod
{\bf j}_{\rm ext}^{\rm (L)} ({\bf q},\omega) \,
\label{1.95}\eeq
and
\beqa
{\bf A}_{\rm ind} ({\bf q},\omega)
&=& {\bf A}({\bf q},\omega) - {\bf A}_{\rm ext} ({\bf q},\omega)
\nonumber \\ [2mm]
&=& \frac{4\pi}{cq^2}\,
\left( \frac{q^2}{q^2 - (\omega/c)^2 \, \epsilon^{\rm (T)}({\bf q}, \omega)}
- \frac{q^2}{q^2 - (\omega/c)^2} \right)
{\bf j}_{\rm ext}^{\rm (T)}({\bf q},\omega)\, .
\label{1.96}\eeqa

The Fourier transforms of the electric field and the magnetic induction
can be calculated directly from the transforms of the potentials by using
Eqs.\ \req{1.75}. We have
\beqa
\ecb({\bf q},\omega) &=&
- {\rm i} {\bf q} \varphi ({\bf q},\omega)
+ {\rm i} \, \frac{\omega}{c} \, {\bf A} ({\bf q},\omega)
\nonumber \\ [2mm]
&=& - {\rm i} \, \frac{4\pi}{\omega} \,
\frac{1}{\epsilon^{\rm (L)} ({\bf q},\omega)} \,
{\bf j}_{\rm ext}^{\rm (L)} ({\bf q},\omega)
\nonumber \\ [2mm]
&& \mbox{}
+ {\rm i} \, \frac{4\pi \omega}{c^2q^2} \,
\frac{q^2}{q^2 - (\omega/c)^2 \,
\epsilon^{\rm (T)}({\bf q}, \omega)} \,
{\bf j}_{\rm ext}^{\rm (T)}({\bf q},\omega)
\label{1.97}\eeqa
and
\beqa
\bcb ({\bf q}, \omega) &=&
{\rm i} {\bf q} \vecprod {\bf A} ({\bf q}, \omega)
\nonumber \\ [2mm]
&=& {\rm i} \, \frac{4\pi}{cq} \,
\frac{q^2}{q^2 - (\omega/c)^2 \,
\epsilon^{\rm (T)}({\bf q}, \omega)} \, \hat{\bf q} \vecprod
{\bf j}_{\rm ext}^{\rm (T)}({\bf q},\omega) \, .
\label{1.98}\eeqa
The Fourier transforms $\ecb_{\rm ext} ({\bf q}, \omega)$
and $\bcb_{\rm ext} ({\bf q}, \omega)$ of fields in vacuum
can be written in the same forms with $\epsilon=1$. Hence, the
transforms of the induced fields are
\beqa
\ecb_{\rm ind} ({\bf q}, \omega) &=&
\ecb ({\bf q}, \omega) - \ecb_{\rm ext} ({\bf q}, \omega)
\nonumber \\ [2mm]
&=&
- {\rm i} \frac{4\pi}{\omega} \left( \frac{1}{\epsilon^{\rm (L)}(
{\bf q}, \omega)} - 1 \right) \, {\bf j}_{\rm ext}^{\rm (L)}({\bf q}, \omega)
\nonumber \\ [2mm]
&& + {\rm i} \frac{4 \pi
\omega}{c^2q^2}\, \left( \frac{q^2}{q^2 - (\omega/c)^2 \,
\epsilon^{\rm (T)}({\bf q}, \omega)} -
\frac{q^2}{q^2 - (\omega/c)^2} \right)
\, {\bf j}_{\rm ext}^{\rm (T)}({\bf q}, \omega) \rule{10mm}{0mm}
\label{1.99}\eeqa
and
\beqa
\bcb_{\rm ind} ({\bf q}, \omega) &=&
\bcb ({\bf q}, \omega) - \bcb_{\rm ext} ({\bf q}, \omega)
\nonumber \\ [2mm]
&=&
{\rm i} \, \frac{4\pi}{cq} \left(
\frac{q^2}{q^2 - (\omega/c)^2 \, \epsilon^{\rm (T)}({\bf q}, \omega)}
-\frac{q^2}{q^2 - (\omega/c)^2} \right)
\hat{\bf q} \vecprod
{\bf j}_{\rm ext}^{\rm (T)}({\bf q},\omega) \, .  \rule{10mm}{0mm}
\label{1.100}\eeqa
The relations
\beq
\dcb({\bf q},\omega) = \epsilon^{\rm (L)}({\bf q},\omega)
\, \ecb({\bf q},\omega)
\label{1.101}\eeq
and
\beqa
\hcb({\bf q},\omega) &=&
\mu^{-1}({\bf q},\omega) \, \bcb({\bf q},\omega)
\nonumber \\ [2mm]
&=&  \left\{ 1 + \left( \frac{\omega}{c q} \right)^2
\left[\rule{0mm}{4mm}\epsilon^{\rm (L)}({\bf q},\omega)
-  \epsilon^{\rm (T)}({\bf q},\omega) \right]\right\}
\bcb({\bf q},\omega)
\label{1.102}\eeqa
complete the definitions of the Fourier transforms of the fields.

The longitudinal and transverse components of the induced current can be
derived from those of the electric field, or from the potentials, as
follows. We notice that the Eqs.\ \req{1.82} can be expressed in terms
of the electric field and the magnetic induction as
\beq
\nablab \dotprod \hat\epsilon^{\rm (L)} \ecb = 4 \pi \rho_{\rm ext},
\qquad
\nablab \vecprod \bcb - \frac{1}{c} \frac{\partial
\hat\epsilon^{\rm (T)} \ecb}{\partial t}
= \frac{4\pi}{c} {\bf j}_{\rm ext}.
\label{1.103}\eeq
These equations are equivalent to the inhomogeneous
Maxwell equations \req{1.67a} and \req{1.67b},
\beq
\nablab \dotprod \ecb = 4 \pi \rho,
\qquad
\nablab \vecprod \bcb - \frac{1}{c} \frac{\partial \ecb}{\partial t}
= \frac{4\pi}{c} {\bf j}\, .
\label{1.104}\eeq
Then, from the first equation in \req{1.103} combined with the equality
\req{1.87}, we can write
\beq
\epsilon^{\rm (L)} ({\bf q},\omega) \, \ecb^{\rm (L)}({\bf q},\omega)
= - {\rm i} \frac{4\pi}{q} \, \hat{\bf q} \,
\rho_{\rm ext}({\bf q},\omega)\,
= - {\rm i} \frac{4\pi}{\omega} \,
{\bf j}_{\rm ext}^{\rm (L)}({\bf q},\omega)\, .
\label{1.105}\eeq
Similarly, the first equation in \req{1.104} implies that
\beq
\ecb^{\rm (L)}({\bf q},\omega) = - {\rm i} \frac{4\pi}{\omega} \, {\bf
j}^{\rm (L)}({\bf q},\omega) \, .
\label{1.106}\eeq
With the aid of Eqs.\ \req{1.48} to \req{1.50}, the
Fourier transform of the second of Eqs.\ \req{1.104} can be expressed as
$$
{\rm i} \, \frac{c}{\omega} \, q^2 \, \hat{\bf q}\vecprod \left[
\hat{\bf q}
\vecprod \ecb^{\rm (T)}({\bf q},\omega) \right] + {\rm i}
\frac{\omega}{c}
\left[
- {\rm i} \, \frac{4\pi}{\omega} \, {\bf j}^{\rm (L)}({\bf
q},\omega) + \ecb^{\rm (T)}({\bf q},\omega) \right] = \frac{4\pi}{c} \,
{\bf j}({\bf q},\omega)
$$
and, recalling that $\hat{\bf q} \vecprod (\hat{\bf q} \vecprod
\ecb^{\rm (T)}) = - \ecb^{\rm (T)}$ because $\ecb^{\rm (T)}$ is
transverse, we can write
\beq
\left( q^2 - \frac{\omega^2}{c^2} \right)
\ecb^{\rm (T)}({\bf q},\omega) = {\rm i} \, \frac{4\pi\omega}{c^2} \,
{\bf j}^{\rm (T)}({\bf q},\omega).
\label{1.107}\eeq
Similarly, the second of Eqs.\ \req{1.103} gives
\beq
\left( q^2 - \frac{\omega^2}{c^2}
\epsilon^{\rm (T)}({\bf q},\omega) \right)
\ecb^{\rm (T)}({\bf q},\omega) = {\rm i} \frac{4\pi\omega}{c^2} \,
{\bf j}_{\rm ext}^{\rm (T)}({\bf q},\omega).
\label{1.108}\eeq
Eqs.\ \req{1.105}-\req{1.108} express the relationship between the Fourier
transforms of the longitudinal and transverse parts of the electric
field $\ecb$ and those of the total and external current densities. It
follows that
\begin{subequations}
\label{1.109}
\beq
{\bf j}_{\rm ind}^{\rm (L)}({\bf q},\omega) =
{\bf j}^{\rm (L)}({\bf q},\omega) -
{\bf j}_{\rm ext}^{\rm (L)} ({\bf q},\omega)= - {\rm i}
\frac{\omega}{4\pi} \left[
\epsilon^{\rm (L)} ({\bf q},\omega) - 1 \right] \, \ecb^{\rm (L)}({\bf
q},\omega)
\label{1.109a}\eeq
and
\beq
{\bf j}_{\rm ind}^{\rm (T)}({\bf q},\omega) =
{\bf j}^{\rm (T)}({\bf q},\omega) -
{\bf j}_{\rm ext}^{\rm (T)} ({\bf q},\omega)= - {\rm i}
\frac{\omega}{4\pi} \left[
\epsilon^{\rm (T)} ({\bf q},\omega) - 1 \right] \, \ecb^{\rm (T)}({\bf
q},\omega).
\label{1.109b}\eeq
\end{subequations}
Expressing the fields in terms of the potentials [see Eq.\ \req{1.75a}]
and using the relations \req{1.74}, these equalities become
\begin{subequations}
\label{1.110}
\beq
{\bf j}_{\rm ind}^{\rm (L)}({\bf q},\omega) = \frac{\omega}{q} \,
\hat{\bf q} \, \rho_{\rm ind} ({\bf q},\omega) =
- \frac{\omega}{4\pi} \left[
\epsilon^{\rm (L)} ({\bf q},\omega) - 1 \right]
{\bf q} \, \varphi ({\bf q},\omega)
\label{1.110a}\eeq
and
\beq
{\bf j}_{\rm ind}^{\rm (T)}({\bf q},\omega) =
\frac{\omega^2}{4\pi c} \left[
\epsilon^{\rm (T)} ({\bf q},\omega) - 1 \right]
{\bf A} ({\bf q},\omega).
\label{1.110b}\eeq
\end{subequations}

\index{electric conductivity}
\index{electric conductivity!longitudinal}
\index{electric conductivity!transverse}
Comparison of Eqs.\ \req{1.109} with Ohm's law, ${\bf j}_{\rm ind} =
\hat{\sigma} \ecb$, leads us to define the {\it longitudinal} and {\it
transverse conductivities} as
\begin{subequations}
\label{1.111}
\beq
\sigma^{\rm (L)}({\bf q},\omega) \equiv
- {\rm i}
\frac{\omega}{4\pi} \left[
\epsilon^{\rm (L)} ({\bf q},\omega) - 1 \right]
\label{1.111a}\eeq
and
\beq
\sigma^{\rm (T)}({\bf q},\omega) \equiv
- {\rm i}
\frac{\omega}{4\pi} \left[
\epsilon^{\rm (T)} ({\bf q},\omega) - 1 \right],
\label{1.111b}\eeq
\end{subequations}
respectively. Thus, the Ohm law takes the form
\begin{subequations}
\label{1.112}
\beq
{\bf j}_{\rm ind}^{\rm (L,T)}({\bf q},\omega) =
\sigma^{\rm (L,T)} ({\bf q},\omega) \, \ecb^{\rm (L,T)}({\bf
q},\omega).
\label{1.112a}\eeq
Evidently,
\beq
\epsilon^{\rm (L,T)} ({\bf q},\omega) = 1 + \frac{\rm 4\pi {\rm
i}}{\omega} \, \sigma^{\rm (L,T)} ({\bf q},\omega).
\label{1.112b}\eeq
\end{subequations}
The properties \req{1.83} imply that
\beq
\left[ \sigma^{\rm (L,T)} ({\bf q},\omega) \right]^\ast =
- \sigma^{\rm (L,T)} (-{\bf q},-\omega).
\label{1.113}\eeq
which in the case of homogeneous isotropic media simplifies to
\beq
\left[ \sigma^{\rm (L,T)} (q,\omega) \right]^\ast =
- \sigma^{\rm (L,T)} (q,-\omega),
\label{1.114}\eeq
\ie, the real (imaginary) parts of $\sigma^{\rm (L,T)} (q,\omega)$ are
odd (even) functions of $\omega$ [cf. Eq.\ \req{1.84}].

In the small-wavelength limit, $q\rightarrow 0$, Eq.\ \req{1.99}
reduces to
\beq
\ecb_{\rm ind} ({\bf 0}, \omega) = - {\rm i} \frac{4\pi}{\omega} \left[
\left( \frac{1}{\epsilon^{\rm (L)}({\bf 0} , \omega)} - 1 \right) \,
{\bf j}_{\rm ext}^{\rm (L)}({\bf 0}, \omega) + \left( \frac{1}{
\epsilon^{\rm (T)}({\bf 0}, \omega)} - 1 \right) \, {\bf j}_{\rm
ext}^{\rm (T)}({\bf 0}, \omega) \right].
\nonumber \eeq
In this limit, the direction of the unit vector $\hat{\bf q}$ is
undefined and, therefore, the decomposition of ${\bf j}_{\rm ext}$ into
its longitudinal and transverse parts is indeterminate. As a matter of fact,
the quantity $\lim_{q\rightarrow 0} \ecb_{\rm ind} (q \hat{\bf q},
\omega)$ has to be independent of the direction of $\hat{\bf q}$. It
then follows that the longitudinal and transverse DFs
coincide at $q=0$,
\beq
\epsilon^{\rm (L)} ({\bf 0} , \omega)
= \epsilon^{\rm (T)} ({\bf 0} , \omega)
\equiv \epsilon(\omega).
\label{1.115}\eeq
We come to the same conclusion from the relation \req{1.90}, which
clearly implies that the difference
\beq
\epsilon^{\rm (L)} -  \epsilon^{\rm (T)} = \left( \frac{c q}{\omega}
\right)^2 (\mu^{-1}-1)
\label{1.116}\eeq
vanishes in the limit where $cq/\omega$ tends to zero.
Hereafter, $\epsilon(\omega)$, the low-$q$ limit of $\epsilon^{\rm
(L,T)} ({\bf q},\omega)$, will be referred to as the {\it optical dielectric
function} (ODF). \index{optical dielectric function}

\index{inverse dielectric function}
The Fourier transform of the induced potential, Eq.\ \req{1.91}, and
those of related quantities, involve the inverse longitudinal DF. As we
will see, for calculation purposes it is convenient to work with the
functions
\begin{subequations}
\label{1.117}
\beq
\eta^{\rm (L)} ({\bf q},\omega) \equiv \frac{1}{\epsilon^{\rm (L)}({\bf
q},\omega)} = \eta_1^{\rm (L)}({\bf q},\omega) - {\rm i}
\eta^{\rm (L)}_2({\bf q},\omega)
\label{1.117a}\eeq
and
\beq
\eta^{\rm (T)} ({\bf q},\omega) \equiv \frac{1}{\epsilon^{\rm (T)}({\bf
q},\omega)} = \eta_1^{\rm (T)}({\bf q},\omega) - {\rm i}
\eta^{\rm (T)}_2({\bf q},\omega) \, ,
\label{1.117b}\eeq
\end{subequations}
where the functions $\eta_2^{\rm (L,T)}({\bf q},\omega)$ are defined as
the negatives of the imaginary parts of $\eta^{\rm (L,T)}({\bf
q},\omega)$, which ensures that they are positive (see Section
\ref{sec1.3}). The functions $\eta_2^{\rm (L,T)}({\bf q},\omega)$ are
called the {\it energy-loss functions}, because they determine the
stopping force exerted by the material on charged projectiles moving
through it (see Section \ref{sec1.4}).
\index{dielectric functions!definitions|)}
\index{Maxwell equations!in material media|)}
\index{Fourier transform|)}


\subsection{Conservation of energy \label{sec1.2.3}}

\index{energy of electromagnetic fields}
The energy density (energy per unit volume) of the electromagnetic
field is \citep{Jackson1975},
\beq
U \equiv \frac{1}{8\pi} \left(\rule{0mm}{4mm}\ecb \dotprod \dcb
+ \bcb \dotprod \hcb \right)\, .
\label{1.118}\eeq
To derive the differential form of the law of energy conservation, we
first note that the work done by the field on the external charges in
a closed volume $V$ per unit time, through the action of the Lorentz
force \req{1.62}, is
\beq
\int_V {\bf j}_{\rm ext}({\bf r},t) \dotprod \ecb ({\bf r},t) \, \d {\bf
r}\, ,
\nonumber \eeq
because the magnetic force, being perpendicular to the velocity of the
charges, does no work. We use the Maxwell equation \req{1.57b}
to write
\beq
\int_V {\bf j}_{\rm ext} \dotprod \ecb
\, \d {\bf r} = \frac{1}{4\pi}
\int_V
\left( c \ecb \dotprod (\nablab \vecprod \hcb)
- \ecb \dotprod \frac{\partial \dcb}{\partial t} \right)
\, \d {\bf r} \, ,
\nonumber \eeq
and, considering the vector equality
\beq
\nablab \dotprod \left( \ecb \vecprod \hcb \right) =
\hcb \dotprod \left( \nablab \vecprod \ecb \right)
- \ecb \dotprod \left(\nablab \vecprod \hcb \right)\, ,
\nonumber \eeq
and the Maxwell equation \req{1.57c}, we find
\beq
\int_V {\bf j}_{\rm ext} \dotprod \ecb
\, \d {\bf r} = \frac{1}{4\pi}
\int_V \left(
- c \nablab \dotprod \left( \ecb \vecprod \hcb \right)
- \hcb \dotprod \frac{\partial \bcb}{\partial t}
- \ecb \dotprod \frac{\partial \dcb}{\partial t} \right)
\, \d {\bf r} \, .
\label{1.119}\eeq
On the other hand, the variation of electromagnetic energy in $V$ per
unit time is
\allowdisplaybreaks{
\beq
\int_V \frac{\partial U}{\partial t} \, \d {\bf r} =
\frac{1}{8\pi} \int_V
\left(
\frac{\partial \ecb}{\partial t} \dotprod \dcb
+ \ecb \dotprod \frac{\partial \dcb}{\partial t}
+ \frac{\partial \bcb}{\partial t} \dotprod \hcb
+ \bcb \dotprod \frac{\partial \hcb}{\partial t}
\right)
\, \d {\bf r} \, .
\nonumber \eeq
Now we add and subtract the last quantity to the right-hand side of
Eq.\ \req{1.119} to write
\beqa
\int_V {\bf j}_{\rm ext} \dotprod \ecb
\, \d {\bf r} &=& - \int_V \frac{c}{4\pi}
\, \nablab \dotprod \left( \ecb \vecprod \hcb \right) \, \d {\bf r}
-\frac{1}{8\pi}
\int_V \left(
2 \hcb \dotprod \frac{\partial \bcb}{\partial t}
+ 2 \ecb \dotprod \frac{\partial \dcb}{\partial t} \right)
\, \d {\bf r}
\nonumber \\ [2mm]
&& \mbox{}
- \int_V \frac{\partial U}{\partial t} \, \d {\bf r} +
\frac{1}{8\pi} \int_V \left(
\ecb \dotprod \frac{\partial \dcb}{\partial t}
+ \frac{\partial \ecb}{\partial t} \dotprod \dcb
+ \bcb \dotprod \frac{\partial \hcb}{\partial t}
+ \frac{\partial \bcb}{\partial t} \dotprod \hcb
\right)
\, \d {\bf r} \, .
\nonumber \eeqa
As the volume $V$ is arbitrary, we can identify the integrands on the
two sides of this equality. This leads to the following differential form
of the energy-conservation law,
\beq
- \frac{\partial U}{\partial t}
- \nablab \dotprod {\bf S}
= {\bf j}_{\rm ext} \dotprod \ecb + \frac{1}{8\pi}
\left(
\ecb \dotprod \frac{\partial \dcb}{\partial t}
- \frac{\partial \ecb}{\partial t} \dotprod \dcb
- \bcb \dotprod \frac{\partial \hcb}{\partial t}
+ \frac{\partial \bcb}{\partial t} \dotprod \hcb
\right),
\label{1.120}\eeq
where \index{Poynting vector}
\beq
{\bf S} = \frac{c}{4\pi} \, \ecb \vecprod \hcb
\label{1.121}\eeq
is the {\it Poynting vector}, which gives the energy flow of the
electromagnetic field (energy that crosses a small surface perpendicular
to ${\bf S}$ per unit surface area and per unit time).

\index{dissipated electromagnetic energy}
The second term on the right-hand side of Eq.\ \req{1.120} describes the
work done by the field on the medium. To prove this assertion, we
calculate the integral of that term over a large volume $\Delta V$
during a suitably long time $\Delta t$,
\beq
\Delta W_{\rm medium} = \int_{\Delta V} \d {\bf r} \int_t^{t+\Delta t} \d
t \,
\frac{1}{8\pi} \left(
\ecb \dotprod \frac{\partial \dcb}{\partial t}
- \frac{\partial \ecb}{\partial t} \dotprod \dcb
- \bcb \dotprod \frac{\partial \hcb}{\partial t}
+ \frac{\partial \bcb}{\partial t} \dotprod \hcb
\right).
\label{1.122}\eeq
In the calculation we express the fields in terms of the potentials, and
recall that the potentials are real [\ie, ${\bf A} (-{\bf q},-\omega) =
{\bf A}^\ast ({\bf q},\omega) $] and that ${\bf A}({\bf q},\omega)$ is
perpendicular to ${\bf q}$ (in the Coulomb gauge). Let us consider the
contribution from the first term,
\beqa
{\cal W}_1 &\equiv& \int_{\Delta V} \d {\bf r} \int_t^{t+\Delta t}
\d t \, \ecb \dotprod \frac{\partial \dcb}{\partial t}
\nonumber \\ [2mm]
&=& \frac{1}{(2\pi)^4} \int_{\Delta V} \d {\bf r} \int_t^{t+\Delta t}
\d t \int \d {\bf q} \int \d \omega \,
\exp\left[ {\rm i} \left( {\bf q} \dotprod {\bf r} - \omega t \right)
\right] \,
\nonumber \\ [2mm]
&& \mbox{} \times
\left( - {\rm i} {\bf q} \varphi ({\bf q},\omega)
+ {\rm i} \, \frac{\omega}{c} \, {\bf A} ({\bf q},\omega) \right)
\int \d {\bf q}' \int \d \omega' \,
\exp\left[ {\rm i} \left( {\bf q}' \dotprod {\bf r} - \omega' t \right)
\right]
\nonumber \\ [2mm]
&& \mbox{} \dotprod
\, (-{\rm i} \omega') \, \epsilon^{\rm (L)} ({\bf q}',\omega')
\left( - {\rm i} {\bf q}' \varphi ({\bf q}',\omega')
+ {\rm i} \, \frac{\omega'}{c} \, {\bf A} ({\bf q}',\omega') \right)
\nonumber \\ [2mm]
&=& \int \d {\bf q} \int \d \omega \,
\, \left( - {\rm i} {\bf q} \varphi ({\bf q},\omega)
+ {\rm i} \, \frac{\omega}{c} \, {\bf A} ({\bf q},\omega) \right)
\nonumber \\ [2mm]
&& \mbox{} \dotprod
\int \d {\bf q}' \int \d \omega' \,
\, (-{\rm i} \omega') \, \epsilon^{\rm (L)} ({\bf q}',\omega')
\left( - {\rm i} {\bf q}' \varphi ({\bf q}',\omega')
+ {\rm i} \, \frac{\omega'}{c} \, {\bf A} ({\bf q}',\omega') \right)
\nonumber \\ [2mm]
&& \mbox{} \times
\frac{1}{(2\pi)^4}  \int_{\Delta V} \d {\bf r} \int_t^{t+\Delta t}
\exp\left[ {\rm i} \left( {\bf q} + {\bf q}'\right) \dotprod {\bf r} \right]
\exp\left[ -{\rm i} \left( \omega + \omega'\right) t \right],
\nonumber \eeqa
where use has been made of the relations \req{1.75a} and \req{1.86a}.
Now we consider that the volume $\Delta V$ has dimensions much larger than
the wavelength and that $\Delta t$ is much longer than the period of the
waves\footnote{Under these circumstances, the integrals of the
exponentials can be approximated as
$$
\int_{-\infty}^\infty \exp({\rm i} u x )\, \d x = 2\pi \, \delta(u),
$$
}, so that we can write
\beqa
{\cal W}_1 &\simeq& \int \d {\bf q}' \int \d \omega' \,
\, \left( {\rm i} {\bf q}' \varphi (-{\bf q}',-\omega')
- {\rm i} \, \frac{\omega'}{c} \, {\bf A} (-{\bf q}',-\omega') \right)
\nonumber \\ [2mm]
&& \mbox{} \dotprod
(-{\rm i} \omega') \, \epsilon^{\rm (L)} ({\bf q'},\omega')
\left( - {\rm i} {\bf q}' \varphi ({\bf q}',\omega')
+ {\rm i} \, \frac{\omega'}{c} \, {\bf A} ({\bf q}',\omega') \right)
\nonumber \\ [2mm]
&=& \int \d {\bf q} \int \d \omega \,
(-{\rm i} \omega) \, \epsilon^{\rm (L)} ({\bf q},\omega)
\left( q^2 \left| \varphi ({\bf q},\omega)\right|^2
+ \frac{\omega^2}{c^2} \, \left| {\bf A} ({\bf q},\omega) \right|^2
\right).
\nonumber \eeqa
The calculation of the last term in \req{1.122} can be performed by
following the same steps. We have
\beqa
{\cal W}_2 &\equiv& \int_{\Delta V} \d {\bf r} \int_t^{t+\Delta t}
\d t \, \frac{\partial \bcb}{\partial t} \dotprod \hcb
\nonumber \\ [2mm]
&=& \frac{1}{(2\pi)^4} \int_{\Delta V} \d {\bf r} \int_t^{t+\Delta t}
\d t \int \d {\bf q} \int \d \omega \,
\exp\left[ {\rm i} \left( {\bf q} \dotprod {\bf r} - \omega t \right)
\right] \, (-{\rm i} \omega)
\,  {\rm i} {\bf q} \vecprod {\bf A} ({\bf q},\omega)
\nonumber \\ [2mm]
&& \mbox{} \dotprod
\int \d {\bf q}' \int \d \omega' \,
\exp\left[ {\rm i} \left( {\bf q}' \dotprod {\bf r} - \omega' t \right)
\right] \dotprod \mu^{-1} ({\bf q'},\omega') \, {\rm i} {\bf q}' \vecprod
{\bf A} ({\bf q}',\omega')
\nonumber \\ [2mm]
&\simeq& \int \d {\bf q}' \int \d \omega' \,
({\rm i} \omega')
\,  {\rm i} (-{\bf q}') \vecprod {\bf A} (-{\bf q}',-\omega')
\, \mu^{-1} ({\bf q'},\omega') \, {\rm i} {\bf q}' \vecprod
{\bf A} ({\bf q}',\omega')
\nonumber \\ [2mm]
&=& \int \d {\bf q} \int \d \omega \,
({\rm i} \omega) q^2
\, \mu^{-1} ({\bf q},\omega) \, \left| {\bf A} ({\bf q},\omega)\right|^2
\nonumber \\ [2mm]
&=& \int \d {\bf q} \int \d \omega \,
({\rm i} \omega) q^2
\, \left\{ 1 - \left( \frac{\omega}{c q} \right)^2
\left[\rule{0mm}{4mm}\epsilon^{\rm (T)}({\bf q},\omega)
-  \epsilon^{\rm (L)}({\bf q},\omega) \right]\right\}
 \, \left| {\bf A} ({\bf q},\omega)\right|^2,
\nonumber \eeqa
where we have used the relation \req{1.90}.

It can be readily verified that the contributions of the second and
third terms in Eq.\ \req{1.122} (including the minus signs) are equal to
${\cal W}_1$ and ${\cal W}_2$, respectively. Hence
\beqa
\Delta W_{\rm medium} &=& \frac{1}{4\pi}
\int \d {\bf q} \int \d \omega \, \left\{
(-{\rm i} \omega) \, \epsilon^{\rm (L)} ({\bf q},\omega)
\left( q^2 \left| \varphi ({\bf q},\omega)\right|^2
+ \frac{\omega^2}{c^2} \, \left| {\bf A} ({\bf q},\omega) \right|^2
\right) \right.
\nonumber \\ [2mm]
&& \mbox{} \left. + ({\rm i} \omega)
\, \left[ q^2 -  \frac{\omega^2}{c^2}
\left(\rule{0mm}{4mm}\epsilon^{\rm (T)}({\bf q},\omega)
- \epsilon^{\rm (L)}({\bf q},\omega) \right)\right] \right\}
 \, \left| {\bf A} ({\bf q},\omega)\right|^2
\nonumber \\ [2mm]
&=&
\int \d {\bf q} \int \d \omega \, \left\{
\frac{-{\rm i} \omega}{4\pi} \, \epsilon^{\rm (L)} ({\bf q},\omega)
\, q^2 \left| \varphi ({\bf q},\omega)\right|^2
+ \frac{-{\rm i} \omega}{4\pi}
\, \epsilon^{\rm (T)}({\bf q},\omega)
\, \frac{\omega^2}{c^2} \, \left| {\bf A} ({\bf q},\omega)\right|^2
\right\} \, . \rule{10mm}{0mm}
\nonumber \eeqa
Notice that the function $\omega q^2 \left| {\bf A} ({\bf
q},\omega)\right|^2$ is odd in $\omega$ and does not contribute to the
integral. Moreover, as the real parts of $\omega \, \epsilon^{\rm
(L,T)}$ are odd functions of $\omega$, we can write
\beqa
\Delta W_{\rm medium}
&=& \int \d {\bf q} \int \d \omega \, \left\{
\left[ {\rm Re} \, \sigma^{\rm (L)} ({\bf q},\omega) \right] \,
\, q^2 \left| \varphi ({\bf q},\omega)\right|^2 \rule{0mm}{6mm}
\right.
\nonumber \\ [2mm]
&& \mbox{} \left. \rule{10mm}{0mm}
+ \left[ {\rm Re} \, \sigma^{\rm (T)} ({\bf q},\omega) \right] \,
\frac{\omega^2}{c^2} \, \left| {\bf A} ({\bf q},\omega)\right|^2
\right\} \, ,
\label{1.123}\eeqa
where $\sigma^{\rm (L,T)} ({\bf q},\omega)$ are the conductivities
defined by Eqs.\ \req{1.111}.

On the other hand, the energy dissipated into the medium can be
calculated as the work done on the induced currents, \ie,
\beqa
\Delta W_{\rm dis} &=& \int_{\Delta V} \d {\bf r} \int_t^{t+\Delta t} \d
t \, {\bf j}_{\rm ind} ({\bf r},t) \dotprod \ecb({\bf r},t)
\nonumber \\ [2mm]
&=& \frac{1}{(2\pi)^4} \int_{\Delta V} \d {\bf r} \int_t^{t+\Delta t}
\d t \int \d {\bf q} \int \d \omega \,
\exp\left[ {\rm i} \left( {\bf q} \dotprod {\bf r} - \omega t \right)
\right]
\nonumber \\ [2mm]
&& \rule{10mm}{0mm} \times
\left(
\sigma^{\rm (L)} ({\bf q},\omega) \ecb^{\rm (L)} ({\bf q},\omega)
+ \sigma^{\rm (T)} ({\bf q},\omega) \ecb^{\rm (T)}({\bf q},\omega) \right)
\nonumber \\ [2mm]
&& \mbox{} \dotprod
\int \d {\bf q}' \int \d \omega' \,
\exp\left[ {\rm i} \left( {\bf q}' \dotprod {\bf r} - \omega' t \right)
\right]
\, \left(
\ecb^{\rm (L)} ({\bf q},\omega)
+ \ecb^{\rm (T)}({\bf q},\omega) \right)
\nonumber \\ [2mm]
&=& \int \d {\bf q}' \int \d \omega' \,
\left[
\sigma^{\rm (L)} (-{\bf q}',-\omega') \ecb^{\rm (L)} (-{\bf q}',-\omega')
\right.
\nonumber \\ [2mm]
&& \mbox{} \left.
+ \sigma^{\rm (T)} (-{\bf q}',-\omega') \ecb^{\rm (T)}(-{\bf
q}',-\omega') \right]
\dotprod \left(
\ecb^{\rm (L)} ({\bf q},\omega)
+ \ecb^{\rm (T)}({\bf q},\omega) \right)
\nonumber \\ [2mm]
&=& \int \d {\bf q}' \int \d \omega' \,
\left[ \sigma^{\rm (L)} (-{\bf q}',-\omega')
\left| \ecb^{\rm (L)} (-{\bf q}',-\omega') \right|^2
\right.
\nonumber \\ [2mm]
&& \mbox{} \left. \rule{10mm}{0mm}
+ \sigma^{\rm (T)} (-{\bf q}',-\omega')
\left| \ecb^{\rm (T)}(-{\bf q}',-\omega') \right|^2 \right].
\nonumber \eeqa
Recalling that the imaginary parts of the conductivities are
odd functions of $\omega$, we can write
\beqa
\Delta W_{\rm dis} &=&
\int \d {\bf q} \int \d \omega \,
\left\{ \left[ {\rm Re} \, \sigma^{\rm (L)} ({\bf q},\omega) \right] \, q^2
\left| \varphi ({\bf q},\omega) \right|^2 \right.
\nonumber \\ [2mm]
&& \mbox{} \left.
+ \left[ {\rm Re} \, \sigma^{\rm (T)} ({\bf q},\omega) \right]
\frac{\omega^2}{c^2}
\left| {\bf A}({\bf q},\omega) \right|^2 \right\}.
\label{1.124}\eeqa
Indeed, this result coincides with \req{1.123}, $\Delta W_{\rm dis} =
\Delta W_{\rm medium}$, showing that the second
term on the right-hand side of the energy-conservation equation
\req{1.120} does represent work performed on the currents induced in the
medium. The energy transferred to the medium in this manner dissipates
irreversibly due to electron collisions and eventually transforms into
heat. The foregoing analysis shows that the energy-conservation equation
\req{1.120} may also be expressed as \citep{Jackson1975}
\beq
\frac{\partial U}{\partial t}
+ \nablab \dotprod {\bf S}
= -\, {\bf j} \dotprod \ecb \, ,
\label{1.125}\eeq
with ${\bf j} = {\bf j}_{\rm ext} + {\bf j}_{\rm ind}$, the total
current.
}


\subsection{Electromagnetic radiation. Optical functions
\label{sec1.2.4}} \index{electromagnetic radiation}

Let us now consider the propagation of electromagnetic fields produced
by distributions of external charges that are localized and electrically
neutral. Far from the charges, the scalar (Coulomb) potential decreases
at least as $r^{-3}$, where $r$ is the distance from the point of
measurement to the center of the charge distribution [see Eqs.\
\req{1.53}].  Moreover, at large $r$ the transverse current also decreases as
$r^{-3}$ [Eq.\ \req{1.55}]. Hence, far away from the external charges,
$\varphi=0$ and ${\bf j}_{\rm ext}^{\rm (T)}=0$, that is, we have a pure
radiation field.

The vector potential of a radiation field satisfies the wave equation
\req{1.82b} with ${\bf j}_{\rm ext} ={\bf 0}$,\index{wave equation!of a radiation field}
\beq
\nabla^2 {\bf A}({\bf r},t) - \frac{1}{c^2}
\frac{\partial^2 [\hat\epsilon^{\rm (T)}\, {\bf
A}({\bf r},t)]}
{\partial t^2} = 0,
\label{1.126}\eeq
which in Fourier space takes the form
\beq
\left[ q^2 -
\frac{\omega^2}{c^2}
\, \epsilon^{\rm (T)} ({\bf q},\omega) \right]
{\bf A}({\bf q},\omega) = 0.
\label{1.127}\eeq

In many practical situations, radiation fields have relatively small
wave numbers, for which the DFs
vary smoothly with $q$ [see, \eg, the DF model given by Eq.\ \req{1.169}
below, or the more realistic DF models described in Section
\ref{sec7.6}] and it is a good approximation to replace $\epsilon^{\rm
(T)}({\bf q},\omega)$ with the ODF, $\epsilon(\omega)$, which amounts to
assuming that $\mu \simeq 1$ (\ie, that magnetic effects are negligible).
Then, the wave equation \req{1.126} becomes
\beq
\nabla^2 {\bf A} - \frac{1}{c^2} \frac{\partial^2
\, \hat\epsilon_{\rm opt} \, {\bf A}} {\partial t^2} = 0,
\label{1.128}\eeq
where $\hat\epsilon_{\rm opt}$ is the ODF operator. Evidently, the
only possible Fourier ``components'' ${\bf A}({\bf q},\omega)$ of the field
are those for which the wavenumber $q$ and the frequency $\omega$
satisfy the dispersion relation, \index{dispersion relation of
electromagnetic radiation}
\beq
q^2 - \frac{\omega^2}{c^2} \, \epsilon(\omega) = 0.
\label{1.129}\eeq
This implies that, for a given angular frequency $\omega$, monochromatic
plane waves of the type\footnote{Since vector potentials are assumed to be
real, \ie, such that ${\bf A}^\ast ({\bf q},\omega) = {\bf A}
(-{\bf q}, -\omega)$ [see Eq.\
\req{1.32}], the Fourier transform of a monochromatic wave is the sum of
a pair of
Fourier terms having opposite wave vectors and frequencies.} \index{monochromatic plane waves}
\beqa
{\bf A}({\bf r},t) &=& {\bf A}_0 \left\{\rule{0mm}{4mm}\!\exp[{\rm i} ({\bf
q}\dotprod {\bf r} - \omega t)] + \exp[-{\rm i} ({\bf
q}\dotprod {\bf r} - \omega t)] \right\}
\nonumber \\ [2mm]
&=& 2{\bf A}_0 \, \cos( {\bf q}\dotprod {\bf r} - \omega t)
\label{1.130}\eeqa
exist only for wave numbers $q$ that satisfy the dispersion relation
\req{1.129}. Notice that, in the Coulomb gauge, the vector
${\bf A}_0$ is perpendicular to ${\bf q}$.
We see that the phase velocity of a plane wave is
\beq
v_{\rm ph}(\omega)= \omega/q =c / \sqrt{\epsilon(\omega)} \, .
\label{1.131}\eeq
In particular, for a plane wave in vacuum, $q=\omega/c$ and $v_{\rm ph} = c$.

\index{complex refractive index}
\index{index of refraction} \index{extinction coefficient}
It is convenient to introduce the {\it complex refractive index} defined by
\beq
N(\omega) = n(\omega) + {\rm i} \kappa(\omega) = \sqrt{\epsilon(\omega)},
\label{1.132}\eeq
where the branch of the square root is the one yielding a non-negative
$n(\omega)$. The real and imaginary parts of $N(\omega)$ are the {\it
index of refraction} $n(\omega)$ and the {\it extinction coefficient}
$\kappa(\omega)$, respectively. They can be expressed in terms of the
real and imaginary parts of the ODF,  $\epsilon(\omega) =
\epsilon_1(\omega) + {\rm i} \epsilon_2(\omega)$, as
\beqa
n(\omega) &=& {\rm Re} \sqrt{\epsilon(\omega)}
= \1o2 \sqrt{2 \epsilon_1(\omega)+ 2 \sqrt{\epsilon_1^2(\omega)
+ \epsilon_2^2(\omega)}},
\label{1.133}\\ [2mm]
\kappa(\omega) &=& {\rm Im} \sqrt{\epsilon(\omega)}
= \frac{\epsilon_2(\omega)}{\sqrt{2 \epsilon_1(\omega)
+ 2 \sqrt{\epsilon_1^2(\omega) +\epsilon_2^2(\omega)}}}\, .
\label{1.134}\eeqa
Conversely, the real and imaginary parts of the complex ODF can be
obtained from $N(\omega)$ as
\beq
\epsilon_1(\omega) = n^2 (\omega) - \kappa^2(\omega)
\qquad \mbox{and} \qquad
\epsilon_2(\omega) = 2 n (\omega) \kappa (\omega).
\label{1.135}\eeq
Notice that $n(\omega)$ and $\kappa(\omega)$ are defined for positive and
negative frequencies. Because $\epsilon(-\omega) = \epsilon^\ast
(\omega)$, the index of refraction (extinction coefficient) is an even
(odd) function of $\omega$.

The dispersion relation \req{1.129} can now be expressed as
\beq
q(\omega) = \frac{\omega}{c} \left[ n(\omega) + {\rm i} \kappa(\omega)
\right],
\label{1.136}\eeq
and the monochromatic plane waves [see Eq.\ \req{1.130}] take the
form
\beqa
{\bf A}({\bf r},t) &=& {\bf A}_0 \left\{\rule{0mm}{4mm}
\exp\left[{\rm i} \left( q(\omega) \hat{\bf q}
\dotprod {\bf r} - \omega t\right) \right] +
\exp\left[- {\rm i} \left( q(-\omega) \hat{\bf q}
\dotprod {\bf r} - \omega t\right) \right] \right\}
\nonumber \\ [2mm]
&=& 2 {\bf A}_0 \cos\left\{\rule{0mm}{4mm} (\omega/c) [n(\omega)
\, \hat{\bf q} \dotprod {\bf r} - c t] \right\} \,
\exp \left[ - (\omega/c)
\kappa(\omega) \,
\hat{\bf q} \dotprod {\bf r} \right].
\label{1.137}\eeqa
From this expression we see that the wave is attenuated exponentially as
it penetrates the medium, and that the phase velocity is $v_{\rm
ph}(\omega) = c/n(\omega)$, according to the definition of the index of
refraction in geometrical optics. As this velocity is a function of
$\omega$, electromagnetic wave packets consisting of Fourier components
of various frequencies change shape as they propagate within the medium.
The effect is most pronounced in frequency regions of ``{\it anomalous
dispersion}'', where $n(\omega)$ varies rapidly over a narrow interval of
frequencies. \index{speed of light!in a dielectric material}

The time-average energy density of the monochromatic plane wave
\req{1.137} is proportional to ${\bf A}^2$, and decreases with the
penetrated thickness, $d = \hat{\bf q} \cdot {\bf r}$, as
$\exp[-2(\omega /c) \kappa d]$. In the quantum description of the
penetration of photons through matter, the attenuation is expressed as
$\exp(- {\cal N} \sigma_{\rm ph} d)$, where $\sigma_{\rm ph}$ is the
molecular photoabsorption cross section for photons of energy $\hbar
\omega$, and ${\cal N}$ is the number of molecules per unit volume. We
are thus led to make the identification \index{photoabsorption cross
section}
\beq
\kappa(\omega) = \frac{{\cal N} c}{2\omega} \sigma_{\rm ph}(\hbar\omega).
\label{1.138}\eeq
For photon energies of the order of a few hundred eV
and larger, the index of refraction is very close to unity
and $\kappa(\omega) \ll 1$, and we can write
\beq
\epsilon_2(\omega) \simeq 2 \kappa(\omega) =
\frac{{\cal N} c}{\omega} \sigma_{\rm ph}(\hbar\omega)
\label{1.139}\eeq
and
\beq
\eta_2(\omega) = {\rm Im} \left( \frac{-1}{\epsilon(\omega)} \right)
= \frac{2 n(\omega) \kappa(\omega)}{[n^2(\omega) + \kappa^2(\omega)]^2}
\simeq 2 \kappa(\omega) =
\frac{{\cal N} c}{\omega} \sigma_{\rm ph}(\hbar\omega).
\label{1.140}\eeq
Thus, at these energies the imaginary part of the ODF (and that of the
inverse ODF) can be obtained from measured or calculated photoabsorption
data. Tabulations of photoelectric cross sections for the elements,
either theoretical or derived from measurements of x-ray attenuation,
are available \citep{Scofield1973, Henke1993, SabbatucciSalvat2016}.


\section{Classical model of the dielectric function \label{sec1.3}}

\index{dielectric functions!classical Lorentz--Drude model|(}
Optical properties of insulators were first modeled by
\citet{Lorentz1909} by
considering that the electrons in the medium react to external
electromagnetic fields as classical damped oscillators. The oscillator
model gives an elementary understanding of optical properties of
materials. Indeed, realistic ODFs of materials, which may be obtained by
combining measurements and elaborate quantum many-body calculations, can
be closely approximated as a set of classical oscillators with
characteristic parameters determined by fitting the known ODF.

Let us consider the simple situation where electromagnetic
fields are spatially homogeneous. For the sake of concreteness, we
assume a molecular material with $Z$ electrons per molecule and molar
mass $A_{\rm m}$ (grams per mol), the number of molecules per unit of
volume is
\beq
{\cal N} = \frac{N_{\rm A} \rho_{\rm m}}{A_{\rm m}} \, ,
\label{1.141}\eeq
where $\rho_{\rm m}$ is the mass density of the material and $N_{\rm A}
= 6.022\, 141 \times 10^{23}$ mol$^{-1}$ is the Avogadro constant. Within
the linear response approximation, the polarization ${\bf P}$ (dipole
moment per unit volume) is parallel to the electric field $\ecb$,
\beq
{\bf P} = \chi_{\rm e} \ecb,
\label{1.142}\eeq
where $\chi_{\rm e}$ is the {\it electric susceptibility}. Under the action of
an electric field $\ecb$, an isolated molecule acquires an
induced dipole moment ${\bf p}_{\rm mol} = \gamma_{\rm mol} \ecb$, where
$\gamma_{\rm mol}$ is the {\it molecular polarizability}. In the
case of fields in a condensed medium with polarization ${\bf P}$, an
atom is acted by an additional field $\ecb_{\rm pol}$ caused by the
polarization of neighboring molecules, which is approximately given by
\citep{Jackson1975}
\beq
\ecb_{\rm pol} = (4\pi/3) {\bf P}.
\label{1.143}\eeq
Thus, for condensed media, we can write
\beq
{\bf P} = {\cal N} \gamma_{\rm mol} \left[ \ecb + (4\pi/3) {\bf
P} \right].
\label{1.144}\eeq
This result can be combined with Eq.\ \req{1.142} to yield
\beq
\chi_{\rm e} = \frac{{\cal N} \gamma_{\rm mol}}{1 - (4\pi/3) {\cal
N}\gamma_{\rm mol}}.
\label{1.145}\eeq
Recalling the definitions \req{1.58a} and \req{1.59},
\beq
\dcb = \ecb + 4\pi {\bf P}, \qquad \dcb = \epsilon \ecb,
\nonumber\eeq
we have
\beq
\epsilon = 1 + 4\pi \chi_{\rm e} = 1 + 3 \, \frac{(4\pi/3) {\cal
N} \gamma_{\rm mol}}{1- (4\pi/3) {\cal N} \gamma_{\rm mol}}.
\label{1.146}\eeq
The inverse relation,
\beq
\frac{4\pi}{3} \, {\cal N}\gamma_{\rm mol} = \frac{\epsilon-1}{\epsilon+2},
\label{1.147}\eeq
is known as the {\it Clausius--Mossotti
equation}\index{Clausius--Mossotti equation}.

The induced atomic dipole moment can be estimated by considering that,
for electric fields that are not too large, the atomic electrons react
as isotropic oscillators with characteristic frequencies $\omega_j$ that
are subject to a friction force proportional to the electron's velocity
(Lorentz oscillators). The equation of motion of an electron (mass
$\me$, charge $-e$) bound by a harmonic force $-\me \omega_j^2 {\bf r}$
under the action of an electric field $\ecb(t)$ (assumed to be constant
over the volume of the atom) and a damping force $-s_j \dot{\bf r}$
is\footnote{Although a time-dependent electric field is always
accompanied by a magnetic field, the magnetic force is much weaker than
the electric force and can be neglected here.}
\beq
\ddot{\bf r} + s_j \dot{\bf r} + \omega_j^2 {\bf r} = - \frac{e}{\me}
\, \ecb(t).
\label{1.148}\eeq
The damping term is intended to account for the energy that is lost by
the oscillator due to radiation emitted because of the electron
acceleration
and to collisions with other electrons; it also helps to remove
mathematical ambiguities that would occur when the dissipative term is missing.
Equation \req{1.148} can be solved by expressing the functions ${\bf
r}(t)$ and $\ecb(t)$ in terms of their Fourier transforms,
\begin{subequations}
\label{1.149}
\beq
{\bf r} (t) = (2\pi)^{-1/2} \int_{-\infty}^\infty \exp(-{\rm i} \omega t)
\, {\bf r}(\omega) \, \d \omega
\label{1.149a}\eeq
and
\beq
\ecb (t) = (2\pi)^{-1/2} \int_{-\infty}^\infty \exp(-{\rm i} \omega t)
\, \ecb (\omega) \, \d \omega,
\label{1.149b}\eeq
\end{subequations}
with
\begin{subequations}
\label{1.150}
\beq
{\bf r} (\omega) = (2\pi)^{-1/2} \int_{-\infty}^\infty \exp({\rm i} \omega t)
\, {\bf r}(t) \, \d t
\label{1.150a}\eeq
and
\beq
\ecb (\omega) = (2\pi)^{-1/2} \int_{-\infty}^\infty \exp({\rm i} \omega t)
\, \ecb (t) \, \d t.
\label{1.150b}\eeq
Notice that both ${\bf r}(t)$ and $\ecb(t)$ are real and, hence,
\beq
{\bf r}(-\omega) = {\bf r}^\ast(\omega)
\qquad \mbox{and} \qquad
\ecb(-\omega) = \ecb^\ast(\omega).
\label{1.150c}\eeq
\end{subequations}
Inserting the expressions \req{1.149} into the equation of motion
\req{1.148} we find
\beq
{\bf r}(\omega) = - \frac{e}{\me} \frac{\ecb(\omega)}{
\omega_j^2 - \omega^2 - {\rm i} s_j \omega}.
\label{1.151}\eeq
The trajectory ${\bf r}(t)$ is obtained by inverting the Fourier
transform, Eq.\ \req{1.149a}.
In the case of a monochromatic (real) wave,
\beq
\ecb(t) = \frac{\ecb_0}{2} \left[
\exp(-{\rm i} \omega_0 t) + \exp({\rm i} \omega_0 t)
\right],
\label{1.152}\eeq
we have
\beq
\ecb(\omega) = (2\pi)^{1/2} \frac{\ecb_0}{2}
\left[ \delta(\omega-\omega_0) +
\delta(\omega+\omega_0) \right],
\label{1.153}\eeq
and
\beqa
{\bf r} (t) &=& - \frac{e}{\me} \, \frac{\ecb_0}{2}
\int \exp(-{\rm i} \omega t) \,\frac{
\delta(\omega-\omega_0) + \delta(\omega+\omega_0)}
{\omega_j^2 - \omega^2 - {\rm i} s_j \omega} \, \d \omega
\nonumber \\ [2mm]
&=&
- \frac{e}{\me} \, \frac{\ecb_0}{2}
\left[ \frac{\exp(-{\rm i} \omega_0 t)}
{\omega_j^2 - \omega_0^2 - {\rm i} s_j \omega_0}
+ \frac{\exp({\rm i} \omega_0 t)}
{\omega_j^2 - \omega_0^2 + {\rm i} s_j \omega_0} \right].
\label{1.154}\eeqa
This exact result shows that in the case a complex monochromatic wave
with positive or negative frequency $\omega$, $\ecb(t) = \ecb_0
\exp(-{\rm i} \omega t)$, we can formally write
\beq
{\bf r}(t) = - \frac{e}{\me} \frac{\ecb_0 \exp(-{\rm i} \omega t)}{
\omega_j^2 - \omega^2 -{\rm i} s_j \omega}.
\label{1.155}\eeq
We see that the electron vibrates with the same frequency as the local
field, but with an amplitude that depends on the frequency.

Let us assume that a molecule consists of a number of sets of
``equivalent'' electrons, each one characterized by the number $f_j$ of
electrons, the resonance frequency $\omega_j$, and the damping constant
$s_j$. The molecular dipole moment can then be expressed as
\beq
{\bf p}_{\rm mol} = \sum_j \left( -e {\bf r}_j \right) f_j =
\left( \frac{e^2}{\me} \sum_j
\frac{f_j}{\omega_j^2 - \omega^2 -{\rm i} s_j \omega} \right)
\ecb_0 \exp(-{\rm i} \omega t),
\label{1.156}\eeq
and the molecular polarizability is
\beq
\gamma_{\rm mol}(\omega) = \frac{e^2}{\me} \sum_j
\frac{f_j}{\omega_j^2 - \omega^2 -{\rm i} s_j \omega},
\label{1.157}\eeq
where the dependence on the frequency of the local field has been
indicated explicitly. Substitution of this result into Eq.\ \req{1.146}
yields a closed expression for the ODF, but with a quite complicated
form. In the case of negligible overlap between resonances, \ie, when
only a single term may contribute to the sum in Eq.\ \req{1.157} for each
$\omega$, it is easily shown that the ODF takes the following form
\beq
\epsilon(\omega) = 1 + \frac{\Omega_{\rm p}^2}{Z}
\sum_j \frac{f_j}{\overline{\omega}_j^2 - \omega^2 -{\rm i} s_j \omega}
\label{1.158}\eeq
with
\beq
\overline{\omega}_j^2 =
\omega_j^2 - \frac{1}{3} \, \frac{f_j}{Z}\, \Omega_{\rm p}^2.
\label{1.159}\eeq
The quantity \index{plasma resonance frequency}
\beq
\Omega_{\rm p} \equiv \sqrt{ 4 \pi \, {\cal N} Z \, \frac{e^2}{\me}}
\label{1.160}\eeq
is the plasma resonance frequency of a free electron gas with the
electron density ${\cal N} Z$ (electrons per unit volume) of the medium.
These expressions are also approximately valid in the case of
overlapping resonances \citep{Sternheimer1952}.
We see that the polarizability of the medium produces a shift of the
resonance frequencies which is proportional to the electronic density;
this shift is sometimes called the {\it Lorenz-Lorentz correction}
\citep{Sternheimer1984}\index{Lorenz--Lorentz correction}.
In the limit of small densities (${\cal N}
\rightarrow 0$), as in thin gases, the resonance frequencies
$\overline{\omega}_j$ tend to the ``natural'' frequencies $\omega_j$ of
the oscillators (which are characteristic of each individual atom or
molecule).

The classical DF for a set of Lorentz oscillators, Eq.\ \req{1.158},
with frequencies $\overline{\omega}_j$ greater than zero and {\it
oscillator strengths} $f_j$ satisfying the sum rule
\beq
\sum_j f_j = Z,
\label{1.161}\eeq
provides a convenient starting point for describing the optical
properties of insulators. Of course, the parameters $f_j$,
$\overline{\omega}_j$ and $s_j$ should be defined so as to approximate
the actual ODF of the considered material, as obtained from experiments
or from quantum calculations. For high frequencies,
\beq
\epsilon(\omega) = 1 - \frac{\Omega_{\rm p}^2}{\omega^2} +
O(\omega^{-3}),
\label{1.162}\eeq
\ie, in the limit $\omega \rightarrow \infty$, the ODF depends only on
the electron density of the medium.

Drude extended the classical ODF model \req{1.158} to the case of
conductors by considering that there
are $f_{\rm ce}$ conducting electrons per molecule that can wander about
the medium \citep{AshcroftMermin1976}. These electrons, being free, should be ascribed a
resonance frequency $\omega_{\rm ce}=0$. It is worth pointing out that
the distinction between ``bound'' and ``free'' electrons has a meaning
only for fields that are constant with time. For alternating fields, the
free electrons do not move arbitrarily far, but oscillate with the
frequency of the electric field. The current density of conducting
electrons, under the combined action of the local electric field and the
damping force $-s_{\rm ce} \dot{\bf r}$ (which now represents their
``friction'' with the ions), is then given by [see Eq.\ \req{1.155}]
\beq
{\bf j}_{\rm ce}(t) = -e \, {\cal N} f_{\rm ce} \frac{\d {\bf r}}{\d
t} = \frac{{\cal N} e^2 f_{\rm ce} }{\me} \, \frac{\ecb_0
\exp(-{\rm i} \omega t)}{ - {\rm i} \omega + s_{\rm ce}}.
\label{1.163}\eeq
From Ohm's law, ${\bf j}_{\rm ce} = \hat{\sigma} \ecb$, we arrive at the
following expression for the conductivity,
\index{electric conductivity}
\beq
\sigma(\omega) = \frac{{\cal N} e^2 f_{\rm ce}}
{\me (s_{\rm ce} - {\rm i} \omega)} = \frac{\Omega_{\rm p}^2}{4\pi Z}
\, \frac{f_{\rm ce}}{s_{\rm ce}-{\rm i} \omega},
\label{1.164}\eeq
which is the result obtained from the Drude theory of metals \citep[see,
\eg,][]{AshcroftMermin1976}.  To be consistent with this qualitative
behavior, we should complement our oscillator model by adding a
``conduction oscillator'' with strength $f_{\rm ce}$ and frequency
$\overline{\omega}_{\rm ce}=0$ and, consequently [from Eq.\ \req{1.159}]
$\omega_{\rm ce}^2 =
f_{\rm ce} \Omega_{\rm p}^2/(3Z)$. Thus, the classical ODF of the
conducting material is
\beq
\epsilon(\omega) = 1 + \frac{\Omega_{\rm p}^2}{Z}
\frac{f_{\rm ce}}{- \omega^2 - {\rm i} s_{\rm ce} \omega}
+ \frac{\Omega_{\rm p}^2}{Z}
{\sum_j}' \frac{f_j}{\overline{\omega}_j^2 - \omega^2 -{\rm i} s_j
\omega}\, ,
\label{1.165}\eeq
where the prime in the sum indicates that the conduction term has been
taken out. The definition \req{1.111a} with this ODF gives
\beqa
\sigma(\omega) &=& - {\rm i} \, \frac{\omega}{4\pi} \, \left[
\epsilon(\omega) - 1 \right] =
\frac{{\cal N} e^2}{\me} \sum_j
\frac{f_j \omega}{s_j \omega + i (\overline{\omega}_j^2 - \omega^2)}
\nonumber \\ [2mm]
&=& \frac{{\cal N} e^2}{\me} \left[ \frac{f_{\rm ce}}{
s_{\rm ce} - {\rm i} \omega} + {\sum_j}'
\frac{f_j \omega}{s_j \omega + i (\overline{\omega}_j^2 - \omega^2)} \right],
\label{1.166}\eeqa
which includes contributions from bound electrons with the usual
resonance-like character. It is seen that the direct-current
conductivity (at $\omega =0$) is real and is determined by the density
of conducting electrons and their friction. Therefore, for sufficiently
small frequencies the conductivity of a conductor is essentially real
(that is, the current is in phase with the electric field) and independent of
frequency. For higher frequencies the conductivity becomes complex and
its variation with $\omega$ is qualitatively described by the simple
expression \req{1.166}.

The formula \req{1.165} is known as the {\it
Lorentz--Drude model} of the ODF. The ODF suffices to describe electric
fields that are spatially homogeneous, \ie, such that their Fourier
transforms vanish for ${\bf q}\ne 0$. Real fields are necessarily
limited in space and, hence, they have finite Fourier components with
$q>0$. Therefore, the Lorentz--Drude model needs to be extended to
account for the dependence of the DF on the wave number $q$.
Unfortunately this is not possible within a classical model of
harmonically bound electrons, which cannot accommodate finite wave
vectors\footnote{We will see below that a finite value of ${\bf q}$
corresponds to a momentum transfer $\hbar {\bf q}$ from the
electromagnetic field to the material. Harmonically bound electrons,
however, have null average momentum and, therefore, they cannot absorb
linear momentum from the field.}, and we must resort to experimental
information or to the results of quantum calculations. To introduce the
$q$-dependence of the DF, we shall proceed in a heuristic manner, taking
as a guide the DF of a gas of electrons that are originally at rest.
\citet{Lindhard1954} derived the DFs of this idealized system using the
random-phase approximation (see Chapter \ref{chapt7}), and found that
for sufficiently large $q$ its longitudinal and transverse DFs are equal
and given by
\beq
\epsilon_{\rm eg}^{\rm (L)}(q,\omega)
= \epsilon_{\rm eg}^{\rm (T)}(q,\omega)
= 1 + \frac{\Omega_{\rm p}^2}{
\displaystyle{(\hbar^2 q^4/4 \me^2) - \omega^2 - {\rm i} s \omega
}}\, ,
\label{1.167}\eeq
where $s$ is a small quantity, which may be interpreted as the
reciprocal of the relaxation time. Comparison with the classical
expression \req{1.165} for free electrons (with
$\overline{\omega}_j=0$),
\beq
\epsilon(\omega) = 1 + \Omega_{\rm p}^2
\, \frac{1}{- \omega^2 - {\rm i} s \omega},
\label{1.168}\eeq
indicates that we may account for the $q$-dependence by simply adding
the quantity $\hbar^2 q^4 / 4 \me^2$ to the denominators in expression
\req{1.165}. Notice that $\hbar^2 q^2/2\me$ is the (non-relativistic)
kinetic energy of an electron with momentum $\hbar q$. By applying this
prescription, the DFs resulting from the classical oscillator model are
expressed as
\beq
\epsilon^{\rm (L)}_{\rm osc} (q, \omega) =
\epsilon^{\rm (T)}_{\rm osc} (q, \omega) =
1 + \frac{\Omega_{\rm p}^2}{Z}
\sum_j \frac{f_j}{\overline{\omega}_j^2 + (\hbar^2 q^4 / 4 \me^2)
- \omega^2 -{\rm i} s_j \omega}.
\label{1.169}\eeq
It should be mentioned that various authors use slightly different
expression for the DFs, with the quantity $\overline{\omega}_j^2 +
\hbar^2 q^4 / 4 \me^2$ in the denominator replaced with expressions like
$(\overline{\omega}_j + \hbar q^2 / 2 \me)^2$ or $\overline{\omega}_j^2
+ \beta_j q^2 +\hbar^2 q^4 / 4 \me^2$, where $\beta_j$ is a constant.
Since we have explicitly disregarded magnetic effects, the longitudinal
and transverse DFs obtained from the classical oscillator model are equal.

In the next Section it is shown that the inverse DF, $\eta(q,\omega) \equiv
1/\epsilon^{\rm (L)}(q,\omega)$, plays a fundamental role in the
classical theory of stopping. In the case of non-overlapping resonances,
straight calculation shows that the inverse of the DF \req{1.169} is
given by the following expression,
\beq
\eta_{\rm osc}(q,\omega) \equiv \frac{1}{\epsilon^{\rm (L)}_{\rm osc} (q,\omega)} = 1 - \frac{\Omega_{\rm p}^2}{Z}
\sum_j \frac{f_j}{\widetilde{\omega}_j^2 + (\hbar^2 q^4 / 4 \me^2)
- \omega^2 -{\rm i} s_j \omega},
\label{1.170}\eeq
with
\beq
\widetilde{\omega}_j^2 = \overline{\omega}_j^2
+ \frac{f_j}{Z} \Omega_{\rm p}^2
= \left\{ \begin{array}{ll} \displaystyle{ \omega_j^2 +
\frac{2}{3} \, \frac{f_j}{Z} \,
\Omega_{\rm p}^2} \rule{10mm}{0mm}
& \mbox{if $\omega_j \ne 0$,} \\ [4mm]
\displaystyle{\frac{f_{\rm ce}}{Z} \, \Omega_{\rm p}^2} & \mbox{for
conduction electrons}. \end{array} \right.
\label{1.171}\eeq
Expression \req{1.170}, albeit derived from a simple classical
approximation, presents the most relevant features of realistic inverse
DFs. With appropriate values of the resonance energies, oscillator
strengths, and damping parameters, it can be used as a convenient
approximation to the actual inverse DF of a material. The classical
expression still has the characteristic form \req{1.165}, apart from the
reversed sign of the summation, but with the resonance frequencies
shifted in an
amount proportional to the electron density of the medium. The shift is
insignificant for low-pressure gases and also for very large resonance
frequencies, such as those corresponding to inner shell excitations. In
the case of metals, it transforms the conducting electron resonance at
$\overline{\omega}_{\rm ce}=0$ into the plasmon line at
$\widetilde{\omega}_{\rm ce} = (f_{\rm ce}/Z)^{1/2} \Omega_{\rm p}$,
which corresponds to the excitation of collective oscillations of the
gas of conduction electrons.
\index{dielectric functions!classical Lorentz--Drude model|)}


\section{Stopping of fast charged particles in a dielectric medium
\label{sec1.4}}

\index{classical stopping theory}\index{Fourier transform}
The classical dielectric theory presented above can be used to describe
the slowing down of fast charged particles in matter. Here we derive the
basic result for the case of homogeneous isotropic media, with DFs
$\epsilon^{\rm (L,T)}(q,\omega)$ that are independent of the direction
of the wave vector ${\bf q}$. As in the foregoing discussion, we assume
that the medium is a compound with $Z$ bound electrons in the molecule
and ${\cal N}$ molecules per unit volume.

Let us consider a projectile of charge $Z_0 e$ that moves with velocity
${\bf v}$ and passes by the origin of coordinates at $t=0$. It is
assumed to follow a straight trajectory, ${\bf r} = {\bf v} t$, with
essentially constant speed. If the projectile charge is positive
(negative), it attracts the electrons (nuclei) and repels the nuclei
(electrons), polarizing the atoms or molecules of the
medium. As the projectile is moving fast, the
polarization is stronger behind the projectile than in front of it. This
inhomogeneous polarization gives rise to an induced electric field
$\ecb_{\rm ind}$ that, in turn, produces a stopping force $Z_0 e \, \ecb_{\rm
ind}$ on the projectile. The stopping power, \ie, the average energy
loss per unit path length, can then be regarded as the result of that
force. Notice that the magnetic force does no work.

We consider the projectile as an external charge injected into the medium.
The associated charge and current distributions are, respectively,
\beq
\rho_{\rm ext}({\bf r},t) = Z_0 e \, \delta ({\bf r} - {\bf v}t)
\qquad \mbox{and} \qquad
{\bf j}_{\rm ext}({\bf r},t) = Z_0 e \, {\bf v} \, \delta ({\bf r} -
{\bf v}t).
\label{1.172}\eeq
The Fourier transforms of these distributions are
\beqa
\rho({\bf q},\omega) &=& (2\pi)^{-2} \int\d {\bf r} \int \d t
\exp[-{\rm i} ({\bf q} \dotprod{\bf r} - \omega t)] \,
Z_0 e \, \delta ({\bf r} - {\bf v}t)
\nonumber \\ [2mm]
&=& (2\pi)^{-2} Z_0 e \int \d t
\exp[-{\rm i} ({\bf q} \dotprod{\bf v} - \omega ) t] \,
\nonumber \\ [2mm]
&=& \frac{Z_0 e}{2\pi} \, \delta({\bf q} \dotprod{\bf v} - \omega )
\label{1.173}\eeqa
and
\beq
{\bf j}({\bf q},\omega)
= \frac{Z_0 e}{2\pi} \, {\bf v} \, \delta({\bf q} \dotprod{\bf v}
- \omega ).
\label{1.174}\eeq
The Fourier transforms of the longitudinal and transverse
parts of the current density are [see Eq.\ \req{1.40a}]
\beqa
&&
{\bf j}^{\rm (L)}_{\rm ext}({\bf q},\omega) = \frac{Z_0 e }{2\pi} \,
(\hat{\bf q} \dotprod {\bf v}) \hat{\bf q}\, \delta( {\bf q} \dotprod
{\bf v} - \omega),
\nonumber \\ [2mm]
&&
{\bf j}^{\rm (T)}_{\rm ext}({\bf q},\omega) =
{\bf j}_{\rm ext}({\bf q},\omega)
-{\bf j}^{\rm (L)}_{\rm ext}({\bf q},\omega) =
\frac{Z_0 e }{2\pi} \, [
{\bf v} - (\hat{\bf q} \dotprod {\bf v}) \hat{\bf q} ]\, \delta( {\bf q}
\dotprod {\bf v} - \omega).
\label{1.175}\eeqa
Introducing these expressions in  Eqs.\ \req{1.97} and \req{1.98}, we
obtain the Fourier transforms of the electromagnetic field,
\beqa
\ecb({\bf q},\omega) &=&
- {\rm i} \, \frac{2 Z_0 e}{\omega} \,
\frac{1}{\epsilon^{\rm (L)} ({\bf q},\omega)} \,
\frac{{\bf q} ({\bf q} \dotprod {\bf v})}{q^2} \,
\delta({\bf q} \dotprod {\bf v} - \omega)
\nonumber \\ [2mm]
&& \mbox{}
+ {\rm i} \, \frac{2 Z_0 e \omega}{c^2} \,
\frac{1}{q^2 - (\omega/c)^2 \, \epsilon^{\rm (T)}({\bf q}, \omega)}
\left( {\bf v} - \frac{{\bf q} ({\bf q} \dotprod {\bf v})}{q^2} \right)
\delta({\bf q} \dotprod {\bf v} - \omega)
\label{1.176}\eeqa
and
\beqa
\bcb ({\bf q}, \omega) &=&
{\rm i} \, \frac{2 Z_0 e}{c} \,
\frac{1}
{q^2 - (\omega/c)^2 \, \epsilon^{\rm (T)}({\bf q}, \omega)} \,
{\bf q} \vecprod {\bf v}
\, \delta({\bf q} \dotprod {\bf v} - \omega).
\label{1.177}\eeqa
Notice that the transforms of the fields in vacuum, $\ecb_{\rm ext}({\bf
q},\omega)$ and $\bcb_{\rm ext}({\bf q},\omega)$, are obtained from
these expressions with $\epsilon^{\rm (L,T)}({\bf q}, \omega)=1$.
Hence, the Fourier transform of the induced electric field is
\beqa
\ecb_{\rm ind}({\bf q},\omega) &=&
\ecb({\bf q},\omega) - \ecb_{\rm ext}({\bf q},\omega)
\nonumber \\ [2mm]
&=& - {\rm i} \frac{2Z_0 e }{\omega} \,
\delta( {\bf q} \dotprod {\bf v} - \omega)
\left[ \left( \frac{1}{\epsilon^{\rm (L)}(
q, \omega)} - 1 \right) \, (\hat{\bf q} \dotprod {\bf v}) \hat{\bf q}
\right.
\nonumber \\ [2mm]
&& \left.- \frac{\omega^2}{c^2}\, \left( \frac{1}{q^2 - (\omega/c)^2 \,
\epsilon^{\rm (T)}(q, \omega)} -
\frac{1}{q^2 - (\omega/c)^2} \right)
\, [{\bf v} - (\hat{\bf q} \dotprod {\bf v}) \hat{\bf q}] \right].
\rule{10mm}{0mm}
\label{1.178}\eeqa
The induced electric field is obtained as the inverse Fourier transform,
\beq
\ecb_{\rm ind} ({\bf r}, t) = (2\pi)^{-2} \int \d {\bf q}
\int \d \omega \,
\exp[ {\rm i} ( {\bf q} \dotprod {\bf r} - \omega t ) ] \,
\ecb_{\rm ind} ({\bf q}, \omega) .
\label{1.179}\eeq

\index{stopping force} \index{stopping power!classical theory}
The stopping force acting on the projectile is $Z_0 e \ecb_{\rm
ind}({\bf v}t, t)$, where $\ecb_{\rm ind}({\bf v}t, t)$ is the induced
electric field at the position of the pro\-jec\-tile. In a small
displacement $\d s\, \hat{\bf v}$ of the projectile, the work made by
the stopping force is $\d W = Z_0 e \, \ecb_{\rm ind}({\bf v}t, t) \cdot
\hat{\bf v} \, \d s$. Since the kinetic energy of the projectile varies
in $\d E = \d W$, the stopping power is
\beq
S \equiv - \frac{\d E}{\d s} = - Z_0 e \;
\hat{\bf v} \dotprod \ecb_{\rm ind}({\bf v} t,t)\, .
\label{1.180}\eeq
Inserting the expression \req{1.179}, with the Fourier transform of the
field given by \req{1.178}, evaluated at the position ${\bf r} = {\bf
v}t$ of the projectile, we have
\beqa
S &=& {\rm i} \, \frac{(Z_0 e)^2}{2 \pi^2 v}
\int \d {\bf q} \int \frac{\d \omega}{\omega} \,
\exp[ {\rm i} ( {\bf q} \dotprod {\bf v} - \omega) t  ] \,
\delta( {\bf q} \dotprod {\bf v} - \omega)
\left[ \left( \frac{1}{\epsilon^{\rm (L)}(
q, \omega)} - 1 \right) \, (\hat{\bf q} \dotprod {\bf v})^2
\right.
\nonumber \\ [2mm]
&& \left.- \frac{\omega^2}{c^2}\, \left( \frac{1}{q^2 - (\omega/c)^2 \,
\epsilon^{\rm (T)}(q, \omega)} -
\frac{1}{q^2 - (\omega/c)^2} \right)
\, [v^2 - (\hat{\bf q} \dotprod {\bf v})^2] \right].
\label{1.181}\eeqa

\begin{figure}[th!] \begin{center}
\vspace*{3mm}
\includegraphics*[width=6.50cm]{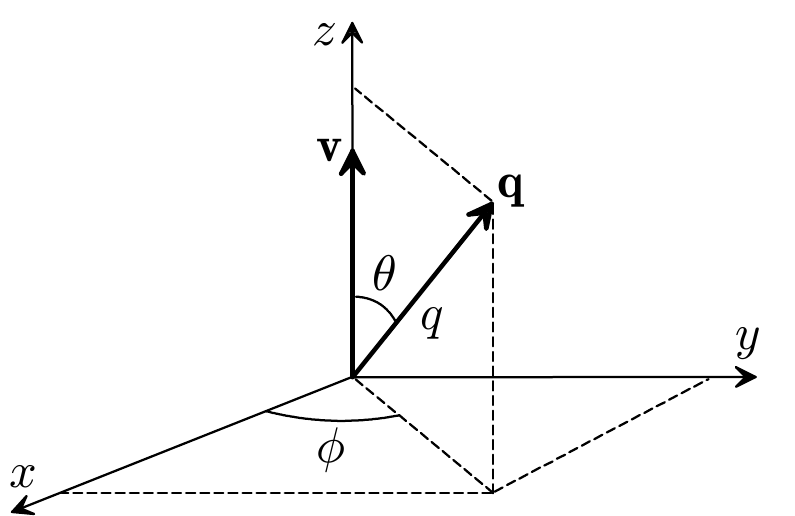}
\caption{Spherical coordinates for the calculation of the stopping
power.
\label{fig1.1}}
\end{center}\end{figure}

\index{spherical polar coordinates}
The integral \req{1.181} can be partially evaluated by expressing ${\bf
q}$ in spherical polar coordinates (Fig.\ \ref{fig1.1}). Taking the
polar axis in the direction of ${\bf v}$, performing the integral over
the azimuthal angle $\phi$, and introducing the variable $\xi\equiv \hat{\bf q}
\cdot \hat{\bf v}$ ($= \cos\theta$), we have
\beqa
S &=& {\rm i} \, \frac{(Z_0 e)^2}{\pi v}
\int_0^\infty q^2 \d q \int_{-1}^{+1} \d \xi \int_{-\infty}^{+\infty}
\frac{\d \omega}{\omega} \,
\delta( q v \xi  - \omega)
\left[ \left( \frac{1}{\epsilon^{\rm (L)}(q, \omega)} - 1 \right) \, (v\xi)^2
\right.
\nonumber \\ [2mm]
&& \left.- \frac{\omega^2}{c^2}\, \left( \frac{1}{q^2 - (\omega/c)^2 \,
\epsilon^{\rm (T)}(q, \omega)} -
\frac{1}{q^2 - (\omega/c)^2} \right)
\, [v^2 - (v\xi)^2] \right].
\label{1.182}\eeqa
Using the relation
\beq
\delta( q v \xi  - \omega) = \frac{1}{qv} \delta\left( \xi -
\frac{\omega}{qv} \right),
\label{1.183}\eeq
and carrying out the integration over $\xi$ (notice that, because
of the delta function, the integrand vanishes unless $-qv < \omega <
qv$), we obtain
\beqa
S &=& {\rm i} \, \frac{(Z_0 e)^2}{\pi v^2}
\int_0^\infty q \d q \int_{-qv}^{+qv}
\frac{\d \omega}{\omega} \,
\left[ \left( \frac{1}{\epsilon^{\rm (L)}(
q, \omega)} - 1 \right) \, (\omega/q)^2
\right.
\nonumber \\ [2mm]
&& \left.- \frac{\omega^2}{c^2}\, \left( \frac{1}{q^2 - (\omega/c)^2 \,
\epsilon^{\rm (T)}(q, \omega)} -
\frac{1}{q^2 - (\omega/c)^2} \right)
\, [v^2 - (\omega/q)^2] \right]
\nonumber \\ [2mm]
&=& {\rm i} \, \frac{(Z_0 e)^2}{\pi v^2}
\int_0^\infty \frac{\d q}{q} \int_{-qv}^{+qv}
\omega \, \d \omega\,
\left[ \left( \frac{1}{\epsilon^{\rm (L)}(
q, \omega)} - 1 \right)
\right.
\nonumber \\ [2mm]
&& \left.- q^2 \left( \frac{1}{q^2 - (\omega/c)^2 \,
\epsilon^{\rm (T)}(q, \omega)} -
\frac{1}{q^2 - (\omega/c)^2} \right)
\, \left( \beta^2 - \frac{\omega^2}{c^2 q^2} \right) \right],
\label{1.184}\eeqa
where $\beta=v/c$. By virtue of the properties \req{1.84}, the values of
the integrand at $-\omega$ and $+\omega$ are complex conjugate of each
other and, therefore,
\beqa
S &=& -2\, \frac{(Z_0 e)^2}{\pi v^2}
\int_0^\infty \frac{\d q}{q} \int_{0}^{qv}
\omega \, \d \omega\, {\rm Im}
\left[ \left( \frac{1}{\epsilon^{\rm (L)}(
q, \omega)} - 1 \right)
\right.
\nonumber \\ [2mm]
&& \left.- q^2 \left( \frac{1}{q^2 - (\omega/c)^2 \,
\epsilon^{\rm (T)}(q, \omega)} -
\frac{1}{q^2 - (\omega/c)^2} \right)
\, \left( \beta^2 - \frac{\omega^2}{c^2 q^2} \right) \right]
\nonumber \\ [2mm]
&=& \frac{2 (Z_0 e)^2}{\pi v^2}
\int_0^\infty \frac{\d q}{q} \int_{0}^{qv}
\omega \, \d \omega
\left[ {\rm Im} \left( \frac{-1}{\epsilon^{\rm (L)} (q, \omega)} \right)
\right.
\nonumber \\ [2mm]
&& \left.+ \left( \beta^2 - \frac{\omega^2}{c^2 q^2} \right)
{\rm Im} \left( \frac{q^2}{q^2 - (\omega/c)^2 \,
\epsilon^{\rm (T)}(q, \omega)} \right)  \right].
\label{1.185}\eeqa
Finally, to facilitate further considerations, it is convenient to
exchange the order of the integrals and write
\beqa
S &=& \frac{2 (Z_0 e)^2}{\pi v^2}
\int_0^\infty \omega \, \d \omega \int_{\omega/v}^\infty
\frac{\d q}{q}
\left[ {\rm Im} \left( \frac{-1}{\epsilon^{\rm (L)} (q, \omega)} \right)
\right.
\nonumber \\ [2mm]
&& \left.+ \beta^2 \left( 1 - \frac{\omega^2}{\beta^2 c^2 q^2} \right)
{\rm Im} \left( \frac{1}{1 - (\omega/cq)^2 \,
\epsilon^{\rm (T)}(q, \omega)} \right)  \right]
\label{1.186}\eeqa
or, equivalently,
\beq
S = \frac{(Z_0 e)^2}{\pi v^2} \; {\rm Im}
\int_0^\infty \omega \, \d \omega \int_{(\omega/v)^2}^\infty
\frac{\d (q^2)}{q^2}
\left[ \frac{-1}{\epsilon^{\rm (L)} (q, \omega)}
+ \frac{c^2 \beta^2 q^2 - \omega^2}
{c^2 q^2 - \omega^2 \,\epsilon^{\rm (T)}(q, \omega)} \right].
\label{1.187}\eeq
This important formula, which was first derived by \citet{Lindhard1954},
separates the contributions of longitudinal and transverse interactions
to the stopping power. It is worth noticing that the stopping power for
non-relativistic projectiles (with $\beta^2 \ll 1$) is determined by the
longitudinal energy-loss function  ${\rm Im} \left[-1/\epsilon_2^{\rm
(L)}\right]$.

Introducing the real and imaginary parts of the transverse DF,
$\epsilon^{\rm (T)} = \epsilon^{\rm (T)}_1 + {\rm i} \epsilon^{\rm
(T)}_2$, we can express the stopping power as
\beqa
S &=& \frac{2 (Z_0 e)^2}{\pi v^2}
\int_0^\infty \omega \, \d \omega \int_{\omega/v}^\infty
\frac{\d q}{q}
\left[ {\rm Im} \left( \frac{-1}{\epsilon^{\rm (L)} (q, \omega)} \right)
+ \beta^2 \left( 1 - \frac{\omega^2}{\beta^2 c^2 q^2} \right)
\left( \frac{\omega}{cq} \right)^2
\rule{0mm}{9mm}
\right.
\nonumber \\ [2mm]
&& \left. \times \frac{\left[ \epsilon_1^{\rm (T)} \right]^2+
\left[ \epsilon_2^{\rm (T)} \right]^2 }
{\left[ 1 - (\omega/cq)^2 \epsilon_1^{\rm (T)} \right]^2 +
[(\omega/cq)^2 \epsilon_2^{\rm T}]^2} \,
{\rm Im} \left( \frac{-1}{\epsilon^{\rm (T)} (q, \omega)} \right)
\right].
\label{1.188}\eeqa
In the special case of a rarefied gas, $\epsilon_1^{\rm (L,T)} \simeq 1$
and $\epsilon_2^{\rm (L,T)} \ll 1$, and the expression \req{1.188} of
the stopping power becomes
\beqa
S & \simeq& \frac{2 (Z_0 e)^2}{\pi v^2}
\int_0^\infty \omega \, \d \omega \int_{\omega/v}^\infty
\frac{\d q}{q}
\left[ {\rm Im} \left( \frac{-1}{\epsilon^{\rm (L)} (q, \omega)} \right)
\right.
\nonumber \\ [2mm]
&& \left.+ \beta^2 \left( 1 - \frac{\beta^2 \omega^2}{c^2 q^2} \right)
\frac{(\omega/cq)^2}{[1 - (\omega/cq)^2]^2} \,
{\rm Im} \left( \frac{-1}{\epsilon^{\rm (T)} (q, \omega)} \right)
\right].
\label{1.189}\eeqa
Hence, the stopping power of low-density gases is determined by the
longitudinal and transverse energy-loss functions, $\eta^{\rm (L,T)}_2
\equiv {\rm Im} \left[-1/\epsilon^{\rm (L,T)}\right]$.


\section{Kramers--Kronig relations and sum rules \label{sec1.5}}
\index{causality principle}
\index{Kramers--Kronig relations|(}

We shall now derive a number of properties of the longitudinal DFs from
the principle of causality and the asymptotic behavior of the DFs at
high frequencies. In the Coulomb gauge, the causality principle implies
that if external charges appear suddenly at a certain time $t_0$ in a
neutral medium, the induced scalar potential vanishes for $t < t_0$. It
is worth pointing out that this form of the causality principle
disregards retardation of the vector potential and, consequently, it is
weaker than the strict (relativistic) causality which requires that
electromagnetic interactions cannot propagate at speeds larger than $c$.
The results from the present study are expected to be valid only under
strict non-relativistic conditions (\ie, when small and moderate
frequencies and wave numbers dominate).

As in Section \ref{sec1.3}, we consider a compound material with $Z$
electrons per molecule and ${\cal N}$ molecules per unit volume, with
the plasma resonance frequency $\Omega_{\rm p}$ defined by Eq.\
\req{1.160}.  Experimental information and elaborate theoretical
(non-relativistic) models indicate that, at frequencies much larger than
the characteristic frequencies $\overline{\omega}_j$ of the material,
the complex ODF admits the following universal asymptotic expansion,
\beq
\lim_{\omega \rightarrow \infty} \epsilon(\omega) = 1 -
\frac{\Omega_{\rm p}^2}{\omega^2} + {\cal O}(\omega^{-3}),
\label{1.190}\eeq
with the term in $\omega^{-1}$ missing. The expansion \req{1.190}
implies that the imaginary part of $\epsilon(\omega)$ decreases faster
than $\omega^{-2}$. Of course, the ODF obtained from the Lorentz--Drude
classical oscillator model [Eq.\ \req{1.169}] does follow this asymptotic
behavior. As that model suggests, the expansion \req{1.190} is also
valid for the longitudinal DF at finite $q$ and frequencies much larger
than $\overline{\omega}_j + \hbar q^2/(2\me)$ \citep{PinesNozieres1989}.
That is,
\beq
\lim_{\omega \rightarrow \infty} \epsilon^{\rm (L)} ({\bf q},\omega) = 1
- \frac{\Omega_{\rm p}^2}{\omega^2} + {\cal O}(\omega^{-3})
\label{1.191}\eeq
and
\beq
\lim_{\omega \rightarrow \infty} \eta^{\rm (L)} ({\bf q},\omega) = 1
+ \frac{\Omega_{\rm p}^2}{\omega^2} + {\cal O}(\omega^{-3}) .
\label{1.192}\eeq

Let us consider an instantaneous charge disturbance at time $t_0$,
\beq
\rho_{\rm ext} ({\bf r}, t) = D({\bf r}) \delta(t-t_0),
\label{1.193}\eeq
where the distribution $D({\bf r})$ is independent of time. Within a
non-relativistic framework, the condition of causality implies that the
induced potential,
\allowdisplaybreaks{
\beq
\varphi_{\rm ind} ({\bf r},t) =
\varphi ({\bf r},t) -
\varphi_{\rm ext} ({\bf r},t),
\label{1.194}\eeq
must vanish identically for $t<t_0$. We have
\beqa
\varphi_{\rm ind} ({\bf r},t) &=&
(2\pi)^{-2} \int \d {\bf q} \int \d \omega \,
\exp\left[ {\rm i} \left( {\bf q} \dotprod {\bf r} - \omega t \right)
\right] \, \varphi_{\rm ind}({\bf q},\omega)
\nonumber \\ [2mm]
&=&
(2\pi)^{-2} \int \d {\bf q} \int \d \omega \,
\exp\left[ {\rm i} \left( {\bf q} \dotprod {\bf r} - \omega t \right)
\right] \, \left[ \varphi({\bf q},\omega) - \varphi_{\rm ext} ({\bf
q},\omega) \right]
\nonumber \\ [2mm]
&=&
(2\pi)^{-2} \int \d {\bf q} \int \d \omega \,
\exp\left[ {\rm i} \left( {\bf q} \dotprod {\bf r} - \omega t \right)
\right] \, \frac{4\pi}{q^2} \left[ \eta^{\rm (L)}({\bf q},\omega) -
1 \right] \rho_{\rm ext}({\bf q},\omega), \rule{15mm}{0mm}
\label{1.195}\eeqa
where use has been made of Eq.\ \req{1.91} and the equivalent
relation for the vacuum.}
The Fourier transform of the disturbance is
\beqa
\rho_{\rm ext}({\bf q},\omega) &=& (2\pi)^{-2} \int \d {\bf r} \int \d t
\, \exp\left[ - {\rm i} \left( {\bf q} \dotprod {\bf r} - \omega t
\right) \right] \, D({\bf r}) \delta (t-t_0)
\nonumber \\ [2mm]
&=&
(2\pi)^{-2} \int \d {\bf r}
\, \exp\left[ - {\rm i} \left( {\bf q} \dotprod {\bf r} - \omega t_0
\right) \right] \, D({\bf r}) = (2\pi)^{-1/2} D({\bf q})
\exp\left( {\rm i} \omega t_0 \right) . \rule{15mm}{0mm}
\label{1.196}\eeqa
Then,
\beqa
\varphi_{\rm ind} ({\bf r},t)
&=& \frac{2}{(2\pi)^{3/2}} \int \d {\bf q}
\exp\left[ {\rm i}
{\bf q} \dotprod {\bf r} \right] \, \frac{D({\bf q})}{q^2}
\nonumber \\ [2mm]
&& \mbox{} \times
\int \d \omega \, \exp\left[ - {\rm i} \omega (t-t_0) \right] \,
\left[ \eta^{\rm (L)}({\bf q},\omega) - 1 \right].
\label{1.197}\eeqa
Since $\varphi_{\rm ind} ({\bf r},t)$ is required to vanish for $t<t_0$
and for any $D({\bf q})$, we must have
\beq
\int_{-\infty}^\infty \d \omega \, \exp\left[ - {\rm i} \omega (t-t_0)
\right] \, \left[ \eta^{\rm (L)}({\bf q},\omega) - 1 \right] =0
\qquad \mbox{for any $t<t_0$.}
\label{1.198}\eeq

\begin{figure}[htb] \begin{center}
\vspace*{3mm}
\includegraphics*[width=6.50cm]{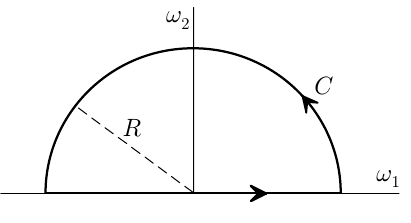}
\caption{Integration contour in the complex $\omega$ plane.\label{fig1.2}}
\end{center}\end{figure}

For a fixed {\bf q}, let us now extend the functions $\exp\left[ - {\rm
i} \omega (t-t_0) \right]$ and $\eta^{\rm (L)}({\bf q},\omega)$
analytically to the complex $\omega$ plane and consider $\omega=\omega_1
+ {\rm i} \omega_2$. For $t<t_0$, the function
$$
\exp\left[ - {\rm
i} \omega (t-t_0) \right] =
\exp\left[ - {\rm i} \omega_1 (t-t_0) \right]
\exp\left[ \omega_2 (t-t_0) \right]
$$
is limited in the upper half plane ($\omega_2 >0$) and increases
exponentially with $|\omega_2|$ in the lower half plane ($\omega_2 <
0$). On the other hand [see Eq.\ \req{1.192}],
for large $\omega$, the function $\eta^{\rm (L)}({\bf q},\omega)-1$
decreases as $\omega^{-2}$. Therefore
\beq
\int_{\rm C} \d \omega \, \exp\left[ - {\rm i} \omega
(t-t_0)
\right] \, \left[ \eta^{\rm (L)}({\bf q},\omega) - 1 \right]
\begin{array}[t]{c} \rightarrow \\ [-2mm] \scriptstyle{ R \rightarrow
\infty} \end{array} 0,
\label{1.199}\eeq
where the integration is along the semicircle shown in Fig.\
\ref{fig1.2}.
Adding the Eqs.\ \req{1.198} and \req{1.199}, it follows
that the integral along the closed contour vanishes,
\beq
\oint_{{\rm C}+{\rm real}\; {\rm axis}} \d \omega \, \exp\left[ - {\rm
i} \omega (t-t_0)
\right] \, \left[ \eta^{\rm (L)}({\bf q},\omega) - 1 \right]
\begin{array}[t]{c} \rightarrow \\ [-2mm] \scriptstyle{ R \rightarrow
\infty} \end{array} 0.
\label{1.200}\eeq
According to Cauchy's integral theorem \citep[see, \eg,][]{Arfken1985},
this relation is satisfied if the function $\eta^{\rm (L)}({\bf
q},\omega) -1$ does not have any poles in the upper half $\omega$
plane. Of course, the function $\eta^{\rm (L)}({\bf q},\omega) -1$ must
have poles in the lower half $\omega$ plane, since otherwise the
integral in Eq.\ \req{1.198}, and the induced scalar potential
\req{1.197}, would vanish also for $t>t_0$.

\index{inverse dielectric function!analyticity}
Summarizing, we have shown that the inverse longitudinal DF is analytic
in the upper half $\omega$ plane. The classical DF model described in
Section \ref{sec1.3} does satisfy this property, which is a direct
consequence of the principle of causality and, therefore, of very
general validity. It should be noted that the DF itself may not be fully
analytical for ${\rm Im} \, \omega \ge 0$, because the DFs of conductors
have a pole at $\omega=0$ [see Eq.\ \req{1.165}].


\subsection{Kramers--Kronig relations \label{sec1.5.1}}

Because of the analyticity of the inverse DF $\eta^{\rm (L)}({\bf
q},\omega)$ for ${\rm Im} \, \omega \ge 0$, we can invoke the Cauchy
theorem to obtain a relation between the real and imaginary parts of
$\eta^{\rm (L)}({\bf q},\omega)$ in the real $\omega$ axis. For an arbitrary
point $z$ in the upper half $\omega$ plane, inside the closed contour
formed by the semicircle C and the real axis, Cauchy's theorem gives
\beq
\eta^{\rm (L)}({\bf q},z) = 1 + \frac{1}{2\pi {\rm i}} \oint
\frac{\eta^{\rm (L)}({\bf q},\omega')-1}{\omega'-z} \d \omega'.
\label{1.201}\eeq
The asymptotic behavior of the inverse DF, Eq.\ \req{1.192}, implies
that at large $\omega$, there is no contribution to the integral from
the semicircle and, therefore,
\beq
\eta^{\rm (L)}({\bf q},z) = 1 + \frac{1}{2\pi {\rm i}}
\int_{-\infty}^\infty
\frac{\eta^{\rm (L)}({\bf q},\omega')-1}{\omega'-z} \d \omega',
\label{1.202}\eeq
where the integral is now along the real axis. Going to the limit where
the point $z$ approaches the real axis from above,
$z=\omega+{\rm i} \xi$, we can write
\beq
\eta^{\rm (L)}({\bf q},\omega) = 1 + \lim_{\xi \rightarrow 0} \frac{1}{2\pi
{\rm i}} \int_{-\infty}^\infty
\frac{\eta^{\rm (L)}({\bf q},\omega')-1}{\omega'-\omega-{\rm i} \xi} \d
\omega'.
\label{1.203}\eeq
Using the formal relation [see Eq.\ \req{B.13}]
\beq
\frac{1}{\omega'-\omega-{\rm i} \xi} = {\cal P} \left( \frac{1}{
\omega'-\omega} \right) + \pi {\rm i} \delta (\omega'-\omega),
\label{1.204}\eeq
where ${\cal P}$ denotes the principal value of the integral,
we obtain
$$
\eta^{\rm (L)}({\bf q},\omega) = 1 + \frac{1}{2\pi {\rm i}}
{\cal P} \int_{-\infty}^\infty
\frac{\eta^{\rm (L)}({\bf q},\omega')-1}{\omega'-\omega} \d \omega' +
\frac{1}{2} \left[\eta^{\rm (L)}({\bf q},\omega)-1 \right].
$$
That is
\beq
\eta^{\rm (L)}({\bf q},\omega) = 1 + \frac{1}{\pi {\rm i}}
{\cal P} \int_{-\infty}^\infty
\frac{\eta^{\rm (L)}({\bf q},\omega')-1}{\omega'-\omega} \d \omega'.
\label{1.205}\eeq
Separating the real and imaginary parts of this equation we can write
\begin{subequations}
\label{1.206}
\beq
\eta_1^{\rm (L)} ({\bf q},\omega) = 1 - \frac{1}{\pi}
{\cal P} \int_{-\infty}^\infty
\frac{\eta_2^{\rm (L)}({\bf q},\omega')}{\omega'-\omega}
\d \omega'
\label{1.206a} \eeq
and
\beq
\eta_2^{\rm (L)} ({\bf q},\omega) = \frac{1}{\pi}
{\cal P} \int_{-\infty}^\infty
\frac{\eta_1^{\rm (L)}({\bf q},\omega')-1}{\omega'-\omega} \d \omega'.
\label{1.206b}\eeq
\end{subequations}
These equalities are known as the {\it Kramers--Kronig dispersion
relations}.

These relations apply to any function of $\omega$ that is analytic for
${\rm Im} \, \omega \ge 0$. The DF obtained from the classical
oscillator model, Eq.\ \req{1.169}, does have this property in the
case of insulators when the damping constants $s_i$ are less than the
corresponding resonance frequencies $\overline{\omega}_j$, because then
the zeros $\omega_\pm$ of the denominators $\overline{\omega}_j^2 +
(\hbar^2 q^4/4\me^2) -\omega^2 - {\rm i} s_j \omega$,
\beq
\omega_\pm = \frac{-{\rm i} s_j}{2} \pm \frac{1}{2}
\sqrt{- s_j^2 + 4\left[ \overline{\omega}_j^2
+ (\hbar^2 q^4/4\me^2)\right]},
\label{1.207}\eeq
lie below the real axis. For a conductor, however, conduction electrons
have the characteristic frequency $\overline{\omega}_{\rm ce}=0$, and
the DF for $q=0$ has a pole at the point $\omega=0$, near which
$\epsilon^{\rm (L)}(0,\omega) \simeq 1 + 4\pi {\rm i} \sigma(0)/\omega$,
where $\sigma(0)$ is the static (longitudinal) conductivity [see Eqs.\
\req{1.112b} and \req{1.165}].
\index{electric conductivity}
This gives an additional term in the Kramers--Kronig
relation for the imaginary part of the DF. The correct result is
obtained readily by noting that the function $\epsilon^{\rm (L)} ({\bf
0},\omega)- 4\pi{\rm i} \sigma(0)/\omega$ is analytic for ${\rm Im} \,
\omega \ge 0$. Therefore,
\beq
\epsilon^{\rm (L)}({\bf q},\omega) - (4 \pi {\rm i} \sigma(0)/\omega)
= 1 + \frac{1}{\pi {\rm i}} {\cal P} \int_{-\infty}^\infty
\frac{\epsilon^{\rm (L)}({\bf q},\omega')
- [4 \pi {\rm i} \sigma(0)/\omega]-1}{\omega'-\omega} \d \omega'.
\nonumber\eeq
Since the conductivity term does not contribute to the integral,
the Kramers--Kronig dispersion relations for the DF, $\epsilon^{\rm
(L)}({\bf q},\omega) = \epsilon^{\rm (L)}_1({\bf q},\omega) + {\rm i}
\epsilon^{\rm (L)}_2({\bf q},\omega)$, read
\begin{subequations}
\label{1.208}
\beq
\epsilon_1^{\rm (L)}({\bf q},\omega) = 1 + \frac{1}{\pi}
{\cal P} \int_{-\infty}^\infty
\frac{\epsilon_2^{\rm (L)}({\bf q},\omega')}{\omega'-\omega}
\d \omega'
\label{1.208a} \eeq
and
\beq
\epsilon^{\rm (L)}_2 ({\bf q},\omega) =
\frac{4 \pi \sigma(0)}{\omega}
- \frac{1}{\pi} {\cal P} \int_{-\infty}^\infty
\frac{\epsilon^{\rm (L)}_1({\bf q},\omega')-1}{\omega'-\omega}
\d \omega',
\label{1.208b}\eeq
\end{subequations}
with the term $4\pi \sigma(0)/\omega$ present only for $q=0$.

For homogeneous isotropic media, the relations \req{1.84} indicate that
$\epsilon_1^{\rm (L)}(q,\omega)$ and $\eta_1^{\rm (L)}(q,\omega)$ are even in
$\omega$, while $\epsilon_2^{\rm (L)} (q,\omega)$ and
$\eta_2^{\rm (L)}(q,\omega)$
are odd. Therefore, the integrals can be transformed to involve only
positive (physically observable) frequencies.  Introducing the change of
variable $\omega' \rightarrow -\omega'$ in the integral over $(-\infty,
0)$, and combining the result with the integral over $(0,\infty)$, we
find
\begin{subequations}
\label{1.209}
\beqa
\eta_1^{\rm (L)}(q,\omega) &=& 1 - \frac{2}{\pi} {\cal P}
\int_{0}^\infty \frac{\omega' \, \eta_2^{\rm (L)}(q,\omega')}
{\omega'^2-\omega^2} \d \omega',
\label{1.209a}\\ [2mm]
\eta_2^{\rm (L)} (q,\omega) &=& \frac{2\omega}{\pi} {\cal P}
\int_0^\infty \frac{\eta_1^{\rm (L)}(q,\omega')-1}
{\omega'^2-\omega^2} \d \omega'.
\label{1.209b}\eeqa
\end{subequations}
Similarly,
\begin{subequations}
\label{1.210}
\beqa
\epsilon_1^{\rm (L)}(q,\omega) &=& 1 + \frac{2}{\pi} {\cal P}
\int_{0}^\infty \frac{\omega'\, \epsilon_2^{\rm (L)}(q,\omega')}
{\omega'^2-\omega^2} \, \d \omega',
\label{1.210a} \\ [2mm]
\epsilon_2^{\rm (L)}(q,\omega) &=&  \frac{4 \pi \sigma(0)}{\omega}
- \frac{2\omega}{\pi} {\cal P}
\int_0^\infty \frac{\epsilon_1^{\rm (L)}(q,\omega')-1}
{\omega'^2-\omega^2} \, \d \omega',
\label{1.210b}\eeqa
\end{subequations}
with the term $4\pi \sigma(0)/\omega$ present only for $q=0$.


\subsection{Sum rules \label{sec1.5.2}}

\index{Kramers--Kronig relations!sum rules derived from}
We can derive useful global properties of the longitudinal DF by
combining the asymptotic behavior, Eqs.\ \req{1.191} and \req{1.192},
with the Kramers--Kronig dispersion relations. Thus, the high-frequency
limit of the Kramers--Kronig relation \req{1.210a} is
\beq
\lim_{\omega \rightarrow \infty} \left( \epsilon_1^{\rm (L)}(q,\omega)
-1 \right) =
- \frac{2}{\pi} \frac{1}{\omega^2}
\int_{0}^\infty \omega' \epsilon_2^{\rm (L)}(q,\omega') \d \omega' +
\cdots
\nonumber\eeq
Comparing with Eq.\ \req{1.191}, and identifying the coefficients of
$\omega^{-2}$, we obtain the so-called {\it $f$-sum rule},
\beq
\int_{0}^\infty \omega \, \epsilon_2^{\rm (L)}(q,\omega) \, \d \omega
= \frac{\pi}{2} \, \Omega_{\rm p}^2.
\label{1.211}\eeq
On the other hand, the high-frequency limit of the second Kramers--Kronig
relation, Eq.\ \req{1.210b}, is
\beq
\lim_{\omega \rightarrow \infty}
\epsilon_2^{\rm (L)} (q,\omega) =  \frac{4 \pi \sigma(0)}{\omega}
+ \frac{2}{\pi} \frac{1}{\omega}
\int_0^\infty \left( \epsilon_1^{\rm (L)}(q,\omega')-1 \right)
\, \d \omega' + \cdots
\nonumber\eeq
Because the large-frequency expansion \req{1.191} has no term in
$\omega^{-1}$, we conclude that
\begin{subequations}
\label{1.212}
\beq
2 \pi^2 \sigma(0) +
\int_0^\infty \left( \epsilon_1^{\rm (L)}(\omega)-1 \right) \,
\d \omega = 0 \qquad \mbox{for $q=0$},
\label{1.212a}\eeq
and
\beq
\int_0^\infty \left( \epsilon_1^{\rm (L)}(q,\omega)-1 \right) \,
\d \omega = 0 \qquad \mbox{for $q>0$}.
\label{1.212b}\eeq
\end{subequations}

Similarly, combining the high-frequency limit of the relations
\req{1.209} with the asymptotic expansion \req{1.192}, we obtain
the equalities
\begin{subequations}
\label{1.213}
\beq
\int_0^\infty \omega \, \eta_2 (q,\omega) \, \d \omega =
\frac{\pi}{2} \Omega_{\rm p}^2
\label{1.213a}\eeq
and
\beq
\int_0^\infty \left[ \eta_1(q,\omega) -1 \right] \, \d \omega
= 0,
\label{1.213b}\eeq
\end{subequations}
which are valid for all $q$.

It is appropriate to mention here that the function \index{generalized
oscillator strength}
\beq
\frac{\d f(q,\omega)}{\d \omega} \equiv
\frac{2 Z}{\pi \Omega_{\rm p}^2} \, \omega \, \eta_2(q,\omega) \,
\label{1.214}\eeq
can be identified with the molecular generalized oscillator strength
(GOS), a property of individual molecules [see Section \ref{sec6.7},
Eq.\ \req{6.251a}].
From Eq.\ \req{1.213a} we see that the GOS satisfies the sum rule
\index{Bethe sum rule}
\beq
\int_0^\infty \frac{\d f(q,\omega)}{\d \omega} \, \d \omega = Z
\qquad \forall q,
\label{1.215}\eeq
which is known as the {\it Bethe sum rule}. When $q=0$ \index{optical
oscillator strength}
the GOS reduces to the optical oscillator strength (OOS), $\d f(\omega)/\d
\omega \equiv \d f(0,\omega)/\d \omega$, and Eq. \req{1.215} becomes the
{\it Thomas-Reiche-Kuhn sum rule} \citep[see, \eg,][]{BransdenJoachain1983},
\beq
\int_0^\infty \frac{\d f(\omega)}{\d \omega} \, \d \omega = Z\, \,
\label{1.216}\eeq
also known as the {\it dipole sum rule}.

The low-frequency limit ($\omega \rightarrow 0$) of the Kramers--Kronig
relations provides other useful sum rules. Thus, Eq.\ \req{1.209a}
implies that
\beq
\eta_1 (q,0) + \frac{2}{\pi}
\int_{0}^\infty \frac{\eta_2(q,\omega)}
{\omega} \, \d \omega = 1 .
\label{1.217}\eeq
In the long-wavelength limit ($q\rightarrow 0$), this result becomes the
{\it perfect-screening sum rule},
\beq
\eta_1(0) + \frac{2}{\pi}
\int_{0}^\infty \frac{\eta_2(\omega)}
{\omega} \, \d \omega = 1 .
\label{1.218}\eeq
For an electric conductor $\eta_1(0)=0$, whereas for an insulator
$\eta_1(0)$ is real and positive. When
expressed in terms of the OOS, this sum rule takes the form
\beq
\eta_1(0) + \frac{\Omega_{\rm p}^2}{Z} \int_0^\infty
\frac{1}{\omega^2} \frac{\d f}{\d \omega}\, \d \omega =1.
\label{1.219}\eeq

These sum rules are consequences from the Kramers--Kronig relations,
which follow from the requirement of ``unretarded'' causality. Hence,
they are expected to hold only in the non-relativistic domain, that is,
when the speeds of the particles in the medium are much less than $c$.
Departures from the sum rules are to be expected for excitations with
frequencies comparable to and larger than $\me c^2/\hbar$, which
normally involve the emission of a single electron with velocity
$\lesssim c$. Quantum calculations for atomic gases indicate that
relativistic corrections reduce the value of the $f$-sum \req{1.211}
\citep{Levinger1957}; the deviation from the non-relativistic value
increases in magnitude with the atomic number of the element, being
about 2.5 \% for the heaviest elements \citep{Salvat2022a}.

\index{Kramers--Kronig relations|)}




\chapter{Schr\"{o}dinger and Dirac wave equations \label{chapt2}}



Quantum theoretical studies of collisions of fast charged particles with
atoms or with other charged particles require solving a wave equation.
In the case of charged particles heavier than the electron, a
non-relativistic study is usually sufficient and the required wave
functions are the solutions of the Schr\"{o}dinger wave equation. When
the particles under consideration are electrons (negatrons or positrons)
or muons, with small or moderate masses and spin $\1o2$, relativistic and
spin effects may be important, even for collisions of projectiles with
relatively low energies. The appropriate wave equation for these
particles is the relativistic Dirac equation. Relativistic effects are
also important in atomic structure calculations because inner-shell
electrons acquire high velocities near the nucleus.

In practical calculations it is expedient to use atomic units (see
Appendix \ref{appC}). In these units, the reduced Planck constant
$\hbar$, the elementary charge $e$ (\ie, the absolute value of the
electron charge), and the mass $\me$ of the electron are taken as unity,
with no dimensions. The atomic unit of length is the Bohr radius
\index{Bohr radius}
\beq
a_0 \equiv \frac{\hbar^2}{\me e^2} = 0.529\, 177\, 211 \; \mbox{\AA},
\label{2.1}\eeq
and the unit of energy is the Hartree energy, \index{Hartree energy}
\beq
E_{\rm h} \equiv \frac{\me e^4}{\hbar^2} = 27.211\, 386\; {\rm eV}.
\label{2.2}\eeq
In atomic units, the speed of light in vacuum is $c=c\hbar/e^2 =
\alpha^{-1} = 137.035\,999$, \ie, the inverse of the fine-structure
constant $\alpha$.

In this Chapter we briefly present the essentials of the Schr\"{o}dinger
and Dirac wave equations for a charged particle of mass $M_0$ and charge
$Z_0e$ in an electromagnetic field represented by the scalar potential
$\varphi({\bf r},t)$ and the vector potential ${\bf A}({\bf r},t)$ (see
Section \ref{sec1.1}). The classical force on the particle is given by the
Lorentz formula
\beq
{\bf F} = Z_0 e \left( \ecb + \frac{1}{c} \, {\bf v} \vecprod \bcb
\right),
\label{2.3}\eeq
where ${\bf v}$ is the velocity of the particle, and
\beq
\ecb = - \nablab \varphi - \frac{1}{c} \,
\frac{\partial {\bf A}}{\partial t}
\qquad \mbox{and} \qquad
\bcb = \nablab \vecprod {\bf A}
\label{2.4}\eeq
are the electric field and the magnetic induction, respectively. The
case of a particle in a central field of force,
\beq
{\bf F} = - \nabla V(r),
\label{2.5}\eeq
which is of importance in atomic and nuclear physics, is studied in
detail. We describe spherical solutions of the Schr\"{o}dinger and Dirac
wave equations, which are important in calculations of atomic and
nuclear structure, and we present corresponding partial-wave expansions
of distorted plane waves, which are used in the description of
collisions of charged particles with atoms.


\section{The Schr\"{o}dinger wave equation\label{sec2.1}}
\index{wave equation!Schr\"{o}dinger}
\index{Schr\"{o}dinger wave equation}

In non-relativistic quantum theory, the wave function of a particle of
mass $M_0$ in a field of force corresponding to the potential $V({\bf
r},t)$ satisfies the time-dependent Schr\"{o}dinger equation
\beq
{\rm i} \hbar \frac{\partial}{\partial t} \Psi({\bf r},t) =
{\cal H} \Psi({\bf r},t),
\label{2.6}\eeq
with the Hamiltonian operator
\beq
{\cal H} = \frac{1}{2M_0} \breve{\bf p}^2 + V({\bf r},t),
\label{2.7}\eeq
where $\breve{\bf p} = - {\rm i} \hbar \nablab$ is the linear momentum
operator\footnote{In the case of operators that are represented with the
same symbols as the corresponding eigenvalues we use the diacritic
``$\breve{\phantom{p}}$'' to denote the operators.}. The Hamiltonian of a
free particle is the kinetic energy operator
\beq
K = \frac{1}{2M_0} \breve{\bf p}^2.
\label{2.8}\eeq
The quantity
\index{Schr\"{o}dinger wave equation!probability density}
\beq
\rho({\bf r},t) = \left| \Psi({\bf r},t) \right|^2
\label{2.9}\eeq
is the {\it probability density} of the particle. From the
time-dependent Schr\"{o}dinger equation \req{2.6},
\beq
{\rm i} \hbar \frac{\partial}{\partial t} \Psi({\bf r},t) = \left[
- \frac{\hbar^2}{2M_0} \nabla^2 + V({\bf r},t) \right]
\Psi({\bf r},t),
\nonumber\eeq
and its complex-conjugate equation,
\beq
- {\rm i} \hbar \frac{\partial}{\partial t} \Psi^\ast({\bf r},t) = \left[
- \frac{\hbar^2}{2M_0} \nabla^2 + V({\bf r},t) \right]
\Psi^\ast({\bf r},t),
\nonumber\eeq
it follows that
\beqa
\frac{\partial \rho}{\partial t} &=& \frac{\partial \Psi^\ast}{\partial t}
\, \Psi + \Psi^\ast  \frac{\partial \Psi}{\partial t}
= \frac{\hbar}{2 {\rm i} M_0} \left[ \left( \nabla^2 \Psi^\ast \right)
\Psi  - \Psi^\ast \left( \nabla^2 \Psi \right) \right]
\nonumber \\ [2mm]
&=& \nablab \dotprod \left\{
\frac{\hbar}{2 {\rm i} M_0} \left[
\left( \nablab \Psi^\ast \right) \Psi
- \Psi^\ast \, \left( \nablab \Psi \right) \right]
\right\},
\nonumber \eeqa
and we obtain the continuity equation
\beq
\frac{\partial \rho}{\partial t} + \nablab \dotprod {\bf j} = 0,
\label{2.10}\eeq
where
\index{Schr\"{o}dinger wave equation!probability current density}
\beq
{\bf j}({\bf r},t) = \frac{\hbar}{2 {\rm i} M_0} \left\{
\Psi^\ast({\bf r},t) \, \left[\nablab \Psi({\bf r},t) \right]
- \left[ \nablab \Psi^\ast({\bf r},t) \right] \Psi({\bf r},t)
\right\}
\label{2.11}\eeq
is the {\it probability current vector}, sometimes called the {\it
probability flux}.

\index{Schr\"{o}dinger wave equation!stationary states}
\index{time-independent Schr\"{o}dinger equation}
If the potential does not depend on time, the time dependence of the
wave function can be factored out by considering stationary states of
the type
\beq
\Psi({\bf r},t) = \psi({\bf r}) \, \exp(- {\rm i} E t/\hbar),
\label{2.12}\eeq
where $E$ is the energy of the particle, and $\psi({\bf r})$ satisfies
the time-independent Schr\"{o}dinger equation
\beq
{\cal H} \psi({\bf r}) = E \psi({\bf r}).
\label{2.13}\eeq
Notice that the probability density of a stationary state is constant
with time and, hence, $\nablab \dotprod {\bf j} =0$.

The Schr\"{o}dinger wave equation for a particle of mass $M_0$ and
electric charge $Z_0e$ in an electromagnetic field described by the scalar
potential $\varphi({\bf r},t)$ and the vector potential ${\bf A}({\bf
r},t)$ has the form \req{2.6} with the Hamiltonian\footnote{The
Hamiltonian \req{2.14} is obtained from that of the free particle,
${\cal H} = \breve{\bf p}^2/(2M_0)$, by applying the minimal coupling
replacements $\breve{\bf p} \rightarrow \breve{\bf p} - (Z_0e/c) {\bf
A}$ and ${\cal H} \rightarrow {\cal H} + Z_0 e \varphi$.}
\beqa
{\cal H} &=& \frac{1}{2M_0} \left[ \breve{\bf p}
- \frac{Z_0e}{c} {\bf A}({\bf r},t)
\right]^2 + Z_0e \, \varphi({\bf r},t) \nonumber \\ [2mm]
&=& \frac{1}{2M_0} \breve{\bf p}^2 - \frac{Z_0e}{2 M_0 c}
\left( \breve{\bf p} \dotprod
{\bf A} + {\bf A} \dotprod \breve{\bf p} \right)
+ \frac{Z_0^2 e^2}{2 M_0 c^2} A^2 + Z_0e \, \varphi.
\label{2.14}\eeqa
As usual, the squared modulus of the wave function defines the
probability density of the particle, Eq.\ \req{2.9}, and the
time-dependent wave equation implies that a continuity equation of the
form \req{2.10} holds with the probability current
\beq
{\bf j}({\bf r},t) = \frac{\hbar}{2 {\rm i} M_0} \left\{
\Psi^\ast({\bf r},t) \, \left[\nablab \Psi({\bf r},t) \right]
- \left[ \nablab \Psi^\ast({\bf r},t) \right] \Psi({\bf r},t)
\right\} - \frac{Z e}{M_0 c} {\bf A} \left| \Psi({\bf r},t) \right|^2,
\label{2.15}\eeq
which contains a term proportional to the vector potential. Evidently,
when ${\bf A} ={\bf 0}$, the particle is acted by the electric force
only, and the associated potential energy is $V = Z_0 e \varphi$. If, in
addition, the scalar potential is constant with time we can consider
stationary states with time-independent wave functions $\psi({\bf r})$
that are solutions of Eq.\ \req{2.13}.


\subsection{Plane waves \label{sec2.1.1}}
\index{Schr\"{o}dinger wave equation!plane waves}
\index{plane waves!periodic boundary conditions}

The Hamiltonian of a free particle ${\cal H} = \breve{\bf p}^2/(2M_0)$ commutes
with the linear momentum operator $\breve{\bf p}$ and, therefore, there is a
basis of common eigenfunctions $\phi_{\bf k}({\bf r})$ of these two
operators,
\beqa
\breve{\bf p} \phi_{\bf k}({\bf r}) &=&
- {\rm i} \hbar \nablab \phi_{\bf k}({\bf r}) =
\hbar {\bf k} \, \phi_{\bf k}({\bf r}),
\label{2.16}\\ [2mm]
{\cal H} \phi_{\bf k}({\bf r}) &=& - \frac{\hbar^2}{2M_0} \nabla^2
\phi_{\bf k}({\bf r}) = E_{\bf k} \phi_{\bf k}({\bf r}),
\label{2.17}\eeqa
with respective eigenvalues
\beq
{\bf p} = \hbar {\bf k}
\qquad \mbox{and} \qquad
E_{\bf k} = \frac{(\hbar k)^2}{2M_0}.
\label{2.18}\eeq
These eigenfunctions are the {\it plane waves}
\beq
\phi_{\bf k}({\bf r}) = (2\pi)^{-3/2} \exp({\rm i}{\bf k}\cdot{\bf r}),
\label{2.19}\eeq
which we have expressed in terms of the wave vector ${\bf k}\equiv{\bf
p}/\hbar$. With the adopted normalization,
\beq
\left< \phi_{{\bf k}'} \left| \phi_{\bf k} \right> \right.
= \int \phi_{{\bf k}'}^\ast({\bf r}) \; \phi_{\bf k} ({\bf r}) \, \d {\bf
r} = \delta ({\bf k}' - {\bf k}).
\label{2.20}\eeq
Notice that the spectra (\ie, the set of eigenvalues) of the
operators $\breve{\bf p}$ and ${\cal H}$ are continuous.

To avoid the difficulties of handling continuous spectra, we consider plane
waves $\phi_{L,{\bf k}}({\bf r})$ obeying periodic boundary conditions
on a cubic box with edges of length $L$ parallel to the axes of the
reference frame. That is, we require that
\beq
\phi_{L,{\bf k}} (x,y,z)= \phi_{L,{\bf k}} (x+L, y+L, z +L).
\label{2.21}\eeq
The solutions of the equation \req{2.17} satisfying
these boundary conditions are the plane waves
\beq
\phi_{L,{\bf k}}({\bf r}) = L^{-3/2} \exp({\rm i}{\bf k}\cdot{\bf r}),
\label{2.22}\eeq
with
\beq
{\bf k} = \frac{2\pi}{L} (n_{x},n_{y},n_{z}),
\label{2.23}\eeq
where $n_{x}$, $n_{y}$ and $n_{z}$ are integers. The normalization
constant in \req{2.22} is such that the probability of finding the
particle within the normalization box equals unity. Notice that the
periodic boundary conditions ensure that the probability current ${\bf
j}({\bf r})$, Eq.\ \req{2.11} takes the same values in opposite faces of
the normalization box, so that the probability fluxes through opposite
faces of the box are equal and, consequently, the probability of finding
the particle within the box remains constant with time.
Evidently $\phi_{L,{\bf k}}({\bf r})$ is an eigenfunction of the
operators $\breve{\bf p}$ and ${\cal H}$ with discrete
eigenvalues $\hbar{\bf k}$ and
\beq
E_{\bf k} = \frac{\hbar^{2}}{2M_0}k^{2} = \frac{\hbar^{2}}{2M_0}
\left(\frac{2\pi}{L}\right)^{2} (n_{x}^{2}+n_{y}^{2}+n_{z}^{2}),
\label{2.24}\eeq
respectively.
The plane waves \req{2.22} are orthonormal, that is,
\beq
\int_{L^3} \phi_{L,{\bf k}'}^\ast ({\bf r}) \,
\phi_{L,{\bf k}}({\bf r}) \, \d {\bf r} = \delta_{{\bf k}',{\bf k}},
\label{2.25}\eeq
where the integral is over the volume of the normalization box and
$\delta_{{\bf k}',{\bf k}}$ is the three-dimensional Kronecker delta,
\index{Kronecker delta}
\beq
\delta_{{\bf k}',{\bf k}} = \left\{
\begin{array}{ll}
1 & \mbox{if ${\bf k}'={\bf k}$}, \\ [1mm]
0 & \mbox{if ${\bf k}' \ne {\bf k}$}.
\end{array} \right.
\nonumber\eeq
The functions $\phi_{L,{\bf k}}({\bf r})$ constitute a complete basis, in
the sense that any function satisfying the periodic boundary conditions
can be expressed as a linear combination of these plane waves (Fourier
series),
\beq
\psi({\bf r}) = \sum_{{\bf k}} \left( \int \phi_{L,{\bf k}}^\ast ({\bf r}')
\, \psi ({\bf r}') \, \d {\bf r}' \right)
\phi_{L,{\bf k}} ({\bf r}).
\label{2.26}\eeq
Hence, the following closure relation is satisfied,
\beq
\sum_{{\bf k}} \phi_{L,{\bf k}}^\ast ({\bf r}')
\phi_{L,{\bf k}} ({\bf r}) = \delta({\bf r} - {\bf r}').
\label{2.27}\eeq

\begin{figure}[tbh] \begin{center}
\includegraphics*[width=6.0cm]{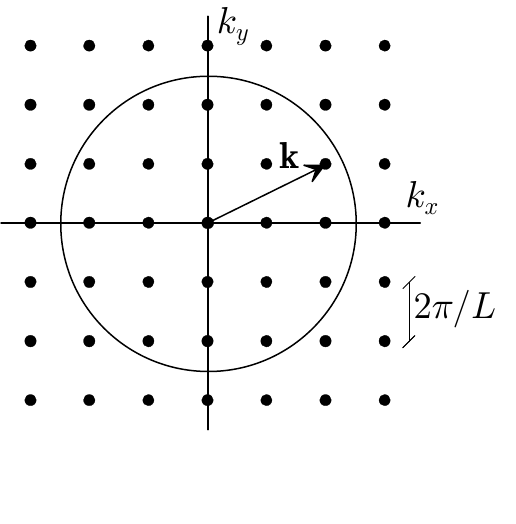} \rule{6mm}{0mm}
\includegraphics*[width=6.5cm]{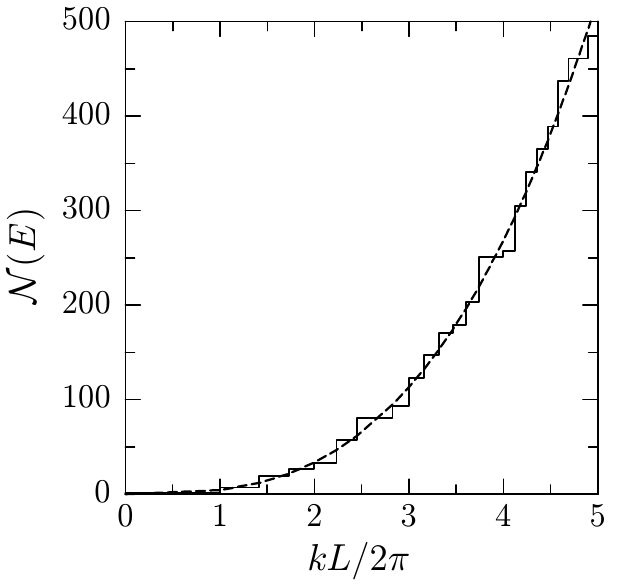}
\caption{\rm Wave vectors of plane waves satisfying periodic boundary
conditions on a cubic box with edge length $L$. The right plot shows the
number of states with energy less than $E=(\hbar k)^2/2 M_0$ (\ie, in
the interior of the sphere of radius $k$) as a function of the
dimensionless quantity $kL/2\pi$; The dashed curve represents the
continuous ``average'' given by the expression \req{2.29}.
\label{fig2.1}}
\end{center} \end{figure}

We can represent each state \req{2.22} by a point in ${\bf k}$-space
with the coordinates given by Eq.\ \req{2.23}, Fig.\ \ref{fig2.1}. The
representative points form a cubic lattice with spacing $2\pi/L$. Hence,
the average number of points per unit volume in ${\bf k}$-space is
\beq
\frac{\d {\cal N}}{\d {\bf k}} = \left(\frac{L}{2\pi} \right)^3.
\label{2.28}\eeq
The states with energy less than $E$ are those having wave vectors
inside a sphere of radius $k=(2M_0 E)^{1/2}/\hbar$. If $k\gg 2\pi/L$,
the number ${\cal N}(E)$ of states with energy less than $E$ can be
estimated by considering the distribution of allowed wave vectors as
continuous. We thus obtain
\beq
{\cal N}(E) = \frac{\d {\cal N}}{\d {\bf k}} \,
\frac{4}{3} \pi k^{3}
= \left(\frac{L}{2\pi}\right)^{3} \frac{4}{3} \pi k^{3} = \frac{L^{3}}{6\pi^{2}} \left(\frac{2M_0}{\hbar^{2}}
\right)^{3/2}E^{3/2}.
\label{2.29}\eeq
This expression gives a kind of average value of
${\cal N}(E)$. When $k$ is not much larger than $2\pi/L$, the discrete
character of the distribution is important (see Fig.\ \ref{fig2.1}) and
cannot be ignored. Actually, the number of states with energy less than
$E$ is a stepwise function (that is, its derivative is either null or
undefined).

\vspace{2mm}

Writing $\d {\bf k} = k^2 \d k \, \d \hat{\bf k}$, where $\hat{\bf k}$
is the unit vector in the direction of ${\bf k}$, we see that the
density of states per unit energy and per unit solid angle (in the
direction $\hat{\bf k}$) is
\index{plane waves!density of states}
\beq
\frac{\d {\cal N}}{\d E \, \d \hat{\bf k}}
= \frac{\d {\cal N}}{\d {\bf k}} \, \frac{k^2 \d k}{\d E} =
\left(\frac{L}{2\pi} \right)^3 \frac{M_0}{\hbar^2} k
= \frac{L^3}{16\pi^3} \left(\frac{2 M_0}{\hbar^2}\right)^{3/2}
\sqrt{E}.
\label{2.30}\eeq
Since there are no privileged directions, the density of states per
unit energy is
\beq
\frac{\d {\cal N}}{\d E }
=\frac{\d {\cal N}}{\d E \, \d \hat{\bf k}} \,  4\pi =
\frac{L^3}{2\pi^2} \, \frac{M_0}{\hbar^2} k
= \frac{L^3}{4\pi^2} \left(\frac{2 M_0}{\hbar^2}\right)^{3/2} \sqrt{E}.
\label{2.31}\eeq


\subsubsection{The ``large box'' limit \label{sec2.1.1.1}}

\index{plane waves!continuum normalization}
With the periodic boundary conditions we assume that the space is
partitioned into cubic boxes of size $L$, each box containing one
particle. This periodic structure has no physical effect when passing to
the limit $L\rightarrow \infty$, in which the wave vector ${\bf k}$
becomes a continuous variable. To study this limiting process, we write
the Fourier series \req{2.26} in the form
\beqa
\psi({\bf r}) &=& \sum_{{\bf k}} \left( \int
L^{-3/2} \exp(-{\rm i}{\bf k}\cdot{\bf r}')
\, \psi({\bf r}') \, \d {\bf r}' \right)
 L^{-3/2} \exp({\rm i}{\bf k}\cdot{\bf r}).
\label{2.32}\eeqa
When increasing the edge $L$ of the normalization box, the amplitude of
the plane waves decreases and the density of states in ${\bf k}$-space,
$\d {\cal N}/\d {\bf k}$, increases. When $L$ is sufficiently
large, we can consider the distribution of allowed ${\bf k}$ values as
continuous and replace the summation over ${\bf k}$ with an integral. Since
the number of states in a volume $\d {\bf k}$ is $(\d  {\cal N}/\d {\bf
k}) \, \d {\bf k}$, we have
\beqa
\psi({\bf r}) &=& \lim_{L\rightarrow\infty}
\sum_{{\bf k}} \left( \int
L^{-3/2} \exp(-{\rm i}{\bf k}\cdot{\bf r}')
\, \psi({\bf r}') \, \d {\bf r}' \right)
 L^{-3/2} \exp({\rm i}{\bf k}\cdot{\bf r})
\nonumber \\ [2mm]
&=& \int \left( \int
L^{-3/2} \exp(-{\rm i}{\bf k}\cdot{\bf r}')
\, \psi({\bf r}') \, \d {\bf r}' \right)
L^{-3/2} \exp({\rm i}{\bf k}\cdot{\bf r}) \, \frac{\d {\cal N}}{\d {\bf
k}} \, \d {\bf k}. \rule{10mm}{0mm}
\label{2.33}\eeqa
That is,
\beq
\psi({\bf r}) = \int \left( \int
(2\pi)^{-3/2} \exp(-{\rm i}{\bf k}\cdot{\bf r}')
\, \psi({\bf r}') \, \d {\bf r}' \right)
(2\pi)^{-3/2} \exp({\rm i}{\bf k}\cdot{\bf r}) \, \d {\bf k}.
\label{2.34}\eeq
This result suggests that in the continuous limit the eigenfunctions of
the linear momentum operator should be normalized in the form
\req{2.20}. Thus, Eq.\ \req{2.34} reduces to the familiar form
\beqa
\psi({\bf r}) &=& \int \left( \int
\phi_{\bf k}^\ast({\bf r}')
\, \psi({\bf r}') \, \d {\bf r}' \right)
\phi_{\bf k}({\bf r}) \, \d {\bf k}.
\label{2.35}\eeqa
Consequently, the limit $L\rightarrow \infty$ can be obtained formally
by replacing the summation over ${\bf k}$ with an integral and multiplying
the discrete plane waves $\phi_{L,{\bf k}}({\bf r})$ by the factor
$(L/2\pi)^{3/2}$, which is the square root of the density of discrete
states per unit volume of ${\bf k}$-space. This same process is used to
obtain the continuous Fourier transform from the discrete Fourier series
\citep[see, \eg,][]{Arfken1985}.

\index{plane waves!density of states}
In the extension to the continuum we have made explicit use of the
density of states, $\d {\cal N}/\d {\bf k}$, which in the limit
$L\rightarrow \infty$ is infinite. The consistency of the process rests
on the fact that the product
\beq
\phi_{L,{\bf k}}({\bf r}) \sqrt{ \frac{\d {\cal N}}{\d {\bf k}}} =
(2\pi)^{-3/2} \exp({\rm i}{\bf k}\cdot{\bf r}) = \phi_{\bf k}({\bf r})
\label{2.36}\eeq
is independent of the volume $L^3$ of the normalization box.
Evidently, in the limit $L \rightarrow \infty$ the density of states has been
absorbed by the normalization constants of the continuum states.
Formally, the density of continuum states $\phi_{\bf k}({\bf r})$
per unit volume of ${\bf k}$-space can be defined by
\beq
\left< \phi_{{\bf k}'} | \phi_{\bf k} \right> = \left( \frac{\d {\cal N}}{\d
{\bf k}} \right)^{-1} \delta({\bf k} - {\bf k}').
\label{2.37}\eeq
This identity shows that, generally, the density of free states is determined
by the adopted normalization. Notice that with the plane waves
$\phi_{\bf k}({\bf r})$ normalized in the form \req{2.19}, the density
of states $\d {\cal N}/\d {\bf k}$ equals unity.


\subsection{Central potentials. Radial wave equation\label{sec2.1.2}}

\index{Schr\"{o}dinger wave equation!central potentials}
The Hamiltonian of a particle of mass $M_0$ in a time-independent central
potential $V(r)$, is [Eq.\ \req{2.7}]
\beq
{\cal H} = -\frac{\hbar^2}{2 M_0} \, \nabla^2 + V(r) =
-\frac{\hbar^2}{2 M_0} \left( \frac{1}{r} \,
\frac{\partial^2}{\partial\,r^2} r - \frac {1}{r^2} L^2 \right) + V(r),
\label{2.38}\eeq
where
\beq
{\bf L} \equiv \hbar^{-1} {\bf r} \vecprod \breve{\bf p}
\label{2.39}\eeq
is the orbital angular momentum operator (in units of $\hbar$) [see Eq.\
\req{B.29e}]. Since ${\bf L}$ commutes with ${\cal H}$, we can construct
simultaneous eigenfunctions of ${\cal H}$, $L^2$ and $L_z$. These
solutions of the Schr\"{o}dinger equation are known as {\it
central-field orbitals} or {\it spherical waves}; they are of the form
\citep{Schiff1968, Merzbacher1970} \index{central-field orbitals}
\beq
\psi_{E\ell m}({\bf r}) = \frac{1}{r} \, P_{E\ell}(r)
\, Y_{\ell m}(\hat{\bf r}),
\label{2.40}\eeq
where the functions $Y_{\ell m}(\hat{\bf r})$, the spherical harmonics
(see Appendix \ref{appB}),
are eigenfunctions of $L^2$ and $L_z$ [with eigenvalues $\ell(\ell+1)$
and $m$, respectively] and the reduced radial function $P_{E\ell}(r)$
satisfies the radial Schr\"{o}dinger equation
\index{radial Schr\"{o}dinger wave equation}
\beq
- \frac{\hbar^2}{2M_0} \, \frac{\d^2 P_{E\ell}}{\d r^2}
+ \left[
\frac{\hbar^2}{2M_0} \, \frac{\ell(\ell+1)}{r^2} + V(r) \right] P_{E\ell}
= E \, P_{E\ell}\, .
\label{2.41}\eeq
The orbital angular momentum quantum number $\ell$ may take
non-negative integer values and, for a given $\ell$, the allowed values
of the magnetic quantum number are $-\ell$, $-\ell+1$, \ldots, $\ell-1$,
and $\ell$ (see Appendix \ref{appB}). The orbitals \req{2.40} have
parity $(-1)^\ell$ under spatial inversion [see Eq.\ \req{B.56d}].

\index{modified Coulomb potentials}
For the sake of simplicity and numerical convenience we shall limit our
considerations to potentials $V(r)$ such that $r V(r)$ is finite for all
$r$. This restriction holds for finite-range potentials, Coulomb
potentials, and any combination of both (the so-called {\it modified
Coulomb potentials}). We can then require that the spatial wave
function $\psi_{E\ell m}({\bf r})$ also be finite in all space. This
implies that the radial wave function $P_{E\ell}(r)$ behaves as
$r^{\ell+1}$ near the origin.

\index{radial Schr\"{o}dinger wave equation!bound states}
When $V(r)$ takes negative values in a certain region, bound states may
exist, where the particle is constrained to move within a limited
volume. The wave functions of bound states have finite norm, and the
corresponding reduced radial functions tend to zero when $r$ goes to
infinity. The radial equation \req{2.41} with the boundary condition
$P_{E\ell}(\infty)=0$ has solutions only for a discrete set of negative
energy eigenvalues. Traditionally, discrete energy levels are identified
by the principal quantum number $n$ and the orbital angular momentum
quantum number $\ell$. For a given $n$, $\ell$ can take values 0, 1,
\ldots $n-1$. Alternatively, instead of $n$, the radial quantum number,
$n_{\rm r} \equiv n-(\ell+1)$, can be used to label the negative energy
levels. The radial quantum number gives the number of nodes of the
radial function, \ie, the zeros of $P_{n\ell}(r)$ other than those at
$r=0$ and $r=\infty$. As Eq.\ \req{2.41} does not depend on the magnetic
quantum number $m$, each energy level $E_{n \ell}$ is at least $2\ell
+1$ times degenerate (the energy levels of a pure Coulomb potential are
also degenerate with respect to $\ell$\/). The wave functions of bound
states (with $E<0$) are normalized by requiring that
\beq
\int \psi_{n'\ell' m'}^\ast({\bf r}) \, \psi_{n\ell m}({\bf r}) \, \d {\bf r} =
\delta_{n',n} \, \delta_{\ell',\ell} \, \delta_{m',m},
\label{2.42}\eeq
where $\delta_{n',n}$ is the Kr\"{o}necker delta. Consequently, the radial
functions are normalized to unity,
\beq
\int_{0}^{\infty} P_{n\ell}^2(r) \d r = 1.
\label{2.43}\eeq

\index{radial Schr\"{o}dinger wave equation!free states}
In the case of free states (with $E>0$), the asymptotic behavior of the
radial function is determined by the Coulomb tail of the potential,
characterized by the parameter
\beq
Z_{\infty} \equiv \frac{1}{e^2} \lim_{r \rightarrow \infty} r \, V(r).
\label{2.44}\eeq
Since the potential tends to zero at large radii, the eigenvalue $E$
represents the kinetic energy of the particle far from the center of
force and we may introduce the corresponding linear momentum $p =
(2M_0E)^{1/2}$, velocity $v=p/M_0$, and wave number
\beq
k \equiv \frac{p}{\hbar} = \sqrt{2M_0 E}/\hbar.
\label{2.45}\eeq
The wave functions of free states will be normalized so that the radial
function has the asymptotic form
\beq
P_{E\ell}(r)
\begin{array}[t]{c}
\sim \\ [-3mm] \scriptstyle{ r \rightarrow \infty}
\end{array}
\sin \left( kr - \ell \frac{\pi}{2} - \eta \ln (2kr) + d_\ell \right),
\label{2.46}\eeq
where \index{Sommerfeld parameter}
\beq
\eta = \frac{Z_\infty e^2}{\hbar v} = \frac{M_0 Z_\infty e^2}{\hbar^2 k}
\label{2.47}\eeq
is the {\it Sommerfeld parameter} ($=0$ for finite-range
potentials) and $d_\ell$ is the phase shift. The radial functions
of free orbitals normalized in this way satisfy the orthonormality condition
\citep[see, \eg,][]{SalvatFernandezVarea2019}
\beq
\int_0^\infty P_{E'\ell'}(r) \, P_{E\ell}(r) \, \d r = \frac{\pi E}{k}
\, \delta(E'-E) \, \delta_{\ell',\ell}
= \frac{\pi}{2} \, \delta(k'-k) \, \delta_{\ell',\ell},
\label{2.48}\eeq
where $\delta(E'-E)$ is the Dirac delta function, and $k'$ is the wave
number of particles with energy $E'$.

\index{radial Schr\"{o}dinger wave equation!free states!phase shifts}
\index{phase-shifts}
The set of phase shifts $d_\ell$ characterizes the scattering of
particles in the potential $V(r)$. Modified Coulomb potentials such that
$Z_\infty \ne 0$ can be regarded as the sum of a Coulomb
potential $V_{\rm C}(r)=Z_\infty e^2/r$ and a finite-range
distortion. The phase shifts for the Coulomb potential are given by
\citep[see, \eg,][and references therein]{Salvat1995}
\beq
\Delta_{\ell} = \arg \Gamma \left( \ell+1+{\rm i}\eta \right),
\label{2.49}\eeq
where $\Gamma(z)$ denotes the complex gamma function. The phase shift
$d_\ell$ is then the sum of the Coulomb phase shift $\Delta_\ell$
and the ``inner'' phase shift $\delta_\ell$, which is only due to
the finite-range distortion of the asymptotic Coulomb potential, $V_{\rm
fr} (r) \equiv V(r)-V_{\rm C}(r)$,
\beq
d_\ell = \delta_\ell + \Delta_\ell\, .
\label{2.50}\eeq
The effect of the Coulomb potential is accounted for by the logarithmic
phase, $-\eta \ln (2kr)$, and the Coulomb phase shift. For a pure
Coulomb potential ($V_{\rm fr} \equiv 0$), $\delta_\ell=0$. Attractive
(repulsive) finite-range potentials give positive (negative) inner phase
shifts. The phase shifts are generally indeterminate in an integer
multiple of $\pi$.

The set of spherical waves (bound and free) constitutes a complete
orthogonal basis of the complex vector space of particle states in the
coordinate representation. Generally, radial wave functions have to be
calculated numerically by solving the radial equation \req{2.41}. The
Fortran subroutine package {\sc radial} \citep{SalvatFernandezVarea2019}
provides highly accurate solutions of the radial equation (that is,
radial functions, energies of bound states, and phase shifts of free
states) for potentials such that $rV(r)$ is finite everywhere.


\subsection{Schr\"{o}dinger distorted plane waves \label{sec2.1.3}}
\index{distorted plane waves!Schr\"{o}dinger wave equation}
\index{Schr\"{o}dinger wave equation!distorted plane waves}

In collision theory, states of free particles (with $E>0$) in the
initial and final channels are described as distorted plane waves
(DPWs), which are solutions of the wave equation for the potential
$V(r)$ that asymptotically behave as a plane wave plus an outgoing
or incoming spherical wave \citep[see, \eg,][]{Joachain1975}.

The Schr\"{o}dinger DPWs of a particle with momentum $\hbar {\bf k}$ in
a potential $V(r)$ {\it of finite range} are solutions of the wave
equation
\beq
\left[ - \frac{\hbar^2}{2 M_0} \, \nabla^2 + V(r) \right]
\psi^{(\pm )}_{\bf k} ({\bf r}) = E \, \psi^{(\pm )}_{\bf k}({\bf r}),
\qquad E=\frac{(\hbar k)^2}{2 M_0},
\label{2.51}\eeq
with the asymptotic behavior
\beq
\psi^{(\pm)}_{\bf k}({\bf r})
\begin{array}[t]{c} \sim \\[-2mm] \scriptstyle{ r
\rightarrow \infty} \end{array}
\phi_{{\bf k}}({\bf r}) + \psi^{(\pm )}_{\rm sc}({\bf r}),
\label{2.52}\eeq
where
$$
\phi_{{\bf k}}({\bf r}) =
\frac{1}{(2\pi)^{3/2}} \, \exp( {\rm i} {\bf k} \dotprod {\bf r})
\eqno{\rm \req{2.36}}$$
is a plane wave, and
\beq
\psi^{(\pm )}_{\rm sc} ({\bf r}) =
\frac{1}{(2\pi)^{3/2}} \, \frac{\exp( \pm
{\rm i} k r)}{r} \, f^{(\pm)}(\hat{\bf k} \dotprod \hat{\bf r})
\label{2.53}\eeq
is an outgoing ($+$) or incoming ($-$) spherical wave\footnote{We recall
that the time-dependent wave function is $\psi^{(\pm )}_{\bf k}({\bf r})
\, \exp(-{\rm i} E t/\hbar)$. The outgoing or incoming character of
the spherical component becomes evident only when considering the
time-dependent factor.}. The function $f^{(\pm )}({\bf k} \cdot {\bf
r})$ is called the {\it scattering amplitude}.
\index{scattering amplitude}
Physically, the DPW
$\psi_{{\bf k}}^{(+)}({\bf r})$ represents a stationary scattering state
with a parallel beam of particles impinging on the scattering center in
the direction $\hat{\bf k}$ and a spherical outgoing wave, modulated by
the scattering amplitude $f^{(+)}(\hat{\bf k} \cdot \hat{\bf r})$,
which describes particles scattered in the direction $\hat{\bf r}$. The
squared modulus of this scattering amplitude equals the scattering
differential cross section (see Section \ref{sec5.1}). The
amplitude $f^{(+)} (\hat{\bf k} \cdot \hat{\bf r})$ of the outgoing
spherical wave admits the following integral representation
[Eq.\ (5.39) in \citet{Joachain1975}]
\beqa
f^{(+)} (\hat{\bf k} \dotprod \hat{\bf r})
&=& - \frac{4 \pi^2 M_0}{\hbar^2} \,
\, \int \phi^\ast_{k \hat{\bf r}}({\bf r}')
\, V(r') \psi^{(+)}_{\bf k}({\bf r}')\, \d {\bf r}'
\nonumber \\ [2mm]
&=& - \frac{(2 \pi)^{1/2} M_0}{\hbar^2} \,
\, \int \exp(-{\rm i} \, k \hat{\bf r} \dotprod {\bf r}')
\, V(r') \psi^{(+)}_{\bf k}({\bf r}')\, \d {\bf r}'.
\label{2.54}\eeqa

\index{distorted plane waves!Schr\"{o}dinger wave equation!partial-wave series}
The DPWs can be expanded in the basis of spherical waves as
\beq
\psi_{{\bf k}}^{(\pm)} ({\bf r}) = \frac{1}{k} \, \sqrt{\frac{2}{\pi}}
\sum_{\ell=0}^\infty  \sum_{m=-\ell}^\ell
{\rm i}^\ell \, \exp \left( \pm {\rm i} \delta_{\ell}
\right) Y_{\ell m}^\ast (\hat{\bf k}) \, \psi_{E \ell m}({\bf r}),
\label{2.55}\eeq
where $\delta_{\ell}$ are the phase shifts (recall that the potential
is assumed to have a finite range). The expansion \req{2.55}
is known as the {\it partial-wave series}. Inserting the expression
\req{2.40} of the spherical waves, we have
\beq
\psi_{{\bf k}}^{(\pm)} ({\bf r}) = \frac{1}{kr} \, \sqrt{\frac{2}{\pi}}
\sum_{\ell} {\rm i}^\ell \, \exp \left( \pm {\rm i} \delta_{\ell}
\right) P_{E\ell}(r)
\sum_{m} Y_{\ell m}^\ast (\hat{\bf k}) \, Y_{\ell m}(\hat {\bf r}).
\label{2.56}\eeq
The angular part can be simplified with the aid of the
addition theorem of the spherical harmonics [Eq.\ \req{B.57}],
\beq
\sum_{m=-\ell}^\ell
Y_{\ell m}^\ast (\hat{\bf k}) \, Y_{\ell m}(\hat {\bf r}) =
\frac{2 \ell +1}{4\pi} \, P_\ell(\cos\theta ), \qquad
\cos\theta =  \hat{\bf k} \dotprod \hat{\bf r},
\label{2.57}\eeq
where $P_\ell(\cos\theta)$ is the Legendre polynomial of degree $\ell$
\citep{Arfken1985,Olver2010},
and $\theta$ is the angle between the vectors ${\bf k}$ and ${\bf r}$.
Making use of the asymptotic form \req{2.46} of the radial functions, we
can readily verify that the DPWs defined by the expansion \req{2.56}
have the required asymptotic behavior \req{2.51} with the
scattering amplitudes given by the following partial-wave expansions
\index{scattering amplitude!partial-wave series}
\begin{subequations}
\label{2.58}
\beq
f^{(+)}(\hat{\bf k} \dotprod \hat{\bf r}) = \frac{1}{2 {\rm i} k} \sum_{\ell}
(2 \ell+1) \left[
{\rm exp}( 2 {\rm i} \delta_{\ell}) - 1 \rule{0mm}{4mm}\right]
P_\ell(\cos\theta)
\label{2.58a}\eeq
and
\beq
f^{(-)}(\hat{\bf k} \dotprod \hat{\bf r}) = \frac{1}{2 {\rm i} k} \sum_{\ell} (-1)^\ell
(2 \ell+1) \left[ 1-
{\rm exp}( - 2 {\rm i} \delta_{\ell}) \rule{0mm}{4mm}\right]
P_\ell(\cos\theta).
\label{2.58b}\eeq
\end{subequations}
From the orthogonality of the radial functions, Eq.\ \req{2.48},
it follows that the DPWs satisfy the orthogonality relation
\beq
\int
\left[ \psi_{{\bf k}'}^{(\pm)} ({\bf r}) \right]^\ast
\psi_{{\bf k}}^{(\pm)} ({\bf r}) \, \d {\bf r}
= \delta({\bf k}'-{\bf k}) \, ,
\label{2.59}\eeq
which implies that the density of DPW states per unit volume in ${\bf
k}$-space is equal to unity (see Section \ref{sec2.1.1.1}).

From the partial-wave series \req{2.56}, and the
properties of the spherical harmonics, it follows that
\beq
\psi_{{\bf k}}^{(-)} ({\bf r}) = \left[ \psi_{-{\bf k}}^{(+)} ({\bf r})
\right]^\ast \qquad \mbox{and} \qquad
f^{(-)}(\hat{\bf k} \dotprod \hat{\bf r}) = \left[
f^{(+)}(- \hat{\bf k} \dotprod \hat{\bf r})\right]^\ast.
\label{2.60}\eeq
The first of these equalities indicates that the DPW $\psi_{{\bf
k}}^{(-)} ({\bf r})$ describes the result of applying the time-reversal
operator on the state corresponding to the DPW $\psi_{-{\bf
k}}^{(+)}({\bf r})$ \citep[see, \eg,][]{Schiff1968}. That is, the DPW
$\psi_{{\bf k}}^{(-)} ({\bf r})$ represents a parallel beam of particles
that {\it emerge} from the center of force in the direction $\hat{\bf
k}$.

Let us consider the continuous family of potentials $a V(r)$, with the
strength parameter $a$ taking values in the interval (0,1). Evidently,
the DPWs for the potentials $aV(r)$ vary continuously with $a$.  In the
limit where $a$ tends to zero, the phase shifts vanish, the radial
functions of free states ($E>0$) are $P_{E\ell}(r)= kr \, j_\ell (kr)$,
where $j_\ell (x)$ are the regular spherical Bessel functions
\citep{Arfken1985}, and the DPWs \req{2.56} reduce to
\beq
\lim_{a \rightarrow 0} \psi^{(\pm)}_{\bf k} ({\bf r})
= \sqrt{\frac{2}{\pi}} \sum_{\ell} {\rm i}^\ell
j_\ell(kr) \sum_{m}
Y_{\ell m}^\ast (\hat{\bf k})  \,
Y_{\ell m}(\hat{\bf r})=(2\pi)^{-3/2} \exp({\rm i} {\bf k} \dotprod
{\bf r}).
\label{2.61}\eeq
With the aid of the addition theorem \req{2.57}, this expression becomes the
familiar Rayleigh expansion of the plane wave,
\beq
\exp({\rm i} {\bf k} \dotprod {\bf r}) = 4\pi \sum_{\ell} {\rm i}^\ell
j_\ell(kr) \sum_{m}
Y_{\ell m}^\ast (\hat{\bf k})  \,
Y_{\ell m}(\hat{\bf r})
=\sum_{\ell} {\rm i}^\ell (2\ell +1) \,
j_\ell(kr) \, P_\ell(\hat {\bf k} \dotprod \hat{\bf r}).
\label{2.62}\eeq
In other words, the DPWs $\psi^{(\pm)}_{\bf k} ({\bf r})$ defined by
Eq.\ \req{2.56} represent states that result from the continuous
transformation of the normalized plane wave obtained by adiabatically
increasing the potential from $0$ to $V(r)$, \ie, varying the strength
parameter $a$ from 0 to 1.

\index{Coulomb distorted plane waves}\index{Kummer function}
In the case of the Coulomb potential, $V_{\rm c}(r)=Z_\infty e^2/r$, the
wave equation \req{2.51} for the DPW $\psi^{(+)}_{\bf k}({\bf r})$ can
be solved analytically by using parabolic coordinates
\citep[see, \eg,][]{Joachain1975}.  The result is
\beq
\psi^{\rm (C,+)}_{{\bf k}} ({\bf r})
= (2\pi)^{-3/2} \exp(-\pi \eta /2) \,
\Gamma (1+{\rm i} \eta) \,
\exp\left( {\rm i} {\bf k} \dotprod {\bf r} \right) \,
_1F_1(-{\rm i} \eta,1 ;{\rm i} [kr - {\bf k} \dotprod {\bf r}]),
\label{2.63}\eeq
where  $\eta = M_0 Z_\infty e^2/(\hbar^2 k)$ is the Sommerfeld parameter
[Eq.\ \req{2.47}] and $_1F_1(a,b,z)$ is the Kummer function,
\citep{AbramowitzStegun1974, Olver2010},
\beq
_1F_1(a;b;z) = 1 + \frac{a}{b} \, z
+ \frac{a(a+1)}{b(b+1)} \, \frac{z^2}{2!}
+ \frac{a(a+1)(a+2)}{b(b+1)(b+2)} \, \frac{z^3}{3!}
+ \ldots
\label{2.64}\eeq
Far from the scattering center, and for large ``transverse distances'',
$D_{\rm tr} = r - \hat{\bf k} \dotprod {\bf r}$,
this DPW has the following asymptotic form,
\beqa
&& \! \! \! \!\! \! \! \! \! \! \! \! \!
\psi_{{\bf k}}^{\rm (C,+)} ({\bf r})
\begin{array}[t]{c} \sim \\ [-2mm] \scriptstyle{ d
\rightarrow \infty} \end{array}
(2\pi)^{-3/2} \, \exp[ {\rm i} {\bf k} \dotprod {\bf r}
+ {\rm i} \eta \ln (kr - {\bf k} \dotprod {\bf r})] \, \left[ 1 +
\frac{\eta^2}{{\rm i}(kr - {\bf k} \dotprod {\bf r})} + \ldots \right]
\nonumber \\ [3mm]
&& \mbox{} + (2\pi)^{-3/2} \,
\frac{\exp[ {\rm
i} k r - {\rm i} \eta \ln(2kr)]}{r} \left[ 1
+ \frac{(1+{\rm i} \eta)^2}{{\rm i}
(kr - {\bf k} \dotprod {\bf r})} + \ldots \right]
f^{\rm (C,+)}(\hat{\bf k} \dotprod \hat{\bf r})
\label{2.65}\eeqa
with the scattering amplitude \index{Coulomb scattering amplitude}
\beq
f^{\rm (C,+)}(\hat{\bf k} \dotprod \hat{\bf r})
= - \eta \exp(2{\rm i}\Delta_0) \,
\frac{\exp[ - {\rm i} \eta \ln(\sin^2(\theta/2))]}{2k\sin^2(\theta/2)},
\label{2.66}\eeq
where $\theta=\arccos(\hat{\bf k} \cdot \hat{\bf r})$ is the angle
between the vectors ${\bf k}$ and ${\bf r}$, and the quantity $\Delta_0
= {\rm arg} \, \Gamma(1 + {\rm i} \eta)$ is the zeroth-order Coulomb
phase shift, Eq.\ \req{2.49}. The asymptotic form \req{2.65} does not
reduce to the expression \req{2.52} characteristic of the DPWs for
finite-range potentials. The long range of the Coulomb potential causes
modifications that, when $D_{\rm tr} \rightarrow \infty$, affect only
the phases of the two terms in the asymptotic form \req{2.65}.
Therefore, these two terms can still be interpreted as describing,
respectively, an incident plane wave and an outgoing scattered wave.

The Coulomb DPW with incoming spherical component, $\psi^{\rm
(C,-)}_{{\bf k}} ({\bf r})$, and the corresponding scattering amplitude
are determined by the formulas \req{2.60}. That is,
\beqa
\psi^{\rm (C,-)}_{{\bf k}} ({\bf r})
&=& \left[ \psi^{\rm (C,+)}_{-{\bf k}} ({\bf r}) \right]^\ast
\nonumber \\ [2mm]
&=& (2\pi)^{-3/2} \exp(-\pi \eta /2) \,
\Gamma (1 - {\rm i} \eta) \,
\exp\left( {\rm i} {\bf k} \dotprod {\bf r} \right) \,
_1F_1({\rm i} \eta; 1; -{\rm i} [kr + {\bf k} \dotprod {\bf r}])
\rule{10mm}{0mm}
\label{2.67}\eeqa
and
\beq
f^{\rm (C,-)}(\hat{\bf k} \dotprod \hat{\bf r})
= \left[ f^{\rm (C,+)}(- \hat{\bf k} \dotprod \hat{\bf r}) \right]^\ast
= - \eta \exp(-2{\rm i}\Delta_0) \,
\frac{\exp[ {\rm i} \eta \ln(\sin^2(\theta/2))]}{2k\sin^2(\theta/2)}.
\label{2.68}\eeq
The Coulomb DPWs and the scattering amplitude can also be
expressed in the form of partial-wave series,
\beq
\psi_{{\bf k}}^{\rm (C,\pm)} ({\bf r}) = \frac{1}{kr}
\, \sqrt{\frac{2}{\pi}}
\sum_{\ell} {\rm i}^\ell \, \exp \left( \pm {\rm i} \Delta_{\ell}
\right) F_{\ell}(\eta, kr)
\sum_{m} Y_{\ell m}^\ast (\hat{\bf k}) \, Y_{\ell m}(\hat {\bf r})
\label{2.69}\eeq
and
\beq
f^{\rm (C, +)}(\hat{\bf k} \dotprod \hat{\bf r})
= \frac{1}{2 {\rm i} k} \sum_{\ell}
(2 \ell+1) \left[
{\rm exp}( 2 {\rm i} \Delta_{\ell}) - 1 \rule{0mm}{4mm}\right]
\, P_\ell(\cos\theta),
\label{2.70}\eeq
where $\Delta_\ell$ are the Coulomb phase shifts \req{2.49}, and
$F_\ell(\eta,kr)$ are the regular Coulomb functions, \ie, the solutions
of the radial Schr\"{o}dinger equation for the Coulomb potential that
are regular at the origin \citep[see, \eg,][]{Salvat1995}.

\index{distorted plane waves!Schr\"{o}dinger wave equation!partial-wave series}
In the general case of a modified Coulomb field, $V(r)$ such that
$\lim_{r\rightarrow \infty} r V(r) = Z_\infty e^2$, the corresponding DPWs are
given by the partial-wave series \req{2.56} with the
total phase shifts $d_{\ell} = \delta_{\ell} +
\Delta_\ell$, which are the sum of the inner and Coulomb phase shifts.
Far from the scattering center, the DPW $\psi_{{\bf k}}^{(+)} ({\bf r})$
has the asymptotic form,
\beqa
&& \! \! \! \!\! \! \! \! \! \! \! \! \!
\psi_{{\bf k}}^{(+)}({\bf r})
\begin{array}[t]{c} \sim \\ [-2mm] \scriptstyle{ d
\rightarrow \infty} \end{array}
(2\pi)^{-3/2} \, \exp[ {\rm i} {\bf k} \dotprod {\bf r}
+ {\rm i} \eta \ln (kr - {\bf k} \dotprod {\bf r})] \, \left[ 1 +
\frac{\eta^2}{{\rm i}(kr - {\bf k} \dotprod {\bf r})} + \ldots \right]
\nonumber \\ [2mm]
&& \mbox{} + (2\pi)^{-3/2} \,
\frac{\exp[ {\rm
i} k r - {\rm i} \eta \ln(2kr)]}{r} \left[ 1
+ \frac{(1+{\rm i} \eta)^2}{{\rm i}
(kr - {\bf k} \dotprod {\bf r})} + \ldots\right]
f^{(+)}(\hat{\bf k} \dotprod \hat{\bf r}),
\label{2.71}\eeqa
with the scattering amplitude [see Eq.\ \req{2.58a}]
\beq
f^{(+)}(\hat{\bf k} \dotprod \hat{\bf r}) = \frac{1}{2 {\rm i} k} \sum_{\ell}
(2 \ell+1) \left[
\exp(2{\rm i}d_{\ell}) - 1 \rule{0mm}{4mm}\right]
P_\ell(\cos\theta).
\label{2.72}\eeq

When the potential $V(r)$ has a Coulomb tail, the partial-wave series
for the scattering amplitude diverges at $\theta=0$ and may converge
very slowly at small angles. The convergence of that series can be
accelerated by adding the Coulomb scattering amplitude, Eq.\ \req{2.66},
and subtracting the partial-wave series \req{2.70}. This yields the
modified series
\beq
f^{(+)}(\theta) = f^{\rm (C,+)}(\theta) +
\frac{1}{2 {\rm i} k} \sum_{\ell} (2\ell +1) \,
{\rm exp}( 2 {\rm i} \Delta_{\ell})
\left[ {\rm exp}( 2 {\rm i} \delta_{\ell}) - 1 \rule{0mm}{4mm}\right]
P_\ell(\cos\theta),
\label{2.73}\eeq
where $\delta_{\ell}$ are the inner phase shifts. This series converges
as rapidly as the partial-wave series of the scattering amplitude for
only the short-range component of the potential.


\section{The Dirac wave equation\label{sec2.2}}
\index{wave equation!Dirac}
\index{Dirac wave equation}

Electrons (mass $\me$, charge $-e$) and positrons (mass $\me$, charge $+
e$) have spin $\1o2$. They are peculiar in that their mass is much
smaller than those of other charged particles. The relativistic wave
equation of spin-$\1o2$ particles is the Dirac equation \citep[see,
\eg,][]{Rose1961, Sakurai1967}
\beq
{\rm i} \hbar \frac{\partial}{\partial t} \Psi({\bf r},t) =
{\cal H}_{\rm D} \Psi({\bf r},t).
\label{2.74}\eeq
The Dirac Hamiltonian for an electron in an external electromagnetic field
described by the scalar potential $\varphi({\bf r},t)$ and the
vector potential ${\bf A}({\bf r},t)$ is
\beq
{\cal H}_{\rm D} = c \widetilde\alphab \dotprod
\left[ \breve{\bf p} + \frac{e}{c} {\bf A} ({\bf r},t)
\right] + \widetilde\beta \, \me c^2 - e \varphi
\label{2.75}\eeq
where $\widetilde{\alphab}=(\alpha_1,\alpha_2,\alpha_3)$ and
$\widetilde{\beta}\equiv \widetilde{\alpha}_4$ are
4$\times$4 anticommuting matrices, \ie, such that
\beq
\left\{ \widetilde\alpha_i , \widetilde\alpha_j \right\}
\equiv \widetilde\alpha_i \widetilde\alpha_j
+ \widetilde\alpha_j \widetilde\alpha_i = 2 \delta_{ij},
\qquad \mbox{$i,j = 1$, 2, 3, 4.}
\label{2.76}\eeq
In the so-called {\it spinor representation}, or {\it Dirac-Pauli
representation}, these matrices are \index{Dirac matrices}
\beq
\widetilde{\alphab} = \left( \begin{array}{cc}
0 & \sigmab \\
\sigmab & 0 \end{array} \right), \qquad
\widetilde{\beta} = \left( \begin{array}{cc}
{\rm I}_2 & 0  \\
0 & -{\rm I}_2 \end{array} \right).
\label{2.77}\eeq
Here, $\sigmab = (\sigma_1, \sigma_2, \sigma_3)$ stands for the familiar
set of {\it Pauli spin matrices} \index{Pauli spin matrices}
\beq
\sigma_1 = \left( \begin{array}{cc}
0 & 1 \\
1 & 0 \end{array} \right), \quad
\sigma_2 = \left( \begin{array}{cc}
0 & -{\rm i} \\
{\rm i} & 0 \end{array} \right), \quad
\sigma_3 = \left( \begin{array}{cc}
1 & 0 \\
0 & -1 \end{array} \right),
		\label{2.78}\eeq
and ${\rm I}_2$ is the 2$\times$2 unit matrix. Explicitly,
\beq
\widetilde{\alpha}_1 = \left( \begin{array}{cccc}
0 & 0 & 0 & 1 \\
0 & 0 & 1 & 0 \\
0 & 1 & 0 & 0 \\
1 & 0 & 0 & 0 \end{array} \right), \quad
\widetilde{\alpha}_2 = \left( \begin{array}{cccc}
0 & 0 & 0 & -{\rm i} \\
0 & 0 & {\rm i} & 0 \\
0 & -{\rm i} & 0 & 0 \\
{\rm i} & 0 & 0 & 0 \end{array} \right), \quad
\widetilde{\alpha}_3 = \left( \begin{array}{cccc}
0 & 0 & 1 & 0 \\
0 & 0 & 0 & -1 \\
1 & 0 & 0 & 0 \\
0 & -1 & 0 & 0 \end{array} \right),
\nonumber \eeq
\beq
\widetilde{\beta} = \left( \begin{array}{cccc}
1 & 0 & 0 & 0 \\
0 & 1 & 0 & 0 \\
0 & 0 & -1 & 0 \\
0 & 0 & 0 & -1 \end{array} \right).
\label{2.79}\eeq
In this representation, the wave function
$\Psi({\bf r},t)$ is a four-component column function or {\it bi-spinor},
\beq
\Psi({\bf r},t) =
\left( \begin{array}{c}
\Psi_1({\bf r},t) \\
\Psi_2({\bf r},t) \\
\Psi_3({\bf r},t) \\
\Psi_4({\bf r},t) \end{array} \right) =
\left( \begin{array}{c}
\Psi_{\rm u} \\
\Psi_{\rm l} \end{array} \right), \quad \mbox{with} \quad
\Psi_{\rm u} =
\left( \begin{array}{c}
\Psi_{\rm 1} \\
\Psi_{\rm 2} \end{array} \right), \quad
\Psi_{\rm l} =
\left( \begin{array}{c}
\Psi_{\rm 3} \\
\Psi_{\rm 4} \end{array} \right).
\label{2.80}\eeq

To derive the differential law of probability conservation, we consider the
Dirac equation \req{2.74}, written in the form
\begin{subequations} \label{2.81}
\beq
{\rm i} \hbar \, \frac{\partial}{\partial t} \Psi = c \left( -{\rm i}
\hbar \nablab +
\frac{e}{c} {\bf A} \right) \dotprod \widetilde\alphab\Psi + \me c^2
\widetilde\beta \Psi - e\varphi \Psi,
\label{2.81a}\eeq
and its Hermitian conjugate equation (that is, the equation obtained by
taking the complex conjugate and replacing each matrix with its transpose)
\beq
- {\rm i} \hbar \, \frac{\partial}{\partial t} \Psi^\dagger = c \left(
+{\rm i} \hbar \nablab +
\frac{e}{c} {\bf A} \right) \dotprod (\Psi^\dagger \widetilde\alphab) + \me c^2
\Psi^\dagger \widetilde\beta - e\varphi \Psi^\dagger \, .
\label{2.81b}\eeq
\end{subequations}
Note that we are assuming that the electromagnetic potentials are real.
Multiplying Eq.\ \req{2.81a} by $\Psi^\dagger$ from the left, and
Eq.\ \req{2.81b} by $\Psi$ from the right and subtracting, we find
\beq
\frac{\partial}{\partial t} (\Psi^\dagger \Psi) =
- \nablab \dotprod (\Psi^\dagger c\widetilde\alphab \Psi).
\label{2.82}\eeq
The definite-positive quantity
\index{Dirac wave equation!probability density}
\beq
\rho({\bf r},t) \equiv \Psi^\dagger({\bf r},t)
\Psi({\bf r},t) =  \sum_{k=1}^{4} \Psi^\ast_k({\bf r},t)
\Psi_k({\bf r},t)
\label{2.83}\eeq
is the probability density, and
\index{Dirac wave equation!probability current density}
\beq
{\bf j}({\bf r},t) \equiv \Psi^\dagger({\bf r},t) c \, \widetilde \alphab \,
\Psi({\bf r},t),
\label{2.84}\eeq
is the probability current vector. With these definitions, \req{2.82}
becomes the probability-conservation equation
\beq
\frac{\partial}{\partial t} \rho({\bf r},t) + \nablab \dotprod {\bf
j}({\bf r},t) = 0.
\label{2.85}\eeq

If the electromagnetic potentials are time-independent, we can consider
stationary states
\index{Dirac wave equation!stationary states}
\beq
\Psi({\bf r},t) = \psi ({\bf r}) \exp(-{\rm i} {\cal W} t/\hbar) =
\left( \begin{array}{c} \psi_{\rm u} ({\bf r}) \\ [1mm]
\psi_{\rm l} ({\bf r}) \end{array} \right)
\exp(-{\rm i} {\cal W} t/\hbar),
\label{2.86}\eeq
where the bi-spinor $\psi({\bf r})$ is a solution of the
{\it time-independent Dirac equation}
\beq
\left[ c \widetilde\alphab\dotprod \left(
\breve{\bf p} + \frac{e}{c} {\bf A} \right) +
\widetilde\beta \me c^2 - e\varphi \right]
\psi({\bf r}) = {\cal W} \psi({\bf r}).
\label{2.87}\eeq
Notice that the eigenvalue ${\cal W}$ is the total energy of the
electron, inclusive of the rest energy $\me c^2$.  In the spinor
representation, this equation takes the form
\beq
\left( \begin{array}{cc} \me c^2-e \varphi & \displaystyle{c
\sigmab\dotprod \left( \breve{\bf p} + \frac{e}{c} {\bf A} \right)} \\
\displaystyle{c \sigmab\dotprod \left( \breve{\bf p} + \frac{e}{c} {\bf A}
\right)} & -\me c^2-e \varphi \end{array} \right)
\left( \begin{array}{c} \psi_{\rm u} ({\bf r}) \\ [4mm]
\psi_{\rm l} ({\bf r}) \end{array} \right) = {\cal W}
\left( \begin{array}{c} \psi_{\rm u} ({\bf r}) \\ [4mm]
\psi_{\rm l} ({\bf r}) \end{array} \right).
\label{2.88}\eeq

A peculiarity of relativistic wave equations is that they have positive
and negative energy eigenvalues. Thus, the Dirac equation for free
particles (\ie, with $\varphi = 0$ and ${\bf A} = {\bf 0}$) has
eigenstates with energies ${\cal W} \ge \me c^2$ and ${\cal W} \le -\me
c^2$. Wave functions with positive energies ${\cal W}=E + \me c^2$
represent states of ordinary electrons (negatrons) with kinetic energy
$E$. According to Dirac's hole theory \citep[see, \eg,][]{Sakurai1967},
states with negative energies ${\cal W} \le - \me c^2$ are not observable
and filled with electrons (Dirac sea); radiative transitions of
electrons in positive-energy states to these negative-energy states are
forbidden by the Pauli exclusion principle. However, an electron in a
state of negative energy $-|{\cal W}|$ can be excited to positive-energy
states (\eg, by absorption of a high-energy photon) leaving a hole in
the Dirac sea.  This hole is observable as a particle of mass $\me$,
charge $+e$, and energy $|{\cal W}|$; this particle is the positron, the
antiparticle of the electron. The operation of charge conjugation
\citep[see, \eg,][]{Rose1961} transforms electron states of energy
${\cal W}$ into positron states of energy $-{\cal W}$. In the case of
positive-energy states, the operator
\beq
K_{\rm D} =  c \widetilde\alphab \dotprod
\breve{\bf p} + \left( \widetilde\beta -1 \right) \me c^2,
\label{2.89}\eeq
which is obtained by subtracting the rest energy $\me c^2$ from the
Hamiltonian of a free particle, can be identified as the kinetic energy
operator of the Dirac theory.


\subsection{Central potentials. Radial wave equations \label{sec2.2.1}}
\index{Dirac wave equation!central potentials}

When ${\bf A}=0$ and the scalar potential is independent of time and
central, the electron is subject to an electric force corresponding to
the potential $V(r) = -e \varphi(r)$, and we can consider {\it spherical
waves}, which are solutions of the time-independent Dirac equation with
well-defined angular momentum.

The Dirac spin operator is \citep{Rose1961, Sakurai1967, Strange1998}
\beq
{\bf S} =\frac{1}{2}
\left( \begin{array}{cc}
\sigmab & 0 \\
0 & \sigmab \\ \end{array} \right) =
\left( \begin{array}{cc}
{\bf S}_{\rm P} & 0 \\
0 & {\bf S}_{\rm P} \\ \end{array} \right),
\label{2.90} \eeq
where ${\bf S}_{\rm P} = \1o2 \sigmab$ is Pauli's non-relativistic spin
operator. Note that all angular momenta are expressed in units of
$\hbar$. The total angular momentum operator, ${\bf J}\equiv {\bf
L}+{\bf S}$, commutes with the Dirac Hamiltonian. It is convenient
to introduce the operator
\beq
{\cal K} \equiv - \widetilde{\beta} \left( 2 {\bf S} \cdot {\bf L} + 1
\right) = - \widetilde{\beta} \left( J^2 - L^2 -S^2 +1 \right),
\label{2.91}\eeq
which commutes with ${\cal H}_{\rm D}$, $J^2$ and $J_z$. Therefore, we
can construct simultaneous eigenfunctions of ${\cal H}_{\rm D}$, $J^2$,
$J_z$ and ${\cal K}$, with respective eigenvalues $W$, $j(j+1)$, $m$ and
$\kappa$. These solutions of the Dirac equation are the {\it
spherical waves} or {\it central-field orbitals}
\index{central-field orbitals}
\beq
\psi_{{\cal W}\kappa m} ({\bf r}) = \frac{1}{r}
\left( \begin{array}{c}
P_{{\cal W}\kappa}(r) \, \Omega_{\kappa, m} (\hat{\bf r}) \\ [2mm]
{\rm i} Q_{{\cal W}\kappa}(r) \, \Omega_{-\kappa, m} (\hat{\bf r})
\end{array} \right),
\label{2.92}\eeq
where $P_{{\cal W}\kappa} (r)$ and $Q_{{\cal W}\kappa}(r)$ are the upper- and
lower-component radial functions, and the spherical spinors
\beq
\Omega_{\kappa m}(\hat{\bf r}) \equiv
\Omega_{j m}^{\ell}(\hat{\bf r})
= \sum_{\mu=\pm 1/2}
\langle \ell, 1/2, m-\mu, \mu | j, m \rangle \;
Y_{\ell,m-\mu} (\hat{\bf r}) \, \chi_\mu
\label{2.93}\eeq
are simultaneous eigenfunctions of the operators $L^2$, $S_{\rm P}^2$,
$J^2$ and $J_{z}$ (where ${\bf J}={\bf L} + {\bf
S}_{\rm P}$ is the non-relativistic total angular momentum) with
eigenvalues $\ell(\ell+1)$, $3/4$, $j(j+1)$ and $m$, respectively.  We
recall that the allowed values of the quantum number $m$ are $-j$,
$-j+1$, \ldots, $j-1$, $j$. The quantities
$\langle\ell,1/2,m-\mu,\mu|j,m\rangle$ are Clebsch--Gordan coefficients,
and the spinors $\chi_\mu$ are eigenstates of $S_{\rm P}^2$ and $S_{{\rm
P} z}$ with eigenvalues $3/4$ and $\mu=\pm 1/2$,
\beq
\chi_{+1/2} = \left( \begin{array}{c} 1 \\ 0 \end{array} \right)
\qquad \mbox{and} \qquad
\chi_{-1/2} = \left( \begin{array}{c} 0 \\ 1 \end{array} \right).
\label{2.94}\eeq

The quantum number $\kappa$ is conventionally used as short-hand
notation for $j$ and $\ell$.  The values of $\kappa$, $j$ and $\ell $
are related by
\begin{subequations}
\label{2.95}
\beq
\kappa = (\ell -j)\, (2j+1) = -(j+\1o2)\, \sigma,
\label{2.95a}\eeq
where $\sigma \equiv -{\rm sign}(\kappa) = - \kappa / |\kappa|$. That is
\beq
\kappa = \left\{
\begin{array}{ll}
-(j+\1o2) \; \; & \mbox{if} \; \; j=\ell+\1o2\, , \\ [2mm]
\; \; \; \; \, j+\1o2 & \mbox{if} \; \; j=\ell-\1o2 \, ,
\end{array} \right.
\qquad \mbox{or} \qquad
\kappa = \left\{
\begin{array}{ll}
-\ell-1 \; \; & \mbox{if} \; \; j=\ell+\1o2\, , \\ [2mm]
\ell  & \mbox{if} \; \; j=\ell-\1o2 \, ,
\end{array} \right.
\label{2.95b}\eeq
and
\beq
j = | \kappa | - \1o2 = \ell + \1o2 \, \sigma \, ,
\label{2.95c}\eeq
\beq
\ell = | \kappa | - \1o2 (1+\sigma) = j - \1o2\, \sigma = \left\{
\begin{array}{ll}
| \kappa | - 1 \; \; & \mbox{if} \; \; \kappa < 0, \\ [2mm]
| \kappa | & \mbox{if} \; \; \kappa > 0\, .
\end{array} \right.
\label{2.95d}\eeq
Notice that
\beq
\kappa(\kappa+1)=\ell(\ell+1).
\label{2.95e}\eeq
An alternative convention is used in x-ray physics, where the angular
momentum quantum numbers are specified by a single numerical label, $i$,
defined by
\beq
i \equiv 2 |\kappa| - \1o2 \left( 1 +\sigma \right) = \left\{
\begin{array}{ll}
2 | \kappa | - 1 \; \; & \mbox{if} \; \; \kappa < 0, \\ [2mm]
2 | \kappa | & \mbox{if} \; \; \kappa > 0\, ,
\end{array} \right.
\label{2.95f}\eeq
which takes positive integer values. The inverse relation is
\beq
\kappa = \1o2 \left[ i - (2i+1)\, \mbox{mod}(i,2) \right] =
\left\{
\begin{array}{ll}
\1o2 \, i & \mbox{if $i$ is even,} \\ [2mm]
- \1o2 (i+1) \; \; & \mbox{if $i$ is odd,}
\end{array} \right.
\label{2.95g}\eeq
\end{subequations}
where ${\rm mod}(i,2) = 0$ if $i$ is even and $=1$ if $i$ is odd.
Table \ref{tab2.1} shows the correspondence between these alternative
representations of angular momentum quantum numbers.

\begin{table}[hbt]
\begin{center}
\caption{
Correspondence between various conventions to designate
angular momentum eigenvalues; ($\ell$, $j$), $\kappa$, spectroscopic
notation (s.n.), and x-ray notation.
\label{tab2.1}}
\vskip 4mm
\begin{tabular}{|r|c|c|c|c|c|c|c|c|c|c|} \hline
\rule{0mm}{4mm}$\ell$ & $0$ & $1$ & $1$ & $2$ & $2$ & $3$ & $3$ & $4$ &
$4$
&
$5$ \\ [1mm]
$j$
& $\frac{1}{2}$
& $\frac{1}{2}$
& $\frac{3}{2}$
& $\frac{3}{2}$
& $\frac{5}{2}$
& $\frac{5}{2}$
& $\frac{7}{2}$
& $\frac{7}{2}$
& $\frac{9}{2}$
& $\frac{9}{2}$
\\ [1mm] \hline \rule{0mm}{4mm}
$\kappa$ & $-1$ & $+1$ & $-2$ & $+2$ &
$-3$ & $+3$ & $-4$ & $+4$ & $-5$ & $+5$ \\ \hline
\rule{0mm}{4mm} s.n. &
 $s_{1/2}$&
 $p_{1/2}$&
 $p_{3/2}$&
 $d_{3/2}$&
 $d_{5/2}$&
 $f_{5/2}$&
 $f_{7/2}$&
 $g_{7/2}$&
 $g_{9/2}$&
 $h_{9/2}$
\\ [1mm]
\rule{0mm}{4mm}&
$s$      &
$\bar{p}$&
$p$      &
$\bar{d}$&
$d$      &
$\bar{f}$&
$f$      &
$\bar{g}$&
$g$      &
$\bar{h}$
\\ [1mm] \hline \rule{0mm}{4mm}
x ray, $i$ &
1 &
2 &
3 &
4 &
5 &
6 &
7 &
8 &
9 &
10 \\ \hline
\end{tabular}
\end{center} \end{table}

Considering the relations \req{2.95c}--\req{2.95d}, and the analytical
expressions of the Clebsch--Gordan coefficients \citep[see,
\eg,][]{Edmonds1960}, the spherical spinors \req{2.93} can be expressed
as
\begin{subequations}
\label{2.96}
\beqa
\Omega_{j=\ell \pm 1/2,m}^\ell (\hat{\bf r},\sigma) &=&
\pm \sqrt{\frac{\ell \pm m + 1/2}{2\ell+1}} \, Y_{\ell, m-1/2}(\hat{\bf
r}) \, \chi_{+1/2} (\sigma)
\nonumber \\ [2mm]
&&
+ \sqrt{\frac{\ell \mp m + 1/2}{2\ell+1}} \, Y_{\ell, m+1/2}(\hat{\bf
r}) \, \chi_{-1/2} (\sigma),
\label{2.96a}\eeqa
or, more compactly,
\beq
\Omega_{\ell\pm1/2,m}^{\ell}(\hat{\bf r})=
\frac{1}{\sqrt{2\ell+1}}
\left( \begin{array}{c}
\pm \sqrt{\ell \pm m + \1o2} \; Y_{\ell,m-1/2}
(\hat{\bf r}) \\ [2mm]
\sqrt{\ell \mp m + \1o2} \; Y_{\ell,m+1/2}
(\hat{\bf r})
\end{array} \right)\, .
\label{2.96b}\eeq
\end{subequations}

\index{radial Dirac wave equation}
It should be noted that the relativistic spherical wave \req{2.92} is not
an eigenfunction of $L^2$; the index $\ell$ used in the spectroscopic
notation is the eigenvalue of the upper-component spinor and serves to
indicate the parity of $\psi_{{\cal W}\kappa m}({\bf r})$. The radial
functions $P_{{\cal W}\kappa}(r)$ and $Q_{{\cal W}\kappa}(r)$ satisfy
the coupled equations \citep{Rose1961}
\beqa
\frac{\d P_{{\cal W}\kappa}}{\d r} & =
& -\, \frac{\kappa}{r} \,P_{W\kappa}
+ \frac{{\cal W}+\me c^2-V}{c\hbar}
\, Q_{{\cal W}\kappa},
\nonumber \\ [2mm]
\frac{\d Q_{{\cal W}\kappa}}{\d r} & =
& \frac{-{\cal W}+\me c^2 +V}{c\hbar} \, P_{{\cal W}\kappa}
+ \frac{\kappa}{r}\, Q_{W\kappa},
\label{2.97}\eeqa
which will be referred to as the {\it radial Dirac equations}.

In the case of a central potential $V(r)$, the radial functions of orbitals
with positive energies ${\cal W}=E+\me c^2$ are given by the radial
equations \req{2.97}. It is convenient to express these equations in terms
of the reduced eigenvalue $E\equiv {\cal W}-\me c^2$ ($W>0$)
\beqa
\frac{\d P_{E\kappa}}{\d r} &=&
- \frac{\kappa}{r} \,P_{E\kappa}
+ \frac{E+2 \me c^2-V}{c\hbar} \, Q_{E\kappa},
\nonumber \\ [2mm]
\frac{\d Q_{E\kappa}}{\d r} &=&
\frac{-E +V}{c\hbar} \, P_{E\kappa}
+ \frac{\kappa}{r}\, Q_{E\kappa},
\label{2.98}\eeqa
where $P_{E\kappa} \equiv P_{{\cal W}\kappa}$ and $Q_{E\kappa} \equiv
Q_{{\cal W}\kappa}$. In the non-relativistic limit ($c\rightarrow
\infty$), the reduced eigenvalues $E$ tend to the eigenvalues of the
radial Schr\"{o}dinger equation \req{2.41} (see below). As in the
non-relativistic theory, when $V(r)$ takes negative values in a certain
region, bound states may exist for a discrete set of negative reduced
eigenvalues ($E<0$). Note that, for attractive potentials that are
strong enough, we may have bound states of ordinary electrons with $E <
-\me c^2$ and ${\cal W}=E+\me c^2 < 0$, that is, deeply bound
positive levels may have negative total energies\footnote{The
feature that characterizes the positive-energy states is the
following. Consider the family of potentials $aV(r)$, where $a$ is a
positive constant. The states of a particle in these potentials vary
continuously with $a$, and reduce to the states of a free particle when
$a \rightarrow 0$. The positive-energy states of the potential
$V(r)$ ($a=1$) are those which, in the limit $a \rightarrow 0$ become
positive-energy states. Similar considerations apply to
negative-energy states.}.

Let us now consider electron states with negative total energy,
${\cal W}=-E-\me c^2$, of a central potential $V(r)=-e \varphi(r)$.
Their radial functions can be calculated by solving the same radial
Dirac equations \req{2.97} as for positive-energy states, but for
the potential $-V(r)$. This is readily seen by introducing the
quantities
\beq
\overline{{\cal W}} \equiv - {\cal W} = E + \me c^2, \qquad
\bar{\kappa} \equiv - \kappa, \qquad
\begin{array}{c}
\overline{P}_{{\cal W}\bar{\kappa}} \equiv Q_{{\cal W}\kappa}, \\ [2mm]
\overline{Q}_{{\cal W}\bar{\kappa}} \equiv P_{{\cal W}\kappa},
\end {array} \qquad
\label{2.99}\eeq
and expressing the radial Dirac equations \req{2.97} as
\beqa
\frac{\d \overline{Q}_{{\cal W}\bar{\kappa}}}{\d r} &=&
+ \frac{\bar{\kappa}}{r}
\,\overline{Q}_{{\cal W}\bar{\kappa}}
+ \frac{- \overline{{\cal W}}+\me c^2-V}{c\hbar} \,
\overline{P}_{{\cal W}\bar{\kappa}},
\nonumber \\ [2mm]
\frac{\d \overline{P}_{{\cal W}\bar{\kappa}}}{\d r} &=&
\frac{\overline{{\cal W}}+\me c^2 +V}{c\hbar} \,
\overline{Q}_{{\cal W}\bar{\kappa}}
- \frac{\bar{\kappa}}{r}\,
\overline{P}_{{\cal W}\bar{\kappa}}.
\nonumber \eeqa
Taking $E$ as the energy variable, and setting
$\overline{P}_{E\kappa} \equiv \overline{P}_{{\cal W}\kappa}$ and
$\overline{Q}_{E\kappa} \equiv \overline{Q}_{{\cal W}\kappa}$, we can
recast these equalities as
\beqa
\frac{\d \overline{P}_{E \bar{\kappa}}}{\d r} &=&
- \frac{\bar{\kappa}}{r}\, \overline{P}_{E\bar{\kappa}}
+ \frac{E + 2 \me c^2 +V}{c\hbar} \,
\overline{Q}_{E\bar{\kappa}}\, ,
\nonumber \\ [2mm]
\frac{\d \overline{Q}_{E \bar{\kappa}}}{\d r} &=&
\frac{- E -V}{c\hbar} \, \overline{P}_{E\bar{\kappa}}
+ \frac{\bar{\kappa}}{r} \,\overline{Q}_{E\bar{\kappa}} \, .
\label{2.100}\eeqa
Indeed, these equations differ from Eqs.\ \req{2.97} only in the sign of
the potential. Hence, the negative-energy orbitals (with ${\cal W}
=-E-\me c^2$) are of the form
\beq
\psi_{{\cal W} \kappa m} ({\bf r}) = \frac{1}{r}
\left( \begin{array}{c}
\overline{Q}_{E \bar{\kappa}}(r) \, \Omega_{\kappa, m}(\hat{\bf r}) \\ [2mm]
{\rm i}
\overline{P}_{E \bar{\kappa}}(r) \, \Omega_{-\kappa, m}(\hat{\bf r})
\end{array} \right),
\label{2.101}\eeq
where the radial functions $\overline{P}_{E \bar{\kappa}}(r)$ and
$\overline{Q}_{E\bar{\kappa}}(r)$ satisfy the radial equations
\req{2.100}. The wave function \req{2.101} represents the
negative-energy state that, when emptied, is observable as a
positron with reduced energy $E$, angular momentum $j=|\kappa|-1/2$ and
magnetic quantum number $-m$. Bound negative-energy states may
exist whenever the potential $-V(r)$ is negative in a sufficiently wide
region.

The operation of charge conjugation \citep{Rose1961} transforms the
negative-energy spherical waves of electrons, Eq.\ \req{2.101}, into the
positive-energy spherical waves of positrons. The latter have the
generic form \req{2.92}, with the radial functions satisfying the radial
Dirac equations \req{2.98} for the potential $V(r)=e \varphi(r)$. In
practical calculations we shall encounter only positive-energy states of
either electrons or positrons.

As in the case of the Schr\"{o}dinger equation, for the sake of
simplicity and numerical convenience, we restrict the study to central
potentials $V(r)$ such that $r V(r)$ is finite for all $r$ (\ie,
finite-range potentials, Coulomb potentials, and modified Coulomb
potentials). Since the spatial wave function $\psi_{E\kappa m}({\bf r})$
must be square summable within a sphere of finite radius,
the radial wave functions $P_{E\kappa}(r)$ and $Q_{E\kappa}(r)$ vanish
at $r=0$.

\index{radial Dirac wave equation!bound states}
When $V(r)$ takes negative values in a certain region, bound states of
Eq.\ \req{2.98} may
exist for a discrete set of negative eigenvalues $E$. Because the wave
functions of bound states have finite norm, the corresponding  radial
wave functions tend to zero when $r$ goes to infinity. Discrete (bound)
energy levels (with $E < 0$), when they exist, are identified by the
principal quantum number $n$ and the quantum number $\kappa$. For a
given $n$, the allowed values of $\kappa$ are $-n, \ldots, -1, 1, 2,
  \ldots, n-1$. As in the non-relativistic theory,
the radial quantum number $n_{\rm r}=n-(\ell +1)$ gives the number of nodes of
$P_{n\kappa}(r)$. Each bound level is, at least, $2j+1$ times degenerate
(the eigenstates of a pure Coulomb potential are degenerate in $\ell$,
that is, states with the same $n$ and $j$ and different values of $\ell$
and $m$ have the same energy). Adequate normalization for bound states is
\beq
\int \psi_{n \kappa m}^\dagger({\bf r}) \,
\psi_{n \kappa m}({\bf r}) \; \d{\bf r} =
\int_{0}^{\infty} \left[ P_{n\kappa}^2(r) + Q_{n\kappa}^2(r) \right] \d r = 1.
\label{2.102}\eeq

\index{radial Dirac wave equation!free states}
The asymptotic behavior of the radial functions of free states (with
$E>0$, is determined by the Coulomb tail of the potential, characterized
by the parameter
$$
Z_{\infty} \equiv \frac{1}{e^2} \lim_{r \rightarrow \infty} r \, V(r).
\eqno{\req{2.44}}$$
Because the potential tends to zero at large radii, $E$ represents the
kinetic energy of the particle far from the center of force and we may
introduce the corresponding linear momentum and velocity, given by
\beq
p = \frac{1}{c} \sqrt{E(E+2\me c^2} \qquad \mbox{and} \qquad
v= c\beta = c \, \frac{\sqrt{E(E+2\me c^2})}{E+\me c^2},
\label{2.103}\eeq
respectively.
Dirac free states will be normalized in such a way that the
radial function $P_{E\kappa}(r)$ asymptotically oscillates with unit amplitude
[cf.\ Eq.\ \req{2.46}],
\beq
P_{E\kappa}(r)
\begin{array}[t]{c}
\sim \\ [-3mm] \scriptstyle{ r \rightarrow \infty}
\end{array}
\sin \left( kr - \ell \frac{\pi}{2} - \eta \ln (2kr) + d_\kappa \right),
\label{2.104}\eeq
where
\beq
k \equiv \frac{p}{\hbar} = \frac{\sqrt{E(E+2\me c^2)}}{c\hbar}
\label{2.105}\eeq
is the relativistic wave number,
\beq
\eta = \frac{Z_\infty e^2}{\hbar v} = \frac{Z_\infty\alpha}{\beta}
\label{2.106}\eeq
is the {\it Sommerfeld parameter} ($=0$ for finite-range potentials), and
$d_\kappa$ is the phase shift. In the case of modified Coulomb
potentials, the phase shifts can be expressed as in the non-relativistic
theory, \ie,
\beq
d_\kappa = \delta_\kappa + \Delta_\kappa,
\label{2.107}\eeq
where $\delta_\kappa$ is the inner phase-shift, attributable to the
finite-range component of the potential, and $\Delta_\kappa$ is the
Dirac phase shift of the Coulomb potential tail \citep{Salvat1995}. The
quantity $\alpha \equiv e^2/(\hbar c) \simeq 1/137$ is the
fine-structure constant.
\index{radial Dirac wave equation!free states!phase shifts}
\index{Sommerfeld parameter}

The radial functions of free spherical waves normalized in the form
\req{2.104} satisfy the orthogonality condition
\citep[see, \eg,][]{Salvat1995}
\beqa
\lefteqn{ \int_0^\infty \left( P_{E'\kappa'} P_{E\kappa} + Q_{E'\kappa'}
Q_{E\kappa} \right) \d r =} \nonumber \\ [2mm]
& = & \frac{\pi E}{k} \;
\delta( E' - E ) \, \delta_{\kappa',\kappa}
=\frac{\pi (E+\me c^2)}{E+2\me c^2} \;
\delta( k' - k ) \, \delta_{\kappa',\kappa},
\label{2.108}\eeqa
where $k'$ is the wave number of particles with energy $E'$. The set of
spherical waves (with positive {\it and} negative energies ${\cal W}=E+\me
c^2$) constitute a complete orthogonal basis of the vector space of
states of an electron.

Except for Coulomb potentials, Dirac radial wave functions have to be
calculated numerically by solving the equations \req{2.98}. The Fortran
subroutine package {\sc radial} \citep{SalvatFernandezVarea2019}
provides highly accurate solutions of the radial equations (that is,
radial functions, energies of bound states, and phase shifts of free
states) for potentials such that $rV(r)$ is finite everywhere.

\index{radial Dirac wave equation!non-relativistic limit}
It is worth noting that in the limit $E-V \ll 2\me c^2$, the radial Eqs.\
\req{2.98} for positive-energy states reduce to
\begin{subequations}
\label{2.109}
\beq
Q_{E\kappa} = \frac{c\hbar}{2\me c^2}
\left( \frac{\kappa}{r} P_{E\kappa} + \frac{\d P_{E\kappa}}{\d r} \right)
\label{2.109a}\eeq
and
\beq
\frac{\d^2 P_{E\kappa}}{\d r^2} = \left[ \frac{\kappa(\kappa+1)}{r^2} -
\frac{2\me }{\hbar^2} (E-V) \right] P_{E\kappa}.
\label{2.109b}\eeq \end{subequations}
Equation \req{2.109a} shows that the lower (``small'') component radial
function $Q_{E\kappa}(r)$ vanishes in the non-relativistic limit
($c\rightarrow \infty$). Using the fact that
$\kappa(\kappa+1)=\ell(\ell+1)$, Eq.\ \req{2.109b} is seen to coincide
with the radial Schr\"{o}dinger equation \req{2.41}, and therefore, in
the non-relativistic limit the upper (``large'') component radial
function $P_{E\kappa}(r)$ reduces to the Schr\"{o}dinger radial
function $P_{E\ell}(r)$.


\subsection{Dirac distorted plane waves \label{sec2.2.2}}
\index{Dirac wave equation!distorted plane waves}
\index{distorted plane waves!Dirac wave equation}

We now consider the positive-energy Dirac DPWs for a central
potential $V(r)$ {\it of finite range}. Notice that modified Coulomb fields, such
that $\lim_{r\rightarrow \infty} rV(r) \ne 0$, are not included in the
present analysis. The case of pure Coulomb fields is studied, \eg, in
\citeauthor{Rose1961}'s (\citeyear{Rose1961}) book. The Dirac DPWs for
an electron with momentum $\hbar {\bf k}$, and kinetic energy
\beq
E=\sqrt{(c\hbar k)^2 + (\me c^2)^2} - \me c^2,
\label{2.110}\eeq
in a pure spin state represented by the Pauli spinor $\chi_\mu$
($\mu=\pm 1/2$) are solutions of the wave equation
\beq
\left[ -{\rm i} \, c \hbar \, \widetilde \alphab \cdot \nablab +
(\widetilde \beta -1) \me c^2 + V(r) \right]
\psi_{{\bf k}\mu}^{(\pm )}({\bf r})
= E \, \psi_{{\bf k}\mu}^{(\pm )}({\bf r})
\label{2.111}\eeq
with the asymptotic behavior
\beq
\psi_{{\bf k}\mu}^{(\pm)}({\bf r})
\begin{array}[t]{c} \sim \\[-3mm] \scriptstyle{ r
\rightarrow \infty} \end{array}
\phi_{{\bf k}\mu}({\bf r}) + \psi_{\rm sc}^{(\pm )}({\bf r}),
\label{2.112}\eeq
where
\beq
\phi_{{\bf k}\mu}({\bf r}) =
\frac{1}{(2\pi)^{3/2}} \, \exp({\rm i} {\bf k}\cdot {\bf r} ) \,
\sqrt{\frac{E+2 \me c^2}{2E+2\me c^2}}
\left( \begin{array}{c} I_2 \\ [2mm]
\displaystyle{\sqrt{\frac{E}{E+2\me c^2}} \,
\sigmab\dotprod\hat{\bf k} }
\end{array} \right) \,
\chi_\mu
\label{2.113}\eeq
is a positive-energy plane wave, and
\beqa
\psi_{\rm sc}^{(\pm )} ({\bf r}) &=&
\frac{1}{(2\pi)^{3/2}} \, \frac{\exp(\pm{\rm i}kr)}{r} \,
\sqrt{\frac{E+2 \me c^2}{2E+2\me c^2}}
\nonumber \\ [2mm]
&& \mbox{} \times
\left( \begin{array}{c} I_2 \\ [2mm]
\displaystyle{\pm \sqrt{\frac{E}{E+2\me c^2}} \,
\sigmab\dotprod\hat{\bf r} }
\end{array} \right) \, {\cal F}^{\rm (\pm)}(\hat{\bf r}, \hat{\bf k}) \,
\chi_\mu
\label{2.114}\eeqa
is an outgoing ($+$) or incoming ($-$) spherical wave.  The
factor ${\cal F}^{\rm (\pm)}(\hat{\bf r}, \hat{\bf k})$, a
2$\times$2 matrix independent of $r$, is the {\it scattering-amplitude
matrix}.

\index{distorted plane waves!Dirac wave equation!partial-wave series}
The Dirac DPWs admit the following {\it partial-wave expansions}
\citep[see][p. 207]{Rose1961}
\beq
\psi_{{\bf k}\mu}^{(\pm)} ({\bf r}) = \frac{1}{k}
\, \sqrt{\frac{E+2 \me c^2}{\pi(E + \me c^2)}}
\sum_{\kappa,m} {\rm i}^\ell \, \exp \left( \pm {\rm i} \delta_{\kappa}
\right) \, \left\{ \left[ \Omega_{\kappa m} (\hat{\bf k})
\right]^\dagger \chi_{\mu} \right\} \psi_{E \kappa m} ({\bf r})\, ,
\label{2.115}\eeq
where $\delta_{\kappa}$ are the Dirac phase shifts, and
$\psi_{E \kappa m}({\bf r})$ are the spherical waves [see Eq.\
\req{2.92}]
\beq
\psi_{E \kappa m} ({\bf r}) = \frac{1}{r}
\left( \begin{array}{c}
P_{E \kappa}(r) \, \Omega_{\kappa, m}(\hat{\bf r}) \\ [2mm]
{\rm i} Q_{E \kappa}(r) \, \Omega_{-\kappa, m}(\hat{\bf r})
\end{array} \right).
\label{2.116}\eeq
With the adopted normalization for free spherical waves, Eq.\
\req{2.104}, the DPWs satisfy the orthogonality relation
\beq
\int
\left[ \psi_{{\bf k}'\mu'}^{(\pm)} ({\bf r}) \right]^\dagger
\psi_{{\bf k}\mu}^{(\pm)} ({\bf r}) \, \d {\bf r}
= \delta({\bf k}'-{\bf k}) \, \delta_{\mu'\mu}\, ,
\label{2.117}\eeq
which implies that the density of positive-energy Dirac DPW states per
unit volume in ${\bf k}$ space is equal to 2 (because of the spin
degeneracy). In the case of the null potential ($V=0$), the radial
functions of free states reduce to regular spherical Bessel functions
\citep{Rose1961} and the DPW becomes the Dirac plane wave $\phi_{{\bf k}
\mu} ({\bf r})$, Eq.\ \req{2.113}.

The DPWs $\psi_{{\bf k}\mu}^{(+)} ({\bf r})$ represent stationary
scattering states, and the scattering-ampli\-tude matrix ${\cal F}^{\rm
(+)}(\hat{\bf r}, \hat{\bf k})$ determines the scattering differential
cross section (see Section \ref{sec5.2}). It can be shown that,
similarly to the Schr\"{o}dinger theory, the DPW $\psi_{{\bf
k}\mu}^{(-)} ({\bf r})$ is obtained by applying the time-reversal
operator \citep{Rose1961, Sakurai1967} to the wave $\psi_{-{\bf k},-\mu}^{(+)}
({\bf r})$ with opposite wave vector and spin.

To simplify the formulas it is convenient to consider a reference frame
with the $z$ axis in the direction of the incident beam, so that ${\bf
k} = k \hat{\bf z}$, and express the scattering-amplitude matrix as a
function of the polar and azimuthal angles, $\theta$ and $\phi$, of the
direction $\hat{\bf r}$. From the expansions \req{2.115}, using the
large-$r$ form \req{2.104} of the radial functions and the expression
\req{2.96b} of the spherical spinors, it follows that
\begin{subequations}
\label{2.118}
\beq
{\cal F}^{\rm (+)}(\hat{\bf r}, \hat{\bf k}) = \left( \begin{array}{cc}
f(\theta) & - \, g(\theta) \, {\rm exp}(-{\rm i}\phi)\\ [2mm]
g(\theta) \, {\rm exp}({\rm i}\phi) & f(\theta) \end{array} \right),
\label{2.118a}\eeq
and
\beq
{\cal F}^{\rm (-)}(\hat{\bf r}, \hat{\bf k}) =
\left( \begin{array}{cc}
f^\ast(\pi-\theta) & g^\ast(\pi-\theta) \, {\rm exp}(-{\rm i}\phi) \\ [2mm]
- \, g^\ast(\pi-\theta) \, {\rm exp}({\rm i}\phi) &
f^\ast(\pi-\theta) \end{array} \right).
\label{2.118b}\eeq
\end{subequations}
The functions $f(\theta)$ and $g(\theta)$ are called the ``direct'' and
``splin-flip'' scattering amplitudes. They admit the following
partial-wave expansions,
\index{direct scattering amplitude}
\begin{subequations}
\label{2.119}
\beq
f(\theta) = \frac{1}{2 {\rm i} k} \sum_{\ell} \left\{
(\ell+1) \left[
{\rm exp}( 2 {\rm i} \delta_{\kappa=-\ell-1}) - 1 \rule{0mm}{4mm}\right]
+ \ell \left[ {\rm exp}(2 {\rm i} \delta_{\kappa=\ell}) - 1 \rule{0mm}{4mm}\right]
\right\} P_\ell(\cos\theta)
\label{2.119a}\eeq
and \index{spin-flip scattering amplitude}
\beq
g(\theta) = \frac{1}{2{\rm i}k}
\sum_{\ell} \left[
{\rm exp}(2 {\rm i} \delta_{\kappa=-\ell-1})
- {\rm exp}( 2 {\rm i} \delta_{\kappa=\ell})
\rule{0mm}{4mm}\right]
P_\ell^1(\cos\theta),
\label{2.119b}\eeq
\end{subequations}
where $P_\ell(\cos\theta)$ and $P_\ell^1(\cos\theta)$ are the Legendre
polynomials and the associated Legendre functions, respectively. Note
that the phase shifts $\delta_{\kappa=-\ell-1}$ and  $\delta_{
\kappa=\ell}$ correspond to $j=\ell+\1o2$ (spin up) and $j=\ell-\1o2$
(spin down), respectively --- see Table \ref{tab2.1}.

In the non-relativistic limit ($c \rightarrow \infty$) the values of the
two phase shifts of orbitals with opposite spins (\ie, with $j=\ell\pm
1/2$) become equal to that of the Schr\"{o}dinger phase shift
\beq
\delta_{\kappa=-\ell-1} = \delta_{\kappa=\ell} = \delta_{\ell}.
\label{2.120}\eeq
Hence, in that limit the direct scattering amplitude reduces to the
Schr\"{o}dinger scattering amplitude, Eq.\ \req{2.58a}, and the
spin-flip amplitude vanishes.


\section{The Pauli equation \label{sec2.3}}
\index{wave equation!Pauli} \index{Pauli wave equation}

The non-relativistic quantum theory of atomic structure requires
introducing the spin of the electron to assign to the ground state of
each atom a shell configuration consistent with the periodic table of
the elements. In the non-relativistic theory spin is introduced as an
internal state variable that has no classical analog and bestows the
electron with a magnetic dipole moment given by
\beq
\mcb = - g_{\rm e} \mu_{\rm B} {\bf S}_{\rm P}\, ,
\label{2.121}\eeq
where $g_{\rm e}=2.002319$ is the gyromagnetic factor, $\mu_{\rm B} = e
\hbar/(2\me c) = 9.274010 \times 10^{-21}$ erg/G is the Bohr magneton,
and ${\bf S}_{\rm P}=\1o2 \sigmab$ is the spin operator. The
classical interaction energy of this magnetic moment with an external
magnetic induction $\bcb$ equals $-\mcb \cdot \bcb$. The
Hamiltonian of an electron in an electromagnetic field represented by
the scalar potential $\varphi({\bf r},t)$ and the vector potential ${\bf
A}({\bf r},t)$ is obtained by adding to the non-relativistic Hamiltonian
\req{2.14} the spin interaction term
\beq
- g_{\rm e} \mu_{\rm B} \, {\bf S}_{\rm P} \dotprod (\nablab \vecprod
{\bf A}) \simeq  - \mu_{\rm B}
\sigmab \dotprod (\nablab \vecprod {\bf A}).
\label{2.122}\eeq
The result is the {\it Pauli} Hamiltonian
\beq
{\cal H}_{\rm P} = \frac{1}{2\me}
\left[ {\bf p} + \frac{e}{c} {\bf A}({\bf r},t)
\right]^2 I_2 - e \, \varphi({\bf r},t) \, I_2  - \mu_{\rm B}
\sigmab \dotprod (\nablab \vecprod {\bf A}),
\label{2.123}\eeq
where $I_2$ is the $2\times 2$ unit matrix. The electron wave functions
are spinors that satisfy the time-dependent Pauli equation
\beq
{\rm i} \hbar \, \frac{\partial}{\partial t} \Psi({\bf r},t)
= {\cal H}_{\rm P} \Psi({\bf r},t).
\label{2.124}\eeq
This equation can also be obtained formally as the non-relativistic
limit of the Dirac equation.

The probability density of the electron is
\beq
\rho({\bf r}) = \Psi^\dagger ({\bf r},t) \, \Psi ({\bf r},t) =
\left| \Psi ({\bf r},t) \right|^2.
\label{2.125}\eeq
From the time-dependent Pauli equation we can derive the following
continuity equation
\beq
\frac{\partial \rho}{\partial t} + \nablab \dotprod {\bf j} = 0,
\label{2.126}\eeq
where the vector
\beqa
{\bf j} ({\bf r},t) &\equiv& \frac{\hbar}{2{\rm i}\me}
\left[ \Psi^\dagger \left( \nablab \Psi \right) - \left( \nablab
\Psi^\dagger \right) \Psi \right]
+ \frac{e}{\me c} \, {\bf A} |\Psi|^2
\label{2.127}\eeqa
is the probability current.

In the case of an electrostatic field (\ie, with ${\bf A} = {\bf
0}$), the Pauli Hamiltonian reduces to the usual non-relativistic form
\beq
{\cal H}_{\rm P} = \left[ \frac{1}{2\me} \, {\bf p}^2
+ V({\bf r},t) \right] I_2,
\label{2.128}\eeq
where
\beq
V({\bf r},t) = - e \varphi({\bf r},t)
\label{2.129}\eeq
is the potential energy of the electron. If, in addition, the scalar
potential is constant with time we can consider stationary states with
time-independent spinor wave functions $\psi$ that are solutions of
the equation
\beq
\left[ \frac{1}{2\me} \, {\bf p}^2
+ V({\bf r}) \right] I_2
\psi = E \psi.
\label{2.130}\eeq
As the Hamiltonian does not depend on the spin, the wave functions can
be expressed as the direct product of a spatial wave function, which
satisfies the Schr\"{o}dinger equation, and a unit spinor,
\beq
\psi = \psi({\bf r}) \chi_{m_{\rm S}}.
\label{2.131}\eeq

To simplify the notation, we may consider the spinors as functions of a
discrete spin variable $\sigma$ (not to be confused with the Pauli
matrices) which can take the values $+1/2$ and $-1/2$. We define the
spin functions of the unit Pauli spinors by
\beq
\left( \begin{array}{c} 1 \\ 0 \end{array} \right) \leftrightarrow
\chi_{+1/2} (\sigma) = \delta_{+1/2, \sigma}, \qquad
\left( \begin{array}{c} 0 \\ 1 \end{array} \right) \leftrightarrow
\chi_{-1/2} (\sigma) = \delta_{-1/2, \sigma},
\label{2.132}\eeq
where $\delta_{\mu,\sigma}$ is the Kronecker delta. The spin
function $\xi(\sigma)$ associated to the spinor $\xi$ is then,
\beq
\xi =  \left( \begin{array}{c} \xi_1 \\ \xi_2 \end{array} \right)
\leftrightarrow
\xi(\sigma) = \xi_1 \chi_{+1/2} (\sigma) +  \xi_2 \chi_{-1/2} (\sigma).
\label{2.133}\eeq
Notice that the scalar product of two spinors,
\beq
\langle \xi \left| \tau \rangle \right. = \xi^\dagger \tau = \xi^*_1
\tau_1 + \xi^\ast_2 \tau_2,
\nonumber \eeq
can be expressed as
\beq
\langle \xi \left| \tau \rangle \right. =
\sum_\sigma \xi^\ast(\sigma) \tau(\sigma).
\label{2.134}\eeq
The solutions of the Pauli equation \req{2.130} can thus be regarded as
functions of the position coordinates ${\bf r}$ and the spin variable
$\sigma$,
\beq
\psi({\bf r},\sigma) = \psi({\bf r}) \, \chi_{m_{\rm S}}(\sigma) .
\label{2.135}\eeq
Evidently, the scalar product of two states takes the form
\beq
\langle \psi_1 \left| \psi_2 \rangle \right. = \sum_\sigma \int \d {\bf
r} \, \psi_1^\ast({\bf r},\sigma) \, \psi_2({\bf r},\sigma).
\label{2.136}\eeq
Defining the composite variable $x \equiv \{ {\bf r},\sigma \}$, we
shall abbreviate the notation by putting
\beq
\psi({\bf r},\sigma) = \psi(x)
\qquad \mbox{and} \qquad
\sum_\sigma \int \d {\bf r}\, f({\bf r},\sigma) = \int \d x \, f(x).
\label{2.137}\eeq
Thus, we will write
\beq
\left< \psi_1 \left| \psi_2 \right> \right. = \int \d x \,
\psi_1^\ast(x) \, \psi_2(x),
\label{2.138}\eeq
which is analogous to the scalar product of two Schr\"{o}dinger wave
functions.


\subsection{Central potentials \label{sec2.3.1}}
\index{Pauli wave equation!central potentials}

In the case of an electron in a central potential $V(r)$, the Pauli
Hamiltonian commutes with the orbital angular momentum ${\bf L}$ and
with the spin angular momentum ${\bf S}_{\rm P}$. Therefore we can
consider a basis of common eigenfunctions of the complete set of
commuting observables ${\cal H}_{\rm P}$, $L^2$, $L_z$, $S_{\rm P}^2$
and $S_{{\rm P}z}$  with corresponding eigenvalues $E$, $\ell(\ell+1)$,
$m_L$, 3/4, and $\mu$. These functions are the spherical waves or
central-field orbitals (in the uncoupled representation)
\index{central-field orbitals}
\beq
\psi_{E\ell m_{\rm L} m_{\rm S}} (x) = \frac{1}{r} \, P_{E\ell}(r) \,
Y_{\ell m_{\rm L}} (\hat{\bf r}) \, \chi_{m_{\rm S}}(\sigma),
\label{2.139}\eeq
where the reduced radial function $P_{E\ell}(r)$ satisfies the radial
Schr\"{o}dinger equation \req{2.41},
\index{radial Schr\"{o}dinger wave equation}
\beq
- \frac{\hbar^2}{2\me} \, \frac{\d^2 P_{E\ell}}{\d r^2}
+ \left[
\frac{\hbar^2}{2\me} \, \frac{\ell(\ell+1)}{r^2} + V(r) \right] P_{E\ell}
= E \, P_{E\ell}\, .
\label{2.140}\eeq

Alternatively, since the total angular momentum ${\bf J} = {\bf L} +
{\bf S}_{\rm P}$
also commutes with ${\cal H}_{\rm P}$, we may consider a basis of
common eigenfunctions of the commuting observables ${\cal H}_{\rm P}$,
$L^2$, $S_{\rm P}^2$, $J^2$, and $J_z$, with respective eigenvalues
$E$, $\ell(\ell+1)$, 3/4, $j(j+1)$, and $m$. These are the spherical
waves (in the coupled representation)
\beq
\psi_{E \ell j m } (x) = \frac{1}{r} \, P_{E\ell}(r) \,
\Omega_{jm}^\ell (\hat{\bf r},\sigma),
\label{2.141}\eeq
where $P_{E\ell}(r)$ is again a solution of Eq.\ \req{2.140}, and
$\Omega_{jm}^\ell (\hat{\bf r},\sigma)$ are the spherical spinors given
by Eq.\ \req{2.96},
\beqa
\Omega_{\ell\pm1/2,m}^{\ell}(\hat{\bf r},\sigma) &=&
\frac{1}{\sqrt{2\ell+1}} \left[
\pm \sqrt{\ell \pm m + \1o2} \; Y_{\ell,m-1/2}
(\hat{\bf r}) \, \chi_{+1/2}(\sigma) \right.
\nonumber \\ [2mm]
&& \mbox{} \left. +
\sqrt{\ell \mp m + \1o2} \; Y_{\ell,m+1/2}
(\hat{\bf r}) \,  \chi_{-1/2}(\sigma) \right] .
\label{2.142}\eeqa

As indicated above, the Pauli equation is obtained as the
non-relativistic limit ($c \rightarrow \infty$) of the Dirac equation.
It has the virtue of introducing the spinor form of the wave function,
but it keeps the spatial part of the Schr\"{o}dinger wave function
unaltered. To obtain first-order relativistic corrections to the wave
function, we should consider the so-called fine-structure Hamiltonian,
which is derived by expanding the Dirac Hamiltonian and retaining terms
of order $E/(\me c^2)$ \citep[see, \eg,][]{BransdenJoachain1983}.
One of these terms is the spin-orbit interaction,
\beq
V_{\rm SO} = - \frac{e \hbar}{2 \me^2 c^2} \, {\bf S}_{\rm P}
\dotprod [(\nablab \varphi) \vecprod {\bf p}],
\label{2.143}\eeq
which accounts for the interaction of the spin with the magnetic field
that is ``seen'' from the reference frame of the electron. If we include
the spin-orbit term, the Hamiltonian of an electron in a central
potential reads
\beq
{\cal H}_{\rm PSO} = \left[ \frac{1}{2\me} \, {\bf p}^2
+ V({\bf r},t) \right] I_2 +
\frac{\hbar^2}{2 \me^2 c^2} \, \frac{1}{r} \, \frac{\d
V}{\d r} \, {\bf L} \dotprod {\bf S}_{\rm P}.
\label{2.144}\eeq
The space of one-electron states then admits a basis of central-field
orbitals of the type (coupled representation)
\beq
\psi_{E \kappa m}(x) =
\frac{1}{r} \, P_{E\kappa}(r) \, \Omega_{\kappa m}(\hat{\bf r}, \sigma),
\label{2.145}\eeq
that are simultaneous eigenfunctions of the mutually commuting operators
${\cal H}_{\rm PSO}$, $L^2$, $S_{\rm P}^2$, $J^2$, and $J_z$, with
corresponding eigenvalues $E$, $\ell(\ell+1)$, 3/4, $j(j+1)$, and $m$.
Notice that the equality ${\bf J} = {\bf L} + {\bf S}_{\rm P}$ implies that
\beq
{\bf L} \dotprod {\bf S}_{\rm P} = \frac{1}{2}
\left[ J^2 - L^2 - S_{\rm P}^2 \right]
\label{2.146}\eeq
and
\beq
{\bf L} \dotprod {\bf S}_{\rm P} \; \Omega_{\kappa m}(\hat{\bf r}, \sigma)
= \frac{1}{2} \left[ j(j+1) - \ell(\ell+1) - \frac{3}{4} \right]
\Omega_{\kappa m}(\hat{\bf r}, \sigma).
\label{2.147}\eeq
The reduced radial functions $P_{E\kappa}(r)$ satisfy the radial
Schr\"{o}dinger equation \index{radial Schr\"{o}dinger wave equation}
\beq
- \frac{\hbar^2}{2\me } \, \frac{\d^2 P_{E\kappa}}{\d r^2} + \left[
\frac{\hbar^2}{2\me } \frac {\ell(\ell+1)}{r^2} + V_{\kappa}(r)
\right] P_{E\kappa}
= E P_{E\kappa}\, ,
\label{2.148}\eeq
with a potential
\beq
V_{\kappa}(r) =  V(r) +
\frac{\hbar^2}{2 \me^2 c^2} \, \frac{1}{r} \, \frac{\d V}{\d r} \,
 \frac{1}{2} \left[ j(j+1) -
\ell (\ell +1) - \frac{3}{4} \right]
\label{2.149}\eeq
that depends on the quantum number $\kappa$ or, equivalently, on $\ell$
and $j$.




\chapter{Atomic structure \label{chapt3}}



\vspace{-5mm}

A realistic description of the structure of atoms is essential to
undertake any theoretical study of interactions of charged particles
with matter. The earliest quantitative estimations of the atomic
electron distribution and the electrostatic potential of atoms were
based on the Thomas--Fermi model, in which the cloud of atomic electrons
is treated semi-classically as a degenerate electron gas obeying
Fermi--Dirac statistics. In this model, the atomic electrons, attracted
by the electrostatic force of the nucleus, occupy a finite volume
because of their mutual repulsion and the effect of the exclusion
principle. More elaborate quantum procedures, based on a self-consistent
solution of the wave equation for the atomic electrons were initiated by
Hartree, and extended by Fock and Slater to account for electron
exchange. Self-consistent atomic models are primarily based on the
approximation of independent electrons in a central potential, with
one-electron orbitals determined from a set of equations that are
derived from the variational principle. Details on self-consistent
atomic models can be found, \eg, in the books by \citet{Hartree1957} and
by \citet{CondonOdabasi1980}.

The state of the art in atomic structure calculations is the
relativistic multi-con\-fi\-gu\-ra\-tion Dirac--Fock method with the
Coulomb--Breit interaction and other quantum-electrodynamics perturbation
corrections \citep[see, \eg,][]{Grant1970,Desclaux1975}.
A basic feature of the underlying physical
picture is that electrons in different orbitals ``feel'' different,
non-local potentials, and this complicates calculations enormously. In
this Chapter we describe simpler atomic models based on the
independent-electron approximation, which facilitate calculations and
provide results that are reasonably accurate for most practical
purposes. After justifying the approximation of independent electrons in
a central potential, we present the non-relativistic statistical
Thomas--Fermi model, which is used to derive a simple analytical
approximation to the atomic electron density. We then describe the
self-consistent Dirac--Hartree--Fock--Slater (DHFS) method, which
incorporates the main relativistic effects. The numerical solution of
the DHFS equations gives fairly realistic atomic electron densities and
for inner bound shells (with ionization energies of the order of 200 eV
and higher) it yields eigenvalues that are closer to the experimental
ionization energies than the Dirac--Fock eigenvalues. Finally, we
describe simple analytical approximations to the Thomas--Fermi and DHFS
potentials that are useful to simplify calculations of elastic
collisions of charged particles with atoms.


\section{The Hamiltonian of atomic systems \label{sec3.1}}

\index{Hamiltonian of atomic systems}
The systems of interest consist of a nucleus with mass $M_0$ and
charge $Ze$, and a set of $N$ ``orbiting'' elementary particles with
respective masses $M_i$ and charges $Z_ie$ ($i=1, 2, \ldots, N$).
Examples of such systems are the isolated atoms or ions, in which the
orbiting particles are electrons with $M_i=\me$ and $Z_i=-1$, and the
exotic atoms, which are atoms where either the nucleus or one of the
atomic electrons is replaced with an elementary particle of another type
(\eg, a muon or a positron). We may also consider the collisions of
charged particles with atoms, where the system includes a free
projectile, with specific mass and charge.

The interactions between the particles of the system are assumed to be
purely electrostatic. Although particles with spin have a magnetic
moment, magnetic interactions are much weaker than the electric ones
and, typically, they are treated as perturbations, \eg, in the
description of the Zeeman effect \citep{BransdenJoachain1983}.
An exception is the spin-orbit interaction of electrons and positrons in
an electric field, which is responsible for the fine structure of atomic
energy spectra. This interaction is a relativistic kinematic effect
and, as such, it is fully accounted for by the Dirac kinetic energy
operator \req{2.89}. In non-relativistic quantum mechanics, the
spin-orbit interaction is frequently included in the Pauli Hamiltonian
(see Section \ref{sec2.3.1}).

\index{atomic number} \index{mass number of the nucleus}
The atomic nucleus is a system of $Z$ protons and $N_{\rm n}$ neutrons,
bound together by the nuclear forces. The number of protons in the
nucleus is usually referred to as the {\it atomic number}, the total
number of nucleons, $A \equiv Z+N_{\rm n}$, is called the {\it mass
number}. The mass
of the nucleus can be estimated from the following empirical formula
\citep{RoyerGautier2006} \index{nuclear mass formula}
\beq
M_0(Z,A) c^2 = Z m_{\rm p}c^2 + N_{\rm n} m_{\rm n} c^2 - B_{\rm nuc},
\label{3.1}\eeq
where $m_{\rm n}=1838.68\; \me $ and $m_{\rm p}=1836.15 \; \me$ are the
masses of the neutron and the proton, respectively, and $B_{\rm nuc}$ is
the nuclear binding energy, given by
\beqa
B_{\rm nuc} &=& (15.7335\times 10^{6} \; {\rm eV})
\left( 1 - 1.6949 \, I^2 \right) A
- (17.8048\times 10^{6} \; {\rm eV})
\left( 1 - 1.0884 \, I^2 \right) A^{2/3}
\nonumber \\ [2mm]
&& \mbox{}
- \frac{3}{5} \, \frac{Z^2 e^2}{R_0} + E_{\rm pair},
\label{3.2}\eeqa
where $I=(N_{\rm n}-Z)/A$ is the relative neutron excess, $R_0 = 1.2181 \,
A^{1/3} \; {\rm fm}$, and
\beq
E_{\rm pair} = \left\{
\begin{array}{ll}
-(11 \times 10^6 \; {\rm eV}) \, A^{-1/2} &
\mbox{for nuclei with odd $Z$ and odd $N_{\rm n}$,} \\ [2mm]
0 & \mbox{for nuclei with odd $A$,} \\ [2mm]
(11 \times 10^6 \; {\rm eV}) \, A^{-1/2}  &
\mbox{for nuclei with even $Z$ and even $N_{\rm n}$.}
\end{array} \right.
\label{3.3}\eeq
The formula \req{3.1}, combined with the total energy of the atomic
electrons, approximates the experimental atomic masses of
naturally occurring isotopes \citep{Coursey2015} with a relative accuracy
better than about $10^{-4}$, which is sufficient for the present purposes.

The effect of the finite size of the nucleus on the atomic structure, as
well as on the collisions of charged particles with atoms, can be
approximately accounted for by using simple models for the distribution
of nuclear charge (\ie, of protons). Stable nuclei are approximately
spherical with proton and neutron densities that vary with the distance
$r$ to the center of the nucleus. These densities are nearly constant
inside the nucleus and fall more or less rapidly to zero at the surface.
A convenient parameterization of the proton density is the Fermi
distribution \citep{Hahn1956}, \index{finite size of nucleus}
\beq
\rho_{\rm p}(r) = \frac{\rho_0}{\exp \left[ (r-R_{\rm n})/ z_{\rm n}
\right] + 1}\, ,
\label{3.4}\eeq
with the ``nuclear radius'' $R_{\rm n}$ and the ``diffuseness'' $z_{\rm
n}$ given by
\beq
R_{\rm n} = 1.07 \, A^{1/3} \, {\rm fm}
\qquad \mbox{and} \qquad
z_{\rm n} = 0.546 \, {\rm fm} \, ,
\label{3.5}\eeq
where the mass number $A$ may be replaced with the atomic weight (mean
relative atomic mass) to describe the average over stable isotopes of
the element.  The constant $\rho_0$, which equals twice the proton
density at $r=R_{\rm n}$, is determined by normalization. The
electrostatic potential for the Fermi distribution has to be calculated
numerically,
\beqa
\varphi_{\rm nuc}(r) &=& e \int \frac{\rho_{\rm p}(r')}{|{\bf r}-{\bf
r}'|}\, \d {\bf r}'
\nonumber \\ [2mm]
&=& \frac{e}{r}
\int_0^r \, \rho_{\rm p}(r') \, 4\pi r'^2\, \d r'
+ e \int_r^\infty \rho_{\rm p}(r')\, 4\pi r' \, \d r'\, .
\label{3.6}\eeqa
For radii much larger than $R_{\rm n}$, $\varphi_{\rm nuc}(r)$ reduces
to the Coulomb potential,
\beq
\varphi_{\rm nuc}(r) = Z e /r.
\label{3.7}\eeq
That is, when the typical distances from the nucleus to the orbiting
particles are much larger than the nuclear radius, the nucleus can be
treated as a point charge,

In calculations of atomic structure, it is commonly assumed
that the nucleus has infinite mass and, consequently, that it remains
fixed at the origin of coordinates. However, the nuclear mass is finite
and the nucleus moves under the actions of the forces from the other
particles of the system, thus affecting the collective motion. Within
the electrostatic approximation, the non-relativistic motion of the
nucleus can be accounted for as follows. \index{finite mass of nucleus}

Let ${\bf R}_0$ and ${\bf R}_i$ denote the position vectors of the
nucleus and of the $i$-th particle, respectively. The classical
Lagrangian function of the system is\footnote{A dot over a symbol
indicates the time derivative, $\dot{\bf R}\equiv \d
{\bf R}/\d t$.}
\beq
L = \frac{M_0}{2} \dot{\bf R}_0^2 + \sum_{i=1}^N
\frac{M_i}{2} \, \dot{\bf R}_i^2
- V ({\bf R}_0, {\bf R}_1, \ldots , {\bf R}_N ),
\label{3.8}\eeq
with the potential energy function
\beq
V ({\bf R}_0, {\bf R}_1, \ldots , {\bf R}_N ) = \sum_{i=1}^N Z_i e
\varphi_{\rm nuc} \left( \left| {\bf R}_i - {\bf R}_0 \right| \right)
+ \sum_{i=1}^{N-1} \sum_{j=i+1}^N
\frac{Z_i Z_j e^2}{ \left| {\bf R}_i - {\bf R}_j\right| }.
\label{3.9}\eeq
To separate the motion of the center of mass (CM) from the relative
motion, we introduce a new set of position variables
\beq
{\bf R}_{\rm CM} \equiv \frac{1}{M_{\rm T}} \left( M_0 {\bf R}_0 +
\sum_{i=1}^N M_i {\bf R}_i \right) , \qquad
{\bf r}_i \equiv {\bf R}_i - {\bf R}_0,
\label{3.10}\eeq
where
\beq
M_{\rm T} = M_0 + \sum_{i=1}^N M_i
\label{3.11}\eeq
is the total mass. Evidently, ${\bf R}_{\rm CM}$ is the position vector
of the CM and ${\bf r}_i$ is the position vector of the $i$-th particle
{\it relative to the nucleus}. The inverse transformation is
\beq
{\bf R}_0 = {\bf R}_{\rm CM} - \frac{1}{M_{\rm T}} \sum_{j=1}^N M_j {\bf
r}_i , \qquad
{\bf R}_i = {\bf R}_{\rm CM} - \frac{1}{M_{\rm T}} \sum_{j=1}^N M_j {\bf r}_j
+ {\bf r}_i.
\label{3.12}\eeq
In the following, the new variables \req{3.10} will be referred to as
{\it relative variables}.

The Lagrange function in terms of the relative variables is
\beqa
L &=& \frac{M_0}{2} \left( \dot{\bf R}_{\rm CM} - \frac{1}{M_{\rm T}}
\sum_{j=1}^N M_j \dot{\bf r}_j \right)^2
\nonumber \\ [2mm]
&& + \sum_{i=1}^N \frac{M_i}{2}\left( \dot{\bf R}_{\rm CM}
- \frac{1}{M_{\rm T}}
\sum_{j=1}^N M_j \dot{\bf r}_j + \dot{\bf r}_i
\right)^2 - V ({\bf r}_1, \ldots , {\bf r}_N )\, ,
\label{3.13}\eeqa
with
\beq
V ({\bf r}_1, \ldots , {\bf r}_N ) =
 \sum_i Z_i e
\varphi_{\rm nuc} \left(r_i\right)
+ \sum_{i=1}^{N-1} \sum_{j=i+1}^N
\frac{Z_i Z_j e^2}{|{\bf r}_i - {\bf r}_j|}\, .
\label{3.14}\eeq
A simple calculation shows that the conjugate momenta of the
relative variables are
\begin{subequations}
\label{3.15}
\beq
{\bf P}_{\rm CM} = \left(
\frac{\partial L}{\partial \dot{X}_{\rm CM}},
\frac{\partial L}{\partial \dot{Y}_{\rm CM}},
\frac{\partial L}{\partial \dot{Z}_{\rm CM}} \right) =
M_{\rm T} \dot{\bf R}_{\rm CM}
\label{3.15a}\eeq
and
\beq
{\bf p}_i = \left(
\frac{\partial L}{\partial \dot{x}_i},
\frac{\partial L}{\partial \dot{y}_i},
\frac{\partial L}{\partial \dot{z}_i} \right) = M_i \dot{\bf r}_i
- \, \frac{M_i}{M_{\rm T}} \sum_{j=1}^N M_j \dot{\bf r}_j .
\label{3.15b}\eeq
\end{subequations}
Recalling that the conjugate momenta of the old variables are
${\bf P}_0 = M \dot{\bf R}_0$ and ${\bf P}_i = M_i
\dot{\bf R}_i$, the relations \req{3.15} imply that
\begin{subequations}
\label{3.16}
\beqa
{\bf P}_{\rm CM} = M_0 \dot{\bf R}_0 +
\sum_{i=1}^N M_i \dot{\bf R}_i = {\bf P}_0 +
\sum_{i=1}^N {\bf P}_i
\label{3.16a}\eeqa
and
\beqa
{\bf p}_i &=& M_i \left( \dot{\bf R}_i - \dot{\bf R}_0 \right)
- \, \frac{M_i}{M_{\rm T}} \sum_{j=1}^N M_j
\left( \dot{\bf R}_j - \dot{\bf R}_0 \right)
\nonumber \\ [2mm]
&=& {\bf P}_i - \frac{M_i}{M_{\rm T}} \left( {\bf P}_0 + \sum_{j=1}^N
{\bf P}_j \right) =
{\bf P}_i - \frac{M_i}{M_{\rm T}} \, {\bf P}_{\rm CM}.
\label{3.16b}\eeqa
\end{subequations}
From these equalities it follows that
\beq
{\bf P}_i = {\bf p}_i + \frac{M_i}{M_{\rm T}} {\bf P}_{\rm CM}
\qquad \mbox{and} \qquad
{\bf P}_0 = \frac{M_0}{M_{\rm T}} {\bf P}_{\rm CM} - \sum_{i=1}^N {\bf
p}_i\, .
\label{3.17}\eeq

Hereafter, we consider the Cartesian components of the position vectors
and their conjugate momenta as quantum operators.
The canonical commutation relations between the Cartesian components
($k,k'=1,2,3$) of the position and momentum operators,
\beq
\left[ \rule{0mm}{4mm}({\bf R}_{i})_k,
({\bf P}_j)_{k'} \right] = {\rm i} \hbar \, \delta_{i,j} \,
\delta_{k,k'} \,,
\label{3.18}\eeq
imply that
\begin{subequations}
\label{3.19}
\beqa
\left[ \rule{0mm}{4mm}({\bf r}_{i})_k, ({\bf p}_i)_k \right]
&=& \left[ \rule{0mm}{4mm}({\bf R}_{i}-{\bf R}_0)_k,
({\bf P}_i)_k - \frac{M_i}{M_{\rm T}} ({\bf P}_0)_k
-\frac{M_i}{M_{\rm T}} \, ({\bf P}_i)_k
\right]
\nonumber \\ [2mm]
&=&
\left( 1 - \frac{M_i}{M_{\rm T}} \right)
\left[ \rule{0mm}{4mm}{\bf R}_{i})_k,
({\bf P}_i)_k \right] +
\frac{M_i}{M_{\rm T}}
\left[ \rule{0mm}{4mm}({\bf R}_0)_k,
({\bf P}_0)_k
\right] = {\rm i} \hbar\, . \rule{10mm}{0mm}
\label{3.19a}\eeqa
Similarly, we find that
\beq
\left[ \rule{0mm}{4mm}({\bf R}_{\rm CM})_k, ({\bf P}_{\rm CM})_k \right] =
{\rm i} \hbar \, ,
\label{3.19b}\eeq
and all the other commutators vanish:
\beq
\left[ \rule{0mm}{4mm}({\bf R}_{\rm CM})_k, ({\bf p}_i)_{k'} \right] =
\left[ \rule{0mm}{4mm}({\bf r}_i)_k, ({\bf P}_{\rm CM})_{k'}
\right] =0.
\label{3.19c}\eeq
\end{subequations}
That is, the conjugate momenta ${\bf p}_i$ of the relative variables ${\bf
r}_i$ also satisfy the canonical commutation relations. As a consequence
\citep[see, \eg,][]{Merzbacher1970}, in coordinate representation, the
corresponding operators have the usual form ${\bf p}_i = - {\rm i}
\hbar \nablab_{{\bf r}_i}$.

The wave functions of the system satisfy the Schr\"{o}dinger equation,
\beq
{\rm i} \hbar \, \frac{\partial}{\partial t} \Psi({\bf R}_{\rm CM}, {\bf
r}_1, \ldots, {\bf r}_N) = {\cal H} \,
\Psi({\bf R}_{\rm CM}, {\bf r}_1, \ldots, {\bf r}_N),
\label{3.20}\eeq
where
\beq
{\cal H} = \frac{1}{2M_0} {\bf P}_0^2 + \sum_{i=1}^N
\frac{1}{2M_i} {\bf P}_i^2
+ V ({\bf R}_0, {\bf R}_1, \ldots , {\bf R}_N )
\label{3.21}\eeq
is the Hamiltonian operator. Introducing the relative variables,
\beqa
{\cal H} &=& \frac{1}{2M_0} \left(
\frac{M_0}{M_{\rm T}}\, {\bf P}_{\rm CM} - \sum_{i=1}^N
{\bf p}_i \right)^2
+ \sum_{i=1}^N \frac{1}{2M_i}
\left( {\bf p}_i + \frac{M_i}{M_{\rm T}}\, {\bf P}_{\rm CM} \right)^2
+ V ({\bf r}_1, \ldots , {\bf r}_N )
\nonumber \\ [2mm]
&=& \frac{1}{2M_{\rm T}} \, {\bf P}_{\rm CM}^2 + \frac{1}{2M_0} \left(
\sum_{i=1}^N {\bf p}_i \right)^2 + \sum_{i=1}^N \frac{1}{2M_i} \,
{\bf p}_i^2 + V ({\bf r}_1, \ldots , {\bf r}_N )
\nonumber \\ [2mm]
&=& \frac{1}{2M_{\rm T}} \, {\bf P}_{\rm CM}^2 +
\sum_{i=1}^N \frac{1}{2}
\left( \frac{1}{M_i} + \frac{1}{M_0} \right) {\bf p}_i^2
+ \frac{1}{M_0}  \sum_{i=1}^{N-1} \sum_{j=i+1}^N {\bf p}_i \dotprod {\bf p}_j
+  V ({\bf r}_1, \ldots , {\bf r}_N )\, .
\nonumber\eeqa
Therefore, in terms of the relative variables, the Hamiltonian operator
reads \index{reduced mass}
\beq
{\cal H} = \frac{1}{2M_{\rm T}} \, {\bf P}_{\rm CM}^2 +
\sum_{i=1}^N \frac{1}{2\mu_i} \, {\bf p}_i^2
+ \frac{1}{M_0}  \sum_{i=1}^{N-1} \sum_{j=i+1}^N {\bf p}_i \dotprod {\bf p}_j
+  V ({\bf r}_1, \ldots , {\bf r}_N ),
\label{3.22}\eeq
where
\beq
\mu_i = \frac{M_iM_0}{M_i+M_0}
\label{3.23}\eeq
is the {\it reduced mass} of the $i$-th particle and the nucleus.

\index{Hamiltonian of the relative motion}
The Hamiltonian \req{3.22} separates the contributions from the motion
of the CM and from the relative motion. The first term indicates that
the CM moves as a free particle with a mass equal to the total mass
$M_{\rm T}$. The remaining terms, which depend only on the position
vectors ${\bf r}_i$, describe the motion of the orbiting particles {\it relative
to the nucleus}. In many practical calculations, the motion of the CM is
left aside, and only the ``relative'' Hamiltonian,
\beq
{\cal H}_{\rm rel} =
\sum_i \frac{1}{2\mu_i} \, {\bf p}_i^2
+  V ({\bf r}_1, \ldots , {\bf r}_N )
+ \frac{1}{M_0} \sum_{i<j} {\bf p}_i \dotprod {\bf p}_j \, ,
\label{3.24}\eeq
is considered. We see that the effects of the finite mass of the nucleus
are 1) the replacement of the actual masses $M_i$ of the particles with
the corresponding reduced masses $\mu_i$, and 2) the occurrence of the
last term, known as the {\it mass-polarization term}, which occurs only
when the number $N$ of particles is larger than one. Because of the
large mass of the nucleus, the mass-polarization term is very small and
it is usually disregarded (it can always be introduced {\it a
posteriori} as a perturbation). Actually, most calculations of atomic
structure are performed by assuming $M_0=\infty$; the effect of the
finite nuclear mass can then be accounted for by merely replacing
$M_i=\me$ with $\mu_i$. In practice, numerical calculation results that
are given in atomic units (see Appendix \ref{appC}) can be readily corrected by
using the following modified atomic units of length and energy:
\beq
a'_0 = a_0 \, \frac{\me}{\mu_e}
\qquad \mbox{and} \qquad
E'_{\rm h} = E_{\rm h} \, \frac{\mu_e}{\me}.
\label{3.25}\eeq

It is interesting to consider the effect of the finite mass of the
nucleus on collisions of charged particles with atoms. Let $M_1$ and
$Z_1 e$ denote, respectively, the mass and charge of the projectile. In
the laboratory reference frame, where the atom is initially at rest at
the origin of coordinates, the projectile moves with momentum
${\bf P}_1$  before the collision. Then, from Eq.\ \req{3.16a}, ${\bf
P}_{\rm CM}= {\bf P}_1$ and, from Eq.\ \req{3.16b} we find that
\index{reduced mass}
\beq
{\bf p}_1 = {\bf P}_1 - \frac{M_1}{M_{\rm T}} \, {\bf P}_1 =
\frac{M_1 M_{\rm atom}}{M_1 + M_{\rm atom}} \, \dot{\bf R}_1,
\label{3.26}\eeq
where $M_{\rm atom}$ is the mass of the target atom. The first factor on
the right-hand side is the reduced mass of the projectile {\it and the
target atom}, which is different from the projectile-nucleus reduced
mass $\mu_1$ [Eq.\ \req{3.23}]. ${\bf p}_1$ is the momentum of the
projectile in the CM frame, a frame that moves with velocity $\dot{\bf
R}_{\rm CM} = {\bf P}_{\rm CM}/M_{\rm T}$ relative to the laboratory
frame (see Section \ref{sec4.2.1}).


\section{Hydrogenic ions \label{sec3.2}}

\index{hydrogenic ions}
The simplest atomic systems are the hydrogenic ions, which consist of a
nucleus of charge $Ze$ and a single bound electron. Because the mass of the
nucleus is much larger than the electron mass, it is natural to describe
the system in a reference frame where the nucleus is at rest at the
origin of coordinates. In addition, as the average distance of the
electron to the nucleus is more than about four orders of magnitude larger than
the nuclear dimensions, we may consider the nucleus as a point charge
and assume that the interaction between the two particles is purely Coulombian.
The non-relativistic time-independent Schr\"{o}dinger wave equation
of the electron then reads
\beq
\left[ - \frac{\hbar^2}{2\me} \nabla^2 - \frac{Z e^2}{r} \right]
\psi({\bf r}) = E \psi({\bf r}).
\label{3.27}\eeq
This equation is solved in most quantum mechanics textbooks \citep[see,
\eg,][]{BransdenJoachain1983, Griffiths1995}. The energy eigenvalues of
the bound states are characterized by the principal quantum number $n$,
\index{hydrogenic ions!energy levels}
\beq
E_n = - \frac{Z^2 e^2}{2 n^2 a_0} =  - \frac{Z^2}{2 n^2} \, E_{\rm h},
\label{3.28}\eeq
where $a_0$ is the Bohr radius and $E_{\rm h}$ is the Hartree energy
(see Appendix \ref{appC}). An infinite denumerable set of bound levels
exist with $n=1$, 2, 3, \ldots The wave functions of bound states are the
central-field orbitals
\beq
\psi_{n \ell m_{\rm L} m_{\rm S}}(x) = \frac{P_{n\ell}(r)}{r} \,
Y_{\ell m_{\rm L}}(\hat{\bf r}) \, \chi_{m_{\rm S}}(\sigma),
\label{3.29}\eeq
where $x = \{ {\bf r}, \sigma\}$, {$Y_{\ell m_{\rm L}}(\hat{\bf r})$ are
the spherical harmonics (Appendix \ref{appB}) with $\ell=0, 1, \ldots,
n-1$, and $m_{\rm L}=-\ell, -\ell +1, \ldots, \ell -1, \ell$. The
reduced radial functions of bound states can be expressed as
\index{hydrogenic ions!radial wave functions}
\beqa
P_{n\ell}(r) &=&
\frac{1}{n(2\ell+1)!}
\left[ \frac{(n+\ell)!}{(n-\ell-1)!}\, \frac{Z}{a_0} \right]^{1/2}
\nonumber \\ [2mm]
&& \mbox{} \times
\left( 2 \frac{Z}{n} \, \frac{r}{a_0} \right)^{\ell+1}
\exp\left( - \frac{Z}{n} \, \frac{r}{a_0}\right) \,
_1F_1 \left( \ell+1-n; 2\ell+2;
 2 \frac{Z}{n} \, \frac{r}{a_0} \right), \rule{12mm}{0mm}
\label{3.30}\eeqa
where $_1F_1(a;b;x)$ is the Kummer function, defined by the power series
\req{2.64}. As the first argument is a negative integer [equal
to minus the radial quantum number $n_{\rm r}=n-(\ell +1)$], the Kummer
function in the expression \req{3.30} is in fact a polynomial of degree
$n_{\rm r}-1$. The radial functions $R_{n\ell}(r) = P_{n\ell}(r) / r$ of
the bound states with the lowest energies are
\begin{subequations}
\label{3.31}
\beqa
R_{10}(r) & = & \left( \frac{Z}{a_0} \right)^{3/2}
2\, \exp(-Zr/a_0), \label{3.31a} \\[2mm]
R_{20}(r) & = & \left( \frac{Z}{a_0} \right)^{3/2}
\frac{1}{\sqrt{2}}
\left[ 1 - \frac{1}{2}\left(\frac{Zr}{a_0}\right) \right]
\exp(-Zr/2a_0), \label{3.31b} \\[2mm]
R_{21}(r) & = & \left( \frac{Z}{a_0} \right)^{3/2}
\frac{1}{2\sqrt{6}} \; \frac{Zr}{a_0}
\exp(-Zr/2a_0).  \label{3.31c}
\eeqa
\end{subequations}
In addition to the usual degeneracy in $m_{\rm L}$ and $m_{\rm S}$, the
hydrogenic energy levels \req{3.28} are also degenerate in $\ell$ (a
peculiarity of the Coulomb potential); the total degeneracy of a
level $E_n$ is $2n^2$. It is interesting to note that the radial wave
functions scale with $Z$, that is,
\beq
R_{n \ell}^{(Z)}(r) = Z^{3/2} R_{n\ell}^{(Z=1)}(Zr),
\label{3.32}\eeq
where $R_{n\ell}^{(Z)}(r)$ is the radial function corresponding to the
indicated value of $Z$.

The expectation value of the potential energy is
\beq
\langle V \rangle \equiv \langle \psi_{n \ell m_{\rm L} m_{\rm S}}(x) \left|
V(r) \right| \psi_{n \ell m_{\rm L} m_{\rm S}}(x) \rangle
= Z e^2 \langle r^{-1} \rangle =  2 E_n,
\label{3.33}\eeq
and the average kinetic energy of the electron is
\beq
\langle K \rangle = \langle {\cal H} \rangle
- \langle V \rangle = - E_n.
\label{3.34}\eeq
These values are in agreement with the virial theorem \citep[see,
\eg,][]{Schiff1968}. \index{virial theorem}

Because their energy levels and wave functions are known analytically,
hydrogenic ions are used in simplified calculations to illustrate specific
aspects of collision theory. They also provide a basis for qualitative
arguments and a justification for semi-empirical formulas.


\section{The independent-electron approximation \label{sec3.3}}

\index{independent-electron approximation}
Let us consider a neutral atom or positive ion of atomic number $Z$ and
$N$ bound electrons, with its center of mass at the origin of
coordinates. We assume that the nucleus is infinitely massive
and that particles interact through instantaneous Coulomb forces. The
Hamiltonian of the system is expressed as the sum of the kinetic
energy operators $\breve{K}_i$ of the individual electrons and
the pairwise electrostatic interactions between the particles,
\beq
{\cal H} = \sum_{i=1}^N \breve {K}_i
+ \sum_{i=1}^N V_{\rm nuc}({\bf r}_i) +
\sum_{i=1}^{N-1} \sum_{j=i+1}^{N}  \frac{e^2}{|{\bf r}_i - {\bf r}_j|},
\label{3.35}\eeq
where ${\bf r}_i$ is the position vector of the $i$-th electron.
$V_{\rm nuc}({\bf r})$ is the nuclear potential (\ie, the electrostatic
interaction energy of an electron at $r$ with the nucleus); for a point
nucleus
\beq
V_{\rm nuc}(r) = - Z e^2 /r.
\label{3.36}\eeq

We wish to calculate the atomic wave function $\Psi$, which depends on the
positions ${\bf r}_i$ and the spin variables $\sigma_i$ of the $N$
electrons and is antisymmetric under the exchange of two electrons, as
required by Pauli's exclusion principle. Because the
direct solution of the wave equation ${\cal H} \Psi = E \Psi$ is
extremely difficult (except for hydrogenic ions
with a single electron) we consider an approximation of independent
electrons, in which atomic electrons are assumed to move independently
in a common local central potential $V(r)$. That is, we replace the
Hamiltonian \req{3.35} with
\beq
{\cal H}_{\rm IEA} = \sum_{i=1}^N \left[ \breve{K}_i + V(r_i)\right],
\label{3.37}\eeq
where $V(r)$ is an effective central potential, which accounts
for the average interaction energy of an electron at $r$ with the
nucleus and with the other electrons. This potential should be selected
in such a way that the difference ${\cal H} - {\cal H}_{\rm IEA}$ is as
small as possible and, hence, it might be introduced {\it a posteriori}
as a perturbation if needed.

The usefulness of the independent-electron approximation stems from the
fact that we can build exact solutions of the wave equation
\beq
{\cal H}_{\rm IEA} \Psi = E_{\rm IEA}  \Psi
\label{3.38}\eeq
in terms of one-electron orbitals that are solutions of the one-electron
wave equation,
\beq
\left[ \breve{K} + V(r)\right] \psi_i (x) = \varepsilon_i \psi_i(x),
\label{3.39}\eeq
where the variable $x$ denotes the position coordinates
${\bf r}$ and the spin variable of the electron (see Section
\ref{sec2.3}).
In practice, we shall consider the atom in its ground state, with the
electrons occupying the $N$ orbitals $\psi_i$ of lower energies.
The atomic wave function $\Psi$ can then be represented as the product
of these orbitals, antisymmetrized to comply with the exclusion
principle. That is, $\Psi$ is described as a {\it Slater determinant},
\beqa
\Psi &=& \frac{1}{\sqrt{N!}} \sum_{P \in S_N} (-1)^P
\psi_{1}(x_{P(1)}) \ldots
\psi_{N}(x_{P(N)})
\nonumber \\ [2mm]
&=& \frac{1}{\sqrt{N!}}\left| \begin{array}{ccc}
\psi_{1}(x_{1}) & \ldots &
\psi_{1}(x_{N}) \\ [2mm]
\vdots & \ddots & \vdots \\ [2mm]
\psi_{N}(x_{1}) & \ldots &
\psi_{N}(x_{N})
\end{array}
\right|,
\label{3.40}\eeqa
where the summation is over the permutations $P$ of $N$ elements and
$P(i)$ denotes the image of the element $i$ by the permutation $P$.
Because the orbitals $\psi_i$ are
orthonormal, the Slater determinant is normalized to unity,
\beq
\left< \Psi \left| \Psi \right> \right. =
\int \d x_1 \ldots \int \d x_N \left| \psi(x_1, \ldots, x_N) \right|^2
= 1,
\label{3.41}\eeq
and it satisfies the wave equation \req{3.38} with the eigenvalue
\beq
E_{\rm IEA}=\varepsilon_{1} + \cdots + \varepsilon_{N}.
\label{3.42}\eeq


\subsection{Non-relativistic independent-electron models
\label{sec3.3.1}}

\index{independent-electron models!non-relativistic}
Although a proper description of atomic structure requires accounting
for relativistic effects, the non-relativistic theory is much simpler to
handle and simplifies certain derivations. Indeed a non-relativistic
formulation is adequate for elements with low atomic numbers and in
situations, such as the degenerate electron gas, where electrons move
with velocities that are much smaller than $c$.

With the non-relativistic one-electron kinetic energy operator,
\beq
\breve{K} = - \, \frac{1}{2\me} \, \breve{\bf p}^2,
\label{3.43}\eeq
the Hamiltonian of the atom, Eq.\ \req{3.35},
is independent of the spin and, consequently, we can consider the
independent-electron model Hamiltonian
\beq
{\cal H}_{\rm IEA} = \sum_{i=1}^N
\left[ - \frac{1}{2\me} \breve{\bf p}_i^2  + V(r_i) \right]
\label{3.44} \eeq
and work with one-electron orbitals in the uncoupled representation,
\beq
\psi_{n\ell m_{\rm L} m_{\rm S}} (x) =
\psi_{n\ell m_{\rm L}} ({\bf r}) \, \chi_{m_{\rm S}}(\sigma)
= \frac{1}{r} \, P_{n\ell}(r) \, Y_{\ell m_{\rm L}} (\hat{\bf r}) \,
\, \chi_{m_{\rm S}}(\sigma),
\label{3.45}\eeq
where $Y_{\ell m_{\rm L}} (\hat{\bf r})$ is a spherical harmonic and the
function $P_{n\ell}(r)$ satisfies the radial Schr\"{o}dinger equation with the
potential $V(r)$,
\beq
- \frac{\hbar^2}{2\me} \, \frac{\d^2 P_{n \ell}}{\d r^2}
+ \left[
\frac{\hbar^2}{2\me} \, \frac{\ell(\ell+1)}{r^2} + V(r) \right] P_{n \ell}
= \varepsilon_{n\ell} \, P_{n\ell}\, .
\label{3.46}\eeq
This equation can be solved to high accuracy, \eg, by means of the
Fortran subroutine package {\sc radial} of
\citet{SalvatFernandezVarea2019}. The set of central-field bound
orbitals is orthonormal,
\beqa
\langle
\psi_{n'\ell' m'_{\rm L} m'_{\rm S}} \left| \psi_{n\ell m_{\rm L} m_{\rm S}}
\rangle \right.
&\equiv& \sum_{\sigma=\pm 1/2} \int \, \d r \; \psi_{n'\ell' m'_{\rm L} m'_{\rm S}}^\ast(x)
\, \psi_{n\ell m_{\rm L} m_{\rm S}}(x)
\nonumber \\ [2mm]
&=& \delta_{n',n} \, \delta_{\ell',\ell} \, \delta_{m'_{\rm L},m_{\rm L}}
\, \delta_{m'_{\rm S},m_{\rm S}}.
\label{3.47}\eeqa
Atomic wave functions $\Psi$ can then be represented as Slater
determinants formed with $N$ orbitals $\psi_{n_i\ell_i m_{{\rm L}i}
m_{{\rm S}i}}(x)$.

The use of central-field orbitals of the type \req{3.45} largely
facilitates the evaluation of matrix elements of spin-independent
operators. Evidently the non-relativistic independent-electron model
describes the ``gross structure'' of the one-electron energy spectrum,
with energy levels $\varepsilon_{n\ell}$ depending on the principal and the
orbital angular momentum quantum numbers.

\index{atomic configuration}
The set of orbitals with equal values of $n$ and $\ell$ is said to
constitute an {\it electron shell}. Usually, electron shells are
indicated by using the notation $(n\ell)$. A shell $(n\ell)$ can
accommodate up to $2(2\ell+1)$ electrons in energy-degenerate orbitals
with different values of the quantum numbers $m_{\rm L}$ and $m_{\rm
S}$. The list of shells $a=(n_a\ell_a)$ that are occupied and their
respective occupancy numbers $\nu_a$ (\ie, the number of orbitals of the
shell that occur in the Slater determinant) define the {\it electron
configuration} of the atom. A configuration is then specified by giving
the list of occupied shells and the corresponding occupancy numbers,
\beq
(n_1\ell_1)^{\nu_1}\, , \, \, (n_2\ell_2)^{\nu_2}\, ,
\, \ldots
\label{3.48}\eeq
A shell with all its orbitals occupied is said to be {\it closed}.
Atomic energy levels corresponding to configurations with open shells
[with $\nu_a < 2(2\ell_a+1)$ electrons] are degenerate; all the states
of such a configuration (which contain $\nu_a$ orbitals with equal
values of $n_a$ and $\ell_a$ and different magnetic
quantum numbers $m_a$ and $m_{{\rm S}a}$) have the same energy.

The atomic electron density is given by
\beq
\rho({\bf r}) = \sum_{i=1}^N \left|
\psi_{n_i\ell_i m_{{\rm L}i} m_{{\rm S}i}} (x) \right|^2.
\label{3.49}\eeq
By virtue of Uns\"{o}ld's theorem \req{B.58}, the electron density of a
closed shell is spherically symmetric. Consequently, we obtain a
spherical average of the electron density by setting
\beq
\rho(r) = \frac{1}{4\pi r^2} \sum_{a} \nu_a P_{n_a \kappa_a }^2(r)
\, ,
\label{3.50}\eeq
where the summation runs over the shells $a=(n_a \kappa_a)$ of the
configuration. For configurations with closed shells the electron
density is spherically symmetric and, consequently, the central-field
approximation is expected to be fairly realistic.


\subsection{Relativistic independent-electron models
\label{sec3.3.2}}

\index{independent-electron models!relativistic}
The non-relativistic formulation presented above is not satisfactory
because it disregards spin effects, other than the double degeneracy of
spatial orbitals, and the fact that atomic electrons acquire large
kinetic energies when they are near the nucleus. It is therefore
pertinent to consider relativistic independent-electron models. The
simplest approach to account for spin and ``kinematical'' relativistic
effects consists of replacing the kinetic energy operator $\breve{K}$ of
an individual electron with the Dirac operator, Eq.\ \req{2.89},
\beq
\breve{K} = c \widetilde{\alphab} \dotprod
\breve{\bf p} + (\widetilde{\beta} -1) \me c^2,
\label{3.51}\eeq
where $\widetilde{\alphab}$, $\widetilde{\beta}$ are the Dirac matrices
in the spinor representation, Eq.\ \req{2.79}. The Hamiltonian of the
atom is thus
\beq
{\cal H} = \sum_{i=1}^N
\left[ c \widetilde{\alphab}_i \dotprod
\breve{\bf p}_i + (\widetilde{\beta}_i
-1) \me c^2 \right]
+ \sum_{i=1}^N V_{\rm nuc}({\bf r}_i) +
\sum_{i=1}^{N-1} \sum_{j=i+1}^{N} \frac{e^2}{|{\bf r}_i - {\bf r}_j|}\, ,
\label{3.52}\eeq
where  $\widetilde{\alphab}_i$, $\widetilde{\beta}_i$ are the Dirac
matrices for the $i$-th electron. The Breit interaction \citep[see,
\eg,][]{Grant1970} and other quantum electrodynamics corrections can be
introduced {\it a posteriori} as perturbations, if required.

Within the relativistic independent-electron approximation, the
Hamiltonian of the atom is approximated as
\beq
{\cal H}_{\rm IEA} = \sum_{i=1}^N
\left[ c \widetilde{\alphab}_i \dotprod
\breve{\bf p}_i + (\widetilde{\beta}_i
-1) \me c^2 + V(r_i) \right].
\label{3.53}\eeq
We can then consider exact solutions of the wave equation
\beq
{\cal H}_{\rm IEA} \Psi = E_{\rm IEA}  \Psi
\label{3.54}\eeq
expressed as Slater determinants build with one-electron orbitals that
satisfy the one-electron wave equation
\beq
\left[ c \widetilde{\alphab}
\dotprod {\bf p} + (\widetilde{\beta}_i -1) \me c^2 + V(r)\right]
\psi_{n\kappa m}(x) = \varepsilon_{n\kappa} \psi_{n\kappa m}(x)
\label{3.55}\eeq
with well defined angular momenta.
These orbitals are spherical waves of the form [see Eq.\ \req{2.92}]
\beq
\psi_{n \kappa m }(x)
= \frac{1}{r} \left( \begin{array}{c} P_{n\kappa}(r) \,
\Omega_{\kappa, m}(\hat{\bf r})\\ [3mm] {\rm i} Q_{n\kappa}(r) \,
\Omega_{-\kappa, m}(\hat{\bf r}) \end{array} \right)\, ,
\label{3.56}\eeq
with radial functions determined by the equations [cf.\ Eqs.\
\req{2.98}],
\beqa
\frac{\d P_{n\kappa}}{\d r} &=&
- \frac{\kappa}{r} \,P_{n\kappa}
+ \frac{\varepsilon_{n\kappa}+2 \me c^2-V}{c\hbar}
\, Q_{n\kappa},
\nonumber \\ [2mm]
\frac{\d Q_{n\kappa}}{\d r} &=&
\frac{-\varepsilon_{n\kappa} +V}{c\hbar} \, P_{n\kappa}
+ \frac{\kappa}{r}\, Q_{n\kappa},
\label{3.57}\eeqa
which can be solved numerically to high accuracy \citep[see,
\eg,][]{SalvatFernandezVarea2019}.
The set of central-field orbitals is orthonormal,
\beq
\langle\psi_{n'\kappa' m'}|\psi_{n\kappa m}\rangle
\equiv \int \psi_{n'\kappa' m'}^\dagger(x)
\, \psi_{n\kappa m}(x) \, \d x =
\delta_{n',n} \, \delta_{\kappa',\kappa} \, \delta_{m',m}.
\label{3.58}\eeq
Slater determinants formed with $N$ orbitals, $\psi_{n_i \kappa_i
m_i}(x)$,
\beq
\Psi =  \frac{1}{\sqrt{N!}}\left| \begin{array}{ccc}
\psi_{n_1 \kappa_1 m_1}(x_{1}) & \ldots &
\psi_{n_1 \kappa_1 m_1}(x_{N}) \\ [2mm]
\vdots & \ddots & \vdots \\ [2mm]
\psi_{n_N \kappa_N m_N}(x_{1}) & \ldots &
\psi_{n_N \kappa_N m_N}(x_{N})
\end{array}
\right|,
\label{3.59}\eeq
satisfy the wave equation \req{3.54} with the eigenvalue
\beq
E_{\rm IEA} = \varepsilon_{n_1 \kappa_1} + \cdots + \varepsilon_{n_N \kappa_N}.
\label{3.60}\eeq

Each one-electron energy level $\varepsilon_{n\kappa}$ is $2j+1$ times
degenerate (we recall that $j=|\kappa| - \1o2$).
The $2j+1$ orbitals $\psi_{n\kappa m}$ that have the same values of the
quantum numbers $n$ and $\kappa$ are said to belong to the same {\it
electron subshell}. Usually, electron subshells are indicated by using the
spectroscopic notation $n\ell_j$ (or, equivalently, $n\kappa$) or the x-ray notation (see Table \ref{tab2.1}).

\begin{table}[ph!]
\caption{Electronic configurations
of neutral atoms in their ground states. Notice that the shells
indicated with $\bar{\ell}$ correspond to $j=\ell - 1/2$ (see Table
\ref{tab2.1}).
\label{tab3.1}}
\begin{center}
\begin{tabular}{rlll} \hline\hline
$Z$ & \multicolumn{2}{c}{Element} \rule{0mm}{4mm}\rule{0mm}{-3mm} &
\multicolumn{1}{c}{Configuration} \\ \hline
 1  & H  & Hydrogen    & $1s^1$                          \\
 2  & He & Helium      & $1s^2$                          \\ \hline
 3  & Li & Lithium     & [He]$2s^1$                      \rule{0mm}{4mm} \\
 4  & Be & Beryllium   & [He]$2s^2$                      \\
 5  & B  & Boron       & [He]$2s^22\bar{p}^1$            \\
 6  & C  & Carbon      & [He]$2s^22\bar{p}^2$            \\
 7  & N  & Nitrogen    & [He]$2s^22\bar{p}^22p^1$        \\
 8  & O  & Oxygen      & [He]$2s^22\bar{p}^22p^2$        \\
 9  & F  & Fluorine    & [He]$2s^22\bar{p}^22p^3$        \\
10  & Ne & Neon        & [He]$2s^22\bar{p}^22p^4$        \\ \hline
11  & Na & Sodium      & [Ne]$3s^1$                      \rule{0mm}{4mm} \\
12  & Mg & Magnesium   & [Ne]$3s^2$                      \\
13  & Al & Aluminium   & [Ne]$3s^23\bar{p}^1$            \\
14  & Si & Silicon     & [Ne]$3s^23\bar{p}^2$            \\
15  & P  & Phosphorus  & [Ne]$3s^23\bar{p}^23p^1$        \\
16  & S  & Sulphur     & [Ne]$3s^23\bar{p}^43p^2$        \\
17  & Cl & Chlorine    & [Ne]$3s^23\bar{p}^43p^3$        \\
18  & Ar & Argon       & [Ne]$3s^23\bar{p}^43p^4$        \\ \hline
19  & K  & Potassium   & [Ar]$4s^1$                      \rule{0mm}{4mm} \\
20  & Ca & Calcium     & [Ar]$4s^2$                              \\
21  & Sc & Scandium    & [Ar]$3\bar{d}^14s^2$                    \\
22  & Ti & Titanium    & [Ar]$3\bar{d}^24s^2$                    \\
23  & V  & Vanadium    & [Ar]$3\bar{d}^34s^2$                    \\
24  & Cr & Chromium    & [Ar]$3\bar{d}^44s^2$                    \\
25  & Mn & Manganese   & [Ar]$3\bar{d}^43d^14s^2$                \\
26  & Fe & Iron        & [Ar]$3\bar{d}^43d^24s^2$                \\
27  & Co & Cobalt      & [Ar]$3\bar{d}^43d^34s^2$                \\
28  & Ni & Nickel      & [Ar]$3\bar{d}^43d^44s^2$                \\
29  & Cu & Copper      & [Ar]$3\bar{d}^43d^54s^2$                \\
30  & Zn & Zinc        & [Ar]$3\bar{d}^43d^64s^2$                \\
31  & Ga & Gallium     & [Ar]$3\bar{d}^43d^64s^24\bar{p}^1$      \\
32  & Ge & Germanium   & [Ar]$3\bar{d}^43d^64s^24\bar{p}^2$      \\
33  & As & Arsenic     & [Ar]$3\bar{d}^43d^64s^24\bar{p}^24p$
\rule{0mm}{-3mm}  \\ \hline \hline
\end{tabular}
\end{center} \end{table}
\addtocounter{table}{-1}

\begin{table}[ph!]
\caption {Continued.
}
\begin{center}
\begin{tabular}{rlll} \hline\hline
$Z$ & \multicolumn{2}{c}{Element} \rule{0mm}{4mm}\rule{0mm}{-3mm} &
\multicolumn{1}{c}{Configuration}  \\ \hline
34  & Se & Selenium    & [Ar]$3\bar{d}^43d^64s^24\bar{p}^24p^2$ \\
35  & Br & Bromine     & [Ar]$3\bar{d}^43d^64s^24\bar{p}^24p^3$ \\
36  & Kr & Krypton     & [Ar]$3\bar{d}^43d^64s^24\bar{p}^24p^4$ \\ \hline
37  & Rb & Rubidium    & [Kr]$5s^1$  \rule{0mm}{4mm}   \\
38  & Sr & Strontium   & [Kr]$5s^2$                    \\
39  & Y  & Yttrium     & [Kr]$4\bar{d}^15s^2$          \\
40  & Zr & Zirconium   & [Kr]$4\bar{d}^25s^2$          \\
41  & Nb & Niobium     & [Kr]$4\bar{d}^35s^2$          \\
42  & Mo & Molybdenum  & [Kr]$4\bar{d}^45s^2$          \\
43  & Tc & Technetium    & [Kr]$4\bar{d}^44d^25s^1$    \\
44  & Ru & Ruthenium     & [Kr]$4\bar{d}^44d^35s^1$                             \\
45  & Rh & Rhodium       & [Kr]$4\bar{d}^44d^45s^1$                             \\
46  & Pd & Palladium     & [Kr]$4\bar{d}^44d^55s^1$                             \\
47  & Ag & Silver        & [Kr]$4\bar{d}^44d^65s^1$                             \\
48  & Cd & Cadmium       & [Kr]$4\bar{d}^44d^65s^2$                             \\
49  & In & Indium        & [Kr]$4\bar{d}^44d^65s^25\bar{p}^1$                   \\
50  & Sn & Tin           & [Kr]$4\bar{d}^44d^65s^25\bar{p}^2$                   \\
51  & Sb & Antimony      & [Kr]$4\bar{d}^44d^65s^25\bar{p}^25p^1$               \\
52  & Te & Tellurium     & [Kr]$4\bar{d}^44d^65s^25\bar{p}^25p^2$               \\
53  & I  & Iodine        & [Kr]$4\bar{d}^44d^65s^25\bar{p}^25p^3$               \\
54  & Xe & Xenon         & [Kr]$4\bar{d}^44d^65s^25\bar{p}^25p^4$               \rule{0mm}{-3mm}  \\ \hline
55  & Cs & Cesium        & [Xe]$6s^1$                                           \rule{0mm}{4mm}   \\
56  & Ba & Barium        & [Xe]$6s^2$                                           \\
57  & La & Lanthanum     & [Xe]$5\bar{d}^16s^2$                                 \\
58  & Ce & Cerium        & [Xe]$4\bar{f}^15\bar{d}^16s^2$                       \\
59  & Pr & Praseodymium  & [Xe]$4\bar{f}^25\bar{d}^16s^2$                       \\
60  & Nd & Neodymium     & [Xe]$4\bar{f}^35\bar{d}^16s^2$                       \\
61  & Pm & Promethium    & [Xe]$4\bar{f}^45\bar{d}^16s^2$                       \\
62  & Sm & Samarium      & [Xe]$4\bar{f}^55\bar{d}^16s^2$                       \\
63  & Eu & Europium      & [Xe]$4\bar{f}^65\bar{d}^16s^2$                       \\
64  & Gd & Gadolinium    & [Xe]$4\bar{f}^64f5\bar{d}^16s^2$                     \\
65  & Tb & Terbium       & [Xe]$4\bar{f}^64f^25\bar{d}^16s^2$                   \\
66  & Dy & Dysprosium    & [Xe]$4\bar{f}^64f^35\bar{d}^16s^2$
                                    \\ \hline\hline
\end{tabular} \end{center} \end{table}
\addtocounter{table}{-1}

\begin{table}[ph!]
\caption {Continued.
}
\begin{center}
\begin{tabular}{rlll} \hline\hline
$Z$ & \multicolumn{2}{c}{Element} \rule{0mm}{4mm}\rule{0mm}{-3mm} &
\multicolumn{1}{c}{Configuration} \\ \hline
67  & Ho & Holmium       & [Xe]$4\bar{f}^64f^45\bar{d}^16s^2$ \\
68  & Er & Erbium        & [Xe]$4\bar{f}^64f^55\bar{d}^16s^2$ \\
69  & Tm & Thulium       & [Xe]$4\bar{f}^64f^65\bar{d}^16s^2$ \\
70  & Yb & Ytterbium     & [Xe]$4\bar{f}^64f^86s^2$           \\
71  & Lu & Lutetium      & [Xe]$4\bar{f}^64f^85\bar{d}^16s^2$ \\
72  & Hf & Hafnium       & [Xe]$4\bar{f}^64f^85\bar{d}^26s^2$ \\
73  & Ta & Tantalum      & [Xe]$4\bar{f}^64f^85\bar{d}^36s^2$ \\
74  & W  & Tungsten      & [Xe]$4\bar{f}^64f^85\bar{d}^46s^2$ \\
75  & Re & Rhenium       & [Xe]$4\bar{f}^64f^85\bar{d}^45d6s^2$                 \\
76  & Os & Osmium        & [Xe]$4\bar{f}^64f^85\bar{d}^45d^26s^2$               \\
77  & Ir & Iridium       & [Xe]$4\bar{f}^64f^85\bar{d}^45d^36s^2$               \\
78  & Pt & Platinum      & [Xe]$4\bar{f}^64f^85\bar{d}^45d^56s^1$               \\
79  & Au & Gold          & [Xe]$4\bar{f}^64f^85\bar{d}^45d^66s^1$               \\
80  & Hg & Mercury       & [Xe]$4\bar{f}^64f^85\bar{d}^45d^66s^2$               \\
81  & Tl & Thallium      & [Xe]$4\bar{f}^64f^85\bar{d}^45d^66s^26\bar{p}^1$     \\
82  & Pb & Lead          & [Xe]$4\bar{f}^64f^85\bar{d}^45d^66s^26\bar{p}^2$     \\
83  & Bi & Bismuth       & [Xe]$4\bar{f}^64f^85\bar{d}^45d^66s^26\bar{p}^26p^1$ \\
84  & Po & Polonium      & [Xe]$4\bar{f}^64f^85\bar{d}^45d^66s^26\bar{p}^26p^2$ \\
85  & At & Astatine      & [Xe]$4\bar{f}^64f^85\bar{d}^45d^66s^26\bar{p}^26p^3$ \\
86  & Rn & Radon         & [Xe]$4\bar{f}^64f^85\bar{d}^45d^66s^26\bar{p}^26p^4$ \\ \hline
87  & Fr & Francium      & [Rn]$7s^1$                           \rule{0mm}{4mm}  \\
88  & Ra & Radium        & [Rn]$7s^2$                            \\
89  & Ac & Actinium      & [Rn]$6\bar{d}^17s^2$                  \\
90  & Th & Thorium       & [Rn]$5\bar{f}^16\bar{d}^17s^2$        \\
91  & Pa & Protactinium  & [Rn]$5\bar{f}^26\bar{d}^17s^2$        \\
92  & U  & Uranium       & [Rn]$5\bar{f}^36\bar{d}^17s^2$        \\
93  & Np & Neptunium     & [Rn]$5\bar{f}^46\bar{d}^17s^2$        \\
94  & Pu & Plutonium     & [Rn]$5\bar{f}^56\bar{d}^17s^2$        \\
95  & Am & Americium     & [Rn]$5\bar{f}^66\bar{d}^17s^2$        \\
96  & Cm & Curium        & [Rn]$5\bar{f}^65f6\bar{d}^17s^2$      \\
97  & Bk & Berkelium     & [Rn]$5\bar{f}^65f^26\bar{d}^17s^2$    \\
98  & Cf & Californium   & [Rn]$5\bar{f}^65f^36\bar{d}^17s^2$    \\
99  & Es & Einsteinium   & [Rn]$5\bar{f}^65f^46\bar{d}^17s^2$    \\ \hline\hline
\end{tabular} \end{center} \end{table}


\index{atomic configuration}
The list of subshells that are occupied and their respective occupancy
numbers (\ie, the number of orbitals of the subshell that occur in the
Slater determinant) define the {\it electron configuration}. A
configuration is specified by giving the occupied subshells $(n\kappa$)
and the corresponding occupancy numbers $\nu$ ($\le 2j+1$). Thus, the
configuration
\beq
(n_1\kappa_1)^{\nu_1}\, , \, \, (n_2\kappa_2)^{\nu_2}\, ,
\, \ldots
\label{3.61}\eeq
consists of shells ($n_1\kappa_1$), ($n_2\kappa_2$), \ldots\ with
$\nu_1$, $\nu_2$, \ldots\ electrons, respectively. The electron
configurations of ground states of neutral atoms are given in Table
\ref{tab3.1}. Notice that ground state configurations typically consist
of a set of {\it closed subshells}, each with $\nu=2j+1$ electrons, and
one or two {\it open subshells} with less than $2j+1$ electrons

The electron density is given by
\beq
\rho(x) = \sum_{i=1}^N \psi^\dagger_{n_i \kappa_i m_i}(x)
\, \psi_{n_i \kappa_i m_i}(x)
\, .
\label{3.62}\eeq
The Uns\"{o}ld theorem, Eq.\ \req{B.58}, implies that the
electron density of a closed subshell is spherically symmetric. We
obtain an spherical average of the electron density by setting
\beq
\rho(r) = \frac{1}{4\pi r^2} \sum_{a} \nu_a \left[
P_{n_a \kappa_a }^2(r)
+ Q_{n_a \kappa_a }^2 (r) \right] \, ,
\label{3.63}\eeq
where the summation runs over the subshells $a=(n_a \kappa_a)$ of the
configuration. In the case of configurations with only closed subshells,
the electron density is spherically symmetric and the central-field
approximation is expected to be quite accurate.


\section{The Thomas--Fermi model  \label{sec3.4}}

\index{Thomas--Fermi model of the atom}
The simplest approach for estimating the atomic electron density
$\rho(r)$ and the potential $V(r)$ of neutral atoms and positive ions is
provided by the Thomas--Fermi model, which is based on statistical and
semiclassical considerations. The atomic electron cloud is described
locally as a degenerate electron gas of density $\rho(r)$ that is
confined by the attraction of the nucleus, represented as a point
charge. Atomic electrons are prevented from falling towards the nucleus
by their mutual repulsion and by the effect of the exclusion principle.
Notice that the Thomas--Fermi model is derived from non-relativistic
arguments and, consequently, it does not account for spin and
relativistic effects.

The electron gas is described within an independent-particle
approximation with single-particle states expressed as spin plane waves
satisfying periodic boundary conditions in a cubic box of side $L$
(see Section \ref{sec2.1.1}),
\beq
\psi_{{\bf k}\mu} (x) = L^{-3} \, \exp({\rm i} {\bf k} \dotprod
{\bf r}) \, \chi_\mu(\sigma),
\label{3.64}\eeq
where ${\bf k}$ is the wave number and $\chi_\mu (\sigma)$ is a
spin function. The assumed spatial
periodicity implies that the allowed wavenumbers are of the form
\beq
{\bf k} = (n_x, n_y, n_x) \frac{2\pi}{L},
\label{3.65}\eeq
where $n_x$, $n_y$, and $n_z$ are integers. The average number of
spatial orbitals per unit volume in ${\bf k}$ space is
\beq
\frac{\d {\cal N}}{\d {\bf k}} = \left( \frac{L}{2\pi} \right)^3.
\label{3.66}\eeq

As dictated by Fermi--Dirac statistics, in the ground state of the gas
the electrons fill all orbitals up to a certain energy $E_{\rm F}$, the
{\it Fermi energy}, corresponding to the wave number $k_{\rm F}$
\beq
E_{\rm F} = \frac{\hbar^2 k_{\rm F}^2}{2\me} \, .
\label{3.67}\eeq
When $k_{\rm F}$ is much larger than $2\pi/L$, the distribution of allowed
wave vectors
can be treated as a continuum. Since each space orbital is occupied by
two electrons with opposite spins, the number of space orbitals occupied
by the electrons in the normalization box is
\beq
\frac{\rho L^3}{2} = \int_{k \le k_{\rm F}} \frac{\d {\cal N}}{\d {\bf k}}
\, \d {\bf k} = \left( \frac{L}{2\pi} \right)^3 \frac{4 \pi k_{\rm F}^3}{3}\, ,
\label{3.68}\eeq
which implies that
\beq
k_{\rm F} = \left( 3 \pi^2 \rho \right)^{1/3}.
\label{3.69}\eeq
The density of space orbitals per unit kinetic energy ($E=\hbar^2
k^2/2\me$) is [see Eq.\ \req{2.31}]
\beq
\frac{\d {\cal N}}{\d E} = \frac{L^3}{4\pi^2}
\left( \frac{2\me}{\hbar^2} \right)^{3/2} \sqrt{E}.
\label{3.70}\eeq
Hence, the average kinetic energy of the electrons in the gas is
\beq
\overline K = \frac{1}{\rho L^3} \int_0^{E_{\rm F}} E\,
\frac{\d {\cal N}}{\d
E} \, \d E = \frac{3\pi^2}{k_{\rm F}^3} \, \frac{1}{2\pi^2}
\left( \frac{2\me}{\hbar^2} \right)^{3/2}
\frac{2}{5} \, E_{\rm F}^{5/2} = \frac{3}{5} \, E_{\rm F}.
\label{3.71}\eeq
Although the expressions \req{3.69} and \req{3.71} are independent of the
volume $L^3$ of the normalization box, they are obtained by assuming
that the distribution of states in ${\bf k}$-space is continuous. This
assumption is valid only when $k_{\rm F} \gg 2\pi/L$ or, equivalently,
when $\rho \gg 8\pi/3 L^3$.


\subsection{The Thomas--Fermi equation\label{sec3.4.1}}

Let us consider the atom or ion as a spherical distribution of electric charge
consisting of a point nucleus with charge $Ze$ (assumed to have an
infinite mass and fixed at the origin of coordinates) and a spherical
electron cloud of density $\rho(r)$ representing $N$ bound electrons,
that is, such that
\beq
\int_{0}^{\infty} \rho(r) 4\pi r^{2} \, \d r = N.
\label{3.72}\eeq

Assuming that the electron charge distribution is rigid, the potential
energy of an electron (external to the atom) at a distance $r$ from the
nucleus is
\beq
V(r) = V_{\rm n}(r) + V_{\rm e}(r),
\label{3.73}\eeq
where
\beq
V_{\rm n}(r) = - \frac{Ze^{2}}{r}
\label{3.74}\eeq
and
\beq
V_{\rm e}(r) =
e^{2} \left[ \frac{1}{r} \int_{0}^{r} \rho(r') 4\pi r'^{2} \, \d r' +
\int_{r}^{\infty} \rho(r') 4\pi r' \, \d r' \right]
\label{3.75}\eeq
are, respectively, the interaction energies with the nucleus and with
the electron cloud. Evidently
\beq
V_{\rm e}(0) =
e^{2} \int_{0}^{\infty} \frac{\rho(r')}{r'} 4\pi r'^{2} \, \d r'
\label{3.76}\eeq
and
\beq
\lim_{r\rightarrow\infty} rV_{\rm e}(r) =
e^{2} \int_{0}^{\infty} \rho(r') 4\pi r'^{2} \, \d r' = e^{2} N.
\label{3.77}\eeq
It is convenient to introduce the {\it screening function} $\Phi(r)$
defined by
\beq
V(r) \equiv - \frac{Ze^{2}}{r} \Phi(r).
\label{3.78}\eeq
Notice that
\beq
\Phi(0) = 1 \qquad \mbox{and} \qquad \Phi(\infty) =
(Z-N)/Z.
\label{3.79}\eeq
The electric charge density, $Ze\delta({\bf r})-e\rho(r)$, and the
electrostatic potential, $V(r)/(-e)$, are related by the Poisson
equation [Eq.\ \req{1.6}]
\beq
\nabla^{2}[-V(r)/e] = -4\pi[Ze\delta({\bf r})-e \rho(r)]
\label{3.80}\eeq
or, using spherical polar coordinates [see Eq.\ \req{B.25}],
\beq
\frac{1}{r} \frac{\d^{2}}{\d r^{2}}[rV(r)] = -4\pi e^{2}\rho(r) \qquad
\mbox{for $r>0$}.
\label{3.81}\eeq

The Thomas--Fermi model of the atom is based on the following two
approximations: \\
$\bullet$ TF1: The potential energy of an {\it atomic} electron at $r$
is given by the expression \req{3.73}, and \\
$\bullet$ TF2: $V(r)$ varies smoothly with $r$. \\
TF1 implies that the potential energy includes the electron
``self-interaction'', that is, $V(r)$ accounts for the screening of the
nuclear charge by the $N$ atomic electrons instead of $N-1$; this
approximation reduces the attraction of atomic electrons by the nucleus
in excess and, consequently tends to increase the atomic volume. The
approximation TF2 means that we can consider a small volume $L^3$ around
a point ${\bf r}$ in which both $V(r)$ and $\rho(r)$ are practically
constant, and we can treat the electrons within $L^3$ as a homogeneous
electron gas.

The maximum energy of an electron at ${\bf r}$ is
\beq
E_{\rm max} = \frac{\hbar^2}{2\me} \, k_{\rm F}^{2} + V(r)
\label{3.82}\eeq
with
\beq
k_{\rm F}(r) = [3\pi^{2}\rho(r)]^{1/3}.
\label{3.83}\eeq
Since the atom is assumed to be in its ground state, $E_{\rm max}$ must
be independent of the position ${\bf r}$ (otherwise the most energetic
electrons would move towards regions with smaller $E_{\rm max}$ values).
In addition, because electrons are bound, we must have $E_{\rm max}\leq
0$.

Combining Eqs.\ \req{3.82} and \req{3.83} we can write
\beq
\rho(r) = \frac{1}{3\pi^{2}} \left(\frac{2\me}{\hbar^{2}}\right)^{3/2}
[E_{\rm max}-V(r)]^{3/2}.
\label{3.84}\eeq
The electron density $\rho(r)$ is not null only in the classically
allowed region where $E_{\rm max} \geq V(r)$. Because the potential
$V(r)$ is attractive (it is negative and increases monotonically with
$r$), the classically allowed region is the interior of a sphere of
radius $r_0$ such that
\beq
E_{\rm max} = V(r_{0}).
\label{3.85}\eeq
Since all the $N$ electrons are inside that sphere,
$V(r_{0})=-(Z-N)e^{2}/r_{0}$ and
\beq
E_{\rm max} = - \frac{(Z-N)e^{2}}{r_{0}}.
\label{3.86}\eeq
Outside the sphere the potential is Coulombian,
\beq
V(r) = - \frac{(Z-N)e^{2}}{r} \qquad (r\geq r_{0}).
\label{3.87}\eeq

Let us introduce the ``maximum kinetic energy'',
\beq
K_{\rm max}(r) \equiv E_{\rm max}-V(r),
\label{3.88}\eeq
and express the electron density, Eq.\ \req{3.84}, as
\beq
\rho(r) = \left\{ \begin{array}{ll}
\displaystyle{\frac{1}{3\pi^{2}}
\left(\frac{2\me}{\hbar^{2}}\right)^{3/2}}
[K_{\rm max}(r)]^{3/2} \rule{5mm}{0mm}
& \mbox{if $K_{\rm max} \geq 0$,} \\ [2mm]
0 & \mbox{if $K_{\rm max}<0$.} \end{array} \right.
\label{3.89}\eeq
On the other hand, the Poisson equation \req{3.81} can be written as
\beq
\frac{1}{r} \frac{\d^{2}}{\d r^{2}}[rK_{\rm max}(r)] = 4\pi e^{2}\rho(r)
\qquad (r>0).
\label{3.90}\eeq
Combining the equations \req{3.89} and \req{3.90} we find that the function
$K_{\rm max}(r)$ satisfies the following differential equation,
\beq
\frac{1}{r} \frac{\d^{2}}{\d r^{2}}[r K_{\rm max}(r)] = \left\{
\begin{array}{ll} \displaystyle{\frac{4e^{2}}{3\pi}
\left(\frac{2\me}{\hbar^{2}}\right)} [K_{\rm max}(r)]^{3/2}
\rule{6mm}{0mm} & \mbox{if $
K_{\rm max} \geq 0$,} \\ \noalign{\smallskip} 0 & \mbox{if $K_{\rm
max}<0$,} \end{array} \right.
\label{3.91}\eeq
with the boundary condition
\beq
\lim_{r\rightarrow 0} r K_{\rm max} (r) = - \lim_{r\rightarrow 0}
rV(r) = Ze^{2}.
\label{3.92}\eeq

Equation \req{3.91} can be simplified by introducing the dimensionless
variable
\beq
x = r/b
\label{3.93}\eeq
with
\beq
b = \frac{(3\pi)^{2/3}}{2^{7/3}} \frac{\hbar^{2}}{\me e^{2}} Z^{-1/3} =
0.88534 \, a_{0} \, Z^{-1/3},
\label{3.94}\eeq
and the function
\beq
\chi(x) = \frac{rK_{\rm max}(r)}{Z e^2}.
\label{3.95}\eeq
The scale parameter $b$ is called the {\it Thomas--Fermi radius}.
\index{Thomas--Fermi radius}
Equation \req{3.91} thus transforms into the {\it Thomas--Fermi
equation}, \index{Thomas--Fermi equation}
\beq
\frac{\d^{2}\chi(x)}{\d x^{2}} = \left\{ \begin{array}{ll}
\displaystyle{\frac{\chi^{3/2}(x)}{x^{1/2}}}
\rule{10mm}{0mm} & \mbox{if $\chi\geq 0$}, \\
\noalign{\smallskip}
0 & \mbox{if $\chi<0$,} \end{array} \right.
\label{3.96}\eeq
and the boundary condition \req{3.92} becomes
\beq
\chi(0) = 1.
\label{3.97}\eeq
Equation \req{3.96} with the condition \req{3.97} admits an infinite
set of solutions, which can be characterized by the value of the
derivative at the origin, $\chi'(0)$. Notice that solutions with
different $\chi'(0)$ values do not intersect, because two solutions that
have identical values at $x=0$ and at another point $x >0$ are
necessarily the same.

Equation \req{3.89} allows obtaining the electron density from the
Thomas--Fermi function,
\beq
\rho(r) = \frac{Z}{4\pi b^{3}} \left[ \frac{\chi(x)}{x} \right]^{3/2}
\qquad \mbox{if $\chi\geq 0$}.
\label{3.98}\eeq
Making use of the differential equation \req{3.96} we can write the more
compact relation
\beq
\rho(r) = \frac{Z}{4\pi b^{3}} \frac{\chi''(x)}{x},
\label{3.99}\eeq
which is valid for all $x$. The number of electrons inside a sphere of
radius $r$ is
\beqa
\int_{0}^{r} \rho(r') 4\pi r'^{2} \, \d r' & = &
Z \int_{0}^{x} \chi''(y)y \, \d y =
Z \left[ y\chi'(y)-\chi(y) \right]_{0}^{x} \nonumber \\[2mm] & = &
Z \left[ x\chi'(x)-\chi(x)+1 \right].
\label{3.100}\eeqa
The potential function \req{3.73},
\beq
V(r) = -\frac{Ze^{2}}{r} +
\left[ \frac{e^{2}}{r} \int_{0}^{r} \rho(r') 4\pi r'^{2} \, \d r' +
e^{2} \int_{r}^{\infty} \rho(r') 4\pi r' \, \d r' \right]
\nonumber \eeq
can be expressed in terms of the Thomas--Fermi function as
\beqa
V(r) & = & \mbox{} - \frac{Ze^{2}}{r} +
\left[ \frac{Ze^{2}}{r} \int_{0}^{x} \chi''(y)y \, \d y +
\frac{Ze^{2}}{b} \int_{x}^{x_{0}} \chi''(y) \, \d y \right] \nonumber
\\[2mm] & = & \mbox{} - \frac{Ze^{2}}{r}
\left[ \chi(x) - x\chi'(x_{0}) \right],
\label{3.101}\eeqa
where $x_0=r_0/b$.
Finally, the screening function defined by Eq.\ \req{3.78} can be
obtained as
\beq
\Phi(r) = - \frac{r}{Ze^{2}} V(r) = \chi(x) - x\chi'(x_{0}).
\label{3.102}\eeq


\subsection{The Thomas--Fermi function for neutral atoms \label{sec3.4.2}}

Numerical calculations show that the solution of Eq.\ \req{3.96} with
$\chi'(0)=-1.58807$ decreases monotonically when $r$ increases and tends
to zero when $r \rightarrow \infty$. For this particular solution
$r_0=\infty$, $\chi'(r_0)=0$, and $r_0 \chi'(r_0)=0$, and from Eqs.\
\req{3.100} and \req{3.102} we have
\beq
\int_{0}^{\infty} \rho(r') 4\pi r'^{2} \, \d r' = Z
\label{3.103}\eeq
and
\beq
\Phi(r) = \chi(x),
\label{3.104}\eeq
respectively. That is, the solution of the Thomas--Fermi equation with
$\chi'(0)=-1.58807$ is the screening function of the neutral atom.
Solutions with $\chi'(0) < -1.58807$ vanish at finite radii $r_0$ and they
can be shown to describe positive ions.

In what follows we shall limit our considerations to the case of neutral
atoms, for which the Thomas--Fermi equation \req{3.96} becomes
\beq
\frac{\d^{2}\chi(x)}{\d x^{2}} =
\frac{\chi^{3/2}(x)}{x^{1/2}} \qquad x \in (0,\infty),
\label{3.105}\eeq
and its solution has the following limiting forms,
\beq
\chi(x) = \left\{ \begin{array}{ll}
1 - 1.58807x + {\cal O}(x^{2}) \rule{5mm}{0mm}& \mbox{if $x\ll 1$} \\
\noalign{\smallskip}
144/x^{3} + {\cal O}(x^{-4}) & \mbox{if $x\gg 1$.} \end{array} \right.
\label{3.106}\eeq
The Thomas--Fermi electron density [Eq.\ \req{3.99}] at small and large
$r$ is
\beq
\rho(r) \sim \left\{ \begin{array}{ll}
1/r^{3/2} \rule{6mm}{0mm}& \mbox{if $r\ll b$,} \\ \noalign{\smallskip}
1/r^{6} & \mbox{if $r\gg b$.} \end{array} \right.
\label{3.107}\eeq
Both expressions are manifestly incorrect. Near the nucleus
$V(r)=-Ze^{2}/r$ and the electron density must remain finite because the
atomic orbitals are nearly hydrogenic. Far from the nucleus the
effective potential energy of an atomic electron reduces to the
interaction energy with the nucleus screened by the other electrons,
that is, $rV(r)\sim -e^{2}$ instead of zero. Consequently, the atomic
orbitals at large $r$ are also hydrogenic and the density $\rho(r)$ must
decrease as a power of $r$ times an exponential, $\sim r^n \exp(-c r)$
[cf.\ Eqs.\ \req{3.31}]. The anomalous behavior at small $r$ is due to
the fast variation of the nuclear potential, while far from the nucleus
the electron density is too small for the statistical model to be valid.
Nonetheless, the Thomas--Fermi model provides reasonable estimates of
the electron density and the potential for a wide range of intermediate
$r$ values for atoms with atomic numbers larger than about 15.

\begin{subequations}
The electron density
\label{3.108}
\beq
\rho(r) = \frac{Z}{4\pi b^{3}} \, \frac{\chi''(x)}{x}
= \frac{Z}{4\pi r} \, \frac{\d^2 \Phi(r)}{\d r^2}
\label{3.108a}\eeq
and the potential
\beq
V(r) = - \, \frac{Ze^{2}}{b} \, \frac{\chi(x)}{x}
= - \, \frac{Z e^2}{r} \, \Phi(r)
\label{3.108b}\eeq
\end{subequations}
are smooth functions of $r$ that have the same form for all neutral
atoms; they differ only by a simple change of scale. Thus, the electron
density of an atom of atomic number $Z$ is
\begin{subequations}
\label{3.109}
\beq
\rho(Z;r) = Z^2 \rho(1;Z^{1/3} r),
\label{3.109a}\eeq
where $\rho(1;r)$ is the density of atoms with $Z=1$. Similarly,
\beq
\Phi(Z;r) = \Phi(1; Z^{1/3} r).
\label{3.109b}\eeq
This feature allows
obtaining analytical expressions for various atomic properties in terms
of only the atomic number $Z$. However, the Thomas--Fermi model is not
able to describe properties that are specific of one atomic species. For
instance, the electron density \req{3.108a} does not exhibit the
oscillations of the density resulting from self-consistent Dirac--Fock
calculations (see below) that arise from contributions of the various
electron subshells.
\end{subequations}

\index{Thomas--Fermi screening function!Lenz--Jensen}
The Thomas--Fermi equation is difficult to solve numerically because of
the singularity at $x=0$, which requires special treatment. The
$\chi(x)$ function was calculated numerically by
\citet{BushCaldwell1931}, \citet{Feynman1949},
and others; a fairly dense tabulation is given by
\citet{CondonOdabasi1980}. However, describing the screening function by
means of a numerical table is not convenient for calculations. Because of
the ability of the Thomas--Fermi model to account for global atomic
properties independently of the atomic number, various analytical
approximations have been proposed. Early calculations of stopping theory
made use of the approximate solution of the Thomas--Fermi equation
derived by \citet{Lenz1932} and \citet{Jensen1932}. With the parameters
determined by a modern calculation \citep{BisterHautala1978}, the
Lenz--Jensen screening function is
\begin{subequations}
\label{3.110}
\beq
\chi_{\rm LJ1} (x) = \exp(-y) \left( 1 + y +0.33443 y^2 + 0.048505 y^3
+0.0026477 y^4 \right)
\label{3.110a}\eeq
with
\beq
y = \sqrt{9.6608 \, r/b}.
\label{3.110b}
\eeq
\end{subequations}
\citet{BisterHautala1978} obtained a more accurate approximation to the
Thomas--Fermi function by considering additional terms in the
Lenz--Jensen series. Their expanded formula reads
\begin{subequations}
\label{3.111}
\beq
\chi_{\rm LJ2} (x) = \exp(-y) \left( 1 + y + \sum_{n=2}^{10} a_n y^n
\right)
\label{3.111a}
\eeq
where
\beq
\begin{array}{lll}
a_2 = 3.2868 \times 10^{-1}, \rule{10mm}{0mm}&
a_3 = 4.2558 \times 10^{-2}, \rule{10mm}{0mm}&
a_4 = 3.3205 \times 10^{-3}, \\ [0mm]
a_5 = 9.4398 \times 10^{-4}, &
a_5 = 9.4440 \times 10^{-5}, &
a_7 = -3.9155 \times 10^{-6}, \\ [0mm]
a_8 = 1.5804 \times 10^{-6}, &
a_9 = -5.7497 \times 10^{-8}, &
a_{10} = 4.3668 \times 10^{-9},
\end{array}
\label{3.111b}\eeq
with
\beq
y = \sqrt{9.2704 \,  r/b}.
\label{3.111c}\eeq
\end{subequations}
These analytical forms are not suited to our purposes. We find
advantageous to use the following approximation, which was obtained by
\citet{Moliere1947} by fitting the numerical Thomas--Fermi screening
function,
\index{Thomas--Fermi screening function!Moli\`{e}re}
\beq
\chi_{\rm M}(x) =
0.1\, \exp({-6x}) + 0.55\, \exp(-1.2x) + 0.35\, \exp(-0.3x).
\label{3.112}\eeq
Values obtained from this formula differ from the exact solution of the
Thomas--Fermi equation \req{3.105} in less than $2\times 10^{-3}$
(absolute difference) for $x<6$. Figure \ref{fig3.1} compares the
numerical Thomas--Fermi screening function and these approximations.
Although the expanded Lenz--Jensen approximation is slightly more
accurate, the Moli\`{e}re formula is better than the original
Lenz--Jensen formula \req{3.110}.  In addition, the expression
\req{3.112} largely facilitates the calculation of elastic collisions of
charged particles with atoms (see Chapter \ref{chapt5}).

\begin{figure}[tbh!] \begin{center}
\includegraphics*[width=10.0cm]{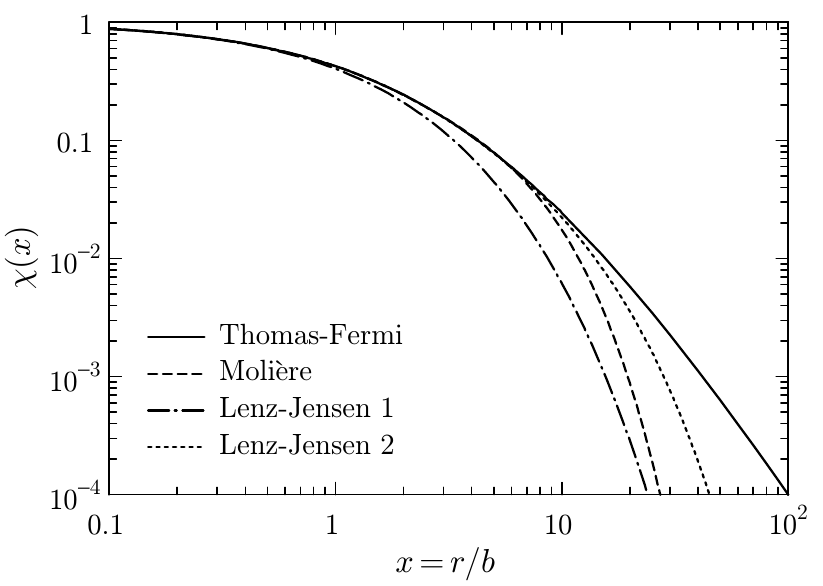}
\caption{
Thomas--Fermi screening function for neutral atoms, as a function of the
reduced radial distance $x=r/b$. The solid curve represents the numerical
solution of the Thomas--Fermi equation. The other curves represent the
Lenz--Jensen approximations, Eqs.\ \req{3.110} (dot-dashed curve) and
\req{3.111} (dotted curve), and the
Moli\`{e}re approximation, Eq.\ \req{3.112} (dashed curve).
\label{fig3.1}}
\end{center}\end{figure}

The electron density corresponding to the Moli\`{e}re screening function
[see Eq.\ \req{3.108a}],
\beq
\rho_{\rm M}(r) = \frac{Z}{4\pi b^2 r}
\left[ 3.6 \exp({-6 \, r/b}) + 0.792\, \exp(-1.2 \, r/b)
+ 0.0315 \exp(-0.3\, r/b) \right],
\label{3.113}\eeq
diverges as $r^{-1}$ near the nucleus and decreases exponentially at
large radii. Realistic electron densities of free atoms obtained from
self-consistent Dirac--Fock calculations (see Section \ref{sec3.5}) are
finite near the nucleus and decrease roughly exponentially at large $r$.
Although the Moli\`{e}re screening function inherits the deficiency
of the Thomas--Fermi model at small $r$, its large-$r$ behavior is more
``natural'' than that of the numerical Thomas--Fermi density and, as a
consequence, the Moli\`{e}re approximation \req{3.112} gives results
more realistic that the Thomas--Fermi function for quantities that are
sensitive to the ``tail'' of the atomic electron distribution.


\subsection{Binding energy of the Thomas--Fermi atom \label{sec3.4.3}}

\index{Thomas--Fermi model of the atom!binding energy}
Because of its universal character, the Thomas--Fermi model is useful to
describe the variation of global atomic properties with the atomic
number. As an interesting example, which is employed in discussions of
the capture of electrons by slow heavy ions (see Section \ref{sec8.9}),
we evaluate here the total binding energy of the Thomas--Fermi atom. We
consider
\beq
\langle E\rangle = \langle K\rangle + \langle V\rangle
\label{3.114}\eeq
with
\beq
\langle K\rangle =
\int_{0}^{\infty} \rho(r) \, \overline{K}(\rho) \, 4\pi r^{2} \, \d r
\label{3.115}\eeq
and
\beq
\langle V\rangle = \int_{0}^{\infty}
\rho(r) \, \left[ -
\frac{Z e^2}{r} + \frac{1}{2}V_{\rm e}(r) \right] \, 4\pi r^{2} \, \d r,
\label{3.116}\eeq
where $\overline{K}(\rho)$ is the average kinetic energy of the
electrons in a gas of density $\rho$, Eq.\ \req{3.71},
\beq
\overline{K}(\rho) = \frac{3}{5} \, E_{\rm F}(\rho) = \frac{3}{5} \,
\frac{\hbar^2}{2\me} \, \left( 3 \pi^2 \rho \right)^{2/3},
\label{3.117}\eeq
and
\beq
V_{\rm e} (r) = V(r) + \frac{Z e^2}{r}
\label{3.118}\eeq
is the electrostatic potential of the electron cloud, which is
multiplied by the factor $\1o2$ to count the interaction between each pair of
electrons only once. Expressing $\rho$ and $V_{\rm e}$ in terms of the
Thomas--Fermi function $\chi$ --Eqs.\ \req{3.98} and \req{3.101}-- we
have
\begin{subequations}
\label{3.119}
\beqa
\langle K\rangle & = & \frac{3}{5} \frac{Z^{2}e^{2}}{b}
\int_{0}^{\infty} \frac{\chi^{5/2}(x)}{x^{1/2}} \, \d x,
\label{3.119a} \\ [2mm]
\langle V\rangle & = & - \frac{1}{2} \frac{Z^{2}e^{2}}{b}
\int_{0}^{\infty} [1+\chi(x)] \frac{\chi^{3/2}(x)}{x^{1/2}} \, \d x.
\label{3.119b}\eeqa
\end{subequations}
The relevant integrals are
\beq
\int_{0}^{\infty} \frac{\chi^{3/2}(x)}{x^{1/2}} \, \d x = - \chi'(0)
\qquad \mbox{and} \qquad
\int_{0}^{\infty} \frac{\chi^{5/2}(x)}{x^{1/2}} \, \d x =
- \frac{5}{7}\chi'(0).
\label{3.120}\eeq
The first of these formulas follows evidently from the Thomas--Fermi
equation \req{3.96}. The second one is not at all trivial; it can be
obtained as a
consequence of the virial theorem (see below), or derived by the
following intricate sequence of integrations by parts. On the one hand,
we have
\begin{subequations}
\label{3.121}
\beqa
\int_0^\infty \frac{\chi^{5/2}(x)}{x^{1/2}} \, \d x &=&
\int_0^\infty \chi(x) \, \chi''(x) \, \d x
= \left[ \chi(x) \chi'(x) \rule{0mm}{4mm} \right]_0^\infty
- \int_0^\infty \chi'^2(x) \, \d x
\nonumber \\ [2mm]
&=& - \chi'(0) - \int_0^\infty \chi'^2(x) \, \d x.
\label{3.121a}\eeqa
On the other hand
\beqa
\int_0^\infty \frac{\chi^{5/2}(x)}{x^{1/2}} \, \d x
&=& \left[ x^{1/2} \chi^{5/2}(x) \rule{0mm}{4mm} \right]_0^\infty
- \int_0^\infty x \left( \frac{5}{2} \, \frac{\chi^{3/2}(x)}{x^{1/2}}\,
\chi'(x) - \frac{1}{2} \frac{\chi^{5/2}(x)}{x^{1/2}} \right) \d x
\nonumber \\ [2mm]
&=& - \frac{5}{2} \int_0^\infty x \, \chi''(x) \, \chi'(x) \, \d x +
\frac{1}{2}
\int_0^\infty \frac{\chi^{5/2}(x)}{x^{1/2}} \, \d x,
\nonumber \eeqa
which gives
\beq
\int_0^\infty \frac{\chi^{5/2}(x)}{x^{1/2}} \, \d x
= - 5 \int_0^\infty x \, \chi''(x) \, \chi'(x) \, \d x.
\nonumber \eeq
The last integral can be transformed by observing that
\beqa
\int_0^\infty x \, \chi''(x) \, \chi'(x) \, \d x
&=& \left[ x \chi'^2(x) \rule{0mm}{4mm} \right]_0^\infty
- \int_0^\infty  \chi'(x) \left[ \chi'(x) + x \chi''(x) \right] \d x
\nonumber \\ [2mm]
&=&
- \int_0^\infty  \chi'^2(x) \, \d x
- \int_0^\infty x \, \chi'(x) \chi''(x) \d x
\nonumber \eeqa
and hence
\beq
\int_0^\infty x \, \chi''(x) \, \chi'(x) \, \d x = - \, \frac{1}{2} \,
\int_0^\infty  \chi'^2(x) \, \d x.
\nonumber \eeq
Therefore,
\beq
\int_0^\infty \frac{\chi^{5/2}(x)}{x^{1/2}} \, \d x
= \frac{5}{2} \int_0^\infty \chi'^2(x) \, \d x.
\label{3.121b}\eeq
\end{subequations}
Combining the equalities \req{3.121a} and \req{3.121b}, we finally
obtain the second formula in Eq.\ \req{3.120}.

Upon replacement of the integrals \req{3.120}, the expressions \req{3.119}
become \index{virial theorem}
\beq
\langle K\rangle = - \frac{3}{7} \frac{Z^{2}e^{2}}{b} \chi'(0),
\qquad
\langle V\rangle = \frac{6}{7} \frac{Z^{2}e^{2}}{b} \chi'(0).
\label{3.122}\eeq
These results show that the virial theorem, $\langle V\rangle = -2
\langle K\rangle$, is satisfied. Inserting the value $\chi'(0)=-1.58807$
of the neutral atom, and the expression \req{3.94} of $b$, we obtain
\beq
\langle E\rangle = \langle K\rangle + \langle V\rangle =
- \langle K\rangle = -0.768745 Z^{7/3} \; E_{\rm h}.
\label{3.123}\eeq
Notice that the average kinetic energy of each atomic electron is
proportional $Z^{4/3}$ and hence the average electron velocity is
proportional to $Z^{2/3}$.


\section{DHFS self-consistent method for atoms \label{sec3.5}}

\index{atomic configuration!ground state}
\index{DHFS self-consistent model}
A more elaborate independent-electron model of atoms is provided by the
Dirac--Hartree--Fock--Slater (DHFS) self-consistent method. The states
of the atom are represented by single Slater determinants built with
Dirac central-field orbitals of the form \req{3.56}. We consider the
ground-state configuration of the atom (see Table \ref{tab3.1}),
specified by giving the occupied subshells $a=(n_a\kappa_a)$ and the
corresponding occupancy numbers $\nu_a$ ($\le 2j_a+1=2|\kappa_a|$);
\beq
(n_1\kappa_1)^{\nu_1}\, , \, \, (n_2\kappa_2)^{\nu_2}\, .
\, \ldots
\label{3.124}\eeq

The orbitals $\psi_{n\kappa m}(x)$ are solutions of the DHFS equations,
\beq
\left[ c\, \widetilde{\alphab} \dotprod {\bf p} +
(\widetilde{\beta} -1) \me c^2
+ V_{\rm DHFS}(r) \right] \psi_{n\kappa m}(x) = \varepsilon_{n\kappa}
\, \psi_{n\kappa m}(x) \, ,
\label{3.125}\eeq
for a common central potential $V_{\rm DHFS}(r)$. These equations are
obtained from the single-configuration Dirac--Fock equations \citep[see,
\eg][]{Grant1970} by replacing the non-local exchange potential with the
local average approximation due to \citet{Slater1951}, which is derived
from the non-relativistic free-electron gas theory. Alternatively, the
DHFS equations result from the non-relativistic Hartree--Fock--Slater
equations \citep[see, \eg,][]{CondonOdabasi1980} when the kinetic energy
operator, ${\bf p}^2/(2\me)$, is replaced with that of Dirac's theory,
Eq. \req{2.89}. Since Eqs.\
\req{3.125} have the form of the Dirac equations for an electron in a
central potential, the  radial functions $P_{n\kappa}(r)$ and
$Q_{n\kappa}(r)$ of the DHFS orbitals satisfy the radial equations
\req{3.57},
\beq
\left. \begin{array}{l}
\displaystyle{\frac{\d P_{n\kappa}}{\d r} =
- \frac{\kappa}{r} \,P_{n\kappa} + \frac{\varepsilon_{n\kappa}-V_{\rm DHFS}+2\me
c^2}{c\hbar} \, Q_{n\kappa}},
\\ [4mm]
\displaystyle {\frac{\d Q_{n\kappa}}{\d r} =
- \frac{\varepsilon_{n\kappa}-V_{\rm DHFS}}{c\hbar} \, P_{n\kappa}}
+ \frac{\kappa}{r}\, Q_{n\kappa}.
\end{array} \right.
\label{3.126}\eeq
The electron density is obtained as
\beq
\rho(r) = \frac{1}{4\pi r^2} \sum_{a} \nu_a \left[
P_{n_a \kappa_a }^2(r)
+ Q_{n_a \kappa_a }^2 (r) \right] ,
\label{3.127}\eeq
where the summation is over occupied subshells. In calculations with
arbitrary configurations, expression \req{3.127} is adopted as the
definition of the local electron density. Thus, for configurations with
open subshells, spherical symmetry of the density (and the potential) is
enforced by considering that the orbitals of an open subshell
$(n_a\kappa_a)$ with $\nu_a$ electrons have a fractional occupancy equal
to $\nu_a / (2j_a + 1)$.

The DHFS potential is given by
\beq
V_{\rm DHFS}(r) =
V_{\rm nuc}(r) + V_{\rm el}(r) + V_{\rm ex}(r)\, .
\label{3.128}\eeq
The term
\beq
V_{\rm nuc}(r) = - e \, \varphi_{\rm nuc}(r)
\label{3.129}\eeq
is the nuclear potential, \ie, the
interaction energy of an electron at $r$ with the nucleus. The finite
size of the nucleus has a small effect on the atomic structure, which
can be approximately accounted for, without complicating the computer
code, by considering the nucleus as a spherical distribution of charge.
In our calculations we use the Fermi distribution \req{3.4},
with its electrostatic potential computed numerically, Eq.\ \req{3.6}.
The second term in expression \req{3.128} is the electronic potential
(\ie, the interaction energy of an electron at $r$ with the whole atomic
electron cloud),
\beqa
V_{\rm el}(r) &=& e^2 \int \frac{\rho(r')}{|{\bf r} - {\bf
r}'|} \, \d {\bf r}'
\nonumber \\ [2mm]
&=& \frac{e^2}{r}
\int_0^r \, \rho(r') \, 4\pi r'^2\, \d r'
+ e^2 \int_r^\infty \rho(r')\, 4\pi r' \, \d r'\, .
\label{3.130}\eeqa
The last term on the right-hand side of Eq.\ \req{3.128} is the
exchange potential, which can be expressed as
\beq
V_{\rm ex}(r) =  C_{\rm ex} \, V_{\rm ex}^{\rm (TF)}(r) \, ,
\label{3.131}\eeq
where $C_{\rm ex}$ is a constant and
\beq
V_{\rm ex}^{\rm (TF)}({\bf r})
= - e^2 (3/\pi)^{1/3} \, [\rho({\bf r})]^{1/3}
\label{3.132}\eeq
is the exchange potential of a free-electron gas of density $\rho({\bf
r})$. \citet{Slater1951}, applied the Thomas--Fermi approximation to the
non-relativistic Hartree--Fock exchange potential (after variation) and
obtained $C_{\rm ex}=3/2$. Later, \citet{KohnSham1965} using arguments
that are equivalent to a variational formulation in which the
non-relativistic Hartree--Fock exchange energy (before variation) is
estimated by means of the Thomas--Fermi approximation, obtained an
exchange potential of the same form but with $C_{\rm ex} = 1$. This
indicates that there is a certain arbitrariness in the value of this
parameter, which has been exploited to seek improved approximations
\citep[see, \eg,][and references therein]{CondonOdabasi1980}. Since the
value $C_{\rm ex}=3/2$ results from applying the Thomas--Fermi
approximation to the exchange potential, the energy eigenvalues
$\varepsilon_{n\kappa}$ keep the physical significance of one-electron
binding energies, as in the Dirac--Fock method (Koopmans' theorem). For
this reason, in the calculations reported below we set $C_{\rm ex}=3/2$.

\index{DHFS self-consistent model!Latter-tail correction}
\index{Latter-tail correction}

Near the nucleus, the exchange potential is outweighed by the nuclear
potential. At intermediate distances, the electron density is relatively
large and the Thomas--Fermi (statistical) approximation to the exchange
potential is nearly correct. However, for large $r$, the electron
density is very small and the statistical approximation is not accurate;
the resulting local exchange potential is so weak there that it cannot
cancel the self-interaction term in the electronic potential. We have,
\beq
\lim_{r\rightarrow \infty} r V_{\rm DHFS}(r) = -(Z-N) e^{2},
\label{3.133}\eeq
which is incorrect. To obtain the correct asymptotic behavior [$=
-(Z-N+1)e^{2}$] it is customary to introduce an {\it ad hoc\/} correction,
known as the Latter tail \citep{Latter1955}, which consists of replacing
$rV_\textrm{DHFS}(r)$ with $-(Z-N+1)e^{2}$ at large distances where the
original DHFS potential becomes higher than this value. The corrected
exchange potential is
\beq
V_{\rm ex}(r) = \left\{
\begin{array}{ll}
C_{\rm ex} V_{\rm ex}^{\rm (TF)}(r)
& \mbox{if $r < r_{\rm Latter}$,}
\\ [2mm]
\displaystyle{-(Z-N-1) \frac{e^2}{r} - V_{\rm nuc}(r) - V_{\rm el}(r)}
\rule{8mm}{0mm}
& \mbox{if $r \ge r_{\rm Latter}$,}
\end{array} \right.
\label{3.134}\eeq
where the cutoff radius $r_{\rm Latter}$ is the outer root of the
equation
\beq
r \left[ V_{\rm nuc}(r) + V_{\rm el}(r)
+ \alpha V_{\rm ex}^{\rm (TF)}(r) \right]  = -(Z-N+1) e^2.
\label{3.135}\eeq
Evidently, this correction yields a continuous exchange potential with the
correct large-$r$ behavior, although the derivative of $V_{\rm ex}(r)$
has a discontinuity at $r_{\rm Latter}$.

Evidently, the DHFS radial equations must be solved numerically. The
Fortran program {\tt DHFS}, which is distributed with the {\sc radial}
subroutine package \citep{SalvatFernandezVarea2019}, computes
one-electron energy eigenvalues, $\varepsilon_{n\kappa}$, radial
functions, and DHFS potentials for neutral atoms and positive ions. The
results are equivalent to those from the similar program developed by
\citet{Liberman1965, Liberman1971}. These programs are available from
the Computer Physics Communications Program Library.

\index{Free-atom boundary conditions}

The {\sc radial} subroutines calculate the one-electron eigenvalues
$\varepsilon_{n\kappa}$ and the radial functions by solving the
equations \req{3.126} with free-atom boundary conditions. That is, with
$P_{n \kappa} (r)$ and $Q_{n \kappa} (r)$ going to zero when $r$ tends
to infinity.


\subsection{Free-atom and Wigner--Seitz boundary conditions
\label{sec3.5.1}}

\index{DHFS self-consistent model!Wigner--Seitz boundary conditions}
\index{Wigner--Seitz boundary conditions}

In many practical cases, atoms are bound in solids and it is of interest
to study what is the effect of aggregation on the atomic orbitals and
ionization energies, at least in an approximate manner. To simplify
arguments, we shall consider a single-element solid of mass density
$\rho_{\rm m}$ (g/cm$^3$). A simple and convenient approach is provided by the
Wigner--Seitz model \citep[see, \eg,][]{Tucker1969}, which assumes that
the atomic charge is fully contained within a sphere of radius $r_{\rm
WS}$ and volume equal to the average atomic volume, \ie,
\beq
\frac{4\pi}{3} \, r_{\rm WS}^3 =A_{\rm m}/(N_{\rm A} \rho_{\rm m}),
\label{3.136}\eeq
where $A_{\rm m}$ is the molar mass or atomic weight (in g/mol) and
$N_{\rm A}$ is Avogadro's constant. Inserting the numerical values of the
constants,
\beq
r_{\rm WS} = 1.3882 \left( A_{\rm m}/ \rho_{\rm m} \right)^{1/3} a_0.
\label{3.137}\eeq
Alternatively, for elemental solids one can set $r_{\rm WS}$ equal to
half the nearest neighbor distance \citep{Kittel1976}; this corresponds
to the so-called muffin-tin model. The potential is spherically
symmetric inside the sphere and constant outside it. To account for the
periodicity of the solid lattice, we would like equivalent orbitals in
neighboring spheres to join smoothly at their surfaces. In a
non-relativistic formulation, this is accomplished (again,
approximately) by requiring that the radial functions $r^{-1} P(r)$ of
odd (even) orbitals vanish (have null derivative) at the surface of the
Wigner--Seitz sphere. In a relativistic calculation, it is not possible
to match simultaneously the large and small radial functions of
corresponding orbitals of neighboring atoms. Nonetheless, the only
orbitals that extend to distances of the order of $r_{\rm WS}$ are those
of outer shells; these are weakly relativistic and, therefore, the small
components of their wave functions at $r_{\rm WS}$ are negligible
[$|Q_{n \kappa}(r_{\rm WS})| \ll |P_{n \kappa}(r_{\rm WS})|$]. We can,
therefore, get quite smooth joining of orbitals of neighboring atoms by
simply matching the large radial functions. The practical recipe is to
require
\beq
\begin{array}{ll}
P_{n \kappa}(r_{\rm WS}) = 0 & \mbox{if $\ell$ is odd}, \\ [3mm]
\displaystyle{\frac{\d }{\d r} \left[ r^{-1} P_{n \kappa}(r_{\rm WS})
\right] = 0}
\rule{10mm}{0mm}
& \mbox{if $\ell$ is even}. \end{array}
\label{3.138}\eeq
Combining these conditions with the radial equations \req{3.126} we
obtain the following boundary values at $r_{\rm WS}$ (apart from a
normalization constant)
\beq
\begin{array}{lll}
P_{n \kappa}(r_{\rm WS}) = 0, \; \; \; & Q_{n \kappa}(r_{\rm WS})
= 1 & \mbox{if $\ell$ is
odd,} \\ [3mm]
P_{n \kappa}(r_{\rm WS}) = 1, & \displaystyle{
Q_{n \kappa}(r_{\rm WS}) =
\frac{c\hbar}{E-V(r_{\rm WS}) + 2\me c^2} \, \frac{1+\kappa}{r_{\rm
WS}}} \rule{5mm}{0mm} & \mbox{if $\ell$ is even.}
\end{array}
\label{3.139}\eeq
These values determine the solution of the DHFS equations unambiguously.


\subsection{Self-consistent solution method\label{sec3.5.2}}

\index{DHFS self-consistent model}
The self-consistent solution of the DHFS equations \req{3.125} is
started by considering the approximate electron density $\rho_0(r)$,
which is obtained from the \citet{Moliere1947} parameterization of the
Thomas--Fermi potential of neutral atoms [Eq.\ \req{3.113}],
\beq
\rho_0 (r) = \frac{N}{4\pi b^2 r}
\left[ 3.6\, {\rm e}^{\displaystyle{-6r/b}}
+ 0.792\, {\rm e}^{\displaystyle{-1.2r/b}}
+ 0.0315\, {\rm e}^{\displaystyle{-0.3r/b}} \right],
\label{3.140}\eeq
where $b=0.88534\,Z^{-1/3}\,a_{0}$ is the Thomas--Fermi radius, Eq.\
\req{3.94}. The
self-consistent solution then proceeds as follows. From the initial
density $\rho_0$ (properly normalized to the number of bound electrons)
we compute a first approximation to the DHFS
potential, Eq.\ \req{3.128}, which we call $V_1$. With this potential,
a set of radial functions and energy eigenvalues is obtained by solving
the radial Dirac equations \req{3.126} for the various subshells. This
yields a new electronic density $\rho_1$ from which we obtain a new
approximation to the DHFS potential, which we indicate as $V_{2{\rm
n}}$. The potential for the next iteration is the weighted average
\beq
V_2 = (1-w) V_1 + w V_{2{\rm n}}.
\label{3.141}\eeq
Inserting this potential into the radial equations, we compute a new set
of radial functions and energy eigenvalues, from which we obtain a new
electron density $\rho_2$, etc. The calculation continues by repeating
these steps and ends when the maximum relative difference between the
potentials in successive iterations is less than the specified
tolerance, $10^{-9}$ in the {\tt DHFS} program of
\citet{SalvatFernandezVarea2019}. At the beginning, the weight $w$ is
set equal to 0.05 and its value is gradually increased as the
calculation progresses, to reach a maximum value of 0.5. This kind of
mixture of the ``new'' and ``old'' potentials is necessary to ensure
convergence for all atoms and positive ions.

Configurations with only one or two electrons ($N=1,2$) in a single subshell
should be handled separately. In these cases we may take $C_{\rm ex}=0$
and set the exchange potential equal to the self interaction energy,
\ie, $N^{-1}$ times the electronic potential. This gives exact results
for the hydrogen atom and one-electron ions in any state. This approach
is also appropriate for helium and two-electron ions
in their ground configuration and in doubly excited symmetric
configurations of the type $(n\kappa)^2$ with $\kappa=-1$, for which the
exchange interaction is negligible \citep{BransdenJoachain1983}.

The program {\tt DHFS} delivers the complete set of results from the
self-consistent calculation, namely, one-electron energy levels,
$\varepsilon_{n\kappa}$, the radial functions $P_{n\kappa}(r)$ and
$Q_{n\kappa}(r)$ of the various subshells, the electron density
$\rho(r)$ and the various components of the self-consistent DHFS
potential $V_{\rm DHFS}(r)$. The program accepts both free-atom and
Wigner--Seitz boundary conditions.

Figure \ref{fig3.2} displays the various components of the
self-consistent DHFS potential (calculated with $C_{\rm ex}=3/2$) of
aluminum ($Z=13$) with a finite nucleus; the corresponding electron
density is shown in Figs.\ \ref{fig3.3} and \ref{fig3.4}. For radii
smaller than $r_{\rm Latter}$, the exchange potential varies smoothly
with $r$; at $r_{\rm Latter}$ the exchange potential changes form (it is
forced to cancel the electron self-interaction) and there is a
discontinuity in the first derivative of $r V_{\rm ex}(r)$.

\begin{figure}[p!] \begin{center}
\includegraphics*[width=11.0cm]{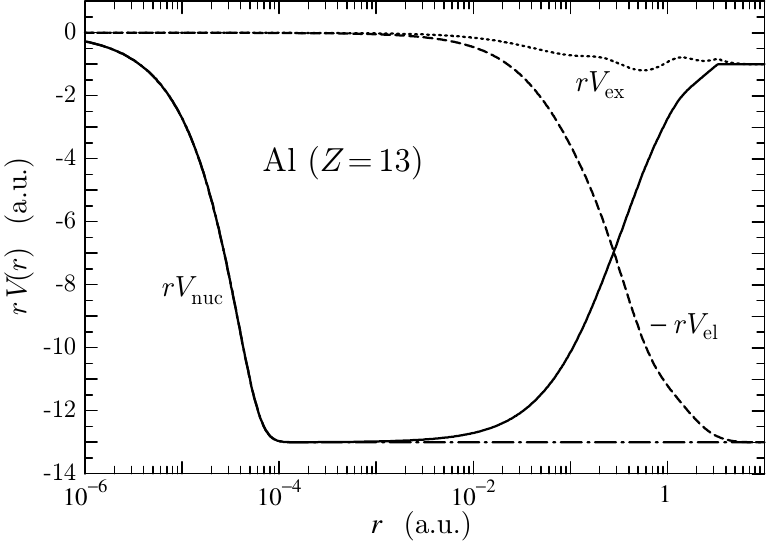}
\caption{
DHFS self-consistent potential $rV_{\rm DHFS}(r)$ of the aluminium atom
(solid curve). The nuclear potential (dot-dashed curve) corresponds to a
finite nucleus, represented by the Fermi distribution, Eq.\ \req{3.4}.
The exchange potential includes the Latter tail correction, Eq.\
\req{3.134}. All quantities are in atomic units (a.u.).
\label{fig3.2}}
\vspace*{7mm}
\includegraphics*[width=7.5cm]{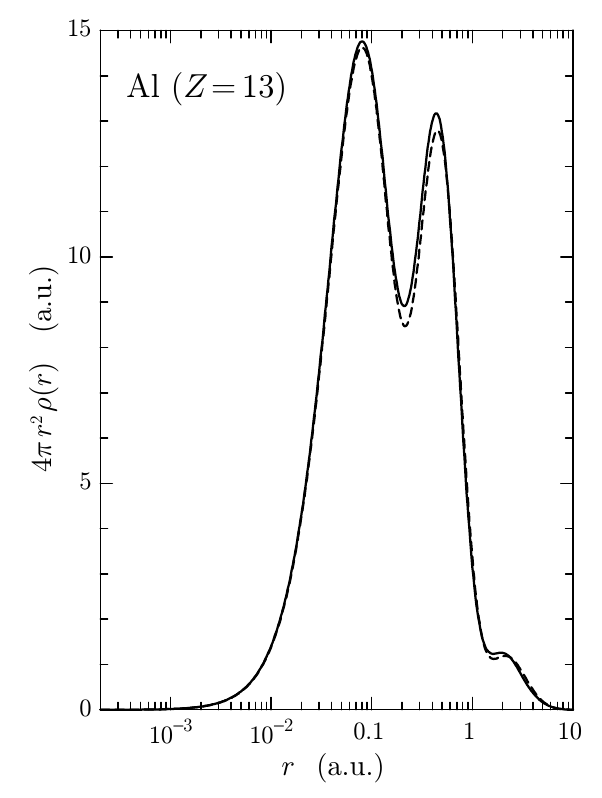} \rule{5mm}{0mm}
\includegraphics*[width=7.5cm]{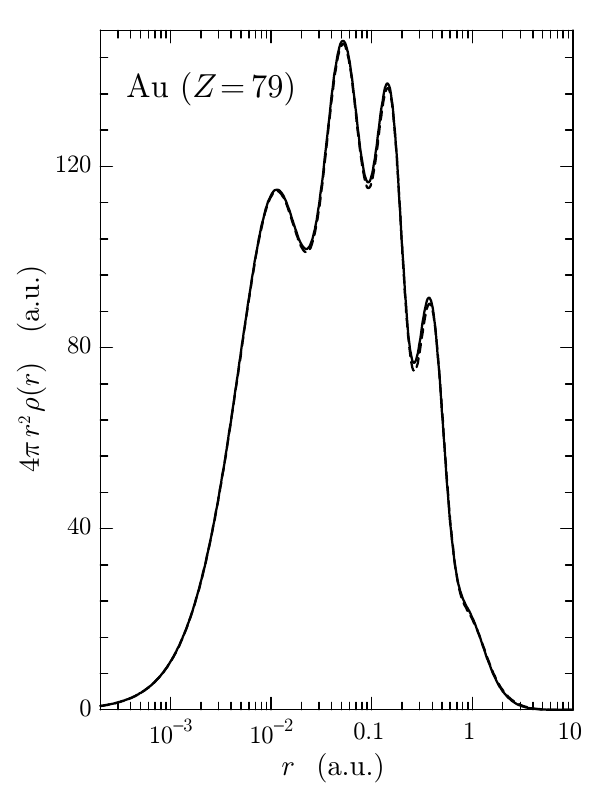}
\caption{
Radial electron densities, $4\pi r^2 \rho(r)$, for the neutral-atom
ground-state configurations of aluminium and gold, obtained with the
DHFS method and calculated using the {\sc mcdf} code (dashed lines). All
quantities are in atomic units (a.u.).
\label{fig3.3}}
\end{center}\end{figure}

Figure \ref{fig3.3} shows the self-consistent radial electron densities
of aluminum and gold ($Z=79$) given by the {\tt DHFS} program, which are
compared with results from the multi-configuration Dirac--Fock program
{\sc mcdf} of Desclaux (\citeyear{Desclaux1975}),
which is considered as the state of
the art in atomic structure calculations. The DHFS method is seen to
provide a better than acceptable description of the atomic ground state.
Electron densities for free atoms and Wigner--Seitz atoms of the
elements aluminum and gold, calculated with the DHFS method, are
displayed in Fig.\ \ref{fig3.4}, together with the Tomas-Fermi densities
of the free atoms. Differences between the electron densities of free
and bound atoms are appreciable only for relatively large radii. Notice
that, as a result of the boundary conditions \req{3.139}, the electron
density at $r = r_{\rm  WS}$ ($\sim 2.7 a_0$ for both elements) is
``flat''. The Thomas--Fermi density is seen to provide a reasonable
qualitative approximation for the bulk of the electron density.

\begin{figure}[tbh!] \begin{center}
\includegraphics*[width=7.3cm]{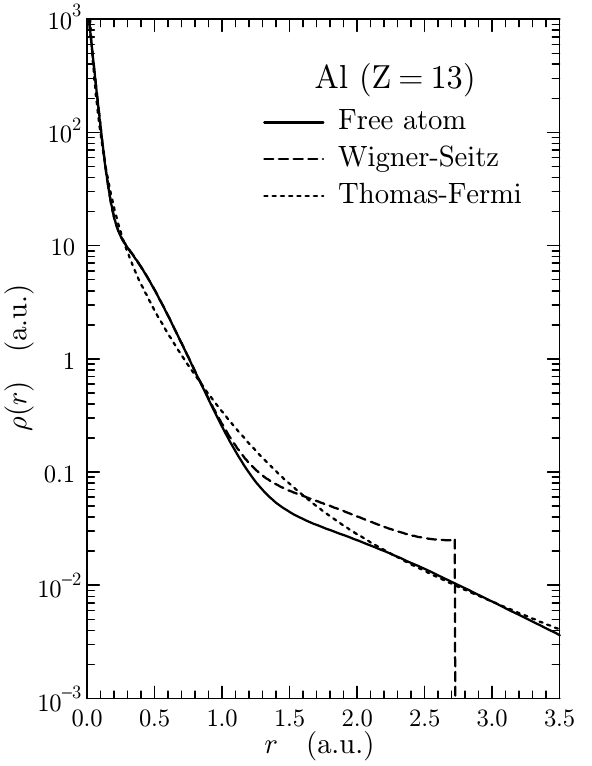} \rule{3mm}{0mm}
\includegraphics*[width=7.3cm]{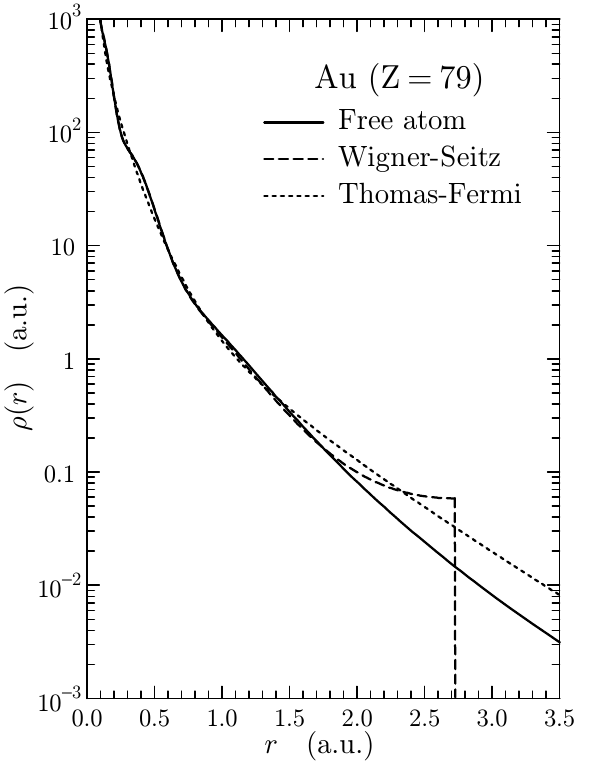}
\caption{
Electron densities for the ground-state configurations of aluminium and
gold atoms under free-atom and Wigner--Seitz boundary conditions,
calculated with the {\tt DHFS} program. The Thomas--Fermi--Moli\`{e}re
electron density, \req{3.113}, is also shown for comparison.
All quantities are in atomic units (a.u.).
\label{fig3.4}}
\end{center}\end{figure}

\index{DHFS self-consistent model!energy eigenvalues}

The absolute values of the DHFS energy eigenvalues, $U_{n\kappa}
= - \varepsilon_{n\kappa}$, of neutral atoms are
compared to the experimental subshell ionization energies
\citep{Carlson1975} in Fig.\ \ref{fig3.5}. As mentioned above, DHFS
eigenvalues keep the significance of one-electron binding energies and,
indeed, they are close to the actual ionization energies for subshells with
$U_{n\kappa}$ larger than about 200~eV. Because of this peculiarity,
the self-consistent DHFS potential is employed in many calculations of
interactions of high-energy radiations with inner-shell electrons
\citep[see, \eg,][]{Pratt1973, BoteSalvat2008, Salvat2022a}.

\index{subshell ionization energies}
\begin{figure}[hp] \begin{center}
\includegraphics*[width=13.0cm]{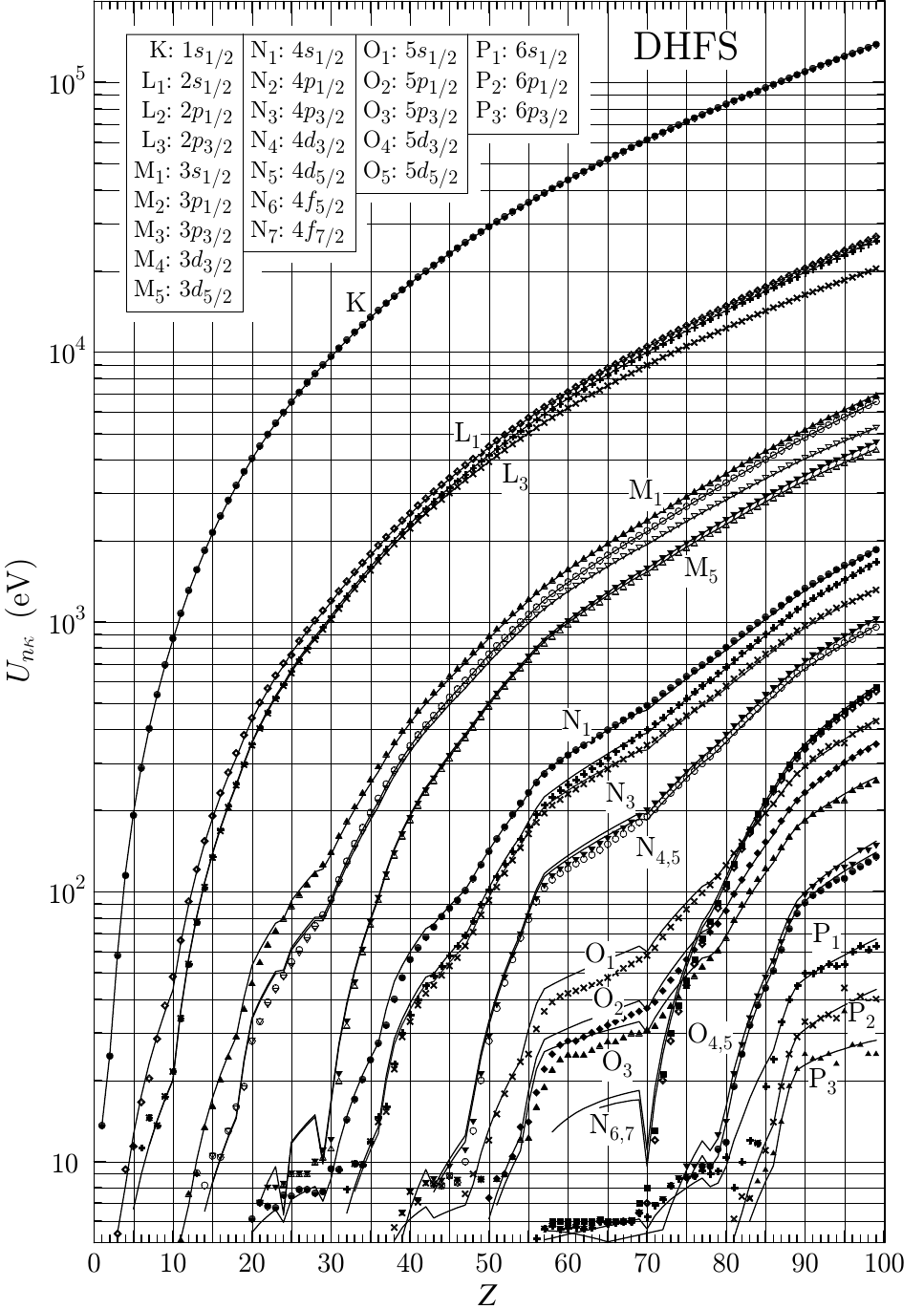}
\caption {\rm
Subshell ionization energies of neutral atoms, in eV. Symbols are
experimental values taken from \citet{Carlson1975}. Lines represent the
absolute values of the one-electron energy levels, $U_{n\kappa} =
- \varepsilon_{n\kappa}$, for
the DHFS potential (with $C_{\rm ex}=3/2)$. Electron subshells are indicated
by using the x-ray notation, in which the value of the principal quantum
number $n=1$, 2, 3, 4, \ldots\ is represented by the respective capital
letters K, L, M, N, \ldots, and angular momenta are indicated by the
numeral subscript $i$ (see Table \ref{tab2.1}).
\label{fig3.5}}
\end{center} \end{figure}


\section{The interaction potential of charged particles and atoms
\label{sec3.6}}

\index{atomic potential|(} \index{atomic screening function} Elastic
collisions of charged particles with neutral atoms are usually described
by means of the {\it static-field approximation}, \ie, as scattering of
the projectile by the electrostatic field of the target atom. Let us
consider collisions of a particle with charge $Z_0e$ and a neutral atom
of the element of atomic number $Z$. To ease calculations, we assume the
atom has a point nucleus and a spherically symmetric cloud of electrons
of density $\rho(r)$. For elements whose ground-state electron
configuration has open shells, a spherical electron density is obtained
after averaging over atomic orientations (or over degenerate states).
Assuming the nucleus at the origin of coordinates, the interaction
potential energy between the projectile at ${\bf r}$ and the target atom
is
\beq
V(r) = Z_0 e \, \varphi_{\rm es}(r) \, ,
\label{3.142}\eeq
where $\varphi_{\rm es}(r)$ is the
electrostatic potential of the atomic charge distribution,
\beq
\varphi_{\rm es}(r) = \frac{Z e}{r} - e
\left( \frac{1}{r} \int_0^r \rho(r') \, 4 \pi
r'^2 \, \d r' + \int_r^\infty \rho(r') \, 4\pi r' \, \d r'
\right).
\label{3.143}\eeq
The electrostatic potential and the electron density
are linked by the Poisson equation [cf.\ Eq.\ \req{3.81}]
\beq
\rho(r) = \frac{1}{4\pi e r} \frac{\d^2}{\d r^2} \,
\left[ r \varphi_{\rm es}(r) \right].
\label{3.144}\eeq
It is customary to write
\beq
\varphi_{\rm es}(r) =  \frac{Z e}{r} \, \Phi(r),
\label{3.145}\eeq
where we have introduced the screening function, $\Phi(r)$,
which describes the electrostatic shielding of the nuclear charge by the
atomic electrons. This function equals unity at $r=0$, where the
projectile ``sees'' the full charge of the nucleus, and decreases
monotonically with $r$, tending to zero at large radii, because the
atomic electrons completely screen the nuclear charge. Notice that
\beq
\rho(r) = \frac{Z}{4\pi r} \frac{\d^2 \Phi(r)}{\d r^2}
\qquad \mbox{and} \qquad
V(r) = \frac{Z_0 Z e^2}{r} \, \Phi(r)\, .
\label{3.146}\eeq

To facilitate calculations of elastic scattering, we use approximate
screening functions given by the following analytical expression
\beq
\Phi(r) = \sum_{i=1}^3 A_i \exp(- a_i r),
\qquad \sum_{i=1}^3 A_i = 1 \, ,
\label{3.147}\eeq
with parameters determined by fitting the numerical screening functions
obtained from different atomic structure calculations. The corresponding
electron density is
\beq
\rho(r) = \frac{Z}{4\pi r} \sum_{i=1}^3 A_i a_i^2 \exp(- a_i r).
\label{3.148}\eeq
and the interaction energy of the charged projectile and the atom is
\index{atomic potential!analytical}
\beq
V(r) = \frac{Z_0 Z e^2}{r} \sum_{i=1}^3 A_i \exp(- a_i r).
\label{3.149}\eeq
Analytical potentials of this kind (a sum of Yukawa-like terms) have
been used by many authors, not only because they lead to a simple
analytical expression of the scattering differential cross section
obtained from the first Born approximation (see Section \ref{sec5.1.2}
below), but also because they give realistic results when used in more
elaborate theoretical formulations. Interaction potentials with similar
expressions are employed in studies of collisions of heavy ions with
atoms \citep[see, \eg,][and references therein]{Ziegler2012}.

In the calculations described below we shall consider the
following screening function models and sets of parameters:

\index{atomic screening function!Thomas--Fermi}
\noindent $\bullet$ {\bf Thomas--Fermi--Moli\`ere (TFM) screening function
} \\
The approximation of \citet{Moliere1947} to the Thomas--Fermi
screening function, Eq.\ \req{3.112}, is of the type \req{3.147} with the
following set of parameters
\beqa
&& A_1 = 0.10, \quad A_2 = 0.55, \quad A_3 = 0.35,
\nonumber \\ [2mm]
&& a_1 = 6.0/b, \quad a_2 = 1.2/b, \quad a_3 = 0.3/b,
\label{3.150}\eeqa
where $b=0.88534 \, Z^{-1/3} \, a_{0}$ is the Thomas--Fermi radius defined by
Eq.\ \req{3.94}.

\begin{figure}[bt!] \begin{center}
\vspace*{3mm}
\includegraphics*[width=7.3cm]{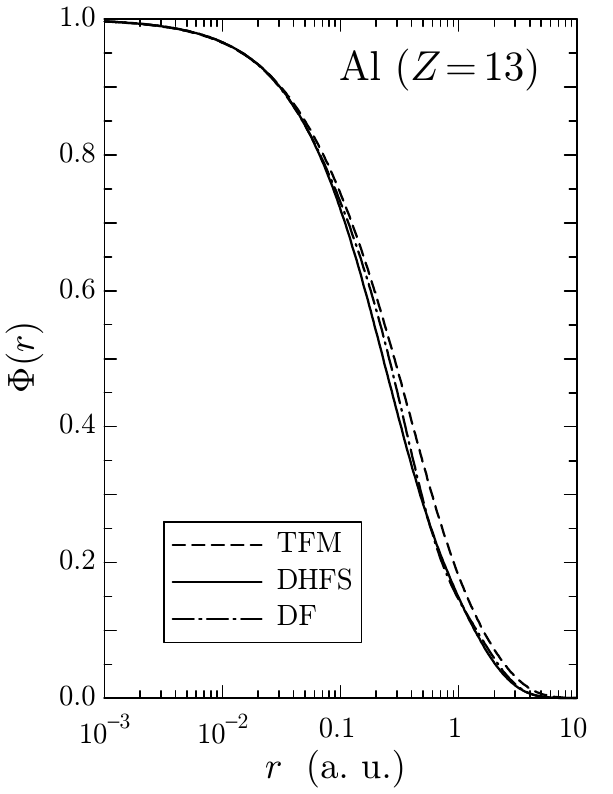} \rule{3mm}{0mm}
\includegraphics*[width=7.3cm]{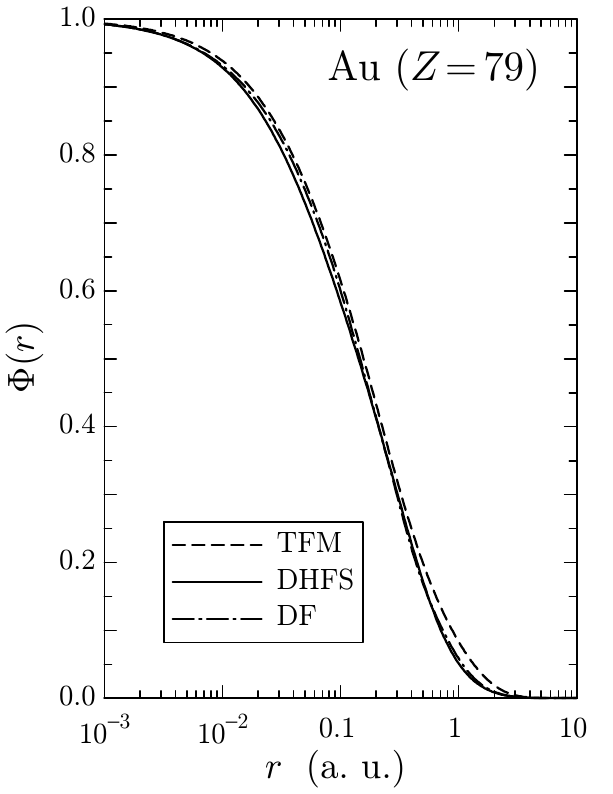}
\caption {\rm TFM and DHFS analytical screening functions of aluminum
and gold atoms. The dot-dashed curves represent the screening function
obtained from the self-consistent Dirac--Fock electron density. Radii in
atomic units (a.u.).
}\label{fig3.6}
\end{center}\end{figure}

\index{atomic screening function!DHFS}
\noindent $\bullet$ {\bf Dirac--Hartree--Fock--Slater (DHFS) screening
function}\\
Realistic screening functions obtained from self-consistent atomic
electron densities are defined numerically. The analytical form
\req{3.147} has been extensively used with parameters determined by
fitting different self-consistent atomic potentials. In the calculations
reported below we use the set of parameters given by \citet{Salvat1987},
which were obtained by fitting the DHFS electron densities of free neutral
atoms with $Z \le 92$. Elements from Z=93 to 103, calculated with the
same strategy, were added in March 1992.
The parameters of the DHFS screening
function are such that the analytical approximation and the numerical
self-consistent DHFS potential yield the same DCSs for scattering of
charged particles at small angles when calculated within the first Born
approximation \citep[see][]{Salvat1987}. The Born approximation is expected
to be accurate for projectiles with sufficiently high energies (see
Chapter \ref{chapt5}).

Figure \ref{fig3.6} displays the TFM and DHFS analytical screening
functions of aluminum and gold atoms. For comparison purposes, the plots
include screening functions obtained from numerical self-consistent
Dirac--Fock electron densities calculated by the computer program of
Desclaux (\citeyear{Desclaux1975}). The
similarity of the TFM and DHFS screening functions confirms the
$Z^{1/3}$ scaling of the Thomas--Fermi model.

\index{atomic screening function!Wentzel} \index{Wentzel potential}
\noindent $\bullet$ {\bf The Wentzel potential}\\
In the first calculation of elastic scattering with the Born
approximation, \citet{Wentzel1927} used the screened potential
\beq
V_{\rm W}(r) = \frac{Z_1 Z e^2}{r} \, \exp(-ar), \qquad a=1/R,
\label{3.151}\eeq
where $R$ is the ``atomic radius'' or ``screening length'', which is
frequently approximated as
\beq
R = 0.88534 \, Z^{-1/3} \, a_{0},
\label{3.152}\eeq
to comply with the Thomas--Fermi scaling. The Wentzel potential has the
generic form \req{3.149} with the one-term screening function $\Phi_{\rm
W}(r)= \exp(-ar)$. The Born approximation for this potential yields a
simple analytical expression for the differential cross section which,
with various values of the ``screening constant'' $a$, has been used in
calculations of multiple elastic scattering \citep[see,
\eg,][]{GoudsmitSaunderson1940, GoudsmitSaunderson1940b,
FernandezVarea1993}.
\index{atomic potential|)}



\chapter{Classical theory of elastic collisions \label{chapt4}}



Elastic collisions of swift charged particles with atoms play a central
role in theoretical radiation physics. By definition, elastic collisions
are interactions that do not cause excitation of the target atom, which
usually is in its ground state. Elastic collisions may produce large
deflections of the projectile and, consequently, have a direct effect on
the structure of trajectories of fast charged particles moving in
material media. In addition, each collision involves an energy transfer
from the projectile to the target atom which manifests as the recoil of
the latter after the interaction and gives rise to the so-called {\it
nuclear} contribution to the stopping power. The recoil of target atoms
is the main cause of radiation damage in solid lattices.

In the present Chapter we first study the scattering of particles by
central potentials by means of the classical non-relativistic trajectory
method, which is expected to be reliable for particles with de Broglie
wavelengths much smaller than the interaction ``radius'' and moving with
speeds much smaller than that of light. The theory is then extended to
the case of collisions of two particles, and we describe the calculation
of the differential cross section (DCS) in the reference frames of the
laboratory (where the target particle is initially at rest) and of the
center of mass. A semi-relativistic extension of the trajectory method
is formulated on the assumption that the interaction in the
center-of-mass frame is represented by a central potential; the DCS in
the laboratory frame is then obtained from the relativistic (Lorentz)
transform of the DCS calculated in the center-of-mass frame. This scheme
qualifies as semi-relativistic, because it accounts for relativistic
kinematics in a rigorous way, but disregards the differences between the
interactions observed from the laboratory and the center-of-mass frames.

\section{Classical scattering of non-relativistic particles by a central
force \label{sec4.1}}

It is instructive to start by studying the scattering of particles of
mass $M$ by a central force associated to the potential $V(r)$ from the
standpoint of non-relativistic classical mechanics
\citep{Goldstein1980}. To simplify theoretical considerations, the
interaction potentials considered hereafter are such that the product $r
V(r)$ is finite for all $r$. This assumption implies that the potential,
and the associated force vanish at $r=\infty$, \ie, when the projectile
is far away from the scattering center.

\index{scattering plane}

Because the force is central, angular momentum is
conserved and the trajectory of the projectile lies on a plane that
contains the center of force, the so-called {\it scattering plane}. To
facilitate the description of the motion, we use polar coordinates
\index{polar coordinates in a plane}
$(r,\varphi)$ on that plane (see Fig.\ \ref{fig4.1}). Introducing the
polar unit vectors
\beq
\hat{\bf e}_r = \hat{\bf r} = (\cos\varphi, \sin\varphi)
\qquad \mbox{and} \qquad
\hat{\bf e}_\varphi = (-\sin\varphi, \cos\varphi),
\label{4.1}\eeq
the position, the velocity, and the acceleration of the projectile are
expressed as
\begin{subequations}
\label{4.2}
\beqa
{\bf r} &=& r \; \hat{\bf e}_r,
\label{4.2a} \\ [2mm]
{\bf v} &=& \dot{\bf r} = \dot r \, \hat{\bf e}_r + r \dot\varphi \,
\hat{\bf e}_\varphi,
\label{4.2b} \\ [2mm]
{\bf a} &=& \dot{\bf v} = (\ddot r - r \dot\varphi^2) \hat{\bf e}_r +
(r \ddot \varphi + 2 \dot r \dot\varphi)
\hat{\bf e}_\varphi \, ,
\label{4.2c}\eeqa
\end{subequations}
where a dot over a quantity denotes the time derivative, \ie,
$\dot{\varphi} \equiv \d \varphi/\d t$. Here we have used the identities
\beq
\frac{\d \hat{\bf e}_r}{\d t} =
\frac{\d \hat{\bf e}_r}{\d \varphi} \, \frac{\d \varphi}{\d t} =
\dot{\varphi} \, \hat{\bf e}_\varphi
\qquad \mbox{and} \qquad
\frac{\d \hat{\bf e}_\varphi}{\d t} =
\frac{\d \hat{\bf e}_\varphi}{\d \varphi} \, \frac{\d \varphi}{\d t}
= - \dot{\varphi} \, \hat{\bf e}_r.
\label{4.3}\eeq
It is worth noticing that the vector product
of the polar unit vectors,
\beq
\hat{\bf e}_r \vecprod \hat{\bf e}_\varphi =
\hat{\bf x} \vecprod \hat{\bf y} =
\hat{\bf z}
\label{4.4}\eeq
is a constant vector, independent of the position of the projectile.

\begin{figure}[htb] \begin{center}
\includegraphics*[width=4.5cm]{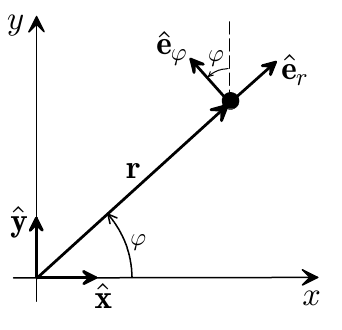}
\caption{
Polar coordinates $(r,\varphi)$ in the plane of scattering. Notice that
$x=r \, \cos\varphi$ and $y=r \, \sin \varphi$.
\label{fig4.1}}
\end{center} \end{figure}

The linear momentum, the angular momentum, and the kinetic energy
of the projectile can thus be expressed as
\beq
{\bf p} \equiv M {\bf v} = M \dot{r} \, \hat{\bf e}_r
+ M r \dot{\varphi} \, \hat{\bf e}_\varphi,
\label{4.5}\eeq
\beq
{\bf L} \equiv {\bf r} \vecprod {\bf p} = r \, \hat{\bf e}_r
\vecprod \left[ M \dot{r} \, \hat{\bf e}_r
+ M r \dot{\varphi} \, \hat{\bf e}_\varphi\right] =
M r^2 \dot{\varphi} \, \hat{\bf e}_r \vecprod \hat{\bf e}_\varphi
= M r^2 \dot{\varphi} \, \hat{\bf z},
\label{4.6}\eeq
and
\beq
K \equiv \frac{p^2}{2 M} = \frac{1}{2} M \dot{r}^2
+ \frac{1}{2} M r^2 \dot{\varphi}^2,
\label{4.7}\eeq
respectively.

The force acting on the projectile is
\beq
{\bf F}({\bf r}) = - \nablab V(r)
= - \frac{\d V(r)}{\d r} \, \hat{\bf e}_r
= F(r) \, \hat{\bf e}_r.
\label{4.8}\eeq
The equation of motion, ${\bf F} = M {\bf a}$, takes the form
\beq
F(r) \, \hat{\bf e}_r
= M \left[ (\ddot r - r \dot\varphi^2) \hat{\bf e}_r +
(r \ddot \varphi + 2 \dot r \dot\varphi)
\hat{\bf e}_\varphi \right]\, ,
\label{4.9}\eeq
which is equivalent to the pair of equations
\begin{subequations}
\label{4.10}
\beqa
&& M ( \ddot{r} - r \dot\varphi^2) = F(r),
\label{4.10a}\\ [2mm]
&& r \ddot\varphi + 2 \dot r \dot \varphi = 0.
\label{4.10b}\eeqa
\end{subequations}
Recalling that ${\bf L} = M r^2 \dot{\varphi} \, \hat{\bf z}$, we have
\beq
\dot{\bf L} = M \left( 2 r \dot{r} \dot{\varphi} + r^2 \ddot{\varphi} \right)
\hat{\bf z}\, .
\label{4.11}\eeq
We thus see that Eq.\ \req{4.10b} expresses the conservation of angular
momentum.

Considering that $L$ is a constant of the motion, from Eq.\ \req{4.6}, we
have
\beq
\dot{\varphi} = \frac{L}{Mr^2}.
\label{4.12}\eeq
Equation \req{4.10a} then becomes
\beq
M \ddot{r} = F(r) + \frac{L^2}{M r^3}\, .
\label{4.13}\eeq
We have thus decoupled the angular and radial motions. The radial
equation \req{4.13} describes the one-dimensional motion of a particle
of mass $M$ under the action of the field force $F(r)$ plus the
``centrifugal force'', $L^2/(Mr^3) = M r \dot{\varphi}^2$. The kinetic
energy \req{4.7} and the total energy $E$ can now be expressed as
\beq
K = \frac{1}{2} M \dot{r}^2
+ \frac{L^2}{2 M r^2},
\label{4.14}\eeq
and
\beq
E \equiv K + V =  \frac{1}{2} M \dot{r}^2
+ \frac{L^2}{2 M r^2} + V(r).
\label{4.15}\eeq
The sum
\beq
V_{\rm rad}(r) \equiv V(r) + \frac{L^2}{2 M r^2}
\label{4.16}\eeq
is the effective potential for the radial motion. The second term on the
right-hand side is called the ``centrifugal potential barrier'', because
\beq
- \frac{d}{\d r} \left( \frac{L^2}{2 M r^2} \right) =
\frac{L^2}{M r^3}\, ,
\label{4.17}\eeq
the centrifugal force. The main features of scattering can be understood
by considering the radial motion under the effective potential combined
with the angular motion, $\dot{\varphi} = L /(Mr^2)$ \citep{Goldstein1980}.


\subsection{Classical trajectories \label{sec4.1.1}}

\index{classical trajectories}
Let us assume that, long before the interaction, the projectile
approaches the center of force with initial energy $E$, momentum ${\bf
p}_{\rm i}$ [$p_{\rm i}=(2ME)^{1/2}$] parallel to the polar axis, and
impact parameter $b$ (see Fig.\ \ref{fig4.2}). It is convenient to
characterize the projectile motion by the magnitude of the angular momentum,
\beq
L=b p_{\rm i}.
\label{4.18}\eeq
Under the action of the force, the projectile follows a continuous
trajectory, \ie, both  $r$, the distance to the projectile from the
center of force, and $\varphi$, the polar angle of the position vector,
evolve with time. We are primarily interested in the deflection angle,
$\vartheta$, defined as the total angle swept by the linear momentum of
the projectile along the complete trajectory. Notice that
$\varphi(t=-\infty)=\pi$ and that $\varphi(t=\infty) = \vartheta$ is the
polar angle of the asymptote of the outgoing trajectory.

\begin{figure}[htb] \begin{center}
\includegraphics*[width=10.5cm]{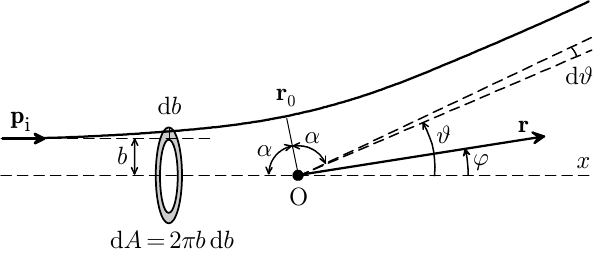}
\caption{
Scattering of particles by a central field.
\label{fig4.2}}
\end{center} \end{figure}

To illustrate the variation of
$\vartheta$ with the impact parameter, Fig.\ \ref{fig4.3} shows
trajectories of electrons ($Z_0=-1$) and positrons ($Z_0=+1$) with an
energy of 500 eV and various angular momenta in the analytical DHFS screened
potential of gold atoms ($Z=79$) ---see Section \ref{sec3.6}---,
\beq
V(r)= \frac{Z_0 Z e^2}{r} \, \Phi_{\rm DHFS}(r).
\label{4.19}\eeq
These trajectories were computed numerically by solving the equation of
motion ${\bf a} = {\bf F}/M$ in Cartesian coordinates by using an
adaptive fourth-order Runge--Kutta method \citep[see,
\eg,][]{DeVries1994} that advances time steps of variable length $\Delta
t$ such that integration errors are less than a prescribed tolerance.
After each integration step the velocity of the particle was
renormalized to ensure that energy is effectively conserved.  To asses
the accuracy of the calculated trajectories, the computer program keeps
track of the time evolution of the angular momentum, which is found to
remain constant in magnitude within the adopted tolerance.

\begin{figure}[p!] \begin{center}
\vspace*{3mm}
\includegraphics*[width=7.6cm]{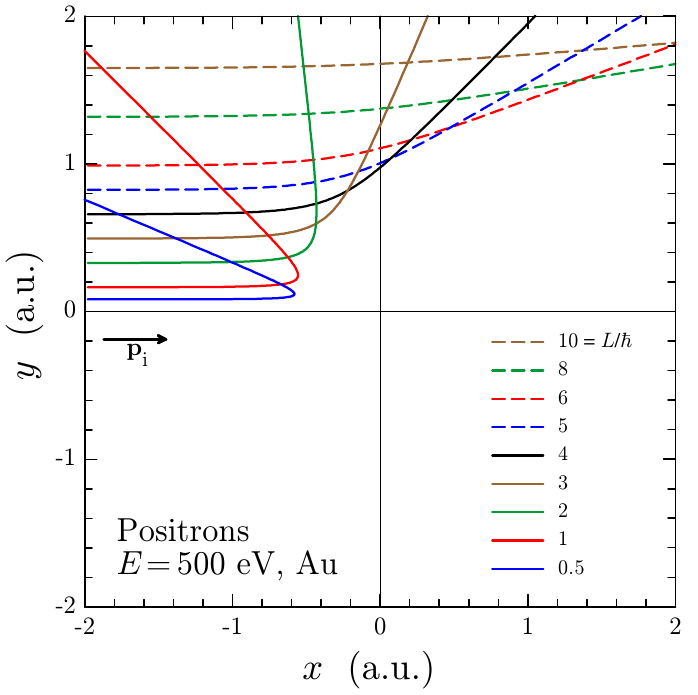} \hfill
\includegraphics*[width=7.6cm]{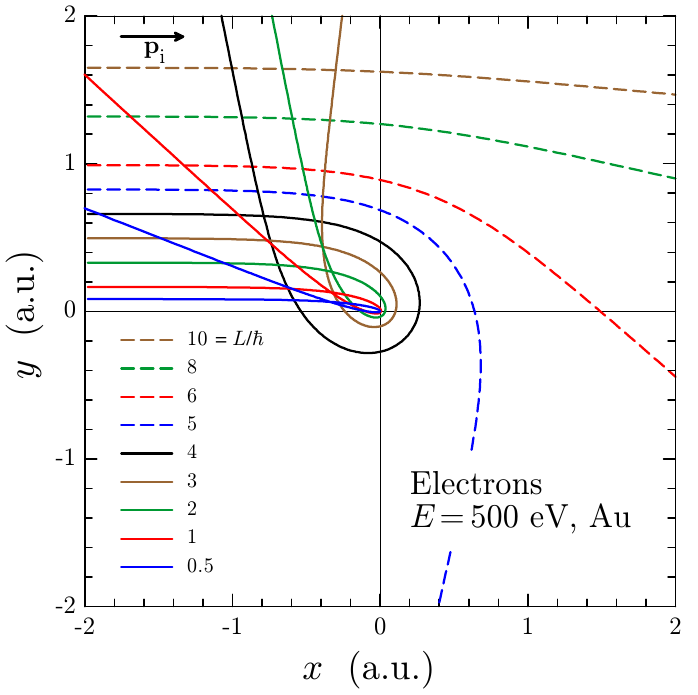}
\caption{Classical trajectories of positrons and electrons with an
energy of 500 eV and the indicated angular momenta in the DHFS screened
potential of the gold atom. Lengths are in atomic units (a.u.).
\label{fig4.3}}
\vspace*{7mm}
\includegraphics*[width=7.6cm]{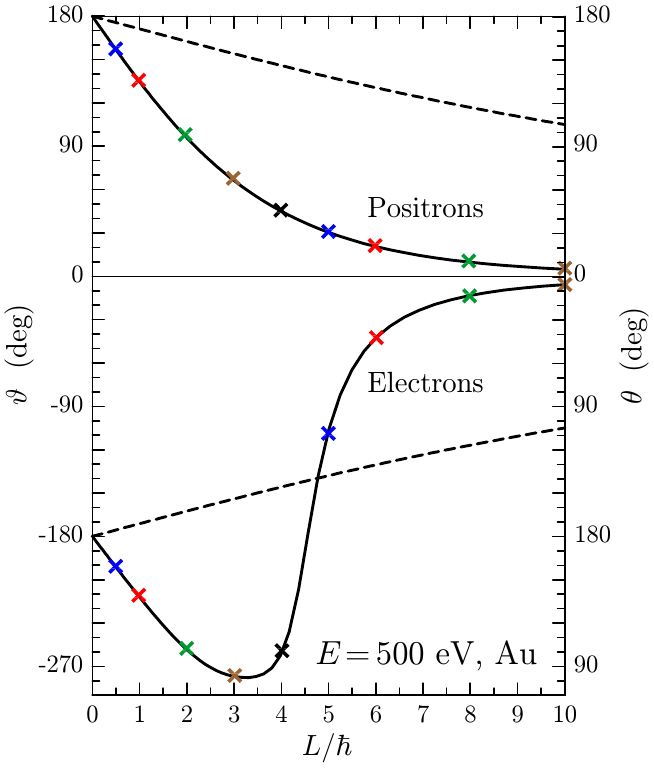}
\caption{Deflection angle as a function of the angular momentum in
the scattering of 500 eV electrons and positrons by the DHFS screened
potential of the gold atom. Crosses correspond to the trajectories
displayed in Fig.\ \ref{fig4.3}. The dashed lines represent the
function $\vartheta(L)$ for scattering of electrons and positrons by the
Coulomb field of the bare nucleus, Eq.\ \req{4.36}. The scale of the
right vertical axis gives the scattering angle $\theta$, Eq.\
\req{4.31}, which is constrained to the interval $(0,\pi)$.
\label{fig4.4}}
\end{center}\end{figure}

In the case of particles with positive charges, such as the positron,
the force is repulsive and induces deflections with positive angles
$\vartheta$. The force on electrons, and other negatively charged
particles, is attractive and causes deflections with negative angles. In
addition, trajectories of projectiles with negative charges and small
impact parameters may loop around the center of force (orbiting
trajectories) and emerge with deflection angles that are less than
$-\pi$. A detailed study of the dependence of the deflection angle on
the impact parameter for a generic central potential is given by
\citet{Goldstein1980}.

Figure \ref{fig4.4} shows the function $\vartheta(L)$ for the same
situations as in Fig.\ \ref{fig4.3}, \ie, scattering of electrons and
positrons with $E=500$ eV by the DHFS screened potential of gold atoms.
In the case of positrons, $\vartheta$ decreases monotonically when $L$
increases (cf.\ Fig.\ \ref{fig4.3}) and there is a one-to-one
correspondence between deflection angle and angular momentum. On the
contrary, the function $\vartheta(L)$ of electrons has a minimum at about
$L=3\hbar$, and for scattering angles less than about 180 degrees the
function $L(\vartheta)$ is double-valued.

With the aid of the constants of motion, the deflection angle
$\vartheta$ can be obtained by quadrature, without having to compute the
whole trajectory. The conservation of angular momentum implies that
[Eq.\ \req{4.12}]
\beq
\dot\varphi = \frac{L}{M r^2} \, .
\label{4.20}\eeq
The total energy
\beq
E = \frac{p^2}{2M} + V(r)
\label{4.21}\eeq
is also conserved. Because $V(r)$ vanishes long before the interaction,
$E=p_{\rm i}^2/(2M)$ and we can write
\beq
p^2(r) = p_{\rm i}^2 - 2MV(r) \, .
\label{4.22}\eeq
Equation \req{4.2b} implies that $v^2 = \dot{r}^2 + r^2 \dot{\varphi}^2$
or, equivalently,
\beq
p^2 = M^2 \dot{r}^2 + M^2 r^2 \dot{\varphi}^2.
\label{4.23}\eeq
This equality, combined with Eqs.\ \req{4.20} and \req{4.22}, gives
\beq
\dot{r} = \pm \, \frac{1}{M} \sqrt{p^2 - \frac{L^2}{r^2}}
=  \pm \, \frac{1}{M} \sqrt{p_{\rm i}^2 - 2 M V(r) - \frac{L^2}{r^2}}
\, ,
\label{4.24}\eeq
where the sign on the right-hand side is $-$ for the part of the orbit
where the particle approaches the scattering center and $+$ when the
particle has passed the point of closest approach ${\bf r}_0$. The
distance of closest approach $r_0$, is determined by the condition
$\dot{r} = 0$, which is met when
\beq
p_{\rm i}^2 - 2M V(r_0) - \frac{L^2}{r_0^2} = 0 \, .
\label{4.25}\eeq
Evidently, $r_0$ is the largest root of this equation, which corresponds
to the outer turning point of the radial motion.

The geometric equation of the trajectory is
\index{trajectory equation}
\beq
\d \varphi =  \frac{\d \varphi}{\d t} \, \frac{\d t}{\d r} \, \d r
= \frac{\dot{\varphi}}{\dot{r}} \d r =
\pm \frac{L/r^2}{\sqrt{p^2(r)-L^2/r^{2}}}\, \d r \, ,
\label{4.26}\eeq
where we have used the equalities \req{4.20} and \req{4.24}.
The deflection angle $\vartheta$ can now be calculated easily by noting
that $2\alpha+\vartheta=\pi$, where $\alpha$ is the angle between the
position vector of the particle at the point of closest approach and the
asymptote of the outgoing trajectory (see Fig.\ \ref{fig4.2}). Because
\beq
\alpha=\int_{r_0}^\infty
\frac{L/r^2}{\sqrt{p^2(r)-L^2/r^{2}}}\, \d r,
\label{4.27}\eeq
we can write \index{classical collisions!deflection angle}
\beq
\vartheta(L) = \pi - 2 \int_{r_0}^\infty
\frac{L/r^2}{\sqrt{p^2(r) -L^2/r^2}}\, \d r.
\label{4.28}\eeq


\subsection{Scattering cross section \label{sec4.1.2}}

In an ideal scattering experiment a parallel beam of projectile
particles with linear momentum ${\bf p}_i$ impinges on the center of
force. We assume that the potential has a finite range and that
the incident beam is laterally homogeneous and
its lateral extension is much larger than the range of the
potential. The beam is then characterized by the current density ${\bf
J}$, a vector parallel to the direction of incidence $\hat{\bf p}_{\rm
i}$ with magnitude equal to the number of incident particles that cross
a small probe surface, perpendicular to the beam and at rest with
respect to the target, per unit time and per unit area of the probe surface.
To describe the experiment, we consider a reference frame
with its origin at the center of force and the $z$ axis in the
direction of the incident beam (Fig.\ \ref{fig4.5}). Direction
vectors $\hat{\bf d}$ are defined by their polar and azimuthal angles,
$\theta$ and $\phi$, respectively,
\beq
\hat{\bf d} = (\sin\theta \, \cos\phi,
\sin\theta \, \sin\phi, \cos\theta).
\label{4.29}\eeq
The polar angle $\theta$ may take values in the interval from 0 to
$\pi$, while the azimuthal angle $\phi$ varies between 0 and $2\pi$.  In
the calculations it is convenient to use the variable $\cos\theta$
instead of the polar angle; note that $\sin\theta = \sqrt{1- \cos^2
\theta}$ is non-negative.

An ideal detector placed far away from the scattering center and
covering a small solid angle $\d \Omega$ around the direction
$(\theta,\phi)$ counts the scattered particles that enter the detector
window. Assuming that the range of the atomic potential is much smaller than the
distance from the scattering center to the detector, the
scattered particles reach the detector moving radially in the direction
$\hat{\bf p}_{\rm f}$ of the final momentum. Let $\dot{N}_{\rm count}$
denote the number of counts per unit time. The scattering {\it
differential cross section} (DCS), per unit solid angle, is defined as
\index{classical scattering!differential cross section}
\beq
\frac{\d \sigma}{\d \Omega} =
\frac{\dot{N}_{\rm count}}{J \, \d \Omega}\, .
\label{4.30}\eeq
The DCS has the dimensions of area/(solid angle); the product $(\d
\sigma/\d \Omega) \d \Omega$ represents the area of a plane surface
that, placed perpendicularly to the incident beam, is hit by as many
projectiles as are scattered into directions within $\d \Omega$.

\begin{figure}[htb] \begin{center}
\includegraphics*[width=6.0cm]{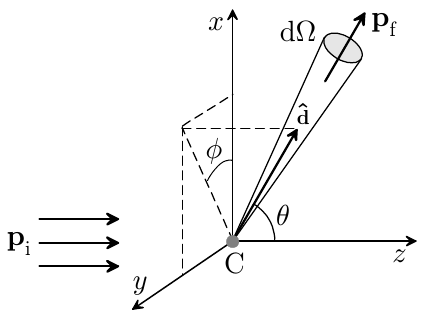}
\caption{
Schematic diagram of a scattering experiment. The incident beam is
parallel to the $z$ axis, and the detector covers a small solid angle
$\d \Omega$ in the direction $\hat{\bf d}$ defined by the polar and
azimuthal angles, $\theta$ and $\phi$. The incident beam is assumed to
be collimated so that only scattered particles can reach the detector.
\label{fig4.5}}
\end{center} \end{figure}

The angle $\theta$ between the directions of motion of the projectile
before and after the interaction is called the {\it (polar) scattering
angle}. Because the system is axially symmetric about the
direction of incidence, the DCS is independent of the azimuthal
scattering angle $\phi$. Hence, to measure the DCS we can use an annular
detector that counts particles that emerge in directions with polar
scattering angle in a small interval $(\theta, \theta + \d \theta)$ and
with any azimuthal angle. The solid angle covered by this detector is
$\d \Omega = 2 \pi \, \sin\theta\, d\theta$.
The detector receives particles whose
outgoing trajectories have the deflection angle $\vartheta = - \theta$
or $\vartheta= \theta$. In addition, for attractive forces, the detector may
receive particles which have circled the center of force and emerge with
deflection angles $\vartheta = \pm \theta$ plus integer multiples of
$2\pi$. Consequently, particles may enter the detector with
multiple deflection angles given by
\beq
\vartheta = \pm \theta + n 2\pi,
\label{4.31}\eeq
where $n$ is an integer. Evidently, the
scattering angle $\theta$ (restricted to the interval from 0 to $\pi$)
corresponding to a given deflection angle $\vartheta$ is determined by
the equality
\beq
\cos\theta = \cos\vartheta .
\label{4.32}\eeq

The DCS can be calculated from knowledge of the function $b(\theta) =
L(\theta)/ p_{\rm i}$, which relates the scattering
angle $\theta$ and the impact parameter $b$.
The number $\dot{N}_{\rm count}$ of scattered particles that enter the
annular detector per unit time is equal to the number of incident
particles that cross a plane perpendicular to the polar axis with impact
parameters between $b$ and $b+\d b$ per unit time (Fig.\ \ref{fig4.2}).
If the function $L(\theta)$ is single-valued,
the classical DCS for scattering in the potential $V(r)$ is
\beq
\frac{\d \sigma}{\d \Omega} =
\frac{J \, 2 \pi b \, \d b}{J \, \d\Omega }
= \frac{b}{\sin\theta} \left| \frac{\d b}{\d \theta} \right|
= \frac{L}{p_{\rm i}^2 \sin\theta} \left| \frac{\d L}{\d \theta} \right|\, .
\label{4.33}\eeq
When the function $L(\theta)$ is multi-valued (as, \eg, in the case of
electron scattering by gold atoms shown in Fig.\ \ref{fig4.4}), the DCS is
the sum of contributions from the various branches of that function.
That is,
\beq
\frac{\d \sigma}{\d \Omega} = \frac{1}{2 p_{\rm i}^2 \sin\theta} \sum_j
\left| \frac{\d L^2}{\d \theta} \right|_{L=L_j}\, ,
\label{4.34}\eeq
where the summation extends over the values $L_j$ of the angular
momentum that give rise to the same value $\theta$ of the scattering
angle. Notice that the DCS may diverge at $\theta=0$ and $\pi$ (because
of the sine in the denominator); the infinite DCSs at forward or
backward scattering angles are referred to as {\it glory scattering}.
The DCS also diverges at the extrema of the function $\theta(L)$; this
situation produces an infinite rise followed or preceded by an abrupt
disappearance of the DCS, which is known as {\it rainbow scattering}
(see the examples in Fig.\ \ref{fig4.9} below).

The {\it total cross section} is defined as the integral of the DCS over
scattering angles,
\beq
\sigma = \int \frac{\d \sigma}{\d \Omega} \, \d \Omega
= 2\pi \int_{-1}^1
\frac{\d \sigma}{\d \Omega} \, \d (\cos\theta).
\label{4.35}\eeq
Note that $\sigma$ has the dimensions of (length)$^2$; it represents the
area of a plane probe surface that, when placed perpendicularly to the beam is
crossed by as many particles as are scattered, irrespective of their
scattering angles. When the interaction potential has a finite range,
the total cross section is finite.


\subsection{Scattering by a Coulomb potential
\label{sec4.1.3}}

\index{classical scattering!Coulomb potential}
Since we are particularly interested in describing the collisions of
charged particles, we consider the case of scattering by the
electrostatic Coulomb interaction of the projectile (charge $Z_0e$) with a point
particle (charge $Ze$) that is fixed at the origin of coordinates,
\beq
V_{\rm C}(r)= \frac{Z_0 Z e^2}{r},
\label{4.36}\eeq
The Coulomb potential is important because it represents the dominant
interaction between charged particles and has interesting scattering
properties. A peculiarity of this potential is that the DCS derived from
the present classical formulation is identical to the exact result from
non-relativistic quantum mechanics.

The point of closest approach is the positive root of the equation [cf.\
Eq.\ \req{4.25}]
\beq
p_{\rm i}^2 r_0^2-2MZ_0Z e^2 r_0-L^2 = 0\, ,
\label{4.37}\eeq
which is given by
\beq
r_0 = \frac{M Z_0 Z e^2}{p_{\rm i}^2} + \sqrt{ \left( \frac{M Z_0 Z
e^2}{p_{\rm i}^2}
\right)^2 + \frac{L^2}{p_{\rm i}^2}} \, .
\label{4.38}\eeq
For the Coulomb potential, the classical formula \req{4.28} for the
deflection angle $\vartheta$ can be evaluated analytically. Introducing
the change of variable $u=r_0/r$, the integral in Eq.\ \req{4.28} is
elementary\footnote{We use the formula
$$
\int \frac{\d u}{\sqrt{c-bu-au^2}} = \frac{1}{\sqrt{a}} \arctan \left(
\frac{2au+b}{2\sqrt{a}\sqrt{c-bu-au^2}} \right), \qquad a,c >0.
$$
},
\allowdisplaybreaks{
\beqa
\vartheta(L) &=& \pi - 2 L \int_0^1
\frac{\d u}{\sqrt{p_{\rm i}^2 r_0^2-2MZ_0 Z e^2 r_0 u-L^2u^2}}
\nonumber \\ [2mm]
&=& \pi - 2 \left[ \arctan \left( \frac{2L^2 u + 2M Z_0 Z e^2
r_0}{2L\sqrt{p_{\rm i}^2 r_0^2 - 2 M Z_0 Z e^2 r_0 u - L^2 u^2}} \right)
\right]_0^1
\nonumber \\ [2mm]
&=& 2 \arctan \left( \frac{M Z_0 Z e^2}{Lp_{\rm i}} \right)
\, .
\label{4.39}\eeqa
When the angular momentum $L$ decreases from $\infty$ to zero, the
deflection angle $\vartheta$ varies from 0 to $\pi$ ($-\pi$) if the
charges of the projectile have equal (opposite) signs. Therefore, the
scattering angle $\theta$ is equal to $|\vartheta|$. The result
\req{4.39} implies that
\beq
L^2(\theta) = \frac{(MZ_0Ze^2)^2}{p_{\rm i}^2} \, \frac{1}{\tan^2(\theta/2)}
= \frac{(MZ_0Ze^2)^2}{p_{\rm i}^2} \,
\frac{\cos^2(\theta/2)}{\sin^2(\theta/2)}\, .
\label{4.40}\eeq
Since the function $L(\theta)$ is single-valued, and
\beq
\frac{\d L^2}{\d \theta} = - \frac{(MZ_0Ze^2)^2}{2p_{\rm i}^2} \,
\frac{\sin\theta}{\sin^4(\theta/2)} ,
\label{4.41}\eeq
we have
\beq
\frac{\d \sigma_{\rm R}}{\d \Omega}
= \frac{1}{2 p_{\rm i}^2 \sin\theta} \left| \frac{\d L^2}{\d \theta} \right|
= \left(\frac{M Z_0Ze^2}{2 p_{\rm i}^2} \right)^2
\frac{1}{\sin^4(\theta/2)} .
\label{4.42}\eeq
\index{classical scattering!Rutherford DCS}This is the famous cross section of
\citeauthor{Rutherford1911}, who derived it in
\citeyear{Rutherford1911}. This formula was the keystone for the
interpretation of the experiments of scattering of alpha particles,
which provided the first experimental evidence of the atomic nucleus.
Notice that the Rutherford DCSs is proportional to the squares of the
charges of the colliding particles, \ie, independent of the signs of
these charges.  }

To facilitate the extension of the theory to the case of collisions of two
charged particles (see Section \ref{sec4.2.1}), we write the Rutherford DCS
as
\beq
\frac{\d \sigma_{\rm R}}{\d \Omega}
= \left( \frac{Z_0Ze^2}{2 v_{\rm i}  p_{\rm i}} \right)^2
\frac{1}{\sin^4(\theta/2)}\, ,
\label{4.43}\eeq
where $v_{\rm i}=p_{\rm i}/M$ is the initial velocity of the projectile
with respect to the center of force. Evidently, the total cross section,
Eq.\ \req{4.35}, is infinite because of the infinite range of the
Coulomb interaction.


\subsection{Scattering by a screened Coulomb potential
\label{sec4.1.4}}

\index{classical scattering!screened Coulomb potential}
Let us now consider the scattering of particles of mass $M$ and charge
$Z_0 e$ with a neutral atom of atomic number $Z$, again assumed to be
held fixed at the origin of the reference frame. In this case, the
interaction is described by a screened Coulomb potential,
\beq
V(r)= \frac{Z_0 Z e^2}{r}\, \Phi(r)\, .
\label{4.44}\eeq

To determine the classical DCS, we have to calculate the function
\req{4.28}
\beq
\vartheta(L) = \pi - 2 \int_{r_0}^\infty
\frac{L r^{-2}}{\sqrt{p_{\rm i}^2 -2M Z_0 Z e^2 \Phi(r) r^{-1}
-L^2 r^{-2}}}\, \d r,
\label{4.45}\eeq
where $r_0$ is the distance of closest approach, given by Eq.\
\req{4.25},
\beq
p_{\rm i}^2 r_0^2-2MZ_0Z e^2 r_0 \Phi(r_0)-L^2 =0\, .
\label{4.46}\eeq
As noted by \citet{Everhart1955}, straight numerical evaluation of
expression \req{4.45} is not convenient, because the second term may
take values close to $\pi$, and the subtraction would magnify errors. We
can obtain an equivalent expression that is less sensitive to numerical
errors by considering the Coulomb potential
\beq
V_0(r)= \frac{Z_0Z e^2}{r}\, \Phi(r_0)\, ,
\label{4.47}\eeq
which coincides with $V(r)$ at $r=r_0$. Recalling that the $\vartheta(L)$
function for this potential is [see Eq.\ \req{4.39}]
\beqa
\vartheta_0 &=& \pi -
2 \int_{r_0}^\infty
\frac{L r^{-2}}{\sqrt{p_{\rm i}^2 -2M Z_0 Z e^2 \Phi(r_0) r^{-1}
-L^2 r^{-2}}}\, \d r
\nonumber \\ [2mm]
&=& 2 \arctan \left( \frac{M Z_0 Z e^2 \Phi(r_0)}{Lp_{\rm i}}
\right)\, ,
\nonumber\eeqa
Eq.\ \req{4.45} can be stated as
\beqa
\vartheta(L) &=&
2 \arctan \left( \frac{M Z_0 Z e^2 \Phi(r_0)}{Lp_{\rm i}} \right)
+ 2 \int_{r_0}^\infty \left(
\frac{L r^{-2}}{\sqrt{p_{\rm i}^2 -2M Z_0 Z e^2 \Phi(r_0) r^{-1}
-L^2 r^{-2}}} \right.
\nonumber \\ [2mm]
&& \mbox{} \rule{10mm}{0mm} \left. -
\frac{L r^{-2}}{\sqrt{p_{\rm i}^2 -2M Z_0 Z e^2 \Phi(r) r^{-1}
-L^2 r^{-2}}} \right) \d r\, .
\label{4.48}\eeqa
Evidently, the integral represents the difference $\vartheta(L) -
\vartheta_0(L)$, between the polar deflections for the potentials
\req{4.44} and \req{4.47}. This difference vanishes for unscreened
Coulomb potentials [\ie, with $\Phi(r)=1$], and its value is small for
screened potentials, so that numerical accuracy is maintained when we
sum the values of the two terms on the right-hand side of Eq.\
\req{4.48}.

The formula \req{4.48} is not yet suited for numerical evaluation,
because the two terms in the integrand diverge at $r=r_0$, although the
integral is finite. To handle this divergence, we change the
integration variable to $u=\sqrt{1-r_0/r}$,
\beqa
\vartheta(L) &=&
2 \arctan \left( \frac{M Z_0 Z e^2 \Phi(r_0)}{Lp_{\rm i}} \right)
\nonumber \\ [2mm]
&& + 4L \int_0^1 \left(
\frac{1}{\sqrt{p_{\rm i}^2 r_0^2-2M Z_0 Z e^2 r_0 \Phi(r_0) [1-u^2]
-L^2 [1-u^2]^{2}}} \right.
\nonumber \\ [2mm]
&& \mbox{} \rule{10mm}{0mm} \left. -
\frac{1}{\sqrt{p_{\rm i}^2 r_0^2 -2M Z_0 Z e^2 r_0 \Phi(r_0/[1-u^2]) [1-u^2]
-L^2 [1-u^2]^{2}}} \right) u\, \d u\, ,
\nonumber\eeqa
and, using the equality \req{4.46}, we have
\beqa
\vartheta(L) &=&
2 \arctan \left( \frac{M Z_0 Z e^2 \Phi(r_0)}{Lp_{\rm i}} \right)
+ 4 \int_0^1 \left(
\frac{1}{\sqrt{C \Phi(r_0)+2-u^{2}}} \right.
\nonumber \\ [2mm]
&& \mbox{} \rule{10mm}{0mm} \left. -
\frac{u}{\sqrt{C \left\{ \Phi(r_0) - \Phi(r_0/[1-u^2]) [1-u^2]
\right\} + 2u^2-u^4}} \right) \d u\, ,
\label{4.49}\eeqa
with
\beq
C = \frac{2MZ_0Z e^2 r_0}{L^2}\, .
\label{4.50}\eeq
Introducing the function
\begin{subequations}
\label{4.51}
\beqa
f(u) &\equiv& \Phi(r_0) -
\frac{\Phi(r_0) - [1-u^2] \, \Phi(r_0/[1-u^2])}{u^2}
\label{4.51a}\\ [2mm]
&=&  r_0 \, \frac{\Phi(r)-\Phi(r_0)}{r-r_0},
\label{4.51b}\eeqa
\end{subequations}
we can write
\beqa
\vartheta(L) &=&
2 \arctan \left( \frac{M Z_0 Z e^2 \Phi(r_0)}{Lp_{\rm i}} \right)
+ 4 \int_0^1 \left\{
\frac{1}{\sqrt{C \Phi(r_0)+2-u^{2}}} \right.
\nonumber \\ [2mm]
&& \mbox{} \rule{10mm}{0mm} \left. -
\frac{1}{\sqrt{C \Phi(r_0) + 2 - u^{2} - C f(u)
}} \right\} \d u\, ,
\label{4.52}\eeqa
where the integrand remains finite at $u=0$ (or $r=r_0$), as desired.
When $u$ increases, the function $f(u)$ decreases tending to zero at
$u=1$. Hence, for values of $u$ near unity, the two terms of the
integrand nearly cancel and the straight evaluation of their
difference would introduce considerable round-off errors. To avoid these
errors, when $f(u)$ is small we use the equivalent expression
\beqa
\vartheta(L) &=&
2 \arctan \left( \frac{M Z_0 Z e^2 \Phi(r_0)}{Lp_{\rm i}} \right)
+ 4 \int_0^1 \frac{\d u}{\sqrt{C \Phi(r_0)+2-u^{2}}}
\nonumber \\ [2mm]
&& \mbox{} \rule{10mm}{0mm} \times \left\{ 1 - \left(
1 - \frac{ C f(u)}
{C \Phi(r_0) + 2 - u^{2}} \right)^{-1/2} \, \right\} ,
\label{4.53}\eeqa
and the binomial expansion of the last factor,
\beq
1 - \left(  1 - x \right)^{-1/2} =
-\frac{1}{2} x
-\frac{1\cdot 3 }{2\cdot 4} x^2
-\frac{1\cdot 3\cdot 5 }{2\cdot 4\cdot 6} x^3 - \cdots
\label{4.54}\eeq

Numerical evaluation of these formulas shows that the deflection angle
$\vartheta$ varies smoothly with $L$. Figure \ref{fig4.4} displays the
functions $\vartheta(L)$ for the case of scattering of electrons and
positrons with $E=500$ eV by the DHFS potential of gold atoms. The plot
also includes the $\vartheta(L)$ functions for scattering by the Coulomb
potential of the bare nucleus, Eq.\ \req{4.36}. For positrons, and for
electrons with angular momenta larger than $\sim 5 \hbar$, the screening
by the atomic electrons makes the deflection angle to decrease (in
absolute value) with $L$ much faster than for Coulomb scattering.

\index{Gauss--Legendre quadrature!adaptive}
In the computer program {\sc elastic} (see Chapter \ref{chapt10})
integrals are evaluated by means of the adaptive Gauss--Legendre
quadrature algorithm (Section \ref{sec10.4.3}),
which allows strict control of numerical errors.
The deflection angle $\vartheta$ is considered as a function of the
variable $\xi = \ln(L^2)$, and it is calculated for a dense, unevenly
spaced grid of $\xi$ values that has a higher density of points where
the function has larger curvature. The function $\vartheta(\xi)$ is then
approximated by the natural cubic spline that interpolates the calculated
table (Section \ref{sec10.4.2}). The DCS is obtained from Eq.\ \req{4.34},
\beq
\frac{\d \sigma}{\d \Omega} = \frac{1}{2 p_{\rm i}^2 \sin\theta}
\sum_j
\left( \left| \frac{\d \vartheta}{\d L^2} \right|_{L=L_j} \right)^{-1}
= \frac{1}{2 p_{\rm i}^2 \sin\theta}
\sum_j
\left( \left| \frac{\d \vartheta}{\d \xi} \exp(-\xi)
\right|_{\xi=\xi_j} \right)^{-1} .
\label{4.55}\eeq
The values $\xi_j$ of $\xi$ corresponding to the scattering angle
$\theta$, as well as the derivatives $\d \vartheta/\d \xi$ are
calculated from the interpolating spline. This procedure is both fast and
accurate.


\subsection{Validity of the classical-trajectory method
\label{sec4.1.5}}

\index{classical scattering!validity}
\citet{Bohr1948} discussed the conditions under which the
non-relativistic classical-trajectory method is expected to be valid
(\ie, consistent with quantum mechanics) on the basis of the following
simple diffraction arguments. He considered the {\it gedanken}
experiment shown in Fig.\ \ref{fig4.6} in which a diaphragm with a
small circular hole of radius $\delta$ filters the incident beam of
projectiles to secure the position of the orbit. Before the diaphragm,
the beam may be represented by a plane wave with the de Broglie
wavelength
\beq
\lambda_{\rm dB} = \frac{2 \pi \hbar}{p_{\rm i}}.
\label{4.56}\eeq
It is assumed that the hole is limited by partly permeable edges so that the
intensity of the transmitted beam, as a function of the lateral distance
from the center of the hole, is Gaussian with standard deviation
$\delta$. According to quantum mechanics, the diaphragm will produce
diffraction and the transmitted beam will diverge with an angular
aperture of the order of \citep[see, \eg,][Section 8.5.2]{BornWolf2002}
\beq
(\Delta \theta)_{\rm diff} = \frac{\lambda_{\rm dB}}{2 \pi \; 2 \delta} =
\frac{\hbar}{p_{\rm i} \, 2 \delta}.
\label{4.57}\eeq

\begin{figure}[htb] \begin{center}
\includegraphics*[width=13.0cm]{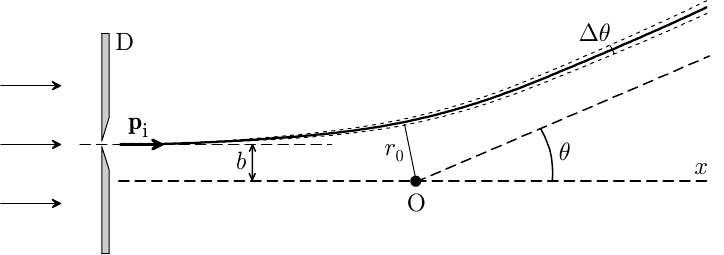}
\caption{
Scattering of particles by a central field. The diaphragm D,
with a small hole, serves to determine the precise localization of the
orbit.
\label{fig4.6}}
\end{center} \end{figure}

Because of the finite size of the hole, there is an uncertainty in the
impact parameter, $\Delta b \sim \delta$, which leads to an additional
uncertainty in the deflection of the transmitted particles caused by the
field. Assuming that the function $L(\theta)$ is single-valued (and,
therefore, that $\theta=|\vartheta|$), the deflection angles produced by
the field will be distributed around the mean value with standard
deviation
\beq
(\Delta \theta)_{\rm field} =
\left| \frac{\d \theta}{\d b} \right|
\delta
= p_{\rm i} \delta
\, \left| \frac{\d \theta}{\d L} \right|
= \frac{p_{\rm i} \delta}{\hbar}
{\cal X},
\label{4.58}\eeq
where
\beq
{\cal X} \equiv
\hbar \left| \frac{\d \theta}{\d L} \right| .
\label{4.59}\eeq

The combined effect of the diffraction by the diaphragm and the field is
to produce, far beyond the scattering center, a diverging beam with
aperture $\Delta \theta$ given by
\beq
(\Delta \theta)^2
= (\Delta \theta)_{\rm diff}^2 + (\Delta \theta)_{\rm field}^2
= \left( \frac{\hbar}{p_{\rm i} \; 2 \delta} \right)^2 +
\left( \frac{p_{\rm i} \delta}{\hbar} \right)^2 {\cal X}^2.
\label{4.60}\eeq
The minimal aperture is obtained for a hole of radius
\beq
\delta_{\rm min} = \frac{\hbar}{p_{\rm i}} \frac{1}{\sqrt{2{\cal X}}},
\label{4.61}\eeq
for which
\beq
(\Delta \theta)_{\rm min}^2 =
\left( \frac{\hbar}{p_{\rm i} \; 2 \delta_{\rm min}} \right)^2 +
\left( \frac{p_{\rm i} \delta_{\rm min}}{\hbar} \right)^2 {\cal X}^2
= {\cal X}.
\label{4.62}\eeq
The aperture $(\Delta \theta)_{\rm min}$ should be small compared with
the mean deflection $\theta$ of the particles due to the field, because
otherwise it will overshadow the orbital deflection. Consequently, the
classical picture will be valid only when
\beq
T_{\rm class}(\theta) \equiv \frac{(\Delta \theta)_{\rm min}}{\theta}
= \frac{\sqrt{{\cal X}}}{\theta} \ll 1.
\label{4.63}\eeq

\noindent $\bullet$ {\bf Scattering by a Coulomb potential} \\
In the case of the Coulomb potential, Eq.\ \req{4.36}, the scattering
law reads [see Eq.\ \req{4.39}]
\beq
\theta(L) = 2 \arctan \left( \frac{p_{\rm i} d_0}{2 L} \right)\,
\qquad \mbox{or} \qquad
L(\theta) = \frac{p_{\rm i} d_0}{2} \, \frac{1}{\tan(\theta/2)},
\label{4.64}\eeq
where
\beq
d_0 \equiv \left| \frac{2 M Z_0 Z e^2 }{p_{\rm i}^2} \right|
\label{4.65}\eeq
is the minimum distance of approach in a head-on collision (with $L=0$)
for a repulsive field. Since the function $\theta(L)$ is single-valued,
\beq
{\cal X} = \hbar \, \left| \frac{\d \theta}{\d L} \right|
= \frac{4\hbar}{p_{\rm i} d_0} \sin^2(\theta/2)
= \frac{2}{|\eta|}\, \sin^2(\theta/2),
\label{4.66}\eeq
where
\beq
|\eta| \equiv \frac{p_{\rm i} d_0}{2 \hbar} = \frac{\pi d_0}{\lambda_{\rm dB}}
= \left| \frac{M Z_0 Z e^2}{\hbar p_{\rm i}} \right|
\label{4.67}\eeq
is the absolute value of the Sommerfeld parameter [cf.\ Eq.\ \req{2.47}].
Hence,\index{Sommerfeld parameter}
\beq
T_{\rm class}(\theta)
= \sqrt{\frac{2}{|\eta|}}
\frac{ \sin(\theta/2)}{\theta}\, .
\label{4.68}\eeq
The second factor on the right-hand side changes smoothly between $\1o2$ and
$1/\pi$ when $\theta$ varies from 0 to $\pi$ and, consequently, the
condition \req{4.63} is satisfied at all angles when the Sommerfeld
parameter is much larger than unity. Therefore, the necessary condition
for the validity of the classical theory of scattering by the Coulomb
field is
\beq
|\eta| \gg 1 \qquad \mbox{or} \qquad  \lambda_{\rm dB} \ll d_0.
\label{4.69}\eeq

\noindent $\bullet$ {\bf Scattering by a screened Coulomb potential} \\
\citet{Bohr1948} also analyzed the validity of the classical theory for
screened Coulomb potentials, taking the Wentzel potential [Eq.\
\req{3.151}]
\beq
V_{\rm W}(r) = \frac{Z_0 Z e^2}{r} \, \exp(-r/R),
\label{4.70}\eeq
as representative of the potentials found in atomic collisions. Since it
is not possible to give an analytical expression for the angular
deflection $\vartheta(b)$ for this potential, Bohr noted that
large-angle interactions correspond to small impact parameters, much smaller
than the atomic radius $R$, for
which screening effects are small and the scattering is dominated by the
unscreened potential of the bare nucleus. Therefore, for large
scattering angles the classical theory is expected to be applicable
when $\lambda_{\rm dB} \ll d_0$.

\index{Wentzel potential}

For angles that are not large, Bohr argued that, because the Wentzel
potential varies appreciably over distances of the order of $R$, it is
to be expected that the classical picture will be valid only when
$\lambda_{\rm dB} \ll R$. In addition, if the hole in the diaphragm has a radius
$\delta$ smaller than $R$, Eq.\ \req{4.57} implies that it will be
impossible to resolve deflections, or scattering angles, smaller than
\beq
\theta_{\rm class} = \frac{\lambda_{\rm dB}}{2 \pi R}
= \frac{\hbar}{p_{\rm i} R} \, .
\label{4.71}\eeq
Hence, the classical calculation may be valid only for scattering angles
larger than $\theta_{\rm class}$. In most cases of practical interest,
$\lambda_{\rm dB} \ll R$ and $\theta_{\rm class} \ll 1$ and, under these
circumstances, quantum calculations (see Section \ref{sec5.1})
confirm that the classical theory is effectively valid for $\theta
\gtrsim \theta_{\rm class}$.

The program {\sc elastic} (Chapter \ref{chapt10}) calculates DCSs for
scattering in screened Cou\-lomb potentials by using the
classical-trajectory method and several quantum approximations. The
ranges of validity of these approaches can then be assessed by simply
comparing the calculation results. As indicated above, {\sc elastic}
builds an accurate representation of the function $\vartheta(L)$ in the
form of a cubic spline, $\vartheta(\xi)$ with $\xi=\ln(L^2)$ from which
the quantity
\beq
\frac{\sqrt{\chi}}{\theta} = \frac{1}{\theta} \sqrt{
\hbar \left| \frac{\d \theta}{\d L} \right| }
= \frac{1}{\theta} \sqrt{
\hbar \left| \frac{\d \theta}{\d \xi} \, 2 \exp(-\xi/2) \right| }
\label{4.72}\eeq
can be readily evaluated. Comparison of classical and quantum results
indicates that the Bohr criterion, Eq.\ \req{4.63}, is far too
restrictive. The classical and quantum DCSs are found to differ by less
than about 2 \% when
\beq
T_{\rm class}(\theta) = \frac{\sqrt{\chi}}{\theta} \lesssim 0.8,
\label{4.73}\eeq
and the smaller the value of $T_{\rm class}(\theta)$ the better the
agreement. In the case of scattering by screened Coulomb potentials, the
condition \req{4.73} is met for angles larger than a certain value
$\theta_{\rm ccl}$ which is determined by {\sc elastic}. The program
delivers the classical DCS only for angles larger than $\theta_{\rm
ccl}$.

\noindent $\bullet$ {\bf Scattering by the potential $C_2 r^{-2}$} \\ In
the relativistic extension of the classical theory, the atomic potential
is replaced with an effective potential [see Eq.\ \req{4.156}] that
contains a term $V_{\rm r1}(r)$ that diverges at the origin as $r^{-2}$.
It is then of interest to investigate the validity of the trajectory
method for the potential
\beq
V_2(r) = \frac{C_2}{r^2}.
\label{4.74}\eeq
The distance of closest approach $r_0$ [see Eq.\ \req{4.25}] is
\beq
r_0 = \frac{1}{p_{\rm i}} \sqrt{2MC_2+ L^2}
\label{4.75}\eeq
and the angular deflection is given by Eq.\ \req{4.28}. With the usual
change of integration variable to $u=r_0/r$, we have
\beqa
\vartheta(L) &=& \pi - 2 L \int_0^1
\frac{\d u}{\sqrt{p_{\rm i}^2 r_0^2-2M C_2 u^2 -L^2u^2}}
\nonumber \\ [2mm]
&=& \pi - \frac{2 L}{\sqrt{2M C_2+L^2}} \left[ \arcsin \left(
\frac{u \sqrt{2M C_2+L^2}}{p_{\rm i} r_0} \right)
\right]_0^1
\nonumber \\ [2mm]
&=& \pi \left( 1 - \frac{L}{\sqrt{2M C_2+L^2}} \right) .
\label{4.76}\eeqa
Interestingly, the function $\vartheta(L)$ is formally independent of
the initial momentum of the projectile. The inverse relation is
\beq
L^2 (\vartheta) = \frac{2M C_2 \,(\pi - \vartheta)^2}{2\pi
\vartheta - \vartheta^2}\, .
\label{4.77}\eeq
Therefore,
\beq
\frac{\d L}{\d \vartheta} =
\pi^2 \left( \frac{2 MC_2}{(2\pi
\vartheta - \vartheta^2)^3} \right)^{1/2}
\label{4.78}\eeq
and
\beq
L \left| \frac{\d L}{\d \vartheta} \right| = 2M \, |C_2|\,
\frac{\pi^2 (\pi - \vartheta)}{(2\pi \vartheta - \vartheta^2)^2}.
\label{4.79}\eeq

For repulsive forces ($C_2 > 0$), the deflection angle
$\vartheta$ is positive and $\le \pi$. Hence, $\theta = \vartheta$ and
\beqa
{\cal X} =
\hbar \left| \frac{\d L}{\d \theta} \right|^{-1}
&=& \left( \frac{\hbar^2}{2MC_2} \right)^{1/2} \, \frac{
\left( 2 \pi \theta - \theta^2 \right)^{3/2}}{\pi^2}.
\label{4.80}\eeqa
The corresponding DCS is
\beq
\frac{\d \sigma}{\d \Omega} = \frac{L}{p_{\rm i}^2 \sin\theta} \left|
\frac{\d L}{\d \theta} \right|
= \frac{2MC_2}{p_{\rm i}^2 \sin\theta} \,
\frac{ \pi^2 (\pi - \theta)}{(2\pi
\theta - \theta^2)^2}\, ,
\label{4.81}\eeq
and the validity of this result is determined by the ratio
\beq
T_{\rm class} = \frac{\sqrt{\cal X}}{\theta} =
\left( \frac{\hbar^2}{2MC_2} \right)^{1/4}
\left[ \frac{\sqrt{2}}{\pi} \,
\frac{ \left( 2 \pi - \theta \right)^{3/4}}{\theta^{1/4}} \right].
\label{4.82}\eeq
The function in square brackets diverges as $\theta^{-1/4}$ at
$\theta=0$, and decreases monotonically when $\theta$ increases to reach
the value $\sqrt{2/\pi}$ at $\theta=\pi$. Consequently, the classical
DCS may not be valid for small scattering angles, and for large angles
it will be correct only if $2M C_2 \gg h^2$.

It is worth noticing that for attractive forces ($C_2 < 0$), the angular
deflection \req{4.76} is defined only for $L>L_{\rm c}$ with the cutoff value
$L_{\rm c}$ given by
\beq
L_{\rm c}^2 = - 2M C_2\, .
\label{4.83}\eeq
The function $\vartheta(L)$ varies monotonically with $L$, going from
$\vartheta(L_{\rm c})=-\infty$ to $\vartheta(\infty)=0$. The angular
momentum corresponding to $\vartheta = - \pi$ is $L(-\pi)=2L_{\rm c}/\sqrt{3}$. Projectiles with
impact parameters less than $b_{\rm c}=L_{\rm c}/p_{\rm i}$ have spiral
orbits that fall to the scattering center after an infinite number of
revolutions. Each value of the scattering angle, $\theta \in (0,\pi)$,
corresponds to a denumerable set of possible deflection angles,
$\vartheta_j=-\theta-j \, 2 \pi$ ($j \ge 0$). Hence, the DCS is the sum
of contributions from all these deflections, which are associated to
angular momenta arbitrarily close to $L_{\rm c}$. That is
\beq
\frac{\d \sigma}{\d \Omega} =
\sum_j \frac{\d \sigma(\vartheta_j)}{\d \Omega}
=  \frac{2M |C_2|}{p_{\rm i}^2 \sin\theta} \,
\sum_j \frac{ \pi^2 (\pi - \vartheta_j)}{(-2\pi
\vartheta_j + \vartheta_j^2)^2}.
\label{4.84}\eeq
Bohr's gedanken experiment is not suited for attractive potentials of
the type \req{4.74}, partly because a diaphragm with a hole at an
impact parameter $b$ larger than $L(-\pi)/p_{\rm i}$, will block
particles with impact parameters close to $b_{\rm c}$ that also
contribute to the DCS at $\theta$. In fact, for a given radius of the
hole, when the impact parameter approaches $b_{\rm c}$ the field
aperture \req{4.58} becomes larger than $\pi$, indicating that the
classical theory may not be valid for any scattering angle.


\section{Non-relativistic classical collisions \label{sec4.2}}

\index{classical collisions}
The theory of scattering by a fixed center of force provides an
acceptable approximation for describing collisions of two particles only
when the projectile has a much smaller mass than the target particle.
This condition holds approximately, for instance, in the case of elastic
collisions of electrons or positrons with atoms (or ions). In other
cases, however, the colliding particles may have comparable masses and a
more elaborate treatment is needed. We now consider collisions of a
``projectile'' with mass $M_1$ and a ``target'' of mass $M_2$, which
interact through a central force. That is, the interaction potential
energy between the two particles has the form $V(|{\bf r}_1 - {\bf
r}_2|)$ where ${\bf r}_1$ and ${\bf r}_2$ are the position vectors of
the projectile and the target, respectively. To facilitate preliminary
arguments, we may assume that the potential has a finite range.

\index{laboratory frame of reference}
The collision experiment is sketched in Fig.\ \ref{fig4.7}. As seen from
the laboratory frame of reference (L), where the target particle (2) is
initially at rest, the projectile particle (1) approaches the target
with velocity ${\bf v}_{1{\rm i}}$. In the plot ${\bf v}_{1{\rm i}}$ is
parallel to the polar $x$ axis of the scattering plane and, therefore,
the linear momenta of the projectile and the target before the collision
are ${\bf p}_{1{\rm i}}=M_1 v_{1{\rm i}} \hat{\bf x}$ and ${\bf
p}_{2{\rm i}}= {\bf 0}$, respectively. After the collision [\ie, long
after the interaction, when $V(r)$ effectively vanishes], the two
particles move outward with velocities ${\bf v}_{1{\rm f}}$ and ${\bf
v}_{2{\rm f}}$ in directions corresponding to the scattering angles
$\theta_{1}$ and $\theta_{2}$, respectively.

The equations of motion are are
\beq
\dot{\bf p}_1 = {\bf F}, \qquad \dot{\bf p}_2 = -{\bf F}
\label{4.85}\eeq
with
\beq
{\bf F} = - \nablab V(r) = - \frac{\d V}{\d r}\, \hat{\bf r}.
\label{4.86}\eeq
where ${\bf r} = {\bf r}_1 - {\bf r}_2$ is the position vector of the
projectile relative to the target. The relative velocity is
\beq
{\bf v} = \dot{\bf r}_1 - \dot{\bf r}_2
= \frac{{\bf p}_1}{M_1} - \frac{{\bf p}_2}{M_2},
\label{4.87}\eeq
and the relative acceleration is
\beq
\dot{\bf v} =  \frac{\dot {\bf p}_1}{M_1} - \frac{\dot {\bf p}_2}{M_2}
= \left( \frac{1}{M_1} + \frac{1}{M_2} \right) {\bf F}.
\nonumber \eeq
Consequently, we have
\beq
\dot{\bf v} = {\bf F} / \mu,
\label{4.88}\eeq
where \index{reduced mass}
\beq
\mu \equiv \frac{M_1M_2}{M_1+M_2}\,
\label{4.89}\eeq
is the {\it reduced mass} of the two particles. That is, the
relative position vector describes the motion of a particle having the
reduced mass under the action of the interaction force.

\index{laboratory frame of reference}
\index{center-of-mass frame of reference}
\begin{figure}[htb] \label{kinem2} \begin{center}
\includegraphics*[scale=0.85]{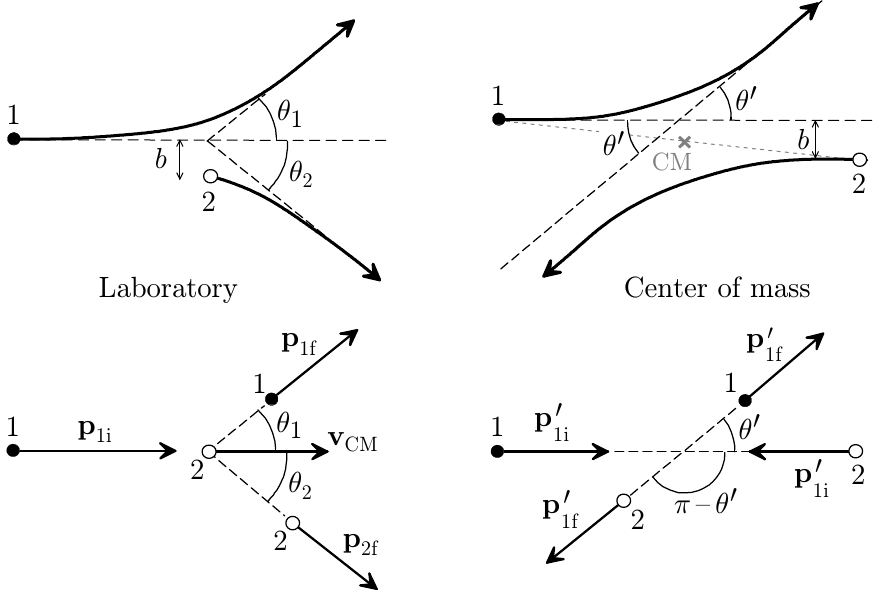}
\caption{
Elastic collisions in the laboratory and in the center of mass
frames. Particle trajectories (top) and momentum diagrams (bottom).
\label{fig4.7}}
\end{center} \end{figure}


\subsection{Relative motion and differential cross section in the CM frame
\label{sec4.2.1}}

\index{classical collisions!relative motion}
\index{cross sections in the center-of-mass frame}
Elastic collisions are examples of two-body processes. It is well known
that the motion of a system of two particles that interact through a
conservative force can be described as the combination of two simpler
motions, namely, that of the center of mass, which moves with constant
velocity ${\bf v}_{\rm CM}$, and the relative motion ${\bf r} (t)$.
Consequently, calculations are easier in the reference frame of the {\it
center of mass} (CM), {\it center of momentum} or {\it barycentric}
frame, \index{center-of-mass frame of reference}
which is a coordinate system with its origin at the center of
mass, its axes parallel to those of the L system, and such that the
total momentum vanishes. In what follows, particle quantities relative
to the CM frame will be indicated by primes, unless the context makes
this distinction unnecessary. As seen from CM, before the interaction
the projectile and the target move toward the origin with momenta
\beq
{\bf p}'_{1{\rm i}} = M_1 \left( {\bf v}_{1{\rm i}} - {\bf v}_{\rm CM} \right)
\qquad \mbox{and} \qquad
{\bf p}'_{2{\rm i}} = M_2 \left( {\bf v}_{2{\rm i}} - {\bf v}_{\rm CM} \right).
\label{4.90}\eeq
By definition, in the CM frame the total
momentum vanishes, ${\bf p}'_{1} + {\bf p}'_{2} =0$, and this implies that
\beq
{\bf v}_{\rm CM} = \frac{ M_1 {\bf v}_{1{\rm i}} + M_2 {\bf
v}_{2{\rm i}}}{M_1+M_2} =  \mu \left( \frac{{\bf v}_{1{\rm i}}}{M_2} +
\frac{{\bf
v}_{2{\rm i}}}{M_1} \right).
\label{4.91}\eeq
Of course, ${\bf v}_{\rm CM} = \dot{\bf r}_{\rm CM}$, where
\beq
{\bf r}_{\rm CM} = \frac{ M_1 {\bf r}_1 + M_2 {\bf
r}_2}{M_1+M_2}\, ,
\label{4.92}\eeq
is the position vector of the center of mass of the two particles
relative to the L frame. Inserting the expression \req{4.91},
Eqs.\ \req{4.90} give
\beq
{\bf p}'_{1{\rm i}} = M_1 {\bf v}'_{1{\rm i}} =
\mu ({\bf v}_{1{\rm i}}-{\bf v}_{2{\rm i}}), \qquad
{\bf p}'_{2{\rm i}} = M_2 {\bf v}'_{2{\rm i}} =
- \mu ({\bf v}_{1{\rm i}}-{\bf v}_{2{\rm i}}).
\label{4.93}\eeq
That is, the initial momenta of the projectile and the target in CM are
${\bf p}'_{1{\rm i}}={\bf p}'_{\rm i}$ and ${\bf p}'_{2{\rm i}}=-{\bf
p}'_{\rm i}$ with
\beq
{\bf p}'_{\rm i} = \mu {\bf v}_{\rm i}\, ,
\label{4.94}\eeq
where ${\bf v}_{\rm i} = {\bf v}_{1{\rm i}}-{\bf v}_{2{\rm i}}$ is the
initial velocity of the projectile relative to the target.
After the collision, the particles move outwards with
momenta ${\bf p}'_{1{\rm f}}$ and ${\bf p}'_{2{\rm f}}$. Because in the
CM frame energy and momentum are conserved, we have ${\bf p}'_{1{\rm
f}}= - {\bf p}'_{2{\rm f}} \equiv {\bf p}'_{\rm f}$ with $p'_{\rm
f}=p'_{\rm i}$, that is, the momenta of the particles before and after
the collision have the same magnitude.
This means that, when observed from the CM frame, the
elastic collision causes only a rotation of the momentum vectors of the
particles (see Fig.\ \ref{fig4.7}). Each collision is then completely
determined by the value of the scattering angle $\theta'$.

The equations of motion in the CM frame are
\beq
\dot{\bf p}'_{1} = {\bf F}
\qquad \mbox{and} \qquad
\dot{\bf p}'_{2} = -{\bf F}
\label{4.95}\eeq
with the force ${\bf F} = - (\d V/\d r') \hat{\bf r}'$ depending only on the
relative position vector, ${\bf r}' = {\bf r}'_{1} - {\bf r}'_{2}$.
The relative velocity is
\beq
{\bf v}' = {\bf v}'_{1} - {\bf v}'_{2} =
\frac{{\bf p}'_{1}}{M_1} -
\frac{{\bf p}'_{2}}{M_2} = \left(
\frac{1}{M_1} + \frac{1}{M_2} \right)
{\bf p}' = \frac{1}{\mu} \, {\bf p}'.
\label{4.96}\eeq
The vector
\beq
{\bf p}' \equiv {\bf p}'_{1} = \mu {\bf v}'
\label{4.97}\eeq
will be referred to as the {\it relative momentum}.
The conservation of energy and angular momentum implies that the
total energy in the CM frame,
\beq
E' = \frac{({\bf p}'_{1})^2}{2M_1} + \frac{({\bf p}'_{2})^2}{2M_2}
+ V(r') = \frac{p'^2}{2\mu} + V(r'),
\label{4.98}\eeq
and the angular momentum in CM,
\beq
{\bf L}' = {\bf r}'_{1} \vecprod {\bf p}'_{1} +
 {\bf r}'_{2} \vecprod {\bf p}'_{2}
= {\bf r}'_{1} \vecprod {\bf p}' -  {\bf r}'_{2} \vecprod {\bf p}'
= {\bf r}' \vecprod {\bf p}'
\label{4.99}\eeq
are constants of the motion. Since ${\bf L}' = \mu {\bf r}' \vecprod
{\bf v}'$, the trajectory ${\bf r}'(t)$ of the relative motion lies on the
plane of scattering, which also contains the trajectories of the two
particles. Before the interaction $V(r)=0$ and the magnitude  of the initial
relative momentum, $p'_{\rm i}=\mu v'_{\rm i}$, determines the total
energy, $E' = p'^2_{\rm i}/(2\mu)$. Hence, Eq.\ \req{4.98} implies that
\beq
p'^2(r') = p'^2_{\rm i} - 2\mu V(r')\, .
\label{4.100}\eeq
Considering polar coordinates in the scattering
plane (see Section \ref{sec4.1}), the conservation of angular momentum
means that [see Eq.\ \req{4.12}]
\beq
\dot{\varphi}' = \frac{L'}{\mu r'^2} .
\label{4.101}\eeq
In addition, the equality $v'^2 = \dot{r}'^2 + r'^2 \dot{\varphi}'^2$
[which follows from Eq.\ \req{4.2b}], combined with the definition
$v'=p'/\mu$ and the results \req{4.100} and \req{4.101}, gives
\beq
\dot{r}' =
\pm \, \frac{1}{\mu} \sqrt{p'^2 - \frac{L'^2}{r'^2}},
\label{4.102}\eeq
where the sign of the square root is $-$ when the particles approach
each other and $+$ when the particles have passed the distance of
closest approach $r'_0$. The latter is determined by the condition
$\dot{r}' = 0$, \ie, as the largest root of the equation
\beq
p'^2(r'_0) - \frac{L'^2}{r'^2_0} =0.
\label{4.103}\eeq
Combining the Eqs.\ \req{4.101} and \req{4.102} we obtain the
equation of the relative motion,
\beq
\d \varphi' = \frac{\d \varphi'}{\d t} \, \frac{\d t}{\d r'} \, \d r'
= \frac{\dot{\varphi}'}{\dot{r}'} \, \d r' = \pm
\frac{L'/r'^2}{\sqrt{p'^2(r')-L'^2/r'^{2}}}\, \d r',
\label{4.104}\eeq
which has the same form as the equation of motion of a particle of mass
$\mu$ in the potential $V(r)$ [cf. Eq.\ \req{4.26}]. Therefore, the DCS in
the CM frame is the same as the DCS for scattering of a particle of mass
$\mu$ by the potential $V(r)$.

For instance, in the case of the Coulomb interaction [Eq.\
\req{4.36}] the DCS in CM is readily obtained from the result
\req{4.43},
\beq
\frac{\d \sigma_{\rm R}}{\d \Omega'}
 = \left(
\frac{Z_0 Z e^2}{ 2 v'_{\rm i} p'_{\rm i}} \right)^2
\frac{1}{\sin^4(\theta'/2)} \, ,
\label{4.105}\eeq
where $v'_{\rm i}=p'_{\rm i}/\mu$ is the relative velocity of the
incident projectile with respect to the target, which has equal values
in the L and CM frames (\ie, $v_{\rm 1i}=v'_{\rm i}$).


\subsection{Cross sections in the L frame \label{sec4.2.2}}

\index{cross sections in the laboratory frame}
Now, we wish to transform the DCS from the CM frame to the L frame,
which is where experimental observations are usually made. We recall
that the coordinate axes of CM are assumed to be parallel to those
of the L frame. For simplicity, we also assume that the velocity ${\bf
v}_{\rm CM}$ of CM is parallel to the $z$ axis of the L frame. Let us
first consider the general situation of a particle moving with
velocities ${\bf v}$ and ${\bf v}'$ in L and CM, respectively. The
vector identity ${\bf v}={\bf v}_{\rm CM}+{\bf v}'$ implies that
\begin{subequations}
\label{4.106}
\beqa
v \sin\theta &=& v' \sin\theta'\, ,
\label{4.106a} \\ [2mm]
v \cos\theta &=& v_{\rm CM} + v' \cos\theta'\, ,
\label{4.106b}\eeqa
\end{subequations}
where $\theta$ and $\theta'$ are the polar angles of the velocity in
L and CM, respectively. As these angles range from 0 to $\pi$,
$\cos\theta$ and $\cos\theta'$ are also unambiguous direction
variables. Because the motion of the CM frame is along the $z$ axis, the
azimuthal angle of the particle velocity in CM is the same as in L,
$\phi'=\phi$.

Equations \req{4.106} imply that
\beq
v = v' \sqrt{1 + \tau^2 + 2 \tau \cos\theta'},
\label{4.107}\eeq
with
\beq
\tau = \frac{v_{\rm CM}}{v'}.
\label{4.108}\eeq
Combining Eqs.\ \req{4.107} and \req{4.106b} we obtain
\beq
\cos\theta = \frac{\tau + \cos\theta'}{\sqrt{1 + \tau^2
+ 2 \tau \cos\theta'}}\, ,
\label{4.109}\eeq
which gives the direction in L that corresponds to the direction
$\cos\theta'$ in CM.

To obtain the inverse relation, we write Eqs.\ \req{4.106} in the form
\begin{subequations}
\beqa
\label{4.110}
v' \sin\theta' &=& v \sin\theta\, ,
\label{4.110a} \\ [2mm]
v' \cos\theta' &=& v \cos\theta - v_{\rm CM} \, .
\label{4.110b}\eeqa
\end{subequations}
Adding the squares, we obtain the quadratic equation
\beq
v'^2 = v^2 + 2 v v_{\rm CM} \cos\theta + v_{\rm CM}^2
\label{4.111}\eeq
with the formal solution
\beq
v = v' \left( \tau \cos\theta \pm \sqrt{ \tau^2 \cos^2 \theta
+ 1 - \tau^2 } \right).
\label{4.112}\eeq
Note that when $\tau <1$ the discriminant $\tau^2 \cos^2 \theta +
1 - \tau^2$ is positive, but only the plus sign gives a valid
solution, with $v$ positive. When $\tau \ge 1$, real solutions exist
only if the discriminant is positive and then the two solutions are
valid. Inserting the expression \req{4.112} into Eq.\ \req{4.110b}, we get
\beq
\cos \theta' = -\tau (1-\cos^2 \theta) \pm \cos\theta
\sqrt{1-\tau^2 (1 - \cos^2 \theta) }\, .
\label{4.113}\eeq
If $\tau \ge 1$, the requirement of a positive discriminant implies that
the angle $\theta$ in L has the upper bound
\beq
\theta_{\rm max} = \arccos \left(\sqrt{1-\frac{1}{\tau^2}}
\right)\, .
\label{4.114}\eeq
The angle in CM corresponding to $\theta_{\rm max}$ is
$\theta'=\arccos(-1/\tau)$.

Figure \ref{fig4.8} is meant to illustrate these transformation
relations for a given value of the velocity $v'$ in CM. When $\tau < 1$,
only the plus sign in expression \req{4.113} is valid; there is a unique
correspondence between the angles $\theta'$ and $\theta$, and the latter
may take any value in the complete interval from 0 to $\pi$. When $\tau
\ge 1$, $\theta$ can only take values from 0 to $\theta_{\rm max}$, and
for each allowed direction $\theta$ in L there are two different
directions $\theta'$ in CM, which correspond to different velocities $v$
in L.

\begin{figure}[htb] \begin{center}
\includegraphics*[scale=0.82]{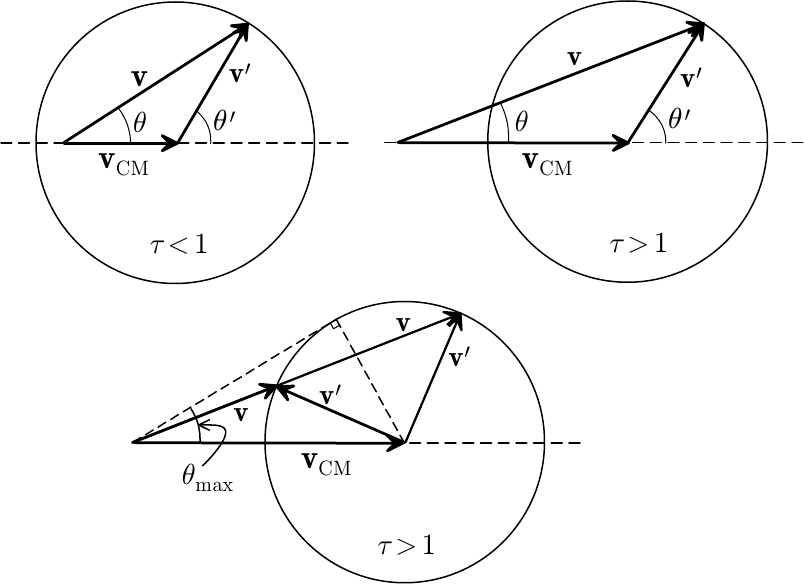}
\caption{
Relationship between the polar direction angles in L and CM, for
$\tau=v_{\rm CM}/v' \lessgtr 1$.
\label{fig4.8}}
\end{center} \end{figure}

The transformation of the DCS from CM to L results from the requirement
that the number of projectile particles scattered to directions within
the solid angle subtended by the detector be the same in both reference
frames. Recalling that the azimuthal angles of the final direction in CM
and L are equal, $\phi_1=\phi'$, we require
\beq
\frac{\d\sigma}{\d\Omega_1}\, \d (\cos\theta_1) \, \d\phi_1 =
\frac{\d\sigma}{\d\Omega'}\, \d(\cos\theta') \, \d\phi'.
\label{4.115}\eeq
That is,
\beq
\frac{\d\sigma}{\d\Omega_1} =
\left| \frac{\d(\cos\theta')}{\d(\cos\theta_1)} \right|
\frac{\d\sigma}{\d\Omega'}.
\label{4.116}\eeq
From Eq.\ \req{4.113},
\beq
\left| \frac{\d(\cos\theta')}{\d(\cos\theta_1)} \right|
= \frac{\left(\tau_1 \cos\theta_1
\pm \sqrt{1 - \tau_1^2 (1-\cos^2 \theta_1)}\; \right)^2}
{\sqrt{1 - \tau_1 (1 - \cos^2 \theta_1)}}\, ,
\label{4.117}\eeq
with
\beq
\tau_1 = \frac{v_{\rm CM}}{v'_{1{\rm f}}} =
\frac{\mu v_{1{\rm i}}/M_2}{\mu v_{1{\rm i}} / M_1}
= \frac{M_1}{M_2},
\label{4.118}\eeq
where use has been made of the equalities $v_{\rm CM}= v'_{\rm 2i} =
p_{\rm i}/M_2$ and
$v'_{1{\rm f}} = p_{\rm i}/M_1$. The DCS obtained by detecting
the scattered projectiles in the L frame is
\beq
\frac{\d\sigma}{\d\Omega_1} =
\frac{\left( \tau_1 \cos\theta_1
\pm \sqrt{1 - \tau_1^2 (1 - \cos^2 \theta_1)}\; \right)^2}
{\sqrt{1 - \tau_1 (1 - \cos^2 \theta_1)}}\,
\frac{\d\sigma(\cos\theta')}{\d\Omega'}\, .
\label{4.119}\eeq
When $M_1 \le M_2$, only the plus sign before the square root must be
considered. For $M_1 > M_2$, the expression on the right-hand side has to be
calculated for both the plus and the minus signs and the two resulting
expressions must be added to obtain the DCS $\d\sigma/\d \Omega_1$ in
the L frame; evidently, in this case the DCS vanishes for scattering
angles $\theta_1$ larger than $\theta_{\rm max}$. When the mass of
the target is much larger than
that of the projectile ($\tau_1 \ll 1$),
\beq
\frac{\d\sigma}{\d\Omega_1} \simeq \frac{\d\sigma}{\d\Omega'},
\label{4.120}\eeq
as it should be.

In the CM frame, after a collision with scattering angle $\theta'$ the
projectile and the target atom move in opposite directions, ($\theta',
\phi'$) and ($\pi - \theta'$, $\phi'+\pi$). Hence, the polar angles
of the recoil directions of the atom in CM and L are related by [see
Eq.\ \req{4.109}]
\beq
\cos\theta_2 =
\frac{\tau_2+\cos(\pi - \theta')}{\sqrt{1+\tau_2^2+2\tau_2\cos(\pi -
\theta')}}\, ,
\label{4.121}\eeq
with
\beq
\tau_2 = \frac{v_{\rm CM}}{v'_{2{\rm f}}} =
\frac{\mu v_{1{\rm i}}/M_2}{\mu v_{1{\rm i}} / M_2}
= 1.
\label{4.122}\eeq
That is,
\beq
\cos \theta_2 = \sqrt{\frac{1-\cos\theta'}{2}} = \sin(\theta'/2) .
\label{4.123}\eeq
Evidently, the polar angle $\theta_2$ of the recoil direction of the
atom in L cannot exceed 90 degrees. When $M_1=M_2$ (\eg, in collisions
of protons with hydrogen ions), $\tau_1=1$ and $\cos \theta_1 =
\cos(\theta'/2)$ so that $\sin(\theta_1+\theta_2)=1$ and, therefore, the
final directions of the two particles in L form a right angle.

For calculations of stopping of charged particles in matter, it is
interesting to express the DCS in terms of the energy loss of the
projectile in the L frame, $W = E_{1{\rm i}} - E_{1{\rm f}}$. The
energy of the projectile after the collision is
\beq
E_{1{\rm f}} = \frac{M_1 v_{1{\rm f}}^2}{2} = \frac{M_1 ({\bf
v}'_{1{\rm f}}+{\bf v}_{\rm CM})^2}{2} =
\frac{M_1 ({v'^2_{1{\rm f}}} + v_{\rm CM}^2 + 2 v'_{1{\rm f}} v_{\rm CM}
\cos\theta')}{2}\, .
\nonumber \eeq
Recalling that $v_{\rm CM} = \mu v_{1{\rm i}}/M_2$ and $v'_{1{\rm f}} =
\mu v_{1{\rm i}} / M_1$, we can write
\beq
E_{1{\rm f}} = E_{1{\rm i}} \, \frac{M_1^2 + M_2^2 +
2 M_1 M_2 \cos\theta'}{(M_1+M_2)^2}\, .
\label{4.124}\eeq
Hence, the energy lost by the projectile is
\beq
W = E_{1{\rm i}} - E_{1{\rm f}}
= \frac{2M_1M_2}{(M_1+M_2)^2} \, E_{1{\rm i}} \, (1-\cos\theta')\, ,
\label{4.125}\eeq
or
\beq
W = W_{\rm max} \, \frac{1-\cos\theta'}{2} = W_{\rm max} \,
\sin^2(\theta'/2)\, ,
\label{4.126}\eeq
where
\beq
W_{\rm max} = \frac{4 M_1 M_2}{(M_1+M_2)^2} \, E_{1{\rm i}}
\label{4.127}\eeq
is the maximum energy loss in a collision. Since in L the dependence of the
DCS on the scattering angle of the projectile is quite involved, it
is advantageous to start from the DCS in CM and write
\index{energy-loss DCS}
\beq
\frac{\d \sigma}{\d W} =
\frac{\d \sigma}{\d \Omega'}
\, \frac{2\pi \sin\theta'\, \d \theta'}{\d W} =
\frac{\d \sigma}{\d \Omega'} \, 2\pi
\left| \frac{\d W}{\d (\cos\theta')} \right|^{-1}
= \frac{4\pi}{W_{\rm max}} \,
\frac{\d \sigma}{\d \Omega'} \, ,
\label{4.128}\eeq
where the last factor is the DCS evaluated at a direction with polar
scattering angle $\theta'$ such that $\cos\theta' = 1 - 2 W/W_{\rm max}$.
In the case of collisions of two charged particles with charges $Z_1 e$
and $Z_2 e$, the DCS in CM is given by the Rutherford formula
\req{4.105} and \index{Thomson cross section}
\beq
\frac{\d \sigma}{\d W} =
\frac{4\pi}{W_{\rm max}} \left(
\frac{Z_1 Z_2 e^2}{ 2 \mu v'^2_{{\rm i}}} \right)^2
\frac{1}{\sin^4(\theta'/2)}
= \frac{2\pi (Z_1 Z_2 e^2)^2}{M_2 v_{1{\rm i}}^2} \,
\frac{1}{W^2}\, .
\label{4.129}\eeq
This result is known as the {\it Thomson formula} \citep{Thomson1912}.
Notice that the energy-loss DCS is inversely proportional to the squared
relative velocity, $v_{1{\rm i}}^2$, and to the mass $M_2$ of the target
particle. Hence, if the slowing down of fast charged particles in matter
were caused by only Coulomb collisions with the individual electrons and
nuclei of the material, the stopping power would be mostly due to
collisions with electrons because of the much larger masses of nuclei.


\section{Relativistic classical collisions
\label{sec4.3}}

\index{classical collisions}
We wish to extend the foregoing study of elastic collisions of charged
particles to the case of projectiles with velocities comparable to that
of light. Because practical methods to compute the DCS are available
only for central potentials, we limit our considerations to interactions
that in the CM frame can be described by a scalar central potential,
$V(r)$, which depends only on the relative position of the particles.
For these potentials we shall obtain the DCS in the CM frame by solving
the relativistic equation for the relative motion, and infer the DCS in
the L frame by means of a Lorentz transformation.

Our approach can be qualified as semi-relativistic, because it
implements strict relativistic kinematics but disregards the fact that
the electromagnetic interaction between the particles (which is
effectively central in the L frame, where the target is at rest)
cannot be central in the CM frame due to the characteristics of the
Lorentz transform of the fields \citep[see, \eg,][]{Jackson1975}. This
theoretical scheme is generally sufficient for describing elastic
collisions of charged particles with kinetic energies (in L) smaller
than their rest energies, for which the interaction is dominated by the
electrostatic force (Coulomb approximation). A rigorous formulation of
the process would require describing the interaction in covariant form,
which would add formidable complications to the theory.

Without losing generality we may assume that $V(r)$ has a finite range,
so that the four-momenta of the particles have definite values before
and after the interaction, that is, when the distance between the
particles exceeds the range of the potential. We will show below that
the trajectories of the colliding particles are in a plane, the
scattering plane. As in previous Sections, we consider that in the L
frame the target particle ($2$) is initially at rest and that the projectile
particle ($1$) impinges in the direction of the $z$ axis with linear
momentum $p_{1{\rm i}}$. We set the $y$-axis in such a way that the
scattering plane is the $z$-$y$ plane. In the L frame, the
energy-momentum four-vectors of the projectile and the target before the
collision are, respectively, (see Fig.\ \ref{fig4.7})
\beq
\underline{ p}_{1{\rm i}} = ({\cal W}_{1{\rm i}} c^{-1}, 0, 0, p_{1{\rm i}})
\quad \mbox{and} \quad
\underline{ p}_{2{\rm i}} = ({\cal W}_{2{\rm i}} c^{-1}, 0, 0, 0),
\label{4.130}\eeq
where ${\cal W}_{j{\rm i}}$ denotes the initial energy of particle $j$,
inclusive of the rest energy, that is,
\beq
{\cal W}_{1{\rm i}} = E_{1{\rm i}} + M_1 c^2
\quad \mbox{and} \quad
{\cal W}_{2{\rm i}} = M_2 c^2,
\label{4.131}\eeq
and $E_{1{\rm i}}$ is the kinetic energy of the projectile before the
interaction. The corresponding four-momenta after the
collision are,
\begin{subequations}
\label{4.132}
\beq
\underline{ p}_{1{\rm f}} =
({\cal W}_{1{\rm f}} c^{-1}, 0, p_{1{\rm f}} \sin\theta_1,
p_{1{\rm f}} \cos\theta_1 )
\label{4.132a} \eeq
and
\beq
\underline{ p}_{2{\rm f}} = ({\cal W}_{2{\rm f}} c^{-1}, 0,
p_{2{\rm f}} \sin\theta_2, p_{2{\rm f}} \cos\theta_2).
\label{4.132b}\eeq
\end{subequations}

The initial four-momenta of the particles in the CM frame
are\footnote{As in previous Sections, the quantities referred to the CM
frame are indicated with primes.} $\underline{p}'_{j} = ({\cal
W}'_{j{\rm i}} c^{-1},{\bf p}'_{j{\rm i}})$. The CM frame, which is
characterized by the property ${\bf p}'_{1} + {\bf p}'_{2} = {\bf 0}$,
moves with respect to the L frame with constant velocity ${\bf v}_{\rm
CM} = v_{\rm CM} \hat{\bf x}$.  Therefore, the transformation from the
L to the CM frame is a boost in the $x$ direction (see Appendix
\ref{appA}, Section
\ref{appA.4}) with velocity ${\bf v}_{\rm CM} =
\beta_{\rm CM} c$ given by Eq.\ \req{A.35},
\beq
\beta_{\rm CM} =
\frac{c p_{1{\rm i}}}{{\cal W}_{1{\rm i}} + {\cal W}_{2{\rm i}}}
= \frac{\beta_1 \gamma_1 M_1 c^2}{M_1 c^2 \gamma_1 + M_2 c^2} \, ,
\label{4.133}\eeq
where $\beta_1$ and $\gamma_1$ pertain to the particle 1
in L. We have
\beq
\gamma_{\rm CM} = \sqrt{\frac{1}{1-\beta_{\rm CM}^2}} =
\sqrt{\frac{({\cal W}_{1{\rm i}}+
{\cal W}_{2{\rm i}})^2}{({\cal W}_{1{\rm i}}+{\cal W}_{2{\rm i}})^2
- c^2 p_{1{\rm i}}^2}}
= \frac{{\cal W}_{1{\rm i}}+{\cal W}_{2{\rm i}}}{s}.
\label{4.134}\eeq
The quantity
\beq
s^2
\equiv c^2 \left( \underline{p}_{1} + \underline{p}_{2} \right)^2
= \left( {\cal W}_{1}+{\cal W}_{2} \right)^2
- c^2 \left( {\bf p}_{1} +{\bf p}_{2} \right)^2
\label{4.135}\eeq
is a Lorentz invariant. In the CM frame, $s$ equals the total
energy of the particles,
\beq
s = {\cal W}'_{1} +{\cal W}'_{2} \, .
\label{4.136}\eeq
Long before the interaction, when the potential effectively vanishes, we have
\beqa
s^2 &=& ({\cal W}_{1{\rm i}} +
{\cal W}_{2{\rm i}})^2 - c^2 p_{1{\rm i}}^2
\nonumber \\ [2mm]
&=& (M_1 c^2 + M_2 c^2)^2 + 2 M_2 c^2 E_{1{\rm i}}.
\nonumber \\ [2mm]
&=& M_1^2 c^4 + M_2^2 c^4 + 2 \gamma_1 M_1 c^2 \, M_2 c^2.
\label{4.137}\eeqa

Before the collision, in the CM frame of reference the colliding
particles have opposite linear momenta,
\beq
{\bf p}'_{1{\rm i}} = - {\bf p}'_{2{\rm i}} \equiv {\bf p}'_{\rm i},
\label{4.138}\eeq
with magnitude [see Eq.\ \req{A.43}]
\begin{subequations} \label{4.139}
\beqa
p'_{\rm i} &=& \beta_{\rm CM} \gamma_{\rm CM} M_2 c =
\frac{M_2 c^2}{s}\, p_{1{\rm i}}
\label{4.139a} \\ [2mm]
&=&
\frac{p_{1{\rm i}}}{\sqrt{
1+ (M_1/M_2)^2 + 2 (M_1/M_2) \gamma_1}}\, .
\label{4.139b}\eeqa
\end{subequations}
The initial energies of the colliding particles in CM are given by
\beq
{\cal W}'_{1{\rm i}} = \sqrt{ M_1^2 c^4 + p'^2_{\rm i} c^2}
\qquad \mbox{and} \qquad
{\cal W}'_{2{\rm i}} = \sqrt{ M_2^2 c^4 + p'^2_{\rm i} c^2}\, .
\label{4.140}\eeq
After the elastic collision, when the interaction energy has vanished,
the particles move away with the same total energies and momenta,
that is,
\beq
{\cal W}'_{1{\rm f}} = {\cal W}'_{1{\rm i}}
\qquad \mbox{and} \qquad
{\cal W}'_{2{\rm f}} = {\cal W}'_{2{\rm i}} \, .
\label{4.141}\eeq


\subsection{Relative motion in the CM frame
\label{sec4.3.1}}
\index{classical collisions!relative motion}

We now consider the equations of motion of the colliding particles in
the CM frame,
\beq
\dot{\bf p}'_{1} = {\bf F}
\quad \mbox{and} \quad
\dot{\bf p}'_{2} = -{\bf F},
\label{4.142}\eeq
where
\beq
{\bf F} = - \nablab' V(r') = - \frac{\d V}{\d r'} \, \hat{\bf r}',
\label{4.143}\eeq
with ${\bf r}' = {\bf r}'_{1} - {\bf r}'_{2}$. The relative velocity,
${\bf v}' = \dot{\bf r}'$, is
\beq
{\bf v}' = {\bf v}'_{1} - {\bf v}'_{2} =
\frac{c^2 {\bf p}'_{1}}{{\cal W}'_{1}} -
\frac{c^2 {\bf p}'_{2}}{{\cal W}'_{2}} = \left(
\frac{c^2}{{\cal W}'_{1}} + \frac{c^2}{{\cal W}'_{2}} \right)
{\bf p}' ,
\label{4.144}\eeq
where ${\bf p}'$ is the linear momentum of the projectile.
Hence, the equation for the relative motion reads
\beq
\dot{\bf v}' = \frac{\d}{\d t} \left[\left(
\frac{c^2}{{\cal W}'_{1}} + \frac{c^2}{{\cal W}'_{2}} \right)
{\bf p}' \right].
\label{4.145}\eeq

As in the non-relativistic study, with the aid of the constants of
motion the deflection angle can be obtained by quadrature. By virtue of
a {\it vis viva} theorem, the quantity
\beq
s \equiv {\cal W}'_{1} + {\cal W}'_{2} + V(r)
\label{4.146}\eeq
is a constant of the motion. In addition, for central forces, the
angular momentum ${\bf L}' = {\bf r}' \vecprod {\bf p}'$ is conserved.
Because
\beq
{\bf r}' \vecprod {\bf v}' = \left(
\frac{c^2}{{\cal W}'_{1}} + \frac{c^2}{{\cal W}'_{2}} \right) {\bf L}',
\label{4.147}\eeq
the conservation of angular momentum implies that the trajectories of
the particles lie in the plane of scattering. The values of the
constants $L'$ and $s$ are determined by the initial conditions;
$L'=b p'_{\rm i}$, where $b$ is the impact parameter (which takes the
same values in L and CM), and $s$ is given by Eq.\ \req{4.137}.

The equality
\beq
s - V(r')
= \sqrt{M_1^2 c^4 + p'^2 c^2} + \sqrt{M_2^2 c^4 + p'^2 c^2}
\label{4.148}\eeq
implies that
\beq
p'^2(r') = \frac{1}{4 c^2}
\left\{ \left[ s - V(r') \right]^2
+ \frac{M_{\rm d}^4 c^8}{\left[ s
- V(r') \right]^2} \right\} -
\frac{M_{\rm s}^2 c^2}{2} \, ,
\label{4.149}\eeq
where
\beq
M_{\rm d}^2 = M_1^2 - M_2^2
\qquad \mbox{and} \qquad
M_{\rm s}^2 = M_1^2 + M_2^2\, .
\label{4.150}\eeq
Naturally, the initial momentum of the particles, given by
Eq.\ \req{4.139}, is such that
\beq
p'^2_{\rm i} = \frac{1}{4 c^2}
\left\{ s^2
+ \frac{M_{\rm d}^4 c^8}{s^2}
\right\} -
\frac{M_{\rm s}^2 c^2}{2} \, .
\label{4.151}\eeq
With obvious rearrangements, expression \req{4.149} becomes
\beqa
p'^2(r') c^2 &=& \frac{1}{4}
\left[ s - V(r') \right]^2
+ \frac{M_{\rm d}^4 c^8}{4\,
\left[ s - V(r') \right]^2}
- \frac{M_{\rm s}^2 c^4}{2}
\nonumber \\ [2mm]
&=& p'^2_{\rm i} c^2
- \frac{s}{2} \, V(r')
\left( 1 - \frac{V(r')}{2 s} \right)
+ \frac{M_{\rm d}^4 c^8}{4 s^2}
\left( \left[ 1 - \frac{V(r')}{s} \right]^{-2} - 1 \right)
\nonumber \\ [2mm]
&=&  p'^2_{\rm i} c^2
- V(r') \, \frac{s^4 -
M_{\rm d}^4 c^8}{2 s^3}
+ V^2(r') \, \frac{s^4 +3 M_{\rm d}^4 c^8}
{4 s^4}
\nonumber \\ [2mm]
&& + \frac{M_{\rm d}^4 c^8}{4 s^2}
\left( \left[ 1 - \frac{V(r')}{s} \right]^{-2} - 1
- 2 \frac{V(r')}{s} - 3
\frac{V^2(r')}{s^2}
\right).
\label{4.152}\eeqa
The constant factors in this formula can be worked out as follows,
\begin{subequations}
\label{4.153}
\beqa
\frac{s^4 - M_{\rm d}^4 c^8}
{2 s^3} &=&
\frac{({\cal W}'_{1{\rm i}}+{\cal W}'_{2{\rm i}})^2 -
({\cal W}'_{1{\rm i}} - {\cal W}'_{2{\rm i}})^2}
{2 ({\cal W}'_{1{\rm i}}+{\cal W}'_{2{\rm i}})}
\nonumber \\ [2mm]
&=& \frac{2 {\cal W}'_{1{\rm i}}{\cal W}'_{2{\rm i}}}
{{\cal W}'_{1{\rm i}}+{\cal W}'_{2{\rm i}}} =
2 \mu_{\rm r} c^2
\label{4.153a}\eeqa
and
\beqa
\frac{s^4 + 3  M_{\rm d}^4 c^8}
{4 s^4} &=&
\frac{({\cal W}'_{1{\rm i}}+{\cal W}'_{2{\rm i}})^2 +
3 ({\cal W}'_{1{\rm i}} - {\cal W}'_{2{\rm i}})^2}
{4 ({\cal W}'_{1{\rm i}}+{\cal W}'_{2{\rm i}})^2}
\nonumber \\ [2mm]
&=& \frac{({\cal W}'_{1{\rm i}})^2 +({\cal W}'_{2{\rm i}})^2 -
{\cal W}'_{1{\rm i}} {\cal W}'_{2{\rm i}}}
{({\cal W}'_{1{\rm i}}+{\cal W}'_{2{\rm i}})^2} = 1 -
\frac{3 \mu_{\rm r}c^2}{s},
\label{4.153b}\eeqa
\end{subequations}
where ${\cal W}'_{1{\rm i}}$ and ${\cal W}'_{2{\rm i}}$ are the initial
total energies of the particles in CM, Eq.\ \req{4.140}. We have
introduced the {\it relativistic reduced mass}, $\mu_{\rm r}$, a
constant defined by \index{reduced mass!relativistic}
\beq
\mu_{\rm r} =  c^{-2}
\frac{{\cal W}'_{1{\rm i}}{\cal W}'_{2{\rm i}}}
{{\cal W}'_{1{\rm i}}+{\cal W}'_{2{\rm i}}}\, .
\label{4.154}\eeq
Thus, after a slight reorganization, Eq.\ \req{4.152} can be written
in the familiar non-relativistic form \req{4.100}
\beq
p'^2(r') = p'^2_{\rm i} - 2 \mu_{\rm r} V_{\rm ef}(r')
\label{4.155}\eeq
with the effective potential given by
\beq
V_{\rm ef}(r') = V(r') + V_{\rm r1}(r') + V_{\rm r2}(r'),
\label{4.156}\eeq
where
\beq
V_{\rm r1}(r') =
- \frac{V^2(r')}{2 \mu_{\rm r} c^2}
\left( 1 - \frac{3\mu_{\rm r}c^2}{s} \right)
\label{4.157}\eeq
and
\beq
V_{\rm r2}(r') =
\frac{(M_1^2 - M_2^2)^2 c^6}
{8 \mu_{\rm r} s^2}
\left( \left[ 1 - \frac{V(r')}{s} \right]^{-2} - 1
- 2 \frac{V(r')}{s} - 3 \frac{V^2(r')}{s^2}
\right).
\label{4.158}\eeq
The terms $V_{\rm r1}(r')$ and $V_{\rm r2}(r')$ are corrections to the
interaction potential that account for the effect of relativistic
kinematics. The first one is proportional to
$V^2(r')$; when $V(r')$ is a screened coulomb potential, $V_{\rm
r1}(r')$ diverges as $r'^{-2}$ at the origin. Considering the binomial
series $(1-x)^{-2}=1+2x+3x^2+\cdots$, the second correction term,
$V_{\rm r2}(r)$, is seen to be of order $(V/s)^3$, and it
vanishes when the projectile and the target particles have the same
mass.

In the non-relativistic limit ($p' \ll M_j c$),
\beq
\mu_{\rm r} \simeq \frac{M_1 M_2}{M_1 + M_2} = \mu \, ,
\label{4.159}\eeq
the familiar reduced mass. In addition,
\beq
{\cal W}'_{j} = \sqrt{M_j^2 c^4 + p'^2 c^2} = M_jc^2 + \frac{p'^2}{2M_j}
- \frac{p'^4}{8 M_j^3c^2} + \cdots,
\label{4.160}\eeq
which implies that
\beq
s = {\cal W}'_{1} + {\cal W}'_{2} \simeq M_1 c^2 + M_2 c^2 + p'^2
\left( \frac{1}{2M_1} + \frac{1}{2M_2}\right) + V(r'),
\label{4.161}\eeq
and, consequently,
\beq
p'^2(r') \simeq 2 \mu \left\{ s - M_1 c^2 - M_2 c^2 \rule{0mm}{4mm}-
V(r') \right\},
\label{4.162}\eeq
where the quantity in curly braces is the total kinetic energy in CM.

It is worth observing that when the mass $M_2$ of the target  tends to
infinity, $\beta_{\rm CM} =0$ and the CM frame coincides with the L
frame. Under these circumstances, ${\bf p}'_{\rm i} = {\bf p}_{1{\rm
i}}$, $s \simeq \infty$, and
\beq
\mu_{\rm r} c^2 \simeq {\cal W}_{1} = \gamma M_1 c^2.
\label{4.163}\eeq
That is, the reduced mass equals the relativistic mass of the
projectile, which increases with $p_{\rm 1i}$.


\subsection{Differential cross section in the CM frame
\label{sec4.3.2}}
\index{collisions in the center-of-mass frame}
\index{cross sections in the center-of-mass frame}

Introducing polar coordinates in the scattering plane (see Section
\ref{sec4.1}), and considering the conservation of angular momentum, Eq.\
\req{4.147} implies that
\begin{subequations}
\label{4.164}
\beq
\dot{\varphi}' = \left( \frac{c^2}{{\cal W}'_{1}} +
\frac{c^2}{{\cal W}'_{2}} \right) \frac{L'}{r'^2} \, .
\label{4.164a}\eeq
In addition, the equality $v'^2 = \dot{r}'^2 + r'^2 \dot{\varphi}'^2$,
together with the results \req{4.144} and \req{4.164a}, gives
\beq
\dot{r}' = \pm \, \left( \frac{c^2}{{\cal W}'_{1}} +
\frac{c^2}{{\cal W}'_{2}} \right) \sqrt{p'^2 - \frac{L^2}{r'^2}},
\label{4.164b}\eeq
\end{subequations}
where the sign on the right-hand side is $-$ when the particles approach
each other, and $+$ when they separate after having passed the point
of closest approach [see Eq.\ \req{4.168} below]. The two equations
\req{4.164} determine the trajectory ${\bf r}'(t')$ of the relative
motion. Notice that the common factor in parenthesis is a function of
only the relative momentum,
\beq
\frac{c^2}{{\cal W}'_{1}} + \frac{c^2}{{\cal W}'_{2}}
=
\frac{1}{M_1 \sqrt{1 +(p'/M_1c)^2}} + \frac{1}{M_2 \sqrt{1
+(p'/M_2c)^2}}\, ,
\label{4.165}\eeq
which for small $p'$ can be approximated by the Taylor expansion
\beqa
\frac{c^2}{{\cal W}'_{1}} + \frac{c^2}{{\cal W}'_{2}}
&=& \frac{1}{\mu_{\rm r}} \left[ 1
- \frac{1}{2} \frac{M_1^3+M_2^3}{(M_1+M_2)^3}
\left( \frac{p'}{\mu_{\rm r} c} \right)^2 \right.
\nonumber \\ [2mm]
&& \mbox{} \left.
+ \frac{1 \cdot 3}{2 \cdot 4} \frac{M_1^5+M_2^5}{(M_1+M_2)^5}
\left( \frac{p'}{\mu_{\rm r} c} \right)^4 - \cdots \right].
\label{4.166}\eeqa
Combining the two equations \req{4.164} we obtain the geometric
equation of the trajectory,
\beq
\d \varphi' = \frac{\d \varphi'}{\d t'} \, \frac{\d t'}{\d r'} \, \d r'
= \frac{\dot{\varphi}'}{\dot{r}'} \, \d r' =
\pm \, \frac{L'/r'^2}{\sqrt{p'^2(r') -L'^2/r'^{2}}}\, \d r'.
\label{4.167}\eeq
The similarity of this formula to the non-relativistic result, Eq.\
\req{4.26} is noteworthy.

The distance of closest approach, $r'_0$, is obtained from the condition
$\dot{r}' = 0$, which gives
\beq
p'^2(r'_0) - \frac{L'^2}{r'^2_0} = 0 \, .
\label{4.168}\eeq
The angle $\alpha'$ between the position vector ${\bf r}'_0$ at the
point of closest approach and the asymptote of the outgoing trajectory
(see Fig.\ \ref{fig4.2}) is given by
\beq
\alpha'=\int_{r'_0}^\infty
\frac{L' r'^{-2}}{\sqrt{p'^2(r')-L'^2 r'^{-2}}}\, \d r'.
\label{4.169}\eeq
Because $2\alpha'+\vartheta'=\pi$, the deflection angle in CM is
\beq
\vartheta'(L') = \pi - 2 \int_{r'_0}^\infty
\frac{L' r'^{-2}}{\sqrt{p'^2(r')-L'^2 r'^{-2}}}\, \d r'\, .
\label{4.170}\eeq

The calculation faces the same difficulties as in the non-relativistic
case, namely, the divergence of the integrand at the turning point and,
in the case of relatively large angular momenta, the near cancellation
of the two terms on the right-hand side of Eq.\ \req{4.170}. In the
case of atomic collisions, where the effective interaction at
intermediate and large radial distances is dominated by the screened
Coulomb field, it is expedient to consider the function
\beq
\Psi(r') \equiv \frac{ r' V_{\rm ef}(r')}{Z_0 Z e^2},
\label{4.171}\eeq
which plays a role similar to that of the screening function,
and apply the calculation strategy developed in Section \ref{sec4.1.4}
[Eqs.\ \req{4.47} to \req{4.52}].  That is, the deflection angle is
obtained as
\beqa
\vartheta'(L') &=&
2 \arctan \left( \frac{\mu_{\rm r} Z_0 Z e^2 \Psi(r'_0)}
{L'p'_{\rm i}} \right)
+ 4 \int_0^1 \left\{
\frac{1}{\sqrt{C \Psi(r'_0)+2-u^{2}}} \right.
\nonumber \\ [2mm]
&& \mbox{} \rule{10mm}{0mm} \left. -
\frac{1}{\sqrt{C \Psi(r'_0) + 2 - u^{2} - C f(u)
}} \right\} \d u\, ,
\label{4.172}\eeqa
where $u=\sqrt{1-r'_0/r'}$,
\beq
C = \frac{2\mu_{\rm r} Z_0Z e^2 r'_0}{L'^2}\, ,
\label{4.173}\eeq
and
\beq
f(u) = r'_0 \, \frac{\Psi(r')-\Psi(r'_0)}{r'-r'_0} \, .
\label{4.174} \eeq
As in the non-relativistic situation, the integrand in Eq.\ \req{4.172}
is finite and varies smoothly over the integration interval. Generally,
the function $f(u)$ is approximately constant for $u < 0.001$, which
corresponds to radii between $r'_0$ and $r'_1=(1+10^{-6}) r'_0$, where
$f(u) \simeq r'_0 \, \d \Psi(r'_0)/\d r'_0$. For radii larger than
$r'_1$, $f(u)$ can be calculated directly from the expression
\req{4.174}. When the effective potential is strongly attractive and the
angular momentum is small, the integral in \req{4.172} diverges (\ie, it
produces numerical overflows); under these circumstances, the program
{\sc elastic} (Chapter \ref{chapt10}) sets the deflection angle equal to
zero, which implies that the DCS at $\theta_0=0$ is infinite [see Eq.\
\req{4.176}]. Conflicting cases are normally apparent in the plots of
the DCS and of the $\vartheta'(L')$ function.

The DCS in the CM frame can be obtained by following the same strategy
as in the non-relativistic study (Section \ref{sec4.1.4}). We recall that the
deflection angle $\vartheta'$ and the scattering angle $\theta'$ [$\in
(0,\pi)$] are related by [see Eqs.\ \req{4.31} and \req{4.32}]
\beq
\cos\theta' = \cos\vartheta'.
\label{4.175}\eeq
The DCS in CM is given by
\beq
\frac{\d \sigma'_{1}}{\d \Omega'} = \frac{1}{2 p'^2_{\rm i} \sin\theta'}
\sum_j \left| \frac{\d L'^2}{\d \theta'} \right|_{L'=L'_j}\, ,
\label{4.176}\eeq
where the summation extends over the values $L'_j$ of the angular
momentum for which the angular deflection corresponds to the scattering
angle $\theta'$. The meaning of this DCS is clarified by using polar
spherical coordinates and assuming that
the beam of projectiles impinges
in the direction of the polar $z$ axis; the DCS expresses the results
from measurements of projectile particles that after the collision
move in the direction $\Omega' = (\theta', \phi')$. We can also consider
the DCS for measurements where the final direction of the target
particle is observed, $\d \sigma'_{2}/\d \Omega'$. Because the particles
$1$ and $2$ move in opposite directions, we have
\beq
\frac{\d \sigma'_{2}(\theta',\phi')}{\d \Omega'} =
\frac{\d \sigma'_{1}(\pi-\theta',\phi'+\pi)}{\d \Omega'} \, .
\label{4.177}\eeq

\index{Gauss--Legendre quadrature!adaptive}
In the computer program {\sc elastic} (see Chapter \ref{chapt10}),
calculations are performed by means of robust algorithms so as to
minimize numerical errors. The integral \req{4.172} is evaluated by
using the adaptive Gauss--Legendre quadrature method (Section
\ref{sec10.4.3.2}), which yields
results with a relative accuracy of the order of $10^{-7}$. The
deflection angle $\vartheta'$ is considered as a function of the variable
$\xi = \ln(L'^2)$, and it is calculated for a dense non-uniform grid of
$\xi$ values. This grid is determined by an adaptive algorithm that sets
a higher density of grid points in regions where the function has larger
curvature, so as to ensure that subsequent interpolation is accurate.
The function $\vartheta'(\xi)$ is replaced with the natural cubic spline
that interpolates the calculated table (see Section \ref{sec10.4.2}).
The DCS is obtained from Eq.\ \req{4.176},
\beq
\frac{\d \sigma'_1}{\d \Omega'} = \frac{1}{2 p'^2_{\rm i} \sin\theta'}
\sum_j
\left( \left| \frac{\d \theta'}{\d L'^2} \right|_{L'=L'_j} \right)^{-1}
= \frac{1}{2 p'^2_{\rm i} \sin\theta'}
\sum_j
\left( \left| \frac{\d \theta'}{\d \xi} \exp(-\xi)
\right|_{\xi=\xi_j} \right)^{-1} .
\label{4.178}\eeq
The values $\xi_j$ of $\xi$ corresponding to a given scattering angle
$\theta'$, as well as the derivatives $\d \theta'/\d \xi$ are
calculated from the interpolating cubic spline.

\begin{figure}[p!] \begin{center}
\includegraphics*[width=7.5 cm]{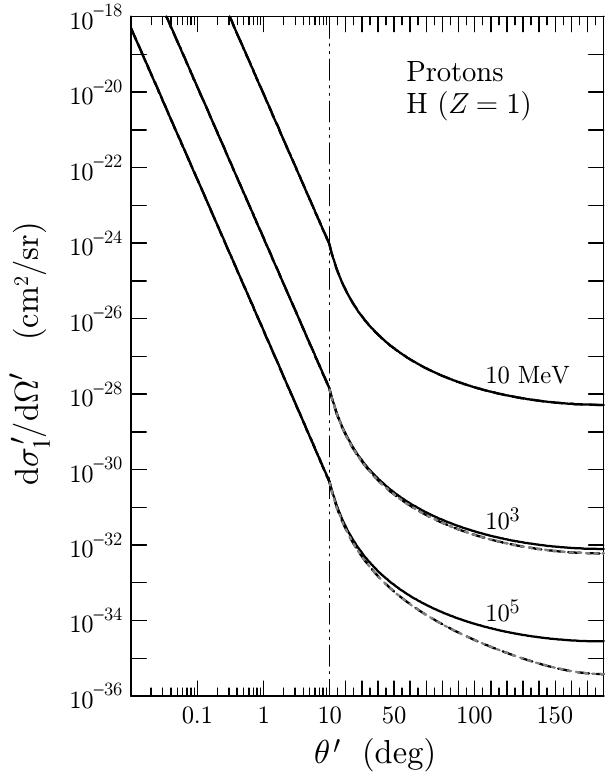} \rule{3mm}{0mm}
\includegraphics*[width=7.5 cm]{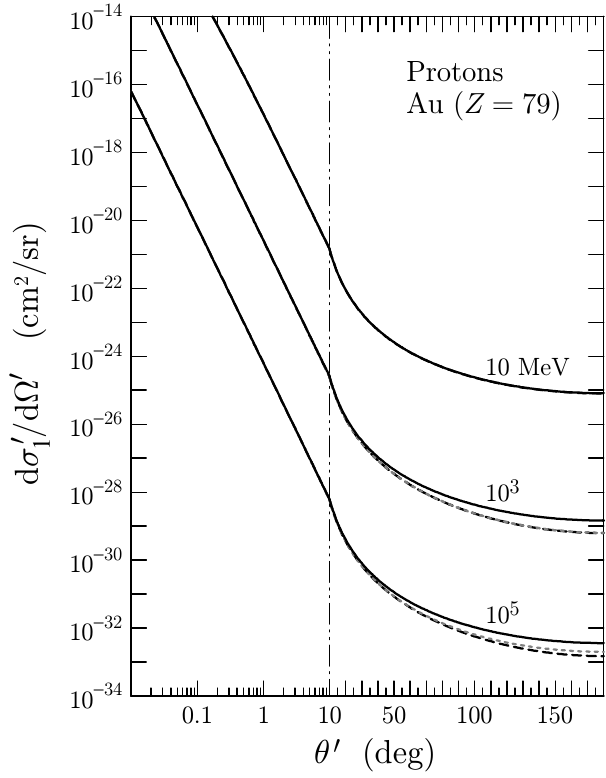} \\ [5mm]
\includegraphics*[width=7.5 cm]{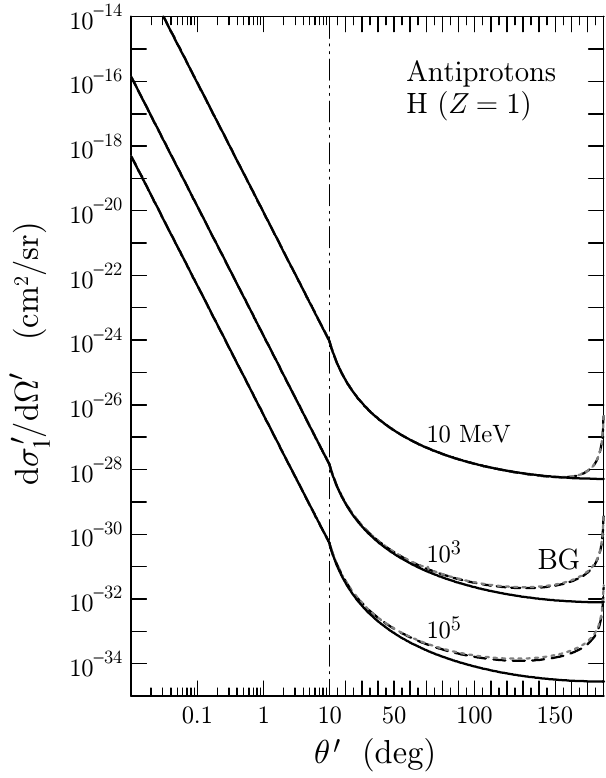} \rule{3mm}{0mm}
\includegraphics*[width=7.5 cm]{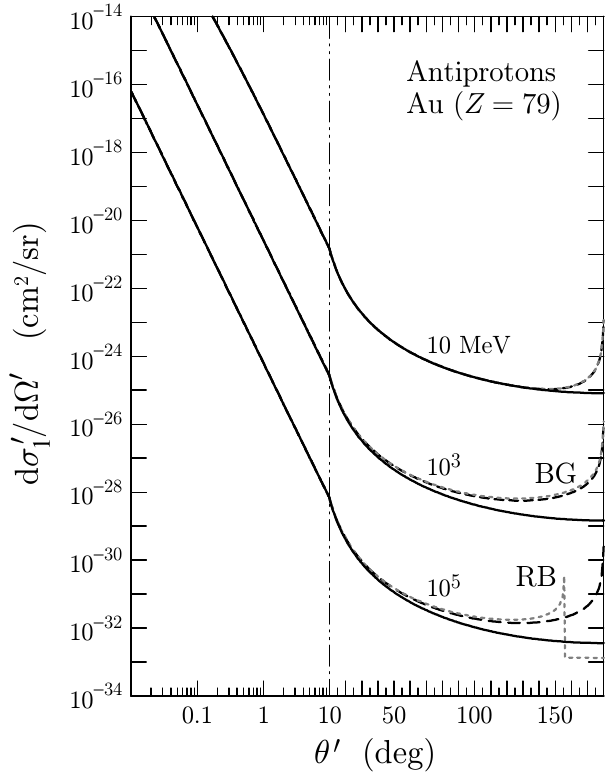}
\caption{
Classical DCSs (in CM) for elastic collisions of protons and antiprotons
with the indicated kinetic energies (in L) with hydrogen and gold atoms.
The curves are results from calculations
with the electrostatic DHFS potential $V(r')$ (solid), with the sum of
$V(r')$ and the first correction (dashed), and  with the full
effective potential \req{4.156} (grey, dotted). The DCSs
for antiprotons show backward glory (BG) and rainbow (RB) structures.
Notice the logarithmic scale for $\theta' < 10^\circ$.
\label{fig4.9}}
\end{center} \end{figure}

Figure \ref{fig4.9} displays the DCSs in CM for collisions of protons
and antiprotons with hydrogen and gold atoms calculated using the
analytical DHFS potential (see Section \ref{sec3.6}). To reveal the
effect of the relativistic correction terms in the effective potential,
Eq.\ \req{4.156}, calculations were performed for the bare DHFS
potential $V(r')$, for the sum $V(r')+V_{\rm r1}(r')$ of the DHFS
potential and the first correction, and for the full effective potential
$V_{\rm ef}(r')$. The correction terms are appreciable at small radii,
where $V(r')$ takes values comparable to $\mu_{\rm r} c^2$. They alter
the divergence of the electrostatic potential at $r'=0$ and may produce
narrow structures in the $\vartheta'(L')$ function, which make the
calculation of the DCS delicate. For instance, the first correction
$V_{\rm r1}(r')$, which is attractive and diverges at $r=0$ as
$r'^{-2}$, may cause that projectiles with small impact parameters are
trapped in spiraling orbits (see Section \ref{sec4.1.5}). This implies
that the function $\vartheta'(L')$ diverges at a certain angular
momentum $L'_{\rm c}$, and it is not defined for angular momenta below
this value. In collisions of antiprotons, the function $\vartheta'(L')$
is negative for intermediate and large angular momenta, with a sharp
minimum where it takes values less than $-\pi$.  Near the minimum,
$\vartheta'(L')$ passes continuously through the value $-\pi$ twice and
hence the DCS is singular at backward angles; this situation is an
example of backward glory scattering \citep[see, \eg,][]{MottMassey1965,
Goldstein1980}. In the case of collisions of $10^5$ MeV antiprotons and
gold atoms described with the effective potential, the value of
$\vartheta'(L')$ at the minimum is $\simeq -155^\circ$, and the DCS
presents a case of rainbow scattering at this angle
\citep{MottMassey1965, Goldstein1980}.

The conditions of validity of this classical semi-relativistic
formulation can be analyzed similarly to the non-relativistic theory
(Section \ref{sec4.1.5}). For calculations performed with only the
electrostatic potential $V(r')$, comparison with results from quantum
theory (see Section \ref{sec5.1}) indicate that the classical approach
is valid for scattering angles $\theta'$ larger than Bohr's limit, Eq.\
\req{4.71},
\vspace*{-1mm}
\beq
\theta'_{\rm class} = \hbar/(p'_{\rm i}R)\, .
\label{4.179}\eeq
However, when considering the full effective potential, Bohr's
diffraction argument does not warrant the validity of the classical
theory. The reason is that the first correction [Eq.\ \req{4.146}] is
attractive and diverges as $r'^{-2}$ at $r'=0$. The function
$\vartheta'(L')$ for the potential $V(r') = C_2 r'^{-2}$ takes values
from $-\infty$ to 0 \citep{Bohr1948} and, hence, the DCS at a given
scattering angle gets contributions from a denumerable set of impact
parameters. Evidently, Bohr's reasoning is not applicable to this
situation.

With due wariness about the validity of the classical method, we observe
that the effect of the relativistic potential terms is appreciable only
for high-energy projectiles, as expected. The effect is negligible for
scattering angles less than, say, about 20 degrees, because orbits with
large impact parameters remain relatively far from the nucleus, where
the relativistic terms are much smaller than $V(r')$, and the
contributions of orbits with small impact parameters are unimportant.
It is worth pointing out that the DCS at large angles is also affected
by the finite size of the nucleus \citep[see][]{SalvatQuesada2020},
which is disregarded in the present study. With regard to transport
calculations of charged particles heavier than the electron, the
relevant feature is that there is a very small probability of having
collisions with large angles for which the effect of the relativistic
correction terms in the effective potential may be appreciable. Thus, in
the case of protons colliding with gold atoms (see Fig.\ \ref{fig4.9})
the total probability of scattering angles larger than 10 degrees in a
collision is about $10^{-6}$, $10^{-8}$ and $10^{-11}$, for $E=10$,
10$^3$, and 10$^5$ MeV, respectively. Consideration of the finite size
of the nucleus reduces the DCS at large angles and, hence, the
probability of large-angle collisions.  Consequently, in transport
calculations of protons and heavier particles, it is generally justified
to utilize DCSs calculated with only the electrostatic potential,
disregarding both the relativistic correction terms and the size of the
nucleus.

\subsection{Collision kinematics in the L frame
\label{sec4.3.3}}

\index{collisions in the laboratory frame}
Let us consider a particle with total energy ${\cal W}'_{1}$ and
momentum ${\bf p}'_{1}$ in CM. The correspondence between the polar angles
$\theta_1$ and $\theta'$ of the direction of the particle in L and CM
is given by the generic transformation formula [Eq.\ \req{A.49} in
Appendix \ref{appA}]
\beq
\tan\theta_1 = \frac{1}{\gamma_{\rm CM}} \,
\frac{\sin\theta'}{\tau_1 + \cos\theta'}, \qquad \tau_1 = v_{\rm
CM}/v'_{1},
\label{4.180}\eeq
where $v'_{1} = c^2 p'_{1} /{\cal W}'_{1}$ is the velocity of the
particle in CM. Recalling that $p'_{1} = p' = \beta_{\rm CM} \gamma_{\rm
CM} M_2 c$,
\beq
\tau_1 = \frac{c \beta_{\rm CM} \, \sqrt{M_1^2 c^4 +
\beta_{\rm CM}^2 \gamma_{\rm CM}^2 M_2^2 c^4}}
{c^2 \, \beta_{\rm CM} \gamma_{\rm CM} M_2 c}
=\sqrt{\left( \frac{M_1}{M_2} \right)^2
(1 - \beta_{\rm CM}^2 ) + \beta_{\rm CM}^2 } \, .
\label{4.181}\eeq
The relation \req{4.180} may be expressed in the equivalent form
\beq
\cos\theta_1 = \frac{\tau_1 + \cos{\theta'}}{\sqrt{
(\tau_1 + \cos{\theta'})^2 + \gamma_{\rm CM}^{-2} \sin^2 \theta'}}\, .
\label{4.182}\eeq
The inverse relation is [see Eq.\ \req{A.58}]
\beq
\cos \theta' = \frac{-\tau_1 \gamma_{\rm CM}^2\sin^2 \theta_1 \pm
\cos\theta_1
\sqrt{\cos^2 \theta_1 +\gamma_{\rm CM}^2 (1-\tau_1^2) \sin^2 \theta_1}}
{\gamma_{\rm CM}^2\sin^2 \theta_1 +\cos^2 \theta_1}\, .
\label{4.183}\eeq
When $\tau_1 < 1$ only the plus sign is valid and $\theta_1$ increases
monotonically with $\theta'$, that is, there is a unique correspondence
between $\theta_1$ and $\theta'$. When $\tau_1 \ge 1$, the angle
$\theta_1$ in the L frame can only take values in the interval from 0 to
the value $\theta_{1,{\rm max}}$ given by Eq.\ \req{A.59},
\beq
\theta_{1,{\rm max}} =
\arccos \left( \sqrt{
\frac{\gamma_{\rm CM}^{2} (\tau_1^2-1)}{\gamma_{\rm CM}^{2} (\tau_1^2-1) + 1 }}
\right).
\label{4.184}\eeq
In addition, if $\tau_1 \ge 1$ each angle $\theta_1 \le \theta_{1,{\rm
max}}$ corresponds to two different angles in CM, which are given by the
formula \req{4.183} with the plus and minus signs. Note that, when $\tau
\ge 1$, $\theta_{1,{\rm max}}$ corresponds to the polar scattering angle
in CM $\theta' = \arccos(- \tau_1^{-1})$, and $\theta_1 =0$ for $\theta'
= 0$ and $\pi$.  Figure \ref{fig4.10} displays the angle $\theta_1$ in L
as a function of the angle $\theta'$ in CM for $\gamma_{\rm CM} =1$ and
different values of the parameter $\tau_1$, which equals the mass ratio
$M_1/M_2$ when $\gamma_{\rm CM}=1$.

\begin{figure}[hptb] \begin{center}
\includegraphics*[width=8.5cm]{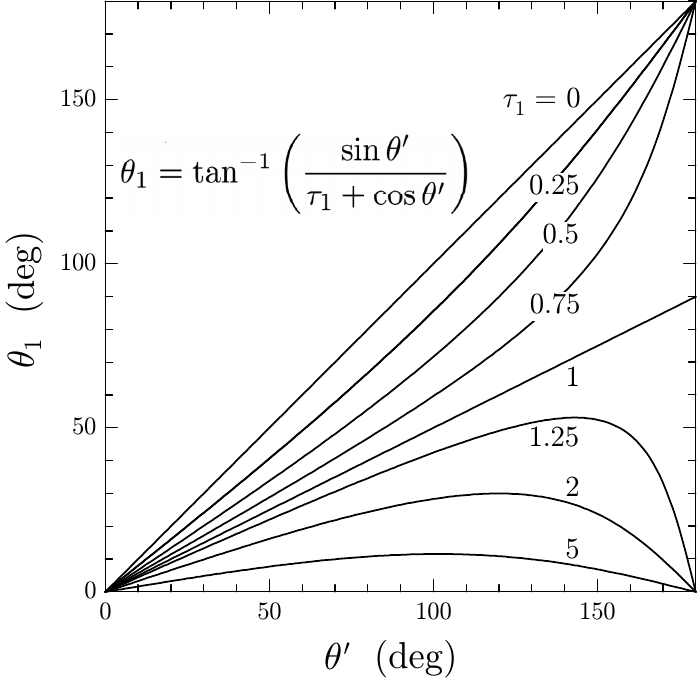}
\vspace*{-2mm}
\caption{Polar angle $\theta_1$ of the direction of motion of
a particle in L as a function of the
polar angle $\theta'$ in CM, evaluated from Eq.\ \req{4.180} with
$\gamma_{\rm CM}=1$ (non-relativistic limit), for the indicated values
of the parameter $\tau_1$ ($=M_1/M_2$ when $\gamma_{\rm CM}=1$).
\label{fig4.10}}
\end{center} \end{figure}

The scattering angle $\theta'$ determines the collision in the CM frame
and, consequently, in any other frame. We wish to express the energy
loss of the projectile in the L frame,  $W=E_{1{\rm i}}-E_{1{\rm f}}$,
in terms of the
scattering angle in CM. We start from the energy-momentum conservation
equation,
\beq
\underline{ p}'_{2{\rm f}} = \underline{ p}'_{1{\rm i}}
+ \underline{ p}'_{2{\rm i}}
- \underline{ p}'_{1{\rm f}}.
\label{4.185}\eeq
Squaring it,
$$
\underline{ p}'_{2{\rm f}}\dotprod \underline{ p}'_{2{\rm f}} =
\underline{ p}'_{1{\rm i}} \dotprod \underline{ p}'_{1{\rm i}} +
\underline{ p}'_{2{\rm i}} \dotprod \underline{ p}'_{2{\rm i}} +
\underline{ p}'_{1{\rm f}} \dotprod \underline{ p}'_{1{\rm f}} +
2 \underline{ p}'_{1{\rm i}} \dotprod \underline{ p}'_{2{\rm i}} -
2 \underline{ p}'_{1{\rm i}} \dotprod \underline{ p}'_{1{\rm f}} -
2 \underline{ p}'_{2{\rm i}} \dotprod \underline{ p}'_{1{\rm f}},
$$
and recalling that $\underline{ p}'_{1{\rm i}} \cdot \underline{
p}'_{1{\rm i}} = (M_1 c)^2$, etc., we have
\beq
(M_1 c)^2 +
 \underline{ p}'_{1{\rm i}} \dotprod \underline{ p}'_{2{\rm i}} -
 \underline{ p}'_{1{\rm i}} \dotprod \underline{ p}'_{1{\rm f}} -
      \underline{ p}'_{2{\rm i}} \dotprod \underline{ p}'_{1{\rm f}} = 0.
\label{4.186}\eeq
We compute the product $ \underline{ p}'_{1{\rm i}} \cdot
\underline{ p}'_{1{\rm f}}$
in CM,
$$
\underline{ p}'_{1{\rm i}} \dotprod \underline{ p}'_{1{\rm f}} =
{\cal W}'_{1{\rm i}} {\cal W}'_{1{\rm f}} c^{-2}
- {\bf p}'_{1{\rm i}} \dotprod
{\bf p}'_{1{\rm f}} =
({\cal W}'_{1{\rm i}})^2 c^{-2} - p'^2_{\rm i} \cos\theta'
= p'^2_{\rm i} ( 1 -\cos\theta') + M_1^2 c^2,
$$
and the other two in L,
$$
\underline{ p}'_{1{\rm i}} \dotprod \underline{ p}'_{2{\rm i}}
= \underline{ p}_{1{\rm i}} \dotprod \underline{ p}_{2{\rm i}} =
{\cal W}_{1{\rm i}} M_2, \qquad
\underline{ p}'_{2{\rm i}} \dotprod \underline{ p}'_{1{\rm f}}
= \underline{ p}_{2{\rm i}} \dotprod \underline{ p}_{1{\rm f}} =
M_2 {\cal W}_{1{\rm f}}.
$$
Introducing these expressions in Eq.\ \req{4.186}, we can write
\beq
{\cal W}_{1{\rm i}} - {\cal W}_{1{\rm f}} = \frac{p'^2_{\rm i}}{M_2}
\left( 1 - \cos\theta'\right).
\label{4.187}\eeq
Finally, using the identity \req{4.139a}, we obtain
\beq
W = {\cal W}_{1{\rm i}} - {\cal W}_{1{\rm f}} =
\frac{M_2 c^4 p_{1{\rm i}}^2}{s^2} \,
\left( 1 - \cos\theta'\right).
\label{4.188}\eeq
We see that the energy transfer $W$ increases with $\theta'$, and
attains its maximum value at $\theta'=\pi$,
\beqa
W_{\rm max} = \frac{2 M_2 c^4 p^2_{1{\rm i}}}{s^2}
&=& \frac{2 M_2 c^2 \, E_{1{\rm i}} (E_{1{\rm i}}
+2M_1 c^2)}
{(M_1 c^2 + M_2 c^2)^2 + 2 M_2 c^2 E_{1{\rm i}}}
\nonumber \\ [2mm]
&=& \frac{2 \beta_1^2 \gamma_1^2 \, M_2 c^2}
{1 + (M_2/M_1)^2 + 2 (M_2/M_1) \gamma_1}.
\label{4.189}\eeqa
Evidently, in the non-relativistic limit ($E_{1{\rm i}} \ll M_1 c^2$) we
regain the result \req{4.127} derived above. The relation \req{4.188}
can then be recast as
\beq
W = W_{\rm max} \,
\frac{1 - \cos\theta'}{2} = W_{\rm max} \, \sin^2(\theta'/2)\, .
\label{4.190}\eeq

In the CM frame the polar angle of the direction of the
target particle after the interaction
is $\pi-\theta'$. In the L frame the target recoils with
kinetic energy $W$ in a direction that lies on the scattering
plane and forms an angle $\theta_2$ with the initial direction of the
projectile. Equation \req{A.49} in Appendix \ref{appA} implies that
\beq
\tan\theta_2 = \frac{1}{\gamma_{\rm CM}} \,
\frac{\sin(\pi -\theta')}{\tau_2 + \cos(\pi - \theta')}
= \frac{1}{\gamma_{\rm CM}} \,
\frac{\sin\theta'}{\tau_2 - \cos\theta'},
\qquad \tau_2 = v_{\rm
CM}/v'_{2},
\label{4.191}\eeq
where $v'_{2} = c^2 p'_{2} /{\cal W}'_{2}$ is the final velocity of the
target particle in CM. Recalling that $p'_{2} = p' = \beta_{\rm CM} \gamma_{\rm
CM} M_2 c$, we have
\beq
\tau_2 = \frac{c \beta_{\rm CM} \, \sqrt{M_2^2 c^4 +
\beta_{\rm CM}^2 \gamma_{\rm CM}^2 M_2^2 c^4}}
{c^2 \, \beta_{\rm CM} \gamma_{\rm CM} M_2 c} =1 .
\label{4.192}\eeq
The Eq.\ \req{4.191} can be expressed in the equivalent form
\beq
\cos\theta_2 = \frac{1 - \cos{\theta'}}{\sqrt{
(1 - \cos{\theta'})^2 + \gamma_{\rm CM}^{-2} \sin^2 \theta'}}\, ,
\label{4.193}\eeq
which shows that the polar angle of the direction of the recoiling atom
cannot exceed 90 degrees.

Let us now consider the variation of the final momenta with the scattering
angles in the L frame. The final momenta of particle $1$ in the L
and CM frames, ${\bf p}_{1{\rm f}}$ and ${\bf p}'_{1{\rm f}}$,
are related by the transformation equations
\beq
p_{1{\rm f}z} = \gamma_{\rm CM} \left(
p'_{\rm i} \cos\theta'+\beta_{\rm CM} {\cal W}'_{1{\rm f}}c^{-1} \right)
\qquad \mbox{and} \qquad
p_{1{\rm f}y} = p'_{\rm i} \sin\theta'.
\label{4.194}\eeq
with ${\cal W}'_{1{\rm f}} =
{\cal W}'_{1{\rm i}}$. Combining these two equalities, to eliminate
$\theta'$,
we have
\beq
\frac{\left(p_{1{\rm f}z} - \beta_{\rm CM} \gamma_{\rm CM}
{\cal W}'_{1{\rm i}}c^{-1}
\right)^2}{\left(\gamma_{\rm CM} p'_{\rm i} \right)^2}
+ \frac{p^2_{1{\rm f}y}}{p'^2_{\rm i}} = 1.
\label{4.195}\eeq
This equation shows that the locus of points on the plane
$(p_{1{\rm f}z},p_{1{\rm f}y})$ that correspond to possible values of the final
momentum ${\bf p}_{1{\rm f}}$ is an ellipse with center at $(\beta_{\rm CM}
\gamma_{\rm CM} {\cal W}'_{1{\rm i}}c^{-1},0)$ and semiaxes $\gamma_{\rm CM}
p'_{\rm i}$ and $p'_{\rm i}$. Introducing the expressions of $\gamma_{\rm CM}$,
Eq.\ \req{4.134}, and the CM momentum, Eq.\ \req{4.139}, the major
semiaxis of the ellipse can be expressed as
\beq
\gamma_{\rm CM} p'_{\rm i} =  \frac{ (E_{1{\rm i}} + M_1 c^2+M_2 c^2) M_2 c^2 }
{(M_1 c^2 + M_2 c^2)^2 + 2 M_2 c^2 E_{1{\rm i}}}\, p_{1{\rm i}}.
\label{4.196}\eeq
Since the ratio of the major to the minor axis is $\gamma_{\rm CM}$,
the ellipse becomes more elongated when the energy of the projectile
increases. In the non-relativistic limit ($\beta_{\rm CM} \rightarrow 0$), the
two semiaxes have the same length, \ie, the ellipse becomes a circle.
The difference between the initial momentum of the target particle and
the major axis of the ellipse is
\beq
p_{1{\rm i}} -
2 \gamma_{\rm CM} p'_{\rm i} =  \frac{ M_1^2 c^4 - M_2^2
c^4}{(M_1 c^2 + M_2 c^2)^2 + 2 M_2 c^2 E_{1{\rm i}}}\, p_{1{\rm i}}.
\label{4.197}\eeq
We see that $p_{1{\rm i}}$ is larger (smaller) than the major axis of
the ellipse if $M_1 > M_2$ ($M_1 < M_2$).

\begin{figure}[htb] \begin{center}
\vspace*{3mm}
\includegraphics*[scale=1.0]{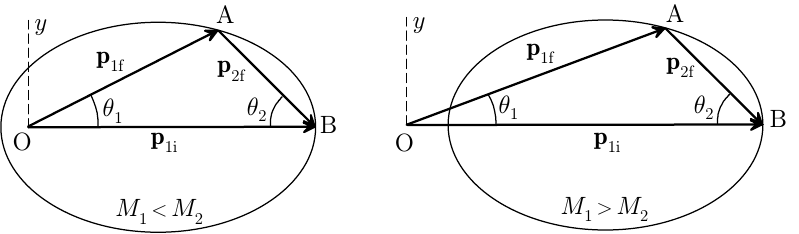}
\caption{
Variation of the final momenta ${\bf p}_{j{\rm f}}$ of the projectile
and the target with the scattering angle $\theta_1$ in the L frame.
}\label{fig4.11}
\end{center} \end{figure}

The relationship between $\theta_1$ and $p_{1{\rm f}}$ can be visualized
by means of the diagrams shown in Fig.\ \ref{fig4.11}. The point O is
the origin of the L frame, whose $z$ axis is pointing in the direction
of the initial momentum ${\bf p}_{1{\rm i}}$. The end point A of the
final momentum ${\bf p}_{1{\rm f}}$ vector lies on the ellipse defined
by Eq.\ \req{4.195}. For $\theta_1=0$, ${\bf p}_{1{\rm f}} = {\bf
p}_{1{\rm i}}$ and, hence, the distance OB is equal to $p_{1{\rm i}}$.
When $M_1 < M_2$, the scattering angle $\theta_1$ varies in the interval
from 0 to $\pi$ and, for each value of $\theta_1$ we have a unique final
momentum $p_{1{\rm f}}$. When $M_1 > M_2$, $\theta_1$ can only vary between 0
and a maximum value $\theta_{1,{\rm max}}$, given by Eq.\ \req{4.184};
for each angle $\theta_1$ in this interval we have two different
possible final momenta (in each allowed direction, we have particles
emerging with two different energies). The cutoff angle $\theta_{1,{\rm
max}}$ corresponds to the direction of the vector ${\bf p}_{1{\rm f}}$
that is tangent to the ellipse. The diagrams in Fig.\ \ref{fig4.11}
show that, for ``soft'' collisions with small scattering angle
$\theta_1$ and small momentum transfer ${\bf p}_{\rm 2f}$, the target
particle $2$ emerges in directions nearly perpendicular to the initial
direction of the projectile.

\subsection{Cross section in the L frame
\label{sec4.3.4}}

\index{collisions in the laboratory frame}
The number of projectile particles emerging after the collision with
direction of flight in the solid angle element $\d \Omega'$ around
$\hat{\bf p}'_{1{\rm f}}$ in CM is the same as the number of those particles
emitted with direction within the corresponding solid angle element $\d
\Omega_1$ around $\hat{\bf p}_{1{\rm f}}$ in L. Therefore, the DCSs in L and CM
are related by
\beq
\frac{\d \sigma_1}{\d \Omega_1} \, \d (\cos\theta_1) \, \d\phi_1 =
\frac{\d \sigma'_1}{\d \Omega'} \,  \d (\cos\theta') \, \d\phi'.
\label{4.198}\eeq
It is worth noticing that this equality implies that the value of the
total cross section $\sigma$ [\ie, the integral of the DCS over
directions, Eq.\ \req{4.35}] is the same in the L and CM frames.
Since the azimuthal angles in L and CM
are equal, $\phi_1=\phi'$, the DCS in L is
\beq
\frac{\d \sigma_1}{\d \Omega_1} =
\left| \frac{\d (\cos\theta')}{\d (\cos\theta_1)}
\right| \frac{\d \sigma'_1}{\d \Omega'}.
\label{4.199}\eeq

\allowdisplaybreaks{
To calculate the derivative of $\cos\theta'$,
we simplify the notation by setting $x=\cos\theta_1$ and $y=\cos
\theta'$, and writing the expression \req{4.183} as
\beq
y = \frac{-\tau_1 \gamma_{\rm CM}^2 ( 1 - x^2) \pm x
\sqrt{x^2 +\gamma_{\rm CM}^2 (1-\tau_1^2) (1-x^2)}}
{\gamma_{\rm CM}^2 (1-x^2)+x^2}\, .
\nonumber\eeq
Then
\beqa
\frac{\d y}{\d x} &=& \frac{
2\tau_1 \gamma_{\rm CM}^2 x \pm
\sqrt{x^2 +\gamma_{\rm CM}^2 (1-\tau_1^2) (1-x^2)}
\pm x \displaystyle{\frac{1}{2} \,
\frac{2x - 2\gamma_{\rm CM}^2 (1-\tau_1^2) x}
{\sqrt{x^2 +\gamma_{\rm CM}^2 (1-\tau_1^2) (1-x^2)}}} }
{\gamma_{\rm CM}^2 (1-x^2)+x^2}
\nonumber \\ [2mm]
&& \mbox{} -
\frac{-\tau_1 \gamma_{\rm CM}^2 ( 1 - x^2) \pm x
\sqrt{x^2 +\gamma_{\rm CM}^2 (1-\tau_1^2) (1-x^2)}}
{\left[\gamma_{\rm CM}^2 (1-x^2)+x^2\right]^2} \,
\left[-2\gamma_{\rm CM}^2 x+2x \right]
\nonumber\\ [2mm]
&=& \frac{
2\tau_1 \gamma_{\rm CM}^2 x \pm
\sqrt{x^2 +\gamma_{\rm CM}^2 (1-\tau_1^2) (1-x^2)}
\pm x \displaystyle{
\frac{x - \gamma_{\rm CM}^2 (1-\tau_1^2) x}
{\sqrt{x^2 +\gamma_{\rm CM}^2 (1-\tau_1^2) (1-x^2)}}} }
{\gamma_{\rm CM}^2 (1-x^2)+x^2}
\nonumber \\ [2mm]
&& \mbox{} +
\frac{-\tau_1 \gamma_{\rm CM}^2 ( 1 - x^2) \pm x
\sqrt{x^2 +\gamma_{\rm CM}^2 (1-\tau_1^2) (1-x^2)}}
{\left[\gamma_{\rm CM}^2 (1-x^2)+x^2\right]^2} \,
2x \left(\gamma_{\rm CM}^2-1 \right)
\nonumber\eeqa
Let us define
\beq
S=\sqrt{x^2 +\gamma_{\rm CM}^2 (1-\tau_1^2) (1-x^2)}
\nonumber\eeq
and write
\beqa
\frac{\d y}{\d x} &=& \frac{1}{\left[\gamma_{\rm CM}^2 (1-x^2)+x^2\right]^2 S}
\nonumber \\ [2mm]
&& \mbox{} \times \left\{ \left(
2\tau_1 \gamma_{\rm CM}^2 x S
\pm S^2 \pm x^2 \left[ 1 - \gamma_{\rm CM}^2 (1-\tau_1^2) \right]
\rule{0mm}{5.5mm} \right) \left[\gamma_{\rm CM}^2 (1-x^2)+x^2\right]
\right.
\nonumber \\ [2mm]
&& \mbox{}  \rule{10mm}{0mm}
+ \left. \left(
-\tau_1 \gamma_{\rm CM}^2 ( 1 - x^2)
S \pm x S^2
\right)
2x \left(\gamma_{\rm CM}^2-1 \right) \rule{0mm}{5.5mm} \right\}.
\nonumber\eeqa
The quantity in curly braces,
\beqa
\left\{ \cdots \right\} &=&
\pm x^2 \left[ 1 - \gamma_{\rm CM}^2 (1-\tau_1^2) \right]
\left[\gamma_{\rm CM}^2 (1-x^2)+x^2\right]
\nonumber \\ [2mm]
&& \mbox{} + S \, 2 \tau_1 x \left\{ \gamma_{\rm CM}^2
\left[\gamma_{\rm CM}^2 (1-x^2)+x^2\right]
- \gamma_{\rm CM}^2 ( 1 - x^2)
\left(\gamma_{\rm CM}^2-1 \right) \right\}
\nonumber \\ [2mm]
&& \mbox{} \pm S^2 \left\{
\left[\gamma_{\rm CM}^2 (1-x^2)+x^2\right] + 2 x^2
\left(\gamma_{\rm CM}^2-1 \right) \right\},
\nonumber\eeqa
can be largely simplified as follows (notice that we add and subtract
various quantities to isolate the final result and to leave a residual
sum of terms that add to zero)
\beqa
\left\{ \cdots \right\} &=&
\pm \gamma_{\rm CM}^2 \tau_1^2 x^2 + \gamma_{\rm CM}^2 2 \tau_1 x S
\pm \gamma_{\rm CM}^2 S^2
\nonumber \\ [2mm]
&& \pm \left\{ \left[ x^2 - \gamma_{\rm CM}^2 (1-\tau_1^2) x^2 \right]
\left[\gamma_{\rm CM}^2 (1-x^2)+x^2\right] - \gamma_{\rm CM}^2 \tau_1^2 x^2
\right.
\nonumber \\ [2mm]
&& \left. \mbox{} \rule{10mm}{0mm} + S^2 \left[
\gamma_{\rm CM}^2 (1-x^2)+x^2 + 2 x^2
\left(\gamma_{\rm CM}^2-1 \right) - \gamma_{\rm CM}^2 \right] \right\}
\nonumber \\ [2mm]
&=& \pm \gamma_{\rm CM}^2 \left[ \tau_1 x \pm S  \right]^2
\nonumber \\ [2mm]
&& \pm \left\{ \left[ x^2 - \gamma_{\rm CM}^2 (1-\tau_1^2) x^2 \right]
\left[\gamma_{\rm CM}^2 (1-x^2)+x^2\right] - \gamma_{\rm CM}^2 \tau_1^2 x^2
\right.
\nonumber \\ [2mm]
&& \left. \mbox{} \rule{10mm}{0mm} +
\left[ x^2 +\gamma_{\rm CM}^2 (1-\tau_1^2) (1-x^2) \right]
\left[ - \gamma_{\rm CM}^2 x^2 + x^2 + 2 x^2
\left(\gamma_{\rm CM}^2-1 \right) \right] \right\}
\nonumber \\ [2mm]
&=& \pm \gamma_{\rm CM}^2 \left[ \tau_1 x \pm S  \right]^2
\nonumber \\ [2mm]
&& \pm \left\{ \left[ x^2 - \gamma_{\rm CM}^2 (1-\tau_1^2) x^2 \right]
\left[\gamma_{\rm CM}^2 (1-x^2)+x^2\right] - \gamma_{\rm CM}^2 \tau_1^2 x^2
\right.
\nonumber \\ [2mm]
&& \left. \mbox{} \rule{10mm}{0mm} +
\left[ x^2 +\gamma_{\rm CM}^2 (1-\tau_1^2) (1-x^2) \right]
x^2 \left(\gamma_{\rm CM}^2-1 \right) \right\}
\nonumber \\ [2mm]
&=& \pm \gamma_{\rm CM}^2 \left[ \tau_1 x \pm S  \right]^2
\nonumber \\ [2mm]
&& \pm \left\{
\gamma_{\rm CM}^2 (1-x^2) x^2 +x^4
- \gamma_{\rm CM}^2 (1-\tau_1^2) x^2
\left[\gamma_{\rm CM}^2 (1-x^2)+x^2\right] - \gamma_{\rm CM}^2 \tau_1^2 x^2
\right.
\nonumber \\ [2mm]
&& \left. \mbox{} \rule{10mm}{0mm} +
\left[ x^2 +\gamma_{\rm CM}^2 (1-\tau_1^2) (1-x^2) \right]
\left(\gamma_{\rm CM}^2 x^2- x^2 \right) \right\}
\nonumber \\ [2mm]
&=& \pm \gamma_{\rm CM}^2 \left[ \tau_1 x \pm S  \right]^2
\nonumber \\ [2mm]
&& \pm \left\{
\gamma_{\rm CM}^2 (1-x^2) x^2 +x^4
- \gamma_{\rm CM}^4 (1-\tau_1^2) x^2 (1-x^2)
- \gamma_{\rm CM}^2 (1-\tau_1^2) x^4
- \gamma_{\rm CM}^2 \tau_1^2 x^2
\right.
\nonumber \\ [2mm]
&& \left. \mbox{} \rule{10mm}{0mm} +
\gamma_{\rm CM}^2 x^4 - x^4
+\gamma_{\rm CM}^4 (1-\tau_1^2) (1-x^2) x^2
-\gamma_{\rm CM}^2 (1-\tau_1^2) (1-x^2) x^2
\right\}
\nonumber \\ [2mm]
&=& \pm \gamma_{\rm CM}^2 \left[ \tau_1 x \pm S  \right]^2
\nonumber \\ [2mm]
&& \pm \left\{
\gamma_{\rm CM}^2 x^2
- \gamma_{\rm CM}^2 (1-\tau_1^2) x^4
- \gamma_{\rm CM}^2 \tau_1^2 x^2
-\gamma_{\rm CM}^2 (1-\tau_1^2) (1-x^2) x^2
\right\}
\nonumber \\ [2mm]
&=& \pm \gamma_{\rm CM}^2 \left[ \tau_1 x \pm S  \right]^2
\pm \left\{
\gamma_{\rm CM}^2 x^2
- \gamma_{\rm CM}^2 \tau_1^2 x^2
-\gamma_{\rm CM}^2 (1-\tau_1^2) x^2
\right\}
\nonumber \\ [2mm]
&=& \pm \gamma_{\rm CM}^2 \left[ \tau_1 x \pm S  \right]^2.
\nonumber\eeqa
Hence
\beq
\frac{\d y}{\d x} = \frac{\pm \gamma_{\rm CM}^2 \left[ \tau_1 x \pm S  \right]^2}
{\left[\gamma_{\rm CM}^2 (1-x^2)+x^2\right]^2 S},
\nonumber\eeq
or, equivalently,
\beq
\frac{\d (\cos \theta')}{\d (\cos \theta_1)}  =
\frac{\pm \gamma_{\rm CM}^2\, \left[\tau_1 \cos\theta_1 \pm \sqrt{\cos^2
\theta_1
+\gamma_{\rm CM}^2(1-\tau_1^2)\sin^2 \theta_1 }\; \right]^2}
{\left( \gamma_{\rm CM}^2\sin^2 \theta_1+\cos^2\theta_1 \right)^2
\sqrt{\cos^2\theta_1+\gamma_{\rm CM}^2 (1-\tau_1^2)\sin^2 \theta_1}} \, .
\label{4.200}\eeq
}
with [see Eq.\ \req{4.181}]
\beq
\tau_1= \frac{v_{\rm CM}}{v'_{1}}
=\sqrt{\left( \frac{M_1}{M_2} \right)^2
(1 - \beta_{\rm CM}^2 ) + \beta_{\rm CM}^2 } \, .
\label{4.201}\eeq
Inserting the result \req{4.200} into Eq.\ \req{4.199}, the DCS in L is
expressed as
\beq
\frac{\d \sigma_1}{\d \Omega_1} =
\frac{\gamma_{\rm CM}^2\, \left[\tau_1 \cos\theta_1 \pm \sqrt{\cos^2
\theta_1
+\gamma_{\rm CM}^2(1-\tau_1^2)\sin^2 \theta_1 }\; \right]^2}
{\left( \gamma_{\rm CM}^2\sin^2 \theta_1+\cos^2\theta_1 \right)^2
\sqrt{\cos^2\theta_1+\gamma_{\rm CM}^2 (1-\tau_1^2)\sin^2 \theta_1}} \,
\frac{\d \sigma'_1}{\d \Omega'}.
\label{4.202}\eeq
If $\tau_1 < 1$ only the plus sign is valid and the scattering angle
$\theta_1$ varies from 0 to $\pi$. When $\tau_1 \ge 1$ the DCS vanishes for
$\theta_1 > \theta_{1,{\rm max}}$, Eq.\ \req{4.184}; for angles
$\theta_1$ less than $\theta_{1,{\rm max}}$,  Eq.\ \req{4.183} yields
two values of $\theta'$ in $(0,\pi)$, the expression on the right-hand
side of Eq.\ \req{4.202} must then be evaluated for these two angles
(with the corresponding plus or minus sign in the numerator),
and the resulting values added up to give the DCS in L.

In calculations of stopping of charged particles in matter, it is
natural to consider the DCS differential in the energy loss of the
projectile, $W = E_{1{\rm i}} - E_{1{\rm f}}$. We have\index{energy-loss
DCS}
\beq
\frac{\d \sigma_1}{\d W} =
\frac{\d \sigma'_1}{\d \Omega'} \, 2\pi
\left| \frac{\d W}{\d (\cos\theta')} \right|^{-1}
= \frac{4\pi}{W_{\rm max}} \, \frac{\d \sigma'_1}{\d \Omega'} \, ,
\label{4.203}\eeq
where the last expression follows from Eqs.\ \req{4.189} and \req{4.190}.
The average energy transfer in a collision is
\beqa
\left< W \right> &=& \int_0^{W_{\rm max}} W \; \frac{1}{\sigma} \,
\frac{\d \sigma_1}{\d W} \, \d W
= \frac{W_{\rm max}}{2 \sigma} \, 2 \pi \int_{-1}^{1} (1-\cos\theta')
\, \frac{\d \sigma'_1}{\d \Omega'} \, \d (\cos \theta')
\nonumber \\ [2mm]
&=& \frac{W_{\rm max}}{2} \, \frac{\sigma_{\rm tr}}{\sigma}\, .
\label{4.204}\eeqa
The quantity
\beq
\sigma_{\rm tr} \equiv 2 \pi \int_{-1}^{1} (1-\cos\theta')
\, \frac{\d \sigma'_1}{\d \Omega'} \, \d (\cos\theta') =
\int (1-\cos\theta')
\, \frac{\d \sigma'_1}{\d \Omega'} \, \d \Omega'
\label{4.205}\eeq
is called the (first) transport cross section (see Section
\ref{sec9.5.2}) and also the {\it momentum-transfer cross section} in
CM. The latter name refers to an interesting property of $\sigma_{\rm
tr}$, which we deduce next. Let us consider the momentum transfer
in an elastic collision,
\beq
\Delta {\bf p}' = {\bf p}'_{\rm i} - {\bf p}'_{\rm f},
\label{4.206}\eeq
where ${\bf p}'_{\rm f}$ is the momentum of the projectile after the
collision. Assuming that the initial momentum is in the direction of the
$z$ axis,
\beq
{\bf p}'_{\rm i} = p'_{\rm i} \, \hat{\bf z}
=  p'_{\rm i} \, (0,0,1),
\label{4.207}\eeq
and, since $p'_{\rm f} = p'_{\rm i}$, we can write
\beq
{\bf p}'_{\rm f} = p'_{\rm i} \left( \sin\theta' \, \cos\phi',
\sin\theta' \, \sin\phi', \cos\theta' \right).
\label{4.208}\eeq
The average momentum transfer in a collision is
\beqa
\left< \Delta {\bf p}'  \right> &\equiv& \int \Delta {\bf p}'  \; \frac{1}{\sigma} \,
\frac{\d \sigma'_1}{\d \Omega'}\, \d \Omega'
\nonumber \\ [2mm]
&=& \frac{p'_{\rm i}}{\sigma}
\int \d \phi'  \int \d \theta' \; \sin\theta'
\left( - \sin\theta' \, \cos\phi',
- \sin\theta' \, \sin\phi',  1- \cos\theta' \right)
\frac{\d \sigma'_1}{\d \Omega'}
\nonumber \\ [2mm]
&=& \frac{p'_{\rm i}}{\sigma}
\, 2\pi \int \d \theta' \; \sin\theta'
\left( 0, 0,  1- \cos\theta' \right)
\frac{\d \sigma'_1}{\d \Omega'}
= p'_{\rm i} \, \hat{\bf z}  \, \frac{1}{\sigma}
\int \left( 1- \cos\theta' \right)
\frac{\d \sigma'_1}{\d \Omega'}\, \d \Omega'.
\nonumber \eeqa
We thus obtain the equality
\beq
\left< \Delta {\bf p}'  \right>
= {\bf p}'_{\rm i} \; \frac{\sigma_{\rm tr}}{\sigma} \, ,
\label{4.209}\eeq
which motivates calling $\sigma_{\rm tr}$ the momentum transfer cross
section.




\chapter{Quantum theory of elastic collisions
\label{chapt5}}



In this Chapter we present the quantum theory of elastic collisions of
charged particles with atoms and ions. We adopt the same assumptions as
in the classical relativistic formulation described in Chapter
\ref{chapt4}. A projectile of mass $M_1$, charge $Z_1 e$, and kinetic
energy $E_1$ collides with a target atom of atomic number $Z$ and mass
$M_2$ initially at rest (in the L frame). As in the classical study,
calculations will be performed in the CM frame where the DCS is
determined by the relative motion of the two particles. We assume that
the interaction between the projectile and the target atom in the CM
frame is described by a scalar central potential $V(r)$ (static-field
approximation). Under this assumption, the relative position of the two
particles obeys the classical equation of motion of a particle with the
relativistic reduced mass $\mu_{\rm r}$, given by Eq.\ \req{4.154}, in
the potential $V(r)$.

It should be mentioned that the static-field approximation gives results
in good agreement with measurements only for particles with sufficiently
high energies, higher than about 1 keV in the case of electrons and
positrons ($\sim 2$ MeV for protons). The static-field approximation
disregards the effect of the dipole polarizability of the target atom
(the electric field of the projectile shifts the atomic charges and the
induced dipole moment acts back on the projectile), which is appreciable
only for slow projectiles and at small scattering angles (large impact
parameters) because atomic polarization is effective only under
slowly-varying electric fields \citep[see, \eg,][and references
therein]{ICRU77}. In addition, results from measurements of elastic
collisions are affected by inelastic interactions, which remove
projectiles from the ``elastic channel''. Finally, in the case of
projectile electrons, elastic collisions are altered by exchange
effects, which result from the indistinguishability of the projectile
and the electrons in the target atom.
The effects of polarization and inelastic
absorption, and exchange in the case of electrons, can be described
approximately by using an {\it optical-model potential}, \ie,
a local potential with an absorptive
imaginary part \citep{Salvat2003}.

We initially consider that projectile particles have zero spin. This
simplification, although questionable for collisions of electrons and
positrons, is quite appropriate for protons and heavier particles whose
magnetic moments are about 1,000 times smaller than that of the
electron. The collisions of heavy particles can then be described by
means of the elementary theory of scattering by a central screened
Coulomb potential
\beq
V(r) = \frac{Z_1 Z e^2}{r} \, \Phi(r),
\label{5.1}\eeq
where $\Phi(r)$ is the atomic screening function (Section \ref{sec3.6}).
Elastic collisions of electrons and positrons with atoms are studied in
Section \ref{sec5.2} in terms of Dirac distorted plane waves. The final
Section of the present Chapter deals with the important case of
collisions of two identical particles, including the special situation
in which the interaction between them is Coulombian.


\section{The wave equation for scattering states \label{sec5.1}}

\index{quantum scattering states}
The wave equation governing the motion of the projectile relative to the
target atom can be obtained by following the familiar heuristic
procedure based on the correspondence principle, which ensures that the
classical laws of motion are regained in the limit where classical
mechanics is applicable \citep[see, \eg,][] {Messiah1999}. We start from
the relativistic equation of motion in the CM frame of reference, Eq.\
\req{4.155} (for simplicity, here we remove the primes in the CM
quantities),
\beq
p^2(r) = p_{\rm i}^2 - 2 \mu_{\rm r} V_{\rm ef}(r),
\label{5.2}\eeq
where $p_{\rm i}$ denotes the initial relative momentum [\ie, the
momentum of the projectile in CM, Eq.\ \req{4.139}],
\beq
p_{\rm i} = \frac{M_2 c}s \,
\sqrt{E_1(E_1+2M_1 c^2)} \qquad \mbox{with} \quad
s^2 = \left[ M_1 c^2 + M_2 c^2\right]^2 + 2 M_2 c^2 \, E_1 .
\label{5.3}\eeq
$\mu_{\rm r}$ is the relativistic reduced mass, defined by Eq.\
\req{4.154},
\beq
\mu_{\rm r} = \frac{\sqrt{M_1^2 c^2 + p_{\rm i}^2}
\sqrt{M_2^2 c^2 + p_{\rm i}^2} }
{\sqrt{M_1^2 c^2 + p_{\rm i}^2} + \sqrt{M_2^2 c^2 + p_{\rm i}^2} },
\label{5.4}\eeq
and $V_{\rm ef}(r)$ is the effective potential [Eq.\ \req{4.156}]
\begin{subequations}
\label{5.5}
\beq
V_{\rm ef}(r) = V(r) + V_{\rm r1}(r) + V_{\rm r2}(r)
\label{5.5a}\eeq
with
\beq
V_{\rm r1}(r) =
- \frac{V^2(r)}{2 \mu_{\rm r} c^2}
\left( 1 - \frac{3\mu_{\rm r}c^2}{s} \right)
\label{5.5b}\eeq
and
\beq
V_{\rm r2}(r) =
\frac{(M_1^2 - M_2^2)^2 c^6}
{8 \mu_{\rm r} s^2}
\left( \left[ 1 - \frac{V(r)}{s} \right]^{-2} - 1
- 2 \frac{V(r)}{s} - 3 \frac{V^2(r)}{s^2}
\right).
\label{5.5c}\eeq
By applying the correspondence rule ${\bf p} \rightarrow - {\rm i} \hbar
\nablab$ to Eq.\ \req{5.2} we obtain the time-independent
wave equation for free states [\ie, states with positive energy $=p_{\rm
i}^2/(2 \mu_{\rm r})$],
\end{subequations}
\beq
\left( - \frac{\hbar^2}{2 \mu_{\rm r}} \, \nabla^2
+ V_{\rm ef}(r) \right) \psi({\bf r})
= \frac{p_{\rm i}^2}{2\mu_{\rm r}} \, \psi({\bf r}) \, ,
\label{5.6}\eeq
which has the same form as the non-relativistic Schr\"odinger equation
that describes the scattering of a particle with the relativistic
reduced mass $\mu_{\rm r}$ and initial momentum $p_{\rm i}$ by the
potential $V_{\rm ef} (r)$. Hence, the wave function $\psi({\bf r})$ and
the scattering DCS can be evaluated by using the methods and
approximations of non-relativistic quantum theory.

To justify the wave equation \req{5.6}, we consider its limiting form
when the mass $M_2$ of the target atom tends to infinity, so that
$\beta_{\rm CM} \simeq 0$ and the CM frame coincides with the L frame,
where the initial momentum of the projectile is ${\bf p}_{\rm 1i}$.
Under these circumstances, ${\bf p}_{{\rm i}} = {\bf p}_{\rm 1i}$,
$s \simeq \infty$,
\beq
\mu_{\rm r} c^2 \simeq {\cal W}_{1} = \gamma M_1 c^2,
\label{5.7}\eeq
\beq
V_{\rm ef}(r) =  V(r) \left[ 1 - \frac{V(r)}{2 \gamma M_1 c^2}
\right],
\label{5.8}\eeq
and the wave equation becomes
\beq
\left( - \frac{\hbar^2}{2 \gamma M_1} \, \nabla^2
+ V(r) \left[ 1 - \frac{V(r)}{2 \gamma M_1 c^2}
\right] \right) \psi({\bf r}) =
\frac{p_{1{\rm i}}^2}{
2\gamma M_1} \, \psi({\bf r}),
\label{5.9}\eeq
which, as expected, coincides with the Klein-Gordon (or relativistic
Scr\"{o}dinger) equation for free states of a particle with mass $M_1$
and initial momentum $p_{1{\rm i}}$ in the electrostatic potential
$V(r)$ \citep[see, \eg,][]{Schiff1968}.  Indeed, our derivation shows
that the second term in the effective Klein-Gordon potential \req{5.8}
has a purely kinematical origin.

In the case of heavy projectiles with masses of the order of, or larger
than the proton mass, $\sim 1836 \, \me$, most of the collisions involve
small scattering angles. The analysis of classical trajectories (see
Section \ref{sec4.1.1}) shows that these angles correspond to
trajectories with moderate and large impact parameters, which progress
through the outer region of the atom where $V(r) \ll s$ and $V(r)
\ll \mu_{\rm r} c^2$. It is therefore appropriate to disregard the
relativistic terms of the effective potential and set
\beq
V_{\rm ef} (r) \equiv V(r).
\label{5.10}\eeq
With this approximation all quantities of interest in the scattering can
be evaluated from the corresponding non-relativistic formulas by
replacing the mass of the projectile with $\mu_{\rm r}$. Moreover, when
the electrostatic potential $V(r)$ is represented in the form
\req{3.149}, \ie, as a sum of Yukawa terms, a good part of the
calculations may be performed analytically. It is worth pointing out
that the relativistic terms of the effective potential are important
only at small $r$, of the order of the nuclear radius and smaller. At
these distances, because of the finite size of the nucleus (see Section
\ref{sec3.1}), the effective electrostatic potential departs from the
Coulomb potential and the present calculations are not expected to be
accurate. In addition, projectile nucleons and heavier nuclei with
small impact parameters interact with the nucleus though short-range
nuclear forces that may induce nuclear reactions. The combined effect of
the finite size of the nucleus and nuclear interactions may be
approximately described by using optical-model potentials \citep[see,
\eg,][and references therein]{BecchettiGreenlees1969, Hodgson1971,
KoningDelaroche2003}. The formulation that follows provides a realistic
description of collisions with impact parameters larger than about the
range of the nuclear interactions, \ie, for moderate and small
scattering angles. The DCS for larger scattering angles, where the effect
of the finite size and structure of the nucleus is appreciable, can be
obtained by combining the present theory with partial-wave
calculations of nuclear scattering with optical-model potentials
\citep{SalvatFernandezVarea2019, SalvatQuesada2020, SalvatHeredia2023}.

A rigorous quantum formulation of the scattering process, analogous to
the classical treatment, would consist in following the time evolution
of wave packets with average position and linear momentum
equivalent to those of the classical process \citep[see,
\eg,][]{GoldbergerWatson1964}. The difficulty of such a calculation
scheme is formidable. Fortunately, equivalent results are obtained by
considering a corresponding stationary process with a parallel and
laterally homogeneous beam of particles incident on the target atom. The
relative motion (in CM) is described by a distorted plane wave, \ie, an
exact solution of the time-independent wave equation for a particle of
mass $\mu_{\rm r}$ and initial linear momentum ${\bf p} = {\bf
p}_{\rm i}$ in the potential $V (r)$, which behaves asymptotically as a
plane wave with wave vector ${\bf k} = {\bf p}/\hbar$ plus an {\it
outgoing} spherical wave (see Section \ref{sec2.1.3}). That is, the
distorted plane wave satisfies the Schr\"{o}dinger wave equation
\beq
\left( - \frac{\hbar^2}{2 \mu_{\rm r}} \nabla^2 + V(r) \right)
\psi_{\bf k}({\bf r}) = \frac{p^2}{2 \mu_{\rm r}} \, \psi_{\bf k} ({\bf r})
\label{5.11}\eeq
and the asymptotic boundary condition
\begin{subequations}
\label{5.12}
\beq
\psi_{\bf k}({\bf r})
\begin{array}[t]{c} \sim \\ [-3mm] \scriptstyle{ r \rightarrow \infty}
\end{array}
(2\pi)^{-3/2} \, \exp({\rm i} {\bf k} \dotprod {\bf r})
+ (2\pi)^{-3/2} \,
\frac{\exp({\rm i} k r)}{r}
f(\hat{\bf k} \dotprod \hat{\bf r}),
\label{5.12a}\eeq
where
\beq
\phi_{\bf k}({\bf r}) =
(2\pi)^{-3/2} \, \exp({\rm i} {\bf k} \dotprod {\bf r})
\label{5.12b}\eeq
is a plane wave with wave vector ${\bf k}$, which represents the
incident beam. The asymptotic spherical component,
\beq
\psi_{\rm sc}({\bf r}) = (2\pi)^{-3/2} \,
\frac{\exp({\rm i} k r)}{r}
f(\hat{\bf k} \dotprod \hat{\bf r}),
\label{5.12c}\eeq
describes the scattered projectiles; its amplitude $f(\hat{\bf k}
\cdot \hat{\bf r})$ is called the {\it scattering amplitude}. Both
the distorted plane wave and the plane wave are normalized so that
\end{subequations}
\beq
\langle \psi_{{\bf k}} \left| \psi_{{\bf k}'} \rangle \right. =
\delta({\bf k} - {\bf k}')
\label{5.13}\eeq
and, consequently, the density of states in ${\bf k}$ space is equal to
unity (see Section \ref{sec2.1.1.1}). It is worth noticing that the constant $E \equiv p^2/(2\mu_{\rm r})$ is
generally different from the kinetic energy of the projectile in CM,
which is given by
\beq
E_1  = {\cal W}_{1} - M_1 c^2
= \sqrt{M_1^2 c^4 + p^2 c^2} - M_1 c^2
= \frac{p^2}{2M_1} - \frac{p^4}{8 M_1^3c^2} + \cdots
\label{5.14}\eeq

To simplify the formulas, hereafter we consider a reference frame with
its $z$ axis in the direction of
the initial momentum (see Fig.\ \ref{fig5.1}), so that ${\bf k} = k
\hat{\bf z}$, where $\hat{\bf z}$ is a unit vector parallel to the $z$
axis. Then, because of the spherical symmetry of the potential, the
distorted plane wave is independent of the azimuthal angle $\phi$ of the
position vector ${\bf r}$, and the scattering amplitude is a function of
the polar angle $\theta$, \ie, the angle between the direction of
incidence and the direction $\hat{\bf r}$ of the emerging
projectile, $\cos\theta = \hat{\bf k} \cdot \hat{\bf r}$.

\begin{figure}[htb] \begin{center}
\includegraphics*[width=6.0cm]{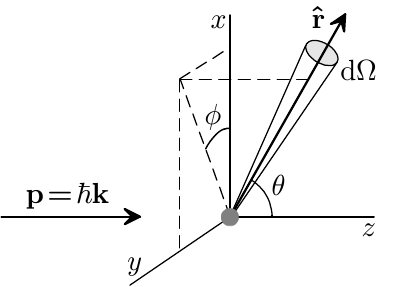}
\caption{
Schematic of a scattering process with the target atom at the origin of
coordinates and the incident beam in the direction of the $z$ axis.
\label{fig5.1}}
\end{center} \end{figure}

The probability density current of the incident wave is [see Eq.\
\req{2.11}]
\beq
{\bf j}_{\rm inc} = \frac{\hbar}{2 {\rm i} \me} \left[
\phi_{\bf k}^{\ast}(\nablab \phi_{\bf k}) -
(\nablab \phi_{\bf k}^{\ast})\phi_{\bf k} \right]
= (2\pi)^{-3} \, \frac{\hbar
{\bf k}}{\me} = (2\pi)^{-3} \, v \hat{\bf z},
\label{5.15}\eeq
which we identify as the current density of projectile
particles. To calculate the probability density current of the
scattered wave,
\beq
{\bf j}_{\rm sc}({\bf r}) = \frac{\hbar}{2 {\rm i} \me}
\left[ \psi_{\rm sc}^{\ast}(\nablab \psi_{\rm sc}) -
(\nablab \psi_{\rm sc}^{\ast})\psi_{\rm sc} \right],
\nonumber \eeq
we express the gradient operator in spherical polar coordinates,
$(r,\theta,\phi)$ [see Fig.\ \ref{figB.1} and Eq.\ \req{B.20}],
\beq
\nablab = \frac{\partial}{\partial r} \hat{\bf r} +
\frac{1}{r} \frac{\partial}{\partial\theta} \hat{\thetab} +
\frac{1}{r\sin\theta} \frac{\partial}{\partial\phi} \hat{\phib},
\label{5.16}\eeq
where $\hat{\bf r},\hat{\thetab},\hat{\phib}$ are unit vectors in the
directions of the trajectories of the vector {\bf r} when the
corresponding coordinate increases, keeping the other two constant. We
have
\beq
{\bf j}_{\rm sc}({\bf r})\dotprod\hat{\bf r} =
\frac{\hbar}{2 {\rm i} \me}
\left[ \psi_{\rm sc}^{\ast}\frac{\partial\psi_{\rm
sc}}{\partial r} - \frac{\partial\psi_{\rm sc}^{\ast}}{\partial
r}\psi_{\rm sc} \right] = (2\pi)^{-3} \frac{\hbar k}{\me}
|f(\theta)|^{2} r^{-2} + O(r^{-3}).
\label{5.17}\eeq
The components of ${\bf j}_{\rm sc}$ in the directions $\hat{\thetab}$
and $\hat{\phib}$ decrease with $r$ as $r^{-3}$.
Hence, ignoring terms of order higher than $r^{-2}$, we can write
\beq
{\bf j}_{\rm sc}({\bf r})\dotprod\hat{\bf r} \sim
(2\pi)^{-3} |f(\theta)|^{2} \, {v} \, r^{-2}.
\label{5.18}\eeq
The number of particles entering the acceptance solid angle $\d \Omega$
of the detector per unit time is
\beq
\dot{N}_{\rm count} = {\bf j}_{\rm sc}({\bf r})\dotprod\hat{\bf r} \;
r^2 \d \Omega.
\label{5.19}\eeq
The scattering differential cross section (DCS) is obtained as [see the
definition \req{4.30}]
\index{quantum scattering!DCS}
\beq
\frac{\d\sigma}{\d\Omega} \equiv
\frac{\dot{N}_{\rm count}}{j_{\rm inc} \, \d \Omega}
= \frac{[{\bf j}_{\rm sc}(r) \dotprod \hat{\bf r}] \, r^{2}}{|{\bf j}_{\rm
inc}|} = |f(\theta)|^{2}.
\label{5.20}\eeq

\subsection{Partial-wave expansion method \label{sec5.1.1}}
\index{quantum scattering!partial-wave expansion method}

The distorted plane wave can be expressed in the form of a partial-wave
series (see Section \ref{sec2.1.3}),
\beq
\psi_{\bf k} ({\bf r}) = (2\pi)^{-3/2}\, \frac{1}{kr}
\sum_{\ell} (2\ell+1) \, {\rm i}^\ell\, \exp \left( {\rm i}
\delta_{\ell} \right) \, P_{E\ell} (r) \,
P_\ell(\hat{\bf k} \dotprod \hat{\bf r}),
\label{5.21}\eeq
where $P_\ell(\hat{\bf k} \cdot \hat{\bf r})$ are the Legendre
polynomials and the functions $P_{E\ell} (r)$ are solutions of the
radial equation
\beq
\left[ - \frac{\hbar^2}{2\mu_{\rm r}} \, \frac{\d^2}{\d r^2} + V(r) +
\frac{\hbar^2}{2\mu_{\rm r}} \, \frac{\ell (\ell+1)}{r^2} \right]
P_{E\ell}(r) = \frac{p^2}{2 \mu_{\rm r}} \, P_{E\ell}(r),
\label{5.22}\eeq
normalized in the form
\beq
P_{E\ell}(r)
\begin{array}[t]{c} \sim \\ [-3mm] \scriptstyle{ r \rightarrow \infty}
\end{array}
\sin \left[ kr - \ell \frac{\pi}{2} + \eta \ln(2kr) + \delta_{\ell}
\right] ,
\label{5.23}\eeq
where $\delta_\ell$ are the scattering phase-shifts, and
\index{Sommerfeld parameter}
\beq
\eta = \frac{\mu_{\rm r}}{\hbar^2 k}
\left[ \lim_{r\rightarrow \infty} r V(r) \right] \, ,
\label{5.24}\eeq
is the Sommerfeld parameter. Purely attractive (repulsive) potentials give
positive (negative) phase shifts. The scattering amplitude is given by
the partial-wave series [see Eq.\ \req{2.58a}]
\begin{subequations}
\label{5.25}
\beqa
f(\theta) &=& \frac{1}{2{\rm i} k} \sum_\ell (2\ell +1)
\left[ \exp(2{\rm i} \delta_\ell) - 1 \right]
P_\ell(\cos\theta)
\label{5.25a}\\ [2mm]
&=& \frac{1}{k} \sum_\ell (2\ell +1)
\exp({\rm i} \delta_\ell) \, \sin (\delta_\ell) \, P_\ell(\cos\theta)
\, .
\label{5.25b}\eeqa
\end{subequations}
The total cross section,
\beq
\sigma = \int \frac{\d \sigma}{\d \Omega} \, \d \Omega = 2 \pi
\int_{-1}^1 |f(\theta)|^{2} \, \d (\cos\theta),
\label{5.26}\eeq
and the momentum transfer cross section,
\beq
\sigma_{\rm tr} = \int (1-\cos\theta) \frac{\d \sigma}{\d \Omega} \, \d
\Omega =2 \pi
\int_{-1}^1 (1-\cos\theta) |f(\theta)|^{2}\, \d (\cos\theta),
\label{5.27}\eeq
can be expressed in terms of the phase shifts. Introducing the expansion
\req{5.25b} of the scattering amplitude and using the orthogonality
relations of the Legendre polynomials
\begin{subequations}
\label{5.28}
\beq
\int_{-1}^1 P_{\ell} (x) \,
P_{\ell'} (x) \, \d x
= \frac{2}{2\ell+1}\,
\delta_{\ell,\ell'} \, ,
\label{5.28a}\eeq
\beq
\int_{-1}^1 x \, P_{\ell} (x) \,
P_{\ell'} (x) \, \d x
= \left\{
\begin{array}{ll}
\displaystyle{
\frac{2 (\ell+1)}{(2\ell+1)(2\ell+3)}
}
 & \mbox{if $\ell'=\ell+1$,} \\ [5mm]
\displaystyle{
\frac{2 \ell}{(2\ell-1)(2\ell+1)} \;
}
 & \mbox{if $\ell'=\ell-1$,}
\end{array} \right.
\label{5.28b}
\eeq
\end{subequations}
we find
\beq
\sigma
= \frac{4\pi}{k^2} \sum_{\ell=0}^\infty (2\ell+1) \sin^2 (\delta_\ell),
\label{5.29}\eeq
and
\beq
\sigma_{\rm tr}
= \frac{4\pi}{k^2} \sum_\ell
(\ell+1) \, \sin^2(\delta_\ell - \delta_{\ell+1}).
\label{5.30}\eeq
Inserting the values $P_{\ell}(1)=1$, Eq.\ \req{B.43a}, in the expansion
\req{5.25b} we see that \index{optical theorem}
\beq
\sigma
= \frac{4\pi}{k} \, {\rm Im} f(0).
\label{5.31}\eeq
This equality shows that the results from exact partial-wave
calculations satisfy the {\it optical theorem} \citep[see, \eg,][]{Schiff1968,
Joachain1975}, which is a direct consequence of the conservation of
probability (the interaction neither produces nor absorbs projectiles).

The difficulty of practical partial-wave calculations increases with the
mass $\mu_{\rm r}$ of the particle. When the projectile is an electron or
positron, the scattering amplitude can be calculated numerically from
its partial-wave expansion \req{5.25} with the phase-shifts obtained by
numerical integration of the radial equation \req{5.22} \citep[see,
\eg,][]{SalvatFernandezVarea2019, Salvat2005}. Unfortunately, for projectiles
heavier than the electron (protons, alphas, ions) partial-wave
calculations are not feasible because of 1) the small de Broglie
wavelengths of these particles, which lengthens the numerical solution
of the radial equation, and 2) the slow convergence of the partial wave
series. In the following Sections we shall describe approximate methods
for computing the DCSs for heavy projectiles.

\index{quantum scattering!Coulomb scattering}
The unscreened Coulomb potential
\beq
V_{\rm C} (r) = \frac{Z_1 Z e^2}{r}
\label{5.32}\eeq
is of fundamental importance to understand global properties of atomic
collisions. The wave equation \req{5.11} for this
potential can be solved analytically by using parabolic coordinates (see
Section \ref{sec2.1.3}). The scattering amplitude for the Coulomb
potential is given by
\beq
f^{\rm (C)}(\theta) = - \eta \, \frac{\Gamma(1+{\rm i}\eta)}
{\Gamma(1-{\rm i}\eta)}\,
\frac{\exp[
- {\rm i} \eta \ln(\sin^2(\theta/2))]}{2k\sin^2(\theta/2)}.
\label{5.33}\eeq
with the Sommerfeld parameter\index{Sommerfeld parameter}
\beq
\eta \equiv \frac{\mu_{\rm r} Z_1 Z e^2}{\hbar^2 k}.
\label{5.34}\eeq
The scattering amplitude \req{5.33} can
also be expressed in the form of a partial-wave series,
\beq
f^{\rm (C)}(\theta) = \frac{1}{2 {\rm i} k}
\sum_{\ell} (2\ell+1) \, \left[ \exp \left( 2 {\rm
i} \Delta_{\ell} \right) - 1 \right] \, P_\ell(\cos\theta)
\label{5.35}\eeq
with the Coulomb phase shifts [Eq.\ \req{2.49}]
\beq
\Delta_\ell = {\rm arg} \, \Gamma(\ell + 1 + {\rm i} \eta).
\label{5.36}\eeq
The DCS for scattering by the Coulomb field is
\beq
\frac{\d \sigma_{\rm C}}{\d \Omega} = \left| f^{\rm (C)}(\theta)
\right|^2
= \frac{\eta^2}{4 k^2 \sin^4(\theta/2)} =
\left(\frac{Z_1Ze^2}{2 v p}\right)^2 \frac{1}{\sin^4(\theta/2)}.
\label{5.37}\eeq
Interestingly, this expression is formally identical to the
classical Rutherford DCS, Eq.\ \req{4.105}. This is a fortunate
peculiarity of the Coulomb potential; for other potentials the classical
and quantum theories yield different DCSs (except when the classical
approximation is valid). \index{Rutherford DCS}

A comment on the relativistic Rutherford formula \req{5.37} is in order.
In principle, the velocity $v$ appearing in that formula is the relative
velocity $v=p/\mu_{\rm r}$ [Eq.\ \req{4.144}] of the two particles in
the CM frame. More accurate calculations that consider the quantized
electromagnetic interaction between the two colliding particles lead to
the result \req{5.37} plus a small relativistic correction, but with the
velocity $v_1$ of the projectile in the L frame instead of $v$.  Hence,
the velocity $v$ in the relativistic Rutherford formula should be set
equal to the velocity of the projectile in L.

\subsection{The plane-wave Born approximation \label{sec5.1.2}}
\index{quantum scattering!plane-wave Born approximation}

The simplest approach for describing elastic collisions is provided by the
plane-wave Born approximation, which is studied in most textbooks on
quantum mechanics. Within this approximation, the states of the
projectile before and after the collision are represented as plane waves
of the type \req{5.12b}, with respective linear momenta ${\bf p}=\hbar
k \hat{\bf z}$ and ${\bf p}'= \hbar {\bf k}'$. The DCS is obtained by
considering the potential $V(r)$ as a perturbation to first order, as
dictated by Fermi's golden rule \citep[see, \eg][]{Baym1974}. We recall
that, with the adopted normalization of the plane waves, Eq.\ \req{5.13},
the density of states per unit volume in ${\bf k}$-space equals unity.
Then, the probability per unit time of a transition from the initial
state ($i$) to final states ($f$) with wave vectors in the volume
element $\d {\bf k}'$ around the final wave vector ${\bf k}'$ is
\beq
\d^2 w_{fi} = \frac{2\pi}{\hbar} \left| T_{fi}^{\rm (B)} \right|^2 \,
\delta(E - E') \, \d {\bf k}'
\label{5.38}\eeq
with the transition matrix element
\beqa
T_{fi}^{\rm (B)} &\equiv& \left. \left. \left<
\phi_{{\bf k}'}({\bf r}) \right| V(r) \right| \phi_{{\bf k}}({\bf r}) \right>
\nonumber \\ [2mm]
&=& \frac{1}{(2\pi)^3} \int
\exp \left( - {\rm i} {\bf k}' \dotprod {\bf r} \right)
V(r) \, \exp \left( {\rm i} {\bf k} \dotprod {\bf r} \right) \, \d {\bf r}.
\label{5.39}\eeqa
From the relation $E'=(\hbar k')^2/(2\mu_{\rm r})$ it follows that
\beq
\d {\bf k}' = k'^2 \d k' \, \d \hat{\bf k}'
= k'^2 \frac{\d k'}{\d E'} \, \d E' \, \d \hat{\bf k}'
= k' \, \frac{\mu_{\rm r}}{\hbar^2} \, \d E' \, \d \hat{\bf k}'
\label{5.40}\eeq
and we can write
\beq
\d^2 w_{fi} = \frac{2\pi \mu_{\rm r}}{\hbar^3} \left| T_{fi}^{\rm (B)} \right|^2 \,
\delta(E - E') \, k' \,
\, \d E' \, \d \hat{\bf k}'.
\label{5.41}\eeq
Integration over the final energy $E'$ gives
\beq
\d w_{fi} = \frac{2\pi \mu_{\rm r}}{\hbar^3} \left| T_{fi}^{\rm (B)} \right|^2 \,
k \, \d \hat{\bf k}'.
\label{5.42}\eeq
Noting that the probability current of the incident wave is [see Eq.\
\req{5.15}]
\beq
{\bf j}_{\rm inc} = (2\pi)^{-3} \frac{\hbar k}{\mu_{\rm r}} \, \hat{\bf k},
\label{5.43}\eeq
we conclude that the DCS is
\beq
\frac{\d \sigma^{\rm (B)}}{\d \Omega} = \frac{1}{\left| {\bf j}_{\rm
inc} \right|} \, \frac{\d w_{fi}}{\d \hat{\bf k}'}
= \frac{(2\pi)^4 \mu_{\rm r}^2}{\hbar^4} \left| T_{fi}^{\rm (B)}
\right|^2.
\label{5.44}\eeq

Evidently, the DCS can be expressed in the familiar form
\beq
\frac{\d \sigma^{\rm (B)}}{\d \Omega} =
\left| f^{\rm (B)}(\theta) \right|^2
\label{5.45}\eeq
with the Born scattering amplitude
\begin{subequations}
\label{5.46}
\beqa
f^{\rm (B)}(\theta)
&\equiv& - \frac{(2\pi)^2 \mu_{\rm r}}{\hbar^2} T_{fi}^{\rm (B)}
= - \frac{\mu_{\rm r}}{2 \pi \hbar^2}
\int \exp \left( - {\rm i} {\bf k}' \dotprod {\bf r} \right)
V(r) \, \exp \left( {\rm i} {\bf k} \dotprod {\bf r} \right) \, \d {\bf r}
\label{5.46a} \\ [2mm]
&=& - \frac{\mu_{\rm r}}{2\pi \hbar^2} \,
\int \exp({\rm i} {\bf q} \dotprod {\bf r})
V(r) \, \d {\bf r} ,
\label{5.46b}\eeqa  \end{subequations}
where $\hbar {\bf q} \equiv {\bf p} - {\bf p}'$ is the momentum
transfer. The global phase factor of the scattering amplitude has been
set in accordance with the result of formal scattering theory
\citep[see,\eg,][]{Joachain1975}. Because the kinetic energies of the projectile before and after
the collision are the same, we have (see Fig.\ \ref{fig5.2})
\beq
q = 2 k \sin(\theta/2) = 2 k \sqrt{\frac{1-\cos\theta}{2}}\, .
\label{5.47}\eeq
For a central potential the integral \req{5.46b} can be expressed in
polar spherical coordinates with the polar axis in the direction
of ${\bf q}$. Then, integration over angles gives
\beqa
f^{\rm (B)}(\theta)
&=& - \frac{\mu_{\rm r}}{2\pi \hbar^2} \,
\int_0^\infty r^2 \, \d r \, 2\pi \int_{-1}^{1} \d (\cos\vartheta)
\exp({\rm i} q r \cos\vartheta) V(r)
\nonumber \\ [2mm]
&=& - \frac{2\mu_{\rm r}}{\hbar^2} \,
\int_0^\infty \frac{\sin(qr)}{qr} \,
V(r) r^2 \, \d r\, .
\label{5.48}\eeqa
As the scattering amplitude is real, the Born approximation does not
satisfy the optical theorem, Eq.\ \req{5.31}.

\begin{figure}[htb] \begin{center}
\includegraphics*[width=6.0cm]{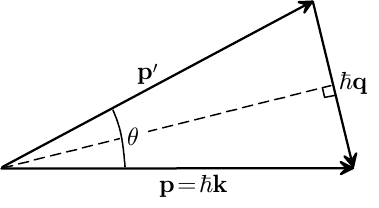}
\caption{
Kinematics of elastic collisions.
\label{fig5.2}}
\end{center} \end{figure}

\index{Wentzel potential}
Because the Born approximation results from a simple first-order
perturbation meth\-od, it is expected to be accurate when the distortion
of the incident plane wave by the scattering potential is small.
Qualitative arguments, based on formal scattering theory, indicate that
the approximation is valid when \citep[see, \eg,][]{Joachain1975}
\beq
\frac{2\mu_{\rm r}}{\hbar^2 k} \left| \int_0^\infty \exp({\rm i} k r) \, V(r) \,
\sin(kr) \, \d r \right| \ll 1.
\label{5.49}\eeq
For the Wentzel potential [Eq.\ \req{3.151}],
\beq
V_{\rm W} (r) = \frac{Z_1 Z e^2}{r} \, \exp(-r/R),
\label{5.50}\eeq
the integral in the relation \req{5.49} can be solved analytically giving
\beq
\frac{\mu_{\rm r}}{\me} \, \left| \frac{Z_1 Z }{ka_0}
\left\{ \arctan(2kR) + \frac{\rm i}{2} \ln[1+4(kR)^2]
\rule{0mm}{4mm}\right\} \right| \ll 1.
\label{5.51}\eeq
For slow projectiles such that $kR \ll 1$, this relation reduces to
$(\mu_{\rm r}/\me) \left| 2Z_1 Z R/a_0\right| \ll 1$, which implies that
the Born approximation may not be valid. For fast particles ($kR \gg
1$), the condition \req{5.49} becomes $(\mu_{\rm r}/\me) \, \left| Z_1 Z
\right| \, \ln(kR)/(ka_0) \ll 1$, so that the approximation is accurate
for projectiles with sufficiently high energies.

Our motivation for considering screened atomic potentials of the form
\req{3.149},
\beq
V(r) = \frac{Z_1 Z e^2}{r} \sum_{i=1}^3 A_i \exp(- a_i r),
\label{5.52}\eeq
is that they lead to the following simple expressions for the
Born scattering amplitude, \index{Born scattering amplitude}
\beqa
f^{\rm (B)}(\theta)
&=& - \frac{2\mu_{\rm r}}{\hbar^2} \, Z_1 Z e^2 \sum_i A_i \frac{1}{q}
\int_0^\infty \sin(qr') \, \exp(-\alpha_i r')
\, \d r'
\nonumber \\ [2mm]
&=& - \frac{2\mu_{\rm r}}{\hbar^2}\, Z_1 Z e^2
\sum_i A_i \, \frac{1}{\alpha_i^2 + q^2},
\label{5.53}\eeqa
and the Born DCS,
\beq
\frac{\d \sigma^{\rm (B)}}{\d \Omega}
= \left| f^{\rm (B)}(\theta) \right|^2
= \left( \frac{2\mu_{\rm r}}{\hbar^2}\, Z_1 Z e^2 \right)^2
\left( \sum_i A_i \, \frac{1}{\alpha_i^2 + q^2} \right)^2.
\label{5.54}\eeq
In the particular case of the Wentzel potential, Eq.\ \req{5.50},
the Born scattering amplitude reads
\beq
f^{\rm (BW)}(\theta)
= - \frac{2\mu_{\rm r}}{\hbar^2}\, Z_1 Z e^2
\, \frac{1}{R^{-2} + q^2},
\label{5.55}\eeq
and the corresponding DCS is \index{Wentzel DCS}
\beq
\frac{\d \sigma^{\rm (BW)}}{\d \Omega}
= \left| f^{\rm (BW)}(\theta) \right|^2
= \left( \frac{2\mu_{\rm r}}{\hbar^2}\, Z_1 Z e^2 \right)^2
\frac{1}{[R^{-2} + 2 k^2 (1-\cos\theta)]^2}.
\label{5.56}\eeq
If we now set $R^{-1}=0$, we obtain the DCS for Coulomb scattering within
the plane-wave Born approximation, \index{Rutherford DCS}
\beq
\frac{\d \sigma^{\rm (B)}_{\rm C}}{\d \Omega}
= \left( \frac{\mu_{\rm r} \, Z_1 Z e^2}{p^2} \right)^2
\frac{1}{(1-\cos\theta)^2}
= \left( \frac{Z_1 Z e^2}{2 v p} \right)^2
\frac{1}{\sin^4(\theta/2)} \, .
\label{5.57}\eeq
We thus see that the Born approximation for the Coulomb potential
yields, again, the classical Rutherford DCS. However, the Born
scattering amplitude for the Coulomb potential differs from the exact
amplitude given by Eq.\ \req{5.33}; the difference is of relevance,
\eg, in the description of collisions of identical particles
(see Section \ref{sec5.3}).

To derive a Born approximation for the phase shifts, we insert the
Rayleigh expansion of the plane waves, Eq.\ \req{2.62}, into the
expression \req{5.46a} of the Born scattering amplitude,
\beqa
f^{\rm (B)}(\theta) & = &
- \frac{8 \pi \mu_{\rm r}}{\hbar^2}
\int \, r^2 \d r \int \d \hat{\bf r}
\left[ \sum_{\ell',m'} {\rm i}^{\ell'}
\, j_{\ell'} (kr) \, Y^\ast_{\ell' m'} (\hat{\bf k}')
Y_{\ell' m'} (\hat{\bf r}) \right]^\ast
V(r)
\nonumber \\ [2mm]
&& \mbox{} \times \left[
\sum_{\ell,m} {\rm i}^\ell
\, j_{\ell} (kr) \, Y^\ast_{\ell m} (\hat{\bf k})
Y_{\ell m} (\hat{\bf r}) \right].
\nonumber \eeqa
Performing the integration over the direction $\hat{\bf r}$,\index{Born
phase shifts}
\beqa
f^{\rm (B)}(\theta) & = &
- \frac{8 \pi \mu_{\rm r}}{\hbar^2} \sum_{\ell m}
\int \, r^2 \d r
\, j_{\ell}^2 (kr) \, V(r) \, Y_{\ell m} (\hat{\bf k}')
Y^\ast_{\ell m} (\hat{\bf k}),
\nonumber \eeqa
and using the addition theorem of the spherical harmonics, Eq.\
\req{B.57}, we obtain
\beqa
f^{\rm (B)}(\theta) & = &
- \frac{2 \mu_{\rm r}}{\hbar^2} \sum_{\ell} (2\ell+1)
\left[ \int
\, j_{\ell}^2 (kr) \, V(r) \, r^2 \d r \right] \,
P_{\ell} (\hat{\bf k}'\dotprod \hat{\bf k})
\nonumber \\ [2mm]
&=& \frac{1}{2{\rm i}k}
\sum_{\ell} (2\ell+1) \, (2{\rm i} \delta_\ell^{\rm (B)})
\, P_\ell(\cos\theta)\, ,
\label{5.58}\eeqa
where
\beq
\delta_\ell^{\rm (B)} = - \frac{2\mu_{\rm r}}{\hbar^2} \, k \int_0^\infty j_\ell^2
(kr') V(r') r'^2 \, \d r'
\label{5.59}\eeq
is the Born approximation for the phase shifts [cf.\ Eqs.\ \req{5.58}
and \req{5.25a}].

A further advantage of using the analytical atomic potentials
\req{5.52} is that their Born phase shifts can be calculated easily
with the aid of the equality [\citeauthor{GradshteynRyzhik2007}
(\citeyear{GradshteynRyzhik2007}), Eq. 6.612.3]
\beq
\int_0^\infty \exp(-\alpha r) j_\ell^2 (k r) r \, dr = \frac{1}{2 k^2}
Q_\ell \left( 1 + \frac{\alpha^2}{2 k^2} \right)\, .
\label{5.60}\eeq
where $Q_\ell(x)$ are the Legendre functions of the second kind
\citep{AbramowitzStegun1974}. We have
\beqa
\delta_\ell^{\rm (B)}
&=& - \frac{2\mu_{\rm r}}{\hbar^2} \, Z_1Z e^2 \sum_i A_i
k \frac{1}{2 k^2}
Q_\ell \left( 1 + \frac{\alpha_i^2}{2 k^2} \right)
\nonumber \\ [2mm]
&=& - \frac{\mu_{\rm r}}{\hbar^2 k} \, Z_1Z e^2 \sum_i A_i
Q_\ell \left( 1 + \frac{\alpha_i^2}{2 k^2} \right).
\label{5.61}\eeqa
The functions $Q_\ell(r)$ can be calculated to high accuracy by using
the calculation scheme described by \citet{FernandezVarea1993}. We start
from the recurrence relation\index{Legendre functions of the second kind}
\beq
(\ell+1) Q_{\ell +1}(x) - (2\ell+1) x Q_\ell(x) + \ell Q_{\ell-1}(x)=0,
\label{5.62}\eeq
and the explicit expressions
\beq
Q_0(x) = \frac{1}{2} \ln \left| \frac{x+1}{x-1} \right|,
\qquad
Q_1(x) = x Q_0(x)-1.
\label{5.63}\eeq
For $x < 1$ the recurrence relation can be applied for increasing values
of $\ell$. However, for $x>1$ the recurrence should only be used for
decreasing values of $\ell$ to avoid the accumulation of round-off
errors. To compute all the $Q_\ell(x)$ for $x>1$ we use the
approximation
\beq
Q_\ell(x) \simeq K_0[\ell \cosh^{-1}(x)],
\label{5.64}\eeq
which holds asymptotically for large values of $\ell$. Here $K_0$ stands
for the modified Bessel function of zeroth order. In the computer
program {\sc elastic} (see Chapter \ref{chapt10}) the expression
\req{5.64} is used to estimate the values of $Q_L(x)$ and $Q_{L+1}(x)$
for an index $L$ that is large enough for Eq.\ \req{5.64} to be
applicable, but not much larger than $80/\cosh^{-1}(x)$ to prevent the
occurrence of computer underflows. Approximate values
$\overline{Q}_\ell(x)$ of the Legendre functions with $\ell = L-1, L-2,
..., 1, 0$ are then generated by applying the recurrence relation
\req{5.62}. The final values of $Q_\ell(x)$ are obtained as
\beq
Q_\ell(x) = \overline{Q}_\ell(x) \left[ Q_0(x)/\overline{Q}_0(x)
\right].
\label{5.65}\eeq


\subsection{Approximate phase shifts and scattering amplitude
\label{sec5.1.3}}

Generally, the Born phase shifts \req{5.59} provide a good approximation
to the actual phase shifts that are small in magnitude, even when the
Born approximation for the scattering amplitude is not accurate. To
allow the calculation of the scattering amplitude from the partial-wave
series \req{5.25} we need an alternative method to obtain the other
phase shifts. We use the semi-classical Wentzel--Kramers--Brillouin (WKB)
approximation, with the \citet{Langer1937} correction, which leads to
the following formula for the phase shifts \citep{Joachain1975}
\index{WKB phase shifts}
\beq
\delta_\ell^{\rm (WKB)} = \frac{1}{2} \left( \ell + \frac{1}{2} \right)
\pi
- k r_0 + \int_{r_0}^\infty \left[ \sqrt{F_\ell(r)} - k \right] \, \d r,
\label{5.66}\eeq
where
\beq
F_\ell (r) = k^2 - \frac{2\mu_{\rm r}}{\hbar^2} V(r) -
\frac{(\ell+1/2)^2}{r^2},
\label{5.67}\eeq
and $r_0$ is the largest zero of $F_\ell(r)$. The WKB
approximation is accurate when the potential $V(r)$ is practically
constant over many wavelengths or, more precisely, when
\citep{Schiff1968}
\beq
\left| \frac{1}{2F_\ell(r)} \, \frac{\d}{\d r} \sqrt{F_\ell(r)} \right|
\ll 1 \qquad \mbox{for $r>r_0$.}
\label{5.68}\eeq
To facilitate numerical calculations, it is
convenient to change the integration variable to $x=1/r$ and write
\beq
\delta_{\ell}^{\rm (WKB)}
= \frac{1}{2} \left( \ell + \frac{1}{2} \right) \pi - kr_{0} +
\int_0^{1/r_{0}} \left[ \sqrt{F_{\ell}(1/x)} - k \right] \, \frac{\d
x}{x^2}\, .
\label{5.69}\eeq
The program {\sc elastic} (Section \ref{sec10.1}) calculates these
integrals by using the adaptive Gauss--Legendre quadrature method
(Section \ref{sec10.4.3.2}) to a relative accuracy of $\sim 10^{-10}$.

It is desirable to get a feel of the accuracy of the WKB and Born
approximations for the phase shifts. In the case of electrons or
positrons, the code system {\sc radial} of
\citet{SalvatFernandezVarea2019}
gives phase shifts obtained from the numerical solution of the radial
equation \req{5.22}, which is integrated by using a robust power-series
solution method. The accuracy of WKB and Born phase shifts can then be
estimated by comparison with the numerical values generated by {\sc
radial}. A similar comparison of approximate and numerical phase shifts
for projectiles heavier than the electron is not possible because of the
already commented difficulty of the numerical calculation of phase
shifts, and it will not be attempted here. As a matter of fact, the
study for electrons is sufficient for our purposes because the WKB
approximation at a given energy is more accurate for the heavier
particles, due to their shorter wave lengths.

Table \ref{tab5.1} shows phase-shifts obtained from the WKB and Born
approximations (as given by the program {\sc elastic}) and
numerical phase shifts calculated with the code {\sc radial}, for the
case of scattering of 10 keV electrons by gold atoms ($Z=79$), with all
phase shifts calculated using the analytical DHFS potential (Section
\ref{sec3.6}). To be consistent with the {\sc radial} results, the
approximate phase shifts were calculated with the reduced mass equal to
the electron mass $\me$ and a linear momentum $p_{\rm i} = (2 \me
E)^{1/2}$. The WKB phase shifts are in fairly good agreement with the
numerical values; relative differences are of the order of 1 \% for
$\ell =0$, and generally they are smaller for larger $\ell$. The Born
phase-shifts of small angular momenta $\ell$ are less accurate than the
WKB phase shifts; the relative differences between the WKB and Born
phases decrease progressively when $\ell$ increases. Beyond a certain
$\ell$, the Born phase shifts give an acceptable approximation to the
actual phase shifts. For projectiles with higher energies we find the
same tendencies, except that the phase shifts decrease in magnitude more
slowly with $\ell$. Accordingly, the convergence of the partial-wave
series slows down (\ie, we need to add more terms to get the sum to a
given relative accuracy) when the energy increases. The differences
between numerical and approximate phase shifts of low orders are the
most relevant, because they amount to adding or subtracting a smooth
component to the DCS, whose effect is magnified at large angles, where
the actual DCS is smaller\footnote{Since the Legendre polynomial
$P_\ell(\cos\theta)$ has $\ell$ zeros nearly uniformly spaced,
an error in the $\ell$-th order phase shift will manifest as an
oscillation of the calculated DCS with respect to the ``correct'' DCS
with nearly $\ell$ crests and troughs of similar amplitude.}.

\begin{table}[htb!]
\caption{\rm
Phase shifts for scattering of 10 keV electrons by the DHFS
potential of gold atoms ($Z=79$). The values in the left column were
computed by the {\sc radial} code. WKB phase shifts were calculated
for orders $\ell \le L=51$. The eikonal phase shifts,
$\delta_\ell^{\rm (eik)} = \chi(b)/2$, with $b=(\ell+1/2)/k$
were calculated from Eq.\ \req{5.98}, \ie, including the Wallace correction.
Adapted from \citet{Salvat2022b}. \label{tab5.1}}
\vskip 3mm
\begin{center}
{\small\tt
\begin{tabular}{|r|ccccc|}
\cline{2-6}
\multicolumn{1}{c}{\rule[-2mm]{0mm}{2mm}} &
\multicolumn{5}{|c|}{$\delta_\ell$} \\ \hline
\multicolumn{1}{|c|}{$\ell$} &
\multicolumn{1}{|c}{\rm numerical} &
\multicolumn{1}{c}{\rm Eqs.\ \req{5.70}} &
\multicolumn{1}{c}{\rm WKB} &
\multicolumn{1}{c}{\rm Born} &
\multicolumn{1}{c|}{\rm eikonal} \\ \hline
  0 & 6.25076E+00 & 6.26033E+00 & 6.26033E+00 & 7.09519E+00 & 1.03652E+01  \\
  1 & 4.85210E+00 & 4.86682E+00 & 4.86682E+00 & 4.45239E+00 & 5.57539E+00  \\
  2 & 3.72213E+00 & 3.73702E+00 & 3.73702E+00 & 3.27638E+00 & 3.86014E+00  \\
  3 & 2.90479E+00 & 2.91414E+00 & 2.91414E+00 & 2.56871E+00 & 2.92794E+00  \\
  4 & 2.32614E+00 & 2.33168E+00 & 2.33168E+00 & 2.08362E+00 & 2.32516E+00  \\
  5 & 1.90282E+00 & 1.90641E+00 & 1.90641E+00 & 1.72640E+00 & 1.89705E+00  \\
  6 & 1.58154E+00 & 1.58402E+00 & 1.58402E+00 & 1.45153E+00 & 1.57569E+00  \\
  7 & 1.33066E+00 & 1.33243E+00 & 1.33243E+00 & 1.23382E+00 & 1.32587E+00  \\
  8 & 1.13061E+00 & 1.13189E+00 & 1.13189E+00 & 1.05782E+00 & 1.12698E+00  \\
  9 & 9.68531E-01 & 9.69443E-01 & 9.69443E-01 & 9.13385E-01 & 9.65866E-01  \\
 10 & 8.35528E-01 & 8.36179E-01 & 8.36179E-01 & 7.93454E-01 & 8.33608E-01  \\
 15 & 4.32699E-01 & 4.32795E-01 & 4.32795E-01 & 4.20759E-01 & 4.32321E-01  \\
 20 & 2.46941E-01 & 2.46926E-01 & 2.46926E-01 & 2.43066E-01 & 2.46832E-01  \\
 25 & 1.50843E-01 & 1.50813E-01 & 1.50813E-01 & 1.49433E-01 & 1.50792E-01  \\
 30 & 9.68752E-02 & 9.68507E-02 & 9.68507E-02 & 9.63097E-02 & 9.68454E-02  \\
 35 & 6.45745E-02 & 6.45571E-02 & 6.45571E-02 & 6.43283E-02 & 6.45556E-02  \\
 40 & 4.42433E-02 & 4.42315E-02 & 4.42315E-02 & 4.41288E-02 & 4.42310E-02  \\
 50 & 2.19469E-02 & 2.19412E-02 & 2.19412E-02 & 2.19184E-02 & 2.19412E-02  \\
 75 & 4.44078E-03 & 4.43988E-03 &   ------    & 4.43947E-03 & 4.43932E-03  \\
100 & 9.76997E-04 & 9.76935E-04 &   ------    & 9.76927E-04 & 9.76527E-04  \\
150 & 5.15943E-05 & 5.15950E-05 &   ------    & 5.15950E-05 & 5.15523E-05  \\
200 & 2.88128E-06 & 2.88181E-06 &   ------    & 2.88181E-06 & 2.87841E-06  \\
250 & 1.65537E-07 & 1.65995E-07 &   ------    & 1.65996E-07 & 1.65742E-07  \\
\hline
\end{tabular}} \end{center} \end{table}

Guided by these observations, and considering that the coefficients in
the Legendre series of the actual scattering amplitude vary smoothly with
$\ell$, we set
\begin{subequations}
\label{5.70}
\beq
\delta_\ell = \left\{
\begin{array}{ll}
\delta^{\rm (WKB)}_\ell \rule{5mm}{0mm} &
\mbox{if $\ell \le L$,} \\ [2mm]
C_\ell \delta^{\rm (B)}_\ell & \mbox{otherwise,}
\end{array} \right.
\label{5.70a}\eeq
where the cutoff order $L$ is the lowest value of $\ell$ for
which either $\delta_\ell^{\rm (B)} < 0.001$ or the relative difference
between the WKB and Born phase shifts is less than 0.001. Evidently,
this prescription implies that only WKB phase shifts of orders $\ell
\le L$ need to be computed. The factor
\beq
C_\ell \equiv  1+ \left( \frac{\delta^{\rm (WKB)}_L}{\delta^{\rm (B)}_L}
- 1 \right) \, \exp \left( - a \, \frac{\ell - L}{L} \right)
\label{5.70b}\eeq
\end{subequations}
is introduced to ensure that the approximate phase shifts $\delta_\ell$
vary smoothly with $\ell$ near the cutoff $L$. The form of this factor,
which has been decided on the basis of the observed variation of the
difference between the WKB and Born phase shifts with $\ell$, is
practically irrelevant as long as it ensures ``continuity'' at $\ell =
L$ and tends to unity at large $\ell$, where the Born approximation to
the phase shifts is expected to be valid. The parameter $a$ is obtained
by requiring that $\delta^{\rm (WKB)}_{L-1}\rule{0mm}{5mm} = C_{L-1}
\delta^{\rm (B)}_{L-1}$, with the proviso that the relative difference of
$\delta_\ell$ and $\delta^{\rm (B)}_\ell$ effectively decreases with
$\ell$. Omission of the factor $C_\ell$ may cause the scattering
amplitude to oscillate about its ``average'' shape with a frequency
depending on the cutoff order $L$ and an amplitude proportional to the
magnitude of the ``discontinuity'' between WKB and Born phase shifts.

As indicated above, the convergence of the partial-wave series may be
very slow. We can alleviate the numerical work by adding the Born
scattering amplitude and subtracting its partial-wave expansion,
\begin{subequations}
\label{5.71}
\beq
f(\theta) = f^{\rm (B)}(\theta) + \sum_{\ell=0}^\infty
{\cal F}_\ell \, P_\ell(\cos\theta)
\label{5.71a} \eeq
with
\beq
{\cal F}_\ell = \frac{1}{2 {\rm i} k} \, (2\ell+1)
\left[ \exp\left( 2 {\rm i} \delta_{\ell}\right) -1
-2{\rm i} \delta_\ell^{\rm (B)} \right].
\label{5.71b}\eeq
\end{subequations}
Since the phase shifts $\delta_\ell$ approximate the Born phase shifts
when $\ell \gtrsim L$, this series converges more rapidly than the
original series \req{5.25}. A further acceleration of convergence for
angles larger than 1 degree is provided by the reduced-series method
described in Section \ref{sec5.2.3}.

This computation scheme is similar to the one adopted by
\citet{Salvat1987c} to describe elastic collisions of electrons by atoms
(including the effects of electron exchange and atomic polarization,
which are disregarded here); a similar calculation strategy was proposed
earlier by \citet{HoerniIbers1953}. Since the calculation of each WKB
phase shift reduces to a single quadrature and the set of Born phase
shifts is generated from a simple recurrence, the whole process is fast
enough to be run interactively on a personal computer. Moreover, the
method is applicable to projectile particles of any type, and it gives
fairly reliable results, provided the velocity of the projectile is
large enough to make the effect of atomic polarization negligible.

At very high energies, the number of phase-shifts needed to ensure
convergence of the partial-wave series \req{5.71} may exceed the allowed
memory storage (500,000 phases in the program {\sc elastic}). For angles
less than 1 degree, when convergence of the partial-wave series is not
reached, the eikonal approximation (Section \ref{sec5.1.4}) is generally
valid, and {\sc elastic} replaces the partial-wave scattering amplitude
and DCS with the corresponding eikonal results. For scattering angles
larger than 1 degree, the program applies the reduced series method,
which causes the partial-wave series to converge with much less terms
(see Section \ref{sec5.2.3}). A limitation of the approximate phase
shifts is found for projectiles with small mass (electrons and
positrons) and high energies; their partial-wave DCS decreases very
rapidly with $\theta$ and the errors in the WKB phase shifts of low
orders produce a visible alteration of the DCS at large angles (see
Fig.\ \ref{fig5.6}).

It is worth noticing that the formula \req{5.69} for the WKB phase
shifts does not assume any specific form of the potential. Therefore,
the WKB phases can be evaluated for the effective relativistic
potential, Eq.\ \req{5.5}. In addition, the phase shifts of the
effective potential $V_{\rm ef}(r)$ approach those of the electrostatic
potential $V(r)$ when the order $\ell$ increases, because the
centrifugal barrier prevents the radial function from seeing the inner
part of the potential, where the relativistic correction terms are appreciable.
Hence, provided the WKB integral \req{5.69} converges, the scattering
amplitude for the effective potential can still be evaluated
approximately by the present scheme, using the WKB phase shifts of
$V_{\rm ef}(r)$ combined with the large-order Born phase shifts of
$V(r)$, Eq.\ \req{5.61}. Unfortunately, for the interesting case of high
energies, the inaccuracies in the WKB phase shifts of low angular
momenta alters the DCS at large angles tending to mask the effect of the
relativistic terms in the effective potential.


\subsection{The eikonal approximation \label{sec5.1.4}}

\index{eikonal approximation|(}
An alternative description of elastic collisions, valid for projectiles
with sufficiently high energies, is provided by the eikonal
approximation \citep{Moliere1947, Schiff1968}, in which the phase of the
scattered wave is obtained from a semi-classical approach under the
assumption of small trajectory deflections.

Let us consider the Schr\"{o}dinger wave equation \req{5.11} for a
free particle of mass $\mu_{\rm r}$ and linear momentum ${\bf p} =
\hbar k$ in a finite-range central potential $V(r)$,
\beq
\left( - \frac{\hbar^2}{2 \mu_{\rm r}} \nabla^2 + V(r) \right)
\psi_{\bf k}({\bf r}) = \frac{\hbar^2 k^2}{2 \mu_{\rm r}} \,
\psi_{\bf k} ({\bf r}).
\label{5.72}\eeq
We express the wave function in the form
\beq
\psi_{\bf k}({\bf r}) =
\sqrt{\rho({\bf r})} \exp[{\rm i} S({\bf r})/\hbar],
\label{5.73}\eeq
where both the phase $S({\bf r})$ and the function $\rho({\bf r})$
are assumed to be real. In addition, $\rho({\bf r})$ is required to be
non-negative. Upon differentiation we have
\beqa
\nablab \psi_{\bf k} &=&
(\nablab \sqrt{\rho}) \exp[{\rm i} S/\hbar]
+ \frac{\rm i}{\hbar} \sqrt{\rho} (\nablab S) \exp[{\rm i} S/\hbar]
\nonumber \\ [2mm]
\nabla^2 \psi_{\bf k} &=&
(\nabla^2 \sqrt{\rho}) \exp[{\rm i} S/\hbar]
+ 2 \frac{\rm i}{\hbar} (\nablab \sqrt{\rho}) \dotprod (\nablab S)
\exp[{\rm i} S/\hbar]
- \frac{1}{\hbar^2} \sqrt{\rho} (\nablab S)^2 \exp[{\rm i} S/\hbar]
\nonumber \\ [2mm]
&& \mbox{}
+ \frac{\rm i}{\hbar} \sqrt{\rho} (\nabla^2 S) \exp[{\rm i} S/\hbar].
\label{5.74}\eeqa
Evidently, $\rho = |\psi_{\bf k}|^2$ is the probability density of the
particle. The probability current vector is
\beq
{\bf j} = \frac{\hbar}{2{\rm i}\mu_{\rm r}} \left[ \psi_{\bf k}^\ast
(\nablab \psi_{\bf k})
+ (\nablab \psi_{\bf k}^\ast) \psi_{\bf k} \right]
= \rho \, (\nablab S)/{\mu_{\rm r}}.
\label{5.75}\eeq
We see that the direction of ${\bf j}$ is that of the gradient of $S$
and, consequently, the vector ${\bf j}$ is perpendicular to the surfaces
of constant $S$.  The quantity $(\nablab S)/\mu_{\rm r}$ can be interpreted
as the ``velocity'' of the probability density.

The wave equation \req{5.72} can be written as
\beqa
&& \! \! \! \! \! \! \! \! \! \! \! \! \! \!
- \frac{\hbar^2}{2 \mu_{\rm r}} \left\{
(\nabla^2 \sqrt{\rho})
+ 2 \frac{\rm i}{\hbar} (\nablab \sqrt{\rho}) \dotprod (\nablab S)
- \frac{1}{\hbar^2} \sqrt{\rho} (\nablab S)^2
+ \frac{\rm i}{\hbar} \sqrt{\rho} (\nabla^2 S) \right\} + V \sqrt{\rho}
\nonumber \\ [2mm]
&=& \frac{\hbar^2 k^2}{2 \mu_{\rm r}} \, \sqrt{\rho} \, .
\label{5.76}\eeqa
Let us now assume that $\hbar$ can be considered as a small quantity
(semi-classical approximation). Then, neglecting terms proportional to
$\hbar$ and to $\hbar^2$, Eq.\ \req{5.76} takes the form
\beq
\frac{1}{2\mu_{\rm r}} \left[ \nablab S({\bf r}) \right]^2 + V({\bf r})
= \frac{\hbar^2 k^2}{2 \mu_{\rm r}}.
\label{5.77}\eeq
The {\it eikonal approximation} consists in expressing the wave function
in the form \req{5.73} with the phase $S({\bf r})$ obtained from Eq.\
\req{5.77}.

\begin{figure}[htb] \begin{center}
\includegraphics*[width=8.0cm]{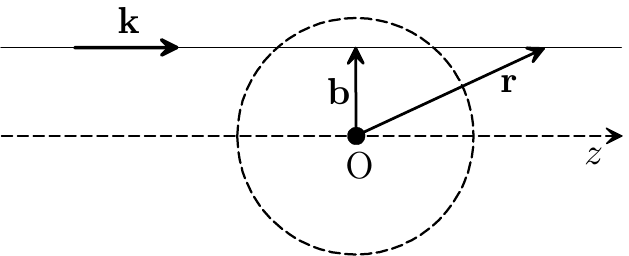}
\caption{Particle trajectory, as considered in the eikonal approximation.
\label{fig5.3}}
\end{center} \end{figure}

To facilitate the calculation of the scattering process, we consider
projectile particles moving initially in the direction of the $z$ axis
(${\bf k} = k \hat{\bf z}$), and we assume that their trajectories are
straight (Fig.\ \ref{fig5.3}). This assumption is justified when the
``kinetic energy'', $E=(\hbar k)^2/(2 \mu_{\rm r})$, is much larger than
$V(r)$ for all points of the trajectory, and it implies that the
calculation results are expected to be valid only for small scattering
angles. Notice that each trajectory is identified by the impact
parameter ${\bf b} = (x,y,0)$. We may then try the following approximate
solution of Eq.\ \req{5.77},
\beq
S({\bf r}) = \hbar \int_{-\infty}^z \left[ k^2 - \frac{2 \mu_{\rm r}}{\hbar^2}
V\left( \sqrt{b^2 + z'^2} \right) \right]^{1/2} \d z' + A,
\label{5.78}\eeq
where the quantity $A$ must be chosen so that $S=\hbar k \, z$ when
$V({\bf r})$ vanishes. Since $E \gg V(r)$, we can put
\beqa
S({\bf r}) &=& \hbar k \int_{-\infty}^z \left[ 1 - \frac{2\mu_{\rm r}}
{\hbar^2 k^2 } V \left( \sqrt{b^2 + z'^2} \right) \right]^{1/2} \d z'
+ A
\nonumber \\ [2mm]
&\simeq& \hbar k \int_{-\infty}^z \left[ 1 - \frac{\mu_{\rm r}}{\hbar^2 k^2 }
V\left( \sqrt{b^2 + z'^2} \right) \right] \d z' + A.
\nonumber \eeqa
Consequently,
\beq
S(z) =
\hbar k \, z -  \frac{\mu_{\rm r}}{\hbar k } \int_{-\infty}^z
V\left( \sqrt{b^2 + z'^2} \right) \d z'.
\label{5.79}\eeq
Thus, the eikonal approximation to the scattering wave function is
\beqa
\psi_{\bf k} ({\bf r}) &=& (2\pi)^{-3/2} \exp \left[ {\rm i}
S({\bf r})/\hbar \right]
\nonumber \\ [2mm]
&=& (2\pi)^{-3/2} \, \exp \left[
{\rm i} k \, z -  \frac{{\rm i} \mu_{\rm r}}{\hbar^2 k } \int_{-\infty}^z
V\left( \sqrt{b^2 + z'^2} \right) \d z' \right],
\label{5.80}\eeqa
with the usual wave-number normalization.

\index{eikonal approximation!scattering amplitude}

Knowing the approximate wave function \req{5.80}, the scattering
amplitude can be calculated by using the formula \req{2.54}
\beq
f^{\rm (eik)}(\theta,\phi) = - \frac{(2\pi)^{1/2}
\mu_{\rm r}}{\hbar^2} \int \d {\bf r} \,
\exp(-{\rm i} {\bf k'}\dotprod {\bf r})\, V({\bf r}) \,
\psi_{\bf k} ({\bf r}),
\label{5.81}\eeq
where ${\bf k}'$ is a vector of modulus $k$ in the direction defined by
the polar angle $\theta$ and the azimuthal angle $\phi$. That is,
\beqa
f^{\rm (eik)}(\theta, \phi) &=& - \frac{\mu_{\rm r}}{2\pi \hbar^2} \int \d {\bf r} \,
\exp\left[- {\rm i} ({\bf k'}\dotprod {\bf r}-kz) \right]
\, V\left( \sqrt{b^2 + z^2} \right)
\nonumber \\ [2mm]
&& \rule{5mm}{0mm} \times
\, \exp \left[
- \frac{{\rm i} \mu_{\rm r}}{\hbar^2 k } \int_{-\infty}^z
V\left( \sqrt{b^2 + z'^2} \right)\, \d z' \right].
\label{5.82}\eeqa
It is worth noticing that, if we remove the last factor, this expression
reduces to that of the plane-wave Born approximation, Eq.\ \req{5.46b}.
That factor accounts for the phase difference between the incident and
the scattered waves, and makes the eikonal approximation generally more
accurate than the Born approximation. The eikonal scattering amplitude
is complex and satisfies the optical theorem (see below), while the Born
scattering amplitude is real and, hence, does not satisfy that theorem.

Because the potential is central, the eikonal scattering amplitude is
independent of the azimuthal angle. Introducing cylindrical coordinates,
${\bf r} = {\bf b} + z \hat{\bf z}$, $\d {\bf r} = b \, \d b \, \d
\phi_b \, \d z$, we can write
\beqa
f^{\rm (eik)}(\theta) &=&
- \frac{\mu_{\rm r}}{2\pi \hbar^2} \int_0^\infty b \, \d b
\left( \int_0^{2\pi}
\d \phi_b
\exp\left[ - {\rm i} ({\bf k'}\dotprod {\bf r}-kz) \right] \right)
\nonumber \\ [2mm]
&& \mbox{} \times
\int_{-\infty}^{\infty}
\, V\left( \sqrt{b^2 + z'^2} \right)
\, \exp \left[
- \frac{{\rm i} \mu_{\rm r}}{\hbar^2 k } \int_{-\infty}^z
V\left( \sqrt{b^2 + z'^2} \right)\, \d z' \right]
\, \d z. \rule{10mm}{0mm}
\label{5.83}\eeqa
Without loosing generality, we can consider that the final direction of
the particle lies in the plane $x$-$z$, so that ${\bf k}' = k
\sin\theta \, \hat{\bf x} + k \cos\theta \, \hat{\bf z}$. We have
$$
{\bf k'}\dotprod {\bf r} -kz = {\bf k}' \dotprod {\bf b} +
({\bf k'}-{\bf k}) \dotprod \hat{\bf z} z
= {\bf k}' \dotprod {\bf b} + k (\cos\theta -1) z.
$$
Recalling that the scattering angle $\theta$ is assumed to be small, the
last term is of the order of $\theta^2$ and can be neglected. We thus have
\beq
{\bf k'}\dotprod {\bf r} -kz \simeq {\bf k}' \dotprod {\bf b} \simeq
kb \theta \cos\phi_b.
\label{5.84}\eeq
The integral over $\phi_b$ is
\beq
\int_0^{2\pi} \d \phi_b \exp\left[- {\rm i} ({\bf k'}\dotprod {\bf
r} -kz) \right] \simeq
\int_0^{2\pi} \d \phi_b \exp( - {\rm i}
kb \theta \cos\phi_b ) =
2\pi \, J_0(kb\theta),
\label{5.85}\eeq
where $J_0(x)$ is the Bessel function of first kind and zeroth order.
The last integral in \req{5.83} is
\beqa
&&
\! \! \! \! \! \! \! \! \!
\! \! \! \! \! \! \! \! \!
\int_{-\infty}^{\infty}
\, V\left( \sqrt{b^2 + z'^2} \right)
\, \exp \left[
- \frac{{\rm i} \mu_{\rm r}}{\hbar^2 k } \int_{-\infty}^z
V\left( \sqrt{b^2 + z'^2} \right)\, \d z' \right]
\, \d z
\nonumber \\ [2mm]
&=&
\left\{
\frac{{\rm i}\hbar^2 k }{\mu_{\rm r}}
\, \exp \left[
- \frac{{\rm i} \mu_{\rm r}}{\hbar^2 k } \int_{-\infty}^z
V\left( \sqrt{b^2 + z'^2} \right)\, \d z' \right]
\right\}_{-\infty}^{\infty}
\nonumber \\ [2mm]
&=&
\frac{{\rm i}\hbar^2 k }{\mu_{\rm r}}
\, \exp \left(
- \frac{{\rm i} \mu_{\rm r}}{\hbar^2 k } \int_{-\infty}^\infty
V\left( \sqrt{b^2 + z'^2} \right)\, \d z' \right).
\nonumber\eeqa
Therefore,
\beq
f^{\rm (eik)}(\theta)
= - {\rm i} k \int_0^\infty J_0(kb\theta) \left\{
\exp \left[
{\rm i} \chi(b) \right] -1 \right\} b \, \d b,
\label{5.86}\eeq
where\index{eikonal approximation!phase function}
\beq
\chi(b) \equiv -\, \frac{2\mu_{\rm r}}{\hbar^2 k } \int_{0}^\infty
V\left( \sqrt{b^2 + z^2} \right)\, \d z
= -\, \frac{2\mu_{\rm r}}{\hbar^2 k } \int_{b}^\infty
V(r) \frac{r\, \d r}{\sqrt{r^2 -b^2}}
\label{5.87}\eeq
is the {\it eikonal phase function}. Introducing the
momentum transfer $\hbar {\bf q} = \hbar{\bf k} - \hbar{\bf k'}$, and
observing that for small angles
\beq
q = |{\bf k} - {\bf k}'| = 2 k \sin(\theta/2) \simeq k \theta\, ,
\label{5.88}\eeq
we can write
\beq
f^{\rm (eik)}(\theta)
= - {\rm i} k \int_0^\infty J_0(qb) \left\{
\exp \left[
{\rm i} \chi(b) \right] -1 \right\} b \, \d b.
\label{5.89}\eeq

\noindent
$\bullet$ {\bf Relationship with the partial-wave expansion}.
\citet{Moliere1947} was the first to use the eikonal approximation
to describe the scattering by a central potential. He pointed out that
the eikonal scattering amplitude can be obtained as the small-angle
limit of the partial-wave expansion \req{5.25a},
$$
f(\theta) = \frac{1}{2{\rm i} k} \sum_\ell (2\ell +1)
\left[ \exp(2{\rm i} \delta_\ell) - 1 \right]
P_\ell(\cos\theta)
$$
with the phase shifts calculated from the WKB approximation, Eq.\
\req{5.66},
$$
\delta_\ell^{\rm (WKB)} = \frac{1}{2} \left( \ell + \frac{1}{2} \right)
\pi
- k r_0 + \int_{r_0}^\infty \left[ \sqrt{F_\ell(r)} - k \right] \, \d r,
$$
where $r_0$ is the outer zero of the function
$$
F_\ell (r) = k^2 - \frac{2\mu_{\rm r}}{\hbar^2} V(r) -
\frac{(\ell+1/2)^2}{r^2}.
$$
To determine the small-angle limit of the partial-wave expansion, we
notice that the WKB phase shifts vanish for $V(r) = 0$ because
\beq
\frac{1}{2} \left( \ell + \frac{1}{2} \right) \pi - k b
= - \int_{b}^\infty \left[ \sqrt{ k^2 -
\frac{(\ell+1/2)^2}{r^2}} - k \right] \, \d r ,
\label{5.90}\eeq
where
\beq
b = \frac{\ell + 1/2}{k}
\label{5.91}\eeq
is the classical impact parameter corresponding to the angular momentum
$\ell + \1o2$; the additional $\1o2$ in the WKB formulas occurs as a
consequence of the fact that the radial motion is limited to
positive values of $r$ \citep{Langer1937}. We can then write
\beqa
\delta_\ell^{\rm (WKB)} &=&
\int_{r_0}^\infty \left[
\sqrt{k^2 - {\cal V}(r) -
\frac{(\ell+1/2)^2}{r^2}} - k \right] \, \d r
- \int_{b}^\infty \left[ \sqrt{ k^2 -
\frac{(\ell+1/2)^2}{r^2}} - k \right] \, \d r
\nonumber \\ [2mm]
&=&
\int_{b}^\infty \left[
\sqrt{K^2(r) - {\cal V}(r)}
- K(r)
\right] \, \d r
\nonumber \\ [2mm]
&& \mbox{}
- \int_{r_0}^{b} \left[ \sqrt{k^2 - {\cal V} (r)
- \frac{(\ell+1/2)^2}{r^2}} - k \right] \, \d r
\label{5.92}\eeqa
with
\beq
{\cal V}(r) = \frac{2\mu_{\rm r}}{\hbar^2} V(r)
\quad \mbox{and} \quad
K^2(r) = k^2 - \frac{(\ell+1/2)^2}{r^2} = k^2 \left( 1 -\frac{b^2}{r^2}
\right).
\label{5.93}\eeq
Small angles correspond to relatively large impact parameters $b$ such
that ${\cal V}(r)$ is much smaller than $K^2(r)$ for $r > b$ and, in
addition, $r_0$ tends to
$b$ when $\ell$ increases. Consequently, for sufficiently large $b$ or
$\ell$, we can neglect the second integral in Eq.\ \req{5.92}, and
approximate the first term in that expression by expanding the square
root in powers of ${\cal V}/K^2$. Neglecting terms of second and higher orders,
we obtain the following eikonal approximation for the phase shifts,
\index{eikonal approximation!phase shifts}
\beq
\delta_\ell^{\rm (eik)} =  - \frac{1}{2} \int_{b}^\infty
\frac{{\cal V}(r)}{K(r)} \, \d r
= - \, \frac{\mu_{\rm r}}{\hbar^2 k_{\rm i}}
\int_{b}^\infty V(r) \, \frac{r \, \d r}{\sqrt{r^2 - b^2}}
= \frac{1}{2} \,  \chi(b),
\label{5.94}\eeq
where $\chi(b)$ is the eikonal phase function, Eq.\ \req{5.87}. The
foregoing derivation indicates that this formula is expected to be valid
(\ie, to give results close to the exact values of the phase shifts) for
large $\ell$'s.

\index{eikonal approximation!scattering amplitude}

Inserting the approximation \req{5.94} into the partial-wave expansion
\req{5.25a} of the scattering amplitude, and replacing the summation
over $\ell= k b$ with an integral over the impact parameter, we have
$$
f(\theta) \simeq \frac{1}{2{\rm i} k} \int_0^\infty \d (kb) \; 2 kb \;
\left\{ \exp[{\rm i} \chi(b)] - 1 \right\}
P_\ell(\cos\theta).
$$
The following approximation to the Legendre polynomials
\citep{Moliere1947}
\beq
P_\ell(\cos\theta) \simeq \sqrt{\frac{\theta}{\sin\theta}}
 J_0 \left( \frac{2
\ell+1}{2} \, \theta \right)
\label{5.95}\eeq
is valid for any $\theta < \pi$ in the limit for large $\ell$, and for
all values of $\ell$ in the limit for small $\theta$. Hence, for small
angles (such that $\sin\theta \simeq \theta$, $q \simeq k \theta$) we
can write
\beq
f(\theta) \simeq - {\rm i} k \int_0^\infty
\left\{ \exp[{\rm i} \chi(b)] - 1 \right\}
J_0( q b)\, b \, \d b,
\label{5.96}\eeq
which is the eikonal scattering amplitude, Eq.\ \req{5.89}.
Consequently, the eikonal approximation can be regarded as the
small-angle limit of the semi-classical WKB approximation.

\noindent
$\bullet$ {\bf Wallace correction}.
\citet{Wallace1971} derived systematic corrections of order $k^{-n}$ to
the eikonal phase. With the first-order Wallace correction included, the
eikonal phase is
\beq
\chi(b)
= -\, \frac{2\mu_{\rm r}}{\hbar^2 k } \int_{b}^\infty
V(r) \left\{ 1 + \frac{\mu_{\rm r}}{\hbar^2 k^2} \left[ V(r) + r \, \frac{\d
V(r)}{\d r} \right] \right\} \frac{ r\, \d r}{\sqrt{r^2 -b^2}}.
\label{5.97}\eeq
In the limit of high energies, the factor in curly braces, the Wallace
correction, tends to unity and \req{5.97} reduces to the expression
\req{5.87} given by Moli\`{e}re. \citet{ByronJoachain1977} have
analyzed the reliability of the eikonal approximation for screened
atomic potentials, and concluded that Wallace's correction does improve
the accuracy of the method.

The formula \req{5.94} with the corrected eikonal phase provides an
improved approximation to the phase shifts,
\beq
\delta_\ell^{\rm (eik)} =
\frac{1}{2} \,  \chi\left( \frac{\ell+1/2}{k} \right).
\label{5.98}\eeq
Eikonal phase shifts for 10 keV electrons scattered by the DHFS
potential of the free gold atom are shown in Table \ref{tab5.1}. Notice
that the relative difference between phase shifts calculated with the
WKB method and with the eikonal approximation decreases rapidly when the
angular momentum increases. Since phase shifts of high orders are
correctly given by the Born approximation with little numerical work
[see Eq.\ \req{5.61}], the eikonal method is not helpful for the
practical evaluation of phase shifts.

\noindent
$\bullet$ {\bf The optical theorem}.
\index{eikonal approximation!optical theorem} \index{optical theorem}
It is interesting to notice that, for particles with sufficiently high
energies, the eikonal approximation (with or without Wallace's
correction) satisfies the optical theorem,
\beq
\sigma^{\rm (eik)} \equiv \int_0^\pi |f^{\rm (eik)}(\theta)|^2 \, 2\pi
\sin(\theta) \,
\d \theta = \frac{4\pi}{k} {\rm Im} f^{\rm (eik)}(0),
\label{5.99}\eeq
which expresses the conservation of the number of particles. This
theorem is useful for verifying the accuracy of numerical calculations.
To prove it, we express $f^{\rm (eik)}(\theta)$ in the form \req{5.89} and
evaluate the integral
\beqa
\int_0^\pi |f^{\rm (eik)}(\theta)|^2 \, 2\pi
\sin \theta
\, \d \theta &=& 2\pi k^2
\int_0^\pi \d \theta \,
\sin \theta
\left\{ \int_0^\infty J_0(qb) \left\{
\exp \left[
{\rm i} \chi(b) \right] -1 \right\} b \, \d b \right\}
\nonumber \\ [2mm]
&& \! \! \! \! \! \! \! \! \times
\left\{ \int_0^\infty J_0(qb') \left\{
\exp \left[
-{\rm i} \chi(b') \right] -1 \right\} b' \, \d b' \right\}
\nonumber \eeqa
\beq
= 2\pi \int_0^\infty \d b \int_0^\infty \d b' \left\{
\int_0^\infty q J_0(qb)\, J_0(qb') \, \d q \right\}\,
b^2 \left\{ \exp \left[ {\rm i} \chi(b) \right] -1 \right\}
\left\{ \exp \left[- {\rm i} \chi(b') \right] -1 \right\},
\nonumber\eeq
where we have used that
\beq
q^2 = 2 k^2 (1-\cos\theta) \; \; \Rightarrow \; \;
k^2 \sin \theta \, \d \theta = q \, \d q,
\nonumber\eeq
and replaced the upper limit ($2k$) in the integral over $q$ with $\infty$.
This replacement is justified only for high energies and, consequently,
we may expect deviations from the theorem at low energies. Using the
equality \citep[][Eq.\ 11.59]{Arfken1985}
\beq
\int_0^\infty q J_0(qb)\, J_0(qb') \, \d q = \frac{1}{b} \,
\delta(b-b'),
\label{5.100}\eeq
we can write
\beqa
&&
\! \! \! \! \! \! \! \! \! \! \! \! \! \! \!
\int_0^\pi |f^{\rm (eik)}(\theta)|^2 \, 2\pi \sin(\theta) \, \d \theta =
2\pi \int_0^\infty \d b \;
b \left\{ \exp \left[ {\rm i} \chi(b) \right] -1 \right\}
\left\{ \exp \left[- {\rm i} \chi(b) \right] -1 \right\}
\nonumber \\ [2mm]
&=& 2\pi \int_0^\infty \d b \;
b \, 2 \left[ 1 - \cos \chi(b) \right].
\nonumber \eeqa
On the other hand, recalling that $J_0(0)=1$, the eikonal scattering
amplitude \req{5.89} at $\theta=0$ ($q=0$) is
\beq
f^{\rm (eik)}(0) = - {\rm i} k \int_0^\infty
\left\{ \exp \left[ {\rm i} \chi(b) \right] -1 \right\} b \, \d b.
\nonumber \eeq
Hence
\beq
\frac{4\pi}{k} {\rm Im} f^{\rm (eik)}(0) =
- 4\pi \int_0^\infty
\left[ \cos \chi(b) -1 \right] b\, \d b,
\nonumber \eeq
which completes the proof of the theorem.

\noindent
$\bullet$ {\bf Practical calculation of the scattering amplitude}.
For atomic potentials of the form \req{5.52}, the eikonal phase
\req{5.97} can be calculated analytically as follows. We first note that
\beq
V(r) + r \, \frac{\d V(r)}{\d r} = -\, Z_1 Z e^2
\sum_i A_i a_i \exp(-a_i r)
\nonumber \eeq
and, consequently,
\beqa
\chi(b)
&=& -\, \frac{2 \mu_{\rm r}}{\hbar^2 k} \int_{b}^\infty \frac{Z_1 Ze^2}{r}
\sum_i A_i \exp(-a_i r)
\left[ 1 - \frac{\mu_{\rm r}}{\hbar^2 k^2} \, Z_1 Ze^2
\sum_j A_j a_j \exp(-a_j r) \right]
\nonumber \\ [2mm]
&& \times \frac{ r\, \d r}{\sqrt{r^2 -b^2}}.
\nonumber \eeqa
Considering the identity [Eq.\ 8.432.3 in \citep{GradshteynRyzhik2007}]
\beq
\int_b^\infty \frac{\exp(-a r)}{\sqrt{r^2-b^2}} \, \d r =
\int_1^\infty \frac{\exp(-a b x)}{\sqrt{x^2-1}} \, \d x =
K_0(a b),
\label{5.101}\eeq
where $K_0(x)$ is the modified Bessel function of the second kind and
zeroth order, we can write
\beq
\chi(b) = -\,
\frac{2 \mu_{\rm r} Z_1 Ze^2}{\hbar^2 k} \,
\sum_i A_i \left\{ K_0(a_i b) -
\frac{\mu_{\rm r} \, Z_1 Ze^2}{\hbar^2 k^2}
\sum_j A_j a_j K_0[(a_i+a_j)b] \right\}.
\label{5.102}\eeq
The evaluation of the expression \req{5.89} thus reduces to a
single quadrature, which must be performed numerically. To ease
the calculation we follow \citet{ZeitlerOlsen1967} and,
instead of directly evaluating the integral \req{5.89}, we introduce the
identity
\beq
(qb) J_0(qb) = \frac{\d }{\d b}
\left[ b J_1(qb) \right],
\label{5.103}\eeq
where $J_1(x)$ is the Bessel function of the first kind and first order.
Integration by parts gives
\beq
f^{\rm (eik)}(\theta)
= - \, \frac{k}{q} \int_0^\infty J_1(qb)
\, \frac{\d \chi(b)}{\d b} \, \exp \left[
{\rm i} \chi(b) \right] \, b \, \d b.
\label{5.104}\eeq
The derivative of the eikonal phase \req{5.102} is
\beqa
\frac{\d \chi(b)}{\d b} &=&
\frac{2 \mu_{\rm r}\, Z_1 Ze^2}{\hbar^2 k} \,
\sum_i A_i \left\{ a_i K_1(a_i b) \rule{0mm}{8mm}\right.
\nonumber \\ [2mm]
&& \mbox{} \left. \rule{10mm}{0mm}
- \frac{\mu_{\rm r} \, Z_1 Ze^2}{\hbar^2 k^2} \sum_j A_j a_j
(a_i+a_j)K_1[(a_i+a_j)b] \right\},
\label{5.105}\eeqa
which follows from the equality $(\d /\d x)K_0(x)=-K_1(x)$, Eq.\
\req{B.63}, where
$K_1(x)$ is the modified Bessel function of the second kind and first
order. The integral \req{5.104} converges rapidly for large values of
$b$ because of the exponentially decaying $K_1(a_ib)$, and the fast
oscillations of the exponential at small $b$ are suppressed by the
$J_1(qb)$ function, which vanishes at $b=0$.  The DCS is given by
\beq
\frac{\d \sigma^{\rm (eik)}}{\d \Omega} = |f^{\rm (eik)} (\theta)|^2.
\label{5.106}\eeq

The eikonal approximation is expected to be accurate for scattering
angles up to about $(k R)^{-1}$ \citep{Moliere1947}, where $R$ is the
atomic radius, Eq.\ \req{3.152}. However, numerical calculations
indicate that for protons and heavier particles the approximation yields
fairly accurate DCSs, practically coincident with those obtained from
classical and partial-wave calculations up to much larger angles, of the
order of
\beq
\theta_{\rm eik} =
\min \left\{ \frac{200}{kR}, 0.1 \pi \right\}\, .
\label{5.107} \eeq
For angles somewhat larger than $\theta_{\rm eik}$, the calculation
presents numerical instabilities and must be discontinued. In the
computer program {\sc elastic} the DCS for $\theta > \theta_{\rm eik}$
is estimated by extrapolation using the empirical formula proposed by
\citet{Salvat2013},
\beq
\frac{\d \sigma^{\rm (eik)}}{\d \Omega}  =
\left(\frac{2 \mu_{\rm r}}{\hbar^2}  \, Z_1 Z e^2\right)^2
\frac{1}{\left[ A + B q^{2/3} + C q^{4/3} +
q^2\right]^2}\, .
\label{5.108}\eeq
with the coefficients $A$, $B$ and $C$ determined by matching the
calculated numerical values of the eikonal DCS and its first and second
derivatives at $\theta=\theta_{\rm eik}$. In the case of proton
collisions, the extrapolated DCS differs by less than about 1 \% from
the classical DCS, which is expected to be accurate for large scattering
angles. Note that for momentum transfers that are large enough, both the
Born DCS and the extrapolated eikonal DCS tend to the Rutherford DCS
\req{5.57}, which decreases rapidly with the scattering angle ($\propto
q^{-4}$).
\index{eikonal approximation|)}

\subsection{Calculation examples \label{sec5.1.5}}

The program {\sc elastic} (see Chapter \ref{chapt10}) computes the DCSs
from the theoretical approximations described above. Figure \ref{fig5.4}
displays the calculated DCSs for collisions of muons ($\mu^{-}$; mass
$M_1= 206.8 \, \me$ and charge $Z_1=-1$) and antimuons ($\mu^{+}$; $M_1=
206.8 \, \me$, $Z_1=+1$) with atoms of oxygen ($Z=8$) and plutonium
($Z=94$), which represent extreme cases of small and large atomic
numbers. The displayed curves correspond to the Born approximation
(Born), the partial-wave expansion method with approximate WKB and Born
phase shifts (pwa), the eikonal approximation, and the DCS obtained from
the classical trajectory method described in Section \ref{sec4.1.4}.
These DCSs were evaluated in the CM frame, for the indicated kinetic
energies $E$ of the projectile in the L frame. All calculations were
performed using the DHFS potential, that is, the relativistic correction
terms in the potential \req{5.5} were disregarded. As shown above, these
terms modify the DCS appreciably only at large scattering angles (see
Fig.\ \ref{fig4.9}), which are unimportant for transport calculations.
The differences between the DCSs for muons and antimuons are small, not
visible in the scale of the plots.

\begin{figure}[p!] \begin{center}
\includegraphics*[width=7.5 cm]{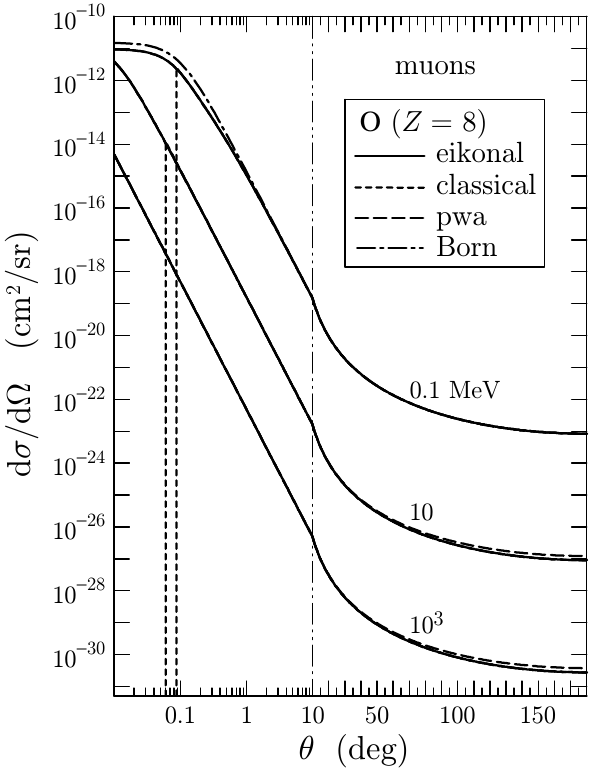} \rule{3mm}{0mm}
\includegraphics*[width=7.5 cm]{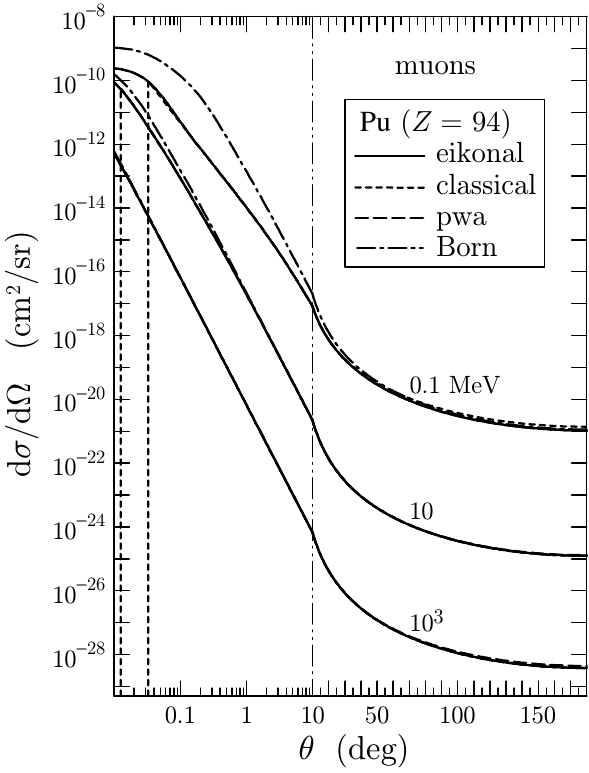} \\ [5mm]
\includegraphics*[width=7.5 cm]{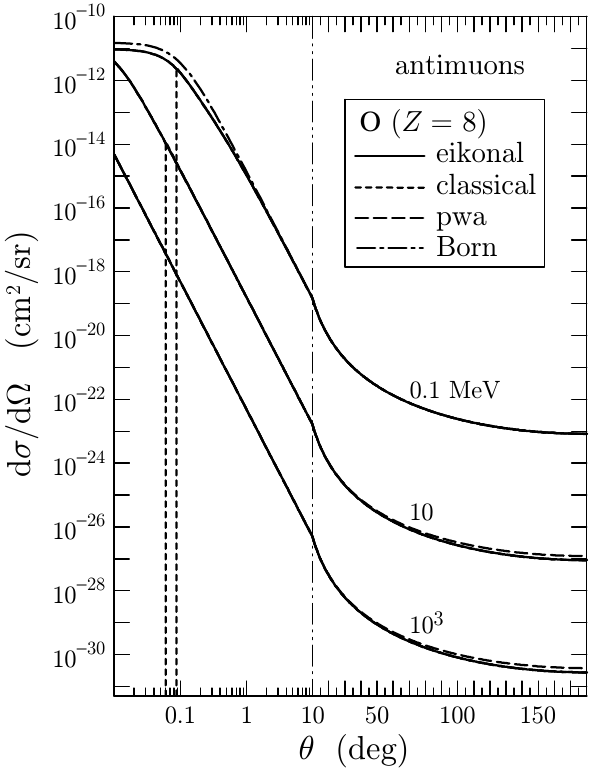} \rule{3mm}{0mm}
\includegraphics*[width=7.5 cm]{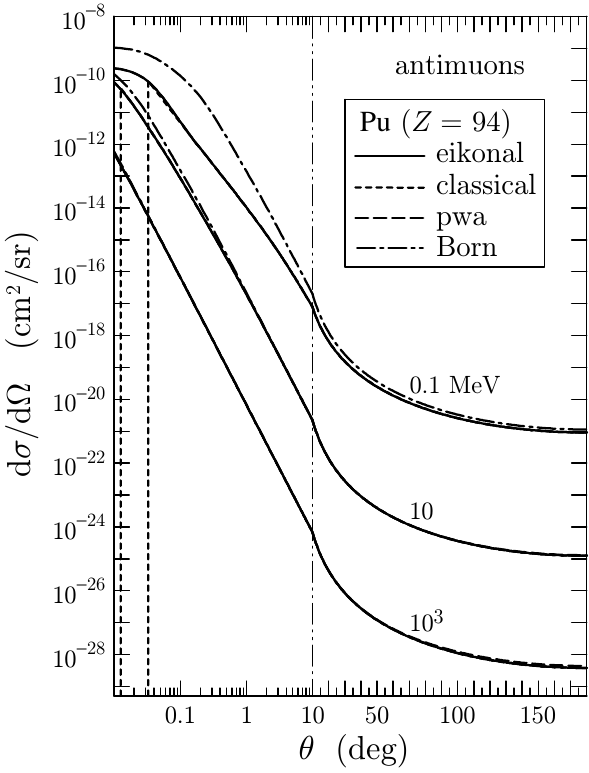}
\caption{
DCSs (in CM) for collisions of muons and antimuons with oxigen and
plutonium atoms, calculated from the approximations indicated in the
legend with the DHFS potential. The labels indicate the kinetic energy
of the projectile (in L). Notice the logarithmic scale for $\theta <
10^\circ$.
\label{fig5.4}}
\end{center} \end{figure}

\begin{figure}[p!] \begin{center}
\includegraphics*[width=7.5 cm]{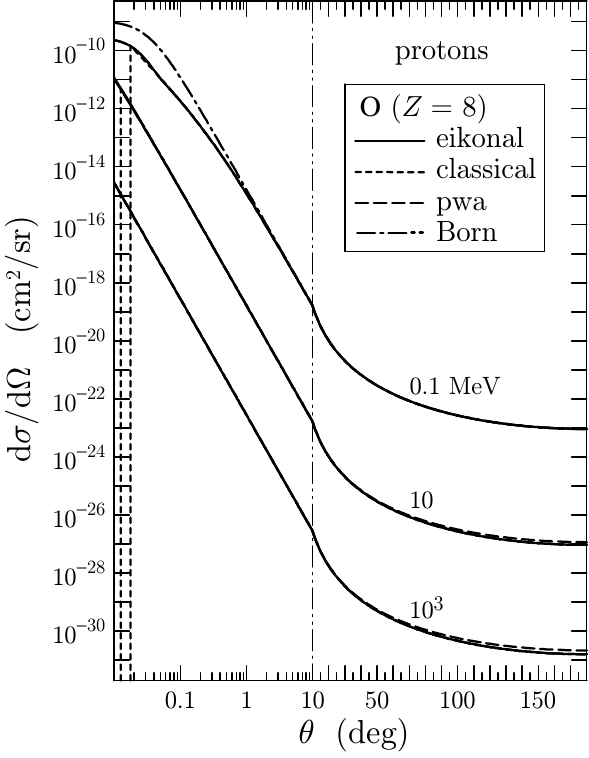} \rule{3mm}{0mm}
\includegraphics*[width=7.5 cm]{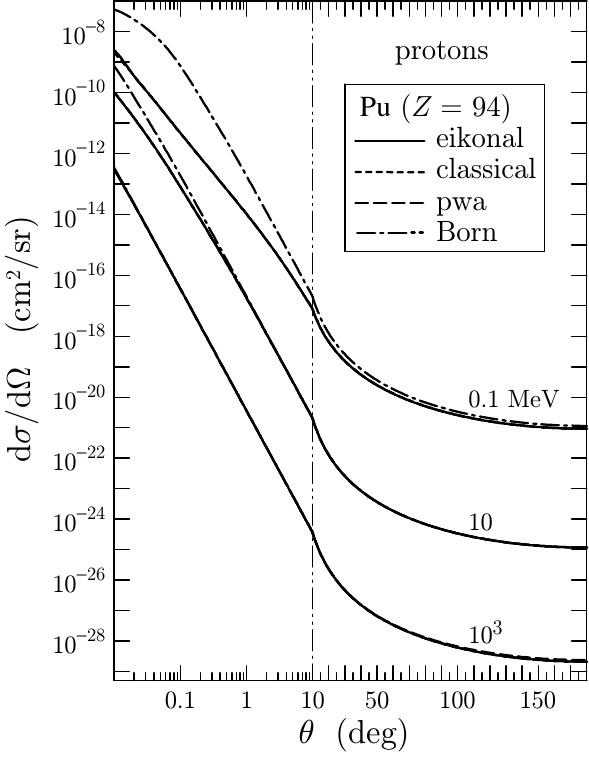} \\ [5mm]
\includegraphics*[width=7.5 cm]{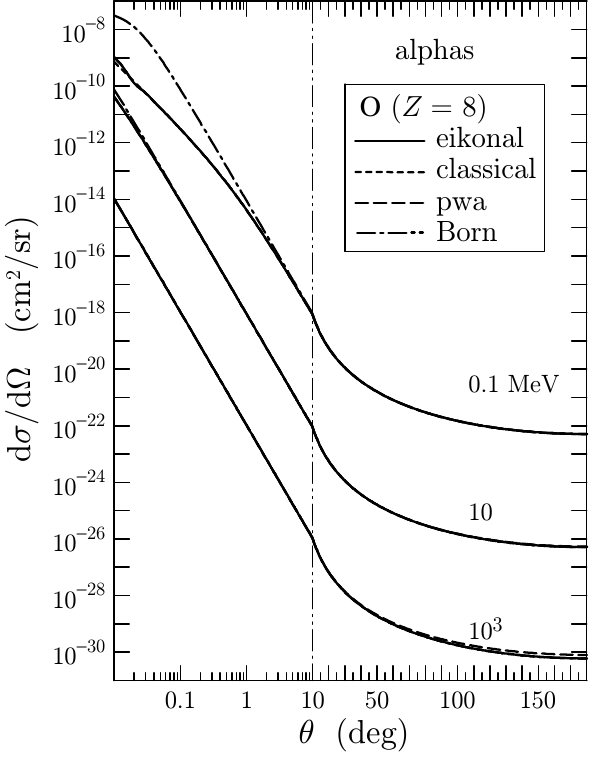} \rule{3mm}{0mm}
\includegraphics*[width=7.5 cm]{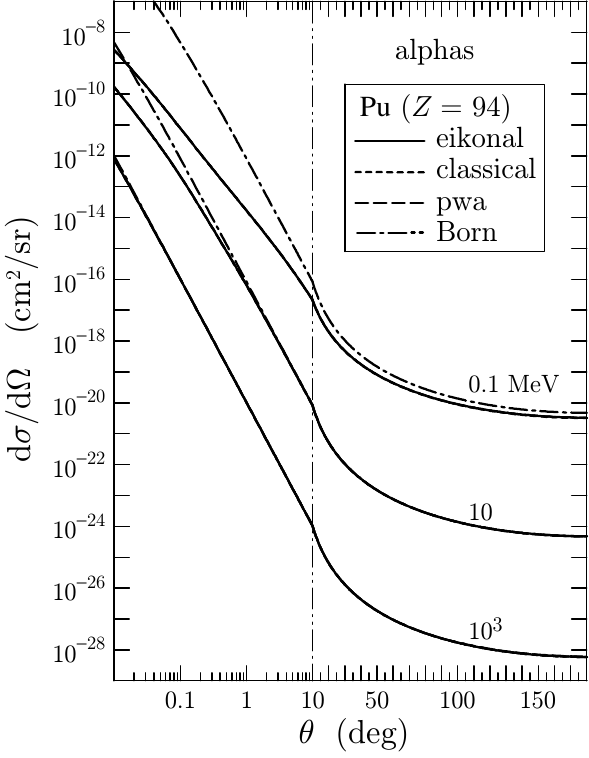}
\caption{
DCSs (in CM) for collisions of protons and alphas with oxigen and
plutonium atoms, calculated from the approximations indicated in the
legend with the DHFS potential. The labels indicate the kinetic energy
of the projectile (in L). Notice the logarithmic scale for $\theta <
10^\circ$.
\label{fig5.5}}
\end{center}\end{figure}

\index{quantum scattering!DCS}
The eikonal approximation is expected to be accurate for scattering
angles less than $\theta_{\rm eik}$, Eq.\ \req{5.107}; for larger
angles, the extrapolation \req{5.108} is used. The partial-wave
scattering amplitude and DCS are found to agree closely with the results
of the eikonal approximation when the first does not face convergence
problems and the latter is accurate, \ie, for sufficiently small
scattering angles. When the partial-wave series do not converge for
scattering angles less than 1 degree, the program replaces the
partial-wave scattering amplitude and the DCS with the eikonal results.

The classical DCS is expected to be valid only for angles such that the
condition \req{4.73} is satisfied. The program {\sc elastic} computes
the quantity
\beq
T_{\rm class} = \frac{\sqrt{\chi}}{\theta}
= \frac{1}{\theta} \sqrt{
\hbar \left| \frac{\d \vartheta}{\d L} \right| }
\label{5.109}\eeq
and it delivers the calculated value of the classical DCS only for
angles larger than a certain value $\theta_{\rm clas}$ for which
either the condition $T_{\rm class} \le 0.8$ holds or the classical DCS
differs by less than 10 \% from the partial-wave (or eikonal) DCS;
otherwise the classical DCS is set to zero.

A striking feature of the numerical results is the close agreement between the
DCSs calculated from the eikonal approximation and from the classical
trajectory method when the ranges of validity of the two approaches
overlap, \ie, when $T_{\rm class} \lesssim 0.8$ for angles $\theta <
\theta_{\rm eik}$. Typically the eikonal and the classical DCSs agree to
three or more digits when the two approximations are valid.
Unfortunately, with the present calculation scheme, it is not
practicable to use the eikonal method with the full effective potential
given by Eq.\ \req{5.5}.

It is worth noticing that the correctness of the Born approximation
improves when the energy increases, as expected. Even when the energy of
the projectile is not high enough to ensure validity of the Born
approximation, the Born DCS does approximate the (extrapolated) eikonal
and classical results at large angles, where the effect of screening is
small (because the Born DCS coincides with the exact quantum result for
unscreened Coulomb potentials).

Figure \ref{fig5.5} displays results from similar calculations for
collisions of protons ($M_1= 1,836 \, \me$, $Z_1=+1$) and alphas ($M_1=
7,294 \, \me$, $Z_1=+2$), again with oxygen and plutonium atoms. The
comparison confirms the general trends observed for muons.


\section{Collisions of electrons and positron with atoms \label{sec5.2}}
\index{quantum scattering!electrons and positrons}

Electrons and positrons (mass $M_1=\me$ and charge $Z_1 e$, $Z_1=\pm 1$)
are peculiar in that they have spin $\1o2$ and their mass is much smaller
than those of other charged particles. The relativistic wave equation of
the electron (and the positron, as well as the muon and its
antiparticle) in an external potential is the Dirac equation (Section
\ref{sec2.2}). Although elastic scattering of electrons can be
accurately calculated by means of numerical Dirac partial-wave analysis
using available computer codes \citep[see, \eg,][] {Salvat2005, ICRU77},
the process provides a unique benchmark to assess the accuracy of the
Born and WKB approximations for the phase shifts. We summarize here the
relevant results and approximations of the theory of scattering of
particles obeying the Dirac equation in central potentials. Because of
the small mass of the electron, we disregard the recoil of the target
atom and perform all calculations in the L frame.

The scattering of electrons (and other Dirac
particles) with linear momentum $p$ by a potential $V(r)$ is
completely described by the direct scattering amplitude, $f(\theta)$,
and the spin-flip scattering amplitude, $g(\theta)$. These are complex
functions of the polar scattering angle $\theta$ determined from the
large-$r$ behavior of the Dirac distorted plane waves, \ie, the
solutions of the Dirac equation for the central potential $V(r)$ that
behave asymptotically as a plane wave plus an outgoing spherical wave
(see Section \ref{sec2.2.2}).
The scattering amplitudes admit the partial-wave expansions \req{2.119},
\begin{subequations}
\label{5.110}
\beq
f^{\rm (D)}(\theta) = \frac{1}{2 {\rm i} k} \sum_{\ell=0}^\infty \left\{
(\ell+1) \left[ \exp\left( 2 {\rm i} \delta_{\kappa=-\ell-1} \right) -
1 \right]
+ \ell \left[ \exp \left( 2 {\rm i} \delta_{\kappa=\ell} \right) - 1
\right]
\right\} \, P_\ell(\cos\theta)
\label{5.110a}\eeq
and
\beq
g^{\rm (D)}(\theta) = \frac{1}{2 {\rm i} k}
\sum_{\ell=0}^\infty \left[
\exp\left( 2 {\rm i} \delta_{\kappa=-\ell-1} \right)
- \exp \left( 2 {\rm i} \delta_{\kappa=\ell}\right)
\right] \, P_\ell^1(\cos\theta),
\label{5.110b}\eeq
\end{subequations}
where $k=p/\hbar$ is the wave number of the projectile, $\delta_\kappa$
are phase shifts, and $P_\ell(\cos\theta)$ and
$P_\ell^1(\cos\theta)$ are Legendre polynomials and associated Legendre
functions, respectively. The DCS for elastic collisions of
spin-unpolarized projectiles is given by
\citep{MottMassey1965, Walker1971}
\beq
\frac{\d \sigma^{\rm (D)}}{\d \Omega} =
|f^{\rm (D)}(\theta)|^2  + |g^{\rm (D)}(\theta)|^2.
\label{5.111}\eeq

The phase shifts $\delta_\kappa$ are determined by the large-$r$
behavior of the radial wave functions $P_{E \kappa}(r)$ and $Q_{E
\kappa}(r)$, which satisfy the radial Dirac equations \citep{Rose1961,
Salvat1995}
\beqa
\frac{\d P_{E \kappa}}{\d r} &=&
-\frac{\kappa}{r} P_{E \kappa}
+\frac{E-V+2 M_1 c^2}{c\hbar} \, Q_{E \kappa},
\nonumber \\ [4mm]
\frac{\d Q_{E\kappa}}{\d r} &=&
- \frac{E-V}{c\hbar} \, P_{E\kappa}
+ \frac{\kappa}{r} Q_{E\kappa}.
\label{5.112}\eeqa
For modified Coulomb potentials $V(r)$ and sufficiently large $r$,
the upper-component radial function $P_{E \kappa}(r)$ oscillates
with constant amplitude, that is,
\beq
P_{E \kappa} (r)
\begin{array}[t]{c}
\sim \\ [-3mm] \scriptstyle{ r \rightarrow \infty}
\end{array}
\sin \left[ kr - \ell \frac{\pi}{2} - \eta \ln (2kr) + \delta_\kappa
\right],
\label{5.113}\eeq
where $\delta_\kappa$ is the phase shift and \index{Sommerfeld parameter}
\beq
\eta=\frac{Z_\infty e^2}{\hbar v}= \frac{Z_\infty \alpha}{\beta}
\label{5.114}\eeq
is the Sommerfeld parameter, Eq.\ \req{2.106}.

It is worth mentioning that the theory summarized here applies to both
electrons and positrons (as well as to muons and antimuons); the only
difference between the particles and their antiparticles is in the sign
of their charges, \ie, the electrostatic interaction with atoms and
positive ions is attractive for electrons and repulsive for positrons.
As in the case of spinless particles, attractive (repulsive) potentials
give positive (negative) phase shifts.

\index{Mott DCS}
\index{Mott DCS!McKinley-Fesbach formula}

The DCS for scattering of electrons and positrons by the Coulomb
potential of a point particle with charge $Ze$, fixed at the origin of
coordinates,
\beq
V(r)= Z_1 Z e^2 /r
\label{5.115}\eeq
is of importance in stopping theory (see Section \ref{sec8.1}). Although
the Dirac phase shifts for this potential are given by known analytical
expressions \citep[see, \eg][]{Rose1961, Salvat1995}, the direct and spin-flip
scattering amplitudes can only be obtained by summing their partial-wave
expansions. The corresponding DCS is known as the Mott DCS
\citep{Mott1929}.
\citet{McKinleyFeshbach1948} have expanded the Mott DCS in powers of
$Z_1 Z e^2$; the result to third order in $Z_1 Z e^2$ is
\beqa
\frac{\d \sigma}{\d \Omega} &=& \left(
\frac{ Z_1Z e^2}{2 v p} \right)^2 \,
\frac{1}{\sin^4(\theta/2)}
\nonumber \\ [2mm]
&& \mbox{} \times \left\{ 1 - \beta^2 \sin^2(\theta/2)
- \pi Z_1 Z \alpha \beta \sin(\theta/2) \,
\left[ 1 - \sin(\theta/2) \right] \rule{0mm}{4mm}\right\},
\label{5.116}\eeqa
where $p=\gamma \beta \, M_1 c$ is the momentum of the projectile and
$\alpha$ is the fine-structure constant. The product of the first two
factors equals the DCS obtained from the relativistic plane-wave Born
approximation [the Rutherford DCS, Eq.\ \req{5.57}]; the second and
third terms in the curly braces arise from the spin of the electron.


\subsection{The Born approximation
\label{sec5.2.1}}

The simplest approach for computing scattering DCSs of Dirac
particles is provided by the plane-wave Born approximation that, as in the
non-relativistic formulation, consists of approximating the wave
functions of the projectile in the initial and final states by Dirac
plane waves and treating the interaction potential as a first-order
perturbation. This Dirac-Born approximation yields the following
expressions of the scattering amplitudes for scattering by a central
potential $V(r)$ \citep{Parzen1950, ICRU77},
\begin{subequations}
\label{5.117}
\beqa
&& f^{\rm (DB)} (\theta) =
\left( \frac{\gamma+1}{2} + \frac{\gamma-1}{2} \cos\theta \right)
f^{\rm (B)} (\theta),
\label{5.117a} \\ [2mm]
&& g^{\rm (DB)} (\theta) =
\frac{\gamma-1}{2} \sin\theta \,
f^{\rm (B)} (\theta),
\label{5.117b}\eeqa
\end{subequations}
where
\beq
\gamma = \frac{E+ M_1c^2}{M_1c^2}
= \sqrt{1 + \left( \frac{p}{M_1c} \right)^2},
\label{5.118}\eeq
and $f^{\rm (B)} (\theta)$ is the
Schr\"{o}dinger-Born scattering amplitude, Eq.\ \req{5.46b},
\beq
f^{\rm (B)}(\theta)
= - \frac{M_1}{2\pi \hbar^2} \,
\int \exp({\rm i} {\bf q} \dotprod {\bf r})
V(r) \, \d {\bf r} .
\nonumber \eeq
The Dirac-Born DCS for spin-unpolarized projectiles is
\beq
\frac{\d \sigma^{\rm (DB)}}{\d \Omega} =
|f^{\rm (DB)}(\theta)|^2  + |g^{\rm (DB)}(\theta)|^2
= \left[ 1-\beta^2 \sin^2(\theta/2)\right] \gamma^2 \,
\left| f^{\rm (B)} (\theta) \right|^2,
\label{5.119}\eeq
with $\beta^2 = 1 - \gamma^{-2}$. The factor $1-\beta^2
\sin^2(\theta/2)$ accounts for the effect of spin, which is mostly
caused by the spin-orbit interaction \citep[see,
\eg][]{BransdenJoachain1983}, and the factor $\gamma^2$ accounts for the
relativistic increase of the projectile mass.

The Dirac-Born scattering amplitudes \req{5.117} can be expressed in
the form of Legendre series. Inserting the expansion \req{5.58} of
$f^{\rm (B)} (\theta)$, and using the following properties of the
Legendre functions,
\begin{subequations} \label{5.120}
\beq
(2\ell +1) \cos\theta \, P_\ell(\cos\theta) =
(\ell+1) P_{\ell +1} (\cos\theta) + \ell P_{\ell -1} (\cos\theta)
\label{5.120a}\eeq
and
\beq
(2\ell +1) \sin\theta \, P_\ell(\cos\theta) =
P_{\ell -1}^1 (\cos\theta) - P_{\ell +1}^1 (\cos\theta),
\label{5.120b}\eeq
\end{subequations}
we obtain
\begin{subequations}
\label{5.121}
\beqa
f^{\rm (DB)}(\theta) & = & \frac{1}{k}
\sum_{\ell=0}^{\infty}
\left[(\ell+1)
\delta_{\kappa=-\ell-1}^{\rm (DB)}
+ \ell \delta_{\kappa=\ell}^{\rm (DB)} \right]
P_{\ell}(\cos\theta),
\label{5.121a}\\[2mm]
g^{\rm (DB)}(\theta) & = & \frac{1}{k}
\sum_{\ell=0}^{\infty}
\left[\delta_{\kappa=-\ell-1}^{\rm (DB)}
-  \delta_{\kappa=\ell}^{\rm (DB)} \right]
P_{\ell}^{1}(\cos\theta),
\label{5.121b}\eeqa
\end{subequations}
with \index{Born phase shifts}
\begin{subequations}
\label{5.122}
\beqa
\delta^{\rm (DB)}_{\kappa = -\ell-1} &=&
\frac{\gamma+1}{2} \delta^{\rm (B)}_\ell +
\frac{\gamma-1}{2} \delta^{\rm (B)}_{\ell+1} \, ,
\label{5.122a} \\ [2mm]
\delta^{\rm (DB)}_{\kappa = \ell} &=&
\frac{\gamma+1}{2} \delta^{\rm (B)}_\ell +
\frac{\gamma-1}{2} \delta^{\rm (B)}_{\ell-1} \, .
\label{5.122b}\eeqa
\end{subequations}
Comparison of Eqs.\ \req{5.121} with the expansions \req{5.110} of the
exact Dirac scattering amplitudes shows that the quantities
$\delta_\kappa^{\rm (DB)}$ can be considered as the phase-shifts
resulting from the Dirac-Born approximation.
\index{quantum scattering!Born approximation}


\subsection{Approximate Dirac phase shifts and scattering amplitudes
\label{sec5.2.2}}

The Dirac-Born phase shifts for potentials of the form \req{5.52} can be
readily evaluated with the aid of the analytical formula \req{5.61}
for the non-relativistic phase shifts. The values so obtained provide a
good approximation to the actual phase shifts of large orders that are
small in magnitude, even when the Born approximation for the
scattering amplitudes is not accurate (see below).

To estimate the phase shifts of small orders, which usually have  larger
absolute values, we take advantage of the fact that the radial Dirac
Eqs.\ \req{5.112} may be reduced to Schr\"{o}dinger form by introducing
the substitution \citep{MottMassey1965, Walker1971}
\beq
P_{E \kappa}(r) = A^{1/2}(r) {\cal P}(r)
\label{5.123}\eeq
with
\beq
A(r)\equiv \frac{E -V(r)+2M_1 c^2}{2 M_1 c^2},
\label{5.124}\eeq
and eliminating the small-component radial function $Q_{E
\kappa}(r)$. The resulting equation is
\beq
\left[ - \frac{\hbar^2}{2M_1} \, \frac{\d^2}{\d r^2} + V_{\rm ef}^{\rm
(D)} (r) +
\frac{\hbar^2}{2M_1} \, \frac{\ell (\ell+1)}{r^2} \right]
{\cal P}(r) = \frac{p^2}{2M_1} \, {\cal P}(r),
\label{5.125}\eeq
where the effective Dirac potential,
\beq
V_{\rm ef}^{\rm (D)} (r)
= V + \frac{1}{2 M_1 c^2} \left\{ V \left(2 E - V \right)
+ (\hbar c)^2
\left[ \frac{\kappa}{r}\frac{A'}{A} +
\frac{3}{4}\left(\frac{A'}{A}\right)^{2}
- \frac{1}{2}\frac{A''}{A} \right] \right\}
\label{5.126}\eeq
depends on the energy and the relativistic quantum number $\kappa$. For
large $r$ values, $A(r)$ becomes a constant, \ie, ${\cal P}$ becomes
proportional to $P_{E \kappa}$, and therefore the phase shifts may be
computed by solving the radial Schr\"{o}dinger equation \req{5.125} as
in the non-relativistic case. In particular, the WKB approximation with
the Langer correction yields [cf.\ Eq.\ \req{5.66}]
\index{quantum scattering!WKB phase shifts}\index{WKB phase shifts}
\beq
\delta_{\kappa}^{\rm (WKB)} =
\frac{1}{2} \left( \ell + \frac{1}{2} \right) \pi - kr_{0} +
\int_{r_{0}}^{\infty} \left[ \sqrt{F_{\kappa}(r)} - k \right] \, \d r,
\label{5.127}\eeq
where
\beq
F_{\kappa}(r) = k^{2}  -
\frac{2M_1}{\hbar^2} V_{\rm ef}^{\rm (D)}(r)
- \frac{(\ell+1/2)^{2}}{r^{2}}
\label{5.128}\eeq
and $r_{0}$ is the largest zero of $F_{\kappa}(r)$. In the program {\sc
elastic} the integrals \req{5.127} are calculated by means of the
adaptive Gauss--Legendre quadrature algorithm (Section
\ref{sec10.4.3.2}) to a relative accuracy of about $10^{-10}$.

\begin{table}[htb!]
\caption{\rm
Dirac phase shifts for scattering of 10 keV electrons by the DHFS
potential of gold atoms ($Z=79$).
\label{tab5.2}}
\vskip 3mm
\begin{center}
{\small\tt
\begin{tabular}{|r|ccc|ccc|}
\cline{2-7}
\multicolumn{1}{c}{\rule[-2mm]{0mm}{2mm}} &
\multicolumn{3}{|c|}{$\delta_{\kappa = -\ell-1}${\rm , spin up}} &
\multicolumn{3}{|c|}{$\delta_{\kappa = \ell}${\rm , spin down}}
\\ \hline
\multicolumn{1}{|c|}{$\ell$} &
\multicolumn{1}{|c}{\rm numerical} &
\multicolumn{1}{c}{\rm WKB} &
\multicolumn{1}{c|}{\rm Born} &
\multicolumn{1}{|c}{\rm numerical} &
\multicolumn{1}{c}{\rm WKB} &
      \multicolumn{1}{c|}{\rm Born}
\\ \hline
  0 & 6.74461E+0 & 6.79607E+0 & 7.18701E+0 & 0.00000E+0 & 0.00000E+0 & 0.00000E+0 \\
  1 & 5.04483E+0 & 5.06634E+0 & 4.51827E+0 & 5.33848E+0 & 5.33310E+0 & 4.55548E+0 \\
  2 & 3.83119E+0 & 3.84728E+0 & 3.32836E+0 & 3.90575E+0 & 3.91894E+0 & 3.34672E+0 \\
  3 & 2.97756E+0 & 2.98732E+0 & 2.61153E+0 & 3.00878E+0 & 3.01778E+0 & 2.62316E+0 \\
  4 & 2.38002E+0 & 2.38572E+0 & 2.11978E+0 & 2.39692E+0 & 2.40231E+0 & 2.12799E+0 \\
  5 & 1.94527E+0 & 1.94892E+0 & 1.75743E+0 & 1.95594E+0 & 1.95947E+0 & 1.76360E+0 \\
  6 & 1.61624E+0 & 1.61878E+0 & 1.47846E+0 & 1.62366E+0 & 1.62612E+0 & 1.48327E+0 \\
  7 & 1.35972E+0 & 1.36153E+0 & 1.25738E+0 & 1.36519E+0 & 1.36695E+0 & 1.26122E+0 \\
  8 & 1.15536E+0 & 1.15666E+0 & 1.07856E+0 & 1.15954E+0 & 1.16081E+0 & 1.08169E+0 \\
  9 & 9.89853E-1 & 9.90785E-1 & 9.31739E-1 & 9.93147E-1 & 9.94062E-1 & 9.34323E-1 \\
 10 & 8.54074E-1 & 8.54741E-1 & 8.09765E-1 & 8.56721E-1 & 8.57375E-1 & 8.11921E-1 \\
 15 & 4.42821E-1 & 4.42921E-1 & 4.30250E-1 & 4.43892E-1 & 4.43989E-1 & 4.31222E-1 \\
 20 & 2.53017E-1 & 2.53003E-1 & 2.48932E-1 & 2.53536E-1 & 2.53521E-1 & 2.49424E-1 \\
 25 & 1.54721E-1 & 1.54691E-1 & 1.53232E-1 & 1.55000E-1 & 1.54970E-1 & 1.53503E-1 \\
 30 & 9.94639E-2 & 9.94391E-2 & 9.88659E-2 & 9.96262E-2 & 9.96010E-2 & 9.90253E-2 \\
 35 & 6.63643E-2 & 6.63466E-2 & 6.61036E-2 & 6.64641E-2 & 6.64462E-2 & 6.62023E-2 \\
 40 & 4.55147E-2 & 4.55026E-2 & 4.53933E-2 & 4.55789E-2 & 4.55668E-2 & 4.54570E-2 \\
 50 & 2.26257E-2 & 2.26199E-2 & 2.25954E-2 & 2.26549E-2 & 2.26491E-2 & 2.26245E-2 \\
 75 & 4.60611E-3 &   ------   & 4.60471E-3 & 4.61155E-3 &   ------   & 4.61014E-3 \\
100 & 1.02005E-3 &   ------   & 1.01997E-3 & 1.02121E-3 &   ------   & 1.02114E-3 \\
150 & 5.45922E-5 &   ------   & 5.45919E-5 & 5.46529E-5 &   ------   & 5.46526E-5 \\
200 & 3.09018E-6 &   ------   & 3.09019E-6 & 3.09357E-6 &   ------   & 3.09358E-6 \\
250 & 1.80387E-7 &   ------   & 1.80388E-7 & 1.80583E-7 &   ------   & 1.80585E-7 \\
300 & 1.07380E-8 &   ------   & 1.07390E-8 & 1.07497E-8 &   ------   & 1.07506E-8 \\
\hline
\end{tabular}} \end{center} \end{table}

The accuracy of WKB and Born approximate phase shifts can be readily
assessed by comparison with numerical phase shifts calculated by the
code system {\sc radial} \citep{SalvatFernandezVarea2019}. Table
\ref{tab5.2} shows such a comparison for 10 keV electrons scattered by
gold atoms, with all phase shifts calculated with the analytical DHFS potential.
The general trends  are the same as for the Schr\"{o}dinger equation
(Section \ref{sec5.1.3}). The WKB phase shifts are in fairly good
agreement with the numerical phases; the relative differences are about
1 \% for $\ell =0$ and decrease
when the order $\ell$ increases. The Born phase shifts are less accurate
than the WKB phases for low $\ell$, but they tend progressively to the
numerical values when $\ell$ increases. Because the situation is similar
to the one found for heavier particles (Section \ref{sec5.1.3}), we use
the recipe given in Eqs.\ \req{5.70} to define the approximate phase
shifts. Thus, for the set of phase shifts with $\kappa < 0$, we define
\begin{subequations}
\label{5.129}
\beq
\delta_{\kappa = - \ell -1} = \left\{
\begin{array}{ll}
\delta^{\rm (WKB)}_{-\ell -1} \rule{5mm}{0mm} &
\mbox{if $\ell < L$,} \\ [2mm]
C_{-\ell-1} \delta^{\rm (DB)}_{-\ell -1} & \mbox{otherwise,}
\end{array} \right.
\label{5.129a}\eeq
where the cutoff value $L$ is the lowest value of $\ell$ for which
either $\delta_{-\ell -1}^{\rm (B)} < 0.001$ or the relative
difference between the WKB and Born phase shifts is less than 0.001. The
factor
\beq
C_{-\ell-1} =  1+ \left( \frac{\delta^{\rm (WKB)}_{-L-1}}{\delta^{\rm
(DB)}_{-L-1}}
- 1 \right) \, \exp \left( -a \, \frac{\ell - L}{L} \right),
\label{5.129b}\eeq
\end{subequations}
with the parameter $a$ determined so that the phase shifts
$\delta_{\kappa=-\ell-1}$ vary smoothly with $\ell$, as described in Section
\ref{sec5.1.3} after Eq.\ \req{5.70}. Phase shifts with positive
$\kappa$ are defined similarly.

The DCS and the scattering amplitudes for high-energy projectiles are
sharply peaked at $\theta =0$ and, consequently, the convergence of the
partial-wave series is quite slow. It can be sped up by adding the
Born scattering amplitudes, Eqs.\ \req{5.117} and subtracting their
partial-wave expansions, Eqs.\ \req{5.121}. We thus obtain
\begin{subequations}
\label{5.130}
\beq
f^{\rm (D)}(\theta) = f^{\rm (DB)}(\theta)
+ \sum_{\ell=0}^\infty {\cal F}_\ell  \, P_\ell(\cos\theta)
\label{5.130a}\eeq
and
\beq
g^{\rm (D)}(\theta) = g^{\rm (DB)}(\theta) +
\sum_{\ell=0}^\infty {\cal G}_\ell \, P_\ell^1(\cos\theta),
\label{5.130b}\eeq
\end{subequations}
where
\begin{subequations}
\label{5.131}
\beqa
{\cal F}_\ell &=&
\frac{1}{2 {\rm i} k} \left\{
(\ell+1) \left[ \exp\left( 2 {\rm i} \delta_{\kappa=-\ell-1} \right) -
1 \right]
+ \ell \left[ \exp \left( 2 {\rm i} \delta_{\kappa=\ell} \right) - 1
\right] \rule{0mm}{4mm}\right\}
\nonumber \\ [0mm]
&& \mbox{} - \frac{1}{k} \left[(\ell+1)
\delta_{\kappa=-\ell-1}^{\rm (DB)}
+ \ell \delta_{\kappa=\ell}^{\rm (DB)} \right]
\label{5.131a}\eeqa
and
\beqa
{\cal G}_\ell &=& \frac{1}{2 {\rm i} k}
\left[
\exp\left( 2 {\rm i} \delta_{\kappa=-\ell-1} \right)
- \exp \left( 2 {\rm i} \delta_{\kappa=\ell}\right)
\rule{0mm}{4mm}\right]
-  \frac{1}{k}
\left[ \delta_{\kappa=-\ell-1}^{\rm (DB)}
- \delta_{\kappa=\ell}^{\rm (DB)}
\right].
\label{5.131b}\eeqa
\end{subequations}


\subsection{The reduced series method  \label{sec5.2.3}}
\index{reduced series method}

At least for scattering angles that are not too small, the convergence
of the partial-wave series can be further accelerated by applying the
``reduced series'' method described by \citet{Yennie1954}, which is
based on the fact that, if a function $f(\theta)$ is strongly peaked at
$\theta=0$, the function $(1-\cos\theta)f(\theta)$ is smoother than
$f(\theta)$, and hence its Legendre expansion may be more rapidly
convergent. Explicitly, to sum up a series of the form
\beq
f(\theta) =
\sum_{\ell=0}^{\infty} {\cal F}_{\ell}^{0} P_{\ell}(\cos\theta)
\label{5.132}\eeq
that is weakly convergent (\ie, ${\cal F}_{\ell}^{0}$
decreases slowly with $\ell$ for large $\ell$), we
consider the transformed series
\beq
\hat{f}(\theta) = (1-\cos\theta)^{n} f(\theta) =
\sum_{\ell=0}^{\infty} {\cal F}_{\ell}^{n} P_{\ell}(\cos\theta)
\label{5.133}\eeq
and compute $f(\theta)$ as $(1-\cos\theta)^{-n}\hat{f}(\theta)$. The
coefficients ${\cal F}_{\ell}^{n}$ are given by
\beq
{\cal F}_{\ell}^{n} = {\cal F}_{\ell}^{n-1} - \frac{\ell+1}{2\ell+3}
{\cal F}_{\ell+1}^{n-1} - \frac{\ell}{2\ell-1} {\cal F}_{\ell-1}^{n-1}
\label{5.134}\eeq
and, for large $\ell$, they decrease more rapidly with $\ell$ than the
coefficients ${\cal F}_{\ell}^{0}$ of the original
series. It can be shown that if
${\cal F}_{\ell}^{n}\approx{\cal O}(\ell^{b})$,
then ${\cal F}_{\ell}^{n+1}\approx{\cal O}(\ell^{b-2})$. In the case of a
series in associated Legendre functions such as $g^{\rm (D)}(\theta)$,
\beq
g(\theta) =
\sum_{\ell=0}^{\infty} {\cal G}_{\ell}^{0} P_{\ell}^1(\cos\theta),
\label{5.135}\eeq
the transformed series is
\beq
\hat{g}(\theta) = (1-\cos\theta)^{n} g(\theta) =
\sum_{\ell=0}^{\infty} {\cal G}_{\ell}^{n} P_{\ell}^{1} (\cos\theta)
\label{5.136}\eeq
with
{\par\nobreak\noindent\vspace*{-\parskip}}
\beq
{\cal G}_{\ell}^{n} = {\cal G}_{\ell}^{n-1} - \frac{\ell+2}{2\ell+3}
{\cal G}_{\ell+1}^{n-1} - \frac{\ell-1}{2\ell-1}
{\cal G}_{\ell-1}^{n-1}.
\label{5.137}\eeq
For large $\ell$, ${\cal G}_{\ell}^{n}$ decreases more rapidly with
$\ell$ than the coefficients ${\cal G}_{\ell}^{0}$.
Again, if ${\cal G}_{\ell}^{n}\approx{\cal O}(\ell^{b})$, then
${\cal G}_{\ell}^{n+1}\approx{\cal O}(\ell^{b-2})$. Of course, the
reduced-series method cannot be used when $\theta\approx 0$.


\subsection{Calculation examples \label{sec5.2.4}}

Figure \ref{fig5.6} shows DCSs for scattering of electrons and positrons
with the indicated energies by aluminum ($Z=13$) and gold ($Z=79$)
atoms, all calculated by the program {\sc elastic} (see Chapter
\ref{chapt10}) for the analytical DHFS potential, Eq.\
\req{3.149}. The plots include results from accurate numerical
partial-wave calculations performed by the {\sc radial} programs
\citep{SalvatFernandezVarea2019}; the accumulated {\it numerical}
relative error of these DCSs is expected to be less than about
$10^{-4}$. Also displayed are the DCSs obtained from the partial-wave
expansion method using approximate WKB and Born phase shifts (pwa).
These are seen to provide a very good approximation to the numerical
data for projectiles with kinetic energy up to about 10 keV.  When the
energy increases, errors in the WKB phase shifts of lower orders
manifest as systematic deviations from the {\sc radial} results at large
angles. Nonetheless, at angles less than about 30$^\circ$ the
partial-wave calculation with approximate phase shifts is seen to agree
closely with the {\sc radial} results, quite irrespectively of the
energy of the projectile.
\index{quantum scattering!DCS}

\begin{figure}[p!] \begin{center}
\includegraphics*[width=7.0 cm]{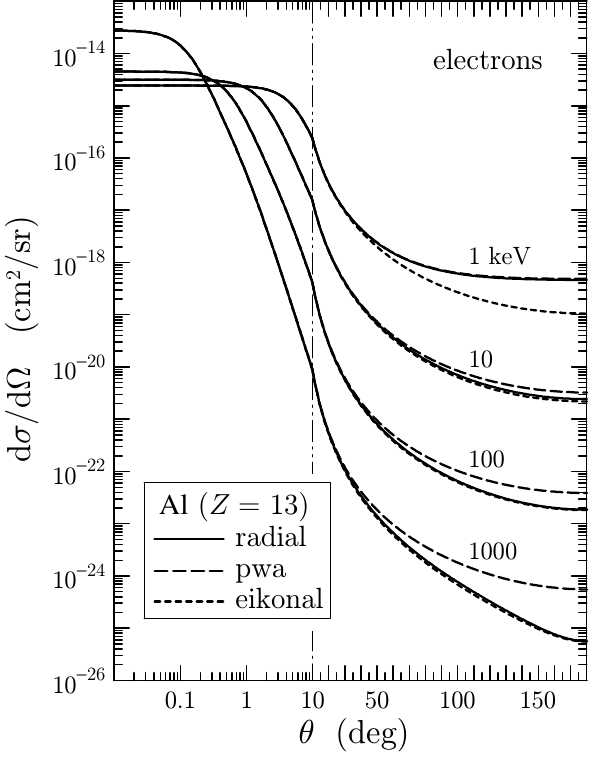} \rule{3mm}{0mm}
\includegraphics*[width=7.0 cm]{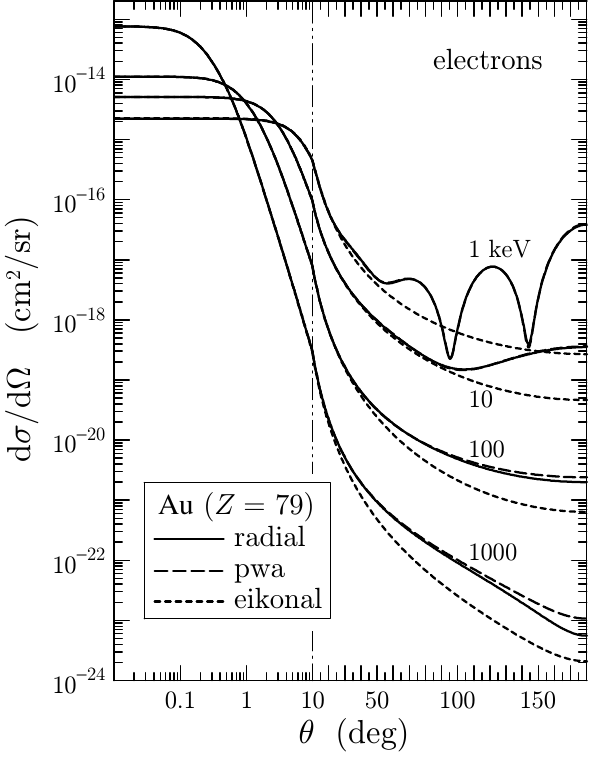} \\ [5mm]
\includegraphics*[width=7.0 cm]{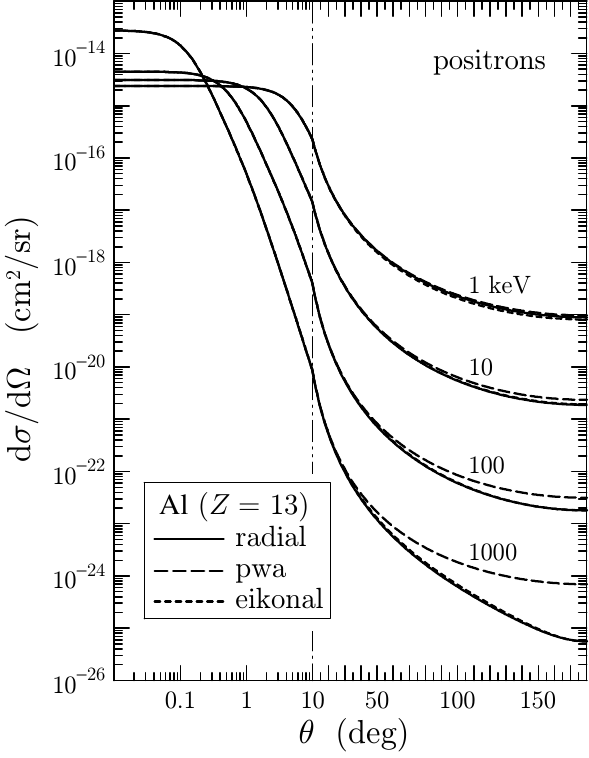} \rule{3mm}{0mm}
\includegraphics*[width=7.0 cm]{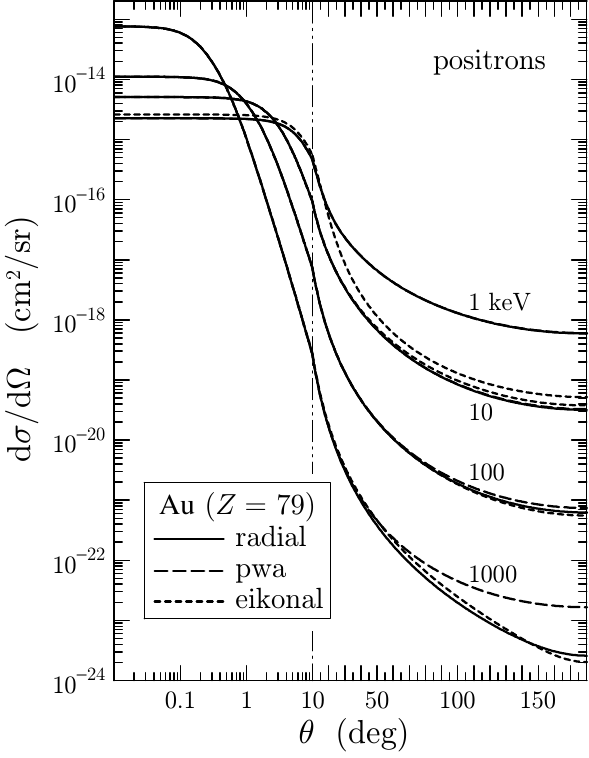}
\caption{
DCS (in the L frame) for elastic collisions of electrons and positrons with
aluminum and gold atoms, calculated
using the analytical DHFS potential. The labels indicate the kinetic
energy of the projectile. Solid curves represent the results
from highly-accurate Dirac partial-wave calculations performed with the
{\sc radial} programs. The dashed curves represent DCSs calculated
with the present partial-wave method with approximate (WKB and Born)
phase shifts. The dotted lines are DCSs obtained from the eikonal
approximation, Eq.\ \req{5.104}.  Notice the logarithmic scale for
$\theta < 10^\circ$.
\label{fig5.6}}
\end{center} \end{figure}

The DCS for spin $\1o2$ projectiles can also be calculated by using the
classical trajectory method (Section \ref{sec4.3.2}) and the eikonal
approximation (Section \ref{sec5.1.4}). These approaches disregard the
effect of the spin of the projectile. To account, at least partially,
for spin effects, the calculated DCSs are multiplied by the factor
$1-\beta^2 \sin^2(\theta/2)$ [see Eq.\ \req{5.118}]. Thus, the eikonal
DCS for Dirac particles is given by
\beqa
\frac{\d \sigma^{\rm (eik)}}{\d \Omega} &=& \left[ 1 - \beta^2
\sin^2(\theta/2) \rule{0mm}{4mm} \right]
\nonumber \\ [2mm]
&& \mbox{} \times
\left\{
\begin{array}{ll}
|f^{\rm (eik)} (\theta)|^2 & \mbox{if $\theta < \theta_{\rm
eik}$,}
\\ [2mm]
\displaystyle{\left(\frac{2 \mu_{\rm r}}{\hbar^2}  \, Z_1 Z e^2\right)^2
\left[ A + B q^{2/3} + C q^{4/3} +
q^2\right]^{-2}} \rule{7mm}{0mm} & \mbox{otherwise.}
\end{array} \right. \rule{10mm}{0mm}
\label{5.138}\eeqa
The close agreement between the eikonal and partial-wave results at
small angles (up to $\sim$30 degrees) is noteworthy. For collisions of
positrons, and of electrons with light-element atoms, the eikonal
approximation is better than acceptable, except for projectile energies
less than $\sim 10$ keV. As in the case of spinless particles, when the
partial-wave series for the scattering amplitude and the DCS do not
converge for angles less than 1 degree, {\sc elastic} replaces them with
the corresponding quantities evaluated from the eikonal approximation.

Incidentally, as the {\sc radial} subroutines
\citep{SalvatFernandezVarea2019} allow the accurate calculation of
electron (and positron) DCSs for any scattering potential, we can use
them to analyze the effect of replacing the numerical DHFS potential
with the analytical approximation \req{3.149}. Figure \ref{fig5.7}
compares DCSs for elastic collisions of electrons with aluminium and
gold atoms calculated by the {\sc radial} subroutines using the
numerical DHFS potential and the potential \req{3.149} with the
parameters given by \citet{Salvat1987}. These results confirm that the
analytical potential is an excellent approximation to the numerical DHFS
potential for computing the scattering of electrons with energies higher
than about 1 keV. A similar degree of agreement is expected for heavier
particles, for which calculation methods with the accuracy of {\sc
radial} are not available.

\begin{figure}[tb!] \begin{center}
\includegraphics*[width=7.5 cm]{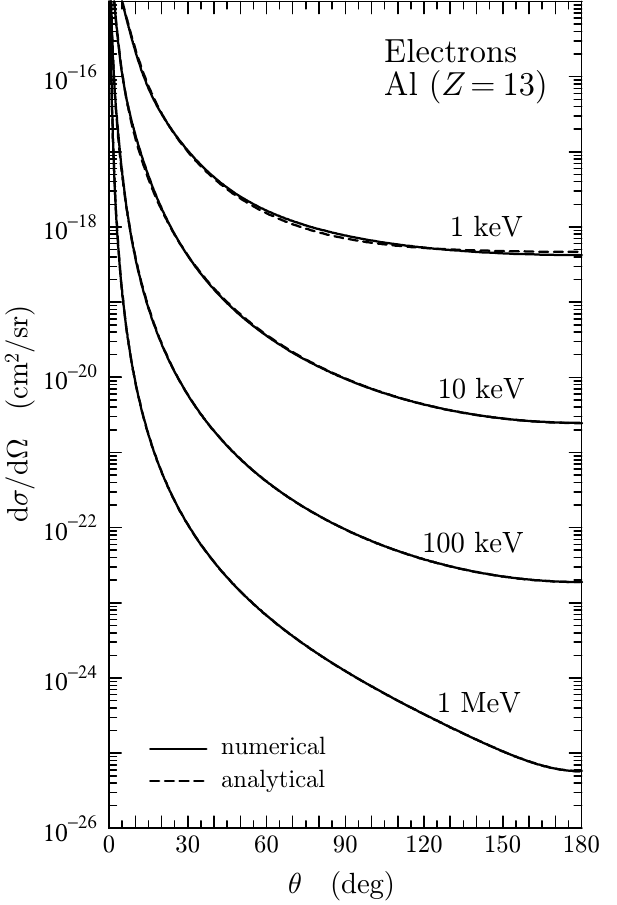} \rule{3mm}{0mm}
\includegraphics*[width=7.5 cm]{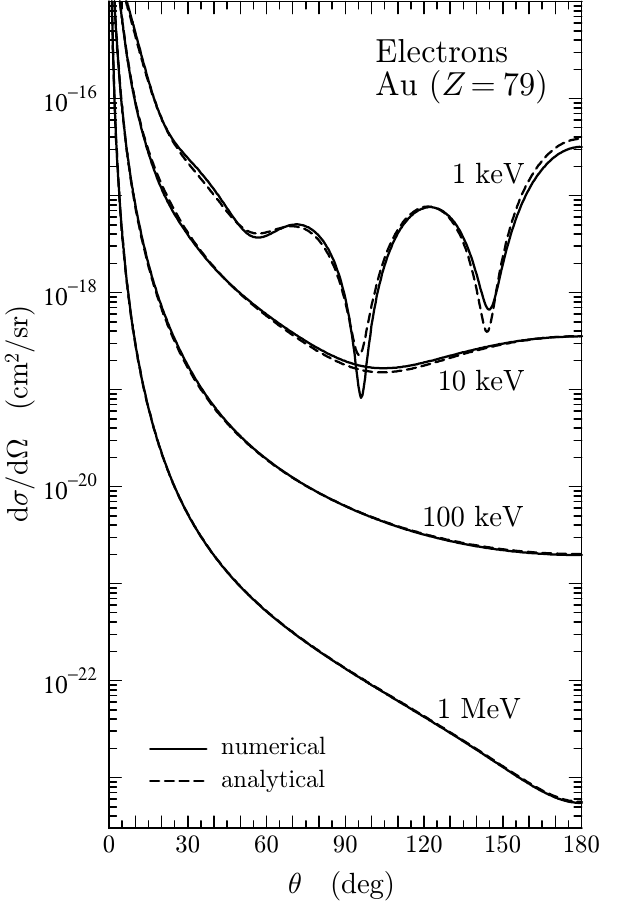}
\caption{
DCS for elastic collisions of electrons with aluminum and gold atoms, calculated by the {\sc radial}
subroutines by using the numerical DHFS potential (solid curves) and its
analytical approximation \req{3.149} with the parameters given by
\citet{Salvat1987} (dashed curves). The labels indicate the kinetic
energy of the projectile.
\label{fig5.7}}
\end{center} \end{figure}


\section{Collisions of identical particles \label{sec5.3}}
\index{collisions of identical particles}

Let us now consider the binary collisions of two identical particles of
mass $M_1$. In the laboratory (L) frame the projectile particle moves
initially with kinetic energy $E_{1{\rm i}}$ and linear momentum ${\bf
p}_{1{\rm i}}$ in the direction of the $z$ axis, and the target particle
is at rest at the origin of the reference frame. As usual, we describe
the collision process in the CM frame where the particles are assumed to
interact through a central force corresponding to a certain potential
$V(r)$, and where the initial linear momentum is [see Eq.\ \req{4.139}]
\beq
{\bf p}'_{\rm i}
= \frac{M_1 c^2 p_{1{\rm i}}}
{\sqrt{2 M_1 c^2 (E_{1{\rm i}}+2M_1 c^2)}} \, \hat{\bf z}
= \sqrt{\frac{M_1}{2} \, E_{1{\rm i}}} \, \hat{\bf z}.
\label{5.139}\eeq
The reduced mass of the two particles [Eq.\ \req{4.154}] is
\beq
\mu_{\rm r} = \frac{1}{2c^2} \sqrt{M_1^2 c^4 + c^2 p'^2_{1{\rm i}}}
= \frac{M_1}{2} \, \frac{E_{1{\rm i}}+M_1 c^2}{M_1 c^2}\, .
\label{5.140}\eeq
After the interaction, the two particles move in opposite directions
and, when they are far from each other, their wave function has
the asymptotic form
\beq
\psi_{\bf k} ({\bf r}) =
(2\pi)^{-3/2} \, \exp({\rm i} {\bf k} \dotprod {\bf r} ) +
(2\pi)^{-3/2} \, \frac{\exp({\rm i} kr )}{r} \, f(\theta),
\label{5.141}\eeq
where ${\bf k} = {\bf p}'_{\rm i}/\hbar$ and $\theta$ is the polar angle
of the direction $\hat{\bf r}$. This wave represents the relative motion
of the two particles. It is important to notice that the exchange
of the two particles is equivalent to inverting the sign of the relative
position vector ${\bf r}$ in the wave function.

In an experiment to measure the collision DCS, a detector covering a
small solid angle $\d \Omega$ about a direction $\hat{\bf r}$ (in CM)
counts {\it all} particles that enter its sensitive volume. That is, the
detector counts both projectiles and recoiling target particles. Hence,
in the evaluation of the scattering DCS, we must account for the
indistinguishability of the colliding particles. If the particles are
bosons their wave function must be symmetrical. If the particles are
spin-$\1o2$ fermions (\eg, electrons, protons, neutrons) its complete
wave function must be antisymmetric. This wave function can be
represented as the direct product of the spatial wave function and a
spin part,
\beq
\psi({\bf r},\sigma_1, \sigma_2) =
\psi({\bf r}) \, \chi_{S,M}(\sigma_1,\sigma_2),
\label{5.142}\eeq
where $\chi_{S,M}(\sigma_1,\sigma_2)$ is an eigenstate of the
total Pauli spin operator, $\breve{\bf S} = \breve {\bf S}_1 + \breve{\bf
S}_2$, that is, an eigenstate of the operators $\breve{S}^2$ and
$\breve{S}_z$ with respective eigenvalues $S(S+1)$ and $M$,
\begin{subequations}
\label{5.143}
\beq
\breve{S}^2 \, \chi_{S,M}(\sigma_1,\sigma_2) = S(S+1)
\, \chi_{S,M}(\sigma_1,\sigma_2)
\label{5.143a}\eeq
and
\beq
\breve{S}_z \, \chi_{S,M}(\sigma_1,\sigma_2) = M
\, \chi_{S,M}(\sigma_1,\sigma_2).
\label{5.143b}\eeq
\end{subequations}
The eigenstates of the total spin are the singlet state
\beq
\chi_{0,0} (\sigma_1,\sigma_2) = \frac{1}{\sqrt{2}}
\left[ \chi_{+1/2}(\sigma_1) \, \chi_{-1/2}(\sigma_2)
- \chi_{-1/2}(\sigma_2) \chi_{+1/2}(\sigma_2) \right],
\label{5.144}\eeq
and the triplet states
\beqa
\chi_{1,1} (\sigma_1,\sigma_2) &=&
\chi_{+1/2}(\sigma_1) \chi_{+1/2}(\sigma_2),
\nonumber \\ [2mm]
\chi_{1,0} (\sigma_1,\sigma_2) &=& \frac{1}{\sqrt{2}}
\left[ \chi_{+1/2}(\sigma_1) \, \chi_{-1/2}(\sigma_2)
+ \chi_{-1/2}(\sigma_2) \chi_{+1/2}(\sigma_2) \right],
\label{5.145}\\ [2mm]
\chi_{1,-1} (\sigma_1,\sigma_2) &=&
\chi_{-1/2}(\sigma_1) \chi_{-1/2}(\sigma_2).
\nonumber \eeqa
Since the singlet state ($S=0$) is antisymmetric under exchange of the
two spins the corresponding spatial wave function must be
symmetric; conversely, as the triplet spin states ($S=1$) are symmetric
the spatial wave function must be antisymmetric.

The symmetric spatial wave function is obtained by symmetrizing the
spatial wave function in \req{5.141},
\beqa
\psi_{S=0} ({\bf r}) &=& \frac{1}{\sqrt{2}} \left[
\psi_{\bf k} ({\bf r}) + \psi_{\bf k} (-{\bf r}) \right]
\nonumber \\ [2mm]
&=&
\frac{(2\pi)^{-3/2}}{\sqrt{2}} \left\{
\exp({\rm i} {\bf k} \dotprod {\bf r})
+ \exp(-{\rm i} {\bf k} \dotprod {\bf r})
+ \frac{\exp({\rm i}kr)}{r} \left[ f(\theta)
+ f(\pi - \theta) \right] \right\}. \rule{17mm}{0mm}
\label{5.146}\eeqa
Similarly, the antisymmetric spatial wave function is
\beqa
\psi_{S=1}({\bf r}) &=& \frac{1}{\sqrt{2}} \left[
\psi_{\bf k} ({\bf r}) - \psi_{\bf k} (-{\bf r}) \right]
\nonumber \\ [2mm]
&=&
\frac{(2\pi)^{-3/2}}{\sqrt{2}} \left\{
\exp({\rm i} {\bf k} \dotprod {\bf r})
- \exp(-{\rm i} {\bf k} \dotprod {\bf r})
+ \frac{\exp({\rm i}kr)}{r} \left[ f(\theta)
- f(\pi - \theta) \right] \right\}. \rule{17mm}{0mm}
\label{5.147}\eeqa
We see that for spin-$\1o2$ fermions the scattering amplitude, which
determines the probability current at the detector, takes the form
\beq
f_{S}(\theta) = f(\theta) + (-1)^S f(\pi-\theta).
\label{5.148}\eeq
For collisions of spinless bosons, the asymptotic behavior of the
(symmetric) spatial wave function is the same as for singlet states of
spin-$\1o2$ fermions. In what follows we consider collisions of
spin-$\1o2$ fermions only.

The DCS in the CM frame is
\beq
\frac{\d\sigma_{S}}{\d\Omega} = \left| f_S(\theta) \right|^2 = \left|
f(\theta) \right|^{2} + \left| f(\pi-\theta) \right|^{2} + (-1)^S 2 {\rm
Re} \left[ f(\theta) f^\ast(\pi-\theta) \right].
\label{5.149}\eeq
The first term on the right-hand side describes projectile particles
that have undergone scattering to angles $(\theta,\phi)$. The second
term corresponds to recoiling target particles from collisions where the
projectile was scattered to the opposite direction ($\pi - \theta, \phi
+ \pi$). The third term accounts for quantum interference effects.
It is worth noticing that the total cross section
\beq
\sigma_S \equiv \int \frac{\d\sigma_{S}}{\d\Omega} \, \d \Omega,
\label{5.150}\eeq
is equal to {\it twice} the number of projectiles that are removed from
the incident beam per unit time and per unit incident current. When the
incident beam is unpolarized, the DCS is obtained by averaging over spin
states,
\beq
\frac{\d\sigma}{\d\Omega} = \frac{1}{4} \frac{\d\sigma_{S=0}}{\d\Omega}
+ \frac{3}{4} \frac{\d\sigma_{S=1}}{\d\Omega} = \left| f(\theta)
\right|^{2} + \left| f(\pi-\theta) \right|^{2} - {\rm Re} \left[
f(\theta) f^\ast(\pi-\theta) \right].
\label{5.151}\eeq

It is useful to express the DCS in the L frame in terms of the energy
transfer $W = E_{1{\rm i}} - E_{1{\rm f}}$, which is related to the
scattering angle in CM by [see Eqs.\ \req{4.189} and \req{4.190}]
\beq
W = W_{\rm max} \sin^2 (\theta/2),
\label{5.152}\eeq
with
\beq
W_{\rm max} = \frac{2 M_1 c^4 p_{1{\rm i}}^2}{2 M_1 c^2 (E_{1{\rm i}} +
2 M_1 c^2)} = E_{1{\rm i}}.
\label{5.153}\eeq
The energy-loss DCS is [see Eq.\ \req{4.203}]
\beq
\frac{\d\sigma}{\d W} =
\frac{4\pi}{W_{\rm max}} \, \frac{\d\sigma}{\d\Omega} \, .
\label{5.154}\eeq
We note that when the particles are distinguishable the maximum allowed
energy loss equals $W_{\rm max}$. However, when the particles are
identical, we must first agree in which of the two particles is
regarded as ``the projectile'' after the collision. Naturally, we
shall consider that the projectile is the particle with the highest energy.
This convention implies that the largest energy transfer is $E_{1{\rm
i}}/2$, and the total cross section defined as
\beq
\sigma \equiv \int_0^{E_{1{\rm i}}/2} \frac{\d\sigma}{\d W} \, \d W
\label{5.155}\eeq
then gives the number of collisions per unit time and per unit incident
current, as usual.


\subsection{Coulomb collisions \label{sec5.3.1}}
\index{collisions of identical particles!Coulomb collisions}

When the interaction between the particles is Coulombian,
\beq
V(r) = \frac{Z_1^2 e^2}{r},
\label{5.156}\eeq
the scattering amplitude is given by Eq.\ \req{5.33},
\beq
f^{\rm (C)}(\theta) = - \eta \, \frac{\Gamma(1+{\rm i}\eta)}
{\Gamma(1-{\rm i}\eta)} \,
\frac{\exp[
- {\rm i} \eta \ln(\sin^2(\theta/2))]}{2k\sin^2(\theta/2)},
\label{5.157}\eeq
where $\eta = \mu_{\rm r} Z_1^2 e^2 /(\hbar^2
k)$ is the Sommerfeld parameter and $k=p'_{\rm i}/\hbar$.
\index{Sommerfeld parameter}
The DCS for collisions of unpolarized identical fermions (in
CM) takes the form
\beq
\frac{\d\sigma}{\d\Omega} = \left(
\frac{Z_1^2 e^{2}}{2 v_{\rm 1i} p'_{\rm i}}\right)^{2}
\left[ \frac{1}{\sin^{4}(\theta/2)} +
\frac{1}{\cos^{4}(\theta/2)} -
\frac{\cos\left\{ 2\eta \, \ln \left[\tan(\theta/2)\right]
\right\} }
{\sin^{2}(\theta/2)\cos^{2}(\theta/2)}\right],
\label{5.158}\eeq
where $v_{\rm 1i}$ is the velocity of the projectile in the L frame [see
the comment after Eq.\ \req{5.37}]. This formula is important because it
describes exchange and interference effects in electron-electron binary
collisions. The analogous DCS for collisions of identical
bosons is obtained by multiplying the last term in the square brackets
by $-2$ [cf.\ Eqs.\ \req{5.151} and Eq.\ \req{5.149} with $S=0$].

Using the relation \req{5.139} and the equality
\req{5.152} we find that the energy-loss DCS corresponding to the
angular DCS \req{5.158} is
\beq
\frac{\d\sigma}{\d W}
= \frac{2 \pi (Z_1^2 e^2)^2}{M_1 v_{\rm 1i}^2}
\left\{ \frac{1}{W^2} +
\frac{1}{(E_{1{\rm i}}-W)^2} -
\frac{1}{W(E_{1{\rm i}}-W)}
\cos\left[ \eta \, \ln \left(\frac{W}{E_{1{\rm i}}-W} \right)
\right] \right\}.
\label{5.159}\eeq
It is worth noticing that this formula applies only to weakly
relativistic particles. Strictly speaking, it is correct only in the
non-relativistic limit ($\gamma \rightarrow 1$).

If instead of the exact Coulomb scattering amplitude,
Eq.\ \req{5.157}, we use the Born approximation [Eq.\ \req{5.55} with
$R^{-1}=0$]
\beq
f^{\rm (CB)} (\theta) =
- \frac{2M_1}{\hbar^2} \, \frac{Z_1^2 e^2}{q^2}
= - \frac{2M_1}{4 p_{\rm i}^2} \,
\frac{Z_1^2 e^2}{\sin^2(\theta/2)}\, ,
\label{5.1601}\eeq
we obtain
\beq
\frac{\d\sigma}{\d W} = \frac{2\pi Z_1^4 e^{4}}{M_1
 v_{1{\rm i }}^{2}} \left[ \frac{1}{W^2} +
\frac{1}{(E-W)^2} - \frac{1}{W(E-W)} \right].
\label{5.161}\eeq

\index{M\o ller cross section!non-relativistic}The correct relativistic
energy-loss DCS for binary collisions of high-energy electrons [Eq.\
\req{6.229}] was derived by \citet{Moller1932} using the methods of
quantum electrodynamics. In the non-relativistic limit, the M{\o}ller
DCS reduces to the result from the Born approximation, Eq.\ \req{5.161}.
When $v_{1{\rm i}} \ll c$, the formula \req{5.159} is more accurate than
the Born approximation \req{5.161}.




\chapter{Quantum theory of inelastic collisions
\label{chapt6}}



The quantum theory of inelastic collisions of charged particles with
isolated atoms has reached a high degree of sophistication
\citep{Bransden1970, Joachain1975}. The most elaborate calculations
provide results in good agreement with measurements, proving that the
theory describes the physical process correctly. However, calculations
are generally difficult and, very frequently, doable only for
projectiles with low energies. In this Chapter we present the elementary
components of the theory of inelastic collisions. We focus on the
plane-wave Born approximation (PWBA), which is the basis of the Bethe
theory of the stopping power. Our main purpose is to derive the basic
formulas describing the excitation of the target atom by impact of
charged particles and the stopping of fast projectiles in thin atomic
gases.  For simplicity, we consider first the non-relativistic version
of the theory. The relevant modifications introduced by relativity are
indicated without proof. They are justified on the basis of the
semi-classical dielectric formalism, which we shall also use to extend
the theory to describe inelastic interactions of charged particles in
dense media (see Chapter \ref{chapt7}).


\section{Collisions of charged particles with atoms \label{sec6.1}}

Let us consider inelastic collisions of a projectile of mass $M_0$ and
charge $Z_0 e$ with a neutral atom of the element $Z$ ($=$ atomic
number) in its ground state. With obvious modifications, the present
study is also valid when the target is a positive ion. We assume that
the interaction between the projectile and the atom is weak and can be
treated as a perturbation to first order. Calculations will be performed
in the laboratory reference frame, where the nucleus of the target atom
(assumed to be pointlike and infinitely massive) is at rest at the
origin of coordinates. The effect of the finite mass of the target may
be accounted for a posteriori by considering the reduced mass
of the projectile and the target atom (see Section \ref{sec3.1}).

Inelastic collisions involve the transfer of energy and linear momentum
from the projectile to the target atom, which result in electronic
excitations of the latter. The interactions of the atomic electrons with
the nucleus, with other electrons, and with the projectile are assumed
to be purely Coulombian (\ie, retardation effects are disregarded). In
addition, to simplify the formulation, atomic states will be described
by means of the independent-electron approximation (see Section
\ref{sec3.3}). That is, atomic electrons are assumed to move
independently in a common central potential $V_{\rm at} (r)$, such as
the self-consistent Hartree-Fock-Slater potential (the non-relativistic
analog of the DHFS potential described in Section \ref{sec3.5}). The
one-electron orbitals are spinors that satisfy the Schr\"{o}dinger
equation
\beq
\left[ \frac{1}{2 \me} \, \breve{\bf p}^2 + V_{\rm at}(r) \right] \psi(x)
= \varepsilon \, \psi(x),
\qquad x \equiv ({\bf r},\sigma).
\label{6.1}\eeq
We will use spherical orbitals of the type (uncoupled representation)
\beq
\psi_{\varepsilon \ell m_{\rm L} m_{\rm S}} (x) =
\frac{P_{\varepsilon \ell}(r)}{r}
\, Y_{\ell m_{\rm L}}(\hat{\bf r}) \, \chi_{m_{\rm S}}(\sigma),
\label{6.2}\eeq
where $Y_{\ell m_{\rm L}}(\hat{\bf r})$ is a spherical harmonic, $\chi_{m_{\rm
S}}(\sigma)$ is a spin function, and the reduced radial function
$P_{\varepsilon \ell}(r)$ satisfies the radial Schr\"{o}dinger equation
(see Section \ref{sec2.1.2})
\beq
- \frac{\hbar^2}{2\me} \, \frac{\d^2 P_{\varepsilon \ell}}{\d r^2}
+ \left[ \frac{\hbar^2}{2\me} \, \frac{\ell(\ell+1)}{r^2} + V_{\rm at}(r)
\right] P_{\varepsilon \ell} = \varepsilon \, P_{\varepsilon \ell}\, .
\label{6.3}\eeq
The atomic potential $V_{\rm at}(r)$ usually has a (Latter) Coulomb tail
[Eqs.\ \req{3.134} and \req{3.135}] and admits an infinite number of
bound orbitals, with discrete negative energy eigenvalues
$\varepsilon_{n\ell}$ characterized by the principal quantum number $n$
and the orbital angular momentum quantum number $\ell$. In addition,
free spherical orbitals exist for any positive energy $\varepsilon$ and
angular momentum $\ell$.

Atomic states $\Psi$ are represented as single Slater determinants,
\beqa
\Psi(1, 2, \ldots, Z) &=& \frac{1}{\sqrt{N!}}\left| \begin{array}{ccc}
\psi_{n_1 \ell_1 m_{{\rm L}1} m_{{\rm S}1}}(x_{1}) & \ldots &
\psi_{n_1 \ell_1 m_{{\rm L}1} m_{{\rm S}1}}(x_{Z}) \\ [2mm]
\vdots & \ddots & \vdots \\ [2mm]
\psi_{n_Z \ell_Z m_{{\rm L}Z} m_{{\rm S}Z}}(x_{1}) & \ldots &
\psi_{n_Z \ell_Z m_{{\rm L}Z} m_{{\rm S}Z}}(x_{Z})
\end{array}
\right|,
\label{6.4}\eeqa
and satisfy the Schr\"{o}dinger equation
\beq
{\cal H}_{\rm IEA}(x_1, \ldots, x_Z)
\Psi(x_1, \ldots, x_Z) = E_{\rm at} \Psi(x_1, \ldots, x_Z),
\label{6.5}\eeq
where
\beq
{\cal H}_{\rm IEA}(x_1, \ldots, x_Z) =
\sum_{j=1}^Z \left[ \frac{1}{2\me} \breve{\bf p}_j^2 + V_{\rm at}(r_j) \right]
\label{6.6}\eeq
is the Hamiltonian of the atomic electrons within the
independent-electron approximation. The energy of the atomic state is
\beq
E_{\rm at} = \varepsilon_{n_1 \ell_1} + \cdots + \varepsilon_{n_Z \ell_Z}.
\label{6.7}\eeq
It is assumed that before the collision the target atom is in its ground
state with the electron configuration\index{atomic configuration}
\beq
(n_1\ell_1)^{\nu_1}\, , \, \, (n_2\ell_2)^{\nu_2}\, ,
\, \ldots
\nonumber \eeq
where $\nu_a$ is the number of electrons in the shell $a=(n_a \ell_a)$.

A collision of the charged projectile with the target atom may induce
transitions of the atom to excited states. In most cases the final state
differs from the initial ground state by a single orbital, which may be
either a bound unoccupied orbital or a free orbital with positive energy
$\varepsilon_b$. To simplify the arguments, bound and free electron
orbitals are assumed to be solutions of the same wave equation
\req{6.3}, \ie, all atomic electrons ``feel'' the same potential $V_{\rm
at}(r)$ in both the initial and the final states. Consequently, the set of
spherical orbitals \req{6.2} constitute an orthonormal basis of
one-electron states.

The Hamiltonian of the system (projectile and target atom) can be
expressed as
\beq
{\cal H}(x_0,x_1,\ldots, x_Z)  = {\cal H}_{\rm IEA}(x_1, \ldots, x_Z)
+ {\cal H}_{\rm p}(x_0) + {\cal H}'(x_0,x_1,\ldots,x_Z),
\label{6.8}\eeq
where ${\cal H}_{\rm IEA}$ is the Hamiltonian of the target atom within
the independent-electron approximation,
\beq
{\cal H}_{\rm p}(0) = \frac{1}{2 M_0} \breve{\bf p}_0^2 =
- \, \frac{\hbar^2}{2 M_0} \, \nabla_0^2
\label{6.9}\eeq
is the Hamiltonian of the (free) projectile, whose variables are
denoted by the subscript ``0'', and
\beq
{\cal H}'(x_0,x_1,\ldots,x_Z) = \frac{Z_0 Z e^2}{r_0} - Z_0 e^2 \
\sum_{j=1}^Z \frac{1}{\left| {\bf r}_0 - {\bf r}_j \right|}
\label{6.10}\eeq
describes the Coulombian interaction of the projectile with the nucleus
and with the $Z$ electrons of the atom. For simplicity we are assuming a
point nucleus. The use of ${\cal H}_{\rm IEA}$ instead of the actual
atomic Hamiltonian, Eq.\ \req{3.24}, implies that electron-correlation
effects are neglected.

In the simplest formulation of the theory, the projectile particle is
assumed to be distinguishable from the atomic electrons and the
interaction ${\cal H}'$ is treated as a first-order perturbation that
induces transitions between the eigenstates $\phi(x_0)\, \Psi(x_1,
\ldots, x_Z)$ of the unperturbed Hamiltonian ${\cal H}_{\rm IEA} + {\cal
H}_{\rm p}$. Since the Hamiltonian ${\cal H}_{\rm p}$ is that of a free
particle, the states of the projectile can be represented as plane
waves\footnote{For the sake of generality we assume that the projectile
has spin. In non-relativistic collisions spin effects are important only
when the projectile is an electron (see Sections \ref{sec6.1.2.1} and
\ref{sec6.6.1}).}
\beq
\phi_{{\bf k},m_{\rm S}}(x_0) = \frac{1}{(2\pi)^{3/2}} \, \exp \left(
{\rm i} {\bf k} \dotprod {\bf r}_0 \right) \, \chi_{m_{\rm S}}(\sigma_0).
\label{6.11}\eeq
This kind of approach is usually referred to as the (first-order)
plane-wave Born approximation (PWBA). The PWBA is valid only for
projectiles with velocities that are much higher than those of the
atomic electrons (see Section \ref{sec6.2}).
\index{quantum inelastic collisions}

A more elaborate theoretical framework,
still based on first-order perturbation theory, is provided by
the (first-order) distorted-wave Born approximation (DWBA), which is
similar to the \citet{Furry1951} representation of quantum electrodynamics.
To formulate the DWBA, we rewrite the Hamiltonian \req{6.8} as
\beq
{\cal H}(x_0,x_1,\ldots, x_Z)  = {\cal H}_{\rm IEA}(x_1, \ldots, x_Z)
+ \left[ {\cal H}_{\rm p}(x_0) + V_{\rm p}(r_0) \right] + {\cal H}''
\label{6.12}\eeq
with ${\cal H}'' = {\cal H}'-V_{\rm p}(r_0)$. That is, we have added and
subtracted a central potential $V_{\rm p}(r_0)$ that depends
only on the coordinates of the projectile. In the DWBA, the states
$\psi(x_0)$ of the projectile before and after the collision are
represented by distorted plane waves,
\beq
\psi^{\pm}_{{\bf k},m_{\rm S}}(x_0) =  \psi^{\pm}_{\bf k}({\bf r}_0) \,
\chi_{m_{\rm S}}(\sigma_0),
\label{6.13}\eeq
where the spatial part is an exact solution of the Schr\"{o}dinger wave
equation
\beq
\left[ - \, \frac{\hbar^2}{2 M_0} \, \nabla^2
+ V_{\rm p}(r_0) \right] \psi^{\pm}_{\bf k}({\bf r}_0)
= E \psi^{\pm}_{\bf k}({\bf r}_0)
\label{6.14}\eeq
of the type \req{2.55}. Asymptotically (\ie, at large radii) the waves
\req{6.13} behave as a plane wave plus a spherical outgoing ($+$) or
incoming ($-$) wave. We can thus consider
\beq
{\cal H}'' (x_0,x_1,\ldots,x_Z) =
\frac{Z_0 Z e^2}{r_0} - Z_0 e^2 \ \sum_{j}
\frac{1}{\left| {\bf r}_0 - {\bf r}_j \right|}
- V_{\rm p}(r_0)
\label{6.15}\eeq
as a perturbation that causes transitions between eigenstates $\psi(x_0)\,
\Psi(x_1, \ldots, x_Z)$ of the ``unperturbed'' Hamiltonian ${\cal H}_{\rm
IEA} + [{\cal H}_{\rm p} + V_{\rm p}]$. The effectiveness of the DWBA
lies in the fact that ${\cal H}''$ can be made much weaker than the
original interaction ${\cal H}'$. Unfortunately, since ${\cal H}'$ does
depend on the coordinates of the atomic electrons, ${\cal H}''$ cannot
be reduced to zero. Nonetheless, it is assumed that, with a proper
choice of the distorting potential $V_{\rm p}$, ${\cal H}''$ can be made
small enough to be treated as a perturbation to first order. Evidently,
when $V_{\rm p}(r) \equiv 0$, the DWBA reduces to the PWBA.


\subsection{The one-active electron approximation \label{sec6.1.1}}
\index{quantum inelastic collisions!one-electron approximation}

Figure \ref{fig6.1} displays the kinematics of the collision. Before the
interaction, the projectile moves with velocity ${\bf v}$, linear
momentum ${\bf p}=\hbar {\bf k}=M_0 {\bf v}$ and kinetic energy $E=\1o2
M_0 v^2$, the corresponding values after the collision are ${\bf v}'$,
${\bf p}'=\hbar {\bf k}'$ and $E'$, respectively. The initial and final
states of the projectile are described as distorted plane waves
$\psi_{{\bf k},m_{\rm S}}^{(\pm)}(x_0)$ of a potential $V_{\rm p}(r_0)$.

\begin{figure}[htb]
\begin{center}
\includegraphics*[width=13.0cm]{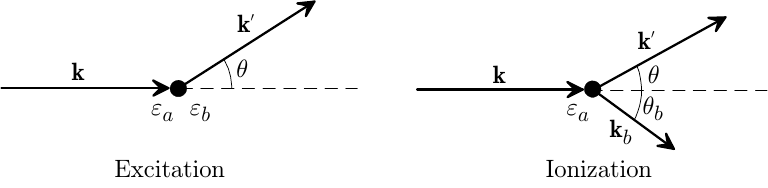}
\caption{Kinematics of inelastic collisions. The quantities $\varepsilon_a$
and $\varepsilon_b$ are the energies of the initial and final states of the
active target electron, respectively; $\theta$ is the polar scattering
angle of the projectile and, in the case of ionizing collisions,
$\theta_b$ is the polar angle of the direction of emission of the
knock-on electron.
\label{fig6.1}}
\end{center}\end{figure}

The interaction with the projectile induces transitions of the target
atom from the initial state $\Psi_{a}$ of energy $E_{{\rm at}a}$
corresponding to the ground-state configuration
\index{atomic configuration}
to a final state
$\Psi_{b}$ of energy $E_{{\rm at}b}$. As indicated above, we
consider that the atomic potential $V_{\rm at}(r)$ describes both the
initial and final states of the atom and, consequently, the atomic wave
function are represented by Slater determinants made of orbitals that
are mutually orthogonal.  Occasionally, the final atomic state may
include free electron orbitals with positive energies, whose spatial
part is a distorted plane wave for the atomic potential, that is, an
exact solution of the wave equation
\beq
\left[- \frac{\hbar^2}{2\me} \, \nabla^2 + V_{\rm at}(r) \right]
\psi^{(\pm)}_{{\bf k}} ({\bf r}) = \varepsilon \,
\psi^{(\pm)}_{{\bf k}} ({\bf r}), \qquad
\varepsilon = \frac{\hbar^2 k^2}{2\me} \, ,
\label{6.16}\eeq
representing an electron emerging in the direction ${\bf k}$.
Notice that free orbitals are orthogonal to all bound orbitals.
We consider that bound orbitals are normalized
to unity and distorted plane waves (of both the projectile and the
atomic electrons) are normalized in wave-vector space,
\ie,
\beq
\int \left[ \psi^{(\pm)}_{{\bf k}'}({\bf r})
\right]^\dagger
\psi^{(\pm)}_{{\bf k}}({\bf r}) \, \d {\bf r}
= \delta({\bf k} - {\bf k}').
\label{6.17}\eeq
With this normalization, the density of free states per unit volume in ${\bf
k}$-space is unity (see Section \ref{sec2.1.1}).

\index{one-active-electron approximation}
As indicated above, the interaction ${\cal H}''$ [or ${\cal H}'$] of the
projectile with the target atom is treated as a first-order perturbation
\citep[see, \eg,][]{Baym1974}. The transition rate is determined by
the transition matrix elements
\beqa
T_{fi} &\equiv&
\left< \psi^{(-)}_{{\bf k'},m'_{\rm S}}(x_0)
\Psi_b (x_1, \ldots, x_Z) \left| {\cal H}''\rule{0mm}{4mm}
\right| \psi^{(+)}_{{\bf k},m_{\rm S}}(x_0) \Psi_a (x_1, \ldots, x_Z)
\right>
\nonumber \\ [2mm]
&=& \left< \psi^{(-)}_{{\bf k'},m'_{\rm S}}(x_0)
\Psi_b (x_1, \ldots, x_Z) \left|
\frac{Z_0 Z e^2}{r_0} - V_{\rm p}(r_0)
\right| \psi^{(+)}_{{\bf k},m_{\rm S}}(x_0) \Psi_a (x_1, \ldots, x_Z)
\right>
\nonumber \\ [2mm]
&+& \left< \psi^{(-)}_{{\bf k'},m'_{\rm S}}(x_0)
\Psi_b (x_1, \ldots, x_Z) \left|
\sum_{j}
\frac{- Z_0 e^2 }{\left| {\bf r}_0 - {\bf r}_j \right|}
\right| \psi^{(+)}_{{\bf k},m_{\rm S}}(x_0)
\Psi_a (x_1, \ldots, x_Z)
\right>. \rule{10mm}{0mm}
\label{6.18}\eeqa
Since atomic orbitals are orthogonal, the first term in this expression
contributes only when $\Psi_b=\Psi_a$ (which corresponds to elastic
scattering). The second term is the matrix element of a one-particle
operator and, by virtue of the Slater--Condon rules \citep{Condon1930},
it is different from
zero only when the atomic states $\Psi_a$ and $\Psi_b$ are equal or
differ by a single orbital. Therefore, the only allowed excitations of
the target atom are single-electron excitations. This is equivalent to
the called {\it one-active-electron approximation}, which consists of
considering only the excitations of a single electron from a bound
orbital $\psi_a$ to an unoccupied (bound or free) orbital $\psi_b$,
whereas the other atomic electrons behave as mere spectators and their
orbitals remain frozen in the course of the interaction. Consequently,
\beqa
T_{fi} &=&
\left< \psi^{(-)}_{{\bf k'},m'_{\rm S}}(x_0) \psi_b (x_1) \left|
\frac{- Z_0 e^2 }{\left| {\bf r}_0 - {\bf r}_1 \right|}
\right| \psi^{(+)}_{{\bf k},m_{\rm S}}(x_0) \psi_a (x_1)
\right>
\label{6.19}\eeqa
when $\Psi_b \ne \Psi_a$, and
\beqa
T_{fi} &=&
\left< \psi^{(-)}_{{\bf k'},m'_{\rm S}}(x_0) \Psi_a (x_1, \ldots, x_Z) \left|
\frac{Z_0 Z e^2}{r_0} - V_{\rm p}(r_0)
\right| \psi^{(+)}_{{\bf k},m_{\rm S}}(x_0) \Psi_a (x_1, \ldots, x_Z)
\right>
\nonumber \\ [2mm]
&& \mbox{}
\nonumber \\ [2mm]
&+&
\left< \psi^{(-)}_{{\bf k'},m'_{\rm S}}(x_0) \Psi_a (x_1, \ldots, x_Z) \left|
\sum_{j}
\frac{- Z_0 e^2 }{\left| {\bf r}_0 - {\bf r}_j \right|}
\right| \psi^{(+)}_{{\bf k},m_{\rm S}}(x_0) \Psi_a (x_1, \ldots, x_Z)
\right> \rule{10mm}{0mm}
\label{6.20}\eeqa
when the target atom remains in its initial state. Notice that distorted
plane waves of initial (final) states have outgoing (incoming) spherical
distortions \citep[see, \eg,][]{BreitBethe1954}. Evidently, the
transition changes the energy of the target atom in $E_{{\rm
at}b}-E_{{\rm at}a} = \varepsilon_b - \varepsilon_a$.


\subsection{Differential cross sections \label{sec6.1.2}}
\index{quantum inelastic collisions!distorted-wave Born approximation}

The differential transition rate for excitation of the active electron
from the orbital $\psi_a$ to an unoccupied bound orbital
$\psi_b$ is given by the Fermi golden rule \citep{Baym1974,Joachain1975}
\beq
\d w_{fi}^{\rm exc} = \frac{2\pi}{\hbar}
\left| T_{fi} \right|^2
\, \delta(E + \varepsilon_a - E' - \varepsilon_b) \, \d {\bf k}',
\label{6.21}\eeq
where we have used that, with the adopted normalization [Eq.\
\req{6.17}], the density of final states of the projectile
per unit volume in ${\bf k}$-space equals unity. The corresponding DCS
is obtained as the ratio of the transition rate to the probability
current of the incident projectile wave,
\beq
{\bf j}_{\rm inc} = (2\pi)^{-3} \frac{\hbar k}{M_0} = (2\pi)^{-3} v \,
\hat{\bf k}.
\label{6.22}\eeq
That is,
\beq
\d \sigma_{fi}^{\rm exc}
= \frac{\d w_{fi}^{\rm exc}}{ \left| {\bf j}_{\rm inc} \right|}
= \frac{(2\pi)^4}{\hbar v}
\left| T_{fi} \right|^2
\, \delta(E + \varepsilon_a - E' - \varepsilon_b) \, \d {\bf k}'.
\label{6.23}\eeq
Using the relation $(\hbar k')^2 = 2M_0 E'$, we have
\beq
\d {\bf k}' = {k'}^2 \, \d k' \, \d \hat{\bf k}'
= {k'}^2 \frac{\d k'}{\d E'} \, \d E'\,\d\hat{\bf k}'
= k'\, \frac{M_0}{\hbar^2} \, \d E'\,\d\hat{\bf k}'
\label{6.24}\eeq
and, therefore,
\beq
\d \sigma_{fi}^{\rm exc} = \frac{(2\pi)^4}{\hbar v} \,
k'\, \frac{M_0}{\hbar^2} \,
|T_{fi}|^2\, \delta(W-\varepsilon_b+\varepsilon_a)
\, \d E' \,\d\hat{\bf k}',
\label{6.25}\eeq
where $W \equiv E - E'$ is the energy lost by the projectile.
Although the delta function, which enforces energy conservation, can be
removed by integration over $E'$, it is more convenient to keep it
and consider the excitation DCS as a function of the angular deflection
$\hat{\bf k}'$ and the energy loss $W = \varepsilon_b -
\varepsilon_a$,
\beq
\frac{\d^2 \sigma_{fi}^{\rm exc}}{\d W \, \d\hat{\bf k}'}
 = \frac{(2\pi)^4}{\hbar v}
\, k'\, \frac{M_0}{\hbar^2} \, \delta(W - \varepsilon_b +
\varepsilon_a) \, |T_{fi}|^2,
\label{6.26}\eeq
where the $T$-matrix element is ``on the energy shell'', \ie, the
initial and final states (of the projectile {\it and} the active
electron) have the same total energy, $E + \varepsilon_a = E' +\varepsilon_b$.

The differential transition rate for ionization (\ie, for transitions
where $\psi_b$ is a free orbital, with $\epsilon_b >0$) is given by the Fermi golden rule
 \citep{Baym1974,Joachain1975}
\beq
\d w^{\rm ion}_{fi} = \frac{(2\pi}{\hbar} \,
|T_{fi}|^2\, \delta(E - E' -\varepsilon_b+\varepsilon_a)
\, \d {\bf k}'\, \d {\bf k}_b \, ,
\label{6.27}\eeq
where $\hbar {\bf k}_b$ is the linear momentum of the ejected electron.
Notice that this expression is valid only when the free orbital $\psi_b$
is normalized in the form \req{6.17}. The corresponding DCS is
\beq
\d \sigma^{\rm ion}_{fi}
= \frac{\d w_{fi}^{\rm ion}}{ \left| {\bf j}_{\rm inc} \right|}
= \frac{(2\pi)^4}{\hbar v} \,
|T_{fi}|^2\, \delta(E - E' -\varepsilon_b+\varepsilon_a)
\, \d {\bf k}'\, \d {\bf k}_b \, .
\label{6.28}\eeq
Using the relation \req{6.24} and the analogous one for ${\bf k}_b$,
\beq
\d {\bf k}_b = k_b \, \frac{\me}{\hbar^2} \, \d
\varepsilon_b \, \d \hat{\bf k}_b,
\label{6.29}\eeq
and performing the integration over $E'$, we get
\beq
\d \sigma^{\rm ion} = \frac{(2\pi)^4}{\hbar v} \,
\, k' \, \frac{M_0}{\hbar^2} \, k_b \,
\frac{\me}{\hbar^2} \,
|T_{fi}|^2\, \d\hat{\bf k}' \, \d \varepsilon_b \, \d\hat{\bf k}_b\, .
\label{6.30}\eeq
That is,
\beq
\frac{\d^3 \sigma^{\rm ion}}{\d W \, \d \hat{\bf k}' \, \d \hat{\bf
k}_b} = \frac{(2\pi)^4}{\hbar v} \,
\, k' \, \frac{M_0}{\hbar^2} \, k_b \,
\frac{\me}{\hbar^2} \,
|T_{fi}|^2,
\label{6.31}\eeq
where use has been made of the fact that $\varepsilon_b = \varepsilon_a + W$.
Hereafter, $T$-matrix elements are on the energy shell.
In many cases (\eg, in calculations of the stopping power of gases)
only the effect of the interactions on the projectile is of interest.
Then, the relevant DCS is obtained by integrating Eq.\ \req{6.31}
over the direction $\hat{\bf k}_b$ of the emitted electron,
\beq
\frac{\d^2 \sigma^{\rm ion}}{\d W \, \d \hat{\bf k}'}  =
\frac{(2\pi)^4}{\hbar v} \,
\, k' \, \frac{M_0}{\hbar^2} \, k_b  \,
\frac{\me}{\hbar^2} \, \int
|T_{fi}|^2 \, \d \hat{\bf k}_b \, .
\label{6.32}\eeq

From now on, we shall limit to consider the doubly-differential cross
sections (DDCSs) given by Eqs.\
\req{6.26} and \req{6.32}. In the derivation of these formulas, we have
assumed transitions from a given initial state $i=$$\{\psi_{{\bf k},
m_{\rm S}}^{(+)}(x_0)$, $\psi_{n_a \ell_a m_{{\rm L}a} m_{{\rm
S}a}} (x) \}$ to a well defined final state $f=$$\{\psi_{{\bf k}',
m'_{\rm S}}^{(-)}(x_0)$, $\psi_{b} (x)\}$. In most practical
cases, the target atoms are randomly oriented, the incident beam is
unpolarized and final magnetic and spin states are not distinguished.
Under these circumstances, the observed DDCS (per electron in the
initial level $\varepsilon_a$) is obtained by averaging
over initial degenerate magnetic and spin states and summing over final
degenerate states. Thus, the observed DDCS for excitation to an empty bound
level $\varepsilon_b$ is given by
\beq
\frac{\d^2 \sigma^{\rm exc}}{\d W \, \d\hat{\bf k}'}
 = \frac{(2\pi)^4}{\hbar v}
\, k'\, \frac{M_0}{\hbar^2} \,
{\cal T}_{fi}^{\rm exc} \, ,
\label{6.33}\eeq
where
\beqa
{\cal T}_{fi}^{\rm exc}
&\equiv& \delta(W - \varepsilon_b + \varepsilon_a) \,
\frac{1}{4(2\ell_a+1)}
\sum_{m_{{\rm L}a}, m_{{\rm S}a}, m_{\rm S}} \; \sum_{m_{{\rm L}b}, m_{{\rm
S}b} , m'_{\rm S}} \left| T_{fi} \right|^2
\nonumber \\ [2mm]
&=& \delta(W - \varepsilon_b + \varepsilon_a) \, \frac{1}{4(2\ell_a+1)}
\sum_{m_{{\rm L}a}, m_{{\rm S}a}, m_{\rm S}} \;
\sum_{m_{{\rm L}b}, m_{{\rm S}b} , m'_{\rm S}}
\nonumber \\ [2mm]
&& \mbox{} \times
\left| \left< \psi^{(-)}_{{\bf k}', m'_{\rm S}}(x_0) \,
\psi_{n_b \ell_b m_{{\rm L}b} m_{{\rm S}b}}({\bf r}) \left|
\frac{- Z_0 e^2 }{\left| {\bf r}_0 - {\bf r} \right|}
\right| \psi^{(+)}_{{\bf k}, m_{\rm S}}(x_0)
\, \psi_{n_a \ell_a m_{{\rm L}a} m_{{\rm S}a}}({\bf r}) \right>
\right|^2. \rule{10mm}{0mm}
\nonumber\eeqa
Using the orthogonality of the unit spinors,
\beq
\left<
\chi_{m_{{\rm S}a}}
\left| \chi_{m_{{\rm S}b}} \right> \right. =
\delta_{m_{{\rm S}a}, m_{{\rm S}b}},
\label{6.34}\eeq
we can evaluate the spin parts and express the $T$-matrix elements in
terms of only the spatial wave functions,
\beqa
{\cal T}_{fi}^{\rm exc}
&=& \delta(W - \varepsilon_b + \varepsilon_a) \,
\frac{1}{2\ell_a+1}
\sum_{m_{{\rm L}a}} \;
\sum_{m_{{\rm L}b}}
\nonumber \\ [2mm]
&& \mbox{} \times
\left| \left< \psi^{(-)}_{{\bf k}'}({\bf r}_0) \,
\psi_{n_b \ell_b m_{{\rm L}b}}({\bf r}) \left|
\frac{- Z_0 e^2 }{\left| {\bf r}_0 - {\bf r} \right|}
\right| \psi^{(+)}_{{\bf k}}({\bf r}_0)
\, \psi_{n_a \ell_a m_{{\rm L}a}}({\bf r})
\right> \right|^2. \rule{10mm}{0mm}
\label{6.35}\eeqa

The observed DDCS for ionization can also be written in the form \req{6.33},
\beq
\frac{\d^2 \sigma^{\rm ion}}{\d W \,\d\hat{\bf k}'}
 = \frac{(2\pi)^4}{\hbar v}
\, k'\, \frac{M_0}{\hbar^2} \,
{\cal T}_{fi}^{\rm ion},
\label{6.36}\eeq
with
\beqa
{\cal T}_{fi}^{\rm ion} &\equiv&
k_b \, \frac{\me}{\hbar^2} \,
\frac{1}{4(2\ell_a+1)}
\sum_{m_{{\rm L}a}, m_{{\rm S}a}, m_{\rm S}} \;
\sum_{m_{{\rm S}b}, m'_{\rm S}}
\, \int \d \hat{\bf k}_b
\nonumber \\ [2mm]
&& \mbox{} \times
\left| \left< \psi^{(-)}_{{\bf k}', m'_{\rm S}}(x_0) \,
\psi^{(-)}_{{\bf k}_b m_{{\rm S}b}}(x) \left|
\frac{- Z_0 e^2 }{\left| {\bf r}_0 - {\bf r} \right|}
\right| \psi^{(+)}_{{\bf k}, m_{\rm S}}(x_0)
\, \psi_{n_a \ell_a m_{{\rm L}a} m_{{\rm S}a}}(x) \right>
\right|^2
\nonumber \eeqa
\vspace*{-4mm}
\beq
= k_b \, \frac{\me}{\hbar^2} \,
\frac{1}{2\ell_a+1}
\sum_{m_{{\rm L}a}} \int \d \hat{\bf k}_b
\left|
\left< \psi^{(-)}_{{\bf k}'} ({\bf r}_0) \, \psi^{(-)}_{{\bf k}_b}
({\bf r}) \left|
\frac{- Z_0 e^2 }{\left| {\bf r}_0 - {\bf r} \right|}
\right| \psi^{(+)}_{{\bf k}} ({\bf r}_0)  \,
\psi_{n_a \ell_a m_{{\rm L}a}} ({\bf r})
\right> \right|^2. \rule{5mm}{0mm}
\label{6.37}\eeq
To facilitate the numerical evaluation of the matrix elements in the
ionization terms, we introduce the partial-wave
expansion of the distorted plane wave, Eq.\ \req{2.55},
\beq
\psi_{{\bf k}_b}^{(-)} ({\bf r})
= \frac{1}{{k}_b} \, \sqrt{\frac{2}{\pi}}
\sum_{\ell_b,m_{{\rm L}b}}
{\rm i}^{\ell_b} \, \exp \left( - {\rm i} \delta_{\varepsilon_b \ell_b}
\right) Y_{\ell_b m_{{\rm L}b}}^\ast (\hat{\bf k}_b) \,
\psi_{\varepsilon_b \ell_b m_{{\rm L}b}}({\bf r}) ,
\label{6.38}\eeq
and write
\beqa
{\cal T}_{fi}^{\rm ion} &=&
\frac{2\me}{\pi \hbar^2 k_b} \, \frac{1}{2\ell_a+1}
\sum_{m_{{\rm L}a}} \int \d \hat{\bf k}_b
 \nonumber \\ [2mm]
 && \mbox{} \times
 \sum_{\ell_b,m_{{\rm L}b}} \, \sum_{\ell'_b,m'_{{\rm L}b}}
{\rm i}^{\ell'_b-\ell_b} \, \exp \left[ {\rm i}
\left( \delta_{\varepsilon_b \ell_b} - \delta_{\varepsilon_b \ell'_b}
\right) \right]
Y_{\ell'_b m'_{{\rm L}b}}^\ast (\hat{\bf k}_b) \,
Y_{\ell_b m_{{\rm L}b}} (\hat{\bf k}_b) \,
\nonumber \\ [2mm]
&& \mbox{} \times
\left< \psi^{(-)}_{{\bf k}'} ({\bf r}_0) \,
\psi_{\varepsilon_b \ell_b m_{{\rm L}b}} ({\bf r}) \left|
\frac{- Z_0 e^2 }{\left| {\bf r}_0 - {\bf r} \right|}
\right| \psi^{(+)}_{{\bf k}} ({\bf r}_0)  \,
\psi_{n_a \ell_a m_{{\rm L}a}} ({\bf r}) \right>
\nonumber \\ [2mm]
&& \mbox{} \times
\left< \psi^{(+)}_{{\bf k}} ({\bf r}_0)  \,
\psi_{n_a \ell_a m_{{\rm L}a}} ({\bf r})
\left|
\frac{- Z_0 e^2 }{\left| {\bf r}_0 - {\bf r} \right|}
\right| \psi^{(-)}_{{\bf k}'} ({\bf r}_0) \,
\psi_{\varepsilon_b \ell'_b m'_{{\rm L}b}} ({\bf r}) \right> .
\label{6.39}\eeqa
Using the orthogonality of the spherical harmonics [Eq.\ \req{B.53}],
we obtain
\beqa
{\cal T}_{fi}^{\rm ion} &=&
\frac{2\me}{\pi \hbar^2 k_b} \,
\frac{1}{2\ell_a+1}
\sum_{m_{{\rm L}a}} \sum_{\ell_b,m_{{\rm L}b}}
\nonumber \\ [2mm]
&& \mbox{} \times
\left|
\left< \psi^{(-)}_{{\bf k}'} ({\bf r}_0) \,
\psi_{\varepsilon_b \ell_b m_{{\rm L}b}} ({\bf r}) \left|
\frac{- Z_0 e^2 }{\left| {\bf r}_0 - {\bf r} \right|}
\right| \psi^{(+)}_{{\bf k}} ({\bf r}_0)  \,
\psi_{n_a \ell_a m_{{\rm L}a}} ({\bf r}) \right>
\right|^2 .
\label{6.40}\eeqa
Apart from a trivial factor and different summation indices, the
ionization matrix elements \req{6.40} have the same form as those for
excitation, Eq.\ \req{6.35}.  That is, when the direction of the ejected
electron is not observed, its final state can be described either as a
distorted plane wave or as a spherical wave. Of course, numerical
calculations are easier with spherical waves.


\subsubsection{Collisions of electrons with atoms \label{sec6.1.2.1}}
\index{quantum inelastic collisions!collisions of electrons}

The expressions for the DDCSs obtained above are appropriate for
describing collisions of spin-$\1o2$ projectiles that are different from
the electron. When the projectile is an electron, it is
indistinguishable from the active target electron and, therefore, they can
undergo re-arrangement collisions (\ie, the projectile and the target
electrons can ``exchange places''). It is natural to consider that the
projectile electron ``sees'' the same atomic potential as the target
electrons. The effect of exchange is then described
by antisymmetrizing the initial and final states in the transition
matrix elements. That is, the transition matrix elements \req{6.19}
are to be replaced with
\beq
T_{fi}^{\rm el} =
\left< \sqrt{2} \, {\cal A}_2 \left[ \psi^{(-)}_{{\bf k}', m'_{\rm S}}
(x_0) \, \psi_b(x_1) \right] \left|
\frac{- Z_0 e^2 }{\left| {\bf r}_0 - {\bf r} \right|}
\right| \sqrt{2} \, {\cal A}_2 \left[ \psi^{(+)}_{{\bf k}, m_{\rm
S}} (x_0) \, \psi_a(x_1) \right] \right> \, ,
\label{6.41}\eeq
where the operator ${\cal A}_2$ is the 2-particle antisymmetrizer,
\beq
{\cal A}_2 \left[ \psi_a(x_0) \psi_b(x_1) \right] \equiv \frac{1}{2!} \left[
 \psi_a(x_0) \psi_b(x_1) - \psi_a(x_1) \psi_b(x_0)\right].
\label{6.42}\eeq
Note that this operator is Hermitian and ${\cal A}_2^2 = {\cal A}_2$. As the
interaction is symmetrical, and the four orbitals are mutually
orthogonal, we have
\beqa
T_{fi}^{\rm el} &=& 2
\left< {\cal A}_2 \left[ \psi^{(-)}_{{\bf k}', m'_{\rm S}} (x_0) \,
\psi_b(x_1) \right] \left|
\frac{- Z_0 e^2 }{\left| {\bf r}_0 - {\bf r} \right|}
\right| \psi^{(+)}_{{\bf k}, m_{\rm S}} (x_0) \, \psi_a(x_1) \right>
\nonumber \\ [2mm]
&=&
\left< \psi^{(-)}_{{\bf k}', m'_{\rm S}} (x_0) \, \psi_b(x_1) \left|
\frac{- Z_0 e^2 }{\left| {\bf r}_0 - {\bf r} \right|}
\right| \psi^{(+)}_{{\bf k},
m_{\rm S}} (x_0) \, \psi_a(x_1) \right>
\nonumber \\ [2mm]
&& \mbox{} -
\left< \psi^{(-)}_{{\bf k}', m'_{\rm S}} (x_1) \, \psi_b(x_0) \left|
\frac{- Z_0 e^2 }{\left| {\bf r}_0 - {\bf r} \right|}
\right| \psi^{(+)}_{{\bf k},
m_{\rm S}} (x_0) \, \psi_a(x_1) \right>.
\label{6.43}\eeqa
The first and second terms in this expression represent direct and
re-arrangement collisions, respectively. In the latter, the interacting
electrons exchange places: the projectile electron is captured to the
orbital $\psi_b$ and the active target electron jumps to the final free
wave $\psi^{(-)}_{{\bf k}', m'_{\rm S}}$.

In the derivation of the electron matrix elements \req{6.43} we have
ignored the presence of the inactive electrons of the target atom. The
result is correct whenever the orbitals of these electrons and the
initial and final orbitals of the target {\it and} the projectile are
mutually orthogonal.


\section{The plane-wave Born approximation \label{sec6.2}}
\index{quantum inelastic collisions!plane-wave Born approximation}

The theory described up to this point corresponds to the so-called
distorted-wave Born approximation (DWBA). The relativistic version of
the DWBA provides a fairly accurate description of the ionization of
inner shells of atoms by electron and positron impact \citep{Segui2003,
Colgan2006, Llovet2014}. However, the calculations are difficult, mostly
because of the slow convergence of the partial-wave series, and the
results are specific to the projectile kind and energy. For the purposes
of stopping theory it is preferable to use the less accurate plane-wave
Born approximation (PWBA), which provides results applicable to any kind
of charged projectile and allows the calculations to be extended up to
arbitrarily high energies.

The PWBA consists of replacing the projectile distorted waves
$\psi^{(\pm)}_{{\bf k}} ({\bf r}_0)$ in the $T$-matrix element
\req{6.35} and \req{6.40} with the corresponding plane waves,
\beq
\phi_{{\bf k}} ({\bf r}_0) = (2\pi)^{-3/2}
\exp({\rm i} {\bf k} \dotprod {\bf r}_0 ).
\label{6.44}\eeq
The $T$-matrix elements then take the form
\beqa
\lefteqn{
T^{\rm pw}_{fi} = \left< \phi_{{\bf k}'}({\bf r}_0) \,
\psi_{b}({\bf r})
\left| \frac{-Z_0 e^2}{\left| {\bf r}_0 - {\bf r} \right|}
\right| \phi_{{\bf k}} ({\bf r}_0) \,
\psi_{a} ({\bf r}) \right>
}
\nonumber \\ [2mm]
&=& (2\pi)^{-3} \left< \exp({\rm i} {\bf k}' \dotprod {\bf r}_0) \,
\psi_{b}({\bf r})
\left| \frac{-Z_0 e^2}{\left| {\bf r}_0 - {\bf r} \right|}
\right| \exp({\rm i} {\bf k} \dotprod {\bf r}_0 ) \,
\psi_{a} ({\bf r}) \right>
\nonumber \\ [2mm]
&=& - \frac{Z_0 e^2}{(2\pi)^{3}} \int \d {\bf r} \,
\psi^\ast_{b}({\bf r}) \,
\psi_{a} ({\bf r})
\int \d {\bf r}_0
\frac{ \exp\left( {\rm i}  {\bf q} \dotprod {\bf r}_0 \right)}
{\left| {\bf r}_0 - {\bf r} \right|},
\nonumber\eeqa
where $\hbar {\bf q} = \hbar({\bf k} - {\bf k}')$ is the momentum
transfer. Noting that the last integral is given by the Bethe formula
\req{B.26}, we can write
\beqa
T^{\rm pw}_{fi}
&=& - \frac{ Z_0 e^2}{2 \pi^2 q^2}
\int \d {\bf r} \,
\psi^\ast_{b}({\bf r}) \,
\psi_{a} ({\bf r}) \,
\exp\left( {\rm i}  {\bf q} \dotprod {\bf r} \right)
\nonumber \\ [2mm]
&=& - \frac{ Z_0 e^2}{2 \pi^2 q^2}
\left< \psi_{b}({\bf r}) \left|
\exp\left( {\rm i}  {\bf q} \dotprod {\bf r} \right) \left| \,
\psi_{a} ({\bf r})
\rule{0mm}{4mm}\right>\right.\right. \, .
\label{6.45}\eeqa

It is worth noticing that elastic collisions are described by the
transition matrix elements \req{6.20} with $\Psi_b=\Psi_a$ and $V_{\rm
p}=0$. Replacing the distorted plane waves by plane waves, these matrix
elements become
\beq
T_{fi}^{\rm pw} = (2\pi)^{-3} \int \d {\bf r}_0
\exp\left( {\rm i}  {\bf q} \dotprod {\bf r} \right) V({\bf r}_0)\, ,
\label{6.46}\eeq
where
\beq
V({\bf r}_0) = \left< \Psi_a (1, \ldots, Z) \left|
\frac{Z_0 Z e^2}{r_0} - \sum_{j}
\frac{Z_0 e^2 }{\left| {\bf r}_0 - {\bf r}_j \right|}
\right| \Psi_a (1, \ldots, Z)
\right>
\label{6.47}\eeq
is the electrostatic potential of the target atom. Evidently, these are
the transition matrix elements \req{5.39} for elastic scattering in the
plane-wave Born approximation.

In what follows, we will consider the case of excitation or ionization
of a shell, $a=(n_a\ell_a)$, with $\nu_a$ electrons. For open
shells with $\nu_a < 2(2\ell_a+1)$ electrons, we assume that the shell
orbitals are all occupied with fractional occupancy, $\nu_a /
[2(2\ell_a+1)]$. The DDCS for excitation of the shell to any
unfilled orbital of the bound level $\varepsilon_{n_b \ell_b}$ is given
by [see Eq.\ \req{6.33}]
\beq
\frac{\d^2 \sigma^{\rm exc}_{a}}{\d W \, \d\hat{\bf k}'}
= \frac{(2\pi)^4}{\hbar v} \, k'\, \frac{M_0}{\hbar^2} \,
{\cal T}^{\rm pw,exc}_{fi} \, ,
\label{6.48}\eeq
where
\beqa
{\cal T}^{\rm pw,exc}_{fi}
&=& \frac{\nu_a}{2\ell_a+1}
\sum_{n_b, \ell_b} \, \frac{2(2\ell_b+1)-\nu_b}{2(2\ell_b+1)} \,
\delta(W - \varepsilon_{n_b \ell_b} + \varepsilon_{n_a \ell_a}) \,
\left( \frac{ Z_0 e^2}{2 \pi^2 q^2} \right)^2
\nonumber \\ [2mm]
&& \mbox{} \times
\sum_{m_a} \; \sum_{m_b} \left|
\left< \psi_{n_b \ell_b m_b}({\bf r}) \left|
\exp\left( {\rm i}  {\bf q} \dotprod {\bf r} \right) \left| \,
\psi_{n_a \ell_a m_a} ({\bf r})
\rule{0mm}{4mm}\right>\right.\right.
\right|^2 ,
\label{6.49}\eeqa
and $\nu_b$ is the number of electrons in the final level.
Similarly the DDCS for ionization of the shell is [see Eq.\ \req{6.36}]
\beq
\frac{\d^2 \sigma^{\rm ion}_a}{\d W \,\d\hat{\bf k}'}
= \frac{(2\pi)^4}{\hbar v} \, k'\, \frac{M_0}{\hbar^2} \,
{\cal T}^{\rm pw,ion}_{fi}
\label{6.50}\eeq
with
\beqa
{\cal T}_{fi}^{\rm pw,ion} &=&
\frac{2\me}{\pi \hbar^2 k_b} \,
\frac{\nu_a}{2\ell_a+1}
\left( \frac{ Z_0 e^2}{2 \pi^2 q^2} \right)^2
\nonumber \\ [2mm]
&& \mbox{} \times
\sum_{m_a}\sum_{\ell_b,m_b}
\left| \left< \psi_{\varepsilon_b \ell_b m_b} \left|
\exp\left( {\rm i}  {\bf q} \dotprod {\bf r} \right)
\rule{0mm}{4mm}\right|
\psi_{n_a \ell_a m_a} ({\bf r})
\right> \right|^2. \rule{10mm}{0mm}
\label{6.51}\eeqa

Introducing the recoil energy $Q$, defined by [cf.\ Eq.\ \req{A.81}],
\beq
Q \equiv \frac{\hbar^2 q^2}{2\me} =
\frac{\hbar^2 ({\bf k} - {\bf k}')^2}{2\me} =
\frac{\hbar^2}{2\me} \left( k^2 + k'^2 - 2 k k' \cos\theta \right),
\label{6.52}\eeq
and noting that
\beq
\d \hat{\bf k}' = 2\pi \, \sin \theta \, \d \theta = 2\pi \,
\frac{\me }{\hbar^2 k k'} \, \d Q
= 2\pi \, \frac{\me}{M_0 v \hbar k'} \, \d Q\, ,
\label{6.53}\eeq
the shell DDCS, including both excitation and ionization, can be written as
\beqa
\frac{\d^2 \sigma_a}{ \d W \, \d Q } &=& \left(
\frac{\d^2 \sigma^{\rm exc}_a}{ \d W \, \d \hat{\bf k}'}
+ \frac{\d^2 \sigma^{\rm ion}_a}{ \d W \, \d \hat{\bf k}'} \right)
\frac{\d \hat{\bf k}'}{\d Q}\rule{10mm}{0mm}
\nonumber \\ [2mm]
&=& \frac{2 \pi Z_0^2 e^4}{\me v^2} \, \frac{1}{WQ}
\, \frac{\d f_a(Q,W)}{\d W}.
\label{6.54}\eeqa
The function $\d f_a(Q,W)/\d W$ is the {\it generalized oscillator
strength} (GOS) of the shell, defined by
\beqa
\lefteqn{
\frac{\d f_a(Q,W)}{\d W} \equiv \frac{W}{Q}
\left( \frac{ Z_0 e^2}{2 \pi^2 q^2} \right)^{-2}
\left[ {\cal T}_{fi}^{\rm pw,exc} + {\cal T}_{fi}^{\rm pw,ion} \right]}
\nonumber \\ [2mm]
&=& \frac{W}{Q} \,
\frac{\nu_a}{2\ell_a+1}  \sum_{n_b, \ell_b} \,
\delta(W - \varepsilon_{n_b \ell_b} + \varepsilon_{n_a \ell_a}) \,
\frac{2(2\ell_b+1)-\nu_b}{2(2\ell_b+1)}
\nonumber \\ [2mm]
&& \mbox{} \rule{5mm}{0mm} \times
\sum_{m_a,m_b}
\left|
\left< \psi_{n_b \ell_b m_b}({\bf r}) \left|
\exp\left( {\rm i}  {\bf q} \dotprod {\bf r} \right) \left| \,
\psi_{n_a \ell_a m_a} ({\bf r})
\rule{0mm}{4mm}\right>\right.\right.
\right|^2
\nonumber \\ [2mm]
&& \mbox{} + \frac{W}{Q} \,
\frac{k_b}{\pi \varepsilon_b} \,
\frac{\nu_a}{2\ell_a+1}
\sum_{m_a}\sum_{\ell_b,m_b}
\left|
\left< \psi_{\varepsilon_b \ell_b m_b} \left|
\rule{0mm}{4mm}\exp\left( {\rm i}  {\bf q} \dotprod {\bf r} \right)
\right|
\psi_{n_a \ell_a m_a} ({\bf r})
\right>
\right|^2,
\label{6.55}\eeqa
where the continuum matrix elements are on the energy shell, \ie, with
\beq
\varepsilon_b = \frac{\hbar^2 k_b^2}{2\me} = \varepsilon_{n_a \ell_a} + W
= W - U_a,
\label{6.56}\eeq
where $U_a \equiv - \varepsilon_{n_a \ell_a}$ is the ionization energy
of the shell.

The DDCS for inelastic collisions with the atom is obtained by adding
the contributions from the different shells of the ground-state
configuration,\index{atomic configuration}

\beq
\frac{\d^2 \sigma}{ \d W \, \d Q }
= \sum_{n_a \ell_a} \frac{\d^2 \sigma_a}{ \d W \, \d Q }
= \frac{2 \pi Z_0^2 e^4}{\me v^2} \, \frac{1}{WQ}
\, \frac{\d f(Q,W)}{\d W} \, ,
\label{6.57}\eeq
where
\beq
\frac{\d f(Q,W)}{\d W} \equiv \sum_{n_a \ell_a} \frac{\d f_a(Q,W)}{\d W}
\label{6.58}\eeq
is the GOS of the atom.

The PWBA is expected to be valid for projectiles with sufficiently high
energies, whose wave functions are only slightly distorted by the
atomic potential. Qualitative arguments indicate that the approximation
is applicable when the speed $v$
of the projectile is much larger than the velocity $u$ of atomic (bound)
electrons \citep{MottMassey1965, Schiff1968}. In the least favorable case
of the K shell ($n=1$), the unscreened hydrogenic
model (Section \ref{sec3.2}) gives
\beq
u^2 = \frac{2 \left< K \right>}{\me} =
\frac{2}{\me} \left( \frac{Z^2}{2n^2} \, \frac{\me e^4}{\hbar^2}
\right),
\label{6.59}\eeq
and the validity condition for the PWBA reads
\beq
\frac{u^2}{v^2} = \frac{Z}{n^2 \beta^2} \, \frac{e^4}{\hbar^2 c^2} =
\frac{Z^2 \alpha^2}{n^2 \beta^2} \ll 1,
\label{6.60}\eeq
where $\alpha = e^2/(\hbar c)\simeq 1/137$ is the fine-structure
constant and $\beta=v/c$. Hence, for excitations of electrons in inner
subshells, the PWBA should be applicable when
\beq
\beta \gg \frac{Z \alpha}{n}.
\label{6.61}\eeq
\citet{Inokuti1971} pointed out that the equivalence of the PWBA and the
impact-parameter approximation \citep{BetheJackiw1997}
implies that the PWBA should be valid when the impact-parameter
approximation is applicable, that is, when
\beq
\beta \gg \frac{\me}{M_0} \, \frac{Z \alpha}{n}.
\label{6.62}\eeq
Additional quantitative indications
on the limits of validity of the PWBA can be obtained by comparing its
results with those from the more elaborate DWBA (see Section
\ref{sec6.1}). \citet{BoteSalvat2008} have performed PWBA and DWBA
calculations of ionization by electron impact for all the subshells of
atoms with $Z=1$ to 99 and concluded that the PWBA is valid (to within
2~\% or so) for electrons with kinetic energies larger than about 30 times
the ionization energy or, invoking again the hydrogenic
formula for the binding energy, when
\beq
v^2 \gtrsim 30 u^2 = 30 \, \frac{Z^2e^4}{n^2 \hbar^2}.
\nonumber \eeq
With evident rearrangements, this relation can be written as
\beq
\beta \gtrsim 6 \, \frac{Z\alpha}{n}\, ,
\label{6.63}\eeq
which agrees with the conditions stated above. Although the latter
result has been numerically verified only for electrons, we may conclude
that it is also valid for heavier projectiles, because their wave
functions are less distorted by the atomic field than those of electrons
(projectiles with larger masses accelerate less than electrons).


\subsection{The generalized oscillator strength \label{sec6.2.1}}
\index{generalized oscillator strength}

An appealing feature of the PWBA is that it can be formulated without
having to specify the details of the electronic structure of the target
system and, hence, it is easily generalized to other targets, such as
ions, molecules, and clusters of atoms. In this Section we describe the
general formulation of the PWBA for collisions of the projectile (mass
$M_0$ and charge $Z_0 e$) with a generic target system that, for the
sake of concreteness, we represent as a molecule with $Z$ bound
electrons. Because nuclei have masses much larger than the electron
mass $\me$, we can assume that the positions of the nuclei in the
molecule remain constant with time and approximate the Hamiltonian of
the molecule in the form
\beq
{\cal H}_{\rm mol} = \sum_{j=1}^Z \frac{1}{2\me} \breve{\bf p}_j^2 + V({\bf
r}_1, \ldots , {\bf r}_Z),
\label{6.64}\eeq
where the summation is over the $Z$ electrons, and the potential energy
$V$ depends only on the relative positions of the electrons and the
nuclei. Let us assume that we know a basis of eigenfunctions $\Psi_n(x_1,
\ldots, x_Z)$ of ${\cal H}_{\rm mol}$ with corresponding eigenvalues
$\varepsilon_n$, which we shall treat as discrete to simplify the
notation. To ease the arguments, we assume that the
molecule is initially in its ground state $\Psi_i$. A collision can
produce the excitation to any final level $\varepsilon_f$. We assume
that molecules are randomly oriented, to make sure that the DDCS does
not depend on the azimuthal scattering angle. Then, the DDCS obtained
from the PWBA can be expressed as [see Eq.\ \req{6.26}]
\beq
\frac{\d^2 \sigma_{fi}}{\d W \, \d\hat{\bf k}'}
= \frac{(2\pi)^4}{\hbar v}
\, k'\, \frac{M_0}{\hbar^2} \, \delta(W - \varepsilon_f +
\varepsilon_i) \, {\cal T}_{fi},
\label{6.65}\eeq
where
\beq
{\cal T}_{fi} =
\sum_{\Psi_f \in \varepsilon_f} |T_{fi}|^2
\label{6.66}\eeq
is the sum over states of the final level of the squared modulus of
the transition matrix elements,
\beq
T_{fi} = \frac{1}{(2\pi)^3}
\left< \exp({\rm i} {\bf k}' \dotprod {\bf r}_0)
\, \Psi_f(x_1, \ldots, x_Z) \left|
- \sum_{j} \frac{Z_0 e^2}{|{\bf r}_0 - {\bf r}_j|}
\right| \exp({\rm i} {\bf k} \dotprod {\bf r}_0)
\, \Psi_i(x_1, \ldots, x_Z) \right>.
\nonumber \eeq
Using the Bethe integral, Eq.\ \req{B.26}, we have
\beq
T_{fi} = - \, \frac{Z_0 e^2}{2\pi^2 q^2}
\left< \, \Psi_f(x_1, \ldots, x_Z) \left|
\sum_{j} \exp({\rm i} {\bf q} \dotprod {\bf r}_j)
\right| \Psi_i(x_1, \ldots, x_Z) \right>,
\label{6.67}\eeq
and, with the aid of the relation \req{6.53}, the DDCS becomes
\beq
\frac{\d^2 \sigma^{\rm (pw)}}{\d W \, \d Q} =
\frac{2\pi \, Z_0^2 e^4}{\me v^2} \, \frac{1}{WQ} \,
\frac{\d f(Q,W)}{\d W} \,
\label{6.68}\eeq
with the GOS defined as
\beqa
\frac{\d f(Q,W)}{\d W} &=& \frac{W}{Q}
\sum_{\Psi_f \in \varepsilon_f}
\delta(W - \varepsilon_f+\varepsilon_i)
\nonumber \\ [2mm]
&& \mbox{} \times
\left|
\left<\Psi_f(x_1, \ldots, x_Z) \left|
\sum_{j=1}^Z \exp({\rm i} {\bf q} \dotprod {\bf r}_j)
\right| \Psi_i(x_1, \ldots, x_Z) \right>
\right|^2,
\label{6.69}\eeqa
where the matrix elements are on the energy shell, that is, with
$\varepsilon_f = \varepsilon_i + W$. This definition is consistent with
that of the atomic GOS obtained from the independent-electron
approximation, Eqs.\ \req{6.55} and \req{6.58}.


\subsubsection{Collisions with electrons at rest \label{sec6.2.1.1}}
\index{quantum inelastic collisions!collisions with electrons at rest}

Collisions of charged particles with free electrons at rest are of
special interest because they admit a simple analytical formulation and
because the associated DCS approximates the DCS of collisions with bound
electrons involving recoil energies much larger than the binding
energies of the target electrons. We wish to calculate the DCS
from the PWBA, by assuming that the projectile is distinguishable from
the target electron. The unperturbed states of both the
projectile and the target electron are free states ($V=0$) which we will
represent as plane waves. Before the collision, the projectile has
kinetic energy $E$ and momentum $\hbar {\bf k}$; the corresponding
quantities after the collision are $E'$ and $\hbar {\bf k}'$. As in the
foregoing study, the
initial and final states of the projectiles are described as plane waves
of the form
$$
\phi_{{\bf k}}({\bf r}) = (2\pi)^{-3/2}
\exp({\rm i} {\bf k} \cdot {\bf r}) \, .
\eqno{\req{6.44}}$$
To keep the analogy with the calculation of collisions with bound
atomic electrons, with initial states normalized to unity, the states of the
target electron need to be described as plane waves
satisfying periodic boundary conditions on a cubic box of side $L$,
\beq
\phi_{L,{\bf k}}({\bf r}) = L^{-3/2}
\exp({\rm i} {\bf k} \cdot {\bf r}) \, ,
\label{6.70}\eeq
which represent one electron in the normalization box. The initial and
final states of the target electron are plane waves with respective
kinetic energies $\epsilon_a=0$ and $\epsilon_b = W$ and wave numbers
${\bf k}_a =0$ and ${\bf k}_b$. Since the operator in Eq.\ \req {6.69} is
independent of the spin, non-vanishing contributions
arise only from transitions that conserve the spin of the target
electron. Therefore, we can disregard the electron spin and express the GOS
of the process as
\beqa
\frac{\d f(Q,W)}{\d W} &=& \frac{W}{Q}
\sum_{{\bf k}_b}
\delta \left( W - \frac{(\hbar k_{\rm b})^2}{2\me} \right)
\left| \left< \phi_{L,{\bf k}_b}({\bf r}) \left|
\exp({\rm i} {\bf q} \dotprod {\bf r})
\right| \phi_{L,{\bf 0}}({\bf r}) \right> \right|^2
\nonumber \\ [2mm]
&=& \frac{W}{Q}
\sum_{{\bf k}_b}
\delta \left( W - \frac{(\hbar k_{\rm b})^2}{2\me} \right)
\left| \frac{1}{L^{3}} \int_{L^3} \exp \left[{\rm i} ({\bf q}-{\bf k}_b)
\dotprod {\bf r} \right] \, \d {\bf r}
\right|^2
\nonumber \\ [2mm]
&=& \frac{W}{Q}
\sum_{{\bf k}_b}
\delta \left( W - \frac{(\hbar k_{\rm b})^2}{2\me} \right)
\left| \delta_{{\bf q},{\bf k}_b} \right|^2.
\nonumber \eeqa
In the limit $L\rightarrow \infty$, the distribution of allowed wave
numbers ${\bf k}_b$ becomes continuous and we obtain
\beq
\frac{\d f(Q,W)}{\d W} = \delta(W-Q),
\label{6.71}\eeq
It is worth observing that the PWBA then yields the following DDCS for
collisions of the projectile with free electrons at rest [Eq.\
\req{6.68}]
\beq
\frac{\d^2 \sigma^{\rm (pw)}}{\d W \, \d Q} =
\frac{2\pi \, Z_0^2 e^4}{\me v^2} \, \frac{1}{WQ} \,
\delta(W-Q).
\label{6.72}\eeq
Because the plane waves are exact solutions of the Schr\"{o}dinger
equation for free electrons, the PWBA and the DWBA lead to the same DDCS
for collisions with free electrons at rest.


\subsubsection{The optical limit \label{sec6.2.1.2}}
\index{optical oscillator strength}

With the aim of determining the limiting value of the GOS at $Q=0$, we
expand the exponentials in the matrix elements \req{6.69},
\beqa
\lim_{Q\rightarrow 0} \frac{\d f(Q,W)}{\d W} &=&
\lim_{Q\rightarrow 0} \frac{W}{Q} \,
\sum_{\Psi_f \in \varepsilon_f}
\delta(W - \varepsilon_i+\varepsilon_f)
\left| \left< \Psi_f \left| Z + {\bf q}
\dotprod \sum_{j} {\bf r}_j
\right| \Psi_i \right> \right|^2.
\nonumber \eeqa
The first term does not contribute because of the orthogonality of the
states $\Psi_n$, and
\beqa
\lim_{Q\rightarrow 0} \frac{\d f(Q,W)}{\d W} &=&
\frac{2\me}{\hbar^2} \, W \,
\sum_{\Psi_f \in \varepsilon_f}
\delta(W - \varepsilon_i+\varepsilon_f)
\left| \hat{\bf q} \dotprod \left< \Psi_f \left|
{\bf D} \right| \Psi_i \right> \right|^2,
\nonumber \eeqa
where
\beq
{\bf D} \equiv  {\sum}_{j} {\bf r}_j
\label{6.73}\eeq
is the dipole-moment operator. We have
\beqa
\left| \left< \Psi_f \left| {\bf D} \right| \Psi_i \right> \right|^2
&\equiv&
\left< \Psi_f \left| {\bf D} \right| \Psi_i \right> \dotprod
\left< \Psi_i \left| {\bf D} \right| \Psi_f \right>
=  \left< \Psi_f \left| D_x \right| \Psi_i \right> \dotprod
\left< \Psi_i \left| D_x \right| \Psi_f \right>
\nonumber \\ [2mm]
&& \mbox{} + \left< \Psi_f \left| D_y \right| \Psi_i \right> \dotprod
\left< \Psi_i \left| D_y \right| \Psi_f \right>
+ \left< \Psi_f \left| D_z \right| \Psi_i \right> \dotprod
\left< \Psi_i \left| D_z \right| \Psi_f \right>
\nonumber \\ [2mm]
&=&
\left| \left< \Psi_f \left| D_x \right| \Psi_i \right> \right|^2
+ \left| \left< \Psi_f \left| D_y \right| \Psi_i \right> \right|^2
+ \left| \left< \Psi_f \left| D_z \right| \Psi_i \right> \right|^2,
\nonumber \eeqa
and we can write
\beq
\left| \hat{\bf q} \dotprod \left< \Psi_f \left|
{\bf D} \right| \Psi_i \right> \right|^2
= \left| \left< \Psi_f \left|
{\bf D} \right| \Psi_i \right> \right|^2 \cos^2 \theta,
\label{6.74}\eeq
where $\theta$ is the angle between the vectors $\hat{\bf q}$ and
$\left< \Psi_f \left| {\bf D} \right| \Psi_i \right> $.
Because the target molecules are randomly oriented, we can replace
$\cos^2 \theta$ with its average over possible orientations,
\beq
\frac{1}{4\pi} \int \cos^2 \theta \; 2 \pi \, \sin\theta\, \d \theta =
\frac{1}{2} \int_{-1}^1 \cos^2\theta \, \d( \cos \theta)  = \frac{1}{3},
\nonumber\eeq
and write
\beq
\left| \hat{\bf q} \dotprod \left< \Psi_f \left|
{\bf D} \right| \Psi_i \right> \right|^2
= \frac{1}{3} \, \left| \left< \Psi_f \left|
{\bf D} \right| \Psi_i \right> \right|^2.
\label{6.75}\eeq
Therefore,
\beq
\lim_{Q\rightarrow 0} \frac{\d f(Q,W)}{\d W} =
\frac{2\me}{3 \hbar^2} \, W \,
\sum_{\Psi_f \in \varepsilon_f}
\delta(W - \varepsilon_i+\varepsilon_f)
\left| \left< \Psi_f \left|
{\bf D} \right| \Psi_i \right> \right|^2.
\label{6.76}\eeq
The last expression is the definition of the {\it dipole} or {\it
optical oscillator strength} (OOS) of the molecule \citep[see,
\eg,][]{BransdenJoachain1983},
\beq
\lim_{Q\rightarrow 0} \frac{\d f(Q,W)}{\d W}  =
\frac{\d f(W)}{\d W}\, .
\label{6.77}\eeq
The oscillator strength is a classical concept (see Chapter
\ref{chapt9}) that was used well before the application of the PWBA to
collision processes by \citet{Bethe1930}; its coincidence at $Q=0$ with
the generalized oscillator strength motivates the name of the latter.


\subsubsection{The Bethe sum rule \label{sec6.2.1.3}}
\index{Bethe sum rule}

The GOS satisfies a sum rule which plays a central role in the derivation
of the Bethe formula for the stopping power of matter for high-energy
charged projectiles (see Section \ref{sec6.9}). To derive that sum rule,
we write the GOS, Eq.\ \req{6.69}, in the form
\beq
\frac{\d f(Q,W)}{\d W} = \frac{1}{Q} \,
\sum_{\Psi_f \in \varepsilon_f}
\delta(W - \varepsilon_i+\varepsilon_f) \, {\cal P}_{fi}
\label{6.78}\eeq
with
\beqa
{\cal P}_{fi} &\equiv& W  \left|
\left<\Psi_f \left|
\sum_{j} \exp({\rm i} {\bf q} \dotprod {\bf r}_j)
\right| \Psi_i \right> \right|^2
\nonumber \\ [2mm]
&=& \left<\Psi_i \left|
\sum_{j} \exp(-{\rm i} {\bf q} \dotprod {\bf r}_j)
\right| \Psi_f \right> \,
\left<\Psi_f \left| \left[ {\cal H}_{\rm mol},
\sum_{j} \exp({\rm i} {\bf q} \dotprod {\bf r}_j)
\right] \right| \Psi_i \right>.
\label{6.79}\eeqa
The operator $\left[ {\cal H}_{\rm mol}, {\sum}_{j} \exp({\rm i} {\bf q}
\dotprod {\bf r}_j) \right]$ acts on the molecular wave functions as
follows,
\beqa
\left[ {\cal H}_{\rm mol}, {\sum}_{j} \exp({\rm i} {\bf q}
\dotprod {\bf r}_j) \right] \Psi
&=& \sum_{j} \frac{1}{2\me} \left[ \breve{\bf p}_j \dotprod
\breve{\bf p}_j ,
\sum_{l} \exp({\rm i} {\bf q}
\dotprod {\bf r}_l) \right] \Psi
\nonumber \\ [2mm]
&=& \sum_{j} \frac{1}{2\me} \left[ \breve{\bf p}_j \dotprod
\breve{\bf p}_j ,
\exp({\rm i} {\bf q} \dotprod {\bf r}_j) \right] \Psi
\nonumber \\ [2mm]
&=& \frac{1}{2\me} \sum_{j}
\left\{ (\hbar q)^2
\exp({\rm i} {\bf q} \dotprod {\bf r}_j)
+ 2 \exp({\rm i} {\bf q} \dotprod {\bf r}_j) \,
\hbar {\bf q} \dotprod
\breve{\bf p}_j \right\} \Psi.
\nonumber \eeqa
Therefore
\beq
\left[ {\cal H}_{\rm mol}, {\sum}_{j} \exp({\rm i} {\bf q}
\dotprod {\bf r}_j) \right] = \frac{1}{2\me} \sum_{j}
\left\{ (\hbar q)^2
\exp({\rm i} {\bf q} \dotprod {\bf r}_j)
+ 2 \exp({\rm i} {\bf q} \dotprod {\bf r}_j) \,
\hbar {\bf q} \dotprod \breve{\bf p}_j \right\},
\label{6.80}\eeq
and we have
\beqa
{\cal P}_{fi} &=& \frac{1}{2\me} \, (\hbar q)^2 \left<\Psi_i \left|
\sum_{j} \exp(-{\rm i} {\bf q} \dotprod {\bf r}_j)
\right| \Psi_f \right> \,
\left<\Psi_f \left| \sum_{l} \exp({\rm i} {\bf q} \dotprod {\bf r}_l)
\right| \Psi_i \right>
\nonumber \\ [2mm]
&& \mbox{} + \frac{\hbar}{2\me} \, \hbar {\bf q} \dotprod
\left<\Psi_i \left|
\sum_{j} \exp(-{\rm i} {\bf q} \dotprod {\bf r}_j)
\right| \Psi_f \right> \, \left<\Psi_f \left| \sum_{l}
\exp({\rm i} {\bf q} \dotprod {\bf r}_l)
\, 2 \breve{\bf p}_j \right| \Psi_i \right>.\rule{10mm}{0mm}
\label{6.81}\eeqa

We can now calculate the integral
\beqa
S_0(Q) &\equiv& \int_0^\infty \frac{\d f(Q,W)}{\d W} \, \d W
\nonumber \\ [2mm]
&=&
\int_0^\infty \frac{1}{Q} \,
\sum_{\Psi_f \in \varepsilon_f}
\delta(W - \varepsilon_f+\varepsilon_i) {\cal P}_{fi} \, \d W
\nonumber \\ [2mm]
&=& \frac{1}{Q} \frac{1}{2\me} \sum_{\Psi_f} \left\{ (\hbar q)^2
\left<\Psi_i \left|
\sum_{j} \exp(-{\rm i} {\bf q} \dotprod {\bf r}_j)
\right| \Psi_f \right> \,
\left<\Psi_f \left| \sum_{l} \exp({\rm i} {\bf q} \dotprod {\bf r}_l)
\right| \Psi_i \right> \right.
\nonumber \\ [2mm]
&& \mbox{} \left. +  2 \hbar {\bf q} \dotprod \left<\Psi_i \left|
\sum_{j} \exp(-{\rm i} {\bf q} \dotprod {\bf r}_j)
\right| \Psi_f \right> \, \left<\Psi_f \left| \sum_{l} \exp({\rm i}
{\bf q} \dotprod {\bf r}_l)
\, \breve{\bf p}_l \right| \Psi_i \right> \right\},
\nonumber\eeqa
where the sum runs over all states. Notice that we have included
a null term with $\Psi_f=\Psi_i$.
Using the completeness property of molecular states,
\beq
\sum_{\Psi_f}  \left| \Psi_f \right> \left<\Psi_f \right| = I
\nonumber\eeq
we can write
\beqa
S_0(Q) &=& \frac{1}{Q} \, \frac{1}{2\me}
\left\{ (\hbar q)^2
\left<\Psi_i \left|
\sum_{j,l} \exp\left[ -{\rm i} {\bf q} \dotprod ({\bf r}_j - {\bf r}_l)
\rule{0mm}{3.5mm} \right]
\right| \Psi_i \right> \right.
\nonumber \\ [2mm]
&& \mbox{} \left. + 2
\hbar {\bf q} \dotprod
\left<\Psi_i \left|
\sum_{j,l} \exp \left[ -{\rm i} {\bf q} \dotprod ({\bf r}_j-{\bf r}_l)
\rule{0mm}{3.5mm} \right]
\, \breve{\bf p}_l \right| \Psi_i \right> \right\}
\nonumber \\ [2mm]
&=& \frac{1}{q^2} \sum_{j,l}
\int \d \tau \, \Psi^\ast_i(\tau)
\exp\left[ -{\rm i} {\bf q} \dotprod ({\bf r}_j - {\bf r}_l)
\rule{0mm}{3.5mm} \right]
\left[ q^2 - 2 {\rm i} {\bf q} \dotprod
\nablab_l \right] \Psi_i(\tau),
\label{6.82}\eeqa
where $\tau$ represents the set of space and spin variables of the
electrons; we have used the definition $Q=(\hbar q)^2/2\me$. Since the
wave function of the ground state can always be taken to be
real\footnote{Because if $\Psi$ is a solution of the time-independent
Schr\"{o}dinger equation with the Hamiltonian \req{6.64}, its complex
conjugate $\Psi^\ast$ also satisfies that equation, and so does their sum
$\Psi +\Psi^\ast$, which is real.}, we can
write
\beq
S_0(Q)
= \frac{1}{q^2} \sum_{j,l}
\int \d \tau \,
\exp\left[ -{\rm i} {\bf q} \dotprod ({\bf r}_j - {\bf r}_l)
\rule{0mm}{3.5mm} \right]
\left[\rule{0mm}{3mm} q^2 - {\rm i} {\bf q} \dotprod
\nablab_l \right] \Psi_i^2(\tau),
\label{6.83}\eeq
Integration by parts of the second term
gives
\beqa
S_0(Q) &=& \frac{1}{q^2} \; \sum_{j \ne l} \left\{
 q^2 \int \d \tau \,
\exp\left[ -{\rm i} {\bf q} \dotprod ({\bf r}_j - {\bf r}_l)
\rule{0mm}{3.5mm} \right] \,
\Psi^2_i(\tau) \right.
\nonumber \\ [2mm]
&& \mbox{} \left. + {\rm i} {\bf q} \dotprod
\int \d \tau \, \left( {\rm i} {\bf q} \right) \,
\exp\left[ -{\rm i} {\bf q} \dotprod ({\bf r}_j - {\bf r}_l)
\rule{0mm}{3.5mm}\right] \,
\Psi^2_i(\tau) \right\}
\nonumber \\ [2mm]
&& \mbox{} + \frac{1}{q^2} \; \sum_{j}
\int \d \tau \, \left\{ q^2 \Psi^2_i(\tau)
- {\rm i} \left[ {\bf q} \dotprod \nablab_j \Psi_i(\tau)
\rule{0mm}{3.5mm}\right]
\Psi_i (\tau) \rule{0mm}{4mm}\right\}
\nonumber \\ [2mm]
&=& \frac{1}{q^2} \; \sum_{j}
\int \d \tau \, \left\{ q^2 \Psi^2_i(\tau)
- {\rm i} \left[ {\bf q} \dotprod \nablab_j \Psi_i(\tau)
\rule{0mm}{3.5mm}\right]
\Psi_i (\tau) \rule{0mm}{4mm}\right\}.
\label{6.84}\eeqa
The integral of the last term,
\beq
\int \d \tau \; \Psi_i(\tau) \, \nablab_I \Psi_i(\tau),
\nonumber\eeq
vanishes because, for a bound state $\Psi_i$, integration by parts gives
the same expression with reversed sign. We thus obtain the celebrated
{\it Bethe sum rule},
\beq
S_0(Q) \equiv \int_0^\infty \frac{\d f(Q,W)}{\d W} \, \d W
= Z \qquad \mbox{for all $Q$.}
\label{6.85}\eeq
In the limit $Q=0$, this important sum rule reduces to the {\it
Thomas-Reiche-Kuhn} sum rule \citep[see, \eg,][]{BransdenJoachain1983}.


\subsection{Generalized oscillator strength of atoms \label{sec6.2.2}}
\index{generalized oscillator strength!calculation for atoms}

We now return to the calculation of the GOS of atoms by considering atomic
wave functions described within the non-relativistic independent-electron
approximation. The relevant matrix elements are of the type [see Eq.\
\req{6.55}]
\beqa
{\cal A}_{ba} &=&
\left< \psi_{\varepsilon_b \ell_b m_b}({\bf r}) \left|
\exp\left( {\rm i}  {\bf q} \dotprod {\bf r} \right) \left| \,
\psi_{n_a \ell_a m_a} ({\bf r})
\rule{0mm}{4mm}\right> \right.\right.
\nonumber \\ [2mm]
&=& \int \d r \int \d \hat{\bf r} \;
P_{\varepsilon_b \ell_b}(r) \, Y_{\ell_b m_b}^\ast (\hat{\bf r})
\; \exp\left( {\rm i}  {\bf q} \dotprod {\bf r} \right) \,
P_{n_a \ell_a}(r) \, Y_{\ell_a m_a} (\hat{\bf r}) .
\label{6.86}\eeqa
Introducing the Rayleigh expansion [cf.\ Eq.\ \req{2.62}],
\beq
\exp({\rm i} {\bf q} \dotprod {\bf r} ) = \sum_{\ell = 0}^\infty
{\rm i}^\ell(2\ell +1) j_\ell(qr) \, \sum_{m=-\ell}^\ell
C^\ast_{\ell m}(\hat{\bf q}) \, C_{\ell m} (\hat{\bf r}) ,
\label{6.87}\eeq
where $j_\ell({x})$ are the spherical Bessel functions and
\beq
C_{\ell m}(\hat{\bf r}) = \sqrt{\frac{4\pi}{2\ell+1}} \,
Y_{\ell m} (\hat{\bf r})
\label{6.88}\eeq
are the components of the Racah spherical tensors, we have
\beqa
{\cal A}_{ba} &=& \sum_\ell {\rm i}^\ell (2\ell+1)
\left( \int
P_{\varepsilon_b \ell_b}(r) \, j_\ell(qr) \, P_{n_a \ell_a}(r) \, \d r
\right)
\nonumber \\ [2mm]
&& \mbox{} \times
\left(\int Y_{\ell_b m_b}^\ast(\hat{\bf r}) \,
C_{\ell m} (\hat{\bf r}) \,
Y_{\ell_a m_a} (\hat{\bf r}) \,  \d \hat{\bf r} \right)
C^\ast_{\ell m}(\hat{\bf q}).
\label{6.89}\eeqa
It is customary to define the radial integrals
\beq
R^\ell_{\varepsilon_b \ell_b; n_a \ell_a} (q) \equiv
\int
P_{\varepsilon_b \ell_b}(r) \, j_\ell(qr) \, P_{n_a \ell_a}(r) \, \d r
\label{6.90}\eeq
and the coefficients
\beqa
c^\ell_m(\ell_b m_b ; \ell_a m_a) &\equiv&
\int Y_{\ell_b m_b}^\ast(\hat{\bf r}) \,
C_{\ell m} (\hat{\bf r}) \,
Y_{\ell_a m_a} (\hat{\bf r}) \,  \d \hat{\bf r}
\nonumber \\ [2mm]
&=& \frac{1}{\sqrt{2\ell_b+1}} \left< \ell_a \ell m_a m
\left| \ell_b m_{{\rm L}_b} \right> \right.
\, \langle \ell_b || C^{(\ell)} || \ell_a \rangle ,
\label{6.91}\eeqa
where $\left< \ell_a \ell m_a m
| \ell_b m_{{\rm L}_b} \right>$ are Clebsch-Gordan
coefficients \citep{Edmonds1960, Rose1995} and
\beq
\langle \ell_b || C^{(\ell)} || \ell_a \rangle =
\sqrt{2\ell_a+1} \left< \ell_a \ell 0 0
\left| \ell_b 0 \right> \right.
\label{6.92}\eeq
are the reduced matrix elements of the Racah tensor $C^{(\ell)}$. With all
this, we can write
\beqa
{\cal A}_{ba} &=& \sum_\ell {\rm i}^\ell (2\ell+1)
R^\ell_{\varepsilon_b \ell_b; n_a \ell_a} (q) \,
c^\ell_m(\ell_b m_b ; \ell_a m_a) \,
C^\ast_{\ell m}(\hat{\bf q}).
\label{6.93}\eeqa

The expression \req{6.55} of the GOS contains the double sum
\beqa
{\cal F}_{ba} &\equiv& \sum_{m_a,m_b}
\left| {\cal A}_{ba} \right|^2 = \sum_{\ell, m} \; \sum_{\ell',m'}
{\rm i}^{-\ell+\ell'} (2\ell+1) (2\ell'+1)
R^\ell_{\varepsilon_b \ell_b; n_a \ell_a} (q) \,
R^{\ell'}_{\varepsilon_b \ell_b; n_a \ell_a} (q) \,
\nonumber \\ [2mm]
&& \mbox{} \times \left( \sum_{m_a,m_b}
c^\ell_m(\ell_b m_b ; \ell_a m_a) \,
c^{\ell'}_{m'}(\ell_b m_b ; \ell_a m_a) \,
\right) C_{\ell m}(\hat{\bf q}) C^\ast_{\ell' m'}(\hat{\bf q}).
\nonumber \eeqa
The properties of the Clebsch-Gordan coefficients imply that
\beq
\sum_{m_a,m_b}
c^\ell_m(\ell_b m_b ; \ell_a m_a) \,
c^{\ell'}_{m'}(\ell_b m_b ; \ell_a m_a) =
\frac{2\ell_a+1}{2\ell+1} \, \langle \ell_a \ell 0 0 | \ell_b 0 \rangle \,
\delta_{\ell,\ell'} \, \delta_{m,m'}.
\label{6.94}\eeq
On the other hand, from the theorem of addition of spherical harmonics
[Eq.\ \req{B.57}], we have
\beq
\sum_m C_{\ell m}(\hat{\bf q})
C^\ast_{\ell m}(\hat{\bf q})=P_\ell(1)=1.
\label{6.95}\eeq
Hence,
\beqa
{\cal F}_{ba} &=& \sum_{\ell} (2 \ell +1) (2\ell_a+1)
\langle \ell_a \ell 0 0 | \ell_b 0 \rangle^2 \,
\left[ R^\ell_{\varepsilon_b \ell_b; n_a \ell_a} (q) \right]^2.
\label{6.96}\eeqa
Introducing this result into expression \req{6.55}, the GOS of a
shell $a=(n_a \ell_a)$ with $\nu_a$ electrons takes the form
\beqa
\frac{\d f_a(Q,W)}{\d W} &\equiv& \frac{W}{Q} \,
\nu_a \sum_{n_b, \ell_b}
\delta(W - \varepsilon_{n_b \ell_b} + \varepsilon_{n_a \ell_a}) \,
\frac{2(2\ell_b+1)-\nu_b}{2(2\ell_b+1)} \,
\nonumber \\ [2mm]
&& \mbox{} \rule{5mm}{0mm} \times
\sum_{\ell} (2 \ell +1)
\langle \ell_a \ell 0 0 | \ell_b 0 \rangle^2 \,
\left[ R^\ell_{n_b \ell_b; n_a \ell_a} (q) \right]^2
\rule{20mm}{0mm}
\nonumber \eeqa
\vspace*{-4mm}
\beq
+ \frac{W}{Q} \, \nu_a \,
\frac{k_b}{\pi \varepsilon_b}
\sum_{\ell_b}
\sum_{\ell}  (2 \ell +1)
\langle \ell_a \ell 0 0 | \ell_b 0 \rangle^2 \,
\left[ R^\ell_{\varepsilon_b \ell_b; n_a \ell_a} (q) \right]^2.
\rule{1mm}{0mm}
\label{6.97}\eeq
It is worth noticing that this GOS is independent of the direction of
the momentum transfer vector, ${\bf q}$, as anticipated.

Incidentally, we can now obtain the OOS
as the $Q\rightarrow 0$ limit of the GOS. Using the
expansion of the spherical Bessel functions for small arguments \citep[see,
\eg,][]{AbramowitzStegun1974},
\beq
j_\ell(x) = \frac{x^\ell}{(2\ell+1)!!}
\left[ 1 - \frac{x^2/2}{1! (2\ell+3)}
+ \frac{(x^2/2)^2}{2! (2\ell+3) (2\ell+5)} - \cdots \right],
\label{6.98}\eeq
we have
\beq
\lim_{q\rightarrow 0}
R_{\varepsilon_b\ell_b;n_a\ell_a}^{\ell}(q) =
\frac{q^\ell}{(2\ell+1)!!} \int_0^\infty
P_{\varepsilon_b\ell_b}(r) \, P_{n_a\ell_a}(r) \, r^\ell \, \d r.
\label{6.99}\eeq
Because of the orthogonality of the initial and final orbitals, terms
with $\ell=0$ do not contribute. The lowest-order non-vanishing
contributions are from the dipole terms ($\ell=1$) and give
\beqa
\frac{\d f_a (W)}{\d W} &\equiv&
\lim_{Q\rightarrow 0} \frac{\d f_a(Q,W)}{\d W} =
\frac{W \, 2\me}{3 \hbar^2}\,
\nu_a \sum_{n_b, \ell_b}
\frac{2(2\ell_b+1)-\nu_b}{2(2\ell_b+1)} \,
\nonumber \\ [2mm]
&& \mbox{} \times \delta(W -\varepsilon_b+\varepsilon_{n_a \ell_a})
\langle \ell_a \ell 0 0 | \ell_b 0 \rangle^2 \,
\left[ R_{n_b\ell_b;n_a\ell_a}^{\rm dip} \right]^2
\nonumber \\ [2mm]
&& \mbox{} + \frac{W \, 2\me}{3 \hbar^2}\, \nu_a \,
\frac{k_b}{\varepsilon_b\pi} \sum_{\ell_b}
\langle \ell_a \ell 0 0 | \ell_b 0 \rangle^2 \,
\left[ R_{\varepsilon_b\ell_b;n_a\ell_a}^{\rm dip} \right]^2
\label{6.100}\eeqa
with
\beqa
R_{\varepsilon_b\ell_b;n_a\ell_a}^{\rm dip} =
\int_0^\infty
P_{\varepsilon_b\ell_b}(r) \, P_{n_a\ell_a}(r) \, r \, \d r.
\label{6.101}\eeqa

The expression \req{6.97} is suited for evaluation of the shell GOS,
which for realistic atomic potentials must be performed numerically. The
only atomic systems for which the GOS is known in analytical form are the
hydrogenic ions with point nuclei (see Section \ref{sec3.2}). The
energy levels $\varepsilon_n$ of a hydrogenic ion
with nuclear charge $Z e$ are given by the Bohr formula, Eq.\
\req{3.28}. Because they depend only on the principal quantum number $n$, they
are degenerate in $\ell$, and represent the so-called gross-structure
energy spectrum. The GOS for excitations from the ground level, $\varepsilon_1$,
to bound levels, including the contributions of all degenerate final
states, is given by \citep{Bethe1930}\index{hydrogenic ions!generalized
oscillator strength}
\begin{subequations}
\label{6.102}
\beqa
\frac{\d f(Q,W)}{\d W} &=& \sum_{n=2}^\infty
\frac{256}{3} \, n^5 \, (n^2 -1) \,
\left[ (n^2-1) + 3 n^2 Q' \right] \,
\left[ (n-1)^2 + n^2 Q' \right]^{n-3} \, \nonumber \\ [2mm]
&& \mbox{} \times \, \left[ (n+1)^2 + n^2 Q' \right]^{-n-3} \, \delta
(W - \varepsilon_n + \varepsilon_1),
\label{6.102a}\eeqa
where
\beq
Q'=\frac{Q}{U_1} \qquad \mbox{and} \qquad
U_1 = \left| \varepsilon_1 \right|.
\label{6.102b}\eeq
\end{subequations}
The GOS for ionization ($W > U_1$) is \citep[see][and
references therein]{Heredia-Avalos2005}
\begin{subequations}
\label{6.103}
\beq
\frac{\d f(Q,W)}{\d W} = \frac{2^7}{U_1} \, W' \, {\cal A}(Q',W') \,
\frac{Q' + W'/3}{\left[ (Q'-W')^2 + 4 Q' \right]^3}\, ,
\label{6.103a}\eeq
with
\beq
{\cal A}(Q',W') = \exp \left[ - \frac{2}{\kappa} \arctan\left(
\frac{2\kappa}{Q'-W'+2} \right) \right]  \left[ 1 - \exp\left( - \,
\frac{2\pi}{\kappa} \right) \right]^{-1},
\label{6.103b}\eeq
where
\beq
W' \equiv \frac{W}{U_1}
\qquad \mbox{and} \qquad
\kappa^2 = W' -1.
\label{16.103c}\eeq
The value of the $\arctan$ function in Eq.\ \req{6.103b} is the one in
the interval ($0,\pi$). Excitations to discrete bound levels (with $W <
U_1$) correspond to negative $\kappa^2$; their contribution to the GOS
can be approximately described as a continuum by extrapolating the
ionization GOS to energy transfers $W$ below $U_1$ as
\citep{Egerton2009}
\beq
{\cal A}(Q',W') = \exp \left[ - \frac{1}{\sqrt{-\kappa^2}} \,
\ln\left( \frac{Q'-W'+2 + 2 \sqrt{-\kappa^2}}{Q'-W'+2 - 2
\sqrt{-\kappa^2}} \right) \right].
\label{6.103e}\eeq
\end{subequations}

The GOS can be represented as a surface on the $(Q,W)$ plane, the
so-called Bethe surface \citep{Inokuti1971}, which contains all
necessary information for describing the effect of inelastic collisions
{\it on the projectile}. A conspicuous feature of the Bethe surface is the
so-called {\it Bethe ridge}, a prominent maximum that peaks near the
line $W=Q$ for energy losses larger than about $2U_a$. This maximum
corresponds to close collisions and has a certain structure that
reflects the distribution of velocities of the electrons in the active
subshell (see Section \ref{sec6.4}). For small recoil energies $Q$, the
Bethe surface is flat, \ie, varies slowly with $Q$.  In the limit $Q
\rightarrow 0$, the GOS equals the OOS.


\subsubsection{The Bethe sum rule for atoms \label{sec6.2.2.1}}
\index{Bethe sum rule!for atoms}

It is interesting to study the contributions of the various electron
shells to the Bethe sum rule. The calculation reveals that the
sum rule is not satisfied by the GOSs of separate shells, although it is
obeyed by the atomic GOS.

Let us consider a closed shell $a=(n_a \ell_a)$ with $\nu_a = 2
(2\ell_a+1)$ electrons and ionization energy $U_a = - \varepsilon_{n_a
\ell_a}$. We wish to calculate the integral
\beqa
S'_0(a;Q) &=& \int \frac{\d f_a(Q,W)}{\d W} \, \d W =
\frac{W}{Q}\,2
 \sum_{n_b,\ell_b}
\sum_{m_a,m_b} \left| \left< \psi_{n_b \ell_b m_b}
\left| \rule{0mm}{4mm}\exp\left( {\rm i}
{\bf q} \dotprod {\bf r} \right) \right| \psi_{n_a \ell_a m_a}
\right> \right|^2
\nonumber \\ [2mm]
&+&
\int_0^\infty \frac{W}{Q}\,
\frac{k_b}{\varepsilon_b \pi} \, 2
\sum_{\ell_b,m_b} \sum_{m_a}
\left| \left<
\psi_{\varepsilon_b \ell_b m_b}
\left| \rule{0mm}{4mm}\exp\left( {\rm i}
{\bf q} \dotprod {\bf r} \right) \right| \psi_{n_a \ell_a m_a}
\right> \right|^2 \, \d W,
\label{6.104}\eeqa
which includes the GOSs of transitions from the active level
$\varepsilon_{n_a\ell_a}$ to {\it all} possible final levels $\varepsilon_b$,
bound and free, even to those corresponding to occupied shells, which
are forbidden by Pauli's exclusion principle.
We first evaluate the quantities
\beq
{\cal P}_{ba} \equiv (\varepsilon_b-\varepsilon_{n_a \ell_a})
\left| \left< \psi_b
\left| \rule{0mm}{4mm}\exp\left( {\rm i}
{\bf q} \dotprod {\bf r} \right) \right| \psi_{n_a \ell_a m_a}
\right> \right|^2
\label{6.105}\eeq
by following the same calculation scheme as in Section \ref{sec6.2.1.3}. That
is, we write
\beq
{\cal P}_{ba} = \left< \psi_{n_a \ell_a m_a}
\left| \rule{0mm}{4mm} \exp\left( -{\rm i}
{\bf q} \dotprod {\bf r} \right) \right| \psi_b \right>
\left< \psi_b
\left| \rule{0mm}{4mm}[  {\cal H}, \exp\left( {\rm i}
{\bf q} \dotprod {\bf r} \right)] \right| \psi_{n_a \ell_a m_a}
\right>,
\label{6.106}\eeq
where ${\cal H}= \hat{\bf p}^2/2\me '+ V_{\rm at}(r)$. Noting that [cf.\
Eq.\ \req{6.80}]
\beqa
[ {\cal H}, \exp\left({\rm i}
{\bf q} \dotprod {\bf r} \right)]
&=& \frac{1}{2 \me} \left\{\rule{0mm}{4mm} (\hbar q)^2
\exp\left({\rm i} {\bf q} \dotprod {\bf r} \right) + 2
\exp\left({\rm i} {\bf q} \dotprod {\bf r} \right)
\, \hbar {\bf q} \dotprod {\bf p} \right\} \, ,
\label{6.107}\eeqa
we have
\beqa
{\cal P}_{ba} &=& Q \left< \psi_{n_a \ell_a m_a}
\left| \rule{0mm}{4mm} \exp\left( -{\rm i}
{\bf q} \dotprod {\bf r} \right) \right| \psi_b \right>
\left< \psi_b
\left| \rule{0mm}{4mm}\exp\left( {\rm i}
{\bf q} \dotprod {\bf r} \right)\right| \psi_{n_a \ell_a m_a}
\right>
\nonumber \\ [2mm]
&& \mbox{} + \frac{1}{\me} \, \hbar {\bf q} \dotprod
\left< \psi_{n_a \ell_a m_a}
\left| \rule{0mm}{4mm} \exp\left( -{\rm i}
{\bf q} \dotprod {\bf r} \right) \right| \psi_b \right>
\left< \psi_b
\left| \rule{0mm}{4mm}\exp\left( {\rm i}
{\bf q} \dotprod {\bf r} \right) {\bf p} \right| \psi_{n_a \ell_a m_a}
\right> \, . \rule{15mm}{0mm}
\label{6.108}\eeqa
Inserting this result into Eq.\ \req{6.104}, and using the completeness
relation of the one-electron spatial states\footnote{The factor in
parenthesis represents the density of free central-field orbitals
per unit energy. Its
reciprocal equals the constant in the normalization relation \req{2.48}.},
\beq
\int_0^\infty \d \varepsilon_b \left( \frac{k_b}{\varepsilon_b \pi}
\right)
\sum_{\ell_b, m_b}
\left| \rule{0mm}{4mm}\psi_{\varepsilon_b\ell_b m_b} \right>
\left< \rule{0mm}{4mm}\psi_{\varepsilon_b\ell_b m_b} \right|
+ \sum_{n_b,\ell_b,m_b} \
\left| \rule{0mm}{4mm}\psi_{n_b \ell_b m_b} \right>
\left< \rule{0mm}{4mm}\psi_{n_b \ell_b m_b} \right| = I,
\label{6.109}\eeq
we obtain
\beqa
S'_0(a;Q) &=&
2 \sum_{m_a} \left(
\left<
\psi_{n_a \ell_a m_a} \left|
\psi_{n_a \ell_a m_a}\rule{0mm}{4mm}
\right>\right. +
\frac{1}{\me} \, \hbar {\bf q} \dotprod
\left<
\psi_{n_a \ell_a m_a}
\left| {\bf p} \, \psi_{n_a \ell_a m_a}\rule{0mm}{4mm}
\right>\right. \right)
\nonumber \\ [2mm]
&=& 2 \sum_{m_a} \left<
\psi_{n_a \ell_a m_a} \left|
\psi_{n_a \ell_a m_a}\rule{0mm}{4mm}
\right>\right. = 2 (2 \ell_a + 1),
\label{6.110}\eeqa
where we have used that $\left< \psi_a | {\bf p} \psi_a\right> =0$, by
parity. This completes the derivation of the Bethe sum rule for a closed
shell,
\beq
S'_0(a;Q) = 2 (2 \ell_a +1).
\label{6.111}\eeq

We now consider the effect of Pauli's exclusion principle, which forbids
transitions of the active electron to bound orbitals that are already
occupied. We express the GOS of the shell $a=(n_a\ell_a)$, Eq.\
\req{6.97}, as
\beqa
\frac{\d f_a(Q,W)}{\d W} &=&
\nu_a \sum_{n_b, \ell_b}
\delta(W - \varepsilon_{n_b \ell_b} + \varepsilon_{n_a \ell_a}) \,
\frac{2(2\ell_b+1)-\nu_b}{2(2\ell_b+1)} \, f_{ba}(Q)
\nonumber \\ [2mm]
&& \mbox{} + \frac{\nu_a}{2(2\ell_a+1)} \,
\frac{\d f_a^{\rm cs}(Q,W)}{\d W} \, {\cal S}(W-U_a),
\label{6.112}\eeqa
where
\beq
f_{ba}(Q) \equiv \frac{W}{Q} \sum_{\ell}
(2\ell+1) \,
\langle \ell_a \ell 0 0 | \ell_b 0 \rangle^2 \,
\left[ R^\ell_{n_b \ell_b; n_a \ell_a} (q) \right]^2,
\label{6.113}\eeq
$\d f_a^{\rm cs}(Q,W)/\d W$ is the GOS for ionization of the closed
shell with $2(2\ell_a+1)$ electrons, and ${\cal S}(x)$ is the unit step function. Notice that the
contribution of transitions to a closed bound level
$\varepsilon_{n_b\ell_b}$ with $\nu_b=2(2\ell_b+1)$ electrons now
vanishes. The quantity $f_{ba}(Q)$ is positive in the case of
excitation to upper levels ($W>0$, $\varepsilon_{n_b\ell_b} >
\varepsilon_{n_a \ell_a}$), and negative for transitions to lower
levels. In addition, the symmetry properties of the Clebsch-Gordan
coefficients and of the radial integrals imply that
\beq
f_{ba}(Q) = - \, \frac{2\ell_b+1}{2\ell_a+1} \, f_{ab}(Q).
\label{6.114}\eeq
The contribution of the shell to the Bethe sum is
\begin{subequations} \label{6.115}
\beqa
S_0(a;Q) &\equiv& \int_0^\infty \frac{\d f_a(Q,W)}{\d W} \, \d W
\nonumber \\ [2mm]
&=& \nu_a
\sum_{n_b,\ell_b} \, \frac{2(2\ell_b+1) - \nu_b}{2(2\ell_b+1)} f_{ba}(Q)
+ \frac{\nu_a}{2(2\ell_a+1)}
\int_{U_a}^\infty \frac{\d f_{a}^{\rm cs}(Q,W)}{\d W} \, \d W
.  \rule{15mm}{0mm}
\label{6.115a}\eeqa
We note that the integral $S'_0(a;Q)$ calculated above by ignoring the
restrictions of the Pauli principle is obtained by setting
$\nu_a=2(2\ell_a+1)$ and $\nu_b=0$,
\beq
S'_0(a;Q) = 2 (2\ell_a+1) \sum_{n_b,\ell_b} f_{ba}(Q)
+ \int_{U_a}^\infty
\frac{\d f_{a}^{\rm cs}(Q,W)}{\d W} \, \d W ,
\label{6.115b}\eeq
\end{subequations}
and we can write
\beqa
S_0(a;Q) &=& \frac{\nu_a}{2(2\ell_a+1)} S'_0(a;Q)
+ \nu_a \sum_{n_b,\ell_b}
\left( \frac{2(2\ell_b+1) - \nu_b}{2(2\ell_b+1)} - 1 \right)  f_{ba}(Q)
\nonumber \\ [2mm]
&=& \frac{\nu_a}{2(2\ell_a+1)} S'_0(a;Q)
- \sum_{n_b,\ell_b} \frac{\nu_a \nu_b}{2} \, \frac{f_{ba}(Q)}{2\ell_b+1}.
\label{6.116}\eeqa
The Bethe sum for the atomic GOS can now be obtained as
\beqa
S_0(Q) &=& \int \frac{\d f(Q,W)}{\d W} \, \d W
= \sum_{n_a,\ell_a} S_0(a;Q)
\nonumber \\ [2mm]
&=& \sum_{n_a,\ell_a} \frac{\nu_a}{2(2\ell_a+1)} \, S'_0(a;Q) -
\sum_{n_a,\ell_a} \sum_{n_b,\ell_b}  \frac{\nu_a \nu_b}{2} \,
\frac{f_{ba}(Q)}{2\ell_b+1}\, .
\nonumber\eeqa
The last summation vanishes because the contributions from pairs of
inverse transitions $a \rightarrow b$ and $b \rightarrow a$ cancel
each other [see Eq.\ \req{6.114}]. The first term in the last expression
is a weighted sum of shell contributions
\req{6.111} {\it ignoring the restrictions by the Pauli principle}. We
thus find that the atomic GOS does satisfy the Bethe sum rule
\beqa
\int_0^\infty \frac{\d f(Q,W)}{\d W} \, \d W &=&
\sum_{n_a,\ell_a} \nu_a = Z .
\label{6.117}\eeqa
However, the sum $S_0(a;Q)$ is not equal to the number of electrons
$\nu_a$ in the shell $a$ because it excludes contributions from
transitions to occupied orbitals. The effect is a net transfer of GOS
from inner to outer shells. The transfer is maximum for $Q=0$, and
decreases with the recoil energy; since transitions to bound levels are
not effective for large $Q$, $S_0(a;Q)$ tends to $\nu_a$ when $Q$ goes to
$\infty$.


\section{Integrated cross sections \label{sec6.3}}
\index{inelastic collisions!integrated cross sections}

The DDCS for inelastic collisions of the projectile is given by Eq.\
\req{6.57},
\beq
\frac{\d^2 \sigma}{ \d W \, \d Q }
= \frac{2 \pi Z_0^2 e^4}{\me v^2} \, \frac{1}{WQ}
\, \frac{\d f(Q,W)}{\d W},
\label{6.118}\eeq
and it is different from zero only within a limited domain of the
$(Q,W)$ plane that is allowed by kinematical constraints. Evidently,
collisions with atoms or molecules in their ground state may only
involve positive energy losses. The recoil energy is defined by Eq.\
\req{A.81},
\beq
Q = \frac{({\bf p} - {\bf p}')^2}{2\me} = \frac{1}{2\me}
\left[ p^2 + p'^2 - 2 pp'\, \cos\theta \right],
\label{6.119}\eeq
where $\theta = \cos^{-1}(\hat{\bf p} \cdot \hat{\bf p}')$ is the polar
scattering angle. For a given energy loss $W$, the allowed
values of $Q$ are limited to the interval ($Q_-, Q_+$) with endpoints
\beq
Q_- = \frac{M_0}{\me} \left( \sqrt{E} - \sqrt{E-W} \right)^2
\qquad \mbox{and} \qquad
Q_+ = \frac{M_0}{\me} \left( \sqrt{E} + \sqrt{E-W} \right)^2
\label{6.120}\eeq
corresponding to $\cos\theta=\pm 1$. When $W \ll E$, $Q_-$ can be
approximated as
\beq
Q_- = \frac{M_0}{\me} \, E \left( 1 - \sqrt{1-\frac{W}{E}} \right)^2
\simeq \frac{M_0}{\me} \, \frac{W^2}{4E} = \frac{W^2}{2\me v^2}.
\label{6.121}\eeq
We observe that $Q_-$ ($Q_+$) is a monotonically increasing (decreasing)
function of $W$. In addition, when the energy loss reaches its maximum
value ($W=E$), $Q_{-}=Q_{+}$. That is, if we represent $Q_{-}$ and
$Q_{+}$ as functions of $W$, we obtain a continuous curve in the $(Q,W)$
plane ---see Fig.\ \ref{fig6.3}--- that defines a single-valued
function of the recoil energy. The equation of this curve is
\beq
W_{\rm m}(Q) = \sqrt{\frac{\me}{M_0}Q}
\left(2\sqrt{E}-\sqrt{\frac{\me}{M_0}Q}\,\right).
\label{6.122}\eeq
For a given value of $Q$, the allowed energy losses are in the interval
($0,W_{\rm m}(Q)$).
The function $W_{\rm m}(Q)$ reaches its maximum value (=$E$) at
$Q=(M_0/\me)E$. When the projectile is an electron or positron ($M_0=\me$),
the diagonal $Q=W$ intersects the curve $W=W_{\rm m}(Q)$ just at its
maximum. For projectiles heavier than the electron, the maximum of the
curve is at the right-hand side of the diagonal (see Fig.\ \ref{fig6.3}).

\begin{figure}[h!]
\begin{center}
\vspace{3mm}
\includegraphics*[width=7.5cm]{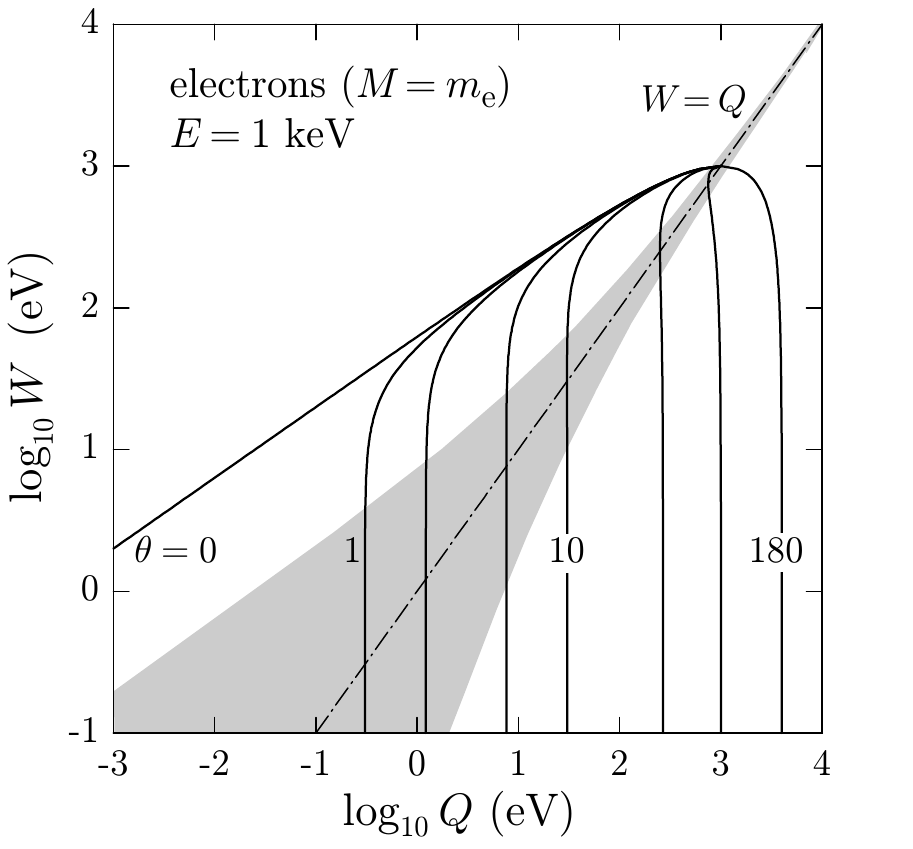} \hfill
\includegraphics*[width=7.5cm]{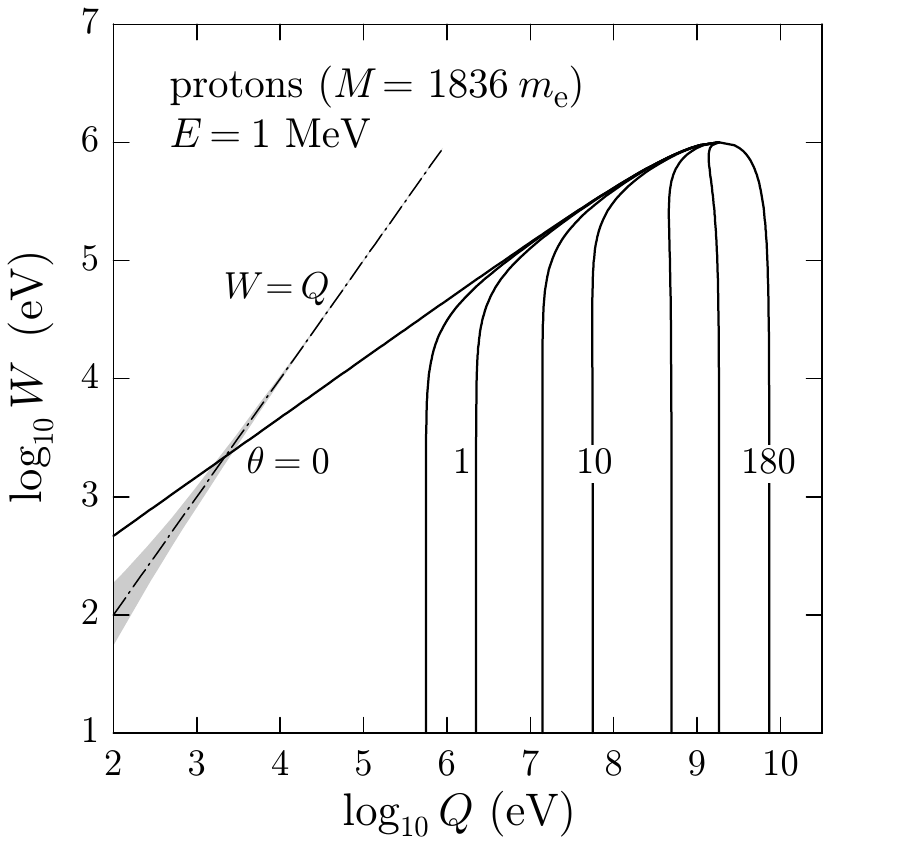}
\caption {\rm Kinematics of inelastic collisions for electrons and
protons with the indicated energies. The curves represent the value of
the recoil energy $Q$ (abscissa) that corresponds to the energy loss $W$
(ordinate) for a given scattering angle. The displayed curves correspond
to scattering angles of 0, 1, 2, 5, 10, 30, 60, and 180 degrees. The
shaded areas represent the region where the GOS of a free-electron gas
with a Fermi energy of 10 eV is different from zero (see Section
\ref{sec6.4.2}).
\label{fig6.3}}
\end{center}\end{figure}


\subsection{Energy-loss DCS and macroscopic cross sections \label{sec6.3.1}}
\index{inelastic collisions!energy-loss DCS}

The energy-loss DCS (\ie, the DCS as a function of only the energy loss
$W$) is obtained by integrating the DDCS over the recoil energy,
\beqa
\frac{\d \sigma}{\d W}
= \int_{Q_{-}}^{Q_{+}}\frac{\d^2 \sigma}{\d W \,\d Q} \, \d Q
= \frac{2 \pi Z_0^2 e^4}{\me v^2} \, \frac{1}{W} \int_{Q_{-}}^{Q_{+}}
\, \frac{\d f(Q,W)}{\d W} \, \frac{1}{Q} \, \d Q \, .
\label{6.123}\eeqa
Notice that the energy-loss DCS is defined only for energy losses that
are less than $E$. For high-energy projectiles and energy losses much
smaller than $E$, $Q_+$ is much larger than $W$, so that the GOS
effectively vanishes at the upper end of the integration interval, and
$Q_-$ is given by Eq.\ \req{6.121}. Hence, the energy-loss DCS for
energy losses much less than $E$ does not depend on the mass $M_0$ of the
projectile. This is not true when $W$ is comparable to $E$, because then
the values of both $Q_-$ and $Q_+$ do depend on $M_0$. Incidentally, the
energy-loss DCS for collisions with free electrons at rest is [see Eq.\
\req{6.72}] \index{Thomson cross section}
\beq
\frac{\d \sigma}{\d W} =
\frac{2\pi \, Z_0^2 e^4}{\me v^2} \, \frac{1}{W^2} \, ,
\label{6.124}\eeq
which coincides with the classical Thomson DCS, Eq.\ \req{4.129}. Again
we find that, in the case of pure Coulomb interactions, the PWBA yields
the same energy-loss DCS as the classical theory.

\index{stopping cross section}
\index{energy straggling cross section} \index{straggling cross section}
For practical purposes, it is useful to introduce the following
integrals (moments) of the energy-loss DCS,
\begin{subequations}
\label{6.125}
\beq
\sigma^{(k)} \equiv
\int_{0}^{E} W^{k} \, \frac{\d\sigma}{\d W} \, \d W\, .
\label{6.125a}\eeq
Notice that $\sigma^{(0)}$ is the total inelastic cross section.
$\sigma^{(1)}$ and $\sigma^{(2)}$ are known as the {\it stopping cross
section} and the {\it energy straggling cross section} (for inelastic
collisions), respectively. Recalling that the probability density of
the energy loss $W$ in a single collision is
\beq
p_1(W) = \frac{1}{\sigma^{(0)}} \, \frac{\d \sigma}{\d W}\, ,
\label{6.125b}\eeq \end{subequations}
we can write
\beq
\sigma^{(k)} =
\sigma^{(0)} \int_{0}^{E} W^{k} \, p_1(W) \, \d W =
\sigma^{(0)} \, \langle W^{k} \rangle\, ,
\label{6.126}\eeq
where $\langle W^{k}\rangle$ is the average value of $W^{k}$ in a
single collision.

\index{macroscopic cross sections}
Let us consider that our fast projectile is moving in a monoatomic gas
of the element of atomic number $Z$ with mass density $\rho_{\rm m}$
(g/cm$^3$). The number of atoms per unit volume is given by Eq.\
\req{1.141},
\beq
{\cal N} = \frac{N_{\rm A} \rho_\textrm{m}}{A_{\rm m}}\, ,
\label{6.127}\eeq
where $N_{\rm A} = 6.022 \, 141 \times 10^{23}$ mol$^{-1}$ is Avogadro's
number, and $A_{\rm m}$ is the molar mass (g/mol) of the element. The
quantities
\beq
\Sigma^{(n)} \equiv {\cal N} \sigma^{(k)}
= \Sigma^{(0)} \, \langle W^{k} \rangle
\label{6.128}\eeq
are called the {\it macroscopic cross sections}; those of orders 0, 1,
and 2 are of importance in particle transport calculations \citep{Salvat2025}.

To clarify the physical meaning of the macroscopic cross section
$\Sigma^{(0)}$, we consider a homogeneous beam of projectiles that
impinges normally on a very thin gas foil of thickness $\d s$. What the
incident projectiles see directly ahead of their path is a uniform
distribution of ${\cal N}\,\d s$ atoms per unit surface. Let $J$ be the
current density of the incident beam. The current density of projectiles
transmitted through the foil without colliding is $J-\d J$, where $\d J
= J \, {\cal N} \sigma^{(0)} \, \d s$ is the number of projectiles that
undergo inelastic collisions per unit time and per unit surface of the
foil (note that ${\cal N}\sigma^{(0)}\,\d s$ is the fractional area
``covered'' by the atomic cross sections). Therefore, the collision
probability per unit path length in the gas is
\beq
\frac{\d J}{J} \, \frac{1}{\d s} = {\cal N} \sigma^{(0)} = \Sigma^{(0)}.
\label{6.129}\eeq

We can now determine the probability density function $p(s)$ of the path
length $s$ of a projectile in the unbounded gas from its current
position to the site of its first inelastic collision. The probability
that the projectile travels a path length $s$ without colliding is
\beq
F(s) = \int_{s}^{\infty} p(s') \, \d s'.
\nonumber\eeq
The probability $p(s)\, \d s$ of having the first collision when the
traveled length is in the interval $(s,s+\d s)$ equals the product of
$F(s)$ (the probability of arrival at $s$ without colliding)
and $\Sigma^{(0)} \,\d s$ (the probability of colliding within
$\d s$). It then follows that
\beq
p(s) = \Sigma^{(0)} \, \int_{s}^{\infty} p(s') \, \d s'.
\label{6.130}\eeq
The solution of this integral equation, with the boundary condition
$p(\infty)=0$, is the familiar exponential distribution
\beq
p(s) = \Sigma^{(0)} \,
\exp\left[-s \, \Sigma^{(0)} \right].
\label{6.131}\eeq
The mean free path $\lambda$ for inelastic collisions in the gas
is defined as the average path length to the first collision:
\beq
\lambda \equiv \langle s\rangle = \int_{0}^{\infty} s \, p(s) \, \d s =
\frac{1}{\Sigma^{(0)}}.
\label{6.132}\eeq
We see that its inverse, $\lambda^{-1}= \Sigma^{(0)}$, is the
probability of collision per unit path length. The {\it electronic (or collision) stopping power} $S$ and the {\it energy straggling
parameter} $\Omega^{2}$ are defined by
\beq
S = \Sigma^{(1)} = \frac{\langle W\rangle}{\lambda}
\label{6.133}\eeq
and
\beq
\Omega^{2} = \Sigma^{(2)} =
\frac{\langle W^{2}\rangle}{\lambda}\, ,
\label{6.134}\eeq
respectively. Evidently, the stopping power gives the average energy
loss per unit path length. The product $\Omega^{2}\,\d s$ is the
variance of the energy distribution of an originally mono-energetic beam
after a short path length $\d s$ (see Section \ref{sec9.4}).

\index{mass stopping power}The macroscopic cross sections are frequently expressed in mass
thickness units, that is, they are defined as
\beq
\Sigma_{\rm m}^{(n)} \equiv
\frac{\Sigma^{(n)}}{\rho_{\rm m}} = \frac{N_{\rm A}}{A_{\rm m}} \,
\sigma^{(n)}.
\label{6.135}\eeq
Division by the factor $\rho_{\rm m}$ has the advantage of weakening the
dependence of the macroscopic cross sections on the mass density of the
material. Thus, in the case of gases, the macroscopic cross sections
$\Sigma_{\rm m}^{(n)}$ are completely independent of the gas density.
The quantity $S/\rho_{\rm m}$ is called the {\it mass stopping power}
and is given in units of eV$\cdot$cm$^2/$g or one of its multiples or
submultiples.


\subsection{Angular differential cross sections \label{sec6.3.2}}
\index{inelastic collisions!angular DCS}

The recoil energy $Q$ is a convenient variable to obtain compact
expressions of the DDCS. For practical purposes, however, one may need
to express the DDCS in terms of the scattering angle. It is advantageous
to measure angular deflections by means of the variable
\beq
\mu \equiv (1-\cos\theta)/2 = \sin^2(\theta/2)\, ,
\label{6.136}\eeq
which is related to the recoil energy by
\beq
Q = \frac{1}{2\me} \left[(p-p')^2 + 4 \, p \, p'  \mu \right].
\label{6.137}\eeq
The DDCS, differential in $W$ and $\mu$, is
\beq
\frac{\d^{2}\sigma}{\d W\,\d \mu} =
\frac{\d^{2}\sigma}{\d W\,\d Q} \frac{\d Q}{\d \mu} =
\frac{\d^{2}\sigma}{\d W\,\d Q} \,
\frac{2p p'}{\me} \, .
\label{6.138}\eeq
The angular DCS is obtained upon integration of the DDCS over $W$,
\beqa
\frac{\d \sigma}{\d \mu} &=& \int_0^{E}
\frac{\d^{2}\sigma}{\d W\,\d Q} \,
\frac{2 p p'}{\me}\, \d W
\nonumber \\ [2mm]
&=&  \frac{2 \pi Z_0^2 e^4}{\me v^2} \, \int_0^{E}
\frac{2 p p'}{\me}\, \frac{1}{WQ}
\, \frac{\d f(Q,W)}{\d W} \, \d W \, .
\label{6.139}\eeqa
Note that the usual angular DCS, per unit solid angle, is
\beq
\frac{\d \sigma}{\d \Omega} = \frac{1}{4\pi} \frac{\d \sigma}{\d \mu}
\, .
\label{6.140}\eeq

The calculation of the integral \req{6.139} is complicated because both
$p'$, $Q$ and the GOS depend on $W$. However, if the energy $E$ of the
projectile is substantially larger than the typical excitation energies
of the atom, we have $p'\simeq p$, and Eq.\ \req{6.137} simplifies to
\beq
Q \simeq \frac{2 p^2}{\me} \ \mu = \frac{4M_0 E}{\me} \, \mu  \, ,
\label{6.141}\eeq
an expression that is independent of $W$. As noted by
\citet{Inokuti1971},
this relation is valid within the region of the ($Q,W$) plane where the
curves of Fig.\ \ref{fig6.3} are straight vertical lines. Under these
circumstances,
\beqa
\frac{\d^{2}\sigma}{\d W\,\d \mu} &\simeq&
\frac{\d^{2}\sigma}{\d W\,\d Q} \frac{\d Q}{\d \mu} =
\frac{2\pi Z_0^2 e^4}{\me v^2} \, \frac{\me} {4M_0 E}
\frac{1}{\mu^2} \, \frac{Q}{W} \, \frac{\d f (Q,W)}{\d W}
\nonumber \\ [2mm]
&=&
\frac{\pi Z_0^2 e^4}{4E^2} \, \frac{1}{\mu^2} \, \frac{Q}{W} \,
\frac{\d f (Q,W)}{\d W} \, .
\label{6.142}\eeqa
If we ignore deviations from this formula, which occur at relatively
large $W$ where the DCS is small, the angular DCS can be obtained as
$$
\frac{\d\sigma}{\d \mu} \simeq
\frac{\pi Z_0^2 e^4}{4E^2} \, \frac{1}{\mu^2}
\, Q \int_0^{E}
\frac{1}{W} \, \frac{\d f (Q,W)}{\d W} \, \d W \, .
$$
As the integrand decreases rapidly with $W$, the upper limit of the
integral can be set to infinity. This leads to the \citet{Morse1932} formula
\beqa
\frac{\d\sigma}{\d \mu} \simeq
\frac{\pi Z_0^2 e^4}{4E^2} \, \frac{1}{\mu^2} \,
S_\textrm{inc}(Q)\, .
\label{6.143}\eeqa
The function \index{incoherent scattering function}
\beq
S_\textrm{inc}(Q) \equiv Q \int_0^\infty \frac{1}{W} \,
\frac{\d f (Q,W)}{\d W} \, \d W
\label{6.144}\eeq
is known as the {\it incoherent-scattering function}
\citep{Inokuti1971}. This function vanishes at $Q=0$ because the
integral is finite, and increases monotonically to reach a saturation
value equal to $Z$, because for large $Q$ the atomic electrons react as
if they were free and at rest and, hence, the GOS can be approximated by
$Z \delta(W-Q)$ [see Section \ref{sec6.4.3}]. The incoherent-scattering
function is employed in approximate calculations of scattering of x-rays
\citep{WallerHartree1929}, electron-positron pair production by gamma
rays, and bremsstrahlung emission by electrons \citep{WheelerLamb1939,
Tsai1974}. The present derivation indicates that such calculations
involve a wealth of simplifications, whose effect on the resulting cross
sections is difficult to estimate.

The incoherent scattering function can be
expressed in terms of only the ground-state wave function $\Psi_0$ of
the target atom or molecule. This is accomplished by following the same
steps as in the derivation of the Bethe sum rule (see Section
\ref{sec6.2.1.3}). From Eqs.\ \req{6.78} and \req{6.79} we can write
\beqa
S_\textrm{inc}(Q) &=& \int_0^\infty \left[
\sum_{\Psi_f \in \varepsilon_f}
\delta(W - \varepsilon_i+\varepsilon_f)
\left| \left<\Psi_f \left|
{\sum}_{j} \exp({\rm i} {\bf q} \dotprod {\bf r}_j)
\right| \Psi_0 \right> \right|^2 \right]\d W
\nonumber \\ [2mm]
&=& {\sum_{\Psi_f}}'
\left| \left<\Psi_f \left|
{\sum}_{I} \exp({\rm i} {\bf q} \dotprod {\bf r}_I)
\right| \Psi_0 \right> \right|^2
\nonumber \\ [2mm]
&=& {\sum_{\Psi_f}}'
\left<\Psi_0 \left|
{\sum}_{j} \exp(-{\rm i} {\bf q} \dotprod {\bf r}_j)
\right| \Psi_f \right>
\left<\Psi_f \left|
{\sum}_{l} \exp({\rm i} {\bf q} \dotprod {\bf r}_l)
\right| \Psi_0 \right>,
\nonumber \eeqa
where the primed summation runs over all excited states.
Adding and subtracting a term with $\Psi_f=\Psi_0$, and using the
completeness of atomic or molecular states,
\beq
\sum_{\Psi_i}
\left| \Psi_i \right>
\left< \Psi_i \right| = I,
\nonumber \eeq
we arrive at the sought result
\beq
S_{\rm inc}(Q) =  \sum_{j=1}^{Z} \sum_{l=1}^{Z}
\left\langle \Psi_{0} \left|
\exp[{\rm i}{\bf q}\dotprod({\bf r}_{l}-{\bf r}_{j})] \right|
\Psi_{0} \right\rangle
- \sum_{j=1}^{Z} \left| \left\langle \Psi_{0}
\left| \exp(-{\rm i}{\bf q}\dotprod{\bf r}_{j}) \right|
\Psi_{0} \right\rangle \right|^{2}.
\label{6.145}\eeq

\index{binary encounter approximation|(}
\index{impulse approximation|(}

\section{Binary-encounter and impulse
approximations \label{sec6.4}}

The numerical calculation of the shell ionization GOS, Eq.\ \req{6.97},
for large momentum transfers $\hbar q$ is extremely difficult because of the
fast oscillations of the radial functions and the slow convergence of
the partial-wave series \req{6.38}. An approximate formula for the
GOS at large $Q$ is provided by the {\it impulse
approximation} (IA), which consists in ignoring the binding forces on
the atomic electrons during the interaction, except for the fact that
the electrons move with a certain distribution of velocities
${\bf v}_a$. The approach of treating the target system as a swarm of
free electrons is also the basis of the {\it classical binary-encounter
approximation} (BEA), which was originally developed by
\citeauthor{Gryzinski1957} (\citeyear{Gryzinski1957, Gryzinski1959})
based on formulas derived by Chandrasekhar for classical collisions of
masses interacting through gravitational forces. The rather involved
formulation of Gryzinski was simplified by \citet{Stabler1964} and
\citet{Vriens1966, Vriens1968}. Here we give a unified derivation of the
IA and the BEA in which the DCS is obtained from the PWBA. Our
approach, which does simplify the derivations, is consistent with the
classical BEA, because the PWBA yields the correct classical Rutherford
DCS for the Coulomb potential.

We assume that the projectile is distinguishable from the target
electrons. Within the PWBA the initial and final states of the
projectile are represented as continuum plane waves,
\beq
\phi_{\bf k}({\bf r}) = (2\pi)^{-3/2}
\exp({\rm i} {\bf k} \cdot {\bf r}),
\label{6.146}\eeq
normalized so that the density of states per unit volume in ${\bf
k}$-space is unity. As in Section \ref{sec6.2.1.1}, the states
of the target electron are described as plane waves satisfying
periodic boundary conditions in a cubic box of side $L$,
\beq
\phi_{L,{\bf k}}({\bf r}) = L^{-3/2}
\exp({\rm i} {\bf k} \cdot {\bf r}) \, ,
\label{6.147}\eeq
which represent one electron in the normalization box. The initial and
final states of the target electron are plane waves with respective
energies $\varepsilon_a$ and $\varepsilon_b$ and wave numbers
${\bf k}_a$ and ${\bf k}_b$. The collision DCS obtained from
Fermi's golden rule is
\beqa
\d \sigma ({\bf k}_a) &=& \sum_{{\bf k}_b} \frac{(2\pi)^4}{\hbar v} \,
|T_{fi}|^2 \, \delta(E - E' - \varepsilon_b + \varepsilon_a ) \,
\d {\bf k}'
\label{6.148}\eeqa
with the transition matrix elements
\beqa
T_{fi} &=& \int \d {\bf r}_0 \, \int_{L^3} \d {\bf r} \,
\phi^\ast_{{\bf k}'} ({\bf r}_0) \,
\phi_{L,{\bf k}_b}^\ast ({\bf r}) \,
\frac{-Z_0 e^2}{|{\bf r}_0 - {\bf r}|} \,
\phi_{{\bf k}} ({\bf r}_0)
\, \phi_{L,{\bf k}_a}({\bf r})
\nonumber \\ [2mm]
&=& \frac{-Z_0 e^2}{(2\pi)^{3}}
\int \d {\bf r}_0 \, \int_{L^3} \d {\bf r} \,
\phi_{L,{\bf k}_b}^\ast ({\bf r}) \,
\frac{\exp\left( {\rm i} {\bf q} \dotprod {\bf r}_0 \right)}
{|{\bf r}_0 - {\bf r}|} \,
\phi_{L,{\bf k}_a}({\bf r})\, ,
\label{6.149}\eeqa
where  $\hbar {\bf q} = \hbar ({\bf k} - {\bf k}')$ is the momentum
transfer and the ${\bf r}$-integral extends over the volume of the
normalization box. Using the Bethe integral, Eq.\ \req{B.26},
we can write
\beqa
T_{fi} &=&
\frac{-Z_0 e^2}{2 \pi^2 L^3 q^2}
\int_{L^3} \d {\bf r}
\, \exp \left[ {\rm i} ({\bf q} + {\bf k}_a -{\bf k}_b) \dotprod {\bf r} \right]
\nonumber \\ [2mm]
&=&
\frac{-Z_0 e^2}{2\pi^2 q^2 } \, \delta_{{\bf q}, {\bf k}_b -{\bf k}_a},
\label{6.150}\eeqa
and the DCS takes the form
\beq
\d \sigma ({\bf k}_a)
= \frac{4 Z_0^2 e^4}{\hbar v}
\sum_{{\bf k}_b} \delta_{{\bf q}, {\bf k}_b -{\bf k}_a}
\, \frac{1}{q^4} \,
\, \delta(E - E'-\varepsilon_b+\varepsilon_a) \, \d {\bf k}'.
\label{6.151}\eeq
The equality $(\hbar k)^2 = 2 M_0 E$ implies that
\beq
\d {\bf k}' = {k'}^2 \, \d k' \, \d \hat{\bf k}'
= {k'}^2 \frac{\d k'}{\d E'} \, \d E'\,\d\hat{\bf k}'
= k'\, \frac{M_0}{\hbar^2} \, \d E'\,\d\hat{\bf k}'
\nonumber\eeq
and, hence, we have
\beq
\frac{\d \sigma ({\bf k}_a)}{\d W \, \d\hat{\bf k}'}
= \frac{4 Z_0^2 e^4}{\hbar v} \, k'\, \frac{M_0}{\hbar^2}
\sum_{{\bf k}_b} \delta_{{\bf q}, {\bf k}_b -{\bf k}_a}
\, \frac{1}{q^4} \,
\, \delta(W-\varepsilon_b+\varepsilon_a) \, ,
\label{6.152}\eeq
where $W=E-E'$ is the energy lost by the projectile.
Taking the limit of expression \req{6.152} when $L \rightarrow \infty$ (large
electron normalization box) the sum over ${\bf k}_b$ becomes an
integral, and we can write
\beq
\frac{\d^2 \sigma ({\bf k}_a)}{\d W \, \d \hat{\bf k}'} =
\frac{4 Z_0^2 e^4 M_0}{ v } \, \int \d {\bf k}_b \;
\delta({\bf q}+{\bf k}_a -{\bf k}_b)
\, \frac{\hbar k'}{\hbar^4 q^4}
\, \delta(W-\varepsilon_b+\varepsilon_a)\, .
\label{6.153}\eeq

In the spirit of the BEA and the IA, we wish to calculate the DCS per target
electron for collisions of the projectile with a swarm of electrons that
move with an isotropic distribution of velocities ${\bf v}_a$. Thus, the
target is characterized by the distribution of electron wave vectors,
${\bf k}_a = \me {\bf v}_a / \hbar$,
\beq
P({\bf k}_a) = \frac{1}{4\pi k_a^2} \, P(k_a),
\label{6.154}\eeq
which is assumed to be normalized to unity. The DCS per electron is given
by
\beqa
\frac{\d^2 \sigma}{\d W \, \d \hat{\bf k}'}
&=&
\int \d {\bf k}_a \; P({\bf k}_a) \;
\frac{\d^2 \sigma({\bf k}_a)}{\d W \, \d\hat{\bf k}'}
\nonumber \\ [2mm]
&=& \frac{4 Z_0^2 e^4 M_0}{ v }
\int \d {\bf k}_a \;  P({\bf k}_a)
\, \int \d {\bf k}_b \;
\delta({\bf q}+{\bf k}_a -{\bf k}_b)
\, \frac{\hbar k'}{\hbar^4 q^4}
\, \delta(W-\varepsilon_b+\varepsilon_a).
\nonumber \eeqa
Integrating first over ${\bf k}_b$, we have
\beqa
\frac{\d^2 \sigma}{\d W \, \d \hat{\bf k}'}
&=& \frac{4 Z_0^2 e^4 M_0}{ v } \, \frac{\hbar k'}{\hbar^4 q^4}
\int \d {\bf k}_a \;  P({\bf k}_a)
\, \delta\left( W- \frac{\hbar^2 ({\bf q}+{\bf k}_a)^2}{2\me}
+ \frac{\hbar^2 {\bf k}_a^2}{2\me} \right)
\nonumber \\ [2mm]
&=& \frac{4 Z_0^2 e^4 M_0}{ v } \, \frac{\hbar k'}{\hbar^4 q^4}
\int \d {\bf k}_a \; P({\bf k}_a)
\, \delta\left[ W- \frac{\hbar^2}{2\me} (q^2+2q k_a \, \cos\theta_a)
\right]\, ,
\nonumber\eeqa
where $\theta_a$ is the angle between the vectors ${\bf q}$ and ${\bf
k}_a$. It is now convenient to express the DCS in terms of the recoil
energy\footnote{Notice that this step is valid only if the distribution
$P({\bf k}_a)$ is isotropic.} [see Eqs.\ \req{6.52} and \req{6.53}]
\beq
Q = \frac{\hbar^2}{2\me} \left( k^2 + k'^2
- 2 k k' \cos\theta \right)
\nonumber\eeq
and write
\beqa
\frac{\d^2 \sigma}{\d W \, \d Q}
&=& \frac{2 \pi Z_0^2 e^4}{\me v^2 } \, \frac{1}{WQ}
\left\{ \frac{W}{Q} \int \d {\bf k}_a \; P({\bf k}_a)
\, \delta\left( W- Q - Q \frac{2 k_a}{q} \, \cos\theta_a \right)
\right\} .
\nonumber\eeqa
By similarity with the result from the PWBA , Eq.\ \req{6.57}, the quantity
in curly braces may be identified as the GOS per electron,
\beq
\frac{\d f(Q,W)}{\d W} =
\frac{W}{Q} \int \d {\bf k}_a \;  P({\bf k}_a) \,
\delta\left( W- Q - Q \frac{2 k_a}{q} \, \cos\theta_a \right).
\label{6.155}\eeq
This is the GOS resulting from both the IA and the BEA. These two
approximations differ only in the way the integral over ${\bf k}_a$ is
evaluated. Indeed, they both lead to the same GOS.

In the IA, the ${\bf k}_a$ integral is calculated partially by using Cartesian
coordinates with the $z$ axis in the direction of ${\bf q}$,
\beqa
\frac{\d f(Q,W)}{\d W} &=&
\frac{W}{Q} \int \d {\bf k}_a \;  P(k_{ax},k_{ay},k_{az})
\, \delta\left( W- Q - Q \frac{2 k_{az}}{q} \right)
\nonumber \\ [2mm]
&=& \frac{W}{Q} \frac{q}{2Q} \int \d k_{ax} \int \d k_{ay} \;
P\left(k_{ax},k_{ay},(W-Q) \frac{q}{2Q} \right).
\nonumber \eeqa
That is,
\beq
\frac{\d f^{\rm (IA)}(Q,W)}{\d W} =
 \frac{W}{Q} \sqrt{\frac{\me}{2\hbar^2 Q}} J \left(
(W-Q) \, \sqrt{\frac{\me}{2\hbar^2 Q}} \right),
\label{6.156}\eeq
where
\beq
J(k_{az}) \equiv \int \d k_{ax} \int \d k_{ay} \;
P\left(k_{ax},k_{ay},k_{az} \right)
\label{6.157}\eeq
is the {\it Compton profile} of the distribution. This definition
implies that \index{Compton profile}
\beq
\int_{-\infty}^\infty J_{n_a,\ell_a}(k_z) \, \d k_z = 1.
\label{6.158}\eeq
Since the velocity distribution is isotropic, we also have
\beq
J(k_z) = 2 \pi \int_{0}^\infty
\rho_{n_a\ell_a}\left( \sqrt{k_z^2 + \kappa^2} \right) \, \kappa\, \d
\kappa,
\label{6.159}\eeq
which shows that the Compton profile is an even function of $k_z$,
\beq
J(k_z) = J_{n_a,\ell_a}(-k_z).
\label{6.160}\eeq
The formula \req{6.156} defines the GOS obtained from the IA. It is expected
to be accurate for excitations involving momentum transfers
$\hbar q$ much larger than the average momentum of the target electrons.
The approximation \req{6.156} is useful, \eg, for extrapolating the
atomic GOS calculated numerically up to a certain large enough $Q$ to
higher recoil energies \citep{Segui2002, BoteSalvat2008}.

In the formulation of the BEA, the integral in Eq.\ \req{6.155} is
evaluated by using spherical coordinates with the polar axis in the
direction of the vector ${\bf q}$. With the assumed spherical symmetry
of the wave vector distribution, Eq.\ \req{6.154}, a trivial integration over
the azimuthal angle gives
\begin{subequations}
\label{6.161}
\beqa
\frac{\d f(Q,W)}{\d W}
&=& \frac{W}{2Q} \int \d k_a \, P(k_a)
\int \d (\cos\theta_a)
\, \delta\left( W- Q - Q \, \frac{2 k_a}{q} \, \cos\theta_a \right)
\rule{12mm}{0mm}
\label{6.161a} \\ [2mm]
&=& \frac{W}{2Q} \int \d k_a \, P(k_a)
\int \d (\cos\theta_a) \, \frac{q}{2 k_a} \, \frac{1}{Q}
\, \delta\left( \cos\theta_a - \cos\theta_1 \right),
\label{6.161b}\eeqa
\end{subequations}
where
\beq
\cos\theta_1 = \frac{W-Q}{Q} \, \frac{q}{2k_a}.
\label{6.162}\eeq
Evidently, this GOS is different from zero only when
$\cos\theta_1$ takes values between $-1$ and 1, \ie, for given $W$ and
$Q$ we have contributions only from $k_a$ values such that
\beq
k_a \ge \frac{\left| W - Q \right|}{2Q} \, q \equiv k_{a,{\rm min}}.
\label{6.163}\eeq
Therefore, the GOS resulting from the BEA is
\beq
\frac{\d f^{\rm (BEA)}(Q,W)}{\d W} = \frac{W}{4Q^2} \, q
\int_{k_{a,{\rm min}}}^\infty
\frac{1}{k_a} \, P(k_a) \, \d k_a.
\label{6.164}\eeq

From Eq.\ \req{6.161a}, inverting the order of the integrals, we find
that
\beq
\int_0^\infty \frac{\d f^{\rm (BEA)} (Q,W)}{\d W}\, \d W
= \frac{1}{2Q} \int_0^\infty \d k_a \, P(k_a)
\int_{-{\rm min}\{1,q/2k_a\}}^1  \d (\cos\theta_a) \:
Q \, \left( 1 + \frac{2 k_a}{q} \, \cos\theta_a \right).
\nonumber \eeq
Hence, when the kinetic energy of the target electrons is bounded
($\varepsilon_a \le \varepsilon_{a, {\rm max}}$), the GOS obtained from
the BEA or from the IA satisfies the Bethe sum rule,
\beq
\int \frac{\d f^{\rm (BEA \, or \, IA)} (Q,W)}{\d W}\, \d W
= \int \d k_a \, P(k_a) = 1,
\label{6.165}\eeq
for all $Q$ larger than $4\varepsilon_{a,{\rm max}}$. For smaller $Q$,
the integral is less than unity. This deviation from the Bethe sum rule
indicates that the BEA and IA should only be used for describing hard
collisions with sufficiently large energy and momentum transfers.


\subsection{Hydrogenic ions \label{sec6.4.1}}
\index{binary encounter approximation!hydrogenic ions}
\index{hydrogenic ions!generalized
oscillator strength!binary encounter approximation}

The first applications of the BEA were aimed at computing cross
sections for ionization of atomic electron shells. The wave-vector
distribution of the electrons in a shell $(n_a\ell_a)$ is
\beq
P({\bf k}_a) = \frac{1}{2\ell_a+1} \sum_{m_a}
\left| {\psi}_{n_a,\kappa_a,m_a} ({\bf k}_a) \right|^2,
\label{6.166}\eeq
where
\beq
{\psi}_{n_a,\kappa_a,m_a} ({\bf k}) \equiv
(2\pi)^{-3/2}
\int \exp(-{\rm i} {\bf k} \dotprod {\bf r}) \,
\psi_{n_a \kappa_a m_a} ({\bf r})  \, \d {\bf r},
\label{6.167}\eeq
\index{Fourier transform}is the Fourier transform of the orbital,
$\psi_{n_a \kappa_a m_a} ({\bf r})$, \ie, the wave function in the
momentum representation.
In the case of hydrogenic ions, the wave functions
${\psi}_{n_a,\ell_a,m_a} ({\bf k_a})$ admit analytical
expressions similar to those of the spatial orbitals \citep[see,
\eg,][]{BetheSalpeter1957, BransdenJoachain1983}, from which we can
readily obtain the wave-number distribution of the electrons in the
shell. In particular, for the ground state ($K$ shell) of hydrogenic
ions with atomic number $Z$ we have
\beq
P(k_a) = \frac{32}{\pi} \left( \frac{Z}{a_0} \right)^5
\frac{k_a^2}{\left( k_a^2 + Z^2 a_0^{-2} \right)^4}\, .
\label{6.168}\eeq

Inserting this expression into the right-hand side of Eq.\ \req{6.164}, the
resulting integral is elementary and gives
\beq
\frac{\d f^{\rm (BEA)}(Q,W)}{\d W} = \frac{2^8}{3\pi} \,
\frac{W Q^{3/2} U_1^{5/2}}{\left[ (Q-W)^2 + 4 QU_1 \right]^3} \,
\label{6.169}\eeq
where $U_1 = \1o2 Z E_{\rm h}$ is the ionization energy of the target
electron. Notice that the last denominator is the same as in the exact
hydrogenic result \req{6.103}.

\vspace{2mm}
\begin{figure}[h!] \begin{center}
\includegraphics*[width=10.5cm]{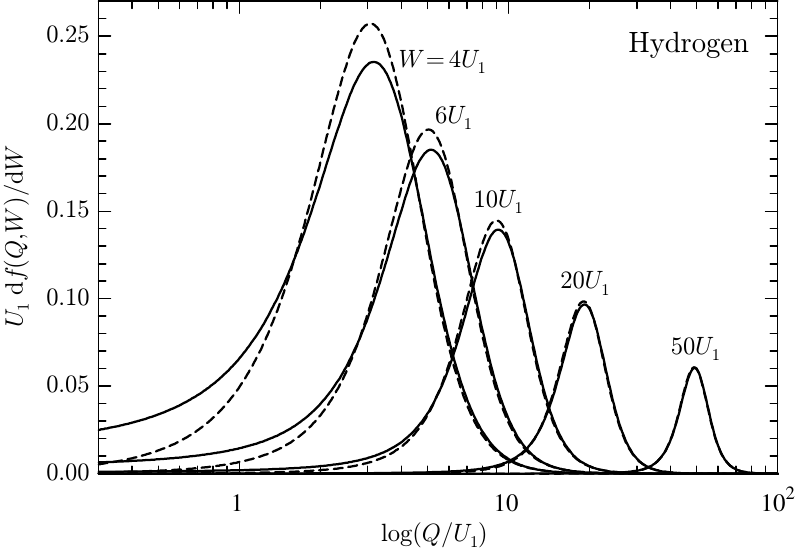}
\caption{GOS of hydrogenic ions in their ground state, for the indicated
values of the energy loss $W$, as a function of the recoil energy $Q$.
The solid curves are calculated from the exact formula \req{6.103}. The
dashed curves are the result from the BEA, Eq.\ \req{6.169}.
\label{fig6.4}}
\end{center} \end{figure}

Figure \ref{fig6.4} compares the exact GOS of hydrogenic ions, Eq.\
\req{6.103}, with the BEA result \req{6.169}. The plotted curves
represent the GOS, $\d f(Q,W)/\d W$, for the indicated values of $W$, as
a function of the recoil energy $Q$. Notice that, when energies are
given in units of the ionization energy $U_1$, the GOS is independent of
the atomic number of the atom or ion. The plot shows that the
accuracy of the BEA improves when the excitation energy increases. For
practical purposes, the BEA can replace the exact formula for $W$
larger than about $50U_1$.


\subsection{Degenerate electron gas \label{sec6.4.2}}
\index{binary encounter approximation!electron gas}

Let us now consider the collisions of a charged projectile in a
degenerate free-electron gas, with electrons filling all states up to
the Fermi level $E_{\rm F}$ with corresponding wave number $k_{\rm F} =
(2 \me E_{\rm F})^{1/2} /\hbar$. This system is of interest because the
plane waves are exact wave functions for both the projectile and the
electrons of the gas and, hence, the BEA and the IA are also exact, in
the sense that they give the same result as the PWBA (and the DWBA).

Because the wave vectors of the electrons in the gas are distributed
uniformly within the Fermi sphere of radius $k_{\rm F}$,
\beq
P(k_a) = \frac{3 k_a^2}{k_{\rm F}^3} \, {\cal S}(k_{\rm F} - k_a),
\label{6.170}\eeq
where ${\cal S}(x)$ is the Heaviside unit step function.
The GOS of the electron gas is [Eq.\ \req{6.164}]
\beq
\frac{\d f^{\rm (BEA)}(Q,W)}{\d W} =
\frac{W}{4Q^2} q \;
\frac{3}{k_{\rm F}^3} \,
\int_{k_{a,{\rm min}}}^\infty
k_a \, \d k_a
=
\frac{W}{4Q^2} \, q \;
\frac{3}{2k_{\rm F}^3} \, \left[ k_{\rm F}^2 - k_{a,{\rm min}}^2
\right] {\cal S}(k_{\rm F} - k_{a,{\rm min}})
\nonumber\eeq
with
$$
k_{a,{\rm min}} = \frac{|W-Q|}{2Q} \, q.
\eqno{\req{6.163}}$$
As the GOS is different from zero only when $k_{a,{\rm min}} \le
k_{\rm F}$, for a given energy loss, the GOS is not null only for recoil
energies in the interval from
\beq
Q_-^{\rm b}  = \left(\sqrt{W+E_{\rm F}} - \sqrt{E_{\rm F}} \right)^2
\qquad \mbox{to} \qquad
Q_+^{\rm b}  = \left(\sqrt{W+E_{\rm F}} + \sqrt{E_{\rm F}} \right)^2 \, .
\label{6.171}\eeq
Equivalently, for a given value of $Q$, the GOS vanishes for energy
transfers $W$ outside the interval limited by the points
\beq
W_-^{\rm b} = Q - 2 \sqrt{Q E_{\rm F}}
\qquad \mbox{and} \qquad
W_+^{\rm b} = Q + 2 \sqrt{Q E_{\rm F}} \, .
\label{6.172}\eeq
Hence the GOS of the electron gas is
\beq
\frac{\d f^{\rm (BEA)} (E_{\rm F}; Q,W)}{\d W} =
\frac{W}{4Q^2} \, \frac{3}{2} \sqrt{\frac{Q}{E_{\rm F}}}
\left[ 1 - \left( \frac{W - Q}{2Q}\right)^2 \frac{Q}{E_{\rm F}}
\right]
\, {\cal W} \left( Q_-^{\rm b}, Q_+^{\rm b}; Q \right),
\label{6.173}\eeq
where ${\cal W}(x_1,x_2;x)$ is the unit window function ($=1$ if $x_1 <
x < x_2$, and $=0$ otherwise). The region of the $(Q,W)$ plane where
the GOS \req{6.173} has non-zero values reduces to a strip, the Bethe
ridge, that extends on both sides of the diagonal $W=Q$, Fig.\
\ref{fig6.5}.

\vspace{2mm}
\begin{figure}[htb] \begin{center}
\includegraphics*[width=8.0cm]{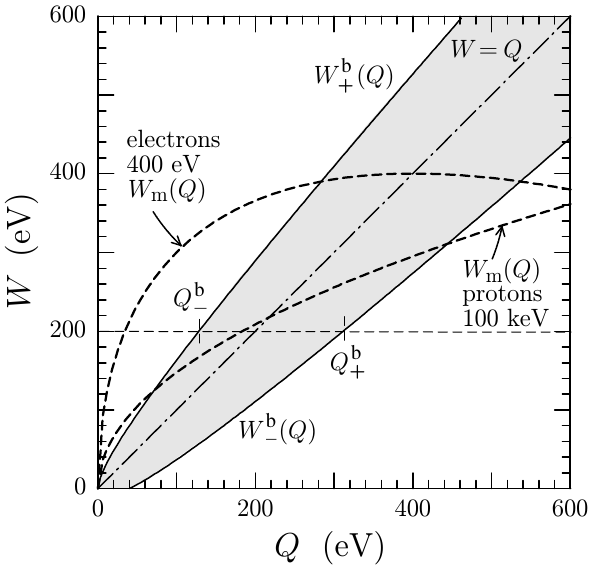}
\caption{
GOS of a degenerate electron gas with Fermi energy $E_{\rm F}=10$ eV. The GOS
is different from zero only on the shaded area, which is limited by the curves
$W_\pm^{\rm b}(Q)$, Eq.\ \req{6.172}. The dashed lines represent the
largest possible energy loss, Eq.\ \req{6.122}, for electrons and
protons with the indicated kinetic energies.
\label{fig6.5}}
\end{center} \end{figure}

The function $W_{\rm m}(Q)$, Eq.\ \req{6.122}, reaches its maximum value
at $Q=(M_0/\me)E$. When the projectile is an electron or positron
($M_0=\me$), the diagonal $Q=W$ intersects the curve $W=W_{\rm m}(Q)$
just at its maximum, and the largest allowed energy loss is $W_{\rm sup}
= E$. For projectiles heavier than the electron, the maximum of the
curve is at the right-hand side of the diagonal (see Figs.\ \ref{fig6.3}
and \ref{fig6.5}). That is, in the case of electrons and positrons (with
$M_0 = \me$) the allowed kinematical domain of the ($Q,W$) plane covers
the whole width of the Bethe ridge when $W \le E-E_{\rm F}$, while for
protons and heavier particles a low-$Q$ portion of the ridge is left out
of the allowed domain. Consequently, the largest possible value of the
energy loss in collisions of heavy particles, $W_{\rm sup}$, is at the
intersection of the curves $W_{\rm m}(Q)$ and $W_-^{\rm b}(Q)$, which
corresponds to the recoil energy
\beq
Q_{\rm sup} = \frac{4}{(1+\me/M_0)^2} \left( \sqrt{\frac{\me}{M_0} \, E} -
\sqrt{E_{\rm F}} \right)^2.
\label{6.174}\eeq
That is,
\beq
W_{\rm sup} = Q_{\rm sup} - 2 \sqrt{Q_{\rm sup} E_{\rm F}}.
\label{6.175}\eeq
If $E$ is much higher than $E_{\rm F}$, we can use the approximate expression
\beq
W_{\rm sup} \simeq \frac{2\me v^{2}}{\left(1+\me/M_0 \right)^{2}},
\label{6.176}\eeq
which corresponds to the intersection of the curve $W=W_{\rm m}(Q)$
and the diagonal $W=Q$. When $M_0=\me$, this expression gives the correct
value, $W_{\rm sup}=E$, for any $E$.

The DDCS per electron in the gas, Eq.\ \req{6.68}, is
\beq
\frac{\d^2 \sigma (E_{\rm F})}{\d W \, \d Q}
= \frac{2 \pi Z_0^2 e^4}{\me v^2 }
\frac{3}{32 E_{\rm F}^{3/2}}
\left[2\, (2E_{\rm F} + W) \, Q^{-5/2} - W^2 Q^{-7/2}
- Q^{-3/2} \right] \, {\cal W} \left( Q_-^{\rm b}, Q_+^{\rm b}; Q
\right).
\label{6.177}\eeq
and the energy-loss DCS is
\beqa
\lefteqn{
\frac{\d \sigma(E_{\rm F})}{\d W} =
\int_{Q_-}^{Q_+}
\frac{\d^2 \sigma (E_{\rm F})}{\d W \, \d Q} \, \d Q
}
\nonumber \\ [2mm]
&=&
\frac{2 \pi Z_0^2 e^4}{\me v^2 }
\frac{3}{32 E_{\rm F}^{3/2}}
\left[- \frac{4}{3}\, (2E_{\rm F} + W) \, Q^{-3/2} + \frac{2}{5}\, W^2 Q^{-5/2}
+ 2 Q^{-1/2} \right]_{Q_1}^{Q_2} \, . \rule{12mm}{0mm}
\label{6.178}\eeqa
The lower limit of the integral is the largest of the values
$Q_-^{\rm b}$ and $Q_-$, and the upper limit is the smallest of
the values $Q_+^{\rm b}$ and $Q_+$, \ie,
\beqa
Q_1 = \max \left\{ Q_-, Q_-^{\rm b} \right\}
\qquad \mbox{and} \qquad
Q_2 = \min \left\{ Q_+, Q_+^{\rm b} \right\} \, .
\label{6.179}\eeqa
The energy-loss DCSs takes different forms for $W$ below and above the
value $W_{\rm c}$ where the curves $W=W_{\rm m}(Q)$ and $W=W_+^{\rm
b}(Q)$ intersect, which is given by
\beq
W_{\rm c} = \frac{4M_0\me}{(M_0+\me)^2} \left[ E-E_{\rm F} -
\sqrt{\frac{\me}{M_0}} \; \frac{M_0 - \me}{\me} \, \sqrt{EE_{\rm F}}
\right]\, .
\label{6.180}\eeq

When the projectile is an electron or positron ($M_0=\me$), $W_{\rm
c}=E-E_{\rm F}$ and the curve $W=W_{\rm m}(Q)$ intersects the two curves
$W=W_+^{\rm b}(Q)$ and $W=W_-^{\rm b}(Q)$ at $W=W_{\rm c}$. For
projectiles heavier than the electron, the largest possible recoil
energy is that of the allowed strip, \ie, $Q_2=Q_+^{\rm b}$ irrespective
of the energy loss (see Fig.\ \ref{fig6.5}). For $W < W_{\rm c}$ the
interval of allowed $Q$ values covers the whole strip, from
$Q_-^{\rm b}$ to $Q_+^{\rm b}$, and we have
\beq
\frac{\d \sigma (E_{\rm F})}{\d W} =
\frac{2 \pi Z_0^2 e^4 }{\me v^2 } \, \frac{1}{W^2}
\left( 1 + \frac{4}{5} \, \frac{E_{\rm F}}{W} \right) =
\frac{2 \pi Z_0^2 e^4 }{\me v^2 } \, \frac{1}{W^2}
\left( 1 + \frac{4}{3} \, \frac{\overline{K}}{W} \right)
\label{6.181}\eeq
for both electrons and heavier projectiles. Here $\overline{K} =
3E_{\rm F}/5$ is the average kinetic energy of an electron of the gas
[Eq.\ \req{3.71}]. For energy transfers larger
than $W_{\rm c}$ the lower limit of the integral \req{6.178} equals
$Q_-$, and the expression of the energy-loss DCS gets more complicated.

Evidently, in the limiting case of electrons at rest
($E_{\rm F} = 0$), the GOS reduces to
\beqa
\frac{\d f^{\rm (BEA)} (0; Q,W)}{\d W}
&=& \frac{W}{2Q} \int \d k_a \, \delta(k_a)
\int \d (\cos\theta_a)
\, \delta\left( W- Q - Q \, \frac{2 k_a}{q} \, \cos\theta_a \right)
\nonumber \\ [2mm]
&=& \frac{W}{2Q}
\int \d (\cos\theta_a)
\, \delta\left( W- Q \right) = \delta(W-Q),
\label{6.182}\eeqa
and the energy-loss DCS becomes the Thomson DCS, Eq.\ \req{4.129},
\beq
\frac{\d \sigma}{\d W} =
\frac{2 \pi Z_0^2 e^4 }{\me v^2 } \, \frac{1}{W^2}.
\label{6.183}\eeq
\index{Thomson cross section}

\vspace{2mm}
\begin{figure}[h!] \begin{center}
\includegraphics*[width=10.5cm]{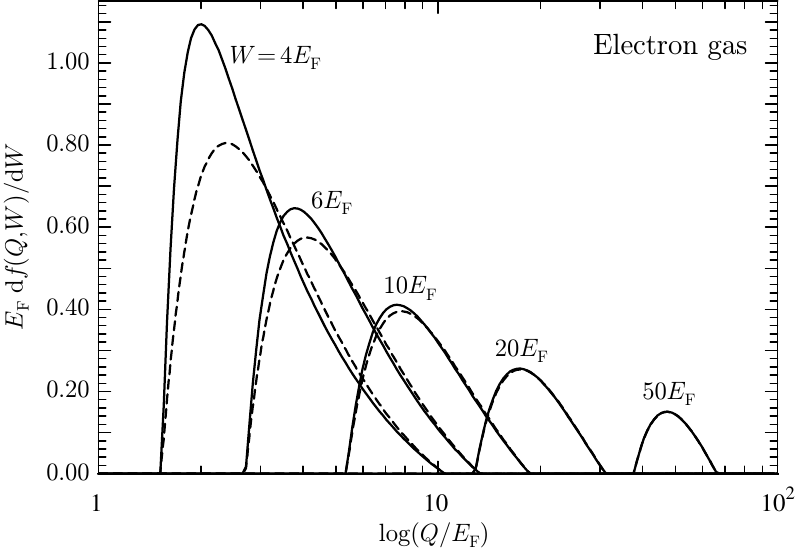}
\caption{GOS of an electron gas with $E_{\rm F} = 10$ eV, for the indicated
values of the energy loss $W$, as a function of the recoil energy $Q$.
The solid curves are calculated from Lindhard's dielectric function by
using the equivalence
\req{6.251a}. The dashed curves are the result from the BEA, Eq.\
\req{6.173}. Notice that all energies are in units of the Fermi energy
$E_{\rm F}$.
\label{fig6.6}}
\end{center} \end{figure}

It should be mentioned that the PWBA does not lead to a realistic GOS
for the electron gas, as evidenced by the fact that the DCS \req{6.181}
diverges at $W=0$ and the interaction cross section is infinite. The
main limitation of the PWBA results from the neglect of collective
effects (plasmon excitations and screening of external charges) which
cannot be accounted for by the independent-electron model underlying the
PWBA. In addition, the BEA neglects the effect of Pauli's exclusion
principle, which forbids transitions of electrons to states with energy
below the Fermi level. A more satisfactory description of the response
of the gas is provided by the Lindhard dielectric function, which is
described in Section \ref{sec7.1}. Figure \ref{fig6.6} compares the
GOS derived from Lindhard's dielectric function [by using the relation
\req{6.251a}] with the result of the BEA. As collective effects are
limited to small excitation energies, the BEA does approach the
realistic GOS for sufficiently large $W$.


\subsection{The structure of the Bethe ridge \label{sec6.4.3}}
\index{Bethe ridge}

The BEA (and the IA) provides a convenient representation of the GOS for
large recoil energies, and puts into evidence that the structure of the
Bethe ridge is determined by the momentum distribution of the target
electrons. Calculations with different targets show that the Bethe ridge
widens when the energy loss $W$ increases. On the other hand, the effect
of electron binding must be small in hard collisions that involve energy
transfers much larger than the characteristic kinetic energies of the
target electrons. Hence, independently of the nature of the
target, the energy-loss DCS at very large $W$ reduces to the Thomson
\index{Thomson cross section}
DCS, Eq.\ \req{6.183}, which corresponds to the GOS $Z \delta(W-Q)$ (a
Bethe ridge with zero width). The apparent conflict between this
feature and the widening of the Bethe ridge with $W$ is resolved by
noticing that $Q$ is not a convenient integration
variable for computing the energy-loss DCS. Equation \req{6.123},
\beqa
\frac{\d \sigma}{\d W}
= \frac{2 \pi Z_0^2 e^4}{\me v^2} \, \frac{1}{W} \int_{Q_{-}}^{Q_{+}}
\, \frac{\d f(Q,W)}{\d W} \, \frac{1}{Q} \, \d Q \, ,
\nonumber \eeqa
clearly suggests considering the GOS as a function of $\ln
Q$ instead of $Q$, because this change of variable in the integral, $\d
Q = Q \, \d (\ln Q)$, cancels out the factor $Q^{-1}$. It turns out
that, when the GOS is considered as a function of $\ln(Q)$,
the width of the Bethe ridge effectively decreases with $W$, as
illustrated in Figs.\ \ref{fig6.4} and \ref{fig6.6}.

Let us verify this assertion for the two examples studied above. The
hydrogenic GOS, Eq.\ \req{6.103a}, takes appreciable values if
\citep{Inokuti1971}
\beq
| Q - W | \lesssim \xi \sqrt{W U_1}\, ,
\label{6.184}\eeq
where $\xi$ is a constant of the order of unity. Hence, the width of the
Bethe ridge increases proportionally to $W^{1/2}$. The inequality
\req{6.184} implies that the GOS is substantial when
\beq
| \ln (Q/U_1) - \ln (W/U_1) |
\lesssim \xi \sqrt{U_1/W}\, .
\label{6.185}\eeq
That is, the width of the ridge in the $\ln Q$ scale is $\sim \xi
(U_1/W)^{1/2}$, which decreases with $W$.

In the case of the electron gas [Eq.\ \req{6.173}] the width of the
Bethe ridge is
\beq
Q_+^{\rm b} - Q_-^{\rm b} = 4 \sqrt{(W+E_{\rm F})E_{\rm F}}
\label{6.186}\eeq
and, evidently, it increases with $W$ and with the Fermi energy $E_{\rm
F}$ of the gas. When $W \gg E_{\rm F}$, the width of the Bethe ridge in the logarithmic plot is
\beq
\ln Q_+^{\rm b} - \ln Q_-^{\rm b} = 2 \ln \left(
\frac{1 + \sqrt{E_{\rm F}/(E_{\rm F}+W)}}{1 - \sqrt{E_{\rm F}/(E_{\rm F}+W)}} \right)
\simeq 4 \sqrt{\frac{E_{\rm F}}{W}}.
\label{6.187}\eeq
Also in this case, the width of the ridge in the $\ln Q$ scale decreases
when $W$ increases.

As a consequence of this tendency, at sufficiently high excitation
energies, the GOS of any target can be approximated by the delta
distribution, $Z \delta(W-Q)$, where $Z$ is the number of electrons in
the target. For projectiles with sufficiently high energies, the maximum
allowed energy loss is determined by the intersect of the curve
$W_\textrm{m} (Q)$, Eq.\ \req{6.122}, with the Bethe ridge ($W=Q$),
which occurs at a point where $W$ has the value
\beq
W_{\rm ridge} = 4 E \left( \sqrt{M_0/\me} + \sqrt{\me/M_0} \right)^{-2}.
\label{6.188}\eeq
Note that, when $M_0=\me$, $W_{\rm ridge}=E$. For heavy projectiles
($M_0 \gg \me$),
\beq
W_{\rm ridge}\simeq 2\me v^2 \, .
\label{6.189}\eeq

\index{binary encounter approximation|)}
\index{impulse approximation|)}


\section{Electron collisions. The Ochkur approximation \label{sec6.5}}
\index{Ochkur approximation|(}

We will now study the effect of exchange in inelastic collisions of
projectile electrons ($Z_0 = -1$, $M_0=\me$) with the electrons in a
closed shell of an atom or ion. The indistinguishability of the
projectile and target electrons leads to observable exchange effects
which, in principle, can be described by anti-symmetrizing the wave
functions of the initial {\it and} final states (see Section
\ref{sec6.1.2.1}). In practice, however, such a modification of the PWBA
is difficult because the projectile plane waves are not orthogonal to
the bound and free orbitals of the active target electron. The only case
where exchange effects can be treated in a consistent way within the
PWBA is that of binary collisions with free electrons, because the
unperturbed wave functions of free electrons are plane waves and these
waves are orthogonal to each other. The energy-loss DCS for binary
collisions with electrons at rest is given by the non-relativistic M\o
ller formula, Eq.\ \req{5.161}.

In the case of collisions with electrons bound to an atom, we can
account partially for exchange effects by using the approximation
proposed by \citeauthor{Ochkur1964} (\citeyear{Ochkur1964,Ochkur1965})
\citep[see also][]{Rudge1968}. Here we present a simple derivation of
Ochkur's approximation, which applies to both excitation and ionization.
Essentially, the approximation consists in ignoring the
non-orthogonality of atomic orbitals and plane waves, and approximating
the transition matrix elements in the form [cf.\ Eq.\ \req{6.43}]
\beq
T_{fi}^{\rm el} \simeq T_{fi}^{\rm dir} - T_{fi}^{\rm exch}\, ,
\label{6.190}\eeq
where
\beq
T_{fi}^{\rm dir}
= \left< \phi_{{\bf k}' m'_{\rm S}}(x_0) \, \psi_{\varepsilon_b \ell_b
m_b m_{{\rm S}b}}(x_1) \left| \frac{e^2}{|{\bf r}_1 - {\bf r}_0|}
\right| \phi_{{\bf k} m_{\rm S}}(x_0) \,
\psi_{n_a \ell_a m_a m_{{\rm S}a}}(x_1)  \right>
\label{6.191}\eeq
and
\beq
T_{fi}^{\rm exch}
= \left< \phi_{{\bf k}' m'_{\rm S}}(x_0) \, \psi_{\varepsilon_b \ell_b
m_b m_{{\rm S}b}}(x_1) \left| \frac{e^2}{|{\bf r}_1 - {\bf r}_0|}
\right| \phi_{{\bf k} m_{\rm S}}(x_1) \,
\psi_{n_a \ell_a m_a m_{{\rm S}a}}(x_0)  \right> \, ,
\label{6.192}\eeq
are the direct and exchange transition-matrix elements, respectively,
with the ``projectile'' states represented as plane waves.

The direct transition-matrix element is
\beqa
T_{fi}^{\rm dir} &=&
\int \d x_0 \int \d x_1 \;
\phi_{{\bf k}' m'_{\rm S}}^\ast(x_0) \, \psi_{\varepsilon_b \ell_b
m_b m_{{\rm S}b}}^\ast(x_1) \, \frac{e^2}{|{\bf r}_1 - {\bf r}_0|} \,
\phi_{{\bf k} m_{\rm S}}(x_0) \,
\psi_{n_a \ell_a m_a m_{{\rm S}a}}(x_1)
\nonumber \\ [2mm]
&=& \frac{e^2}{(2\pi)^3} \, \delta_{m'_{\rm S},m_{\rm S}}
\delta_{m_{{\rm S}b},m_{{\rm S}a}}
\int \d {\bf r}_0 \int \d {\bf r}_1
\frac{\exp \left( {\rm i} {\bf q}
\dotprod {\bf r}_0 \right)}{|{\bf r}_1 - {\bf r}_0|} \,
\psi_{\varepsilon_b \ell_b m_b}^\ast({\bf r}_1) \,
\psi_{n_a \ell_a m_a}({\bf r}_1)\, ,
\nonumber \eeqa
where ${\bf q} = {\bf k} - {\bf k}'$. Using the Bethe integral formula
\req{B.26}, we obtain
\beqa
T_{fi}^{\rm dir} &=&
\frac{e^2}{2\pi^2} \, \frac{1}{q^2}
\, \delta_{m'_{\rm S},m_{\rm S}}
\delta_{m_{{\rm S}b},m_{{\rm S}a}}
\int \d {\bf r}_1 \; \exp \left( {\rm i} {\bf q} \dotprod {\bf r}_1 \right)
\, \psi_{\varepsilon_b \ell_b m_b}^\ast({\bf r}_1) \,
\psi_{n_a \ell_a m_a}({\bf r}_1). \rule{10mm}{0mm}
\label{6.193}\eeqa

The exchange transition-matrix element can be transformed as follows,
\beqa
T_{fi}^{\rm exch} &=&
\int \d x_0 \int \d x_1 \;
\phi_{{\bf k}' m'_{\rm S}}^\dagger(x_0) \, \psi_{\varepsilon_b \ell_b
m_b m_{{\rm S}b}}^\dagger(x_1) \,
\frac{e^2}{|{\bf r}_1 - {\bf r}_0|} \,
\phi_{{\bf k} m_{\rm S}}(x_1) \,
\psi_{n_a \ell_a m_a m_{{\rm S}a}}(x_0)
\nonumber \\ [2mm]
&=& \frac{e^2}{(2\pi)^3}  \int \d {\bf r}_1 \;
\left[ \psi_{\varepsilon_b \ell_b m_b}^\ast({\bf r}_1)
 \exp({\rm i} {\bf k} \dotprod {\bf r}_1) \,
\chi_{m_{{\rm S}b}}^\dagger
 \chi_{m_{\rm S}} \right]
\nonumber \\ [2mm]
&& \mbox{} \times
\int \d {\bf r}_0 \;
\left[  \exp(-{\rm i} {\bf k}' \dotprod {\bf r}_0) \,
\psi_{n_a \ell_a m_a}({\bf r}_0)
\, \chi_{m_{\rm S}'}^\dagger \chi_{m_{\rm S}a} \right]
\frac{1}{|{\bf r}_1 - {\bf r}_0|} \, .
\label{6.194}\eeqa
Let us consider first the integral over ${\bf r}_0$,
\beq
X({\bf r}_1) \equiv
\int \d {\bf r}_0
\exp(-{\rm i} {\bf k}' \dotprod {\bf r}_0)
\psi_{n_a \ell_a m_a}({\bf r}_0)
\, \frac{1}{|{\bf r}_1 - {\bf r}_0|} \, .
\label{6.195}\eeq
\index{Fourier transform}Introducing the Fourier transform of the bound orbital
$\psi_{n_a \ell_a m_a}$,
\beq
\widetilde{\psi}_{n_a \ell_a m_a}({\bf k}_0) \equiv (2\pi)^{-3/2}
\int  \d {\bf r}_0\, \exp(-{\rm i} {\bf k}_0 \cdot {\bf r}_0) \,
\psi_{n_a \ell_a m_a}({\bf r}_0)\, ,
\label{6.196}\eeq
we have
\beq
\psi_{n_a \ell_a m_a}({\bf r}_0) = (2\pi)^{-3/2}
\int \d {\bf k}_0 \, \exp({\rm i} {\bf k}_0 \cdot {\bf r}_0) \,
\widetilde{\psi}_{n_a \ell_a m_a} ({\bf k}_0)
\label{6.197}\eeq
and
$$
X({\bf r}_1) = (2\pi)^{-3/2}
\int \d {\bf k}_0 \,
\widetilde{\psi}_{n_a \ell_{{\rm L}a} m_a} ({\bf k}_0)
\int \d {\bf r}_0
\frac{\exp[{\rm i} ({\bf k}_0-{\bf k}') \dotprod {\bf r}_0]}
{|{\bf r}_1 - {\bf r}_0|} \, ,
$$
which, with the aid of the Bethe integral \req{B.26}, can be written as
\beq
X({\bf r}_1) = 4 \pi (2\pi)^{-3/2}
\, \exp(-{\rm i}{\bf k}'\dotprod{\bf r}_1)
\int \d {\bf k}_0 \,
\widetilde{\psi}_{n_a \ell_a m_a} ({\bf k}_0) \,
\frac{\exp({\rm i} {\bf k}_0\dotprod {\bf r}_1)}
{|{\bf k}_0-{\bf k}'|^2} \, .
\label{6.198}\eeq

Now, we recall that the PWBA provides the dominant term of the expansion
of the exact transition matrix in powers of $k^{-2}$ \citep[see,
\eg,][]{Joachain1975}. Assuming that the Ochkur approximation gives
the correct value of the exchange transition matrix for high energies,
we can expand $T_{fi}^{\rm exch}$ as a power series in $k^{-2}$ and
keep only the lowest-order term. It is legitimate to disregard
higher-order terms, because they are of the same order as the
corrections introduced by the second and higher Born approximations.
The average squared momentum of the target electron is
\beq
\hbar^2 \left< {\bf k}_0^2 \right> \equiv \hbar^2 \int k_0^2 \left|
\widetilde{\psi}_{n_a \ell_a m_a}({\bf k}_0) \right|^2
 \, \d {\bf k}_0\, ,
\label{6.199}\eeq
and can be expressed as
\beq
\hbar^2 \left< k_0^2 \right> =
2 \me \langle K \rangle_{n_a\ell_a},
\label{6.200}\eeq
where $\langle K \rangle_{n_a\ell_a}$ is the kinetic energy of the
target electron in its initial state. For hydrogenic ions, $\langle K
\rangle_{n_a\ell_a}$ equals the ionization energy $U_a = -\varepsilon_{n_a
\ell_a}$, by virtue of the virial theorem. \index{virial theorem}
In general, $\langle K
\rangle_{n_a\ell_a}= \xi U_a$, where $\xi$ is of the
order of unity. With the approximation
\beq
\frac{1}{|{\bf k}_0 - {\bf k}' |^2} = \frac{1}{ \left< k_0^2
\right> + k'^2}\, \left[ 1 + \frac{k_0^2 - \left< k_0^2
\right> - 2 {\bf k}_0 \dotprod {\bf k}'}{ \left< k_0^2 \right> +
k'^2} \right]^{-1} \simeq \frac{1}{ \left< k_0^2 \right> + k'^2}\, ,
\label{6.201}\eeq
and the equality \req{6.197}, Eq.\ \req{6.198} leads to
\beq
X({\bf r}_1) = 4 \pi \, \frac{1}{ \left< k_0^2 \right> + k'^2}
\, \exp(-{\rm i}{\bf k}'\dotprod{\bf r}_1)
\, {\psi}_{n_a \ell_{{\rm L}a} m_a} ({\bf r}_1) \, .
\label{6.202}\eeq
Introducing this result into Eq.\ \req{6.194}, we get
\beqa
T_{fi}^{\rm exch} &=&
\frac{e^2}{(2\pi)^3}  \int \d {\bf r}_1 \;
\left[ \psi_{\varepsilon_b \ell_b m_b}^\ast({\bf r}_1) \,
 \exp({\rm i} {\bf k} \dotprod {\bf r}) \,
\chi_{m_{{\rm S}b}}^\dagger
 \chi_{m_{\rm S}} \right]
\nonumber \\ [2mm]
&& \mbox{} \times
 4 \pi \, \frac{1}{ \left< k_0^2 \right> + k'^2}
\, \exp(-{\rm i}{\bf k}'\dotprod{\bf r}_1)
\, {\psi}_{n_a \ell_{{\rm L}a} m_a} ({\bf r}_1)
\, \chi_{m_{\rm S}'}^\dagger \chi_{m_{\rm S}a}
\nonumber \\ [2mm]
&=&  \frac{e^2}{2\pi^2}
\, \frac{1}{ \left< k_0^2 \right> + k'^2}
\int \d {\bf r}_1 \;  \exp({\rm i} {\bf q} \dotprod {\bf r}_1) \,
\psi_{\varepsilon_b \ell_b m_b}^\ast({\bf r}_1)
\, {\psi}_{n_a \ell_a m_a} ({\bf r}_1)
\nonumber \\ [2mm]
&& \mbox{} \times
\chi_{m_{{\rm S}b}}^\dagger
\left( \chi_{m_{\rm S}} \chi_{m_{\rm S}'}^\dagger  \right)
\chi_{m_{{\rm S}a}}\, .
\label{6.203}\eeqa

The complete $T$-matrix element for electron collisions is
\beqa
T_{fi}^{\rm el} &=& T_{fi}^{\rm dir} - T_{fi}^{\rm exch}
\nonumber \\ [2mm]
&=& {\cal F}_{ba} \, \delta_{m_{{\rm S}b},m_{{\rm S}a}}
\, \delta_{m'_{\rm S},m_{\rm S}} - {\cal G}_{ba} \,
\chi_{m_{{\rm S}b}}^\dagger
\left( \chi_{m_{\rm S}} \chi_{m_{\rm S}'}^\dagger  \right)
\chi_{m_{{\rm S}a}} \, ,
\label{6.204}\eeqa
where
\beq
{\cal F}_{ba} = \frac{e^2}{2\pi^4}
\, \frac{1}{q^2}
\left<  \psi_{\varepsilon_b \ell_b m_b} \left|
\exp \left( {\rm i} {\bf q} \dotprod {\bf r} \right) \left|
\psi_{n_a \ell_a m_a} \right>\right.\right.
\label{6.205}\eeq
and
\beq
{\cal G}_{ba} = \frac{q^2}{ \left< k_0^2 \right> + k'^2} \,
{\cal F}_{ba} \, .
\label{6.206}\eeq
With these expressions, the sum over spins that occurs in
the definition of the DDCS and the GOS can be readily evaluated,
\beqa
{\cal T}_{fi}^\textrm{Ochkur} &=& \frac{1}{4}
\sum_{m_{{\rm S}a}, m_{{\rm S}b}} \; \; \sum_{m_{\rm S},m'_{\rm S}} \;
\left| T_{fi}^{\rm el} \right|^2 = \frac{1}{4}
\sum_{m_{{\rm S}a}, m_{{\rm S}b}} \; \; \sum_{m_{\rm S},m'_{\rm S}} \;
\left| T_{fi}^{\rm dir} - T_{fi}^{\rm exch}  \right|^2
\nonumber \\ [2mm]
&=& \frac{1}{4} \sum_{m_{{\rm S}a}, m_{{\rm S}b}} \; \;
\sum_{m_{\rm S},m'_{\rm S}} \;
\left(  {\cal F}_{ba} \, \delta_{m_{{\rm S}b},m_{{\rm S}a}}
\, \delta_{m'_{\rm S},m_{\rm S}} - {\cal G}_{ba} \,
\chi_{m_{{\rm S}b}}^\dagger
\left( \chi_{m_{\rm S}} \chi_{m_{\rm S}'}^\dagger  \right)
\chi_{m_{{\rm S}a}}  \right)
\nonumber \\ [2mm]
&& \mbox{} \times
\left(  {\cal F}_{ba}^\ast \, \delta_{m_{{\rm S}b},m_{{\rm S}a}}
\, \delta_{m'_{\rm S},m_{\rm S}} - {\cal G}_{ba}^\ast \,
\chi_{m_{{\rm S}a}}^\dagger
\left( \chi_{m_{\rm S}'} \chi_{m_{\rm S}}^\dagger  \right)
\chi_{m_{{\rm S}b}}  \right)
\nonumber \\ [2mm]
&=& |{\cal F}_{ba}|^2 + |{\cal G}_{ba}|^2 -
\frac{1}{2} \left( {\cal F}_{ba} {\cal G}_{ba}^\ast +
{\cal F}_{ba}^\ast {\cal G}_{ba} \right)\, ,
\label{6.207}\eeqa
where we have used the closure relation for spin states,
$$
\sum_{m_\textrm{S}} \chi_{m_{\rm S}} \chi^\dagger_{m_{\rm S}'} = I_2.
$$
That is,
\beqa
{\cal T}_{fi}^\textrm{Ochkur} &=&  |{\cal F}_{ba}|^2 + |{\cal G}_{ba}|^2
- \1o2 \left( {\cal F}_{ba} {\cal G}_{ba}^\ast +
{\cal F}_{ba}^\ast {\cal G}_{ba} \right)
\nonumber \\ [2mm]
&=& \left[ 1 + \left(\frac{q^2}{ \left< k_0^2 \right> + k'^2}\right)^2
- \frac{q^2}{ \left< k_0^2 \right> + k'^2}
\right] |{\cal F}_{ba}|^2.
\label{6.208}\eeqa
Or, in terms of the recoil energy, $Q \equiv (\hbar q)^2/2\me$,
\beqa
{\cal T}_{fi}^\textrm{Ochkur}
&=& \left[ 1 + \left(\frac{Q}{ \langle K \rangle_{n_a\ell_a}
+ E - W} \right)^2
- \frac{Q}{ \langle K \rangle_{n_a\ell_a} + E - W}
\right] |{\cal F}_{ba}|^2. \rule{10mm}{0mm}
\label{6.209}\eeqa
The first term in this expression corresponds to direct transitions, the
second term describes exchange transitions, and the last one arises from
interference between direct and exchange transitions.

When the projectile particle is a positron, or a distinguishable
electron, the transition matrix elements obtained from the PWBA are
$T_{fi} = {\cal F}_{ba} \, \delta_{m_{{\rm S}b},m_{{\rm S}a}} \,
\delta_{m'_{\rm S},m_{\rm S}}$, and the sum over spins is
\beq
{\cal T}_{fi}^{\rm pw} = \frac{1}{4}
\sum_{m_{{\rm S}a}, m_{{\rm S}b}} \; \; \sum_{m_{\rm S},m'_{\rm S}} \;
\left| T_{fi}^{\rm el} \right|^2
=  |{\cal F}_{ba}|^2 \, .
\label{6.210}\eeq
Hence, the DDCS resulting from the Ochkur approximation is obtained by
simply multiplying the DDCS derived within the PWBA by the factor in
square brackets on the right-hand side of Eq.\ \req{6.209},
\beq
\frac{\d^2 \sigma^{\rm Ochkur}_a}{ \d W \, \d Q } = \left[ 1 +
\left(\frac{Q}{ \langle K \rangle_{n_a\ell_a} + E - W}\right)^2
- \frac{Q}{\langle K \rangle_{n_a\ell_a} + E-W} \right]
\frac{\d^2 \sigma_a}{ \d W \, \d Q }\, .
\label{6.211}\eeq
This result is valid for both excitation and ionization. Transitions of
the target electron to discrete bound levels $\varepsilon_b$ are allowed
whenever the energy loss $W = \varepsilon_b - \varepsilon_a$ is less than the
kinetic energy of the projectile. Ionizing collisions are allowed only
for energy transfers larger than the shell ionization energy $U_a=-\varepsilon_a$; after the interaction we have two free electrons
with energies $E'=E-W$ and $E_\textrm{s} = W-U_a$. Since these
two electrons are indistinguishable, we will consider the fastest one as
the ``primary''. Because of this convention, the maximum allowed energy
loss is $W_{\rm max} = (E + U_a)/2$. Evidently, when $W=W_{\rm
max}$ the primary and secondary electrons have the same energy,
$E'=E_{\rm s}$.

The Ochkur approximation yields the following DDCS for collisions
of an electron with a free electron at rest,
\index{M\o ller cross section}
\beq
\frac{\d^{2}\sigma^{\rm Ochkur}}{\d W\d Q} =
\frac{2\pi e^{4}}{\me v^2} \, \frac{1}{WQ}
\left( 1 + \frac{Q^2}{(
E - W )^2 } - \frac{Q}{E - W}
\right) \, \delta(W-Q)\, .
\label{6.212}\eeq
This result coincides with the non-relativistic M\o ller DDCS, Eq.\
\req{5.161}. That is, the Ochkur approximation is exact (within the
non-relativistic PWBA) for interactions of a non-relativistic electron
with a free electron at rest.
\index{Ochkur approximation|)}


\section{The relativistic plane-wave Born approximation \label{sec6.6}}
\index{relativistic Born approximation}

The relativistic PWBA for collisions of charged particles with atoms was
first formulated by \citeauthor{Bethe1930}
(\citeyear{Bethe1930,Bethe1932}) and it was later recast in a somewhat more
transparent form by \citet{Fano1963}. The theory is based on an
independent-electron model of the atom, with electron orbitals that are
solutions of the Dirac equation for a self-consistent atomic potential
(\eg, the DHFS potential). Aside from the necessary use of relativistic
kinematics (see Appendix \ref{appA}), the relativistic formulation
differs from its non-relativistic version in that the interaction
between the projectile (0) and the active target electron (1) consists
of the instantaneous (longitudinal) Coulomb interaction plus a
transverse term that accounts for the exchange of virtual photons
between the projectile and the target electron. The transverse term is
obtained by describing the interaction through the quantized
electromagnetic field as a perturbation to second order \citep[see,
\eg,][]{BetheSalpeter1957}. Specifically, in the case of spin-$\1o2$
projectiles different from the electron, the interaction is represented
by the effective Hamiltonian \citep{Fano1963}
\beq
{\cal H}_{\rm int}(0,1) =
- \frac{Z_0 e^2}{2\pi^2}
\int \d {\bf q} \, \left( \frac{1}{q^2} - \frac{
\widetilde{\alphab}_0 \dotprod \widetilde{\alphab}_1 -
(\widetilde{\alphab}_0 \dotprod \hat{\bf q})
(\widetilde{\alphab}_1 \dotprod \hat{\bf q})}{q^2 - (W/\hbar c)^2} \right)
\exp\left[ {\rm i} {\bf q} \dotprod \left({\bf r}_1 -
{\bf r}_0 \right) \right].  \rule{5mm}{0mm}
\label{6.213}\eeq
where $\widetilde{\alphab}_i$ and $\widetilde{\beta}_i$ are the Dirac
matrices of the two particles. The first term in this expression is the
unretarded Coulomb interaction,
\beq
- \frac{Z_0 e^2}{2\pi^2}
\int \d {\bf q} \, \frac{1}{q^2} \,
\exp\left[ {\rm i} {\bf q} \dotprod \left({\bf r}_1 -
{\bf r}_0 \right) \right] =
- \frac{Z_0 e^2}{ |{\bf r}_1 - {\bf r}_0|},
\label{6.214}\eeq
as we can readily verify by expressing the integral in spherical
coordinates with the polar axis in the direction of the vector ${\bf r}
= {\bf r}_1 - {\bf r}_0$.  The second term in \req{6.213} represents the
transverse interaction, \ie, the effect of the exchange of a virtual
photon between the projectile and the active target electron.

Following \citet{Fano1963} the recoil energy $Q$ is defined as the
kinetic energy of an electron with linear momentum equal to the momentum
transfer $\hbar {\bf q}= {\bf p} - {\bf p}'$, that is,
\beq
Q(Q+2\me c^2)= (c\hbar q)^2 = c^2 (p^2 + p'^2 - 2 p p' \, \cos\theta)\, ,
\label{6.215}\eeq
where $\theta$ is the polar scattering angle (see Fig.\ \ref{figA.2}).
More explicitly,
\beq
Q = \sqrt{ (c\hbar q)^2 + \me^2 c^4} \, - \me c^2 .
\label{6.216}\eeq
For a given energy loss $W$, the allowed recoil energies
are limited to the interval $(Q_-,Q_+)$ with [Eq.\ \req{A.69}]
\beq
Q_{\pm} = \sqrt{ \left[ \sqrt{E(E+2M_0c^2)} \pm
\sqrt{(E-W)(E-W+2M_0c^2)}\right]^2 + \me^2
c^4 } - \me c^2\, .
\label{6.217}\eeq
When $W \ll E$,
\beq
Q_- \simeq \frac{W^2}{2 \me c^2 \, \beta^2},
\label{6.218}\eeq
where $\beta=v/c$.
Conversely, for a given recoil energy $Q< Q_+$, the maximum allowed
energy loss is [Eq.\ \req{A.75}]
\beq
W_{\rm m}(Q) = E + M_0c^2 - \sqrt{\left[ \sqrt{E(E+2M_0 c^2)} -
\sqrt{Q(Q+2\me c^2)} \right]^2 + M_0^2 c^4}\, .
\label{6.219}\eeq
When the kinetic energy of the projectile is much larger than $W_{\rm
m}(Q)$, we can use the approximation [Eq.\ \req{A.76}]
\beq
W_{\rm m}(Q) \simeq \beta \sqrt{Q(Q+2\me c^2)}= \hbar q v \, .
\label{6.220}\eeq

The relativistic PWBA may be formulated in complete analogy with the
non-rel\-a\-tiv\-is\-tic theory \citep{Fano1963, BoteSalvat2008}, with the
projectile wave functions represented by Dirac plane waves. For
concreteness, we consider inelastic collisions of projectiles with mass
$M_0$ and charge $Z_0 e$ with atoms of the element of atomic number $Z$,
under the usual conditions of spin-unpolarized projectiles and randomly
oriented atoms. \citet{BoteSalvat2008} expressed the DDCS as the sum of
contributions from longitudinal (L) and transverse (T) interactions,
\begin{subequations}
\label{6.221}
\beq
\frac{\d^2 \sigma}{\d W \, \d Q} = \frac{\d^2 \sigma^{\rm L}}{\d
W \, \d Q} + \frac{\d^2 \sigma^{\rm T}}{\d W \, \d Q} \, ,
\label{6.221a}\eeq
with
\beqa
\frac{\d^2 \sigma^{\rm L}}{\d W \, \d Q}
&=& \frac{2\pi Z_0^2 e^4}{\me v^2} \,
\frac{2\me c^2}{WQ(Q+2\me c^2)}
\nonumber \\ [2mm]
&& \mbox{} \times
\left\{ 1 - \frac{4(E+M_0c^2)W - W^2 + Q(Q+2\me c^2)}{4\,
(E+M_0 c^2)^2} \right\}
\frac{\d f (Q,W)}{\d W} \rule{15mm}{0mm}
\label{6.221b}\eeqa
and
\beqa
\frac{\d^2 \sigma^{\rm T}}{\d W \, \d Q}
&=& \frac{2\pi Z_0^2 e^4}{\me v^2} \,
\frac{2\me c^2 W}{[Q(Q+2\me c^2)-W^2]^2} \,
\nonumber \\ [2mm]
&& \mbox{} \times
\left( \beta^2 \sin^2\theta_{\rm r}
+ \left\{ \frac{Q(Q+2\me c^2)- W^2 }{2(E+ M_0 c^2)^2}\,
\right\} \right) \frac{\d g (Q,W)}{\d W}\, , \rule{10mm}{0mm}
\label{6.221c}\eeqa
where $\theta_{\rm r}$ is the recoil angle, that is, the angle between the
momentum transfer ${\bf q}$ and the momentum $\hat{\bf p}$ of
the projectile before the interaction (Fig.\ \ref{figA.2}),
\beq
\beta^2 \sin^2\theta_\textrm{r} = \beta^2
- \frac{W^2}{Q(Q+2\me c^2)} \left\{ 1 + \frac{ Q(Q+2\me c^2)-W^2}
{2 W (E+M_0 c^2)} \right\}^2 \, .
\label{6.221d}\eeq
\end{subequations}
The functions $\d f(Q,W)/\d W$ and $\d g(Q,W)/\d W$ are the longitudinal
GOS and the transverse GOS (TGOS), respectively. In the non-relativistic
limit ($c \rightarrow 0$) the transverse DDCS vanishes and the
longitudinal DDCS reduces to the non-relativistic form \req{6.55}.

For projectiles heavier than the electron ($M_0 \gg \me$), and also for
fast electrons and positrons, the quantities in curly braces in Eqs.\
\req{6.221} are very close to unity or null. Then the DDCS takes the
simplified form
\beqa
\frac{\d^2 \sigma_{\rm PWBA}}{\d Q \, \d W}
&=& \frac{2\pi Z_0^2 e^4}{\me v^2} \,
\left[ \frac{2\me c^2}{WQ(Q+2\me c^2)} \, \frac{\d f(Q,W)}{\d W} \right.
\nonumber \\ [2mm]
&+& \left.\left( \beta^2 - \frac{W^2}{Q(Q+2 \me c^2)} \right)
\frac{ 2\me c^2 W } {[Q(Q+2\me c^2) - W^2 ]^2} \,
\frac{\d g(Q,W)}{\d W} \right], \rule{10mm}{0mm}
\label{6.222}\eeqa
which is used in virtually all practical calculations.

As in the non-relativistic case, the atomic GOSs are obtained as the sum
of contributions from the various electron subshells $a=(n_a \kappa_a)$
of the ground state configuration of the target atom,
\index{atomic configuration}
\beq
\frac{\d f (Q,W)}{\d W} = \sum_a \frac{\d f_a (Q,W)}{\d W}\, ,
\qquad
\frac{\d g (Q,W)}{\d W} = \sum_a \frac{\d g_a (Q,W)}{\d W} \, .
\label{6.223}\eeq
In the limit $Q \rightarrow 0$, both the longitudinal and the transverse
GOS reduce to the dipole oscillator strength
\beq
\frac{\d f (0,W)}{\d W} = \frac{\d g (0,W)}{\d W} =
\frac{\d f (W)}{\d W}.
\label{6.224}\eeq
For recoil energies $Q$ much larger than
the ionization energy, $U_a = -\varepsilon_{n_a\kappa_a}$, the GOS and
the TGOS differ from zero only in the vicinity of the line $W\sim Q$,
the Bethe ridge, because the target electrons then react as if they were
essentially free and at rest. The GOS and the TGOS for a free stationary
target electron are
\beq
\frac{\d f^{\rm free}(Q,W)}{\d W} = Z \, \delta(Q-W) \quad \mbox{and} \quad
\frac{\d g^{\rm free}(Q,W)}{\d W} = Z \, \delta(Q-W),
\label{6.225}\eeq
respectively. It follows that, for large enough recoil energies, the GOS
and the TGOS satisfy the Bethe sum rule. However, relativistic effects
cause slight deviations from the Bethe sum rule that are appreciable, of
the order of $-2.5$ \%, for high-$Z$ elements and small recoil energies
\citep{Cohen2003}.

\citet{Salvat2022a} have calculated extensive numerical tables of the
GOS and the TGOS of all subshells of neutral atoms ($Z=1$--99) using the
independent-electron model with the DHFS self-consistent potential (see
Section \ref{sec3.5}). Figures \ref{fig6.7} and \ref{fig6.8} display the
GOS and TGOS of helium (two electrons in the closed $1s_{1/2}$ shell),
and of the $2p_{3/2}$ subshell of silicon, respectively. The structure
of the Bethe surfaces is generally more complex for subshells with
higher angular momenta. The calculated dipole oscillator strength is such
that
\beq
\frac{\d f (W)}{\d W} = \frac{\me c}{2 \pi^2 \hbar e^2} \,
\sigma_{\rm ph}^{\rm dipole} (W),
\label{6.226}\eeq
where $\sigma_{\rm ph}^{\rm dipole} (W)$ is the atomic cross section for
absorption of photons of energy $W$ calculated within the dipole
approximation. \citet{SabbatucciSalvat2016} have computed accurate
photoabsorption cross sections for all the elements using the DHFS
atomic potential. The theory of the photoeffect
\citep{SabbatucciSalvat2016}, as well as the relativistic PWBA
\citep{BoteSalvat2008}, describes excited states as if they were
stationary, \ie, the finite lifetime of excited states is neglected. As
a result, the calculated GOSs and photoabsorption cross sections
represent excitations to bound states as a series of discrete resonances
(delta functions), while the actual functions vary continuously with
$W$. This defect of the theory can be partially corrected by convolving
the calculated GOS with a Lorentzian distribution having a full-width at
half maximum equal to the line width of atomic levels
\citep{SabbatucciSalvat2016}. Such a correction is normally ignored
because it produces changes in the stopping power and mean free path
that are much smaller than the uncertainties arising from the
approximations in the atomic states.

\begin{figure}[p!]
\begin{center}
\includegraphics*[width=10.0cm] {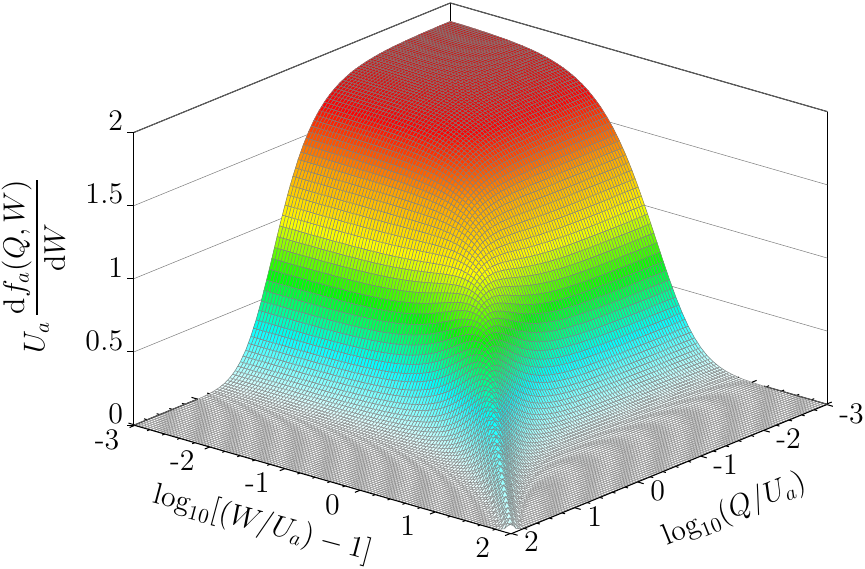}  \\ [2mm]
\includegraphics*[width=10.0cm] {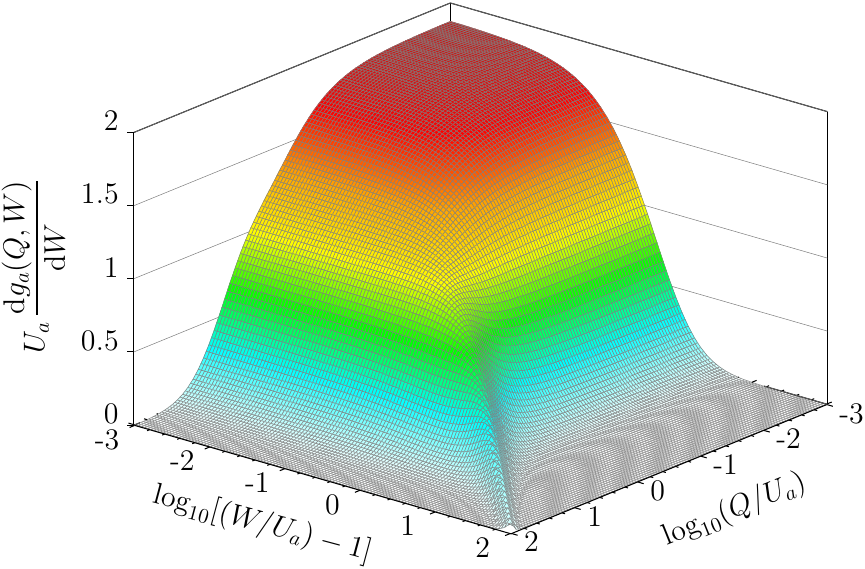}  \\ [2mm]
\includegraphics*[width=7cm] {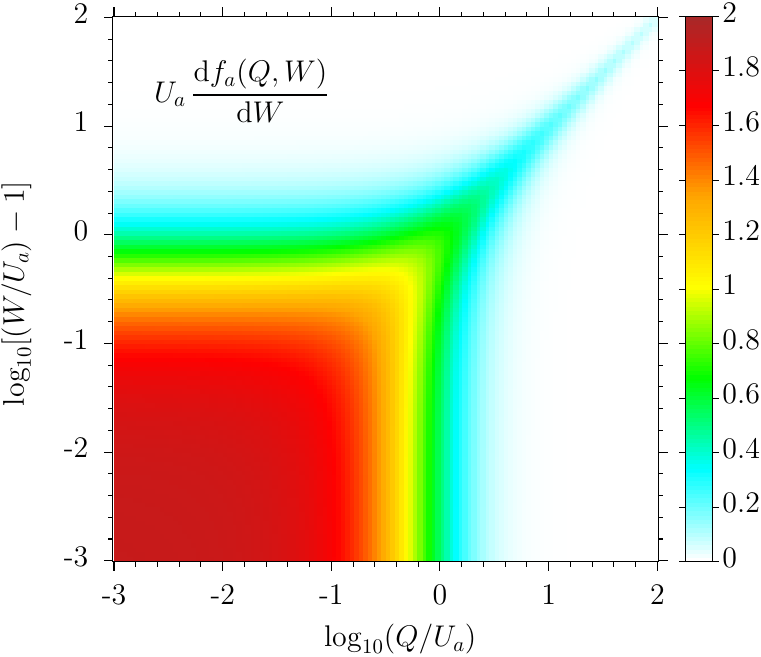} \rule{7mm}{0mm}
\includegraphics*[width=7cm]{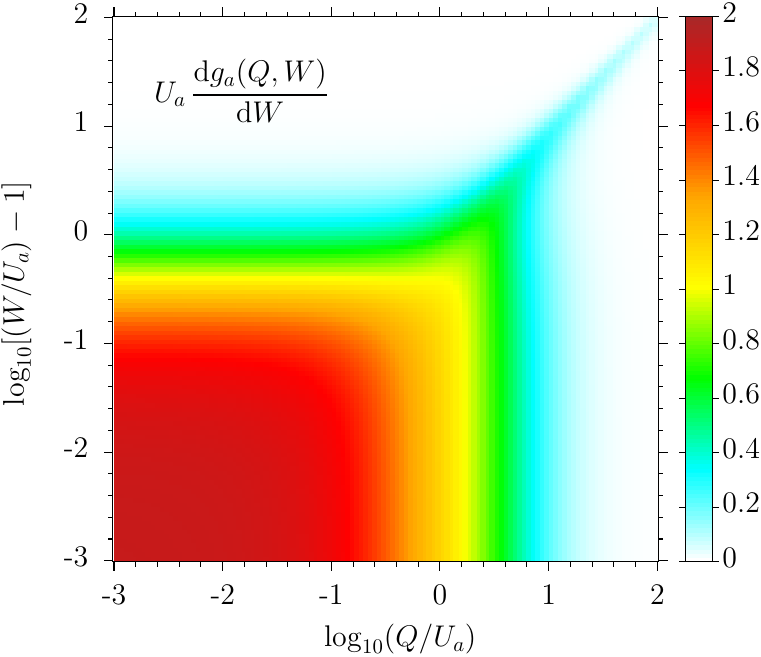}
\caption{Relativistic longitudinal and transverse GOSs for the
ionization of the helium atom ($Z=2$, two electrons in the closed 1$s_{1/2}$
shell), calculated by \citet{Salvat2022a}.
Energies are expressed in units of the shell ionization energy $U_a =
- \varepsilon_{n_a\kappa_a}$. The GOSs are displayed as 3-dimensional
surfaces (top and middle plots) and as color-level diagrams (bottom).
\label{fig6.7}}
\end{center}\end{figure}

\begin{figure}[p!]
\begin{center}
\includegraphics*[width=10.0cm] {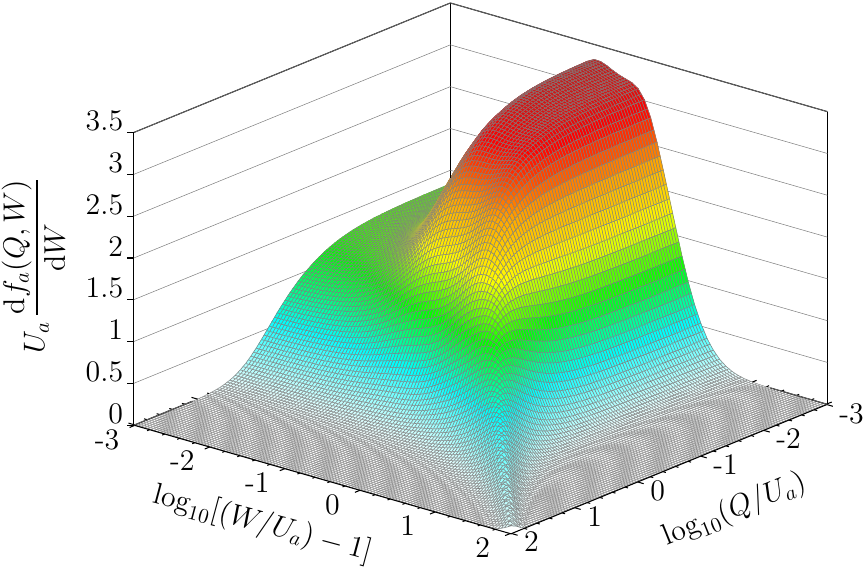}  \\ [2mm]
\includegraphics*[width=10.0cm] {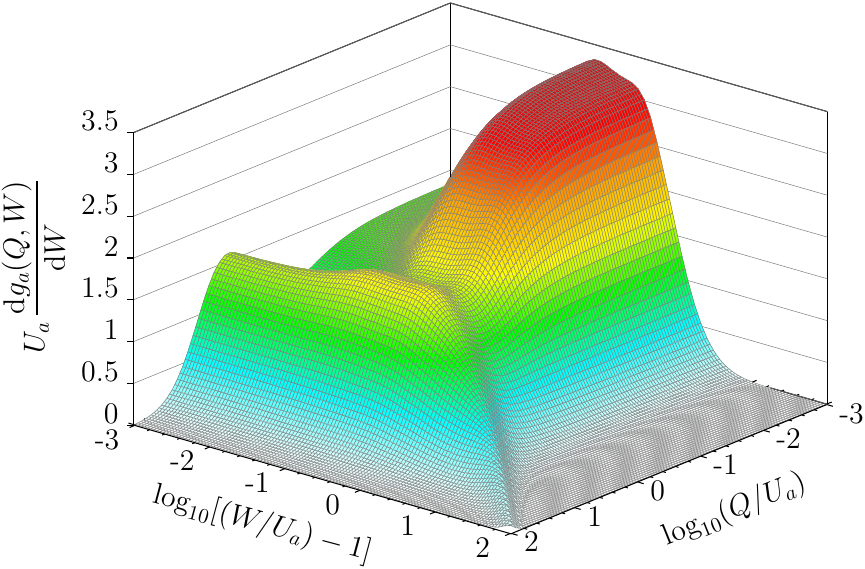}  \\ [2mm]
\includegraphics*[width=7cm] {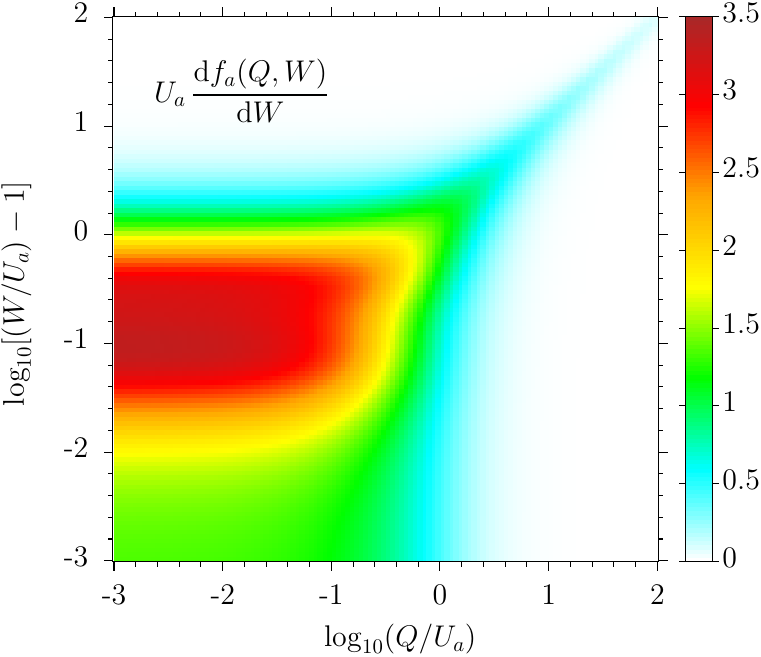} \rule{7mm}{0mm}
\includegraphics*[width=7cm]{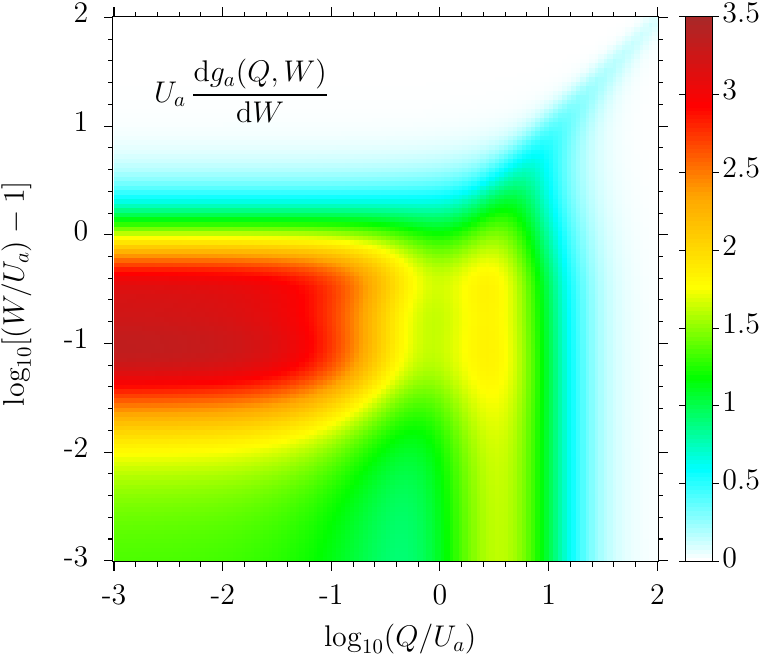}
\caption{Relativistic longitudinal and transverse GOSs for the
ionization of the 2$p_{3/2}$
subshell of the silicon atom ($Z=14$), calculated by
\citet{Salvat2022a}.
Energies are expressed in units of the subshell ionization energy $U_a =
- \varepsilon_{n_a\kappa_a}$. The GOSs are displayed as 3-dimensional
surfaces (top and middle plots) and as color-level diagrams (bottom).
\label{fig6.8}}
\end{center}\end{figure}

The atomic energy-loss DCS is obtained as [cf.\ \req{6.123}]
\beq
\frac{\d \sigma}{\d W}
= \int_{Q_{-}}^{Q_{+}}\frac{\d^2 \sigma}{\d W \,\d Q} \, \d Q
\, ,
\label{6.227}\eeq
where $Q_-$ is given by Eq.\ \req{6.217}.  As in the non-relativistic
theory, for high-energy projectiles and energy losses much smaller than
$E$, $Q_+$ is much larger than $W$ and the GOSs effectively vanish for
$Q > Q_+$. Then the upper limit of the integral can be replaced by
$\infty$. Hence, the energy-loss DCS for energy losses much less than
$E$ does not depend on the mass $M_0$ of the projectile. Of course, this
is not true when $W$ is comparable to $E$, because then the values of
both $Q_-$ and $Q_+$ do depend on $M_0$

\index{total inelastic cross section} \index{stopping cross section}
\index{energy straggling cross section} \index{straggling cross section}
The total inelastic cross section, $\sigma^{(0)}$, the stopping cross
section, $\sigma^{(1)}$, and the energy-straggling cross section,
$\sigma^{(2)}$, are given by [cf.\ Eq.\ \req{6.125a}]
\beq
\sigma^{(k)} \equiv
\int_{0}^{E} W^{k} \, \frac{\d\sigma}{\d W} \, \d W\, .
\label{6.228}\eeq
The relations \req{6.223} imply that these integrated cross sections can
be expressed as sums of the corresponding contributions from the various
electron subshells,
\beq
\sigma^{(k)} = \sum_a \sigma^{(k)}_a \qquad
\mbox{with} \qquad \sigma^{(k)}_a \equiv
\int_{0}^{E} W^{k} \, \frac{\d\sigma_a}{\d W} \, \d W\, ,
\label{6.229}\eeq
where $\d\sigma_a/\d W$ is the energy-loss DCS for excitations of the
$a$ subshell.

Figure \ref{fig6.9} displays the integrated cross sections
$\sigma^{(0)}$ and $\sigma^{(1)}$ for inelastic collisions of protons
with aluminium and gold atoms as functions of the kinetic energy of the
projectile. The atomic cross sections (solid curves) were obtained as the
sums of contributions from individual electron subshells, also displayed
in the plots, which were evaluated from the GOS tables computed with the
self-consistent DHFS potential \citep{Salvat2022a}. Notice that, because
of the extra factor in the definition \req{6.228}, the relative
contributions of inner subshells to the stopping cross section are
larger than those to the total inelastic cross section.

\begin{figure}[p!]
\begin{center}
\includegraphics*[width=7.50cm] {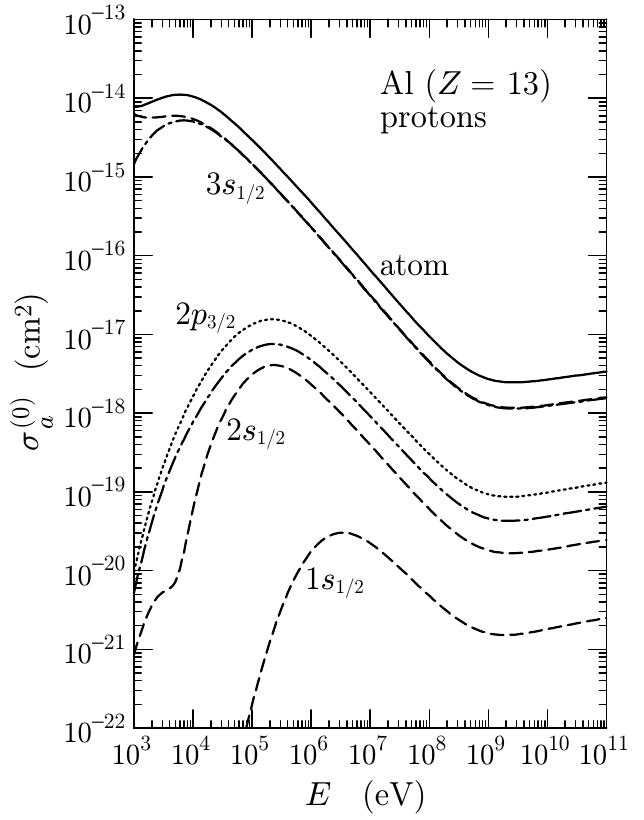} \rule{5mm}{0mm}
\includegraphics*[width=7.50cm] {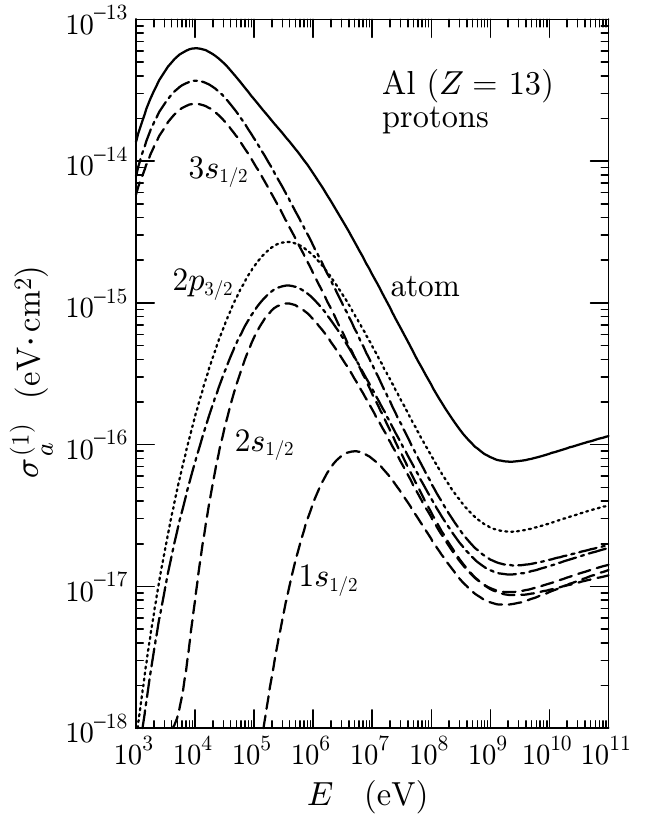} \\ [5mm]
\includegraphics*[width=7.50cm] {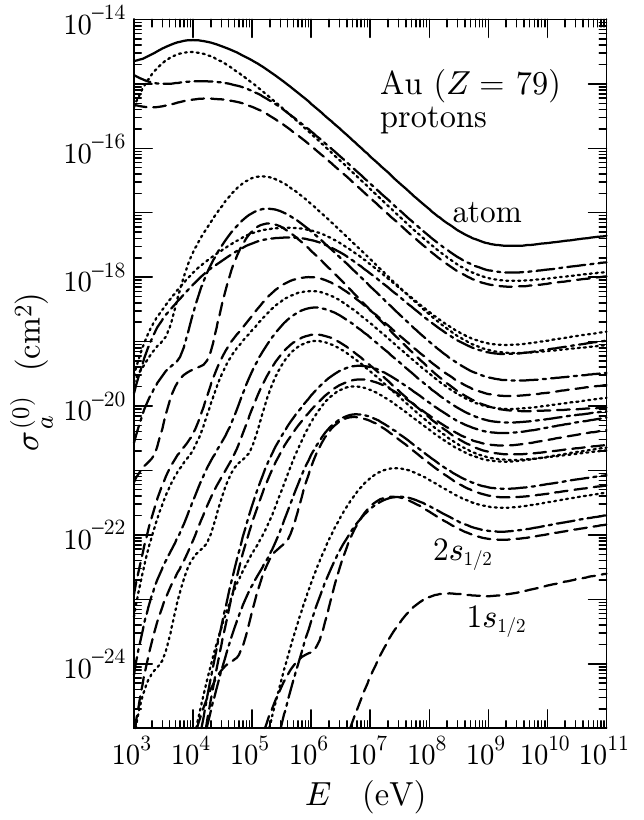} \rule{5mm}{0mm}
\includegraphics*[width=7.50cm] {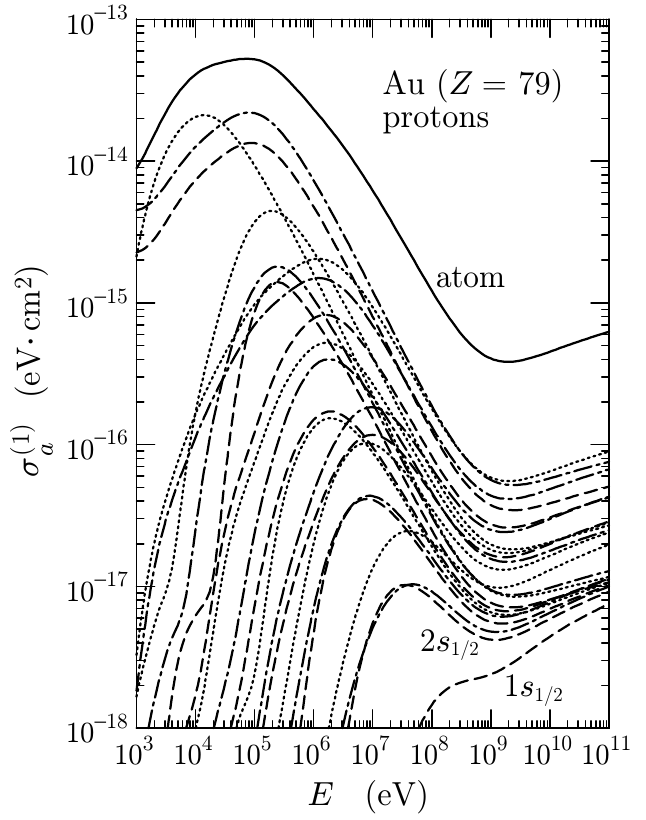}
\caption{Atomic total and stopping cross sections (solid curves) for
	inelastic
collisions of protons with aluminium (top) and gold (bottom) atoms,
calculated from the relativistic PWBA with GOSs computed with the
self-consistent DHFS potential. The
contributions of the various electron subshells are also displayed.
\label{fig6.9}}
\end{center}\end{figure}


\subsection{Collisions of electrons \label{sec6.6.1}}
\index{collisions of electrons}

Collisions of electrons ($Z_0=-1$, $M_0=\me$) with atoms differ from
those of heavier particles in that the projectile is indistinguishable
from the atomic electrons (see Section \ref{sec6.5}) and, consequently,
interactions are affected by exchange effects (\ie, re-arrangement
collisions and interference between direct and exchange transitions). As
in the non-relativistic theory, it is difficult to make allowance of
exchange within the PWBA, because the projectile plane waves are not
orthogonal to the bound and free orbitals of the active electron.
Exchange corrections can be calculated consistently only in the case of
collisions with free electrons at rest (because both the projectile and
the target are then described by plane waves); the corresponding
exchange-corrected PWBA DCS for projectile electrons was derived by
\citet{Moller1932}. The common approach to account for the effect
of electron exchange on the stopping power \citep{RohrlichCarlson1954,
ICRU37} of electrons and positrons is to consider that, for energy
transfers $W$ much larger than the ionization energy $U_a$, the target
electrons can be regarded as free and at rest.

In the case of collisions of non-relativistic electrons, exchange
effects can be described approximately by using the Ochkur
approximation (Section \ref{sec6.5}), which consists of
multiplying the GOS by the factor
\beq
F_\textrm{Ochkur} (Q,W) =
1+\left(\frac{Q}{\langle K \rangle+ E-W}\right)^{2}
-\frac{Q}{\langle K \rangle +E-W}\, ,
\label{6.230}\eeq
where $\langle K \rangle$ is the kinetic energy of the target
electron.
Notice that the Ochkur factor \req{6.230} reduces to unity when $Q=0$,
\ie, exchange effects are negligible for small-$Q$ collisions. This
peculiarity does make sense because, in a semi-classical picture,
interactions with small momentum transfers correspond to distant
collisions (with large impact parameters) in which the two particles
retain their identities. In ionizing collisions, we have two
(indistinguishable) free electrons in the final state, and it is natural
to consider the fastest as the ``primary''. Consequently, the largest
allowed energy loss in ionizations of a subshell $a=(n_a\kappa_a)$ is
\beq
W_{\rm max} = (E+U_a)/2 .
\label{6.231}\eeq
In the case of excitation to bound levels, the ``primary'' electron is
the one that remains free after the collisions and, therefore, the
maximum energy loss is $W_{\rm max}=E$.

The relativistic energy-loss DCS for collisions with a free electron at
rest was calculated by \citet{Moller1932} using the methods of quantum
electrodynamics. It is given by the formula,
\index{M\o ller cross section}
\beqa
\frac{\d \sigma_{\rm M\o ller}}{\d W} & = &
\frac{2 \pi e^4}{\me v^2} \frac{1}{W^{2}}
\, F_{\rm M\o ller} (W) \, ,
\label{6.232}\eeqa
where
\beq
F_{\rm M\o ller} (W) =
1  + \left( \frac{W}{E-W} \right)^{2}
- \frac{(1-b_0)W}{E-W} + \frac{b_0 W^{2}}{E^{2}}
\label{6.233}\eeq
with
\beq
b_0 = \left( \frac{E}{E+\me c^{2}} \right)^{2}
= \left( \frac{\gamma-1}{\gamma} \right)^{2} = \left( 1 -
\sqrt{1-\beta^2} \right)^2\, ,
\label{6.234}\eeq
and $\gamma= (1 -\beta^2)^{-1/2}$.

Let us note that the DDCS for collisions with free stationary electrons
($\langle K \rangle=0$) can be expressed in the form
\beqa
\frac{\d^2 \sigma_{\rm M\o ller}}{\d Q \, \d W} & = &
\frac{2 \pi e^4}{\me v^2} \frac{1}{W^{2}}
\, F_{\rm OM} (Q,W) \, \delta(W-Q)\, ,
\label{6.235}\eeqa
with
\beq
F_{\rm OM} (Q,W) =
1  + \left( \frac{Q}{\langle K \rangle+E-W} \right)^{2}
- \frac{(1-b_0)Q}{\langle K \rangle+E-W} +
\frac{b_0Q^{2}}{(\langle K \rangle+E)^{2}}\, .
\label{6.236}\eeq
The last expression differs from the M\o ller
factor \req{6.233} in that the numerators contain $Q$ instead of $W$. We have
introduced this seemingly arbitrary replacement to get an expression
analogous to $F_\textrm{Ochkur}(Q,W)$, Eq.\ \req{6.230}, and such that
$F_{\rm OM} (Q,W)$ reduces to unity for $Q=0$. Of course, this
formal modification does not alter the DDCS for collisions with free
electrons at rest. In the non-relativistic limit ($b_0\rightarrow
0$), the factor $F_\textrm{OM}(Q,W)$ reduces to the Ochkur factor.

In the case of binary collisions of an electron with a free electron
at rest, considering the two electrons as distinguishable, the PWBA
yields the DDCS [see Eqs.\ \req{6.221} and \req{6.225}]
\beqa
\frac{\d^2 \sigma_\textrm{free}}{\d Q \d W}
&=& \left( \frac{2\pi e^4}{\me
v^2} \, \frac{1}{W^2} \right) \, F_{\rm bin}^0(W)
\, \delta(Q-W)\,
\label{6.237}\eeqa
with
\beq
F_{\rm bin}^0(W) = 1 - \frac{(2E-W+4 \me c^2)W}{2(E+\me c^2)^2}\, .
\label{6.238}\eeq

To account for exchange effects in electron-atom collisions, we multiply
the DDCS \req{6.221} by correcting factors. In the case of low-$Q$
excitations of a subshell $a=(n_a\kappa_a)$, with $Q \le
U_a$, we will multiply the DDCS for longitudinal
interactions by the factor \req{6.230}, but leave the DDCS for
transverse interactions unchanged, because these are distant
interactions for which exchange effects are expected to be negligible.
That is,
\begin{subequations}
\label{6.239}
\beq
\frac{\d^2 \sigma_a (\textrm{e}^-)}{\d W \, \d Q} =
F_{\rm Ochkur} (Q,W) \, \frac{\d^2 \sigma^{\rm L}_a}{\d
W \, \d Q} + \frac{\d^2 \sigma^{\rm T}_a}{\d W \, \d Q}
\qquad
\mbox{for $Q<U_a$.}
\label{6.239a}\eeq
For large-$Q$ ionizations, with $Q>U_a$,
exchange effects are accounted by setting
\beq
\frac{\d^2 \sigma_a (\textrm{e}^-)}{\d W \, \d Q} =
\frac{F_{\rm OM}(Q,W)}{F_{\rm bin}^0 (W)} \left( \frac{\d^2
\sigma^{\rm L}_a}{\d W \, \d Q} + \frac{\d^2 \sigma^{\rm T}_a}{\d W \,
\d Q}\right) \qquad \mbox{for $Q>U_a$\, .}
\label{6.239b}\eeq
\end{subequations}
Note that for very large recoil energies the subshell GOS reduces to the delta
function, $(2j_a+1) \delta(W-Q)$, and the DDCS \req{6.239b} reduces to
the M\o ller DDCS \req{6.232}, multiplied by the number of electrons in
the subshell.


\subsection{Collisions of positrons \label{sec6.6.2}}
\index{collisions of positrons}

In collisions of positrons ($Z_0=+1$, $M_0=\me$), the (active)
electron-positron pair can undergo annihilation followed by recreation
of a new pair \citep[see, \eg,][]{Sakurai1967}, a process that coexists
with ordinary scattering. Positron collisions are also affected by
exchange effects because of the indistinguishability of the target
electron from the electrons in virtual states of negative energy (the
Dirac sea). The DCS for collisions of positrons of kinetic
energy $E$ with free electrons at rest was derived by
\citet{Bhabha1936} by using the methods of quantum electrodynamics.
It is given by \index{Bhabha cross section}
\beq
\frac{\d \sigma_{\textrm{Bhabha}}}{\d Q \, \d W} =
\frac{2 \pi e^4}{\me v^2} \, \frac{1}{W^{2}}
\, F_{\rm Bhabha} (W) \, \delta(W-Q) ,
\label{6.240}\eeq
where
\beq
F_{\rm Bhabha} (W) = 1 - b_{1}\frac{W}{E} +
b_{2}\left(\frac{W}{E}\right)^{2} - b_{3}\left(\frac{W}{E}\right)^{3} +
b_{4}\left(\frac{W}{E}\right)^{4}
\label{6.241}\eeq
with
\beqa
b_{1} & = & \left( \frac{\gamma-1}{\gamma} \right)^{2}
\frac{2(\gamma+1)^{2}-1}{\gamma^{2}-1}, \nonumber \\[2mm]
b_{2} & = & \left( \frac{\gamma-1}{\gamma} \right)^{2}
\frac{3(\gamma+1)^{2}+1}{(\gamma+1)^{2}}, \nonumber \\[2mm]
b_{3} & = & \left( \frac{\gamma-1}{\gamma} \right)^{2}
\frac{2\gamma(\gamma-1)}{(\gamma+1)^{2}}, \nonumber \\[2mm]
b_{4} & = & \left( \frac{\gamma-1}{\gamma} \right)^{2}
\frac{(\gamma-1)^{2}}{(\gamma+1)^{2}}\, .
\label{6.242}\eeqa
The maximum allowed energy loss is $W_{\rm max} = E$, because the
positron and the electron are distinguishable. Note that in the
non-relativistic limit ($\gamma \rightarrow 1$) $b_i=0$.

We see that, in collisions of positrons with free electrons, exchange
and annihilation-recreation effects are appreciable only for relativistic projectiles (with $\gamma
> 1$) and for collisions with relatively large energy transfers. That
is, most of the effect occurs for large-$W$ collisions, for which the
Bhabha DCS is expected to provide a fairly good approximation. In the
case of collisions with electrons bound in atoms and ions,
we give a partial account of exchange and annihilation-recreation
by simply multiplying the DDCS for high-$W$ excitations,
for which $Q \simeq W$, by a factor that yields the Bhabha DCS
\req{6.240} at large $Q$, leaving the DDCS for low- and intermediate $Q$
unchanged. Specifically, for excitations of a subshell $a=(n_a\kappa_a)$
we will set
\beq
\frac{\d^2 \sigma_a (\textrm{e}^+)}{\d W \, \d Q} = \left\{
\begin{array}{ll}
\displaystyle{
\frac{\d^2 \sigma_a}{\d W \, \d Q}  }
& \mbox{if $Q<U_a$\, ,} \\ [6mm]
\displaystyle{\frac{F_{\rm Bhabha}(W)}{F_{\rm bin}^0 (W)} \,
\frac{\d^2 \sigma_a}{\d W \, \d Q} \; \; \; } & \mbox{if
$Q> U_a$\, ,}
\end{array}\right.
\label{6.243}\eeq
where the DDCSs on the right-hand side is given by Eqs. \req{6.221}, and
$F_{\rm bin}^0(W)$ is defined by Eq.\ \req{6.238}.


\section{Semi-classical dielectric theory \label{sec6.7}}
\index{semi-classical dielectric theory|(}

\index{Fourier transform}
The classical dielectric approach presented in Section \ref{sec1.4}
describes the slowing down of a swift charged particle as a continuous
process, while in reality the energy loss is the result of multiple
discrete interactions. Semi-classical arguments \citep[see,
\eg,][]{Jackson1975} indicate that the variables $\omega$ and ${\bf q}$,
which were introduced merely as variables of the Fourier transformation,
can be assigned a physical meaning by considering that $W=\hbar \omega$
and $\hbar {\bf q}$ represent, respectively, the energy loss and the
momentum transfer in one interaction. In accordance with this
interpretation, these variables are subject to the constraints of
energy and linear momentum conservation [Eqs.\ \req{6.215} to
\req{6.220}]. This semi-classical picture constitutes the link between
the classical and quantum formulations. Indeed, in the case of thin
atomic gases the semi-classical dielectric formalism should be
consistent with the results of perturbative quantum-mechanical
calculations.

Let us consider the case of an isotropic material, with DFs
$\epsilon^{\rm (L,T)} (q,\omega) = \epsilon_1^{\rm (L,T)} + {\rm i}
\epsilon_2^{\rm (L,T)}$ depending on only the magnitude of ${\bf q}$.
The classical stopping power for projectiles with charge $Z_0 e$ and
velocity $v$ is given by Eq.\ \req{1.186},
\beqa
S_{\rm cl} &=& \frac{2 (Z_0  e)^2}{\pi v^2}
\int_0^\infty \omega \, \d \omega \int_{\omega/v}^\infty
\frac{\d q}{q}
\left[ {\rm Im} \left( \frac{-1}{\epsilon^{\rm (L)} (q, \omega)} \right)
\right.
\nonumber \\ [2mm]
&& \left.+ \beta^2 \left( 1 - \frac{\omega^2}{\beta^2 c^2 q^2} \right)
{\rm Im} \left( \frac{1}{1 - (\omega/cq)^2 \,
\epsilon^{\rm (T)}(q, \omega)} \right)  \right],
\label{6.244}\eeqa
which we will express in terms of the single-interaction variables $Q$
[Eq.\ \req{6.216}] and $W=\hbar \omega$. Assuming that $W \ll E$, the
lower limit of the integral over $Q$ is [see Eq.\ \req{6.218}],
\beq
\sqrt{(c \hbar \omega/v)^2 - \me^2 c^4} - \me c^2 \simeq
\frac{W^2}{2 \me c^2 \beta^2} \simeq Q_- \, .
\label{6.245}\eeq
In addition, the upper limits of the integrals can be replaced with
$W_{\rm max}$ and $Q_+$, respectively, because the integrand tends to
zero at large $W$ or $Q$ and, hence, it practically vanishes outside the
kinematically allowed domain. We can thus write
\beqa
S_{\rm cl} &=& {\cal N} \, \frac{2 \pi Z_0^2 e^4}{\me v^2} \, \frac{2 Z}
{\pi (\hbar \Omega_{\rm p})^2}
\int_0^{W_{\rm max}} \d W \; W \int_{Q_-}^{Q_+}
\d Q  \; \frac{2(Q+\me c^2)}{Q(Q+2\me c^2)}
\left[
{\rm Im} \left( \frac{-1}{\epsilon^{\rm (L)} (Q, W)} \right)
\right.
\nonumber \\ [2mm]
&& \left.+ \beta^2 \left( 1 - \frac{W^2}{\beta^2 Q(Q+2\me c^2} \right)
 {\rm Im} \left( 1 - \frac{W^2}{Q(Q+2\me c^2)}
\; \epsilon^{\rm (T)}(Q, W) \right)^{-1}
\right], \rule{10mm}{0mm}
\label{6.246}\eeqa
where \index{plasma resonance frequency}
\beq
\Omega_{\rm p} = \sqrt{4 \pi \, {\cal N} Z \, \frac{e^2}{\me}}
\label{6.247}\eeq
is the plasma resonance frequency of an electron gas with the average
electron density ${\cal N} Z$ of the medium, Eq.\ \req{1.160}.

To reveal the connection between the PWBA and the dielectric theory, we
consider a low-density gas, for which $\epsilon_1^{\rm (L,T)} \simeq 1$
and $\epsilon_2^{\rm (L,T)} \ll 1$. The semi-classical expression
\req{6.246} of the stopping power for such a material simplifies to
\beqa
S_{\rm cl} &=& {\cal N} \, \frac{2 \pi Z_0^2 e^4}{\me v^2} \, \frac{2 Z}
{\pi (\hbar \Omega_{\rm p})^2}
\int_0^{W_{\rm max}} \d W \; W \int_{Q_-}^{Q_+}
\d Q  \frac{ 2(Q+\me c^2)}{Q(Q+2\me c^2)}
\left[ {\rm Im} \left( \frac{-1}{\epsilon^{\rm (L)} (Q, W)} \right)
\right.
\nonumber \\ [2mm]
&& \! \! \! \! \! \! \! \! \left.
+ \beta^2 \left( 1 - \frac{W^2}{\beta^2 Q(Q+2\me c^2)} \right)
\frac{W^2Q(Q+2\me c^2)}{\left[Q(Q+2\me c^2) - W^2 \right]^2}
{\rm Im} \left( \frac{-1}{\epsilon^{\rm (T)} (Q, W)} \right)
\right]. \rule{10mm}{0mm}
\label{6.248}\eeqa
We recall that in the PWBA, and in other quantum formulations, the
stopping power is obtained as the following integral of the DDCS (see
Section \ref{sec6.3})
\beq
S_{\rm PWBA} = {\cal N} \int_0^{W_{\rm max}} \d W \,
W \int_{Q_-}^{Q_+} \d Q \;
\frac{\d^2 \sigma_{\rm PWBA}}{\d Q \, \d W}.
\label{6.249}\eeq
The similarity of these two expressions allows inferring a formula for
the semi-classical DDCS. From the definition of the stopping power,
$$
S_{\rm cl} = {\cal N} \int_0^{W_{\rm max}} \d W \,
W \int_{Q_-}^{Q_+} \d Q \;
\frac{\d^2 \sigma_{\rm cl}}{\d Q \, \d W},
$$
we conclude that
\beqa
\frac{\d^2 \sigma_{\rm cl}}{\d Q \, \d W}
&=& \frac{2 \pi Z_0^2 e^4}{\me v^2} \,
\frac{2 Z}{\pi (\hbar \Omega_{\rm p})^2} \,
\frac{2(Q+\me c^2)}{Q(Q+2\me c^2)} \,
\left[ {\rm Im} \left( \frac{-1}{\epsilon^{\rm (L)} (Q, W)} \right)
\right.
\nonumber \\ [2mm]
&&  \! \! \! \! \! \! \! \! \! \! \! \! \! \! \! \! \! \!
\left.+ \beta^2 \left( 1 - \frac{W^2}{\beta^2 Q(Q+2\me c^2)} \right)
\frac{ W^2 \, Q(Q+2\me c^2)}{\left[ Q(Q+2\me c^2) - W^2 \right]^2} \;
{\rm Im} \left( \frac{-1}{\epsilon^{\rm (T)} (Q, W)} \right)
\right]. \rule{15mm}{0mm}
\label{6.250}\eeqa
Making the identifications
\begin{subequations}
\label{6.251}
\beq
\frac{\d f(Q,W)}{\d W}
= W \, \frac{2Z}{\pi (\hbar \Omega_{\rm p})^2} \,
{\rm Im} \left( \frac{-1}{\epsilon^{\rm (L)}(Q, W)} \right)
= W \, \frac{2Z}{\pi (\hbar \Omega_{\rm p})^2} \,
\eta_2^{\rm (L)}(Q, W)
\label{6.251a}\eeq
and
\beq
\frac{\d g(Q,W)}{\d W}
= W \, \frac{2Z}{\pi (\hbar \Omega_{\rm p})^2} \,
{\rm Im} \left( \frac{-1}{\epsilon^{\rm (T)}(Q, W)} \right)
= W \, \frac{2Z}{\pi (\hbar \Omega_{\rm p})^2} \,
\eta_2^{\rm (T)}(Q, Wa),
\label{6.251b}\eeq
\end{subequations}
we can write
\beqa
\frac{\d^2 \sigma_{\rm cl}}{\d Q \, \d W}
&=& \frac{2 \pi Z_0^2 e^4}{\me v^2} \, \left[
\frac{2(Q+\me c^2)}{W Q(Q+2\me c^2)} \; \frac{\d f(Q,W)}{\d W}
\right.
\nonumber \\ [2mm]
&&  \! \! \! \! \! \! \! \! \! \! \! \! \! \! \! \!
\left.+ \beta^2 \left( 1 - \frac{W^2}{\beta^2 Q(Q+2\me c^2)} \right)
\frac{ 2 (Q+\me c^2) \, W^2}{\left[ Q(Q+2\me c^2) - W^2 \right]^2} \;
\frac{\d g(Q,W)}{\d W}
\right]. \rule{10mm}{0mm}
\label{6.252}\eeqa
Now, as the integrand decreases rapidly with $Q$, we can replace the
factor $Q+\me c^2$ in the numerators with $\me c^2$ to obtain
\index{generalized oscillator strength!and dielectric functions}
\index{dielectric functions!and generalized oscillator strengths}
\beqa
\frac{\d^2 \sigma_{\rm cl}}{\d Q \, \d W}
&=& \frac{2 \pi Z_0^2 e^4}{\me v^2} \, \left[
\frac{2\me c^2}{W Q(Q+2\me c^2)} \; \frac{\d f(Q,W)}{\d W}
\right.
\nonumber \\ [2mm]
&&  \! \! \! \! \! \! \! \! \! \! \! \! \! \! \! \!
\left.+ \beta^2 \left( 1 - \frac{W^2}{\beta^2 Q(Q+2\me c^2)} \right)
\frac{ 2\me c^2 \, W^2}{\left[ Q(Q+2\me c^2) - W^2 \right]^2} \;
\frac{\d g(Q,W)}{\d W} \right], \rule{10mm}{0mm}
\label{6.253}\eeqa
which is identical to the result \req{6.222} from the relativistic PWBA.

Because the extension of the PWBA to dense materials is beyond our
capabilities, we assume that the equivalence of the quantum perturbation
theory and the semi-classical dielectric formalism, which we have just
proved for thin gases, also holds for any amorphous material.
Accordingly, the integrand in Eq.\ \req{6.246} is identified as the
semi-classical DCS,
\beqa
\frac{\d^2 \sigma}{\d Q \, \d W} &=&
\frac{2 \pi Z_0^2 e^4}{\me v^2}
\, \frac{2 Z}{\pi (\hbar \Omega_{\rm p})^2} \,
\frac{2(Q+\me c^2)}{Q(Q+2\me c^2)}
\left[ {\rm Im} \left( \frac{-1}{\epsilon^{\rm (L)} (Q, W)} \right)
\rule{0mm}{7mm}\right.
\nonumber \\ [2mm]
&&  \! \! \! \! \! \! \! \! \! \! \! \! \! \! \! \!
\left.+ \beta^2 \left( 1 - \frac{W^2}{\beta^2 Q(Q+2\me c^2)} \right)
 {\rm Im} \left( 1 - \frac{W^2}{Q(Q+2\me c^2)}
\; \epsilon^{\rm (T)}(Q, W) \right)^{-1}
\right]. \rule{10mm}{0mm}
\label{6.254}\eeqa
Introducing the longitudinal and transverse GOSs defined by Eqs.\
\req{6.251}, this DDCS can be expressed as
\beq
\frac{\d^2 \sigma}{\d Q \, \d W} =
\frac{\d^2 \sigma_0}{\d Q \, \d W} +
\frac{\d^2 (\Delta \sigma)_{\rm pol}}{\d Q \, \d W}
\label{6.255}\eeq
with
\beqa
\frac{\d^2 \sigma_0}{\d Q \, \d W}
&=& \frac{2\pi Z_0^2 e^4}{\me v^2} \left( 1 + \frac{Q}{\me c^2} \right)
\left[ \frac{2\me c^2}{WQ(Q+2\me c^2)} \, \frac{\d f(Q,W)}{\d W} \right.
\nonumber \\ [2mm]
&+& \left. \beta^2 \left( 1 - \frac{W^2}{\beta^2 Q(Q+2 \me c^2)} \right)
\frac{ 2\me c^2 W } {[Q(Q+2\me c^2) - W^2 ]^2} \,
\frac{\d g(Q,W)}{\d W} \right]\, ,
\rule{15mm}{0mm}
\label{6.256}\eeqa
and
\beqa
\frac{\d^2 (\Delta \sigma)_{\rm pol}}{\d Q \, \d W} &=&
\frac{2 \pi Z_0^2 e^4}{\me v^2} \,
\frac{2 Z}{\pi (\hbar \Omega_{\rm p})^2} \,
\frac{2(Q+\me c^2)}{Q(Q+2\me c^2)} \, \beta^2
\left( 1 - \frac{W^2}{\beta^2 Q(Q+2 \me c^2)} \right)
\nonumber \\ [2mm]
&& \mbox{} \times \left[ {\rm Im} \left( 1 - \frac{W^2}{Q(Q+2\me c^2)}
\; \epsilon^{\rm (T)}(Q, W) \right)^{-1} \right.
\nonumber \\ [2mm]
&& \mbox{} \left. - \frac{W^2 \, Q(Q+2\me c^2)}{\left[ Q(Q+2\me c^2)
- W^2 \right]^2} \;
{\rm Im} \left( \frac{-1}{\epsilon^{\rm T}(Q,W)} \right)
\rule{0mm}{7mm}\right].
\label{6.257}\eeqa
The first term on the right-hand side of Eq.\ \req{6.255} represents the
DDCS of an ideal ``unpolarizable'' material, in which the
electromagnetic field of the projectile is the same as in vacuum, while
the second term can be regarded as a correction accounting for the
effect of the dielectric polarization of the medium, the so-called Fermi
density effect. It is worth noticing that (aside from a factor $1 +
Q/\me c^2 \simeq 1$) the result \req{6.256} is
formally identical to that of the PWBA with the GOSs of the material,
which are defined by Eqs.\ \req{6.251}. The dielectric formalism not
only completes the quantum perturbative approach by providing the
polarization correction, but also allows accounting for collective
plasmon-like excitations, which are alien to the PWBA and are naturally
described within the quantum dielectric theory (see Chapter
\ref{chapt7}).

In most practical calculations, the DDCS of the unpolarizable material
and the polarization correction are considered separately. Because $Q
\ll \me c^2$ for the most probable excitations (for which also $W \ll
\me c^2$), we set
\beqa
\frac{\d^2 \sigma_0}{\d Q \, \d W}
&=& \frac{2\pi Z_0^2 e^4}{\me v^2}
\left[ \frac{2\me c^2}{WQ(Q+2\me c^2)} \, \frac{\d f(Q,W)}{\d W} \right.
\nonumber \\ [2mm]
&+& \left. \beta^2 \left( 1 - \frac{W^2}{\beta^2 Q(Q+2 \me c^2)} \right)
\frac{ 2\me c^2 W } {[Q(Q+2\me c^2) - W^2 ]^2} \,
\frac{\d g(Q,W)}{\d W} \right]\, ,
\rule{20mm}{0mm}
\label{6.258}\eeqa
which is formally identical to the DDCS obtained from the PWBA, Eq.\
\req{6.222}, except for the fact that now the GOSs pertain to the
material, Eqs.\ \req{6.251}. The
polarization correction is appreciable only for fast projectiles, with
velocity $v$ comparable to $c$, for which
\beq
Q_- \simeq \sqrt{(W/\beta)^2 + \me^2 c^4} - \me c^2 \simeq
\frac{W^2}{2 \me c^2 \beta^2}\, .
\label{6.259}\eeq
It follows that the value of $Q_-$ is much smaller than $W$ ($Q_- \simeq
1$ eV for $W \simeq 1$ keV). On the other hand, a photon of energy $W$ in
vacuum has a momentum $\hbar q = W/c$ and, hence, transitions
corresponding to the emission of bare photons would be located on the
vacuum photon line,
\beq
Q = \sqrt{W^2 + \me^2 c^4} - \me c^2
\qquad \mbox{or} \qquad
W=\sqrt{Q(Q+2\me c^2)},
\label{6.260}\eeq
which lies outside the kinematically allowed region (see Figs.\
\ref{figA.3}). Because the polarization correction \req{6.257} almost
diverges at the photon line, the relevant recoil energies are small and,
consequently, we can replace the DFs on the right-hand side of Eq.
\req{6.257} with their values at $Q=0$, \ie, by the ODF $\epsilon(W)$.
High-$Q$ transverse interactions are properly accounted for by the
unpolarized DDCS \req{6.258}, because the response of target electrons
to high-$Q$ excitations is dominated by their inertia. That is, the
polarization effect alters the DDCS only for low-$Q$ transverse
interactions and the correction is completely determined by the ODF.
Heretofore, all calculations of the density effect have utilized this
approximation \citep[see, \eg,][]{ Fano1963}.

With the $Q$-dependence of the DFs removed, Eq.\ \req{6.257} becomes
\beqa
\frac{\d^2 (\Delta \sigma)_{\rm pol}}{\d Q \, \d W} &=&
\frac{2 \pi Z_0^2 e^4}{\me v^2} \,
\frac{2 Z}{\pi (\hbar \Omega_{\rm p})^2} \,
\frac{2(Q+\me c^2)}{Q(Q+2\me c^2)} \, \beta^2
\left( 1 - \frac{W^2}{\beta^2 Q(Q+2 \me c^2)} \right)
\nonumber \\ [2mm]
&& \mbox{} \times
\left[ {\rm Im} \left( 1 - \frac{W^2}{Q(Q+2\me c^2)}
\; \epsilon(W) \right)^{-1} \right.
\nonumber \\ [2mm]
&& \mbox{} \rule{5mm}{0mm} \left.
- \frac{W^2 \, Q(Q+2\me c^2)}{\left[ Q(Q+2\me c^2)
- W^2 \right]^2} \;
{\rm Im} \left( \frac{-1}{\epsilon(W)} \right) \right].
\label{6.261}\eeqa
We point out that the value of this expression decreases rapidly with
$Q$, that is, the most probable scattering angles $\theta$ are small.
The polarization correction to the energy-loss DCS is
\beq
\frac{\d (\Delta \sigma)_{\rm pol}}{\d W} = \int_{Q_-}^{\infty}
\frac{\d^2 (\Delta \sigma)_{\rm pol}}{\d Q \, \d W} \, \d Q,
\label{6.262}\eeq
where we have replaced the upper limit of the integral, $Q_+$, with
infinity, because the integrand practically vanishes for $Q>Q_+$.
Changing the integration variable to
\beq
x = \frac{W^2}{\beta^2 Q (Q+2\me c^2)},
\label{6.263}\eeq
we have [notice that $x(Q_-) =1$]
\beqa
\frac{\d (\Delta \sigma)_{\rm pol}}{\d W} &=&
\frac{2 \pi Z_0^2 e^4}{\me v^2} \,
\frac{2 Z}{\pi (\hbar \Omega_{\rm p})^2}
\int_{0}^1 \frac{\d x}{x} \, \beta^2 (1-x)
\left[ {\rm Im} \left( \frac{1}{1 - \beta^2 x \, \epsilon(W)}\right) \right.
\nonumber \\ [2mm]
&& \mbox{} \left. - \frac{\beta^2 x}{(1-\beta^2 x)^2} \;
{\rm Im} \left( \frac{-1}{\epsilon(W)} \right) \right].
\label{6.264}\eeqa
The integral of the second term in square brackets is elementary,
\beq
\int_0^1 \frac{ \beta^4 (1- x)}{(1-\beta^2
x)^2} \, \d x
= \left[ \frac{\beta^2-1}{1 - \beta^2 x} - \ln(1-\beta^2 x)
\right]_0^1 = \ln \left( \frac{1}{1-\beta^2} \right) - \beta^2,
\label{6.265}\eeq
and the integral of the first term can be obtained as
\begin{subequations}
\label{6.266}
\beqa
&& \! \! \! \! \! \! \! \! \! \! \! \! \! \!
\beta^2 \, {\rm Im}
\int_0^1 \frac{(1-x)}{1-\beta^2 \epsilon x} \, \frac{\d x}{x}
= \beta^2 \, {\rm Im} \left[ \ln x + \left( \frac{1}{\beta^2 \epsilon}
-1 \right) \ln (1-\beta^2 \epsilon x) \right]_0^1
\nonumber \\ [2mm]
&=& {\rm Im} \left[ - \left( \beta^2 - \frac{1}{\epsilon} \right)
\ln \left(1-\beta^2 \epsilon\right) \right] \,
\label{6.266a} \\ [2mm]
&=& {\rm Im} \left[ - \left( \beta^2 - \frac{\epsilon_1
- {\rm i}\epsilon_2}{|\epsilon|^2} \right) \ln \left( 1-\beta^2
\epsilon_1 - {\rm i} \beta^2 \epsilon_2 \right) \right]
\nonumber \\ [2mm]
&=& \left( \beta^2 - \frac{\epsilon_1}{|\epsilon|^2} \right) \arctan
\! \left( \frac{\beta^2 \epsilon_2}{1-\beta^2 \epsilon_1} \right)
- \frac{\epsilon_2}{|\epsilon|^2} \ln \left[ (1-\beta^2
\epsilon_1)^2 + \beta^4 \epsilon_2^2 \right]^{1/2}. \rule{15mm}{0mm}
\label{6.266b}\eeqa
\end{subequations}
Hence
\beqa
\frac{\d (\Delta \sigma)_{\rm pol}}{\d W} &=&
\frac{2 \pi Z_0^2 e^4}{\me v^2} \,
\frac{2 Z}{\pi (\hbar \Omega_{\rm p})^2}
\left\{
\left( \beta^2 - \frac{\epsilon_1}{|\epsilon|^2} \right) \arctan
\! \left( \frac{\beta^2 \epsilon_2}{1-\beta^2 \epsilon_1} \right)
\right.
\nonumber \\ [2mm]
&& \mbox{} \left.
- \frac{\epsilon_2}{|\epsilon|^2} \ln \left[ (1-\beta^2
\epsilon_1)^2 + \beta^4 \epsilon_2^2 \right]^{1/2}
- \frac{\epsilon_2}{|\epsilon|^2}
\left[ \ln \left( \frac{1}{1-\beta^2} \right) - \beta^2 \right]
\right\}. \rule{10mm}{0mm}
\label{6.267}\eeqa
In calculations it may be convenient to express the polarization
correction in terms of the inverse DF, $\eta=\epsilon^{-1}= \eta_1 -
{\rm i} \eta_2$, and the dipole oscillator strength [see Eqs.\
\req{6.251}]. We have
\beqa
\frac{\d (\Delta \sigma)_{\rm pol}}{\d W} &=&
\frac{2 \pi Z_0^2 e^4}{\me v^2} \, \frac{1}{W} \, \frac{\d f(W)}{\d W}
\left\{
\left( \frac{\beta^2 - \eta_1}{\eta_2} \right)
\arctan \! \left( \frac{\beta^2 \eta_2}{\eta_1 (\eta_1 -\beta^2) +
\eta_2^2} \right) \right.
\nonumber \\ [2mm]
&& \mbox{} \left.
- \frac{1}{2} \ln \left[ \frac{[\eta_1(\eta_1 -\beta^2)+ \eta_2^2]^2
+ \beta^4 \eta_2^2}{(\eta_1^2+\eta_2^2)^2} \right]
- \left[ \ln \left( \frac{1}{1-\beta^2} \right) - \beta^2 \right]
\right\}. \rule{10mm}{0mm}
\label{6.268}\eeqa

In what follows we will concentrate on the DDCS for collisions with the
``unpolarizable'' material, Eq.\ \req{6.258}. The polarization correction
which, as we have shown, alters only the low-$Q$ transverse interactions,
will be further analyzed in Section \ref{sec8.2}.
\index{semi-classical dielectric theory|)}


\section{Interactions with large recoil energies \label{sec6.8}}
\index{inelastic collisions!with large recoil energies}

When the energy transfer $W$ is much larger than the ionization energy
$U_a = -\varepsilon_a$ of the active subshell, the target electron
reacts as if it were essentially free and at rest. Let $W_a$ be a
moderately large energy loss (of the order of, say, $10^3 \, U_a$) such
that the integral of the OOS on the interval $(W_a, \infty)$ is
negligible in comparison with the number $2j_a+1$ of electrons in the
subshell. Then, for energy losses larger than $W_a$, and also for recoil
energies larger than this value, the Bethe surface is different from
zero only in the vicinity of the Bethe ridge, and the GOS and the TGOS
can be approximated by the delta function, $(2j_a+1) \, \delta(W-Q)$,
that is, collisions with $Q>W_a$ can be described as binary collisions
with free electrons at rest. The corresponding DDCS (per target electron) is
obtained from Eqs.\ \req{6.221} with the GOSs set equal to
$\delta(W-Q)$,
\beqa
\frac{\d \sigma_\textrm{free}}{\d Q \, \d W} =
\left( \frac{2\pi Z_0^2 e^4}{\me
v^2} \, \frac{1}{W^2} \right) \, F_{\rm bin}(W) \, \delta(W-Q)
\label{6.269}\eeqa
with
\beq
F_{\rm bin}(W) = 1 -
\frac{\left[(2E-W+2 M_0 c^2)\me c^2 +M_0^2c^4+\me^2 c^4\right]W}
{2\me c^2(E+M_0 c^2)^2} \, .
\label{6.270}\eeq
The quantity in parenthesis in Eq.\ \req{6.269} is the non-relativistic
Thomson energy-loss DCS [Eq.\ \req{6.124}], but with the relativistic speed $v=\beta c$.
Hence, the factor $F_{\rm bin}(W)$ accounts for the remaining
relativistic corrections, which are appreciable only for binary
collisions of high-energy projectiles with large energy transfers [see
Eq.\ \req{6.273}].

The maximum allowed energy loss is determined by the intersect of the curve
$W_\textrm{m} (Q)$, Eq.\ \req{6.219}, with the Bethe ridge ($W=Q$), which
occurs at a point where $W$ has the value\footnote{$W_{\rm ridge}$ is the
maximum energy transfer in collisions with free electrons at rest,
which is also given by Eq.\ \req{4.189}.}
\beq
W_{\rm ridge} = \frac{2\me c^2 \beta^2}{1-\beta^2} \, R \qquad
\mbox{with} \qquad
R \equiv
\left[ 1+\left(\frac{\me}{M_0}\right)^2+ \frac{2}{\sqrt{1 - \beta^2}}
\, \frac{\me}{M_0} \right]^{-1}\, .
\label{6.271}\eeq
Note that, when $M_0=\me$, $W_{\rm ridge}=E$.
For heavy projectiles ($M_0 \gg \me$) with kinetic
energies much less than their rest energy $M_0c^2$, $R \sim 1$ and
\beq
W_{\rm ridge}\simeq \frac{2\me c^2 \beta^2}{1-\beta^2}\, .
\label{6.272}\eeq
For energies of the order of $M_0c^2$ or larger, the complete
expression \req{6.271} should be used.

The kinematical factor \req{6.270} can be written as
\beq
F_{\rm bin}(W) = 1 - \beta^2 \frac{W}{W_{\rm ridge}} +
\frac{1-\beta^2}{2 M_0^2 c^4} \, W^2 \, .
\label{6.273}\eeq
The contribution of large-$Q$ interactions, with $Q>W_a$, to the
energy-loss DCS of a closed subshell $a=(n_a\kappa_a)$ is
\beqa
\frac{\d \sigma_{a,Q>W_a}^\textrm{free}}{\d W}
&\simeq& {\cal C}_a
\, \frac{1}{W^2} \left[
1 - \beta^2 \frac{W}{W_{\rm ridge}} +
\frac{1-\beta^2}{2 M_0^2 c^4} \, W^2 \right]
{\cal S}(W_\textrm{ridge} - W_a)\, ,
\label{6.274}\eeqa
where ${\cal S}(x)$ is the unit step function, and
\beq
{\cal C}_a = \frac{2 \pi Z_0^2 e^4}{\me v^2} \, (2j_a+1).
\label{6.275}\eeq
The corresponding contributions to the integrated cross
sections,
$$
\sigma_{a,Q>W_a}^{(k)} = \int_{W_a}^{W_{\rm ridge}} W^k \,
\frac{\d \sigma_{a,Q>W_a}^\textrm{free}}{\d W} \, \d W,
$$
are
\begin{subequations}
\label{6.276}
\beq
\sigma_{a,Q>W_a}^{(0)} = {\cal C}_a
\left[ - \frac{1}{W} - \beta^2 \,\frac{\ln W}{W_{\rm ridge}}
+ \frac{1-\beta^2}{2M_0^2 c^4} \, W
\right]_{W_a}^{W_\textrm{ridge}}\, ,
\label{6.276a}\eeq
\beq
\sigma_{a,Q>W_a}^{(1)} = {\cal C}_a
\left[ \ln W - \beta^2 \, \frac{W}{W_{\rm ridge}}
+ \frac{1-\beta^2}{2M_0^2 c^4} \, \frac{W^2}{2}
\right]_{W_a}^{W_\textrm{ridge}} \,
\label{6.276b}\eeq
and
\beq
\sigma_{a,Q>W_a}^{(2)} = {\cal C}_a
\left[ W - \beta^2 \, \frac{W^2}{2W_{\rm ridge}}
+ \frac{1-\beta^2}{2M_0^2 c^4} \, \frac{W^3}{3}
\right]_{W_a}^{W_\textrm{ridge}} \, .
\label{6.276c}\eeq
\end{subequations}
Note that these formulas apply only when $W_{\rm ridge}>W_a$; otherwise
$\sigma^{(n)}_{a,Q>W_a}=0$.

\vspace*{5mm}
\noindent $\bullet$ {\bf Cross sections for electrons}

\noindent
In the case of projectile electrons, the energy-loss DCS per electron
in the subshell $a$ for collisions with $Q>W_a$ is given by the M\o ller
formula, Eq.\ \req{6.232},
\begin{subequations}
\label{6.277}
\beq
\frac{\d \sigma_{a,Q>W_a}^\textrm{free}}{\d W} =      {\cal C}_a
\, \frac{1}{W^{2}} \, F_{\rm bin}^{(-)} (W)
\label{6.277a}\eeq
with
\beq
F_{\rm bin}^{(-)} (W) =
 1  + \left( \frac{W}{E-W} \right)^{2}
- \frac{(1-b_0)W}{E-W} + \frac{b_0 W^{2}}{E^{2}} \, .
\label{6.277b}\eeq
\end{subequations}
The contributions of these collisions to the total, stopping and
energy-straggling cross sections are given by the following analytical
expressions,
\begin{subequations}
\label{6.278}
\beqa
\sigma_{a,Q>W_a}^{(0)}  (\textrm{e}^-) & = & {\cal C}_a \,
\left[ -\frac{1}{W} + \frac{1}{E-W}
+ \frac{1-b_0}{E}
\ln \left( \frac{E-W}{W}\right) + \frac{b_0 W}{(E)^{2}}
\right]_{W_a}^{W_{\rm max}} , \rule{10mm}{0mm}
\label{6.278a}\eeqa
\beqa
\sigma_{a,Q>W_a}^{(1)}  (\textrm{e}^-)& = & {\cal C}_a  \,
\left[ \ln W + \frac{E}{E-W}
+ (2-b_0)\ln(E-W) + \frac{b_0 W^{2}}{2(E)^{2}}
\right]_{W_a}^{W_{\rm max}} ,  \rule{10mm}{0mm}
\label{6.278b}\eeqa
and
\beqa
\sigma_{a,Q>W_a}^{(2)} (\textrm{e}^-) & = & {\cal C}_a  \,
\left[ (2-b_0)W + \frac{2 E^2-W^2}{E-W}
\right. \nonumber \\[2mm] & & \left. \rule{25mm}{0mm}
+ (3-b_0)E \ln(E-W) + \frac{b_0 W^3}{3E^2}
\right]_{W_a}^{W_{\rm max}}.  \rule{13mm}{0mm}
\label{6.278c}\eeqa
\end{subequations}
Of course, these formulas apply only when $W_{\rm max}>W_a$; otherwise
$\sigma^{(n)}_{a,Q>W_a}=0$.

\vspace*{5mm}
\noindent $\bullet$ {\bf Cross sections for positrons}

\noindent
The DDCS for high-$Q$ collisions, with $Q>W_a$, of positrons with
an electron in the subshell $a$ is given by the Bhabha formula
\req{6.240},
\begin{subequations}
\label{6.279}
\beq
\frac{\d \sigma_{a,Q>W_a}^\textrm{free}}{\d W} =
{\cal C}_a  \, \frac{1}{W^{2}} \, F_{\rm bin}^{(+)} (W)
\label{6.279a}\eeq
with
\beq
F_{\rm bin}^{(+)} (W) = 1 - b_{1}\frac{W}{E} +
b_{2}\left(\frac{W}{E}\right)^{2} - b_{3}\left(\frac{W}{E}\right)^{3} +
b_{4}\left(\frac{W}{E}\right)^{4}.
\label{6.279b}\eeq
\end{subequations}
The contributions of these collisions to the total, stopping and
energy-straggling cross sections can then be calculated analytically and
are
\begin{subequations}
\label{6.280}
\beq
\sigma^{(0)}_{a,Q>W_a} (\textrm{e}^+) = {\cal C}_a
\left[ -\frac{1}{W} - b_{1}\frac{\ln W}{E} + b_{2}\frac{W}{E^{2}} -
b_{3}\frac{W^{2}}{2E^{3}} + b_{4}\frac{W^{3}}{3E^{4}}
\right]_{W_a}^{E},
\label{6.280a}\eeq
\beq
\sigma^{(1)}_{a,Q>W_a} (\textrm{e}^+) = {\cal C}_a
\left[ \ln W - b_{1}\frac{W}{E} + b_{2}\frac{W^{2}}{2E^{2}} -
b_{3}\frac{W^{3}}{3E^{3}} + b_{4}\frac{W^{4}}{4E^{4}}
\right]_{W_a}^{E},
\label{6.280b}\eeq
and
\beq
\sigma^{(2)}_{a,Q>W_a} (\textrm{e}^+) = {\cal C}_a
\left[ W - b_{1}\frac{W^2}{2E} + b_{2}\frac{W^{3}}{3E^{2}} -
b_{3}\frac{W^{4}}{4E^{3}} + b_{4}\frac{W^{5}}{5E^{4}}
\right]_{W_a}^{E},
\label{6.280c}\eeq
\end{subequations}
if $E>W_a$; otherwise they all vanish.


\section{High-energy stopping power formula \label{sec6.9}}
\index{Bethe stopping power formula}

To close this Chapter, we are going to derive an asymptotic
(high-energy) formula for the electronic stopping power of a thin gas for
high-energy charged particles. The gas is assumed to consist of ${\cal
N}$ neutral atoms or molecules per unit volume, each with $Z$ bound
electrons.

We start by considering the case of projectiles much heavier than the
electron ($M_0 \gg \me$) that move with kinetic energy $E$ in the gas.
We wish to calculate the stopping power of the gas,
\beq
S \equiv {\cal N} \sigma^{(1)} =
{\cal N} \int_0^{W_{\rm max}} \d W \, W \int_{Q_-}^{Q_+} \d Q \,
\frac{\d^2 \sigma}{\d Q \, \d W},
\label{6.281}\eeq
with the DDCS obtained from the PWBA [Eq.\ \req{6.222}], or from the
semi-classical DDCS \req{6.258} of the unpolarized
medium\footnote{Notice that here we disregard the polarization effect,
which will be considered in Chapter \ref{chapt9}.}
\beqa
\frac{\d^2 \sigma}{\d Q \, \d W}
&=& \frac{2\pi Z_0^2 e^4}{\me v^2} \,
\left[ \frac{2\me c^2}{WQ(Q+2\me c^2)} \, \frac{\d f(Q,W)}{\d W} \right.
\nonumber \\ [2mm]
&+& \left.\left( \beta^2 - \frac{W^2}{Q(Q+2 \me c^2)} \right)
\frac{ 2\me c^2 W } {[Q(Q+2\me c^2) - W^2 ]^2} \,
\frac{\d g(Q,W)}{\d W} \right]. \rule{10mm}{0mm}
\label{6.282}\eeqa
We assume that the energy $E$ of the projectile is much higher than the
binding energies of the electrons in the target atoms or molecules, so
as to take advantage of the topological properties of the Bethe surface
and the sum rules obeyed by the GOSs.

\index{Bethe stopping power formula!derivation|(}
Following \citet{Fano1963}, we evaluate the integrals in Eq.\
\req{6.281} approximately by considering various ranges of $Q$. Although
the impact parameter is not defined in the quantum formulation, the
classical picture suggests that large (small) momentum transfers roughly
correspond to small (large) impact parameters; indeed, interactions with
small and large recoil energies are frequently referred to as distant
and close interactions, respectively.

\noindent $\bullet$ {\bf Low-$Q$ range}. This is the range where the
GOS and the TGOS can be approximated by the OOS; it extends up to a
certain recoil energy $Q_1$ that is much less than the energy $E$ of the
projectile. The contribution of these low-$Q$ excitations to the
stopping power is
\beqa
S_{\rm low} &=& {\cal N} \int_0^{W_{\rm max}} \d W \int_{Q_-}^{Q_1} \d Q \,
W \, \frac{\d^2 \sigma}{\d Q \, \d W}
\nonumber \\ [2mm]
&\simeq& \frac{2\pi Z_0^2 e^4}{\me v^2} \,{\cal N}
\int_0^{\infty} \d W  \, \frac{\d f(W)}{\d W}
\int_{Q_-}^{Q_1} \d Q \, \left[ \frac{1}{Q}  \right.
\nonumber \\ [2mm]
&+& \left.\left( \beta^2 - \frac{W^2}{Q(Q+2 \me c^2)} \right)
\frac{ 2\me c^2 W^2} {[Q(Q+2\me c^2) - W^2 ]^2} \right].
\label{6.283}\eeqa
Since the OOS decreases rapidly with $W$, we have replaced the largest
allowed energy loss $W_{\rm max}$ with infinity and we have dropped a
relativistic correction of order $Q/2 \me c^2$; notice also that
\beq
Q_- \simeq W^2/(2 \me c^2 \beta^2).
\label{6.284}\eeq
The integral of the transverse
term can be reduced to a simpler form by introducing the ``recoil
angle'' $\vartheta_{\rm r}$ defined by
\beq
\cos^2 \vartheta_{\rm r} = \frac{Q_- (Q_- + 2 \me c^2)}{Q (Q+ 2 \me c^2)}
= \frac{W^2 /\beta^2}{Q (Q+ 2 \me c^2)},
\label{6.285}\eeq
and by replacing the upper limit $Q_1$ with infinity. We have
\beqa
S_{\rm low} &\simeq& \frac{2\pi Z_0^2 e^4}{\me v^2} \,{\cal N}
\int_0^{\infty} \d W  \, \frac{\d f(W)}{\d W}
\left[ \ln \left( \frac{2 \me c^2 \beta^2 Q_1}{W^2} \right)  \right.
\nonumber \\ [2mm]
&& \mbox{} \left. + \int_0^1 \frac{ \beta^4 (1- \cos^2
\vartheta_{\rm r})}{(1-\beta^2
\cos^2\vartheta_{\rm r})^2} \, \d (\cos^2 \vartheta_{\rm r})
\right],
\nonumber \eeqa
and, using the result \req{6.265},
\beq
S_{\rm low} = \frac{2\pi Z_0^2 e^4}{\me v^2} \,{\cal N}
\int_0^{\infty} \d W  \, \frac{\d f(W)}{\d W}
\left[ \ln \left( \frac{2 \me c^2 \beta^2 Q_1}{W^2}  \right)
+ \ln \left( \frac{1}{1-\beta^2} \right) - \beta^2 \right].
\label{6.286}\eeq
Because we have extended the integral over $Q$ of the transverse
term to infinity, we have included at least part of the stopping power
of transverse interactions with intermediate and large $Q$ values.
Finally, using the dipole sum rule, we can write
\beqa
S_{\rm low} &=& \frac{2\pi Z_0^2 e^4}{\me v^2} \, {\cal N} Z
\left[ \ln \left( \frac{2 \me v^2\,  Q_1}{I^2}  \right)
+ \ln \left( \frac{1}{1-\beta^2} \right) - \beta^2 \right].
\label{6.287}\eeqa
where the quantity $I$, defined by\index{mean excitation
energy}\index{mean excitation potential}
\beq
Z \ln I = \int_0^\infty \ln W \, \frac{\d f(W)}{\d W} \, \d W,
\label{6.288}\eeq
is the so-called {\it mean excitation energy} (or {\it mean excitation
potential}) of the material. Empirical values of the mean excitation
energies for many materials of interest in radiation metrology and
dosimetry are compiled in the \citet{ICRU37}.

\noindent $\bullet$ {\bf Intermediate-$Q$ range}. For recoil energies
between $Q_1$ and a certain value $Q_2$, which we assume much larger
than the binding energies of the atomic electrons, the maximum allowed
energy transfer $W_{\rm m}(Q)$ is large and we can consider that the
practical totality of the GOS is contained in the interval from 0 to
$W_{\rm m}(Q)$. We disregard the contribution of
transverse interactions, which has already been accounted for in $S_{\rm
low}$. The contribution of longitudinal interactions with intermediate
$Q$ values is
\beqa
S_{\rm int} &=&
\frac{2\pi Z_0^2 e^4}{\me v^2} \,
{\cal N} \int_0^{W_{\rm max}} \d W \, \int_{Q_1}^{Q_2} \d Q \,
\frac{1}{Q} \, \frac{\d f(Q,W)}{\d W} \, ,
\label{6.289}\eeqa
where we have dropped a correction of order $Q/2\me c^2$. After
replacing $W_{\rm max}$ with $\infty$. Exchanging the order of the
integrals, and using the Bethe sum rule, we obtain
\beqa
S_{\rm int} &=&
\frac{2\pi Z_0^2 e^4}{\me v^2} \,
{\cal N} Z \, \ln \left( \frac{Q_2}{Q_1} \right).
\label{6.290}\eeqa

\noindent $\bullet$ {\bf High-$Q$ range}. Excitations with recoil
energies larger than $Q_2$ can be described as collisions with free
electrons at rest. For these excitations the GOS and the TGOS are
different from 0 only when $W \simeq Q$, and they can be approximated as
\beq
\frac{\d f(Q,W)}{\d W} =
\frac{\d g(Q,W)}{\d W} \simeq Z \, \delta(W-Q).
\label{6.291}\eeq
For projectiles heavier than the electron, and for $Q>Q_2$, the
atomic DDCS obtained from the PWBA with these GOSs [Eqs.\ \req{6.269}
and \req{6.273}] reads
\beq
\frac{\d^2 \sigma_{\rm free}}{\d Q\, \d W} =
\frac{2\pi Z_0^2 e^4}{\me v^2} Z \left( \frac{1}{W^2}
- \frac{\beta^2}{W_{\rm ridge} W} + \frac{1-\beta^2}{2 M_0^2 c^4} \right)
\delta(Q-W)
\, ,
\label{6.292}\eeq
where
\beq
W_{\rm ridge} = \frac{2\me c^2 \, \beta^2 \, R}{1-\beta^2}
\qquad \mbox{with} \qquad
R = \left[ 1 + \left(\frac{\me}{M_0} \right)^2 +
\frac{2}{\sqrt{1-\beta^2}} \, \frac{\me}{M_0} \right]^{-1},
\label{6.293}\eeq
is the maximum allowed energy loss.
The contribution of high-$Q$ excitations to the stopping power is
\beqa
S_{\rm high} &=& {\cal N}
\int_{Q_2}^{W_{\rm ridge}} \d W \int_{Q_-}^{Q_+} \d Q \,
W \, \frac{\d^2 \sigma_{\rm free}}{\d Q \, \d W}
\nonumber \\ [2mm]
&=& \frac{2\pi Z_0^2 e^4}{\me v^2} \, {\cal N} Z
\int_{Q_2}^{W_{\rm ridge}}
\left( \frac{1}{W}- \frac{\beta^2}{W_{\rm ridge}}
+ \frac{1-\beta^2}{2 M_0^2 c^4} \, W\right) \, \d W
\nonumber \\ [2mm]
&=& \frac{2\pi Z_0^2 e^4}{\me v^2} \, {\cal N} Z
\left\{ \ln \left( \frac{W_{\rm ridge}}{Q_2} \right)
- \frac{\beta^2}{W_{\rm ridge}} \left( W_{\rm ridge} - Q_2 \right)
\rule{0mm}{7mm}\right.
\nonumber \\ [2mm]
&& \left. + \frac{1-\beta^2}{2M_0^2 c^4} \, \frac{1}{2} \left[
W_{\rm ridge}^2 - Q_2^2 \right]
\right\}.
\label{6.294}\eeqa
Neglecting terms of the order of $Q_2/2\me c^2$ and smaller, for
consistency with the intermediate-$Q$ case, we obtain
\beqa
S_{\rm high} &=& \frac{2\pi Z_0^2 e^4}{\me v^2} \, {\cal N} Z
\left[ \ln \left(\frac{2 \me c^2 \beta^2}{Q_2} \right)
+ \ln \left( \frac{1}{1-\beta^2} \right)
- \beta^2 \right.
\nonumber \\ [2mm]
&& \left. + \ln(R) + \frac{\me^2}{M_0^2} \beta^4 \gamma^2 R^2 \right].
\label{6.295}\eeqa
For projectiles much heavier than the electron, $R \simeq 1$ and we can
write
\beqa
S_{\rm high} &=& \frac{2\pi Z_0^2 e^4}{\me v^2} \, {\cal N} Z
\left[ \ln \left(\frac{2 \me c^2 \beta^2}{Q_2} \right)
+ \ln \left( \frac{1}{1-\beta^2} \right)
- \beta^2 \right].
\label{6.296}\eeqa
The last two terms represent the
contribution of high-$Q$ transverse interactions which, within the
present approximation, is seen to be identical to the corresponding
contribution of low- and intermediate-$Q$ interactions [see Eq.\
\req{6.287}].

Finally, the stopping power is obtained as the sum of these results,
\beq
S = S_{\rm low} + S_{\rm int} + S_{\rm high},
\label{6.297}\eeq
which gives
\beq
S = \frac{4\pi Z_0^2 e^4}{\me v^2} \, {\cal N} Z
\left[ \ln \left(\frac{2 \me v^2}{I} \right)
+ \ln \left( \frac{1}{1-\beta^2}  \right) - \beta^2 + \frac{1}{2} \,
f(\gamma) \right]
\label{6.298}\eeq
with \index{Bethe stopping power formula}
\beq
f(\gamma) = \ln(R) + \left( \frac{\me}{M_0}\,
\frac{\gamma^2-1}{\gamma} \, R \right)^2.
\label{6.299}\eeq
Equation \req{6.298} is the celebrated {\it Bethe formula} for the
stopping power, which was first obtained by \citet{Bethe1932}. It is
worth noticing that the formula is derived from the relativistic PWBA,
that is, from a first-order perturbation approach which is valid only
when the speed of the projectile is much larger than the speeds of the
electrons bound in the atom (see Section \ref{sec6.2}). The accuracy of
the formula worsens progressively when the energy $E$ decreases (even
assuming that the PWBA remains accurate) because the kinematically
allowed domain in the $Q$-$W$ plane reduces and some of the assumptions
in the above derivation lose validity.

\index{restricted stopping power}

Incidentally, we can consider the stopping power restricted to energy
losses that are less than a given cutoff $W_{\rm c}$, which  is still
larger than $Q_2$, the lower limit of the high-$Q$ domain. The
restricted stopping power is a useful concept in practical dosimetry
studies, and a basic component of Landau's energy-straggling theory
(Section \ref{sec9.4.2}). The high-$Q$ contribution to the restricted
stopping power for heavy particles (for which $R\simeq 1$) can be
calculated by following the same steps as in the derivation of Eq.\
\req{6.294} with the value $W_{\rm c}$ replacing $W_{\rm ridge}$ in the
upper limit of the integral,
\beqa
S_{\rm high} (W<W_{\rm c})
&=& \frac{2\pi Z_0^2 e^4}{\me v^2} \, {\cal N} Z
\int_{Q_2}^{W_{\rm c}}
\left( \frac{1}{W}- \frac{\beta^2}{W_{\rm ridge}}
+ \frac{1-\beta^2}{2 M_0^2 c^4} \, W\right) \, \d W
\nonumber \\ [2mm]
&=& \frac{2\pi Z_0^2 e^4}{\me v^2} \, {\cal N} Z
\left\{ \ln \left(\frac{W_{\rm c}}{Q_2} \right)
- \frac{\beta^2}{W_{\rm ridge}} \left( W_{\rm c} - Q_2 \right)
+ \frac{1-\beta^2}{2M_0^2 c^4} \, \frac{1}{2} \left( W_{\rm c}^2 - Q_2^2
\right) \right\}
\nonumber \\ [2mm]
&\simeq& \frac{2\pi Z_0^2 e^4}{\me v^2} \, {\cal N} Z
\left\{ \ln \left(\frac{W_{\rm c}}{Q_2} \right) - \beta^2 \frac{W_{\rm
c}}{W_{\rm ridge}} \right\} ,
\label{6.300}\eeqa
where we have neglected terms of the order of $Q_2/\me c^2$ and
$\me^2/M_0^2$. Hence, the restricted stopping power is
\index{Bethe stopping power formula!restricted stopping power}
\beqa
S (W< W_{\rm c}) &=& S_{\rm low} + S_{\rm int} + S_{\rm high} (W< W_{\rm c})
\nonumber \\ [2mm]
&& \!  \! \! \! \! \! \! \! \! \! \! \! \! \! \! \! \! \! \! \! \! \!
\! \! \!
\mbox{} = \frac{2 \pi Z_0^2 e^4}{\me v^2} \, {\cal N} Z
\left[ \ln \left(\frac{2 \me v^2 W_{\rm c}}{I^2} \right)
+ \ln \left( \frac{1}{1-\beta^2}  \right) -  \beta^2
\left( 1 + \frac{W_{\rm c}}{W_{\rm ridge}} \right) \right].
\rule{10mm}{0mm}
\label{6.301}\eeqa

\vspace*{2mm}

\noindent $\bullet$ {\bf Electrons and positrons}.
The Bethe formula, in the form \req{6.298}, is not suited for electrons
and positrons, because the interactions of these particles are affected
by exchange and annihilation-recreation effects (see Sections
\ref{sec6.6.1} and \ref{sec6.6.2}). For high-energy projectiles, these
effects modify the DDCS for collisions with relatively large recoil
energies, but have a negligible influence on the low- and
intermediate-$Q$ interactions. Hence, the correct formula for the
stopping power of high energy electrons and positrons can be obtained by
simply considering the appropriate DDCSs for close, high-$Q$
interactions, which are given by the M\o ller and Bhabha formulas [Eqs.
\req{6.277a} and \req{6.279a}. Thus, in the case of electrons, the
contribution of high-$Q$ interactions to the stopping power is given by
\beqa
S_{\rm high}^{(-)} &=& {\cal N}
\int_{Q_2}^{E/2} \d W \int_{Q_-}^{Q_+} \d Q \,
W \, \frac{\d^2 \sigma_{\rm M\o ller}}{\d Q \, \d W}
\nonumber \\ [2mm]
&=&  \frac{2\pi Z_0^2 e^4}{\me v^2} \, {\cal N} Z
\int_{Q_2}^{E/2} \frac{1}{W} \, F_{\rm M\o ller}(W) \, \d W.
\label{6.302}\eeqa
Similarly, for positrons we have
\beq
S_{\rm high}^{(+)} =  \frac{2\pi Z_0^2 e^4}{\me v^2} \,
{\cal N} Z \int_{Q_2}^{E} \frac{1}{W} \, F_{\rm Bhabha}(W) \, \d W.
\label{6.303}\eeq
These integrals are given by expressions \req{6.278b} and \req{6.280b}.
In the case of high-energy projectiles with $E \gg Q_2$, the resulting
expressions can be somewhat simplified. As in the case of heavy
particles, the stopping power is obtained by adding those expressions
and the contributions from the low- and intermediate-$Q$ ranges that are
given above. The result is\index{Bethe stopping power formula!electrons
and positrons}
\beq
S^{(\pm)} = \frac{4\pi Z_0^2 e^4}{\me v^2} \, {\cal N} Z
\left[ \ln \left(\frac{2 \me v^2}{I} \right)
+ \ln \left( \frac{1}{1-\beta^2}  \right) - \beta^2 + \frac{1}{2} \,
f^{(\pm)}(\gamma) \right],
\label{6.304}\eeq
where
\beq
f^{(-)}(\gamma) = \frac{2\gamma^2-1}{\gamma^2}
+ \frac{1}{8} \left( \frac{\gamma-1}{\gamma} \right)^2
- \left[ 4 - \left( \frac{\gamma-1}{\gamma} \right)^2 \right] \ln 2
- \ln(\gamma+1)
\label{6.305}\eeq
for electrons, and
\beq
f^{(+)}(\gamma) = \frac{\gamma^2-1}{12\gamma^2} \left( 1
- \frac{14}{\gamma+1}
- \frac{10}{(\gamma+1)^2}
- \frac{4}{(\gamma+1)^3} \right)
- \ln 2 - \ln(\gamma+1)
\label{6.306}\eeq
for positrons. Here the superscripts $(-)$ and $(+)$ indicate projectile
electrons and positrons, respectively.

\begin{figure}[h!]
\begin{center}
\includegraphics*[width=9.5cm]{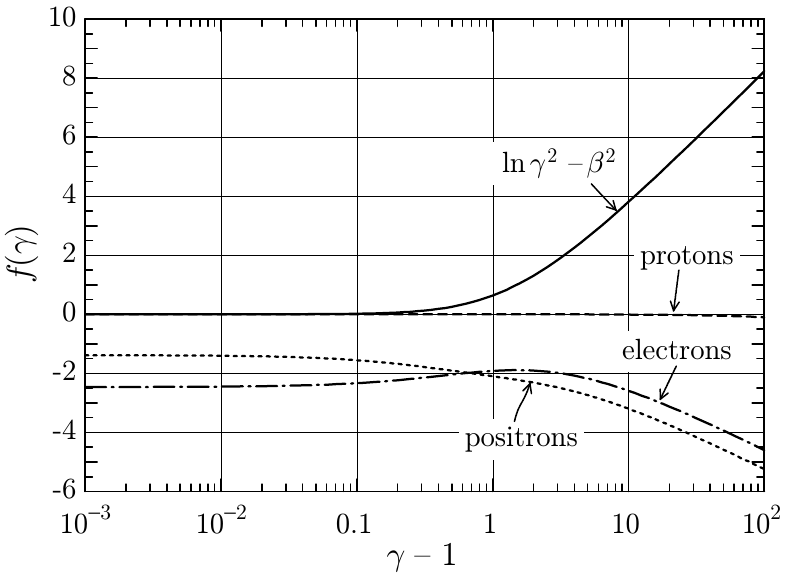}
\caption{Energy-dependent terms in the Bethe
stopping power formula, Eq.\ \req{6.307}. The quantity $\gamma-1$ is the
kinetic energy of the projectile in units of its rest energy.
\label{fig6.10}}
\end{center}\end{figure}

Summarizing, the Bethe stopping power formula can be written as
\beq
S (E) = \frac{4\pi Z_0^2 e^4}{\me v^2} \, {\cal N} Z
\left[ \ln \left(\frac{2 \me v^2}{I} \right)
+ \ln \left( \frac{1}{1-\beta^2} \right) - \beta^2 + \frac{1}{2}
f(\gamma) \right],
\label{6.307}\eeq
where the function $f(\gamma)$ is given by Eqs.\ \req{6.305},
\req{6.306}, and \req{6.299} for electrons, positrons, and heavier
particles, respectively (see Fig.\ \ref{fig6.10}), and the last term is
a relativistic correction. This formula is
applicable to any charged projectile with sufficiently high velocity,
irrespective of its mass and charge. An interesting and very useful
feature of the Bethe formula is that the material is completely
characterized by its average electron density, ${\cal N} Z$, and the
mean excitation energy $I$ defined by Eq.\ \req{6.288}.

\index{mean excitation energy}\index{mean excitation potential}
\index{additivity approximation}
As mentioned above, empirical $I$ values for elemental materials,
derived from various combinations of experimental information and
theoretical considerations, have been compiled by the ICRU
\citep{ICRU37}, Fig.\ \ref{fig6.11}. The ICRU recommended $I$ values are
used in most radiation dosimetry and transport studies.
The mean excitation energy of compounds,
alloys, or mixtures may be estimated by invoking the additivity
approximation. Let us consider a compound whose molecules consist of
$n_j$ atoms of the element of atomic number $Z_j$. According to the
additivity approximation, the GOS of a molecule is the
sum of the GOSs of the atoms and, consequently, the $I$ value of the
compound is given by
\beq
\left( \sum_j n_j Z_j \right) \ln I =
\sum_j n_j Z_j \ln(I_j),
\label{6.308}\eeq
where $I_j$ denotes the mean excitation energy of the element $Z_j$.
Since the additivity approximation neglects the effect of aggregation on
the atomic GOS, the $I$ value resulting from Eq.\ \req{6.308} may differ
appreciably from the ``true'' mean excitation energy of the material. A
better estimate of the $I$ value can only be obtained either from
stopping measurements or from knowledge of the optical dielectric
function of the material (see Chapter \ref{chapt7}).

\begin{figure}[h!]
\begin{center}
\includegraphics*[width=10.5cm]{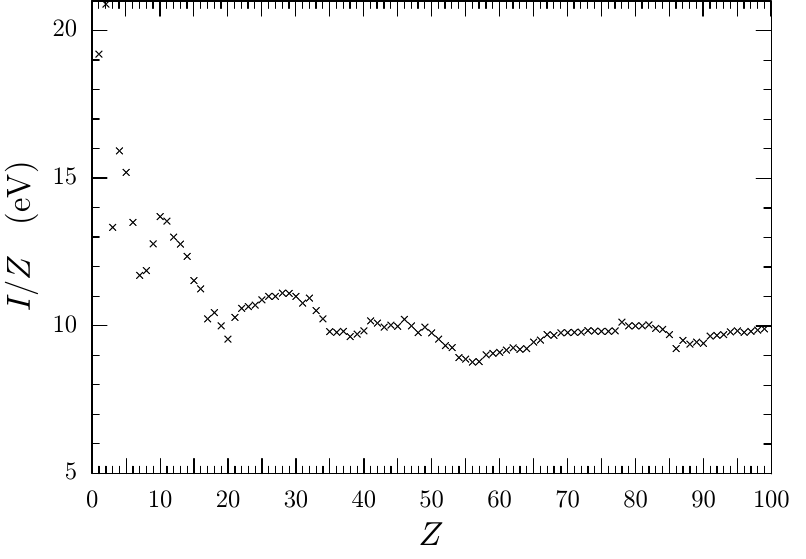}
\caption{Empirical $I$ values for elemental materials recommended by
the ICRU \citep{ICRU37}. The plotted quantity is the ratio $I/Z$ as a
function of the atomic number $Z$.
\label{fig6.11}}
\end{center}\end{figure}

In the non-relativistic limit ($\gamma \rightarrow 1, \beta \rightarrow
0$), the Bethe formula \req{6.307} simplifies to
\beqa
S_{\rm nr} &=& \frac{4\pi Z_0^2 e^4}{\me v^2} \, {\cal N} Z \left[
\ln \left(\frac{2 \me v^2}{I} \right) + \frac{1}{2} \, f(1) \right]
\nonumber \\ [2mm]
&=& \frac{4\pi Z_0^2 e^4}{\me v^2} \, {\cal N} Z \;
\ln \left(\frac{2 \me v^2}{I} \, a \right)
\label{6.309}\eeqa
with
\beq
a = \exp \left[ \frac{1}{2} f(1) \right] = \left\{
\begin{array}{ll}
\displaystyle{\sqrt{\exp(1)/2^5}}
\rule{5mm}{0mm} & \mbox{for electrons,}
\\ [3mm]
\displaystyle{1/2}
& \mbox{for positrons,} \\ [3mm]
\displaystyle{\frac{M_0}{M_0 + \me}}
& \mbox{for heavier projectiles.}
\end{array} \right.
\label{6.310}\eeq

For projectiles with low energies, the Bethe formula \req{6.307} gives
negative stopping powers, a clear indication of the fact that we are
outside its range of validity. In approximate calculations [\eg, of the
CSDA range --- see Section \req{sec9.4.1}] it is desirable to extend
the formula to lower energies, preventing the occurrence of negative
stopping powers. To obtain a suitable extended formula, we follow
\citet{Rao-SahibWittry1974} and consider the reciprocal of the
non-relativistic Bethe stopping power, Eq.\ \req{6.309}, as a function
of the variable $x = \beta^2$ which, aside from a multiplicative
constant, is
$$
h(x) = \frac{x}{\ln(Ax)} \qquad \mbox{with} \qquad A = \frac{2a \, \me c^2}{I}.
$$
This function has an inflexion point at $x_{\rm c} = \exp(2)/A$, where
$$
h(x_{\rm c}) = x_{\rm c}/2 \quad \mbox{and} \quad
h'(x_{\rm c}) = 1/4.
$$
To avoid negative stopping powers, for $x$ values less than $x_{\rm c}$
we replace the function $h(x)$ with the first two terms of its Taylor
expansion about $x_{\rm c}$ [notice that $h''(x_{\rm c})=0$],
$$
h(x) \simeq \frac{x_{\rm c}}{2} + \frac{1}{4} \left(x-x_{\rm c} \right)
= \frac{x_{\rm c} + x}{4} \, .
$$
That is, we consider the modified function
$$
h_{\rm mod} (x) = \left\{ \begin{array}{ll}
x/\ln(A x) & \mbox{if $x > x_{\rm c}$,} \\ [2mm]
(x + x_{\rm c})/4  \rule{5mm}{0mm} & \mbox{if $x \le x_{\rm c}$,}
\end{array} \right.
$$
which is positive for $x\in(0,x_{\rm c})$. Notice that $h_{\rm mod}(x)$
and its first and second derivatives are continuous. It is thus
reasonable to modify the non-relativistic Bethe formula, by making the
replacement
\index{Bethe stopping power formula!low-energy extrapolation}
\beq
\frac{1}{\beta^2} \ln \left( \frac{2a \, \me c^2}{I} \, \beta^2 \right)
\; \; \rightarrow \; \;
\frac{4}{\beta^2 + \beta^2_{\rm c}}
\label{6.311}\eeq
when $\beta^2 < \beta^2_{\rm c}$, where
\beq
\beta^2_{\rm c} = \exp(2) \, \frac{I}{2a \, \me c^2}
\label{6.312}\eeq
and $\ln a = f(1)/2$. This replacement can also be applied to the
relativistic formula \req{6.307}, after adding and subtracting the
quantity $\ln a$. Thus, for $\beta < \beta_{\rm c}$, we take
\beq
S = \frac{4\pi Z_0^2 e^4}{\me v^2} \, {\cal N} Z
\left[ \frac{4 \beta^2}{\beta^2 + \beta^2_{\rm c}} - \frac{1}{2} f(1)
+ \ln \left( \frac{1}{1-\beta^2} \right) - \beta^2 + \frac{1}{2}
f(\gamma) \right],
\label{6.313}\eeq
which extrapolates the original Bethe formula to low energies with
continuity of the function and its first and second derivatives. In the
low-energy limit, the extended formula \req{6.313} gives a finite value
of the stopping power, while experimental results indicate that $S$
vanishes when the speed of the projectile tends to zero (see Fig.\
\ref{fig7.12}). Nonetheless,
the extended formula yields fairly realistic stopping powers for
projectiles with speeds well below $\beta_{\rm c} c$, where the Bethe
formula is manifestly incorrect.
\index{Bethe stopping power formula!derivation|)}


\subsection{The shell correction \label{sec6.9.1}}
\index{shell correction|(}

It is customary to extend the validity of the Bethe formula \req{6.307}
to lower energies of the projectile by adding a term $C/Z$, the
so-called {\it shell correction}, which represents the difference
between the ``exact'' stopping power calculated from the PWBA and the
result from the asymptotic formula \req{6.307}. The corrected Bethe
formula reads
\beq
S = \frac{4\pi Z_0^2 e^4}{\me v^2} \, {\cal N} Z
\left[ \ln \left(\frac{2 \me v^2}{I} \right)
+ \ln \left( \frac{1}{1-\beta^2} \right) - \beta^2 + \frac{1}{2}
f(\gamma) - \frac{C}{Z} \right].
\label{6.314}\eeq
The shell correction $C/Z$ thus accounts for the various approximations
involved in the derivation of the Bethe formula, namely, \\
1) it was assumed that all electron subshells contribute to the stopping
power irrespective of their ionization energies, \\
2) the lowest allowed recoil energy $Q_-$ was replaced with the approximate
value given by \req{6.284}, \\
3) any variation of the GOS with $Q$ for interactions with recoil energies
smaller than $Q_2$ (low- and intermediate-$Q$ ranges) was disregarded,
and \\
4) close (high-$Q$) excitations were represented as collisions with free
electrons at rest, thus neglecting the finite width of the Bethe ridge
\citep{Fano1963}.

\citet{Walske1952, Walske1956} used hydrogenic wave functions to obtain
shell corrections for K and L shells; \citet{KhandelwalMerzbacher1966}
and \citet{Bichsel1983} performed similar calculations for M shells.
\citet{Bonderup1967} obtained an atomic shell correction from stopping
powers calculated using the local-plasma approximation of
\cite{LindhardScharff1953} (see Section \ref{sec7.5.1}). More recently,
\citet{Bichsel2002} determined the corrections for the inner shells of
aluminum and silicon by direct integration of the non-relativistic GOSs
of \citet{Manson1972}. These calculations are essentially non
relativistic and they predict that, for projectiles with kinetic
energies much higher than the ionization energy of the K shell, the
shell correction $C/Z$ decreases when the energy of the projectile
increases, tending to zero in the high-energy limit. References on
available sources of theoretical and empirical shell corrections are
given in the \citet{ICRU37}.

\begin{figure}[h]
\begin{center}
\vspace*{3mm}
\includegraphics*[width=10.5cm]{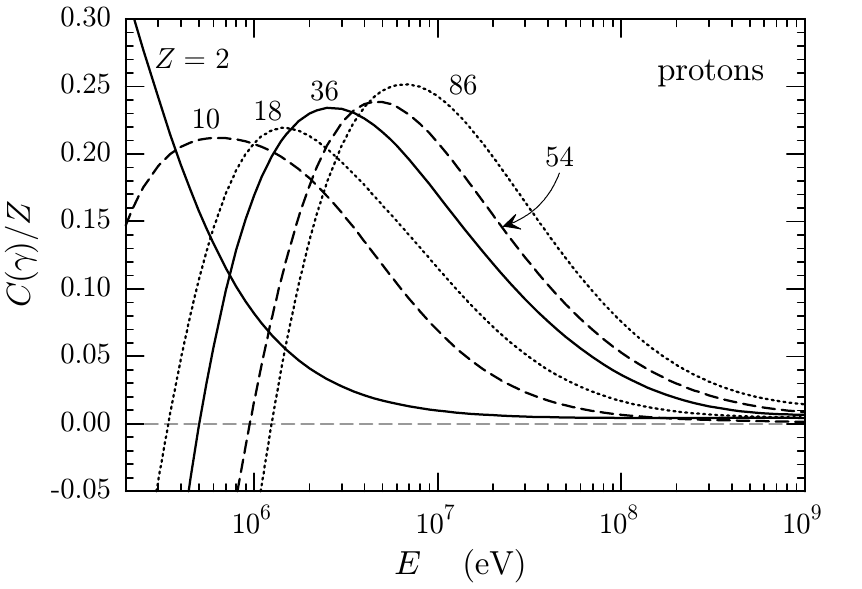}
\caption{Shell correction $C'_0/Z$ to the Bethe formula of the stopping
power of noble gases for protons, as functions of the kinetic energy of
the projectile. The displayed values were calculated from DHFS
relativistic GOSs \citep{Salvat2022a}.
\label{fig6.12}}
\end{center}
\vspace*{-5mm}
\end{figure}

Aside from the approximations listed above, the derivation of the Bethe
stopping power formula is not completely satisfactory because it ignores
relativistic deviations from the Bethe sum rule \citep[see,
\eg,][]{Cohen2003}. In reality, the integral $S_0(Q)$ in Eq.\ \req{6.85}
is slightly less than the atomic number $Z$ for low $Q$ excitations,
although it approaches $Z$ when $Q$ increases \citep{BoteSalvat2008,
Salvat2022a}. A rigorous shell correction can only be obtained from a
comparison of the stopping power predicted by the Bethe formula with the
result ${\cal N} \sigma^{(1)}$ calculated numerically from GOS tables.
The latter calculation is unwieldy and delicate because of the massive
interpolations involved in the integration of the DDCS.

\begin{figure}[h!]
\begin{center}
\includegraphics*[width=7.9cm]{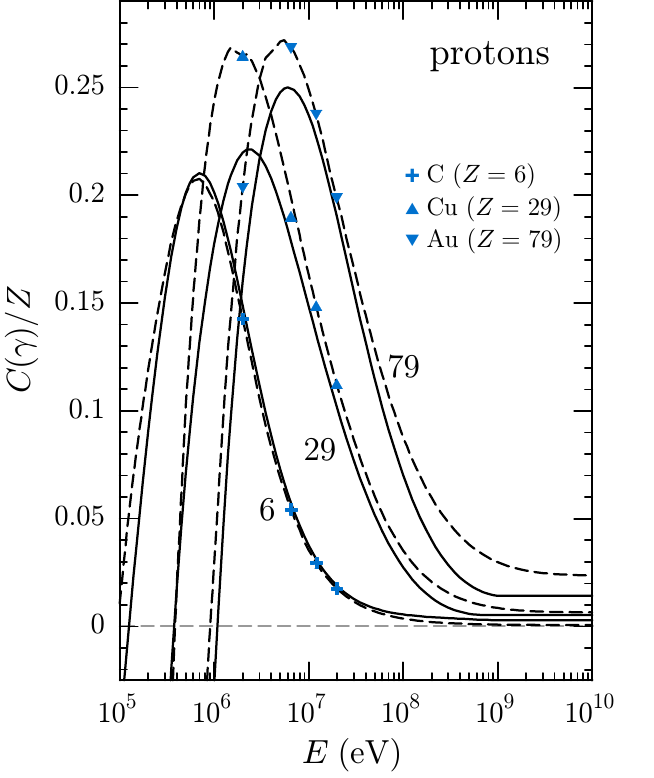} \rule{1mm}{0mm}
\includegraphics*[width=7.9cm]{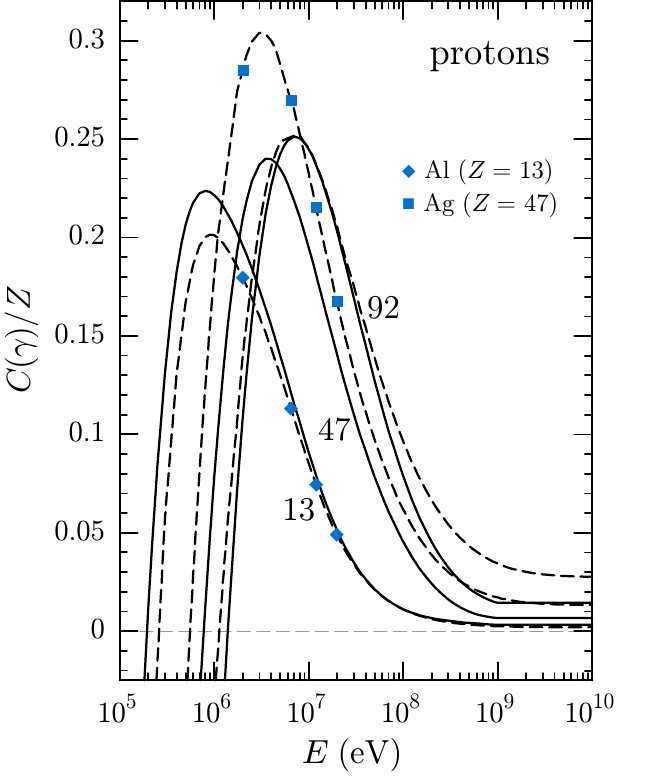}
\caption{Shell corrections $C(\gamma)/Z$ to the asymptotic formula of the
stopping cross section for inelastic collisions of protons with atoms of
the elements with the indicated atomic numbers, as functions of the
kinetic energy of the projectile. Solid curves are the results from DHFS
calculations, symbols are Bichsel's semi-empirical shell corrections given in
the \citet{ICRU37}, and the dashed curves were generated with
the program {\sc best} \citep{BergerBichsel1994}. Adapted from
\citet{Salvat2022a}.
\label{fig6.13}}
\end{center}\end{figure}

Calculations of the stopping power of atomic gases for protons using the
relativistic DDCS given by Eqs.\ \req{6.221} with the GOS tables
computed from the DHFS self-consistent potential have been performed
recently by \citet{Salvat2022a}. Their results show that the stopping
power obtained from the relativistic PWBA for projectiles much heavier
than the electron can be expressed in the form
of the Bethe formula,
\beqa
S (E) &=&
\frac{4 \pi Z_1^2 e^4}{\me v^2} \, {\cal N} Z \,
\left\{
\ln \left(\frac{2 \me v^2}{I'} \right)
+ \ln \left( \frac{1}{1-\beta^2}\right) - \beta^2
+ \frac{1}{2} \, f(\gamma) \right.
\nonumber \\ [2mm]
&& \left. + \frac{S_0-Z}{2Z} \left[ \ln (\beta^2 \gamma^2) - \beta^2
\right]- \frac{C(\gamma)}{Z}
\right\},
\label{6.315}\eeqa
where the first term in the second line is a relativistic correction
with
\beq
S_0 = \int_0^\infty \frac{\d f(W)}{\d W}\, \d W \,
\label{6.316}\eeq
and with an ``effective'' $I$ value, $I'$, slightly larger than the
one resulting from the conventional definition, Eq.\ \req{6.288}. The
shell correction $C(\gamma)/Z$ for free atoms was obtained as the
difference between the stopping power calculated by direct integration
of the DDCS and the prediction of the Bethe formula, Eq.\ \req{6.315},
without the shell correction. It is worth noticing that the stopping
cross section obtained from the PWBA is determined by the speed and
the squared charge of the projectile; it does not depend on neither
the mass nor the sign of the charge of the projectile. This feature
implies that the shell correction, considered as a function of
$\gamma$ or of the projectile speed, is valid for any kind of charged
projectile heavier than the electron. Since the main contributions to
the shell correction originate from inner electron subshells, which
are assumed to be described accurately by the DHFS model, the shell
correction calculated for free DHFS atoms is expected to be applicable
also to dense materials and compounds.  Figure \ref{fig6.12} shows the
shell correction calculated for protons in noble gases
\citep{Salvat2022a}.  Figure \ref{fig6.13} compares these shell
corrections with those obtained from the program {\sc best} of
\citet{BergerBichsel1994}, which was used to generate the high-energy
stopping power tables given in the \citet{ICRU49}. Atomic shell
corrections for electrons and positrons have also been calculated by
the author by using the same methodology as for heavier projectiles,
\ie, $C(\gamma)/Z$ was derived from the difference between the stopping
cross sections computed from the DDCS resulting from the DHFS-model
GOSs and from the asymptotic Bethe formula without the shell correction.
Tables of these DHFS shell corrections for electrons, positrons, and
heavier projectiles, and for all elements ($Z=1$ to 99), are included in
the database of the {\sc sbethe} program (see Chapter \ref{chapt10} and
\citeauthor{SalvatAndreo2023}, and \citeyear{SalvatAndreo2023}).

\begin{figure}[t!]
\begin{center}
\includegraphics*[width=7.25cm]{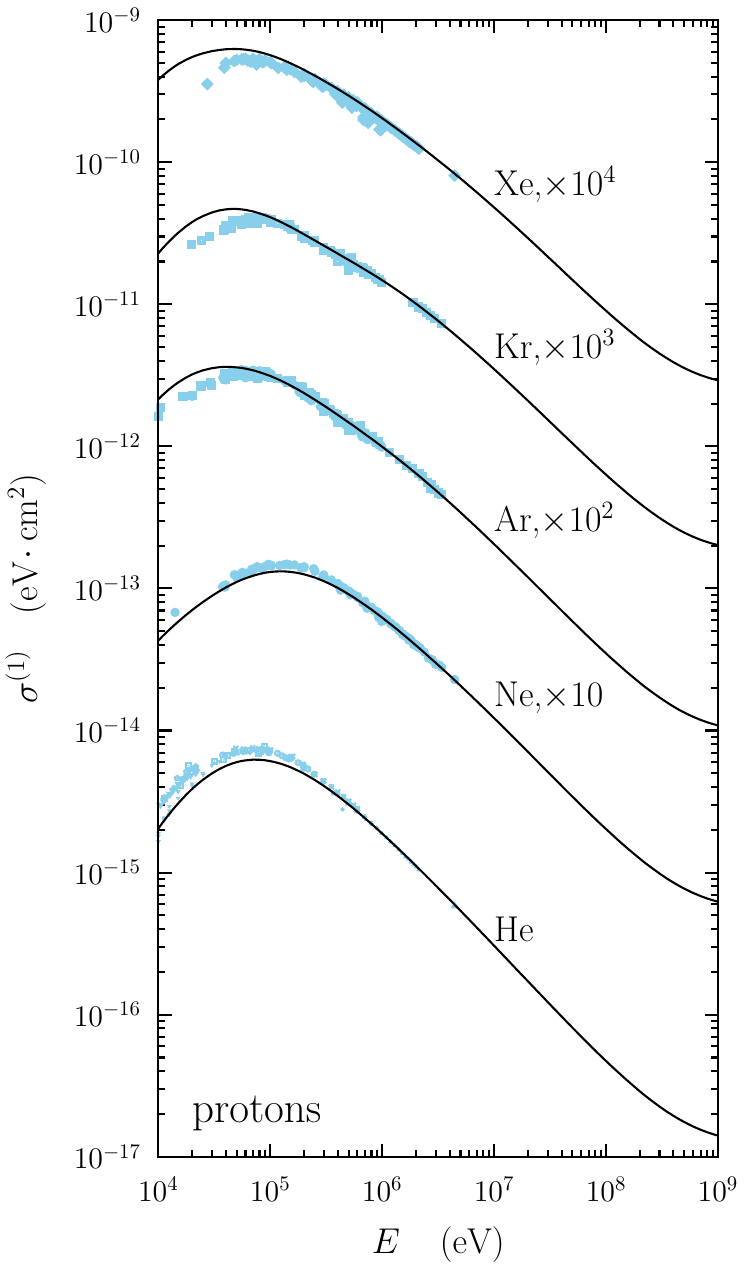}
\includegraphics*[width=7.25cm]{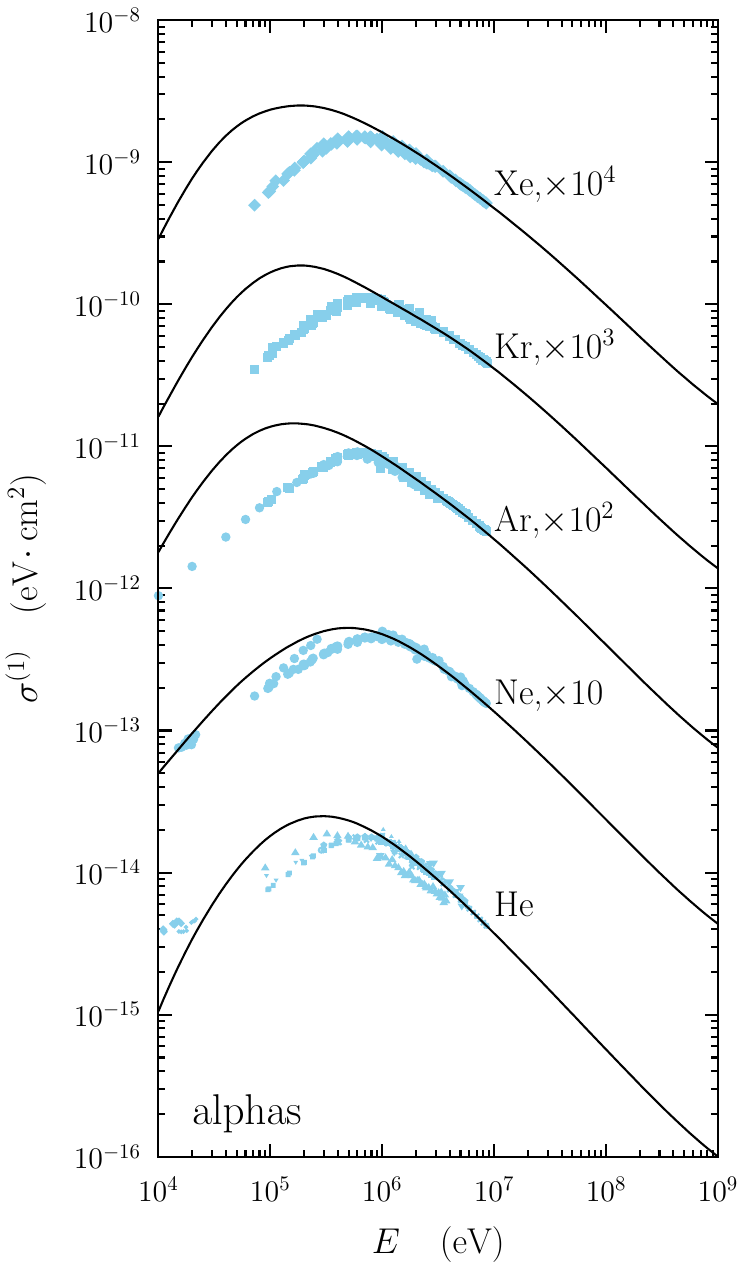}
\caption{Stopping cross sections of protons and alpha particles in noble
gases, calculated from the relativistic PWBA with the DHFS
self-consistent potential (solid curves). Symbols represent experimental
data from the IAEA stopping-power database.
\label{fig6.14}}
\vspace*{-4mm}
\end{center}\end{figure}

It is worth recalling that the relativistic PWBA, as well as the
dielectric theory described in Section \ref{sec6.7}, are expected to be
valid only for charged projectiles with sufficiently large energies (see
Section \ref{sec6.2}). Figure \ref{fig6.14} compares stopping cross
sections calculated from the PWBA for collisions of protons and alpha
particles with atoms of the noble gases \citep{Salvat2022a} with
experimental data from the IAEA online database
(\citeauthor{Montanari2017}, \citeyear{Montanari2017};
\url{https://www-nds.iaea.org/stopping}).
Since the PWBA with the DHFS potential is expected to provide realistic
results for atoms with closed-subshell configurations, this comparison
indicates that the results from the PWBA are reliable for protons and
alphas with kinetic energies higher than about 0.75~MeV and 5~MeV,
respectively. It is interesting that these values coincide with the
estimated validity limits of the corrected Bethe formula \req{6.315} for
protons and alphas \citep{Salvat2022c}. This coincidence lends support
to the applicability to arbitrary materials of atomic shell corrections
derived from free-atom DHFS calculations.

\index{shell correction|)}




\chapter{Inelastic collisions in a local electron gas. Dielectric theory
\label{chapt7}}
\index{dielectric functions!of the electron gas|(}

Excitations of condensed media differ from those of isolated atoms due
to atomic aggregation effects: the orbitals of electrons bound in outer
subshells are affected by the presence of neighboring atoms, which also
has an influence on the wave functions of free electrons moving with
small and moderate speeds. Aggregation causes modifications of the
excitation spectrum which manifest in the so-called x-ray absorption
fine structure of the photoabsorption cross section \citep[see,
\eg,][]{RehrAlbers2000}. In addition, electrons that are in weakly bound
orbitals can react collectively through quantized oscillation modes
called {\it plasmons}. The simplest system presenting collective
excitations is the degenerate electron gas, which provides an idealized
model for the conduction electrons in metals. More importantly, the
dielectric function (DF) of the electron gas is a fundamental component
of the so-called optical-data models of the DF of real materials.

The dielectric formalism (Chapter \ref{chapt1} and Section \ref{sec6.7})
provides a convenient description of inelastic collisions of high-energy
charged particles in material media. The main hindrance of the formalism
is the extreme difficulty of the calculation of the dielectric function
(DF) of a given material from first principles, which is caused by the
complexity of the quantum structure of the material as a many-particle
system. In principle, the DF can be calculated by using the methods of
density functional theory \citep[see, \eg,][]{Martin2004} which provide
electronic wave functions obtained by means of a self-consistent method. In
practice, however, only partial results for a limited number of solids
have been reported to date. A quantity that has been calculated using
this kind of approach is the optical dielectric function (ODF)
\citep[see, \eg.,][]{Werner2009}. Unfortunately, the calculation method
still involves approximations, and calculations are feasible only for
low and moderate excitation energies, $W \equiv \hbar \omega$, up to
about 100 eV. An alternative approach to calculate the ODF is provided
by the real-space Green's function formalism utilized in studies of
x-ray absorption fine structure \citep[see, \eg,][]{RehrAlbers2000}.

In the present Chapter we first consider the stopping of charged
particles in a degenerate homogeneous electron gas. We derive the DF of
the gas by
means of the random phase approximation, the so-called Lindhard DF. We
then consider the modification known as the Mermin DF, which accounts
for the finite life time of the gas excitations, and an ad hoc extension
of the Mermin DF that includes a gap in the excitation spectrum to mimic
the response of valence electrons in semiconductors and insulators. We
also consider a general strategy to build the complete DF,
$\epsilon(Q,W)$, from the limited theoretical and experimental
information that may be available for a material. We limit our
considerations to isotropic materials or solids with cubic lattices
whose DFs do not depend on the direction of the wave vector ${\bf q}$.
We concentrate on the so-called {\it optical-data models}, that are the
most elaborate schemes employed in practical calculations of the
stopping of charged particles in matter. The first component of an
optical-data model is the optical oscillator strength (OOS), which is
built by assembling available experimental and theoretical information
on optical constants [\ie, the refractive index $n(W)$ and the
extinction coefficient $\kappa(W)$, which we consider as functions of
the excitation energy $W=\hbar \omega$]. The DF $\epsilon(Q,W)$ is
obtained by extending the OOS to finite recoil energies by means of a
suitable algorithm. For low and moderate frequencies, the extension
algorithms most frequently used are based on the Lindhard and Mermin DFs
of the electron gas.


\section{Lindhard dielectric functions of an electron gas
\label{sec7.1}}
\index{Lindhard dielectric function|(}

\citet{Lindhard1954} obtained analytical expressions for the longitudinal
and transverse DFs of a degenerate electron gas (a homogeneous gas of
free electrons at zero temperature) of density $\rho^0$ (electrons per
unit volume) using the random-phase approximation, an elementary
quantum-mechanical many-body first-order perturbation theory. Because of
the fundamental importance of the Lindhard DFs, we provide here a
detailed derivation of it.

\subsection{Derivation of the Lindhard DF \label{sec7.1.1}}

We consider the {\it jellium} model of a solid conductor, which consists
of the non-interacting electron gas and a uniform jelly background of
positive charge such that the average spatial charge density equals zero.
Within an independent-electron approximation, the electrostatic
potential of the positive background cancels with that of the electron
gas, and the one-electron Hamiltonian of the gas reduces to that of a
free particle
\beq
{\cal H}_0 = \frac{\breve{\bf p}^2}{2 \me} = - \, \frac{\hbar^2}{2 \me} \,
\nabla^2.
\label{7.1}\eeq
We wish to determine the response of the gas to an electromagnetic field
described by the potentials $\varphi({\bf r},t)$ and ${\bf A}({\bf
r},t)$ in the Coulomb gauge. We recall that, for a transverse vector
potential $\breve{\bf p} \cdot {\bf A} = {\bf A} \cdot \breve{\bf
p}$. The interaction energy of an electron with the field is given by
the non-relativistic Hamiltonian\footnote{For the sake of simplicity
here we ignore the interaction of the electron spin with the magnetic
field, although we consider that the spatial orbitals are doubly
occupied by electrons with opposite spins.} [see Eq.\ \req{2.14}]
\beq
{\cal H}' = -e \, \varphi({\bf r},t) + \frac{e}{\me c} \, {\bf A}
\dotprod \breve{\bf p} = -e \, \varphi({\bf r},t) - {\rm i} \, \frac{e
\hbar}{\me c} \, {\bf A} \dotprod \nablab ,
\label{7.2}\eeq
which will be considered as a perturbation to first order. Notice that
we have disregarded a term proportional to $A^2$, which should be
considered in calculations to second order of perturbation.

The unperturbed one-electron spatial orbitals can be represented as plane
waves satisfying periodic boundary conditions in a cubic box of side
$L$,
\beq
\phi^0_{\bf k} ({\bf r},t) =
L^{-3/2} \, \exp \left[{\rm i} ({\bf k} \dotprod
{\bf r} - \omega_{\bf k} t) \right],
\label{7.3}\eeq
where ${\bf k}$ is the electron wave number, and
\beq
E_{\bf k} = \frac{\hbar^2 k^2}{2 \me} \equiv \hbar \omega_{\bf k}
\label{7.4}\eeq
is the kinetic energy of the electron. The density of spatial orbitals
(number of states per unit volume) in {\bf k}-space is [Eq.\ \req{2.28}]
\beq
\frac{\d {\cal N}}{\d {\bf k}} = \left( \frac{L}{2\pi} \right)^3.
\label{7.5}\eeq
In the ground state of the gas, all orbitals with energies less than the
Fermi energy, $E_{\rm F}$ are occupied. We can write
\beq
E_{\rm F} = \frac{\hbar^2 k_{\rm F}^2}{2\me} ,
\label{7.6}\eeq
where
\beq
k_{\rm F} = \left( 3\pi^2\rho^0 \right)^{1/3}
\label{7.7}\eeq
is the Fermi wave number (\ie, the wave number of an electron at the
Fermi level).

The perturbed orbitals can be expressed as
\beq
\phi_{{\bf k}}({\bf r},t) = \phi^0_{{\bf k}}({\bf r},t)
+\exp(st) \, \phi^1_{{\bf k}}({\bf r},t) + \cdots,
\label{7.8}\eeq
where $\exp(st) \, \phi^1_{\bf k}({\bf r},t)$ is the correction to first
order in the external field. To ensure that at $t=-\infty$ these
orbitals reduce to the unperturbed states, we multiply the correction by
a factor $\exp(st)$, where $s$ is a small positive constant which should
be set to zero at the end of the calculation. The effect of this factor
is to switch the correction on adiabatically in the remote past. With a
certain {\it ad hoc} modification (see Section \ref{sec7.2}), the theory
can be extended to account for the finite relaxation time of electronic
excitations, and then the quantity $s$ can be identified with the
reciprocal of the relaxation time. The perturbed orbitals obey the
time-dependent Schr\"{o}dinger equation
\beq
{\rm i} \hbar \, \frac{\partial}{\partial t}
\, \phi_{\bf k}({\bf r},t)
= \left[{\cal H}_0 + {\cal H}' \right] \, \phi_{\bf k}({\bf r},t),
\label{7.9}\eeq
which implies that the first-order correction $\exp(st) \, \phi^1_{\bf
k}({\bf r},t)$ satisfies the equation (neglecting terms of higher orders)
\index{Fourier transform}
\beqa
\left( {\rm i} \hbar \frac{\partial}{\partial t} - {\cal H}_0 \right)
\exp(st) \, \phi^1_{\bf k}({\bf r},t) &=& {\cal H}' \,
\phi^0_{\bf k}({\bf r},t).
\label{7.10}\eeqa
We consider the Fourier transform of the correction to the wave function,
\beq
\overline{\phi}^1_{\bf k} ({\bf q},\omega)
= (2\pi)^{-2} \int \d {\bf r} \int \d t \;
\exp\left[ - {\rm i} \left( {\bf q} \dotprod {\bf r} - \omega t \right)
\right] \; \exp(st) \, \phi^1_{\bf k}({\bf r},t),
\nonumber\eeq
and write
\beq
\exp(st) \, \phi^1_{\bf k} ({\bf r},t)
= (2\pi)^{-2} \int \d {\bf q} \int \d \omega \;
\exp\left[ {\rm i} \left( {\bf q} \dotprod {\bf r} - \omega t \right)
\right] \, \overline{\phi}^1_{\bf k}({\bf q},\omega),
\label{7.11}\eeq
Introducing this expression and the Fourier transforms of the
potentials, Eq.\ \req{7.10} takes the form
\beqa
&& \! \! \! \! \! \! \! \! \! \! \! \! \! \! \! \!
(2\pi)^{-2} \int \d {\bf q} \int \d \omega \, \left\{
\hbar (\omega+{\rm i} s) - \frac{\hbar^2 q^2}{2 \me} \right\}
\exp\left[ {\rm i} \left( {\bf q} \dotprod {\bf r} - \omega t \right)
\right] \, \overline{\phi}^1_{\bf k} ({\bf q},\omega)
\nonumber \\ [2mm]
&=& (2\pi)^{-2} \int \d {\bf q}' \int \d \omega' \,
\left\{-e\, \varphi({\bf q}',\omega')
+ \frac{e\hbar}{\me c}  {\bf A}({\bf q}',\omega') \dotprod
{\bf k} \right\}
\nonumber \\ [2mm]
&& \mbox{} \times
\exp\left[ {\rm i} \left( {\bf q}' \dotprod {\bf r} - \omega' t \right)
\right] \, L^{-3/2} \, \exp \left[{\rm i} ({\bf k} \dotprod
{\bf r} - \omega_{\bf k} t) \right].
\nonumber \eeqa
Changing variables to ${\bf q}={\bf q}'+{\bf k}$ and
$\omega=\omega'+\omega_{\bf k}$ in the integrals on the right-hand
side, we see that
\beq
\left\{ \frac{\hbar^2
q^2}{2 \me} - \hbar (\omega + {\rm i} s) \right\}
\overline{\phi}^1_{\bf k} ({\bf q},\omega)
= L^{-3/2} \left\{e \varphi({\bf q}-{\bf k},\omega-\omega_{\bf k})
- \frac{e\hbar}{\me c}  {\bf A}({\bf q}-{\bf k},\omega-\omega_{\bf k})
\dotprod {\bf k}
\right\} ,
\nonumber \eeq
and, consequently,
\beqa
\overline{\phi}^1_{\bf k} ({\bf q},\omega)
&=& \frac{2 \me e}{\hbar^2 L^{3/2}} \, \frac{
\varphi({\bf q}-{\bf k},\omega-\omega_{\bf k})
- (\hbar/\me c)  {\bf A}({\bf q}-{\bf k},\omega-\omega_{\bf k})
\dotprod {\bf k}}{q^2 - 2 \me (\omega+{\rm i} s) /\hbar} \, .
\rule{15mm}{0mm}
\label{7.12}\eeqa


\subsubsection{Induced charge and current densities \label{sec7.1.1.1}}

The perturbed charge density of electrons in the gas is
\beq
\rho({\bf r},t) = - e \, 2 \sum_{\bf k} \left|
\phi_{\bf k}({\bf r},t) \right|^2
\equiv
-e \rho^0 + \rho_{\rm ind} ({\bf r},t),
\label{7.13}\eeq
where the summation is over spatial orbitals that are filled, the factor
2 accounts for the spin degeneracy, and the quantity $\rho_{\rm ind}$ is
the induced electron charge density. To first order in the external field,
\beq
\rho({\bf r},t) = -e \, 2
\sum_{\bf k} \left|
\phi^0_{\bf k}({\bf r},t)
+ \exp(st) \, \phi^1_{\bf k}({\bf r},t) \right|^2,
\label{7.14}\eeq
and, to the same order, we have
\beq
\rho_{\rm ind} ({\bf r},t) = -e \, 2
\sum_{\bf k} \left\{ \left[
\phi^0_{\bf k}({\bf r},t) \right]^\ast
\exp(st) \, \phi^1_{\bf k}({\bf r},t) +
\phi^0_{\bf k}({\bf r},t)
\left[ \exp(st) \, \phi^1_{\bf k}({\bf r},t) \right]^\ast
\right\}.
\label{7.15}\eeq
Let us consider the quantity
\beqa
\lefteqn{X \equiv \sum_{\bf k}
\left[ \phi^0_{\bf k}({\bf r},t) \right]^\ast
\exp(st) \, \phi^1_{\bf k}({\bf r},t)
= \sum_{\bf k}
L^{-3/2} \, \exp \left[-{\rm i} ({\bf k} \dotprod
{\bf r} - \omega_{\bf k} t) \right] }
\nonumber \\ [2mm]
&& \mbox{} \; \; \; \; \times
(2\pi)^{-2} \int\d {\bf q}' \int \d \omega' \; \exp\left[{\rm i} ({\bf
q}'\dotprod {\bf r}-\omega' t)\right]
\nonumber \\ [2mm]
&& \mbox{} \; \; \; \; \times
\frac{2 \me e}{\hbar^2 L^{3/2}}
\, \frac{
\varphi({\bf q}'-{\bf k},\omega'-\omega_{\bf k})
- (\hbar/\me c)  {\bf A}({\bf q}'-{\bf k},\omega'-\omega_{\bf k})
\dotprod {\bf k}}{q'^2 - 2 \me (\omega'+{\rm i} s) /\hbar}.
\nonumber \eeqa
Changing the integration variables to ${\bf q} = {\bf q}'-{\bf k}$ and
$\omega = \omega' - \omega_{\bf k}$, we have
\beqa
X &=& (2\pi)^{-2} \int\d {\bf q} \int \d \omega \;
\exp \left[ {\rm i} ({\bf q} \dotprod
{\bf r} - \omega t) \right]
\frac{2 \me e}{\hbar^2 L^{3}}
\sum_{\bf k} \frac{\varphi({\bf q},\omega)
- (\hbar/\me c)  {\bf A}({\bf q},\omega)
\dotprod {\bf k}}{q^2 + 2 {\bf q} \dotprod {\bf k}
- 2 \me (\omega+{\rm i} s) /\hbar} .
\nonumber \eeqa
Passing to the limit $L \rightarrow \infty$, the summation over ${\bf
k}$ can be replaced with an integral,
\beq
\sum_{\bf k} \cdots \rightarrow \int \d {\bf k} \, \left( \frac{L}{2\pi}
\right)^3 \, f({\bf k}) \, \cdots ,
\label{7.16}\eeq
where the first factor in the integrand is the density of spatial orbitals
per unit volume in ${\bf k}$-space, and $f({\bf k}) = {\cal S }(k_{\rm
F}-k)$ is the probability that the orbital $\phi^0_{\bf k}$
is occupied. Notice that the integral
\beq
\int \d {\bf k} \, \left( \frac{L}{2\pi} \right)^3 \, f({\bf k})
= \left( \frac{L}{2\pi} \right)^3
\int_0^{k_{\rm F}} 4\pi \, k^2 \d k
= \left( \frac{L}{2\pi} \right)^3 4\pi \, \frac{k_{\rm F}^3}{3}
= \frac{1}{2} L^3 \rho^0,
\label{7.17}\eeq
gives the number of occupied space orbitals. We can thus write
\beqa
X &=& (2\pi)^{-2} \int\d {\bf q} \int \d \omega \;
\exp \left[ {\rm i} ({\bf q} \dotprod
{\bf r} - \omega t) \right]
\nonumber \\ [2mm]
&& \mbox{} \times
\frac{2 \me e}{\hbar^2 (2\pi)^3} \int \d {\bf k} \, f({\bf k})\,
\frac{\varphi({\bf q},\omega)
- (\hbar/\me c)  {\bf A}({\bf q},\omega)
\dotprod {\bf k}}{q^2 + 2 {\bf q} \dotprod {\bf k}
- 2 \me (\omega+{\rm i} s) /\hbar} .
\nonumber \eeqa
Inserting this result into Eq.\ \req{7.15}, we see that the Fourier
transform of the induced charge density is \index{Fourier transform}
\beqa
\rho_{\rm ind} ({\bf q},\omega) &=& - \frac{4 \me e^2}{\hbar^2 (2\pi)^{3}}
\left\{
\int \d {\bf k} \, f({\bf k}) \, \frac{\varphi({\bf q},\omega)
- (\hbar/\me c)  {\bf A}({\bf q},\omega)
\dotprod {\bf k}}{q^2 + 2 {\bf q} \dotprod {\bf k}
- 2 \me (\omega+{\rm i} s) /\hbar} \right.
\nonumber \\ [2mm]
&& \mbox{} + \left.
\int \d {\bf k} \, f({\bf k}) \, \frac{\varphi^\ast({\bf q},\omega)
- (\hbar/\me c)  {\bf A}^\ast({\bf q},\omega)
\dotprod {\bf k}}{q^2 + 2 {\bf q} \dotprod {\bf k}
- 2 \me (\omega - {\rm i} s) /\hbar} \right\}.
\label{7.18}\eeqa
Since the potentials are real [\ie, $\varphi(- {\bf q},- \omega) =
\varphi^\ast({\bf q}, \omega)$, and similarly for the vector
potential], we can write
\beqa
\rho_{\rm ind} ({\bf q},\omega) &=& - \frac{\me e^2}{2 \hbar^2 \pi^{3}}
\int \d {\bf k} \, f({\bf k}) \,
\left\{\rule{0mm}{6mm}
\frac{1}{q^2 + 2 {\bf q} \dotprod {\bf k} - 2 \me (\omega+{\rm i} s) /\hbar}
\right.
\nonumber \\ [2mm]
&& \mbox{} \left. + \frac{1}{q^2 - 2 {\bf q} \dotprod {\bf k}
+ 2 \me (\omega+{\rm i} s) /\hbar} \right\}
\left[ \varphi({\bf q},\omega)
- (\hbar/\me c)  {\bf A}({\bf q},\omega)
\dotprod {\bf k} \rule{0mm}{4mm}\right]. \rule{12mm}{0mm}
\label{7.19}\eeqa
The contribution of the term with the vector potential is null. This
feature can be proved by noting that, for a fixed ${\bf q}$, we can
evaluate the integral
\beq
I \equiv \int \d {\bf k} \, f({\bf k}) \,
\left\{
\frac{1}{q^2 + 2 {\bf q} \dotprod {\bf k} - 2 \me (\omega+{\rm i} s) /\hbar}
+ \frac{1}{q^2 - 2 {\bf q} \dotprod {\bf k} + 2 \me (\omega+{\rm i} s) /\hbar}
\right\} ({\bf A} \dotprod {\bf k})
\nonumber \eeq
using spherical coordinates with the polar $z$ axis in the direction of
${\bf q}$ and the $x$ axis in the direction of ${\bf A}$ (recall that
in the Coulomb gauge ${\bf A}$ is transverse). Thus,
${\bf k} = k(\sin\theta \cos\phi, \sin\theta \sin\phi, \cos\theta)$,
${\bf A}=(A,0,0)$, and
\beqa
I &=& \int_0^{k_{\rm F}} k^2 \d k \int_0^\pi \sin\theta \, \d \theta
\int_0^{2\pi} \d \phi \; Ak \sin\theta \cos\phi
\nonumber \\ [2mm]
&& \mbox{} \times
\left\{
\frac{1}{q^2 + 2 q k \cos \theta - 2 \me \omega /\hbar}
+ \frac{1}{q^2 - 2 q k \cos\theta + 2 \me \omega /\hbar}
\right\},
\nonumber \eeqa
where the integral over $\phi$ evidently vanishes. Hence, setting $\xi=\cos\theta$, we have
\beqa
\rho_{\rm ind} ({\bf q},\omega) &=&
- \varphi({\bf q},\omega) \, \frac{\me e^2}{\pi^2 \hbar^2}
\int_0^{k_{\rm F}} \d k \, k^2 \int_{-1}^1 \d \xi \left(
\frac{1}{q^2 + 2 qk\xi - 2 \me (\omega+{\rm i} s)
/\hbar}
\right.
\nonumber \\ [2mm]
&& \mbox{} \left.
+ \frac{1}{q^2 - 2qk\xi + 2 \me (\omega+{\rm i} s) /\hbar}
\right).
\label{7.20}\eeqa

The external field also causes a variation of the current density of
electrons in the gas. The perturbed charge current density is [see Eq.\
\req{2.15}],
\beqa
{\bf j} ({\bf r},t) &=& - e \, 2 \sum_{\bf k}
\left\{     \frac{\hbar}{2 {\rm i} \me}  \left(\rule{0mm}{4mm}
\left[ \phi_{\bf k}({\bf r},t) \right]^\ast  \left[
\nablab \phi_{\bf k}({\bf r},t) \right]
- \left[ \nablab \phi_{\bf k}({\bf r},t) \right]^\ast
\left[ \phi_{\bf k}({\bf r},t) \right] \right)
\right.
\nonumber \\ [2mm]
&& \mbox{} \left. + \frac{e}{\me c}
\left| \phi_{\bf k}({\bf r},t) \right|^2 {\bf A}({\bf r},t)
\rule{0mm}{4mm}\right\}
\nonumber \\ [2mm]
&=& {\bf j}^0 ({\bf r},t) + {\bf j}_{\rm ind} ({\bf r},t),
\label{7.21}\eeqa
where the factor 2 accounts for the spin degeneracy. The unperturbed
charge current density
\beqa
{\bf j}^0 ({\bf r},t) &=& {\rm i} \, \frac{e \hbar}{\me}
\sum_{\bf k} \left\{
(\phi_{\bf k}^0)^\ast \left( \nablab \phi_{\bf k}^0 \right)
-\rule{0mm}{4mm} \left( \nablab \phi_{\bf k}^0 \right)^\ast
\phi_{\bf k}^0 \right\}
\label{7.22}\eeqa
vanishes because terms with ${\bf k}$ and $-{\bf k}$ cancel each other.
The charge current density induced by the external field is,
to first order,
\beqa
{\bf j}_{\rm ind} ({\bf r},t) &=&
\sum_{\bf k}
\left\{
{\rm i} \, \frac{e \hbar}{\me} \, \exp(st) \left[ \rule{0mm}{4mm}
(\phi_{\bf k}^1)^\ast (\nablab \phi_{\bf k}^0)
+ (\phi_{\bf k}^0)^\ast (\nablab \phi_{\bf k}^1)
- (\nablab \phi_{\bf k}^1)^\ast \phi_{\bf k}^0
- (\nablab \phi_{\bf k}^0)^\ast \phi_{\bf k}^1 \right]
\right.
\nonumber \\ [2mm]
&& \mbox{} \left.
- \frac{2 e^2}{\me c} L^{-3}
{\bf A}
\rule{0mm}{4mm}\right\},
\label{7.23}\eeqa
where we have used that $|\phi_{\bf k}^0|^2 = L^{-3}$.
Taking the limit $L\rightarrow \infty$, the summation over ${\bf k}$
can be replaced with an integral [see Eq.\ \req{7.16}] to give
\beqa
{\bf j}_{\rm ind} ({\bf r},t) &=&
\left( \frac{L}{2\pi} \right)^3 \int \d {\bf k} \, f({\bf k})
\left\{
{\rm i} \, \frac{e \hbar}{\me} \, \exp(st) \left[ \rule{0mm}{4mm}
(\phi_{\bf k}^1)^\ast (\nablab \phi_{\bf k}^0)
+ (\phi_{\bf k}^0)^\ast (\nablab \phi_{\bf k}^1)
\right. \right.
\nonumber \\ [2mm]
&& \mbox{} \left. \left. \rule{0mm}{4mm}
- (\nablab \phi_{\bf k}^1)^\ast \phi_{\bf k}^0
- (\nablab \phi_{\bf k}^0)^\ast \phi_{\bf k}^1 \right]
- \frac{2 e^2}{\me c} L^{-3} {\bf A}
\rule{0mm}{4mm}\right\}
\label{7.24}\eeqa
Writing the correction $\exp(st) \, \phi_{\bf k}^1({\bf r},t)$ in terms of
its Fourier transform, Eq.\ \req{7.11}, and using the expression
\req{7.3} of the unperturbed orbitals, we have \index{Fourier transform}
\beqa
\left[ \exp(st) \phi_{\bf k}^1\right]^\ast (\nablab \phi_{\bf k}^0)
&=& \phi_{\bf k}^0({\bf r},t) \, (2\pi)^{-2}
\int \d {\bf q} \int \d \omega \;
\exp\left[ - {\rm i} \left( {\bf q} \dotprod {\bf r} - \omega t \right)
\right] \, \overline{\phi}^{1\ast}_{\bf k}({\bf q},\omega) \, ({\rm i}{\bf k}),
\nonumber \eeqa
\beqa
(\phi_{\bf k}^0)^\ast \left[ \exp(st) \nablab \phi_{\bf k}^1 \right]
&=& \left[ \phi_{\bf k}^0({\bf r},t) \right]^\ast \; (2\pi)^{-2}
\int \d {\bf q} \int \d \omega \;
\exp\left[ {\rm i} \left( {\bf q} \dotprod {\bf r} - \omega t \right)
\right] \, \overline{\phi}^{1}_{\bf k}({\bf q},\omega) \, ({\rm i}{\bf q}),
\nonumber \eeqa
\beqa
\left[ \exp(st) \nablab \phi_{\bf k}^1 \right]^\ast \phi_{\bf k}^0
&=&  \phi_{\bf k}^0({\bf r},t) \; (2\pi)^{-2}
\int \d {\bf q} \int \d \omega \;
\exp\left[ - {\rm i} \left( {\bf q} \dotprod {\bf r} - \omega t \right)
\right] \, \overline{\phi}^{1\ast}_{\bf k}({\bf q},\omega) \, (-{\rm i}{\bf q}),
\nonumber \eeqa
\beqa
(\nablab \phi_{\bf k}^0)^\ast \exp(st) \phi_{\bf k}^1
&=&  \left[ \phi_{\bf k}^0({\bf r},t) \right]^\ast \; (2\pi)^{-2}
\int \d {\bf q} \int \d \omega \;
\exp\left[ {\rm i} \left( {\bf q} \dotprod {\bf r} - \omega t \right)
\right] \, \overline{\phi}^{1}_{\bf k}({\bf q},\omega) \, (-{\rm i}{\bf k}).
\nonumber \eeqa
Combining these expressions, we can evaluate the following integral
\beqa
{\bf I} &\equiv& {\rm i}
\int \d {\bf k} \, f({\bf k}) \, \exp(st)
\left[ \rule{0mm}{4mm}
(\phi_{\bf k}^1)^\ast (\nablab \phi_{\bf k}^0)
+ (\phi_{\bf k}^0)^\ast (\nablab \phi_{\bf k}^1)
- (\nablab \phi_{\bf k}^1)^\ast \phi_{\bf k}^0
- (\nablab \phi_{\bf k}^0)^\ast \phi_{\bf k}^1 \right]
\nonumber \\ [2mm]
&=& - L^{-3/2} \int \d {\bf k}  \, f({\bf k}) \, (2\pi)^{-2}
\nonumber \\ [2mm]
&& \mbox{} \times
\left[
\int \d {\bf q} \int \d \omega \;
\exp\left\{\rule{0mm}{4mm}
- {\rm i} \left[ ({\bf q}-{\bf k}) \dotprod {\bf r} -
(\omega-\omega_{\bf k}) t \right] \right\}
\, \overline{\phi}^{1\ast}_{\bf k}({\bf q},\omega) \, ({\bf k}+{\bf q}) \right.
\nonumber \\ [2mm]
&& \mbox{} \rule{5mm}{0mm} \left. +
\int \d {\bf q} \int \d \omega \;
\exp\left\{\rule{0mm}{4mm} {\rm i} \left[ ({\bf q}-{\bf k}) \dotprod {\bf r} -
(\omega-\omega_{\bf k}) t \right] \right\}
\, \overline{\phi}^{1}_{\bf k}({\bf q},\omega) \, ({\bf q}+{\bf k})
\right].
\label{7.25}\eeqa
By applying evident changes of variables we get
\beqa
{\bf I}&=& - L^{-3/2} \int \d {\bf k}  \, f({\bf k}) \, (2\pi)^{-2}
\nonumber \\ [2mm]
&& \mbox{} \times
\left[
\int \d {\bf q}' \int \d \omega' \;
\exp\left[
{\rm i} \left({\bf q}' \dotprod {\bf r} -
\omega' t \right) \right]
\, \overline{\phi}^{1\ast}_{\bf k}({\bf k}-{\bf q}',\omega_{\bf k}-\omega') \,
(2{\bf k}-{\bf q}') \right.
\nonumber \\ [2mm]
&& \mbox{} \rule{5mm}{0mm} \left. +
\int \d {\bf q}' \int \d \omega' \;
\exp\left[\rule{0mm}{4mm}
{\rm i} \left({\bf q}' \dotprod {\bf r} -
\omega' t \right) \right]
\, \overline{\phi}^{1}_{\bf k}({\bf k}+{\bf q}',\omega_{\bf k}
+\omega') \, (2{\bf k}+{\bf q}') \right]
\nonumber \\ [2mm]
&=& \, (2\pi)^{-2} \int \d {\bf q}' \int \d \omega' \;
\exp\left[
{\rm i} \left({\bf q}' \dotprod {\bf r} -
\omega' t \right) \right] \left\{ - L^{-3/2} \int \d {\bf k}  \, f({\bf k})
\, \exp(st)
\right.
\nonumber \\ [2mm]
&& \mbox{} \times \left.
\left[\rule{0mm}{4mm}
\overline{\phi}^{1\ast}_{\bf k}({\bf k}-{\bf q}',\omega_{\bf k}-\omega') \,
(2{\bf k}-{\bf q}') +
\overline{\phi}^{1}_{\bf k}({\bf k}+{\bf q}',\omega_{\bf k} +\omega') \, (2{\bf
k}+{\bf q}') \right]
\rule{0mm}{5mm}\right\}.
\nonumber \eeqa
From Eq.\ \req{7.12}, recalling that the scalar and vector potentials are
real, we have
\beqa
\overline{\phi}^{1\ast}_{\bf k}({\bf q},\omega) &=&
 \frac{2 \me e}{\hbar^2 L^{3/2}} \, \frac{
\varphi^\ast({\bf q}-{\bf k},\omega-\omega_{\bf k})
- (\hbar/\me c)  {\bf A}^\ast({\bf q}-{\bf k},\omega-\omega_{\bf k})
\dotprod {\bf k}}{q^2 - 2 \me (\omega-{\rm i} s) /\hbar}
\nonumber \\ [2mm]
&=& \frac{2 \me e}{\hbar^2 L^{3/2}} \,
\frac{ \varphi(-{\bf q}+{\bf k},-\omega+\omega_{\bf k})
- (\hbar/\me c)  {\bf A}(-{\bf q}+{\bf k},-\omega+\omega_{\bf k})
\dotprod {\bf k}}{q^2 - 2 \me (\omega-{\rm i} s) /\hbar}\, .
\rule{15mm}{0mm}
\label{7.26}\eeqa
Now we can write
\beqa
{\bf I} &=& (2\pi)^{-2} \int \d {\bf q}' \int \d \omega' \;
\exp\left[
{\rm i} \left({\bf q}' \dotprod {\bf r} -
\omega' t \right) \right] \left\{ - L^{-3/2} \int \d {\bf k}  \, f({\bf k})
\right.
\nonumber \\ [2mm]
&& \mbox{} \times \left.
\left[\rule{0mm}{4mm}
\overline{\phi}^{1\ast}_{\bf k}({\bf k}-{\bf q}',\omega_{\bf k}-\omega') \,
(2{\bf k}-{\bf q}') +
\overline{\phi}^{1}_{\bf k}({\bf k}+{\bf q}',\omega_{\bf k} +\omega') \, (2{\bf
k}+{\bf q}') \right]
\rule{0mm}{5mm}\right\}
\nonumber \\ [2mm]
&=& (2\pi)^{-2} \int \d {\bf q}' \int \d \omega' \;
\exp\left[
{\rm i} \left({\bf q}' \dotprod {\bf r} -
\omega' t \right) \right] \left\{ - \frac{2 \me e}{\hbar^2 L^{3}} \,
\int \d {\bf k}  \, f({\bf k})
\right.
\nonumber \\ [2mm]
&& \mbox{} \times \left.
\left[\rule{0mm}{4mm}
\frac{ \varphi({\bf q}',\omega')
- (\hbar/\me c)  {\bf A}({\bf q}',\omega')
\dotprod {\bf k}}{q'^2 -2 {\bf q}'\dotprod {\bf k}
+ 2 \me (\omega'+{\rm i} s) /\hbar}
\, (2{\bf k}-{\bf q}')
\right. \right.
\nonumber \\ [2mm]
&& \mbox{} \rule{5mm}{0mm} \left. \left. +
\frac{ \varphi({\bf q}',\omega')
- (\hbar/\me c)  {\bf A}({\bf q}',\omega')
\dotprod {\bf k}}{q'^2 +2 {\bf q}'\dotprod {\bf k}
- 2 \me (\omega'+{\rm i} s) /\hbar}
\, (2{\bf k}+{\bf q}') \right]
\rule{0mm}{5mm}\right\}.
\label{7.27}\eeqa
With these results, the induced charge current can be expressed as
\beqa
{\bf j}_{\rm ind} ({\bf r},t) &=& \left( \frac{L}{2\pi} \right)^3
\left( \frac{e \hbar}{\me} \, {\bf I} -
\frac{2 e^2}{\me c} L^{-3} \int \d {\bf k} \, f({\bf k})
\, {\bf A}({\bf r},t) \right).
\label{7.28}\eeqa \index{Fourier transform}
We see that the Fourier transform of the induced charge current is
\beqa
{\bf j}_{\rm ind} ({\bf q},\omega) &=&
- \frac{e^2}{4 \pi^3 \me c} \,
{\bf A}({\bf q},\omega) \int \d {\bf k}  \, f({\bf k})
\nonumber \\ [2mm]
&& - \frac{e^2}{4 \pi^3 \hbar} \,
\int \d {\bf k}  \, f({\bf k})
\left[\rule{0mm}{4mm}
\frac{ \varphi({\bf q},\omega)
- (\hbar/\me c)  {\bf A}({\bf q},\omega)
\dotprod {\bf k}}{q^2 +2 {\bf q}\dotprod {\bf k}
- 2 \me (\omega+{\rm i} s) /\hbar}
\, (2{\bf k}+{\bf q})
\right.
\nonumber \\ [2mm]
&& \mbox{} \rule{5mm}{0mm} \left. +
\frac{ \varphi({\bf q},\omega)
- (\hbar/\me c)  {\bf A}({\bf q},\omega)
\dotprod {\bf k}}{q^2 -2 {\bf q}\dotprod {\bf k}
+ 2 \me (\omega+{\rm i} s) /\hbar}
\, (2{\bf k}-{\bf q}) \right].
\label{7.29}\eeqa
The integrals in this expression can be made more explicit by using
spherical coordinates with the polar $z$ axis in the direction of ${\bf
q}$ and the $x$ axis in the direction of ${\bf A}$. This implies that
${\bf q}=(0,0,q)$, ${\bf k} = k(\sin\theta \cos\phi, \sin\theta
\sin\phi, \cos\theta)$, ${\bf A}=(A,0,0)$, and
\beq
{\bf A} \dotprod {\bf k} =  A k \sin\theta \cos\phi, \qquad
{\bf q} \dotprod {\bf k} = qk \cos\theta.
\label{7.30}\eeq
We thus have
\beqa
\lefteqn{
{\bf j}_{\rm ind} ({\bf q},\omega) =
- \frac{e^2}{4 \pi^3 \me c} \,
{\bf A}({\bf q},\omega)
\int_0^{k_{\rm F}} \d k \, k^2
\int_{-1}^1 \d (\cos\theta) \int_0^{2\pi} \d \phi }
\nonumber \\ [2mm]
&& - \frac{e^2}{4 \pi^3 \hbar} \int_0^{k_{\rm F}} \d k \, k^2
\int_{-1}^1 \d (\cos\theta) \int_0^{2\pi} \d \phi
\left[\rule{0mm}{4mm}
\frac{ \varphi - (\hbar/\me c)   A k \sin\theta \cos\phi
}{q^2 +2 qk \cos\theta - 2 \me (\omega+{\rm i} s) /\hbar}
\, (2{\bf k}+{\bf q})
\right.
\nonumber \\ [2mm]
&& \mbox{} \rule{5mm}{0mm} \left. +
\frac{ \varphi - (\hbar/\me c)  A k \sin\theta \cos\phi
}{q^2 -2 qk \cos\theta + 2 \me (\omega+{\rm i} s) /\hbar}
\, (2{\bf k}-{\bf q})
\right].
\label{7.31}\eeqa
We notice that
\begin{subequations}
\label{7.32}
\beqa
\int_0^{2\pi} \d \phi \, {\bf k} &=&
\int_0^{2\pi} \d \phi \, k (\sin\theta \cos\phi, \sin\theta
\sin\phi, \cos\theta) = 2\pi k (0,0,\cos\theta)
\nonumber \\ [1mm]
&=& 2 \pi k \, \cos\theta \, \hat{\bf q},
\label{7.32a}\eeqa
\beq
\int_0^{2\pi} \d \phi \, {\bf A} \dotprod {\bf k} =
\int_0^{2\pi} \d \phi \, Ak \sin\theta \cos\phi = 0,
\label{7.32b}\eeq
and
\beqa
\int_0^{2\pi} \d \phi \, ({\bf A} \dotprod {\bf k}) {\bf k} &=&
\int_0^{2\pi} \d \phi \, Ak^2 \sin\theta \cos\phi
\, (\sin\theta \cos\phi, \sin\theta
\sin\phi, \cos\theta)
\nonumber \\ [2mm]
&=& \pi A k^2 (\sin^2\theta,0,0)
= \pi \, \sin^2\theta \, k^2 {\bf A}.
\label{7.32c}\eeqa
\end{subequations}
Hence, writing $\xi=\cos\theta$, we have
\beqa
\lefteqn{
{\bf j}_{\rm ind} ({\bf q},\omega) =
- \frac{e^2}{2 \pi^2 \me c} \,
{\bf A}({\bf q},\omega)
\int_0^{k_{\rm F}} \d k \, k^2
\int_{-1}^1 \d \xi }
\nonumber \\ [2mm]
&& - \frac{e^2}{4 \pi^3 \hbar} \int_0^{k_{\rm F}} \d k \, k^2
\int_{-1}^1 \d \xi
\left[\rule{0mm}{4mm}
\frac{ \varphi (4\pi k \xi \hat{\bf q} + 2 \pi {\bf q})
- (\hbar/\me c) 2\pi (1 - \xi^2) k^2 {\bf A}}
{q^2 +2 qk \xi - 2 \me (\omega+{\rm i} s) /\hbar}
\right.
\nonumber \\ [2mm]
&& \mbox{} \rule{5mm}{0mm} \left. +
\frac{ \varphi (4\pi k \xi \hat{\bf q} - 2 \pi {\bf q})
- (\hbar/\me c) 2\pi (1 - \xi^2) k^2 {\bf A}}
{q^2 -2 qk \xi + 2 \me (\omega+{\rm i} s) /\hbar}
\right].
\label{7.33}\eeqa
Consequently, the transverse and longitudinal components of the induced charge
current are, respectively,
\beqa
{\bf j}_{\rm ind}^{\rm (T)} ({\bf q},\omega) &=&
\frac{e^2}{2 \pi^2 \me c } \,
\int_0^{k_{\rm F}} \d k \, k^2 \int_{-1}^1 \d \xi
\left(\rule{0mm}{4mm}
\frac{k^2(1 - \xi^2)}{q^2 +2 qk \xi - 2 \me (\omega+{\rm i} s) /\hbar} \right.
\nonumber \\ [2mm]
&& \mbox{} \rule{5mm}{0mm} \left.
+ \frac{ k^2 (1 - \xi^2)}{q^2 -2qk\xi
+ 2 \me (\omega+{\rm i} s) /\hbar} - 1 \right)
{\bf A}({\bf q},\omega)
\label{7.34}\eeqa
and
\beqa
{\bf j}_{\rm ind}^{\rm (L)} ({\bf q},\omega)
&=&  - \, \frac{\bf q}{q^2} \, \frac{e^2}{2 \pi^2 \hbar} \,
\varphi({\bf q},\omega)
\int_0^{k_{\rm F}} \d k \, k^2 \int_{-1}^1 \d \xi
\left(\rule{0mm}{4mm}
\frac{2qk\xi + q^2}{q^2 +2 qk\xi - 2 \me (\omega+{\rm i} s) /\hbar}
\right.
\nonumber \\ [2mm]
&& \mbox{} \rule{5mm}{0mm} \left.
+ \frac{2 qk\xi - q^2}{q^2
-2 qk\xi + 2 \me (\omega+{\rm i} s) /\hbar}
\right).
\label{7.35}\eeqa
It is worth noticing that the longitudinal component of the induced charge
current and the induced charge density \req{7.20} are related by
\beq
(\omega + {\rm i} s) \rho_{\rm ind} ({\bf q},\omega) = {\bf q} \dotprod
{\bf j}_{\rm ind}^{\rm (L)} ({\bf q},\omega),
\label{7.36}\eeq
which in the limit $s \rightarrow 0$ reduces to the continuity equation
\req{1.74a}.


\subsubsection{Dielectric functions \label{sec7.1.1.2}}

The longitudinal DF of the electron gas (eg) can be derived from the
relation \req{1.110a},
\beq
\rho_{\rm ind} ({\bf q},\omega) = - \, \frac{q^2}{4\pi} \left[ \epsilon^{\rm
(L)}_{\rm eg} (q,\omega) -1 \right] \varphi({\bf q},\omega),
\nonumber \eeq
which, combined with Eq.\ \req{7.20}, implies that
\beqa
\epsilon^{\rm (L)}_{\rm eg} (q,\omega) -1  &=& - \, \frac{4\pi}{q^2} \, \frac{
\rho_{\rm ind} ({\bf q},\omega)}{\varphi ({\bf q},\omega)}.
\label{7.37}\eeqa
Hence,
\beqa
\epsilon^{\rm (L)}_{\rm eg} (q,\omega) -1
&=& \frac{4 \me e^2}{\pi \hbar^2 q^{2}}
\int_0^{k_{\rm F}} \d k \, k^2 \int_{-1}^1 \d \xi
\left(
\frac{1}{q^2 + 2 qk\xi - 2 \me (\omega+{\rm i} s)
/\hbar} \right.
\nonumber \\ [2mm]
&& \mbox{} \left.
+ \frac{1}{q^2 - 2 qk\xi + 2 \me (\omega+{\rm i} s) /\hbar} \right)
\nonumber \\ [2mm]
&=& \frac{4 \me e^2}{\pi \hbar^2 4 z^2 k_{\rm F}^2}
\int_0^{1} \d \tau \, \tau^2 k_{\rm F}^3 \int_{-1}^1 \d \xi
\left(
\frac{1}{4z^2k_{\rm F}^2 + 4 z \tau k_{\rm F}^2 \xi -
2 u' 2 z k_{\rm F}^2 } \right.
\nonumber \\ [2mm]
&& \mbox{} \left.
+ \frac{1}{4z^2k_{\rm F}^2 - 4 z \tau k_{\rm F}^2 \xi +
2 u' 2 z k_{\rm F}^2 }  \right)
\nonumber \\ [2mm]
&=& \frac{\chi^2}{4 z^3}
\int_0^1 \d \tau \, \tau^2 \int_{-1}^1 \d \xi
\left(
\frac{1}{\tau \xi + z - u'}
- \frac{1}{ \tau \xi - z - u'} \right),
\label{7.38}\eeqa
where we have introduced the dimensionless variables
\beq
z = \frac{q}{2 k_{\rm F}}, \qquad
u' = \frac{\me (\omega + {\rm i} s)}{\hbar q k_{\rm F}} ,
\qquad
\tau = \frac{k}{k_{\rm F}},
\label{7.39}\eeq
and the constant
\beq
\chi^2= \frac{\me e^2}{\pi \hbar^2 k_{\rm F}}\, .
\label{7.40}\eeq
The integrals in Eq.\ \req{7.38} can now be evaluated analytically. The
integral over $\xi$ is elementary, and the following simple
transformations give
\beqa
\epsilon^{\rm (L)}_{\rm eg} (q,\omega) - 1
&=& \frac{\chi^2}{4 z^3}
\int_0^1 \d \tau \, \tau
\left[ \ln( \tau \xi + z - u')
- \ln ( \tau \xi - z - u')\rule{0mm}{4mm}
\right]_{-1}^{1}
\nonumber \\ [2mm]
&=& \frac{\chi^2}{4 z^3}
\int_0^1 \d \tau \, \tau
\left[ \ln \left( \frac{\tau + z - u'}
{\tau - z + u'} \right)
+ \ln \left( \frac{ \tau + z + u'}{\tau - z - u'}
\right) \right]
\nonumber \\ [2mm]
&=& \frac{\chi^2}{4 z^3}
\int_0^1 \d \tau \, \tau
\left[ \ln \left( \tau + z - u' \right)
+ \ln \left( \tau + z + u' \right) \rule{0mm}{4mm}\right]
\nonumber \\ [2mm]
&& \mbox{} - \, \frac{\chi^2}{4 z^3}
\int_0^1 \d \tau \, \tau
\left[ \ln \left( \tau - z + u' \right)
+ \ln \left( \tau - z - u' \right) \rule{0mm}{4mm}\right]
\nonumber \\ [2mm]
&=& \frac{\chi^2}{4 z^3}
\int_0^1 \d \tau \, \tau
\left[ \ln \left( \tau + z - u' \right)
+ \ln \left( \tau + z + u' \right) \rule{0mm}{4mm}\right]
\nonumber \\ [2mm]
&& \mbox{} + \frac{\chi^2}{4 z^3}
\int_{-1}^0 \d \tau' \, \tau'
\left[ \ln \left( \tau' + z - u' \right)
+ \ln \left( \tau' + z + u' \right) \rule{0mm}{4mm}\right]
\nonumber \\ [2mm]
&=& \frac{\chi^2}{4 z^3}
\int_{-1}^1 \d \tau \, \tau
\left[ \ln \left( \tau + z - u' \right)
+ \ln \left( \tau + z + u' \right) \rule{0mm}{4mm}\right].
\nonumber\eeqa
Hereafter the function ${\rm ln}(u)$ represents the principal
value of the logarithm of the complex number $u$,
\beq
\ln (u) = \ln|u| + {\rm i} \phi, \qquad -\pi < \phi \le \pi.
\label{7.41}\eeq
The remaining integrals are evaluated by using the formula
\beq
\int_{-1}^1  x \ln (x+a) \, \d x
= \frac{1-a^2}{2} \, \ln \left( \frac{a+1}{a-1} \right) +a,
\label{7.42}\eeq
which gives
\beqa
\epsilon^{\rm (L)}_{\rm eg} (q,\omega) - 1
&=& \frac{\chi^2}{4 z^3}
\left[
\frac{1-(z - u')^2}{2} \, \ln \left( \frac{z - u'+1}{z - u' -1}
\right) + z - u' \right.
\nonumber \\ [2mm]
&& \mbox{} \left.
+ \frac{1-(z + u')^2}{2} \, \ln \left( \frac{z + u'+1}{z + u' -1}
\right) + z + u'
\right]
\nonumber \\ [2mm]
&=& \frac{\chi^2}{4 z^3} \, 4z
\left[
\frac{1-(z - u')^2}{8z} \, \ln \left( \frac{z - u'+1}{z - u' -1}
\right) \right.
\nonumber \\ [2mm]
&& \mbox{} \left.
+ \frac{1-(z + u')^2}{8z} \, \ln \left( \frac{z + u'+1}{z + u'-1}
\right) + \frac{1}{2} \right].
\nonumber \eeqa
Finally, we can write
\beqa
\epsilon^{\rm (L)}_{\rm eg} (q,\omega) &=& 1
+ \frac{\chi^2}{ z^2} \left[ \frac{1}{2} +
\frac{1-(z - u')^2}{8z} \, \ln \left( \frac{z - u'+1}{z - u' -1}
\right) \right.
\nonumber \\ [2mm]
&& \mbox{} \left.
+ \frac{1-(z + u')^2}{8z} \, \ln \left( \frac{z + u'+1}{z + u'-1}
\right) \right].
\label{7.43}\eeqa
\allowdisplaybreaks{
The transverse DF can be calculated from the general relation
\req{1.110b},
\beq
{\bf j}_{\rm ind}^{\rm (T)} (q,\omega) =
\frac{\omega^2}{4\pi c}
\left[ \epsilon^{\rm (T)}_{\rm eg} (q,\omega) -1 \right]
{\bf A}({\bf q},\omega),
\nonumber \eeq
which combined with the expression \req{7.34} of the induced transverse
current gives
\beqa
\epsilon^{\rm (T)}_{\rm eg} (q,\omega) -1 &=&
\frac{4\pi c}{\omega^2} \, \frac{e^2}{2 \pi^2 \me c } \,
\int_0^{k_{\rm F}} \d k \, k^2 \int_{-1}^1 \d \xi
\left(\rule{0mm}{4mm}
\frac{k^2(1 - \xi^2)}{q^2 +2 qk \xi - 2 \me (\omega+{\rm i} s) /\hbar} \right.
\nonumber \\ [2mm]
&& \mbox{} \rule{5mm}{0mm} \left.
+ \frac{ k^2(1 - \xi^2)}{q^2 -2qk\xi
+ 2 \me (\omega+{\rm i} s) /\hbar} - 1 \right).
\label{7.44}\eeqa
Introducing the variables \req{7.39}, we have
\beqa
\epsilon^{\rm (T)}_{\rm eg} (q,\omega) -1
&=& \frac{2 e^2}{\pi \me \omega^2} \,  k_{\rm F}^3
\int_0^{1} \d \tau \, \tau^2 \left[ -2 + \frac{\tau^2}{4z}
\int_{-1}^1 \d \xi \left(\rule{0mm}{4mm}
\frac{1 - \xi^2}{\tau \xi + z - u'}
- \frac{1 - \xi^2}
{\tau \xi - z - u'} \right) \right].
\nonumber \eeqa
The integrals over $\xi$ can be evaluated by using the formula
\beq
\int_{-1}^1 \frac{1-x^2}{a x + b} \, \d x
= \frac{2 b}{a^2} + \frac{a^2 - b^2}{a^3}
\, \ln\left(\frac{b + a}{b-a}\right),
\label{7.45}\eeq
which gives
\beqa
\epsilon^{\rm (T)}_{\rm eg} (q,\omega) -1
&=& \frac{2 e^2}{\pi \me \omega^2} \,  k_{\rm F}^3
\int_0^{1} \d \tau \, \tau^2 \left\{ -1
+ \frac{\tau^2 - (z-u')^2}{4\tau z} \, \ln\left(
\frac{z-u'+\tau}{z-u'-\tau} \right) \right.
\nonumber \\ [2mm]
&& \mbox{} \left.
- \frac{\tau^2 - (z+u')^2}{4\tau z} \, \ln\left(
\frac{z+u'-\tau}{z+u'+\tau} \right)
\right\}.
\nonumber \eeqa
Finally, using the formula
\beq
\int_{-1}^1 x\,(x^2 - a^2)\,\ln ( a - x)  \, \d x = -\frac{1}{4} \,
\left(1-a^2\right)^2\, \ln \left(\frac{a+1}{a-1}\right) - \frac{a}{6}
+ \frac{a^3}{2},
\label{7.46}\eeq
and introducing the plasma resonance angular frequency
\index{plasma resonance frequency}
\beq
\omega_{\rm p} \equiv \sqrt{\frac{4 e^2}{3\pi \me} \, k_{\rm F}^3}
= \sqrt{\frac{4\pi e^2}{\me} \,  \rho^0}\, ,
\label{7.47}\eeq
we obtain
\beqa
\epsilon^{\rm (T)}_{\rm eg} (q,\omega)
&=& 1 - \, \frac{\omega_{\rm p}^2}{\omega^2}
\left\{ \frac{3}{8} \left(z^2 + 3 u'^2 +1 \right)
- \frac{3}{32 z} \left[ 1 - (z-u')^2 \right]^2 \, \ln\left(
\frac{z-u'+1}{z-u'-1} \right) \right.
\nonumber \\ [2mm]
&& \mbox{} \left.
-  \frac{3}{32 z} \left[ 1 - (z+u')^2 \right]^2 \, \ln\left(
\frac{z+u'+1}{z+u'-1} \right)
\right\}.
\label{7.48}\eeqa


\subsection{Practical calculation of Lindhard's DF \label{sec7.1.2}}

\allowdisplaybreaks{
The basic parameters entering the Lindhard formulas are the plasma
resonance angular frequency, $\omega_{\rm p}$,
\beq
\omega_{\rm p}^2 \equiv 4\pi \rho^0 \frac{e^2}{\me} = \frac{4 e^2}{3\pi \me} \,
k_{\rm F}^3 ,
\label{7.49}\eeq
and the Lindhard parameter
\beq
\chi^2 \equiv \frac{\me e^2}{\pi \hbar^2 k_{\rm F}}
= \frac{3}{16} \,\frac{\hbar^2 \omega_{\rm p}^2}{E_{\rm F}^2}\, .
\label{7.50}\eeq
In practical calculations it is convenient to characterize the electron
gas by means of the plasmon energy,
\beq
W_{\rm p} = \hbar \omega_{\rm p} \, ,
\label{7.51}\eeq
from which other relevant parameters are obtained as
\beq
k_{\rm F} = \left(
\frac{3\pi}{4} \right) ^{1/3} \left( \frac{W_{\rm p}}{E_{\rm h}}
\right)^{2/3} a_0^{-1}
= 1.330\, 670\, 04 \left( \frac{W_{\rm p}}{E_{\rm h}} \right)^{2/3}
a_0^{-1}\, ,
\label{7.52}\eeq
\beq
E_{\rm F} = \frac{1}{2} \left(
\frac{3\pi}{4} \right) ^{2/3} \left( \frac{W_{\rm p}}{E_{\rm h}}
\right)^{4/3} E_{\rm h}
= 0.885\, 341\, 38 \left( \frac{W_{\rm p}}{E_{\rm h}} \right)^{4/3}
E_{\rm h}\, ,
\label{7.53}\eeq
and
\beq
\chi^2 = \frac{3}{16} \,\frac{W_{\rm p}^2}{E_{\rm F}^2}
= \left( \frac{4}{3\pi^4} \right)^{1/3}
\left( \frac{E_{\rm h}}{W_{\rm p}} \right)^{2/3}
= 0.239\, 210\, 23 \left( \frac{E_{\rm h}}{W_{\rm p}} \right)^{2/3},
\label{7.54}\eeq
where $a_0$ is the Bohr
radius and $E_{\rm h}$ is the Hartree
energy (see Appendix \ref{appC}).

It is also convenient to express the Lindhard DFs in terms of the
dimensionless variables \citep{Lindhard1954, Ritchie1959}
\beq
z \equiv \frac{q}{2k_{\rm F}}, \qquad
x \equiv \frac{\hbar \omega}{E_{\rm F}}, \qquad \mbox{and} \qquad
u \equiv \frac{\hbar (\omega+{\rm i}
s)}{E_{\rm F}} = x + {\rm i} y\, .
\label{7.55}\eeq
The longitudinal DF of the electron gas is given by
\beq
\epsilon_{\rm eg}^{\rm (L)} (\omega_{\rm p}, s; q,\omega)
= 1 + \frac{\chi^2}{z^2} \, f(z,u),
\label{7.56}\eeq
with
\beqa
f(z,u)=\frac{1}{2} &+&
\frac{1}{8z} \left[ 1 - \left( z - u/4z \right)^2 \right]
\ln \left( \frac{z-u/4z+1}{z-u/4z-1} \right) \nonumber \\ [1mm]
&+& \frac{1}{8z} \left[ 1 - \left( z + u/4z \right)^2 \right]
\ln \left( \frac{z+u/4z+1}{z+u/4z-1} \right),
\label{7.57}\eeqa
and the transverse DF of the gas is
\beq
\epsilon_{\rm eg}^{\rm (T)} (\omega_{\rm p}, s; q, \omega) =
1 - \frac{\omega_{\rm p}^2}{\omega^2} \,  g(z,u)
= 1 - \frac{16 \chi^2}{3x^2} \,  g(z,u) \, ,
\label{7.58}\eeq
with
\beqa
g(z,u) &=&
\frac{3}{8} \left[ z^2 + 3 (u/4z)^2+1 \right]
-\frac{3}{32 z} \left[
1 - \left( z - u/4z \right)^2 \right]^2
\ln \left( \frac{z-u/4z+1}{z-u/4z-1} \right)
\nonumber \\ [1mm]
&& \mbox{} - \frac{3}{32 z}  \left[ 1 - \left( z + u/4z \right)^2 \right]^2
\ln \left( \frac{z+u/4z+1}{z+u/4z-1} \right).
\label{7.59}\eeqa
Because the parameter $s$ was introduced so as to switch the
perturbation on adiabatically; the limit $s \rightarrow 0$ should be
taken at the end of the calculation. Notice that values
of the plasma resonance frequency $\omega_{\rm p}$ and the damping
parameter $s$ are indicated explicitly as arguments of the DFs. In the
following, various DF models will be distinguished by the set of
parameters listed as arguments.

When $|z+u/4z| \gg 1$ and $|z-u/4z| \gg 1$ we can expand the complex
functions $f(z,u)$ and $g(z,u)$ in the forms\footnote{We use the
expansions
$$
\ln \left( \frac{u+1}{u-1} \right) = 2 u^{-1} + \frac{2}{3}u^{-3}
+ \frac{2}{5}u^{-5}+ \frac{2}{7}u^{-7} + \frac{2}{9}u^{-9} + \cdots
\qquad  |u| \ge 1,\; \; u \ne \pm 1,
$$
$$
(1-u^2) \ln \left( \frac{u+1}{u-1} \right) = - 2 u + \frac{4}{3}u^{-1}
+ \frac{4}{3\cdot 5}u^{-3}+ \frac{4}{5\cdot 7}u^{-5}
+ \frac{4}{7\cdot 9}u^{-7} + \cdots
\qquad  |u| \ge 1,\; \; u \ne \pm 1,
$$
and
$$
(1-u^2)^2 \ln \left( \frac{u+1}{u-1} \right) = 2 u^3 - \frac{10}{3}u
+ \frac{16}{3\cdot 5}u^{-1} + \frac{16}{3\cdot 5\cdot 7}u^{-3}
+ \frac{16}{5\cdot 7\cdot 9}u^{-5} + \cdots
\qquad  |u| \ge 1,\; \; u \ne \pm 1.
$$
}
\beqa
f(z,u) &=& \frac{1}{2} + \frac{1}{8z}
\left[ - 2(z-u/4z) + \frac{4}{3(z-u/4z)} + \frac{4}{3\cdot 5(z-u/4z)^3}
\right.
\nonumber \\ [1mm]
&& \mbox{} \left.
+ \frac{4}{5\cdot 7(z-u/4z)^5} - \cdots \right]
\nonumber \\ [1mm]
&& \! \! \! \! \! \! \! \! \! \! \! \! \! \! \! \! \! \! \!
\mbox{} + \frac{1}{8z}
\left[ - 2(z+u/4z) + \frac{4}{3(z+u/4z)} + \frac{4}{3\cdot 5(z+u/4z)^3}
+ \frac{4}{5\cdot 7(z+u/4z)^5} - \cdots \right]
\nonumber \eeqa
\beq
= \frac{16z^2}{16 z^4-u^2} \left[ \frac{1}{3}
+ \frac{z^2 + 3 (u/4z)^2}{3\cdot 5[z^2-(u/4z)^2]^2}
+ \frac{z^4 + 10 z^2 (u/4z)^2+5 (u/4z)^4}{5\cdot 7[z^2-(u/4z)^2]^4} + \cdots
\right]
\label{7.60}\eeq
and
\beqa
g(z,u) &=& \frac{3}{8} \left[ z^2 + 3 (u/4z)^2+1 \right]
\nonumber \\ [1mm]
&& \! \! \! \! \! \! \! \! \! \! \! \! \! \! \! \! \! \! \! \!  \! \! \!
\mbox{} - \frac{3}{32z}
\left[ 2(z-u/4z)^3 - \frac{10}{3}\, (z-u/4z)
+ \frac{16}{3\cdot 5(z-u/4z)} + \frac{16}{3\cdot 5 \cdot 7(z-u/4z)^3}
+ \cdots \right]
\nonumber \\ [1mm]
&& \! \! \! \! \! \! \! \! \! \! \! \! \! \! \! \! \! \! \! \!  \! \! \!
\mbox{} - \frac{3}{32z}
\left[ 2(z+u/4z)^3 - \frac{10}{3} \, (z+u/4z)
+ \frac{16}{3\cdot 5(z+u/4z)}+ \frac{16}{3 \cdot 5 \cdot 7(z+u/4z)^3}
+ \cdots \right]
 \nonumber \\ [1mm]
&& \! \! \! \! \! \! \! \! \! \! \! \! \! \! \! \! \! \! \! \! \! \! \!
\! \!
= 1 - \frac{1}{5[z^2-(u/4z)^2]}
- \frac{z^2 + 3 (u/4z)^2}{5\cdot 7[z^2-(u/4z)^2]^3}
-  \frac{z^4 + 10 z^2 (u/4z)^2+5 (u/4z)^4}
{5 \cdot 7 \cdot 3[z^2-(u/4z)^2]^5} + \cdots
\nonumber \\
\label{7.61}\eeqa

For small $z$ and finite $u$ (such that $z \ll 1$ and $|u/4z| \gg 1$) we have
\beq
f(z,u) \simeq - \frac{1}{3} \left( \frac{4z}{u} \right)^2
- \frac{1}{5} \left( \frac{4z}{u} \right)^4
- \frac{1}{7} \left( \frac{4z}{u} \right)^6 - \cdots\,
\label{7.62}\eeq
and
\beq
g(z,u) \simeq 1 + \frac{1}{5} \left( \frac{4z}{u} \right)^2
+ \frac{3}{35} \left( \frac{4z}{u} \right)^4
+ \frac{5}{105} \left( \frac{4z}{u} \right)^6
+ \cdots\, .
\label{7.63}\eeq
Hence, for small $q$,
\begin{subequations}
\label{7.64}
\beq
\epsilon_{\rm eg}^{\rm (L)} (\omega_{\rm p}, s; q,\omega)
= 1 + \frac{\chi^2}{z^2} \,
f(z,u) \simeq 1 - \chi^2
\left[ \frac{16}{3u^2} + \frac{256 z^2}{5 u^4} +
\frac{4096 z^4}{7u^6} + \cdots \right]
\label{7.64a}\eeq
and
\beq
\epsilon_{\rm eg}^{\rm (T)} (\omega_{\rm p}, s; q,\omega)
= 1 - \frac{16 \chi^2}{3x^2} g(z,u)
\simeq 1 - \frac{\chi^2}{x^2}
\left[ \frac{16}{3} + \frac{256 z^2}{15 u^2} +
\frac{4096 z^4}{35u^4} + \cdots \right].
\label{7.64b}\eeq
\end{subequations}
That is,
\beq
\epsilon_{\rm eg}^{\rm (L)} (\omega_{\rm p}, s; 0,\omega) =
1 - \frac{\omega_{\rm p}^2}{(\omega+{\rm i} s)^2}
\qquad \mbox{and} \qquad
\epsilon_{\rm eg}^{\rm (T)} (\omega_{\rm p}, s; 0,\omega) =
1 - \frac{\omega_{\rm p}^2}{\omega^2}\, .
\label{7.65}\eeq

On the other hand, for large $z$ such that $z \gg 1$ and $|u/4z| \ll z$,
\beq
f(z,u) \simeq \frac{1}{3} + \frac{1}{15} z^{-2}
+ \frac{1}{35} z^{-4} + \frac{1}{63} z^{-6} + \cdots\,
\label{7.66}\eeq
and
\beq
g(z,u) \simeq 1 - \frac{1}{5} z^{-2} - \frac{1}{35} z^{-4}
- \frac{1}{105} z^{-6} + \cdots\, .
\label{7.67}\eeq
Thus the DFs for large $z$ become
\beq
\epsilon_{\rm eg}^{\rm (L)} (\omega_{\rm p}, s; q,\omega)
\simeq 1 + \frac{\chi^2}{z^2}
\left[ \frac{1}{3} + \frac{1}{15} z^{-2}
+ \frac{1}{35} z^{-4} + \frac{1}{63} z^{-6} + \cdots \right]
\label{7.68}\eeq
and
\beq
\epsilon_{\rm eg}^{\rm (T)} (\omega_{\rm p}, s; q,\omega)
\simeq 1 - \frac{16 \chi^2}{3x^2}
\left[  1 - \frac{1}{5} z^{-2} - \frac{1}{35} z^{-4}
- \frac{1}{105} z^{-6} + \cdots \right]\, .
\label{7.69}\eeq

The derivation of the Lindhard DF is purely
non-relativistic and, consequently, it applies only to electron gases
such that $E_{\rm F} \ll \me c^2$ and to excitations with $\hbar \omega
\ll \me c^2$. \citet{Lindhard1954} also calculated the DFs for a gas of
electrons that are originally at rest using electron wave functions that
are solutions of the Dirac equation for a free electron. He found that
the longitudinal and transverse DFs of a gas of electrons at rest are
equal and given by
\beqa
\epsilon_{\rm eg}^{\rm (L)}(\omega_{\rm p}, s; q,\omega)
&=& \epsilon_{\rm eg}^{\rm (T)}(\omega_{\rm p}, s; q,\omega)
\nonumber \\ [2mm]
&=& 1 + \frac{\omega_{\rm p}^2}{
\displaystyle{
\frac{\hbar^2 q^4}{4 \me^2} - \left( 1 + \frac{\hbar^2 q^2}{2\me^2 c^2}
\right) (\omega + {\rm i} s)^2
+ \frac{\hbar^2 (\omega+{\rm i} s)^4}{4 \me^2 c^4}}}\, .
\label{7.70}\eeqa
Notice that this simple calculation accounts only for relativistic
kinematical effects; a consistent relativistic calculation should
include the possibility of electron-positron pair production and
annihilation \citep{Jancovici1962}. Nonetheless, Lindhard's result for
the relativistic gas of electrons at rest should also be approximately
valid for
excitations of a real nonrelativistic electron gas (that is, such that
$E_{\rm F} \ll \me c^2$) in the limit of large frequencies and wave
numbers. In the non-relativistic limit ($c\rightarrow \infty$),
Lindhard's result, Eq.\ \req{7.70}, simplifies to
\beq
\epsilon_{\rm eg}^{\rm (L)}(\omega_{\rm p}, s; q,\omega)
= \epsilon_{\rm eg}^{\rm (T)}(\omega_{\rm p}, s; q,\omega)
= 1 + \frac{\omega_{\rm p}^2}{
\displaystyle{
\frac{\hbar^2 q^4}{4 \me^2} - (\omega + {\rm i} s)^2}} \, ,
\label{7.71}\eeq
which is the form adopted in Section \ref{sec1.3} to extend the
Lorentz-Drude model of the ODF to finite wave numbers [see Eq.\
\req{1.169}]. Indeed, the optical limit ($q \rightarrow 0$) of the
Lindhard DF equals the ODF of a classical oscillator with null resonance
frequency and damping constant $s/2$, Eq.\ \req{1.168}.
}


\subsection{Undamped Lindhard dielectric functions \label{sec7.1.3}}

The Lindhard DFs for the electron gas are obtained as
the $s \rightarrow 0$ limit of the expressions \req{7.56} and
\req{7.58}. The analytical expressions of the DFs in this limit follow
readily by considering the equalities
\beqa
&& \!  \! \! \! \! \! \! \! \! \! \! \! \! \!
\ln \left( \frac{z\pm (x+{\rm i     } y)/4z+1}{z\pm (x+{\rm i} y)/4z-1}
\right) =  \ln \left( \frac{(z\pm x/4z)^2 -1 +(y/4z)^2 \mp {\rm i} y/2z}
{(z\pm x/4z-1)^2+(y/4z)^2} \right)
\nonumber \\ [2mm]
&=& \ln \left| \frac{z\pm (x+{\rm i} y)/4z+1}{z\pm (x+{\rm i} y)/4z-1}
\right| + {\rm i} \arg \left( (z\pm x/4z)^2 -1 +(y/4z)^2
\mp {\rm i} y/2z \right) \rule{10mm}{0mm}
\label{7.72}\eeqa
and using the property
\beq
\lim_{s \rightarrow 0} \ln ( - |u| \pm {\rm i} s) =
\ln|u|\pm{\rm i} \pi.
\label{7.73}\eeq
We find different situations in the following regions of the $(z,x)$
plane (see Fig.\ \ref{fig7.1}): \\ [2mm]
\rule{5mm}{0mm} Region a:\; $z+x/4z \le 1, \quad -1 < z-x/4z \le 1$.\\
\rule{5mm}{0mm} Region b:\; $z+x/4z > 1, \quad -1 < z-x/4z \le 1$.\\
\rule{5mm}{0mm} Region c$_1$: $z+x/4z > 1, \qquad \quad \; \; z-x/4z \le -1$.\\
\rule{5mm}{0mm} Region c$_2$: $z+x/4z > 1, \qquad \quad \; \; z-x/4z > 1$.\\

\begin{figure}[tbh!]
\begin{center}
\includegraphics*[width=9cm]{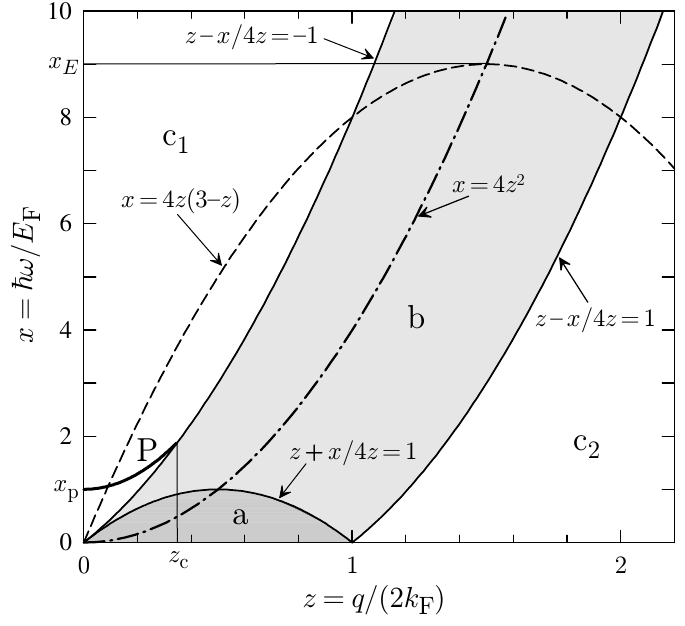}
\caption {\rm Regions of the $(z,x)$ plane where different expressions
of the functions $f_2(z,x)$ and $g_2(z,x)$ apply. Regions a and b, the
so-called Lindhard continuum, represent electron-hole excitations; the
line P in region  c$_1$ corresponds to plasmon excitation. Adapted from
\citet{Salvat2003}.
\label{fig7.1}}
\end{center}
\end{figure}

In the limit of small $s$, the longitudinal Lindhard DF takes the
form
\beq
\epsilon_{\rm eg}^{\rm (L)} (\omega_{\rm p}; q,\omega)
\equiv \epsilon_{\rm eg}^{\rm (L)} (\omega_{\rm p}, 0; q,\omega)
= 1 + \frac{\chi^2}{z^2} \left[
f_1(z,x) + {\rm i} f_2(z,x) \right],
\label{7.74}\eeq
where
\beqa
f_1(z,x)=\frac{1}{2} &+&
\frac{1}{8z}       \left[ 1 - \left( z - x/4z \right)^2 \right]
\ln\left| \frac{z-x/4z+1}{z-x/4z-1} \right| \nonumber \\ [2mm]
&+& \frac{1}{8z} \left[ 1 - \left( z + x/4z \right)^2 \right]
\ln\left| \frac{z+x/4z+1}{z+x/4z-1} \right|.
\label{7.75}\eeqa
The function $f_2(z,x)$ takes different expressions
on the regions of the $(z,x)$ plane shown in Fig.\ \ref{fig7.1}:
\beqa
f_2(z,x) &=& \frac{\pi x}{8z}
\rule{3mm}{0mm} \mbox{if $z+x/4z \le 1$ (region a),}
\nonumber \\ [2mm]
&=& \frac{\pi}{8z} \left[ 1 - \left( z-x/4z\right)^2 \right]
\rule{3mm}{0mm}
\mbox{if $-1 < z-x/4z< 1$ and $z+x/4z > 1$ (region b),}
\nonumber \\ [2mm]
&=& s \, h_{\rm L}(z,x)
\rule{3mm}{0mm}
\mbox{if $z-x/4z \le -1$ or
$z-x/4z\ge 1$ (regions $c_1$ and $c_2$),}
\label{7.76}\eeqa
where we have retained $s$ as a factor that multiplies a function
$h_{\rm L}(z,x)$; in the considered limit $s \rightarrow 0$, the
exact form of that function is irrelevant.

Similarly, in the limit $s \rightarrow 0$, the transverse Lindhard
DF reads
\beq
\epsilon_{\rm eg}^{\rm (T)} (\omega_{\rm p}; q,\omega) \equiv
\epsilon_{\rm eg}^{\rm (T)} (\omega_{\rm p}, 0; q,\omega)
= 1 - \frac{16 \chi^2}{3 x^2}
\left[ g_1(z,x) + {\rm i} g_2(z,x) \right]
\label{7.77}\eeq
with
\beqa
g_1(z,x) &=&
\frac{3}{8} \left[ z^2 + 3 (x/4z)^2+1 \right]
-\frac{3}{32 z} \left[
1 - \left( z - x/4z \right)^2 \right]^2
{\rm ln}\left| \frac{z-x/4z+1}{z-x/4z-1} \right|
\nonumber \\ [2mm]
&& \mbox{} -\frac{3}{32 z} \left[ 1 - \left( z + x/4z \right)^2 \right]^2
{\rm ln} \left| \frac{z+x/4z+1}{z+x/4z-1} \right|,
\label{7.78}\eeqa
and
\beqa
g_2(z,x) &=& - \, \frac{3\pi x}{16 z} \left[1-z^2-(x/4z)^2 \right]
\rule{3mm}{0mm} \mbox{if $z+x/4z \le 1$ (region a),}
\nonumber \\ [2mm]
&=& - \, \frac{3\pi}{32z} \left[ 1 - \left( z - x/4z\right)^2
\right]^2
\rule{3mm}{0mm}
\mbox{if $-1 < z-x/4z< 1$ and $z+x/4z > 1$ (region b),}
\nonumber \\ [2mm]
&=& s \, h_{\rm T} (z,x)
\rule{3mm}{0mm}
\mbox{if $z-x/4z \le -1$ or
$z-x/4z\ge 1$ (regions $c_1$ and $c_2$),}
\label{7.79}\eeqa
where, again, the exact form of the function $h_{\rm T}(z,x)$ is
irrelevant in the limit $s=0$.

\index{plasmon excitation}
The energy-loss function (ELF)
\beq
\eta_{2,{\rm eg}}^{\rm (L)}(\omega_{\rm p}; q,\omega) \equiv
{\rm Im} \left( \frac{-1}{\epsilon_{\rm eg}^{\rm (L)}
(\omega_{\rm p}; q,\omega)} \right)
= \frac{\chi^2 z^2 f_2(z,x)}{\left[ z^2 + \chi^2 f_1(z,x)
\right]^2 + \chi^4 f_2^2 (z,x)}
\label{7.80}\eeq
plays a central role in stopping theory: charged projectiles can produce
excitations of the gas only if $\eta_{2,{\rm eg}}^{\rm (L)}(q,\omega)
\ne 0$. When $s \rightarrow 0$ this function takes positive values in
the region a+b, the so-called Lindhard continuum, which corresponds to
energy and momentum transfers that can be absorbed by individual
electrons of the gas (see Section \ref{sec6.4.2}). The ELF vanishes in
regions c$_1$ and c$_2$ except on a resonance line in region c$_1$ where
the denominator of Eq.\ \req{7.80} vanishes. This line is defined by the
implicit equation
\beq
F(z,x) \equiv z^2 + \chi^2 f_1(z,x) = 0
\label{7.81}\eeq
and corresponds to the excitation of free oscillations of the gas, \ie, to
{\it plasmon excitation}. In region c$_1$ we have
\beq
{\rm Im} \left( \frac{-1}{\epsilon_{\rm eg}^{\rm (L)}
(\omega_{\rm p}; q,\omega)} \right)
= z^2 \lim_{s\rightarrow 0} \,
\frac{\chi^2 s h_{\rm L} (z,x)}{F^2(z,x)
+ \chi^4 s^2 h_{\rm L}^2 (z,x)}
= z^2 \, \pi \delta\left[ F(z,x) \right]
\nonumber \eeq
where $\delta(x)$ is the Dirac delta function. Using the property
\req{B.10}, we can write
\beq
{\rm Im} \left( \frac{-1}{\epsilon_{\rm eg}^{\rm (L)}
(\omega_{\rm p}; q,\omega)} \right)
= \pi z^2 \left( \left| \frac{\partial
F(z,x)}{\partial x} \right|_{x_0}\right)^{-1} \delta(x-x_0),
\nonumber \eeq
where $x_0$ is the root of Eq.\ \req{7.81}. That is,
\beq
{\rm Im} \left( \frac{-1}{\epsilon_{\rm eg}^{\rm (L)}
(\omega_{\rm p}; q,\omega)} \right)
= \frac{\pi z^2}{\chi^2} \left( \left| \frac{\partial
f_1(z,x)}{\partial x} \right|_{x_0}\right)^{-1}
\delta\left( x- x_0 \right)
\label{7.82}\eeq
with
\beqa
\frac{\partial
f_1(z,x)}{\partial x}
&=&
\frac{1}{16z^2} \left[ ( z - x/4z )
-\ln\left| \frac{z-x/4z+1}{z-x/4z-1} \right|
\right.
\nonumber \\ [2mm]
&& \mbox{} \times \left.
- ( z + x/4z )
\ln\left| \frac{z+x/4z+1}{z+x/4z-1} \right| \right]\, . \rule{10mm}{0mm}
\label{7.83}\eeqa
The plasmon cutoff wave number $z_{\rm c}$ (see Fig.\ \ref{fig7.1}) is
defined by the entrance of the plasma resonance line into the Lindhard
continuum, \ie, as the root of the equation
\beq
F\left(\rule{0mm}{4mm}z_{\rm c},4z_{\rm c}(z_{\rm c}+1)\right)=0.
\label{7.84}\eeq
For small $z$, we can use use the expansion \req{7.62} and write
\beq
F(z,x) \simeq z^2 - x_{\rm p}^2 \left[
\left( \frac{z}{x} \right)^2
+\frac{48}{5} \left( \frac{z}{x} \right)^4 \right]
\label{7.85}\eeq
where $x_{\rm p}=W_{\rm p}/E_{\rm F}$. Solving the equation $F(z,x_0)=0$
we obtain
\beq
x_0^2 \simeq x_{\rm p}^2 + \frac{48}{5} \, z^2
\nonumber \eeq
or, equivalently
\beq
\omega_0^2 = \omega_{\rm p}^2 + \frac{6}{5} \, \frac{E_{\rm F}}{\me} \,
q^2.
\label{7.86}\eeq
At $z=0$ we have
\beq
{\rm Im} \left( \frac{-1}{\epsilon_{\rm eg}^{\rm (L)}(\omega_{\rm p};
0,\omega)} \right)
= \frac{\pi}{2} \, x_{\rm p} \, \delta(x-x_{\rm p}) = \frac{\pi}{2} \,
\omega_{\rm p} \, \delta(\omega - \omega_{\rm p}) .
\label{7.87}\eeq

Similarly, the transverse ELF is
\beq
\eta_{2,{\rm eg}}^{\rm (T)}(\omega_{\rm p}; q,\omega) \equiv
{\rm Im} \left( \frac{-1}{\epsilon_{\rm eg}^{\rm (T)}
(\omega_{\rm p}; q,\omega)} \right)
= \frac{ - 3 x^2 \, 16 \chi^2 \, g_2(z,x)}{\left[
3x^2 - 16 \chi^2 \, g_1(z,x) \right]^2
+ \left[ 16 \chi^2 \, g_2 (z,x) \right]^2}.
\label{7.88}\eeq
This function vanishes in regions c$_1$ and c$_2$ except on a resonance
line in region c$_1$ where the denominator vanishes. This line is
defined by the implicit equation
\beq
G(z,x) \equiv 3 x^2 - 16 \chi^2 \, g_1(z,x) = 0.
\label{7.89}\eeq
Hence, in region c$_1$ we have
\beqa
{\rm Im} \left( \frac{-1}{\epsilon_{\rm eg}^{\rm (T)}
(\omega_{\rm p}; q,\omega)} \right)
&=&  3x^2 \lim_{s\rightarrow 0} \,
\frac{16 \chi^2 \, s \, h_{\rm T}(z,x)}{G^2(z,x)
+ \left[16 \chi^2 \, s \, h_{\rm T} (z,x) \right]^2}
= 3 x^2 \, \pi \delta\left[ G(z,x) \right]
\nonumber \\ [2mm]
&=& 3 \pi x^2 \left( \left| \frac{\partial
G(z,x)}{\partial x} \right|_{x'_0}\right)^{-1} \delta(x-x'_0),
\nonumber \eeqa
where $x'_0$ is the root of Eq.\ \req{7.89}. That is,
\beq
{\rm Im} \left( \frac{-1}{\epsilon_{\rm eg}^{\rm (T)}
(\omega_{\rm p}; q,\omega)} \right)
= 3 \pi x^2 \left( \left| 6 x - 16 \chi^2
\frac{\partial g_1(z,x)}{\partial x} \right|_{x'_0}\right)^{-1}
\delta\left( x- x'_0 \right)
\label{7.90}\eeq
with
\beqa
\frac{\partial g_1(z,x)}{\partial x}
&=& \frac{6x}{32 z^2} -
\frac{3}{32z^2} \left[ 1 - ( z - x/4z )^2 \right] ( z - x/4z )
\, \ln\left| \frac{z-x/4z+1}{z-x/4z-1} \right|
\nonumber \\ [2mm]
&& \mbox{} +
\frac{3}{32z^2} \left[ 1 - ( z + x/4z )^2 \right] ( z + x/4z )
\, \ln\left| \frac{z+x/4z+1}{z+x/4z-1} \right| \, .
\label{7.91}\eeqa
The plasmon cutoff wave number $z'_{\rm c}$ (see Fig.\ \ref{fig7.1}) is
defined by the entrance of the plasma resonance line into the Lindhard
continuum, \ie, as the root of the equation
\beq
G\left(\rule{0mm}{4mm}z_{\rm c},4z_{\rm c}(z_{\rm c}+1)\right)=0.
\label{7.92}\eeq
For small $z$, we can use use the expansion \req{7.63} and write
\beq
G(z,x) \simeq 3x^2 - 3 x_{\rm p}^2 \left[1 + \frac{1}{5}
\left( \frac{4z}{x} \right)^2 \right].
\label{7.93}\eeq
Solving the equation $G(z,x'_0)=0$ we obtain
\beq
x'^2_0 \simeq x_{\rm p}^2 + \frac{16 z^2}{5x_{\rm p}^2}
\nonumber \eeq
or, equivalently,
\beq
\omega'^2_0 = \omega_{\rm p}^2 + \frac{4}{10} \, \frac{E_{\rm
F}^3}{W_{\rm p}^2} \, \frac{q^2}{\me} \, .
\label{7.94}\eeq
At $z=0$ we have
\beq
{\rm Im} \left( \frac{-1}{\epsilon_{\rm eg}^{\rm (T)}(\omega_{\rm p}; 0,\omega)} \right)
= \frac{\pi}{2} \, x_{\rm p} \, \delta(x-x_{\rm p}) = \frac{\pi}{2} \,
\omega_{\rm p} \, \delta(\omega - \omega_{\rm p}) .
\label{7.95}\eeq
It is worth noticing that the transverse plasma line typically has less
dispersion than the longitudinal line, that is, the ``slope'' of the
transverse resonance line in the $(z,x)$ plane is smaller than that of
the longitudinal line (see Fig.\ \ref{fig7.2}).

\begin{figure}[tbh!]
\begin{center}
\includegraphics*[width=10.0cm]{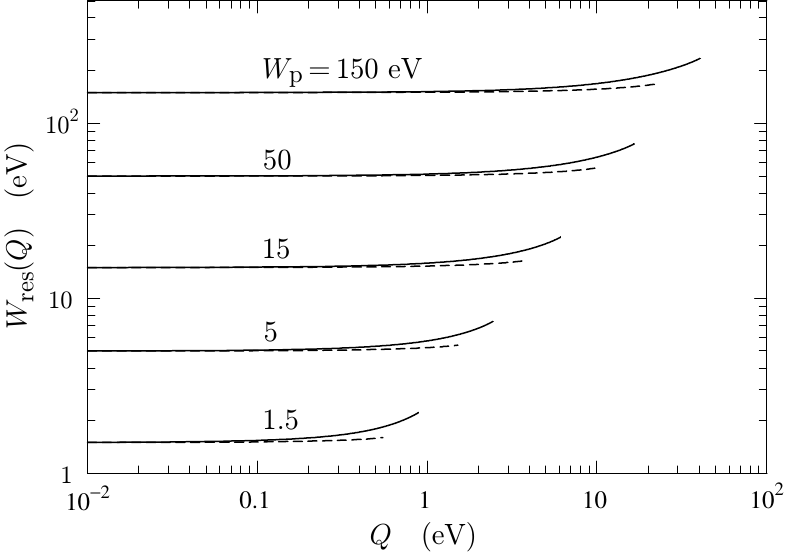}
\caption{Plasmon resonance curves for electron gases with the indicated
plasma energies $W_{\rm p}$, obtained from Lindhard's DFs. The solid
and dashed curves represent longitudinal and transverse
plasmon lines, respectively.
\label{fig7.2}}
\end{center}\end{figure}


\subsection{The plasmon-pole approximation \label{sec7.1.4}}
\index{plasmon-pole approximation}

In applications where low-$q$ excitations dominate, it may be
sufficiently accurate to represent the inverse DF of an electron gas by
means of the so-called {\it plasmon-pole approximation}
\citep{Ritchie1976, Echenique1979},
\beq
\eta_{\rm pp}(\omega_{\rm p}, s; q,\omega) \equiv 1 -
\frac{\omega_{\rm p}^2}{\omega_{\rm r}^2(q)
- \omega(\omega+{\rm i} s)} \, ,
\label{7.96}\eeq
where
\beq
\omega_{\rm r}^2 (q) = \omega_{\rm p}^2 + \beta_{\rm pp}^2 q^2
+ (\hbar^2 q^4 /4 \me^2)
\label{7.97}\eeq
with
\beq
\beta_{\rm pp}^2 = \frac{6}{5} \, \frac{E_{\rm F}}{\me}.
\label{7.98}\eeq
At small $q$ this DF model approximates the Mermin DF (see below) and,
when $s\rightarrow 0$, it yields a dispersion relation for plasmon
excitations in agreement with the result of Lindhard's theory. The term
$\hbar^2 q^4 /4 \me^2$ in expression \req{7.97} is introduced in an {\it
ad hoc} fashion to account for large-$q$ electron-hole excitations.

In the limit of no damping ($s=0$), the plasmon-pole inverse DF becomes
\beqa
\eta_{{\rm pp}}(\omega_{\rm p}; q,\omega) = 1 -
\frac{\omega_{\rm p}^2}{
\omega_{\rm r}^2(q) - \omega^2 }
- {\rm i} \, \lim_{s\rightarrow 0} \,
\frac{\omega_{\rm p}^2 \, s\omega}{
\left[ \omega_{\rm r}^2(q) - \omega^2 \right]^2 + s^2 \omega^2}\, .
\label{7.99}\eeqa
The ELF takes a relatively simple form,
\beqa
\eta_{2,{\rm pp}}(\omega_{\rm p}; q,\omega) )
&=& \pi \omega_{\rm p}^2 \, \delta \left[ \omega_{\rm p}^2
+ \beta_{\rm pp}^2 q^2
+ (\hbar^2 q^4 /4 \me^2) - \omega^2 \right]
\nonumber \\ [2mm]
&=& \frac{\pi}{2} \, \frac{\omega_{\rm p}^2}{\omega_{\rm r}(q)}
\, \delta[\omega-\omega_{\rm r}(q)] \, ,
\label{7.100}\eeqa
which in the optical limit reduces to
\beq
\eta_{2,{\rm pp}}(\omega_{\rm p}; \omega) =
\frac{\pi}{2} \,
\omega_{\rm p} \, \delta(\omega-\omega_{\rm p}).
\label{7.101}\eeq
Evidently, the only possible excitations lay on the line
$\omega=\omega_{\rm r}(q)$ of the $(q,\omega)$ plane.
\index{Lindhard dielectric function|)}


\section{Mermin dielectric functions of an electron gas \label{sec7.2}}
\index{Mermin dielectric function}

In a real material, electronic excitations decay due to collisions with
a finite lifetime $\tau$, which in the relaxation-time approximation is
assumed to be independent of $\omega$ \citep[see,
\eg,][]{KliewerFuchs1969}. It is tempting to account for this finite
lifetime by simply interpreting the constant $s$ in the Lindhard DF as
the reciprocal of the plasmon lifetime $\tau$ derived, \eg, from
electron energy-loss measurements. However, the use of a complex
frequency $\omega+{\rm i} s$ in expressions \req{7.56} to \req{7.59}
leads to a conflict with the conservation of the local number of
electrons \citep{Mermin1970}, which surely originates from the incorrect
normalization of the perturbed electron wave functions \req{7.8}. A generalization of the longitudinal
Lindhard DF function which introduces a finite plasmon lifetime
$\tau=1/s$, reduces to the classical optical limit, and still conserves
the local electron number was derived by \citet{Mermin1970}. The
longitudinal DF proposed by Mermin is
\beqa
\epsilon^{\rm (L)}(\omega_{\rm p}, s; q,\omega) &=&
\nonumber \\ [1mm]
&& \! \! \! \! \! \! \! \! \! \! \! \! \!  \! \! \! \! \! \! \! \! \!
\! \! \! \! \! \!\mbox{}
1 + \frac{\omega+{\rm i} s}{\omega} \,
\frac{ \left[ \epsilon_{\rm eg}^{\rm (L)}
(\omega_{\rm p}, s; q,\omega) -1 \right]}
{1 + ({\rm i}s/\omega) \left[ \epsilon_{\rm eg}^{\rm (L)}
(\omega_{\rm p}, s; q,\omega) -1 \right]
/\left[ \epsilon_{\rm eg}^{\rm (L)}
(\omega_{\rm p}; q,0) -1 \right] } \, , \rule{10mm}{0mm}
\label{7.102}\eeqa
where $\epsilon_{\rm eg}^{\rm (L)}(\omega_{\rm p}, s; q,\omega)$ is the
Lindhard longitudinal DF, Eq.\ \req{7.56}, depending on the complex
frequency $\omega+{\rm i} s =(x+{\rm i} y) E_{\rm F}$. Equivalently, we
can write
\beq
\epsilon^{\rm (L)}(\omega_{\rm p}, s; q,\omega) =
1 + \frac{\chi^2}{z^2} \,
\frac{x + {\rm i} y}{x} \, \frac{f(z,x+{\rm i} y)}
{1 + ({\rm i}y/x) \, f(z,x+{\rm i}y) / f(z,0)} \, ,
\label{7.103}\eeq
where $f(z,0)$ is the Lindhard static function, Eq.\ \req{7.93}. It is worth
noting that the optical limit of the Mermin DF [see Eq.\ \req{7.65}],
\beq
\lim_{q\rightarrow 0} \epsilon^{\rm (L)}(\omega_{\rm p}, s; q,\omega) =
1 + \frac{\chi^2}{z^2} \,
\frac{ x + {\rm i} y}{\omega} \, f(0,x+{\rm i} y)
= 1 - \frac{\omega_{\rm p}^2}{\omega (\omega+{\rm i} s )}\, ,
\label{7.104}\eeq
coincides with the classical result, Eq.\ \req{1.168}. Moreover, in the
static limit the Mermin DF equals the Lindhard DF,
\beq
\lim_{\omega\rightarrow 0} \epsilon^{\rm (L)}
(\omega_{\rm p}, s; q,\omega)
= \epsilon_{\rm eg}^{\rm (L)}(\omega_{\rm p}; q,0)
= 1 + \frac{\chi^2}{z^2} \, f(z,0),
\label{7.105}\eeq
and, consequently, the Mermin modification does not alter the screening
of charges predicted by the (undamped) Lindhard DF, which is determined
by the DF in the static limit \citep{AshcroftMermin1976}. Clearly, the factor
$(\omega+{\rm i}s)/\omega$ in Eq.\ \req{7.102} restores the classical
optical limit. On the other hand, the Mermin DF, has no singularities in
the upper half of the complex $\omega$ plane (${\rm Im} \, \omega > 0$) and,
therefore, it satisfies the Kramers--Kronig relations \req{1.210}. That
is, the Mermin DF is consistent with the principle of unretarded
causality (Section \ref{sec1.5}).

In an alternative derivation of the transverse DF of the electron gas,
\citet{KliewerFuchs1969} obtained,
\beq
\epsilon_{\rm rt}^{\rm (T)} (\omega_{\rm p}, s; q,\omega) =
1 - \frac{16 \chi^2}{3x(x+{\rm i}y)} \,  g(z,x+{\rm i}y) \, ,
\label{7.106}\eeq
which differs from Lindhard's transverse DF, Eq.\ \req{7.58}, in that a
factor $\omega$ in the denominator has been
replaced with $\omega+{\rm i} s$. The DF
\req{7.106} can be considered as the transverse DF in the relaxation-time (rt)
approximation. Indeed, the optical limit of expression \req{7.106}
coincides with the classical result \req{1.168}. However, the DF
\req{7.106} does not satisfy the Kramers--Kronig dispersion relations
[Eqs.\ \req{1.210}]. This defect can be corrected by introducing a
low-$\omega$ correction analogous to the one employed in the Mermin
longitudinal DF \req{7.103},
\beq
\epsilon^{\rm (T)} (\omega_{\rm p}, s; q,\omega) =
1 - \frac{16 \chi^2}{3x(x+{\rm i}y)} \,  \frac{g(z,x+{\rm i}y)}
{1 + ({\rm i}y/x) \, f(z,x+{\rm i}y) / f(z,0)} \, ,
\label{7.107}\eeq
which removes the singularities of the transverse DF from the upper half
of the complex $\omega$ plane. This {\it ad hoc} modification of the
Lindhard formula gives a transverse DF that has the classical optical
limit, Eq.\ \req{7.104}, and satisfies the Kramers--Kronig relations
\req{1.210}.


\section{DF for valence electrons in semiconductors and
insulators \label{sec7.3}}
\index{gapped dielectric function}
\index{dielectric functions!of semiconductors and insulators}

The Lindhard and Mermin DFs are appropriate for describing the
dielectric response of conduction electrons in free-electron-like
materials such as aluminum. The main characteristic of these materials
is that the function $\eta_2^{\rm (L)}(q,\omega)$ for small frequencies
decreases with $\omega$ and tends monotonically to zero at $\omega=0$,
reflecting the fact that even small energy transfers are able to produce
excitations of electrons that are close enough to the Fermi level. The
situation is different for semiconductors and insulators, which exhibit
a gap in the electronic excitation spectrum. Consequently their energy
loss function vanishes for frequencies less than $\omega_{\rm g} =
E_{\rm g} /\hbar$, where $E_{\rm g}$ denotes the gap energy. To describe
the DF of these materials, we need a model of the ELF with a
well-defined threshold at $\omega_g$.

\citet{LevineLouie1982} suggested a simple transformation that converts
the Lindhard DF into a form with a gap in its ELF without altering
relevant features of the original Lindhard DF such as the high-frequency
analytical behavior and unretarded causality (as expressed by the
Kramers--Kronig relations), and still admits a relatively simple
analytical representation. Here we present a similar transformation of
the Mermin DFs, Eqs.\ \req{7.103} and \req{7.107}, which introduces a
frequency gap $\omega_{\rm g}$.

The transformation takes a simpler and more plausible form when
expressed in terms of the inverse DF, $\eta(q,\omega) = \eta_1 - {\rm i}
\eta_2$. Following \citet{LevineLouie1982}, the inverse longitudinal DF
for a semiconductor or insulator with the energy gap $\hbar \omega_{\rm
g}$ at frequencies $\omega$ above $\omega_{\rm g}$ is expressed as
\beq
\eta^{\rm (L)}(\omega_{\rm p}, s, \omega_{\rm g}; q, \omega)
\equiv \eta^{\rm (L)}(\omega_{\rm p}, s; q, \overline{\omega})
= \frac{1}{\epsilon^{\rm (L)}(\omega_{\rm p}, s; q, \overline{\omega})}
\qquad \mbox{for $\omega > \omega_{\rm g}$,}
\label{7.108}\eeq
with the ``shifted'' frequency $\overline{\omega}$ given by
\beq
\overline{\omega} = \sqrt{\omega^2 - \omega_{\rm g}^2},
\label{7.109}\eeq
where the positive branch of the square root is implied. For the sake
of simplicity, we use the same symbol for the original and the new DFs,
which are distinguished by the number of their arguments. For
frequencies below $\omega_{\rm g}$ we set
\beq
\eta^{\rm (L)}_2(\omega_{\rm p}, s, \omega_{\rm g}; q, \omega) \equiv 0
\qquad \mbox{for $\omega \le \omega_{\rm g}$.}
\label{7.110}\eeq
Because Eqs.\ \req{7.108} and \req{7.110} define the function
$\eta_2^{\rm (L)}(\omega_{\rm p}, s, \omega_{\rm g}; q, \omega)$ for all
$\omega \ge 0$, the function $\eta_1^{\rm (L)}(\omega_{\rm p}, s,
\omega_{\rm g}; q, \omega)$ can be obtained from the Kramers--Kronig
relation \req{1.209a},
\beqa
\eta_1^{\rm (L)}(\omega_{\rm p}, s, \omega_{\rm g}; q, \omega)
&=& 1 - \frac{2}{\pi} {\cal P}
\int_{0}^\infty \frac{\omega'}
{\omega'^2-\omega^2} \,
\eta_2^{\rm (L)}(\omega_{\rm p}, s, \omega_{\rm g}; q, \omega')
\, \d \omega'.
\nonumber \eeqa
With a change of variables, we can write
\beq
\eta_1^{\rm (L)}(\omega_{\rm p}, s, \omega_{\rm g}; q, \omega)
= 1 - \frac{2}{\pi} {\cal P}
\int_{0}^\infty \frac{\overline{\omega'}}
{\overline{\omega'}^2-\omega^2+\omega_{\rm g}^2} \,
\eta_2^{\rm (L)}(\omega_{\rm p}, s; q, \overline{\omega'})
\, \d \overline{\omega'} .
\label{7.111}\eeq
Because the original DF satisfies the Kramers--Kronig relations, for
$\omega > \omega_{\rm g}$ we have
\beq
\eta_1^{\rm (L)}(\omega_{\rm p}, s, \omega_{\rm g}; q, \omega)
= \eta_1^{\rm (L)}(\omega_{\rm p}, s; q, \overline{\omega}) ,
\label{7.112}\eeq
in accordance with the definition \req{7.108}. The real part of the
inverse DF for $\omega \le \omega_{\rm g}$ is now determined by Eq.\
\req{7.111}, which gives
\beq
\eta_1^{\rm (L)}(\omega_{\rm p}, s, \omega_{\rm g}; q, \omega)
= \eta_1^{\rm (L)}(\omega_{\rm p}, s; q, {\rm i} \, \widetilde{\omega})
\qquad \mbox{for $\omega \le \omega_{\rm g}$}
\label{7.113}\eeq
with
\beq
\widetilde{\omega} = \sqrt{\omega_{\rm g}^2 - \omega^2}.
\label{7.114}\eeq
Because the imaginary part of the Mermin DF vanishes for
imaginary frequencies, the definition \req{7.110} can be written
similarly,
\beq
\eta_2^{\rm (L)}(\omega_{\rm p}, s, \omega_{\rm g}; q, \omega)
= \eta_2^{\rm (L)}(\omega_{\rm p}, s; q, {\rm i} \, \widetilde{\omega})
\qquad \mbox{for $\omega \le \omega_{\rm g}$}.
\label{7.115}\eeq
Since the derivation rests only on the fact that the original DF
satisfies the Kramers--Kronig relations, the same transformation can be
employed to introduce the frequency gap in the transverse Mermin DF.

Summarizing, the gapped Mermin DFs can be expressed in the compact, and
convenient form
\begin{subequations}
\label{7.116}
\beq
\epsilon^{\rm (L,T)}(\omega_{\rm p}, s, \omega_{\rm g}; q, \omega) =
\epsilon^{\rm (L,T)}(\omega_{\rm p}, s;
q, \overline{\omega}),
\label{7.116a}\eeq
where the  ``shifted'' frequency is defined as
\beq
\overline{\omega} \equiv
\left\{
\begin{array}{ll}
\sqrt{\omega^2 - \omega_{\rm g}^2} \rule{5mm}{0mm}&
\mbox{if $\omega \ge \omega_{\rm g}$,} \\ [2mm]
{\rm i} \sqrt{\omega_{\rm g}^2 - \omega^2} &
\mbox{if $\omega < \omega_{\rm g}$,}
\end{array} \right.
\label{7.116b}\eeq
\end{subequations}
with the positive values of the square roots implied. Evidently, in the
limit $\omega_{\rm g} =0$ these DFs reduce to the Mermin DFs,
$\epsilon^{\rm (L,T)} (\omega_{\rm p}, s; q, \omega)$. The gapped Mermin
DFs \req{7.116} satisfy the Kramers--Kronig relations and the set of sum
rules derived from them. Moreover, all these properties are retained
when the gap frequency $\omega_{\rm g}$ is allowed to vary with $q$.

For practical calculation purposes, we note that the optical limit
($q=0$) of the real and imaginary parts of the inverse
longitudinal DF read
\begin{subequations}
\label{7.117}
\beqa
\eta_1^{\rm (L)}(\omega_{\rm p}, s, \omega_{\rm g}; \omega) &=&
1 - \frac{
\omega_{\rm p}^2 \left( \omega_{\rm p}^2 - \omega^2 + \omega_{\rm g}^2
\right)}{\left(
 \omega_{\rm p}^2 - \omega^2 + \omega_{\rm g}^2 \right)^2 + s^2
\left( \omega^2 - \omega_{\rm g}^2 \right)} \quad \quad
\mbox{if $\omega \ge \omega_{\rm g}$}, \rule{10mm}{0mm}
\label{7.117a} \\ [2mm]
&=& 1 - \frac{\omega_{\rm p}^2}
{\omega_{\rm p}^2 + \omega_{\rm g}^2 - \omega^2
+ s \sqrt{\omega_{\rm g}^2-\omega^2}} \quad \quad \rule{7mm}{0mm}
\mbox{if $\omega < \omega_{\rm g}$},
\label{7.117b}\eeqa
\end{subequations}
and
\beq
\eta_2^{\rm (L)}(\omega_{\rm p}, s, \omega_{\rm g}; \omega) =
\frac{ \omega_{\rm p}^2 \, s \sqrt{\omega^2-\omega_{\rm g}^2}}
{(\omega_{\rm p}^2 - \omega^2 + \omega_{\rm g}^2)^2 +
s^2 (\omega^2-\omega_{\rm g}^2)} \,
{\cal S} (\omega - \omega_{\rm g}) \, .
\label{7.118}\eeq
The ODF is
\begin{subequations}
\label{7.119}
\beqa
\epsilon (\omega_{\rm p}, s, \omega_{\rm g}; 0, \omega)
&=&  1 - \frac{\omega_{\rm p}^2}
{ \sqrt{\omega^2 - \omega_{\rm g}^2}
\left( \sqrt{\omega^2 - \omega_{\rm g}^2} + {\rm i} s \right) }
\qquad \mbox{if $\omega > \omega_{\rm g}$}, \rule{10mm}{0mm}
\label{7.119a} \\
&=&  1 + \frac{\omega_{\rm p}^2}
{\sqrt{\omega_{\rm g}^2 - \omega^2}
\left( \sqrt{\omega_{\rm g}^2 - \omega^2} +  s \right) }
\qquad \; \mbox{if $\omega \le \omega_{\rm g}$}.
\label{7.119b}\eeqa
\end{subequations}
In particular, the static long-wavelength limit of the DF is
\beq
\epsilon(\omega_{\rm p}, s, \omega_{\rm g}; 0, 0)
= 1 + \frac{\omega_{\rm p}^2}
{\omega_{\rm g} \left( \omega_{\rm g} +  s \right) } \, .
\label{7.120}\eeq
It can be easily verified that the OELF $\eta_2
(\omega_{\rm p}, s, \omega_{\rm g}; 0, \omega)$ attains its maximum
value at a frequency $\omega_{\rm max}$ given by
\begin{subequations}
\label{7.121}
\beq
\omega_{\rm max} = \sqrt{
\omega_{\rm g}^2 + \frac{2\omega_{\rm p}^2 - s^2 +
\sqrt{\left( 2\omega_{\rm p}^2 - s^2 \right)^2 + 12 \omega_{\rm
p}^4}}{6}} \, .
\label{7.121a}\eeq
Conversely, the resonance frequency $\omega_{\rm p}$ for which the
ELF has its maximum at $\omega_{\rm max}$ is
\beq
\omega_{\rm p} = \sqrt{
\sqrt{\left( s^2+4 \omega_{\rm max}^2 -4 \omega_{\rm g}^2 \right)
\left( \omega_{\rm max}^2 - \omega_{\rm g}^2\right)}
-\omega_{\rm max}^2 + \omega_{\rm g}^2}\, .
\label{7.121b}\eeq
\end{subequations}
Notice that the formulas \req{7.117} to \req{7.121} with $\omega_{\rm g}=0$
are also valid for the DFs of Mermin and for the
Lorentz-Drude oscillator model, Eq.\ \req{1.168} with $\Omega_{\rm p} =
\omega_{\rm p}$.


\noindent $\bullet$ {\bf The gapped Lindhard DF} \\
The above strategy to introduce a gap in the Mermin DF can also be
applied to the undamped Lindhard DF. The resulting DF,
\begin{subequations}
\label{7.122}
\beq
\epsilon_{\rm eg}^{\rm (L,T)}(\omega_{\rm p}, \omega_{\rm g}; q, \omega) =
\epsilon_{\rm eg}^{\rm (L,T)}(\omega_{\rm p}; q, \overline{\omega})
\label{7.122a}\eeq
with
\beq
\overline{\omega} \equiv
\left\{
\begin{array}{ll}
\sqrt{\omega^2 - \omega_{\rm g}^2} \rule{5mm}{0mm}&
\mbox{if $\omega \ge \omega_{\rm g}$,} \\ [2mm]
{\rm i} \sqrt{\omega_{\rm g}^2 - \omega^2} &
\mbox{if $\omega < \omega_{\rm g}$,}
\end{array} \right.
\label{7.122b}\eeq
\end{subequations}
finds an important application in the optical-data models described in
the following Sections. It is worth noticing that the corresponding optical
energy-loss function is
\beq
{\rm Im} \left( \frac{-1}{\epsilon_{\rm eg}^{\rm (L)}(\omega_{\rm p},
\omega_{\rm g}; 0,\omega)} \right)
= \frac{\pi}{2} \,
\omega_{\rm p} \, \delta(\omega - \omega_{\rm res})
\label{7.123}\eeq
with
\beq
\omega_{\rm res} = \sqrt{\omega_{\rm p}^2 + \omega_{\rm g}^2}.
\label{7.124}\eeq


\section{Stopping of charged particles in an electron gas
\label{sec7.4}}
\index{stopping power of the electron gas}

Let us consider a charged particle of mass $M_0$ and charge $Z_0e$
moving with kinetic energy $E$ in an electron gas characterized by the
plasma resonance energy $W_{\rm p}=\hbar \omega_{\rm p}$. In preparation
for the calculations that follow, we describe the response of
the gas by means of the Mermin DFs with a damping energy parameter
$\Gamma = \hbar s$, $\epsilon_{\rm eg}^{\rm (L,T)}(W_{\rm p}, \Gamma; Q,
W)$, and the undamped ($\Gamma=0$) Lindhard DFs, $\epsilon_{\rm eg}^{\rm
(L,T)}(W_{\rm p}; Q, W)$. Hereafter we regard the DFs as functions of
the recoil energy $Q$,
\beq
Q(Q+2\me c^2)=(c\hbar q)^2
\label{7.125}\eeq
and the energy transfer $W$. The calculations that follow are readily
generalized to the gapped DFs.

The interactions of the projectile with the gas are described by means
of the semi-classical dielectric formalism (see Section \ref{sec6.7}).
We disregard the dielectric polarization of the gas [Eq.\ \req{6.264}]
which is appreciable only for projectiles with kinetic energies of the
order of, and higher than the rest energy $M_0c^2$. A general
description of the effect of polarization is given in Section
\ref{sec8.2}. The DDCS per electron in the gas is then [see Eq.\
\req{6.256}]
\beqa
&&  \! \! \!  \! \! \! \! \!
\frac{\d^2 \sigma}{\d Q \, \d W}
= \frac{2\pi Z_0^2 e^4}{\me v^2} \,
\left[ \frac{2\me c^2}{WQ(Q+2\me c^2)} \,
\frac{\d f_{\rm eg}(Q,W)}{\d W} \right.
\nonumber \\ [2mm]
&+& \left.\left( \beta^2 - \frac{W^2}{Q(Q+2 \me c^2)} \right)
\frac{ 2\me c^2 W } {[Q(Q+2\me c^2) - W^2 ]^2} \,
\frac{\d g_{\rm eg}(Q,W)}{\d W} \right]\, ,
\rule{10mm}{0mm}
\label{7.126}\eeqa
with the GOSs [Eqs.\ \req{6.251}]
\begin{subequations}
\label{7.127}
\beq
\frac{\d f_{\rm eg}(Q,W)}{\d W} =
W \,
\frac{2}{\pi W_{\rm p}^2} \,
{\rm Im} \left( \frac{-1}{
\epsilon^{\rm (L)}_{\rm eg} (W_{\rm p}, \Gamma; Q, W)} \right)
\label{7.127a}\eeq
and
\beq
\frac{\d g_{\rm eg}(Q,W)}{\d W} =
W \,
\frac{2}{\pi W_{\rm p}^2} \,
{\rm Im} \left( \frac{-1}{\epsilon^{\rm (T)}_{\rm eg} (W_{\rm p}, \Gamma;
Q, W)} \right) .
\label{7.127b}\eeq
\end{subequations}
To obtain the energy-loss DCS,
\beq
\frac{\d \sigma}{\d W} = \int_{Q_-}^{Q_+}
\frac{\d^2 \sigma}{\d Q \, \d W} \, \d W,
\label{7.128}\eeq
\index{Gauss--Legendre quadrature!adaptive}we must perform a numerical
quadrature. In the case of the Mermin DF, the integral is evaluated
numerically by means of the adaptive Gauss--Legendre method (Section
\ref{sec10.4.3.1}), which delivers results to a relative accuracy better
than 0.01 \%.

The calculation for the undamped Lindhard DF is somewhat tricky, because
of the resonance character of plasmon excitations. Our computer program
{\sc stopping} (see Chapter \ref{chapt10}) first determines the longitudinal and transverse plasmon lines,
$W=W_{\rm res}(Q)$, by solving Eqs.\ \req{7.81} and \req{7.89} for a
grid of $Q$ values that is dense enough to allow accurate cubic spline
interpolation (Section \ref{sec10.4.2}). The roots of these equations are obtained by the
bisection method. Figure \ref{fig7.2} displays plasma resonance curves,
$W_{\rm res}(Q)$, for electron gases with various plasma resonance energies
$W_{\rm p}$; notice that transverse plasmons have weaker dispersion.
The GOS of each plasma line is represented in the form [see Eqs.\
\req{7.82} and \req{7.90}]
\beq
\left[\frac{\d f(Q,W)}{\d W} \right]_{\rm plasmon} =
F_{\rm res}(Q) \, \delta[W-W_{\rm res}(Q)],
\label{7.129}\eeq
where $F_{\rm res}(Q)$ is the GOS of the line. The contribution of
plasmon excitations to the energy-loss DCS is obtained easily from this
expression, with appropriate interpolation by natural cubic splines.
The $Q$-integration over the Lindhard
continuum can be performed by using the adaptive Gauss--Legendre method.
However, to reduce the numerical effort, for energy transfers larger
than $100 \, W_{\rm p}$ the GOSs are replaced with the analytical
expression obtained from the binary-encounter approximation [Eq.\
\req{6.173}],
\beq
\frac{\d f^{\rm (BEA)} (E_{\rm F}; Q,W)}{\d W}
= \frac{W}{4Q^2} \, \frac{3}{2} \sqrt{\frac{Q}{E_{\rm F}}}
\left[ 1 - \left( \frac{W - Q}{2Q}\right)^2 \frac{Q}{E_{\rm F}}
\right]
\, {\cal W} \left( Q_-^{\rm b}, Q_+^{\rm b}; Q \right),
\label{7.130}\eeq
with
\beq
Q_\pm^{\rm b}  = \left(\sqrt{W+E_{\rm F}} \pm \sqrt{E_{\rm F}} \right)^2
\, ,
\label{7.131}\eeq
where $E_{\rm F}$ is the Fermi energy of the gas.
When the integration interval $(Q_-,Q_+)$ covers the whole width of the
Lindhard continuum, the integral may be evaluated analytically, and the
contribution of electron-hole excitations to the energy-loss DCS can be
approximated as the product of the non-relativistic energy-loss DCS
[Eq.\ \req{6.181}] and a relativistic correction factor (see Section
\ref{sec6.8}),
\beq
\left[ \frac{\d \sigma} {\d W} \right]_{\rm eh} =
\frac{2 \pi Z_0^2 e^4 }{\me v^2 } \, \frac{1}{W^2}
\left( 1 + \frac{4}{5} \, \frac{E_{\rm F}}{W} \right) F_{\rm bin}(W).
\label{7.132}\eeq

It is interesting to notice here that the PWBA (see Section
\ref{sec6.4.2}) is only capable of describing
electron-hole excitations of the electron gas. It does not account for
plasmon excitations, and also fails to describe the screening of
external charges by the gas. The latter effect,
combined with Pauli's exclusion principle, causes the energy-loss DCS to
vanish when $W$ goes to zero, in contrast with the divergence of the
PWBA energy-loss DCS at $W=0$ [see Eq.\ \req{6.181}].  The reason for
the superiority of the dielectric formalism can be found in the
derivation of the Lindhard DF (Section \ref{sec7.1.1}), which relates
the charge and current induced in the gas to the total, {\it external
plus induced}, scalar and vector potentials, while the PWBA accounts
only for the effect of the {\it external} scalar potential.

To give a feel of the similarities and differences between the
descriptions obtained from the Lindhard and Mermin DFs, Fig.\
\ref{fig7.3} shows the energy-loss DCSs calculated from these DFs for
protons ($Z_0=1$, $M_0=\me$) of various kinetic energies in an electron
gas with $W_{\rm p} = 15$~eV (and $W_{\rm g}=0$), with a damping
parameter in the Mermin DF $\Gamma=0.75$~eV, which grossly corresponds
to the observed plasmon loses in metallic aluminum. The damping of the
Mermin DF causes a little increase of the energy-loss DCS at small
energy transfers, as well as a slight displacement of the upper end of
the energy-loss spectrum towards higher values. However, the results
from the two DFs agree closely for energy transfers $W$ larger than
about $W_{\rm p}$ and not too close to the higher end of the spectrum,
that is, where the integration over $Q$ includes the bulk of the
Lindhard continuum (or the Bethe ridge in PWBA terminology). The
consistency of the two calculations, which utilize quite different
numerical methods, indicates that our treatment of the discrete plasmon
resonances in Lindhard's DFs is correct.

\begin{figure}[bth!]
\begin{center}
\includegraphics*[width=8.0cm]{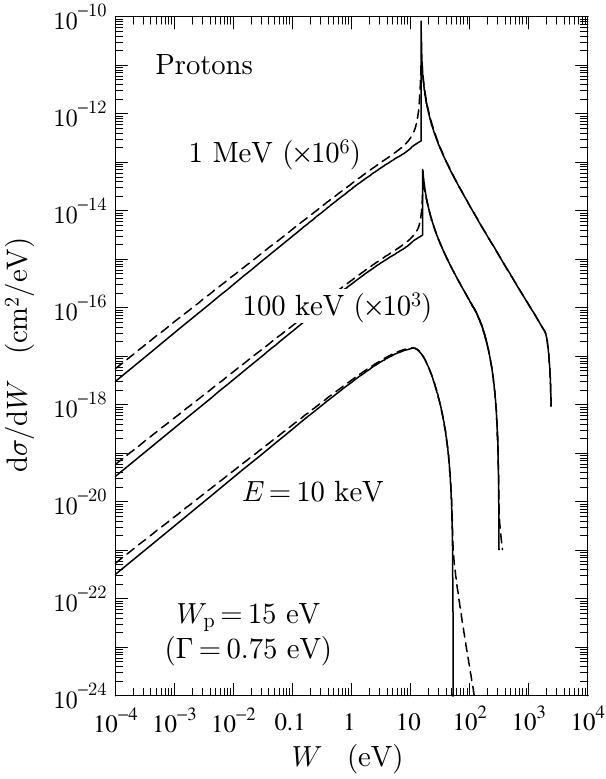}
\caption{Energy-loss DCSs for protons of the indicated kinetic energies
in an electron gas with $W_{\rm p}=15$ eV. Solid
curves are calculated by using the Lindhard DF; the dashed curves
are the results obtained by considering the Mermin dielectric function with
damping parameter $\Gamma = 0.75$ eV.
\label{fig7.3}}
\end{center}\end{figure}
 		 	
A more quantitative assessment of the influence of the DF model on the
calculations of inelastic interactions of charged particles with an
electron gas is provided by the integrals of the energy-loss DCS and the
corresponding macroscopic cross sections. We consider the mean free path,
\begin{subequations}
\label{7.133}
\beq
\lambda =  \frac{1}{\rho^0 \sigma^{(0)}}, \qquad
\sigma^{(0)} = \int_0^{W_{\rm max}} \frac{\d \sigma}{\d W} \, \d W,
\label{7.133a}\eeq
the stopping power,
\beq
S = \rho^0 \sigma^{(1)}, \qquad
\sigma^{(1)} = \int_0^{W_{\rm max}} W \, \frac{\d \sigma}{\d W} \, \d W,
\label{7.133b}\eeq
and the energy-straggling parameter
\beq
\Omega^2 = \rho^0 \sigma^{(2)}, \qquad \sigma^{(2)} = \int_0^{W_{\rm max}}
W^2 \, \frac{\d \sigma}{\d W} \, \d W,
\label{7.133c}\eeq
\end{subequations}
of positrons ($Z_0=1$, $M_0=\me$) and protons ($Z_0=1$, $M_0=1836 \, \me$)
in the gas. Notice that $\rho^0$ denotes the electron
density (electrons per unit volume) of the gas, which is determined by
the plasma resonance energy [see Eqs.\ \req{7.49} and \req{7.51}],
\beq
\rho^0 = \frac{\me}{4\pi \, e^2 \hbar^2} \, W_{\rm p}^2.
\label{7.134}\eeq

\begin{figure}[p!]
\begin{center}
\includegraphics*[width=7.5cm]{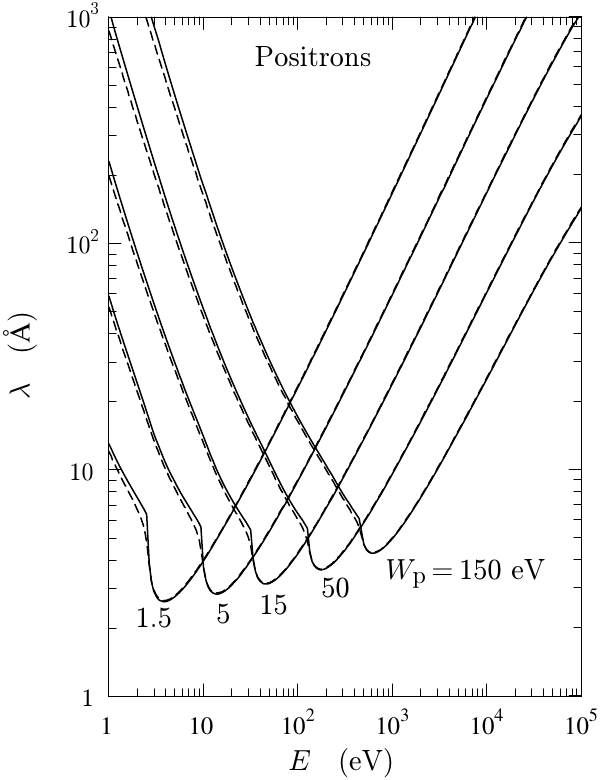} \rule{3mm}{0mm}
\includegraphics*[width=7.5cm]{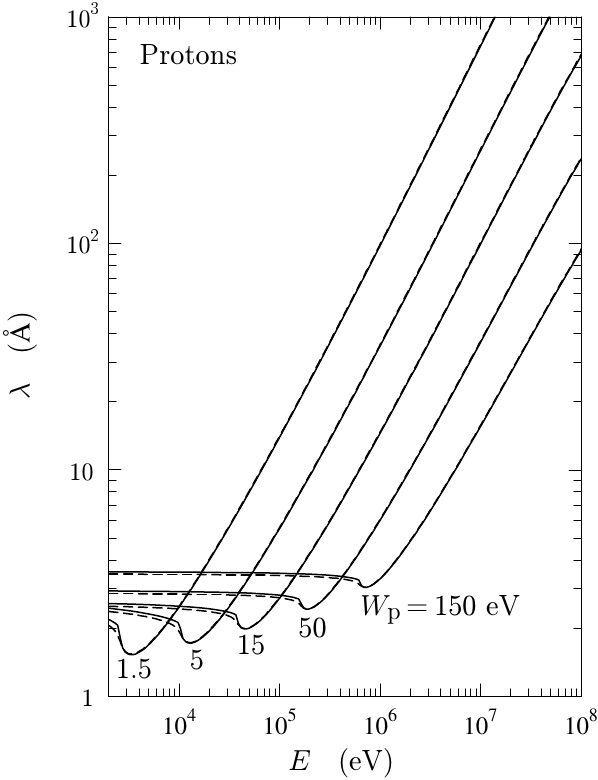} \\ [5mm]
\includegraphics*[width=7.5cm]{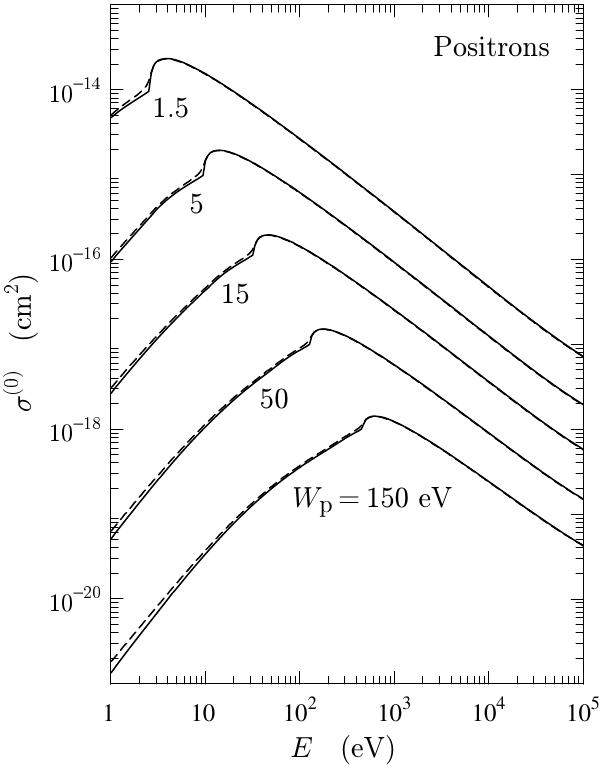} \rule{3mm}{0mm}
\includegraphics*[width=7.5cm]{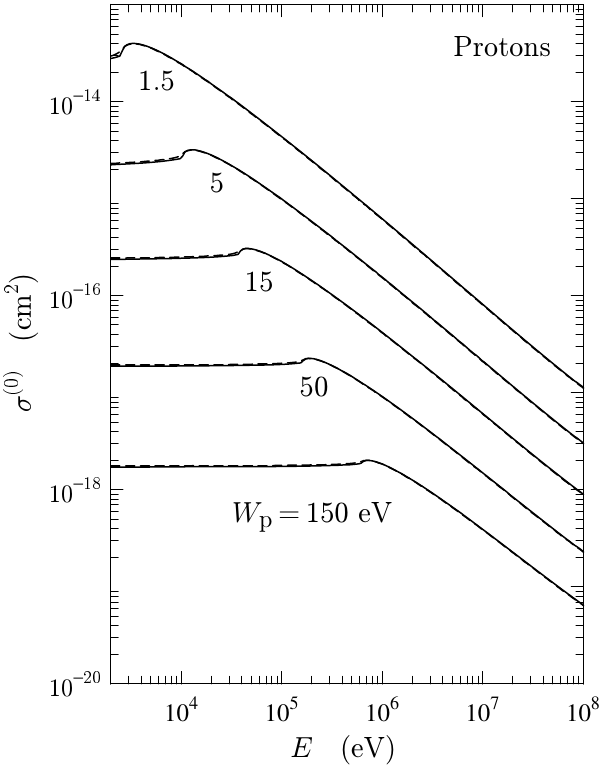}
\caption{Mean free paths $\lambda$ and total cross sections
$\sigma^{(0)}$ for positrons and protons in electron gases with the
indicated plasma resonance energies, $W_{\rm p}$, calculated from the
Lindhard DF (solid curves) and from the Mermin DF with $\Gamma=
0.05 W_{\rm p}$ (dashed curves), as functions of the kinetic energy
of the projectile.
\label{fig7.4}}
\end{center}\end{figure}

\begin{figure}[p!]
\begin{center}
\includegraphics*[width=7.5cm]{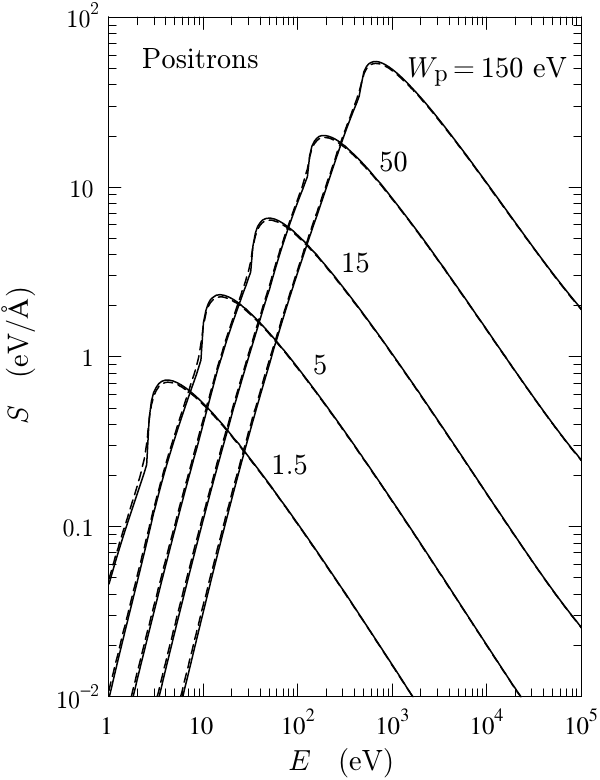} \rule{3mm}{0mm}
\includegraphics*[width=7.5cm]{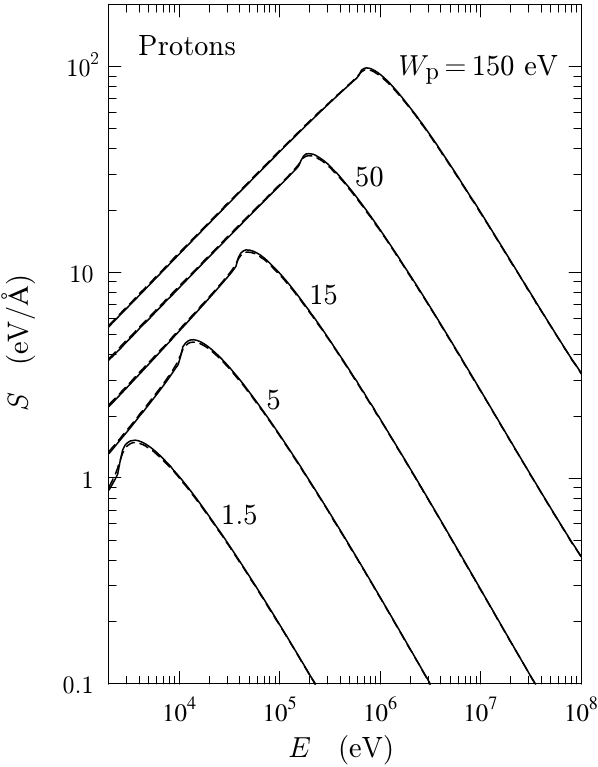} \\ [5mm]
\includegraphics*[width=7.5cm]{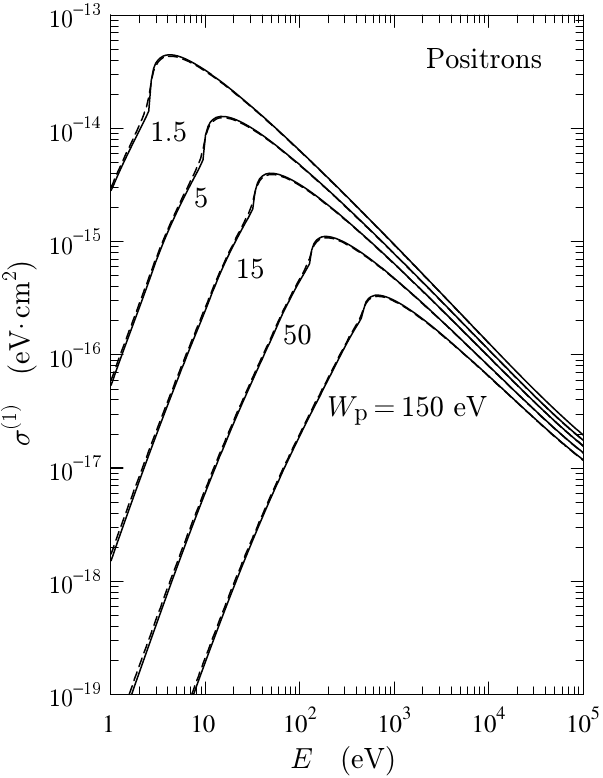} \rule{3mm}{0mm}
\includegraphics*[width=7.5cm]{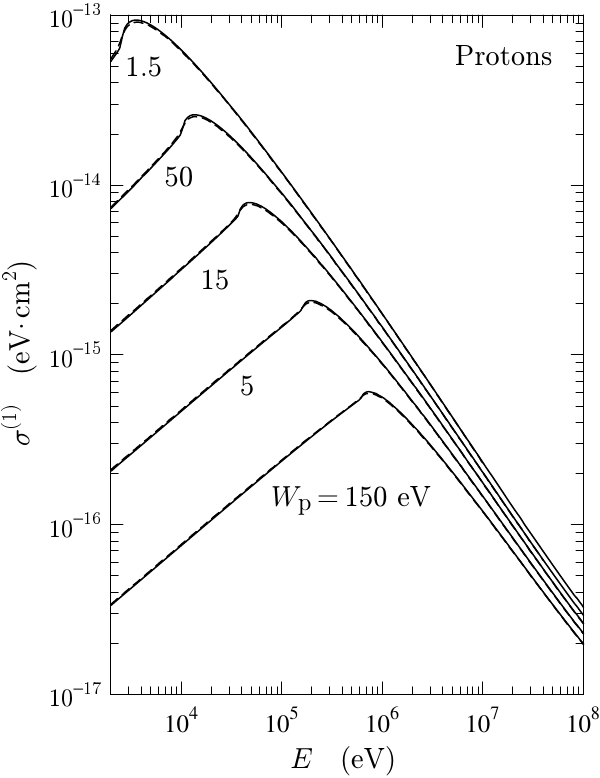}
\caption{Stopping powers $S$ and stopping cross sections
$\sigma^{(1)}$ for positrons and protons in electron gases with the
indicated plasma resonance energies, $W_{\rm p}$, calculated from the
Lindhard DF (solid curves) and from the Mermin DF with $\Gamma=
0.05 W_{\rm p}$ (dashed curves), as functions of the kinetic energy
of the projectile.
\label{fig7.5}}
\end{center}\end{figure}

\begin{figure}[p!]
\begin{center}
\includegraphics*[width=7.5cm]{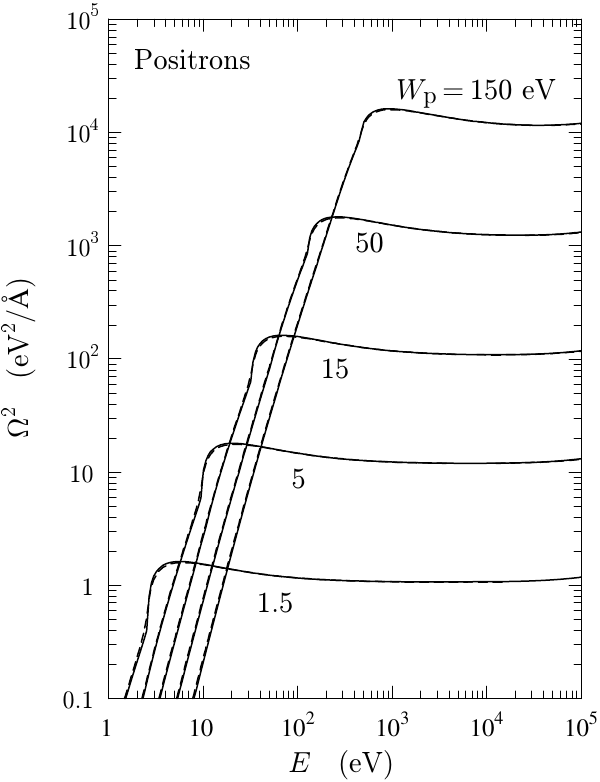} \rule{3mm}{0mm}
\includegraphics*[width=7.5cm]{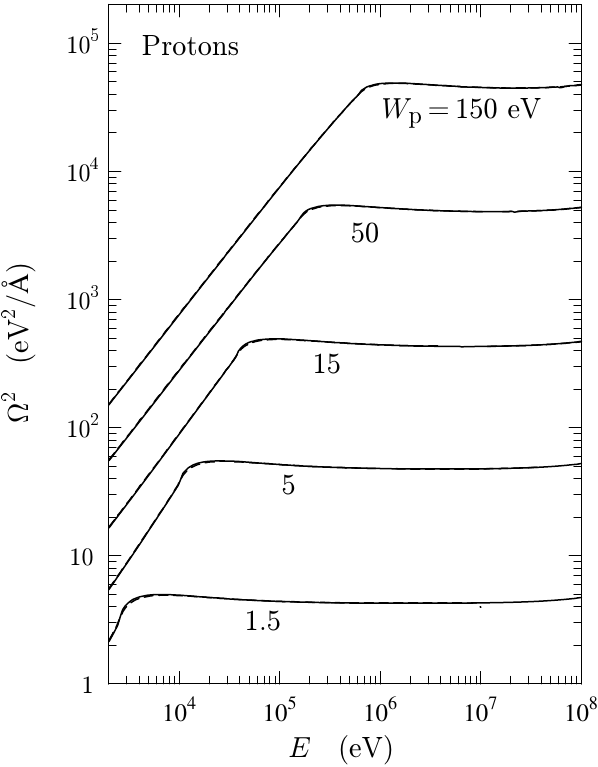} \\ [5mm]
\includegraphics*[width=7.5cm]{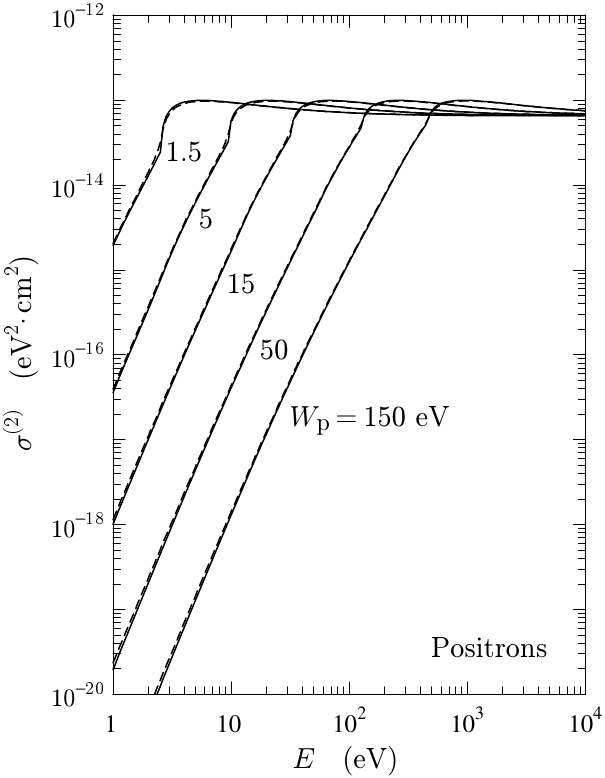} \rule{3mm}{0mm}
\includegraphics*[width=7.5cm]{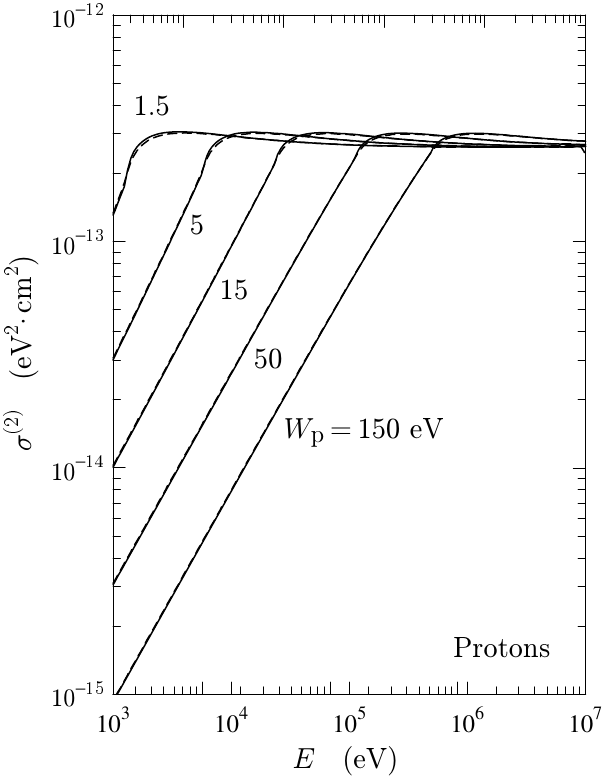}
\caption{Energy-straggling parameters $\Omega^2$ and cross sections
$\sigma^{(2)}$ for positrons and protons in electron gases with the
indicated plasma resonance energies, $W_{\rm p}$, calculated from the
Lindhard DF (solid curves) and from the Mermin DF with $\Gamma=
0.05 W_{\rm p}$ (dashed curves), as functions of the kinetic energy
of the projectile.
\label{fig7.6}}
\end{center}\end{figure}

\noindent Figures \ref{fig7.4} to \ref{fig7.6} display the quantities
	\req{7.133} for positrons and protons in electron gases with various
	plasma resonance energies as functions of the kinetic energy of the
	projectile, calculated from the Lindhard and Mermin DFs.  The damping
	energy parameter $\Gamma$ of the Mermin DFs was set equal to $0.05
	W_{\rm p}$. For the sake of clarity, in the case of positrons we have
	not accounted for the Bhabha correction [Eq.\ \req{6.236}].
	Differences between the results from the Lindhard and Mermin DFs are
	consistent with the tendency shown in Fig.\ \ref{fig7.3}.

The calculation of the energy-loss DCS and its integrals can be
simplified by noticing that transverse contributions are appreciable
only for high-energy projectiles. Then, we can disregard the dispersion
of the transverse plasmon line, which is indeed small (see Fig.\
\ref{fig7.2}). In addition, the contribution of high-$Q$ transverse
interactions can be accounted for with good approximation by replacing
the transverse GOS with the longitudinal GOS.


\section{The optical oscillator strength of materials \label{sec7.5}}
\index{optical oscillator strength of materials}

For concreteness, we consider a compound material with $Z$ electrons per
molecule and ${\cal N}$ molecules per unit volume. The plasma resonance
frequency corresponding to the average electron density in the material is
\index{plasma resonance frequency}
\beq
\Omega_{\rm p} = \sqrt{4\pi {\cal N} Z \frac{e^2}{\me}}.
\label{7.135}\eeq
We concentrate on the modeling of the longitudinal DF, which describes
the optical properties of the material and the stopping of
non-relativistic charged particles. The transverse DF, which occurs in
the description of the energy loss of relativistic charged particles,
could be treated in essentially the same way as the longitudinal DF.
However, stopping calculations make explicit use of the fact that the
longitudinal and transverse DFs are identical for both low and large $Q$
values, and the transverse DF is seldom needed.

A fundamental quantity in the classical theory of stopping of charged
particles in matter (see Chapter \ref{chapt6}) is the {\it dipole} or
{\it optical oscillator strength} (OOS) distribution,
\beq
F(W) \equiv \d f(W)/\d W\, ,
\label{7.136}\eeq
which is the optical limit ($Q \rightarrow 0$) of the generalized
oscillator strength (GOS) [see Section \ref{sec6.6}, Eq.\ \req{6.224}]
\beq
F(W) = \lim_{Q \rightarrow 0} \frac{\d f (Q,W)}{\d W}\, .
\label{7.137}\eeq
It is worth recalling that the OOS of atoms naturally splits into
contributions from the various electron shells, Eq.\ \req{6.223}.

Calculations of relativistic GOSs of atoms \citep{BoteSalvat2008} and of
the atomic photoeffect \citep[see, \eg][]{SabbatucciSalvat2016} show
that the OOS is proportional to the cross section for absorption of
unpolarized photons of energy $W$ calculated within the dipole
approximation,
\beq
F(W) =
\frac{\me c}{2 \pi^2 \hbar e^2} \,
\sigma_{\rm ph}^{\rm dipole}(W).
\label{7.138}\eeq
The cause of this connection between the OOS, a property of interactions
of charged particles, and the photoabsorption cross section is the fact
that the interaction of the projectile with the target electrons is
through the electromagnetic field. It should also be noticed that the
equality \req{7.138} holds only when the OOS and $\sigma_{\rm ph}^{\rm
dipole}(W)$ are calculated by using first-order perturbation theory.
That is, deviations may occur when higher-order (quadrupole, octupole,
\ldots) terms are appreciable.

Equivalently, instead of the OOS we can use the optical energy-loss
function (OELF) of the material (see Section \ref{sec1.2.4}),
\beq
\eta_2 (W) \equiv {\rm Im} \left( \frac{-1}{\epsilon(W)} \right)
= \frac{2 n(W)\kappa(W)}{
[n^2(W) + \kappa^2(W)]^2}\, ,
\label{7.139}\eeq
which is related to the OOS by [Eq.\ \req{6.251a}]
\beq
\eta_2 (W) = \frac{\pi (\hbar \Omega_{\rm p})^2}{2 Z} \, \frac{1}{W} \,
F(W).
\label{7.140}\eeq
It is convenient to regard the OELF as the primary optical function
because it is analytic for ${\rm Im}\, W \ge 0$ (see Section
\ref{sec1.5}) and, therefore, we can calculate from it the real part of
the inverse optical DF, $\eta_1(W)$, by using the Kramers--Kronig
relation \req{1.209a}. That is, knowledge of the OELF determines the
full inverse ODF, $\eta(W)=\eta_1(W)- {\rm i} \, \eta_2(W)$.

In practical numerical calculations, we shall use the OOS defined by a
table of values for a suitably spaced grid of excitation energies, $\{
W_i, F(W_i) \}$, which is built by assembling available theoretical and
experimental information on optical constants. The value of the OOS or
the OELF at an arbitrary $W$ is typically evaluated by linear log-log
interpolation or extrapolation on this table.


\subsection{The local-plasma approximation \label{sec7.5.1}}
\index{optical oscillator strength of materials!local-plasma
approximation}\index{local-plasma approximation}

In elementary studies, and when experimental optical data or
photoabsorption cross sections are not available, the OOS can be
estimated from knowledge of only the local density of electrons in the
medium by means of the {\it local-plasma approximation} (LPA), which was
first proposed by \citet{LindhardScharff1953} to estimate atomic or
molecular OOSs. In essence, this approximation assumes that under the
action of external electromagnetic fields, the electrons in a volume
element $\d {\bf r}$ of the medium, where the electron density is
$\rho({\bf r})$, react in the same way as if they were in a
free-electron gas of the same density.

We recall that the undamped Lindhard DF of an electron gas
yields the following OOS per electron in the gas [Eq.\ \req{7.95}],
\beq
\frac{\d f_{\rm eg}(\hbar\omega_{\rm p}; W)}{\d W}
= \frac{2}{\pi (\hbar\omega_{\rm p})^2}
\, W \, \eta_{2,{\rm eg}} (\hbar\omega_{\rm p} ; W)
= \delta(W-\hbar\omega_{\rm p}),
\label{7141}\eeq
where
\beq
\omega_{\rm p}(\rho) = \sqrt{4 \pi \rho e^2/\me}
\label{7.142}\eeq
is the plasma resonance frequency of an electron gas with density
$\rho$. Hence, according to the LPA picture, the OOS of the material
can be approximated as \citep{JohnsonInokuti1983}
\beq
F_{\rm LPA}(W) = \int \d {\bf r} \, \rho({\bf r}) \,
\delta\left[W - \tau \hbar \omega_{\rm p}(\rho({\bf r})) \right].
\label{7.143}\eeq
The parameter $\tau$, which defines a rescaling of the energy axis, is
introduced empirically with the purpose of ensuring that the OOS
\req{7.143}
yields, through the definition \req{6.288},
a value of the mean excitation
energy $I$ that agrees with empirical estimates (see Fig.\
\ref{fig6.11}). Qualitative considerations \citep{LindhardScharff1953,
Tung1988} indicate that the value of $\tau$ should be between 1 and $2$.
The integral in \req{7.143} extends over the volume of the unit cell for
solids, or over the volume of an atom or molecule in gases. As the
integral of the density over this volume equals the number of electrons,
the OOS \req{7.143} satisfies the dipole sum rule \req{1.216},
irrespective of the value of $\tau$.

The electron density of free atoms can be obtained from DHFS
self-consistent calculations (see Section \ref{sec3.5}). The same
method, with electron orbitals satisfying Wigner--Seitz boundary
conditions (Section \ref{sec3.5.1}), can be used to estimate the
electron density of elemental solids. These atomic electron densities
are spherically symmetric (in case of electronic configurations with
open subshells, \index{atomic configuration}
a spherical average is usually assumed). We may take
advantage of this symmetry to simplify the calculation of the integral
\req{7.143} as follows. Using the properties of the delta function, we can
write
\beqa
\delta\left[W - \tau \, \hbar \omega_{\rm p}(\rho(r))\right] &=&
\delta\left(W - \tau\,\sqrt{\frac{4\pi \hbar^2 e^2}{\me}\, \rho(r)}
\right)
\nonumber \\ [2mm]
&=& \frac{1}{\tau} \sum_i \sqrt{\frac{\me}{\pi \hbar^2 e^2} \rho(r_i)}\;
\frac{1}{\left|\rho'(r_i) \right|}\,
\delta(r-r_i),
\label{7.144}\eeqa
where the summation is over the roots $r_i$ of the equation
\beq
\frac{W}{\tau} - \sqrt{\frac{4\pi \hbar^2 e^2}{\me} \rho(r_i)} =0.
\label{7.145}\eeq
As the atomic electron density $\rho(r)$ decreases monotonically with
$r$, this equation normally has a single root. We thus have,
\beqa
F_{\rm LPA}(W) &=& 4\pi
\int \d r \,/ r^2 \, \rho(r) \,
\delta[W - \tau \hbar \omega_{\rm p}(\rho(r))]
\nonumber \\ [2mm]
&=& \frac{4}{\tau} \, \sqrt{\frac{\pi \me}{\hbar^2 e^2}} \sum_i \; r_i^2\,
\frac{[\rho(r_i)]^{3/2}}{\left|\rho'(r_i) \right|}\, .
\label{7.146}\eeqa
The LPA tends to approximate the gross features of optical functions
obtained from experiments and from more elaborate theoretical
calculations, although the numerical values may differ substantially
from accurate data. Figure \ref{fig7.7} displays OELFs obtained from the
LPA for solid aluminum and silicon, using DHFS atomic electron densities
calculated with Wigner--Seitz boundary conditions. Since the electron
density is finite within the Wigner--Seitz sphere, the OOS presents a
lower edge that corresponds to the density at the surface of the sphere.
The adopted values of $\tau$ were 1.31 and 1.41 for Al and Si,
respectively.

\begin{figure}[hp!] \begin{center}
\includegraphics*[width=7.25cm]{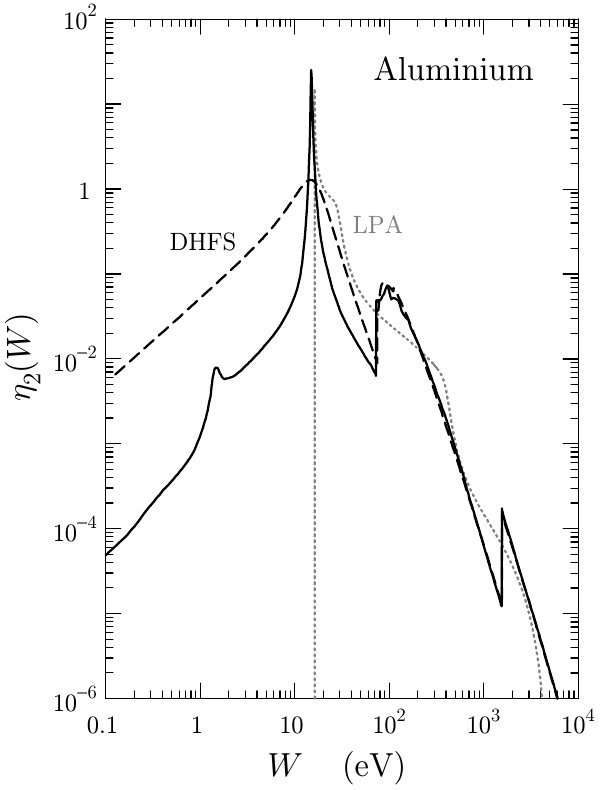} \rule{5mm}{0mm}
\includegraphics*[width=7.25cm]{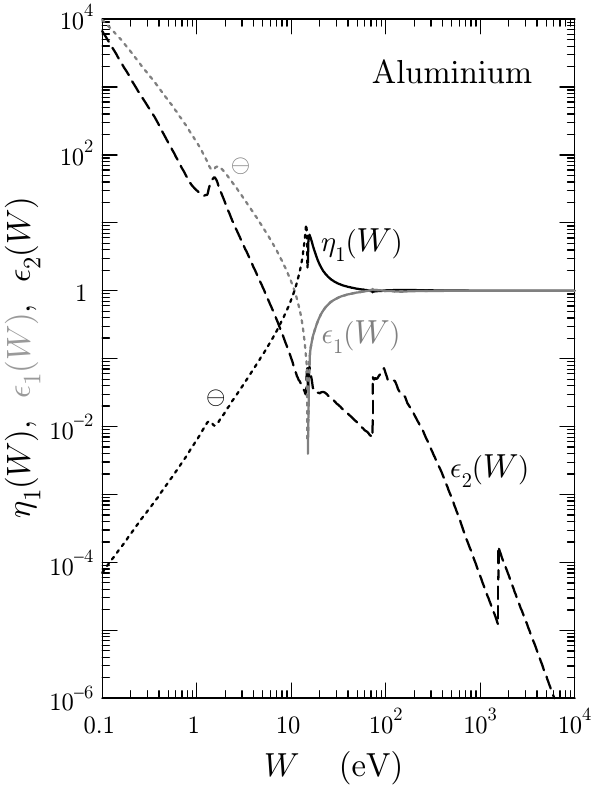} \\ [5mm]
\includegraphics*[width=7.25cm]{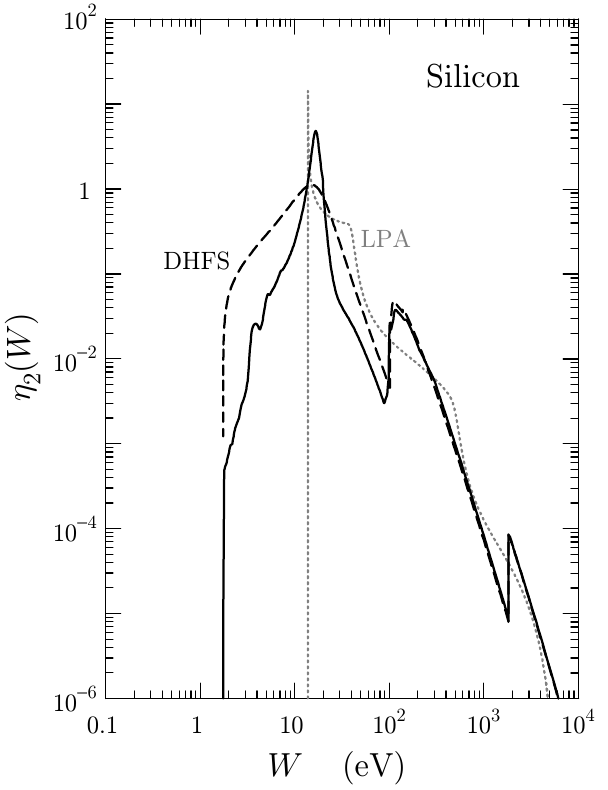} \rule{5mm}{0mm}
\includegraphics*[width=7.25cm]{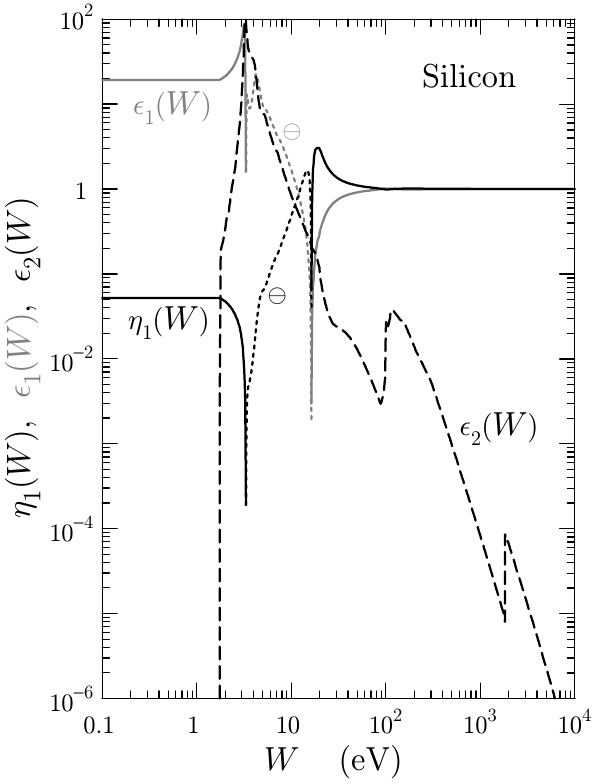}
\caption{Optical DFs of aluminum and silicon. The solid curves in the
left plots show the OELFs obtained from experimental optical data and
atomic photoabsorption cross sections; the dashed curves are OELFs
derived from the DHFS-model OOSs (Section \ref{sec7.5.2}); and the dotted
gray curves are OELFs obtained from the local-plasma
approximation, Eq.\ \req{7.143}, with the DHFS electron densities of
Wigner--Seitz atoms. The dotted curves in the right plots represent
negative values of the functions.
\label{fig7.7}}
\end{center}\end{figure}


\subsubsection{The OOS of Thomas--Fermi atoms \label{sec7.5.1.1}}

\index{optical oscillator strength of Thomas--Fermi atoms}
The simplest theoretical framework to obtain approximate atomic electron
densities is the Thomas--Fermi statistical model (see Section
\ref{sec3.4}), which is built on the same basic assumption as the LPA,
namely, that electrons in the atom behave as if they were part of a
homogeneous electron gas with the local electron density $\rho(r)$ that
is bound by the screened Coulomb potential of the nucleus. This
approach, commonly referred to as the {\it statistical model} (sm), is
useful, \eg, for understanding the variation with $Z$ of basic stopping
properties.

Because the numerical Thomas--Fermi electron density is inconvenient to
work with, in the calculations we shall use the analytical approximation
proposed by \citet{Moliere1947} [Eq.\ \req{3.113}]
\beq
\rho(r) = \frac{Z}{4\pi b^{2} r}
\left[ 3.6 \exp({-6 \, r/b}) + 0.792 \exp(-1.2 \, r/b)
+ 0.0315 \exp(-0.3\, r/b) \right],
\label{7.147}\eeq
where $b=0.88534 a_0 Z^{-1/3}$ is the Thomas--Fermi radius [Eq.\
\req{3.94}].
As the Thomas--Fermi--Moli\`{e}re
density decreases monotonically with $r$, the atomic OOS resulting from the
statistical model can be expressed as
\beq
F_{\rm sm} (Z,\tau;W) =
\frac{4}{\tau} \, \sqrt{\frac{\pi \me}{\hbar^2 e^2}} \; r_0^2\,
\frac{[\rho(r_0)]^{3/2}}{\left|\rho'(r_0) \right|}\, \qquad \mbox{with}
\qquad
\rho(r_0) = \frac{\me}{4\pi \hbar^2 e^2} \, \frac{W^2}{\tau^2}.
\label{7.148}\eeq
Introducing the reduced variable $x=r/b$, and the Moli\`{e}re TF
density, we can write
\beq
F_{\rm sm} (Z,\tau;W)  =
\frac{1}{\tau} \, \sqrt{\frac{4 \me}{\hbar^2 e^2} \, Z b^3} \; H(x_0)
= \frac{1}{\tau} \, \frac{3\pi}{2^{5/2}} \, \frac{\hbar^2}{\me e^4}
\; H(x_0) \, ,
\label{7.149}\eeq
where
\beq
H(x) = \left( x \sum_{n} A_n a_n^2 \exp(-a_n x) \right)^{3/2}
\left[ \sum_{n} A_n a_n^2 \left( x^{-1} + a_n
\right) \exp(-a_n x) \right]^{-1}
\label{7.150}\eeq
and $x_0$ is the root of the equation
\beq
\frac{1}{x_0} \sum_{n} A_n a_n^2 \exp(-a_n x_0) =
\frac{b^3}{Z}
\frac{\me}{\hbar^2 e^2} \, \frac{W^2}{\tau^2}
= \frac{(3\pi)^{2}}{2^{7} \tau^2}
\, \left( \frac{W}{Z E_{\rm h}} \right)^2,
\label{7.151}\eeq
where $E_{\rm h}$ is the Hartree energy (Appendix \ref{appC}).

Because the TF model has a length scale proportional to $Z^{-1/3}$, the
electron density scales like $Z^2$ [Eq.\ \req{3.109a}],
\beq
\rho(Z;r) = Z^2 \rho(1;Z^{1/3} r),
\label{7.152}\eeq
and the plasma resonance energy $W_{\rm p}(r) = \hbar \omega_{\rm
p}(\rho(r))$ scales like $Z$,
\beq
W_{\rm p}(Z; {\bf r}) \equiv \sqrt{\frac{4\pi \hbar e^2}{m} \,
\rho(Z ; {\bf r})} =
Z \, W_{\rm p}(1 ; Z^{1/3} {\bf r}).
\label{7.153}\eeq
These scaling properties imply that the atomic OOS is a universal
function of the variable $W/Z$. We have
\beqa
&& \! \! \! \! \! \! \! \! \! \! \! \! \! \! \! \! \! \! \!
F_{\rm sm}(Z, \tau ; W)
= Z^2 \int \d {\bf r} \, \rho(1; Z^{1/3} {\bf r}) \,
\delta\left[W - \tau Z W_{\rm p}(1; Z^{1/3} {\bf r}) \right]
\nonumber \\ [2mm]
&& = Z \int \d {\bf r}' \, \rho(1; {\bf r}') \,
\delta\left[W - \tau Z W_{\rm p}(1; {\bf r}') \right]
\quad [\mbox{changing variable to ${\bf r}'=Z^{1/3}{\bf r}$}]
\nonumber \\ [2mm]
&& = \frac{1}{\tau} \int \d {\bf r}' \, \rho(1; {\bf r}') \,
\delta\left[\frac{W}{\tau Z} - W_{\rm p}(1; {\bf r}') \right]
\nonumber \\ [2mm]
&& = \frac{1}{\tau} F_{\rm sm}\left( 1,1; \frac{W}{\tau Z} \right),
\label{7.154}\eeqa
where $F_{\rm sm}\left(1, 1; W \right)$ is the statistical-model OOS
calculated with $Z=1$ {\it and} $\tau=1$. The OOS \req{7.154} has been
utilized by \citet{JacksonMcCarthy1972} in a generic calculation of the
$Z_1^3$ correction to the Bethe stopping power [see Section
\ref{sec8.4}].

As indicated above, the statistical-model OOS satisfies the $f$-sum rule
\beq
\int_0^\infty F_{\rm sm} (Z,\tau;W)  \, \d W = Z.
\label{7.155}\eeq
The mean excitation energy $I$, the primary parameter occurring in the
Bethe formula for the stopping power (see Section \ref{sec6.9}), is
defined by [Eq. \req{6.288}]
\beq
\ln I = \frac{1}{Z}
\int_0^\infty \ln W \, F_{\rm sm} (Z,\tau;W)
\, \d W.
\label{7.156}\eeq
The $I$ value obtained from the statistical model is
\beqa
\ln I &=& \frac{1}{Z \tau}
\int_0^\infty \ln(W) \, F_{\rm sm} \left(1,1; \frac{W}{Z\tau} \right)  \, \d W
\nonumber \\ [2mm]
&=&
\int_0^\infty \ln(W'Z\tau) \, F_{\rm sm} \left( 1,1; W' \right) \, \d W'
\qquad [\mbox{changing variable to $W'=W/(Z\tau)$}]
\nonumber \\ [2mm]
&=& \left[ \ln(Z\tau) + \ln I^{(1)}_{\rm sm} \right]
= \ln \left( Z \tau I^{(1)}_{\rm sm}
\right),
\label{7.157}\eeqa
where $I^{(1)}_{\rm sm}$ is the mean excitation energy calculated for
$Z=1$ and $\tau=1$,
\beq
\ln I^{(1)}_{\rm sm} =
\int_0^\infty \ln(W) F_{\rm sm} \left( 1,1; W \right)  \, \d W.
\label{7.158}\eeq
A numerical integration gives $I^{(1)}_{\rm sm} = 6.322$ eV.  Thus, the
statistical model yields $I = Z \tau \times  6.322$ eV. With the value
$\tau=2^{1/2}$ we obtain $I/Z = 8.941$ eV, which is reasonably close to
the value 8.9 eV obtained by \citet{LindhardScharff1953} from the
electron density of the Thomas--Fermi atom instead of the Moli\`{e}re
approximation to it. These authors also considered the atomic electron density
obtained from the Lenz--Jensen screening function [Eq.\ \req{3.110a}].
With the value $\tau=2^{1/2}$, the Lenz--Jensen model gives $I/Z=10.7$ eV,
quite close to the average of the empirical $I$ values which is
about $10.5$ eV [see Fig.\ \ref{fig6.11}]. In
practical applications, the value of $\tau$ can be fixed by requiring
that the statistical model OOS reproduces the empirical $I$ value of the
material. This requirement gives
\beq
\tau = \frac{I}{Z \, I^{(1)}_{\rm sm}}.
\label{7.159}\eeq
As indicated above, the resulting value of $\tau$ usually lies between 1
and 2.


\subsection{OOSs calculated from the DHFS potential
\label{sec7.5.2}}
\index{optical oscillator strength of materials!DHFS shell contributions}

A better approach to obtain approximate OOSs of materials is to use
theoretical OOSs resulting from numerical self-consistent calculations
for free atoms. \citet{Salvat2022a} have calculated extensive
numerical tables of the GOS and the TGOS of all subshells of neutral
atoms ($Z=1$--99) using the independent-electron model with the DHFS
self-consistent potential (see Section \ref{sec3.5}).

At excitation energies higher than a few hundred eV, the refractive
index tends to unity and the extinction coefficient is small. Under
these circumstances we can use the identities \req{1.140} and
\req{1.214} to conclude that
\beq
F(W) \simeq \frac{\me c}{2 \pi^2 \ \hbar e^2} \,
\sigma_{\rm ph}(W),
\label{7.160}\eeq
where $\sigma_{\rm ph}(W)$ is the molecular cross section for
photoelectric absorption of photons of energy $W$. The recipes
\req{7.138}
and \req{7.160} are not equivalent. The identity \req{7.138} is justified
by exact results from the PWBA at $Q=0$ and it may be expected to be
generally valid. Consequently, Eq.\ \req{7.160} should be used only when
the dipole approximation for the photoeffect is applicable, that is,
only when the wavelength of the photon, $\lambda_{\rm ph}= 2\pi \hbar
c/E$ is much larger than the atomic radius \citep[see, \eg,][page
181]{BransdenJoachain1983}. The differences between $\sigma_{\rm
ph}(W)$ and its dipole approximation may introduce small departures from
the $f$-sum rule, which add to those arising from relativistic
corrections.

The calculated atomic OOSs account for transitions of individual
electrons from their initial orbitals to final orbitals with positive
energy (ionization) and to unoccupied bound orbitals with negative
energies (excitation). The contribution from excitations to bound states
is described as a series of discrete resonances (delta functions).
Because the subshell ionization energies $U_a = -\epsilon_{n_a\kappa_a}$
obtained from the DHFS potential differ slightly from the experimental
ionization energies (see Fig.\ \ref{fig3.5}), we have prepared a
database of subshell OOSs that are shifted in energy to the correct
(empirical) ionization energies given by \citet{Carlson1975}. This
database contains OOS tables for the subshells of the elements ($Z=1$ to
99) that cover the whole range of excitation energies where the OOS is
appreciable. For the purposes of stopping calculations, excitations to
bound atomic levels must be taken into account to ensure that the
resulting OOS are compliant with the dipole or $f$-sum rule. Because the
fine details of the excitation spectrum are not of much importance, the
contribution of discrete excitations to the OOS have been represented
approximately by extending the ionization OOS to excitation energies
below the ionization threshold. Explicitly, the OOS of the subshell
$a=(n_a\kappa_a)$ of an atom with atomic number $Z$ was obtained as
\beq
F_a(Z;W) = \left\{
\begin{array}{ll}
F_a^{\rm ion} (Z;W) & \mbox{if $U_a \le W$,} \\ [2mm]
F_a^{\rm ion} (Z;U_a) \rule{5mm}{0mm} &
\mbox{if $U'_a \le W < U_a$,} \\ [2mm]
0 & \mbox{if $W < U'_a$,}
\end{array} \right.
\label{7.161}\eeq
where $F_a^{\rm ion} (Z;W)$ is the OOS for ionization of the subshell,
as calculated from the DHFS potential \citep{BoteSalvat2008,
Salvat2022a}, $U_a$ is the binding energy of the subshell, and the
cutoff energy $U'_a$ is such that the product $(U_a-U'_a) F_a^{\rm ion}
(Z; U_a)$ equals the sum of OOSs for excitations to discrete levels. For
the outmost subshells the cutoff energy so defined may be less than
$0.5\, U_a$; in this case, the recipe \req{7.161} is modified by setting
$U'_a \simeq 0.5 U_a$, and defining the constant OOS in the interval
$(U'_a,U_a)$ so that the $f$-sum value of the subshell is preserved.

In the case of a monoatomic gas of the element with atomic number $Z$,
the OOS can be approximated in terms of the subshell OOSs in the
database as
\beq
F_{\rm atom}(Z;W) = \sum_a F_a(Z;W) \, ,
\label{7.162}\eeq
where the summation runs over the various electron subshells of the atom
in its ground state configuration. This atomic OOS deviates slightly
from the dipole sum rule,
\beq
\int_0^\infty F_{\rm atom}(Z;W) \, \d W = Z,
\label{7.163}\eeq
because of relativistic corrections \citep{Levinger1957, CohenLeung1998}.
The theoretical study of \citet{Cohen2003} and the numerical
calculations by \citet{Salvat2022a} show that the integral
\req{7.163} of the OOS is slightly less than $Z$; the deviation increases
monotonically with the atomic number and is about 2.5 \% for einsteinium
($Z=99$). Since the dipole sum rule is instrumental in the derivation of
the stopping power formula (Section \ref{sec6.9}), we re-normalize the
OOS to fulfill that sum rule. This modification ensures that the GOS
obtained with the optical-data model described below does satisfy the
Bethe sum rule, which is known to be correct for sufficiently large
recoil energies. In addition, to assure agreement with stopping powers
obtained from the high-energy Bethe formula, we shall rescale the
excitation energies so as to reproduce the empirical value of the mean
excitation energy $I$. That is, we consider the OOS
\beq
F(Z;W) = a_1 a_2 F_{\rm atom}(Z;a_2W)
\label{7.164}\eeq
with the constants $a_1$ and $a_2$ determined from the conditions
\begin{subequations}
\label{7.165}
\beq
Z = \int_0^\infty F(Z;W) \, \d W
= a_1 \int_0^\infty F_{\rm atom}(Z;W') \, \d W'
\qquad [W'=a_2 W]
\label{7.165a}\eeq
and
\beqa
\ln I &=& \frac{1}{Z} \int_0^\infty \ln W \, F(Z; W) \, \d W
\nonumber \\ [2mm]
&=& \frac{1}{Z} \, a_1 \int_0^\infty \ln (W'/a_2) \,
F_{\rm atom}(Z; W') \, \d W' \qquad [W'=a_2W]
\nonumber \\ [2mm]
&=& - \ln a_2 + \frac{a_1}{Z} \int_0^\infty \ln W' \,
F_{\rm atom}(Z; W') \, \d W'.
\label{7.165b}\eeqa
\end{subequations}

The recipe \req{7.162} is not suited for molecular gases and condensed
materials, because the wave functions of electrons in outer subshells
are strongly affected by atomic aggregation. In addition, the presence
of neighboring atoms modifies the final-state orbitals of the active
electron. These complications give rise to the near-edge and extended
x-ray absorption fine structure \citep[see, \eg,][]{RehrAlbers2000},
which manifests as generally weak oscillatory wiggles of the OOS and of the
photoabsorption cross section, which average to a smooth atom-like
shape. Because of the lack of more accurate models, the free-atom
subshell OOS \req{7.161} will be used to approximate the
contribution of inner subshells with binding energies larger than about
50 eV, but the OOS of electrons in weakly bound outer subshells
must be modeled independently.

When more reliable theoretical or empirical information is not
available, an OOS model suited for stopping calculations can be built
from the subshell OOSs $F_a(Z;W)$ in the database as
follows. For the sake of generality, let us consider the case of a
compound with molecules consisting of $n_i$ atoms of the element with
atomic number $Z_i$ ($i=1$,2, \ldots). The contributions from inner
subshells with binding energies $U_a$ larger than a certain threshold
value $W_{\rm th}$ of the order of 50 eV are approximated by the
free-atom form \req{7.161}.
The OOS of electrons in outer subshells with binding energies $U_a <
W_{\rm th}$ is represented by a single damped oscillator
with resonance energy $W_r$ and damping constant $\Gamma$. In the case
of insulators and semiconductors, an energy gap $W_{\rm g}$ may be introduced
by considering the frequency shift of \citet{LevineLouie1982} (see
Section \ref{sec7.3}). Explicitly, we set
\beq
F_{\rm out}(W) = f_{\rm out} \, \frac{2 \Gamma}{\pi} \, \frac{W
\sqrt{W^2 - W_{\rm g}^2}}{\left( W_{\rm r}^2 + W_{\rm g}^2 -W^2
\right)^2+ \Gamma^2 (W^2-W_{\rm g}^2)} \, {\cal S}(W-W_{\rm g}),
\label{7.166}\eeq
where $f_{\rm out}$ is the effective number of electrons in outer
subshells of the atoms in the molecule. That is, the model OOS is
obtained as
\beq
F(W) = F_{\rm out}(W) +
{\sum}_i n_i \left( {\sum}'_a F_a(Z_i;W) \right) \, \Theta(W-W_{\rm th}),
\label{7.167}\eeq
where the first summation extends over the various elements $Z_i$ in the
molecule, and the second summation runs over the inner subshells of the
element $Z_i$. Notice that the OOSs of inner subshells are truncated at
$W_{\rm th}$.

\index{DHFS-model OOS}
The effective number $f_{\rm out}$ of electrons in outer
subshells is obtained by requiring that the  dipole sum rule is satisfied,
\ie,
\beq
f_{\rm out} = Z - \int_{W_{\rm th}}^\infty \left[
{\sum}_i n_i \left( {\sum}'_a F_a(Z_i;W) \right) \right]
\d W,
\label{7.168}\eeq
where $Z$ is the total number of electrons in a molecule. The resonance
energy $W_{\rm r}$ of outer electrons is set equal to the plasma
resonance energy
of an electron gas with the average density of outer electrons,
\beq
W_{\rm r} = \sqrt{\hbar^2 \Omega^2_{\rm p} f_{\rm out}/Z}.
\label{7.169}\eeq
The gap energy $W_{\rm g}$ is defined by the user, possibly guided by
experimental information, \eg, from electron energy-loss experiments.
Finally the damping constant $\Gamma$ is fixed by requiring that the mean
excitation energy $I$ obtained from the model OOS,
\beq
\ln I = \frac{1}{Z} \int_0^\infty \ln (W) \, F(W) \, \d
W\, ,
\label{7.170}\eeq
coincides with the empirical $I$ value of the material \citep[as given,
\eg, in the][]{ICRU37}. The resulting OOS has a realistic appearance for
large energy transfers $W$, it satisfies the $f$-sum rule, and it yields
the adopted empirical $I$ value. In the following it will be referred to
as the {\it DHFS-model OOS}.


\subsection{Empirical OOSs derived from optical data \label{sec7.5.3}}
\index{optical oscillator strength of materials!empirical}

The most elaborate optical-data models use OOS [or OELFs $\eta_2(W)$, as
defined by Eq.\ \req{7.139}] obtained from experimental information on
optical constants. The required information can be inferred from
measurements with monochromatized synchrotron radiation \citep[see,
\eg,][]{Shiles1980}. The main source of measured optical data is the
{\it Handbook of Optical Constants of Solids} \citep{Palik1985,
Palik1991, Palik1998}, which includes tables of optical constants for a
number of metals, semiconductors and insulators.  These tables contain
the refractive index $n(W)$ and the extinction coefficient $\kappa(W)$
measured with different methods, frequently by various groups and with
various degrees of accuracy. They cover a range of excitation energies
$W=\hbar \omega$ from about $10^{-3}$ eV up to an upper energy that
depends on the material, typically about 100 eV. Alternatively,
empirical OELFs of solids can be derived from electron energy-loss
measurements \citep[see, \eg,][]{Egerton2011}. For solids for which
experimental information is not available, approximate ODFs could be
obtained from density-functional theory calculations \citep[see,
\eg,][]{Werner2009}.

\index{additivity approximation} \begin{subequations} \label{7.171}
The tables of photon interaction coefficients of \citet{Henke1993,
Henke2010}, which were obtained from a compilation of experimental data
and calculations, cover the energy range from 30 eV to 30 keV for the
elements with $Z=1$ to 92. The atomic photoabsorption cross sections,
$\sigma_{\rm ph}(Z;W)$, derived from these tables may be used to
estimate the molecular cross section by means of the additivity
approximation. The corresponding OOS is [see Eq.\ \req{7.138}]
\beq
F(W) = \frac{\me c}{2 \pi^2 \hbar e^2} \,
\sum_{i} n_i \, \sigma_{\rm ph}(Z_i;W),
\label{7.171a}\eeq
where the summation is over the elements in a molecule. The empirical
OOS can thus be extended up to energies of the order of 10 keV. At these
energies and above, the refractive index approaches unity and the OOS
can be approximated by calculated atomic OOSs [Eq.\ \req{7.162}],
\beq
F(W) = \sum_i n_i F_{\rm atom}(Z_i;W).
\label{7.171b}\eeq
Unfortunately atomic calculations disregard aggregation effects. A
better solution would be using OOSs derived from measured photoelectric
cross sections or electron energy-loss spectra, which naturally include
aggregation effects, but they are not generally available.
\end{subequations}

It should be borne in mind that optical data obtained from both
experimental measurements and theoretical calculations are affected by
considerable distortions which arise, respectively, from experimental
uncertainties and from the adopted theoretical approximations. These
distortions may become evident when comparing results from different
experiments and calculations, and also as departures from the sum rules
obeyed by the optical functions (see Section \ref{sec1.5.2}). The most
relevant of these are the $f$-sum rule,
\beq
\int_0^\infty W \, \eta_2 (W) \, \d W =
\frac{\pi}{2} (\hbar \Omega_{\rm p})^2,
\label{7.172}\eeq
and the perfect-screening sum rule,
\beq
\eta_1(0) + \frac{2}{\pi}
\int_{0}^\infty \frac{\eta_2(W)}
{W} \, \d W = 1 .
\label{7.173}\eeq
For an electric conductor $\eta_1(0)=0$, whereas for an insulator
$\eta_1(0)$ is real and positive. In practice, to satisfy these sum
rules and to ensure consistency with the Kramers--Kronig relations,
sensible and generally small modifications to the collected data may
have to be introduced \citep[see, \eg, ][]{Shiles1980}.

Figure \ref{fig7.7} displays the optical functions for solid aluminum
and silicon, a conductor and a semiconductor. The OELFs were obtained by
combining the refractive indexes and extinction coefficients given in
the Handbook by \citet{Palik1985, Palik1991, Palik1998} with calculated
OOSs \citep{Salvat2022a} at higher energies. The empirical OELF is
renormalized so that the $f$-sum rule is satisfied. The functions
$\eta_1(W)$ displayed in the plots were obtained from the composite OELF
$\eta_2(W)$ by using the Kramers--Kronig relation \req{1.209a},
\beq
\eta_1(W) = 1 - \frac{2}{\pi} {\cal P}
\int_{0}^\infty \frac{W' \, \eta_2(W')}
{W'^2-W^2} \d W',
\label{7.174}\eeq
Notice that this calculation scheme implies that the perfect-screening
sum rule \req{7.173} is automatically satisfied (apart from accumulated
numerical errors). For the sake of completeness, the plots also include
the real and imaginary parts of the ODF,
\beq
\epsilon(W) = \frac{1}{\eta(W)}
= \epsilon_1(W) + {\rm i} \, \epsilon_2(W)\, .
\label{7.175}\eeq
For comparison purposes, the plots also display the OELFs obtained from
calculated atomic DHFS-model OOSs (Section \ref{sec7.5.2}).

The curves labeled LPA in Fig.\ \ref{fig7.7} are the results from the
local-plasma approximation (see Section \ref{sec7.5.1}), calculated from
DHFS self-consistent atomic electron densities computed under
Wigner--Seitz boundary conditions (Section \ref{sec3.5.1}), to account
approximately for the effect of aggregation on the electron density. It
is interesting to notice that the structure of the OELF resulting from
the LPA with the DHFS electron density grossly resembles that of the
empirical function, bending visibly near the absorption edges of inner
subshells. This structure is not present in the OELFs calculated from
the LPA with the Thomas--Fermi electron density, which does not account
for shell effects \citep{JohnsonInokuti1983}.

\index{DHFS-model OOS}

To illustrate the difficulties encountered in building the optical DFs
of materials, we consider the case of aluminum oxide, Al$_2$O$_3$. The
refractive index and the extinction coefficients of this material were
tabulated by \citet{Gervais1991} in the second volume of the Handbook of
Optical Constants \citep{Palik1991} for photon energies up to 100 eV. In
the third volume of that Handbook \citep{Palik1998},
\citet{TropfThomas1998} extended the tables by adding results from new
measurements by various authors, including information on the anisotropy
of the crystalline state, and extending the energy range of the
tabulation up to 140 eV. However, the various sets of data in the latter
tables differ considerably and, consequently, it is not easy to combine
them to get optical functions that vary continuously with energy. After
analyzing various options, we adopted the values in the tables of Tropf
and Thomas in the low-energy range up to 40 eV, were we took the
refractive index and the extinction coefficient of the ordinary ray,
disregarding the small optical anisotropy. For energies below 27 eV,
where Tropf and Thomas provide data from three different sources, we
gave preference to the set of values from \citet{Tomiki1993}, which are
systematically higher than the other two sets. From this composite data
set, we determined the OELF for photon energies up to 40 eV. The OELF
for energies $W$ higher than 121 eV, the aluminum L$_1$ absorption
edge, was derived from calculated DHFS-model OOSs
\citep{Salvat2022a} using the identity \req{7.171b}. The gap
between 40 and 120 eV, was filled by adding an extra grid point at 70
eV, and the corresponding value of $\eta_2(W)$ was determined so that
the $f$-sum rule is satisfied. The function $\eta_1(W)$ was calculated
from $\eta_2(W)$ by using the Kramers--Kronig relation \req{7.174}.

\index{DHFS-model OOS}

\begin{figure}[hp!] \begin{center}
\includegraphics*[width=7.25cm]{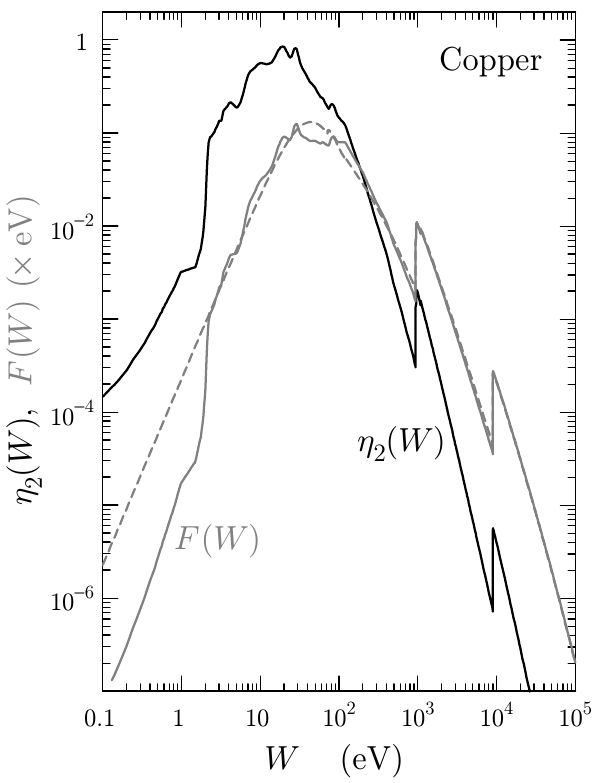} \rule{5mm}{0mm}
\includegraphics*[width=7.25cm]{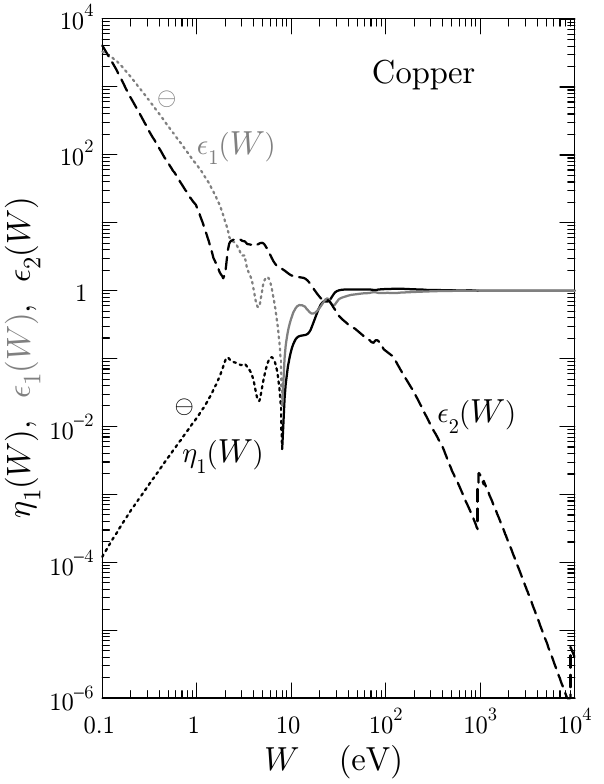} \\ [5mm]
\includegraphics*[width=7.25cm]{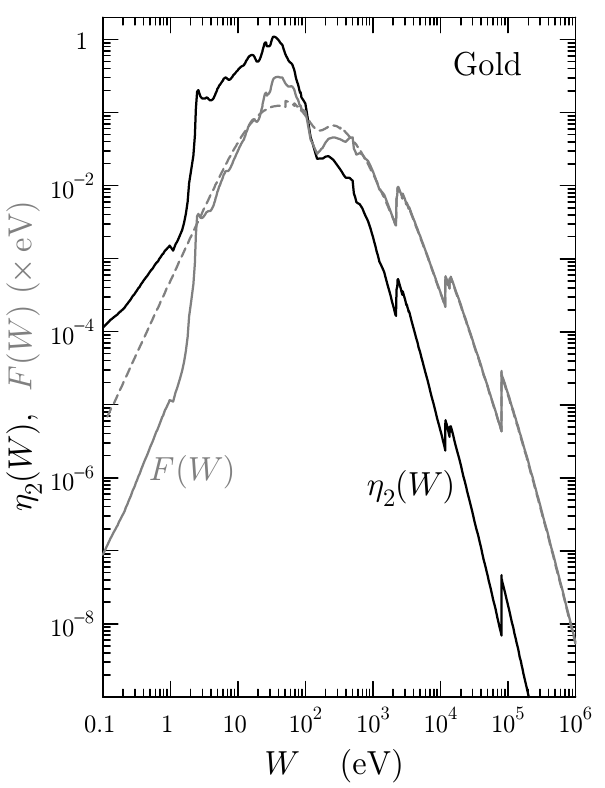} \rule{5mm}{0mm}
\includegraphics*[width=7.25cm]{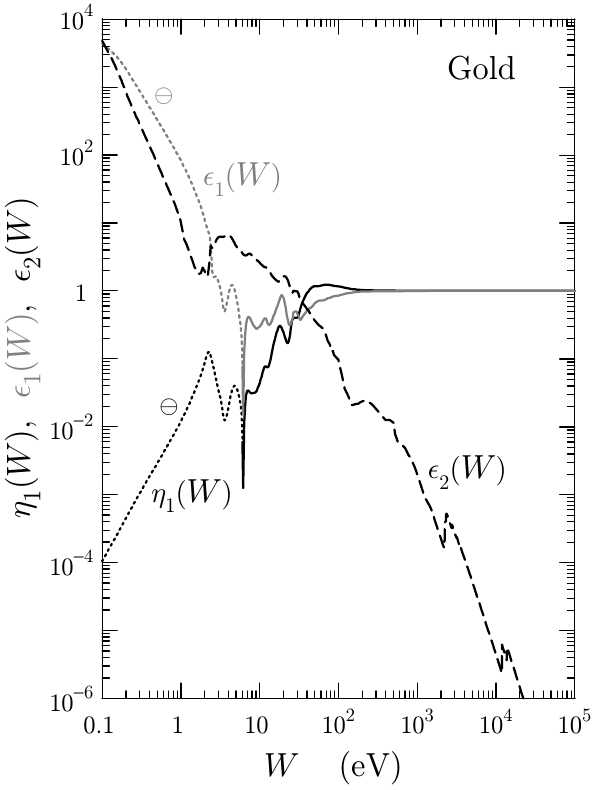}
\caption{Optical DFs of solid copper and gold. The dotted curves in the
right plots represent negative values of the functions. The dashed
curves in the left plots are OOSs built from subshell OOS calculated
from the DHFS potential, Eq.\ \req{7.162}.
\label{fig7.8}}
\end{center}\end{figure}

\begin{figure}[hp!] \begin{center}
\includegraphics*[width=7.25cm]{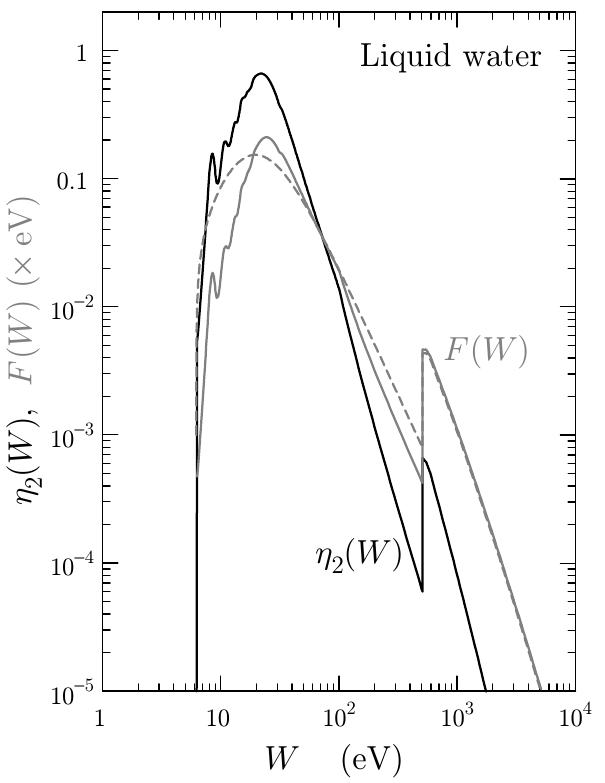} \rule{5mm}{0mm}
\includegraphics*[width=7.25cm]{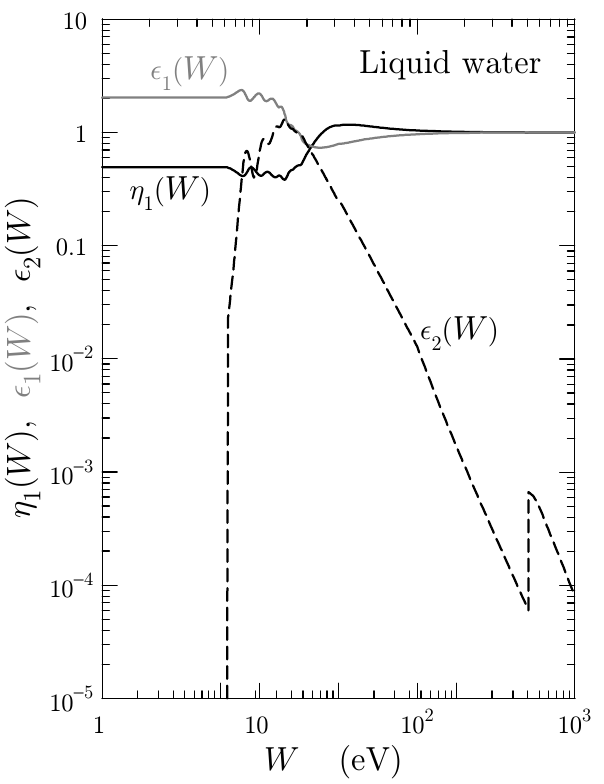} \\ [5mm]
\includegraphics*[width=7.25cm]{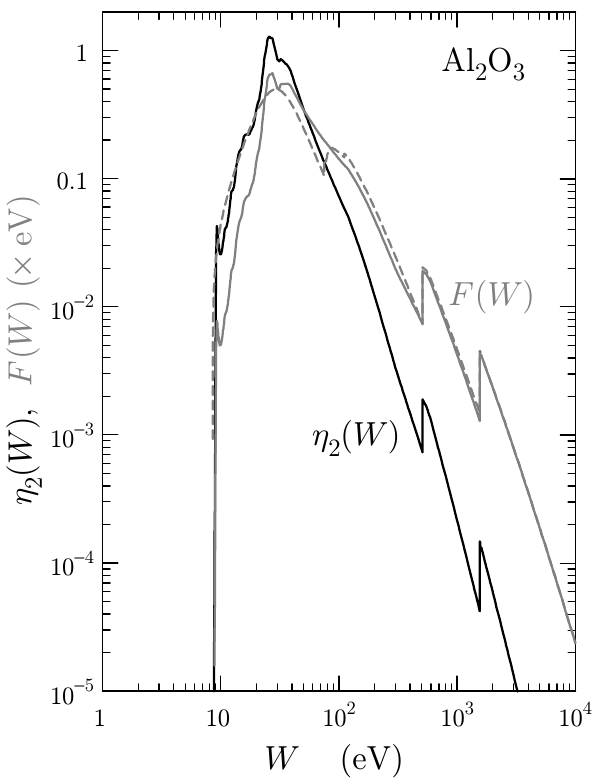} \rule{5mm}{0mm}
\includegraphics*[width=7.25cm]{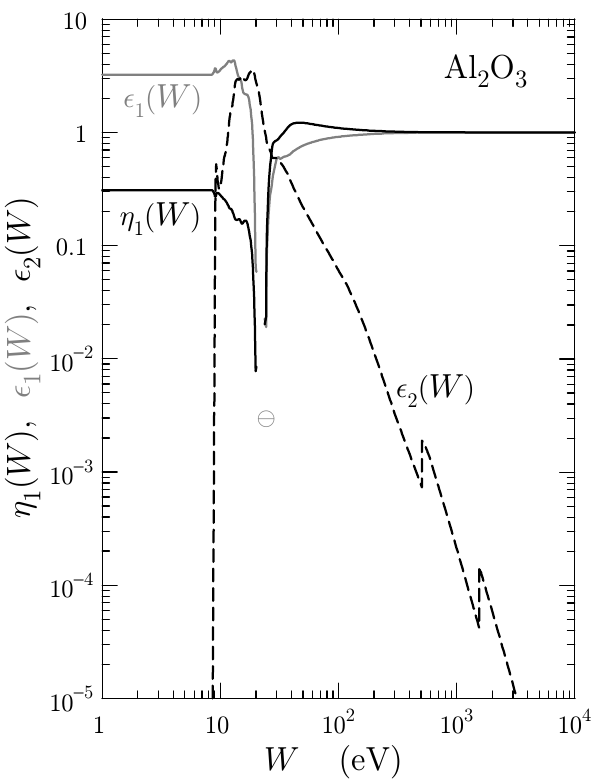}
\caption{Optical DFs of liquid water and alumina (Al$_2$O$_3$). The
dotted curves in the right plots represent negative values of the
functions. The dashed curves in the left plots represent the DHFS-model
OOSs, Eq.\ \req{7.162}.
\label{fig7.9}}
\end{center}\end{figure}

We have collected OOSs obtained in similar ways for a set of materials
consisting of aluminium \citep{Shiles1980}, silicon \citep{Bichsel1988},
titanium, copper, gold, aluminium oxide, liquid water, diamond,
magnesium oxide, and beryllium. As indicated above, to ensure
consistency with theory, the empirical OOS was renormalized so that the
$f$-sum rule is exactly satisfied. Figures \ref{fig7.8} and \ref{fig7.9}
display the optical functions of copper, gold, alumina and liquid water,
derived from the empirical OOSs.

Because of the unavailability of reliable optical information for many
materials, we frequently rely on OOSs tables build from DHFS-model OOSs
as described in Section \ref{sec7.5.2}.  Figures \ref{fig7.8} and
\ref{fig7.9} compare these OOSs with the empirical OOSs of copper, gold,
aluminium oxide, and liquid water. The close agreements for energy
transfers higher than about 100 eV does not imply reliability because
both models utilize the same calculated atomic OOSs.


\section{Optical-data models of the DF
\label{sec7.6}}
\index{optical-data models}

Practical calculations of stopping powers of materials for charged
particles, as well as elaborate Monte Carlo simulations of the transport
of charged particles in matter, are based on {\it optical-data models}.
An optical-data model combines an empirical OOS, $F(W) = \d f(W)/\d W$,
with physically motivated extension algorithms to define the generalized
oscillator strength (GOS), $\d f(Q,W)/ \d W$, for finite recoil energies
$Q$. Equivalently, the model may start from the OELF $\eta_2(W)$ to
build the complex inverse DF, $\eta(Q,W)$. The resulting DF model should
satisfy the relevant sum rules (see Section \ref{sec1.5.2}) and mimic
the conspicuous features of the GOS and the DF. Evidently, the
reliability of the results from calculations with an optical-data model
is determined by the adequacy of the PWBA; even with complete knowledge of
the actual OOS of a material, the predictions of the model will not
be reliable when the kinetic energy of the charged projectile is too low
for the PWBA to be valid.


\subsection{The DF model of Penn \label{sec7.6.1}}
\index{optical-data models!Penn model}

\citet{Penn1987} extended the central idea underlying the LPA to build
the DF, not from the electron density, which in general is known only
approximately, but from the empirical OELF, which is expected to
represent the optical properties of the medium dependably. More
specifically, in Penn's model, the medium is considered as a continuous
admixture of electron gases with plasma resonance energies $W'=\hbar
\omega_{\rm p}$ and weights determined by the OOS, $F(W')$. The response
of the electron gas is described by means of the undamped Lindhard DF
(see Section \ref{sec7.1.3}),
\beq
\eta_{\rm eg}(W'; Q,W) \equiv
\frac{1}{\epsilon_{\rm eg}(W'; Q,W)} =
\eta_{1,{\rm eg}} (W'; Q, W) - {\rm i}
\eta_{2,{\rm eg}} (W'; Q, W).
\label{7.176}\eeq
We recall that the GOS (per electron in the gas)
is [see Eq.\ \req{6.251a}]
\beq
\frac{\d f_{\rm eg}(W'; Q ,W)}{\d W}
= \frac{2W}{\pi W'^2} \, \eta_{\rm 2,eg} (W'; Q, W)),
\label{7.177}\eeq
and the corresponding OOS is [Eq. \req{7.95}]
\beq
\frac{\d f_{\rm eg}(W';W)}{\d W}
= \frac{2W}{\pi W'^2}
{\rm Im} \left( \frac{-1}{\epsilon_{\rm eg}(W'; W)}\right)
= \delta(W-W').
\label{7.178}\eeq
From the obvious equality
\beq
\frac{\d f(W)}{\d W} = \int_0^\infty F(W') \,
\delta(W-W') \, \d W' \, ,
\nonumber\eeq
it follows that
\beq
\frac{\d f(W)}{\d W} = \int_0^\infty F(W') \,
\frac{\d f_{\rm eg}(W';W)}{\d W} \, \d W' \, .
\label{7.179}\eeq
The Penn model for the GOS results from the assumption that the identity
\req{7.179} is generalizable to non-zero recoil energies (\ie, to finite wave
numbers). That is, the model GOS is defined as
\beq
\frac{\d f(Q,W)}{\d W} = \int_0^\infty F(W') \,
\frac{\d f_{\rm eg}(W';Q,W)}{\d W} \, \d W' \, .
\label{7.180}\eeq
We can regard the expression on the right-hand side as the weighted
admixture of electron gases that reproduces the adopted OOS when $Q
\rightarrow 0$. For large $Q$ and $W$ the model GOS exhibits a Bethe ridge
(\ie, a peak at $W \simeq Q$) with a physically plausible shape. In
addition, when the empirical OOS satisfies the dipole sum rule, the
model GOS satisfies the Bethe sum rule, Eq.\ \req{6.85}.

The Penn model may also be formulated in terms of the OELF.
Using the identity \req{6.251a} we can
express the OOS and the GOSs in terms of the corresponding ELFs, and
write Eq.\ \req{7.180} in the form
$$
\frac{2W}{\pi (\hbar \Omega_{\rm p})^2} \, \eta_{\rm 2} (Q,W) =
\int_0^\infty
\frac{2W'}{\pi (\hbar \Omega_{\rm p})^2} \, \eta_{\rm 2} (W')
\, \frac{2W}{\pi W'^2} \, \eta_{\rm 2,eg}(W';Q,W) \, \d W' \, .
$$
Hence, the model ELF is
\beq
\eta_2(Q,W)
= \frac{2}{\pi} \int_0^\infty
\eta_2(W') \, \frac{1}{W'}
\, \eta_{2,{\rm eg}} (W'; Q, W)
\, \d W'.
\label{7.181}\eeq
The real part of the inverse DF can be obtained from the Kramers--Kronig
relation \req{1.209a},
\beqa
\eta_1 (Q, W) &=& 1 -
\frac{2}{\pi} {\cal P} \int_0^\infty \frac{W''\,
\eta_2(Q, W'')}{W''^2-W^2} \, \d W''
\nonumber \\ [2mm]
&=& 1 - \frac{2}{\pi} {\cal P} \int_0^\infty
\frac{W''}{W''^2-W^2}
\left( \frac{2}{\pi} \int_0^\infty
\eta_2(W') \, \frac{1}{W'}
\, \eta_{2,{\rm eg}} (W'; Q, W'')
\, \d W' \right) \, \d W''.
\nonumber\eeqa
Inverting the order of the integrals, and recalling that the Lindhard DF
satisfies the Kramers--Kronig relations \req{1.209}, we have
\beqa
1 - \eta_1 (Q, W)  &=&
\frac{2}{\pi} \int_0^\infty
\eta_2(W') \, \frac{1}{W'} \left(
\frac{2}{\pi} {\cal P} \int_0^\infty \frac{W''
\eta_{2,{\rm eg}} (W'; q, W'')}
{W''^2-W^2}  \, \d W'' \right)
\, \d W'
\nonumber \\ [2mm]
&=& \frac{2}{\pi} \int_0^\infty
\eta_2(W') \, \frac{1}{W'} \left[
1 - \eta_{1,{\rm eg}} (W'; q, W')
\right] \, \d W'.
\label{7.182} \eeqa
That is, the Penn model provides the complex longitudinal inverse DF,
$\eta(Q,W) = \eta_1(Q,W) + {\rm i} \eta_2(Q,W)$, from knowledge of only
the OELF.

Naturally, the Penn model can also be considered to set the transverse
GOS and DF. We have
\beq
\frac{\d g(Q,W)}{\d W} = \int_0^\infty F(W') \,
\frac{\d g_{\rm eg}(W';Q,W)}{\d W} \, \d W' ,
\label{7.183}\eeq
where
\beq
\frac{\d g_{\rm eg}(W'; Q ,W)}{\d W}
= \frac{2}{\pi W_{\rm p}^2} \, W \,
{\rm Im} \left( \frac{-1}{\epsilon_{\rm eg}^{\rm (T)}
(W'; Q, W)}\right),
\label{7.184}\eeq
is the one-electron TGOS of the electron gas with resonance energy $W'$
obtained from Lindhard's undamped transverse DF, Eq.\ \req{7.77}. Of course,
the identities \req{7.181} and \req{7.182} also hold for the transverse DF.

It is worth noticing that the Penn model is consistent with the basic
assumptions of the LPA (see Section \ref{sec7.5.1}); the two can be
combined to calculate an approximation to the GOSs and the DFs from
knowledge of only the local electron density of the material.

The physical implications of the Penn model are revealed vividly when
considering the calculation of the DDCS for collisions of charged
particles [Eq. \req{6.222}]
\beqa
\frac{\d^2 \sigma}{\d Q \, \d W}
&=& \frac{2\pi Z_0^2 e^4}{\me v^2}
\left[ \frac{2\me c^2}{WQ(Q+2\me c^2)} \, \frac{\d f(Q,W)}{\d W} \right.
\nonumber \\ [2mm]
&+& \left. \beta^2 \left( 1 - \frac{W^2}{\beta^2 Q(Q+2 \me c^2)} \right)
\frac{ 2\me c^2 W } {[Q(Q+2\me c^2) - W^2 ]^2} \,
\frac{\d g(Q,W)}{\d W} \right]\, .
\rule{10mm}{0mm}
\label{7.185}\eeqa
Here we disregard the polarization correction, which will be considered
separately in Section \ref{sec8.2}. Introducing the expressions
\req{7.180} and \req{7.183} of the GOS and the TGOS, we find that the DDCS
obtained from Penn's model is
\beq
\frac{\d^2 \sigma}{\d Q \, \d W}
= \int_0^\infty F(W') \,
\frac{\d^2 \sigma_{\rm eg}(W')}{\d Q \, \d W} \, \d W'
\label{7.186}\eeq
where $\d^2 \sigma_{\rm eg}(W')/\d Q \, \d W$ is the one-electron DDCS
for collisions of the projectile in a homogeneous electron gas with
plasma resonance energy $W'$. In the terminology of stopping theory
(Chapter \ref{chapt9}), Penn's approach amounts to considering the
medium as a continuous set of ``oscillators'' (electron gas components)
with resonance energies $W'$ and strength $F(W')$. The response of each
oscillator is described by the Lindhard DFs of the undamped electron
gas. Evidently, the integrated cross sections [see Eq.\ \req{6.124}]
\beq
\sigma^{(n)} = \int_0^E  W^n \left( \int_{Q_-}^{Q_+}
\frac{\d^2 \sigma}{\d Q \, \d W} \, \d Q \right) \d W,
\label{7.187}\eeq
can be calculated as
\beq
\sigma^{(n)} = \int_0^\infty F(W') \, \sigma_{\rm eg}^{(n)}(W')
\, \d W',
\label{7.188}\eeq
where $\sigma_{\rm eg}^{(n)}(W')$ is the integrated cross section for
excitations of the electron gas with resonance energy $W'$. In the case
of insulators or semiconductors, $\sigma_{\rm eg}^{(n)}(W')$ will be
calculated by using the gapped form of the Lindhard DF [see
Section \ref{sec7.3}, Eqs. \req{7.122}].

Calculations with the Penn model are delicate because of the zero width
of the plasmon line and the sharp peak in the entrance of that line into
the Lindhard continuum. A practical alternative \citep[see, \eg,][]{
Abril1998, Da2014} consists in representing the DF
as a linear combination of Mermin DFs [Eq.\ \req{7.103}], with
coefficients obtained by fitting the empirical OELF. This scheme
produces a smooth continuous DF, which is easier to handle numerically,
and provides a realistic representation of the DF for valence electrons.
However, the mean free path, the stopping power, and the
energy-straggling parameter calculated from the Mermin DF do not differ
much from those obtained by using the Penn model with the OOS
corresponding to the Mermin DF. This is illustrated in Fig.\
\ref{fig7.10}, which shows that the differences in the mean free path and
stopping power calculated by the two methods are of the order of a few
percent at low energies, and decrease when the energy of the projectile
increases. We prefer to base our calculations on the Penn model with the
undamped Lindhard DF, with the aid of an efficient numerical algorithm
for computing the integrated cross sections of the electron gas.

\begin{figure}[htb!] \begin{center}
\includegraphics*[width=7.70cm]{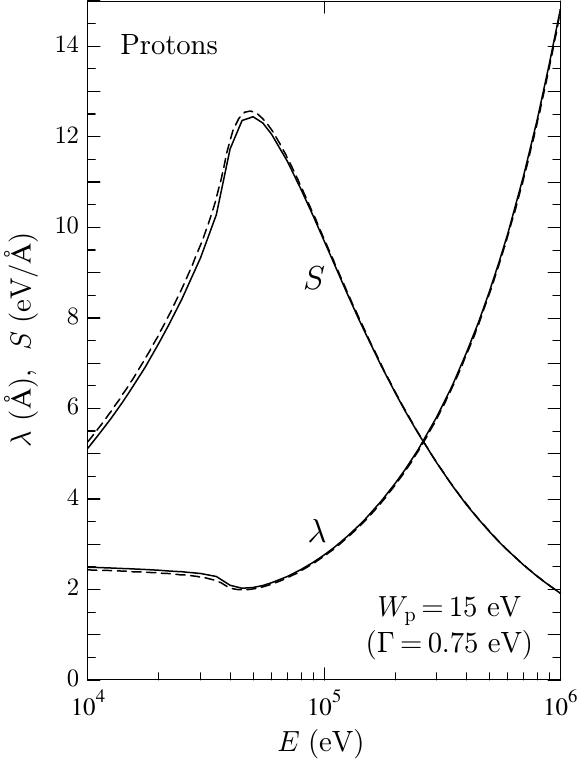}
\caption{Mean free path and stopping power of protons in an electron gas
with $W_{\rm p}=15$ eV, as functions of the kinetic energy $E$ of the
projectile, calculated from the Mermin DF with $\Gamma=0.75$
eV (dashed curves) and from the Penn model with the OOS of the
Mermin DF (solid curves).
\label{fig7.10}}
\end{center}\end{figure}

The Penn model of the GOS and the DF is acceptable for conduction and
valence electrons, which respond through collective plasmon-like
excitations at small $Q$ and through individual electron-hole
excitations (binary collisions) at moderate and large $Q$. However, as
noted by \citet{FernandezVarea1993b, FernandezVarea2005}, the Lindhard
DF is not appropriate for describing excitations of electrons in inner
atomic subshells. The contribution to the GOS of an inner subshell
presents an abrupt threshold at $W \sim U_a$, where $U_a$ is the
ionization energy of the subshell. This feature cannot be reproduced by
using the GOSs of Lindhard or Mermin as extension algorithms, because
the corresponding GOSs allow excitations with energy transfers below the
ionization threshold. In addition, the gapped form of the Mermin DF,
Section \ref{sec7.3}, gives an OOS that decreases too slowly with $W$.
Consequently, we shall use the Lindhard GOS of the electron gas to
extend the OOS of valence electrons only, that is, for resonance
energies $W'$ from 0 up to a certain cut-off energy $W_{\rm c}$  of the
order of 80 eV (corresponding to a frequency $W_{\rm c}/\hbar$ in the
extreme ultraviolet), which should be selected slightly below the first
excitation threshold of inner sub-shells that is clearly distinguishable
in the OOS.


\section{Inelastic collisions of charged particles in dense media
\label{sec7.7}}

\index{inelastic collisions in dense media}
\index{optical-data models!Lindhard--Liljequist extension scheme}

Considered as an extension algorithm, the plasmon-pole oscillator model
[Eq.\ \req{7.100}] is easy to implement, but it has the
inconvenience of shifting the ionization threshold of inner subshells by
the impact of electrons and positrons to roughly twice the ionization
energy of the subshell, because the allowed kinematical domain
intersects the resonance line \req{7.97} only when $E\gtrsim 2 W'$. To
describe excitations of inner subshells, in our calculations we shall
use the {\it delta oscillator model}, which was introduced by
\citet{Liljequist1983} and can be considered as an adaptation of the
plasmon-pole model. The GOS of a delta oscillator with resonance energy
$W'$ is defined by
\beq
\frac{\d f_\delta(W';Q,W)}{\d W} = \delta \left[ W-W_{\rm r}(W',Q) \right]
\label{7.189}\eeq
with the resonance line given by
\beq
W_{\rm r} (W';Q) \equiv \max \left\{ W', Q \right\}.
\label{7.190}\eeq
That is,
\beq
\frac{\d f_\delta(W';Q,W)}{\d W} = \delta(W-W') \, {\cal S}(W'-Q)
+ \delta(W-Q) \, {\cal S}(Q-W').
\label{7.191}\eeq
The corresponding inverse DF is
\beq
\eta_{\delta}(W'; Q,W)
= 1 - \frac{W'^2}{W_{\rm r}^2(W',Q) - W^2}
- {\rm i} \,  \frac{\pi}{2} \,
\frac{W'^2}{W_{\rm r}(W';Q)}
\, \delta[W-W_{\rm r}(W';Q)]\, .
\label{7.192}\eeq
The rationale for this model is to mimic the global features of the
atomic GOS obtained from the plane-wave Born approximation
\citep{FernandezVarea1993b, FernandezVarea2005}, and to ensure the
correct threshold energies for ionization of inner shells. Low-$Q$
excitations of the oscillator have a resonance-like character (with
$W=W'$), while the only allowed high-$Q$ excitations occur at the line
$W=Q$, which represents binary collisions with free electrons at rest.
Since the longitudinal and transverse GOSs coincide at small and large
$Q$, we assume that the TGOS of the delta oscillator is also given by
Eq.\ \req{7.191}.

The Lindhard--Liljequist extension scheme consist in
using a predefined OOS and considering a cutoff energy loss $W_{\rm c}$
of the order of 50 eV (usually coinciding with an absorption edge).
Excitations of valence electrons with
resonance energies $W'<W_{\rm c}$ are described by using the Lindhard
GOSs as extension algorithm, and interactions with oscillators having
resonance energies larger than $W_{\rm c}$ are modeled by means of the
Liljequist delta oscillator. The GOS model built from this
Lindhard--Liljequist extension scheme is given by
\beqa
&& \! \! \! \! \! \! \! \! \! \! \! \! \! \!
\frac{\d f_{\rm LL}(Q,W)}{\d W}
= \int_0^{W_{\rm c}} F(W') \,
\frac{\d f_{\rm eg}(W';Q,W)}{\d W} \, \d W'
\nonumber \\ [2mm]
&& \mbox{} +
\left[ F(W) \, {\cal S}(W-Q) +
\left( \int_{W_{\rm c}}^W F(W') \, \d W' \right) \delta(W-Q)
\right] {\cal S} (W-W_{\rm c}), \rule{10mm}{0mm}
\label{7.193}\eeqa
and the corresponding TGOS is
\beqa
&& \! \! \! \! \! \! \! \! \! \! \! \! \! \!
\frac{\d g_{\rm LL}(Q,W)}{\d W}
= \int_0^{W_{\rm c}} F(W') \,
\frac{\d g_{\rm eg}(W';Q,W)}{\d W} \, \d W'
\nonumber \\ [2mm]
&& \mbox{} +
\left[ F(W) \, {\cal S}(W-Q) +
\left( \int_{W_{\rm c}}^W F(W') \, \d W' \right) \delta(W-Q)
\right] {\cal S} (W-W_{\rm c}). \rule{10mm}{0mm}
\label{7.194}\eeqa

The program {\sc stopping} (see Chapter \ref{chapt10}) performs
calculations of inelastic collisions by using an optical-data model of
the OOS with
the Lindhard--Liljequist extension scheme. The DDCS for collisions of
charged particles heavier than the electron is calculated from Eq.\
\req{6.222}, which describes collisions in an unpolarizable medium. The
dielectric polarization of the material is considered separately (see
Section \ref{sec8.2}). The
energy-loss DCS is obtained as the integral of the DDCS over
kinematically allowed recoil energies. Integrated cross sections
$\sigma^{(k)}$ (see Section \ref{sec6.3}) can be calculated as integrals
over resonance energies $W'$ of the integrated cross sections
$\sigma_1^{(k)}(W')$ of individual oscillators,
\beq
\sigma^{(k)} =
\int_0^\infty F(W') \, \sigma_1^{(k)}(W') \, \d W'.
\label{7.195}\eeq
With this scheme, the need of evaluating the GOS is avoided.

\index{inelastic collisions in dense media!integrated cross sections}
Integrated cross sections for
excitations of Lindhard oscillators are calculated as described in
Section \ref{sec7.4}. To obtain the integrated cross sections of the
delta oscillator, we consider separately the contributions of close
($Q>W'$) and distant ($Q<W'$) interactions. The energy-loss DCS of
distant interactions is
\beqa
\frac{\d \sigma_{\rm 1,dist}(W')}{d W}
&=& \frac{2\pi Z_0^2 e^4}{\me v^2} \,  \delta(W-W') \int_{Q_-}^{W'}
\left[ \frac{2\me c^2}{WQ(Q+2\me c^2)} \right.
\nonumber \\ [2mm]
&+& \left. \beta^2 \left( 1 - \frac{W^2}{\beta^2 Q(Q+2 \me c^2)} \right)
\frac{ 2\me c^2 W } {[Q(Q+2\me c^2) - W^2 ]^2} \,
\right] \d Q.
\rule{10mm}{0mm}
\label{7.196}\eeqa
The integral of the first term is elementary. That of the second term,
is evaluated by introducing the variable
\beq
x = \frac{W^2}{\beta^2 Q (Q + 2 \me c^2)}
\label{7.197}\eeq
so that
\beqa
\lefteqn{\int_{Q_-}^{W'}
\beta^2 \left( 1 - \frac{W^2}{\beta^2 Q(Q+2 \me c^2)} \right)
\frac{ 2\me c^2 W } {[Q(Q+2\me c^2) - W^2 ]^2} \, \d Q}
\nonumber \\ [2mm]
&\simeq& \int_{0}^{1}
\beta^2 (1-x)
\frac{ 2\me c^2 W }{1 - \beta^2 x} \,
\frac{\beta^2}{W^2 2(Q + \me c^2)} \, \d x
\simeq \frac{1}{W} \int_{0}^{1}
\frac{\beta^4(1-x)} {1 - \beta^2 x} \, \d x
\nonumber \eeqa
where we have used that $x \simeq 1$ at $Q=Q_-$ and, as the
integrand decreases rapidly with $Q$, we have replaced the upper limit
$W'$ of the integral with $\infty$. The last expression results from
considering that, for transverse interactions, $Q \ll 2 \me c^2$. Hence,
using the result \req{6.265}], we have
\beqa
\frac{\d \sigma_{\rm 1,dist}(W')}{d W}
&=& \frac{2\pi Z_0^2 e^4}{\me v^2} \,  \frac{\delta(W-W')}{W} \left\{
\left[ \ln \left( \frac{Q}{Q+2\me c^2} \right) \right]_{Q_-}^{W'}
+ \ln \left( \frac{1}{1-\beta^2} \right) - \beta^2 \right\}.
\nonumber \\
\label{7.198}\eeqa
The energy-loss DCS for close collisions with a delta oscillator is
given by the expressions derived in Section \ref{sec6.8},
\beq
\frac{\d \sigma_{\rm 1,clo}(W')}{\d W} =
\frac{2 \pi Z_0^2 e^4 }{\me v^2 } \, \frac{1}{W^2} \, F_{\rm bin}(W).
\label{7.199}\eeq
The integrated cross sections of a delta oscillator,
\beq
\sigma_1^{(k)}(W') = \int W^k \left[
\frac{\d \sigma_{\rm 1,dis}(W')}{\d W}
+ \frac{\d \sigma_{\rm 1,clo}(W')}{\d W} \right] \d W,
\label{7.200}\eeq
can then be evaluated analytically.

\index{inelastic collisions in dense media!polarization correction}
The polarization of the medium, which affects only the low-$Q$
transverse interactions, is accounted for by adding to the energy-loss
DCS of the unpolarized material the correction given by Eq.\
\req{6.268},
\beqa
\frac{\d (\Delta \sigma)_{\rm pol}}{\d W} &=&
\frac{2 \pi Z_0 e^4}{\me v^2} \, \frac{1}{W} \, \frac{\d f(W)}{\d W}
\left\{
\left( \frac{\beta^2 - \eta_1}{\eta_2} \right)
\arctan \! \left( \frac{\beta^2 \eta_2}{\eta_1 (\eta_1 -\beta^2) +
\eta_2^2} \right) \right.
\nonumber \\ [2mm]
&& \mbox{} \left.
- \frac{1}{2} \ln \left[ \frac{[\eta_1(\eta_1 -\beta^2)]^2 + \beta^4
\eta_2^2}{(\eta_1^2+\eta_2^2)^2} \right]
- \left[ \ln \left( \frac{1}{1-\beta^2} \right) - \beta^2 \right]
\right\}. \rule{10mm}{0mm}
\label{7.201}\eeqa
In the {\sc stopping} program this correction is applied by simply
modifying the energy-loss DCS for distant interactions of each active
oscillator. That is, the correction is introduced only to the
oscillators that are effectively excited, \ie, those with energies
$\lesssim W_{\rm ridge}$ [see Eq.\ \req{6.271}].

The Lindhard--Liljequist extension scheme, combined with a realistic OOS
model provides a fairly accurate
description of the stopping of fast charged particles heavier than the
electron. The theoretical framework (\ie, the dielectric formalism and
the PWBA) is supposed to be appropriate for projectiles with
sufficiently high energies, and we expect it to lose validity at low
energies because of the influence of various effects that are not
included in the theory. The qualitative discussion in Section
\ref{sec6.2} indicates that the PWBA and related formulations are
appropriate for projectiles with velocities larger than those of the
electrons in the K shell of the heaviest atoms in the material. In
practice this validity limit is far too restrictive, because low-energy
projectiles are able to excite valence and conduction electrons only.
Hence, we may expect the theory to be approximately valid for
projectiles with velocities larger than those of the electrons in the
valence and conduction bands, which is much lower than the velocity of
electrons in the K shell.


\subsection{Collisions of electrons and positrons \label{sec7.7.1}}

The DDCS for collisions of electrons and positrons is obtained by
following the schemes described in Sections \ref{sec6.6.1} and
\ref{sec6.6.2}. However, here we disregard exchange corrections for
distant interactions (\ie, plasmon excitations of Lindhard oscillators
and low-$Q$ excitations of delta oscillators), which are expected to be
small and would complicate the calculations.

\index{collisions of positrons}
\noindent $\bullet$ {\bf Positron collisions}.
The modifications required for positrons reduce to multiplying the DDCS
for close interactions (electron-hole interactions in the case of
Lindhard oscillators) by the Bhabha factor \req{6.241},
\beq
F_{\rm Bhabha} (W) = 1 - b_{1}\frac{W}{E} +
b_{2}\left(\frac{W}{E}\right)^{2} - b_{3}\left(\frac{W}{E}\right)^{3} +
b_{4}\left(\frac{W}{E}\right)^{4}\, .
\label{7.202}\eeq

\index{collisions of electrons}
\noindent $\bullet$ {\bf Electron collisions}. The calculation of
electron collisions is more delicate owing to the stronger influence of
exchange effects and to the restrictions implied by Pauli's exclusion
principle. The account for the effect of exchange in close collisions,
we multiply the DDCS by the Ochkur-M\o ller factor [see Eq.\ \req{6.239b}],
\beq
\frac{\d^2 \sigma ({\rm e}^-)}{\d Q \, \d W} =
\frac{F_{\rm OM}(Q,W)}{F_{\rm bin}^0(W)} \,
\frac{\d^2 \sigma}{\d Q \, \d W},
\label{7.203}\eeq
where
\beq
F_{\rm OM} (Q,W) =
1  + \left( \frac{Q}{E+\langle K \rangle-W} \right)^{2}
- \frac{(1-b_0)Q}{E+\langle K \rangle-W} +
\frac{b_0Q^{2}}{(E+\langle K \rangle)^{2}}\, ,
\label{7.204}\eeq
and
\beq
F_{\rm bin}^0(W) = 1 - \frac{(2E-W+4 \me c^2)W}{2(E+\me c^2)^2}\, .
\label{7.205}\eeq
with $b_0 = [E/(E+\me c^2]$.
The quantity $\langle K \rangle$ is the average kinetic energy of the
target electrons.

To estimate the average kinetic energy of electrons represented by delta
oscillators, we recall that the hydrogenic model gives $\langle K
\rangle = U_a$, the ionization energy of the target electron. On the
other hand, the average resonance energies $W'$ of the oscillators,
weighted by the OOS \req{7.161} of the $1s$ shell of hydrogen is,
$$
\int W' F_a(Z=1; W') \, \d W' = 1.86 \, U_a.
$$
It is then reasonable to regard the resonance energy $W'$ of an
oscillator as a kind of binding energy and to set $\langle K \rangle =
\xi W'$, where $\xi$ is a parameter of the order of unity. Following
\citet{FernandezVarea2005}, we set $\xi=1$ and express the energy-loss DCS
for close collisions with the oscillator as
\beqa
\frac{\d \sigma_{\rm 1,clo}(W')}{\d W} &=&
\frac{2 \pi Z_0^2 e^4 }{\me v^2 } \, \frac{1}{W^2} \,
\left[
1  + \left( \frac{W}{E+W'-W} \right)^{2}
- \frac{(1-b_0)W}{E+W'-W} +
\frac{b_0W^{2}}{(E+W')^{2}}
\right]. \nonumber \\
\label{7.206}\eeqa
After the collision, the kinetic energy of the recoiling target electron
is $W-W'$, while that of the projectile is $E-W$. The allowed energy
transfers lie in the interval from $W'$ up to a certain maximum value
$W_{\rm max}$ such that the projectile and the target electrons have the
same final kinetic energy. That is,
\beq
W_{\rm max} = \1o2 \left( E + W' \right).
\label{7.207}\eeq

The average kinetic energy of electrons represented by Lindhard
oscillators is that of an electron gas $\langle K \rangle = 3E_{\rm
F}/5$ [Eq.\ \req{3.71}]. In the spirit of the LPA (Section
\ref{sec7.5.1}), the material medium is regarded as an inhomogeneous
electron gas with a constant (\ie, position-independent) Fermi level,
which we set to zero. Consequently, the electrons move in a potential
$V({\bf r}) = - E_{\rm F}({\bf r})$, and their average ``binding
energy'' is about $-E_{\rm F} + \langle K \rangle = - 2 E_{\rm F}/5$. By
similarity with the delta
oscillator, the maximum allowed energy loss should be about $0.5 [E + 2
E_{\rm F}/5]$. For projectiles with kinetic energy near $E_{\rm F}$ we
may allow a larger maximum loss because the main effect of the Pauli
principle is already accounted for by the Lindhard GOS. The program {\sc
stopping} sets
\beq
W_{\rm max} = {\rm min} \left\{ E, \1o2 \left( E + 2E_{\rm F}/5
\right) \right\},
\label{7.208}\eeq
which yields results in reasonable agreement with experimental data at
low energies.

\index{optical-data models!inelastic mean free path}
\index{optical-data models!stopping power}
\index{optical-data models!energy-straggling parameter}
Figure \ref{fig7.11} shows mean free paths, stopping powers, and
energy-straggling parameters (as defined in Section \ref{sec6.3.1}) for
electrons and positrons in copper
calculated by the program {\sc stopping} with the Lindhard--Liljequist
extension scheme using the empirical OOS plotted in Fig.\ \ref{fig7.8}. Notice
that lengths are expressed in mass-thickness units (g/cm$^2$), that is,
the quantities plotted are $\lambda \rho_{\rm m}$, $S /\rho_{\rm m}$ and
$\Omega^2/\rho_{\rm m}$, where $\rho_{\rm m}$ is the mass density of the
material.  Apparently, the integrated cross sections of positrons are
larger than those of electrons. This is mostly a consequence of our
convention of considering the projectile as the faster of the two
electrons after a collision.

\begin{figure}[htb!] \begin{center}
\includegraphics*[width=7.20cm]{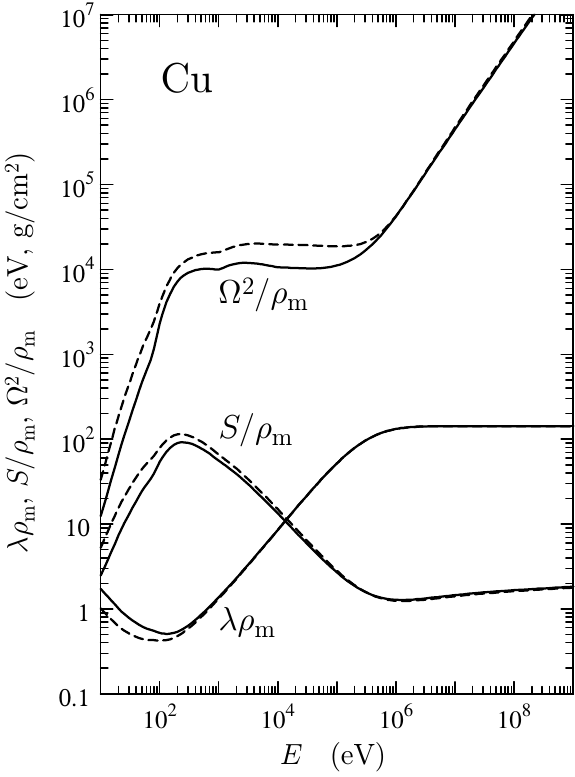}
\caption{Mean free path, $\lambda$, stopping power $S$, and
energy-straggling parameter $\Omega^2$ of electrons (solid curves) and
positrons (dashed curves) in copper, calculated by the program {\sc
stopping} with the empirical OOS derived from optical data
(as plotted in Fig.\ \ref{fig7.8}) and the Lindhard--Liljequist
extension scheme.
\label{fig7.11}}
\end{center}\end{figure}


\subsection{Comparison with experimental data \label{sec7.7.2}}

\begin{figure}[p!] \begin{center}
\includegraphics*[width=7.0cm]{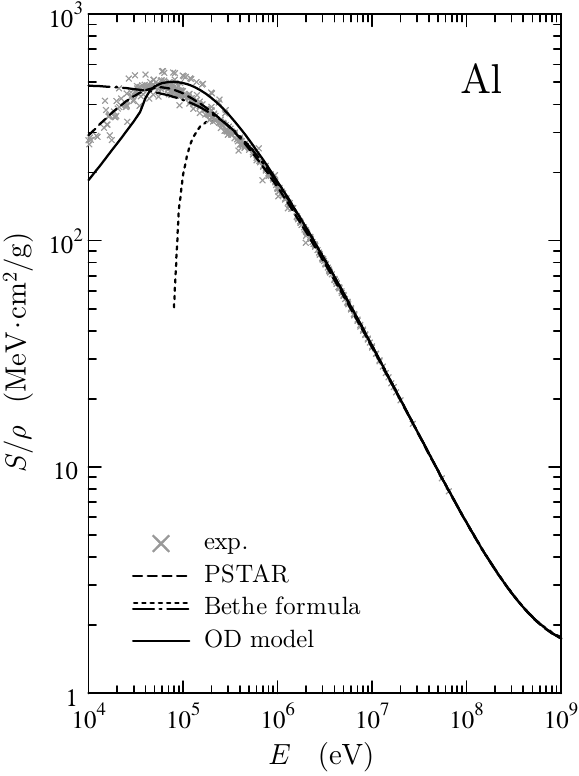} \rule{5mm}{0mm}
\includegraphics*[width=7.0cm]{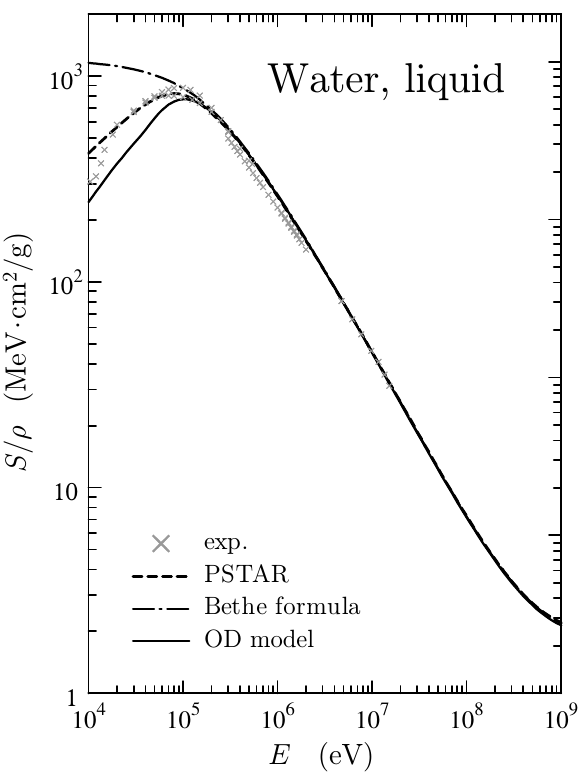} \\ [5mm]
\includegraphics*[width=7.0cm]{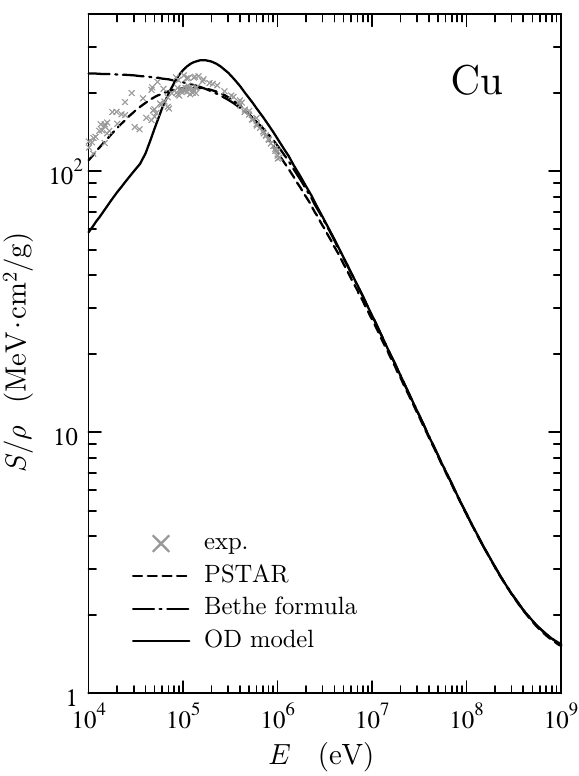} \rule{5mm}{0mm}
\includegraphics*[width=7.0cm]{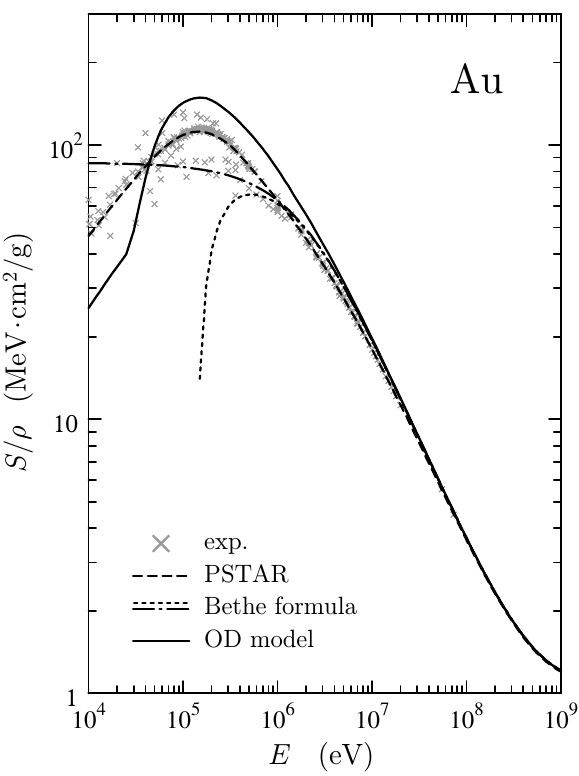}
\caption{Stopping powers of aluminium, liquid water, copper, and gold
for protons, as functions of the kinetic energy $E$ of the projectile.
The solid curves represent calculation results from
optical-data model calculations. The dot-dashed curves
represent results from the Bethe formula with the low-energy
extrapolation \req{6.313}; the doted curves are calculated from
the Bethe formula, Eq.\ \req{6.307}. Symbols are experimental data, and the
dashed curves are results from the {\sc pstar} program.
	\label{fig7.12}}
\end{center}\end{figure}

\begin{figure}[p!] \begin{center}
\includegraphics*[width=7.20cm]{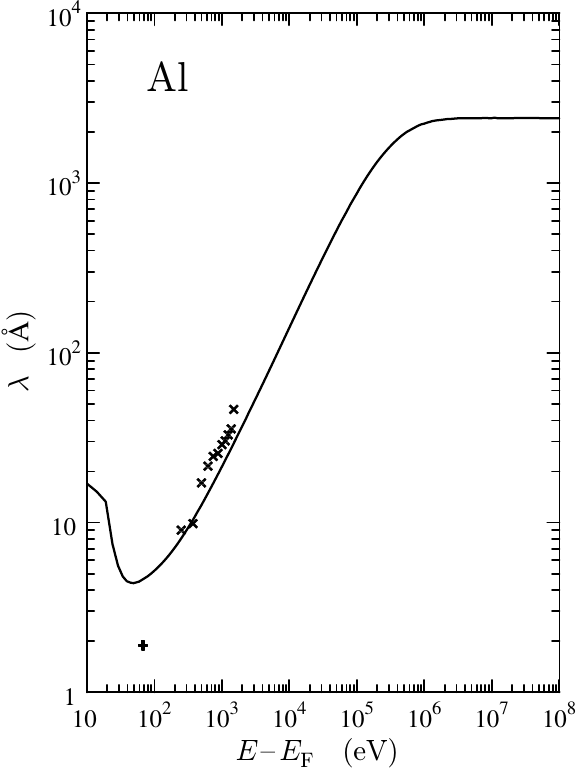} \rule{5mm}{0mm}
\includegraphics*[width=7.20cm]{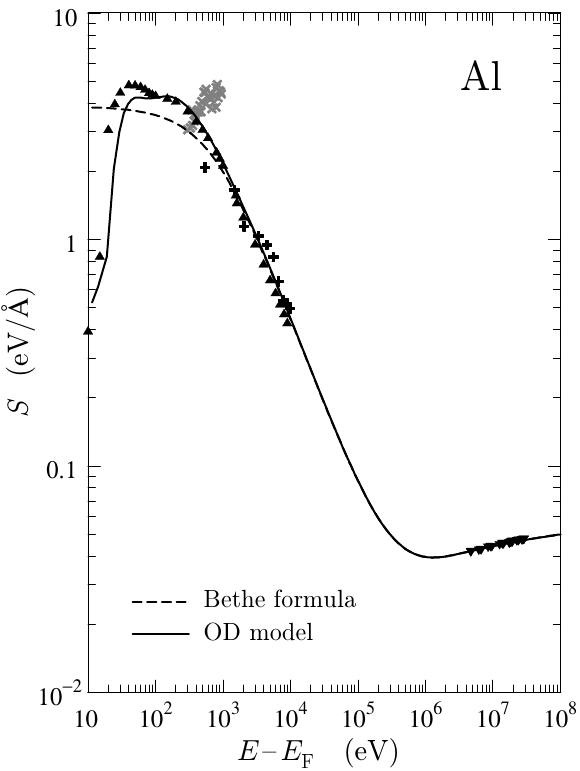} \\ [5mm]
\includegraphics*[width=7.20cm]{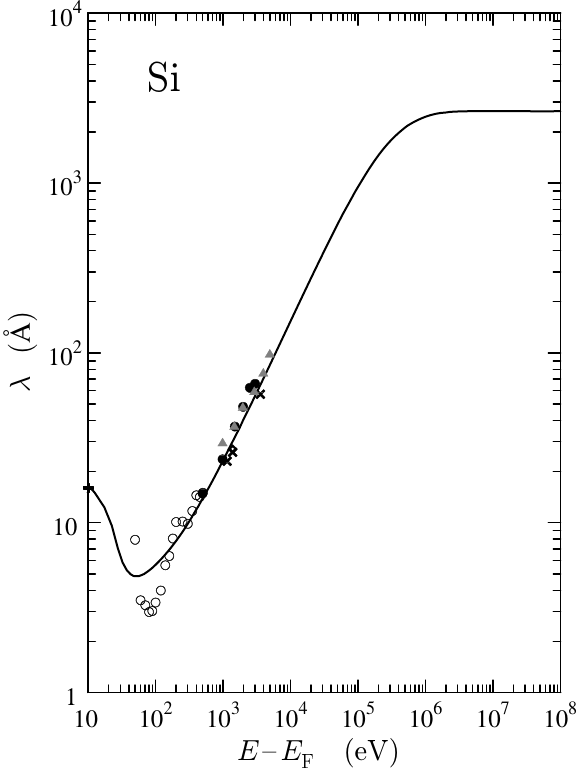} \rule{5mm}{0mm}
\includegraphics*[width=7.20cm]{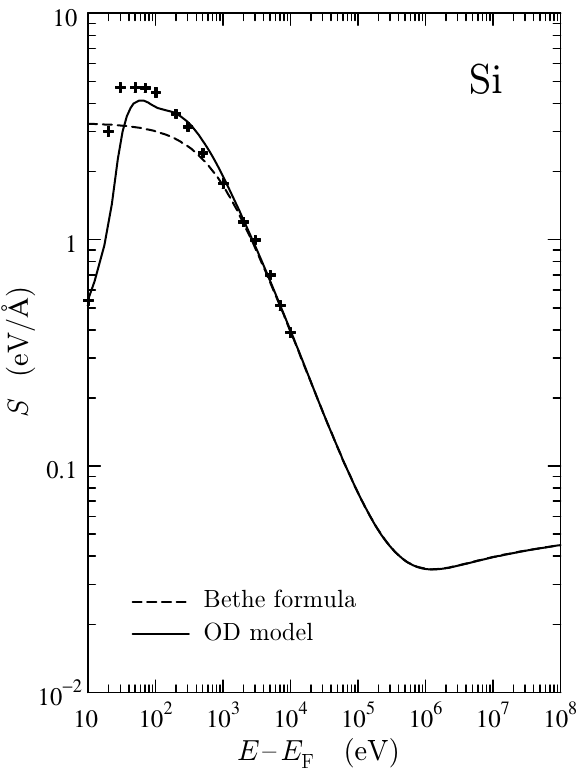}
\caption{Mean free paths and stopping powers of electrons in aluminium
($E_{\rm F}=11.7$ eV) and silicon ($E_{\rm F}=12.5$ eV).
Experimental data from various authors, compiled by
\citet{FernandezVarea2005}.
\label{fig7.13}}
\end{center}\end{figure}

\begin{figure}[p!] \begin{center}
\includegraphics*[width=7.20cm]{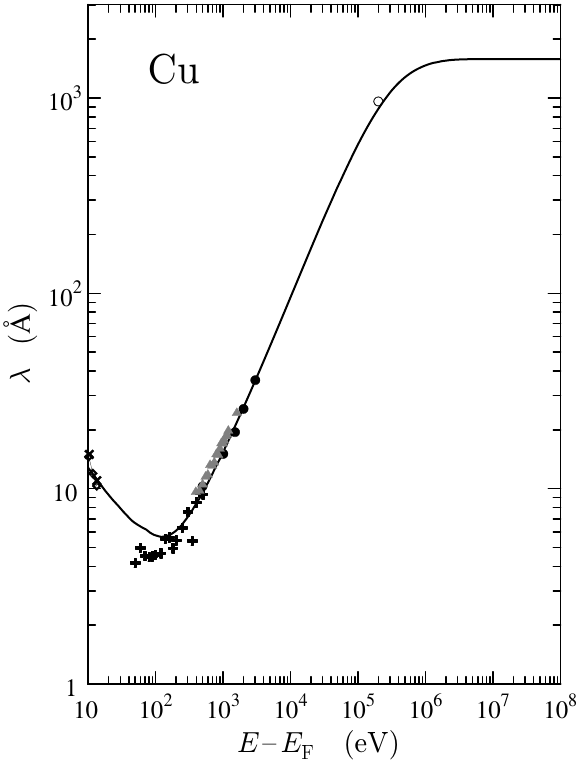} \rule{5mm}{0mm}
\includegraphics*[width=7.20cm]{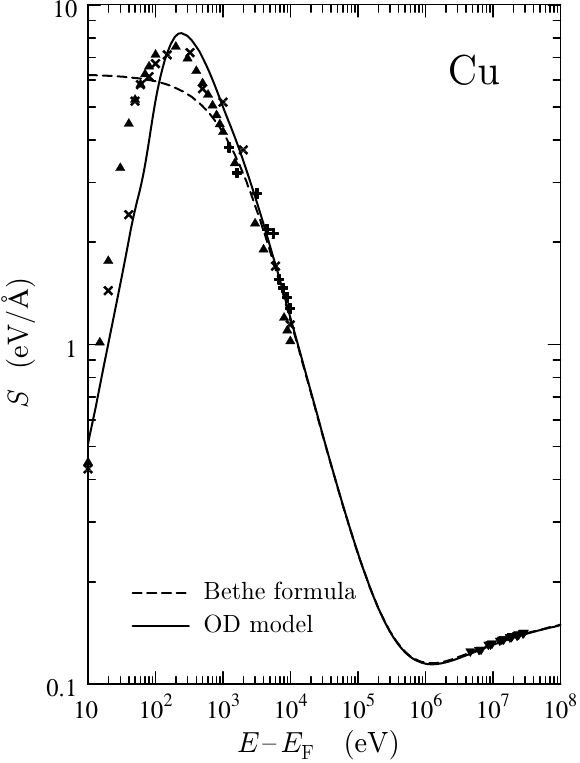} \\ [5mm]
\includegraphics*[width=7.20cm]{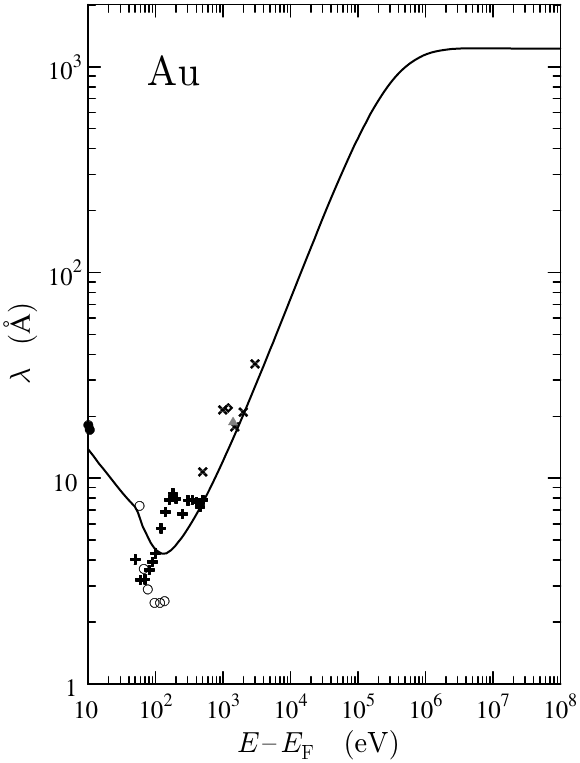} \rule{5mm}{0mm}
\includegraphics*[width=7.20cm]{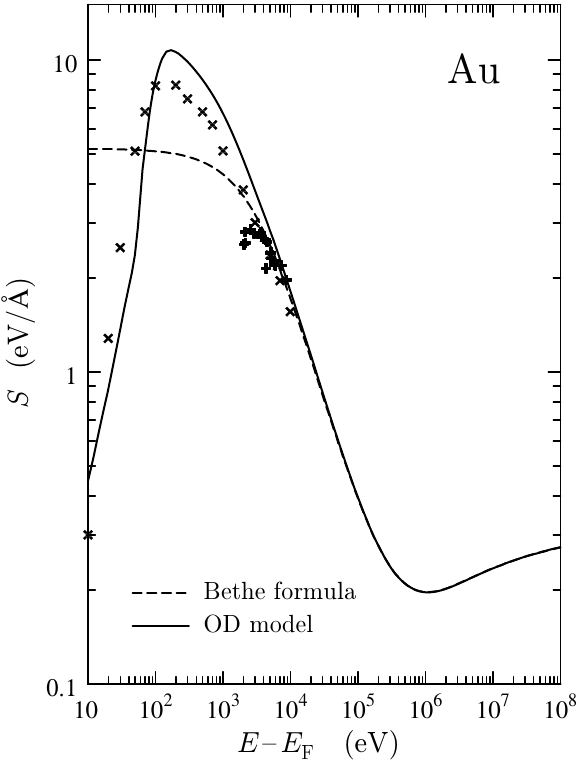}
\caption{Mean free paths and stopping powers of electrons in copper
($E_{\rm F}=7$ eV) and gold ($E_{\rm F}=5.5$ eV). Experimental data
from various authors, compiled by \citet{FernandezVarea2005}.
\label{fig7.14}}
\end{center}\end{figure}

\index{optical-data models!inelastic mean free path}
\index{optical-data models!stopping power}
To give a feel of the reliability of the Lindhard--Liljequist extension
scheme, we shall compare results from calculations using
empirical OOSs [obtained as described in Section \ref{sec7.5.3}] with
experimental inelastic data. The considered quantities are electronic
stopping powers for
protons, and mean free paths and stopping powers for electrons. Figure
\ref{fig7.12} displays the stopping powers of aluminium, liquid water,
copper and gold for protons as functions of the kinetic energy $E$ of
the projectile.
Symbols represent experimental data from various groups, taken from the
IAEA online database on ``Electronic Stopping Power of Matter for
Ions''\footnote{This database is available from the IAEA web site,
\url{https://www-nds.iaea.org/stopping}. The data used here
were downloaded in March 2022.} which contains an exhaustive collection
of measured stopping powers \citep{Montanari2017}. The plots also
include stopping powers generated by the program {\sc pstar}
\citep{Berger1992, ICRU49}, which uses semi-empirical fitting formulas
based on theoretical trends that provide satisfactory fits to
experimental stopping-power data. The dot-dashed curves represent
stopping powers calculated from the Bethe formula with the low-energy
extrapolation, Eq.\  \req{6.313}, and with the mean excitation energy
$I$ obtained from Eq.\ \req{6.288} using the same OOSs employed in the
optical-data model calculations. The plots for aluminium and gold
include results from the Bethe formula without the low-energy
extrapolation, Eq.\ \req{6.307}. In spite of the ad-hoc character of
that extrapolation, it provides stopping powers with the right order of
magnitude even for the lower energies in the plots.

The solid curves in Fig.\ \ref{fig7.12} are results from calculations for
protons with the Lindhard--Liljequist extension scheme and the empirical
OOS derived from optical data. They are in close
agreement with experimental data (and the PSTAR curves) for energies
higher than about 100 keV for water, 1 MeV for aluminium, 10 MeV for
copper and 50 MeV for gold. Below these limits, the difference between
theory and experiment tends to increase when the kinetic energy of the
projectile decreases.  A part of the difference is due to the
simplifications in the optical-data model GOS of inner subshells, where
we are disregarding both the variation of the GOS with $Q$ for low-$Q$
interactions and the finite width of the Bethe ridge. Differences are
also caused by non-linear effects (\ie, contributions beyond the PWBA)
which are known to increase in importance when the velocity of the
projectile decreases \citep[see, \eg,][]{Echenique1986}.

Figures \ref{fig7.13} and \ref{fig7.14} display inelastic mean free paths
and stopping powers of electrons in aluminium, silicon, copper, and gold
calculated from the present optical-data model, together with
experimental data compiled by \citet{FernandezVarea2005}. Typically, the
energies of electrons are referred to the Fermi level $E_{\rm F}$ of the
material, which we have set equal to the Fermi energy of an electron gas
with the density of the conduction band; in the case of silicon, the
density of the gas was estimated by assuming four valence electrons per
atom. The experimental data have relatively large uncertainties (which
are evident from the discrepancies between data sets from different
sources), especially at energies of about 100 eV and below, where
$\lambda$ and $S$ attain their extreme values, because of the
difficulties arising from the competition between elastic and inelastic
collisions, which is stronger for electrons and positrons than for
protons and heavier particles. Globally, calculated results are in
satisfactory agreement with measurements. Unfortunately, available
experimental data are very scarce above a few hundred keV. The
calculated inelastic mean free paths saturate at high energies because
the ``interaction'' time ($\sim R/v$), and hence the interaction
probability, becomes independent of $E$ for ultrarelativistic
projectiles.

In conclusion, we have shown that the Lindhard--Liljequist extension
scheme, combined with empirical OOSs derived from optical data,
provides a better than acceptable description of inelastic
collisions of projectiles with sufficiently high energies. However,
inaccuracies are to be expected, not only because of the failure of the
PWBA at low energies, but also from the simplicity of the model and from
inaccuracies in the adopted optical functions. While the gapped Lindhard
DF is expected to be a fairly reliable extension algorithm for
excitations of valence and conduction electrons, the delta oscillator
does not account for relevant details of the GOS of inner electron
subshells. Since electrons in inner subshells react essentially as in
free atoms, their contributions to the integrated cross sections can be
estimated more reliably from the relativistic PWBA with GOSs calculated
with the DHFS self-consistent potential \citep{BoteSalvat2008,
Salvat2022a}.




\chapter{Stopping theory  \label{chapt8}}

The slowing down of fast charged particles in matter, a central problem
in radiation physics, has been under study since the discovery of the
atomic nucleus by Rutherford in the early 1900's. The first theoretical
description of the process was given by \citet{Bohr1913, Bohr1915} who
derived a formula for the stopping power on the basis of classical
mechanics and dielectric theory. The analogue quantum formula was
obtained by \citet{Bethe1930, Bethe1932}, and the connection between the
classical and quantum stopping-power formulas was studied by
\citet{Bloch1933}.  The effect of the polarization of the material (or
density effect) was first considered by \citet{Fermi1940} and studied
later by \citeauthor{Sternheimer1952} (\citeyear{Sternheimer1952}) and
others \citep{Sternheimer1982, Sternheimer1984}. The  study of the
Barkas effect, or $Z_1^3$ correction to the stopping power, was
initiated by the classical dielectric approach of Ashley, Ritchie and
Brandt (\citeyear{Ashley1972}). For a thorough review of the subject,
see \citet{Ahlen1980}.

In this Chapter we describe the various components of the Bethe--Bloch
stopping power formula, following basically the chronological order in
which they were introduced. To give a feel of historical evolution of
the subject, calculations are presented similarly to the original
publications, except for those results that follow directly from the
derivations in previous Chapters. We concentrate on the stopping effect
of interactions causing electronic excitations of the material, the {\it
electronic stopping}. High-energy electrons and positrons slow down also
due to the emission of bremsstrahlung \citep[see,
\eg,][]{HaugNakel2004}, the so-called {\it radiative stopping}, whose
calculation is considered in Sections \ref{sec8.6} and \ref{sec8.7} for
electrons/positrons and muons, respectively. We also describe the
process of {\it nuclear stopping}, which results from energy transfers
in elastic collisions with atoms (Section \ref{sec8.8}). Finally, the
effect of the capture of electrons by low-energy positive ions on the
stopping power is briefly considered in Section \ref{sec8.9}.


\section{Classical stopping power \label{sec8.1}}
\index{stopping power!Bohr formula|(}
\index{classical stopping power}

A fast charged particle moving in a material medium interacts with the
atomic electrons and nuclei. The classical Thomson formula, Eq.\
\req{4.129}, shows that the energy-loss DCS for binary collisions with
free electrons or nuclei is inversely proportional to the mass of the
target particle. This feature implies that the stopping of fast charged
particles is largely due to interactions with the electrons; collisions
with nuclei have a much smaller stopping effect, the so-called {\it
nuclear stopping}, that is significant only for projectiles with very low
energies and light elements (Section \ref{sec8.8}). On the other hand,
the energy-loss DCS is approximately proportional to $W^{-2}$ and,
therefore, the average energy loss is generally much smaller than the
kinetic energy of the projectile. The effect of multiple collisions with
small energy transfers is equivalent to that of a ``stopping force'',
which we usually call the {\it stopping power} (although this is not an
appropriate name to refer to a force).

We describe here the early theoretical calculation of the electronic
stopping power made by \citet{Bohr1913, Bohr1915}. The situation to be
considered is that of a fast particle of mass $M_1$ and charge $Z_1 e$
moving with velocity $v$ through an homogeneous and isotropic
material\footnote{ In the present Chapter we adhere to the usual
convention in stopping theory and designate the mass and charge of the
projectile by $M_1$ and $Z_1e$, respectively.}. For the sake of
concreteness, we assume a molecular material with $Z$ electrons per
molecule and ${\cal N}$ molecules per unit volume [Eq.\ \req{1.141}].
The material is characterized by its optical oscillator strength (OOS)
density,
\beq
\frac{\d f(\omega)}{\d \omega} = \frac{2Z}{\pi \Omega_{\rm p}^2} \,
\omega \, {\rm Im} \left( \frac{-1}{\epsilon(\omega)} \right),
\label{8.1}\eeq
where $\epsilon(\omega)$ is the optical DF and
\index{plasma resonance frequency}
\beq
\Omega_{\rm p} = \sqrt{4\pi\, {\cal N} Z \, (e^2/\me )}
\label{8.2}\eeq
is the plasma resonance frequency of a free electron gas with the
average electron density ${\cal N} Z$ of the material. The OOS is
assumed to satisfy the dipole sum rule
\beq
\int_0^\infty \frac{\d f(\omega)}{\d \omega} \, \d \omega = Z,
\label{8.3}\eeq
which suggests that the quantity $[\d f(\omega)/\d \omega] \, \d
\omega$ represents the number of electrons per molecule that have
resonance frequencies between $\omega$ and $\omega+\d \omega$.

In the classical form of the theory, ``close'' collisions are treated as
binary collisions between the projectile and the electrons in the
material, while energy transfers in ``distant'' interactions are
described by considering the electrons as classical harmonic oscillators
acted by the electric field of the projectile (the magnetic induction
does no work). It is assumed that the energy-loss caused by interactions
with electrons is much smaller than the kinetic energy $E$ of the
particle. This assumption is generally valid for particles much heavier
than the electron ($M_1 \gg \me$). In addition, we will disregard
deflections of the particle trajectory caused by collisions with nuclei
and electrons. Under these circumstances, the velocity ${\bf v}$ remains
essentially constant along a trajectory length similar to the range of
the interaction fields. To simplify the formulas, we consider that the
projectile moves along the $z$ axis of the laboratory reference frame,
where the material is at rest, with velocity ${\bf v}= v \hat{\bf z}$,
and that it passes by the origin of coordinates at $t=0$.  That is, the
projectile follows the trajectory ${\bf r} = {\bf v} t = (0,0,vt)$.


\subsection{Energy transfer in Coulomb collisions \label{sec8.1.1}}
\index{energy transfer in Coulomb collisions}

Let us first study the slowing down of the projectile by assuming it is
caused only by collisions with the electrons in the material. If the
velocity $v$ of the projectile is assumed to be much higher than the
typical velocities of the electrons in their atomic orbits (but still
non-relativistic), the target electrons can be treated as free and
initially at rest in the laboratory frame.

We limit our considerations to projectiles much heavier than the electron
($M_1 \gg \me$), so that the center-of-mass (CM) frame practically
coincides with the rest frame of the projectile. Evidently, the velocity
of the electron in the CM frame is $\simeq -{\bf v}$. The classical DCS
in the CM frame is given by the Rutherford formula, Eq.\ \req{4.105},
\index{Rutherford cross section}
\beq
\frac{\d \sigma_{\rm R}}{\d \Omega} = \left( \frac{Z_1 e^2}{2\me v^2}\right)^2
\frac{1}{\sin^4(\theta/2)},
\label{8.4}\eeq
where $\theta$ is the scattering angle in CM. The angle $\theta$ and the
impact parameter $b$ are related by Eq.\ \req{4.39}, that is,
\begin{subequations}
\label{8.5}
\beq
\tan(\theta/2) = \frac{Z_1 e^2}{\me v^2\, b}
\label{8.5a}\eeq
or, equivalently,
\beq
\sin^2(\theta/2) = \frac{(Z_1 e^2/\me v^2)^2}{b^2+(Z_1 e^2/\me v^2)^2}
\, .
\label{8.5b}\eeq
\end{subequations}
Notice that the impact parameter $b$ has the same values in the
laboratory and in the CM frames.
As discussed in Section \ref{sec4.1.5}, the classical
Rutherford DCS is valid (\ie,
consistent with the DCS obtained from a quantum calculation) when the
absolute value of the Sommerfeld parameter $\eta$ is much larger than
unity, that is, when \index{Sommerfeld parameter}
\beq
|\eta| =  \left| \frac{Z_1 e^2}{\hbar v} \right| \gg 1.
\label{8.6}\eeq
Strictly speaking, the formulas \req{8.4} and \req{8.5} are applicable
only to projectiles with velocities much smaller than the speed of
light. The main consequence of using these non-relativistic formulas will be
discussed below (see Section \ref{sec8.1.4}).

The energy transfer to the target electron in a collision is [Eq.\
\req{4.126}]
\beq
W = W_{\rm max} \, \sin^2(\theta/2),
\label{8.7}\eeq
where
\beq
W_{\rm max} = \frac{4 M_1 \me}{(M_1 + \me)^2} \, E =
\frac{2 \me v^2}{(1 + \me/M_1)^2}
\label{8.8}\eeq
is the maximum allowable energy transfer in a collision [Eq.\ \req{4.127}],
which occurs when $\theta=\pi$, \ie, in collisions where the electron
reverses its momentum in CM.
Equations \req{8.5} and \req{8.7} imply the following one-to-one
correspondence between the energy transfer and the impact parameter
\beq
W(b) = W_{\rm max} \,
\frac{(Z_1 e^2/\me v^2)^2}{b^2+(Z_1 e^2/\me v^2)^2}.
\label{8.9}\eeq
For the heavy particles considered here, we may use the approximations
\beq
W_{\rm max} \simeq 2 \me v^2
\qquad \mbox{and} \qquad
W (b) \simeq \frac{2 (Z_1 e^2)^2}{\me v^2} \, \frac{1}{b^2+(Z_1 e^2/\me
v^2)^2}.
\label{8.10}\eeq

Assuming the classical collision model is applicable, we can calculate
the stopping power as follows. When the projectile crosses a thin
material foil of thickness $\d z$, it interacts with electrons with all
possible impact parameters. Electrons with impact parameter $b$ ``see''
an appreciable field during a time interval $\Delta t \sim b/v$. In
reality these electrons are in
bound states with a certain frequency of motion $\omega'$. If $\Delta t$
is much shorter than the orbital period of the electron, $2\pi/ \omega'$
(\ie, at small impact parameters), the target electron can be considered
as free. In the opposite situation, when the time $\Delta t$ is longer
than the orbital period, the target electron responds adiabatically,
that is, its orbit stretches slowly as the particle approaches the
origin of coordinates and returns to normal when the particle moves
away, without absorbing any amount of energy \citep[see,
\eg,][]{Griffiths1995}. That is, ``collisions'' with
impact parameter greater than a certain adiabatic cutoff, of the order of
\beq
b_{\rm ad} = \frac{v}{\omega'},
\label{8.11}\eeq
do not contribute to the electronic stopping power. We thus have
\beq
\left[ - \, \frac{\d E}{\d z}\right]_{\rm col}
= {\cal N} Z \int_0^{b_{\rm ad}}
W(b) \, 2\pi b \, \d b.
\label{8.12}\eeq
Inserting the second expression in Eq.\ \req{8.10}, we obtain
\beq
\left[ - \, \frac{\d E}{\d z}\right]_{\rm col}
= {\cal N} Z \,  2\pi \frac{(Z_1 e^2)^2}{\me v^2}
\ln \left[ \frac{b_{\rm ad}^2+(Z_1 e^2/\me v^2)^2}
{(Z_1 e^2/\me v^2)^2} \right].
\label{8.13}\eeq
For high-energy particles $b_{\rm ad} \gg \left| Z_1 \right| e^2/(\me
v^2)$, and we can write
\beq
\left[ - \, \frac{\d E}{\d z}\right]_{\rm col}
\simeq {\cal N} Z \,  2\pi \frac{(Z_1 e^2)^2}{\me v^2}
\ln \left[ \left( \frac{b_{\rm ad}}
{b_{\rm min}} \right)^2 \right]
\label{8.14}\eeq
with
\beq
b_{\rm min} = \frac{\left| Z_1 \right| e^2}{\me v^2 }
= |\eta| \, \frac{\hbar}{p}\, .
\label{8.15}\eeq
For the purpose of calculating the stopping power, this approximation is
equivalent to setting
\beq
W(b) = \frac{2 (Z_1 e^2)^2}{\me v^2} \, \frac{1}{b^2}\, ,
\label{8.16}\eeq
and considering that only impact parameters larger than $b_{\rm min}$
(but less than $b_{\rm ad}$) are allowed. We thus obtain
\beq
\left[ - \, \frac{\d E}{\d z}\right]_{\rm col} \simeq
\frac{4 \pi Z_1^2 e^4}{\me v^2} \, {\cal N} Z \,
\ln \left( \frac{v}{\omega'} \,
\frac{\me v^2 }{\left| Z_1 \right| e^2} \right).
\label{8.17}\eeq
Evidently, the validity of this formula is questionable because of the
ad-hoc introduction of the adiabatic limit to the impact parameter.


\subsection{Energy transfer to classical oscillators \label{sec8.1.2}}
\index{energy transfer to classical oscillators}

Classical binary collisions with sufficiently large impact parameters
involve energy transfers $W(b)$ that are less than the typical
excitation energies of the material. Consequently, these ``distant''
excitations are affected by binding effects. To compute the energy
transfer to the material, with account of electron binding, we follow
\citet{Bohr1913, Bohr1915} and consider the electrons in the medium as isotropic
harmonic oscillators with characteristic frequencies $\omega_j$ that are
subject to the electromagnetic field of the passing particle. In the
quantum formulation, the oscillator has an infinite number of discrete
levels, $\epsilon_n = (n+3/2) \hbar \omega_j$, with corresponding
excitation energies $\Delta E_n = n \hbar \omega_j$. Apparently, a
microscopic formulation with classical oscillators, which can absorb
arbitrary amounts of energy, is incompatible with a discrete excitation
spectrum. It is a fortunate fact that the classical formulation does
provide the correct average energy transfer to an oscillator that is
sufficiently far from the trajectory of the particle. That is, the
classical theory is unable to describe individual interactions but
yields the correct contribution of the oscillator to the stopping power.

\begin{figure}[hbt] \begin{center}
\includegraphics*[scale=0.80]{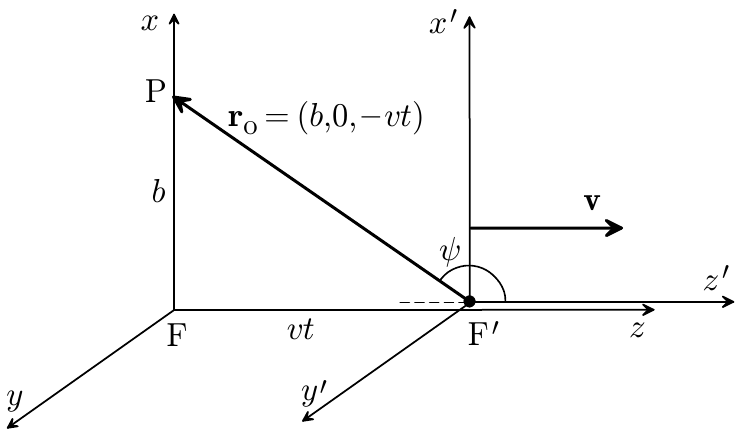}
\caption{
Laboratory reference frame (F) and rest frame (F') of a particle moving
with constant velocity ${\bf v}=v \hat{\bf z}$ in F. The particle is
assumed to pass the origin of coordinates of F at $t=0$.
}\label{fig8.1}
\end{center} \end{figure}

Hence, we can calculate the energy transfer to distant oscillators
classically, as the work done on them by the electromagnetic force. As
before, we assume that the charged particle follows the trajectory ${\bf
r} = vt \, \hat{\bf z}$ and we consider a target electron at the
position $(b,0,0)$, which corresponds to the impact parameter $b$, under
the action of the electromagnetic field of the passing particle.  The
electromagnetic field of the projectile at the position of the electron
can be obtained either from the Li\'{e}nard-Wiechert formulas of the
retarded fields [see Eqs.\ \req{1.28}] or as the Lorentz transform of
the Coulomb field in the rest frame of the projectile \cite[see,
\eg,][Section 11.10]{Jackson1975}. We are interested in the values of
the electromagnetic field at points on the plane $z=0$.  Because the
field is axially symmetric about the $z$ axis, it suffices knowing it at
the points P $(b,0,0)$ of the $x$ axis (Fig.\ \ref{fig8.1}). The
components of the electric and magnetic fields at P are
\index{electromagnetic field!of a moving charge}
\beq
\begin{array}{ll}
\displaystyle{
{\cal E}_x (b,t) = \frac{Z_1 e \, \gamma b}{(b^2+\gamma^2 v^2 t^2)^{3/2}},} &
{\cal B}_x (b,t) = 0, \\ [2mm]
{\cal E}_y (b,t) = 0,  &
\displaystyle{
{\cal B}_y (b,t) =  \frac{Z_1 e\, \beta \gamma b}{(b^2+\gamma^2 v^2
t^2)^{3/2}} =
\beta {\cal E}_x(b,t),}
\\ [2mm]
\displaystyle{
{\cal E}_z (b,t) = - \frac{Z_1 e \, \gamma vt}{(b^2+\gamma^2 v^2 t^2)^{3/2}},}
\rule{10mm}{0mm} &
{\cal B}_z (b,t) = 0,
\end{array}
\label{8.18}\eeq
where
\beq
\beta = v/c \qquad \mbox{and} \qquad
\gamma=\sqrt{ \frac{1}{1-\beta^2}} \, .
\label{8.19}\eeq
We can write the electric field more compactly as [see Eq.\ \req{1.26a}]
\beq
\ecb(b,t) = \frac{Z_1 e}{r_{\rm o}^3 \gamma^2} \,
\frac{{\bf r}_{\rm o}}{\left( 1 - \beta^2 \sin^2 \psi \right)^{3/2}}\, ,
\label{8.20}\eeq
where ${\bf r}_{\rm o}$ is the vector from the position of the charge to
the observation point P and $\psi=\cos^{-1}(\hat{\bf r}_{\rm o} \cdot
\hat{\bf v})$. The electric field is radial (\ie, parallel to ${\bf
r}_{\rm o}$) but differs from the isotropic field of a charge at rest;
the field along the direction of motion ($\psi=0$) is decreased by a
factor $\gamma^{-2}$, while the field in the transverse direction
($\psi=\pi/2$) is larger by a factor of $\gamma$. Figure \ref{fig8.2}
displays the components of the electric field at the observation point P
as functions of the position $vt$ of the passing charged particle.
Notice that the fields at P are appreciable over a time interval of the
order of
\beq
\Delta t = \frac{b}{\gamma v}\, .
\label{8.21}\eeq

\begin{figure}[hbt] \begin{center}
\includegraphics*[scale=0.75]{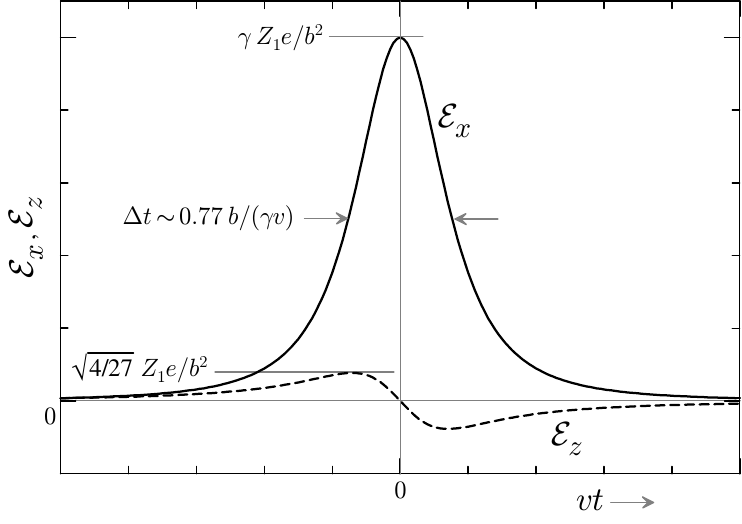}
\caption{
Components of the electric field of a passing charged particle at a point
P with impact parameter $b$ (Fig.\ \ref{fig8.1}) as functions of the
particle position $vt$, with characteristic quantities.
}\label{fig8.2}
\end{center} \end{figure}

For large enough impact parameters $b$, the force on the electron is
small and the amplitude of the oscillatory motion is much smaller than
$b$ throughout the interaction. Hence, we can neglect the variation of
that force with the position of the electron, similarly to the so-called
dipole approximation that is employed in studies of emission and
absorption of radiation by atoms \citep[see,
\eg,][]{BransdenJoachain1983}. Initially (\ie, at $t=-\infty$) the
electron is at rest in its equilibrium position. The displacement ${\bf
u}_j(t)$ of the electron at time $t$ from its equilibrium position is
determined by the force equation
\beq
\ddot{\bf u}_j + s_j \dot{\bf u}_j + \omega_j^2 {\bf u}_j = - \frac{e}{\me}
\, \ecb(b,t).
\label{8.22}\eeq \index{Fourier transform}
Frequency-Fourier analysis of ${\bf u}_j(t)$ and $\ecb(b,t)$ gives [see
Eq.\ \req{1.151}]
\beq
{\bf u}_j(\omega) = - \frac{e}{\me} \, \frac{\ecb(b,\omega)}{
\omega_j^2 - \omega^2 - {\rm i} s_j \omega}.
\label{8.23}\eeq
The rate of work done on the oscillator by the field of the projectile
is
\beq
\frac{\d W}{\d t} = \int {\bf j}_{\rm osc}({\bf r},t) \dotprod \ecb(b,t) \,
\d {\bf r}\, ,
\label{8.24}\eeq
where ${\bf j}_{\rm osc}({\bf r},t)=-e \, \dot{\bf u}_j \, \delta({\bf r}
-{\bf u}_j)$ is the current density of the oscillator. The total work done
along the complete trajectory of the projectile is
\beq
W (b,\omega_j) = \int_{-\infty}^\infty
\frac{\d W}{\d t} \, \d t = -e \int_{-\infty}^\infty
\dot{\bf u}_j(t) \dotprod \ecb(b,t) \, \d t.
\label{8.25}\eeq
Expressing the vectors in terms of their Fourier transforms,
\beqa
W (b,\omega_j) &=& -e \int_{-\infty}^\infty
(2\pi)^{-1} \left[ -{\rm i} \int_{-\infty}^\infty
\d \omega \, \omega \, \exp(-{\rm i} \omega t)
\, {\bf u}_j(\omega) \right] \dotprod \left[
\int_{-\infty}^\infty \d \omega' \, \exp(-{\rm i} \omega' t)
\, \ecb (b,\omega') \right] \d t
\nonumber \\ [2mm]
&=& {\rm i} e \int_{-\infty}^\infty \d \omega  \int_{-\infty}^\infty
\d \omega' \, \omega \,
{\bf u}_j (\omega) \dotprod
\ecb(b,\omega') \, \delta(\omega+\omega')
= {\rm i} e \int_{-\infty}^\infty
\d \omega \, \omega \, {\bf u}_j(\omega) \dotprod
\ecb (b,-\omega)
\nonumber \\ [2mm]
&=& {\rm i} e \int_{-\infty}^\infty \d \omega \,
\omega \, {\bf u}_j(\omega) \dotprod \ecb^\ast (b,\omega).
\label{8.26}\eeqa
Since the vectors ${\bf u}_j(t)$ and $\ecb(b,t)$ are real, we can recast
this expression as an integral over positive frequencies,
\beqa
W (b,\omega_j) &=&
{\rm i} e \int_{-\infty}^0 \d \omega \,
\omega \, {\bf u}_j(\omega) \dotprod \ecb^\ast (b,\omega)
+ {\rm i} e \int_0^\infty \d \omega \,
\omega \, {\bf u}_j(\omega) \dotprod \ecb^\ast (b,\omega)
\nonumber \\ [2mm]
&=& - {\rm i} e \int_0^\infty \d \omega' \,
\omega' \, {\bf u}_j(-\omega') \dotprod \ecb^\ast (b,-\omega')
+ {\rm i} e \int_0^\infty \d \omega \,
\omega \, {\bf u}_j(\omega) \dotprod \ecb^\ast (b,\omega)
\nonumber \\ [2mm]
&=& - {\rm i} e \int_0^\infty \d \omega \,
\omega \, {\bf u}_j^\ast(\omega) \dotprod \ecb (b,\omega)
+ {\rm i} e \int_0^\infty \d \omega \,
\omega \, {\bf u}_j(\omega) \dotprod \ecb^\ast (b,\omega)
\nonumber \\ [2mm]
&=& 2 e \, {\rm Re} \left( {\rm i} \int_0^\infty \d \omega \,
\omega \, {\bf u}_j(\omega) \dotprod \ecb^\ast (b,\omega) \right).
\label{8.27}\eeqa
Introducing the expression \req{8.23}
\beqa
W (b,\omega_j) &=& - \frac{2e^2}{\me} \, {\rm Re}
\left( {\rm i} \int_0^\infty \d \omega \,
\omega \, \frac{\ecb(b,\omega)}{
\omega_j^2 - \omega^2 - {\rm i} s_j \omega}
\dotprod \ecb^\ast (b,\omega) \right)
\nonumber \\ [2mm]
&=& \frac{2e^2}{\me} \,
\int_0^\infty \d \omega \, \left|\ecb (b,\omega)\right|^2
\frac{s_j \omega^2}{
(\omega_j^2 - \omega^2)^2 + s_j^2 \omega^2}\, .
\label{8.28}\eeqa
Assuming that the damping constant $s_j$ is much smaller than the
resonance energy $\omega_j$, which is usually the case, the integrand
has a sharp peak at $\omega_j$ with an approximately Lorentzian shape.
Then, $\ecb (b,\omega)$ can be approximated by its value at
$\omega=\omega_j$ and we can write
\beqa
W (b,\omega_j) &\simeq& \frac{2e^2}{\me} \,  \left|\ecb (b,\omega_j)\right|^2
\int_0^\infty \d \omega \, \frac{s_j \omega^2}{
(\omega_j^2 - \omega^2)^2 + s_j^2 \omega^2}
\nonumber \\ [2mm]
&=& \frac{2e^2}{\me} \,  \left|\ecb (b,\omega_j)\right|^2
\int_0^\infty \d x \, \frac{x^2}{
[(\omega_j/s_j)^2 - x^2]^2 + x^2}
\label{8.29}\eeqa
with $x=\omega/s_j$. The integral equals $\pi/2$ independently of the value of
$\omega_j/s_j$. Hence
\beq
W (b,\omega_j) \simeq
\frac{\pi e^2}{\me} \,  \left|\ecb (b,\omega_j)\right|^2 .
\label{8.30}\eeq
It is worth noticing that this expression of the energy absorbed by a
non-relativistic oscillator is valid not only for the field of a passing
charged particle, but also for any radiation pulse that is essentially
constant within the atomic volume.

The Fourier transform of the electric field \req{8.18} is
\beqa
\ecb(b,\omega) &=&
(2\pi)^{-1/2} \int_{-\infty}^\infty \d t \, \exp({\rm i} \omega t)
\, \left[ Z_1 e \, \frac{\gamma b \, \hat{\bf x} - \gamma vt \, \hat{\bf
z}}{(b^2 + \gamma^2 v^2 t^2)^{3/2}}\right]
\nonumber \\ [2mm]
&=& \frac{Z_1 e}{b^2} \, \frac{b}{\gamma v}
(2\pi)^{-1/2} \int_{-\infty}^\infty \d x \, \exp[{\rm i} (\omega b
/\gamma v) x]
\, \frac{\gamma \, \hat{\bf x} - x \, \hat{\bf
z}}{(1 + x^2)^{3/2}}
\label{8.31}\eeqa
with $x=\gamma v t/b$. Noting that the real and imaginary parts of the
terms in the integrand
have definite parities, the integrals in this expression are of the
types
\begin{subequations}
\label{8.32}
\beq
\int_{-\infty}^\infty \d x \, \frac{\exp({\rm i} \xi x)}{(1 + x^2)^{3/2}}
= 2 \int_0^\infty \d x \, \frac{\cos(\xi x)}{(1 + x^2)^{3/2}} =
2\xi \, K_1(\xi)
\label{8.32a}\eeq
and
\beq
\int_{-\infty}^\infty \d x \, \frac{x \exp({\rm i} \xi x)}{(1 + x^2)^{3/2}}
= 2 {\rm i} \int_0^\infty \d x \, \frac{x \sin(\xi x)}{(1 + x^2)^{3/2}}
= 2 {\rm i} \, \xi \, K_0(\xi),
\label{8.32b}\eeq
\end{subequations}
where $K_n(x)$ are modified Bessel functions of orders $n=0$ and 1 (see
Section \ref{appB.3} in Appendix \ref{appB}). We can then write
\beqa
\ecb(b,\omega) &=&
\frac{Z_1 e}{v b}\, \left( \frac{2}{\pi} \right)^{1/2}
\left[ \, \xi K_1(\xi) \, \hat{\bf x}
- {\rm i}  \frac{1}{\gamma} \, \xi K_0(\xi) \, \hat{\bf z}  \right]
\label{8.33}\eeqa
with
\beq
\xi = \frac{\omega b}{\gamma v}\, .
\label{8.34}\eeq
Inserting the expression \req{8.33} into the right-hand side of Eq.\
\req{8.30}, we find that the energy transfer to the oscillator is
\beq
W (b,\omega_j) =
\frac{2 Z_1^2 e^4}{\me v^2} \; \frac{1}{b^2}
\left[ \xi_j^2 K_1^2(\xi_j) + \frac{1}{\gamma^2} \xi_j^2 K_0^2(\xi_j)
\right], \qquad \xi_j = \frac{\omega_j b}{\gamma v} \, .
\label{8.35}\eeq
The factor before the square bracket is just the approximate result for
classical collisions with free electrons \req{8.16}. Introducing the
limiting forms of the modified Bessel functions for small and large
arguments [see Eqs.\ \req{B.60} and \req{B.62} in Appendix \ref{appB}], we
obtain the following  limiting expressions
\beq
W (b,\omega_j) \simeq
\frac{2 Z_1^2 e^4}{\me v^2} \; \frac{1}{b^2}
\left\{
\begin{array}{ll}
1 & \mbox{if $\xi_j \ll 1$,} \\ [2mm]
\displaystyle{\left( 1 + \frac{1}{\gamma^2} \right) \frac{\pi}{2} \,
\xi_j
\exp(-2\xi_j)} \rule{7mm}{0mm}
& \mbox{if $\xi_j \gg 1$.}
\end{array} \right.
\label{8.36}\eeq
Because $\xi_j=b/b_{\rm ad}$ with $b_{\rm ad} = \gamma v /\omega_j$ [the
relativistic version of Eq.\ \req{8.11}], this result shows that for $b
\ll b_{\rm ad}$ the energy transfer is approximately the same as in
collisions with free electrons at rest, Eq.\ \req{8.16}. Additionally,
for $b \gg b_{\rm ad}$ the energy transfer decreases exponentially to
zero; this behavior justifies the qualitative adiabatic argument used to
estimate $b_{\rm ad}$ in the previous Section.

The contribution to the stopping power of oscillators with impact
parameters larger than a given cutoff value $a$ is
\beqa
\left[ - \, \frac{\d E}{\d z} \right]_{b>a} &=& 2 \pi {\cal N} \,
\int_0^\infty
\d \omega_j \, \frac{\d f(\omega_j)}{\d \omega_j} \int_{a}^\infty
W (b,\omega_j) \, b \, \d b
\nonumber \\ [2mm]
&=& 2 \pi {\cal N} \, \int_0^\infty \d \omega_j \,
\frac{\d f(\omega_j)}{\d \omega_j}
\frac{2 Z_1^2 e^4}{\me v^2} \left( \frac{1}{b^2} \right)
\int_{a}^\infty
\left[ \xi_j^2 K_1^2(\xi_j) + \frac{1}{\gamma^2} \xi_j^2 K_0^2(\xi_j) \right]
\, b \, \d b
\nonumber \\ [2mm]
&=& \frac{4\pi Z_1^2 e^4}{\me v^2} \, {\cal N} \,
\int_0^\infty \d \omega_j \, \frac{\d f(\omega_j)}{\d \omega_j}
\int_{\xi_a}^\infty
\left[ K_1^2(\xi) + \frac{1}{\gamma^2} K_0^2(\xi) \right]
\, \xi \, \d \xi
\label{8.37}\eeqa
with $\xi_a = \omega_j a/(\gamma v)$. Since the integrand decreases
rapidly to zero for large $b$, the upper limit of the second integral has been
set to $\infty$. The integral over $\xi$ can be solved in closed form
with the aid of the formulas \req{B.69} and \req{B.70}, which give
\beqa
\left[ - \, \frac{\d E}{\d z} \right]_{b>a} &=&
\frac{4\pi Z_1^2 e^4}{\me v^2} \, {\cal N} \,
\int_0^\infty \d \omega_j \, \frac{\d f(\omega_j)}{\d \omega_j}
\nonumber \\ [2mm]
&& \mbox{} \times \left\{ \xi_a K_1(\xi_a) K_0(\xi_a)
- \frac{v^2}{2c^2} \xi_a^2 \left[ K_1^2(\xi_a) - K_0^2(\xi_a) \right]
\right\}.
\label{8.38}\eeqa

In situations of practical interest with high-energy particles (see
below) $\xi_a \ll 1$ and we can use the limiting expressions \req{B.64}
to simplify \req{8.38} as follows
\beqa
\left[ - \, \frac{\d E}{\d z} \right]_{b>a}
&=& \frac{4\pi Z_1^2 e^4}{\me v^2} \, {\cal N} \,
\int_0^\infty \d \omega_j \, \frac{\d f(\omega_j)}{\d \omega_j}
\left\{ - \ln(\xi_a/2) - g - \frac{v^2}{2c^2} \right\}
\nonumber \\ [2mm]
&=& \frac{4\pi Z_1^2 e^4}{\me v^2} \, {\cal N} \,
\int_0^\infty \d \omega_j \, \frac{\d f(\omega_j)}{\d \omega_j}
\left\{ \ln\left(\frac{2 \exp(-g) \gamma v}{a \omega_j} \right)
- \frac{v^2}{2c^2} \right\},
\label{8.39}\eeqa
where $g=0.5772$ is Euler's constant. \index{Euler constant}
Using the dipole sum rule \req{8.3}, we have
\beqa
\left[ - \, \frac{\d E}{\d z} \right]_{b>a}
&=& \frac{4\pi Z_1^2 e^4}{\me v^2} \, {\cal N} \, Z
\left\{ \ln\left(\frac{2 \exp(-g) \, v}{a \overline{\omega}} \right)
+ \frac{1}{2} \, \ln \left( \frac{1}{1-\beta^2} \right)
- \frac{1}{2} \, \beta^2 \right\},
\label{8.40}\eeqa
where $\overline{\omega}$ is an average resonance frequency defined by
\index{mean excitation energy}
\index{mean excitation potential}
\beq
Z \ln \overline{\omega} = \int_0^\infty \ln(\omega) \,
\frac{\d f(\omega)}{\d \omega} \,  \d \omega.
\label{8.41}\eeq
Evidently, the definition \req{6.285} implies that $\hbar
\overline{\omega}=I$, the mean excitation energy of the Bethe theory.


\subsection{Bohr's classical stopping-power formula \label{sec8.1.3}}

In the foregoing analysis we have treated interactions of the particle with
electrons at small and moderate impact parameters $b$ as
binary collisions. Interactions with electrons that are far
from the trajectory of the particle are more appropriately represented
by considering the electrons in the material as classical oscillators.
We have already shown that the two calculations yield equivalent results
for impact parameters that are not too large. It is then reasonable to
combine the two approaches by introducing an intermediate cutoff impact
parameter $a$ that separates ``close'' and ``distant'' collisions, that is,
collisions with impact parameters smaller and larger than $a$.

The contribution of close collisions to the stopping power is [see the
second expression in Eq. \req{8.10}]
\beqa
\left[ - \, \frac{\d E}{\d z} \right]_{b<a}
&=& {\cal N} Z \int_0^a W(b) \, 2\pi \, b \d b
\nonumber \\ [2mm]
&=& \frac{2\pi Z_1^2 e^4}{\me v^2} \, {\cal N} \, Z \,
\ln \left[ \frac{a^2+(Z_1 e^2/\me v^2)^2}
{(Z_1 e^2/\me v^2)^2} \right].
\label{8.42}\eeqa
Generally $a \gg Z_1 e^2/(\me v^2)$, and we can write
\beq
\left[ - \, \frac{\d E}{\d z} \right]_{b<a}
= \frac{4\pi Z_1^2 e^4}{\me v^2} \, {\cal N} \, Z \,
\ln \left( a \, \frac{\me v^2}{|Z_1| \,  e^2} \right).
\label{8.43}\eeq

The stopping power $S$ is obtained by adding the contributions from close
and distant collisions,
$$
S = - \, \frac{\d E}{\d z} =
\left[ - \, \frac{\d E}{\d z} \right]_{b<a}
+ \left[ - \, \frac{\d E}{\d z} \right]_{b>a}.
$$
This gives \index{Bohr's stopping power formula}
\beq
S =
\frac{4\pi Z_1^2 e^4}{\me v^2} \, {\cal N} \, Z
\left[ \ln\left(\frac{2 \exp(-g) \, \me v^3}{|Z_1| \, e^2 \,
\overline{\omega}} \right)
+ \frac{1}{2} \,
\ln \left( \frac{1}{1-\beta^2} \right) - \frac{1}{2} \, \beta^2
\right],
\label{8.44}\eeq
which is the classical formula for the stopping power, first derived by
\citet{Bohr1913, Bohr1915}. It is interesting to observe that the only
characteristics of the material entering this formula are the total
average electron density ${\cal N} Z$ and the mean frequency
$\overline{\omega}$. For a given speed $v$ of the projectile, the
classical stopping power is independent of the mass of the particle and
an even function of its charge $Z_1$, \ie, particles with equal masses
and opposite charges experience the same stopping force.


\subsection{Relativistic effects in close collisions \label{sec8.1.4}}

\index{stopping power!Bohr formula!relativistic close collisions}

The Bohr formula differs from the Bethe formula \req{6.295} in that it
contains only one half of the relativistic correction, $\ln (\gamma^2) -
\beta^2$, which is correctly predicted by the PWBA. The derivation of
the Bethe formula (Section \ref{sec6.9}) shows that exactly one half of
this correction is due to distant (low-$Q$) collisions and the other
half is due to close (high-$Q$) collisions. The Bohr formula gives the
correct relativistic correction for distant interactions, however the
correction for close collisions is missing because these collisions were
described by using non-relativistic physics.

To get the correct relativistic form of the stopping-power formula, we
slightly modify Bohr's argument. For projectiles much heavier than the
electron ($M_1 \gg \me$) the center-of-mass (CM) frame practically
coincides with the rest frame of the projectile, and the velocity of the
electron in the CM frame is $\simeq -{\bf v}$. As discussed by
\citet{JacksonMcCarthy1972}, the DCS in CM is essentially the same as
the DCS for scattering of an electron that moves with speed $v$ in the
Coulomb potential of the projectile; this approach is also followed by
\citet{LindhardSorensen1996} in their study of the stopping power of
high-energy projectiles. The scattering of an electron by a fixed point
charge $Z_1 e$ is described by the Mott DCS, which is obtained from the
exact solution of the Dirac equation for the Coulomb potential (see
Section \ref{sec5.2}). We consider the \citet{McKinleyFeshbach1948}
expansion of the Mott DCS, Eq.\ \req{5.116},
\beqa
\frac{\d \sigma}{\d \Omega} &=& \left( \frac{Z_1 e^2}{2 v p} \right)^2
\frac{1}{\sin^4(\theta/2)}
\nonumber \\ [2mm]
&& \mbox{} \times \left\{ 1 - \beta^2 \sin^2(\theta/2)
+ \pi Z_1 \alpha \beta \sin(\theta/2) \,
\left[ 1 - \sin(\theta/2) \right] \rule{0mm}{4mm}\right\},
\label{8.45}\eeqa
where $p=\beta\gamma \me c$ is the momentum of an electron moving with
the speed of the projectile and $\alpha = e^2/(\hbar c)\simeq 1/137$ is the
fine-structure constant. The relativistic correction that is
missing in Bohr's stopping-power formula is obtained by using
relativistic kinematics and by replacing the Rutherford DCS \req{8.4} with
the approximate DCS,
\beq
\frac{\d \sigma}{\d \Omega} = \left( \frac{Z_1 e^2}{2vp}\right)^2
\frac{1}{\sin^4(\theta/2)} \left[ 1 - \beta^2 \sin^4(\theta/2) \right],
\label{8.46}\eeq
obtained by discarding the $Z_1^3$ term of the McKinley--Feshbach
DCS. The
DCS \req{8.46} is also obtained from the Born approximation for spin
$\1o2$ particles [cf.\ Eq.\ \req{5.119}]. Since the spin
correction, $-\beta^2 \sin^2(\theta/2)$, decreases when the energy loss
decreases, the equality \req{8.5b} with the electron relativistic mass
$\gamma \me$,
\beq
\sin^2(\theta/2) = \frac{(Z_1 e^2/\gamma \me v^2)^2}
{b^2+(Z_1 e^2/\gamma \me v^2)^2},
\label{8.47}\eeq
is still valid for impact parameters that are sufficiently large (\ie,
for small scattering angles in CM).

After the collision, the target electron recoils with the kinetic energy
\beq
W = W_{\rm max} \, \sin^2(\theta/2) = W_{\rm max} \,
\frac{1-\cos\theta}{2},
\label{8.48}\eeq
where [see Eq.\ \req{4.189}]
\beq
W_{\rm max} = \frac{2 \beta^2 \gamma^2 \, \me c^2}{
1 + 2 (\me/M_1) \gamma + (\me/M_1)^2} \simeq
2 \beta^2 \gamma^2 \, \me c^2
\label{8.49}\eeq
is the largest allowed energy transfer. Hence, the energy-loss DCS is
\beq
\frac{\d \sigma}{\d W} = \frac{2\pi \, \sin \theta \, \d \theta}{\d W}
\, \frac{\d \sigma}{\d \Omega} =
\frac{2\pi Z_1^2 e^4}{\me v^2} \, \frac{1}{W^2}
\left( 1 - \beta^2 \frac{W}{W_{\rm max}} \right),
\label{8.50}\eeq
which differs from the DCS underlying Bohr's formulation in that 1)
$v=\beta c$ is the relativistic velocity and 2) it includes the spin
correction factor. Equations \req{8.47} and \req{8.48} imply the
following relationship between the energy loss and the impact parameter
$b$,
\beq
W(b) = W_{\rm max} \, \frac{(Z_1 e^2/\gamma \me v^2)^2}
{b^2+(Z_1 e^2/\gamma \me v^2)^2}\, ,
\label{8.51}\eeq
Then, for binary collisions with electrons that are not too close to the
projectile's trajectory (for which spin effects are expected to be
small) we can write
\beq
W(b) = W_{\rm max} \, \frac{b_{\rm min}^2}{b^2 + b_{\rm min}^2}
\qquad \mbox{with} \qquad
b_{\rm min} =  \frac{|Z_1| \, e^2}{\gamma \, \me v^2}.
\label{8.52}\eeq
We can now select the cutoff impact parameter $a$ such that $a \gg
b_{\rm min}$ and, hence,
\beq
\frac{W(a)}{W_{\rm max}} \simeq \frac{b_{\rm min}^2}{a^2} \ll 1.
\nonumber\eeq
The contribution to the stopping power of collisions with impact
parameters less than $a$ is given by
\beq
\left[ - \, \frac{\d E}{\d z} \right]_{b<a} =
{\cal N} Z \int_{W(a)}^{W_{\rm max}} W \,
\frac{\d \sigma}{\d W} \, \d W
= \frac{2\pi Z_1^2 e^4}{\me v^2} \, {\cal N} Z
\left[ \ln W - \beta^2 \frac{W}{W_{\rm max}}
\right]_{W(a)}^{W_{\rm max}}.
\label{8.53}\eeq
For projectiles with sufficiently high energies we have
\beqa
\left[ - \, \frac{\d E}{\d z} \right]_{b<a}
& \simeq & \frac{2\pi Z_1^2 e^4}{\me v^2} \, {\cal N} Z
\left[ \ln \left( \frac{a^2}{b_{\rm min}^2} \right) - \beta^2
\right]
\nonumber \\ [2mm]
&=& \frac{2\pi Z_1^2 e^4}{\me v^2} \, {\cal N} Z
\left[ 2 \, \ln \left( a \,
\frac{\gamma \, \me v^2}{|Z_1| e^2} \right)
- \beta^2 \right]\rule{25mm}{0mm}
\nonumber \\ [2mm]
&=& \frac{4\pi Z_1^2 e^4}{\me v^2} \, {\cal N} Z
\left[ \ln \left( a \,
\frac{\me v^2}{|Z_1| e^2} \right) + \frac{1}{2} \ln \left(
\frac{1}{1-\beta^2} \right)
- \frac{1}{2} \, \beta^2 \right].
\rule{10mm}{0mm}
\label{8.54}\eeqa
The first term in this expression coincides with Bohr's non-relativistic
result, Eq.\ \req{8.43}. The second and third terms give the correct
relativistic contribution of close interactions to the
stopping power [cf.\ Eq.\ \req{6.304}] thus amending a deficiency of
the non-relativistic classical formula of Bohr. The ``relativistic'' Bohr
formula now reads  \index{Bohr's relativistic stopping power formula}
\beqa
S &=& \left[ - \, \frac{\d E}{\d z} \right]_{b<a}
+ \left[ - \, \frac{\d E}{\d z} \right]_{b>a}
\nonumber \\ [2mm]
&=& \frac{4\pi Z_1^2 e^4}{\me v^2} \, {\cal N} \, Z
\left[ \ln\left(\frac{2 \exp(-g) \, \me v^3}{|Z_1| \, e^2 \,
\overline{\omega}} \right)
+ \ln \left( \frac{1}{1-\beta^2} \right) - \beta^2
\right].
\label{8.55}\eeqa

\index{stopping power!Bohr formula|)}


\section{Dielectric polarization of the medium \label{sec8.2}}
\index{dielectric polarization}
\index{Fermi density effect}

In reality, the mass stopping power is not strictly proportional to the
density of the material for two reasons. First, although the OOS is
frequently considered as an atomic property, it is clear that the
resonance frequencies of weakly bound (valence and conduction) electrons
depend somehow on the state of aggregation of the material. That is, the
OOS and the mean frequency $\overline{\omega}$ are characteristics of
each material phase, defined by its chemical composition {\it and}
microscopic structure. Thus, water vapor, liquid water and ice have
different OOSs and $\overline{\omega}$ values. Secondly, the stopping
power also depends on the density of the material because the
electromagnetic field of the projectile is modified by the electric
polarization of the medium. As pointed out by \citet{Fermi1940} when a
charged projectile traverses a given mass thickness of material, the
energy loss is larger in a rarefied medium than in a dense one. The
reduction of the stopping power that arises from the larger polarizability
of the denser medium (assuming that the two media have the same OOS!) is
known as the {\it Fermi density effect}, a misnomer for what should be
called the {\it dielectric polarization effect} .

The dependence of the stopping power on the density and structure of the
material can be accounted for within the classical dielectric formalism
(Section \ref{sec1.4}), which assumes complete knowledge of the
longitudinal and transverse DFs of the material, $\epsilon^{\rm
(L,T)}({\bf q},\omega)$, as functions of the wave vector and the
frequency of the field. Evidently, differences between the responses of
media with the same composition but in different phases (solid, liquid,
or gas) are embodied in the DFs. For the sake of concreteness (and
consistency) we limit our considerations to the case of homogeneous
isotropic media, with DFs that are independent of the direction of the
wave vector.

As in previous Sections, we consider a projectile of charge $Z_1 e$ that
passes by the origin of coordinates at $t=0$ moving with velocity ${\bf
v}=(0,0,v)$ in the direction of the $z$ axis. The charge and current
distributions associated with the projectile are, respectively,
\beq
\rho_{\rm ext}({\bf r},t) = Z_1 e \, \delta ({\bf r} - {\bf v}t)
\qquad \mbox{and} \qquad
{\bf j}_{\rm ext}({\bf r},t) = Z_1 e \, {\bf v} \, \delta ({\bf r} -
{\bf v}t).
\label{8.56}\eeq
As shown in Section \ref{sec1.4},
the Fourier transforms of the electromagnetic field of the projectile
are [see Eqs.\ \req{1.176} and \req{1.177}] \index{Fourier transform}
\beqa
\ecb({\bf q},\omega) &=&
- {\rm i} \, \frac{2 Z_1 e}{\omega} \,
\frac{1}{\epsilon^{\rm (L)} ({\bf q},\omega)} \,
\frac{{\bf q} ({\bf q} \dotprod {\bf v})}{q^2} \,
\delta({\bf q} \dotprod {\bf v} - \omega)
\nonumber \\ [2mm]
&& \mbox{}
+ {\rm i} \, \frac{2 Z_1 e \omega}{c^2} \,
\frac{1}{q^2 - (\omega/c)^2 \, \epsilon^{\rm (T)}({\bf q}, \omega)}
\left( {\bf v} - \frac{{\bf q} ({\bf q} \dotprod {\bf v})}{q^2} \right)
\delta({\bf q} \dotprod {\bf v} - \omega)
\label{8.57}\eeqa
and
\beqa
\bcb ({\bf q}, \omega) &=&
{\rm i} \, \frac{2 Z_1 e}{c} \,
\frac{1}
{q^2 - (\omega/c)^2 \, \epsilon^{\rm (T)}({\bf q}, \omega)} \,
{\bf q} \vecprod {\bf v}
\, \delta({\bf q} \dotprod {\bf v} - \omega).
\label{8.58}\eeqa
We recall that the transforms of the fields in vacuum, $\ecb_{\rm ext}({\bf
q},\omega)$ and $\bcb_{\rm ext}({\bf q},\omega)$, are obtained from
these expressions with $\epsilon^{\rm (L,T)}({\bf q}, \omega)=1$.

The frequency-Fourier transforms of the fields at a point ${\bf r}$ are
given by
\beqa
\ecb ({\bf r},\omega) & \equiv & \frac{1}{(2\pi)^{3/2}} \int \d {\bf q} \,
\exp({\rm i} {\bf q} \dotprod {\bf r}) \, \ecb({\bf q},\omega)
\nonumber \\ [2mm]
&=& \frac{2Z_1 e}{(2\pi)^{3/2}} \int \d {\bf q} \,
\exp({\rm i} {\bf q} \dotprod {\bf r}) \,
\left[
- {\rm i} \, \frac{1}{\omega} \,
\frac{1}{\epsilon^{\rm (L)} ({\bf q},\omega)} \,
\frac{({\bf q} \dotprod {\bf v}) {\bf q} }{q^2}
\right.
\nonumber \\ [2mm]
&& \mbox{} + \left.
{\rm i} \, \frac{\omega}{c^2} \,
\frac{1}{q^2 - (\omega/c)^2 \, \epsilon^{\rm (T)}({\bf q}, \omega)}
\left( {\bf v} - \frac{({\bf q} \dotprod {\bf v}) {\bf q}}{q^2} \right)
\right]
\delta({\bf q} \dotprod {\bf v} - \omega)
\label{8.59}\eeqa
and
\beqa
\bcb ({\bf r},\omega) &=& \frac{2 Z_1 e}{(2\pi)^{3/2}} \int \d {\bf q} \,
\exp({\rm i} {\bf q} \dotprod {\bf r}) \,
\left(
{\rm i} \, \frac{1}{c} \,
\frac{{\bf q} \vecprod {\bf v}}
{q^2 - (\omega/c)^2 \, \epsilon^{\rm (T)}({\bf q}, \omega)}
\right)
\delta({\bf q} \dotprod {\bf v} - \omega). \rule{15mm}{0mm}
\label{8.60}\eeqa


\subsection{The optical approximation \label{sec8.2.1}}

Because these expressions can only be evaluated numerically, Fermi
considered the case of distant excitations for which the electromagnetic
field set up by the projectile varies slowly within the atomic volume.
Then, only the long-wavelength (small-$q$) components of the field are
important, for which the DFs $\epsilon^{\rm (L,T)}(q,\omega)$ can be
approximated by the ODF $\epsilon(\omega)$, similarly to the dipole
approximation used in atomic physics. Evidently, the field has
axial symmetry around the $z$ axis and, consequently, we only
need to determine it at the points ${\bf r} =(b,0 ,0)$ in the
$x$ axis, where we have
\beqa
\ecb(b,\omega) &=& \frac{2Z_1 e}{(2\pi)^{3/2}} \int \d {\bf q} \,
\exp({\rm i} q_x b) \,
\left[
- {\rm i} \, \frac{1}{\omega} \,
\frac{1}{\epsilon(\omega)} \,
\frac{q_z v {\bf q} }{q^2}
\right.
\nonumber \\ [2mm]
&& \mbox{} + \left.
{\rm i} \, \frac{\omega}{c^2} \,
\frac{1}{q^2 - (\omega/c)^2 \, \epsilon(\omega)}
\left( {\bf v} - \frac{q_z v {\bf q}}{q^2} \right)
\right]
\delta(q_z v - \omega)
\label{8.61}\eeqa
and
\beqa
\bcb(b,\omega) &=& \frac{2 Z_1 e}{(2\pi)^{3/2}} \int \d {\bf q} \,
\exp({\rm i} q_x b) \,
\left(
{\rm i} \, \frac{1}{c} \,
\frac{(q_y,-q_x,0) v}
{q^2 - (\omega/c)^2 \, \epsilon(\omega)}
\right)
\delta(q_z v - \omega).
\label{8.62}\eeqa
The integral over $q_z$ can be done immediately, giving
\beqa
\ecb(b,\omega) &=& \frac{2{\rm i}Z_1 e}{(2\pi)^{3/2} v}
\int_{-\infty}^\infty \d q_x \int_{-\infty}^\infty \d q_y \,
\exp({\rm i} q_x b) \,
\left[
- \frac{1}{\epsilon(\omega)} \,
\frac{(q_x,q_y,\omega/v) }{q_x^2+q_y^2+(\omega/v)^2}
\right.
\nonumber \\ [2mm]
&& \mbox{} + \left.
\frac{\omega}{c^2} \,
\frac{1}{q_x^2+q_y^2+\mu^2}
\left( (0,0,v) - \frac{\omega (q_x,q_y,\omega/v)}
{q_x^2+q_y^2+(\omega/v)^2} \right)
\right]
\label{8.63}\eeqa
and
\beqa
\bcb(b,\omega) &=& \frac{2 {\rm i} Z_1 e}{(2\pi)^{3/2} v}
\int_{-\infty}^\infty \d q_x \int_{-\infty}^\infty \d q_y \,
\exp({\rm i} q_x b) \,
\left( \frac{v}{c} \,
\frac{(q_y,-q_x,0)}
{q_x^2+q_y^2+\mu^2}
\right)
\label{8.64}\eeqa
with the complex quantity
\beq
\mu^2 = \frac{\omega^2}{v^2} - \frac{\omega^2}{c^2} \, \epsilon(\omega)
= \frac{\omega^2}{v^2} \left[ 1 -\beta^2 \epsilon(\omega) \right].
\label{8.65}\eeq
In preparation for the integration over $q_y$ we note that the various
terms in the integrands have definite parity under reflections on the
plane $q_y=0$ and, therefore, only the even terms contribute. That is,
\beqa
\ecb(b,\omega) &=& \frac{2{\rm i}Z_1 e}{(2\pi)^{3/2} v}
\int_{-\infty}^\infty \d q_x \int_{-\infty}^\infty \d q_y \,
\exp({\rm i} q_x b) \,
\left[
- \frac{1}{\epsilon(\omega)} \,
\frac{(q_x,0,\omega/v) }{q_x^2+q_y^2+(\omega/v)^2}
\right.
\nonumber \\ [2mm]
&& \mbox{} + \left.
\frac{\omega}{c^2} \,
\frac{1}{q_x^2+q_y^2+\mu^2}
\left( (0,0,v) - \frac{\omega (q_x,0,\omega/v)}
{q_x^2+q_y^2+(\omega/v)^2} \right)
\right]
\label{8.66}\eeqa
and
\beqa
\bcb(b,\omega) &=& \frac{2 {\rm i} Z_1 e}{(2\pi)^{3/2} v}
\int_{-\infty}^\infty \d q_x \int_{-\infty}^\infty \d q_y \,
\exp({\rm i} q_x b) \,
\left( \frac{v}{c} \,
\frac{(0,-q_x,0)}
{q_x^2+q_y^2+\mu^2}
\right).
\label{8.67}\eeqa
It is now clear that ${\cal E}_y=0$ and ${\cal B}_x = {\cal B}_z =0$,
that is, the electric field is in the $x$-$z$ plane, and the magnetic
field is parallel or anti-parallel to the $y$ axis.

The component of the electric field along the direction of the
projectile is
\beqa
{\cal E}_z &=& \frac{2{\rm i}Z_1 e}{(2\pi)^{3/2} v}
\int_{-\infty}^\infty \d q_x \int_{-\infty}^\infty \d q_y \,
\exp({\rm i} q_x b) \,
\left[
- \frac{1}{\epsilon(\omega)} \,
\frac{\omega/v }{q_x^2+q_y^2+(\omega/v)^2}
\right.
\nonumber \\ [2mm]
&& \mbox{} + \left.
\frac{\omega}{c^2} \,
\frac{1}{q_x^2+q_y^2+\mu^2}
\left( v - \frac{\omega^2/v}
{q_x^2+q_y^2+(\omega/v)^2} \right)
\right].
\label{8.68}\eeqa
After some rearrangements, we write
\beq
{\cal E}_z = - \frac{2{\rm i}Z_1 e \omega}{(2\pi)^{3/2} v^2}
\left( \frac{1}{\epsilon(\omega)} - \frac{v^2}{c^2} \right)
\int_{-\infty}^\infty \d q_x \, \exp({\rm i} q_x b)
\int_{-\infty}^\infty \d q_y \,
\frac{1}{q_x^2+q_y^2+\mu^2}.
\label{8.69}\eeq
The integral over $q_y$ equals $\pi/(q_x^2 + \mu^2)^{1/2}$ and we have
\beq
{\cal E}_z = - \frac{{\rm i}Z_1 e \omega}{(2\pi)^{1/2} v^2}
\left( \frac{1}{\epsilon(\omega)} - \frac{v^2}{c^2} \right)
\int_{-\infty}^\infty \d q_x \, \frac{\exp({\rm i} q_x b)}
{\sqrt{q_x^2+\mu^2}}.
\label{8.70}\eeq
Finally, the remaining integral can be evaluated with the aid of formula
\req{B.67} in Appendix \ref{appB},
\beq
\int_{-\infty}^\infty \d q_x \, \frac{\exp({\rm i} q_x b)}
{\sqrt{q_x^2+\mu^2}}
= 2 \int_{0}^\infty \d q_x \, \frac{\cos(q_x b)}
{\sqrt{q_x^2+\mu^2}} = 2 K_0 (\mu b),
\label{8.71}\eeq
to give
\beqa
{\cal E}_z (b,\omega) &=& - \frac{{\rm i}Z_1 e \omega}{v^2} \,
\left(\frac{2}{\pi}\right)^{1/2}
\left( \frac{1}{\epsilon(\omega)} - \frac{v^2}{c^2} \right)
K_0(\mu b)
\nonumber \\ [2mm]
&=& - {\rm i} \left(\frac{2}{\pi}\right)^{1/2}
\frac{Z_1 e}{v} \, \frac{1}{\epsilon(\omega)} \,
\frac{v}{\omega} \, \mu^2  K_0(\mu b),
\label{8.72}\eeqa
where $\mu$ is the square root of the quantity \req{8.65} such that
${\rm Re}(\mu) > 0$.

The radial component of the electric field is
\beqa
{\cal E}_x (b,\omega) &=& \frac{2{\rm i}Z_1 e}{(2\pi)^{3/2} v}
\int_{-\infty}^\infty \d q_x \int_{-\infty}^\infty \d q_y \,
\exp({\rm i} q_x b) \,
\left[
- \frac{1}{\epsilon(\omega)} \,
\frac{q_x}{q_x^2+q_y^2+(\omega/v)^2}
\right.
\nonumber \\ [2mm]
&& \mbox{} - \left.
\frac{\omega}{c^2} \,
\frac{1}{q_x^2+q_y^2+\mu^2}
\left( \frac{\omega q_x}
{q_x^2+q_y^2+(\omega/v)^2} \right)
\right]
\nonumber \\ [2mm]
&=& - \frac{2{\rm i}Z_1 e}{(2\pi)^{3/2} v}
\, \frac{1}{\epsilon(\omega)}
\int_{-\infty}^\infty \d q_x
\, \exp({\rm i} q_x b) \, q_x
\int_{-\infty}^\infty \d q_y \,
\frac{1}{q_x^2+q_y^2+\mu^2} \, .
\label{8.73}\eeqa
Performing the integral over $q_y$, we have
\beq
{\cal E}_x (b,\omega) = - \frac{2{\rm i}Z_1 e}{(2\pi)^{3/2} v}
\, \frac{1}{\epsilon(\omega)} \, \pi
\int_{-\infty}^\infty \d q_x
\, \exp({\rm i} q_x b) \, q_x
\, \frac{1}{\sqrt{q_x^2+\mu^2}}.
\label{8.74}\eeq
The remaining integral is of the type given by Eq.\ \req{B.67},
\beq
\int_{-\infty}^\infty \d q_x
\, \frac{q_x \, \exp({\rm i} q_x b)}{\sqrt{q_x^2+\mu^2}}
= 2 {\rm i} \int_{-\infty}^\infty \d q_x
\, \frac{q_x \, \sin(q_x b)}{\sqrt{q_x^2+\mu^2}}
= 2 {\rm i} \, \mu \, K_1(\mu b)\, ,
\label{8.75}\eeq
and we finally obtain
\beq
{\cal E}_x (b,\omega)
= \left(\frac{2}{\pi}\right)^{1/2}
\frac{Z_1 e}{v} \,
\frac{1}{\epsilon(\omega)}  \, \mu \, K_1(\mu b).
\label{8.76}\eeq

The non-vanishing component of the magnetic field is
\beqa
{\cal B}_y (b,\omega) &=& - \frac{2 {\rm i} Z_1 e}{(2\pi)^{3/2} v}
\int_{-\infty}^\infty \d q_x
\exp({\rm i} q_x b) \, q_x
\left( \frac{v}{c} \,
\int_{-\infty}^\infty \d q_y \,
\frac{1}{q_x^2+q_y^2+\mu^2} \right)
\nonumber \\ [2mm]
&=& - \frac{2 {\rm i} Z_1 e}{(2\pi)^{3/2} v} \int_{-\infty}^\infty
\d q_x \exp({\rm i} q_x b) \, q_x
\left( \frac{v}{c} \, \pi
\frac{1}{\sqrt{q_x^2+\mu^2}} \right),
\label{8.77}\eeqa
and using the equality \req{8.75} we find
\beqa
{\cal B}_y (b,\omega)
&=& \left(\frac{2}{\pi}\right)^{1/2}  \frac{Z_1 e}{v} \,
\frac{v}{c} \, \mu \, K_1(\mu b)
\nonumber \\ [2mm]
&=&
\beta \, \epsilon(\omega) \, {\cal E}_x(b,\omega).
\label{8.78}\eeqa

With these results we can evaluate the stopping power due to
interactions with molecules at impact parameters $b \ge a$. Following
\citet{Jackson1975}, we first determine the energy transfer to a
molecule at the impact parameter $b$, which is obtained by adding the
expressions \req{8.27} of the energy transfer to individual
oscillators,
\beq
W_{\rm m}(b) = 2 e \int_0^\infty \frac{\d f(\omega_j)}{\d \omega_j}
\, {\rm Re} \left( {\rm i} \int_0^\infty \omega \, {\bf u}_j(\omega)
\dotprod \ecb^\ast(b,\omega) \, \d \omega \right) \d \omega_j.
\label{8.79}\eeq
Inverting the order of the integrals, we write
\beq
W_{\rm m}(b) = {\rm Re} \left[ - 2 {\rm i}
\int_0^\infty \omega \,
\left( - e \int_0^\infty \frac{\d f(\omega_j)}{\d \omega_j}
{\bf u}_j(\omega) \, \d \omega_j \right)
\dotprod \ecb^\ast(b,\omega) \, \d \omega \right].
\label{8.80}\eeq
The quantity in parentheses is the Fourier transform of the
molecular dipole moment, \index{Fourier transform}
\beq
{\bf p}_{\rm mol} (\omega) =
 - e \int_0^\infty \frac{\d f(\omega_j)}{\d \omega_j}
{\bf u}_j(\omega) \, \d \omega_j,
\label{8.81}\eeq
which we express in terms of the dielectric function as [see Eq.\
\req{1.86c}]
\beq
{\bf p}_{\rm mol} (\omega) = \frac{\bf P}{\cal N} =
\frac{\epsilon(\omega)-1}{4\pi {\cal
N}} \, \ecb (b,\omega).
\label{8.82}\eeq
Hence,
\beqa
W_{\rm m}(b) &=& {\rm Re} \left[ - 2 {\rm i}
\int_0^\infty \omega \,
\frac{\epsilon(\omega)-1}{4\pi {\cal
N}} \, \left| \ecb (b,\omega) \right|^2 \, \d \omega \right]
\nonumber \\ [2mm]
&=& {\rm Re} \left[ - 2 {\rm i}
\int_0^\infty \omega \,
\frac{\epsilon(\omega)}{4\pi {\cal
N}} \, \left| \ecb (b,\omega) \right|^2 \, \d \omega \right].
\label{8.83}\eeqa
Now the contribution to the stopping power of molecules with impact
parameter larger than $a$ is
\beqa
\left[ - \, \frac{\d E}{\d z} \right]_{b\ge a} &=&
2 \pi {\cal N} \int_a^\infty W_{\rm m}(b) \, b \, d b
\nonumber \\ [2mm]
&=& \int_a^\infty
{\rm Re} \left[ - {\rm i}
\int_0^\infty \omega \,
\epsilon(\omega) \,
\left| \ecb(b,\omega) \right|^2
\d \omega \right] \, b \, d b
\nonumber \\ [2mm]
&=& \int_a^\infty \left(
\int_0^\infty \omega \,
{\rm Im}[\epsilon(\omega)] \,
\left| \ecb(b,\omega) \right|^2
\d \omega \right) \, b \, d b.
\label{8.84}\eeqa
Introducing the expressions \req{8.72} and \req{8.76} of the components of
the electric field,
\beqa
\left| \ecb (b,\omega) \right|^2 &=&
{\cal E}_x (b,\omega) \, {\cal E}_x^\ast (b,\omega)
+ {\cal E}_z (b,\omega) \, {\cal E}_z^\ast (b,\omega)
\nonumber \\ [2mm]
&=& \frac{2 Z_1^2 e^2}{\pi v^2}
\, \frac{1}{\left| \epsilon(\omega) \right|^2}
\left[ K_1(\mu b) \, K_1(\mu^\ast b)
+ \frac{v^2}{\omega^2} \, \mu \mu^\ast
K_0(\mu b) \, K_0(\mu^\ast b) \right] \mu \mu^\ast, \rule{10mm}{0mm}
\label{8.85}\eeqa
and reversing the order of the integrations, Eq.\ \req{8.84} becomes
\beqa
\left[ - \, \frac{\d E}{\d z} \right]_{b\ge a}
&=& \frac{2 Z_1^2 e^2}{\pi v^2} \,
\int_0^\infty \omega \,
\frac{{\rm Im}[\epsilon(\omega)]}{\left| \epsilon(\omega) \right|^2}
\, {\cal X} \, \d \omega
\label{8.86}\eeqa
with the real quantity
\beqa
{\cal X} &\equiv& \mu \mu^\ast \int_a^\infty \left[
 K_1(\mu b) \, K_1(\mu^\ast b)
+ \frac{v^2}{\omega^2} \, \mu \mu^\ast
K_0(\mu b) \, K_0(\mu^\ast b) \right] \, b \, \d b
\nonumber \\ [2mm]
&=& \mu \mu^\ast a \left[ \frac{ \mu^\ast K_1(\mu a) \, K_0(\mu^\ast a)- \mu
K_0(\mu a) \, K_1(\mu^\ast a)}{(\mu^\ast)^2 - \mu^2} \right.
\nonumber \\ [2mm]
&& \mbox{} \left.
- \frac{v^2}{\omega^2} \mu \mu^\ast
\frac{ \mu K_1(\mu a) \, K_0(\mu^\ast a)- \mu^\ast
K_0(\mu a) \, K_1(\mu^\ast a)}{(\mu^\ast)^2-\mu^2}\right],
\label{8.87}\eeqa
where we have used the equalities \req{B.71} and \req{B.72}
in Appendix \ref{appB}. Since
\beqa
{\cal X} &=& \beta^2 a \frac{\mu \mu^\ast}{(\mu^\ast)^2-\mu^2}
\left\{ \epsilon(\omega) \mu^\ast K_1(\mu a) \, K_0(\mu^\ast a)-
\epsilon^\ast(\omega) \mu K_0(\mu a) \, K_1(\mu^\ast a) \right\}
\nonumber \\ [2mm]
&=& \frac{\beta^2 a}{(\mu^\ast)^2 - \mu^2}
\left\{ \epsilon(\omega) \left[1-\beta^2 \epsilon^\ast(\omega) \right]
\mu K_1(\mu a) \, K_0(\mu^\ast a) \right.
\nonumber \\ [2mm]
&& \mbox{} \left. -
\epsilon^\ast(\omega)  \left[1-\beta^2 \epsilon(\omega) \right]
\mu^\ast K_0(\mu a) \, K_1(\mu^\ast a) \right\}
\nonumber \\ [2mm]
&=& \frac{1}{2 {\rm i} \; {\rm Im} \epsilon(\omega)}
\left\{ - 2 {\rm i} \; {\rm Im} \left(
\rule{0mm}{4mm}\epsilon^\ast(\omega)
\left[1-\beta^2 \epsilon(\omega) \right] \mu^\ast a \,
K_0(\mu a) \, K_1(\mu^\ast a) \right) \right\},
\nonumber \eeqa
we have
\beqa
{\cal X} &=& \frac{\left| \epsilon(\omega) \right|^2}{{\rm Im}
[\epsilon(\omega)]}
\left\{ - {\rm Im} \left[
\left( \frac{1}{\epsilon(\omega)} -\beta^2 \right)
\mu^\ast a \,
K_0(\mu a) \, K_1(\mu^\ast a) \right] \right\}.
\label{8.88}\eeqa
Hence, \index{Fermi's distant stopping power}
\beq
\left[ - \, \frac{\d E}{\d z} \right]_{b\ge a} =
\frac{2}{\pi} \, \frac{Z_1^2 e^2}{v^2} \;
{\rm Re} \int_{0}^\infty
{\rm i} \, \omega \, \mu^\ast a \, K_1(\mu^\ast a) \,
K_0(\mu a)
\left( \frac{1}{\epsilon(\omega)} - \beta^2 \right)
\, \d\omega.
\label{8.89}\eeq
This result, which was first obtained by \citet{Fermi1940} using a
different calculation scheme (see the next Section), accounts for
the effect of electric polarization in excitations of ``distant''
oscillators. Numerical calculations of expression \req{8.89} with
realistic ODFs are quite complicated and do not give additional insight
on the physical process; a more efficient calculation scheme will be
presented below.

The polarization effect is associated to the occurrence of complex
arguments in the modified Bessel functions, which come from the presence
of the ODF in the definition \req{8.65},
$$
\mu^2 = \frac{\omega^2}{v^2} \left[ 1 - \beta^2 \epsilon(\omega)
\right].
\eqno{\req{8.65}}$$
To elucidate this fact, we note that Eq.\ \req{8.86} can be rewritten as
\beqa
\left[ - \, \frac{\d E}{\d z} \right]_{b\ge a}
&=& \frac{4\pi Z_1^2 e^4}{\me v^2} \,
\int_0^\infty \frac{\d f (\omega)}{\d \omega}
\, {\cal X} \, \d \omega,
\label{8.90}\eeqa
where we have used the identity \req{8.1}. If, in addition, now we set
$\epsilon(\omega) \simeq 1$, $\mu \simeq \omega/(\gamma v)$ and the
quantity ${\cal X}$ becomes
\beq
{\cal X} \simeq \int_{\xi_a}^\infty \left[
 K_1^2 (\xi)
+ \frac{1}{\gamma^2} \,
K_0^2(\xi) \right] \, \xi \, \d \xi
\label{8.91}\eeq
with $\xi_a = \omega a/(\gamma v)$. Evidently, expression \req{8.90} then
reduces to the formula \req{8.37}. Consequently, polarization is fully
accounted by the factor $\epsilon(\omega)$ in the definition \req{8.65}.
Since the ODF is multiplied by $\beta^2$, we see that polarization is
important only for projectiles with high energies.


\subsection{Cherenkov radiation \label{sec8.2.2}}
\index{Cherenkov radiation}

It is interesting to consider the alternative derivation of the formula
\req{8.89} given by \citet{Fermi1940}, which shows that the formula does
include energy losses associated to the emission of Cherenkov radiation.
Indeed, Fermi calculated the energy lost by the projectile per unit time
in distant interactions with impact parameter $b \ge a$ as the energy
flow of the electromagnetic field through the surface of a cylinder of
radius $a$ around the trajectory of the projectile. Since the Poynting
vector is
\beq
{\bf S} (a,t) = \frac{c}{4\pi} \, \ecb(a,t) \vecprod \bcb(a,t) =
\frac{c}{4\pi} \, \left( - {\cal E}_z {\cal B}_y, 0,
{\cal E}_x {\cal B}_y \right),
\label{8.92}\eeq
the contribution of distant interactions to the stopping power is given by
\beq
\left[ - \, \frac{\d E}{\d z} \right]_{b\ge a} = \frac{1}{v} \, \frac{\d
E}{\d t} = - \frac{c}{4\pi\, v} \int_{-\infty}^\infty
{\cal E}_z(a,z) {\cal B}_y(a,z) \, 2\pi\, a \, \d z.
\label{8.93}\eeq
Because the particle is assumed to follow the trajectory ${\bf r}= {\bf
v} t$, the integral over $z$ at the instant $t$ is equivalent to the
integral of the fields at $z=0$ over all time. Setting $z=vt$ we can
write
\beq
\left[ - \, \frac{\d E}{\d z} \right]_{b\ge a} =
- \frac{ca}{2} \int_{-\infty}^\infty
{\cal E}_z(a,t) {\cal B}_y(a,t) \, \d t.
\label{8.94}\eeq
Expressing the fields in the usual form,
\beq
\left( \begin{array}{l}
{\cal E}_z(a,t) \\
{\cal B}_y(a,t) \end{array} \right)
= \frac{1}{(2\pi)^{1/2}} \int_{-\infty}^\infty
\d \omega \, \exp(-{\rm i} \omega t)
\left( \begin{array}{l}
{\cal E}_z(a,\omega) \\
{\cal B}_y(a,\omega) \end{array} \right),
\label{8.95}\eeq
and recalling that they are real, \ie, ${\cal E}_z(a,-\omega) ={\cal
E}_z^\ast (a,\omega)$, we find
\beq
\left[ - \, \frac{\d E}{\d z} \right]_{b\ge a} =
- ca \, {\rm Re} \int_{0}^\infty
{\cal E}_z(a,\omega) {\cal B}_y^\ast (a,\omega) \, \d\omega.
\label{8.96}\eeq
Inserting the expressions \req{8.72} and \req{8.78}, we obtain again
Fermi's formula \req{8.89},
\beq
\left[ - \, \frac{\d E}{\d z} \right]_{b\ge a} =
\frac{2}{\pi} \, \frac{Z_1^2 e^2}{v^2} \;
{\rm Re} \int_{0}^\infty
{\rm i} \, \omega \, \mu^\ast a \, K_1(\mu^\ast a) \,
K_0(\mu a)
\left( \frac{1}{\epsilon(\omega)} - \beta^2 \right)
\, \d\omega,
\nonumber \eeq
which gives the energy transfer per unit path length to radial distances
larger than $a$.

Let us consider the limit of large distances, $\mu a  \gg 1$,
where the modified Bessel functions can be approximated by their
asymptotic form given by Eq.\ \req{B.60} in Appendix \ref{appB}. The
non-vanishing components of the electromagnetic field at those large
distances become
\beq
{\cal E}_x (b,\omega)
= \left(\frac{2}{\pi}\right)^{1/2}
\frac{Z_1 e}{v} \,
\frac{1}{\epsilon(\omega)}  \, \mu \, K_1(\mu b)
\simeq \frac{Z_1 e}{v} \,
\frac{1}{\epsilon(\omega)}  \, \sqrt{\frac{\mu}{b}}
\, {\rm e}^{-\mu b},
\label{8.97}\eeq
\beq
{\cal E}_z (b,\omega) =
{\rm i} \left(\frac{2}{\pi}\right)^{1/2}
\frac{Z_1 e \omega}{c^2} \,
\left[ 1 - \frac{1}{\beta^2 \epsilon(\omega)} \right] K_0(\mu b)
\simeq {\rm i} \frac{Z_1 e \omega}{c^2} \,
\left[ 1 - \frac{1}{\beta^2 \epsilon(\omega)} \right]
\, \frac{{\rm e}^{-\mu b}}{\sqrt{\mu b}},
\label{8.98}\eeq
and
\beq
{\cal B}_y (b,\omega)
= \left(\frac{2}{\pi}\right)^{1/2}  \frac{Z_1 e}{c} \,
\mu \, K_1(\mu b)
\simeq  \frac{Z_1 e}{c} \,
\sqrt{\frac{\mu}{b}} \, {\rm e}^{-\mu b}.
\label{8.99}\eeq
Introducing these expressions into Eq.\ \req{8.96} we obtain the energy
transfer to large distances per unit path length of the projectile,
\beq
\left[ - \, \frac{\d E}{\d z} \right]_{b\ge a} =
 \frac{Z_1^2 e^2}{c^2} \;
{\rm Re} \int_{0}^\infty \omega
\left[ 1 - \frac{1}{\beta^2 \epsilon(\omega)} \right]
\left( - {\rm i} \sqrt{\frac{\mu^\ast}{\mu}} \right)
\, {\rm e}^{-(\mu+\mu^\ast) a}
\, \d\omega.
\label{8.100}\eeq
When $\mu = (\omega/v) \left[ 1 - \beta^2 \epsilon(\omega)
\right]^{1/2}$ has a positive real part, which is generally the case,
the exponential factor makes the expression to decrease rapidly with
distance, and all the energy is deposited near the projectile's path.
This is not true when $\mu$ is purely imaginary, because then the
exponential is unity and the expression is independent of $a$.  In this
very exceptional case, part of the energy escapes to large distances in
the form of {\it Cherenkov radiation}. $\mu$ can be purely imaginary only when
$\epsilon(\omega)$ is real (\ie, when there is no absorption) and
positive, and $\beta^2 \epsilon(\omega) > 1$, that is
\beq
{\rm Im} [\epsilon(\omega)] \simeq 0
\qquad \mbox{and} \qquad
v > \frac{c}{\sqrt{\epsilon(\omega)}}.
\label{8.101}\eeq
Hence, emission of Cherenkov radiation of frequency $\omega$ is
possible only when absorption is weak and the speed of the projectile is
larger than the phase velocity of electromagnetic radiation of that
frequency.

If Cherenkov radiation of frequency $\omega >0$ is emitted, $\beta^2
\epsilon(\omega) >1$ and we can assume that
$\epsilon(\omega)=\epsilon_1 + {\rm i} \epsilon_2$ ($\omega >0$) has a
very small imaginary part ($\epsilon_2 \ll \epsilon_1$). Then,
\beq
\mu = \frac{\omega}{v} \sqrt{1 - \beta^2 \epsilon_1 - {\rm i}
\beta^2 \epsilon_2} = \left| \mu \right| \, \exp\left[ \frac{{\rm i}}{2}
\arctan \left( \frac{\epsilon_2}{1 - \beta^2 \epsilon_1} \right)
\right],
\label{8.102}\eeq
and we see that the phase of $\mu$ equals $-\pi/2$. That is,
\beq
\mu = - {\rm i} \left| \mu \right| .
\label{8.103}\eeq
Therefore, $\mu^\ast / \mu \sim -1$ and expression \req{8.100} gives
\beq
\left[ - \, \frac{\d E}{\d z} \right]_{\rm rad} =
\frac{Z_1^2 e^2}{c^2} \;
\int_{0}^\infty \omega
\left[ 1 - \frac{1}{\beta^2 \epsilon(\omega)} \right] \d\omega.
\label{8.104}\eeq
The integrand in this expression gives the frequency distribution of the
emitted radiation. The direction of propagation of this radiation
is that of the Poynting vector \req{8.92}, which forms with the
direction of the emitting particle an angle $\theta_{\rm C}$ given by
\beq
\tan \theta_{\rm C} = \frac{-{\cal E}_z}{{\cal E}_x}.
\label{8.105}\eeq
Inserting the expressions \req{8.97} and \req{8.98} of the distant fields,
we find
\beq
\tan \theta_{\rm C} = \sqrt{\beta^2 \epsilon(\omega) -1}
\label{8.106}\eeq
or, equivalently
\beq
\cos \theta_{\rm C} = \frac{1}{\beta \sqrt{\epsilon(\omega)}}.
\label{8.107}\eeq
We see that Cherenkov radiation is emitted only when $\theta_{\rm C}$ is
a physical angle with cosine less than unity. Because the ODF generally
varies with frequency, radiations of different frequencies are emitted
at different angles. Evidently, the radiation is linearly
polarized in the plane defined by the direction of observation and the
trajectory of the projectile.

The foregoing study shows that the energy released as Cherenkov
radiation is part of the contribution of ``distant'' oscillators to the
stopping power, and it is fully accounted for by the Fermi formula
\req{8.89}.


\subsection{Density-effect correction to the stopping power
\label{sec8.2.3}}
\index{stopping power!density effect correction}
\index{density-effect correction}

A consistent description of the stopping of charged particles in
materials is provided by the semi-classical dielectric formalism
(Section \ref{sec6.7}), where the response of the material is
determined by the longitudinal and transverse DFs, $\epsilon^{\rm
(L,T)}(q,\omega)$, as functions of the wave number and the frequency of
the field. Again, we limit our considerations to the case of homogeneous
isotropic media, with DFs that are independent of the direction of the
wave vector ${\bf q}$.

To determine the polarization or density-effect correction to the
stopping power, we consider the classical formula [Eq.\ \req{1.189}],
which we write in the form
\beq
S = S_0 + (\Delta S)_{\rm pol},
\label{8.108}\eeq
where
\beqa
S_0 &=& \frac{2 (Z_1 e)^2}{\pi v^2}
\int_0^\infty \omega \, \d \omega \int_{\omega/v}^\infty
\frac{\d q}{q}
\left[ {\rm Im} \left( \frac{-1}{\epsilon^{\rm (L)}} \right)
\right.
\nonumber \\ [2mm]
&& \left.
+ \left( \beta^2 - \frac{\omega^2}{c^2 q^2} \right)
\frac{(cq/\omega)^2}{\left[ (cq/\omega)^2 - 1
\right]^2} \,
{\rm Im} \left( \frac{-1}{\epsilon^{\rm (T)}} \right)
\right].
\label{8.109}\eeqa
is the stopping power of an ``unpolarizable'' medium having the same DFs
as the actual material, and $(\Delta S)_{\rm
pol}$ is the polarization correction,
\beq
(\Delta S)_{\rm pol} \equiv S - S_0 =
\frac{2 (Z_1 e)^2}{\pi v^2}
\int_0^\infty \omega \, {\cal Y} (\omega) \, \d \omega
\label{8.110}\eeq
with
\beqa
{\cal Y} (\omega) &=&
\int_{\omega/v}^\infty
\frac{\d q}{q} \,
\left( \beta^2 - \frac{\omega^2}{c^2 q^2} \right)
\nonumber \\ [2mm]
&& \mbox{} \times \left[
{\rm Im} \left( \frac{q^2}{q^2 - (\omega/c)^2 \,
\epsilon^{\rm (T)}(q, \omega)} \right)
- \frac{(cq/\omega)^2}{\left[ (cq/\omega)^2 - 1
\right]^2} \,
{\rm Im} \left( \frac{-1}{\epsilon^{\rm (T)}(q,\omega)} \right)
\right]. \rule{15mm}{0mm}
\label{8.111}\eeqa
As shown in Section \ref{sec6.7}, $S_0$ can be regarded as the
stopping power given by the PWBA with the GOSs obtained from the
DFs, Eqs.\ \req{6.251} and \req{6.256}.

The function ${\cal Y}(\omega)$ can be calculated approximately as
follows. We first observe that the polarization correction is expected to
be appreciable only for fast projectiles with $\beta$ close to unity.
Consequently, the lower limit of the
integral \req{8.111} is generally small and, since the integrand almost
diverges at the photon line $q=\omega/c$, we can replace the DFs with the
ODF, $\epsilon(\omega)$. Hence,
\beqa
{\cal Y} (\omega) &\simeq&
\int_{\omega/v}^\infty
\frac{\d q}{q} \,
\beta^2 \left[ 1 - \left( \frac{\omega}{\beta c q} \right)^2 \right]
\nonumber \\ [2mm]
&& \mbox{} \times \left[
{\rm Im} \left( \frac{1}{1 - (\omega/cq)^2 \,
\epsilon(\omega)} \right)
- \frac{(\omega/cq)^2}{\left[ 1-(\omega/cq)^2 \right]^2} \,
{\rm Im} \left( \frac{-1}{\epsilon(\omega)} \right)
\right], \rule{10mm}{0mm}
\label{8.112}\eeqa
and, changing the integration variable to
\beq
x = \left[ \omega/(c \beta q) \right]^2,
\label{8.113}\eeq
we can write
\beqa
{\cal Y} (\omega) &=& \frac{1}{2}
\int_{0}^1
\frac{\d x}{x} \, \beta^2 (1 - x ) \left[
{\rm Im} \left( \frac{1}{1 - \beta^2 x \,
\epsilon(\omega)} \right)
- \frac{\beta^2 x}{\left( 1-\beta^2 x \right)^2} \,
{\rm Im} \left( \frac{-1}{\epsilon(\omega)} \right)
\right]. \rule{10mm}{0mm}
\label{8.114}\eeqa
These integrals are evaluated in Section \ref{sec6.7} [Eqs.\
\req{6.265} and \req{6.266}]. The result is
\beq
{\cal Y}(\omega) =
\frac{1}{2} \left\{
{\rm Im} \left[ - \left( \beta^2 - \frac{1}{\epsilon} \right)
\ln \left(1-\beta^2 \epsilon\right) \right]
- \left[ \ln \left( \frac{1}{1-\beta^2} \right) - \beta^2 \right]
{\rm Im} \left( \frac{-1}{\epsilon(\omega)} \right) \right\}.
\rule[-6mm]{0mm}{0mm}
\label{8.115}\eeq

\citet{Fano1956} derived a compact formula for the high-energy density
effect correction to the stopping power by using contour integration in
the complex plane. He started from a result equivalent to \req{8.112}
and showed that the correction \req{8.110} can be expressed as
\beq
(\Delta S)_{\rm pol} =  - \, \frac{2 \pi Z_1^2 e^4}{\me v^2} \,
{\cal N} \, Z \, \delta_{\rm F}
\label{8.116}\eeq
with
\beq
\delta_{\rm F} =
\frac{2 }{\pi \Omega_{\rm p}^2} \int_{0}^{\infty} \omega \;
{\rm Im} \left( \frac{-1}{\epsilon(\omega)} \right)
\ln \left( 1 + \frac{\ell^{2}}{\omega^{2}} \right) \, \d \omega
- \frac{\ell^{2}}{\Omega_{\rm p}^{2}} \left( 1-\beta^{2} \right),
\label{8.117}\eeq
where the frequency $\ell$ is a real-valued function of $\beta^2$
defined as the positive root of the equation
\beq
1 - \beta^2 \epsilon({\rm i} \ell)=0.
\label{8.118}\eeq
As pointed out by \citet{InokutiSmith1982}, to simplify the numerical
evaluation of $\ell$ it is best to write this equation in the form
\beq
1 - \beta^2 = 1 - \frac{1}{\epsilon({\rm i} \ell)},
\label{8.119}\eeq
and express the right-hand side as an integral involving only
$\epsilon(\omega)$ for real values of $\omega$. The sought expression is
obtained by using the equality \req{1.209a}, which gives
\beq
1 - \frac{1}{\epsilon({\rm i} \ell)} = \frac{2}{\pi} \int_0^\infty
\d \omega \; \frac{\omega}{\omega^2 + \ell^2} \, {\rm Im} \left(
\frac{-1}{\epsilon(\omega)} \right).
\label{8.120}\eeq

For calculation purposes, it is advantageous to express the
density-effect correction in terms of the OOS. We have
\beq
\delta_{\rm F} \equiv \frac{1}{Z} \int_{0}^{\infty} \frac{\d
f(W)}{\d W}
\ln \left( 1 + \frac{L^{2}}{W^{2}} \right) \, \d W
- \frac{L^{2}}{\hbar^2 \Omega_{\rm p}^{2}} \left( 1-\beta^{2} \right),
\label{8.121}\eeq
where $L$ is the positive root of the equation
\beq
{\cal F}(L) \equiv \frac{1}{Z} \hbar^2
\Omega_{\rm p}^{2} \int_{0}^{\infty}
\frac{1}{W^{2}+L^{2}} \frac{\d f(W)}{\d W} \, \d W =
1 - \beta^{2}.
\label{8.122}\eeq
The function ${\cal F}(L)$ decreases monotonically with $L$, and hence,
the root $L(\beta^{2})$ exists only when $1-\beta^{2}<{\cal F}(0)$;
otherwise it is $\delta_{\rm F}=0$. Therefore, the function
$L(\beta^{2})$ starts with zero at $\beta^{2}=1-{\cal F}(0)$ and grows
monotonically with increasing $\beta^{2}$.

\begin{figure}[h!] \begin{center}
\includegraphics*[width=10.5cm]{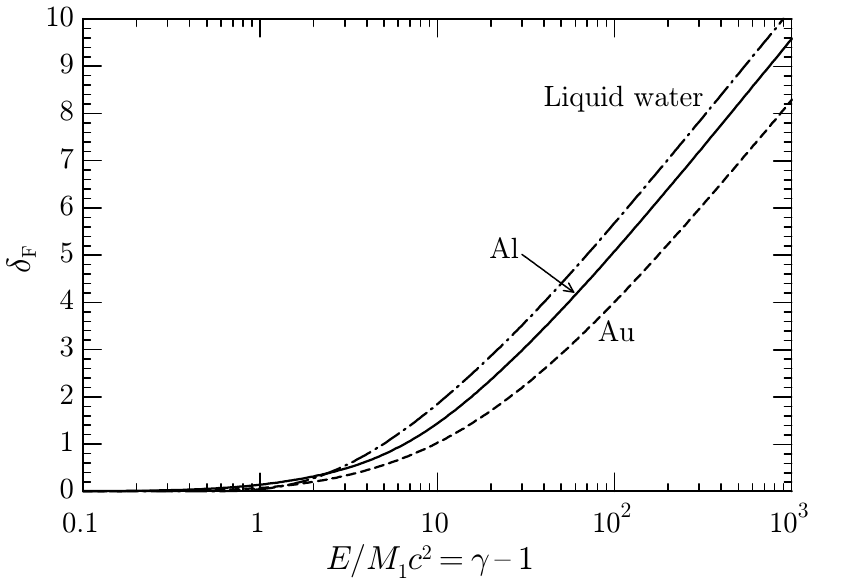}
\caption{
Density effect corrections for metallic aluminium and gold, and for liquid
water, as functions of the kinetic energy of the projectile in units of
$M_1 c^2$.
}\label{fig8.3}
\end{center} \end{figure}

The stopping power $S_0$ of the unpolarizable medium for high-energy
projectiles can be calculated from the Bethe formula \req{6.314}. The
stopping power of the polarizable material is obtained by adding the
density-effect correction,
\beq
S = \frac{4\pi Z_1^2 e^4}{\me v^2} \, {\cal N} Z
\left[ \ln \left(\frac{2 \me v^2}{I} \right)
+ \ln \left( \frac{1}{1-\beta^2} \right) - \beta^2 + \frac{1}{2}
f(\gamma) - \frac{C}{Z} - \frac{1}{2} \delta_{\rm F} \right].
\label{8.123}\eeq
In the limit of very high energies ($\beta \rightarrow 1$), the $L$
value resulting from Eq.\ \req{8.122} is much larger than the
characteristic atomic excitation energies ($L\gg W$) and it can be
approximated as $L^2 = \hbar^2 \Omega_{\rm p}^2/(1-\beta^2)$. Then,
using the dipole sum rule and the definition
$$
Z \ln I = \int_0^\infty \ln(W) \, \frac{\d f(W)}{\d W} \, \d W,
\eqno{\req{6.288}}$$
we obtain
\beq
\delta_{\rm F} \simeq \ln \left( \frac{\hbar^2
\Omega_{\rm p}^2}{(1-\beta^2)I^2}
\right) - 1, \qquad \mbox{when $\beta \rightarrow 1$}.
\label{8.124}\eeq
Inserting this expression into Eq.\ \req{8.123}, and assuming that the
shell correction vanishes in the high-energy limit, we see that the stopping
power for ultra-relativistic projectiles is completely determined by
${\cal N} Z$, the average density of electrons in the material. Figure
\ref{fig8.3} shows the density-effect correction $\delta_{\rm F}$ for
aluminium, gold, and liquid water, calculated from the OOSs displayed in
Figs.\ \ref{fig7.7}, \ref{fig7.8}, and \ref{fig7.9}, respectively.


\section{The Bloch or $Z_1^4$ correction \label{sec8.3}}
\index{stopping power!Bloch $Z_1^4$ correction}
\index{Bloch $Z_1^4$ correction}

The Bohr stopping power formula [see Section \ref{sec8.1.4}, Eq.\
\req{8.55}] for an elemental material of atomic number $Z$,
\beq
S_{\rm Bohr} =
\frac{4\pi Z_1^2 e^4}{\me v^2} \, {\cal N} \, Z
\left[ \ln\left(\frac{2 \exp(-g)\, \me v^3}{|Z_1| \, e^2 \,
\overline{\omega}} \right)
+ \ln \left( \frac{1}{1-\beta^2} \right)
- \beta^2 \right],
\label{8.125}\eeq
was derived from classical arguments that are expected to be valid
(that is coincident with the results from exact quantum
calculations) when the absolute value of the Sommerfeld parameter
\index{Sommerfeld parameter}
\beq
|\eta | = \frac{|Z_1| e^2}{\hbar v} = \frac{|Z_1| \alpha}{\beta}
\label{8.126}\eeq
is much larger than unity. $\alpha = e^2/(\hbar c)\simeq 1/137$ is the
fine-structure constant.

On the other hand, the Bethe stopping power formula \req{8.123},
excluding the shell and polarization corrections and noting that
$f(\gamma) \simeq 0$ (Fig.\ \ref{fig6.10}) for projectiles much heavier
than the electron,
\beq
S_{\rm Bethe} = \frac{4\pi Z_1^2 e^4}{\me v^2} \, {\cal N} Z
\left[ \ln \left(\frac{2 \me v^2}{I} \right)
+ \ln \left( \frac{1}{1-\beta^2} \right) - \beta^2 \right] ,
\label{8.127}\eeq
is the result from the quantum PWBA, a first-order perturbation
calculation, or from the equivalent semi-classical dielectric formalism.
The PWBA is valid when the speed $v$ of the projectile is
much larger than the velocities of atomic (bound) electrons (see
Section \ref{sec6.2}) or, equivalently, when
\beq
Z \alpha \ll \beta.
\label{8.128}\eeq

\index{stopping logarithm}
Recalling that $\hbar
\overline{\omega}=I$, the two formulas are seen to differ only in their
logarithmic terms,
\begin{subequations}
\label{8.129}
\beq
L_{\rm Bohr} =
\ln\left(\frac{2 \exp(-g)\, \me v^3 \hbar}{|Z_1| \, e^2 \,
I} \right)
\label{8.129a}\eeq
and
\beq
L_{\rm Bethe} =
\ln \left(\frac{2 \me v^2}{I} \right),
\label{8.129b}\eeq
\end{subequations}
and the difference
\beq
L_{\rm dif} = L_{\rm Bohr} - L_{\rm Bethe} =
\ln \left( \frac{\exp(-g)}{\eta} \right),
\label{8.130}\eeq
is determined by the Sommerfeld parameter and the Euler constant
\index{Euler constant}\index{Sommerfeld parameter}
$g=0.5772$. \citet{Bloch1933} derived a comprehensive formula for
the high-energy stopping power that practically coincides with Bohr's
formula when $\eta^2 \gg 1$ and approaches the Bethe formula in the limit
of small values of $\eta^2$. Here we present a simple derivation of the
Bloch correction due to \citet{LindhardSorensen1996}. The considerations
that follow are non-relativistic; our aim here is to derive the usual Bloch
correction to the stopping power and to clarify its physical meaning.

In the classical collision model all energy transfers are assumed to be
caused by collisions with electrons, and the stopping power is given by
[see Eqs.\ \req{8.12} and \req{8.7}],
\beq
\left[ - \, \frac{\d E}{\d z} \right]_{\rm col} = {\cal N} Z \int W(b) \,
2\pi b\, \d b = {\cal N} Z \, W_{\rm max}
\int_0^{b_{\rm ad}} \sin^2[\theta(b)/2] \,  2 \pi b \, \d b ,
\label{8.131}\eeq
where $W_{\rm max} \simeq 2 \me v^2$ [Eq.\ \req{8.8}] is the maximum
energy transfer in a
collision, $\theta(b)$ is the polar scattering angle in the CM frame
corresponding to the impact parameter $b$, which is given by [Eq.\
\req{8.5b}]
\beq
\sin^2[\theta(b)/2] = \frac{Z_1^2 e^4}{(\me v^2 \, b)^2 + Z_1^2 e^4},
\label{8.132}\eeq
and the upper limit of the integral is the classical adiabatic impact
parameter.

Recalling that the classical DCS can be expressed as [see Eq.\
\req{4.33}]
\beq
\frac{\d \sigma}{\d \Omega} = \frac{2\pi b \, \d b}{\d \Omega}
\label{8.133}\eeq
and the definition of the transport cross section (in the CM frame),
\beq
\sigma_{\rm tr}
= \int (1 - \cos\theta) \,
\frac{\d \sigma}{\d \Omega} \, \d \Omega
= 4 \pi \int_0^\infty \sin^2[\theta(b)/2] \, b \, \d b,
\label{8.134}\eeq
we see that the integral in Eq.\ \req{8.131} represents the contribution
to the transport cross section of collisions with impact parameters up
to a certain adiabatic limit $b_{\rm ad}$. It is then natural to
introduce the quantity,
\beq
\sigma_{\rm tr}^{\rm cl} (b_{\rm ad})
\equiv 4 \pi \int_0^{b_{\rm ad}} \sin^2[(\theta(b)/2] \, b \, \d b.
\label{8.135}\eeq
To allow comparison of the classical picture with a quantum description
of the collision process, it is convenient to replace the impact
parameter $b$ with the angular momentum $\ell \hbar = \me v \, b$ of the
electron in the reference frame of the projectile, and write
\beq
\sigma_{\rm tr}^{\rm cl} (\ell_{\rm ad}) =
\frac{4 \pi}{\me^2 v^2} \int_0^{\ell_{\rm ad}} \sin^2[(\theta(\ell)/2] \,
\ell \, \d \ell
\label{8.136}\eeq
with $\ell_{\rm ad} = \me v \, b_{\rm ad} /\hbar$ and
\beq
\sin^2[\theta(\ell)/2] = \frac{Z_1^2 e^4}{\ell^2 \hbar^2 v^2 + Z_1^2 e^4}
= \frac{\eta^2}{\ell^2 + \eta^2} \, .
\label{8.137}\eeq
We thus have
\beq
\sigma_{\rm tr}^{\rm cl} (\ell_{\rm ad}) = \frac{4\pi Z_1^2 e^4}{\me^2 v^4}
\int_0^{ \ell_{\rm ad}}
\frac{\ell}{\ell^2 +\eta^2}\, \d \ell ,
\label{8.138}\eeq
and we can express the classical stopping power as
\beq
\left[ - \, \frac{\d E}{\d z} \right]_{\rm col} =
{\cal N} Z \, \frac{W_{\rm max}}{2} \, \sigma_{\rm
tr}^{\rm cl} (\ell_{\rm ad}) = \frac{4\pi Z_1^2 e^4}{\me v^2} \,
{\cal N} Z \, L_{\rm tr}^{\rm cl}
\label{8.139}\eeq
with
\beq
L_{\rm tr}^{\rm cl} = \int_0^{\ell_{\rm ad}}
\frac{\ell}{\ell^2 +\eta^2}\, \d \ell = \frac{1}{2} \,
\ln\left( \frac{\ell_{\rm ad}^2 + \eta^2}{\eta^2} \right) \simeq
\ln\left( \frac{\ell_{\rm ad}}{\eta} \right).
\label{8.140}\eeq
Upon comparison with the Bohr formula, Eq.\ \req{8.125}, we conclude that
the correct adiabatic limit $\ell_{\rm ad}$ is
\beq
\ell_{\rm ad} = \frac{2 \, \exp(-g) \, \me v^2}{I}.
\label{8.141}\eeq
Indeed, this gives
\beq
L_{\rm tr}^{\rm cl} \simeq
\ln\left(\frac{\exp(-g)\, \me v^3 \hbar}{|Z_1| \, e^2 \,
I} \right) = L_{\rm Bohr}.
\label{8.142}\eeq

\citet{LindhardSorensen1996} considered the analogous calculation of the
stopping power based on the transport cross section obtained from the
exact quantum description of (non-relativistic) Coulomb scattering. The
transport cross section for Coulomb scattering can be written as [see
Eq.\ \req{5.30}]
\beq
\sigma_{\rm tr}^{\rm qu} = \frac{4\pi\, \hbar^2}{\me^2 v^2}
\sum_{\ell=0}^\infty (\ell+1) \, \sin^2(\Delta_\ell - \Delta_{\ell+1})
\label{8.143}\eeq
with the Coulomb phase shifts \req{5.36},
\beq
\Delta_\ell = \arg \Gamma(\ell + 1 - {\rm i} \eta).
\label{8.144}\eeq
The property
\beq
\Delta_{\ell +1} = \Delta_\ell + \arctan(-\eta/\ell)
\label{8.145}\eeq
implies that
\beq
\sin^2(\Delta_\ell - \Delta_{\ell+1}) =
\frac{\eta^2}{(\ell+1)^2 + \eta^2}
\nonumber\eeq
and, consequently,
\beq
\sigma_{\rm tr}^{\rm qu} = \frac{4\pi\, \hbar^2}{\me^2 v^2} \, \eta^2
\sum_{\ell=0}^\infty \frac{\ell+1}{(\ell+1)^2 + \eta^2}
= \frac{4\pi Z_1^2 e^4}{\me^2 v^4}
\sum_{\ell=0}^\infty \frac{\ell+1}{(\ell+1)^2 + \eta^2}\, .
\label{8.146}\eeq
Comparison with Eq.\ \req{8.138} shows that the quantity
\beq
L_{\rm tr}^{\rm qu} \equiv \sum_{\ell=0}^{\infty}
\frac{\ell+1}{(\ell+1)^2 + \eta^2}
\label{8.147}\eeq
is analogous to the classical logarithm $L_{\rm
tr}^{\rm cl}$, Eq.\ \req{8.140}.

On the other hand, when a first-order perturbation approach is valid,
$\eta \ll 1$, and we can disregard $\eta^2$ in expression \req{8.147},
which becomes
\beq
L_{\rm tr}^{\rm pert} \equiv \sum_{\ell=0}^{\ell'_{\rm ad}}
\frac{1}{\ell+1} \simeq \int_1^{\ell'_{\rm ad}} \frac{\d \ell}{\ell} =
\ln\left(\ell'_{\rm ad} \right).
\label{8.148}\eeq
We have introduced a perturbative adiabatic limit $\ell'_{\rm ad}$,
which is determined by identifying the quantity \req{8.148}
with the Bethe logarithm,
\beq
L_{\rm tr}^{\rm pert} = L_{\rm Bethe} =
\ln \left(\frac{2 \me v^2}{I} \right).
\label{8.149}\eeq
The desired correction is obtained as the difference
\beq
Z_1^2 L^{\rm nr}_2 \equiv L_{\rm tr}^{\rm qu} - L_{\rm tr}^{\rm pert}
= \sum_{\ell=0}^{\infty}
\frac{\ell+1}{(\ell+1)^2 + \eta^2}
- \sum_{\ell=0}^{\ell'_{\rm ad}} \frac{1}{\ell+1}.
\label{8.150}\eeq
Since the difference of the unrestricted
summations converges, we introduce a negligible error by setting
\beq
Z_1^2 L^{\rm nr}_2 = \sum_{\ell=0}^\infty \left(
\frac{\ell+1}{(\ell+1)^2 + \eta^2} - \frac{1}{\ell+1} \right)
= - \eta^2 \sum_{n=1}^\infty \frac{1}{ n(n^2+\eta^2)}\, .
\label{8.151}\eeq
This is the {\it Bloch non-relativistic correction} to the Bethe formula,
which should be calculated by considering the Sommerfeld parameter, Eq.\
\req{8.126}, with the relativistic speed $v=\beta c$ to account, at
least partially, for relativistic kinematical effects.

The series \req{8.151} converges quite slowly. The calculation of the
correction can be simplified by using the approximations
\begin{subequations} \label{8.152}
\beqa
Z_1^2 L^{\rm nr}_2 &\simeq& - \eta^2\left( 1.20205683-1.03691432 \eta^2
+ 1.00756875 \eta^4 \right.
\nonumber \\ [2mm]
&& \left. -0.98268318 \eta^6 + 0.78219262 \eta^8 \right)
\label{8.152a}\eeqa
for $\eta^2 < 0.1$, and
\beq
Z_1^2 L^{\rm nr}_2 = - g - \ln \eta - \frac{1}{12 \eta^2}
- \frac{1}{120 \eta^4} - \frac{1}{252 \eta^6} - \frac{1}{240 \eta^8}
\label{8.152b}\eeq
\end{subequations}
for $\eta^2 > 10$. The relative error of these approximations is less
than $10^{-7}$. Figure \ref{fig8.4} shows the Bloch correction as a
function of the Sommerfeld parameter $\eta=|Z_1| \alpha/\beta$. Notice
that $\eta$ equals $|Z_1|\alpha$ at very high energies and increases
when the energy of the projectile decreases.

\begin{figure}[hbt] \begin{center}
\includegraphics*[width=10.5cm]{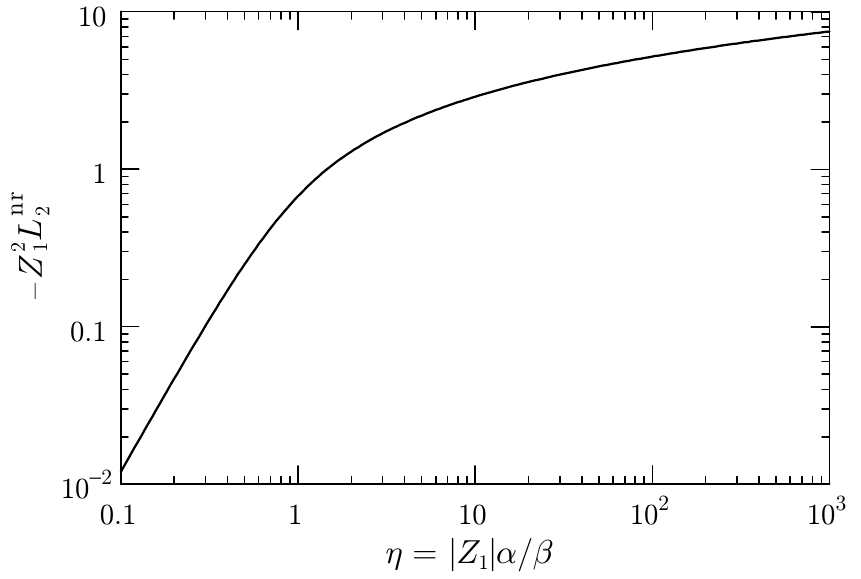}
\caption{
Negative of the non-relativistic Bloch correction as a function of the
Sommerfeld parameter $\eta$.
}\label{fig8.4}
\end{center} \end{figure}

\index{Bethe--Bloch formula}
The corrected formula
\beq
S_{\rm Bethe-Bloch} = \frac{4\pi Z_1^2 e^4}{\me v^2} \, {\cal N} Z
\left[ \ln \left(\frac{2 \me v^2}{I} \right)
+ \ln \left( \frac{1}{1-\beta^2} \right) - \beta^2 + Z_1^2 L^{\rm nr}_2 \right]
\label{8.153}\eeq
is known as {\it the Bethe--Bloch stopping power formula}. For very
small values of $\eta$ the Bloch correction is negligible and we
regain the Bethe formula; for $\eta \gg 1$, $Z_1^2 L_2$ approaches the
value $-g - \ln \eta$ and the resulting formula is Bohr's classical
result (with the full relativistic correction).

\subsection{The Lindhard--S\o rensen correction \label{sec8.3.1}}

\index{stopping power!Lindhard--S\o rensen relativistic correction}
\index{Lindhard--S\o rensen relativistic correction}

\citet{LindhardSorensen1996} have extended their non-relativistic
calculation of the Bloch correction to account for relativistic effects.
They replaced the non-relativistic transport cross section \req{8.143}
with its relativistic generalization expressed in terms of the
Dirac-Coulomb phase shifts instead of the Schr\"{o}dinger--Coulomb phase
shifts \req{8.144}, which amounts to using the correct relativistic
theory of elastic collisions of electrons with pointlike charged
projectiles. The relativistic extension of the Bloch correction was
derived as the difference between the exact transport cross section and
its perturbative expansion. This procedure is justified only for
projectiles with low and intermediate energies, say up to $E \sim 10 M_1
c^2$. At higher energies, the finite size of the charge distribution of
the projectile becomes relevant, and one should modify the theory by
replacing the Dirac-Coulomb phase shifts with the corresponding phase
shifts for scattering of electrons by the projectile's finite-size
charge distribution. While the Dirac-Coulomb phase shifts admit an
analytical expression, the relativistic phase shifts for scattering of
electrons by other potentials have to be calculated numerically.

The stopping power formula with the Lindhard--S\o rensen correction
reads
\beq
S = \frac{4\pi Z_1^2 e^4}{\me v^2} \, {\cal N} Z
\left[ \ln \left(\frac{2 \me v^2}{I} \right)
+ \ln \left( \frac{1}{1-\beta^2} \right) - \beta^2 +
\Delta L^{\rm LS} \right] \, .
\label{8.154} \eeq
The results from numerical calculations of the relativistic correction
$\Delta L^{\rm LS}$ to the stopping power for
projectiles with small charges ($|Z_1| \le 2$) and energies less than
about $10 M_1 c^2$ are closely approximated by the following analytical
expression
\beq
\Delta L^{\rm LS}_{\rm point} = \left ( \frac{1+A}{1+1.92 \,
(\gamma-1)^{1.41}} - A \right) Z_1^2 L^{\rm nr}_2,
\label{8.155}\eeq
where $\gamma-1=E/(M_1c^2)$, and $A=180.20$ for $Z_1=+1$ (protons,
deuterons, tritons, antimuons), $A=-178.34$ for $Z_1=-1$ (antiprotons,
muons), $A=90.59$ for $Z_1=+2$ (alphas), and $A=-88.73$ for $Z_1=-2$.
Values from this empirical formula agree closely with numerical results
calculated from the theory of \citet{LindhardSorensen1996} for pointlike
projectiles, the absolute differences being normally less than about
5$\times 10^{-4}$. Figure \ref{fig8.5} displays the Lindhard--S\o
rensen correction for point projectiles with charge number $Z_1$ equal
to $\pm1$ and $\pm 2$; the approximation \req{8.155} is seen to closely
approximate the numerical results calculated with the Dirac--Coulomb
phase shifts.

\begin{figure}[h] \begin{center}
\includegraphics*[width=8.5cm]{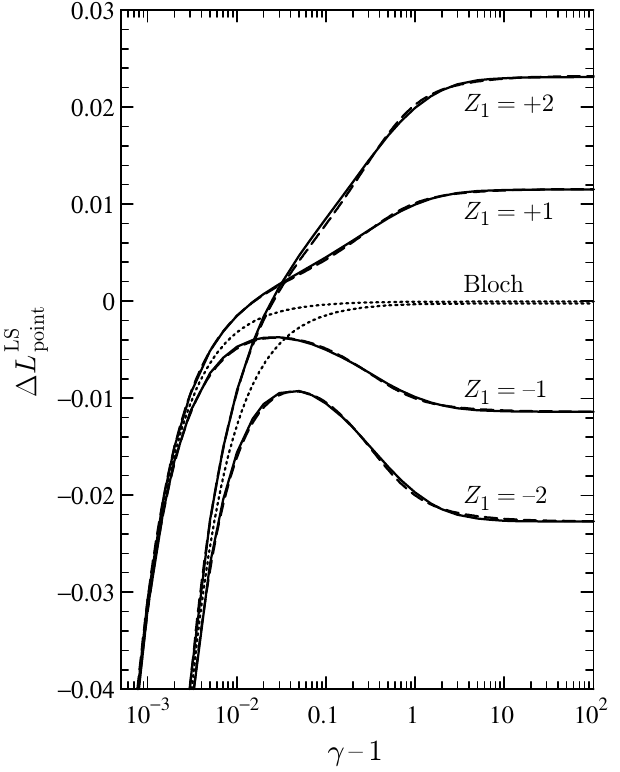}
\caption{
Lindhard--S\o rensen correction for point projectiles of the indicated
charge numbers. The solid curves are the results from accurate numerical
calculations. The dashed curves represent the analytical approximation
\req{8.155}. The dotted curves show the non-relativistic Bloch
correction, Eq.\ \req{8.151}.
}\label{fig8.5}
\end{center} \end{figure}

Although the correction \req{8.155} reduces to the Bloch form, $Z_1^2
L_2^{\rm nr}$, for slow projectiles, it also accounts for part of the
$Z_1^3$ and higher-order relativistic corrections. As noted by
\citeauthor{LindhardSorensen1996}, the angular momenta that give sizable
contributions to the correction $\Delta L^{\rm LS}$ correspond to impact
parameters much smaller than the atomic radii. Therefore, we consider
that the Lindhard--S\o rensen correction effectively accounts for the
$Z_1^3$ correction for close collisions.


\section{The Barkas or $Z_1^3$ correction \label{sec8.4}}
\index{stopping power!Barkas $Z_1^3$ correction|(}
\index{Barkas $Z_1^3$ correction|(}

The Bethe--Bloch formula \req{8.153} is even in $Z_1$ and, consequently,
it predicts the same stopping powers for particles and their
antiparticles. However, there is experimental evidence of appreciable
differences between the stopping powers for particles with the same
masses and opposite charges.  The origin of these differences is called
the Barkas effect, which introduces a correction term proportional to
$Z_1^3$ in the formula of the stopping power. As indicated above, the part of
the Barkas effect arising from close collisions is already accounted for
by the Lindhard--S\o rensen correction. Here we describe the
relativistic calculation of the Barkas correction for distant
interactions by \citet{JacksonMcCarthy1972}, which parallels a
non-relativistic formulation by \citet{Ashley1972}.

As in previous Sections, we consider distant interactions of the
swift particle with harmonically bound electrons at the impact parameter
$b$. The classical formula for the stopping power is based on the
assumption that the electric field of the projectile is constant within
the volume swept by the target electron. We can go a step further by
considering the variation of the electric field with the displacement
${\bf u}$ of the electron from its equilibrium position, which is given
by [see Eqs.\ \req{8.18}]
\beq
\ecb(b,t) = Z_1 e \, \frac{\gamma (b +u_x) \, \hat{\bf x}
- \gamma (vt-u_z) \, \hat{\bf z}}
{[(b+u_x)^2 + \gamma^2 (v t-u_z)^2]^{3/2}}\, .
\label{8.156}\eeq
Following \citep{Ashley1972}, we will study the motion of
the target electron in the electric field obtained by expanding the
expression \req{8.156} for small displacements and retaining first-order
terms only. That is, we take
\beq
\ecb(b,t) = \ecb_0(b,t) + \Delta \ecb(b,t),
\label{8.157}\eeq
where
\beq
\ecb_0(b,t) = Z_1 e \, \frac{\gamma b \, \hat{\bf x} - \gamma vt \, \hat{\bf
z}}{(b^2 + \gamma^2 v^2 t^2)^{3/2}}
\label{8.158}\eeq
is the electric field at the equilibrium position and
\beqa
\Delta \ecb(b,t) &=&
\frac{\partial \ecb}{\partial u_x} \, u_x +
\frac{\partial \ecb}{\partial u_z} \, u_z =
\frac{Z_1 e}{(b^2 + \gamma^2 v^2 t^2)^{5/2}}
\nonumber \\ [2mm]
&& \mbox{} \times \left\{
\left[ (\gamma^3 v^2 t^2 -2 \gamma b^2) u_x + 3 \gamma^3 vtb \, u_z
\rule{0mm}{4mm}\right] \hat{\bf x}
\right.
\nonumber \\ [2mm]
&& \mbox{} \; \left. + \left[
3b\gamma v t \, u_x + (\gamma b^2 -2 \gamma^3 v^2 t^2 ) u_z
\rule{0mm}{4mm}\right] \hat{\bf z} \right\}
\label{8.159}\eeqa
is the first-order correction.

We wish to calculate the energy transferred to an electron that is
harmonically bound at the point $(b,0,0)$ (see Fig.\ \ref{fig8.1}) and
initially at rest. The displacement ${\bf u}(t)$ of the target electron
at time $t$ from its equilibrium position is determined by the force
equation
\beq
\ddot{\bf u} + \omega_j^2 {\bf u} = - \frac{e}{\me} \, \ecb(b,t),
\label{8.160}\eeq
where $\omega_j$ is the characteristic frequency of the oscillator and
$\ecb(b,t)$ is the electric field acting on the electron. We express the
electron trajectory in the form
\beq
{\bf u}(t) = {\rm Im} \, \frac{e}{\me \omega_j} \int_{-\infty}^t
\exp[-{\rm i} \omega_j (t-t')] \, \ecb(b,t')  \, \d t',
\label{8.161}\eeq
which does satisfy the differential equation \req{8.160} with the
initial conditions ${\bf u}(-\infty) = 0$ and $\dot{\bf u}(-\infty)=0$.
At time $t$ the oscillator has absorbed an energy
\beqa
W(b,\omega_j,t) &=& \frac{\me}{2} \left[ \dot{\bf u}^2(t) + \omega_j^2 {\bf
u}^2(t) \rule{0mm}{4mm} \right]
\nonumber \\ [2mm]
&=& \frac{\me}{2} \left\{
\left( {\rm Im} \, \frac{-{\rm i} e}{\me} \int_{-\infty}^t
\exp[-{\rm i} \omega_j (t-t')] \, \ecb(b,t')  \, \d t' \right)^2 \right.
\nonumber \\ [2mm]
&& \mbox{} \left. + \omega_j^2
\left( {\rm Im} \, \frac{e}{\me \omega_j} \int_{-\infty}^t
\exp[-{\rm i} \omega_j (t-t')] \, \ecb(b,t')  \, \d t' \right)^2 \right\}
\nonumber \\ [2mm]
&=& \frac{e^2}{2\me} \left|
\int_{-\infty}^t
\exp({\rm i} \omega_j t') \, \ecb(b,t')  \, \d t' \right|^2 .
\label{8.162}\eeqa
The energy absorbed along the complete (unperturbed) trajectory of the
projectile is
\beq
W (b,\omega_j) \equiv W(b,\omega_j,\infty) =
\frac{\pi e^2}{\me} \left| \ecb(b,\omega_j) \right|^2 .
\label{8.163}\eeq
In agreement with the result obtained above [see Eq.\ \req{8.30}].

Assuming that the displacement of the electron is small, $|{\bf u}| \ll
(b^2 + \gamma^2 v^2 t^2)^{1/2}$, we can approximate the electric field
in the form \req{8.157}. The term $\Delta \ecb(b,t)$ may be calculated
by iteration as follows. We start from the zeroth-order
displacement ${\bf u}(t)$ obtained from Eq.\ \req{8.161} with
$\ecb_0(b,t)$ replacing $\ecb(b,t)$,
\beq
{\bf u}(t) = {\rm Im} \, \frac{e}{\me \omega_j} \int_{-\infty}^t
\exp[-{\rm i} \omega_j (t-t')] \, \ecb_0(b,t')  \, \d t'.
\label{8.164}\eeq
Inserting this expression into Eq.\ \req{8.159} we calculate the field
correction to first order in the displacement. The corrected field can
then be used to obtain the electron trajectory ${\bf u}(t)$ to first
order, and so on. Thus, to determine the first-order correction to the
electric field we can use the displacement \req{8.164},
\beqa
{\bf u}(t) &=& \frac{Z_1 e^2}{\me \omega_j} \, {\rm Im} \, \int_{-\infty}^t
\exp[-{\rm i} \omega_j (t-t')] \,
\frac{\gamma b \, \hat{\bf x} - \gamma vt' \, \hat{\bf
z}}{(b^2 + \gamma^2 v^2 t'^2)^{3/2}}\, \d t'
\nonumber \\ [2mm]
&=& \frac{Z_1 e^2}{\me b^2 \omega_j} \, \frac{b}{\gamma v} \,
{\rm Im} \, \int_{-\infty}^x
\exp[-{\rm i} \xi (x-y)] \,
\frac{\gamma \, \hat{\bf x} - y \, \hat{\bf
z}}{(1 + y^2)^{3/2}}\, \d y\, ,
\label{8.165}\eeqa
where we have introduced the variables
\beq
x= \gamma v t/b, \qquad y= \gamma v t' /b, \qquad  \mbox{and} \qquad
\xi=\omega_j b/(\gamma v).
\label{8.166}\eeq
Hence,
\begin{subequations}
\label{8.167}
\beq
u_x(t) = - \frac{Z_1 e^2}{\me v \, \omega_j b} \,
F_1(\xi,x)
\label{8.167a}\eeq
and
\beq
u_z(t) = \frac{Z_1 e^2}{\me v \, \omega_j b} \, \frac{1}{\gamma} \,
F_2(\xi,x)
\label{8.167b}\eeq
\end{subequations}
with
\begin{subequations}
\label{8.168}
\beq
F_1(\xi,x) = \int_{-\infty}^y
\frac{\sin[\xi (x-y)]}{(1 + y^2)^{3/2}}\, \d y
\label{8.168a}\eeq
and
\beq
F_2(\xi,y) = \int_{-\infty}^x
\frac{y \, \sin[\xi (x-y)]}{(1 + y^2)^{3/2}}\, \d y\, .
\label{8.168b}\eeq
\end{subequations}

\index{Fourier transform}
The Fourier transforms of the two terms of the electric field \req{8.157}
are [see Eq.\ \req{8.33}]
\beq
\ecb_0(b,\omega) = \frac{Z_1 e}{bv} \left( \frac{2}{\pi} \right)^{1/2}
\left[  \xi K_1 (\xi) \, \hat{\bf x}
- \frac{\rm i}{\gamma} \, \xi
K_0 (\xi) \, \hat{\bf z} \right]
\label{8.169}\eeq
and
\beqa
\Delta \ecb (b,\omega) &=&
(2\pi)^{-1/2} \int_{-\infty}^\infty \d t \, \exp({\rm i} \omega t)\, \frac{Z_1e}{
(b^2 + \gamma^2 v^2 t^2)^{5/2}}
\nonumber \\ [2mm]
&& \mbox{} \times \left\{
\left[ (\gamma^3 v^2 t^2 -2 \gamma b^2) u_x + 3 \gamma^3 vtb \, u_z
\rule{0mm}{4mm}\right] \hat{\bf x}
\right.
\nonumber \\ [2mm]
&& \mbox{} \; \left. + \left[
3b\gamma v t \, u_x + (\gamma b^2 -2 \gamma^3 v^2 t^2 ) u_z
\rule{0mm}{4mm}\right] \hat{\bf z} \right\}.
\label{8.170}\eeqa
Introducing the variables \req{8.166},
\beqa
\Delta \ecb (b,\omega) &=& \frac{Z_1 e}{b^2 \gamma v}
(2\pi)^{-1/2} \int_{-\infty}^\infty \d x \, \frac{\exp({\rm i} \xi x)}{
(1 + x^2)^{5/2}}
\nonumber \\ [2mm]
&& \mbox{} \times \left\{
\gamma \left[ (x^2 -2) u_x + 3 x \gamma \, u_z
\rule{0mm}{4mm}\right] \hat{\bf x}
+ \left[
3x \, u_x + (1 -2 x^2 ) \gamma u_z
\rule{0mm}{4mm}\right] \hat{\bf z} \right\},
\nonumber \eeqa
and inserting the expressions \req{8.167},
\beqa
\Delta \ecb (b,\omega) &=& \frac{Z_1^2 e^3}{b^3 \omega_j \, \gamma \me v^2}
\, (2\pi)^{-1/2} \int_{-\infty}^\infty \d x \,
\frac{\exp({\rm i} \xi x)}{(1 + x^2)^{5/2}}
\nonumber \\ [2mm]
&& \mbox{} \times \left\{
\gamma \left[ (2 - x^2) F_1(\xi,x) + 3 x \, F_2(\xi,x)
\rule{0mm}{4mm}\right] \hat{\bf x}
\right.
\nonumber \\ [2mm]
&& \mbox{} \; \left.
+ \left[
- 3x \,  F_1(\xi,x) + (1 -2 x^2 )  \, F_2(\xi,x)
\rule{0mm}{4mm}\right] \hat{\bf z} \right\}.
\label{8.171}\eeqa

Now we can obtain the energy transfer from Eq.\ \req{8.163}. Noting
that, to fist order in the field correction,
\beq
\left| \ecb_0 + \Delta \ecb \right|^2
\simeq \left| \ecb_0 \right|^2
+ 2 \, {\rm Re} [ \ecb_0]
\, {\rm Re} [\Delta \ecb]
+ 2 \, {\rm Im} [ \ecb_0]
\, {\rm Im} [\Delta \ecb],
\label{8.172}\eeq
we can write
\beqa
W(b,\omega_j) &=&
\frac{\pi e^2}{\me} \left| \ecb_0(b,\omega_j)
+ \Delta \ecb (b,\omega_j)
\right|^2
\nonumber \\ [2mm]
&=& \frac{2 Z_1^2 e^4}{\me v^2}  \, \frac{1}{b^2}
\left[  \xi^2 K_1^2 (\xi)
+ \frac{1}{\gamma^2} \, \xi^2 K_0^2 (\xi) \right]
\nonumber \\ [2mm]
&& \mbox{} +
\frac{2 Z_1^3 e^6}{\me^2 v^3} \, \frac{1}{b^4 \omega_j}
\left[ - \xi K_1 (\xi) G_1(\xi)
+ \frac{1}{\gamma^2} \, \xi K_0 (\xi) G_0(\xi) \rule{0mm}{4mm}\right],
\label{8.173}\eeqa
with
\begin{subequations}
\label{8.174}
\beq
G_1(\xi) = \int_{-\infty}^\infty
\d x \, \frac{\cos(\xi x)}{(1 + x^2)^{5/2}}
\left[ (x^2-2) F_1(\xi,x) - 3 x \, F_2(\xi,x)
\rule{0mm}{4mm}\right]
\label{8.174a}\eeq
and
\beq
G_0(\xi) =
\int_{-\infty}^\infty \d x \, \frac{\sin(\xi x)}{(1 + x^2)^{5/2}}
\left[
3x \,  F_1(\xi,x) - (1 -2 x^2 )  \, F_2(\xi,x)
\rule{0mm}{4mm}\right].
\label{8.174b}\eeq
\end{subequations}
The first term in the expression \req{8.173} is the known zeroth-order
result [cf.\ Eq.\ \req{8.35}], the second term represents the
Barkas-effect correction.

The stopping power of distant oscillators, with
impact parameters larger than the cutoff value $a$, is
\beqa
\left[ - \, \frac{\d E}{\d z} \right]_{b>a} &=& 2 \pi {\cal N} \,
\int_{0}^\infty \d \omega_j \,
\frac{\d f(\omega_j)}{\d \omega_j} \int_{a}^\infty W (b,\omega_j) \, b \, \d b
\nonumber \\ [2mm]
&=& \left[ - \, \frac{\d E}{\d z} \right]_{b>a}^{(0)}
+ \left[ - \, \frac{\d E}{\d z} \right]_{b>a}^{(1)}.
\label{8.175}\eeqa
Proceeding as in Section \ref{sec8.1.2}, the contribution of the zeroth
order term can be expressed in the form
$$
\left[ - \, \frac{\d E}{\d z} \right]_{b>a}^{(0)} =
\frac{4\pi Z_1^2 e^4}{\me v^2} \, {\cal N} \,
\int_{0}^\infty \d \omega_j \, \frac{\d f(\omega_j)}{\d \omega_j}\rule{7cm}{0mm}
$$
$$
\mbox{} \times \left\{ \xi_a K_1(\xi_a) K_0(\xi_a)
- \frac{v^2}{2c^2} \xi_a^2 \left[ K_1^2(\xi_a) - K_0^2(\xi_a) \right]
\right\},
\eqno{\req{8.38}}$$
where $\xi_a = \omega_j a/(\gamma v)$. The contribution of the $Z_1^3$
term is
\beqa
\left[ - \, \frac{\d E}{\d z} \right]_{b>a}^{(1,{\rm osc})}
&=& 2 \pi {\cal N} \, \int_0^\infty \d \omega_j \,
\frac{\d f(\omega_j)}{\d \omega_j} \int_{a}^\infty \d b \, b
\nonumber \\ [2mm]
&& \mbox{} \times
\left\{
\frac{2 Z_1^3 e^6}{\me^2 v^3} \, \frac{1}{b^4 \omega_j}
\left[ - \xi K_1 (\xi) G_1(\xi)
+ \frac{1}{\gamma^2} \, \xi K_0 (\xi) G_0(\xi) \rule{0mm}{4mm}\right]\right\}
\nonumber \\ [2mm]
&=& 4 \pi \, \frac{Z_1^3 e^6}{\me^2 v^3} \, {\cal N} \,
\frac{1}{\gamma^2 v^2}
\int_0^\infty \d \omega_j \,
\frac{\d f(\omega_j)}{\d \omega_j} \, \omega_j \int_{a}^\infty \d b
\nonumber \\ [2mm]
&& \mbox{} \times
\left\{
\frac{\gamma^2 v^2}{b^3 \omega_j^2}
\left[ - \xi K_1 (\xi) G_1(\xi)
+ \frac{1}{\gamma^2} \, \xi K_0 (\xi) G_0(\xi) \rule{0mm}{4mm}\right]\right\}
\nonumber \\ [2mm]
&=& 4 \pi \, \frac{Z_1^3 e^6}{\gamma^2 \me^2 v^5} \, {\cal N} \,
\int_0^\infty \d \omega_j \,
\frac{\d f(\omega_j)}{\d \omega_j} \, \omega_j
\nonumber \\ [2mm]
&& \mbox{} \times \int_{\xi_a}^\infty \frac{\d
\xi}{\xi^2} \left[ - K_1 (\xi) G_1(\xi)
+ \frac{1}{\gamma^2} \, K_0 (\xi) G_0(\xi) \rule{0mm}{4mm}\right].
\nonumber\eeqa
That is,
\beqa
\left[ - \, \frac{\d E}{\d z} \right]_{b>a}^{(1,{\rm osc})}
&=& \frac{4 \pi \, Z_1^3 e^6}{\gamma^2 \me^2 v^5} \, {\cal N} \,
\int_0^\infty \d \omega_j \,
\frac{\d f(\omega_j)}{\d \omega_j} \, \omega_j
\left[ I_1 (\xi_a) + \frac{1}{\gamma^2} \, I_2 (\xi_a) \rule{0mm}{4mm}\right]
\label{8.176}\eeqa
with
\begin{subequations}
\label{8.177}
\beq
I_1(\xi_a) = - \int_{\xi_a}^\infty \frac{1}{\xi^2} \, K_1(\xi) \,
G_1(\xi) \, \d \xi
\label{8.177a}\eeq
and
\beq
I_2(\xi_a) = \int_{\xi_a}^\infty \frac{1}{\xi^2} \, K_0(\xi) \, G_0(\xi)
\, \d \xi,
\label{8.177b}\eeq
\end{subequations}

\index{Gauss--Legendre quadrature!adaptive}
Since we have been unable to find published numerical tables, we have
computed the functions $I_1(\xi_a)$ and $I_2(\xi_a)$ numerically as
follows. First the functions $G_1(\xi)$ and $G_0(\xi)$ were evaluated
from expressions \req{8.168} and \req{8.174} by using the adaptive
20-point Gauss--Legendre quadrature method (Section \ref{sec10.4.3.2})
to perform the integrations on
$x$ and $y$, for a grid of $\xi$ values suitably spaced to allow
accurate natural cubic spline interpolation (see Section
\ref{sec10.4.2}). Then the integrals
$I_1(\xi_a)$ and $I_2(\xi_a)$ were computed from Eqs.\ \req{8.177}, with
$G_1(\xi)$ and $G_0(\xi)$ replaced with their interpolating splines, by
using the adaptive Gauss--Legendre method.  The numerical results are
displayed in Fig.\ \ref{fig8.6}; they are generally accurate to four or
five digits.

\begin{figure}[h!] \begin{center}
\includegraphics*[width=7.5cm]{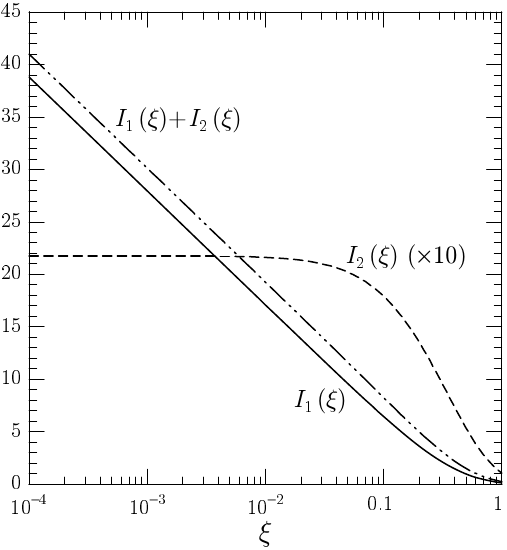} \rule{1mm}{0mm}
\includegraphics*[width=7.9cm]{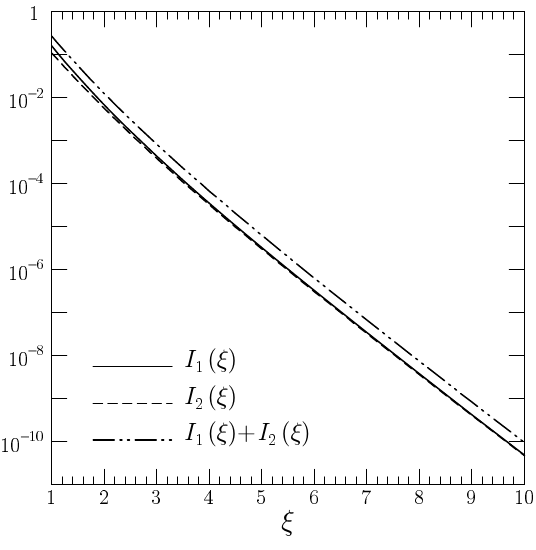}
\caption{Calculated $I_1(\xi)$ and $I_2(\xi)$ functions, as functions of
the dimensionless variable $\xi=\omega a/\gamma v$. Notice the
logarithmic scale of the vertical axes in the right plot.
\label{fig8.6}}\end{center} \end{figure}

\begin{table}[th!]
\begin{center}
\caption{ Parameters of the analytical approximations to the functions
$I_1(\xi)$ and $I_2(\xi)$, Eqs.\ \req{8.178}. Numbers in parentheses
indicate powers of ten.
\label{tab8.1}}
\vskip 4mm
\begin{tabular}{|r|c|c|c|c|} \hline
$i$ & $p_{1i}$ & $p_{2i}$ & $p_{3i}$ & $p_{4i}$\rule[-2.0mm]{0mm}{3mm}\\ \hline
$1$ & $+1.796676(+1)$ & $-3.062544(+0)$ & $-1.018033(-3)$ & $-2.798361(+1)$ \\
$2$ & $+4.825194(+1)$ & $+1.787672(+1)$ & $+1.080448(-1)$ & $+3.964525(+0)$ \\
$3$ & $+4.661380(+1)$ & $-4.100572(+1)$ & $+1.568923(+0)$ & $-5.706711(-1)$ \\
$4$ & $+2.269807(+1)$ & $+4.652797(+1)$ & $-2.779110(+0)$ & $+2.652851(-2)$ \\
$5$ & $+5.409180(+0)$ & $-2.728692(+1)$ & $+8.079930(+0)$ & $-6.442942(-4)$ \\
$6$ & $+5.116173(-1)$ & $+8.070905(+0)$ & $-7.244712(+0)$ & $+8.003948(-6)$ \\
$7$ & $--$            & $-9.587701(-1)$ & $--           $ & $-4.016094(-8)$
\\ \hline
\end{tabular}

\vskip 3mm
\begin{tabular}{|r|c|c|c|c|} \hline
$i$ & $q_{1i}$ & $q_{2i}$ & $q_{3i}$ & $q_{4i}$ \rule[-2.0mm]{0mm}{3mm} \\ \hline
$1$ & $+2.177590(+0)$ & $+3.768512(+0)$ & $+7.431717(-5)$ & $-2.808307(+1)$ \\
$2$ & $+5.689823(-1)$ & $-2.135702(+1)$ & $+6.662051(-2)$ & $+3.977364(+0)$ \\
$3$ & $+1.038828(+0)$ & $+4.601060(+1)$ & $+2.142710(+0)$ & $-5.714987(-1)$ \\
$4$ & $+2.877808(-4)$ & $-4.714780(+1)$ & $-7.407167(+0)$ & $+2.655846(-2)$ \\
$5$ & $--$            & $+2.551946(+1)$ & $+2.327532(+1)$ & $-6.449171(-4)$ \\
$6$ & $--$            & $-7.102352(+0)$ & $-3.497742(+1)$ & $+8.010890(-6)$ \\
$7$ & $--$            & $+8.042960(-1)$ & $+1.950319(+1)$ & $-4.019306(-8)$
\\ \hline
\end{tabular}
\end{center} \end{table}

In order to facilitate numerical calculations, the tabulated functions have been
fitted by the following analytical expressions which, with the
parameters given in Table \ref{tab8.1}, approximate the numerical
results with an accuracy better than $0.1$ \% for $\xi$ from zero up to
$\sim$ 15.
\begin{subequations}
\label{8.178}
For $x < 10^{-9}$,
\beq
I_1(\xi) = (3\pi/2) \ln(3/8\xi), \qquad I_2(\xi) = 2.17758\, .
\label{8.178a}\eeq
For $10^{-9} \le \xi < 0.25$,
\beqa
I_1(\xi) &=&
\frac{3\pi}{2} \ln \left( \frac{3}{8\xi} \right)-2 \pi \xi^2
\left[p_{11}+ p_{12} \, \ln \xi + p_{13} (\ln\xi)^2 \right.
\nonumber \\ [2mm]
&& \mbox{} \left.
+ p_{14} (\ln\xi)^3 + p_{15} (\ln\xi)^4 + p_{16} (\ln\xi)^5
\right],
\label{8.178b}\eeqa
\beq
I_2(\xi) =
q_{11} -2 \pi \xi^2
\left[q_{12}+ q_{13} (\ln \xi)^2 + q_{14} (\ln\xi)^4 \right].
\label{8.178c}\eeq
For $0.25 \le \xi < 2$,
\beqa
I_1(\xi) &=&
\left(p_{21}+ p_{22} \xi^{-1/2} + p_{23} \xi^{-1} + p_{24} \xi^{-3/2}
+ p_{25} \xi^{-2}
\right.
\nonumber \\ [2mm]
&& \mbox{} \left.
+p_{26} \xi^{-5/2} + p_{27} \xi^{-3} \right) \xi^{-5/8},
\label{8.178d}\eeqa
\beqa
I_2(\xi) &=&
\left(q_{21}+ q_{22} \xi^{-1/2} + q_{23} \xi^{-1}
+ q_{24} \xi^{-3/2} + q_{25} \xi^{-2}
\right.
\nonumber \\ [2mm]
&& \mbox{} \left.
+q_{26} \xi^{-5/2} + q_{27} \xi^{-3} \right) \xi^{-1} \exp(-3\xi/2).
\label{8.178e}\eeqa
For $2 \le \xi < 15$,
\beq
I_1(\xi) =
\left( p_{31}+ p_{32} \xi^{-1} + p_{33} \xi^{-2} + p_{34} \xi^{-3}
+ p_{35} \xi^{-4} + p_{36} \xi^{-5}
\right) \exp(-2\xi),
\label{8.178f}\eeq
\beqa
I_2(\xi) &=&
\left( q_{31}+ q_{32} \xi^{-1} + q_{33} \xi^{-2} + q_{34} \xi^{-3}
+ q_{35} \xi^{-4} + q_{36} \xi^{-5} \right.
\nonumber \\ [2mm]
&& \mbox{} \left. + q_{37} \xi^{-6}  \right) \exp(-2\xi).
\label{8.178g}\eeqa
For $15 \le \xi \lesssim 50$,
\beq
I_1(\xi) =\exp
\left( p_{41}+ p_{42} \xi^{1} + p_{43} \xi^{2} + p_{44} \xi^{3}
+ p_{45} \xi^{4} + p_{46} \xi^{5}+ p_{47} \xi^{6} \right),
\label{8.178h}\eeq
\beqa
I_2(\xi) &=& \exp
\left( q_{41}+ q_{42} \xi^{1} + q_{43} \xi^{2} + q_{44} \xi^{3}
+ q_{45} \xi^{4} + q_{46} \xi^{5} + q_{47} \xi^{6}  \right).
\label{8.178i}\eeqa
\end{subequations}
For $\xi > 50$ the two functions are less than $10^{-40}$ and decrease
exponentially; the computer program sets them to zero.

Expressed in terms of the energy transfer $W$, the zeroth-order
and the Barkas-correction contributions to the stopping power
of distant oscillators are, respectively,
\beqa
\left[ - \, \frac{\d E}{\d z} \right]_{b>a}^{(0)} &=&
\frac{4\pi Z_1^2 e^4}{\me v^2} \, {\cal N} \,
\int_0^\infty \d W \, \frac{\d f(W)}{\d W}
\nonumber \\ [2mm]
&& \mbox{} \times \left\{ \xi K_1(\xi) K_0(\xi)
- \frac{1}{2} \, \beta^2 \, \xi^2 \left[ K_1^2(\xi) - K_0^2(\xi)
\right]
\right\}
\label{8.179}\eeqa
and
\beq
\left[ - \, \frac{\d E}{\d z} \right]_{b>a}^{(1,{\rm osc})}
= \frac{4 \pi \, Z_1^2 e^4}{\me v^2} \, {\cal N} \,
\frac{Z_1 \alpha}{\gamma^2 \beta^3 \me c^2}
\int_0^\infty \d W \, \frac{\d f(W)}{\d W} \, W
\left[ I_1 (\xi) + \frac{1}{\gamma^2} \, I_2 (\xi)
\rule{0mm}{4mm}\right],
\label{8.180}\eeq
where $\xi = W a/(\gamma v \hbar)$ and $\alpha=e^2/(\hbar c)$ is the
fine-structure constant.
For sufficiently fast projectiles, and moderate values of the cutoff
impact parameter $a$, $\xi$ is much less than unity, and the integral
in Eq.\ \req{8.179} can be simplified as in Section \ref{sec8.1.2} to
give the result \req{8.40} of the Bohr theory.

\index{energy-loss DCS!Barkas correction to the}

From the expression \req{8.180} of the $Z_1^3$ correction to the
stopping power of distant oscillators we can identify the following
Barkas correction to the energy-loss DCS,
\beq
\frac{\d \, (\Delta \sigma)_{\rm B,osc}}{\d W} =
\frac{4 \pi \, Z_1^2 e^4}{\me v^2} \,
\frac{ Z_1 \alpha}{\gamma^2 \beta^3 \, \me c^2} \,
\frac{\d f(W)}{\d W}
\left[ I_1 (\xi) + \frac{1}{\gamma^2} \, I_2 (\xi)
\rule{0mm}{4mm}\right] \, {\cal S}(W_{\rm max}-W),
\label{8.181}\eeq
with $\xi=W a/(\gamma v\, \hbar)$. We have added a step function to make
explicit that the largest energy loss in distant interactions is assumed
to be equal to $W_{\rm max}$, Eq.\ \req{8.49}, the highest allowed
energy transfer in close collisions. The correction \req{8.181} can be
added to the energy-loss DCS obtained from the PWBA, to account
partially for non-linear effects in a form similar to the second-order
Born approximation.

It is customary to express the Barkas correction to the stopping power
in the form
\beq
\left[ - \, \frac{\d E}{\d z} \right]_{b>a}^{{\rm (1,osc)}}
= \frac{4 \pi \, Z_1^2 e^4}{\me v^2} \, {\cal N} Z \,
\Delta L^{\rm B}(a)
\label{8.182}\eeq
where
\beqa
\Delta L^{\rm B}(a) =
\frac{Z_1 \alpha}{\gamma^2 \beta^3 \me c^2} \, \frac{1}{Z}
\int_0^{W_{\rm max}} \d W \, \frac{\d f(W)}{\d W} \, W
\left[ I_1 (\xi) + \frac{1}{\gamma^2} \, I_2 (\xi)
\rule{0mm}{4mm}\right]
\rule{10mm}{0mm}
\label{8.183}\eeqa
is the correction to the stopping logarithm.

The cutoff impact parameter $a$ now remains an essential parameter of
the theory, which is selected on the basis of qualitative arguments.
\citet{JacksonMcCarthy1972} suggested taking
\beq
a= \sqrt{\frac{\hbar}{2 \me \omega}}, \qquad \xi
= \frac{W a}{\gamma v\, \hbar}
= \frac{1}{\gamma \beta} \sqrt{\frac{W}{2 \me c^2}} \, .
\label{8.184}\eeq
This $a$ value is the magnitude of the dipole oscillator strength of the
harmonic oscillator, and a measure of the amplitude of the classical
motion of an oscillator with energy $\hbar\omega$. \citet{Lindhard1976},
based on considerations of stopping by an electron gas, proposed using a
value of the order of the impact parameter that corresponds to an
angular momentum of the order of $\hbar$. More precisely, Lindhard's
cutoff impact parameter is
\beq
a = \exp(-g) \, \frac{\hbar}{\me v}, \qquad
\xi = \frac{W a}{\gamma v\, \hbar}
= \frac{\exp(-g)}{\gamma \beta^2} \, \frac{W}{\me c^2}\, ,
\label{8.185}\eeq
where $g=0.5772$ is the Euler constant\index{Euler constant},
and $\exp(-g) = 0.5616$. In the case of protons in aluminum,
\citet{Ashley1991b} found that the calculated correction to the stopping
power with \citeauthor{Lindhard1976}'s $a$ value agrees reasonably with
experimental results.

In the following we adopt Lindhard's $a$ value, which has the advantage
of being independent of the energy transfer.
Because of the approximate character of the Barkas correction, it is expedient
to allow a certain flexibility and define the cutoff impact parameter as
\beq
a = C_{\rm B} \exp(-g) \, \frac{\hbar}{\me v} \, , \qquad
\xi =  C_{\rm B}
\frac{\exp(-g)}{\gamma \beta^2} \, \frac{W}{\me c^2}\, ,
\label{8.186}\eeq
where $C_{\rm B}$ is a dimensionless constant, of the order of unity, to
be determined empirically for each material.  A systematic comparison
with results from measurements of the stopping power of elemental
materials for protons in the IAEA database (\citeauthor{Montanari2017},
\citeyear{Montanari2017};
\url{https://www-nds.iaea.org/stopping}) indicates that a
value of $C_{\rm B}$ between 1 and 10 (increasing roughly with the
atomic number $Z$) yields stopping powers in reasonable agreement with
the experiments \citep{Salvat2022c}. In the calculations we set $C_{\rm
B} = \max \{ 1, \overline{Z} / 10 \}$, where $\overline{Z}$ is the
average atomic number of the material.

It is worth noticing that the Barkas correction \req{8.183} depends
explicitly on the OOS. Previous calculations of this correction used
OOSs obtained from the LPA \citep{Ashley1972, JacksonMcCarthy1972,
ICRU49}, whose reliability is questionable.  Figure \ref{fig8.7} shows
the Barkas correction for protons in solid aluminium, silicon, copper,
and gold, as functions of the kinetic energy of the projectile. The
displayed corrections were computed for the indicated values of the
coefficient $C_{\rm B}$, Eq.\ \req{8.186}, with the empirical OOSs
derived from optical data, and with the DHFS-model OOS, Eq.\ \req{7.162}
(see Figs.\ \ref{fig7.7} and \ref{fig7.8}). The plotted results indicate
that the two OOS models yield nearly equivalent results for energies
larger than about 1 MeV. We conclude that, the Barkas correction can be
estimated by using the DHFS-model OOSs when more reliable optical
information is not available.

\begin{figure}[hp!] \begin{center}
\includegraphics*[width=7.20cm]{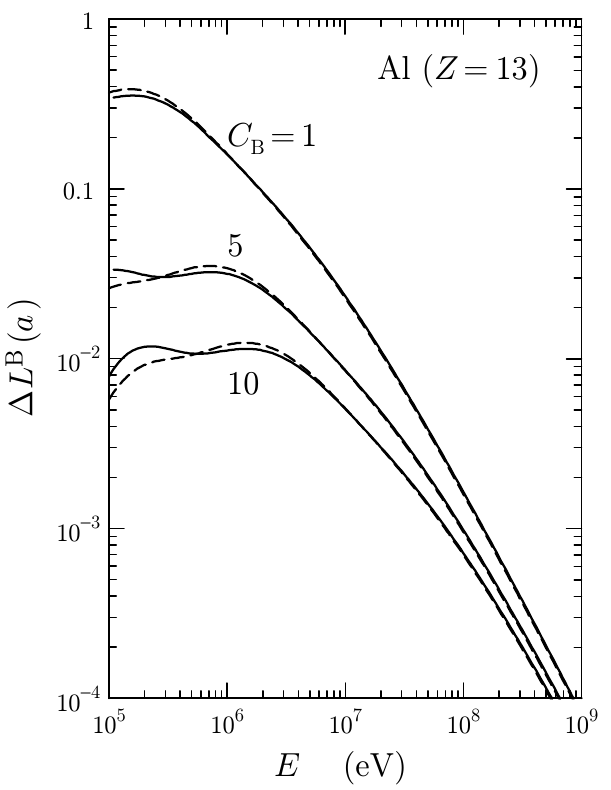} \rule{5mm}{0mm}
\includegraphics*[width=7.20cm]{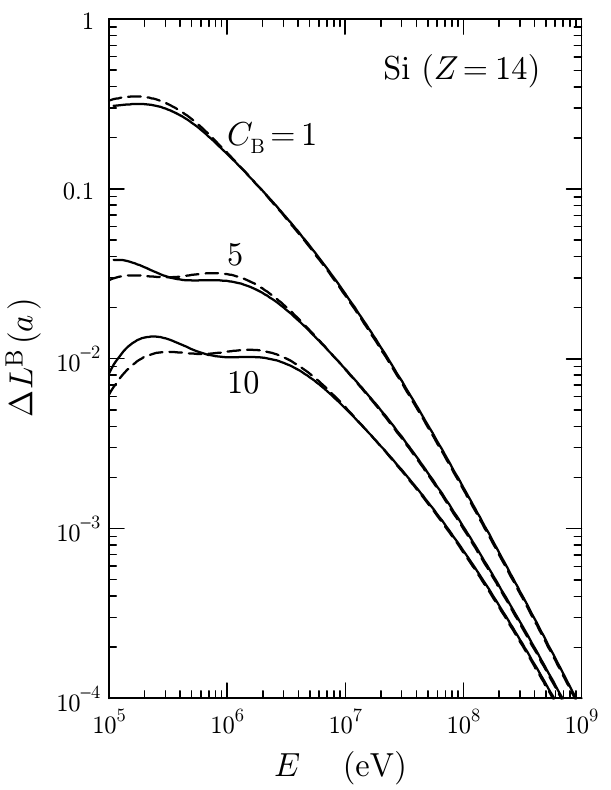} \\ [5mm]
\includegraphics*[width=7.20cm]{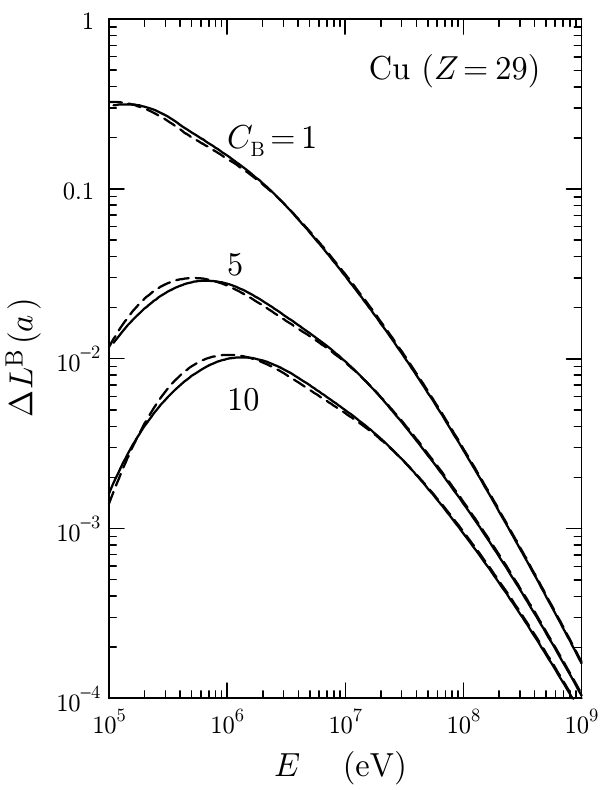} \rule{5mm}{0mm}
\includegraphics*[width=7.20cm]{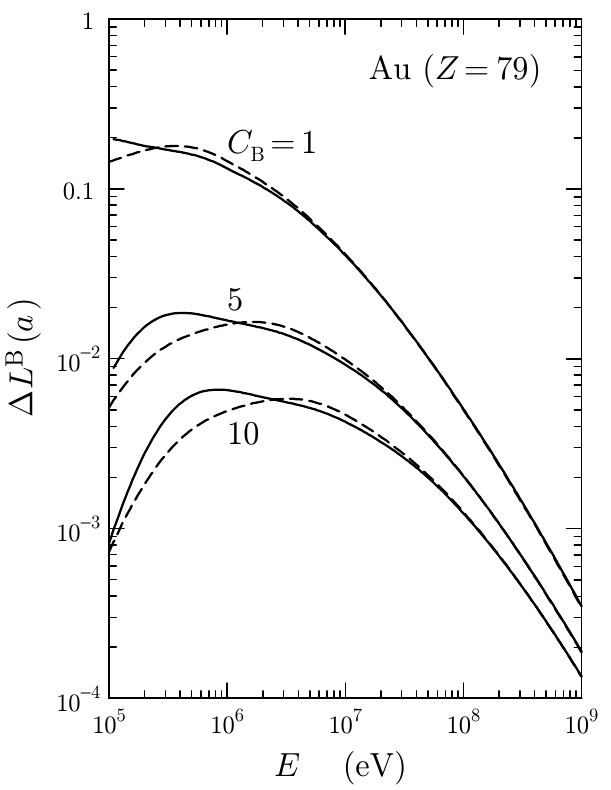}
\caption{Barkas correction for protons in solid aluminium,
silicon, copper, and gold, calculated with the empirical OOSs derived
ìfrom optical data (solid curves) and with the DHFS-model OOSs (dashed
curves), for the indicated values of the parameter $C_{\rm B}$, Eq.\
\req{8.186}, as functions of the proton energy.
\label{fig8.7}}
\end{center}\end{figure}

\index{Barkas $Z_1^3$ correction|)}

\index{stopping power!Barkas $Z_1^3$ correction|)}


\section{Corrected Bethe formula for the electronic stopping
\label{sec8.5}}
\index{stopping power!corrected Bethe formula|(}
\index{corrected Bethe formula}
\index{Bethe formula}
Adding to the Bethe formula [Eq.\ \req{6.307}] the density-effect
correction [Eq.\ \req{8.121}], the shell correction (Section
\ref{sec6.9.1}), the Lindhard--S\o rensen correction [Eq. \req{8.155}],
and the Barkas correction [Eq.\ \req{8.183}], we obtain the {\it
corrected Bethe stopping power formula},
\beqa
S (E) = - \, \frac{\d E}{\d z} &=&
\frac{4 \pi Z_1^2 e^4}{\me v^2} \, {\cal N} Z \,
\left[ \ln\left(\frac{2 \me v^2}{I} \right)
+ \ln \left( \frac{1}{1-\beta^2} \right) - \beta^2  + \frac{1}{2} \,
f(\gamma) \right.
\nonumber \\ [2mm]
&& \mbox{} \left.
- \frac{1}{2} \, \delta_{\rm F} - \frac{C_{\rm mod}(\gamma)}{Z} +
\Delta L^{\rm LS} + \Delta L^{\rm B}(a)
\right].
\label{8.187}\eeqa

\index{stopping logarithm}
Following the tradition, the corrected Bethe formula can be written in
condensed form as
\beq
S (E) = \frac{4 \pi Z_1^2 e^4}{\me v^2} \, {\cal N} Z
\left[ L_B + \Delta L^{\rm B}(a) + \Delta L^{\rm LS} \right],
\label{8.188}\eeq
where the quantity in square brackets is called the {\it stopping
logarithm}. The zeroth-order term is
\beq
L_B =  \ln\left(\frac{2 \me v^2}{I} \right) + \ln \gamma^2
- \frac{\gamma^2-1}{\gamma^2} + \frac{1}{2} \, f(\gamma)
- \frac{C_{\rm mod}(\gamma)}{Z}
- \frac{1}{2} \, \delta_{\rm F}
\label{8.189}\eeq
with
\begin{subequations}
\label{8.190}
\beq
f(\gamma) = \ln(R) + \left( \frac{\me}{M_1}\,
\frac{\gamma^2-1}{\gamma} \, R \right)^2, \qquad
R = \left[ 1 + \left(\frac{\me}{M_1} \right)^2 +
2\gamma \, \frac{\me}{M_1} \right]^{-1}
\label{8.190a}\eeq
for particles much heavier than the electron [Eq.\ \req{6.299}],
\beq
f(\gamma) = \frac{2\gamma^2-1}{\gamma^2}
+ \frac{1}{8} \left( \frac{\gamma-1}{\gamma} \right)^2
- \left[ 4 - \left( \frac{\gamma-1}{\gamma} \right)^2 \right] \ln 2
- \ln(\gamma+1)
\label{8.190b}\eeq
for electrons [Eq.\ \req{6.305}],
\beq
f(\gamma) = \frac{\gamma^2-1}{12\gamma^2} \left( 1
- \frac{14}{\gamma+1}
- \frac{10}{(\gamma+1)^2}
- \frac{4}{(\gamma+1)^3} \right)
- \ln 2 - \ln(\gamma+1)
\label{8.190c}\eeq
for positrons [Eq.\ \req{6.306}],
\end{subequations}
and
\beq
\frac{C_{\rm mod}(\gamma)}{Z} = \frac{C(\gamma)}{Z}
- \frac{S_0-Z}{2Z} \left[ \ln (\beta^2 \gamma^2) - \beta^2
\right],
\label{8.191}\eeq
where the second term is a relativistic correction, cf.\ Eq.\
\req{6.315} \citep{Salvat2022a}.
In the calculations reported below, and in the {\sc sbethe} program
(Chapter \ref{chapt10}), the stopping logarithm $L_B$ is obtained by
adding to the Bethe logarithm,
\beq
L_0 = \ln\left(\frac{2 \me v^2}{I} \right)
+ \ln \left(\frac{1}{1-\beta^2} \right) - \beta^2 \, ,
\label{8.192}\eeq
the function $f(\gamma)/2$, the tabulated DHFS shell correction and the
density-effect correction derived from the OOS for the DHFS potential,
Eq.\ \req {8.121}.

The Barkas correction, $\Delta L^{\rm B}$, is calculated from Eq.\
\req{8.183} also with the OOS for the DHFS potential and with Lindhard's
$a$ value [Eq.\ \req{8.186} with $C_B=\max\{1,Z/10\}$].  Finally, the
Lindhard--S\o rensen correction, $\Delta L^{\rm LS}$, is evaluated by
using the parameterization in Eq.\ \req{8.155}.

It is worth stressing the fact that the theory leading to the
corrected Bethe formula is essentially based on the linear-response
approximation, with the Barkas and Lindhard--S\o rensen corrections to
account for higher-order effects. In the case of an electron gas,
non-linear effects increase in importance when the velocity of the
projectile decreases, causing substantial differences between the
stopping powers obtained from the present dielectric formalism and from
non-linear theory \citep{Echenique1986} for projectiles with velocities
of the order of or less than the Fermi velocity, $v_{\rm F} =\hbar
k_{\rm F}/\me$.

The derivation of the Barkas and Lindhard--S\o rensen corrections starts
from the assumption that the projectile follows essentially a straight
trajectory, which is certainly not the case for electrons and positrons.
As a matter of fact, for these particles, the negative Barkas correction
is found to exceed the value of the Bethe logarithm at low energies.
Consequently, in calculations for projectile electrons and positrons we
will ignore the Barkas and Lindhard--S\o rensen corrections. As shown in
Figs.\ \ref{fig7.13} and \ref{fig7.14}, optical-data model calculations
without these corrections yield fairly realistic values of the mean free
path and the stopping power even for low-energy electrons and positrons.

Figures \ref{fig8.8} and \ref{fig8.9} display calculated stopping logarithms
\beq
L_{\rm total} = L_0 + \frac{1}{2} \, f(\gamma)
- \frac{C_{\rm mod}(\gamma)}{Z}
- \frac{1}{2} \, \delta_{\rm F} + \Delta L^{\rm B}(a) + \Delta L^{\rm LS}
\label{8.193}\eeq
for protons, antiprotons, electrons and positrons in solid aluminium and
gold given by the program {\sc sbethe} \citep{SalvatAndreo2023}. The
mean excitation energy $I$ and the corrections $\delta_{\rm F}$ and
$\Delta L^{\rm B}(a)$ are determined by the OOS of the material. In the
adopted approach, $I$ is considered as an empirical parameter (=166 eV
for aluminium and 790 eV for gold), which is used to build the DHFS OOS
model (see Section \ref{sec7.5.2}) of the material and, consequently,
determines its stopping properties.  Similar calculations for protons in
aluminium were presented by \citet{Ashley1991b}. The Barkas correction
calculated here from realistic OOSs is expected to be more reliable than
the values used in the \citet{ICRU49}, which were derived from a
formulation by \citet{Ashley1972, Ashley1973} using the OOS obtained
from the LPA (Section \ref{sec7.5.1}) with the atomic electron density
of the Lenz--Jensen model. The calculations by \citeauthor{Ashley1973}
were made consistent with measurement by fixing the value of the LPA
parameter $\tau$ so as to reproduce the correction derived from
experimental stopping power data. Similar calculations of the Barkas
correction, also using the Lenz--Jensen atomic electron density, were
presented by \citet{JacksonMcCarthy1972}.

\begin{figure}[hp!] \begin{center}
\includegraphics*[width=7.20cm]{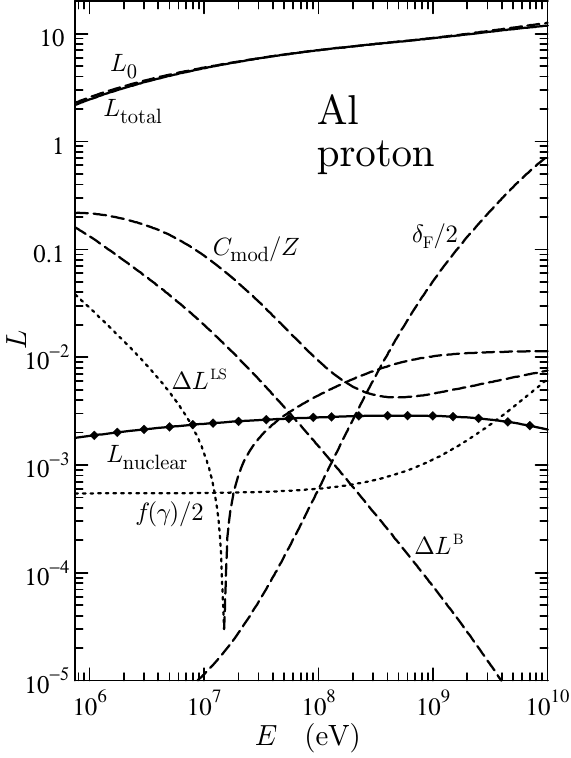} \rule{5mm}{0mm}
\includegraphics*[width=7.20cm]{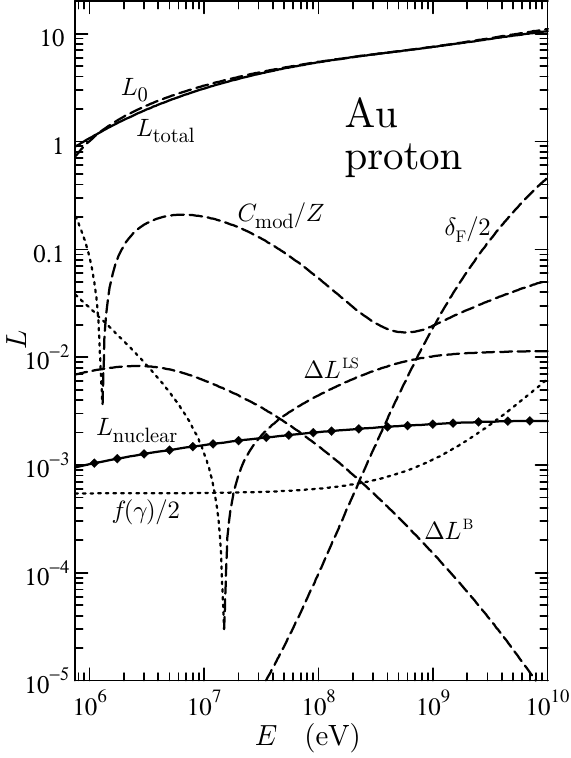} \\ [5mm]
\includegraphics*[width=7.20cm]{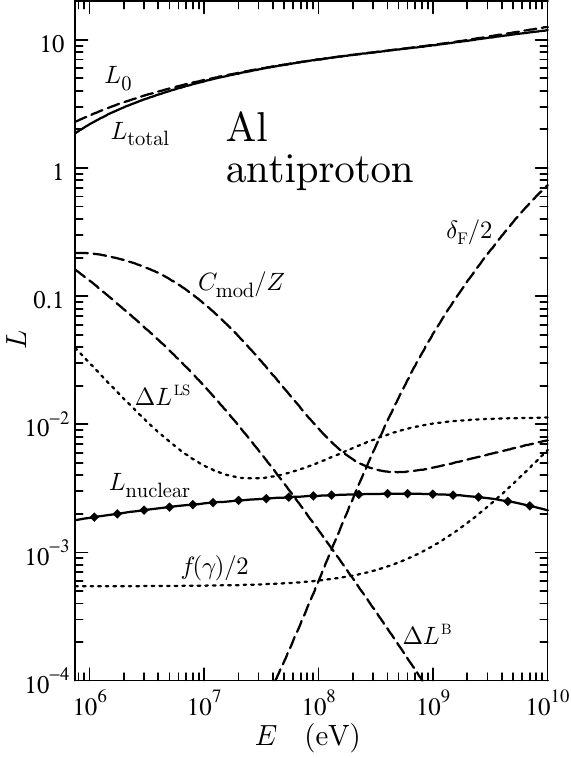} \rule{5mm}{0mm}
\includegraphics*[width=7.20cm]{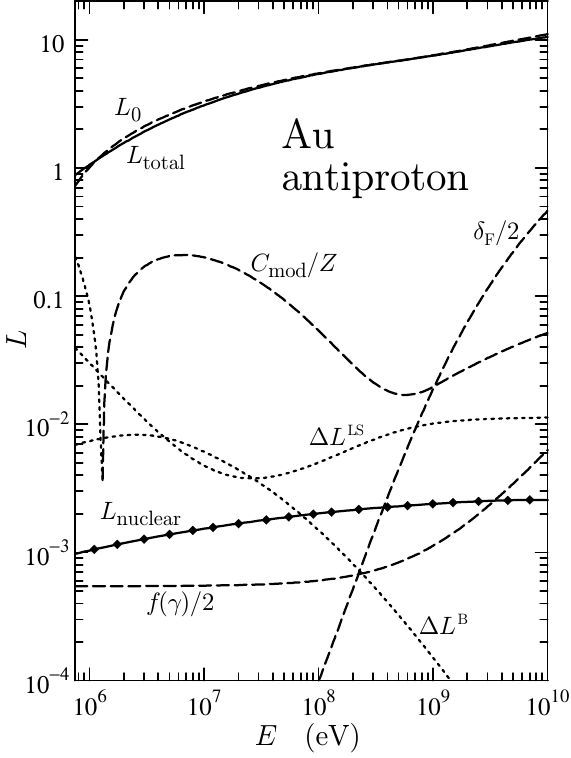}
\caption{Stopping logarithms for protons and antiprotons in aluminium
	and gold. The plotted quantities are: total logarithm, $L_{\rm
	total}$ (solid curves), Eq.\ \req{8.193}; Bethe logarithm, $L_0$, Eq.\ \req{8.192};
	$\1o2 f(\gamma)$ function, Eq.\ \req{8.190a}; modified shell
	correction, $C_{\rm mod}/Z$, Eq.\ \req{8.191}; density-effect
	correction, $\delta_{\rm F}/2$, Eq.\ \req{8.121}; Barkas correction,
	$\Delta L^{\rm B}$, Eq.\ \req{8.183}; Lindhard--S\o rensen correction,
	$\Delta L^{\rm LS}$, Eq.\ \req{8.155} and nuclear stopping logarithm,
	Eq.\ \req{8.229} (diamonds). The various contributions to the
stopping logarithm are shown as dashed curves (positive values) or as dotted
	curves (negative values).
\label{fig8.8}}
\end{center}\end{figure}

\begin{figure}[hp!] \begin{center}
\includegraphics*[width=7.20cm]{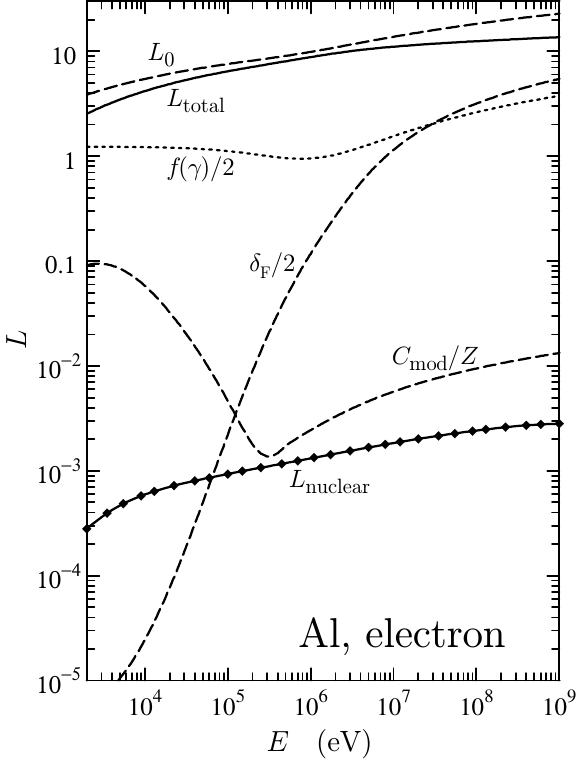} \rule{5mm}{0mm}
\includegraphics*[width=7.20cm]{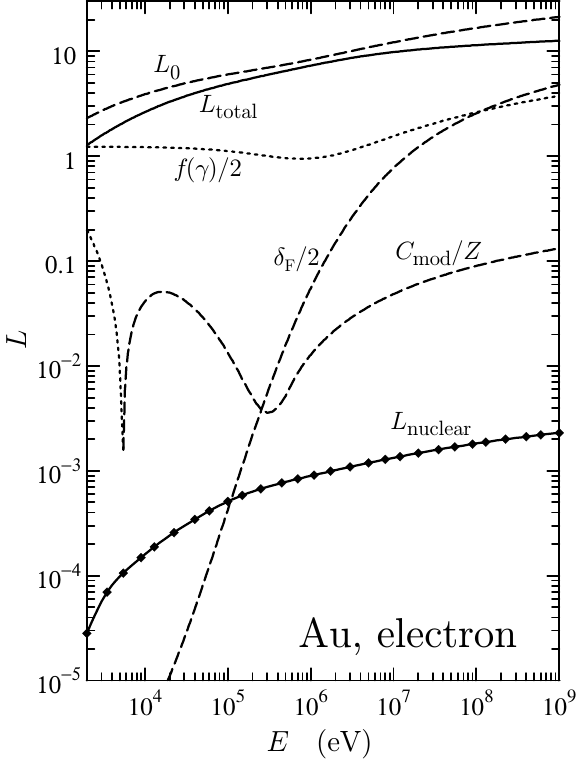} \\ [5mm]
\includegraphics*[width=7.20cm]{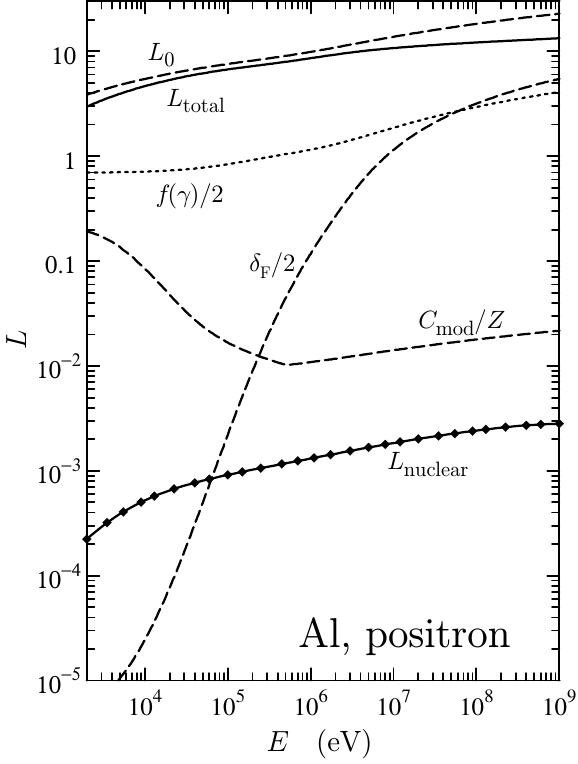} \rule{5mm}{0mm}
\includegraphics*[width=7.20cm]{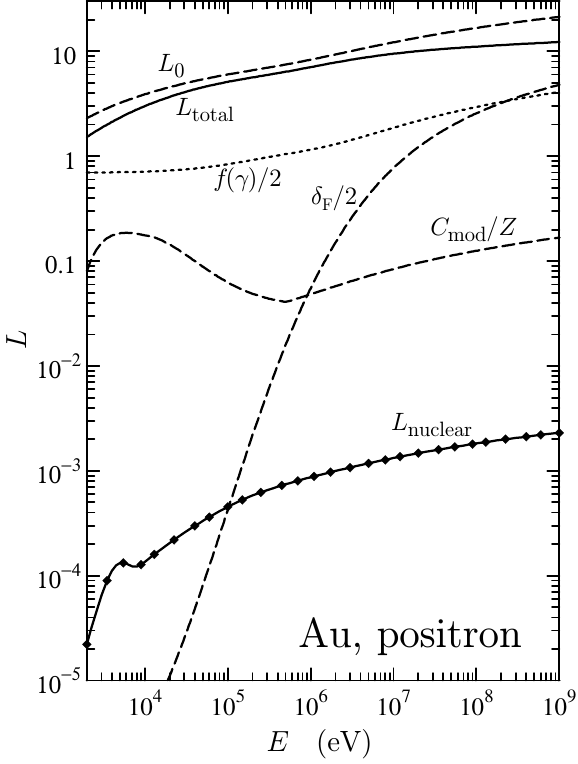}
\caption{Stopping logarithms for electrons and positrons in aluminium
	and gold. The plotted quantities are:  total logarithm, $L_{\rm
	total}$, Eq.\ \req{8.193} (solid curves); Bethe logarithm, $L_0$,
	Eq.\ \req{8.192};
	$\1o2 f(\gamma)$ function, Eq.\ \req{8.190b} or \req{8.190c}; modified shell
	correction, $C_{\rm mod}/Z$, Eq.\ \req{8.191}; density-effect
	correction, $\delta_{\rm F}/2$, Eq.\ \req{8.121};
	and nuclear stopping logarithm, Eq.\ \req{8.229} (diamonds).
	The various contributions to the stopping
	logarithm are shown as dashed curves (positive values) or as dotted
	curves (negative values).
\label{fig8.9}}
\end{center}\end{figure}

A recent detailed study of the reliability of the corrected Bethe
formula \citep{Salvat2022c} has shown that the formula \req{8.187}, with
the shell correction derived from the relativistic PWBA by
\cite{Salvat2022a}, the Lindhard--S\o rensen correction \req{8.155}, and
the density-effect and Barkas corrections calculated from the DHFS-model
OOS (Section \ref{sec7.5.2}) by using the empirical $I$ values from the
\citet{ICRU37}, yields stopping powers of elemental materials in close
agreement with experimental data from the IAEA database
(\citeauthor{Montanari2017}, \citeyear{Montanari2017};
\url{https://www-nds.iaea.org/stopping})
for protons and alpha particles with kinetic
energies higher than 0.75 MeV and 5 MeV, respectively. In the case of
electrons and positrons, the corrected Bethe formula (with the
density-effect correction, and with the adequate $f(\gamma)$ function
and shell correction) is expected to be reliable for projectiles with
kinetic energies higher than about 500 eV. The calculations by
\citet{Salvat2022c} indicate that optical-data models based on the
free-electron-gas theory, or using even simpler extension algorithms
(see Chapter \ref{chapt7}), are not capable of giving realistic estimates of
the shell correction, which is important for projectiles with
intermediate and low energies. Consequently, stopping powers derived
from optical-data models are not expected to be highly accurate.
Nonetheless, optical-data models are useful for practical calculations
and Monte Carlo simulations because they provide the most realistic DCSs
available for inelastic collisions of charged particles in condensed
matter. A convenient practice is to use stopping powers calculated from
the corrected Bethe formula, and to renormalize the DCSs obtained from
an optical-data model to reproduce the adopted stopping power.

\index{stopping power!corrected Bethe formula|)}


\subsection{Low-energy extrapolation \label{sec8.5.1}}

\index{corrected Bethe formula!low-energy extrapolation}
\index{low-energy extrapolation}

As alread mentioned, the results from the corrected Bethe formula,
calculated from the {\sc sbethe} approach, closely approximate the
measured stopping powers for protons and alpha particles with kinetic
energies higher than a value $E_{\rm cut}$ of the order of 0.75~MeV and
5~MeV respectively. The very limited experimental information available
on the stopping of low energy electrons, suggests that the formula is
also valid for electrons, and positrons, with kinetic energies higher
than $E_{\rm cut} \sim 2$~keV.  In order to permit the approximate
calculation of particle ranges of low-energy projectiles, it is
convenient to extrapolate the predictions of the Bethe formula to
energies lower than $E_{\rm cut}$. Our aim here is not to give reliable
stopping powers for low-energy projectiles, but simply to permit
estimating the order of magnitude of their ranges.

Inspection of available experimental data for protons and alphas
\citep{Salvat2022c} indicates that the stopping power of these particles
has a wide maximum at an energy $E_{\rm max}$ of about 50~keV for
protons and 0.8~MeV for alphas \citep{Salvat2022c}. The stopping power
for electrons has a similar energy dependence, with a maximum at $E_{\rm
max}\sim 100$~eV.

The program {\sc sbethe} calculates the electronic stopping power from
the corrected Bethe formula \req{8.187} for energies higher than $E_{\rm
cut}$, which the program tentatively sets equal to 2~keV for electrons and
positrons, 150~keV for muons and antimuons, 0.75~MeV for protons and
antiprotons, and 5~MeV for alpha particles. The calculated values are
extrapolated to lower energies by using the analytical form
\beq
S_{\rm in}(E) = \left\{
\begin{array}{ll}
	\displaystyle{	\exp \left[ \rule{0mm}{5mm}
	A - B \left( \ln t \right)^{1.5} \right]}
& \mbox{if $E_{\rm max} \le E \le E_{\rm cut}$,}
\\ [3mm]
\displaystyle{ 1.5 \, \exp(A) \, \sqrt{t}
	\left( 1 - \frac{t}{3} \right)} &
\mbox{if $E \le E_{\rm max}$,}
\end{array} \right.
\label{8.194}\eeq
with $t = E/E_{\rm max}$, and the parameters $A$ and $B$ determined by
requiring continuity of $S_{\rm in}(E)$ and its derivative at $E=E_{\rm
cut}$. The program may increase the value of the cutoff energy $E_{\rm
cut}$ so that $E_{\rm cut} > 1.05 E_{\rm max}$ and the parameter $B$ is
larger than 0.05. The energy dependence at low energies, $S_{\rm in}
\propto \sqrt{E}$, is in accordance with the theory of the free-electron
gas (which predicts that the stopping power of slow projectiles is
proportional to their velocity \citep{Lindhard1954}).

A great effort has been done over the years to measure the stopping
cross sections of elementary solids for protons and alphas. The
available experimental information is included in the exhaustive IAEA
online database\footnote{This database can be downloaded from the IAEA
web site, \url{https://www-nds.iaea.org/stopping}. The data
used in the present work were downloaded in March 2022.} on ``Electronic
Stopping Power of Matter for Ions'' \citep{Montanari2017}.
Following \citet{Salvat2022c}, to improve the description of the
stopping powers of elemental solids for protons and alphas with low
energies ($E<E_{\rm cut}$), and also to validate the corrected Bethe
formula at intermediate energies, we parameterize the measured stopping
power for low-energy protons and alphas by using the empirical formula
adopted in the \citet{ICRU49}, and attributed there to
Andersen and Ziegler,
\beq
S_{\rm low}(E) = \frac{s_1(T) \, s_2(T)}{s_1(T) +s_2(T)}
\label{8.195}\eeq
with
\begin{subequations} \label{8.196}
\beq
s_1 (T) = a_1 T^{a_2}
\label{8.196a}\eeq
and
\beq
s_2 (T) = \frac{a_3}{T} \, \ln \left( 1 + \frac{a_4}{T} + a_5 T
\right)
\label{8.196b}\eeq
\end{subequations}
where $a_1$, \ldots, $a_5$ are adjustable parameters. The variable $T$
is the kinetic energy $E$ of the projectile in units of keV for protons
and in units of MeV for alphas. The parameters $a_i$ were determined
from a least-squares fit of the function \req{8.195} to the stopping
powers $S_{\rm exp}(E_k)$ given in the IAEA database with $E_k \in (10$
keV, $2 E_{\rm cut}$). They were obtained, for each material (for which
enough measured data were available) and projectile kind (protons and
alphas), by minimizing the function
\beq
\chi_2^2(a_1, \ldots, a_5) = \sum_k \left[ S_{\rm low}(E_k) - S_{\rm
exp}(E_k) \right]^2
\label{8.197}\eeq
where the summation is over the data with energies $E_k$ in the interval
($0, 2E_{\rm cut}$). The minimization was performed by using the simplex
method of \citet{NelderMead1965}.

Generally, the fitted formula \req{8.195} and the corrected Bethe formula
\req{8.187} (with the ICRU $I$ value) do not
yield the same values at $E_{\rm cut}$. To obtain a continuous function
of the projectile's kinetic energy $E$, with a gradual transition from
the low-$E$ fit to the high-energy Bethe formula \req{8.187}, we set
\beq
S_{\rm in}(E) = \left\{
\begin{array}{ll}
S_{\rm low}(E) & \mbox{if $E \le E_{\rm cut}$,} \\ [2mm]
S_{\rm mix}(E) \rule{5mm}{0mm}
	& \mbox{if $E_{\rm cut} < E \le 2 E_{\rm cut}$,} \\ [2mm]
S_{\rm cB}(E) & \mbox{if $E> 2 E_{\rm cut}$,}
\end{array} \right.
\label{8.198}\eeq
with
\beq
S_{\rm mix}(E) =
\left[ 1-\left(\frac{E}{E_{\rm cut}}-1\right)
\left( 1- \frac{S_{\rm cB}(2E_{\rm cut})}{S_{\rm low}(2E_{\rm cut})}
\right) \right]
S_{\rm low}(E)
\label{8.199}\eeq
where $S_{\rm cB}(E)$ is the value obtained from the corrected Bethe
formula \req{8.187}.

\begin{figure}[h!] \begin{center}
\includegraphics*[width=7.8cm]{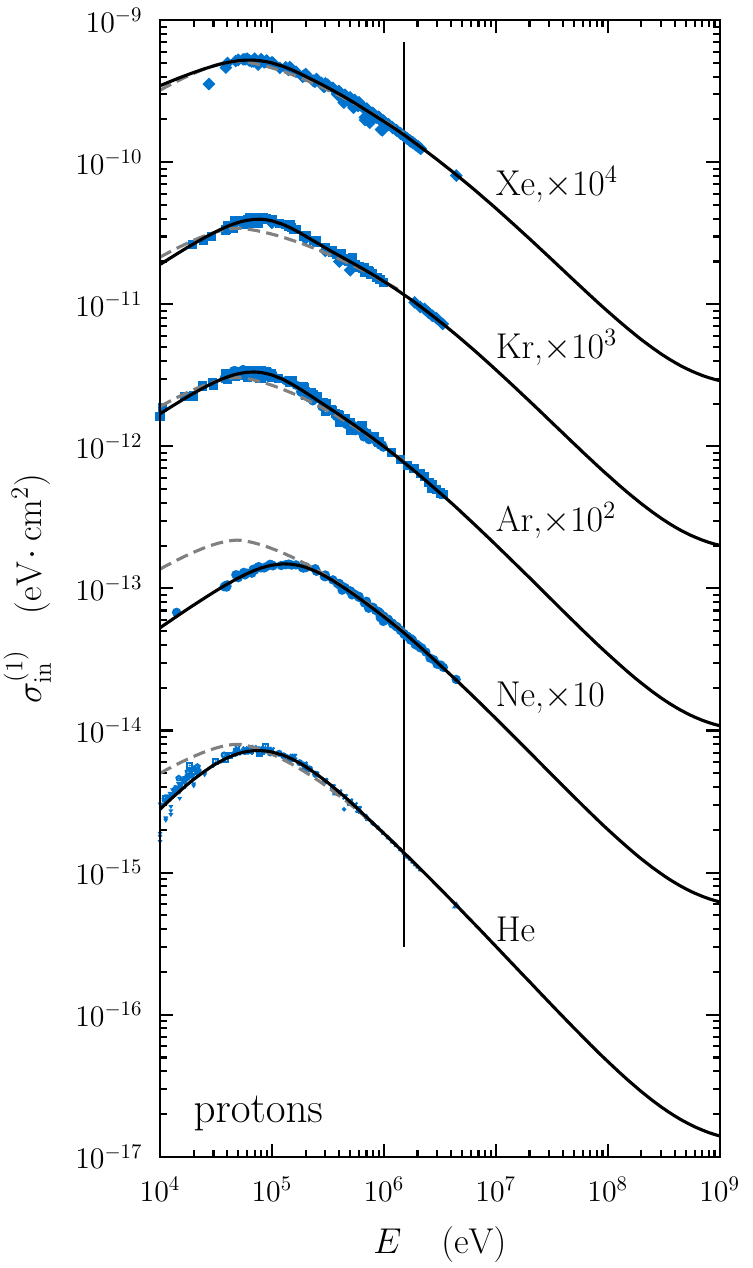} \rule{0mm}{0mm}
\includegraphics*[width=7.8cm]{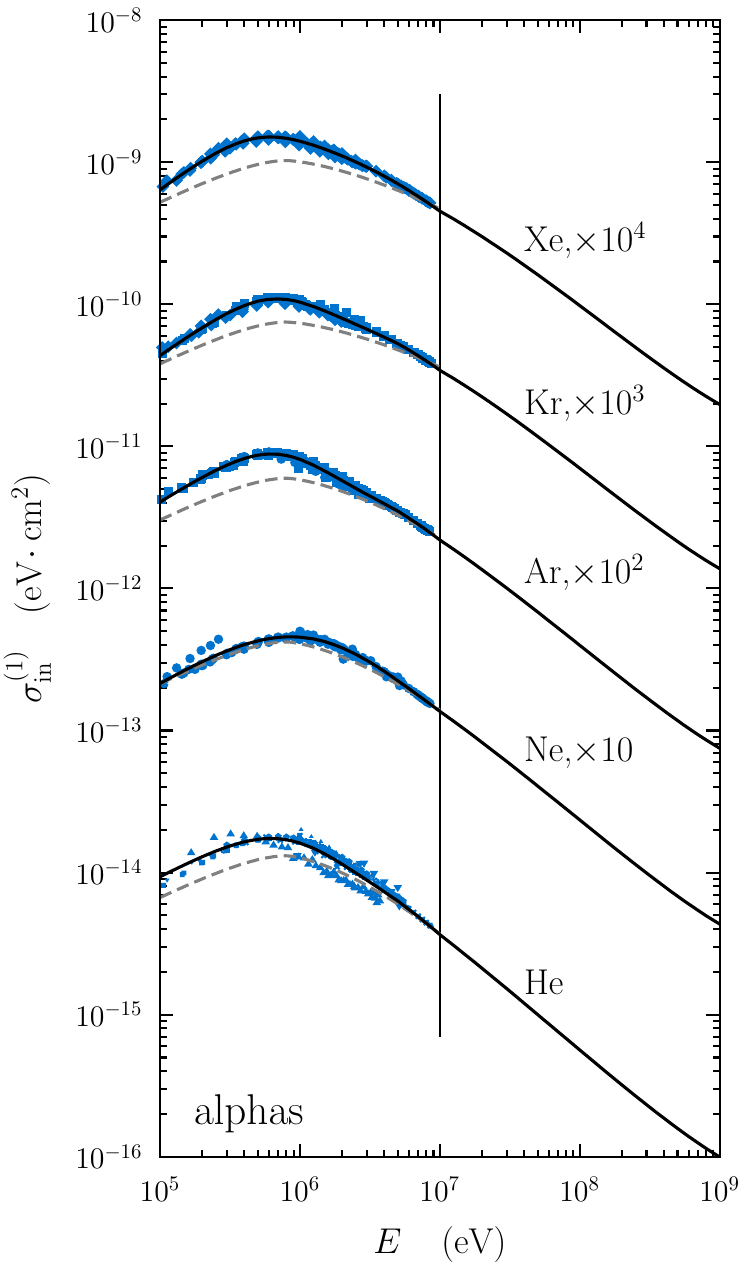}
\caption{
Electronic stopping cross sections of noble gases for protons (left) and
	alpha particles (right) as functions of the kinetic energy of the
	projectile, multiplied by the indicated powers of 10 to improve
	visibility. Solid black curves are obtained from the fitted formula
	\req{8.195}; dashed gray curves were calculated from the extrapolation
	formula \req{8.194}. Symbols represent experimental data from the IAEA
	database. Other details are explained in the text.
\label{fig8_10}}\end{center} \end{figure}

Electronic stopping cross sections of noble gases for protons and alpha
particles calculated by the {\sc sbethe} program are compared with
results from measurements in Fig.\ \ref{fig8_10}. The experimental data
were taken from the IAEA online database on ``Electronic Stopping Power
of Matter for Ions'' \citep{Montanari2017}. The vertical lines are at
the energy $2E_{\rm cut}$ above which the corrected Bethe formula is
applied. Below this energy, the plotted values were generated from the
extrapolation formula \req{8.192} (gray dashed curves) and from the fitted
formula \req{8.195} (solid curves). In spite of the simplicity of the
extrapolation formula, its results follow the global trends of the
experimental data.

The electronic stopping cross section of water molecules (liquid and
ice) for protons and alpha particles are shown in Fig. \ref{fig8_11},
which compares results from calculations according to Eq.\ \req{8.195}
with experimental values from the IAEA database. The vertical lines
indicate the cutoff energy $E_{\rm cut}$.  The transition from $S_{\rm
low}(E)$ to $S_{\rm cB}(E)$, is described by the interpolation given by
Eq.\ \req{8.197} the joining between the low-energy formula \req{8.195}
and the high-energy corrected Bethe formula \req{8.187} is almost
imperceptible.

\begin{figure}[h!] \begin{center}
\includegraphics*[width=8.5cm]{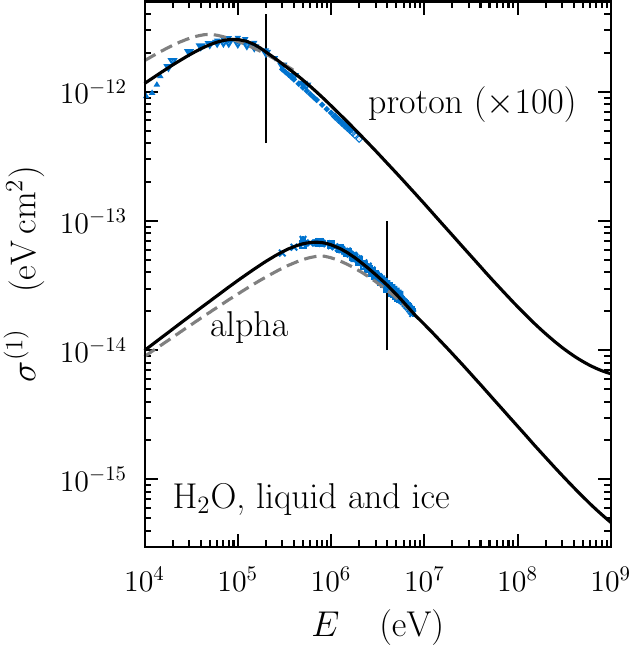}
\caption{
Electronic stopping cross sections of water molecules for protons
	($\times 100$) and alpha particles as functions of the kinetic energy
	of the projectile. The vertical lines indicate the $E_{\rm cut}$
	values adopted (=200 keV for protons and 4 MeV for alphas). Solid
	black curves are obtained from the fitted formula \req{8.195} for
	$E<E_{\rm cut}$ and from the interpolation scheme \req{8.198}; dashed
	gray curves were calculated from the extrapolation formula \req{8.194}.
	Symbols represent experimental data from the IAEA database.
\label{fig8_11}}\end{center} \end{figure}

Figure \ref{fig8_12} compares stopping powers of metallic aluminium,
silicon, copper, and gold for projectile electrons calculated by the
{\sc sbethe} program with experimental data from
\citet{Garber1971}, \citet{AlAhmad1983}, \citet{Luo1991},
\citet{Joy2008}, and \citet{MacPherson1998}. The
dashed portion of the curves are results from the extrapolation
\req{8.194}, which yields realistic values of the electronic stopping power
for electron energies down to about 100~eV.

\begin{figure}[ph!] \begin{center}
\includegraphics*[width=7.8cm]{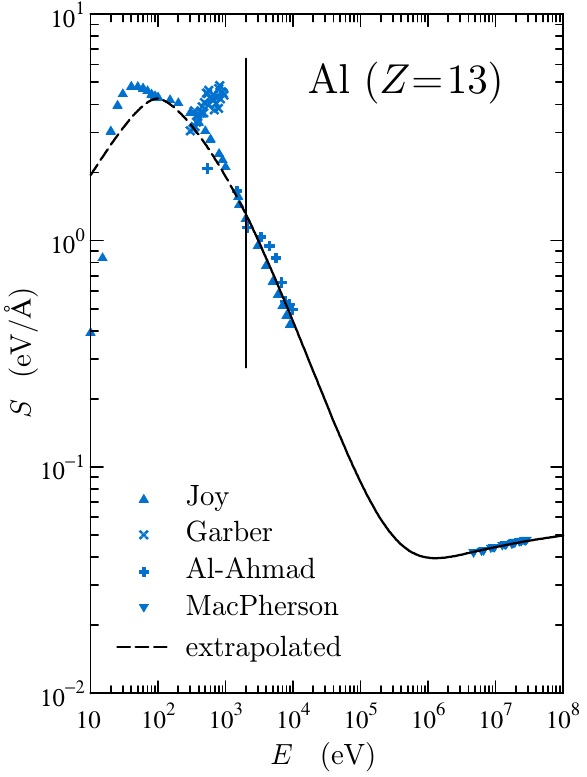} \rule{0mm}{0mm}
\includegraphics*[width=7.8cm]{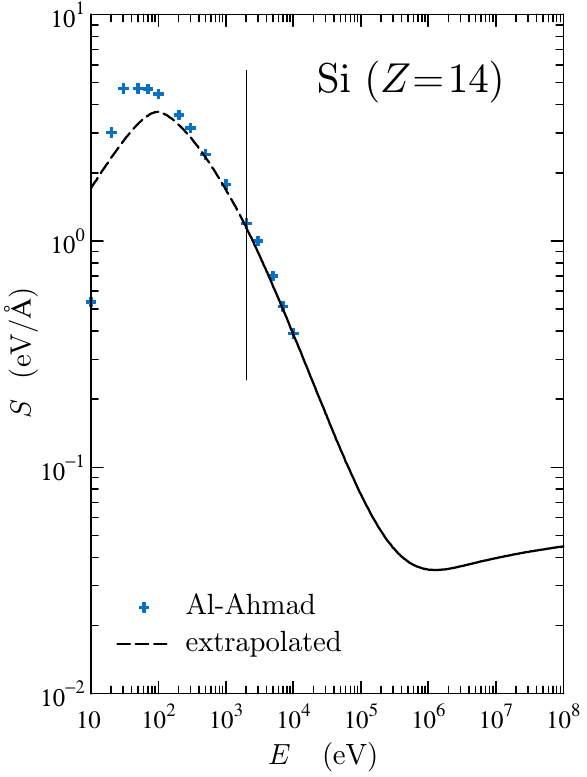}
\includegraphics*[width=7.8cm]{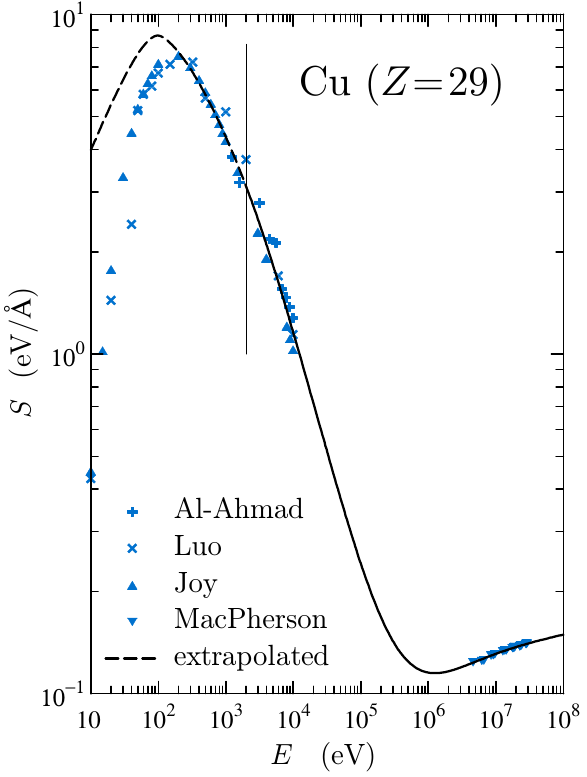} \rule{0mm}{0mm}
\includegraphics*[width=7.8cm]{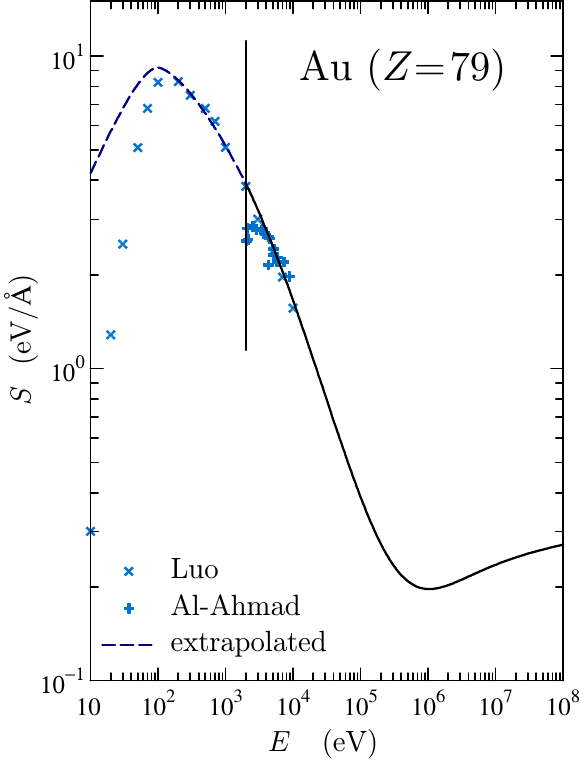}
\caption{
Electronic stopping powers of solid aluminium, silicon, copper, and gold
for electrons, as functions of the kinetic energy $E$.  The curves are
results from {\sc sbethe}. Symbols represent experimental data from the
indicated references.
\label{fig8_12}}\end{center} \end{figure}


\subsection{Corrected Bethe formula for compound materials
\label{sec8.5.2}}

The corrected Bethe formula \req{8.187} is applicable to arbitrary
materials, including compounds and mixtures of various elements. Let us
consider a compound whose molecules consist of $n_j$ atoms of the
element of atomic number $Z_j$ ($j=1,2, \ldots$). The mean excitation
energy of the compound can be estimated by using the additivity
approximation, \ie, by assuming that the molecular cross section can be
approximated as the sum of atomic cross sections of the atoms in a
molecule. The OOS of a molecule is then the sum of the OOSs of its atoms
and, consequently, the $I$ value of the compound is given by
\beq
Z \ln I =
{\sum}_j n_j Z_j \ln(I_j) \; \; \; \mbox{with} \; \; \;
Z = {\sum}_j n_j Z_j \, ,
\label{8.200}\eeq
where $I_j$ denotes the mean excitation energy of the element with
atomic number $Z_j$. Since the additivity approximation neglects the
effect of aggregation on the atomic OOSs, the $I$ value resulting from
Eq.\ \req{8.200} may differ appreciably from the ``true'' mean excitation
energy of the material. A better estimate of the $I$ value can only be
obtained either from stopping measurements or from knowledge of the
OOS of the material.

As discussed by \citet{Salvat2022c}, the atomic shell correction
obtained from the PWBA is valid for arbitrary materials because the main
contribution to that correction arises from inner electron subshells,
which are only slightly affected by aggregation effects. The modified
shell correction $C_{\rm mod}(\gamma)/Z$ of the compound, obtained from
the additivity approximation, is given by
\beq
\frac{C_{\rm mod}(\gamma)}{Z}  = \frac{1}{Z} {\sum}_j n_j \, Z_j \,
\frac{C_{{\rm mod},j}(\gamma)}{Z_j},
\label{8.201}\eeq
where the quantity $C_{{\rm mod},j}(\gamma)/Z_j$ is the modified
shell correction for the
elemental material of atomic number $Z_j$. The cutoff impact parameter
$a$, which determines the Barkas correction, may be estimated from Eq.\
\req{8.186} with
\beq
C_{\rm B} = \max \{ 1, \overline Z/10 \}
\; \; \; \mbox{where} \; \; \; \overline{Z} = Z \left( {\sum}_j n_j
\right)^{-1} = \left( {\sum}_j n_j Z_j \right) \left( {\sum}_j n_j
\right)^{-1}.
\label{8.202}\eeq

As indicated above, the electronic stopping power obtained from the
corrected Bethe formula is completely determined by the adopted value of
the mean excitation energy $I$. By default, the program {\sc sbethe}
sets $I=I_{\rm ICRU}$, the $I$ value recommended in the \citet{ICRU37},
which yields results in close agreement with available experimental
stopping powers \citep{Salvat2022c}. In order to allow analyzing the
dependence of the calculated stopping power on this parameter, the user
is allowed to modify its default value.

Notice that Eq.\ \req{8.201} was incorrect in the original article of
\citet{SalvatAndreo2023} and in previous versions of the software: the
factor $n_j$ was missing in the Eq.\ while in the program we missed the
ratio $Z_j/Z$. This fact made the results of our calculations for
compounds wrong. This annoying feature was discovered after an
inconsistency of some results for compounds (for various compound
materials the results from the code where plagued with ``not-a-number''
messages), which was kindly pointed out by Dr Huang Yu.

\section{Radiative stopping power for electrons and po\-si\-trons
\label{sec8.6}}
\index{radiative stopping power for electrons and positrons}

The corrected Bethe formula describes the so-called {\it electronic stopping
power} of a material for high-energy charged particles, that is, the
energy loses caused by interactions that produce electronic excitations
of the material. In the case of protons and heavier particles, the
slowing down is almost completely due to these interactions; elastic
collisions with nuclei have an appreciable stopping effect only for
projectiles with relatively small velocities (see Section \ref{sec8.8}).

Electrons and positrons, because of their small mass, experience large
accelerations when they penetrate the electrostatic field of an atom, or
of an electron, and, as a result, they emit bremsstrahlung (braking
radiation). A thorough review of the theory and experimental
measurements of bremsstrahlung emission is given in the monograph by
\citet{HaugNakel2004}. The process is responsible for the so-called {\it
radiative stopping power}, which dominates the stopping power for
high-energy electrons and positrons. We give here only a brief
description of the calculation of the radiative stopping power because
the process is beyond the framework of collision theory.

In each bremsstrahlung event, an electron with kinetic energy $E$
emits a photon of energy $W$, which may take values in the interval
from 0 to $E$. As a good approximation, we can ignore the recoil of the
target particle (nucleus or electron) and consider that the energy of
the projectile after the collision is $E-W$. Notice that an electron can
be brought to rest by emission of a single bremsstrahlung photon. The
relevant information on the radiative process is provided by the atomic
energy-loss DCS, differential in only the energy $W$ of the emitted
photon \citep[see][and references therein]{ICRU37}.  Theoretical
considerations \citep{BetheHeitler1934, Tsai1974} show that the
DCS for bremsstrahlung emission in the field of an atom of atomic number
$Z$ can be expressed in the form
\beq
\frac{\d\sigma_{\rm rad}}{\d W} =
\frac{Z^2}{\beta^2} \, \frac{1}{W} \, \chi(Z,E;\kappa),
\label{8.203}\eeq
where $\kappa$ is the reduced photon energy,
\beq
\kappa \equiv W/E,
\label{8.204}\eeq
which takes values between 0 and 1. The quantity
\beq
\chi(Z,E; \kappa) \equiv (\beta^2/Z^2) W
\frac{\d\sigma_{\rm rad}}{\d W}
\label{8.205}\eeq
is known as the ``scaled'' bremsstrahlung DCS; for atoms of a given element $Z$,
it varies smoothly with $E$ and $\kappa$. \citet{SeltzerBerger1985,
SeltzerBerger1986} produced extensive tables of the scaled DCS for all
the elements ($Z=$1--99) and for electron energies from 1 keV to 10 GeV.
They tabulated the scaled DCSs for emission in the (screened) field of
the nucleus (electron-nucleus bremsstrahlung) and in the field of atomic
electrons (electron-electron bremsstrahlung) separately, as well as
their sum, the total scaled DCS. The electron-nucleus bremsstrahlung DCS
was calculated by combining analytical high-energy theories with results
from partial-wave calculations by \citet{Pratt1977}
for bremsstrahlung emission in screened atomic fields and energies below
2 MeV. The scaled DCS for electron-electron bremsstrahlung was obtained
from the theory of \citet{Haug1975} combined with a screening correction
that involves Hartree-Fock incoherent scattering functions. Seltzer and
Berger's scaled DCS tables constitute the most reliable theoretical
representation of bremsstrahlung energy spectra available at present.

The total atomic cross section for bremsstrahlung emission is infinite
due to the divergence of the DCS \req{8.197} at $W=0$ (the so-called
infrared divergence), which is associated with the null mass of the
photon. Nevertheless, the radiative stopping cross section,
\beq
\sigma_{\rm rad}^{(1)} (E) \equiv \int_{0}^{E}
W \frac{\d\sigma_{\rm rad}}{\d W} \, \d W
= \frac{Z^2}{\beta^2} E \, \int_0^1 \chi(Z,E;\kappa) \, \d
\kappa\, ,
\label{8.206}\eeq
is finite. The radiative stopping power (\ie, the average energy
radiated per unit path length) is
\beq
S_{\rm rad}(E) = {\cal N} \sigma_{\rm rad}^{(1)}(E),
\label{8.207}\eeq
where ${\cal N}$ is the number of atoms per unit volume.
The tables of Seltzer and Berger include the quantity
\beq
\phi_{\rm rad} (Z, E) \equiv \frac{1}{Z^2 \alpha r_{\rm e}^2 (E+\me c^2)}
\int_0^E W \frac{\d\sigma_{\rm rad}}{\d W} \, \d W,
\label{8.208}\eeq
where $\alpha$ is the fine-structure constant and $r_{\rm e}=\alpha^2
a_0$ is the classical electron radius. The stopping power of electrons
with kinetic energy $E$ can then be calculated easily by interpolation
of the Seltzer and Berger tables.

In the case of compounds (or mixtures), the molecular DCS is obtained
from the additivity approximation, \ie, as the sum of the DCSs of all
the atoms in a molecule. Consider a compound whose molecules consist of
$n_i$ atoms of the element $Z_i$. The molecular DCS is
\beq
\frac{\d\sigma_{\rm rad,mol}}{\d W} = \frac{1}{\beta^2 W}
\sum_i n_i Z_i^2 \, \chi (Z_i,E; \kappa).
\label{8.209}\eeq
The radiative stopping power of the compound is
\beq
S_{\rm rad}(E) = {\cal N} \, \alpha r_{\rm e}^2 (E+\me c^2)
\sum_i n_i Z_i^2 \, \phi_{\rm rad} (Z_i, E),
\label{8.210}\eeq
where ${\cal N}$ is the number of molecules per unit volume.

The radiative DCS and the stopping power for positrons are generally
smaller than those for electrons because positrons are repelled by the
nucleus and, therefore, experience less acceleration than electrons with
the same energy. Owing to the lack of more detailed calculations, the
atomic DCS for positrons is obtained by multiplying the electron DCS by
a $\kappa$-independent factor, \ie,
\beq
\frac{\d\sigma_{\rm rad}^{(+)}}{\d W} =
F_{\rm p}(Z,E) \, \frac{\d\sigma_{\rm rad}^{(-)}}{\d W}.
\label{8.211}\eeq
The factor $F_{\rm p}(Z,E)$ is set equal to the ratio of the radiative
stopping powers for positrons and electrons, which has been calculated
by \citet{Kim1986} \citep[cf.][]{BergerSeltzer1982}. In the
calculations we use the following analytical approximation
\beqa
F_{\rm p}(Z,E) &=& 1 - \exp (
- 1.2359\times 10^{-1}\, t
+ 6.1274\times 10^{-2}\, t^{2}
- 3.1516\times 10^{-2}\, t^{3}
\nonumber \\ [2mm]
&& \mbox{}
+ 7.7446\times 10^{-3}\, t^{4}
- 1.0595\times 10^{-3}\, t^{5}
+ 7.0568\times 10^{-5}\, t^{6}
\nonumber \\ [2mm]
&& \mbox{}
- 1.8080\times 10^{-6}\, t^{7}),
\label{8.212}\eeqa
where
\beq
t = \ln \left( 1 + \frac{10^{6}}{Z^{2}} \frac{E}{\me c^{2}} \right).
\label{8.213}\eeq
Expression \req{8.212} reproduces the values of $F_{\rm p}(Z,E)$
tabulated by \citet{Kim1986} to an accuracy of about 0.5\%.
Correspondingly, the radiative stopping power for positrons is
calculated as
\beq
S_{\rm rad}^{(+)}(E) = {\cal N} \, \alpha r_{\rm e}^2 (E+\me c^2)
\sum_i n_i Z_i^2 \, F_{\rm p}(Z_i,E) \, \phi_{\rm rad} (Z_i, E).
\label{8.214}\eeq

Figure \ref{fig8.13} compares the radiative stopping powers of electrons
and positrons in aluminium and gold, with the corresponding
electronic and nuclear stopping powers. The nuclear stopping power, \ie,
the effect of the energy transfers in elastic collisions (see
the following Section), is seen to be negligible for these particles.
Stopping is dominated by electronic excitations at low energies, while
Bremsstrahlung emission prevails at high energies. Electronic and
radiative processes contribute equal amounts to the stopping power for
projectiles with a critical kinetic energy, which is about 50 MeV for
aluminium and near 10 MeV for gold. The critical energy decreases when
the atomic number of the material increases.

\begin{figure}[hp!] \begin{center}
\includegraphics*[width=7.20cm]{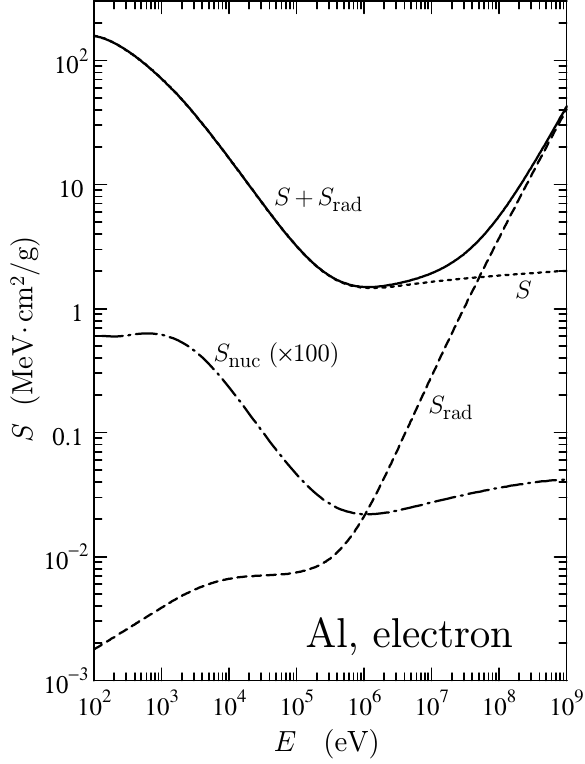} \rule{5mm}{0mm}
\includegraphics*[width=7.20cm]{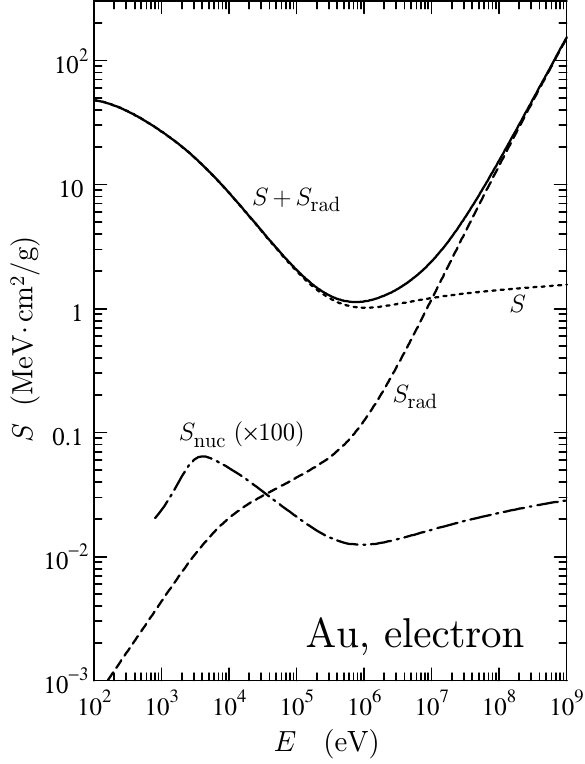} \\ [5mm]
\includegraphics*[width=7.20cm]{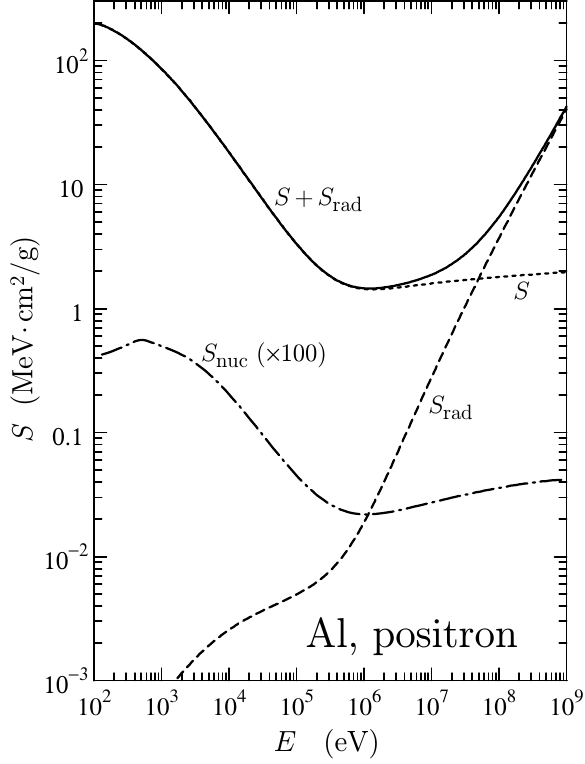} \rule{5mm}{0mm}
\includegraphics*[width=7.20cm]{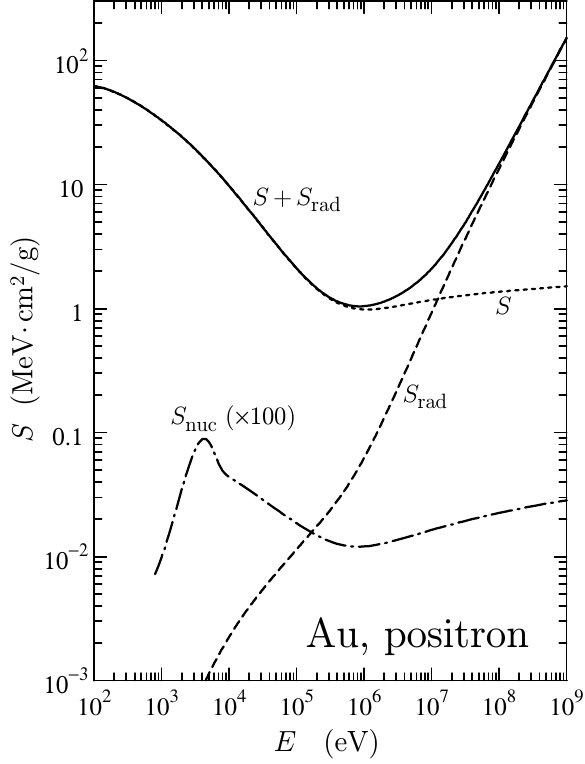}
\caption{Stopping powers for electrons and positrons in aluminium and
gold. The plotted stopping powers are: electronic, Eq.\ \req{8.187}
(dotted curves); radiative, Eq.\ \req{8.210} or \req{8.214} (dashed);
nuclear, Eq.\ \req{8.228}, multiplied by 100 (dot-dashed); and
electronic plus radiative (solid).
\label{fig8.13}}
\end{center}\end{figure}


\section{Radiative stopping power for muons \label{sec8.7}}

As in the case of electrons, the stopping power for high-energy muons is
affected by radiative contributions. In this Section we use the
numerical results given by \citet{Groom2001}. The radiative mass
stopping power (\ie, the radiative stopping power divided by the mass
density of the material) of muons can be expressed as
\beq
\frac{S_{\rm rad} (E)}{\rho} = E_{\rm T}
\left[ b_{\rm br}(E_{\rm T})+b_{\rm pair}(E_{\rm T})
+ b_{\rm photo}(E_{\rm T}) \right],
\label{8.215}\eeq
where $E_{\rm T}= E + m_\mu c^2$ is the total energy of the projectile
($m_\mu c^2 = 105.6583755$ MeV is the muon's rest energy), $b_{\rm
br}(E_{\rm T})$, $b_{\rm pair}(E_{\rm T})$, and $b_{\rm photo}(E_{\rm T})$ are
contributions from the processes of bremsstrahlung emission,
electron-positron pair production, and photonuclear interactions,
respectively. These quantities vary very slowly with energy, tending to
constant values at high energies.

The article of \citet{Groom2001} contains tables of the quantities $b_{\rm
br}(E_{\rm T})$, $b_{\rm pair}(E_{\rm T})$, and $b_{\rm photo}(E_{\rm
T})$, and their sum
\beq
b_{\rm total} (E_{\rm T}) = b_{\rm br}(E_{\rm T}) + b_{\rm pair}(E_{\rm T})
+ b_{\rm photo}(E_{\rm T})
\label{8.216}\eeq
for a grid of 16 total energies, $E_{{\rm T},j}$, the same for all
elements, which covers the interval from 500 MeV (where the radiative
stopping power effectively vanishes) up to 100 TeV, and for a set of 62
elementary materials with atomic numbers $Z$. We have
generated tables for the whole periodic system by means of natural cubic
spline interpolation in $Z$ (for fixed $E_{\rm T}$). The database of
{\sc sbethe} contains the original files of \citet{Groom2001} for the
elements included in that publication and the files generated by
spline interpolation of $b_{\rm br}$, $b_{\rm pair}$, and $b_{\rm
photo}$, with $b_{\rm total}$ obtained from Eq.\ \req{8.216}, for the rest
of elements. This set of files covers all the atomic numbers from hydrogen
($Z=1$) to einsteinium ($Z=99$).

The {\sc sbethe} program uses only the total coefficient $b_{\rm
tot}(E_{{\rm T},j},Z)$ for elementary materials; the quantity
\beq
d(E_{\rm T},Z) = \frac{A_{\rm m}(Z)}{N_{\rm A}}\, E_{\rm T} \, b_{\rm tot}(E_{\rm T},Z),
\label{8.217}\eeq
where $A_{\rm m}(Z)$ is the molar mass of the element with atomic
number $Z$ and $N_{\rm A} = 6.02214076$ $\times 10^{23}$ mol$^{-1}$ is
Avogadro's constant, can be regarded as an atomic cross section. In the
case of a compound consisting of $n_i$ atoms of the element $Z_i$ ($i=1,
\ldots, N$), the corresponding molecular cross section is obtained as
the sum of cross sections of the atoms in a molecule,
\beq
d_{\rm compound}(E_{\rm T}) = \sum_{i=1}^N n_i d(E_{\rm T},Z_i),
\label{8.218}\eeq
and the radiative stopping power at a certain kinetic energy $E$ is
calculated as
\beq
S_{\rm rad} (E) = {\cal N} d_{\rm compound}(E_{\rm T})
\label{8.219}\eeq
by means of natural cubic spline interpolation in energy of a precalculated
table of the molecular cross section $d_{\rm compound}(E_{\rm T})$ at
the 16 total energies of the tables' grid.

For the sake of consistency with the calculation by Groom \etal, aside
from the radiative stopping power described above, we include a
correction term to the electronic stopping power expressed as [see Eq.\
(13) in \citet{Groom2001}]
\beq
\Delta S_{\rm in} = {\cal N} r_{\rm e}^2 \, \me c^2 \, \alpha Z \left[
\ln \left( \frac{2 E_{\rm T}}{m_\mu c^2} \right)
- \frac{1}{3} \, \ln \left( \frac{2 W_{\rm max}}{\me c^2} \right)
\right] \ln^2 \left( \frac{2 W_{\rm max}}{\me c^2} \right),
\label{8.220}\eeq
where $W_{\rm max}$ is given by Eq.\ \req{8.8}, and $Z$ is
the total number of electrons in a molecule, Eq.\ \req{8.200}.


\section{Nuclear stopping \label{sec8.8}}
\index{nuclear stopping}

Because nuclei have masses much larger than the electron mass, the
average energy transfer in collisions of charged projectiles with nuclei
is much smaller than in collisions with electrons. Nonetheless,
elastic collisions with atoms (\ie, nuclei screened by the atomic
electrons) involve finite energy transfers and contribute to the
stopping power. The effect is known as {\it nuclear stopping}.

Let us consider the collisions of a charged projectile (mass $M_1$,
charge $Z_1 e$) moving with kinetic energy $E$ with an atom of atomic
number $Z$ initially at rest. For simplicity, the mass of the atom may
be estimated as $M_2 = A_{\rm m}/N_{\rm A}$, where $A_{\rm m}$ is the
atomic molar mass (gr/mol) and $N_{\rm A}$ is the Avogadro number. From the
kinematics of elastic collisions (Section \ref{sec4.3.3}), the energy
transfer $W$ is conveniently expressed in terms of the scattering angle
$\theta$ in CM [Eqs.\ \req{4.190} and \req{4.189}]
\beq
W = W_{\rm max} \,
\frac{1 - \cos\theta}{2} = W_{\rm max} \, \sin^2(\theta/2)\, ,
\label{8.221}\eeq
where
\beq
W_{\rm max} = \frac{2M_2c^2 E (E+ 2 M_1 c^2)}{(M_1c^2 + M_2 c^2)^2 + 2
M_2c^2 E}
\label{8.222}\eeq
is the maximum energy transfer, which occurs in head-on collisions.

The stopping effect of elastic collisions is described by the
energy-loss DCS given by [Eq.\ \req{4.203}]
\beq
\frac{\d \sigma}{\d W} = \frac{4\pi}{W_{\rm max}} \, \frac{\d \sigma}{\d
\Omega},
\label{8.223}\eeq
where the last factor is the angular DCS in CM at a direction with polar
scattering angle
\beq
\theta= \cos^{-1} \left(1 - 2 \, \frac{W}{W_{\rm max}} \right).
\label{8.224}\eeq
The nuclear stopping cross section is
\beqa
\sigma_{\rm nuc}^{(1)} &=& \int_0^{W_{\rm max}} W \,
\frac{\d \sigma}{\d W} \, \d W
\nonumber \\ [2mm]
&=& \frac{W_{\rm max}}{2} \, 2\pi \int_{-1}^{1} (1-\cos\theta)
\, \frac{\d \sigma}{\d \Omega} \, \d (\cos\theta)
\nonumber \\ [2mm]
&=& \frac{W_{\rm max}}{2} \, \sigma_{\rm tr}\, ,
\label{8.225}\eeqa
where
\beq
\sigma_{\rm tr} = \int (1-\cos\theta)
\, \frac{\d \sigma}{\d \Omega} \, \d \Omega
\label{8.226}\eeq
is the transport cross section in the CM frame. That is, the stopping cross
section is proportional to the transport cross
section, and their ratio equals half the maximum energy transfer in the
L frame.

In the case of compounds (or mixtures), the molecular cross section for
nuclear stopping can be evaluated by means of the additivity
approximation, that is, as the sum of the cross sections of all the
atoms in a molecule. Consider a compound whose molecules consist of
$n_i$ atoms of the element $Z_i$. The molecular stopping cross section
is
\beq
\sigma_{\rm nuc}^{(1)} = \sum_i n_i \,
\sigma_{\rm nuc}^{(1)}(Z_i)\, ,
\label{8.227}\eeq
where $\sigma_{\rm nuc}^{(1)}(Z_i)$ is the stopping cross section of
atoms of the element with atomic number $Z_i$, Eq.\ \req{8.213}.

The nuclear stopping power of the material is
\beq
S_{\rm nuc} = {\cal N} \sigma_{\rm nuc}^{(1)} \, .
\label{8.228}\eeq
Generally $S_{\rm nuc}$ is much smaller than the electronic stopping
power, except for heavy projectiles with very low energies and
materials with light elements.

To give a feel of the relative importance of nuclear stopping, Figs.\
\ref{fig8.8} and \ref{fig8.9} show the ``nuclear stopping logarithm''
$L_{\rm nuc}$ defined as
\beq
L_{\rm nuc} = \left(
\frac{4 \pi Z_1^2 e^4}{\me v^2} \, {\cal N} Z \right)^{-1}
S_{\rm nuc}
\label{8.229}\eeq
for protons, antiprotons, electrons, and positrons in aluminium and
gold. The $L_{\rm nuc}$ values in the plots (curves with diamonds) were
calculated by the program {\sc stopping} from DCSs for elastic
scattering evaluated by means of the eikonal approximation (Section
\ref{sec5.1.4}) with the analytical DHFS
screened atomic potential given by
Eq.\ \req{3.149}. For protons and antiprotons in aluminium, the relative
contribution of nuclear stopping to the total stopping power is of the
order of $10^{-2}$ at $E= 100$ keV, and decreases with increasing energy
to about $10^{-4}$ at 10 GeV. The relative contribution of nuclear
stopping decreases when the atomic number $Z$ of the target atom
increases. In the case of electrons and positrons, Fig.\
\ref{fig8.13},
the relative contribution of nuclear stopping is less than about $10^{-4}$ at
any energy. Consequently, for electrons and positrons nuclear stopping
is generally neglected. It may be noted that the eikonal approximation
is not accurate for electrons and positrons with energies below about 50
keV, where elastic DCSs should be calculated by the partial-wave
expansion method. The maximum of $L_{\rm nuc}$ for positrons in gold at
$E \sim 8$ keV is an artifact introduced by the eikonal approximation.


\section{Electron capture by positively charged ions
\label{sec8.9}}
\index{electron capture by positive ions}

A fundamental assumption in the theory of stopping is that the projectile
behaves as
a traveling point particle with constant charge. This is not true for
slow positive ions, which capture electrons from the medium and loose
them through a complex dynamical process. Since within the first PWBA
the energy-loss DCS is proportional to the squared charge of the
projectile, the effect of electron capture (and other low-energy
corrections) may be accounted for by considering an effective charge
$Z_1^\ast$ of the projectile, which may be defined in terms of the ratio
of the measured stopping power $S_{\rm meas}(E)$ and the calculated
stopping power of a projectile with unit charge, $S(Z_1=1;E)$,
\beq
Z_1^\ast = \sqrt{\frac{S_{\rm meas}(E)}{S(Z_1=1;E)}} \, .
\label{8.230}\eeq
Experimental evidence gives support to Bohr's suggestion that the
orbital velocity of bound electrons is the dominant parameter of the
process, and that an ion gets stripped of all its electrons that (in
their bound orbitals) have orbital velocities smaller than the velocity
of the ion \cite[see, \eg,][and references therein]{Ahlen1980}. In the
case of heavy ions with charges $Z_1 \gg 1$ and moderate velocities, for
which a sufficiently large number of electrons is captured, arguments
based on the Thomas--Fermi model lead to the semi-empirical formula
\beq
Z_1^\ast = Z_1 \left[ 1 - \exp\left( \frac{-v}{v_0 Z_1^{2/3}} \right)
\right],
\label{8.231}\eeq
where $v_0 = e^2/\hbar$ is the atomic unit of velocity, \ie, the velocity
of the electron in the ground state of the hydrogen atom. The formula
\req{8.231} is widely used to estimate effective ion charges $Z_1^\ast$.
For light ions, with small $Z_1$, it is more natural to consider that
the capture process is ruled by the velocity of an electron bound to the
ion in the ground state, which is $v_0 Z_1$ (see Section \ref{sec3.2}).
Thus, for light ions we consider
\beq
Z_1^\ast = Z_1 \left[ 1 - \exp\left( \frac{-v}{v_0 Z_1} \right)
\right] =  Z_1 \left[ 1 - \exp\left( \frac{-\beta}{\alpha Z_1} \right)
\right],
\label{8.232}\eeq
where $\alpha = v_0/c$ is the fine structure constant. Equivalently, in
terms of the non-relativistic kinetic energy $E$ of the ion,
\beq
Z_1^\ast = Z_1 \left[ 1 - \exp\left( \frac{-1}{\alpha Z_1}
\sqrt{ \frac{2E}{M_1c^2} }\right) \right].
\label{8.233}\eeq
The occurrence of electron capture, considered simply as a modification
of the effective charge of the ion, modifies the calculated cross
sections for slow projectiles. If, for the sake of concreteness, we
assume that the effect is appreciable when $Z_1^\ast < 0.98 Z_1$, we
conclude that calculations ignoring electron capture are valid for
protons with energies higher than about 400 keV, and for alphas with $E
\gtrsim 6.5$ MeV. \citet{Ziegler1999} mentions that these limits may be
somewhat too high, and that protons may not really be able to capture
electrons because the size of the bound electron orbital is larger than
the interatomic distances of most solids. The current consensus for
protons is that $Z_1^\ast = Z_1$, at least for condensed materials.




\chapter{Transport theory \label{chapt9}}



In the present Chapter we consider the transport of charged particles in
matter.  Transport processes can be described by using two different
methodologies, namely the conventional numerical solution of the
transport equation by, \eg, finite-difference techniques
\citep{CaseZweifel1967}, and Monte Carlo simulation methods
\citep{Berger1963, Jenkins1988}. Monte Carlo simulation consists of
generating random trajectories of particles through the material system
according to the DCSs adopted for the various interaction mechanisms
\citep[see, \eg,][and references therein]{Salvat2025}.

As compared with alternative deterministic finite-difference methods,
Monte Carlo simulation has several distinct advantages. Firstly, it can
describe arbitrary interaction processes, including those with cross
sections that are non-continuous functions of the physical variables
(\eg, the atomic photoelectric effect, whose total cross section
presents sharp absorption edges). Secondly, Monte Carlo simulation can
easily follow particles through material systems with complex
geometries, where deterministic methods would find great difficulties
even to define the appropriate boundary conditions. Finally, the
stochastic nature of Monte Carlo methods permits a straightforward
evaluation of statistical (class A) uncertainties of simulation results,
while finite-difference methods allow only rough estimations of
accumulated numerical errors. Although Monte Carlo codes have reached a
high degree of sophistication; simulation suffers from the drawback of
requiring very large computation times, particularly for fast charged
particles which experience a very large number of interactions before
being brought to rest.

On the other hand, formulations based on the transport equation provide
a deeper insight into the physics of the process. Unfortunately,
numerical difficulties for solving the transport equation pose severe
limits to the scattering models and geometries that can be considered.
It is only after introducing drastic simplifications that approximate
solutions can be obtained \citep{SpencerFano1954, Zheng-MingBrahme1993}.
Nevertheless, certain partial approximate solutions provide useful
information on specific aspects of the transport process such as the
energy loss distributions or the angular distributions. These solutions
are also the basis of the so-called condensed simulation algorithms,
which are employed in most Monte Carlo codes for the simulation of
high-energy charged particles.

In the first Sections of this Chapter we introduce basic radiometric
concepts and quantities employed for the description of radiation
fields, and we present the linear transport equation, which is obtained
from simple balance arguments. The rest of the Chapter is devoted to the
derivation of specific approximate solutions of the transport equation
that are useful to devise efficient Monte Carlo transport algorithms for
high-energy charged particles. Energy-straggling theories
\citep{Landau1944, BlunckLeisegang1950}, which give the energy
distribution of charged particles after traveling a given path length,
disregarding position and direction variables, are presented in Section
\ref{sec9.4}, which includes a simple algorithm for the numerical
calculation of the ``exact'' energy distribution. The final Sections are
devoted to the multiple scattering theories of \citet{Moliere1947,
Moliere1948}, \citet{GoudsmitSaunderson1940, GoudsmitSaunderson1940b},
Fermi--Eyges, and \citet{Lewis1950}, which yield the angular
distribution of charged particles after traveling a given path length in
an unlimited (infinite) medium.


\section{Radiation fields \label{sec9.1}}

\index{radiation fields|(}A {\it radiation field} is an ensemble of
energetic particles of various types (photons, electrons, positrons,
protons, neutrons, alphas, etc.).  Each particle moves in a direction
specified by the unit vector $\Omegab$ with radiant energy $E$, which is
the energy of the particle excluding its rest energy. That is, for
*massive particles, $E$ is the kinetic energy and, in the case of
photons, $E$ is simply the energy. In general, particles can be in
different ``inner'' states: photons can have various polarizations,
electrons and positrons can be in different spin states, etc. In the
following we shall assume that radiation fields are unpolarized and,
consequently, that all definitions and equations include an implicit
average over possible inner states of the particles. Because most
radiometric quantities are additive, definitions will be given for
radiation fields with particles of one type only. For mixed fields, with
particles of different types, the definitions contain an implicit
summation over the different types of particles.

In this Section we essentially follow the terminology of the
\citet{ICRU85} on fundamental quantities and units for ionizing
radiation.  However, the sequence of definitions of  radiometric
quantities presented here is opposite to the one adopted in the ICRU
Report 85. We start from the microscopic properties of the radiation
field and we obtain macroscopic (average) quantities by integration. The
discrete nature of radiation fields is thus incorporated in a natural
and rigorous way.

\index{radiation fields!particle density}
A complete description of the radiation field is provided by the {\it
particle density}, $n (t;{\bf r},$ $ E, \Omegab)$. The quantity $n (t;{\bf
r}, E, \Omegab) \, \d {\bf r} \, \d E \, \d \Omegab$ is the number of
particles in the volume element $\d {\bf r}$ at ${\bf r}$ that move with
directions within the solid angle element $\d \Omegab$ about the
direction $\Omegab$ and have energies in the interval $(E, E + \d E)$ at
time $t$. Direction vectors are usually specified using spherical
coordinates, \ie, by giving the polar angle $\theta$ and the azimuthal
angle\index{polar angles} $\phi$, $\Omegab = (\theta, \phi)$; see Fig.\
\ref{fig9.1}.
Because $|\Omegab|=1$, each direction vector defines a point on the
surface of the unit sphere. Notice that the time $t$ is
regarded as a parameter, and may be removed when considering
stationary processes.

\begin{figure}[htb!] \begin{center}
\vspace*{3mm}
\includegraphics*[width=6cm]{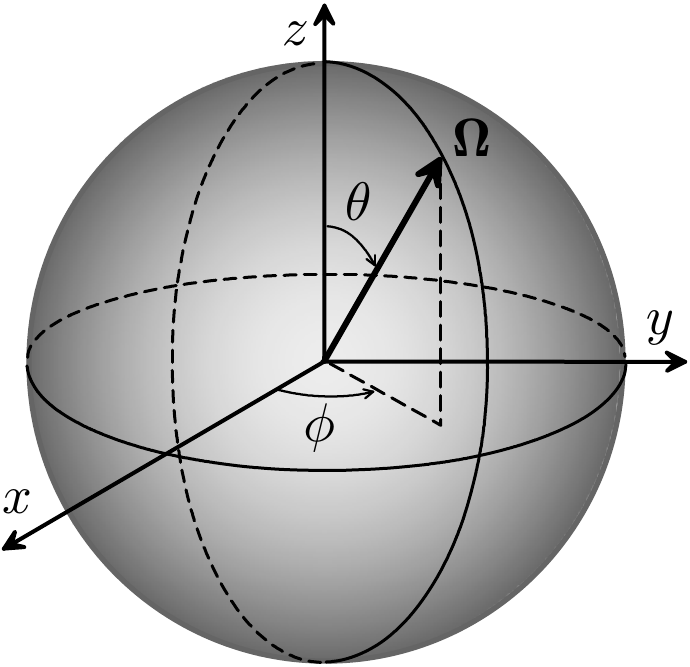}
\caption{
Unit direction vector $\Omegab$ in spherical coordinates. The polar
angle $\theta$ takes values in the interval $[0,\pi]$, while the
azimuthal angle $\phi$ varies between 0 and $2\pi$.
\label{fig9.1}}
\end{center}\end{figure}

A classical particle that moves freely with constant
velocity ${\bf v}_1$ has the associated particle density
\beq
n (t;{\bf r}, E, \Omegab) = \delta({\bf r} - {\bf r}_1^0 - {\bf v}_1 t) \,
\delta(\Omegab - \hat{\bf v}_1) \delta(E-E_1)\, ,
\label{9.1}\eeq
where ${\bf r}_1^0$ is the initial position vector (at $t=0$), $E_1$
is the kinetic energy of the particle, and $\delta(x)$ is the Dirac
delta distribution (Appendix \ref{appB}). Notice that
\beq
\delta({\bf r}-{\bf r}') = \delta(x-x') \, \delta(y-y') \, \delta(z-z')
\quad \mbox{and} \quad
\delta(\Omegab-\Omegab') = \delta(\cos\theta-\cos\theta') \,
\delta(\phi-\phi').
\label{9.2}\eeq

\index{radiation fields!trajectory picture}
In the case of classical particles, their instantaneous positions and
velocities can be determined with arbitrarily high accuracy. The
evolution of radiation fields, however, is governed by the laws of
quantum mechanics, which imply that particles cannot be assigned
definite values of position {\it and} linear momentum. Indeed, results from
measurements of these two quantities yield random values with standard
deviations satisfying Heisenberg's uncertainty relation,
\beq
\Delta r_i \, \Delta p_i \ge \frac{\hbar}{2} \qquad (i=x,y,z).
\label{9.3}\eeq
Thus, a particle with well-defined momentum necessarily has a poorly defined
position. Interactions with the atoms of the medium can be regarded as
position measurements \citep{Schiff1968}, which determine the position
vector ${\bf r}$ of the particle with an uncertainty $\Delta r$ of the
order of the atomic radius. According to the Thomas--Fermi model
(Section \ref{sec3.4}) the atomic radius is of the order of $Z^{-1/3}
a_0$, where $a_0$ is the Bohr radius.  After the interaction, the
components of the linear momentum have uncertainties $\Delta p \sim
\hbar /(2 \Delta r) = \hbar Z^{1/3}/(2 a_0)$. In transport studies, {\bf
r} and {\bf p} are considered as {\it independent} variables, that is,
quantum correlations between position and momentum are disregarded. This
simplification is acceptable only when the momentum of the particle is
much larger than its uncertainty, \ie, $p \gg \Delta p$. Hence, the
trajectory picture underlying transport calculations and Monte Carlo
simulation is expected to be valid for electrons and positrons with
kinetic energies much higher than $(\Delta p)^2/(2 \me)\sim Z^{2/3}
E_{\rm h}$, where $E_\textrm{h}$ is the Hartree energy (Appendix
\ref{appC}).  Thus, the trajectory picture is applicable to electrons
with kinetic energy higher than about 100 eV for oxygen ($Z=8$), and 500
eV for gold ($Z=79$).


\subsection{Current density and flux density \label{sec9.1.1}}

\index{radiation fields!angular current density}
\index{flux density}
The {\it angular current density} of a radiation field is defined by
\beq
{\bf j} (t; {\bf r}, E, \Omegab) \equiv \Omegab \, v(E) \,
n (t; {\bf r}, E, \Omegab),
\label{9.4}\eeq
where $v(E)$ is the speed of particles with energy $E$. To clarify the
physical meaning of the angular current density, we recall that an {\it
oriented} surface element of area $\d A$ with normal unit vector
$\hat{\bf n}$ is represented by the vector $\d {\bf A}= \hat{\bf n} \,
\d A$. Let us consider a surface element $\d {\bf A}$ at the position
${\bf r}$ and a particle that at the time $t$ is at the point ${\bf
r}_1$ moving with velocity ${\bf v}_1=v_1 \Omegab_1$; in a time interval
$\d t$, the particle will cross the surface element $\d {\bf A}$ at
${\bf r}$ only if the point ${\bf r}_1$ is within the space region swept
by the surface $\d {\bf A}$ when it is translated a distance $-{\bf v}_1
\d t$ (see Fig.\ \ref{fig9.2}a); the volume of this region is $|{\bf v}_1
\cdot \d {\bf A}| \d t = |\hat{\bf n} \cdot \Omegab_1| \, v_1 \d
t\, \d A$. Hence, the quantity
\beq
d N \equiv
{\bf j} (t; {\bf r}, E, \Omega) \, \dotprod \, \d {\bf A} \, \d E
\, \d \Omegab \, \d t
=
n (t; {\bf r}, E, \Omegab) \, \d E \, \d \Omegab \, \left[ (\hat{\bf n}
\dotprod \Omegab) \rule{0mm}{4mm} v(E) \, \d t \, \d A  \right] \,
\label{9.5}\eeq
is the net number of particles in the ranges $\d E$ and $\d \Omegab$
around $E$ and $\Omegab$
that cross the surface element $\d {\bf A}$ placed at ${\bf r}$ during
the time interval $\d t$. Note that particles moving in directions
$\Omegab$ such that $\hat{\bf n} \cdot \Omegab >0$ ($<0$)
give positive (negative) contributions.

\begin{figure}[htb!] \begin{center}
\includegraphics*[width=12cm]{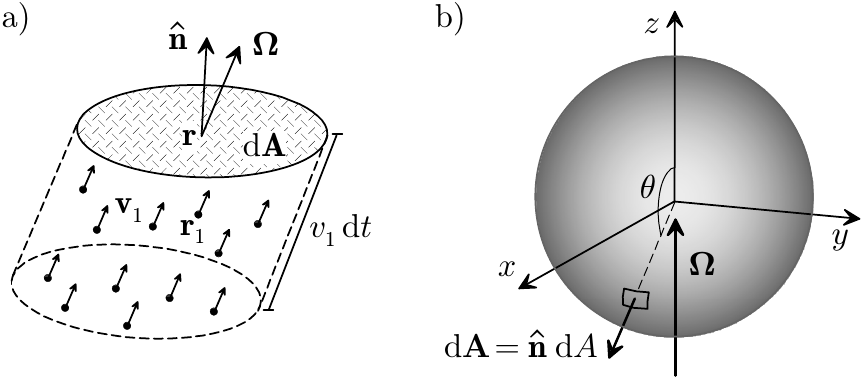}
\caption{
Surfaces in a radiation field. a) A surface element $\d {\bf A}$ in a
field of particles moving with the same velocity ${\bf v}_1$. In a time
interval $\d t$ the surface element is crossed by the particles that
initially are within the volume that would be swept by the surface if it
were moving with velocity $-{\bf v}_1$. b) Schematic diagram of the
coordinate system used to calculate the integral of Eq.\ \req{9.7} over the
surface of the sphere.
\label{fig9.2}}
\end{center}\end{figure}

In most practical cases, the radiation field is present only during a
finite time interval, \ie, $n(t;{\bf r},E,\Omegab)=0$ when $t\rightarrow
\pm \infty$, and one is usually interested in total quantities,
integrated over time. Let us consider a finite volume ${\cal V}$ limited by a
closed surface ${\cal S}$; an element of this surface is represented by
a vector $\d {\bf A} \equiv \hat{\bf n} \, \d A$, where $\hat{\bf n}$ is
the outward normal to the surface. The total number of particles that
enter the volume ${\cal V}$ through its limiting surface is
\beqa
N_{\rm in} ({\cal S}) &=& -
\int_0^\infty \d E \int \d \Omegab
\int_{-\infty}^{\infty} \d t \int_{\cal S} \d {\bf A} \dotprod
{\bf j} (t; {\bf r}, E, \Omegab)
{\cal S} (- \hat{\bf n} \dotprod \Omegab)
\nonumber \\ [2mm]
&=& - \int_0^\infty \d E \int \d \Omegab
\int_{-\infty}^{\infty} \d t  \int_{\cal S} \d A \,
(\hat{\bf n} \dotprod \Omegab) v(E) \,
n (t; {\bf r}, E, \Omegab)
{\cal S} (- \hat{\bf n} \dotprod \Omegab),\rule{10mm}{0mm}
\label{9.6} \eeqa
where we have introduced the unit step function ${\cal S}(x)$ ($=1$ if $x
>0$, $=0$ otherwise) to count only particles that enter the volume (\ie,
with directions $\Omegab$ such that $\hat{\bf n} \cdot \Omegab < 0$).
If we remove this step function, particles that enter and leave the
volume would not contribute. The expression (note the sign!)
$$
- \int_0^\infty \d E \int \d \Omegab
\int_{-\infty}^{\infty} \d t  \int_{\cal S} \d A \,
(\hat{\bf n} \dotprod \Omegab) v(E) \,
n (t; {\bf r}, E, \Omegab)
$$
gives the net number of particles absorbed in
${\cal V}$, \ie, the number of entering particles that are absorbed
minus the number of particles generated in the volume. When ${\cal S}$
is a sphere of very small radius $r_\textrm{s}$ (such that the particle
density is nearly constant within the sphere), the surface integral in
\req{9.6} can be easily evaluated by using spherical polar coordinates
with the origin at the center of the sphere and the polar axis along the
direction of $\Omegab$ (see Fig.\ \ref{fig9.2}b), so that $\hat{\bf n}
\cdot \Omegab = \cos\theta$. We have
\beqa
- \int_{\cal S} \d A \, (\hat{\bf n} \dotprod \Omegab) \,
n (t; {\bf r}, E, \Omegab) {\cal S}(- \hat{\bf n} \dotprod \Omegab)
&=& 2\pi r_\textrm{s}^2 \int_{\pi/2}^{\pi} \d\theta \, \sin\theta
\, (-\cos\theta) \, n (t; {\bf r}, E, \Omegab)
\nonumber \\ [2mm]
&=& \pi r_\textrm{s}^2 \, n (t; {\bf r}, E, \Omegab).
\label{9.7}\eeqa
Hence
\beq
N_{\rm in} ({\rm sphere}) =  \pi r_\textrm{s}^2 \, \int_0^\infty \d E \,
\int \d \Omegab \,
\int_{-\infty}^{\infty} \d t \, v(E) \,
n (t; {\bf r}, E, \Omegab),
\label{9.8}\eeq
where $\pi r_\textrm{s}^2$ is the cross-sectional area of the sphere.

Analogously, the total radiant energy that enters the
volume ${\cal V}$ through its surface is given by
\beq
R_{\rm in} ({\cal S}) = -
\int_0^\infty \d E \int \d \Omegab
\int_{-\infty}^{\infty} \d t \int_{\cal S} \d {\bf A} \dotprod
{\bf j} (t; {\bf r}, E, \Omegab) \, E \,
{\cal S} (-\hat{\bf n} \dotprod \Omegab).
\label{9.9}\eeq
In the case of a small sphere with radius $r_{\rm s}$,
\beq
R_{\rm in} ({\rm sphere}) =  \pi r_{\rm s}^2 \, \int_0^\infty \d E
\int \d \Omegab \int_{-\infty}^{\infty} \d t \, v(E) \,
n (t; {\bf r}, E, \Omegab) \, E.
\label{9.10}\eeq

The angular current density is awkward to handle because of its vector
nature. A more convenient quantity is the {\it angular flux density},
which is defined as the magnitude of the angular current
density,\index{radiation fields!angular flux density}
\beq
\Phi (t; {\bf r}, E, \Omegab) \equiv
v(E) \, n(t; {\bf r}, E, \Omegab).
\label{9.11}\eeq
Notice that $\Phi(t; {\bf r}, E, \Omegab) \, \d E \, \d \Omegab
\, \d A \, \d t$ is the number of particles in the ranges $\d E$ and $\d
\Omegab$ that in the time interval $\d t$ cross a small surface
element $\d {\bf A}$ placed at ${\bf r}$ and perpendicular to $\Omegab$,
\ie, $\d {\bf A} = \Omegab \, \d A$. The SI unit of angular flux
density is (J$\cdot$m$^2\cdot$s)$^{-1}$. The integral of the
angular flux density over directions is the {\it flux
density},\index{flux density}
\beq
\Phi(t; {\bf r}, E) \equiv \int \d \Omegab \,
\Phi(t; {\bf r}, E, \Omegab).
\label{9.12}\eeq
That is, $\Phi(t; {\bf r}, E)\, \d A \, \d E \, \d t$ gives the number
of particles with energies in $(E,E+\d E)$ that cross a small surface of
area $\d A$, placed at $r$ and perpendicular to the direction of motion
{\it of each particle}, during the time interval $\d t$. Also, according
to Eq.\ \req{9.8}, $\Phi(t; {\bf r}, E)\, \d A \, \d E \, \d t$ is the
number of particles incident on a small sphere of cross-sectional area
$\d A$, centered at ${\bf r}$, with energies in $(E,E+\d E)$ during the
time $\d t$.


\subsection{Radiometric quantities \label{sec9.1.2}}

The angular flux density, $\Phi (t; {\bf r}, E, \Omegab) = v(E) \,
n(t; {\bf r}, E, \Omegab)$, provides the most detailed description of a
radiation field. Other radiometric quantities can be expressed as
integrals of the angular flux density over appropriate ranges of
its arguments. As mentioned above, we follow the terminology of the
\citet{ICRU85} for radiometric quantities. However, we adopt a
more explicit notation, where the dependence of a quantity on the
variables $t$, ${\bf r}$, $E$, and $\Omegab$ (when applicable) is
indicated in the list of arguments.

In the ICRU Report 85, the angular flux density, \req{9.11}, is denoted by
$\dot{\Phi}_{{\bf \Omega},E}$ and is called the {\it distribution of the
scalar particle radiance with respect to energy}. The quantity
\beq
\Psi(t; {\bf r}, E, \Omegab) \equiv E
\Phi (t; {\bf r}, E, \Omegab)
\label{9.13}\eeq
[$\dot{\Psi}_{{\bf \Omega},E}$ in the ICRU notation] is the {\it
distribution of the scalar energy radiance with respect to energy}. The
ICRU Report 85 also considers vector quantities, which are expressed as
the product of a scalar function and the direction vector $\Omegab$.
Thus, from the scalar quantities \req{9.11} and \req{9.13} we can
define the {\it distribution of the vector particle radiance with
respect to energy},
\beq
\Omegab \; \Phi (t; {\bf r}, E, \Omegab) =
{\bf j} (t; {\bf r}, E, \Omegab),
\label{9.14}\eeq
and the {\it distribution of the vector energy radiance with respect to
energy},
\beq
\Omegab \,
\Psi(t; {\bf r}, E, \Omegab) = \Omegab \, E
\Phi (t; {\bf r}, E, \Omegab).
\label{9.15}\eeq
Although vector quantities are useful in transport theory, they are of
limited interest in radiation dosimetry because most radiation effects
are independent of the particle direction.

\index{radiation fields!fluence}  \index{fluence}
The {\it fluence} $\Phi({\bf r})$ at the point ${\bf r}$ is defined as
\beq
\Phi({\bf r}) \equiv \frac{\d N_{\rm in}}{\d A},
\label{9.16}\eeq
where $\d N_{\rm in}$ is the total number of particles incident on a
small sphere ${\cal S}$ of cross-sectional area $\d A$, centered at
${\bf r}$. The unit of fluence in the SI is m$^{-2}$. Equations
\req{9.8} and \req{9.11} imply that
\beqa
\Phi({\bf r}) &=&
\int_{-\infty}^{\infty} \d t
\int_0^\infty \d E \int \d \Omegab \,
\Phi (t; {\bf r}, E, \Omegab)
\nonumber \\ [2mm]
&=&
\int_{-\infty}^{\infty} \d t
\int_0^\infty \d E \int \d \Omegab \,
v(E) \, n(t; {\bf r}, E, \Omegab).
\label{9.17}\eeqa
\index{radiation fields!average fluence in a volume}
The {\it average fluence} in a finite volume ${\cal V}$ is
\beq
\overline{\Phi} \equiv
\frac{1}{{\cal V}} \int_{\cal V} \d {\bf r} \, \Phi({\bf r}).
\label{9.18}\eeq
That is,
\beq
\overline{\Phi} =
\frac{1}{{\cal V}} \int_{\cal V} \d {\bf r} \, \int_0^\infty \d E \,
\int \d \Omegab \,
\int_{-\infty}^{\infty} \d t \, v(E)
n(t; {\bf r}, E, \Omegab).
\label{9.19}\eeq

To reveal the physical significance of $\overline{\Phi}$, we
consider a field of classical particles with density [cf.\ Eq.\ \req{9.1}]
\beq
n (t;{\bf r}, E, \Omegab) = \sum_i \delta[{\bf r} - {\bf r}_{i}^0 -
{\bf v}_i (t-t_i^0)] \, \delta(\Omegab - \hat{\bf v}_i) \, \delta(E-E_i),
\label{9.20}\eeq
where ${\bf v}_i$ and $E_i$ are the velocity and the
energy of the $i$-th particle, and ${\bf r}_i^0$ and $t_i^0$ denote the
position and time at which that particle underwent its last interaction.
The average fluence in a volume ${\cal V}$ is
\beqa
\overline{\Phi} &=& \frac{1}{\cal V}
\int_{\cal V} \d {\bf r} \int_0^\infty \d E \int \d \Omegab
\int_{-\infty}^{\infty} \d t \,
v(E) \sum_i \delta[{\bf r} - {\bf r}_{i}^{0} -
{\bf v}_i (t-t_i^0)] \, \delta(\Omegab - \hat{\bf v}_i) \, \delta(E-E_i)
\nonumber \\ [2mm]
&=& \frac{1}{\cal V} \sum_i
\int_{\cal V} \d {\bf r}
\int_{-\infty}^{\infty} \d t \, v_i \,
\delta[{\bf r} - {\bf r}_{i}^{0} + {\bf v}_i (t-t_i^0)] \, .
\label{9.21}\eeqa
The integral over ${\bf r}$ is now trivial: the delta function
$\delta[{\bf r} - {\bf r}_{i}^0 + {\bf v}_i (t-t_i^0)]$ simply implies that a
particle contributes only when its trajectory intersects the volume
${\cal V}$.  Moreover, the product $\d t \, v_i$ is the path length $\d
s$ that particle $i$ travels {\it within the volume} ${\cal V}$ during
the time $\d t$. Hence,
\beq
\overline{\Phi} = \frac{1}{\cal V}
\sum_i (\mbox{path length of particle $i$ in ${\cal V}$}).
\label{9.22}\eeq
That is, the average fluence gives the total path length of particles
per unit volume. It has dimensions of (surface)$^{-1}$ and is expressed
in m$^{-2}$ or a multiple of it. The equality \req{9.22} is valid
also for real radiation fields, provided only that particles follow
classical trajectories, ${\bf r}_i(t)={\bf r}_i^{0} + {\bf v}_i \,
(t-t_i^0)$,
between consecutive interactions. Expression \req{9.22} can be used to
estimate the average fluence in Monte Carlo simulations. The
efficiency of the simulation decreases when the detector volume ${\cal V}$
is reduced, because less and less particles contribute to the fluence
score. That is, Monte Carlo codes are only able to provide average
fluences in {\it finite} volumes.

\index{radiation fields!energy fluence}
The {\it energy fluence} is defined by
\beqa
\Psi ({\bf r}) &\equiv&
\int_{-\infty}^{\infty} \d t \int_0^\infty \d E
\int \d \Omegab \, E\, \Phi (t; {\bf r}, E, \Omegab)
\nonumber \\ [2mm]
&=& \int_0^\infty \d E \, \int \d \Omegab \,
\int_{-\infty}^{\infty} \d t \, E\, v(E) \,
n(t; {\bf r}, E, \Omegab).
\label{9.23}\eeqa
Note that
\beq
\Psi ({\bf r}) = \frac{\d R_{\rm in}}{\d A},
\label{9.24}\eeq
where $\d R_{\rm in}$ is the total radiant energy of particles incident on a
small sphere of cross-sectional area $\d A$, centered at ${\bf r}$.
\index{radiation fields!average energy fluence in a volume}
The {\it average energy fluence} in a finite volume ${\cal V}$ is
\beq
\overline{\Psi} \equiv
\frac{1}{{\cal V}} \int_{\cal V} \d {\bf r} \, \Psi({\bf r}).
\label{9.25}\eeq
That is,
\beq
\overline{\Psi} =
\frac{1}{{\cal V}} \int_{\cal V} \d {\bf r} \, \int_0^\infty \d E \,
\int \d \Omegab \,
\int_{-\infty}^{\infty} \d t \, E \, v(E) \, n(t; {\bf r}, E, \Omegab).
\label{9.26}\eeq
A derivation parallel to that of Eq.\ \req{9.22} leads to
\beq
\overline{\Psi} = \frac{1}{\cal V}
\sum_{i,k} E_{ik} \times (\mbox{path length of particle $i$ with energy
$E_{ik}$ in ${\cal V}$}),
\label{9.27}\eeq
where the subscript $k$ denotes the different energies acquired by a
particle along its trajectory.

The {\it fluence rate}, $\Phi(t;{\bf r})$, and the {\it energy-fluence
rate} $\Psi(t;{\bf r})$ are defined as
\beq
\Phi(t;{\bf r}) \equiv
\int \d \Omegab \int_0^\infty \d E \, \Phi (t; {\bf r}, E, \Omegab)
= \int \d \Omegab \int_0^\infty \d E \,
v(E) \, n(t; {\bf r}, E, \Omegab) \, ,
\label{9.28}\eeq
and
\beq
\Psi(t;{\bf r}) \equiv \int \d \Omegab
\int_0^\infty \d E \, \Psi (t; {\bf r}, E, \Omegab)
= \int \d \Omegab \int_0^\infty \d E \,  E\,
v(E) \, n(t; {\bf r}, E, \Omegab) ,
\label{9.29}\eeq
respectively. The physical significance of these quantities is made
clear by considering a small sphere of cross-sectional area $\d A$
centered at ${\bf r}$. Then, $\Phi(t;{\bf r}) \, \d A$ and
$\Psi(t;{\bf r}) \, \d A$  are, respectively, the number of particles
and the radiant energy incident on the sphere per unit time.

For a monoenergetic radiation field composed of particles with radiant
energy $E_0$,
\beq
n(t; {\bf r}, E, \Omegab) = n(t; {\bf r}, \Omegab)\, \delta(E-E_0),
\label{9.30}\eeq
the fluence rate and the energy-fluence rate are, respectively,
\beq
\Phi(t;{\bf r}) = v(E_0) \, \overline{n}(t;{\bf r})
\qquad \mbox{and} \qquad
\Psi(t;{\bf r}) = E_0 \, v(E_0) \, \overline{n}(t;{\bf r}) \, ,
\label{9.31}\eeq
where
\beq
\overline{n}(t;{\bf r}) = \int \d \Omegab \,
n(t; {\bf r}, \Omegab)
\label{9.32}\eeq
is the particle number density (\ie, the number of particles per unit
volume).

The {\it particle radiance} $\Phi(t;{\bf r},{\bf \Omega})$ is defined by
\beqa
\Phi (t;{\bf r},{\bf \Omega}) &\equiv&
\int_0^\infty \d E \, \Phi(t;{\bf r},E,{\bf \Omega})
= \int_0^\infty \d E \,
v(E) \, n(t; {\bf r}, E, \Omegab),
\label{9.33}\eeqa
and the {\it energy radiance} is
\beqa
\Psi(t;{\bf r},{\bf \Omega}) &\equiv&
\int_0^\infty \d E \, \Psi (t;{\bf r},E,{\bf \Omega})
= \int_0^\infty \d E \, E\,
v(E) \, n(t; {\bf r}, E, \Omegab).
\label{9.34}\eeqa
The quantities $\Phi(t;{\bf r},{\bf \Omega}) \, \d A$ and
$\Psi(t;{\bf r},{\bf \Omega}) \, \d A$  are, respectively,
the number of particles and the radiant energy that enter a small sphere
of cross sectional area $\d A$ per unit time and per unit solid angle in
the direction $\Omegab$.

\index{radiation fields!distribution of fluence with respect to energy}
\index{fluence!distribution with respect to energy}

The {\it distribution of fluence with respect to energy} is defined as
\beq
\Phi({\bf r},E) \equiv \int_{-\infty}^{\infty} \d t
\int \d \Omegab \, \Phi(t;{\bf r},E,{\bf \Omega})
= \int_{-\infty}^{\infty} \d t \int \d \Omegab \,
v(E) \, n(t; {\bf r}, E, \Omegab),
\label{9.35}\eeq
and the {\it distribution of energy fluence with respect to energy} is
\beq
\Psi({\bf r},E) \equiv \int_{-\infty}^{\infty} \d t
\int \d \Omegab \, \Psi(t; {\bf r}, E, \Omegab)
= \int_{-\infty}^{\infty} \d t \int \d \Omegab \, E \,
v(E) \, n(t; {\bf r}, E, \Omegab)\, .
\label{9.36}\eeq
The quantities $\Phi({\bf r},E) \, \d E \, \d A$ and $\Psi({\bf
r},E) \, \d E \, \d A$ are, respectively, the number of particles and the
radiant energy that enter a small sphere of cross sectional area $\d A$
with energies in the interval $(E, E+\d E)$.

The distribution of fluence with respect to energy $\Phi({\bf r},E)$ is
used in dosimetry to calculate important quantities such as the absorbed
dose and the kerma \citep{ICRU85}. In a Monte Carlo simulation we can
only determine the {\it average distribution of fluence with respect to
energy} in a finite volume ${\cal V}$,
\index{radiation fields!average distribution of fluence with respect to energy}
\beq
\overline{\Phi}(E) \equiv \frac{1}{{\cal V}} \int_{\cal V} \d {\bf r}
\,
\Phi({\bf r},E)
= \frac{1}{{\cal V}}
\int_{\cal V} \d {\bf r} \int_{-\infty}^{\infty} \d t \int \d \Omegab \,
v(E) \, n(t; {\bf r}, E, \Omegab).
\label{9.37}\eeq
In the limit of small volumes, $\overline{\Phi}(E)$ reduces to
$\Phi({\bf r},E)$. Arguments similar to those employed in the
derivation of Eq.\ \req{9.22} show that
\beq
\overline{\Phi}(E) \, \Delta E \simeq
\int_{E}^{E+\Delta E} \overline{\Phi}(E') \, \d E'
= \frac{1}{\cal V}
\sum_{i} \left(
\begin{array}{l}
\mbox{path length of particle $i$ while it is} \\
\mbox{inside ${\cal V}$ and its radiant energy $E_i$ }\\
\mbox{is in the interval $(E, E+\Delta E)$}
\end{array} \right). \rule[-10mm]{0mm}{0mm}
\label{9.38}\eeq
That is, ${\cal V} \overline{\Phi}(E)\, \Delta E$ is the total path
length in ${\cal V}$ of field particles having radiant energies in the
interval $(E, E+\Delta E)$. In shower theory the quantity ${\cal V}
\overline{\Phi}(E)$ (which represents the path length covered by
particles of energy $E$, per unit energy range) is called the {\it
differential track length} (in ${\cal V}$).

Let us consider a small volume ${\cal V}$ around the position ${\bf r}$,
and an energy interval $(E,E+\Delta E)$. We have
\beq
\Phi({\bf r}) = \lim_{{\cal V} \rightarrow 0} \overline{\Phi} \, ,
\label{9.39}\eeq
and
\beq
\Phi({\bf r},E) = \lim_{\Delta E \rightarrow 0} \frac{1}{\Delta E}
\lim_{{\cal V} \rightarrow 0} \overline{\Phi} (E) \, \Delta E \, .
\label{9.40}\eeq
We thus see that $\Phi({\bf r})$ gives the path length of field
particles per unit volume, and $\Phi({\bf r},E)$ is the path length of
field particles per unit volume and per unit energy range.

\index{radiation fields!flux of particles across a surface}
The {\it flux} $N_{\cal S}(t)$ of particles across a surface ${\cal S}$
is defined as the net number of particles that cross the surface per unit
time,
\beqa
N_{\cal S}(t) &\equiv&
\int_0^\infty \d E \int \d \Omegab
\int_{\cal S} \d A \, (\hat{\bf n} \dotprod \Omegab)
\Phi(t; {\bf r}, E, \Omegab)
\nonumber \\ [2mm]
&=& \int_0^\infty \d E \int \d \Omegab
\int_{\cal S} \d A \, (\hat{\bf n} \dotprod \Omegab)
v(E) \, n (t; {\bf r}, E, \Omegab),
\label{9.41}\eeqa
where $\hat{\bf n}$ is the unit vector normal to the surface. The {\it
energy flux} $R_t({\cal S})$ is the net radiant energy that flows across
the surface per unit time,
\beqa
R_{\cal S}(t) &\equiv&
\int_0^\infty \d E \int \d \Omegab
\int_{\cal S} \d A \, (\hat{\bf n} \dotprod \Omegab) E\,
v(E) \, n (t; {\bf r}, E, \Omegab).
\label{9.42}\eeqa
The number of particles and
the radiant energy that cross the surface during the time interval
$(t_1,t_2)$ are
\beq
N_{\cal S} = \int_{t_1}^{t_2} \d t \, N_{\cal S}(t)
\qquad \mbox{and} \qquad
R_{\cal S} = \int_{t_1}^{t_2} \d t \, R_{\cal S}(t),
\label{9.43}\eeq
respectively.
\index{radiation fields|)}


\section{Interactions with material media \label{sec9.2}}

Particles in a material medium undergo interactions of various kinds.
In each interaction, a particle makes a transition from the initial
``state'' $E, \Omegab$ to a new state $E', \Omegab'$ and, in certain
cases, secondary particles are emitted. We consider only media that are
amorphous, homogeneous and isotropic, made by spherically symmetrical
atoms or randomly oriented molecules with uniform number density and
random positions. Diffraction and channeling effects resulting from
coherent scattering by several centers are assumed to be negligible and,
consequently, the theory is applicable only to amorphous materials and
to polycrystalline solids. For the sake of concreteness, let us assume
that individual interaction events are characterized by the energy
transfer, $W\equiv E-E'$, and the angular deflection $\Omega =
(\vartheta,\varphi)$. The polar scattering angle $\vartheta$
is the angle between the directions of motion immediately before and
after the interaction, \ie, $\cos\vartheta=\Omegab\cdot \Omegab'$.
To describe the transport of primary particles, we only need to specify the
double-differential cross sections (DDCS), differential in the energy
loss and the angular deflection, for the various
interaction mechanisms. Note that, because of the assumed isotropy of
the medium, the DDCSs are independent of the azimuthal scattering angle
$\varphi$.

\index{differential inverse mean free path}
The macroscopic total DDCS, or {\it differential inverse mean free path}
(DIMFP), of a particle with energy $E$ is defined by
\beq
\mu(E; W, \Omegab \rightarrow \Omegab') \equiv
\mu(E; W, \cos\vartheta)
= {\cal N} \, \frac{\d^2 \sigma(E)}{\d W \, \d \Omega},
\label{9.44}\eeq
where ${\cal N}$ is the number of atoms (or molecules) per unit volume
and the factor $\d^2 \sigma(E)$ $/[\d W \, \d \Omega]$ is the total DDCS per
atom (or molecule), \ie, the sum of DDCSs for all relevant interaction
processes of the considered type of particle. The inverse mean free path,
\beq
\mu(E) \equiv 2\pi \int_{-1}^{1} \d (\cos \vartheta) \int \d W \;
\mu(E; W,\cos\vartheta) = {\cal N} \, \sigma(E),
\label{9.45}\eeq
gives the interaction probability per unit path length (see Section
\ref{sec6.3.1}).
The quantity $\sigma(E)$ is the total cross section per atom (or molecule).
To clarify the significance of the DIMFP, consider a very large number
$N$ of particles moving in an infinite medium, all them with the same
energy $E$. In a short interval of time $\d t$, each particle travels a
distance $\d s = v \, \d t$. The average number of particles that
experience collisions during this time interval is $N \, \mu(E) v(E) \, \d t$.
Hence, the product $\mu(E) v$ is the interaction probability per unit
time. The function
\beq
p_1(W,\cos\vartheta) \equiv \frac{1}{\mu(E)} \,
\mu(E; W ,\cos\vartheta) =
\frac{2\pi}{\sigma(E)} \, \frac{\d^2 \sigma(E)}{\d W \, \d \Omega}
\label{9.46}\eeq
is the joint probability density (normalized to unity) of the energy loss $W$
and the angular deflection $\cos\vartheta$ of a particle in individual
interactions. Hence, the average number $\d N$ of particles that in $\d
t$ experience collisions with energy loss and angular deflection in the
respective intervals $(W, W+\d W)$ and $(\cos\vartheta, \cos\vartheta + \d
\cos\vartheta)$ is
\beq
\d N = [N \mu(E) \, v(E) \, \d t] \, p_1(W,\cos\vartheta) \,
\d W \, \d (\cos\vartheta).
\label{9.47}\eeq

Interactions of primary particles may cause the emission of secondary
particles within the material. For instance, photon interactions and
inelastic interactions of charged particles produce fast secondary
electrons (the so-called {\it delta rays})\index{delta rays}. Heavy ions may induce
nuclear reactions and the release of fast nuclear fragments and gamma
rays. Even elastic collisions (\eg, of protons with hydrogen atoms) may
originate fast recoiling nuclei.

To account for the generation of secondaries by a particle of a given type
it is convenient to
introduce a DIMFP for the emission of each type of secondary particles. For
example, the generation of secondary electrons by interactions of
protons is described by a DIMFP,
$\mu_{\rm sec}(E; E_{\rm sec}, \Omegab \rightarrow \Omegab_{\rm sec})$,
defined as the probability density per unit path length for a proton of energy
$E$ that moves in the direction $\Omegab$ to generate a secondary
electron which leaves the interaction site with energy $E_{\rm sec}$ in
the direction $\Omegab_{\rm sec}$. The DIMFP $\mu_{\rm sec}(E; E_{\rm
sec}, \Omegab \rightarrow \Omegab_{\rm sec})$ is determined by the DDCS and
the kinematics of inelastic collisions. For instance, for primary
protons with very high energy $E$, the atomic electrons may be
considered as approximately free and at rest. Then, in each inelastic
collision, a secondary electron is emitted with energy $E_{\rm
sec}=W$ and momentum $q \Omegab_{\rm sec}$ equal to the momentum
transfer,
\beq
q \Omegab_{\rm sec} = p \Omegab - p' \Omegab',
\label{9.48}\eeq
where $p$ and $p'$ are the magnitudes of the linear momentum of the
projectile before and after the interaction. Taking the dot product of
this equation by $\Omegab$, we have
\beq
\Omegab \dotprod \Omegab' = \frac{p - q \, \Omegab \dotprod \Omegab_{\rm
sec}}{p'}.
\label{9.49}\eeq
Under those circumstances,
\beqa
\mu_{\rm sec}(E; E_{\rm sec}, \Omegab\rightarrow \Omegab_{\rm sec}) &=&
{\cal N} 2\pi
\int \d W \int \d(\cos\vartheta) \frac{\d \sigma_{\rm in}(E)}{\d W \,
\d \Omega}
\nonumber \\ [2mm]
&& \mbox{} \times \delta(E_{\rm sec} - W) \, \delta(\cos\vartheta - \Omegab
\dotprod \Omegab'),
\label{9.50}\eeqa
where $\d^2 \sigma_{\rm in}(E)/[\d W \, \d \Omega]$ is the DDCS for
inelastic collisions of the primary particles.

Charged particles may emit bremsstrahlung photons. The DIMFP for
emission of a brems\-strah\-lung photon with energy $E_{\rm br}$ in the
direction $\Omegab_{\rm br}$ is given by
\beqa
\mu_{\rm br}(E; E_{\rm br}, \Omegab\rightarrow \Omegab_{\rm br}) &=&
{\cal N} 2\pi
\int \d W \int \d(\cos\vartheta) \frac{\d \sigma_{\rm rad}(E)}{\d W \,
\d \Omega}
\nonumber \\ [2mm]
&& \mbox{} \times \delta(E_{\rm br} - W) \, \delta(\cos\vartheta - \Omegab
\dotprod \Omegab_{\rm br}),
\label{9.51}\eeqa
where $\d^2 \sigma_{\rm rad}(E)/[\d W \, \d \Omega]$ is the DDCS for
emission of bremsstrahlung photons by primary particles of energy $E$,
which is considered as a function of the ``intrinsic'' direction
$\Omega=(\vartheta,\varphi)$ of the photon. Here we are assuming that
the primary particle is not deflected in the radiative event.


\section{The transport equation \label{sec9.3}}
\index{Boltzmann transport equation|(}
\index{transport equation|(}
For the sake of concreteness, in this Section we consider that primary
particles are fast electrons and we disregard the generation of
secondary radiations other than electrons. We will derive the transport
equation, also known as the {\it linear Boltzmann equation}, which
determines the angular flux density $\Phi$ of electrons within the
material bodies of the system under the assumption that forces between
transported electrons are negligible (hence, these electrons propagate
independently of each other). With obvious modifications, the theory can
be extended to other types of primary particles (positrons, photons,
neutrons, protons, alphas, and heavier charged particles) as well as to
include various types of secondary radiation (\eg, x-rays generated by
electrons).

The dominant interactions of electrons are elastic collisions (el), inelastic
collisions (in), and brems\-strah\-lung emission (rad). The total DDCS
is
\beq
\frac{\d^2 \sigma(E)}{\d W \, \d \Omega}
= \frac{\d \sigma_{\rm el}(E)}{\d \Omega} \, \delta (W)
+\frac{\d^2 \sigma_{\rm in}(E)}{\d W \, \d \Omega}
+\frac{\d \sigma_{\rm rad}(E)}{\d W} \, \frac{1}{2\pi} \,
\delta(\cos\vartheta-1).
\label{9.52}\eeq
As before, we consider that primery particles are not deflected in
bremsstrahlung events. It should be mentioned that the bremsstrahlung
DDCS diverges at $W=0$ and, consequently, the total brems\-strah\-lung
cross section is infinite. This causes numerical difficulties, which can
be avoided by simply setting the bremsstrahlung DDCS to zero for energy
losses less than a certain cutoff value $W_{\rm c}$, small enough so
that the neglected radiative events do not contribute appreciably to the
stopping power.

Although we shall be mainly concerned with geometries consisting of
locally homogeneous bodies, we may assume that the DIMFP is a function
of the position ${\bf r}$. This makes the formulation valid also for
material systems that are inhomogeneous.

We consider that primary electrons are injected into the material
structure from an external source (\eg, an electron accelerator)
characterized by the {\it source strength function} $Q(t;{\bf r}, E,
\Omegab)$, which is defined in such a way that $Q(t;{\bf r}, E, \Omegab)
\, \d t \, \d {\bf r} \, \d E \, \d \Omegab$ is the number of
electrons that are injected into the volume element $\d {\bf r}$ at
${\bf r}$ with energy between $E$ and $E+\d E$ and direction of motion
$\Omegab$ within the solid angle element $\d \Omegab$ during the
time interval between $t$ and $t +\d t$.

The transport equation is obtained by requiring conservation of the
number of electrons within a small volume element of phase space.  Let
us consider a small space volume ${\cal V}$ with surface ${\cal S}$
about the point ${\bf r}$, and let $N(t)$ be the number of electrons
that at time $t$ are within ${\cal V}$ and have energies and directions
in $\d E$ and $\d \Omegab$,
\beq
N(t) = \d E \, \d \Omegab \int_{\cal V} \d {\bf r} \; n(t;
{\bf r}, E, \Omegab),
\label{9.53}\eeq
where $n (t;{\bf r}, E, \Omegab)$ is the particle density (see Section
\ref{sec9.1}). After a short time interval $\d t$, this number changes in
\beq
\d N = \d t \, \d E \, \d \Omegab \int_{\cal V} \frac{\partial n(t;
{\bf r}, E, \Omegab)}{\partial t} \, \d {\bf r}.
\label{9.54}\eeq
On the other hand, we have the balance relation
\beqa
\d N &=&
\mbox{} - \d t \, \d E \, \d \Omegab \int_{\cal S} {\bf j}
(t;{\bf r}, E, \Omegab)\dotprod \hat{\bf n} \, \d A \nonumber \\
[2mm]
&& \mbox{} -
\d t \, \d E \, \d \Omegab \int_{\cal V} n
(t; {\bf r}, E, \Omegab)\, v(E) \, \mu (E) \,
\d
{\bf r} \nonumber \\ [2mm]
&& \mbox{} +
\d t \, \d E \, \d \Omegab \int_{\cal V} \int \! \int
n(t; {\bf r}, E', \Omegab')\, v(E')\,
\mu(E'; E'-E, \Omegab'\rightarrow \Omegab) \,
\d E'\, \d \Omegab' \, \d {\bf r} \nonumber \\ [2mm]
&& \mbox{} +
\d t \, \d E \, \d \Omegab \int_{\cal V} \int \! \int
n(t; {\bf r}, E', \Omegab')\, v(E')\,
\mu_{\rm sec}(E'; E, \Omegab'\rightarrow \Omegab) \,
\d E'\, \d \Omegab' \, \d {\bf r} \nonumber \\ [2mm]
&& \mbox{} +
\d t \, \d E \, \d \Omegab\, \int_{\cal V} Q
(t; {\bf r}, E, \Omegab)\, \d {\bf r},
\label{9.55}\eeqa
where ${\bf j} (t;{\bf r}, E, \Omegab)$ is the angular current density,
Eq.\ \req{9.4}, and $\hat{\bf n}$ indicates the outward normal to $\d
A$. The first term on the right-hand side of this equation accounts for
those electrons that enter or leave ${\cal V}$ with no change in
velocity. The second term gives the number of electrons that are
scattered into energies and directions outside the ranges $\d E$ and $\d
\Omegab$, the third term corresponds to electrons that are scattered
into these ranges. The fourth term accounts for the generated secondary
electrons and, finally, the last term gives the number of electrons
injected by the source.

Using Gauss' theorem and the definition \req{9.4}, we can write
\beq
\int_{\cal S} {\bf j} (t;{\bf r}, E, \Omegab)\dotprod \hat{\bf n}
\, \d A = \int_{\cal V} \nablab \dotprod {\bf j}(t; {\bf r}, E, \Omegab
) \, \d {\bf r} = \int_{\cal V} v(E) \, \Omegab \dotprod
\nablab n(t; {\bf r}, E, \Omegab) \, \d {\bf r}.
\label{9.56}\eeq
Then, from relations \req{9.54} and \req{9.55}, upon noting that the
volume ${\cal V}$ is arbitrary, we obtain the {\it Boltzmann transport
equation}\index{Boltzmann transport equation}
\beqa
&&\frac{\partial n(t; {\bf r}, E, \Omegab)}{\partial t}
+ v(E) \, \Omegab \dotprod \nablab n(t; {\bf r}, E, \Omegab)
+ n(t; {\bf r}, E, \Omegab)\, v(E)\, \mu(E)
\nonumber  \\ [2mm]
&&\rule{5mm}{0mm} = \int \! \int
n(t; {\bf r}, E', \Omegab')\, v(E')\,
\mu(E'; E'-E, \Omegab\dotprod \Omegab') \,
\d E' \, \d \Omegab' \nonumber \\ [2mm]
&&\rule{5mm}{0mm} + \int \! \int n(t; {\bf r}, E', \Omegab')
\, v(E')\, \mu_{\rm sec}(E'; E, \Omegab\dotprod
\Omegab') \, \d E' \, \d \Omegab' +  Q(t; {\bf r}, E, {\bf
\Omega}). \rule{13mm}{0mm}
\label{9.57}\eeqa
Usually, this equation is expressed in terms of the
angular flux density, Eq.\ \req{9.11},
$$
\Phi(t; {\bf r}, E, \Omegab) \equiv v(E) n(t; {\bf r}, E, \Omegab),
$$
as
\beqa
&& \frac{1}{v(E)} \, \frac{\partial \Phi(t; {\bf r}, E, \Omegab)
}{\partial t} + \Omegab \dotprod \nablab \Phi(t; {\bf r}, E, {\bf
\Omega}) + \Phi(t; {\bf r}, E, \Omegab)\, \mu(E)
\nonumber \\ [2mm]
&& \rule{5mm}{0mm}
= \int \! \int \Phi(t; {\bf r}, E', \Omegab')\, \mu(E'; E'-E,
\Omegab'\rightarrow \Omegab) \, \d E' \, \d \Omegab'
\nonumber \\ [2mm]
&& \rule{5mm}{0mm} + \int \! \int
\Phi(t; {\bf r}, E', \Omegab')\,
\mu_{\rm sec}(E'; E, \Omegab'\rightarrow \Omegab)
\, \d E' \, \d \Omegab' +
Q(t; {\bf r}, E, \Omegab). \rule{10mm}{0mm}
\label{9.58}\eeqa
As a direct consequence of the assumption that transported electrons do not
interact among them, the transport equation is linear. To obtain a
complete solution of this equation it is then necessary to
specify the electron flux at some initial time $t=0$, and the electron
sources. In vacuum, $\mu \equiv 0$ and the electrons propagate in
straight lines, so that the outer vacuum acts as a {\it perfect
absorber}. The
boundary condition at an interface between a diffusing medium and a
perfect absorber is
\beq
\Phi(t; {\bf r}, E, \Omegab) =0 \qquad \mbox{for $\Omegab
\dotprod \hat{\bf n} < 0$} \, ,
\label{9.59}\eeq
where $\hat{\bf n}$ is the normal to the interface entering the
perfect absorber.

It is clear that the numerical solution of the transport equation
\req{9.58}, with its six independent variables, is a problem of
formidable difficulty. Some simplification is obtained when the geometry
of the source and the material structure has some sort of symmetry that
allows reducing the number of variables. Alternatively, we can integrate
the equation over some of the variables to obtain simpler equations for
the integrated flux density.

For the sake of simplicity, in the following, the term accounting for
the emission of secondary electrons will be considered as included in
the interaction term, \ie, the second term on the right-hand side of
Eq.\ \req{9.58} will be dropped. With this simplification, the transport
equation reduces to
\beqa
&& \frac{1}{v(E)} \, \frac{\partial \Phi(t; {\bf r}, E, \Omegab)
}{\partial t} + \Omegab \dotprod \nablab \Phi(t; {\bf r}, E, {\bf
\Omega}) + \Phi(t; {\bf r}, E, \Omegab)\, \mu(E)
\nonumber \\ [2mm]
&& \rule{5mm}{0mm}
= \int \! \int \Phi(t; {\bf r}, E', \Omegab')\, \mu(E';
E'-E, \Omegab'\rightarrow \Omegab) \, \d E' \, \d \Omegab'
+ Q(t; {\bf r}, E, \Omegab). \rule{10mm}{0mm}
\label{9.60}\eeqa

The multiple scattering theories described below assume an ideal point
source that emits $N_0$ electrons from the origin at $t=0$ with well defined
initial energy $E_0$ and direction of movement $\Omegab_0$,
\beq
Q(t; {\bf r}, E, \Omegab) = N_0\, \delta(t) \, \delta({\bf r}) \,
\delta(E-E_0) \, \delta(\Omegab-\Omegab_0).
\label{9.61}\eeq
In this case, we can remove the source term from the transport equation,
\beqa
&& \frac{1}{v(E)} \, \frac{\partial \Phi(t; {\bf r}, E, \Omegab)
}{\partial t} + \Omegab \dotprod \nablab \Phi(t; {\bf r}, E, {\bf
\Omega}) + \Phi(t; {\bf r}, E'-E, \Omegab)\, \mu(E)
\nonumber \\ [2mm]
&& \rule{5mm}{0mm}
= \int \! \int \Phi(t; {\bf r}, E', \Omegab')\, \mu(E';
E, \Omegab'\rightarrow \Omegab) \, \d E' \, \d \Omegab',
\label{9.62}\eeqa
and define the characteristics of the source through the
initial condition
\beq
\Phi(t=0; {\bf r}, E, \Omegab) = N_0 \, \delta({\bf r})
\, \delta(E-E_0) \, \delta(\Omegab-\Omegab_0).
\label{9.63}\eeq

In many practical situations, it is not necessary to study the
time-evolution of the electron flux. A trivial example is that of a
stationary source, the electron flux is then constant with $t$ and
satisfies the {\it stationary transport equation}
\beqa
&&\Omegab \dotprod \nablab \Phi({\bf r}, E, \Omegab)
+ \Phi({\bf r}, E, \Omegab)\, \mu(E) \nonumber \\ [2mm]
&&\rule{10mm}{0mm} =
\int \! \int \Phi({\bf r}, E', \Omegab')\, \mu(E'; E'-E,
\Omegab\dotprod \Omegab') \, \d E' \, \d \Omegab'
+ Q({\bf r}, E, \Omegab) ,
\rule{10mm}{0mm}
\label{9.64}\eeqa
where the argument $t$ has been dropped in the functions $Q$ and
$\Phi$.

For arbitrary sources, we may limit to consider the total number of
emitted electrons
\beq
Q_{\rm tot} ({\bf r}, E, \Omegab) \equiv \int_{-\infty}^{\infty}
Q(t; {\bf r}, E, \Omegab) \, \d t,
\label{9.65}\eeq
and the time-integrated flux
\beq
\Phi_{\rm tot}({\bf r}, E, \Omegab) \equiv
\int_{-\infty}^{\infty} \Phi(t; {\bf r}, E, \Omegab) \, \d t.
\label{9.66}\eeq
Assuming that electrons are emitted from the source only during a finite
time interval (as it happens for real sources), upon integration of Eq.\
\req{9.60} over $t$ from $-\infty$ to $\infty$, we obtain the {\it
time-integrated transport equation}
\beqa
&&\Omegab \dotprod \nablab \Phi_{\rm tot}({\bf r}, E, \Omegab)
+ \Phi_{\rm tot}({\bf r}, E, \Omegab)\,
\mu(E) \nonumber \\ [2mm]
&&\rule{10mm}{0mm} =
\int \! \int \Phi_{\rm tot} ({\bf r}, E', \Omegab')\, \mu(
E'; E'-E, \Omegab\dotprod \Omegab') \, \d E' \, \d {\bf
\Omega}' + Q_{\rm tot}({\bf r}, E, \Omegab), \rule{15mm}{0mm}
\label{9.67}\eeqa
which is formally equivalent to Eq.\ \req{9.64}. As a consequence, the
solution of a stationary problem and that of a time-integrated problem
are obtained by solving the same equation. Again, this is a
consequence of the assumption that interactions between transported
electrons are negligible.
\index{Boltzmann transport equation|)}
\index{transport equation|)}


\section{Energy straggling \label{sec9.4}}

\index{energy straggling|(}
Let us consider charged particles of mass $M_1$, charge $Z_1e$ with
initial kinetic energy $E_0$ moving in an infinite medium. Each
particle undergoes discrete interactions and, after a path length $s$,
its energy is reduced to a value $E \le E_0$. Since both the
number of interactions and the energy loss in each interaction are
random variables, the energy $E$ at the end of the step is a random
variable, which follows a certain distribution $\Phi(s; E)$. This
distribution is usually referred to as the {\it energy-straggling
distribution}.

Charged particles lose energy preferentially through inelastic
collisions. The allowed energy transfers $W$ in inelastic collisions
range from the first excitation energy of the target atom (which, for
simplicity, we set equal to zero) to a maximum value $W_{\rm max}$,
which depends on the mass of the projectile [Eq.\ \req{9.49}]. Since
energy is mostly transferred to atomic electrons, $W_{\rm max} \simeq E$
for projectile electrons and positrons, and $W_{\rm max} \ll E$ for
heavy charged particles (with $M \gg \me$).

In the case of fast electrons and positrons, the radiative stopping
power (Section \ref{sec9.6}) is important; it exceeds the collision
stopping power for projectiles with sufficiently high energies. In each
radiative event the energy of the projectile decreases in an amount
equal to the energy $W$ of the emitted photon, which may take values
from $0$ to $E$. Because of the divergence of the bremsstrahlung DCS at
$W=0$ [see Eq.\ \req{9.197}], the total cross section for bremsstrahlung
emission is infinite.  Temporarily, to put aside difficulties associated
with infinite cross sections, we introduce a small cutoff energy $W_{\rm
cr}$, of the order of a few tens of eV, and ignore the emission of
photons with smaller energies, which have a negligible stopping effect.

The atomic energy-loss DCS for a projectile with kinetic energy $E$ is
the sum of the collision and bremsstrahlung DCSs,
\beq
\frac{\d \sigma(E)}{\d W} =
\frac{\d \sigma_{\rm col}(E)}{\d W} \, {\cal S}(W_{\rm max} -W) + \frac{\d
\sigma_{\rm rad}(E)}{\d W} \, {\cal S}(E -W) \,
{\cal S}(W - W_{\rm cr}) \, .
\label{9.68}\eeq
The probability that the projectile loses an energy between
$W$ and $W + \d W$ in an individual interaction is $p_1(E;W) \, \d W$,
with the probability density
\beq
p_1(E;W) = \frac{1}{\sigma(E)} \, \frac{\d \sigma(E)}{\d W},
\label{9.69}\eeq
where
\beq
\sigma(E) = \int_0^{W_{\rm max}} \frac{\d \sigma_{\rm in}(E)}{\d W}\, \d W
+ \int_{W_{\rm cr}}^E \frac{\d \sigma_{\rm rad}(E)}{\d W}\, \d W
 \label{9.70}\eeq
is the total atomic cross section for energy-loss interactions.
The mean free path $\lambda(E)$ between energy-loss events is given by
\beq
\lambda^{-1}(E) = {\cal N} \sigma(E),
\label{9.71}\eeq
and the corresponding DIMFP can be expressed as
\beq
\mu(E; W) \equiv {\cal N} \frac{\d \sigma(E)}{\d W}
= \lambda^{-1}(E) \, p_1(E;W).
\label{9.72}\eeq

The transport equation for the energy straggling distribution is
obtained by integrating Eq.\ \req{9.62} over spatial coordinates
${\bf r}$ and over directions $\Omegab$. Considering that the DIMFP
$\mu(E;W)$ vanishes for $W<0$ and $W>E$, and replacing the time
variable by the path length, $\d s = v(E) \d t$, the transport equation
can be written in the form
\beq
\frac{\partial  \Phi(s; E)}{\partial s}
= \int_0^{\infty} \Phi(s; E+W) \, \mu(E+W;W) \, \d W
- \Phi(s, E) \lambda^{-1}(E),
\label{9.73}\eeq
with the initial condition
\beq
\Phi(s=0; E) = \delta(E-E_0).
\label{9.74}\eeq
It is convenient to consider the straggling distribution as a function
of the accumulated energy loss $\Delta = E_0 - E$, which satisfies the
equation
\beq
\frac{\partial  \Phi (s; \Delta)}{\partial s} =
\int_0^{\infty} \left[
\Phi(s; \Delta-W) \mu(E_0 - \Delta + W; W)
- \Phi(s;\Delta) \, \mu(E-\Delta;W) \rule{0mm}{4mm}\right]\, \d W
\label{9.75}\eeq
with
\beq
\Phi(s=0; \Delta) = \delta(\Delta).
\label{9.76}\eeq

For small enough path lengths $s$, for which energy losses are
relatively small, $\Delta \ll E_0$, we can ignore the dependence of the
DIMFP on the energy and write
\beq
\frac{\partial  \Phi (s; \Delta)}{\partial s} =
\int_0^{\infty} \left[
\Phi(s; \Delta-W) - \Phi(s;\Delta) \rule{0mm}{4mm} \right]
\mu(E_0;W) \, \d W\, .
\label{9.77}\eeq
Although the solution of this equation is still difficult, it is trivial
to determine the first moments of the energy-loss distribution
\beq
\langle\Delta^{n}\rangle \equiv \int_{0}^{\infty} \Delta^{n}
\Phi(s;\Delta) \, \d\Delta.
\label{9.78}\eeq
We have
\beqa
\frac{\d}{\d s} \langle\Delta^{n}\rangle & = &
\int_{0}^{\infty} \d \Delta
\int_0^{\infty} \d W \,  \Delta^{n} \left[
\Phi(s; \Delta-W) - \Phi(s;\Delta) \rule{0mm}{4mm} \right]
\mu(E_0;W)
\nonumber \\[1mm]
& = & \int_{0}^{\infty} \d \Delta'
\int_0^{\infty} \d W \,  (\Delta'+W)^{n}
\Phi(s; \Delta') \, \mu(E_0;W)
- \langle \Delta^n \rangle
\int_0^{\infty} \d W \, \mu(E_0;W)
\nonumber \\ [2mm]
&=& \sum_{k=1}^{n} \frac{n!}{k!(n-k)!} \langle \Delta^{n-k} \rangle
\int_{0}^{\infty} W^{k} \mu(E_0;W) \, \d W.
\nonumber \eeqa
That is,
\beq
\frac{\d}{\d s} \langle\Delta^{n}\rangle
= \sum_{k=1}^{n} \frac{n!}{k!(n-k)!} \, \langle \Delta^{n-k} \rangle
\lambda^{-1}(E_0) \, \langle W^{k} \rangle_1,
\label{9.79}\eeq
where
\beq
\langle W^{k} \rangle_1 \equiv
\int_{0}^{W_{\rm max}} W^{k} p_1(E_0; W)
\, \d W
\label{9.80}\eeq
is the expectation value of $W^k$ in a single interaction. The
equations for the first and second moments,
\beq
\frac{\d}{\d s} \langle\Delta\rangle
= \lambda^{-1}(E_0) \, \langle W \rangle_1 \equiv S(E_0),
\eeq
\label{9.81}and
\beq
\frac{\d}{\d s} \langle\Delta^{2}\rangle
= 2 \langle \Delta \rangle
\lambda^{-1}(E_0) \, \langle W \rangle_1
+ \lambda^{-1}(E_0) \, \langle W^{2} \rangle_1 =
2 \langle \Delta \rangle S(E_0)
+ \Omega^2(E_0),
\label{9.82}\eeq
can be readily integrated and give
\beq
\langle\Delta\rangle = S(E_0) \, s
\label{9.83}\eeq
and
\beq
\langle\Delta^{2}\rangle =
[S(E_0) \, s]^2 + \Omega^2(E_0) \, s.
\label{9.84}\eeq
The quantities
\beq
S(E) \equiv
 \lambda^{-1} (E) \, \langle W \rangle_1 = {\cal N}
\int_{0}^{W_{\rm max}} W \frac{\d \sigma(E)}{\d W} \, \d W
\label{9.85}\eeq
and
\beq
\Omega^2(E) \equiv
 \lambda^{-1} (E) \,  \langle W^2 \rangle_1 = {\cal N}
\int_{0}^{W_{\rm max}} W^{2}  \frac{\d \sigma(E)}{\d W} \, \d W
\label{9.86}\eeq
are the {\it stopping power} and the {\it energy straggling parameter},
respectively. Evidently, the stopping power is the average energy loss
per unit path length. We note that the variance of the energy loss
distribution is
\beq
{\rm var}(\Delta) =
\langle\Delta^{2}\rangle - \langle\Delta\rangle^{2} =
\Omega_{\rm s}^{2}(E_0) s.
\label{9.87}\eeq
Hence, the energy straggling parameter equals the
variance increase per unit path length.


\subsection{The continuous slowing down approximation \label{sec9.4.1}}
\index{energy straggling!continuous slowing down approximation}
\index{continuous slowing down approximation}
In cases where energy straggling is small [that is, when $\sqrt{{\rm
var}(\Delta)} \ll \langle\Delta\rangle$], and also for certain
theoretical considerations, the continuous slowing-down approximation
(CSDA) may be useful. In this approximation, energy straggling is
completely neglected. The projectile particle is assumed to lose energy
continuously along its path and the slowing-down process is completely
characterized by the stopping power $S(E)$.

The usefulness of the CSDA stems from the fact that it sets a strict
correspondence between path length and energy loss. The CSDA range of a
particle with kinetic energy $E$ is given by \index{CSDA range}
\index{range of charged particles} \index{energy-range curve}
\beq
R(E) = \int_{E_{\rm abs}}^{E} \frac{\d E'}{S(E')},
\label{9.88}\eeq
where $E_{\rm abs}$ is the ``absorption'' energy, \ie, the energy at
which the particle is assumed to be effectively absorbed in the medium.
If a particle starts its trajectory with kinetic energy $E$, the
energy loss $\Delta_s$ after a path length $s$ is determined by the equation
\beq
s = \int_{E-\Delta_s}^{E} \frac{\d E'}{S(E')} = R(E) - R(E-\Delta_s).
\label{9.89}\eeq
Hence, to calculate the energy loss as a function of the path length,
$\Delta_s$, we only need to know the CSDA range as a function of energy,
$R(E)$, which in early studies used to be defined graphically by means of
an {\it energy-range curve}.

The energy deposited into the medium per unit path length can be estimated as
\beq
D(z)=S(E(z)),
\label{9.89.1}\eeq
where $E(z)$ is the energy of the projectile after traveling a path
length $z$, \ie, such that $R(E(z))=R(E_0)-z$. The quantity $D(z)$ can
be identified with the CSDA depth-dose distribution calculated with
elastic scattering effects neglected.

The energy loss distribution corresponding to the CSDA is
\beq
\Phi_{\rm CSDA}(s;\Delta) = \delta(\Delta-\Delta_s).
\label{9.90}\eeq
The CSDA is expected to be approximately valid only for heavy charged
particles (mesons, protons, and heavier ions), because for these
particles the largest energy loss $W_{\rm max}$ is much less than E, and
the contribution of bremsstrahlung emission to the stopping power is
small. This is not true for electrons and positrons, because these
particles can lose a large fraction of their kinetic energy in a single
collision or radiative event.  Nevertheless, the CSDA formulas
\req{9.88} and \req{9.89} yield fairly accurate estimates of the
average particle range and of the average path length that particles
have to travel to lose an average energy $\Delta_s$, even for electrons
and positrons.

There is a special situation in which the straggling distribution can be
evaluated analytically in terms of only the first and second moments of
the energy-loss DCS. It occurs in the case of heavy particles and path
lengths $s$ much larger than the mean free path $\lambda(E_0)$ for
energy-loss interactions, but still small enough to have $\Delta_s \ll
E_0$. Then, the energy loss $\Delta$ is the result of a large number of
interactions, each of which causing an energy loss $W$ much smaller than
$\Delta$. Under these circumstances, the central limit theorem
\citep[see, \eg,][]{Cramer1962} applies, and it implies that the
straggling distribution is Gaussian. Since the mean and variance of the
energy loss are given by Eqs.\ \req{9.83} and \req{9.87}, we have
\beq
\Phi_{\rm CSDA2}(s;\Delta) = \frac{1}{\sqrt{2\pi s \Omega^2(E_0)}}
\exp\left(- \, \frac{[\Delta - s S(E_0)]^2}{2 s \Omega^2(E_0)} \right) .
\label{9.91}\eeq
Note that this distribution applies only when
\beq
\sqrt{s \Omega^2(E_0)} \ll s S(E_0) \ll E_0
\label{9.92}\eeq
because otherwise there would be a finite probability for energy losses
that are either negative or larger than the initial energy of the
particle. Although the conditions may look too restrictive, the
distribution \req{9.91} is useful for describing the energy straggling
caused by soft interactions of charged particles (including electrons
and positrons) in mixed Monte Carlo algorithms \citep{Salvat2025}.


\subsection{The Landau distribution \label{sec9.4.2}}
\index{energy straggling!Landau distribution|(}
\index{Landau distribution}
\citet{Landau1944} obtained an approximate energy-straggling distribution,
$\Phi_{\rm L}(s; \Delta)$, under the following assumptions: \\ [2mm]
{\bf L1}: Energy losses due to bremsstrahlung emission are negligible.
\\ [2mm]
{\bf L2}: The path length $s$ is small enough to ensure that the average energy
loss along the trajectory segment is much less than the initial energy
$E_0$, so that the DIMFP can be considered to remain essentially
constant (\ie, independent of the energy of the particle). \\ [2mm]
{\bf L3}: We can select a certain cutoff energy loss, $W_{\rm c}$, much smaller
than $E_0$, such that the DIMFP for ``hard'' energy-loss events with
$W>W_{\rm c}$ can be approximated by the Thomson formula [Eq.\
\req{4.129}],
\beq
\mu(W) = {\cal N} \frac{\d \sigma_{\rm R} }{\d W} =
{\cal N} Z \, \frac{2\pi Z_0^2 e^4}{\me v^2} \, \frac{1}{W^2}
\qquad \mbox{for $W>W_{\rm c}$},
\label{9.93}\eeq
where $Z$ is the number of electrons per atom or molecule. \\ [2mm]
{\bf L4}: The maximum energy loss in a single collision is much larger than the
cutoff energy loss, $W_{\rm max} \gg W_{\rm c}$. \\ [2mm]
{\bf L5}: The average energy loss per unit path length caused by
``soft'' individual interactions with $W<W_{\rm c}$ can be calculated
from the Bethe formula for the restricted stopping power, Eq.\
\req{6.301},
\beq
S (W<W_{\rm c}) =
\int_0^{W_{\rm c}} W \mu(W) \, \d W =
{\cal N} Z \, \frac{2\pi Z_0^2 e^4}{\me v^2} \, \ln\frac{W_c}{W_{\rm 1}},
\label{9.94}\eeq \index{restricted stopping power}
with
\beq
\ln W_1 = \ln \left( \frac{(1-\beta^2) I^2}{2\me v^2} \right)
+ \beta^2,
\label{9.95}\eeq
where $I$ is the mean excitation energy of the material, Eq.\
\req{6.288}.

Since energy losses are assumed to be
negligible (L2), the transport equation takes the form \req{9.77},
\beq
\frac{\partial  \Phi_{\rm L}(s; \Delta)}{\partial s} = \int_0^\infty \left[
\Phi_{\rm L}(s; \Delta-W)
- \Phi_{\rm L}(s;\Delta) \right] \, \mu(W) \, \d W.
\label{9.96}\eeq
Although the maximum energy loss $W_{\rm max}$ in a single collision is
finite, assumption L3 implies that for large energy losses the DIMFP
decreases rapidly with $W$. Hence, we can assume that the expression
\req{9.93} of the DIMFP is valid for $W>W_{\rm max}$ and replace the
finite value $W_{\rm max}$ with infinity. Then, the variable $\Delta$
appears in Eq.\ \req{9.96} only as the argument of the
energy-straggling distribution, which can be determined by applying the
Laplace transformation \citep{Arfken1985, AbramowitzStegun1974}. The
Laplace transform of the function $\Phi_{\rm L}(s;\Delta)$ with respect to
$\Delta$ is defined by
\beq
\widetilde{\Phi}_{\rm L}(s;q) \equiv \int_0^\infty \Phi_{\rm L}(s;\Delta) \exp(-
q \Delta ) \, \d \Delta.
\label{9.97}\eeq
The transformation can be inverted to express $\Phi_{\rm L}(s;\Delta)$
in terms of $\widetilde{\Phi}_{\rm L}(s;q)$ as follows
\beq
\Phi_{\rm L}(s;\Delta) \equiv \frac{1}{2\pi {\rm i}}
\int_{c-{\rm i} \infty}^{c+{\rm i} \infty}
\widetilde{\Phi}_{\rm L}(s;q) \exp( q \Delta ) \, \d q,
\label{9.98}\eeq
where $c$ is a real positive value, that is, the integration is along a
straight line parallel to the imaginary axis and at the right of the
latter. Multiplying both sides of Eq.\ \req{9.96} by $\exp( - q
\Delta)$ and integrating over $\Delta$, we obtain
\beq
\frac{\partial  \widetilde{\Phi}_{\rm L}(s; q)}{\partial s} =
- \widetilde{\Phi}_{\rm L}(s;q) \int_0^\infty \left[
1-\exp(-qW) \right] \, \mu(W) \, \d W.
\label{9.99}\eeq
The boundary condition \req{9.76} [$\Phi_{\rm L}(s=0;\Delta) = \delta(\Delta)$]
implies that
\beq
\widetilde{\Phi}_{\rm L}(s=0;q) = 1.
\label{9.100}\eeq
Integrating Eq.\ \req{9.99} with this condition gives
\beq
\widetilde{\Phi}_{\rm L}(s; q) = \exp\left\{ -s
\int_0^\infty \left[
1-\exp(-qW)\right] \, \mu(W) \, \d W\right\}.
\label{9.101}\eeq
Inserting this result in expression \req{9.98} we obtain
\beq
\Phi_{\rm L}(s;\Delta) = \frac{1}{2\pi {\rm i}}
\int_{c-{\rm i} \infty}^{c+{\rm i} \infty}
\exp\left\{ q \Delta -s
\int_0^\infty \left[
1-\exp(-qW)\right] \, \mu(W) \, \d W\right\}
\, \d q.
\label{9.102}\eeq

With the DIMFP model adopted by Landau, Eqs.\ \req{9.93} and \req{9.94}, the
integral over $W$ can be evaluated analytically. To this end, we split
the integral in two parts, with limits from 0 to $W_{\rm c}$ and
from $W_{\rm c}$ to $\infty$,
\beq
\int_0^\infty \left[ 1-\exp(-qW)\right] \, \mu(W) \, \d W = J_1 + J_2
\label{9.103}\eeq
with
\begin{subequations}
\label{9.104}
\beqa
J_1 &=& \int_0^{W_{\rm c}} \left[ 1-\exp(-qW)\right] \, \mu(W) \, \d W,
\label{9.104a}\\ [2mm]
J_2 &=& \int_{W_{\rm c}}^\infty \left[ 1-\exp(-qW)\right] \, \mu(W) \, \d W.
\label{9.104b}\eeqa
\end{subequations}
Introducing the expression \req{9.93}, the second of these integrals becomes
\beq
J_2 = {\cal N} Z \, \frac{2\pi Z_0^2 e^4}{\me v^2}
\int_{W_{\rm c}}^\infty \frac{1-\exp(-qW)}{W^2} \, \d W.
\label{9.105}\eeq
Integrating by parts, and assuming that contributions to the integral
arise only from $q$ values such that $qW_{\rm c} \ll 1$, we get
\beqa
\int_{W_{\rm c}}^\infty \frac{1-\exp(-qW)}{W^2} \, \d W
&=& \frac{1-\exp(-qW_{\rm c})}{W_{\rm c}} +
q \int_{W_{\rm c}}^\infty \frac{\exp(-qW)}{W} \, \d W
\nonumber \\ [2mm]
&\simeq& q + q \int_{pW_{\rm c}}^\infty \frac{\exp(-z)}{z} \, \d z.
\nonumber\eeqa
That is,
\beqa
\frac{1}{q} \int_{W_{\rm c}}^\infty \frac{1-\exp(-qW)}{W^2} \, \d W
&\simeq& 1 + \int_{qW_{\rm c}}^\infty \frac{\exp(-z)}{z} \, \d z
\nonumber \\ [2mm]
&=& 1 + \int_{qW_{\rm c}}^1 \frac{1}{z} \, \d z
+ \int_{1}^\infty \frac{\exp(-z)}{z} \, \d z
+ \int_{0}^1 \frac{\exp(-z)-1}{z} \, \d z
\nonumber\eeqa
The sum of the last two integrals is $-g$, where $g  = 0.5772$ is
Euler's constant. Hence,
\beq
J_2 = {\cal N} Z \, \frac{2\pi Z_0^2 e^4}{\me v^2} q \left[
1-g- \ln (qW_{\rm c}) \right].
\label{9.106}\eeq
Assuming, as before, that $qW_{\rm c} \ll 1$, the integral $J_1$ can be
approximated as
\beq
J_1 \simeq q \int_0^{W_{\rm c}} W \, \mu(W) \, \d W
- \frac{1}{2} q^2 \int_0^{W_{\rm c}} W^2 \, \mu(W) \, \d W.
\label{9.107}\eeq
The integrals on the right-hand side are, respectively, the stopping
power and the energy-straggling parameter for soft collisions [cf.\
Eqs\ (\req{9.85} and \req{9.86}]. Landau assumed that the second
integral was negligible, and he approximated the first integral
according to the assumption L5 [Eq.\ \req{9.94}],
\beq
J_1 = q
{\cal N} Z \, \frac{2\pi Z_0^2 e^4}{\me v^2} \, \ln\frac{W_{\rm c}}{W_1}
\label{9.108}\eeq
Combining these results, we have
\beq
\int_0^\infty \left[ 1-\exp(-qW)\right] \, \mu(W) \, \d W =
{\cal N} Z \, \frac{2\pi Z_0^2 e^4}{\me v^2} q \left[
1-g - \ln (qW_1) \right].
\label{9.109}\eeq
Introducing this expression in Eq.\ \req{9.102} we obtain the Landau
distribution,
\beq
\Phi_{\rm L}(s;\Delta) = \frac{1}{2\pi {\rm i}}
\int_{c-{\rm i} \infty}^{c+{\rm i} \infty}
\exp\left\{ q \Delta - s
{\cal N} Z \, \frac{2\pi Z_0^2 e^4}{\me v^2} q \left[
1-g - \ln (qW_1) \right] \right\}
\, \d q.
\label{9.110}\eeq

This distribution can be cast in a more convenient form by
introducing the quantity
\beq
\xi \equiv s \,
{\cal N} Z \, \frac{2\pi Z_0^2 e^4}{\me v^2},
\label{9.111}\eeq
which has the dimensions of energy, and by changing the integration
variable to $u=\xi q$. We can thus write
\beq
\Phi_{\rm L}(s;\Delta) = \frac{1}{\xi} \varphi(\lambda)
\label{9.112}\eeq
with
\beq
\varphi(\lambda) \equiv
\frac{1}{2\pi {\rm i}}
\int_{c-{\rm i} \infty}^{c+{\rm i} \infty}
\exp(u \ln u + \lambda u) \, \d u
\label{9.113}\eeq
and
\beq
\lambda \equiv \frac{\Delta}{\xi} -
\left[ \ln \left( \frac{\xi}{W_1} \right)+1-g \right].
\label{9.114}\eeq
Thus, the Landau distribution is expressed as the product of $1/\xi$ by
a universal function $\varphi_{\rm L}(\lambda)$ of the dimensionless
variable $\lambda$. For numerical calculation, it is more convenient to use
the following equivalent form of the integral [Eq.\ (24) in
\citet{Landau1944}]
\beq
\varphi(\lambda) = (1/\pi) \int_0^\infty
\exp(-t \ln t - \lambda t) \sin(\pi t) \, \d t.
\label{9.115}\eeq
The Landau distribution, calculated from this expression, is plotted in
Fig.\ \ref{fig9.3}. The distribution has the shape of an asymmetric bell, with
its most probable value at $\lambda = -0.2228$, and with a long tail in the
positive side, which tends to $1/\lambda^2$ for $\lambda\rightarrow
\infty$. The approximation
\begin{subequations}
\label{9.116}
\beq
\varphi(\lambda) \simeq \lambda^{-2} + 10 \, \lambda^{-3}
\label{9.116a}\eeq
differs from the exact distribution by less than 0.5 \% for $\lambda >
200$. For negative $\lambda$, $\varphi(\lambda)$ decreases rapidly
with $|\lambda|$, and for $\lambda < -3.0$ the approximation
\beq
\varphi(\lambda) \simeq \frac{1}{\sqrt{2\pi}}
\exp\left( \frac{|\lambda|-1}{2} - \exp(|\lambda|-1) \right),
\label{9.116b}\eeq
\end{subequations}
differs from the exact result by less than 0.5 \%.

\begin{figure}[htb!] \begin{center}
\vspace*{3mm}
\includegraphics*[width=11cm]{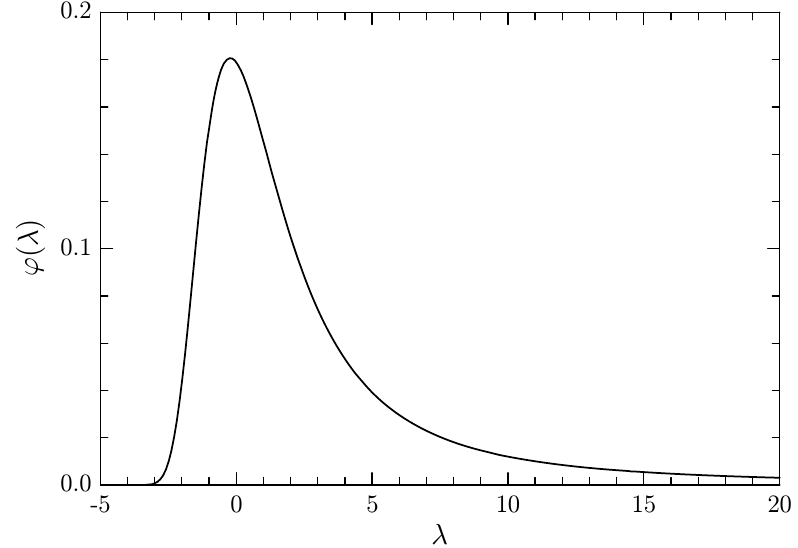}
\includegraphics*[width=11cm]{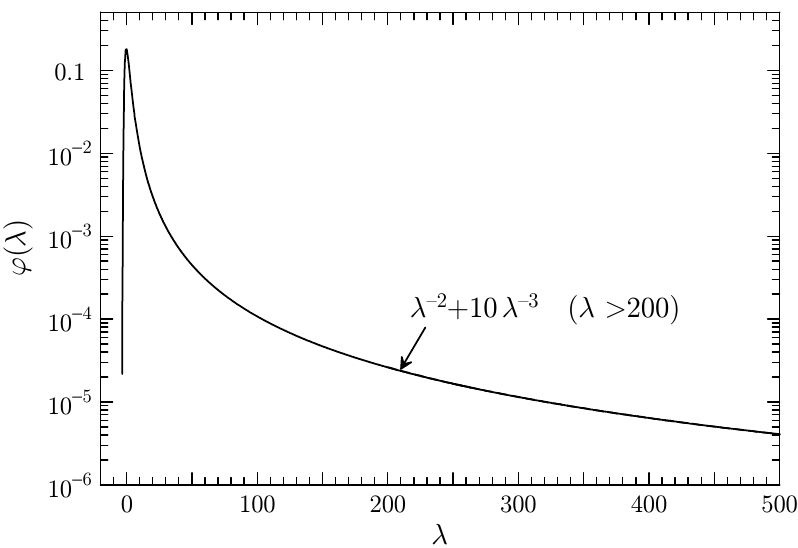}
\caption{
Landau energy-straggling distribution.
\label{fig9.3}}
\end{center}\end{figure}

The most probable energy loss is obtained from Eq.\ \req{9.114} with
$\lambda=-0.2228$, which gives
\beq
\Delta_{\rm p} = \xi \left[ \ln \left( \frac{2 \me c^2 \, \beta^2
\gamma^2 \, \xi}{I^2} \right) - \beta^2 + 0.200 \right].
\label{9.117}\eeq

\index{energy straggling!Landau distribution|)}


\subsection{The distribution of Blunk and Leisegang \label{sec9.4.3}}
\index{energy straggling!Blunk--Leisegang distribution}
\index{Blunk--Leisegang distribution}

The Landau distribution is derived under the assumption L5, which
implies that the energy straggling associated to soft collisions is
neglected. \citet{BlunckLeisegang1950} introduced this straggling
component by keeping the second-order term in Eq.\ \req{9.107},
\beq
J_1 = p {\cal N} Z \, \frac{2\pi Z_0^2 e^4}{\me v^2} \ln\frac{W_1}{W_{\rm c}}
- \1o2 p^2 \Omega^2_{\rm soft},
\label{9.118}\eeq
where
\beq
\Omega^2_{\rm soft} =
\int_0^{W_{\rm c}} W^2 \, \mu(W) \, \d W,
\label{9.119}\eeq
is the energy straggling parameter of the soft interactions. With the
second-order term included, the straggling distribution [see Eq.\
\req{9.102}] becomes
\beqa
\Phi_{\rm BL}(s;\Delta) &=& \frac{1}{2\pi {\rm i}}
\int_{c-{\rm i} \infty}^{c+{\rm i} \infty}
\exp\left\{ p \Delta - s
{\cal N} Z \, \frac{2\pi Z_0^2 e^4}{\me v^2} p \left[
1 - g - \ln (pW_1) \right] \right\}
\nonumber \\ [2mm]
&& \mbox{} \times \exp\left(+ \1o2 p^2 \, s \Omega^2_{\rm soft}\right)
\, \d p.
\label{9.120}\eeqa
Setting $c=0$ and $p={\rm i} x$, the distribution
$\Phi_{\rm BL}(s;\Delta)$ is expressed as a Fourier integral
\beqa
\Phi_{\rm BL}(s;\Delta) &=& \frac{1}{2\pi}
\int_{-\infty}^{\infty}
\exp\left\{ {\rm i} x \Delta - s
{\cal N} Z \, \frac{2\pi Z_0^2 e^4}{\me v^2} p \left[
1 - g - \ln (i xW_1) \right] \right\}
\nonumber \\ [2mm]
&& \mbox{} \times \exp\left(- \1o2 x^2 \, s \Omega^2_{\rm soft}\right)
\, \d x.
\label{9.121}\eeqa
Recalling that \citep{GradshteynRyzhik2007}
\beq
\frac{1}{\sqrt{2\pi}} \int_{-\infty}^\infty \exp(-{\rm i} k x) \,
\exp(-a x^2) \, \d x = \frac{1}{a \sqrt{2}} \exp(-k^2/ 4 a^2),
\nonumber\eeq
and using the convolution theorem of Fourier analysis
\citep[see,eg,][]{Arfken1985}, we can write \index{Fourier
transform!convolution theorem}
\beq
\Phi_{\rm BL}(s;\Delta) = \int_{-\infty}^\infty
\Phi_{\rm L}(s;\Delta-x) \, \frac{1}{\sqrt{2\pi s \Omega^2_{\rm
soft}}} \exp\left(- \, \frac{x^2}{2 s \Omega^2_{\rm soft}} \right)
\, \d x.
\label{9.122}\eeq
That is, the Blunck--Leisegang distribution results from the convolution of
the Landau distribution and the normal distribution with zero mean and
variance $s \Omega^2_{\rm soft}$.

The Landau and Blunck--Leisegang distributions are approximately valid
for electrons and positrons, for which the maximum value $W_{\rm max}$
of the energy loss $W$ in a single collision is equal to $E_0/2$ and
$E_0$, respectively. For heavier projectiles, $W_{\rm max}$ may be
considerably less that $E_0$ [see Eqs.\ \req{4.127} and \req{4.189}], so
that the assumption L4 is not valid and the replacement of $W_{\rm max}$
with $\infty$ in Eq.\ \req{9.96} is not permissible.
\citet{Vavilov1957} followed a method similar to Landau's to determine
the energy-straggling distribution using a simple analytical DIMFP that
vanishes for $W>W_{\rm max}$, thus removing the assumption L4. Vavilov's
distribution provides a more realistic description of energy straggling
for charged particles heavier than the electron. In particular, when
$W_{\rm max} \ll \xi$ [see Eq.\ \req{9.111}], the Vavilov distribution
becomes a normal distribution, which is the result we would expect from
the central-limit theorem [\ie, the total energy loss resulting from
multiple collisions with very small energy losses is a normal random
variate, cf.\ Eq.\ \req{9.91}]. A modification of the Vavilov
distribution to account for the straggling of soft interactions, similar
to the Blunck--Leisegang formulation, is described by
\citet{BichselSaxon1975}.


\subsection{Numerical solution of the energy-straggling equation
\label{sec9.4.4}}
\index{energy straggling!numerical distributions|(}

The energy-straggling theories of \citet{Landau1944},
\citet{BlunckLeisegang1950} and \citet{Vavilov1957} are of limited
accuracy because of the approximate nature of the underlying energy-loss
DCSs. With today's computational power, energy distributions from
realistic energy-loss DCSs can be obtained by solving the transport
equation numerically. \citet{BichselSaxon1975} describe a numerical
algorithm in which the energy-loss distribution after $n$ interactions
is obtained as the $n$-fold convolution of the single-scattering
energy-loss distribution $p_1(E;\Delta)$; the straggling distribution is
then evaluated by considering that the number $n$ of interactions along
the path length $s$ follows a Poisson distribution. This method,
however, ignores the variation of the DIMFP with the energy $E$ of the
projectile particle and, therefore, it is able to calculate only
straggling distributions for small path lengths.

The transport equation \req{9.73},
\beq
\frac{\partial  \Phi(s; E)}{\partial s}
= \int_0^{\infty} \Phi(s; E+W) \, \mu(E+W;W) \, \d W
- \Phi(s, E) \lambda^{-1}(E),
\label{9.123}\eeq
could be solved by using conventional finite-difference methods.
However, the implementation of such methods is not trivial, because
energy-loss DIMFPs vary rapidly in certain $W$ intervals. Here we
describe a numerical solution algorithm that, although approximate, is
applicable to arbitrary interaction models. This algorithm is inspired
by a more general method used by \citet{FathersRez1984} for solving the
full transport equation \req{9.60}. Our algorithm is much simpler than
theirs because straggling calculations do not involve spatial
variables.

The first stage in the formulation of the numerical algorithm is to
transform Eq.\ \req{9.123} into a system of linear equations by
discretizing the $E$ variable. We introduce a partition of the energy
interval $(0,E_0)$ into set of $N$ subintervals ($e_i, e_{i+1}$), whose
endpoints $e_i$ are such that $0=e_1 < e_2 < \cdots < e_{N} < E_0 <
e_{N+1}$ and
\beq
e_{i+1} = e_i + r^{N-i} D,
\label{9.124}\eeq
where $D$ is a constant, defined by the user, and $r$ is determined
so that
\beq
e_1=0\qquad \mbox{and} \qquad e_{N} = E_0-\1o2 D.
\label{9.125}\eeq
Such a partition exists provided only that $D < E_0/(N-0.5)$.
Note that the width of the $i$-th subinterval is
$$
H_i=e_{i+1}-e_i = r^{N-i} D,
$$
and, therefore, the grid spacing diminishes as $e_i$ increases, that is,
the energy resolution is higher near the initial energy (Fig.\
\ref{fig9.4}). The $i$-th
endpoint energy is the sum of a geometric sequence,
$$
e_i = \left( r^{N-i+1} + r^{N-i} + \cdots + r^{N-2} + r^{N-1} \right)
D = \frac{r^{N-i+1} - r^{N}}{1-r} D,
$$
that is,
\beq
e_i = \frac{1-r^{-i+1}}{r-1} \, r^{N} D.
\label{9.126}\eeq
To allow a simple physical interpretation of the solution algorithm, we
will assume that transported particles can only move with discrete
energies $E_i^{\rm d}$, and that the allowed ``energy levels'' are the
midpoints of the partition subintervals, that is,
\beq
E_i^{\rm d} = \frac{e_i + e_{i+1}}{2} = e_i + \frac{1}{2} \, r^{N-i} D
= \left( \frac{1 - r^{-i+1}}{r-1} + \frac{1}{2} \,
r^{-i} \right) r^N D.
\label{9.127}\eeq
The highest energy level is required to be $E_N^{\rm d} = E_0$ [see Eq.\
\req{9.125}], \ie, the mid point of the subinterval ($e_N,e_{N+1}$),
whose width is $D$. This requirement fixes the geometric factor $r$,
which is the root of the equation
\beq
E_N^{\rm d} = \left(  \frac{r^N - r}{r-1} + \frac{1}{2} \right) D.
\label{9.128}\eeq
Evidently, the parameter $D$ defines the ``energy resolution''
of the discrete levels near $E_N^{\rm d}$. This parameter, together
with the number $N$ of energy levels, determines the accuracy of the
numerical solution.

\begin{figure}[htb] \begin{center}
\vspace*{3mm}
\includegraphics*[width=14cm]{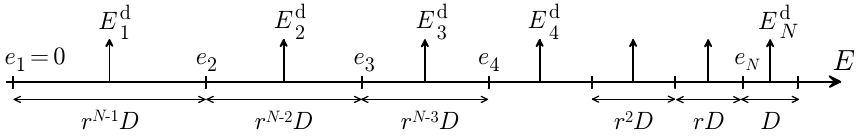}
\caption{
Partition of the energy interval $(0, E_N^{\rm d})$, showing the subinterval
endpoints $e_i$ and the discrete energy levels $E_i^{\rm d}$.
\label{fig9.4}}
\end{center}\end{figure}

Let $\Phi_i (s)$ denote the probability that a particle, after
traveling a path length $s$, is in level $E_i^{\rm d}$. This quantity
is related to the actual continuous distribution, $\Phi(s,E)$,
through
\beq
\Phi_i (s) = \int_{e_i}^{e_{i+1}}
\Phi(s,E) \, \d E \simeq  \Phi(s,E_i^{\rm d}) H_i.
\label{9.129}\eeq
The
stopping integrals on the right-hand side of Eq.\ \req{9.123} can be expressed
as follows,
\beq
\int \d E' \, \Phi (s,E') \, \mu (E';E'-E_i^{\rm d})
= \sum_{j=i+1}^{N} \, T_{i,j} \, \Phi_j (s)
\label{9.130}\eeq
with
\beq
T_{i,j} = \frac{1}{\Phi_j (s)}
\int_{e_j}^{e_{j+1}} \d E' \, \Phi (s,E') \, \mu
(E';E'-E_i^{\rm d}).
\label{9.131}\eeq
Note that the quantity $T_{ij}$ represents the probability per unit path
length that a particle in the energy level $E_j^{\rm d}$ makes a transition to
the level $E_i^{\rm d}$. As the stopping DIMFP is a rapidly varying function of
the energy loss, energy discretization may introduce quite sizeable
errors when the spacing between levels is not small enough. In
particular, the stopping power (average energy loss per unit
path length) derived from the transition probabilities,
\beq
S_{\rm disc}(E_i^{\rm d}) =
\sum_{j=0}^{i-1} \, T_{j,i} \, (E_i^{\rm d}-E_j^{\rm d}),
\label{9.132}\eeq
may differ considerably from its actual value,
\beqa
S_1(E_i^{\rm d}) &=&
\int_{0}^{E_i^{\rm d}} (E_i^{\rm d}-E') \, \mu
(E_i^{\rm d};E_i^{\rm d}-E') \, \d E'
\nonumber \\ [2mm]
&=& \sum_{j=0}^{i-1}
\int_{e_j}^{e_{j+1}} (E_i^{\rm d}-E') \, \mu
(E_i^{\rm d};E_i^{\rm d}-E') \, \d E'
\nonumber \\ [2mm]
&& +
\int_{e_{i}}^{E_i^{\rm d}} (E_i^{\rm d}-E') \, \mu
(E_i^{\rm d};E_i^{\rm d}-E') \, \d E'.
\label{9.133}\eeqa
Since the stopping power is the most relevant magnitude to ensure a
proper description of energy losses, it is convenient to define the
transition rates as
\beq
T_{j,i} \equiv \frac{1}{E_i^{\rm d}-E_j^{\rm d}} \times \left\{
\begin{array}{ll}
\displaystyle{
\int_{e_j}^{e_{j+1}} (E_i^{\rm d}-E') \, \mu
(E_i^{\rm d};E_i^{\rm d}-E') \, \d E'} \rule{10mm}{0mm} & \mbox{if $j<i-1$,} \\ [4mm]
\displaystyle{
\int_{e_{i-1}}^{E_{i}^{\rm d}} (E_i^{\rm d}-E') \, \mu
(E_i^{\rm d};E_i^{\rm d}-E') \, \d E'} & \mbox{if $j=i-1$} .
\end{array} \right.
\label{9.134}\eeq
With this definition, Eq.\ \req{9.132} gives exact values of the stopping
power for the energies $E_i^{\rm d}$, irrespective of the number of energy
levels. Moreover, in the limit where all $H_i$ are small, the transition
rates obtained from Eqs.\ \req{9.131} and \req{9.134} coincide. To be
consistent, the stopping mean free path must also be calculated from the
transition rates
\beq
\lambda_{\rm st}^{-1} (E_i^{\rm d}) = \sum_{j=0}^{i-1} \, T_{j,i}.
\label{9.135}\eeq
Stopping interactions are thus completely characterized by the
transition rates $T_{j,i}$ ($0 \le j<i$). This discrete scheme is
equivalent to a continuous-energy description with energy integrals
evaluated by the trapezoidal rule; the advantage of using discrete
energy levels is the straightforward interpretation of the transition
rates $T_{j,i}$.

With all this, the transport equation \req{9.123} can be expressed as
\beq
\frac{\d  \Phi_i (s)}{\d s} =
\sum_{j=i+1}^{N} T_{i,j} \Phi_j (s) -
\lambda_{\rm st}^{-1} (E_i^{\rm d}) \, \Phi_i (s)
\label{9.136}\eeq
or, in matrix form,
\beq
\frac{\d }{\d s} \Phib(s) = {\bf T} \Phib (s)\, ,
\label{9.137}\eeq
where $\Phib$ is an $N$-dimensional vector (column matrices) and ${\bf T}$
is an $N\times N$ matrix,
\beq
\Phib (s) =
\left( \begin{array}{c}
\Phi_1 (s) \\
\vdots \\
\Phi_N (s)
\end{array} \right),
\qquad
{\bf T} \equiv \left( \begin{array}{cccc}
-\lambda^{-1}_{\rm st} (E_1^{\rm d}) & T_{1,2} & T_{1,3} & \ldots \\
0 & -\lambda^{-1}_{\rm st} (E_2^{\rm d}) & T_{2,3} & \ldots \\
0 & 0 & -\lambda^{-1}_{\rm st} (E_3^{\rm d})  & \ldots \\
\vdots & \vdots & \vdots & \ddots  \end{array} \right).
\label{9.138}\eeq
Since $E_0 = E_{N}$, the boundary condition \req{9.74} becomes
\beq
\Phi_i(0) =  \delta_{i,N}.
\label{9.139}\eeq
The discrete energy distribution after a path length $s$
can now be obtained as
\beq
\Phib(s) = \exp({\bf T} s) \, \Phib(0).
\label{9.140}\eeq
When $s$ is much less than the value $\Lambda \equiv
{\rm min}\{\lambda_{\rm st} (E_i^{\rm d}); i=1, 2, \ldots, N \}$,
the matrix $\exp(s{\bf T})$ can be evaluated from the
Taylor expansion,
\beq
\exp(s{\bf T}) = I + s{\bf T} + \frac{1}{2!} (s{\bf T})^2 +
\frac{1}{3!} (s{\bf T})^3 + \cdots
\label{9.141}\eeq
Otherwise, we can consider a path length $s'=2^{-n} s$ such that $s'\sim
0.1 \Lambda$, evaluate $\exp(s'{\bf T})$ from its Taylor expansion,
which converges rapidly, and obtain $\exp(s{\bf T})$ by squaring $n$
times the matrix $\exp(s'{\bf T})$. This calculation is trivial to
program and runs quite fast, even on modest personal computers and for
energy grids of relative large dimensions, with $N\sim 2,000$ levels. To
compare with experimental or simulated continuous energy distributions,
the discrete distribution $\Phi_i(s)$ may be transformed into a
histogram with bars of width $H_i$ and height $\Phi_i (s)/H_i$.

\begin{figure}[hp!] \begin{center}
\vspace*{3mm}
\includegraphics*[width=14cm]{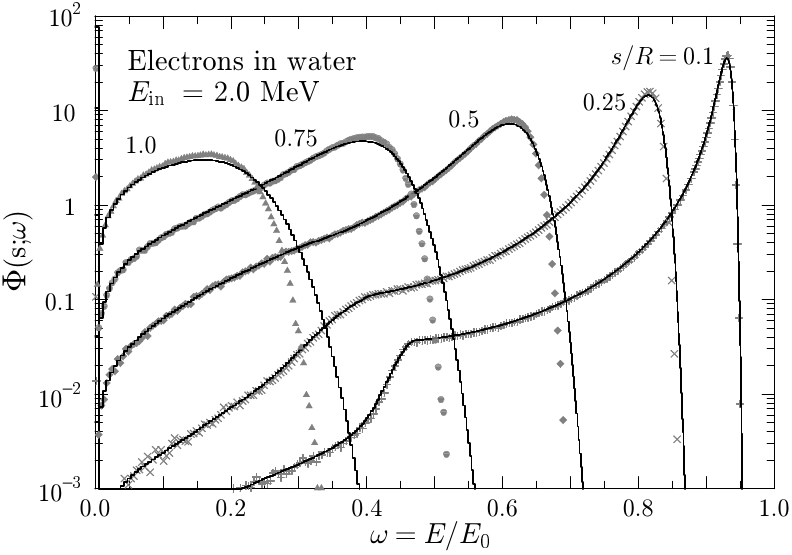} \\ [5mm]
\includegraphics*[width=14cm]{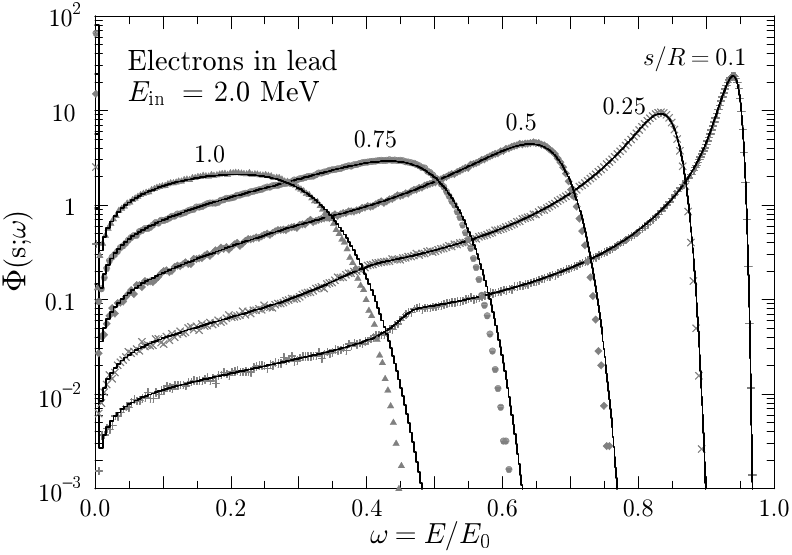}
\caption{
Energy distributions of 2 MeV electrons in water and lead after
traveling the indicated path lengths $s$, expressed in units of the CSDA
range $R$.  Symbols represent results from {\sc penelope} simulations.
Histograms are distributions calculated with the numerical algorithm
described in the text, with $D=50$ eV and $N=1,000$ energy levels.
\label{fig9.5}}
\end{center}\end{figure}

This numerical algorithm has been employed to calculate energy
distributions of electrons and positrons using the energy-loss DCSs
implemented in the Monte Carlo code {\sc penelope} \citep{Salvat2025}.
An obvious advantage of this algorithm, as compared with the usual
energy-straggling theories, is that it accounts for energy losses caused
by both inelastic collisions and radiative events. Figure \ref{fig9.5}
compares calculated energy distributions of electrons in water and lead
with results from simulations by {\sc penelope} of exactly the same
processes. Apart from statistical uncertainties, the {\sc penelope}
results represent ``exact'' energy distributions. The discrete
distributions were calculated with $D=50$ eV and $N=1,000$ energy
levels. The calculation of the transition probabilities $T_{j,i}$ for
lead took about 3 minutes on a Intel Core i7M 2~GHz processor; the
evaluation of the distribution $\Phi_i(s)$ from Eq.\ \req{9.140} was
performed in less than 3 seconds. The numerical distributions agree
reasonably with the simulation results; only slight differences are seen
in the high-energy part of the distributions, reflecting the distortion
of the low-energy-loss portion of the energy-loss DCS introduced by
energy discretization. It is worth mentioning that the calculated
distributions give average energies that agree almost exactly (again,
apart from statistical uncertainties) with the values resulting from the
simulation. In any case, this simple numerical algorithm is far more
accurate than conventional energy-straggling theories, and accounts for
the totality of energy-loss processes. Therefore, it could be adopted in
high-energy Monte Carlo transport codes for describing energy-loss
events with a class I (complete grouping) scheme. In this case, only
distributions for relatively small path lengths, say $s\sim 0.1 R$,
would be required; and it is at these moderate path lengths where the
numerical algorithm is more accurate.  \index{energy
straggling!numerical distributions|)}

\index{energy straggling|)}


\vspace*{-3mm}

\section{Multiple scattering \label{sec9.5}}

\index{multiple scattering|(}

A basic ingredient of condensed Monte Carlo algorithms is the angular
distribution of charged particles after traveling a certain path length
$s$ in an infinite, homogeneous medium. This distribution can be
obtained from various multiple-scattering theories with different
degrees of complexity. We shall start from a somewhat simplified
situation in which fast charged particles with kinetic energy $E$ travel
a path length $s$ which is so short that their average energy loss is
negligible compared to $E$. This is the framework of the multiple
scattering theories of \citet{GoudsmitSaunderson1940} and
\citet{Moliere1948}. After describing these theories, we will present
the Fermi--Eyges small-angle approach and the elaborate formulation of
\citet{Lewis1950}, which combine multiple scattering with energy-loss
processes treated within the CSDA.

Angular deflections of the trajectories of transported particles are
caused by elastic collisions and, to a lesser extent, by inelastic
interactions. We consider that particles are not deflected in radiative
(bremsstrahlung emission) events. The effective scattering DCS includes
contributions from elastic and inelastic collisions,
\beq
\frac{\d \sigma}{\d \Omega} =
\frac{\d \sigma_{\rm el}}{\d \Omega} + \int
\frac{\d^2 \sigma_{\rm in}}{\d W \, \d \Omega} \, \d W,
\label{9.142}\eeq
and, for isotropic
materials, it is a function of only the polar scattering angle $\theta$.
The mean free path $\lambda$ between scattering events is given by
\beq
\lambda = 1/({\cal N}\sigma),
\label{9.143}\eeq
where
\beq
\sigma \equiv \int
\frac{\d \sigma}{\d \Omega} \, \d \Omega
\label{9.144}\eeq
is the total scattering cross section. If particles move initially along
the direction $\hat{\bf z}$ of the $z$ axis, the angular
distribution $f_{1}(\theta)$ after a single scattering event is
\beq
f_{1}(\theta) = \frac{1}{\sigma} \frac{\d\sigma}{\d \Omega}.
\label{9.145} \eeq
Note that the probability of scattering into directions within the solid
angle element $\d \Omegab$ about the direction $\Omegab=(\theta,\phi)$
is equal to $f_{1}(\theta) \, \d\Omegab$. It is convenient to write
$f_{1}(\theta)$ in the form of a Legendre series \citep[see,
\eg,][]{Arfken1985}
\beq
f_{1}(\theta) = \sum_{\ell=0}^{\infty} \frac{2\ell+1}{4\pi} F_{\ell}
\, P_{\ell}(\cos\theta),
\label{9.146} \eeq
where $P_{\ell}(\cos\theta)$ are the Legendre polynomials
(Appendix \ref{appB}) and
\beq
F_{\ell} \equiv 2\pi \int_{-1}^{1} P_{\ell}(\cos\theta) f_{1}(\theta)
\, \d(\cos\theta) = \langle P_\ell (\cos\theta) \rangle_1,
\label{9.147} \eeq
where the subscript ``1'' indicates values for a single
collision.

The derivations that follow are easier when the single-scattering
angular distribution is expressed in terms of the spherical harmonics,
$Y_{lm}(\theta,\phi)$ (Section \ref{appB.2}).
Using the relation \req{B.56b}, the single-scattering angular
distribution can be expressed as
\beq
f_{1}(\cos\theta) =
f_{1}(\hat{\bf z} \rightarrow \Omegab) =
\sum_{\ell} \sqrt{\frac{2\ell+1}{4\pi}} F_{\ell}
\, Y_{\ell 0}(\Omegab).
\label{9.148} \eeq
Because the angular distribution is axially symmetric, the summation
contains only terms with $m=0$.

Now, let us consider a particle that is moving in a direction $\Omegab_0
= (\theta_0,\phi_0)$ before scattering, and let $\theta'$ and $\phi'$
denote the polar and azimuthal scattering angles in the collision. The
angular distribution after the collision is given by an expansion like
\req{9.148}, but with the axes of the reference frame rotated so that
the polar axis $\hat{\bf z}$ points along the direction $\Omegab_0$. The
addition theorem of spherical harmonics [Eq.\ \req{B.57}] implies that
\beq
Y_{\ell 0}(\Omegab') =
\sqrt{\frac{4\pi}{2\ell+1}}
\sum_m Y_{lm}^\ast(\Omegab_0) \, Y_{\ell m} (\Omegab),
\label{9.149}\eeq
where $\Omegab'=(\theta', \phi')$. Hence, the angular distribution after
the collision is given by
\beq
f_{1}(\Omegab_0 \rightarrow \Omegab) =
\sum_{\ell} \sqrt{\frac{2\ell+1}{4\pi}} F_{\ell}
\, Y_{\ell 0}(\Omegab')
= \sum_{\ell,m} F_{\ell}
\; Y_{lm}^\ast(\Omegab_0) \, Y_{\ell m} (\Omegab).
\label{9.150}\eeq
The property \req{B.56c} implies that when $\Omegab_0 = \hat{\bf z}$
this expression reduces to the form \req{9.148}. Thus, the DIMFP of a
particle that moves in the direction $\Omegab_0$ is given by
\beq
\mu(\Omegab_0 \rightarrow \Omegab)
= \frac{1}{\lambda} \, f_{1}(\Omegab_0 \rightarrow \Omegab)
= \frac{1}{\lambda} \, \sum_{\ell,m} F_{\ell}
\; Y_{lm}^\ast(\Omegab_0) \, Y_{\ell m} (\Omegab).
\label{9.151}\eeq
Since the coefficients $F_\ell$ are real, the property \req{B.56a}
implies that
\beq
\mu(\Omegab_0 \rightarrow \Omegab) = \mu(\Omegab \rightarrow \Omegab_0),
\label{9.152}\eeq
reflecting the fact that the DCS depends only on the cosine of the
scattering angle, $\cos\theta'= \Omegab_0\cdot \Omegab$.


\subsection{The folding theorem \label{sec9.5.1}}
\index{folding theorem}
The multiple-scattering theory of Goudsmit and Saunderson is based on a
fundamental property of Legendre functions, the so-called {\it folding
theorem}, which allows expressing the angular distribution after a given
number of collisions in a simple form. We start by giving a simple
derivation of this theorem.

Let us assume that particles move in a medium where they can undergo
scattering events of two different types, say $a$ and $b$, with single
scattering angular distributions $f_1^{(a)}(\Omegab)$ and
$f_1^{(b)}(\Omegab)$ respectively. For instance, these distributions may
correspond to scattering by two atoms of different elements. Now
consider that a particle, which initially moves in the direction of the
$z$ axis, $\hat{\bf z} = (\theta=0,\phi)$, suffers a collision of type
$a$ followed by a collision of type $b$.

After the first collision, the probability of finding
the particle moving in the direction $\Omegab_1$ is given by
\beq
f_1^{(a)}(\hat{\bf z} \rightarrow \Omegab_1) = \sum_{\ell}
\sqrt{\frac{2\ell+1}{4\pi}} F^{(a)}_{\ell}
\, Y_{\ell 0}(\Omegab_1).
\label{9.153} \eeq
A particle that is scattered in the direction
$\Omegab_1$, flies in this direction until the collision $b$
takes place, and the angular distribution of the
particle after the $b$ scattering is
\beq
f_1^{(b)}(\Omegab_1 \rightarrow \Omegab) =
\sum_{\ell, m } F^{(b)}_{\ell}
Y_{lm}^\ast(\Omegab_1) \, Y_{\ell m} (\Omegab).
\label{9.154} \eeq
The angular distribution of particles after undergoing the two collisions is
obtained by integrating over possible intermediate directions $(\Omegab_1)$,
\beqa
&& \! \! \! \! \! \! \! \! \! \! \!
f_2^{(a,b)}(\Omegab) = \int \d \Omegab_1 \;
f_1^{(a)}(\hat{\bf z} \rightarrow \Omegab_1) \,
f_1^{(b)}(\Omegab_1 \rightarrow \Omegab)
\nonumber \\ [2mm]
&=& \int \d \Omegab_1
\left[ \sum_{\ell'}
\sqrt{\frac{2\ell'+1}{4\pi}} F^{(a)}_{\ell'}
\, Y_{\ell' 0}(\Omegab_1)\right]
\left[ \sum_{\ell, m } F^{(b)}_{\ell}
Y_{lm}^\ast(\Omegab_1)
Y_{\ell m} (\Omegab) \right].
\nonumber \eeqa
Using the orthogonality of the spherical harmonics, we obtain
\beq
f_2^{(a,b)}(\Omegab) = \sum_{\ell}
\sqrt{\frac{2\ell+1}{4\pi}} F^{(a)}_{\ell}
F^{(b)}_{\ell} Y_{\ell 0} (\Omegab)\, .
\label{9.155}\eeq
The distribution $f_2^{(a,b)}$ is said to be the result of ``folding''
the distributions $f_1^{(a)}$ and $f_1^{(b)}$. This operation is usually
denoted by the symbol $\otimes$,
$$
f_2^{(a,b)} \equiv f_1^{(a)} \otimes f_1^{(b)}.
$$
The result \req{9.155} constitutes the {\it folding theorem}. It
follows from the derivation that the order in which the collisions $a$
and $b$ occur is irrelevant, \ie, $f_1^{(a)}\otimes
f_1^{(b)}=f_1^{(b)}\otimes f_1^{(a)}$.


\subsection{The Goudsmit--Saunderson distribution \label{sec9.5.2}}

\index{multiple scattering!Goudsmit--Saunderson distribution}
\index{Goudsmit--Saunderson distribution}

Let us now consider that particles start off from a certain position,
which we select as the origin of our reference frame, moving in the
direction of the $z$ axis. Considering the Legendre expansion of the
single-scattering distribution given by Eq.\ \req{9.148}, it follows from
the folding theorem that the angular distribution after exactly $n$
scattering events is
\beq
f_{n}(\Omegab) = \sum_{\ell=0}^{\infty} \sqrt{\frac{2\ell+1}{4\pi}}
(F_{\ell})^{n} Y_{\ell 0} (\Omegab).
\label{9.156} \eeq
We recall that the inverse mean free path $\lambda^{-1}={\cal N}\sigma$
gives the
collision probability per unit path length. Since the energy is assumed
to remain constant, the
probability distribution of the number $n$ of collisions after a path
length $s$ is Poissonian with mean $\langle n \rangle = s/\lambda$,
\ie,
\beq
P(n) = \exp(-s/\lambda) \frac{(s/\lambda)^{n}}{n!}.
\label{9.157} \eeq
The angular distribution after a path length $s$ can then be expressed
as
\beqa
\Phi_{\rm GS}(s;\Omegab) &=& \sum_{n=0}^{\infty} P(n)
f_{n}(\Omegab) = \sum_{\ell=0}^{\infty}
\sqrt{\frac{2\ell+1}{4\pi}} \left[ \exp(-s/\lambda) \sum_{n=0}^{\infty}
\frac{(s/\lambda)^{n}}{n!} (F_{\ell})^{n} \right] Y_{\ell 0}(\Omegab)
\nonumber \\ [2mm]
&=& \sum_{\ell=0}^{\infty}
\sqrt{\frac{2\ell+1}{4\pi}} \left[ \exp(-s/\lambda)
\exp(sF_\ell/\lambda) \rule{0mm}{4mm}\right] Y_{\ell 0}(\Omegab).
\nonumber \eeqa

Introducing the transport mean free paths $\lambda_\ell$ defined by
\beq
\lambda_\ell^{-1} \equiv \frac{1-F_\ell}{\lambda} =
\frac{2\pi}{\lambda} \int_{-1}^{1} \left[1-P_{\ell}(\cos\theta)
\right] f_{1}(\theta) \, \d(\cos\theta),
\label{9.158}\eeq
the multiple-scattering distribution can be finally written in the form
\beq
\Phi_{\rm GS}(s;\Omegab) \equiv \Phi_{\rm GS}(s;\theta)
= \sum_{\ell=0}^{\infty}
\frac{2\ell+1}{4\pi} \exp(-s /\lambda_\ell) P_{\ell}(\cos\theta),
\label{9.159} \eeq
which is the expansion derived by \citet{GoudsmitSaunderson1940}.
Because the single-scattering DCSs are assumed to be independent of the
azimuthal angle $\phi$, the multiple-scattering distribution is axially
symmetric, \ie, it depends only on the polar angle $\theta$ of the final
direction.  The value of $F_{\ell}$ [see Eq.\ \req{9.147}] decreases
with $\ell$ owing to the increasingly faster oscillations of
$P_{\ell}(\cos\theta)$ and, hence, $\lambda_{\ell}^{-1}$ increases
monotonically with $\ell$ and tends to $\lambda^{-1}$ when $\ell$ goes
to infinity.

Figure \ref{fig9.6} (a) shows Goudsmit--Saunderson distributions of 15
MeV electrons in lead for different path lengths $s$, given as
multiples of the mean free path $\lambda$. In these calculations we used
the elastic DCS obtained by means of the partial-wave expansion method
with approximate phase-shifts for the DHFS potential, as described in
Section \ref{sec5.2.2}. These distributions are not very realistic
because we are neglecting the deflections caused by inelastic collisions
as well as the energy loss of the electrons along the path length. We
see that the angular distribution widens progressively when the number
of collisions $\langle n \rangle = s/\lambda$ increases. 

\begin{figure}[htb] \begin{center}
\includegraphics*[width=7cm]{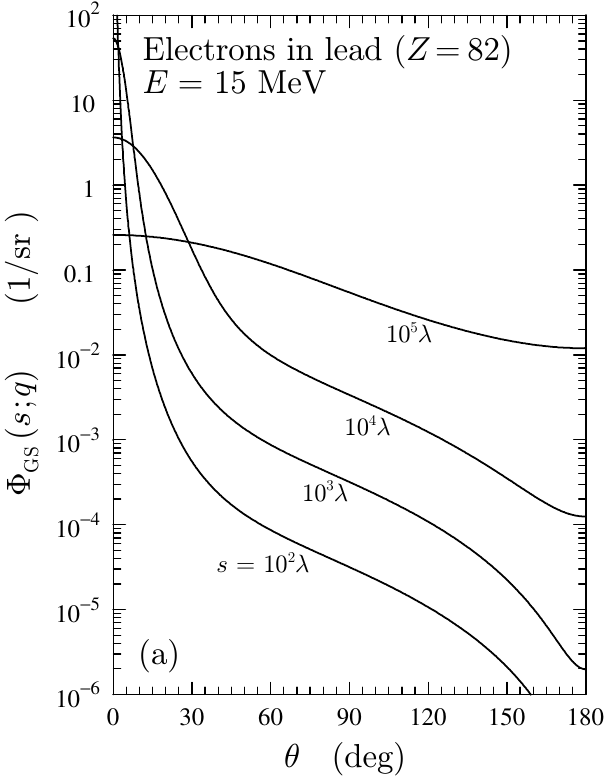} \rule{6mm}{0mm}
\includegraphics*[width=7cm]{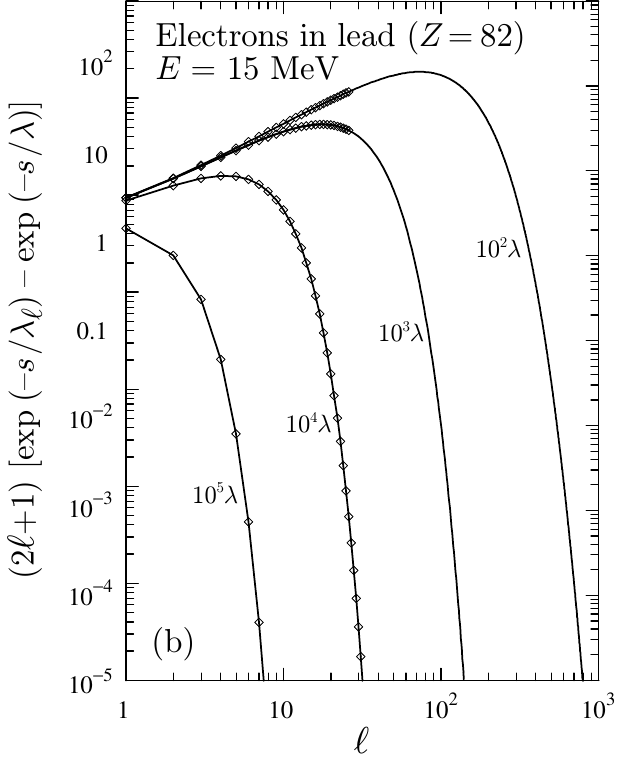}
\caption{
(a) Goudsmit--Saunderson angular distributions, Eq.\ \req{9.159} with
	the calculated DCS for the DHFS analytical potential
	\citep{Salvat1987}, of 15 MeV electrons in lead for the indicated path
	lengths $s$, expressed in units of the mean free path $\lambda$. (b)
	Coefficients $(2\ell+1)$ $\times [ \exp(-s/ \lambda_\ell)-\exp(-s/
	\lambda)]$ of the Goudsmit--Saunderson series, Eq.\ \req{9.167}, for
	the same path lengths.
\label{fig9.6}}
\end{center} \end{figure}

It is common to consider the $\ell$-th {\it transport cross section},
defined by
\beq
\sigma_\ell \equiv \sigma \, 2\pi
\int_{-1}^{1} \left[1-P_{\ell}(\cos\theta)
\right] f_{1}(\theta) \, \d(\cos\theta) =
\int_{-1}^{1} \left[1-P_{\ell}(\cos\theta)
\right] \frac{\d \sigma}{\d \Omega} \, \d \Omegab\, ,
\label{9.160}\eeq
so that $\lambda_\ell^{-1} ={\cal N} \sigma_\ell$. Introducing the
expressions \req{B.41} of the Legendre Polynomials, we have
\beqa
\sigma_{1} &=& 2\pi \int_{-1}^{1}
\left[1-\cos\theta
\right] f_{1}(\theta) \, \d(\cos\theta) =
\langle 1- \cos\theta \rangle_1,
\nonumber \\ [2mm]
\lambda_{1}^{-1} &=& \frac{\langle 1- \cos\theta \rangle_1}{\lambda},
\label{9.161} \eeqa
and
\beqa
\sigma_{2} &=& 2\pi \int_{-1}^{1}
\left[1-\frac{1}{2} \left( 3 \cos^2 \theta - 1 \right) \right]
f_{1}(\theta) \, \d(\cos\theta) = \frac{3}{2}
\langle 1- \cos^2\theta \rangle_1,
\nonumber \\ [2mm]
\lambda_{2}^{-1} &=& \frac{3
\langle 1- \cos^2\theta \rangle_1}{2 \lambda}.
\label{9.162} \eeqa
$\lambda_{1}^{-1}$ gives a measure of the average angular deflection per
unit path length. By analogy with the stopping power, which is the mean
energy loss per unit path length [Eq.\ \req{9.85}], the quantity $2
\lambda_{1}^{-1}$ is sometimes called the {\it scattering
power}\footnote{When small angles dominate, $\left< 1 - \cos\theta
\right>_1 \simeq
\left< \theta^2 \right>_1/2$ and $2 \lambda_{1}^{-1} \simeq
\left< \theta^2 \right>_1/ \lambda$.}. Notice that $\sigma_1$
is the momentum-transfer cross section [see Eq.\ \req{4.205}].

From the orthogonality of the Legendre polynomials, it follows that
\beq
\langle P_{\ell}(\cos\theta) \rangle_{\rm GS} \equiv 2\pi \int_{-1}^{1}
P_{\ell}(\cos\theta) \Phi_{\rm GS}(\theta;s) \, \d(\cos\theta) =
\exp(-s/\lambda_{\ell}).
\label{9.163} \eeq
For $\ell = 0$ this formula gives
\beq
2\pi \int_{-1}^{1}
\Phi_{\rm GS}(s;\theta) \, \d(\cos\theta) = 1,
\label{9.164}\eeq
\ie, the probability density function is normalized to
unity. For $\ell =1$ and 2, Eq.\ \req{9.163} implies that
\begin{subequations} \label{9.165}
\beq
\langle \cos\theta \rangle_{\rm GS} = \exp(-s / \lambda_{1})
\label{9.165a} \eeq
and
\beq
\langle \cos^{2}\theta \rangle_{\rm GS} =
\frac{1}{3} \left[ 1+2\exp(-s / \lambda_{2}) \right],
\label{9.165b} \eeq
\end{subequations}
respectively

The Goudsmit--Saunderson expansion \req{9.159} and the results
\req{9.163} and \req{9.165} are exact. To compute these quantities for
a given single-scattering DCS, which usually is available only in
numerical form, we have to evaluate the transport coefficients
\req{9.147} and the transport mean free paths $\lambda_{\ell}$, Eq.\
\req{9.158}. The number of terms needed to get convergence of the
Goudsmit--Saunderson series increases as the path length decreases. In
the case of small path lengths, the convergence of the series can be
improved by separating the contribution from particles that have had no
collisions. The angular distribution of unscattered
particles is
\begin{subequations} \label{9.166}
\beq
\frac{\delta(\cos\theta-1)}{\pi}
= \sum_{\ell=0}^\infty \frac{2\ell+1}{2} \, g_\ell
\, P_\ell (\cos\theta)
\label{9.166a}\eeq
with
\beq
g_\ell = \frac{1}{\pi}
\int_{-1}^1 \delta(\cos\theta-1) \, P_\ell(\cos\theta) \, \d
(\cos\theta) = \frac{1}{2\pi},
\label{9.166b}\eeq
\end{subequations}
where we have taken into account that the delta function is at the
endpoint of the interval $[-1,1]$ of the variable $\cos\theta$ [see Eq.\
\req{B.16}]. Hence, we can write
\beqa
\Phi_{\rm GS}(s; \Omegab) &\equiv& \exp(-s/\lambda)
\frac{\delta(\cos\theta-1)}{\pi}
\nonumber \\ [2mm]
&& + \sum_{\ell=0}^{\infty}
\frac{2\ell+1}{4\pi} \left[ \exp(-s/\lambda_{\ell})
- \exp(-s/\lambda) \rule{0mm}{4mm}\right]
P_{\ell}(\cos\theta). \rule{10mm}{0mm}
\label{9.167} \eeqa
Figure \ref{fig9.6} (b) displays the coefficients $(2\ell +1)
[\exp(-s/\lambda_\ell) - \exp(-s/\lambda)]$ of the Legendre series, Eq.\
\req{9.167}, for the same path lengths as in Fig.\ \ref{fig9.6} (a)
(notice that for $\ell =0$, the coefficient is equal to unity). It is
clear that the convergence of the series \req{9.167} improves rapidly
when $s$ increases. In the limit $s\rightarrow \infty$, the first term
in Eq.\ \req{9.167} vanishes, only the term $\ell=0$ contributes to
the summation and the distribution becomes isotropic, $\Phi_{\rm
GS}(s;\theta) \rightarrow (4\pi)^{-1}$.


\subsubsection{Calculation of the Legendre coefficients
\label{sec9.5.2.1}}

\index{computer code!{\sc gosan}}

The Legendre coefficients
\beq
F_\ell =
2\pi \int_{-1}^{1} P_{\ell}(\cos\theta) \,
f_{1}(\theta) \, \d(\cos\theta)
\label{9.168}\eeq
can be calculated by using the numerical algorithm described by
\citet{Negreanu2005}, which is implemented in the Fortran code {\sc
gosan} (see Section \ref{sec10.1}). The DCS is specified by means of a
table with a suitably spaced grid of angles, and its value at any angle
is evaluated by natural cubic spline interpolation (see Section
\ref{sec10.4.2}). The single-scattering distribution $f_1(\theta)$ is
considered as a function of the variable $y=\cos\theta$.  A partition of
the integration interval, $y_1=-1 < y_2 < \ldots < y_{n} <y_{n+1} = 1$,
is established such that, within each subinterval ($y_j$, $y_{j+1}$) the
DCS varies by less than a factor of about 9.
Then,
\beq
F_\ell = 2\pi \sum_{j=1}^n \int_{y_j}^{y_{j+1}}
P_{\ell}(y) \, f_{1}(y) \,
\d y,
\label{9.169}\eeq
and the integrals on the right-hand side involve only Legendre
polynomials and a slowly varying function. Therefore, they can be
evaluated to very high accuracy using the $N$-point Gauss--Legendre
quadrature formula \citep{AbramowitzStegun1974}
\beq
\int_{a}^{b} f(y) \, \d y =
\frac{b-a}{2} \, \sum_{i=1}^{N} \, w(x_i) \,
f(y_i) \quad \mbox{with} \quad
y_i = \frac{b+a}{2} + \frac{b-a}{2} x_i ,
\label{9.170}\eeq
where the abscissas $x_i$ ($-1<x_i<1$) are the $N$ zeros of the Legendre
polynomial $P_{N}(x)$ and the corresponding weights $w(x_i)$ are
\beq
w(x_i) = \frac{2}{(1-x_i^2)(P'_{N}(x_i))^2}.
\label{9.171}\eeq
The difference $R_N(f)$ between the real value of the integral and the
result from the formula \req{9.170} is
\beq
R_N(f) = \frac{(b-a)^{(2N+1)} \, (N!)^4}{(2N+1) [(2N)!]^3} \, f^{(2N)} (\xi),
\label{9.172}\eeq
where $f^{(2N)}(y)$ denotes the $2N$-th derivative of the integrand and
$\xi$ is a point inside the interval $[a,b]$. Notice that the formula
yields the exact values of the integrals of polynomials with degrees up
to $2N-1$.

\index{Legendre polynomials!recurrence relation}

The key feature here is that the abscissas and weights of the
Gauss--Legendre formula can be calculated rapidly and accurately for
arbitrary $N$. In our calculations we use an adapted version of the
subroutine {\tt gauleg} of {\it Numerical Recipes} \citep{Press1992},
which delivers the abscissas and weights of the $N$-point
Gauss--Legendre formula, for a given value of $N$. The number $N_{\rm
L}$ of Legendre coefficients that are to be calculated is selected in
advance; the maximum number of coefficients that can be considered in
the program {\sc gosan} is $N_{\rm L}^{\rm max} = 15,000$ (although this
value can be increased by editing the source file). The
number $N$ of points in the quadrature formula is determined as
\beq
N = \max \left\{500, \min (N_{\rm L}/2, N_{\rm L}^{\rm max}/2)+10
\right\}.
\label{9.173}\eeq
Assuming that the function $f_1(y)$ can be accurately approximated by a
polynomial of degree 20 or less within each subinterval, the integrands
for the coefficient $F_\ell$ of highest order $\ell = N_{\rm L}$ are
polynomials of degree less than $2N$ and, therefore, the Gauss-Legendre
formula should essentially yield exact values for all the coefficients
with $\ell \le N_{\rm L}$. Moreover, for the evaluation of these
coefficients we only need the values of the single-scattering
distribution $f_1(y)$ and of the polynomials $P_\ell (y)$ at the points
$y_i$ corresponding to the abscissas $x_i$. The Legendre polynomials are
generated by using their recurrence relation
\citep{AbramowitzStegun1974}
\beqa
&& P_0(x)=1, \quad P_1(x)=x
\nonumber \\ [2mm]
&& P_{\ell}(x) = \frac{1}{\ell} \left[
(2\ell-1) x P_{\ell-1}(x) - (\ell-1) P_{\ell -2}(x) \right].
\label{9.174}\eeqa
Thus, all the Legendre coefficients $F_\ell$ with $\ell \le N_{\rm L}$ can be
calculated {\it simultaneously}; the algorithm can be coded in a very
compact form, as three nested {\sc do} loops with a small number of
arithmetic operations. Notice that $F_0 = 1$ and that the value of
$F_\ell$ decreases monotonically with $\ell$ tending to zero for large
$\ell$ because of the rapid oscillations of the Legendre polynomials.
Because the single-scattering distribution $f_1(\theta)$ is sharply peaked
at $\theta=0$, the coefficients $F_\ell$ are all positive. In
practice, the number of coefficients that can be calculated
accurately is limited by the accumulated errors of the integration
algorithm, which manifest by giving negative values to some coefficients
with large $\ell$. The program considers the occurrence of a negative
coefficient, say with $\ell=\ell_{\rm max}+1$, as an indication of
relevant numerical errors and terminates the summation of the Legendre
series.

The accuracy of the calculated $F_{\ell}$ coefficients can be readily
verified by comparing the original DCS and the DCS ``reconstructed'' by
summing the Legendre expansion,
\beq
\frac{\d \sigma}{\d \Omega} = \sigma \; \sum_{\ell=0}^{\ell_{\rm max}}
\frac{2\ell+1}{4\pi} F_{\ell} \, P_{\ell}(\cos\theta).
\label{9.175}\eeq
With the DCSs obtained from the partial-wave expansion method for
electrons and positrons (Section \ref{sec5.2}), the relative differences
between the original DCS and the reconstructed DCS with $N_{\rm L}
\simeq 5000$ terms are negligible at small angles; they are only
appreciable for electrons with energies larger than about 1 MeV and at
relatively large angles \citep{Negreanu2005}. These differences can be
reduced by adding more terms in the Legendre expansion. It is worth
noticing that the single-scattering distribution is the most
unfavourable case; the Goudsmit--Saunderson series converge faster (\ie,
with a smaller number of terms) when the path length increases. The
program {\sc gosan} computes the multiple-scattering angular distribution
as [see Eq.\ \req{9.167}]
\beqa
\Phi_{\rm GS}(s; \Omegab) &\equiv& \exp(-n_{\rm av})
\frac{\delta(\cos\theta-1)}{\pi}
\nonumber \\ [2mm]
&& + \sum_{\ell=0}^{\ell_{\rm max}}
\frac{2\ell+1}{4\pi} \left\{ \exp[- n_{\rm av} (1-F_\ell)]
\rule{0mm}{4mm}
- \exp(-n_{\rm av}) \right\}
P_{\ell}(\cos\theta), \rule{15mm}{0mm}
\label{9.176} \eeqa
where $n_{\rm av} = s/\lambda$ is the average number of collisions in
the path length $s$. This result shows that the multiple-scattering
distribution is determined by the transport coefficients
\beq
G_\ell \equiv 1 - F_\ell =
2\pi \int_{-1}^{1} \left[1-P_{\ell}(\cos\theta)
\right] f_{1} (\theta) \, \d(\cos\theta).
\label{9.177}\eeq


\subsection{The Wentzel DCS model \label{sec9.5.3}}
\index{multiple scattering!Goudsmit--Saunderson distribution!Wentzel
model}
\index{Wentzel DCS}

In general, the inverse transport mean free paths $\lambda_\ell$, eq.\
\req{9.158} must be calculated numerically, and because of the fast
oscillations of Legendre polynomials of large orders, the calculation is
far from trivial. We shall consider here the Goudsmit--Saunderson
distribution for the following DCS model,
\beq
\frac{\d \sigma^{\rm (W)}(\theta)}{\d \Omega} = \sigma^{\rm (W)}
\frac{A(1+A)}{\pi (2A + 1 - \cos\theta)^2},
\label{9.178}\eeq
because the corresponding inverse transport mean free paths are given by
an analytical formula, which can be used to check the accuracy of
numerical algorithms used for calculation of the $\lambda_\ell^{-1}$
values for arbitrary DCSs.  Notice that the DCS  \req{9.178} has the
same angular dependence as the Wentzel DCS, Eq.\ \req{5.56}, which
describes the scattering by a screened Coulomb potential [$= Z_1 Z e^2
\exp(-\alpha r)/r$] within the plane-wave Born approximation (Section
\ref{sec5.1.2}). Here, however, $\sigma^{\rm (W)}$ and $A$ are
considered as adjustable parameters that may be determined, \eg, by
fitting a realistic DCS obtained from partial-wave calculations.

The factor
\beq
f_1^{\rm (W)} (\theta) = \frac{A(1+A)}{\pi (2A + 1 - \cos\theta)^2}
\label{9.179}\eeq
is the normalized angular distribution for single scattering and,
therefore, the total cross section is equal to $\sigma^{\rm (W)}$. The
transport coefficients for the Wentzel DCS \req{9.178} are
\beqa
G_\ell^{\rm (W)} &=&
2\pi \int_{-1}^{1} \left[1-P_{\ell}(\cos\theta)
\right] f_{1}^{\rm (W)} (\theta) \, \d(\cos\theta)
\nonumber \\ [2mm]
&=& 1 - 2 A(1+A) \int_{-1}^{1}
\frac{P_{\ell}(\cos\theta)}{(2A+1-\cos\theta)^2}\,
\d(\cos\theta).
\nonumber\eeqa
Using the formula (7.228) of \citet{GradshteynRyzhik2007} and the
properties of the Legendre functions, we obtain the result
\index{Legendre functions of the second kind}
\beq
G_\ell^{\rm (W)} =
1 - \ell \left[ Q_{\ell-1}(1+2A) - (1+2A) Q_\ell (1+2A) \right],
\label{9.180}\eeq
where $Q_\ell(x)$ are Legendre functions of the second kind
\citep{AbramowitzStegun1974}. These functions can be calculated
easily from their recursion relation [see Eqs.\ \req{5.62} to
\req{5.65}]. The first and second transport coefficients are
\beq
G_1^{\rm (W)} = 2A
\left[ (1+A) \ln \left( \frac{1+A}{A} \right) -1 \right]
\label{9.181}\eeq
and
\beq
G_2^{\rm (W)} = 6A (1+A)
\left[ (1+2A) \ln \left( \frac{1+A}{A} \right) -2 \right].
\label{9.182}\eeq

Given a realistic DCS, with total cross section $\sigma$ and first
transport cross section $\sigma_1$, we can set $\sigma^{\rm (W)}=\sigma$
and determine the ``screening'' parameter $A$ by requiring that
$\lambda_1^{\rm (W)} = 1/({\cal N} \sigma_1)$. The Wentzel DCS, Eq.\
\req{9.178}, then describes an elastic-scattering process that has the
same mean free path and the same single-scattering average deflection as
the real process. Because the shape of the Wentzel DCS is physically
plausible, it yields multiple-scattering distributions that resemble
those of the realistic DCS. The resemblance improves when the path length
increases. For relatively large path lengths, when the average number of
interactions is larger than about 20 (the multiple-scattering regime),
repeated scattering smears out the details of the DCS and the angular
distributions obtained from the realistic DCS and the Wentzel DCS agree
closely.


\subsection{The Moli\`{e}re multiple-scattering theory
\label{sec9.5.4}}
\index{multiple scattering!Moli\`{e}re distribution|(}
\index{Moli\`{e}re distribution}
\citet{Moliere1948} \citep[see also][]{Bethe1953} derived an approximate
analytical expression for the multiple-scattering angular distribution
that is used in several popular Monte Carlo codes.
\citet{FernandezVarea1993} showed that the Moli\`{e}re theory is
essentially equivalent to the Goudsmit--Saunderson theory for a Wentzel
model with specific values of the total cross section $\sigma^{\rm (W)}$
and the parameter $A$. The DCS underlying Moli\`{e}re's theory is the
Wentzel DCS, Eq.\ \req{9.178}, with screening parameter and total
cross section given by
\beq
A^{\rm (M)} = \frac{1}{4} \frac{(\hbar c)^2}{E(E+2M_1 c^2)} \,
\frac{1.13 + 3.76 Z^2 / (137\, \beta)^2}{(0.885 Z^{-1/3} a_0)^2},
\label{9.183}\eeq
and
\beq
\sigma^{\rm (M)} = \frac{(ZZ_1e^2)^2}{\beta^2 \, E (E+2 M_1 c^2)}\,
\frac{\pi}{A^{\rm (M)}(1+A^{\rm (M)})},
\label{9.184}\eeq
where $Z$ is the atomic number of the target atoms, and $M_1$, $Z_1 e$,
and $E$ are, respectively, the mass, the charge and the kinetic energy of
the transported particles. These expressions were determined by
Moli\`{e}re by fitting the elastic DCS obtained from the eikonal
approximation with the parameterization \req{3.149}--\req{3.150} of the
Thomas--Fermi atomic potential. In the following, the Wentzel DCS with
Moli\`{e}re's parameters will be referred to as the Wentzel--Moli\`{e}re
DCS.

The Moli\`{e}re theory is devised to describe multiple scattering
distributions of high-energy particles, for which the screening
parameter $A$ is small. Let us consider the case $A \ll 1$ and
set $x=1+2A$. From the relation \citep{GradshteynRyzhik2007}
\beq
Q_{\ell}(x) = \frac{1}{2} P_{\ell}(x) \ln \left( \frac{x+1}{x-1} \right)
- \sum_{m=1}^{\ell} \frac{1}{m} P_{m-1}(x) P_{\ell-m}(x),
\label{9.185} \eeq
and the limiting form of the Legendre polynomials
\beq
P_{\ell}(1+2A) = 1 + \ell(\ell+1)A + O(\ell^{4}A^{2}),
\label{9.186} \eeq
it follows that
\beq
Q_{\ell}(1+2A) = \left[ 1+\ell(\ell+1)A \right] \left[ \frac{1}{2}
\ln \left( \frac{1+A}{A} \right) - \Phi(\ell) \right] + \ell(\ell+1)A
+{\cal O}(l^{4}A^{2}),
\label{9.187} \eeq
where $\Phi(\ell)$ is the partial harmonic series,
\beq
\Phi(\ell) \equiv \sum_{m=1}^{\ell} \frac{1}{m}.
\label{9.188} \eeq
We replace $\Phi(\ell)$ with the approximation \citep{Moliere1948}
\beq
\Phi(\ell) = g + \ln (\ell+1/2) + \frac{1}{24(\ell+1/2)^{2}},
\label{9.189} \eeq
where $g=0.577\, 215\, 665$ is Euler's constant.
Introducing these results in expression \req{9.180}, keeping only the
dominant terms, we obtain the Moli\`ere approximation to the transport
coefficients,
\beq
G_{\ell}^{\rm (M)}
= \ell (\ell+1) A \left[ \ln \left( \frac{1+A}{A}
\right) + 1 - 2 g - 2\ln (\ell+1/2) \right] \, .
\label{9.190} \eeq
In particular, we have
  \beqa
G_{1}^{\rm (M)} & = &
2A \left[ \ln \left( \frac{1+A}{A} \right) - 0.965 \right],
\label{9.191} \\[2mm]
G_{2}^{\rm (M)} & = &
6A \left[ \ln \left( \frac{1+A}{A} \right) - 1.987 \right].
\label{9.192} \eeqa
Comparing these approximations with the exact expressions given by
Eqs.\ \req{9.181} and \req{9.182}, we
see that the approximations introduced up to this point are not serious,
provided $A\ll 1$. However, from Eq.\ \req{9.186} it is clear that the
approximation $G_{\ell}^{\rm (M)}$ will fail for $\ell$ values such that
$\ell^{2}A\sim 1$. Actually, when $\ell$ increases from 0 to $\infty$,
the right-hand side of Eq.\ \req{9.190} first increases from 0 up to a
maximum value $\sim (1+A)\exp(-2 g)$, which is reached when
$\ell=\ell_{\rm max}\sim (1+A^{-1})^{1/2} \exp(-g)$, and for larger
values of $\ell$ it decreases monotonically and eventually becomes
negative. As the ``exact'' transport coefficients $G_{\ell}^{\rm (W)}$,
Eq.\ \req{9.180}, increase monotonically with $\ell$ and tend to unity when $\ell$ goes to $\infty$, it is
clear that we should limit the use of the approximation given by Eq.\
\req{9.190} to path lengths $s$ that are large enough to make sure that
the Goudsmit--Saunderson series, Eq.\ \req{9.159}, for the
Wentzel-Moli\`{e}re DCS model, converges with less than $\ell_{\rm max}$ terms.

Following \citet{Moliere1948}, we assume that the number of terms in the
Goudsmit--Saunderson series
$$
\Phi_{\rm M}(s;\theta)
= \sum_{\ell=0}^{\infty}
\frac{2\ell+1}{4\pi} \exp\left[ -(s / \lambda^{\rm (M)}) \, G^{\rm (M)}_\ell
\right] \, P_{\ell}(\cos\theta)
\eqno{\req{9.159}}$$
is large enough, so that the
summation can be replaced with an integral, and we use the following
approximation
\beq
P_{\ell}(\cos\theta) \simeq
\left( \frac{\theta}{\sin\theta} \right)^{1/2} J_{0}([\ell+1/2]\theta),
\label{9.193} \eeq
where $J_{0}$ is the Bessel function of the first
kind. Assuming that $s\gg\lambda^{\rm (M)} = \left[{\cal N} \sigma^{\rm
(M)}\right]^{-1}$, we have
\beqa
\Phi_{\rm M}(s;\theta) & = &
\frac{1}{2\pi} \left( \frac{\theta}{\sin\theta} \right)^{1/2}
\int_{0}^{\infty} (\ell+1/2)
\exp \left[ -(s/\lambda^{\rm (M)})G_{\ell}^{\rm (M)} \right]
J_{0}([\ell+1/2]\theta) \, \d\ell \nonumber \\[2mm]
& = & \frac{1}{2\pi} \left( \frac{\theta}{\sin\theta} \right)^{1/2}
\int_{0}^{\infty} y \exp[-h(y)] J_{0}(y\theta) \, \d y,
\label{9.194} \eeqa
where $y=\ell+1/2$ and
\beq
h(y) \equiv \frac{s}{\lambda^{\rm (M)}} \, G_{\ell}^{\rm (M)} =
\frac{s}{\lambda^{\rm (M)}} \,
(y^{2}-1/4) A \left[ \ln \left(\frac{1+A}{A} \right) + 1 -2 g
- 2\ln y \right].
\label{9.195} \eeq
Introducing the parameters
\beq
\chi_{\rm c}^{2} \equiv \frac{s}{\lambda^{\rm (M)}} \, 4A, \qquad
b \equiv \ln \left[ \frac{s}{\lambda^{\rm (M)}} (1+A) \right] + 1 -
2 g,
\label{9.196} \eeq
we can write
\beq
h(y) = \frac{1}{4} \chi_{\rm c}^{2} (y^{2}-1/4) \left[ b-\ln \left(
\frac{1}{4} \chi_{\rm c}^{2} y^{2} \right) \right],
\label{9.197} \eeq
and, changing to the variable $u=\chi_{\rm c}y$,
\beqa
\Phi_{\rm M}(s;\theta) & = &
\frac{1}{2\pi} \left( \frac{\theta}{\sin\theta} \right)^{1/2}
\frac{1}{\chi_{\rm c}^{2}} \nonumber \\[2mm]
& & \times \int_{0}^{\infty} u \exp \left[ - \left( \frac{u^{2}}{4} -
\frac{\chi_{\rm c}^{2}}{16} \right)
\left( b-\ln (u^{2}/4) \right) \right]
J_{0}(u\theta/\chi_{\rm c}) \, \d u.
\label{9.198} \eeqa
To facilitate the evaluation of this integral, Moli\`ere set
\beq
b = B - \ln B, \qquad \omega = uB^{1/2} = \chi_{\rm c}B^{1/2}y
\label{9.199} \eeq
so that he could write
\beqa
\Phi_{\rm M}(s;\theta) & = &
\frac{1}{2\pi} \left( \frac{\theta}{\sin\theta} \right)^{1/2}
\frac{1}{\chi_{\rm c}^{2}B} \int_{0}^{\infty} \omega \exp
\left[ \frac{\chi_{\rm c}^{2}}{16} \left( B-\ln(\omega^{2}/4) \right)
\right] \nonumber \\[2mm]
& & \times \exp \left[ \frac{\omega^{2}}{4B}
\ln (\omega^{2}/4)-\omega^{2}/4 \right]
J_{0} \left( \frac{\omega\theta}{\chi_{\rm c}B^{1/2}} \right) \,
\d\omega.
\label{9.200} \eeqa

Now, we note that $e^{B}/B=e^{b}\sim s/\lambda^{\rm (M)}$, the average
number of collisions in the path length $s$. In practice $B$ takes
values much larger than unity [\eg, $B\approx 3.6$ for
$s=10\lambda^{\rm (M)}$, which is a considerably small path length] and
hence, for the values of $\omega$ that effectively contribute to the
integral in Eq.\ \req{9.200}, we can write
\beq
\exp \left[ \frac{\omega^{2}}{4B} \ln (\omega^{2}/4) \right] \simeq
1 + \frac{1}{B} \frac{\omega^{2}}{4} \ln (\omega^{2}/4)
+ \frac{1}{2!} \frac{1}{B^{2}} \left( \frac{\omega^{2}}{4} \ln
(\omega^{2}/4) \right)^{2}
\label{9.201} \eeq
and
\beq
\exp \left[ \frac{\chi_{\rm c}^{2}}{16} \left( B - \ln(\omega^{2}/4)
\right) \right] \simeq \exp (\chi_{\rm c}^{2}B/16).
\label{9.202} \eeq
With all this, we obtain
\beq
\Phi_{\rm M}(s;\theta) =
\frac{1}{2\pi} \left( \frac{\theta}{\sin\theta} \right)^{1/2}
\frac{\exp(\chi_{\rm c}^{2}B/16)}{\chi_{\rm c}^{2}B}
\left[ f^{(0)}(\vartheta) + \frac{1}{B}  f^{(1)}(\vartheta) +
\frac{1}{B^{2}} f^{(2)}(\vartheta) \right],
\label{9.203} \eeq
where
\beq
\vartheta \equiv \frac{\theta}{\chi_{\rm c}B^{1/2}}
\label{9.204} \eeq
and
\beq
f^{(n)}(\vartheta) \equiv \frac{1}{n!} \int_{0}^{\infty} \omega
\exp(-\omega^{2}/4) \left( \frac{\omega^{2}}{4} \ln (\omega^{2}/4)
\right)^{n} J_{0}(\vartheta\omega) \, \d\omega.
\label{9.205} \eeq
This is the distribution obtained by \citet{Moliere1948}
and \citet{Bethe1953}. The first term in the sum \req{9.203} is the
Gaussian distribution
\beq
f^{(0)}(\vartheta) = 2 \exp(-\vartheta^{2}).
\label{9.206} \eeq
The functions $f^{(1)}(\vartheta)$ and $f^{(2)}(\vartheta)$ have been
tabulated by \citet{Bethe1953}. The results from an accurate numerical
calculation are displayed in Fig.\ \ref{fig9.7}. For $\vartheta$ values up to
about 10, $f^{(1)}(\vartheta)$ and $f^{(2)}(\vartheta)$ can be obtained
by cubic spline interpolation from a numerical table (see Section
\ref{sec10.4.2}). For $\vartheta\geq 10$, the expressions
\begin{subequations}
\label{9.207}
\beqa
f^{(1)}(\vartheta) & \simeq & 2\vartheta^{-4}
\left( 1-5\vartheta^{-2} \right)^{-4/5}, \label{9.207a} \\[2mm]
f^{(2)}(\vartheta) & \simeq & 16\vartheta^{-6} (\ln \vartheta +
g - 3/2) \left( 1+9\vartheta^{-2} \right) - 38\vartheta^{-8},
\label{9.207b} \eeqa
\end{subequations}
are accurate to within 0.01 percent.

\begin{figure}[htb] \begin{center}
\includegraphics*[width=9.5cm]{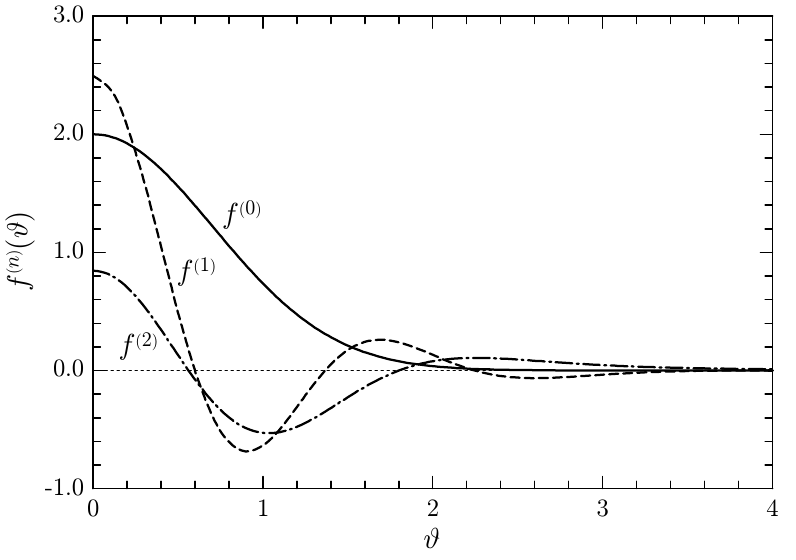}
\caption{
Moli\`{e}re functions $f^{(n)}(\vartheta)$, Eq.\ \req{9.205}.
\label{fig9.7}}
\end{center} \end{figure}

With the screening parameter given by Eq.\ \req{9.183}, and assuming it
to be much less than unity, we find
\begin{subequations}
\label{9.208}
\beqa
\chi_{\rm c}^{2} & = &
s {\cal N} 4 \pi \frac{(ZZ'e^{2})^{2}}{(p\beta c)^{2}}, \label{9.208a}
\\[2mm]
b & = &
\ln \left( s {\cal N} \pi \frac{(ZZ'e^{2})^{2}}{(p\beta c)^{2}A^{\rm (M)}}
\right) + 1 - 2 g,
\label{9.208b} \eeqa
\end{subequations}
which coincide exactly with the parameters used by \citet{Moliere1948}.
Therefore, the original form of the Moli\`ere theory is nothing more
than an analytical approximation to the exact Goudsmit--Saunderson
distribution for the Wentzel--Moli\`{e}re DCS model.

The mathematical approximations introduced in the derivation of the
Moli\`ere distribution, Eq.\ \req{9.203}, put certain limits on its
range of validity. First, the limiting form of the transport
coefficients given by Eq.\ \req{9.190} is valid only when the screening
parameter $A$ is small. This limits the applicability of the theory to
high energies, for which $\lambda_{1}^{\rm (M)}\gg\lambda^{\rm (M)}$.
Secondly, we must have $b>1$ [otherwise, the parameter
$B$, see Eq.\ \req{9.199}, is not defined]. This means that the path
length $s$ must be larger than about $4\lambda^{\rm (M)}$ [see Eq.\
\req{9.196}], \ie, the scattering must be at least plural. Actually, a
restriction of this sort was to be expected from the very beginning
since, when $s\sim\lambda^{\rm (M)}$, the Goudsmit--Saunderson series is
slowly convergent and contributions from terms with
$\ell>\ell_{\rm max}$ [see the discussion after Eq.\ \req{9.192}] may
not be negligible. As $G_{\ell}^{\rm (M)}$ is not adequate for these
high order terms, the whole theory fails when $s\sim\lambda^{\rm (M)}$.
Finally, the approximation given by Eq.\ \req{9.193} is very accurate
for small angles, it remains valid for intermediate angles, and breaks
down for values of $\theta$ near $180$ deg where the factor
$(\theta/\sin\theta)^{1/2}$ diverges. This divergence is not important
when $s\ll\lambda_{1}^{\rm (M)}$, since then the angular distribution is
strongly peaked in the forward direction and the only effect of the
divergence is a very narrow peak in the backward direction with a
negligible area. Undesirable effects of this divergence become prominent
when $s\simeq \lambda_{1}^{\rm (M)}$. Under these circumstances, the
distribution \req{9.198} shows a conspicuous peak in the backward
direction and therefore differs appreciably from the ``exact'' angular
distribution, which tends to the isotropic distribution when
$s\gg\lambda_{1}$. To avoid this anomalous behavior, we should limit to
path lengths such that the Gaussian part, Eq.\ \req{9.206}, of the
Moli\`ere distribution has a width less than $\sim$ 1 radian
\citep{Bethe1953}
or, equivalently, such that $\chi_{\rm c}^{2}B\leq 1$. In conclusion,
the Moli\`ere distribution, Eq.\ \req{9.203}, gives a good
approximation to the Goudsmit--Saunderson distribution for the Wentzel
model when the conditions
\beq
\lambda^{\rm (M)} \ll \lambda_{1}^{\rm (M)} \quad {\rm and} \quad
4\lambda^{\rm (M)} < s < \lambda_{1}^{\rm (M)}
\label{9.209}\eeq
are simultaneously fulfilled. For path lengths in this range, the
Moli\`{e}re $B$ parameter can be estimated by using the formula
\beq
B \simeq \frac{\ln(1.345 \, X)+X-2g}{1-1/(1.345 \, X)}
\qquad \mbox{with} \qquad
X=\ln \left[ \frac{s}{\lambda^{\rm (M)}} (1+A) \right],
\label{9.210}\eeq
which approximates the exact solution of Eq.\ \req{9.199} with an
accuracy better than 0.2~\%.
\index{multiple scattering!Moli\`{e}re distribution|)}
\index{multiple scattering|)}


\section{Multiple scattering with energy loss. \label{sec9.6}}
\index{multiple scattering with energy loss|(}

Elastic collisions coexist with energy-loss processes that
slow down the projectile along its path. The only attempts to account
for the effect of energy loss on the multiple scattering distribution
are based on the CSDA. As shown in Section \ref{sec9.4.1}, the
CSDA determines a correspondence between the traveled path length $s$ and
the associated energy loss $\Delta_s$, which is defined by
$$
s = \int_{E-\Delta_s}^{E} \frac{\d E'}{S(E')} = R(E) - R(E-\Delta_s),
\eqno{\req{9.89}}$$
where $E$ is the initial energy of the projectile (at $s=0$). Thus,
characteristic scattering quantities can be regarded as functions of
either the energy $E$ or the path length $s$. For example,
the average number of scattering events undergone by the
particle along the path length $s$ is
\beq
\langle n \rangle = \int_0^s \frac{\d s'}{\lambda(s')} =
\int_{E-\Delta_s}^E \frac{1}{\lambda(E')} \, \frac{\d E'}{S(E')}.
\label{9.211}\eeq
Notice that $\lambda$ here is first regarded as a function of the path
length $s$ and, afterward, as a function of the ``local'' energy
$E'=E-\Delta_{s'}$.

Again, we consider a moving charged particle in an infinite medium that
starts from the origin of our reference frame and moves in the direction
of the $z$-axis. We wish to determine the probability density $\Phi(s;
{\bf r},\Omegab)$ of finding the particle at the position ${\bf
r}=(x,y,z)$, moving in the direction given by the unit vector
$\Omegab=(\theta,\phi)$ after having traveled a path length $s$. The
diffusion equation for this problem is
\beq
\frac{\partial \Phi}{\partial s} + \Omegab \cdot \nablab \Phi =
\int \left[ \Phi(s;{\bf r},\Omegab')-\Phi(s;{\bf r},\Omegab)
\right] \mu(s; \Omegab\rightarrow \Omegab') \, \d\Omegab',
\label{9.212} \eeq
where $\mu(s;\Omegab \cdot \Omegab')$ is the differential
inverse mean free path of a particle having energy $E-\Delta_s$. Before
giving a formal solution of this problem, we will describe a
simpler approach that is valid for particles with very high energies.


\subsection{The Fermi--Eyges distribution \label{sec9.6.1}}
\index{Fermi--Eyges theory}
\index{multiple scattering with energy loss!Fermi--Eyges distribution|(}
As seen above, the DCS for scattering of high-energy charged
particles is dominated by small angular deflections such that $\sin
\theta \simeq \theta$ and $\cos\theta \simeq 1 - \theta^2/2$. Early
studies of multiple scattering took advantage of this feature for
deriving approximate analytical solutions of the transport equation.
Here we present an approximation first proposed by Fermi
\citep{RossiGreisen1941}, who assumed that particles move with constant
energy, and latter extended by \citet{Eyges1948} to include energy
losses within the CSDA.

Fermi noted that, in small angle processes, particles move nearly
parallel to the $z$ axis for a long while and, if the average
deflection angle remains small, we can set $s \simeq z$. In addition, he
considered the projections of the particle trajectories on the plane
$x-z$. Let $F(z;x,\theta_x)$ denote the probability density for
particles at a ``depth'' $z$ to have the projected lateral displacement $x$ and
the direction of motion making a projected angle $\theta_x$ with the $z$
axis. Because the system has axial symmetry around the $z$ axis, the
same distribution describes the process projected on the $y-z$ plane.
Considering these two projected processes as independent, we can write
\beq
\Phi(s; {\bf r}, \Omegab) = F(z; x, \theta_x) \, F(z;y,\theta_y).
\label{9.213}\eeq
The first benefit of this factorization is a reduction of the
dimensionality of the problem.

Under the assumption of small scattering angles, the distribution $F(z;
x, \theta_x)$ has a very narrow bell shape with its maximum at $x=0$ and
$\theta_x = 0$. Consequently, as the probability of having large
projected angles is negligible, we can consider that the variables
$x$ and $\theta_x$ take values from $-\infty$ to $\infty$. Evidently,
the distribution is symmetrical under inversion, that is,
\beq
F(z; x, \theta_x) = F(z; - x, \theta_x)
= F(z; x, -\theta_x).
\label{9.214}\eeq
Expressing the direction vector $\Omegab = (\theta, \phi)$ in
Cartesian components,
\beq
\Omegab= (\cos\phi \, \sin\theta, \sin\phi \, \sin\theta , \cos\theta),
\label{9.215}\eeq
the projected angle $\theta_x$ is determined by
\begin{subequations}
\label{9.216}
\beq
\cos \theta_x = \frac{\Omegab - (\Omegab\dotprod \hat{\bf y}) \hat{\bf
y}}
{\left| \Omegab - (\Omegab\dotprod \hat{\bf y}) \hat{\bf y} \right|} \, \dotprod
\hat{\bf z} = \frac{\cos\theta}{\sqrt{1- \sin^2 \phi \sin^2 \theta}}.
\label{9.216a}\eeq
Similarly,
\beq
\cos \theta_y = \frac{\Omegab - (\Omegab\dotprod \hat{\bf x}) \hat{\bf
x}}
{\left| \Omegab - (\Omegab\dotprod \hat{\bf x}) \hat{\bf x} \right|} \, \dotprod
\hat{\bf z} = \frac{\cos\theta}{\sqrt{1- \cos^2 \phi \sin^2 \theta}}.
\label{9.216b}\eeq
\end{subequations}
When the angles $\theta$, $\theta_x$, and $\theta_y$ are small, these
relations reduce to
\begin{subequations}
\label{9.217}
\beq
\theta_x^2 = \left( 1 - \sin^2 \phi \right) \theta^2 =
\cos^2 \phi \; \theta^2
\label{9.217a}\eeq
and
\beq
\theta_y^2 = \left( 1 - \cos^2 \phi \right) \theta^2 =
\sin^2 \phi \; \theta^2.
\label{9.217b}\eeq
\end{subequations}

We now consider the variation of the distribution $F(z;x,\theta_x)$
caused by a thin layer of material at depths between $z$ and $z + \Delta
z$. Let $\rho_{\Delta z}(\theta_x) \, \d \theta_x$ be the probability
that a particle traversing the thickness $\Delta z$ will be deflected a
projected angle between $\theta_x$ and $\theta_x + \d \theta_x$ as the
result of scattering events in $\Delta z$. Since the DCS is independent
of the azimuthal angle, the distribution $\rho_{\Delta z}(\theta_x)$ is
symmetrical, centered at $\theta_x=0$, and, by definition, normalized to
unity , \ie,
\begin{subequations}
\label{9.218}
\beq
\rho_{\Delta z}(-\theta_x) = \rho_{\Delta z}(\theta_x), \qquad
\int \rho_{\Delta z}(\theta_x) \, \d \theta_x = 1, \qquad
\int \theta_x \,\rho_{\Delta z}(\theta_x) \, \d \theta_x = 0.
\label{9.218a}\eeq
Additionally, in the small angle approximation, the second moment of
$\rho_{\Delta z} (\theta_x)$,
\beq
\langle \theta_x^2 \rangle_{\Delta z}  = \int \theta_x^2 \,
\rho_{\Delta z}(\theta_x) \, \d \theta_x,
\nonumber \eeq
can be obtained with the aid of the relation \req{9.217a},
\beq
\langle \theta_x^2 \rangle_{\Delta z} = \left[\frac{1}{2\pi} \int_0^{2\pi}
\left( 1 - \sin^2 \phi \right) \, \d \phi \right]
\langle \theta^2 \rangle_{\Delta z}
= \frac{1}{2} \langle \theta^2 \rangle_{\Delta z},
\nonumber \eeq
where $\langle \theta^2 \rangle_{\Delta z}$ is the average for the
actual multiple scattering process in a path length $\Delta z$,
\beq
\langle \theta^2 \rangle _{\Delta z}
= \Delta z \, {\cal N}
\int_{-\infty}^\infty \theta^2 \, \frac{\d \sigma}{\d \Omega} \, \d
\Omega \simeq \Delta z \, {\cal N} \;
2 \int_{-\infty}^\infty (1-\cos\theta) \, \frac{\d \sigma}{\d \Omega} \,
\d \Omega.
\nonumber \eeq
Hence,
\beq
\langle \theta_x^2 \rangle_{\Delta z}
= \frac{\Delta z}{\lambda_1},
\label{9.218b}\eeq
\end{subequations}
where $\lambda_1$ is the first transport mean free path, Eq.\
\req{9.161}, which here is considered as a function of the traveled path
length $z$.

The scattering in the layer $\Delta z$ modifies the function
$F(z;x,\theta_x)$ mostly because particles traveling at an angle
$\theta_x$ undergo an additional lateral displacement $\theta_x \,
\Delta z$. We first consider the change in the spatial distribution, by
neglecting the effect of scattering on the angular distribution (which is
of second order, see below), and write
\begin{subequations}
\label{9.219}
\beq
F(z+ \Delta z;x,\theta_x) = F(z;x - \theta_x \, \Delta z,\theta_x)
= F(z;x,\theta_x) - \theta_x \, \Delta z \,
\frac{\partial F}{\partial x} \, .
\label{9.219a}\eeq
On the other hand, the change in the angular distribution, calculated by
disregarding the variation of the spatial distribution, is
\beq
F(z + \Delta z ;x,\theta_x) = F(z;x,\theta_x) + \int_{-\infty}^\infty
F(z;x,\theta'_x) \, \rho_{\Delta z}(\theta_x-\theta'_x) \, \d \theta'_x.
\nonumber \eeq
Because $\rho_{\Delta z}(\theta_x-\theta'_x)$ is different from zero
only for very small values of the argument, we can expand $F(z; x,
\theta'_x)$ as a Taylor series about $\theta_x$. Then, keeping only terms
up to second order and using the properties \req{9.218}, we have
\beq
F(z + \Delta z ;x,\theta_x) = F(z;x,\theta_x) +
\frac{1}{2} \, \langle \theta_x^2 \rangle_{\Delta z}
\, \frac{\partial^2 F}{\partial \theta_x^2}
= F(z;x,\theta_x) + \frac{\Delta z}{2 \lambda_1}
\, \frac{\partial^2 F}{\partial \theta_x^2}.
\label{9.219b}\eeq
\end{subequations}
The complete variation $\Delta F$ of $F(z; x, \theta_x)$ is obtained as
the sum of the
changes caused independently by spatial and angular variations,
\beq
\Delta F(z ;x,\theta_x) = - \theta_x \, \Delta z \,
\frac{\partial F}{\partial x}
+ \frac{\Delta z}{2 \lambda_1}
\, \frac{\partial^2 F}{\partial \theta_x^2}.
\nonumber \eeq
Hence, the diffusion equation for the distribution $F(z; x,\theta_x)$ is
\citep{RossiGreisen1941}
\beq
\frac{\partial F}{\partial z} = - \theta_x \,
\frac{\partial F}{\partial x}
+ \frac{1}{2 \lambda_1}
\, \frac{\partial^2 F}{\partial \theta_x^2}.
\label{9.220}\eeq

Following \citet{Eyges1948}, with the aid of the CSDA, we consider the
transport mean free path as a function of the path length $z$. To solve
the equation \req{9.220}, we introduce the Fourier transform
\index{Fourier transform}
\beq
\widetilde{F}(z;u,v) = \frac{1}{2\pi}
\int_{-\infty}^\infty \d x \int_{-\infty}^\infty \d \theta_x \,
F(z;x,\theta_x) \, \exp\left[ -{\rm i} \left( x u + \theta_x v \right)
\right]
\label{9.221}\eeq
and its inverse,
\beq
F(z;x,\theta_x) = \frac{1}{2\pi}
\int_{-\infty}^\infty \d u \int_{-\infty}^\infty \d v \,
\widetilde{F}(z;u,v) \, \exp\left[ {\rm i} \left( x u + \theta_x v \right)
\right].
\label{9.222}\eeq
Inserting the latter into Eq.\ \req{9.220}, after simple manipulations,
we find that the function $\widetilde{F}(z;u,v)$ satisfies the equation
\beq
\frac{\partial \widetilde{F}(z;u,v)}{\partial z} = u\,
\frac{\partial \widetilde{F}(z;u,v)}{\partial v}
- \frac{v^2}{2 \lambda_1} \, \widetilde{F}(z;u,v).
\label{9.223}\eeq
Changing variables to
\beq
z'=z, \quad u'=u, \quad v'=z+ \frac{v}{u},
\label{9.224}\eeq
the chain rule gives
\beq
\frac{\partial \widetilde{F}}{\partial z} =
\frac{\partial \widetilde{F}}{\partial z'}
\frac{\partial z'}{\partial z}
+ \frac{\partial \widetilde{F}}{\partial u'}
\frac{\partial u'}{\partial z}
+ \frac{\partial \widetilde{F}}{\partial v'}
\frac{\partial v'}{\partial z}
=
\frac{\partial \widetilde{F}}{\partial z'}
+ \frac{\partial \widetilde{F}}{\partial v'}
\nonumber \eeq
and
\beq
\frac{\partial \widetilde{F}}{\partial v} =
\frac{\partial \widetilde{F}}{\partial v'} \, \frac{1}{u},
\nonumber \eeq
and Eq.\ \req{9.223} takes the form
\beq
\frac{\partial \widetilde{F}}{\partial z'}
= - \, \frac{u'^2(v'-z')^2}{2 \lambda_1(z')} \, \widetilde{F}.
\label{9.225}\eeq
This equation admits the formal solution
\beq
\widetilde{F}(z';u',v') = C(v') \exp \left( - u'^2 \int_k^{z'}
\frac{(v'-\zeta)^2}{2 \lambda_1(\zeta)} \, \d \zeta \right),
\nonumber \eeq
where the function $C(v')$ is an ``integration constant'' and $k$ is a
finite value not yet specified. Expressed in terms of the Fourier
variables, this solution reads
\beq
\widetilde{F}(z;u,v) = C\left( z+v/u \right) \exp \left( - u^2 \int_k^{z}
\frac{(z+v/u-\zeta)^2}{2 \lambda_1(\zeta)} \, \d \zeta \right).
\label{9.226}\eeq
Imposing the boundary condition
\beq
F(0; x, \theta_x) = \delta(x) \, \delta(\theta_x)
\qquad \Rightarrow \qquad
\widetilde{F}(0; u,v)= \frac{1}{2\pi}\, ,
\label{9.227}\eeq
we conclude that
\beq
C\left( z+v/u \right) = \frac{1}{2\pi} \,
\exp \left(  u^2 \int_k^{0}
\frac{(z+v/u-\zeta)^2}{2 \lambda_1(\zeta)} \, \d \zeta \right)
\nonumber \eeq
and
\beq
\widetilde{F}(z;u,v) = \frac{1}{2\pi} \,
\exp \left( - u^2 \int_0^{z}
\frac{(v/u+z-\zeta)^2}{2 \lambda_1(\zeta)} \, \d \zeta \right).
\label{9.228}\eeq

With the definitions
\begin{subequations}
\label{9.229}
\beqa
A_0(z) &=& \int_0^{z}  \frac{1}{2 \lambda_1(\zeta)} \, \d \zeta,
\label{9.229a} \\ [2mm]
A_1(z) &=& \int_0^{z}  \frac{z-\zeta}{2 \lambda_1(\zeta)} \, \d \zeta,
\label{9.229b}\\ [2mm]
A_2(z) &=& \int_0^{z}  \frac{(z-\zeta)^2}{2 \lambda_1(\zeta)} \, \d \zeta,
\label{9.229c}\eeqa
\end{subequations}
we can write
\beq
\widetilde{F}(z;u,v) = \frac{1}{2\pi} \,
\exp \left[ - \left( A_0 v^2 + 2 A_1 v u + A_2 u^2 \right) \right].
\label{9.230}\eeq
Inserting this result into the right-hand side of Eq.\ \req{9.222} and
evaluating the integrals\footnote{We use the formula \citep[][Eq.\
3.323.2]{GradshteynRyzhik2007},
$$
\int_{-\infty}^\infty \exp\left( - a x^2 - bx -c \right) \d x =
\sqrt{\frac{\pi}{a}} \exp\left( \frac{b^2-4ac}{4a} \right)
\qquad \qquad {\rm Re}\, a >0,
$$
which is valid for complex coefficients.},
we finally obtain the {\it Fermi--Eyges distribution},
\beq
F(z;x,\theta_x) = \frac{1}{4\pi\sqrt{B}}
\exp \left( - \frac{A_2 \theta_x^2 - 2 A_1 \theta_x x + A_0 x^2}{4B}
\right),
\label{9.231}\eeq
where
\beq
B=A_0 A_2 - A_1^2.
\label{9.232}\eeq
Integration of the distribution \req{9.231} over $x$ gives the
projected angular distribution irrespective of the lateral displacement,
\begin{subequations}
\label{9.233}
\beq
F(z;\theta_x) = \int_{-\infty}^\infty F(z;x,\theta_x) \, \d x =
\frac{1}{\sqrt{4\pi A_0}}
\exp \left( - \frac{\theta_x^2}{4A_0} \right).
\label{9.233a}\eeq
Similarly, integration over $\theta_x$ gives the distribution of lateral
displacements, irrespective of the projected angle,
\beq
F(z;x) = \int_{-\infty}^\infty F(z;x,\theta_x) \, \d \theta_x =
\frac{1}{\sqrt{4\pi A_2}}
\exp \left( - \frac{x^2}{4A_2} \right).
\label{9.233b}\eeq
\end{subequations}
These two distributions are Gaussian and normalized to unity, as expected.

When the energy loss in the path length $z$ is small, the transport mean
free path remains practically constant and the definitions \req{9.229}
give
\beq
A_0= \frac{z}{2 \lambda_1}, \quad
A_1= \frac{z^2}{4 \, \lambda_1}, \quad
A_2= \frac{z^3}{6 \, \lambda_1}, \quad
B=\frac{z^4}{48 \, \lambda_1^2}.
\label{9.234}\eeq
Then, the distribution \req{9.231} simplifies to
\beq
F(z;x,\theta_x) = \frac{\sqrt{3}}{\pi} \, \frac{\lambda_1}{z^2}
\exp \left[ -  \frac{2 \lambda_1}{z} \left( \theta_x^2 - 3 \theta_x \,
\frac{x}{z} + 3 \, \frac{x^2}{z^2} \right) \right],
\label{9.235}\eeq
which is the result given by \citet{RossiGreisen1941}. In this case, it
can be easily verified that the function \req{9.235} verifies the differential
equation \req{9.220} and the boundary condition \req{9.227}. The
distributions \req{9.233} of the projected angle and the lateral
displacement become
\begin{subequations}
\label{9.236}
\beq
F(z;\theta_x) =
\sqrt{\frac{\lambda_1}{2 \pi z}}\,
\exp \left( - \frac{\lambda_1}{2z} \, \theta_x^2 \right)
\label{9.236a}\eeq
and
\beq
F(z;x) =
\sqrt{\frac{3 \lambda_1}{2\pi z^3}}
\exp \left( - \frac{3 \lambda_1}{2z^3} \, x^2\right),
\label{9.236b}\eeq
\end{subequations}
respectively.

The Fermi--Eyges distribution is employed for modeling particle beams in
hadrontherapy. Monte Carlo simulations with such beam models require
generating random values of the variables $x$ and $\theta_x$ from their
joint distribution \req{9.231}. To simplify the sampling, it is
convenient to write the distribution in terms of new variables $r_1$ and
$r_2$ defined by
\beq
x=\sqrt{2 A_2}\,  r_1, \qquad \mbox{and} \qquad
\theta_x = \sqrt{\frac{2A_1^2}{A_2}}\;  r_1+
\sqrt{\frac{2B}{A_2}}\; r_2.
\label{9.237}\eeq
The Jacobian of the transformation is
\beq
\left|\frac{\partial
(x,\theta_x)}{\partial (r_1,r_2)}\right| = \left| \begin{array}{cc}
\sqrt{2 A_2} & 0 \\ [3mm]
\displaystyle{\sqrt{\frac{2A_1^2}{A_2}}} &
\displaystyle{\sqrt{\frac{2B}{A_2}}}
\end{array} \right| = 2 \sqrt{B}.
\label{9.238}\eeq
The probability density function of the new variables is
\beq
F(z;r_1,r_2) = F(z;x,\theta_x) \,
\left|\frac{\partial
(x,\theta_x)}{\partial (r_1,r_2)}\right|
\nonumber \\ [2mm]
= \frac{1}{2\pi} \exp \left( - \frac{r_1^2 + r_2^2}{2} \right),
\nonumber \eeq
or
\beq
F(z;r_1,r_2) = \frac{1}{\sqrt{2\pi}}\, \exp(-r_1^2/2)
\times \frac{1}{\sqrt{2\pi}}\, \exp(-r_2^2/2).
\label{9.239}\eeq
Hence, $r_1$ and $r_2$ are independent random variables with standard
normal distribution, that is, their probability density function is the
Gaussian with null mean and unit variance. Values of $r_1$ and $r_2$ can
be sampled by means of the Box--M\"{u}ller method \citep[see,
\eg,][]{Salvat2025}, which uses two random numbers to produce a pair of
independent normal deviates. The values of the displacement $x$ and the
projected angle $\theta_x$ are then obtained from the transformation
equations \req{9.237}.
\index{multiple scattering with energy loss!Fermi--Eyges distribution|)}


\subsection{The Lewis theory \label{sec9.6.2}}
\index{multiple scattering with energy loss!Lewis theory|(}
\index{Lewis theory}

A formal theory of multiple scattering with energy loss was formulated
by \citet{Lewis1950} as an extension of the Goudsmit and Saunderson
theory (Section \ref{sec9.5.2}). Lewis' theory includes energy loss
through the CSDA, and it describes both angular deflections and spatial
displacements.

As usual, we consider a fast charged particle that starts from the
origin of the reference frame moving in the direction of the
$z$-axis. We wish to solve the diffusion equation
\beq
\frac{\partial \Phi_{\rm L}}{\partial s} + \Omegab \cdot \nablab \Phi_{\rm L} =
\int \left[ \Phi_{\rm L}(s;{\bf r},\Omegab')-\Phi_{\rm L}(s;{\bf r},\Omegab)
\right] \mu(s; \Omegab\rightarrow \Omegab') \, \d\Omegab',
\label{9.240} \eeq
with the energy-dependent differential inverse mean free path
$\mu(s;\Omegab \cdot \Omegab')$ considered as a function of the traveled path
length. We write [see Eq.\ \req{9.151}]
\beq
\mu(s;\Omegab \rightarrow \Omegab') = {\cal N} \sigma(s) \,
\sum_{\ell,m} F_{\ell}(s)
\; Y_{lm}^\ast(\Omegab) \, Y_{\ell m} (\Omegab')
\label{9.241}\eeq
with [Eq.\ \req{9.147}]
\beq
F_\ell (s) = \frac{1}{\sigma(s)} \int P_\ell (\cos\theta) \, \frac{\d
\sigma}{\d \Omega} \, \d \Omegab\, .
\label{9.242}\eeq
Equation \req{9.240} has to be solved under the boundary condition
\beq
\Phi_{\rm L}(0;{\bf r},\Omegab) = \delta({\bf r})\, \delta(\Omegab)
=(1/\pi)\delta({\bf r})\delta(1-\cos\theta).
\label{9.243}\eeq
Notice that this distribution is normalized to unity, because
the delta function is located at the end of the $\cos\theta$
interval and, therefore, its integral equals 1/2.

Following Lewis, we expand Eq.\ \req{9.240} in
spherical harmonics in $\Omegab$,
\beq
\Phi_{\rm L}(s;{\bf r},\Omegab) = \sum_{\ell,m} \Phi_{\ell m}(s;{\bf r}) \,
Y_{\ell m} (\Omegab), \qquad
\Phi_{\ell m}(s;{\bf r}) \equiv \int \d \Omegab \,
Y_{\ell m}^\ast (\Omegab) \, \Phi_{\rm L}(s;{\bf r},\Omegab).
\label{9.244}\eeq
Multiplying Eq.\ \req{9.240} by $Y_{\ell m}^\ast (\Omegab)$ and
integrating over $\Omegab$, we obtain
\beqa
&& \! \! \! \! \! \! \! \! \! \! \! \! \! \! \! \! \!
\int \d \Omegab \, Y_{\ell m}^\ast (\Omegab)
\left( \frac{\partial \Phi_{\rm L}}{\partial s} + \Omegab \cdot
\nablab \Phi_{\rm L} \right) =
\int \d \Omegab \, Y_{\ell m}^\ast (\Omegab) \,
\nonumber \\ [2mm]
&& \mbox{} \rule{10mm}{0mm} \times
\int  \d\Omegab' \left[ \Phi_{\rm L}(s;{\bf r},\Omegab')
\mu(s; \Omegab'\rightarrow \Omegab)
-\Phi_{\rm L}(s;{\bf r},\Omegab)
\mu(s; \Omegab\rightarrow \Omegab') \right], \rule{10mm}{0mm}
\label{9.245} \eeqa
where we have used the symmetry \req{9.152} of the DIMFP.
The left-hand side can be expanded with the help of the closure relation
of spherical harmonics [Eq.\ \req{B.54}],
\beqa
&& \! \! \! \! \! \! \! \! \! \! \! \! \! \! \!
\int \d \Omegab \, Y_{\ell m}^\ast (\Omegab)
\left( \frac{\partial \Phi_{\rm L}}{\partial s} + \Omegab \cdot \nablab
\Phi_{\rm L} \right)
\nonumber \\ [2mm]
&=& \frac{\partial}{\partial s} \int \d \Omegab \, Y_{\ell m}^\ast (\Omegab)
\Phi_{\rm L}(s; {\bf r},\Omegab) + \int \d \Omegab \, Y_{\ell m}^\ast (\Omegab)
\int \d \Omegab' \delta(\Omegab - \Omegab') \;
\Omegab \cdot \nablab \Phi_{\rm L}(s; {\bf r}, \Omegab')
\nonumber \\ [2mm]
&=& \frac{\partial}{\partial s} \Phi_{\ell m}(s; {\bf r})
+ \int \d \Omegab \, Y_{\ell m}^\ast (\Omegab)
\int \d \Omegab'
\sum_{\lambda,\mu} Y_{\lambda\mu}^\ast (\Omegab')
Y_{\lambda\mu}(\Omegab) \;
\Omegab \cdot \nablab \Phi_{\rm L}(s; {\bf r}, \Omegab')
\nonumber \\ [2mm]
&=& \frac{\partial}{\partial s} \Phi_{\ell m}(s; {\bf r})
+ \sum_{\lambda,\mu} {\bf Q}_{\ell m,\lambda\mu} \dotprod
\nablab \Phi_{\lambda\mu}(s; {\bf r}),
\label{9.246}\eeqa
where
\beq
{\bf Q}_{\ell m,\lambda\mu} =
\int \d \Omegab \, Y_{\ell m}^\ast (\Omegab)
\; \Omegab \; Y_{\lambda\mu}(\Omegab)
\label{9.247}\eeq
is a three-dimensional vector. The latter integrals can be readily
calculated by noting that the Cartesian components of $\Omegab
=(\theta,\phi)$ can be expressed as
\beq
\Omegab = \sqrt{\frac{4\pi}{3}} \left(
\frac{1}{\sqrt{2}} \left[Y_{1,-1}(\Omegab)
- Y_{1,1}(\Omegab) \right],
\frac{\rm i}{\sqrt{2}} \left[Y_{1,-1}(\Omegab)
+ Y_{1,1}(\Omegab) \right],
Y_{1,0}(\Omegab) \right),
\label{9.248}\eeq
and using the well-known formula for the integral of three spherical
harmonics \cite[see, \eg,][]{Edmonds1960}.

To evaluate the right-hand side of Eq.\ \req{9.245}, we introduce the
expansion \req{9.241} and make use of the definitions \req{9.244},
\beqa
&& \! \! \! \! \! \! \! \! \! \! \! \! \! \! \!
\int \d \Omegab \, Y_{\ell m}^\ast (\Omegab) \,
\int  \d\Omegab' \left[ \Phi_{\rm L}(s;{\bf r},\Omegab')-\Phi_{\rm L}(s;{\bf r},\Omegab) \right]
\mu(s; \Omegab\rightarrow \Omegab')
\nonumber \\ [2mm]
&=& {\cal N} \sigma(s) \int \d \Omegab \, Y_{\ell m}^\ast (\Omegab) \,
\int  \d\Omegab' \left[ \Phi_{\rm L}(s;{\bf r},\Omegab')-\Phi_{\rm L}(s;{\bf r},\Omegab) \right]
\,  \sum_{\lambda,\mu} F_{\lambda}(s)
\; Y_{\lambda\mu}^\ast(\Omegab') \, Y_{\lambda\mu} (\Omegab)
\nonumber \\ [2mm]
&=& {\cal N} \sigma(s) \sum_{\lambda,\mu} F_{\lambda}(s)
\int \d \Omegab \, Y_{\ell m}^\ast (\Omegab) \,
\int  \d\Omegab' \Phi_{\rm L}(s;{\bf r},\Omegab')
\; Y_{\lambda\mu}^\ast(\Omegab') \, Y_{\lambda\mu} (\Omegab)
\nonumber \\ [2mm]
&& \mbox{} -  {\cal N} \sigma(s)\sum_{\lambda,\mu} F_{\lambda}(s)
\int \d \Omegab \, Y_{\ell m}^\ast (\Omegab) \,
\int  \d\Omegab' \Phi_{\rm L}(s;{\bf r},\Omegab) \,
\; Y_{\lambda\mu}^\ast(\Omegab) \, Y_{\lambda\mu} (\Omegab')
\nonumber \\ [2mm]
&=&  {\cal N} \sigma(s)
\sum_{\lambda,\mu} F_{\lambda}(s) \, \Phi_{\lambda\mu}(s;{\bf r})
\int \d \Omegab \, Y_{\ell m}^\ast (\Omegab) \, Y_{\lambda\mu} (\Omegab)
\nonumber \\ [2mm]
&& \mbox{} - \Phi_{\ell m}(s;{\bf r})
\int  \d\Omegab'  {\cal N} \sigma(s) \left\{ \sum_{\lambda,\mu} F_{\lambda}(s)
\; Y_{\lambda\mu}^\ast(\Omegab) \, Y_{\lambda\mu} (\Omegab') \right\}.
\label{9.249} \eeqa
Using the orthonormality of the spherical harmonics, and recalling that
the expansion in curly braces represents the normalized
single-scattering angular distribution, we obtain
\beqa
&& \! \! \! \! \! \! \! \! \! \! \! \! \! \! \!
\int \d \Omegab \, Y_{\ell m}^\ast (\Omegab) \,
\int  \d\Omegab' \left[ \Phi_{\rm L}(s;{\bf r},\Omegab')-\Phi_{\rm L}(s;{\bf r},\Omegab) \right]
\mu(s; \Omegab\rightarrow \Omegab')
\nonumber \\ [2mm]
&=&  {\cal N} \sigma(s)
F_{\ell}(s) \Phi_{\ell m}(s;{\bf r}) -  \Phi_{\ell m}(s;{\bf r}) \, {\cal
N} \sigma(s)
= -  \Phi_{\ell m}(s;{\bf r})  {\cal N} \sigma(s)
\left[ 1 - F_\ell (s) \right]. \rule{10mm}{0mm}
\label{9.250} \eeqa
Therefore the transport equation \req{9.240} is equivalent to the
system of equations
\beq
\frac{\partial}{\partial s} \Phi_{\ell m}(s; {\bf r})
+ \sum_{\lambda,\mu} {\bf Q}_{\ell m,\lambda\mu} \dotprod
\nablab \Phi_{\lambda\mu}(s; {\bf r}) = -
 \Phi_{\ell m}(s;{\bf r})  {\cal N} \sigma(s)
\left[ 1 - F_\ell (s) \right].
\label{9.251}\eeq
Introducing the inverse transport mean free paths [cf.\ Eq.\ \req{9.158}]
\beq
\lambda_\ell^{-1} (s) \equiv \frac{1 - F_\ell(s)}{\lambda}
=  {\cal N}
\int [1-P_\ell(\cos\theta) ] \, \frac{\d \sigma(s)}{\d \Omega} \, \d
\Omegab,
\label{9.252}\eeq
we can express the Eqs.\ \req{9.251} in the form \citep{Lewis1950}
\beq
\frac{\partial}{\partial s} \Phi_{\ell m}(s; {\bf r}) +
 \lambda_\ell^{-1} (s) \, \Phi_{\ell m}(s;{\bf r}) = -
\sum_{\lambda,\mu} {\bf Q}_{\ell m,\lambda\mu} \dotprod
\nablab \Phi_{\lambda\mu}(s; {\bf r}).
\label{9.253}\eeq
The initial conditions satisfied by the $\Phi_{\ell m}$ coefficients are
\beq
\Phi_{\ell m}(0;{\bf r})= \delta({\bf r}) \, \delta_{m0} Y_{\ell
0}(\hat{\bf z}) = \sqrt{\frac{2\ell +1}{4\pi}} \delta_{m0} \delta({\bf
r}).
\label{9.254}\eeq


\subsubsection{Angular distribution \label{sec9.6.2.1}}

Integrating Eqs.\ \req{9.253} over the spatial coordinates, we
obtain the equations for the Legendre coefficients of the angular
distribution,
\beq
\Phi_{\rm L}(s; \Omegab) = \int\d {\bf r} \, \Phi_{\rm L}(s; {\bf r}, \Omegab).
\label{9.255}\eeq
Upon integration by parts, the integral over ${\bf r}$ of the right-hand
side of Eq.\ \req{9.253} is seen to vanish because the function
$\Phi_{\lambda\mu}(s; {\bf r})$ tends to zero when $r \rightarrow
\infty$. Then we have
\beq
\frac{\partial}{\partial s} \Phi_{\ell m}(s) +
\lambda_\ell^{-1}(s) \, \Phi_{\ell m}(s)= 0, \qquad
\Phi_{\ell m}(s) = \int \d {\bf r} \; \Phi_{\ell m}(s;{\bf r}) .
\label{9.256}\eeq
Moreover, equations with different $\ell,m$ are uncoupled. Since the
initial distribution \req{9.254} contains only components with $m=0$,
we lose all terms with $m\ne 0$, and find
\beq
\frac{\partial}{\partial s} \Phi_{\ell}(s) +
\lambda_\ell^{-1}(s) \, \Phi_{\ell}(s) = 0,
\label{9.257}\eeq
with the initial condition
\beq
\Phi_\ell(0) = \sqrt{\frac{2\ell +1}{4\pi}}.
\label{9.258}\eeq
Evidently, the solution is
\beq
\Phi_\ell (s) = \sqrt{\frac{2\ell +1}{4\pi}} \exp\left( - \int_0^s
\lambda_\ell^{-1} (s) \, \d s \right).
\label{9.259}\eeq
Hence, the angular distribution is given by
\beq
\Phi_{\rm L}(s; \Omegab) = \sum_\ell \Phi_\ell(s) \, Y_{\ell 0} (\Omegab) =
\sum_l \frac{2\ell +1}{4\pi}
\, k_\ell(s) \, P_\ell(\cos\theta)
\label{9.260}\eeq
with
\beq
k_\ell(s)
= \exp \left( - \int_0^s \lambda_\ell^{-1} (s') \, \d s' \right)
= \exp \left( -
\int_{E-\Delta_s}^E \frac{1}{\lambda_\ell (E')} \, \frac{\d
E'}{S(E')} \right)
\label{9.261}\eeq
In the last expression, $\lambda_\ell$ and the stopping power $S$ are
considered as
functions of the kinetic energy of the projectile. Evidently, the result
\req{9.260} reduces to the distribution of Goudsmit and Saunderson, Eq.\
\req{9.159}, when the energy loss is negligible.

For small path lengths, the Legendre expansion \req{9.260} converges
very slowly, mainly due to the contribution of unscattered particles
which is proportional to $\delta(\cos\theta -1)$. It is then
advantageous to separate the contribution from particles that have
undergone no collisions,
\beqa
\Phi_{\rm L}(s;\Omegab) &\equiv& \exp(-\langle n \rangle )
\frac{\delta(\cos\theta-1)}{\pi}
\nonumber \\ [2mm]
&& + \sum_{\ell=0}^{\infty}
\frac{2\ell+1}{4\pi} \left[ \exp(-k_{\ell})
- \exp(-\langle n \rangle ) \right]
P_{\ell}(\cos\theta)\, ,
\label{9.262} \eeqa
where
\beq
\langle n \rangle = \int_0^s \frac{\d s'}{\lambda(s')} =
\int_{E-\Delta_s}^E \frac{1}{\lambda(E')} \, \frac{\d E'}{S(E')}
\label{9.263}\eeq
is the average number of collisions undergone by the particle along the
path length $s$.

The average values of the Legendre polynomials take simple expressions,
\beq
\langle P_{\ell}(\cos\theta) \rangle_{\rm L} \equiv 2\pi \int_{-1}^{1}
P_{\ell}(\cos\theta) \Phi_{\rm L}(s;\theta) \, \d(\cos\theta) =
k_{\ell}(s).
\label{9.264} \eeq
In particular,
\beq
\langle \cos\theta \rangle_{\rm L} = k_{1}(s),
\label{9.265} \eeq
and
\beq
\langle \cos^{2}\theta \rangle_{\rm L} =
\frac{1}{3} \left[ 1+2 k_{2}(s) \right].
\label{9.266} \eeq

\citet{Negreanu2005} showed that angular distributions of electrons
transmitted through thin foils predicted by the Lewis theory are in good
agreement with measurements by \citet{Hanson1951} and by other authors.
In their comparisons, \citet{Negreanu2005} used elastic cross sections
and stopping powers given by the Monte Carlo code {\sc penelope}.


\subsubsection{Spatial moments and correlation functions
\label{sec9.6.2.2}}

Let us now consider the moments of the Lewis distribution,
Eq.\ \req{9.244}. Specifically, we would like to calculate the following
quantities,
\beq
\langle F({\bf r}) \, Y_{l,m}^\ast(\Omegab) \rangle \equiv \int \d {\bf r}
\, F({\bf r})
\int \d \Omegab \; Y_{l,m}^\ast(\Omegab) \, \Phi_{\rm L}(s;{\bf r},\Omegab),
\label{9.267}\eeq
where $F({\bf r})$ is a function of only the spatial coordinates.
Inserting the expansion \req{9.244}, we can write
\beq
\langle F({\bf r}) \, Y_{l,m}^\ast(\Omegab) \rangle = \int \d {\bf r}
\, F({\bf r}) \, \Phi_{\ell m} (s;{\bf r}).
\label{9.268}\eeq
Recalling that $Y_{00}(\Omegab) = (4\pi)^{-1/2}$, we see that purely
spatial moments, $\langle F({\bf r}) \rangle$, are determined by the
Legendre coefficient $\Phi_{00}$. Unfortunately, Eqs.\ \req{9.253}
cannot be solved for $\Phi_{00}$ because the equations for the various
$\Phi_{\ell m}$ are coupled. We shall therefore limit ourselves to
calculating the low-order moments of the spatial distribution and
correlation functions with the angles. The following derivation is taken
from \citet{KawrakowBielajew1998}, which generalizes the one sketched
by \citet{Lewis1950}.

We introduce the quantities
\beq
h_{\ell, m}^{p,q} (s) \equiv
\langle x^p z^q \, Y_{l,m}^\ast(\Omegab) \rangle
= \int \d {\bf r} \, x^p z^q \, \Phi_{\ell, m} (s; {\bf r}),
\label{9.269}\eeq
Note that, because the problem is invariant under rotations around the
$z$ axis, we only need to consider one of the transverse coordinates,
$x$ or $y$.
Multiplication of Eqs.\ \req{9.253} by $x^p z^q$, insertion of the values
of ${\bf Q}_{\ell m, \lambda \mu}$, and occasional integration by parts,
gives the following set of equations \citep{KawrakowBielajew1998},
\beqa
\left( \frac{\partial }{\partial s} + \lambda_\ell^{-1} \right)
h_{\ell, m}^{p,q}
&=&
\frac{p}{2} \left[
A_{m}^{\ell} h_{\ell-1, m-1}^{p-1,q}
+ A_{m+1}^{\ell+1} h_{\ell+1, m+1}^{p-1,q}
- A_{-m}^{\ell} h_{\ell-1, m+1}^{p-1,q}
- A_{-m+1}^{\ell+1} h_{\ell+1, m-1}^{p-1,q} \right]
\nonumber \\ [2mm]
&& \mbox{} + q \left[ B_{m}^{\ell} h_{\ell-1, m}^{p,q-1}
+ B_{m}^{\ell+1} h_{\ell+1, m}^{p,q-1} \right],
\label{9.270}\eeqa
with
\beq
A^\ell_m = \sqrt{\frac{(\ell + m)(\ell + m -1)}{4\ell^2 -1}} \qquad
\mbox{and} \qquad
B^\ell_m = \sqrt{\frac{(\ell + m)(\ell - m )}{4\ell^2 -1}}.
\label{9.271}\eeq
The corresponding boundary conditions are
\beq
h_{\ell, m}^{p,q} (0) = \sqrt{\frac{2\ell+1}{4\pi}} \, \delta_{q,0}
\, \delta_{p,0} \, \delta_{m,0}.
\label{9.272}\eeq
These equations can be solved in ascending order in $p$ and $q$.

To illustrate the solution method, we consider the case of the moments
 $\langle z^q \rangle$ of the longitudinal displacement and the
correlation function of $z^q$ and $\cos\theta$. We have
\beq
\langle z^q \rangle = \sqrt{4\pi} \, \langle z^q Y_{0,0} (\Omegab) \rangle
= \sqrt{4\pi} \, h_{0,0}^{0,q}(s)
\label{9.273}\eeq
and
\beq
\langle z^q \cos\theta \rangle = \sqrt{\frac{4\pi}{3}} \, \langle z^q
Y_{1,0} (\Omegab) \rangle
= \sqrt{\frac{4\pi}{3}} \, h_{1,0}^{0,q}(s).
\label{9.274}\eeq
From Eq.\ \req{9.270},
\beq
\left( \frac{\partial }{\partial s} + \lambda_\ell^{-1} \right) h_{\ell, 0}^{0,q}
= q  \left[ \frac{\ell}{\sqrt{4\ell^2-1}} \, h_{\ell-1, 0}^{0,q-1}
+ \frac{\ell+1}{\sqrt{4(\ell+1)^2-1}}\,  h_{\ell+1, 0}^{0,q-1} \right]
\label{9.275}\eeq
with the boundary conditions
\beq
h_{\ell, 0}^{0,q} (0) = \sqrt{\frac{2\ell+1}{4\pi}} \, \delta_{q,0}.
\label{9.276}\eeq
The first ($q=0$) of Eqs.\ \req{9.275} gives
\beq
h_{\ell, 0}^{0,0} (s) = \sqrt{\frac{2\ell+1}{4\pi}} \; k_\ell(s),
\label{9.277}\eeq
with $k_\ell (s)$ given by Eq.\ \req{9.261}.
The second equation ($q=1$) reads
\beqa
\left( \frac{\partial }{\partial s} + \lambda_\ell^{-1} \right) h_{\ell, 0}^{0,1}
&=& \frac{\ell}{\sqrt{4\ell^2-1}} \, h_{\ell-1, 0}^{0,0}
+ \frac{\ell+1}{\sqrt{4(\ell+1)^2-1}}\,  h_{\ell+1, 0}^{0,0}
\nonumber \\ [2mm]
&=&  \frac{\ell}{\sqrt{4\ell^2-1}} \,
\sqrt{\frac{2\ell-1}{4\pi}} \; k_{\ell-1}(s)
+ \frac{\ell+1}{\sqrt{4(\ell+1)^2-1}}\,
\sqrt{\frac{2\ell+3}{4\pi}} \; k_{\ell+1}(s)
\nonumber \\ [2mm]
&=&  \frac{1}{\sqrt{4\pi(2\ell+1)}} \left[ \ell k_{\ell-1}(s)
+ (\ell+1) k_{\ell+1}(s) \right],
\label{9.278}\eeqa
and its solution is
\beq
h_{\ell, 0}^{0,1} (s) =  \frac{1}{\sqrt{4\pi(2\ell+1)}}  \; k_\ell(s)
\int_0^s \frac{\d s'}{k_\ell(s')} \left[ \ell k_{\ell-1}(s') + (\ell+1)
k_{\ell+1}(s') \right]
\label{9.279}\eeq
Therefore,
\begin{subequations}
\label{9.280}
\beq
\langle z \rangle = \sqrt{4\pi} \, h_{0,0}^{0,1}(s) =
\int_0^s \d s' \, k_{1}(s'),
\label{9.280a}\eeq
and
\beq
\langle z \cos\theta \rangle = \sqrt{\frac{4\pi}{3}} \, h_{1,0}^{0,1}(s)
= \frac{1}{3}  \; k_1(s)
\int_0^s \frac{\d s'}{k_1(s')} \left[ 1 + 2 k_{2}(s') \right],
\label{9.280b}\eeq
where we have used that $\lambda_0^{-1} =0$ and $k_0(s)=1$.

Kawrakow and Bielajew (1998) used this method to derive
the following additional results:
\beq
\langle z^2 \rangle = \frac{2}{3} \int_0^s \d s' \, k_1(s')
\int_0^{s'} \frac{\d s''}{k_1(s'')} \left[ 1 + 2 k_{2}(s'') \right],
\label{9.280c}\eeq
\beq
\langle x^2 + y^2 \rangle = \frac{4}{3} \int_0^s \d s' \, k_1(s')
\int_0^{s'} \frac{\d s''}{k_1(s'')} \left[ 1 - k_{2}(s'') \right],
\label{9.280d}\eeq
\beq
\langle x \sin\theta\cos\phi + y \sin\theta\sin\phi\rangle = \frac{2}{3}
\, k_1(s)
\int_0^{s} \frac{\d s'}{k_1(s')} \left[ 1 - k_{2}(s') \right],
\label{9.280e}\eeq
\beqa
\langle x z \sin\theta\cos\phi \rangle &=& \frac{1}{15}
\, k_1(s)
\int_0^{s} \frac{\d s'}{k_1(s')}
\nonumber \\ [2mm]
&& \mbox{} \times \int_0^{s'} \d
s''  \left\{ 5 k_1(s'') + \frac{k_2(s')}{k_2(s'')}
[k_1(s'')- 6 k_{3}(s'')] \right\}, \rule{15mm}{0mm}
\label{9.280f}\eeqa
\beqa
\langle x^2 z \rangle &=& \frac{1}{15} \int_0^s \d s'
\, k_1(s')
\int_0^{s'} \frac{\d s''}{k_1(s'')}
\nonumber \\ [2mm]
&& \mbox{} \times \int_0^{s''} \d
s'''  \left\{ 9.k_1(s''') + \frac{k_2(s'')}{k_2(s''')}
[8 k_1(s''')- 18 k_{3}(s''')] \right\}.  \rule{15mm}{0mm}
\label{9.280g}\eeqa
\end{subequations}
It is worth observing that the quantities \req{9.280} are
completely determined by the values of the coefficients $k_\ell(s)$ of
lower orders, \ie, by the transport mean free paths
$\lambda_{\ell} (E)$. Additionally, they are independent
of the mean free path $\lambda$.
\index{multiple scattering with energy loss!Lewis theory|)}
\index{multiple scattering with energy loss|)}




\chapter{Computer programs
\label{chapt10}}



\index{Fortran programs|(} In this Chapter we describe two Fortran
codes that calculate cross sections for elastic
interactions and stopping powers of charged particles in matter by
using the theoretical models described in previous Chapters. These
programs are stand-alone and can be run interactively on modest personal
computers. They provide detailed quantitative information on the
interaction processes, which is useful for revealing differences between
various approaches and also for illustrating fundamental aspects of the
theory.

\index{computer code!{\sc elastic}}
The program {\sc elastic} computes differential and integrated cross
sections for elastic collisions of charged particles with atoms by using
the classical and quantum-mechanical approaches presented in Chapters
\ref{chapt4} and \ref{chapt5}. To alleviate the numerical work without
sacrificing the reliability of the calculated DCSs, the program uses
interaction potentials expressed as a sum of Yukawa-like terms (Section
\ref{sec3.6}). A summary description of {\sc elastic} has been published
recently by \citet{Salvat2022b}; the program, complemented with a Java
graphical user interface, is available from the Computer Physics
Communications Program Library.

\index{computer code!{\sc sbethe}} The {\sc sbethe} program gives the
stopping power calculated from the corrected Bethe formula, Eq.\
\req{8.187}, with the DHFS-model OOS (Section \ref{sec7.5.2}) and the
shell correction derived from the DHFS model for electrons, positrons,
and charged particles much heavier than the electron (Section
\ref{sec6.9}). {\sc Sbethe} accounts for the dielectric polarization of
the medium (Section \ref{sec8.2}). The stopping power for low-energy
particles is obtained from either the extrapolation formula \req{8.194}
or the fitted formula given by Eqs.\ \req{8.195} and \req{8.198}. The
{\sc sbethe} program was published by \citet{SalvatAndreo2023}; the
computer code and the database files are available from the Computer
Physics Communications Program Library.

The interaction models implemented in the programs have been tailored to
simplify numerical calculations. Their most relevant characteristic is
that the calculation involves only interpolations of tabulated functions
and integrations of functions of a single variable. Generally, we use
linear or linear log-log interpolation, except for a few smooth
functions which require high accuracy and are interpolated by using
natural cubic splines. Integrals of tabulated functions are frequently
evaluated by integrating the interpolating function. Integration of
functions given by an analytical formula are performed by using an
adaptive Gauss--Legendre quadrature algorithm. Numerical methods for
determining interpolation tables, various interpolation schemes, and
\index{Gauss--Legendre quadrature!adaptive}
Gauss--Legendre integration are described in the Complement at the end
of this Chapter. The computer codes generate multiple output files with
tables of functions, which can be visualized by means of a plotting
program. I recommend using {\sc gnuplot}, which is small in size,
\index{gnuplot}
available for various platforms (including Linux and Windows) and free;
this software package can be downloaded from the distribution sites
listed at the {\sc gnuplot} homepage, \url{http://www.gnuplot.info}.

\section{The program {\sc elastic} \label{sec10.1}}
\index{computer code!{\sc elastic}|(}

The program {\sc elastic} calculates DCSs for elastic collisions of
charged particles with atoms in the CM frame using the theory and
approximations described in Chapters \ref{chapt4} and \ref{chapt5}.
Specifically, the program delivers the DCSs calculated from the
relativistic trajectory method (Section \ref{sec4.3}), the Born
approximation (Section \ref{sec5.1.2}), the partial-wave expansion
method with Born and WKB phase shifts (Section \ref{sec5.1.3}) and the
eikonal approximation (Section \ref{sec5.1.4}). Optionally, in the case
of electrons and muons, the Born approximation and the partial-wave
expansion method can be replaced with their Dirac versions (see Section
\ref{sec5.2}).

Accurate calculations of the scattering of electrons and positrons by a
potential $V(r)$ given in tabular form may be performed by using
dedicated numerical procedures \citep[see, \eg,][and references
therein]{Salvat2005, SalvatFernandezVarea2019} for solving the radial
wave equation and summing up the partial-wave series. These procedures
are, generally, very demanding and, in addition, they cannot be used for
calculating the scattering of particles heavier than the electron.
{\sc Elastic} avoids most of the computational
burden by using approximate interaction potentials of the form
\req{3.149},
\beq
V(r) = \frac{Z_0 Z e^2}{r} \sum_{i=1}^3 A_i \exp(-a_i r),
\label{10.1}\eeq
which allow performing a good part of the calculations analytically.
With these potentials the evaluation of the DCS is reduced to a set of
integrals of functions of a single variable.

\index{Gauss--Legendre quadrature!adaptive}
As mentioned above, the program utilizes robust numerical methods. In
particular, most integrals are calculated by using an adaptive algorithm
based on the 20-point Gauss--Legendre quadrature formula, complemented
with a bisection scheme that allows a strict control of integration
errors (see Section \ref{sec10.4.3}). In calculations of the classical
DCS, the
deflection function $\vartheta(L)$ in the center-of-mass (CM)
frame\footnote{Here $\theta$ and $\vartheta$ are the scattering and
deflection angles in the CM frame, which were denoted with a prime in
Chapter \ref{chapt4}.} is computed to a relative accuracy of $\sim
10^{-7}$ for a dense grid of angular momenta, which is defined in an
adaptive way, and the table is interpolated by using log-lin natural
cubic spline interpolation (Section \ref{sec10.4.2}).  Because the
calculation of the angular deflection for a given angular momentum is
relatively fast, and to ensure accuracy of the DCS for small and large
angles, the table extends to very small and very large angular momenta,
from $\sim 10^{-7} \hbar$ to $\sim 10^8 \hbar$. When the integration
routine cannot attain the required accuracy, of at lest three decimal
places, the program issues an error message, and when the estimated
relative error of the DCS is larger than 1 \%, the classical DCS is set
to zero.

\index{Gauss--Legendre quadrature!adaptive} The eikonal DCS is evaluated
by using the expression \req{5.102} of the eikonal phase and the Zeitler
and Olsen formula \req{5.104} for the scattering amplitude. The integral
in the latter formula is calculated by using the adaptive
Gauss--Legendre quadrature method; the results are generally accurate to
four or more digits for scattering angles less than the practical cutoff
$\theta_{\rm eik} \sim 200(kR)^{-1}$, Eq.\ \req{5.107}. The WKB phase
shifts are also calculated with the Gauss--Legendre adaptive algorithm
to a relative accuracy of about $10^{-10}$. In the summation of
partial-wave series, the reduced-series method (Section \ref{sec5.2.3})
is applied for scattering angles larger than 1$^\circ$. In spite of all
this care for numerical accuracy, partial-wave results for high-energy
projectiles tend to be affected by the error of low-order phase-shifts
introduced by the WKB approximation. This kind of error manifests as a
high-frequency oscillation, normally with small amplitude, of the DCS
about an apparently correct smooth curve; the effect is clearly revealed
by plotting the relative difference of DCSs obtained from the
partial-wave method and from the eikonal approximation, which yields a
DCS that varies smoothly with the scattering angle. For projectiles with
very high energies, it may be impossible to achieve convergence of the
partial-wave series, not only because of the limited memory storage
allowed but also due to accumulated round-off errors. When such a case
is identified, partial-wave results are not delivered by the program
(the corresponding DCS is set to zero).

The user can select the projectile particle (the allowed options are
electrons, positrons, muons, antimuons, protons, antiprotons, and
alphas), the atomic number of the target atom (from $Z=1$ to 99), and
the parameterization of the atomic potential (see Section \ref{sec3.6}).
To allow studying the effect of the relativistic terms of the effective
potential [Eqs.\ \req{5.5}], classical trajectories and WKB phase shifts
may be evaluated by using either of the following potentials: \\ [2mm]
1) the bare electrostatic potential $V(r)$, \\ [2mm]
2) the potential with the second-order correction
\beq
V(r) - \frac{V^2(r)}{2\mu_{\rm r} c^2} \left( 1 - \frac{3
\mu_{\rm r} c^2}{\cal S} \right),
\label{10.2}\eeq
which reduces to the Klein-Gordon potential, Eq.\ \req{5.8}, for a
target atom with infinite mass, or \\ [2mm]
3) the full effective potential \req{5.5a}. \\ [2mm]
However, the Born DCS and phase shifts and the eikonal DCS are always
calculated with only the electrostatic potential $V(r)$ because their
calculation algorithms rest on the assumed analytical form \req{10.1} of
the potential.

{\sc elastic} generates an output file, named {\tt dcs.dat}, that
contains a table of DCSs in CM obtained by means of the classical
trajectory method, the Born approximation, the eikonal approximation,
and the partial-wave expansion method with approximate phase shifts. The
DCS is given for a predefined grid of 1,000 scattering angles, which is
densely spaced to allow accurate interpolation. The output file {\tt
dcs.dat} also contains the corresponding values of the total cross
section,
\beq
\sigma = \int \frac{\d \sigma}{\d \Omega} \, \d \Omega\, ,
\label{10.3}\eeq
and the first and second transport cross sections, \\ [2mm]
\beq
\sigma_1 = \int \left[ 1 - P_1(\cos\theta ) \right]
\frac{\d \sigma}{\d \Omega} \, \d \Omega
= \int (1 - \cos\theta )\frac{\d \sigma}{\d \Omega} \, \d \Omega
\label{10.4}\eeq
and \\ [2mm]
\beq
\sigma_2 = \int \left[ 1 - P_2(\cos\theta ) \right]
\frac{\d \sigma}{\d \Omega} \, \d \Omega
= \int \frac{3}{2} (1 - \cos^2\theta )
\frac{\d \sigma}{\d \Omega} \, \d \Omega .
\label{10.5}\eeq
Because the classical and partial-wave DCSs may present calculation
artifacts in difficult cases, these integrals are calculated by linear
log-log interpolation of the DCS tables, a method that is not highly
accurate but is generally robust against numerical fluctuations.

{\sc Elastic} also generates various files with results from
intermediate stages of the calculations and complementary information:
\\ [2mm]
$\bullet$ {\tt potential.dat}: table with values of the components of
the potential \req{5.5} used in the calculation.
\\ [2mm]
$\bullet$ {\tt thl.dat} and {\tt thl-spline.dat}: table of calculated
values of the $\vartheta(L)$ function, which determines the classical
DCS \req{4.176}, and of its interpolating spline for a denser grid of
$L$ values. The goodness of the interpolation can be verified by
plotting the contents of these two files (see Fig.\ \ref{fig10.1}, lower
plot). \\ [2mm]
$\bullet$ {\tt Schro-phase-shifts.dat} or {\tt Dirac-phase-shifts.dat}:
WKB and Born phase shifts, and the combined phase-shifts [Eqs.\
\req{5.70} or \req{5.129}] used in the calculation of the partial-wave
series of the scattering amplitudes.
\\ [2mm]
$\bullet$ {\tt scatamp.dat}: Complex scattering amplitudes from the
eikonal approximation, and from the partial-wave expansion method
whenever phase shifts can be calculated and the Legendre series do
converge to the required accuracy.
\\ [2mm]
$\bullet$ {\tt dcs-lab-angle.dat} and {\tt dcs-lab-energy.dat}: DCSs in
the L frame, per unit solid angle and differential in the energy loss
$W$ [Eqs.\ \req{4.202} and \req{4.203}, respectively], calculated from
the eikonal approximation, and from the partial-wave expansion method
when it is feasible. At the end of file {\tt dcs-lab-energy.dat},
the program gives the stopping cross section,
\beq
\sigma_{\rm st} \equiv \int_0^{W_{\rm max}} W \,
\frac{\d \sigma_1}{\d W} \, \d W \, ,
\label{10.6}\eeq

\noindent
evaluated from the eikonal and the partial-wave DCSs. This quantity
determines the slowing down of the projectile caused by elastic
collisions, the nuclear stopping power (see Section \ref{sec8.8}). The
files {\tt dcs-lab-angle.dat} and {\tt dcs-lab-energy.dat} are generated
only when the CM frame moves with respect to the L frame (\ie, when
$M_2$ is finite).

It is pertinent to bear in mind that {\sc elastic} implements various
massive numerical procedures (sum of Legendre series with up to several
hundred thousand terms, adaptive integration requiring a large number of
function evaluations, \ldots) whose accuracy is difficult to estimate.
Normally the program identifies conflicting situations and issues
warning messages when a calculation does not converge to the required
accuracy. Because accumulated round-off errors manifest as noise of the
calculated quantities, the calculation results are expected to be
correct only when the $\vartheta(L)$ function, the phase shifts of
orders $\ell \gtrsim 10$, and the DCSs vary smoothly with the
corresponding variable or index.

\begin{table}[th!]
\caption{Example of input data file for the {\sc elscat} program.
The scale lines are not part of the file.}
\label{tab10.1}
\begin{center}
\begin{verbatim}
C...+....1....+....2....+....3....+....4....+....5....+....6....+....7..
6         Projectile (1=e-, 2=e+, 3=mu-, 4=mu+, 5=p+, 6=p-, 7=alpha)
1         Wave equation (1=Schrod, 2=Schrod M=infty, 3=Dirac)
3         Potential (1=V(r), 2=+1st corr, 3=+1st+2nd corrs)
1         Screening model (1=DHFS, 2=TFM, 3=Wentzel)
79        Atomic number
1e8       Kinetic energy in LAB, as many lines as needed...
1e9
C...+....1....+....2....+....3....+....4....+....5....+....6....+....7..
\end{verbatim}
\end{center} \end{table}

\begin{figure}[p!] \begin{center}
\includegraphics*[width=15 cm]{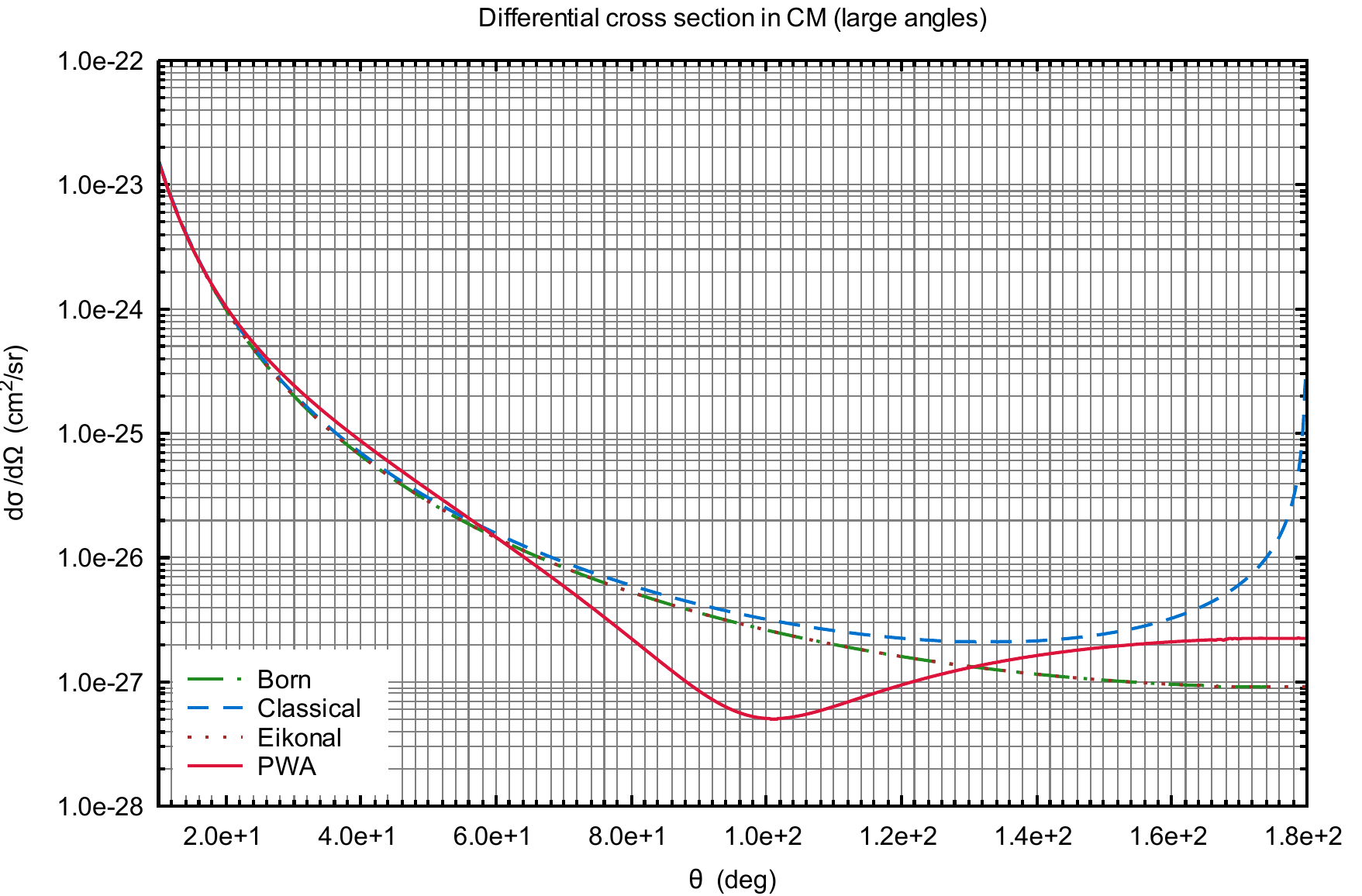}
\includegraphics*[width=15 cm]{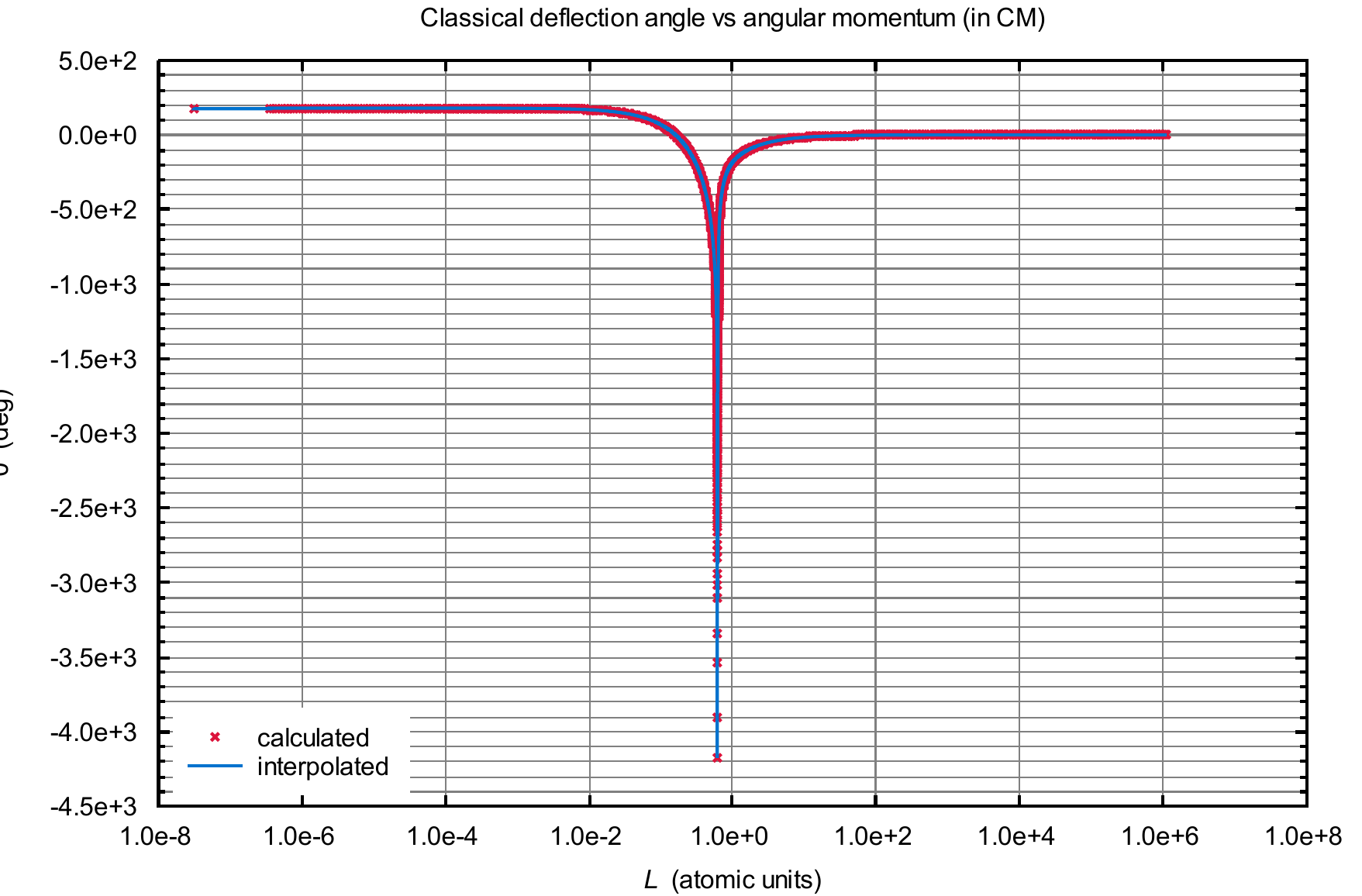}
\caption{ Results from {\sc elscat} for collisions of 100 MeV
antiprotons with gold atoms, using the full effective DHFS potential, as
specified in the example input file shown in Table \ref{tab10.1}.
\label{fig10.1}}
\end{center} \end{figure}

\begin{figure}[p!] \begin{center}
\includegraphics*[width=15 cm]{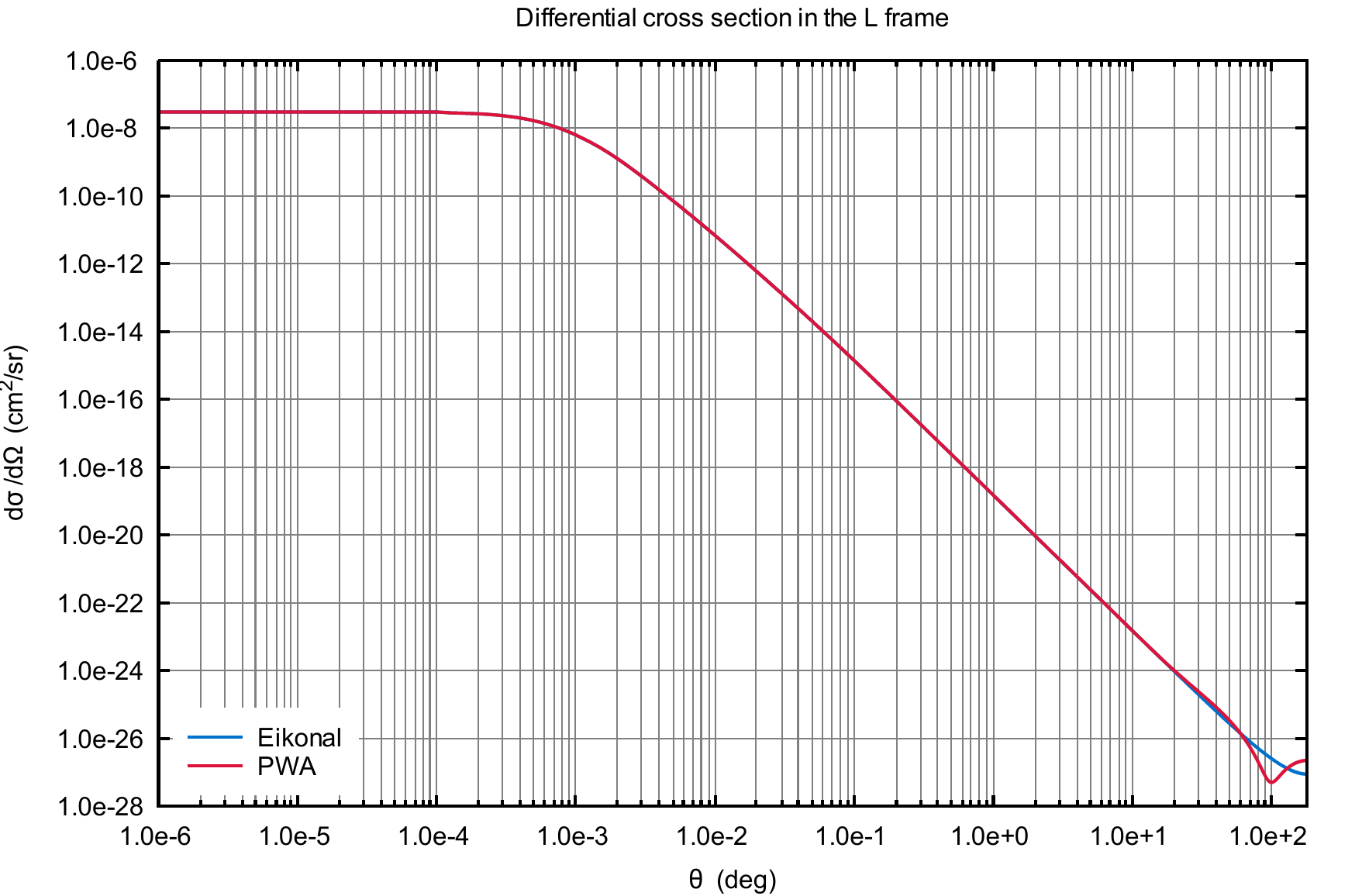}
\includegraphics*[width=15 cm]{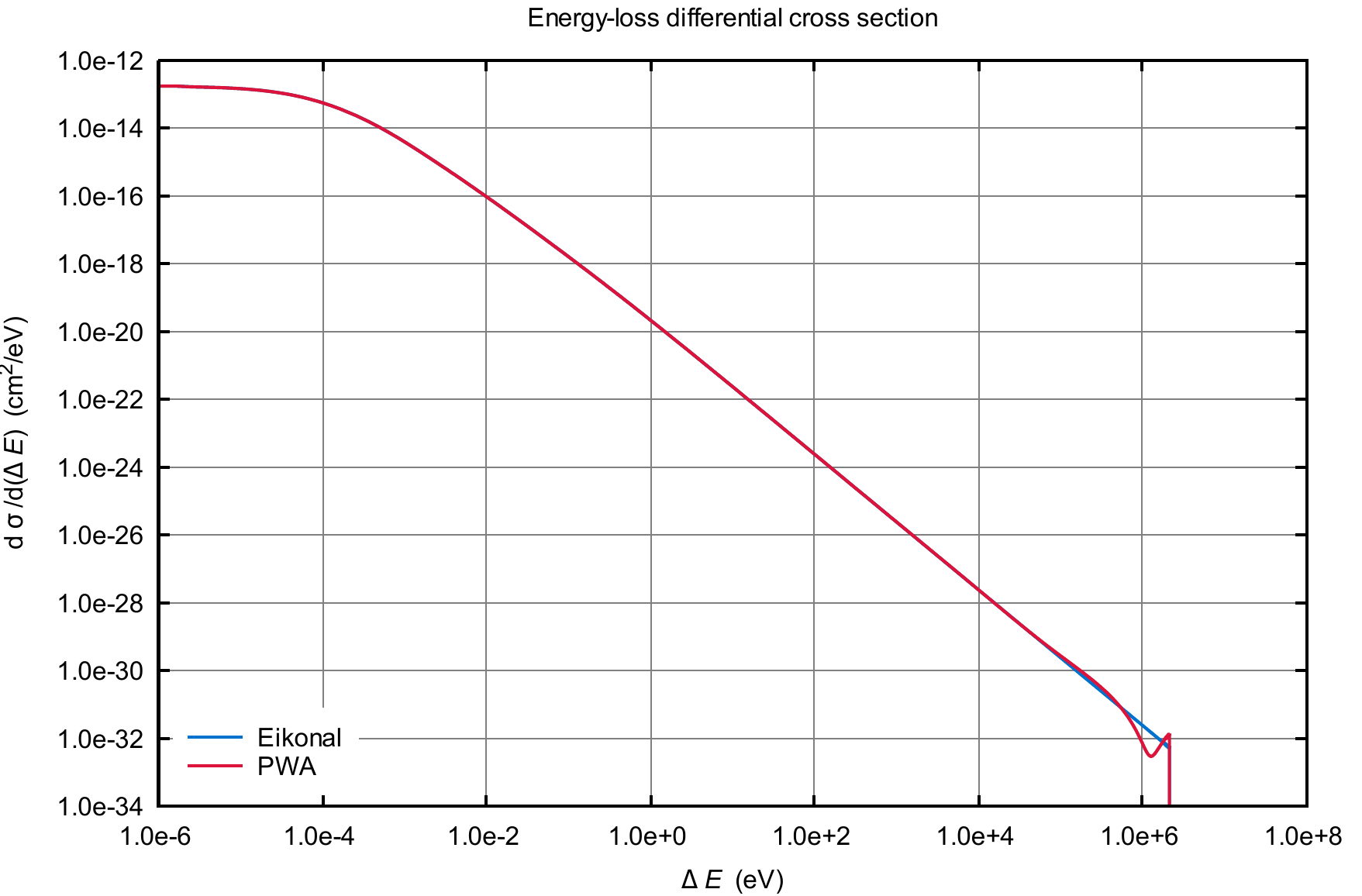}
\caption{ DCSs for collisions of 100 MeV antiprotons with gold atoms in
the L frame, calculated by {\sc elscat} for the full effective DHFS
potential, as specified in the example input file shown in Table
\ref{tab10.1}.
\label{fig10.2}}
\end{center} \end{figure}

{\sc Elastic} is devised to run interactively; typical calculation times
on an Intel Core i7 CPU at 1.99 GHz are less than about 5 seconds.
The user may enter the
definition data of the problem from the keyboard, in response to prompts
from the program, which are self-explanatory. Alternatively the program
can read data from an input file and be run in batch mode. The input
file provides the same information that would be entered from the
keyboard. Each line in this file contains numerical values (in free
format) followed by a brief text description (a reminder to the
user, not read by the program). Table \ref{tab10.1} shows an example of
input file for antiprotons colliding with gold atoms. The information in
the file is the following: \\ [2mm]
$\bullet$ 1st line: Kind of projectile, an integer, with the possible
values 1 (electron), 2 (positron), 3 (muon), 4 (antimuon), 5 (proton), 6
(antiproton), and 7 (alpha particle). \\ [2mm]
$\bullet$ 2nd line: Wave equation. An integer {\tt IWEQ} with one of the
values\\
\rule{5mm}{0mm}{\tt IWEQ}
\begin{minipage}[t]{13cm}
= 1, Schrodinger equation, \\
= 2, Schrodinger equation with $M_2=\infty$, \\
= 3, Dirac equation with $M_2=\infty$ (for electrons and muons only),
\end{minipage} \\ [2mm]
$\bullet$ 3rd line: Potential model. An integer,\\
\rule{5mm}{0mm} {\tt IVEF} \begin{minipage}[t]{13cm}
= 1, the atomic potential, $V(r)$, \\
= 2, the sum $V(r) + V_{\rm r1}(r)$, \\
= 3, the full effective potential, $V_{\rm eff} = V(r) + V_{\rm r1}(r)
+ V_{\rm r2}(r)$.
\end{minipage} \\
For the Dirac equation ({\tt IWEQ}=3), the program ignores the input
value and sets {\tt IVEF=1}. \\ [2mm]
$\bullet$ 4th line: Screening model. An integer with possible values 1
(DHFS), 2 (TFM), and 3 (Wentzel). \\ [2mm]
$\bullet$ 5th line: Atomic number of the target atom, an integer.\\ [2mm]
$\bullet$ 6th and following lines: Kinetic energy of the projectile in
the L frame, in eV. A single real value in each line, as many lines as
needed.

The program overwrites the output files at the end of the calculation of
each energy. To avoid loosing the generated information, a duplicate of
the {\tt dcs.dat} file is produced with the name {\tt dcs-xpxxxexx.dat}
derived from the character string {\tt x.xxxexx} of the energy value in
exponential format. The calculation of the set of DCSs for each
specific case (\ie, target atom, projectile type, and kinetic energy of
the projectile in L) takes only a few seconds on a personal computer,
quite independently of the details of the case. Partial
results from the calculation defined by the input file shown in Table
\ref{tab10.1} are displayed in Figs.\ \ref{fig10.1} and \ref{fig10.2}.
These plots were generated by using the {\sc gnuplot} scripts included
in the distribution package.

\index{computer code!{\sc elastic}|)}
\index{gnuplot}
\index{Goudsmit--Saunderson distribution}

The distribution bundle includes the Fortran program {\sc gosan}, which
calculates Gouds\-mit--Saun\-der\-son multiple scattering distributions
from the DCSs generated by {\sc elastic} by using the numerical
algorithm described in Section \ref{sec9.5.2.1}. {\sc Gosan} reads the
DCS table from the output file {\tt dcs.dat} of {\sc elastic},
calculates the coefficients $F_\ell$ of its Legendre expansion [Eq.\
\req{9.168}], and then calculates the multiple-scattering distribution
$\Phi_{\rm GS}(s;\Omegab)$ [Eq.\ \req{9.176}] for the path lengths $s$
selected by the user. The path length is specified by entering the
average number of collisions, $n_{\rm av} = s/\lambda$. The program
produces three output files, named {\tt GS-coefs.dat}, {\tt GS-pdf.dat},
and {\tt GS-dcs.dat} that contain, respectively, a table of coefficients
$F_\ell$, a table of the calculated multiple scattering distribution,
and the table of the DCS (as read from the file {\tt dcs.dat}) and the
DCS recalculated from its Legendre expansion. The contents of the output
files can be visualized by using the provided {\sc gnuplot} script {\tt
gosan.gnu}. Notice that the files {\tt GS-coefs.dat} and {\tt
GS-pdf.dat} are overwritten at the end of the calculation of each new
path length.


\section{The program {\sc sbethe} \label{sec10.3}}
\index{computer code!{\sc sbethe}|(}

The Fortran program {\sc sbethe}  calculates the
electronic stopping power of a material for fast charged particles from
the corrected Bethe formula, Eq.\ \req{8.187}, with the various
corrections computed as described in Chapter \ref{chapt8}, and the
extrapolation \req{8.194} or the low-energy fitted formula \req{8.195}
with the joining recipe \req{8.198}. The {\sc sbethe} program utilizes a
database of subshell OOSs and atomic shell corrections obtained from
PWBA calculations with the DHFS self-consistent potential of free
neutral atoms \citep[Section \ref{sec7.5.2}][]{BoteSalvat2008,
Salvat2022a, Salvat2022c}. 

The database associated to the {\sc sbethe} program is contained in the
subdirectory {\tt ./sdbase} that contains the following 599 ASCII files.  The
string {\tt zz} in the file names denotes the atomic number of the
element, two digits: \\
$\circ$ 99 files {\tt oos-zz.tab} with tables of the subshell OOSs that were
extracted from the database of GOSs calculated with the DHFS
potential. \\
$\circ$ 99 files {\tt shcorr-zz.tab} with tables of the atomic
shell correction $C(\gamma)/Z$ (see Subsection \ref{sec6.9.1}) for protons
and other projectiles heavier than the electron. \\
$\circ$ 99 files {\tt eshcorr-zz.tab} with tables of the atomic
shell correction $C(\gamma)/Z$ for electrons. \\
$\circ$ 99 files {\tt pshcorr-zz.tab} with tables of the atomic
shell correction $C(\gamma)/Z$ for positrons. \\
$\circ$ 99 files {\tt pdebr-zz.p08} with tables of scaled bremsstrahlung
DCSs, $\chi(Z,E;\kappa)$, and integrated cross sections, $\phi_{\rm
rad}(Z,E)$, for electrons (Section \ref{sec8.6}). \\
$\circ$ 99 files {\tt rmuonzz.tab} with tables of the quantities $b_{\rm
br}$, $b_{\rm pair}$, $b_{\rm photo}$, and $b_{\rm total}$ (see Section
\ref{sec8.7}) for a grid of 16 total energies (the same for all the
elements). Obtained by interpolation in $Z$ of the tables given in
\citet{Groom2001}. \\
$\circ$ {\tt pdatconf.p14}, this file contains a list of ground-state
configurations, and ``experimental'' subshell ionization energies
\citep{Carlson1975} of free atoms of the elements.\\
$\circ$ {\tt pdcompos.pen} contains composition data and physical
parameters for the materials listed in {\tt material-list.txt}, taken
from the database of the ESTAR program of \citet{Berger1992}. \\
$\circ$ {\tt shparams.tab} lists the numerical values of the
energy-independent parameters in the asymptotic formulas
\citet{Salvat2022a} for
the integrated cross sections of the subshells of neutral DHFS atoms. \\
$\circ$ {\tt atparams.tab} lists the numerical values of the
energy-independent parameters in the asymptotic formulas for
the integrated cross sections of neutral DHFS atoms. \\
$\circ$ {\tt exp-param.tab} parameters of the analytical formulas
\req{8.195} for protons and alphas in those elemental materials with
sufficient stopping data in the IAEA database to ensure consistency of
the fits.

The program {\sc sbethe} runs interactively, input data are entered from the
keyboard following the program prompts, which are self-explanatory. The
program starts by asking the name {\tt mname} of the material, an
alphanumeric string of up to 15 characters.  If a file with the name
{\tt mname.mat} exists in the working directory, the program reads the
material parameters (composition, mass density and mean excitation
energy) and the OOS from that file.  Otherwise, {\sc sbethe} asks for
the parameters of the material, and builds the OOS table by using the
DHFS subshell OOSs in the database. To minimize the amount of input
information, the program can read the material characteristics from the
file {\tt pdcompos.pen}; the list of predefined materials with their
identifying numbers is given in the file {\tt material-list.txt} and in
Table \ref{tab10.2}. Once
the material parameters and the OOS table are set, the program writes
them in the output file {\tt mname.mat}, which will be read directly in
future runs for that material. The user can select the kind of projectile
particle among the default options (electrons, positrons, negative muons,
antimuons, protons, antiprotons, and alphas), or enter the charge and
mass of the desired projectile.

The electronic stopping powers obtained from the corrected Bethe formula
are determined by the adopted values of the mean excitation energy $I$.
By default, the program sets $I=I_{\rm ICRU}$, the $I$ value recommended
in the ICRU Reports 37 (\citeyear{ICRU37}) and 90 (\citeyear{ICRU90}). The
results shown in Figs.\ \ref{fig8.8} to \ref{fig8.13} were
generated with this default $I$ value. In order to permit analyzing the
dependence of the calculated stopping power on the adopted mean
excitation energy, the user is allowed to change the proposed value of
this parameter. With the default $I$ value and for projectiles with $E >
E_{\rm cut}$, the results from the program are in close agreement with
the ICRU recommended stopping powers \citep{ICRU37, ICRU49, ICRU90}.

The {\sc sbethe} program generates tables of the stopping power and
related quantities for a nearly logarithmic grid of kinetic energies of
the projectile with 66 points per decade. The output of {\sc sbethe}
consists of the following formatted text files: \\ [2mm]
$\bullet$ {\tt OOS.dat}: optical oscillator strength $F(W) \equiv \d
f(W)/ \d W$ of the material, as a function of the excitation energy $W$,
calculated from the DHFS-model atomic subshell OOSs with the adopted $I$
value, as described in Section \ref{sec4.1}. \\ [2mm]
$\bullet$ {\tt stplog.dat}: table of the electronic stopping cross
section per atom or molecule, $\sigma^{(1)}_{\rm in} \equiv S_{\rm in}
/{\cal N}$, calculated from the corrected Bethe formula \req{8.187}, the
Bethe ``logarithm''
\beq
L_0 \equiv \ln \left(\frac{2 \me v^2}{I} \right)
+ \ln \gamma^2 - \beta^2 ,
\label{10.11}\eeq
the function $f(\gamma)/2$, and the corrections
$C_{\rm mod}(\gamma)/Z$, $\delta_{\rm F}/2$,
$\Delta L^{\rm LS}$ and $\Delta L^{\rm B}(a)$. 

\addtocounter{table}{+1}
\newpage
\noindent {\small {\bf Table 10.2}: \rule[-2.5mm]{0mm}{0mm}List of the
280 pre-defined materials
included in the {\tt pdcompos.pen} file, with their identifying numbers
\citep[adapted from][]{Berger1992}.\label{tab10.2}}
\vspace*{3mm}
\addtolength{\baselineskip}{-0.86mm}
\hrule
\vspace*{-4mm}
\begin{verbatim}
 ***  ELEMENTS (id. number = atomic number):
   1  Hydrogen              34  Selenium              67  Holmium
   2  Helium                35  Bromine               68  Erbium
   3  Lithium               36  Krypton               69  Thulium
   4  Beryllium             37  Rubidium              70  Ytterbium
   5  Boron                 38  Strontium             71  Lutetium
   6  Amorphous carbon      39  Yttrium               72  Hafnium
   7  Nitrogen              40  Zirconium             73  Tantalum
   8  Oxygen                41  Niobium               74  Tungsten
   9  Fluorine              42  Molybdenum            75  Rhenium
  10  Neon                  43  Technetium            76  Osmium
  11  Sodium                44  Ruthenium             77  Iridium
  12  Magnesium             45  Rhodium               78  Platinum
  13  Aluminium             46  Palladium             79  Gold
  14  Silicon               47  Silver                80  Mercury
  15  Phosphorus            48  Cadmium               81  Thallium
  16  Sulfur                49  Indium                82  Lead
  17  Chlorine              50  Tin                   83  Bismuth
  18  Argon                 51  Antimony              84  Polonium
  19  Potassium             52  Tellurium             85  Astatine
  20  Calcium               53  Iodine                86  Radon
  21  Scandium              54  Xenon                 87  Francium
  22  Titanium              55  Cesium                88  Radium
  23  Vanadium              56  Barium                89  Actinium
  24  Chromium              57  Lanthanum             90  Thorium
  25  Manganese             58  Cerium                91  Protactinium
  26  Iron                  59  Praseodymium          92  Uranium
  27  Cobalt                60  Neodymium             93  Neptunium
  28  Nickel                61  Promethium            94  Plutonium
  29  Copper                62  Samarium              95  Americium
  30  Zinc                  63  Europium              96  Curium
  31  Gallium               64  Gadolinium            97  Berkelium
  32  Germanium             65  Terbium               98  Californium
  33  Arsenic               66  Dysprosium            99  Einsteinium

 ***  COMPOUNDS AND MIXTURES (in alphabetical order):
 100  Acetone
 101  Acetylene
 102  Adenine
 103  Adipose tissue (ICRP)
 104  Air, dry (near sea level)
 105  Alanine
 106  Aluminum oxide
 107  Amber
 108  Ammonia
 109  Aniline
 110  Anthracene
 111  B-100 bone-equivalent plastic
 112  Bakelite
 113  Barium fluoride
 114  Barium sulfate
 115  Benzene
 116  Beryllium oxide
 117  Bismuth germanium oxide
 118  Blood (ICRP)
 119  Bone, compact (ICRU)
 120  Bone, cortical (ICRP)
 121  Boron carbide
 122  Boron oxide
 123  Brain (ICRP)
 124  Butane
 125  N-butyl alcohol
 126  C-552 air-equivalent plastic
 127  Cadmium telluride
 128  Cadmium tungstate
 129  Calcium carbonate
 130  Calcium fluoride
 131  Calcium oxide
 132  Calcium sulfate
 133  Calcium tungstate
 134  Carbon dioxide
 135  Carbon tetrachloride
 136  Cellulose acetate, cellophane
 137  Cellulose acetate butyrate
 138  Cellulose nitrate
 139  Ceric sulfate dosimeter solution
 140  Cesium fluoride
 141  Cesium iodide
 142  Chlorobenzene
 143  Chloroform
 144  Concrete, portland
 145  Cyclohexane
 146  1,2-dichlorobenzene
 147  Dichlorodiethyl ether
 148  1,2-dichloroethane
 149  Diethyl ether
 150  N,n-dimethyl formamide
 151  Dimethyl sulfoxide
 152  Ethane
 153  Ethyl alcohol
 154  Ethyl cellulose
 155  Ethylene
 156  Eye lens (ICRP)
 157  Ferric oxide
 158  Ferroboride
 159  Ferrous oxide
 160  Ferrous sulfate dosimeter solution
 161  Freon-12
 162  Freon-12b2
 163  Freon-13
 164  Freon-13b1
 165  Freon-13i1
 166  Gadolinium oxysulfide
 167  Gallium arsenide
 168  Gel in photographic emulsion
 169  Pyrex glass
 170  Glass, lead
 171  Glass, plate
 172  Glucose
 173  Glutamine
 174  Glycerol
 175  Graphite
 176  Guanine
 177  Gypsum, plaster of Paris
 178  N-heptane
 179  N-hexane
 180  Kapton polyimide film
 181  Lanthanum oxybromide
 182  Lanthanum oxysulfide
 183  Lead oxide
 184  Lithium amide
 185  Lithium carbonate
 186  Lithium fluoride
 187  Lithium hydride
 188  Lithium iodide
 189  Lithium oxide
 190  Lithium tetraborate
 191  Lung (ICRP)
 192  M3 wax
 193  Magnesium carbonate
 194  Magnesium fluoride
 195  Magnesium oxide
 196  Magnesium tetraborate
 197  Mercuric iodide
 198  Methane
 199  Methanol
 200  Mixed wax
 201  Ms20 tissue substitute
 202  Muscle, skeletal (ICRP)
 203  Muscle, striated (ICRU)
 204  Muscle-equivalent liquid, with sucrose
 205  Muscle-equivalent liquid, without sucrose
 206  Naphthalene
 207  Nitrobenzene
 208  Nitrous oxide
 209  Nylon, du Pont elvamide 8062
 210  Nylon, type 6 and type 6/6
 211  Nylon, type 6/10
 212  Nylon, type 11 (rilsan)
 213  Octane, liquid
 214  Paraffin wax
 215  N-pentane
 216  Photographic emulsion
 217  Plastic scintillator (vinyltoluene based)
 218  Plutonium dioxide
 219  Polyacrylonitrile
 220  Polycarbonate (makrolon, lexan)
 221  Polychlorostyrene
 222  Polyethylene
 223  Polyethylene terephthalate (mylar)
 224  Polymethyl methacrilate (lucite, perspex, plexiglass)
 225  Polyoxymethylene
 226  Polypropylene
 227  Polystyrene
 228  Polytetrafluoroethylene (teflon)
 229  Polytrifluorochloroethylene
 230  Polyvinyl acetate
 231  Polyvinyl alcohol
 232  Polyvinyl butyral
 233  Polyvinyl chloride
 234  Polyvinylidene chloride (saran)
 235  Polyvinylidene fluoride
 236  Polyvinyl pyrrolidone
 237  Potassium iodide
 238  Potassium oxide
 239  Propane
 240  Propane, liquid
 241  N-propyl alcohol
 242  Pyridine
 243  Rubber, butyl
 244  Rubber, natural
 245  Rubber, neoprene
 246  Silicon dioxide
 247  Silver bromide
 248  Silver chloride
 249  Silver halides in photographic emulsion
 250  Silver iodide
 251  Skin (ICRP)
 252  Sodium carbonate
 253  Sodium iodide
 254  Sodium monoxide
 255  Sodium nitrate
 256  Stilbene
 257  Sucrose
 258  Terphenyl
 259  Testes (ICRP)
 260  Tetrachloroethylene
 261  Thallium chloride
 262  Tissue, soft (ICRP)
 263  Tissue, soft (ICRU four-component)
 264  Tissue-equivalent gas (methane based)
 265  Tissue-equivalent gas (propane based)
 266  Tissue-equivalent plastic (A-150)
 267  Titanium dioxide
 268  Toluene
 269  Trichloroethylene
 270  Triethyl phosphate
 271  Tungsten hexafluoride
 272  Uranium dicarbide
 273  Uranium monocarbide
 274  Uranium oxide
 275  Urea
 276  Valine
 277  Viton fluoroelastomer
 278  Water, liquid
 279  Water vapour
 280  Xylene
\end{verbatim} \hrule
\addtolength{\baselineskip}{+0.86mm}

\noindent
$\bullet$ {\tt stp.dat}: table of the electronic stopping power
calculated from the corrected Bethe formula, Eq.\ \req{8.187}, with and
without the shell correction (useful for visualizing the effect of the
shell correction). \\ [2mm]
$\bullet$ {\tt stp-low.dat}: table of the electronic stopping power
obtained from the low-energy extrapolation, Eq.\ \req{8.194}, of results
from the corrected Bethe formula. For protons and alpha particles in
elements with enough experimental stopping-power data, the program
gives preference to the fitted analytical formula \req{8.195}.
\\ [2mm]
$\bullet$ {\tt lstp.dat}: tables of the electronic stopping power
(including the low-energy extrapolation), the radiative stopping power
(null for projectiles heavier than the muon), and their sum, the
total stopping power, $S(E)=S_{\rm in}(E)+S_{\rm rad}(E)$, all in eV/\AA.
The fifth column is
the CSDA range $R(E)$, in cm, Eq.\ \req{9.88} with $E_{\rm abs}$ equal to
the lowest energy in the table or plot. \\  [2mm]
$\bullet$ {\tt mstp.dat}: table of the electronic, radiative and total
mass stopping powers, $S(E)/\rho$ (in MeV cm$^2$/g), and the CSDA
range times the mass density of the material (in g/cm$^2$). \\  [2mm]
$\bullet$ {\tt depth-dose.dat}: table of CSDA depth-dose distributions, Eq.\
\req{9.89.1}, for projectiles with various initial energies. \\ [2mm]
$\bullet$ {\tt asymptotic.dat}: table of inelastic total, stopping, and
energy-straggling molecular cross sections, calculated from the
asymptotic formulas, given by Eqs.\ (89), (96), and (105) in the article
of \citet{Salvat2022a}, for DHFS atoms.
The values in this table are not expected to be realistic; they are
provided only to reveal the limitations of the uncorrected asymptotic
formulas. \\ [2mm]
$\bullet$ {\tt PENstp.dat}: table of the electronic
mass stopping power, $S_{\rm in}(E)/\rho$ (in MeV cm$^2$/g), with the
energy grids used by the simulation codes {\sc penelope} and {\sc
penhan} \citep{Salvat2025, SalvatHeredia2023}. These tables can be used
to replace the default stopping power tables in the material data files
generated by the codes, which are less reliable than the tables produced
by {\sc sbethe}. This replacement requires having a flexible file editor
that allows selecting and replacing a text block.

Results from {\sc sbethe} for projectile protons and electrons in copper
(material identification number 29) are displayed in Figs.\ \ref{fig10.5}
to \ref{fig10.8}, which are screen shots of the plots generated with the
provided {\sc gnuplot} scripts. Similar data for muons and protons in
liquid water are shown in Figs \ref{fig10.9} and \ref{fig10.10}. Of course,
the scripts will work only when {\sc gnuplot} is installed on the
computer. When the extension ``{\tt .gnu}'' is associated to {\sc
gnuplot}, a script can be executed by simply clicking the mouse with the
pointer on the script icon.

\index{computer code!{\sc sbethe}|)}


\section{Distribution package and installation \label{sec10.4}}
\index{computer code!{\sc elastic}}
\index{computer code!{\sc sbethe}}

The programs {\sc elastic} and {\sc sbethe} are
distributed as ancillary material in a single compressed zip file. The
contents of that file consists of a single directory
that contains a {\tt readme.txt} file with operation instructions, the {\tt
material-list.txt} file with the list of predefined materials in the
file 'pdcompos.pen' and the corresponding identification num-

\begin{figure}[t!] \begin{center}
\includegraphics*[width=15.0cm]{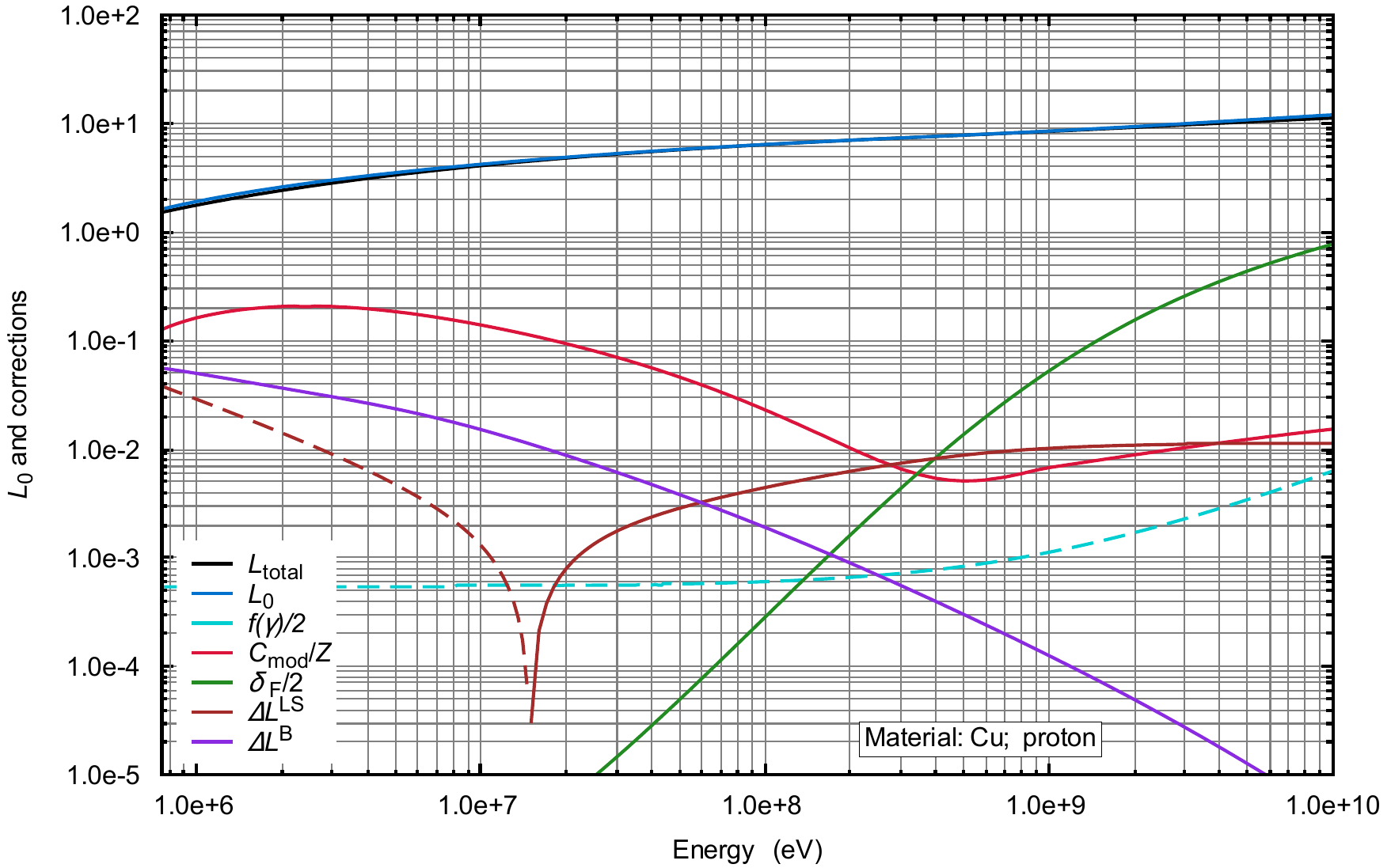}
\caption{Results from {\sc sbethe} for protons in solid copper,
	plotted with {\sc gnuplot} by using the provided script. Terms in the
	corrected Bethe formula \req{8.187}, as functions of the kinetic energy
	of the projectile. The dashed curves represent negative values.
\label{fig10.5}}\end{center} \end{figure}

\begin{figure}[b!] \begin{center}
\includegraphics*[width=15.0cm]{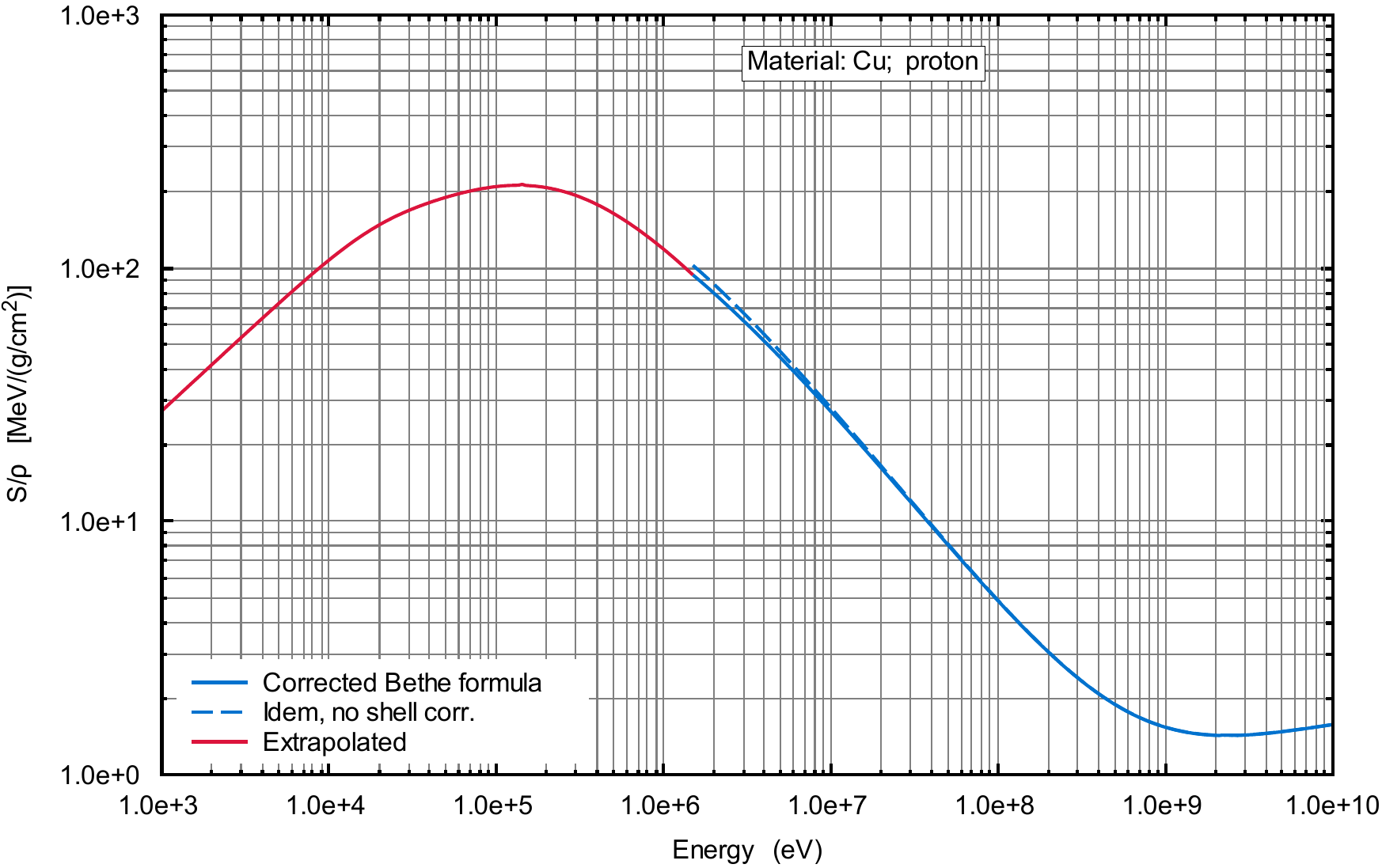}
\caption{Results from {\sc sbethe} for protons in solid copper,
	plotted with {\sc gnuplot} by using the provided script. Mass stopping
	power as a function of the kinetic energy	of the projectile.
\label{fig10.6}}
\end{center} \end{figure}

\begin{figure}[t!] \begin{center}
\includegraphics*[width=15.0cm]{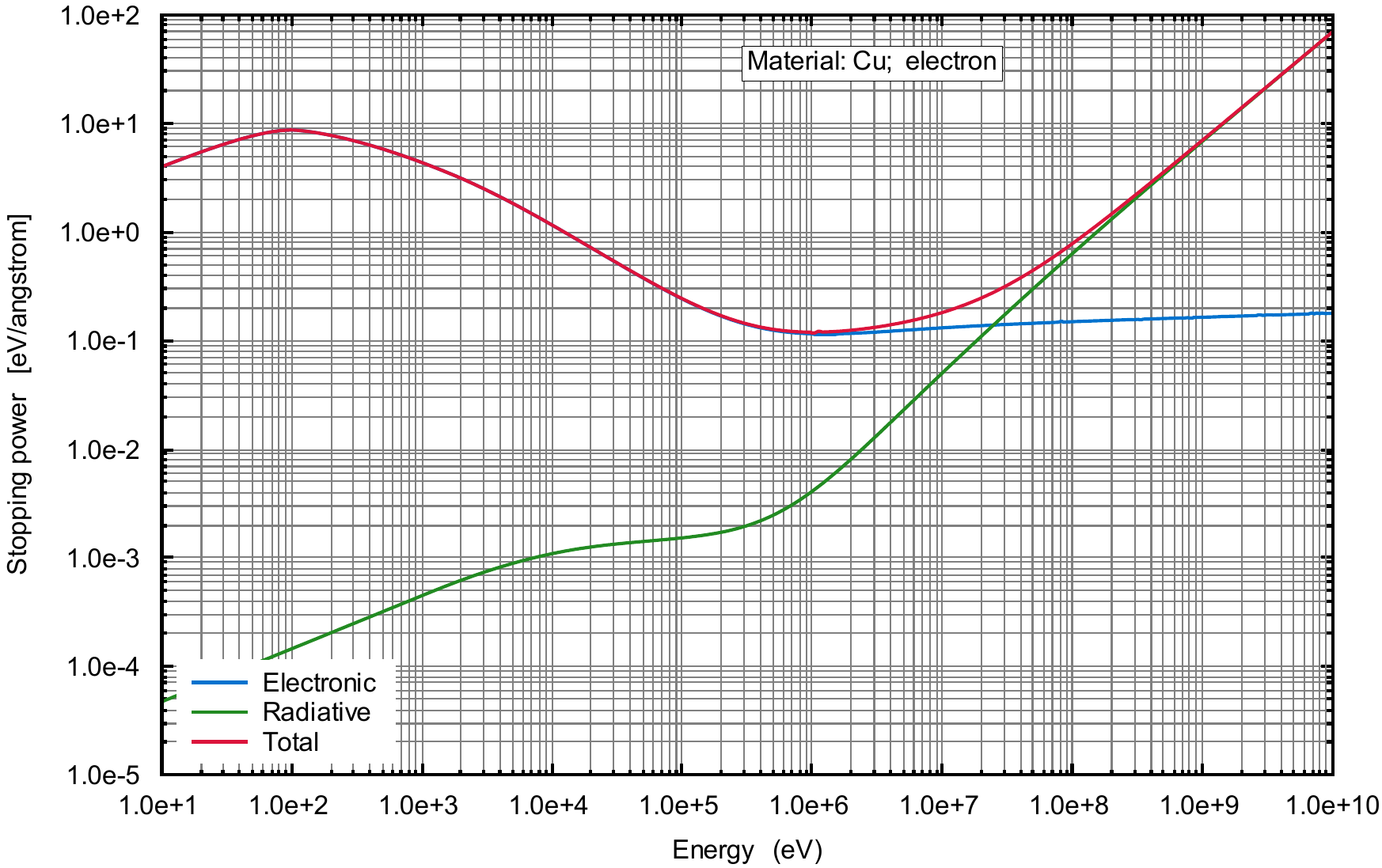}
\caption{Results from {\sc sbethe} for electrons in solid copper,
	plotted with {\sc gnuplot} by using the provided script. Electron
	stopping powers as functions of the kinetic energy of the projectile.
\label{fig10.7}}\end{center} \end{figure}

\begin{figure}[b!] \begin{center}
\includegraphics*[width=15.0cm]{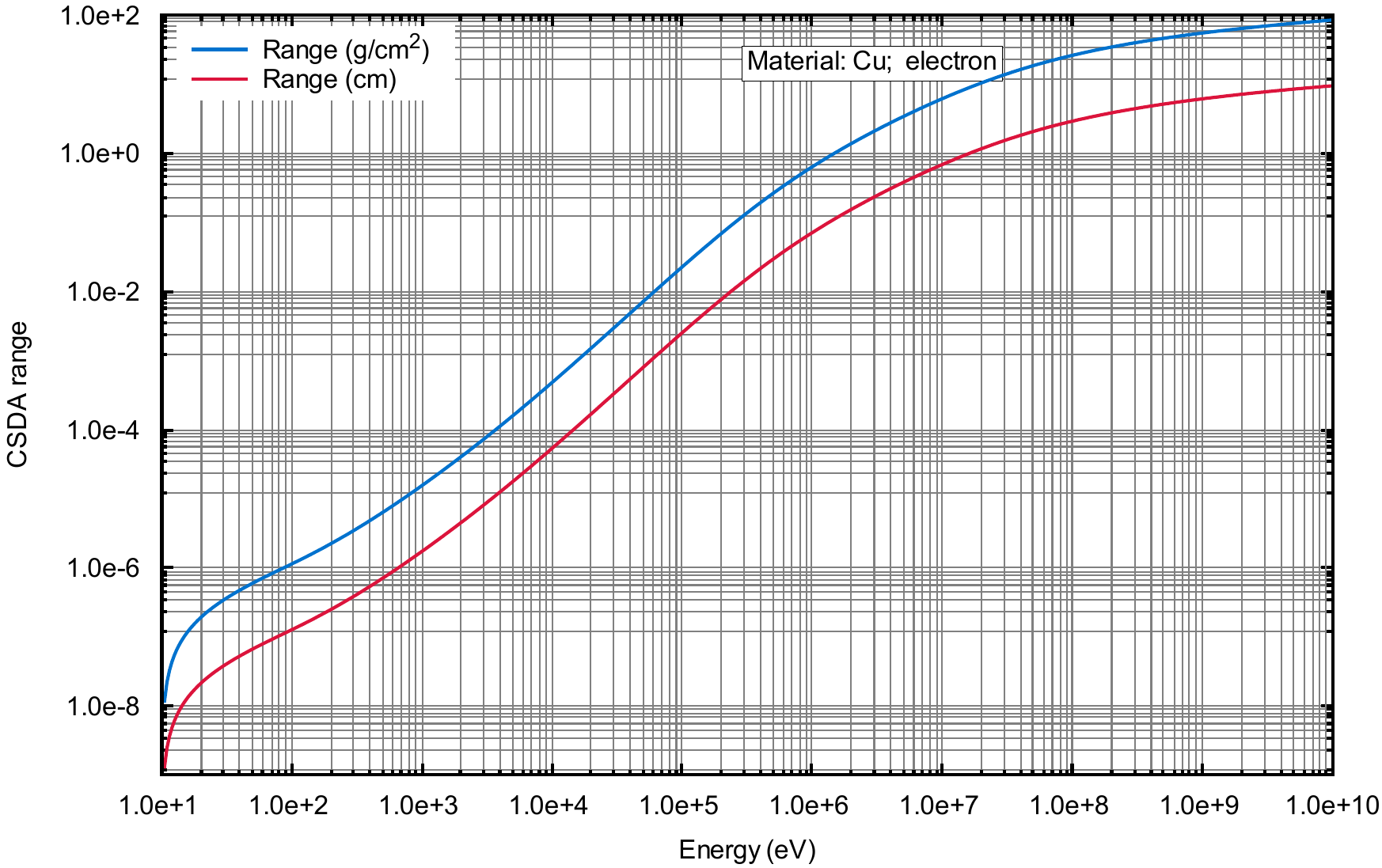}
\caption{Results from {\sc sbethe} for electrons in solid copper,
	plotted with {\sc gnuplot} by using the provided script. Electron
	range as a function of the kinetic energy	of the projectile.
\label{fig10.8}}\end{center} \end{figure}

\begin{figure}[t!] \begin{center}
\includegraphics*[width=15.0cm]{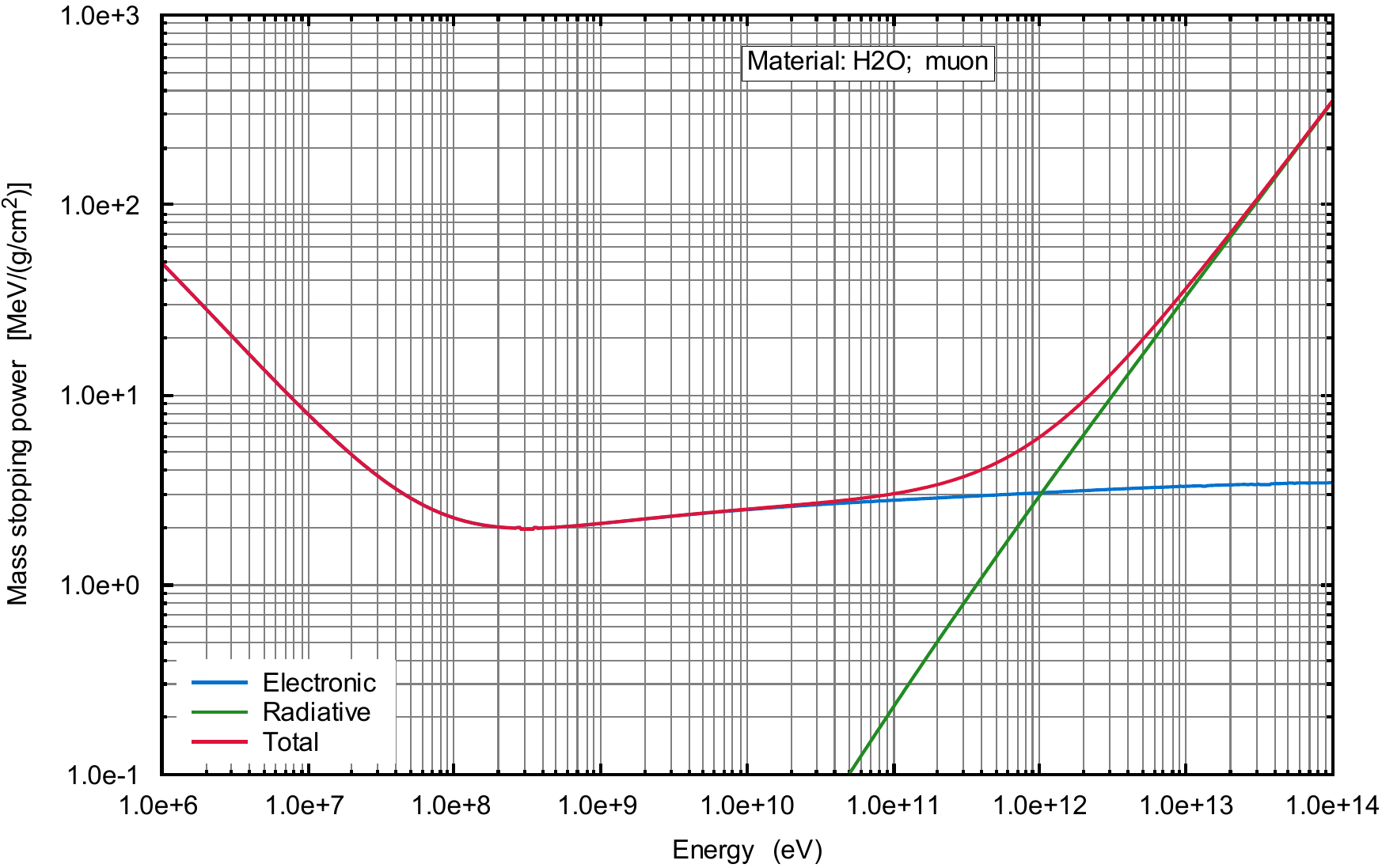}
\caption{Results from {\sc sbethe} for muons in liquid water,
	plotted with {\sc gnuplot} by using the provided script. Muon
	stopping powers as functions of the kinetic energy of the projectile.
\label{fig10.9}}\end{center} \end{figure}

\begin{figure}[b!] \begin{center}
\includegraphics*[width=15.0cm]{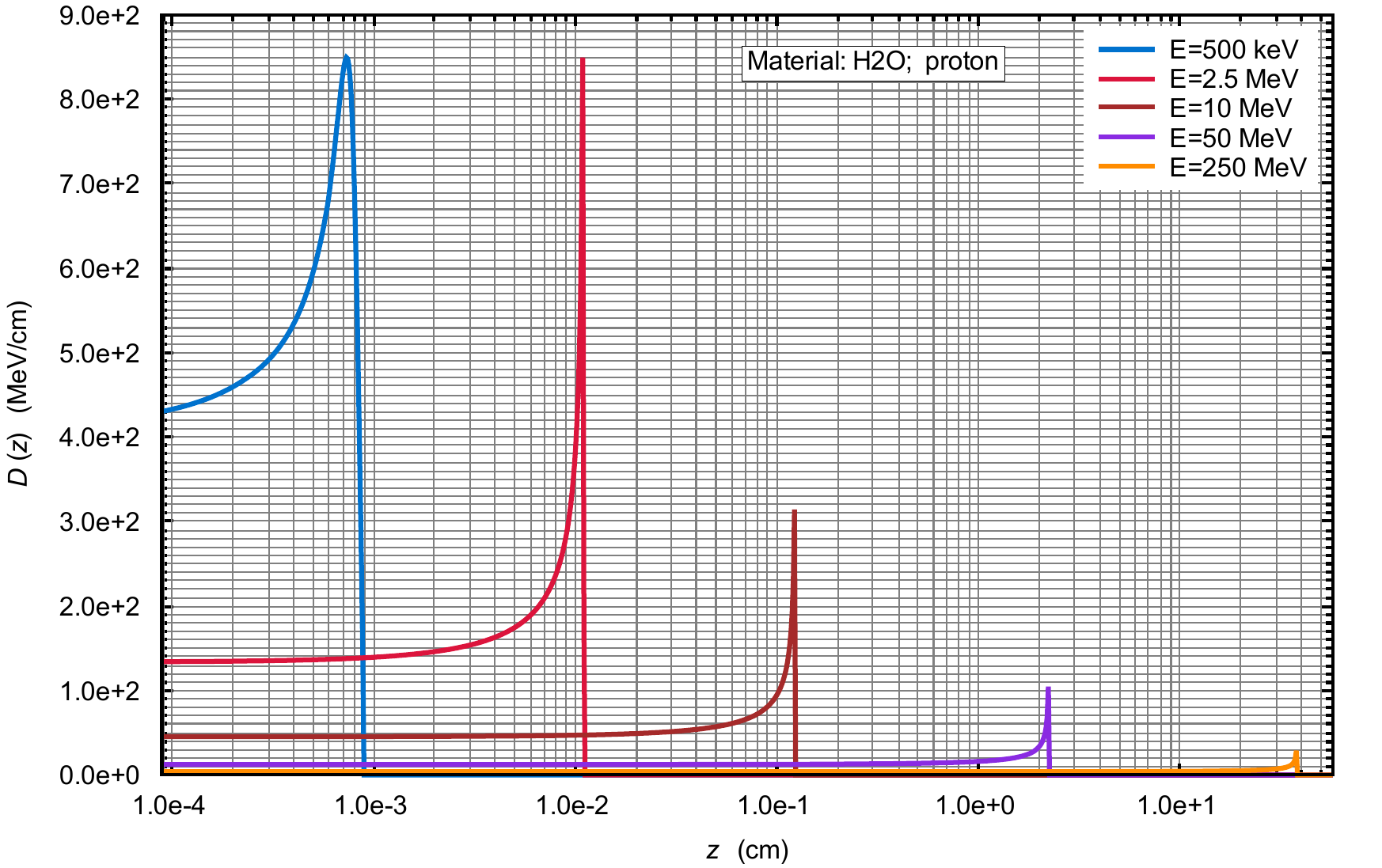}
\caption{Results from {\sc sbethe} for protons in liquid water, plotted
	with {\sc gnuplot} by using the provided script. Proton CSDA depth
	dose (Eq.\ \req{9.89.1}, no elastic scattering) for projectiles with
	the indicated kinetic energies.
\label{fig10.10}}\end{center} \end{figure}

\clearpage

\noindent bers, and two subdirectories named {\tt ./elastic} and {\tt
./sbethe}. The directory {\tt ./elastic} contains the source file of the
program, an example of input file, and various {\sc gnuplot} scripts.
The subdirectory {\tt ./gosan} includes the source file of the Fortran
program {\sc gosan} and an associated {\sc gnuplot} script. The directory {\tt
./sbethe} contains the program's source file, five {\sc gnuplot}
scripts, the subdirectory {\tt ./doc} with the manual of the program,
and the database {\tt ./sdbase} consisting of the 599 files listed
above, which include the DHFS atomic shell corrections
\citep{Salvat2022a}, the scaled bremsstrahlung DCSs of Seltzer and
Berger (\citeyear{SeltzerBerger1985, SeltzerBerger1986}) (Section
\ref{sec8.6}) for projectile electrons, and the radiative corrections to
the stopping power of muons derived from the tables given by
\citet{Groom2001}.

To install the package on your computer, expand the contents of the zip file
into the hard disc, keeping the directory structure
unchanged. To get the executable binary files of the three programs,
their source files have to be compiled in the corresponding directories.
Notice that the {\sc sbethe} program assumes that the
required databases are in a subdirectory of the directory where the
executable file is run.

\index{gnuplot}

The output files are in a format ready for visualization with a
plotting program. The program directories contain
a number of {\sc gnuplot} scripts for plotting the contents of relevant
output files. These scripts have filenames with the extension ``{\tt
.gnu}''. The script {\tt script-name.gnu} plots the contents of the
output file with the same name and the extension ``{\tt .dat}''.
Evidently, the scripts work only when {\sc gnuplot} is installed on the
computer. In Windows platforms, if the extension ``{\tt .gnu}'' is
associated to {\sc wgnuplot.exe}, each script can be run by simply
clicking on its icon. The provided scripts use the wxt terminal, which
works well on Microsoft Windows. If you are a Linux user and the wxt
terminal is not available on your computer, you may change the terminal
by editing the script files.

\index{Fortran programs|)}


\section{Complement. Numerical methods\label{sec10.4}}

\index{numerical methods|(}
As indicated above, the programs {\sc elastic} and {\sc sbethe} make a
fairly intensive use of interpolation and integration. In this Section
we describe generic tools that are utilized in the programs for these
tasks.


\subsection{Numerical tables: interpolation and integration
\label{sec10.4.1}}

\index{interpolation} \index{numerical integration}
\index{numerical methods!interpolation}
A function $f(x)$ may be defined by means of a numerical table, \ie, a
list of $N$ values $x_i$ of the variable and the corresponding function
values $f_i =f(x_i)$ ($i=1, 2, \ldots, N$). This type of representation
is convenient for functions that are difficult to calculate, and also
when the table values are taken from an external source or obtained from
measurements.

We assume that the grid points $x_i$ are given in non-decreasing order
($x_1 \le x_2 \le \ldots \le x_N$), but they are not required to be
uniformly spaced. Repeated grid points are used for describing
discontinuities of the function. When the interval $(x_1,x_N)$ contains
discontinuities, each continuous subinterval is treated separately.  In
the following we consider that the table represents a continuous
function. The values $f(x)$ of the function represented by the discrete
table are evaluated by using a suitable interpolation scheme. When the
function has small curvature we use linear interpolation,
\beq
f(x) = f_j + \frac{x-x_j}{x_{j+1}-x_j} \, (f_{j+1}-f_{j})
\qquad \mbox{if $x_j \le x \le x_{j+1}$.}
\label{10.10}\eeq
If the function varies smoothly, and the grid points are conveniently spaced
we use four-point Lagrange interpolation of the four data points that
are closer to $x$ in the table,
\beqa
f(x) &=&
\frac{(x-x_{j}) (x-x_{j+1})(x-x_{j+2})}
{(x_{j-1}-x_{j})(x_{j-1}-x_{j+1})(x_{j-1}-x_{j+2})} \, f_{j-1}
\nonumber \\ [2mm]
&+& \frac{(x-x_{j-1}) (x-x_{j+1})(x-x_{j+2})}
{(x_{j}-x_{j-1})(x_{j}-x_{j+1})(x_{j}-x_{j+2})} \, f_{j}
\nonumber \\ [2mm]
&+& \frac{(x-x_{j-1}) (x-x_{j})(x-x_{j+2})}
{(x_{j+1}-x_{j-1})(x_{j+1}-x_{j})(x_{j+1}-x_{j+2})} \, f_{j+1}
\nonumber \\ [2mm]
&+& \frac{(x-x_{j-1}) (x-x_{j})(x-x_{j+1})}
{(x_{j+2}-x_{j-1})(x_{j+2}-x_{j})(x_{j+2}-x_{j+1})} \, f_{j+2}
\qquad \mbox{if $x_j \le x \le x_{j+1}$,} \rule{15mm}{0mm}
\label{10.11}\eeqa
except when the point $x$ is in the first or the last subinterval, where
we take $j=2$ or $j=N-2$, respectively. Lagrange interpolation should
not be used when the tabulated values are affected by appreciable
uncertainties because the interpolating cubic polynomial may wiggle
widely magnifying the errors.

Let us assume that a function $f(x)$ can be calculated accurately but
its calculation is very time consuming. To speed up calculations with
that function, we can first tabulate it over a sufficiently wide
interval and then obtain the required $f(x)$ values rapidly by
interpolation of the pre-calculated table. Evidently, the accuracy of
the interpolated values is determined by the spacing of the grid points
where the function is tabulated. A graph of the table, with $x$ as
abscissa and $f(x)$ as ordinate, is useful to determine the appropriate
interpolation scheme. Functions that have narrow peaks (such as the DCS
for elastic collisions of high energy particles with atoms) or that
extend over large intervals covering several decades (as a table of
stopping powers) are visualized more clearly by using a logarithmic
scale for either the abscissa, or the ordinate, or both. Needless to
say, logarithmic axes may be used only when the variable or the function
take positive values. For instance,
the function $\exp[-(ax^2+bx+c)]$ becomes a parabola when plotted with a
logarithmic ordinate axis and, hence, Lagrange interpolation of
$\ln[f(x)]$ will be highly accurate, even with a sparse grid. In the
programs we use linear or four-point Lagrange interpolation of $f(x)$ or
$\ln[f(x)]$ as a function of $x$ or $\ln x$. We use the combination of
linear and logarithmic  axes for the abscissa and the ordinate that
yields the ``smoother'' plot of the function, \ie, that can be better
approximated by straight or polynomial segments. The grid points $x_i$
may be determined adaptively, starting by a uniform or logarithmic grid
with a moderate number of points, and adding new points where the
existing grid seems to give larger interpolation errors.

Given a table of a function $f(x)$, integrals of the types
\index{numerical methods!integration}
\beq
\int_A^B x^n f(x) \, \d x,  \qquad
\int_A^B \ln x \; x^n \; f(x) \, \d x, \ldots
\label{10.12}\eeq
with $(A,B) \in (x_1,x_N)$
can be approximated by integrating the interpolating function.
When using linear-linear interpolation, the contribution of each
subinterval $(x_i,x_{i+1})$ can be evaluated analytically using the
familiar antiderivatives
\beq
\int x^a \d x = \left\{
\begin{array}{ll}
\displaystyle{\frac{x^{a+1}}{a+1}} \rule{5mm}{0mm}
& \mbox{if $a \ne -1$,} \\ [3mm]
\ln x & \mbox{if $a = -1$,}
\end{array} \right.
\label{10.13}\eeq
and
\beq
\int x^a \; \ln x \; \d x = \left\{
\begin{array}{ll}
\displaystyle{\frac{x^{a+1}}{a+1} \left( \ln x - \frac{1}{a+1} \right)}
\rule{5mm}{0mm}
& \mbox{if $a \ne -1$,} \\ [3mm]
\1o2 \ln^2 x & \mbox{if $a = -1$,}
\end{array} \right.
\label{10.14}\eeq
If both the variable and function values of a table are positive, we
can use linear log-log interpolation for a fast evaluation of the
integrals \req{10.12}. The interpolating function in
the interval between the points $x_1$ and $x_2$, where the function has
respective values $f_1$ and $f_2$, is defined by
\beq
\ln[f(x)] = \ln f_1 + \frac{\ln f_2 - \ln f_1}{\ln x_2 - \ln x_1}
\, (\ln x - \ln x_1)
\nonumber\eeq
or, more directly,
\beq
f(x) = \frac{f_1}{x_1^a} \, x^a
\qquad \mbox{with} \qquad a = \frac{\ln(f_2/f_1)}{\ln(x_2/x_1)}.
\label{10.15}\eeq
Hence, if $a\ne -1$,
\beq
\int_{x_1}^{x_2} f(x) \, \d x
= \frac{f_1}{x_1^a} \left[ \frac{x^{a+1}}{a+1} \right]_{x_1}^{x_2}
= \frac{1}{a+1} \left( f_2 x_2 - f_1 x_1
\right)
\label{10.16}\eeq
and
\beqa
\int_{x_1}^{x_2} f(x) \, \ln x \, \d x
&=& \frac{f_1}{x_1^a} \left[ \frac{x^{a+1}}{a+1}
\left( \ln x - \frac{1}{a+1} \right) \right]_{x_1}^{x_2}
\nonumber \\ [2mm]
&=& \frac{1}{a+1} \left[ f_2 x_2 \left( \ln x_2 - \frac{1}{a+1} \right)
- f_1 x_1 \left( \ln x_1 - \frac{1}{a+1} \right) \right].
\rule{15mm}{0mm}
\label{10.17}\eeqa
These formulas, complemented with similar ones for the exceptional case
$a=-1$, allow the stepwise calculation of the integrals \req{10.12}.

In our computer programs, the preparation and manipulation of
interpolation tables and the calculation of certain integrals are
performed by the following Fortran subroutines.  Notice that the
physical dimensions of all input-output arrays must be defined in the
calling program.

\noindent 1) {\tt
SUBROUTINE TABLEF(FCT,XL,XU,X,Y,TOL,ERR,NPM,NFIX,NU,NP,MODE)}
\\ [2mm]
\noindent
This subroutine builds a table of the external function {\tt FCT(X)}
(provided by the user) in the interval ({\tt XL},{\tt XU}). The grid
consists of a first subgrid with {\tt NU} equally spaced points and {\tt
NP-NU} additional points that concentrate in regions where the function
has the largest ``curvature''. Optionally a number {\tt NFIX} of {\tt
X}-values can be fixed; they are to be entered in the first {\tt NFIX}
positions of the input array {\tt X(.)}. {\tt TOL} is the tolerance; the
subroutine returns when the table is such that the largest relative error
of the interpolation is less than {\tt TOL}.
\\ [2mm]
\parbox[t]{2.1cm}{INPUT:}
\parbox[t]{13.2cm }{
{\tt FCT} = name of the external function. \\
{\tt XL,XU} = end points of the considered interval. \\
{\tt TOL} = tolerance, desired relative error of the interpolation. \\
{\tt NPM} = physical dimension of arrays {\tt X} and {\tt Y}. \\
{\tt NFIX} = number of fixed points. Their abscissas must be entered as
\\
\rule{5mm}{0mm}the first {\tt NFIX} elements of the array {\tt X}. A
duplicated value is considered \\
\rule{5mm}{0mm}as a discontinuity. \\
{\tt NU} = number of points in the initial uniform subgrid. \\
{\tt NP} =  desired number of points in the table. Must be larger
than {\tt NFIX+NU}.
}

\noindent
\rule{2.1cm}{0mm}
\parbox[t]{13.2cm }{
{\tt MODE} = interpolation mode: \\
\rule{10mm}{0mm}  $\le 0$ or $>$7, linear, \\
\rule{10mm}{0mm}  1, lin-log linear, \\
\rule{10mm}{0mm}  2, log-lin linear, \\
\rule{10mm}{0mm}  3, log-log linear, \\
\rule{10mm}{0mm}  4, four-point Lagrange, \\
\rule{10mm}{0mm}  5, lin-log four-point Lagrange, \\
\rule{10mm}{0mm}  6, log-lin four-point Lagrange, \\
\rule{10mm}{0mm}  7, log-log four-point Lagrange.
} \\ [2mm]
\parbox[t]{2.1cm}{OUTPUT:}
\parbox[t]{13.2cm }{
{\tt X(1:NP),Y(1:NP)} = generated arrays of abscissas and function
values. \\
{\tt ERR} = estimate of the interpolation error; relative error if
$\ln(${\tt FCT}$)$ is interpolated, absolute error otherwise. \\
{\tt  NP} = number of points in the generated grid. It may be less
than the \\
\rule{5mm}{0mm}input value, because the subroutine stops adding grid
points as soon \\
\rule{5mm}{0mm}as the required tolerance is reached.
}

\vspace*{3mm}

\allowdisplaybreaks{
\noindent 2) {\tt
FUNCTION FINTRP(XC,X,Y,NP,MODE)}
\\ [2mm]
\noindent
Interpolation (and extrapolation) in a table {\tt X(1:NP)}, {\tt Y(1:NP)}
of a function {\tt Y(X)} with {\tt NP} data points. The abscissas {\tt
X(I)} must be in non-decreasing order. A duplicated abscissa with the
corresponding values at the lower and higher sides, is
treated as a discontinuity. To prevent undesirable extrapolations, the
function values outside the definition interval may be set to zero
by adding discontinuities at the beginning and at the end of the
table.
\\ [2mm]
\parbox[t]{2.1cm}{INPUT:}
\parbox[t]{13.2cm }{
{\tt XC} = value of the variable. \\
{\tt X(1:NP)} = array of variable values (in increasing order). \\
{\tt Y(1:NP)} = corresponding function values. \\
{\tt N} =  number of points in the table. \\
{\tt MODE} = interpolation mode (see the codes in subroutine {\tt
TABLEF}).
} \\ [2mm]
\parbox[t]{2.1cm}{OUTPUT:}
\parbox[t]{13.2cm }{At output, {\tt FINTRP} is the value of the function
at {\tt XC}.
}
}

\noindent The logarithm of a zero or negative abscissa ({\tt} $<
10^{-99}$) will abort the program. In log interpolations of the
function, negative function values are replaced temporarily
with $10^{-99}$.

\vspace*{3mm}

\allowdisplaybreaks{
\noindent 3) {\tt
SUBROUTINE TCLEAN(X,Y,ERCUT,ERRM,NP,NPS,MODE)}
\\ [2mm]
\noindent
This subroutine cleans a table of function values {\tt X(I)}, {\tt
Y(I)}. The error at each grid point is estimated by temporarily removing
the point and considering the interpolated value from the rest of the
table. Points with errors less than the tolerance {\tt ERCUT} are
permanently removed.
\\ [2mm]
\parbox[t]{2.1cm}{INPUT:}
\parbox[t]{13.2cm }{
{\tt X(1:NP)} = array of variable values (in increasing order). \\
{\tt Y(1:NP)} = corresponding function values. \\
{\tt ERCUT} = desired accuracy (or tolerable error; relative error if $\ln
{\tt Y}$ \\
\rule{5mm}{0mm}is interpolated, absolute error otherwise). \\
{\tt NP} =  number of points in the table. \\
{\tt NPS} = desired size of the cleaned table, points with the smaller
relative
}

\noindent
\rule{2.1cm}{0mm}
\parbox[t]{13.2cm }{
\rule{5mm}{0mm}(log) or absolute (lin) errors are removed from the table. \\
{\tt MODE} = interpolation mode (see the codes in subroutine {\tt
TABLEF}).
} \\ [2mm]
\parbox[t]{2.1cm}{OUTPUT:}
\parbox[t]{13.2cm }{
{\tt X(1:NP)}, {\tt Y(1:NP)} = abscissas and function values in the
cleaned table. \\
{\tt NPS} =  number of points in the cleaned table. \\
{\tt ERRM} = largest interpolation error of the cleaned table (relative
error \\
\rule{5mm}{0mm}if $\ln {\tt Y}$ is interpolated, absolute error otherwise).
}
}

\allowdisplaybreaks{
\noindent 4) {\tt
SUBROUTINE MERGE2(X1,Y1,X2,Y2,X,Y,N1,N2,NPM,N,MODE)}
\\ [2mm]
\noindent
This subroutine merges two tables ({\tt X1}, {\tt Y1}) and ({\tt X2},
{\tt Y2}) of two functions to produce a table ({\tt X}, {\tt Y}) of the
sum of these functions, with the abscissas in increasing order. A
discontinuity of a function is described by giving the abscissa twice,
with the function values at the lower and upper sides. The merged table
maintains the discontinuities of the original tables.
\\ [2mm]
\parbox[t]{2.1cm}{INPUT:}
\parbox[t]{13.2cm }{
{\tt X1(1:N1)}, {\tt Y1(1:N1)} = arrays of variable values (in increasing
order), \\
\rule{5mm}{0mm}and corresponding values of the first function. }
\\ [2mm]
\parbox[t]{2.1cm}{\rule{10mm}{0mm}}
\parbox[t]{13.2cm }{
{\tt N1} =  number of points in the first table. \\
{\tt X2(1:N2)}, {\tt Y2(1:N2)} = arrays of variable values (in increasing
order), \\
\rule{5mm}{0mm}and corresponding values of the second function. \\
{\tt N1} =  number of points in the second table. \\
{\tt NPM} =  physical dimension of arrays {\tt X} and {\tt Y}. It is
assumed that \\
\rule{5mm}{0mm} {\tt NPM} $\ge$ {\tt N1}$+${\tt N2}. \\
{\tt MODE} = interpolation mode (see the codes in subroutine {\tt
TABLEF}).
} \\ [2mm]
\parbox[t]{2.1cm}{OUTPUT:}
\parbox[t]{13.2cm }{
{\tt X(1:N)}, {\tt Y(1:N)} = abscissas and function values in the
merged table. \\
{\tt N} =  number of points in the merged table.
}
}

\allowdisplaybreaks{
\noindent 5) {\tt FUNCTION SMOMLL(X,Y,XL,XU,NP,MOM,ILOG)}
\\ [2mm]
\noindent
This function operates with positive definite functions (\ie,
probability density functions). It calculates integrals (moments) of a
tabulated function, ${\tt Y}= f({\tt X})$, over the interval ({\tt
XL},{\tt XU}) by using linear log-log interpolation of the input table.
The values of both the variable and the function are assumed to be
non-negative. Of course, a global shift ${\tt X} \rightarrow {\tt X} +
a$ allows considering also functions that extend to negative {\tt X}.
\\ [2mm]
\parbox[t]{2.1cm}{INPUT:}
\parbox[t]{13.2cm }{
{\tt X(1:NP)} = array of variable values (in increasing order). \\
{\tt Y(1:NP)} = corresponding function values (must be non-negative).
\\
{\tt NP} =  number of points in the table. \\
{\tt XL}, {\tt XU} = limits of the integration interval. \\
{\tt MOM} = moment order. \\
{\tt ILOG} = include the optional $\ln {\tt X}$ factor in the
integrand if {\tt ILOG=1}.
} \\ [2mm]
\noindent
\parbox[t]{2.1cm}{OUTPUT:}
\parbox[t]{13.2cm }{At output, {\tt SMOMLL} is the value of the integral
$$
\int_{\tt XL}^{\tt XU} x^{\tt MOM} f(x) \, \d x
\qquad \mbox{or} \qquad
\int_{\tt XL}^{\tt XU} x^{\tt MOM} \, \ln x \, f(x) \, \d x,
$$
depending on the input value of {\tt ILOG}.
}
}


\subsection{Cubic spline interpolation\label{sec10.4.2}}
\index{numerical methods!cubic spline interpolation|(}
\index{interpolation!cubic splines|(}

When a tabulated function is known to vary smoothly, interpolation
by natural cubic splines is generally appropriate. The following
presentation of spline interpolation is adapted from the book of
\citet{Maron1982}.

Suppose that a function $f(x)$ is given in numerical form, \ie, as a
table of values
\beq
f_i = f(x_i) \qquad (i=1,\ldots , N).
\label{10.18}\eeq
The points (knots) $x_i$ do not need to be equispaced, but we
assume that they are in (strictly) increasing order
\beq
x_1 <x_2< \cdots <x_N.
\label{10.19}\eeq

A function $\varphi(x)$ is said to be an interpolating cubic spline
if \\
1) It reduces to a cubic polynomial within
each interval [$x_i$,$x_{i+1}$], \ie, if $x_i \leq x \leq x_{i+1}$
\beq
\varphi(x)=a_i + b_i x + c_i x^2 + d_i x^3 \equiv p_i(x)
\qquad (i=1, \ldots, N-1).
\label{10.20}\eeq
2) The polynomial $p_i(x)$ matches the values of $f(x)$ at the
end-points of the $i$-th interval,
\beq
p_i(x_i) = f_i, \quad p_i(x_{i+1}) = f_{i+1} \qquad (i=1, \ldots, N-1),
\label{10.21}\eeq
so that $\varphi(x)$ is continuous in [$x_1$,$x_N$]. \\
3) The first and second derivatives of $\varphi(x)$ are continuous
in [$x_1$,$x_N$]
\begin{subequations}
\label{10.22}
\beqa
p'_i(x_{i+1}) &=& p'_{i+1}(x_{i+1}) \qquad (i=1, \ldots, N-2),
\label{10.22a} \\ [2mm]
p''_i(x_{i+1}) &=& p''_{i+1}(x_{i+1}) \qquad (i=1, \ldots, N-2).
\label{10.22b}\eeqa
\end{subequations}
Consequently, the curve $y=\varphi(x)$ interpolates the tabulated values
\req{10.18} and has a continuously turning tangent.

To obtain the spline coefficients $a_i$, $b_i$, $c_i$, $d_i$
($i=1, \ldots, N-1$) we start from the fact that $\varphi''(x)$ is
linear in [$x_i$,$x_{i+1}$]. Introducing the quantities
\beq
h_i \equiv x_{i+1}-x_i \qquad (i=1, \ldots, N-1)
\label{10.23}\eeq
and
\beq
\sigma_i = \varphi''(x_i) \qquad (i=1, \ldots, N),
\label{10.24}\eeq
we can write the obvious identity
\beq
p''_i(x) = \sigma_i \frac{x_{i+1}-x}{h_i} +
\sigma_{i+1} \frac{x-x_i}{h_i} \qquad
(i=1, \ldots, N-1).
\label{10.25}\eeq
Notice that $x_{i+1}$ must be larger than $x_i$ in order to have $h_i >
0$. Integrating Eq.\ \req{10.25} twice with respect to $x$ gives for
$i=1,\ldots, N-1$
\beq
p_i(x) =
\sigma_i \frac{(x_{i+1}-x)^3}{6 h_i} +
\sigma_{i+1} \frac{(x-x_i)^3}{6 h_i} +
A_i (x-x_i) + B_i (x_{i+1}-x),
\label{10.26}\eeq
where $A_i$ and $B_i$ are constants. These can be determined by
introducing the expression \req{10.26} into Eqs.\ \req{10.21}; this
gives the pair of Eqs.\
\beq
\sigma_i \frac{h_i^2}{6} + B_i h_i = f_i \qquad {\rm and} \qquad
\sigma_{i+1} \frac{h_i^2}{6} + A_i h_i = f_{i+1}.
\label{10.27}\eeq
Finally, solving for $A_i$ and $B_i$ and substituting the result in
\req{10.26}, we obtain
\beq
\left. \begin{array}{lcl}
p_i(x) & = & \displaystyle{\frac{\sigma_i}{6} \left[
\frac{(x_{i+1}-x)^3}{h_i}
- h_i (x_{i+1}-x) \right] + f_i \frac{x_{i+1}-x}{h_i}} \\ [4mm]
& & \displaystyle{\mbox{}+ \frac{\sigma_{i+1}}{6} \left[
\frac{(x-x_i)^3}{h_i} - h_i (x-x_i) \right] + f_{i+1}
\frac{x-x_i}{h_i}}. \end{array} \right.
\label{10.28}\eeq

To be able to use $\varphi(x)$ to approximate $f(x)$, we must find the
second derivatives $\sigma_i$ ($i=1, \ldots, N$). To this end, we impose
the conditions \req{10.22}. Differentiating \req{10.28} gives
\beq
p'_i(x) = \frac{\sigma_i}{6} \left[
- \frac{3(x_{i+1}-x)^2}{h_i} + h_i \right] +
\frac{\sigma_{i+1}}{6} \left[ \frac{3(x-x_i)^2}{h_i}
- h_i \right] + \delta_i,
\label{10.29}\eeq
where
\beq
\delta_i = \frac{y_{i+1}-y_i}{h_i}.
\label{10.30}\eeq
Hence,
\begin{subequations}
\label{10.31}
\beq
p'_i(x_{i+1}) = \sigma_i \frac{h_i}{6} + \sigma_{i+1} \frac{h_i}{3} +
\delta_i,
\label{10.31a}\eeq
\beq
p'_i(x_{i}) = - \sigma_i \frac{h_i}{3} - \sigma_{i+1} \frac{h_i}{6} +
\delta_i,
\label{10.31b)}\eeq
and, similarly,
\beq
p'_{i+1}(x_{i+1}) = - \sigma_{i+1} \frac{h_{i+1}}{3} - \sigma_{i+2}
\frac{h_{i+1}}{6} + \delta_{i+1}.
\label{10.31c}\eeq
\end{subequations}
Replacing \req{10.31a} and \req{10.31c} in \req{10.22a}, we find
\beq
h_i \sigma_i + 2 (h_i + h_{i+1}) \sigma_{i+1} + h_{i+1} \sigma_{i+2} = 6
\left( \delta_{i+1}-\delta_i \right) \qquad (i=1, \ldots, N-2).
\label{10.32}\eeq

The system of equation \req{10.32} is linear in the $N$ unknowns
$\sigma_i$ ($i=1, \ldots, N$). However, since it contains only $N-2$
equations, it is under-determined. This means that we need either to add
two additional (independent) equations or to fix arbitrarily two of the
$N$ unknowns.  The usual practice is to adopt {\it end-point strategies}
that introduce constraints on the behavior of $\varphi(x)$ near $x_1$
and $x_N$. An end-point strategy fixes the values of $\sigma_1$ and
$\sigma_N$, yielding an $(N-2)\times (N-2)$ system in the variables
$\sigma_i$ ($i=2, \ldots, N-1$). The resulting system is, in matrix
form,
\beq
\left( \begin{array}{ccccccc}
H_2 & h_2 & 0 & \cdots & 0 & 0 & 0 \\
h_2 & H_3 & h_3 & \cdots & 0 & 0 & 0 \\
0 & h_3 & H_4 & \cdots & 0 & 0 & 0 \\
\vdots & \vdots & \vdots & \ddots & \vdots & \vdots & \vdots \\
0 & 0 & 0 & \cdots & H_{N-3} & h_{N-3} & 0 \\
0 & 0 & 0 & \cdots & h_{N-3} & H_{N-2} & h_{N-2} \\
0 & 0 & 0 & \cdots & 0 & h_{N-2} & H_{N-1}
\end{array} \right)
\left( \begin{array}{c}
\sigma_2 \\  \sigma_3 \\
\sigma_4 \\  \vdots \\
\sigma_{N-3} \\  \sigma_{N-2} \\
\sigma_{N-1}   \end{array} \right) =
\left( \begin{array}{c}
D_2 \\  D_3 \\  D_4 \\  \vdots \\  D_{N-3} \\  D_{N-2} \\
D_{N-1} \end{array} \right) ,
\label{10.33}\eeq
where
\beq
H_i = 2(h_{i-1} + h_{i}) \qquad (i=2, \ldots, N-1)
\label{10.34}\eeq
and
\beq
\left. \begin{array}{lclr}
D_2&=&6 (\delta_{2} - \delta_{1}) - h_1 \sigma_1, & \\ [2mm]
D_i&=&6 (\delta_{i} - \delta_{i-1}) & (i=3, \ldots, N-2), \\ [2mm]
D_{N-1}&=&6 (\delta_{N-1} - \delta_{N-2}) - h_{N-1} \sigma_N. &
\end{array} \right.
\label{10.35}\eeq
($\sigma_1$ and $\sigma_N$ are removed from the first and last
equations, respectively). The matrix of coefficients is symmetric,
tridiagonal and diagonally dominant (the larger coefficients are in the
diagonal), so that the system \req{10.33} can be easily (and accurately)
solved by Gauss elimination. The spline coefficients $a_i$, $b_i$,
$c_i$, $d_i$ ($i=1, \ldots, N-1$) ---see Eq.\ \req{10.20}--- can then be
obtained by expanding the expressions \req{10.28}:
\beq
\left. \begin{array}{lcl}
a_i &=& \displaystyle{\frac{1}{6h_i} \left[ \sigma_i x_{i+1}^3 -
\sigma_{i+1} x_i^3 + 6 \left( f_i x_{i+1} - f_{i+1} x_i \right) \right]
+ \frac{h_i}{6} \left( \sigma_{i+1} x_i - \sigma_i x_{i+1} \right),} \\
[3mm]
b_i &=& \displaystyle{ \frac{1}{2h_i} \left[ \sigma_{i+1} x_i^2 -
\sigma_i x_{i+1}^2 + 2 \left( f_{i+1} - f_i \right) \right] +
\frac{h_i}{6} \left( \sigma_i - \sigma_{i+1} \right),} \\ [3mm]
c_i &=& \displaystyle{ \frac{1}{2h_i} \left( \sigma_i x_{i+1} -
\sigma_{i+1} x_i \right),} \\ [3mm]
d_i &=& \displaystyle{ \frac{1}{6h_i} \left( \sigma_{i+1} - \sigma_i
\right)}.
\end{array} \right.
\label{10.36}\eeq
In order to allow moderate extrapolation we extend the spline to the
intervals $(-\infty,x_1)$ and $(x_N,\infty)$ by means of quadratic
polynomials that match the values of the spline and its first and second
derivatives at the end-points. That is, we set
\begin{subequations}
\label{10.37}
\beq
\varphi(x) = (a_1+d_1x_1^3) + (b_1-3d_1x_1^2) x + (c_1+3d_1x_1)  x^2
\qquad \mbox{if $x \le x_1$,}
\label{10.37a}\eeq
and
\beqa
\varphi(x) &=& (a_{N-1}+d_{N-1}x_N^3)
+ (b_{N-1}-3d_{N-1}x_N^2) x
\nonumber \\ [2mm]
&& \mbox{} + (c_{N-1}+3d_{N-1}x_N)  x^2
\qquad \qquad \mbox{if $x \ge x_N$,}
\label{10.37b}\eeqa
\end{subequations}
We thus obtain a continuous function with continuous first and second
derivatives, \ie, a cubic spline in ($-\infty$,$\infty$).

When accurate values of $f''(x)$ are known, the best strategy is to set
$\sigma_1 = f''(x_1)$ and $\sigma_N = f''(x_N)$, since this will
minimize the spline interpolation errors near the end-points.
Unfortunately, the exact values $f''(x_1)$ and $f''(x_N)$ are not always
available.
The so-called {\it natural spline} corresponds to taking $\sigma_1 =
\sigma_N = 0$. It results in a $y=\varphi(x)$ curve with the shape that
would be taken by a flexible rod (such as a draftsman's spline) if it
were bent around pegs at the knots but allowed to maintain its natural
(straight) shape outside the interval [$x_1$,$x_N$].

The accuracy of the spline interpolation is mainly determined by the
density of knots in the regions where $f(x)$ has strong variations. For
constant, linear, quadratic and cubic functions the interpolation errors
can be reduced to zero by using the exact values of $\sigma_1$ and
$\sigma_N$. Evidently, in the case of quadratic or cubic polynomials the
natural spline may introduce considerable errors near the end-points. It
is also important to keep in mind that a cubic polynomial has, at most, one
inflexion point. As a consequence, we should have at least a knot
between each pair of inflexion points of $f(x)$ to ensure proper
interpolation. Special care must be taken when interpolating functions
that have a practically constant value in a partial interval, since the
spline tends to wiggle instead of staying constant. In this particular
case, it may be more convenient to use linear interpolation in the
interval where the function is constant.

Obviously, the interpolating cubic spline $\varphi(x)$ can be used not
only to obtain interpolated values of $f(x)$ between the knots, but also
to calculate integrals such as
\beq
\int_a^b f(x) \, x^k \, \d x \simeq
\int_a^b \varphi(x) \, x^k \, \d x
\label{10.38}\eeq
analytically. In addition, it is worth recalling that derivatives of
$\varphi(x)$ other than the first one may differ significantly from
those of $f(x)$.

To obtain the interpolated value $\varphi(x_{\rm c})$ ---see Eq.\
\req{10.20}--- of $f(x)$ at the point $x_{\rm c}$, we must first determine
the interval [$x_i$,$x_{i+1}$] that contains the point $x_{\rm c}$. To
reduce the effort to locate the point, we use the following binary
search algorithm \\ [2mm]
\rule{6mm}{0mm} \parbox[t]{13cm}{
1. Set $i=$1 and $j=N$. \\
2. Set $k=[(i+j)/2]$, where $[x]$ indicates the integer part of
$x$. \\
3. If $x_{k} < x_{\rm c}$ set $i=k$, otherwise set $j=k$. \\
4. If $j-i >1$ go to step 2. \\
5. Deliver $i$.
} \\ [4mm]
\noindent
Notice that the maximum delivered value of $i$ is $N-1$. Hence, the
cases $x < x_{1}$ and $x>x_{N}$ (where the quadratic extrapolation
applies) should be considered separately.

\vspace*{3mm}

Our programs use cubic spline interpolation of various functions.  The
source files include the following subprograms that define and use
interpolating cubic splines. Notice that the physical dimensions of all
input-output arrays must be defined in the calling program.

\noindent 1) {\tt SUBROUTINE SPLINE(X,Y,A,B,C,D,S1,SN,N)}
\\ [1mm]
This subroutine determines the coefficients of a piecewise cubic spline
that interpolates the input table of variable and function values
defined as a double array {\tt (X,Y)} of dimension {\tt N}. Duplicated
abscissas are considered as discontinuities; a separate spline is used
for each interval between consecutive discontinuities, with ``natural''
end-point shape (null second derivative) at the discontinuities. \\
[1mm]
\parbox[t]{2.1cm}{INPUT:}
\parbox[t]{13.2cm }{
{\tt X(1:N)} = grid points $x_i$, must be in non decreasing order. \\
{\tt Y(1:N)} = corresponding function values, $y_i = f(x_i)$.\\
{\tt S1}, {\tt SN} =  $\sigma_1$ and $\sigma_N$, second derivatives at
$x_1$ and $x_N$, respectively. \\
\rule{5mm}{0mm} The natural cubic spline is considered when ${\tt S1}=0$
and {\tt SN=0}.\\
{\tt N} = $N$, number of grid points.
} \\  [1mm]
\parbox[t]{2.1cm}{OUTPUT:}
\parbox[t]{13.2cm }{
{\tt A}(1:N), {\tt B(1:N)}, {\tt C(1:N)}, {\tt D(1:N)} = arrays of spline coefficients.
} \\ [1mm]
\noindent OTHER SUBPROGRAMS USED: subroutine {\tt SPLIN0}.

\noindent The interpolating cubic polynomial in the {\tt I}-th interval,
from {\tt X(I)} to {\tt X(I+1)}, is
\beq
\varphi(x) = {\tt A(I)} + x ({\tt B(I)}+x ({\tt C(I)} +x {\tt D(I)}))).
\label{10.39}\eeq

\vspace*{3mm}

\noindent 2) {\tt SUBROUTINE SPLIN0(X,Y,A,B,C,D,S1,SN,N)}
\\ [1mm]
\noindent
Initialization of cubic spline interpolation of tabulated data. It is
assumed that the function and its first two derivatives are continuous;
duplicated abscissas are not allowed. \\ [1mm]
\parbox[t]{2.1cm}{INPUT:}
\parbox[t]{13.2cm }{
{\tt X(1:N)} = grid points $x_i$, must be in strictly increasing order.
\\
{\tt Y(1:N)} = corresponding function values, $y_i = f(x_i)$.\\
{\tt S1}, {\tt SN} =  $\sigma_1$ and $\sigma_N$, second derivatives at
$x_1$ and $x_N$, respectively. \\
\rule{5mm}{0mm} The natural cubic spline is considered when ${\tt S1}=0$
and {\tt SN=0}.\\
{\tt N} = $N$, number of grid points.
} \\  [1mm]
\parbox[t]{2.1cm}{OUTPUT:}
\parbox[t]{13.2cm }{
{\tt A(1:N)}, {\tt B(1:N)}, {\tt C(1:N)}, {\tt D(1:N)} = arrays of spline
coefficients.
} \\ [1mm]
\noindent This subroutine is standalone.

\vspace*{3mm}

\noindent 3) {\tt SUBROUTINE FINDI(XC,X,N,I)} \\ [1mm]
\noindent
This subroutine finds the interval ({\tt X(I)},{\tt X(I+1)})
that contains the value {\tt XC} by using the binary search algorithm.
\\ [1mm]
\parbox[t]{2.1cm}{INPUT:}
\parbox[t]{13.2cm }{
{\tt XC} = point to be located. \\
{\tt X(1:N)} = grid points $x_i$. \\
{\tt N} = $N$, number of grid points.
} \\  [1mm]
\parbox[t]{2.1cm}{OUTPUT:}
\parbox[t]{13.2cm }{
{\tt I} = interval index.
}

\vspace*{3mm}

\noindent 4) {\tt FUNCTION SPLVAL(XC,X,A,B,C,D,N)} \\ [1mm]
\noindent
This function evaluates a cubic spline at the point {\tt XC}; quadratic
extrapolation is used for points outside the interval $(x_1,x_N)$.
\\ [2mm]
\parbox[t]{2.1cm}{INPUT:}
\parbox[t]{13.2cm }{
{\tt XC} = spline argument. \\
{\tt X(1:N)} = grid points $x_i$, must be in non decreasing order.
}

\noindent
\rule{21mm}{0mm}
\parbox[t]{13.2cm }{
{\tt A(1:N)}, {\tt B(1:N)}, {\tt C(1:N)}, {\tt D(1:N)} = arrays of spline
coefficients. \\
{\tt N} = $N$, number of grid points.
} \\ [1mm]
\parbox[t]{2.1cm}{OUTPUT:}
\parbox[t]{13.2cm }{
{\tt SPLVAL} = value of the spline function at {\tt XC}.
}

\vspace*{2mm}

\index{numerical integration!cubic splines}
\noindent 5) {\tt FUNCTION SPLINT(X,A,B,C,D,XL,XU,N,NPOW,ILOG)}  \\ [2mm]
\noindent
This function calculates the integral of a cubic spline function
$\varphi(x)$ multiplied by a power of $x$ and, optionally, by $\ln x$. \\ [2mm]
\parbox[t]{2.1cm}{INPUT:}
\parbox[t]{13.2cm }{
{\tt X(1:N)} = spline grid points $x_i$. \\
{\tt A(1:N)}, {\tt B(1:N)}, {\tt C(1:N)}, {\tt D(1:N)} = arrays of spline
coefficients. \\
{\tt XL} and {\tt XU}= lower and upper limits of the integral. \\
{\tt N} = $N$, number of spline grid points. \\
{\tt NPOW} = power of $x$ in the integrand. \\
{\tt ILOG} = optional logarithm:
} \\ [-1mm]
$$
\begin{array}{ll}
\mbox{if {\tt ILOG} $=1$,} \; \; \; \; &  {\tt SPLINT} =
\displaystyle{\int_{\tt XL}^{\tt XU} \varphi(x) \, x^{\tt NPOW} \, \ln x
\, \d x,}
\\ [5mm]
\mbox{if {\tt ILOG} $\ne 1$,} & {\tt SPLINT} =
\displaystyle{\int_{\tt XL}^{\tt XU} \varphi(x) \, x^{\tt NPOW} \, \d
x.}
\end{array}
$$
\parbox[t]{2.1cm}{OUTPUT:}
\parbox[t]{13.2cm }{
{\tt SPLINT} = value of the integral.
}

\index{numerical methods!cubic spline interpolation|)}
\index{interpolation!cubic splines|)}

\subsection{Numerical quadrature \label{sec10.4.3}}

In many cases, we need to calculate integrals of the form
\beq
\int_{A}^{B} f(z) \, \d z,
\label{10.40}\eeq
where the integrand is coded as an external function subprogram, which
gives nominally exact values. These integrals are evaluated by using the
Fortran external function {\tt SUMGA}, which implements the twenty-point
Gauss--Legendre quadrature method with an adaptive bisection scheme to
allow for error control. This procedure is comparatively fast and is
able to deal even with functions that have integrable singularities
located at the endpoints of the interval $[A,B]$, a quite exceptional
feature. It integrates functions $f(z)$ with localized structures such
as narrow peaks or valleys more efficiently than non-adaptive
algorithms. In addition, the Fortran function {\tt SUMGA} can be readily
extended to integrate functions of a complex variable $z$.

\subsubsection{Gauss--Legendre integration \label{sec10.4.3.1}}
\index{numerical methods!Gauss--Legendre quadrature}
\index{Gauss--Legendre quadrature}

We use the twenty-point Gauss--Legendre quadrature formula
\citep[see, \eg,][]{AbramowitzStegun1974}, given by
\beq
\int_{a}^{b} f(z) \, \d z =
\frac{b-a}{2} \sum_{i=1}^{20} w_{i} f(z_{i})
\label{10.41}\eeq
with
\beq
z_{i} = \frac{b-a}{2} x_{i} + \frac{b+a}{2} = a + \frac{b-a}{2} \left( 1
+ x_i \right).
\label{10.42}\eeq
The abscissa $x_{i}$ ($-1<x_{i}<1$) is the $i$-th zero of the Legendre
polynomial $P_{20}(x)$, the weights $w_{i}$ are defined as
\beq
w_{i} =
\frac{2}{\left(1-x_{i}^{2}\right) \left[ P'_{20}(x_{i}) \right]^{2}}.
\label{10.43}\eeq
The numerical values of the abscissas and weights are given in Table
\ref{tab10.3}. The difference between the exact value of the integral and the
right-hand side of Eq.\ \req{10.41} is
\beq
\Delta_{20} = \frac{(b-a)^{41}(20!)^{4}}{41\,(40!)^{3}} f^{(40)}(\xi),
\label{10.44}\eeq
where $\xi$ is a point in the interval $[a,b]$.

\begin{table}[hbt]
\begin{center}
\caption{\rm Abscissas and weights for twenty-point Gauss--Legendre
integration.
\label{tab10.3}}
\vspace*{4mm}
{\tt
\begin{tabular}{|c|c|} \hline\hline
$\pm x_{i}$ & $w_{i}$ \\ \hline
7.6526521133497334D-02&    1.5275338713072585D-01 \\
2.2778585114164508D-01&    1.4917298647260375D-01 \\
3.7370608871541956D-01&    1.4209610931838205D-01 \\
5.1086700195082710D-01&    1.3168863844917663D-01 \\
6.3605368072651503D-01&    1.1819453196151842D-01 \\
7.4633190646015079D-01&    1.0193011981724044D-01 \\
8.3911697182221882D-01&    8.3276741576704749D-02 \\
9.1223442825132591D-01&    6.2672048334109064D-02 \\
9.6397192727791379D-01&    4.0601429800386941D-02 \\
9.9312859918509492D-01&    1.7614007139152118D-02 \\  \hline\hline
\end{tabular}} \end{center} \end{table}

The Gauss--Legendre method gives an estimate of the integral of $f(z)$ over the
interval $[a,b]$, which is obtained as a weighted sum of function values
at fixed points inside the interval. We point out that \req{10.41} is an
open formula, \ie, the value of the function at the endpoints of the
interval is never required. Owing to this fact, function {\tt SUMGA} can
integrate functions that are singular at the endpoints. As an example,
the integral of $f(x)=x^{-1/2}$ over the interval [0,1] is correctly
evaluated. This would not be possible with a method based on a closed
formula (\ie, one that uses the values of the integrand at the interval
endpoints).

\subsubsection{Adaptive bisection \label{sec10.4.3.2}}
\index{Gauss--Legendre quadrature!adaptive}

\index{numerical methods!Gauss--Legendre quadrature!adaptive bisection}
The Fortran function {\tt SUMGA} exploits the fact that the error
$\Delta_{20}$, Eq.\ \req{10.44}, of the calculated integral decreases
when the interval length is reduced. Thus, halving the interval and
applying the Gauss--Legendre method to each of the two subintervals gives a much
better estimate of the integral, provided only that the function $f(x)$
is smooth enough over the initial interval. Notice that the error
decreases by a factor of about $2^{-40}$(!).

The algorithm implemented in {\tt SUMGA} is as follows. The integration
interval $(A,B)$ is successively halved so that each iteration gives a
doubly finer partition of the initial interval. We use the term
``$n$-subinterval'' to denote the subintervals obtained in the $n$-th
iteration. In each iteration, the integrals over the different
$n$-subintervals are evaluated by the Gauss--Legendre method, Eq.\
\req{10.41}. Consider that the integral over a given $n$-subinterval is
$S_{1}$. In the following iteration, this $n$-subinterval is halved and
the integrals over each of the two resulting $(n+1)$-subintervals are
evaluated, giving values $S_{1a}$ and $S_{1b}$. If
$S'_{1}=S_{1a}+S_{1b}$ differs from $S_{1}$ in less than the selected
tolerance, $S'_{1}$ is the sought value of the integral in the
considered $n$-subinterval; the value $S'_{1}$ is then accumulated and
this $n$-subinterval is no longer considered in subsequent iterations.
Each iteration is likely to produce new holes (eliminated subintervals)
in the regions where the function is smoother and, hence, the numerical
effort progressively concentrates in the regions where $f(x)$ has
stronger variations. The calculation terminates when the exploration of
the interval $(A,B)$ has been successfully completed or when a clear
indication of an anomalous behavior of $f(x)$ is found (\eg, when there
is a persistent increase of the number of remaining $n$-subintervals in
each iteration). In the second case a warning message may optionally be
printed in an output file, and control is returned to the calling
program.
\index{numerical methods|)}


%
   \newpage
   \appendix
   \addtocontents{toc}{\protect\vspace{1cm}}


\appendix
\chapter{Collision kinematics \label{appA}}

To cover the energy range of interest in radiation transport
studies, particle collisions must be described by using relativistic
kinematics. The conservation of energy and momentum, together with the
transformation properties of these quantities under Lorentz
transformations, yield general relations that are independent of the
nature of the particles and of their mutual interactions. This Appendix
contains general considerations and elementary derivations of
kinematical relations that are useful for describing the collisions of
relativistic particles.


\section{Lorentz transformations \label{appA.1}}

\index{Lorentz transformation} \index{four vectors}
An event is described by a four-vector $\underline{r}$, \ie, by the set of
four space-time coordinates, $\{ t, x, y, z \}$, relative to a certain
reference frame $F$,
\beq
\underline{r} = (r_0,r_1,r_2,r_3)=(ct,{\bf r}),
\label{A.1}\eeq
where $c$ is the speed of light in vacuum. We denote four-vectors using
underlined italic letters, ordinary vectors and scalar quantities are
represented by bold and italic letters, respectively. The space part of
a four-vector $\underline{r}$ is an ordinary vector, ${\bf
r}=(r_1,r_2,r_3)=(x,y,z)$. The Minkowski product of two four-vectors,
$\underline{r}$ and $\underline{s}$, is defined by
\index{four vectors!inner product of}
\beq
\underline{r} \dotprod \underline{s} \equiv
r_0 s_0 - r_1 s_1 - r_2 s_2 - r_3 s_3
= r_0 s_0 - {\bf r} \dotprod {\bf s}.
\label{A.2}\eeq
The squared ``norm'' of a four-vector,
\beq
\underline{r} \dotprod \underline{r} \equiv r_0^2 - r_1^2 - r_2^2 - r_3^2 =
c^2 t^2 - r^2,
\label{A.3}\eeq
is invariant under Lorentz transformations \citep[see, \eg,][Chapter
11]{Jackson1975}. These are real linear transformations that relate the
space-time coordinates of an event as observed from two inertial
reference frames. Let us consider a second reference frame $G$ whose
axes are parallel to those of $F$ and that moves with respect to frame
$F$ with velocity ${\bf v}_G=v_G \hat{\bf z}$ parallel to the $z$ axis;
quantities measured from the $G$ frame will be denoted by primes. For
simplicity, we assume that the origins of the two frames coincide at
$t=t'=0$. The coordinates $\underline{r}'=(ct',{\bf r}')$ of the event
\req{A.1} in $G$ are given by
\begin{subequations}
\label{A.4}
\beqa
x' &=& x, \quad y'=y,
\label{A.4a} \\ [2mm]
z' &=& \gamma_G (z - \beta_G ct),
\label{A.4b} \\ [2mm]
ct' &=& \gamma_G ( ct - \beta_G z),
\label{A.4c}\eeqa
\end{subequations}
with
\beq
\beta_G \equiv \frac{v_G}{c} \qquad {\rm and} \qquad
\gamma_G \equiv \sqrt{\frac{1}{1-\beta_G^2}}.
\label{A.5}\eeq
Equations \req{A.4} represent the Lorentz transformation corresponding
to a {\it boost} in the $z$ direction. The inverse transformation is
\begin{subequations}
\label{A.6}
\beqa
z &=& \gamma_G (z' + \beta_G ct'),
\label{A.6a} \\ [2mm]
ct &=& \gamma_G ( ct' +\beta_G z').
\label{A.6b}\eeqa
\end{subequations}

When the velocity ${\bf
v}_G=v_G\hat{\bf u}_G$ of $G$ with respect to $F$ is arbitrary, the
transformation equations are
\begin{subequations}
\label{A.7}
\beqa
ct' &=& \gamma_G ( ct - \beta_G r_\parallel),
\label{A.7a} \\ [2mm]
r_\parallel' &=& \gamma_G (r_\parallel - \beta_G ct),
\qquad {\bf r}'_\perp = {\bf r}_\perp,
\label{A.7b}\eeqa
where
\beq
r_\parallel = {\bf r} \dotprod \hat{\bf v}_G
\qquad \mbox{and} \qquad
{\bf r}_\perp = {\bf r} - ({\bf r} \dotprod \hat{\bf v}_G )
\hat{\bf v}_G
\label{A.7c}\eeq
are the components of ${\bf r}$ parallel and perpendicular,
respectively, to ${\bf v}_G$. In matrix form,
\beq
\left( \begin{array}{c}
ct' \\ r'_\parallel \end{array} \right) =
\left( \begin{array}{cc}
\gamma_G & -\gamma_G \beta_G \\
-\gamma_G \beta_G & \gamma_G \end{array} \right)
\left( \begin{array}{c}
ct \\ r_\parallel \end{array} \right), \qquad {\bf r}'_\perp = {\bf
r}_\perp\, .
\label{A.7d}\eeq
\end{subequations}
Equivalently, we can write
\beqa
{\bf r}' &=& {\bf r} - ({\bf r} \dotprod \hat{\bf v}_G ) \hat{\bf v}_G +
\gamma_G ({\bf r} \dotprod \hat{\bf v}_G - \beta_G c t) \hat{\bf v}_G,
\nonumber \\ [2mm]
ct' &=& \gamma_G ( ct - \beta_G {\bf r} \dotprod \hat{\bf v}_G).
\label{A.8}\eeqa
The inverse transformation is obtained by reversing the sign of
$\hat{\bf v}_G$ and interchanging primed and unprimed quantities.

The consequences of the Lorentz transformation are apparent in
measurements of length (\ie, distance between the positions of two
simultaneous events) and time intervals (separating two events that
occur in the same location). Consider a bar of rest length $L'$ that
moves with velocity ${\bf v}_{\rm G}$ parallel to it; when its rear
end passes the origin ($t=0$), its front end is at a point $L$ such
that [from Eq.\ \req{A.4b}] $L=L'/\gamma_{\rm G}$. That is, objects
moving with velocity ${\bf v}_{\rm G}$ are contracted in the direction
of motion by a factor $\gamma_{\rm G}^{-1}$ (FitzGerald--Lorentz
contraction). Similarly, assume a clock at rest at the origin of the
frame $G$ ($z'=0$) that ticks with period $T'$. Equation \req{A.6b}
implies that an observer in $F$ sees the clock ticking with a period $T$
longer than the period $T'$ of the clock at rest: $T=T' \gamma_{\rm G}$
(time dilation).

It is clear that $\underline{r}'^2 = \underline{r}^2$, \ie, the squared
norm of four-vectors is an invariant under Lorentz transformations. This
invariance implies that the speed of light $c$ is the same in all
inertial frames.  If $\underline{r}$ and $\underline{s}$ are
four-vectors, so is their sum $\underline{r}+\underline{s}$. Since
$(\underline{r} + \underline{s})^2 = \underline{r}^2 + 2 \underline{r}
\cdot \underline{s} + \underline{s}^2$ is an invariant, the product
of four-vectors, $\underline{r} \cdot \underline{s}$ is also an
invariant.


\section{The energy-momentum four-vector \label{appA.2}}

\index{energy-momentum four-vector} \index{four-momentum}
The energy and the linear momentum of a particle form a four-vector,
called the {\it energy-momentum four-vector} (or {\it four-momentum}), given by
\beq
\underline{p} = ({\cal W}c^{-1},{\bf p}),
\label{A.9}\eeq
where ${\cal W}$ is the total energy (including the rest energy) and
${\bf p}$ is the linear momentum. The components of the four-momentum
transform in the same way as the coordinates. Thus, in the case of a
boost with velocity ${\bf v}_G$,
\beq
\left( \begin{array}{c}
{\cal W}'c^{-1} \\ p'_\parallel \end{array} \right) =
\left( \begin{array}{cc}
\gamma_G & -\gamma_G \beta_G \\
-\gamma_G \beta_G & \gamma_G \end{array} \right)
\left( \begin{array}{c}
{\cal W} c^{-1} \\ p_\parallel \end{array} \right), \qquad {\bf
p}'_\perp = {\bf p}_\perp\, .
\label{A.10}\eeq
The four-momentum of a particle at rest is
\begin{subequations}
\label{A.11}\beq
\underline{p}^0 = (Mc,{\bf 0}),
\label{A.11a}\eeq
where $M$ is the (rest) mass and $c p^0_0 = Mc^2$ is the rest energy
of the particle\index{rest energy of a particle}. The four-momentum of a
particle that moves with velocity ${\bf v}$ can be obtained as the
transform of $\underline{p}^0$ under a boost with velocity $-{\bf v}$,
which gives
\beq
\underline{p} = Mc \, \gamma (1,\beta \hat{\bf v}),
\label{A.11b}\eeq
\end{subequations}
where
\beq
\beta \equiv \frac{v}{c} \qquad {\rm and} \qquad
\gamma \equiv \sqrt{\frac{1}{1-\beta^2}} \, .
\label{A.12}\eeq
That is,
\beq
{\bf p} = \beta \, \gamma M c\, \hat{\bf v}
\qquad \mbox{and} \qquad
{\cal W} = \gamma Mc^2 = \frac{cp}{\beta}.
\label{A.13}\eeq
The rest mass $M$ of a particle determines the invariant length of its
energy-momentum,
\beq
\underline{p} \dotprod \underline{p} =
{\cal W}^2 c^{-2} - {\bf p}^2 = (M c)^2.
\label{A.14}\eeq
Consequently, \index{relativistic total energy}
\beq
{\cal W} = \sqrt{(cp)^2 + M^2 c^4}
= Mc^2 \sqrt{1 + \left( \frac{p}{Mc}\right)^2} \, .
\label{A.15}\eeq

\index{relativistic kinetic energy}
The kinetic energy $E$ of a massive particle ($M\neq 0$) is defined as
\beq
E \equiv {\cal W} - M c^{2}= (\gamma-1) Mc^2\, .
\label{A.16}\eeq
The magnitude of the momentum is given by
\beq
(cp)^{2} = {\cal W}^2 - M^2c^4 = (\gamma \beta\, Mc^2)^2 = E(E+2Mc^{2}).
\label{A.17}\eeq
Evidently, $\gamma$ is the total energy in units of the rest
energy,
\begin{subequations}
\label{A.18}
\beq
\gamma = \frac{\cal W}{Mc^2} =
\sqrt{1 + \left( \frac{p}{Mc}\right)^2} =
\frac{E+Mc^2}{Mc^2}\, ,
\label{A.18a}\eeq
and
\beq
\beta = \frac{cp}{\cal W} = \left[ 1 + \left( \frac{Mc}{p} \right)^2
\right]^{-1/2} =
\sqrt{\frac{E(E+2Mc^2)}{(E+Mc^2)^2}}\, .
\label{A.18b}\eeq
\end{subequations}

\index{relativistic velocity}
The velocity of the particle can be expressed in terms of its linear
momentum as
\beq
{\bf v} = \frac{c^2 {\bf p}}{{\cal W}} =
\left[ 1 + \left( \frac{p}{Mc} \right)^2 \right]^{-1/2}
\frac{{\bf p}}{M} = \frac{{\bf p}}{\gamma M}.
\label{A.19}\eeq
Also,
\beq
E = M c^2 \left[ \sqrt{1 + \left( \frac{p}{Mc} \right)^2} - 1 \right]
= Mc^2(\gamma -1) = M v^2 \, \frac{\gamma^2}{\gamma+1}.
\label{A.20}\eeq


\section{Addition of velocities \label{appA.3}}

\index{addition of velocities}
From the Lorentz transformation of coordinates we can derive the law of
addition of velocities. Let ${\bf v}$ and ${\bf v}'$ denote the
velocities of a particle as seen from the reference frames $F$ and $G$,
respectively. The frame $G$ moves with velocity ${\bf v}_G$ with respect
to $F$. As seen from $F$, the particle moves from ${\bf
r}$ to ${\bf r} + \d {\bf r}$ in a time interval $\d t$. Using the
transformation relations \req{A.8},
\beqa
\d {\bf r} &=& \d {\bf r}' - (\d {\bf r}' \dotprod \hat{\bf v}_G )
\hat{\bf v}_G +
\gamma_G (\d {\bf r}' \dotprod \hat{\bf v}_G + \beta_G c \d t') \hat{\bf v}_G,
\nonumber \\ [2mm]
c \d t &=& \gamma_G ( c \d t' + \beta_G \d {\bf r}' \dotprod
\hat{\bf v}_G),
\label{A.21}\eeqa
we obtain
\beqa
{\bf v} = \frac{\d {\bf r}}{\d t } &=&
\frac{\d {\bf r}' - (\d {\bf r}' \dotprod \hat{\bf v}_G )
\hat{\bf v}_G +
\gamma_G (\d {\bf r}' \dotprod \hat{\bf v}_G + \beta_G c \d t')
\hat{\bf v}_G}{\gamma_G \d t' [ 1 + c^{-1} \beta_G
(\d {\bf r}'/\d t') \dotprod \hat{\bf v}_G]}
\nonumber \\ [2mm]
&=& \frac{{\bf v}' - ( {\bf v}' \dotprod \hat{\bf v}_G )
\hat{\bf v}_G +
\gamma_G ({\bf v}' \dotprod \hat{\bf v}_G + c \beta_G )
\hat{\bf v}_G}{\gamma_G [ 1 + c^{-1} \beta_G {\bf v}' \dotprod
\hat{\bf v}_G]}.
\label{A.22}\eeqa
Since the components of the velocities ${\bf v}$ and ${\bf v}'$ parallel
and perpendicular to ${\bf v}_G$ are
\beq
\left.
\begin{array}{ll}
v_\parallel = {\bf v} \dotprod \hat{\bf v}_G \\
v'_{\parallel} = {\bf v}' \dotprod \hat{\bf v}_G \end{array} \right\}
\qquad \mbox{and} \qquad
\left\{ \begin{array}{l}
{\bf v}_\perp = {\bf v} - ({\bf v} \dotprod \hat{\bf v}_G ) \hat{\bf
v}_G \\
{\bf v}'_{\perp} = {\bf v}' - ({\bf v}' \dotprod \hat{\bf v}_G )
\hat{\bf v}_G \end{array}\right. \, ,
\label{A.23}\eeq
the result \req{A.22} can be expressed as
\beq
{\bf v} =
\frac{{\bf v}'_\perp +
\gamma_G (v'_\parallel + c \beta_G)
\hat{\bf v}_G}{\gamma_G [ 1 +  c^{-1}\beta_G \, {\bf v}' \dotprod
\hat{\bf v}_G]}.
\label{A.24}\eeq
That is,
\beq
v_\parallel = \frac{v'_{\parallel} + c \beta_G}{1 + c^{-1} \beta_G
\, {\bf v}' \dotprod \hat{\bf v}_G}, \qquad {\bf v}_\perp = \frac{{\bf
v}'_{\perp}}{\gamma_G (1 +  c^{-1} \beta_G
\, {\bf v}' \dotprod \hat{\bf v}_G)}\, .
\label{A.25}\eeq
This is the famous relativistic law of addition of velocities.
The inverse transformation is obtained by interchanging primed and
unprimed quantities and changing the sign of $\beta_G$,
\beq
v'_\parallel = \frac{v_{\parallel} - c \beta_G }{1 -  c^{-1}\beta_G \,
{\bf v} \dotprod \hat{\bf v}_G},
\qquad {\bf v}'_\perp = \frac{{\bf
v}_{\perp}}{\gamma_G (1 -  c^{-1}\beta_G \,
{\bf v} \dotprod \hat{\bf v}_G)}\, .
\label{A.26}\eeq
In the non-relativistic limit ($c\rightarrow \infty$) these results
reduce to the familiar relation ${\bf v}={\bf v}' + {\bf
u}_G$.

From the law of addition of velocities, it readily follows that
\beqa
v^2 &=& v_\parallel^2 + v_\perp^2 =
\frac{(v'_{\parallel} + c\beta_G)^2 + (1- \beta_G^2) v_\perp'^2
}{(1 +  c^{-1} \beta_G \, {\bf v}' \dotprod \hat{\bf v}_G)^2}
\nonumber \\ [2mm]
&=&\frac{(v' \cos \chi + c \beta_G)^2 + (1-\beta_G^2)  v'^2 \sin^2\chi}
{(1 +  c^{-1}\beta_G \, v' \cos\chi)^2},
\label{A.27}\eeqa
where $\chi$ is the angle between the vectors ${\bf v}'$ and ${\bf
v}_G$. After simple manipulations, we obtain
\beq
v^2 = c^2 - \, \frac{c^2 - v'^2}{\gamma^2 (1 +  c^{-1}\beta_G \, v'
\cos\chi)^2}\, .
\label{A.28}\eeq
When the speed $v'$ in the $G$ frame equals $c$, this formula
implies that $v=c$ {\it irrespective of the direction of} ${\bf v}'$.
That is, the speed of light is the same in all inertial frames, in
accordance with Einstein's principle of special relativity. Moreover, if
$v'$ and $v_G$ are less than $c$, $v$ is also less than $c$.

\section{Laboratory and center-of-mass frames \label{appA.4}}

\index{laboratory frame of reference}
Let us consider a reaction,
\beq
1 + 2 \rightarrow 3 + 4 + ...
\label{A.29}\eeq
in which a system of two particles, $1$ and $2$, transforms into a
set of particles, $3$, $4$, etc. Here we use the term ``particle'' in a
generic sense; our particles may be composite systems as long as their
internal states remain stable when the interactions with the other
particles have ceased. Reactions are frequently observed from the
{\it laboratory} (L) frame, where the particle $2$ (the target) is at rest
before the interaction. We assume that the initial momentum of the
projectile (particle $1$) is in the direction of the $z$ axis. Thus, the
four-momenta of the two particles before the interaction are
\begin{subequations}
\label{A.30}
\beq
\underline{p}_{1} = ({\cal W}_{1} c^{-1},{\bf p}_{1}) =
({\cal W}_1 c^{-1},0,0, p_{1})
\label{A.30a}\eeq
and
\beq
\underline{p}_{2} = ({\cal W}_{2} c^{-1},{\bf p}_{2}) =
(M_2 c,0,0,0),
\label{A.30b}\eeq
\end{subequations}
where $M_i$ is the mass of particle $i$.
On the other hand, to simplify theoretical \index{center-of-mass frame
of reference} \index{barycentric frame of reference}
calculations, it is convenient to work in the {\it center of mass} (CM),
{\it center of momentum} or {\it barycentric} frame, which is a
coordinate system with its axes parallel to those of the L system and
such that the total momentum vanishes,
\beq
{\bf P}'_{\rm t} \equiv {\bf p}'_{1} + {\bf p}'_{2}
= {\bf p}'_{3} + {\bf p}'_{4} + \cdots = {\bf 0}.
\label{A.31}\eeq
Hereafter, quantities relative to the CM system will be indicated by
primes. The four-momenta of the initial particles in the CM frame are
$\underline{p}'_{1} = ({\cal W}'_{1} c^{-1},{\bf p}'_{1})$ and
$\underline{p}'_{2} = ({\cal W}'_{2} c^{-1},{\bf p}'_{2})$,
respectively. The CM frame moves with respect to the laboratory
frame with a certain velocity ${\bf v}_{\rm CM} = v_{\rm CM} \hat{\bf
z}$ and, therefore, the transformation from the CM frame to the L
frame is a boost in the $z$ direction. Thus, the components of a
four-momentum in the L and CM frames are related by
\beq
\begin{array}{l}
p_{x} = p'_{x}, \quad p_{y}=p'_{y}, \\ [2mm]
p_{z} = \gamma_{\rm CM} ( p'_{z} + \beta_{\rm CM} {\cal W}'c^{-1}), \\ [2mm]
{\cal W} = \gamma_{\rm CM} ({\cal W}' + \beta_{\rm CM} c p'_{z} ).
\end{array}
\label{A.32}\eeq
with
\beq
\beta_{\rm CM} = \frac{v_{\rm CM}}{c},\qquad
\gamma_{\rm CM} = \frac{1}{\sqrt{1 - \beta_{\rm CM}^2}}.
\label{A.33}\eeq
Note that here we are considering the transformation \req{A.6}, the inverse of
\req{A.4}. To determine the velocity $v_{\rm CM}$, we apply the transformation
\req{A.32} to the total four-momentum in CM, $\underline{P}'_{\rm t} = ({\cal
W}'_{1}c^{-1}+{\cal W}'_{2} c^{-1}, 0)$,
\beq
p_{1} = \gamma_{\rm CM} \beta_{\rm CM}
({\cal W}'_{1}+{\cal W}'_{2}) c^{-1}, \qquad
{\cal W}_{1} + {\cal W}_{2} = \gamma_{\rm CM} ({\cal W}'_{1}+{\cal W}'_{2}).
\label{A.34}\eeq
These equations give
\beq
\beta_{\rm CM} =
\frac{c p_{1}}{{\cal W}_{1} + {\cal W}_{2}}
= \frac{\sqrt{E_1(E_1+2 M_1 c^2)}}{E_1+M_1c^2+M_2 c^2}\, ,
\label{A.35}\eeq
where $E_1={\cal W}_1 - M_1 c^2$ is the kinetic energy of the projectile
in L.

The square of the total four-momentum of the particles before the
reaction,
\beq
s^2 \equiv c^2 (\underline{p}_1 + \underline{p}_2)^2 = ({\cal W}_1+{\cal W}_2)^2
- c^2 ({\bf p}_1 +{\bf p}_2)^2 ,
\label{A.36}\eeq
is a Lorentz invariant. Evidently, $s$ is the total energy in the CM
frame,
\beq
s = {\cal W}'_{1}+{\cal W}'_{2}.
\label{A.37}\eeq
In the L frame, $s^2$ is expressed as
\beqa
s^2 &=& ({\cal W}_{1} + {\cal W}_2)^2 - c^2 p_{1}^2
\nonumber \\ [2mm]
&=& (M_1 c^2 + M_2 c^2)^2 + 2 M_2 c^2 E_{1}.
\label{A.38}\eeqa
Because energy and momentum are conserved in the reaction,
$\underline{p}_1+\underline{p}_2 =
\underline{p}_3+\underline{p}_4+\cdots$, we also have
\beq
s^2 = c^2 (\underline{p}_3 + \underline{p}_4 + \cdots)^2
= \left( {\cal W}'_{3}+{\cal W}'_{4} + \cdots \right)^2 .
\label{A.39}\eeq
The first equality in \req{A.35} implies that
\beq
\gamma_{\rm CM} = \sqrt{\frac{1}{1-\beta_{\rm CM}^2}} =
\sqrt{\frac{({\cal W}_{1}+{\cal W}_{2})^2}{({\cal W}_{1}+{\cal W}_{2})^2
- c^2 p_{1}^2}}
= \frac{{\cal W}_{1}+{\cal W}_{2}}s
\label{A.40}\eeq
and
\beq
\beta_{\rm CM} \gamma_{\rm CM} =
\frac{c p_{1}}s.
\label{A.41}\eeq

The initial three-momentum in CM is defined by
\beq
{\bf p}' \equiv {\bf p}'_{1} = - {\bf p}'_{2},
\label{A.42}\eeq
and its magnitude can be calculated as the Lorentz transform of
${\bf p}_2 =0$,
\beq
p' = p'_{2} = \beta_{\rm CM} \gamma_{\rm CM} M_2 c = \frac{M_2 c^2
p_1}s.
\label{A.43}\eeq
The initial energies of the colliding particles in CM are given by
\begin{subequations}
\label{A.44}
\beqa
{\cal W}'_{1} &=& \sqrt{ M_1^2 c^4 + p'^2 c^2}
= \sqrt{ M_1^2 c^4 + \beta_{\rm CM}^2 \gamma_{\rm CM}^2 M_2^2 c^4}\, ,
\label{A.44a}\\ [2mm]
{\cal W}'_{2} &=& \sqrt{ M_2^2 c^4 + p'^2 c^2}
= \sqrt{ M_2^2 c^4 + \beta_{\rm CM}^2 \gamma_{\rm CM}^2 M_2^2 c^4}
\, .
\label{A.44b}\eeqa
\end{subequations}

\index{threshold energy of a reaction}
For a reaction to be possible, the kinetic energy of the projectile in L
cannot be smaller than a certain threshold value $E_{\rm L,th}$. The
threshold energy is such that all particles in the final channel are at
rest in the CM frame. From the equalities
\beq
s^2 = (M_1 c^2 + M_2 c^2)^2 + 2 M_2 c^2 E_{\rm L,th}
= (M_3 c^2+ M_4 c^2+ \cdots)^2,
\label{A.45}\eeq
we obtain
\beq
E_{\rm L,th} = \frac{c^2}{2 M_2} \left[ (M_3+M_4 + \cdots)^2
- (M_1 + M_2)^2 \right].
\label{A.46}\eeq
As an application of this result, consider the creation of an
electron-positron pair ($M_3=M_4=\me$) by a photon ($M_1=0$) in the
field of a target charged particle of mass $M_2$. The threshold  energy
for this process is
\beq
E_{\rm L,th} = \frac{c^2}{2 M_2} \left[ (2\me + M_2)^2- M_2^2 \right]
= c^2 \frac{2\me^2 + 2\me M_2}{M_2}.
\label{A.47}\eeq
If the target particle is a nucleus ($M_2\gg \me$), $E_{\rm L,th} \sim
2\me c^2$. When the pair production occurs in the field of an electron
(triplet production), $M_2=\me$ and $E_{\rm L,th}=4 \me c^2$.

\section{The transformation of angles \label{appA.5}}

\index{relativistic transformation of angles |(}
In collision theory we frequently need to transform cross sections
between two reference frames $F$ and $G$, with $G$ moving with velocity
${\bf v}_G$ relative to $F$. Cross sections are usually expressed in
terms of the energy and the direction of motion of particles in each
reference frame. Let ${\bf v}$ and ${\bf v}'$ denote the velocity of a
particle as measured from $F$ and $G$, respectively. For simplicity, we
assume that the axes of the two frames are parallel, with the $z$ and
$z'$ axes in the direction of the relative velocity ${\bf v}_G$, and
with the $y$ and $y'$ axes perpendicular to the velocity of the particle
in such a way that ${\bf v} \cdot \hat{\bf x} \ge 0$ (see Fig.\
\ref{figA.1}). Thus the direction of motion is completely determined by
the polar angle, $\theta$ or $\theta'$. Since a polar angle $\theta$
takes values in the interval $(0,\pi)$, the value of $\cos\theta$
defines the direction of motion unambiguously.

To facilitate the transformation of cross sections, it is convenient to
write the relations \req{A.25} in the equivalent forms
\beq
v\sin\theta = \frac{1}{\gamma_G} \frac{v' \sin\theta'}{1 +
c^{-2} v_Gv' \cos\theta'},\qquad
v \cos\theta = \frac{v' \cos\theta' + v_G}{1 +
c^{-2} v_Gv' \cos\theta'}.
\label{A.48}\eeq
We thus see that the polar angles of the velocity of the particle in $F$ and
$G$ are related by
\beq
\tan\theta = \frac{1}{\gamma_G} \, \frac{\sin\theta'}{\tau + \cos\theta'}
\qquad \mbox{with} \qquad \tau = v_G/v' \, .
\label{A.49}\eeq
Because $\theta$ and $\theta'$ are in $(0,\pi)$, this identity implies
that $\cos\theta$ and $\tau + \cos\theta'$ have the same sign. We can
thus write the equivalent equation
\beq
\cos\theta = \frac{\gamma_G(\tau + \cos{\theta'})}{\sqrt{
\sin^2 \theta' + \gamma_G^2 (\tau + \cos{\theta'})^2}}.
\label{A.50}\eeq

\begin{figure}[hbt] \begin{center}
\includegraphics*[scale=0.75]{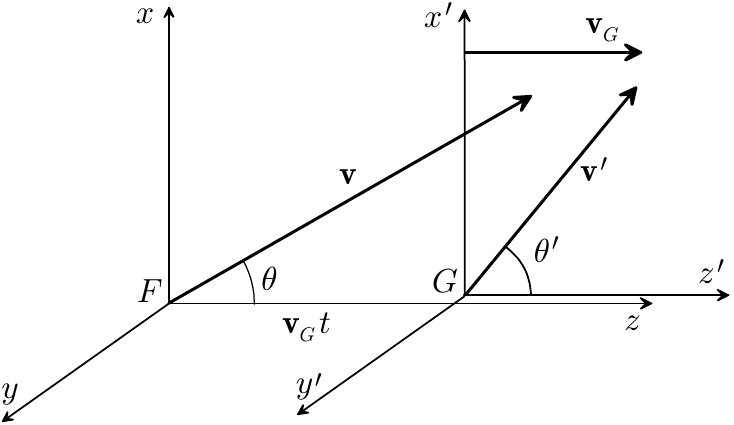}
\caption{
Velocities of a particle in two Lorentz frames. Notice that the polar
axes, $z$ and $z'$, are parallel to the relative velocity ${\bf v}_G$, and
the velocities ${\bf v}$ and ${\bf v}'$ lie on the $x-z$ plane. The polar
angles $\theta$ and $\theta'$ may only take values in the interval
$(0,\pi)$.
}\label{figA.1}
\end{center} \end{figure}

To derive the inverse relation without loosing control of possible sign
ambiguities, we start from the inverse of the transformation \req{A.48},
\beq
v' \sin\theta' = \frac{v\,\sin\theta}{\gamma_G\,\left(1 - c^{-2} v_G v
\cos\theta\right)} \, , \qquad
v' \cos\theta' = \frac{v\,\cos\theta - v_G}{ 1 - c^{-2} v_G
v\cos\theta}\, ,
\label{A.51}\eeq
and, adding the squares, we write
\beq
v'^2 = \frac{\gamma_G^{-2} v^2 \sin^2\theta + v^2 \cos^2\theta+ v_G^2
- 2 v_G v\,\cos\theta}{\left(1 - c^{-2} v_G v\cos\theta\right)^2}.
\label{A.52}\eeq
Making use of the relation
\beq
v'^2 = c^2 \beta'^2 = c^2 \left( 1 - \gamma'^{-2} \right),
\nonumber \eeq
the equality \req{A.52} can be formulated as the
quadratic equation
\beq
\left( \frac{v}{v'} \right) ^2 \left[ \gamma'^2 \gamma_G^{-2}
+ \beta_G^{2} \,\cos^2\theta
\right] - 2\, \frac{v}{v'} \tau \cos\theta +
\gamma'^2 \left(\tau^2 - 1 \right)= 0,
\label{A.53}\eeq
which admits the formal solution
\beq
\frac{v}{v'} = \frac{\tau \, \cos\theta \pm \sqrt{\Delta}}{
\gamma'^2 \gamma_G^{-2} + \beta_G^{2} \cos^2\theta}
\label{A.54}\eeq
with the discriminant
\beqa
\Delta &=& \tau^2 \cos^2 \theta -
\left[ \gamma'^2 \gamma_G^{-2} + \beta_G^{2}  \,\cos^2\theta
\right] \gamma'^2 \left(\tau^2-1 \right)
\nonumber \\ [2mm]
&=& \gamma'^4 \gamma_G^{-4} \left[ \cos^2 \theta + (1-\tau^2) \gamma_G^2
\, \sin^2 \theta \right] .
\label{A.55}\eeqa
Since
\beq
\tau^2 = \frac{v_G^2}{v'^2}
= \frac{1 - \gamma_G^{-2}}{1-\gamma'^{-2}}\, ,
\label{A.56}\eeq
when $\tau < 1$ the discriminant $\Delta$ is larger than $\tau^2
\cos^2 \theta$ and only the plus sign in expression \req{A.54} gives a
valid solution with $v \ge 0$. When $\tau \ge 1$, real solutions exist
only if the discriminant is positive and then the two solutions are
valid. Inserting the expression \req{A.54} into the second of Eqs.\
\req{A.51}, we get
\beq
\cos\theta' = \frac{\tau \gamma_G^{-2} (\cos^2 \theta - \gamma'^2) \pm
\sqrt{\Delta} \, \cos\theta}{\gamma'^2 \gamma_G^{-2} \mp
\tau \sqrt{\Delta} \, \beta'^2 \cos\theta},
\label{A.57}\eeq
which after some rearrangements can be written as
\beq
\cos \theta' =
\frac{-\tau \gamma_G^2\sin^2 \theta \pm \cos\theta
\sqrt{\cos^2 \theta+(1-\tau^2)\gamma_G^2  \sin^2 \theta}}
{\gamma_G^2\sin^2 \theta+\cos^2 \theta}.
\label{A.58}\eeq
As indicated above, when $\tau < 1$ only the plus sign is valid. If
$\tau \ge 1$, the requirement of a positive discriminant, \ie,
$\cos^2 \theta + (1-\tau^2) \gamma_G^2 \sin^2\theta \ge 0$,
implies that the angle $\theta$ in $F$ has the upper bound,
\beq
\theta_{{\rm max}} = \arccos \left(\sqrt{\frac{1}{1 +
\gamma_G^{-2}(\tau^2 - 1)^{-1}}} \right)\, .
\label{A.59}\eeq
The direction $\theta_{\rm max}$ in $F$ corresponds to $\theta'=\arccos
(-\tau^{-1})$ in $G$. Note that when $\tau=1$ the largest angle in $F$,
$\theta_{\rm max}=\pi/2$, occurs for $\theta' = \pi$.
\index{relativistic transformation of angles |)}


\section{Kinematics of inelastic collisions \label{appA.6}}

\index{inelastic collisions! kinematics of |(}
We consider here the kinematics of inelastic interactions of fast
charged particles (projectiles) of mass $M$ moving with velocity ${\bf
v}$ with a target atom or molecule. For simplicity, we will work in
the laboratory frame, where the target is at rest. Our study is focused
on the effect of inelastic interactions on the projectile.

Let ${\bf p}=\hbar{\bf k}$ and $E$ be the momentum and
the kinetic energy of the projectile just before an inelastic collision,
the corresponding quantities after the collision are denoted,
respectively, by ${\bf
p'}=\hbar{\bf k}'$ and $E'=E-W$; where $W$ is the energy lost by the
projectile. We recall that the
kinetic energy and the momentum of a free particle that moves with
velocity ${\bf v}$ are, respectively
\beq
E = (\gamma-1) M c^2 \quad \mbox{and} \quad {\bf p} = \beta \gamma Mc
\hat{\bf v}\, ,
\label{A.60}\eeq
where
\beq
\beta = \frac{v}{c} = \sqrt{\frac{\gamma^2-1}{\gamma^2}} =
\sqrt{\frac{E(E+2Mc^2)}{(E+Mc^2)^2}}
\label{A.61}\eeq
is the speed in units of $c$ and
\beq
\gamma = \sqrt{\frac{1}{1-\beta^2}} =
\frac{E+Mc^2}{Mc^2}
\label{A.62}\eeq
is the total energy in units of the rest energy of the particle. Note
that $E$ and $p$ are related by
\beq
(cp)^2 = (E+Mc^2)^2 - M^2 c^4 = E(E+2Mc^2)\, .
\label{A.63}\eeq

For projectile particles other than electrons, the maximum
energy loss in an inelastic collision is $W_{\rm max}=E$. In the case of ionization
by electron impact, owing to the indistinguishability between the
projectile and the ejected electron, the maximum energy loss is $W_{\rm
max} \simeq E/2$ (see Section \ref{sec6.5}).  The momentum transfer in the collision is ${\bf
p}-{\bf p'} \equiv \hbar {\bf q}$; note that ${\bf q}={\bf k} - {\bf
k}'$ is the momentum transfer in units of $\hbar$. It is customary to
introduce the {\it recoil energy} $Q$ defined by
\beq
Q(Q+2 \me c^{2}) = (c \hbar q)^{2} =
c^{2} \left( p^{2}+p'^{2}-2pp'\cos\theta \right),
\label{A.64}\eeq
where $\me$ is the electron mass and $\theta=\arccos(\hat{\bf p}
\cdot \hat{\bf p}')$ is the scattering angle (see Fig.\ \ref{figA.2}).
Equivalently, we can write
\beq
Q = \sqrt{(c \hbar q)^2 + \me^2 c^4 }- \me c^2.
\label{A.65}\eeq
When the collision is with a free electron at rest, the energy loss is
completely transformed into kinetic energy of the recoiling electron,
\ie, $Q=W$. For collisions with bound electrons, the relation $Q\simeq
W$ still holds for hard ionizing collisions (that is, when the energy
transfer $W$ is much larger than the ionization energy of the target
electron so that binding effects are negligible).

\begin{figure}[htb]
\begin{center}
\vspace{3mm}
\includegraphics*[width=7.0cm]{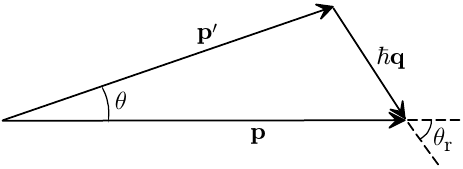}
\caption {\rm Momentum transfer and scattering angles in inelastic
collisions.
\label{figA.2}}
\end{center}\end{figure}

Equation \req{A.64} relates the energy loss, the scattering angle and the
recoil energy. The curves displayed in Fig.\ \ref{figA.3} represent the
recoil energy as a function of the energy loss for given scattering
angles. It is worth noting that for energy transfers that are much less
than the energy of the projectile, the curves are nearly vertical
straight lines. That is, when the energy loss is small, the recoil
energy $Q$ is a function of only the scattering angle ($Q$ is
independent of $W$). This behavior changes when the energy loss
increases, because each curve with $\theta \lesssim 60$ deg approaches
smoothly the $\theta=0$ curve. All curves converge to a single point
when the energy loss reaches its maximum allowed value
$W_\textrm{max}=E$.

\begin{figure}[h!]
\begin{center}
\vspace{3mm}
\includegraphics*[width=7.5cm]{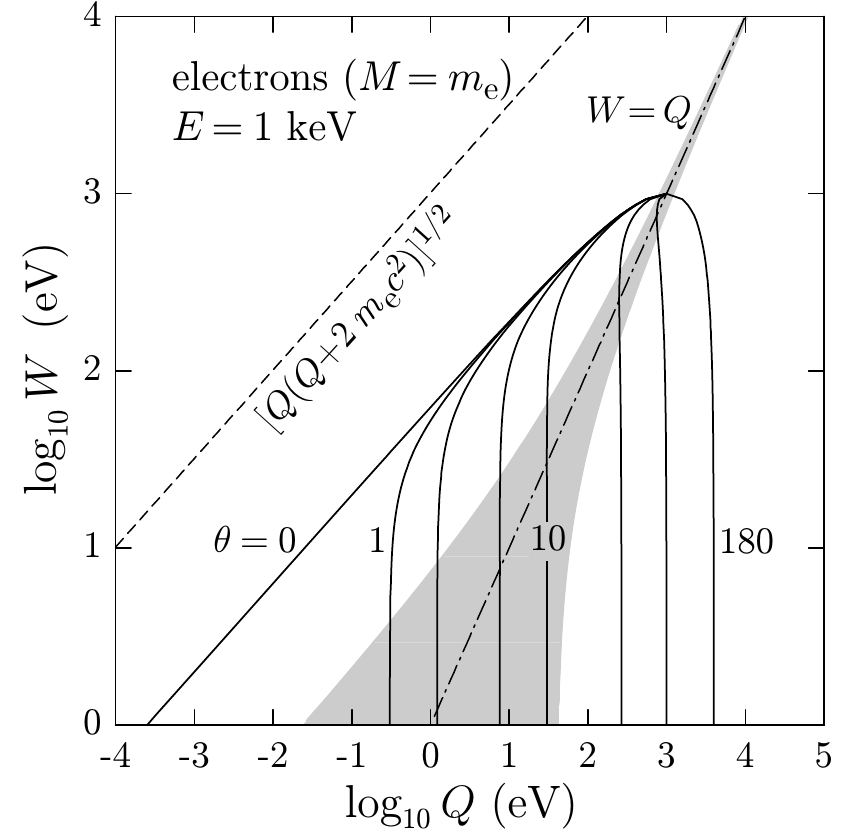} \hfill
\includegraphics*[width=7.5cm]{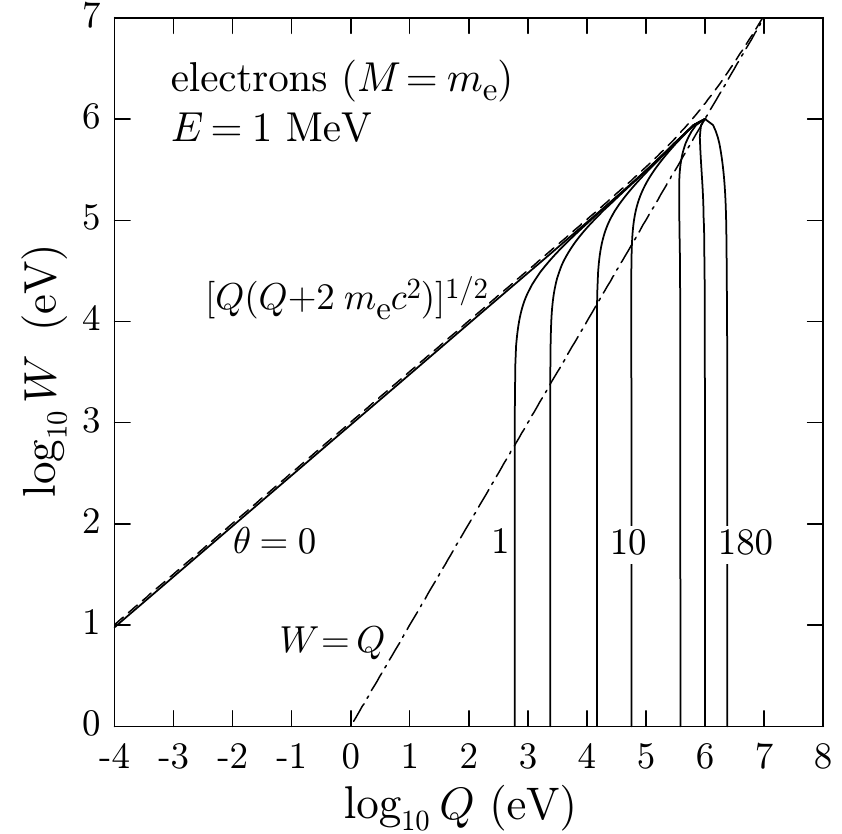} \\
\includegraphics*[width=7.5cm]{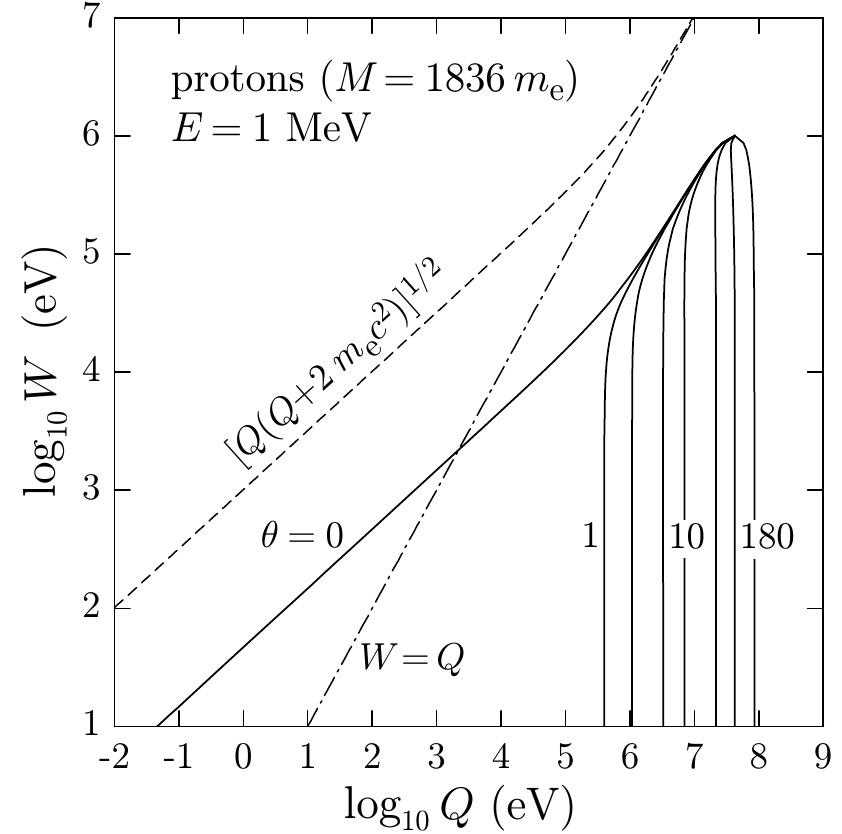} \hfill
\includegraphics*[width=7.5cm]{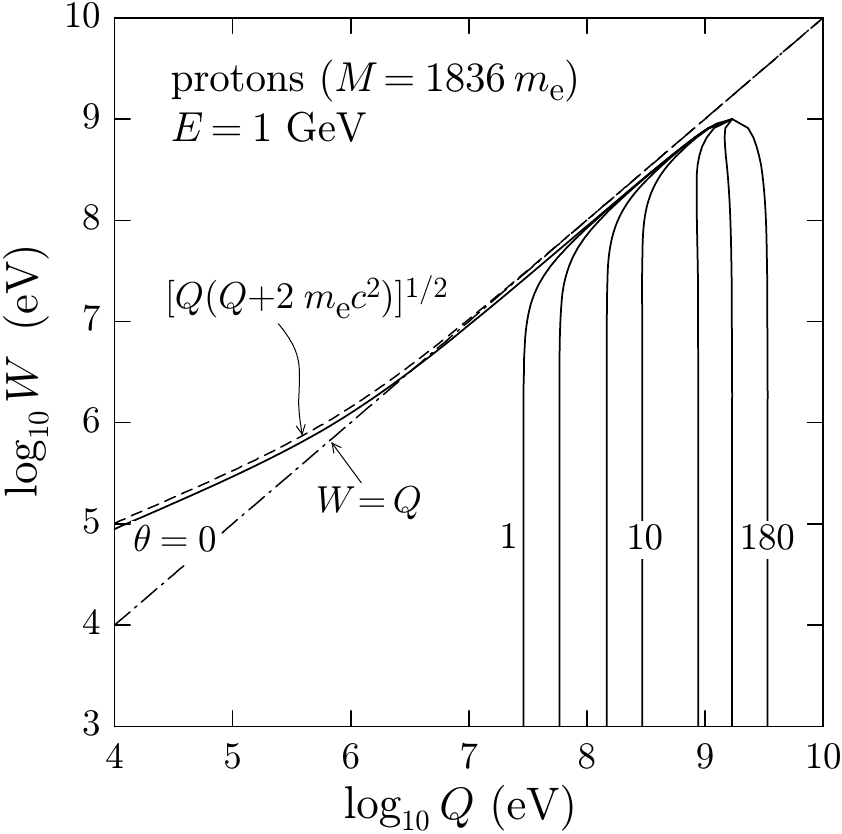}
\caption {\rm Kinematics of inelastic collisions for electrons and
protons with the indicated kinetic energies. The curves represent the value of
the recoil energy $Q$ (abscissa) that corresponds to the energy loss $W$
(ordinate) for a given scattering angle. The displayed curves correspond
to scattering angles of 0, 1, 2, 5, 10, 30, 60, and 180 degrees. The
shaded area in the top-right plot represents the possible excitations of
a free-electron gas with a Fermi energy of 10 eV.
\label{figA.3}}
\end{center}\end{figure}

From the definition of the momentum transfer $\hbar {\bf q}$ (see Fig.\
\ref{figA.2}), we can obtain expressions of the {\it scattering angle}
$\theta$ and the {\it recoil angle} $\theta_{\rm r}$ (\ie, the angle
between ${\bf p}$ and ${\bf q}$) in terms of the energy loss $W$ and the
recoil energy $Q$. Squaring the identity ${\bf p} - {\bf p}' = \hbar
{\bf q}$, we have
$$
(cp)^2 + (cp')^2 - 2 (cp) (cp') \cos\theta = (c \hbar q)^2,
$$
and it follows that
\beqa
\cos\theta &=& \frac{(cp)^2 + (cp')^2 - (c \hbar q)^2}{2(cp)(cp')}
\nonumber \\ [2mm]
&=& \frac{ E(E+2M c^2) + (E-W)(E-W+2M c^2) - (c \hbar q)^2}
{2\sqrt{ E(E+2M c^2) \; (E-W)(E-W+2M c^2)}}\, .
\label{A.66}\eeqa
Similarly, taking the square of the  identity ${\bf p} - \hbar {\bf q} =
{\bf p}'$, we obtain
\beqa
\cos\theta_{\rm r} &=& \frac{(cp)^2 - (cp')^2 + (c \hbar q)^2}
{2(cp)(c \hbar q)}
\nonumber \\ [2mm]
&=&
\frac{ E(E+2M c^2) - (E-W)(E-W+2M c^2) + (c q)^2}
{2\sqrt{ E(E+2M c^2)} \; (c \hbar q)}
\nonumber \\ [2mm]
&=&
\frac{W}{\beta (c \hbar q)} \left( 1 + \frac{ (c \hbar q)^2-W^2}
{2 W (E+M c^2)} \right)\, .
\label{A.67}\eeqa
For heavy ($M\gg \me$) high-energy projectiles and collisions such that
$Q\ll E$ and $W \ll E$,
\beq
\cos \theta_{\rm r} \simeq \frac{W}{\beta (c \hbar q)}\, .
\label{A.68}\eeq

For a given energy loss $W$, the kinematically allowed recoil energies lie in the interval $Q_- < Q <
Q_+$, with end points given by Eq.\ \req{A.64} with $\cos\theta =+1$
and $-1$, respectively. That is
\beqa
Q_{\pm} &= &\sqrt{ (cp \pm cp')^2 + \me^2 c^4} - \me c^2
\nonumber \\ [4mm]
&=& \sqrt{
\left[ \sqrt{E(E+2Mc^2)} \pm \sqrt{(E-W)(E-W+2Mc^2)}\right]^2 + \me^2
c^4 } - \me c^2\, . \rule{10mm}{0mm}
\label{A.69}\eeqa
Notice that, for $W < E$, $Q_{+}$ is larger than $W$. When
$W\ll E$, expression \req{A.69} is not suitable for evaluating $Q_-$
since it involves the subtraction of two similar quantities. In this
case, it is more convenient to use the approximate relation
\beqa
cp-cp' & \simeq &
c \left( \frac{\d p}{\d E} \, W
- \frac{1}{2} \, \frac{\d^{2}p}{\d E^{2}} \, W^{2}
+ \frac{1}{6} \, \frac{\d^{3}p}{\d E^{3}} \, W^{3} \right)
\nonumber \\[2mm]
& = & \frac{W}{\beta} \left[ 1 + \frac{1}{2\gamma(\gamma+1)} \,
\frac{W}{E} + \frac{1}{2(\gamma+1)^{2}} \left( \frac{W}{E} \right)^{2}
\right]
\label{A.70}\eeqa
and calculate $Q_-$ as
\beq
Q_{-} =
\me c^{2} \left[ \sqrt{ \left(\frac{cp-cp'}{\me c^{2}} \right)^{2} +
1} - 1 \right]
\simeq
\me c^{2} \left( x - \frac{x^{2}}{2} + \frac{x^{3}}{2} \right)\, ,
\label{A.71}\eeq
with
\beq
x \equiv \frac{1}{2} \left( \frac{cp-cp'}{\me c^{2}} \right)^{2} \simeq
\frac{W^{2}}{2\beta^{2}\,\me^{2}c^{4}}
\left[ 1 + \frac{1}{2\gamma(\gamma+1)} \,
\frac{W}{E} + \frac{1}{2(\gamma+1)^{2}} \left( \frac{W}{E} \right)^{2}
\right]^{2}\, .
\label{A.72}\eeq
Thus, for $W \ll E$,
\beq
Q_{-}(Q_{-}+2\me c^{2}) = (cp-cp')^{2} \simeq W^{2}/\beta^{2}\, ,
\label{A.73}\eeq
and
\beq
\cos \theta_{\rm r} \simeq \frac{W}{\beta (c \hbar q)} \simeq
\sqrt{\frac{Q_{-}(Q_{-}+2\me c^{2})}{Q(Q+2\me c^{2})}}\, .
\label{A.74}\eeq

\begin{figure}[thb]
\begin{center}
\vspace{3mm}
\includegraphics*[width=7.7cm]{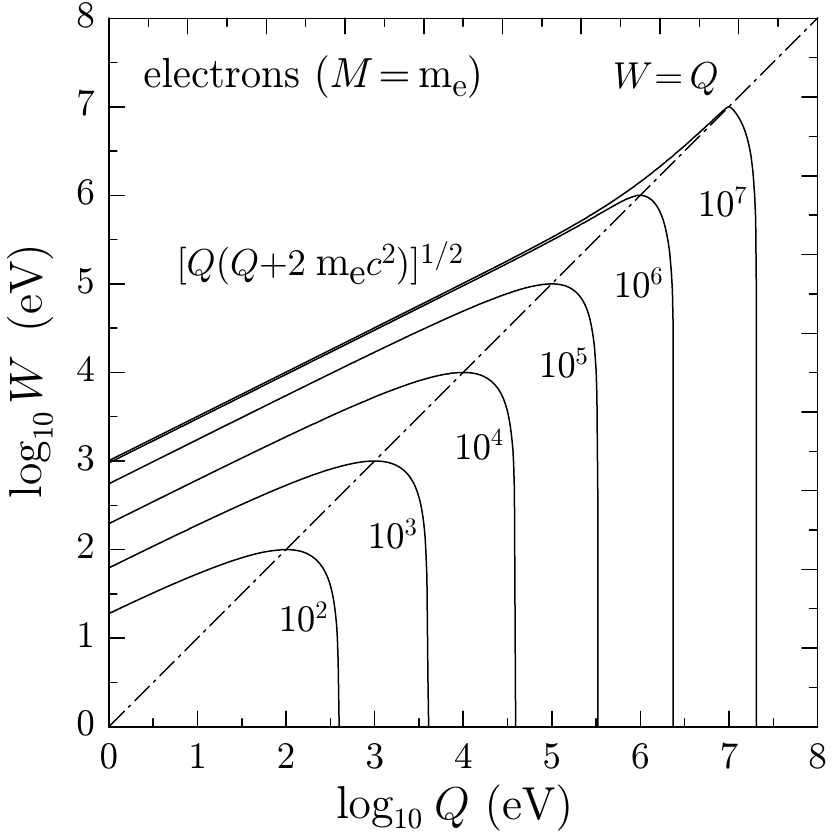}\hfill
\includegraphics*[width=7.7cm]{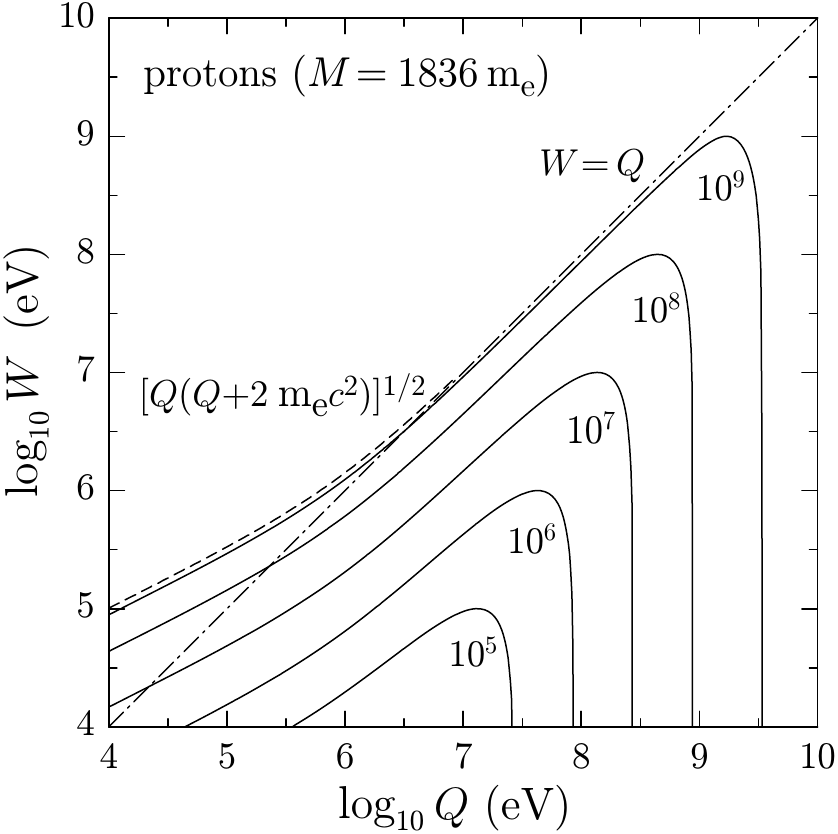}
\caption {\rm Domains of kinematically allowed transitions in the
$(Q,W)$ plane for electrons/positrons (left) and protons (right). The
curves represent the maximum allowed energy loss $W_{\rm m}(Q)$, given
by Eq.\ \req{A.75}, for projectiles with the indicated kinetic
energies (in eV). When $E$ increases, $W_{\rm m}(Q)$ approaches the
vacuum photon line, $W_0(Q) =[Q(Q+2\me c^2)]^{1/2}$, which is an absolute
upper bound for the allowed energy losses.
\label{figA.4}}
\end{center}\end{figure}

From \req{A.69}, it is clear that the curves $Q=Q_-(W)$ and
$Q=Q_+(W)$ intersect at $W = E$. Thus, they define a single continuous
function $W=W_{\rm m}(Q)$ in the interval $0 < Q < Q_+(0)$. By solving
the equations $Q=Q_{\pm}(W_{\rm m})$ we obtain
\beq
W_{\rm m}(Q) = E + Mc^2 - \sqrt{\left[ \sqrt{E(E+2Mc^2)} -
\sqrt{Q(Q+2\me c^2)} \right]^2 + M^2 c^4}\, .
\label{A.75}\eeq
Evidently, for given values of $E$ and $Q$ [$<Q_+(0)$], the only
kinematically allowed values of the energy loss are those in the
interval $0 < W < W_{\rm m}(Q)$ (see Fig.\ \ref{figA.4}). When $E \gg
W_{\rm m}(Q)$, the expression \req{A.75} reduces to [see Eq.\ \req{A.73}]
\beq
W_{\rm m}(Q) \simeq \beta \sqrt{Q(Q+2\me c^2)}\, .
\label{A.76}\eeq
The next-order approximation,
\beq
W_{\rm m}(Q) \simeq \beta \sqrt{Q(Q+2\me c^2)}- \frac{1}{2} \left[
\gamma M c^2 + \beta \sqrt{Q(Q+2\me c^2)} \right]
\left( \frac{\sqrt{Q(Q+2\me c^2)}}{\gamma^2 M c^2} \right)^2 \,
\label{A.77}\eeq
gives results with 9 or more digits correct for $Q$ values up to about
$10^{-4}E$.

The plots in Figs.\ \ref{figA.3} and \ref{figA.4} reveal a noteworthy
difference between the allowed domains for electrons and protons. For
electrons and positrons ($M=\me$) the curves $W_{\rm m}(Q)$ have their
maxima at $Q=E$, just at the intersections of these curves with the
diagonal of the $(Q,W)$ plane. In the case of protons, and heavier ions,
the maxima of the curves occur at recoil energies larger than $E$, \ie,
on the right-hand side of the diagonal.  It is also worth observing that the
intersection of the curve $W_{\rm m}(Q)$ with the diagonal $W=Q$, the
Bethe ridge, occurs at a point where the energy loss is
\beq
W_{\rm ridge} = 2 (\gamma^2-1) \, \me c^2 \, R
\label{A.78}\eeq
with
\beq
R = \left( 1 + \frac{\me^2}{M^2} +
2\gamma \, \frac{\me}{M} \right)^{-1}.
\label{A.79}\eeq
For electrons and positrons, $W_{\rm ridge}=E$. For heavy
projectiles ($M \gg \me$) with energies much smaller than their rest
energy $Mc^2$, $R \simeq 1$ and
\beq
W_{\rm ridge} \simeq 2 (\gamma^2-1)
\me c^2 = 2 \, \frac{E (E+2 Mc^2)}{M^2 c^4} \, \me c^2 .
\label{A.80}\eeq

In the non-relativistic regime, the recoil energy is
\beq
Q \equiv (\hbar q)^{2}/2\me \, ,
\label{A.81}\eeq
and the limits of the interval ($Q_-^\textrm{nr},Q_+^\textrm{nr})$ of
allowed recoil energies are
\beq
Q_{\pm}^\textrm{nr} = \frac{M}{\me} \left[ \sqrt{E} \pm \sqrt{E-W} \right]^{2}
\, .
\label{A.82}\eeq
The maximum energy loss for a given value of $Q$ $[< Q_+^{\rm nr}(0)]$ is
\beq
W_\textrm{m}^\textrm{nr} (Q) = \sqrt{\frac{\me}{M} \, Q} \left( 2 \sqrt{E}
- \sqrt{\frac{\me}{M} \, Q} \right) \, .
\label{A.83}\eeq
Note that $W_\textrm{m}^\textrm{nr}(Q)$ increases without limit when the
energy $E$ of the projectile increases. This is in sharp contrast with
the relativistic case; Eq.\ \req{A.76} shows that $[Q(Q+2\me c^2)]^{1/2}$ is
an upper bound for $W_\textrm{m}(Q)$.

\index{vacuum photon line} \index{photon line}
For a given energy loss $W$, the quantity
\beq
\hbar q_- \equiv c^{-1} \sqrt{Q_-(Q_-+2\me c^2)}\, ,
\label{A.84}\eeq
is the minimum value of the momentum transfer in an inelastic collision,
which occurs when $\theta=0$. $\hbar q_-$ is always larger than $W/c$.
When the energy of the projectile increases, $\beta \rightarrow 1$ and
$\hbar q_-$ decreases approaching (but never reaching) the value $W/
c$. It is worth recalling that a photon of energy $W$ in vacuum has a
linear momentum $\hbar q = W/c$ and, hence, interactions consisting of
emission or absorption of bare photons would be located on the line
\beq
W_0(Q) = c\hbar q = \sqrt{Q(Q+2\me c^2)}
\label{A.85}\eeq
of the ($Q$,$W$) plane, the so-called {\it vacuum photon line}. This line lies
outside the kinematically allowed region, \ie, the ``recoil'' energy of
the photon is less than $Q_-$ (see Figs.\ \ref{figA.3} and
\ref{figA.4}).
Therefore, when the target is a single atom, the emission of photons by
the projectile is not possible\footnote{In a condensed medium,
ultra-relativistic projectiles can emit real photons (Cherenkov
radiation) under certain, very restricting circumstances (Section
\ref{sec8.2.2}).}.
\index{inelastic collisions! kinematics of |)}



\chapter{Mathematical formulas \label{appB}}

\section{The Dirac delta distribution \label{appB.1}}

\index{Dirac delta function|(}
The {\it Dirac delta distribution} $\delta(x)$ is defined by the property
\beq
\int_a^b f(x) \delta(x-x_0) \, \d x = \left\{ \begin{array}{ll}
f(x_0) \rule{3mm}{0mm}& \mbox{if $a<x_0<b$,} \\ [4mm]
0 & \mbox{if $x_0<a$ \'{o} $x_0>b$,}
\end{array} \right.
\label{B.1}\eeq
where $f(x)$ is an arbitrary function of the scalar variable $x$ that is
continuous at the point $x_0$. A more intuitive definition of the delta
distribution is obtained by considering the set of step functions
\beq
F_\epsilon (x) = \left\{ \begin{array}{cl}
\displaystyle{\frac{1}{2\epsilon}}  \rule{3mm}{0mm}&
\mbox{if $-\epsilon<x<\epsilon$} \\ [4mm]
0 & \mbox{if $|x|>\epsilon$,} \end{array} \right.
\label{B.2}\eeq
with $\epsilon>0$. In the limit $\epsilon\rightarrow 0$, the function
$F_\epsilon(x)$ extends over an ``infinitesimal'' interval about $x=0$
and, if $a<0$ and $b>0$,
\beq
\int_a^b f(x) \left( \lim_{\epsilon\rightarrow 0} F_\epsilon(x) \right)
\,
\d x
= f(0) \lim_{\epsilon\rightarrow 0} \int_a^b  F_\epsilon(x) \, \d x
= f(0),
\label{B.3}\eeq
for any continuous function $f(x)$. Consequently, we can set
\beq
\delta(x) = \lim_{\epsilon\rightarrow 0} F_\epsilon(x).
\label{B.4}\eeq

Evidently, $\delta(x)$ is not an ordinary function because being null
almost everywhere (\ie, except on a set of null measure) its integral
is not zero. The Dirac delta is an example of {\it distribution} or
{\it generalized function}. Normally, a distribution can be represented as a
limit of a sequence of functions that are well-behaved, \ie, continuous
with continuous derivatives \citep[see, \eg,][]{DenneryKrzywicki1996}.
The following are alternative representations of the delta distribution
\begin{subequations}
\label{B.5}
\beqa
\delta(x)
\label{B.5a}&=&
\lim_{\sigma \rightarrow 0} \frac{1}{\sigma\sqrt{2\pi}} \exp\left( -
\frac{x^2}{2\sigma^2} \right),
\\ [2mm]
\label{B.5b}
&=& \lim_{\epsilon \rightarrow 0} \frac{\epsilon}{\pi (x^2+\epsilon^2)},
\eeqa
\end{subequations}
which are limiting forms of positive-definite functions with
symmetric bell shapes centered at $x=0$ and of unit area. Notice that
the delta function is symmetric
\beq
\delta(x)= \delta(-x).
\label{B.6}\eeq
Starting from a representation in terms of well-behaved functions,
such as \req{B.5a} or \req{B.5b}, we can define the derivative of the
delta distribution,
\begin{subequations}
\label{B.7}
\beqa
\delta'(x) =\frac{\d }{\d x} \delta(x)
&\equiv&
\lim_{\sigma \rightarrow 0} \frac{-x}{\sigma^3\sqrt{2\pi}} \exp\left( -
\frac{x^2}{2\sigma^2} \right)
\label{B.7a} \\ [2mm]
&=& \lim_{\epsilon \rightarrow 0} \frac{-2 \epsilon x}{\pi
(x^2+\epsilon^2)^2},
\label{B.7b}
\eeqa
\end{subequations}
which, evidently is also a distribution.

The delta distribution can also be obtained as limiting forms of
functions that are not definite positive, or by means of integrals:
\begin{subequations}
\label{B.8}
\beqa
\delta(x)
&=& \lim_{c\rightarrow \infty} \frac{\sin(cx)}{\pi x}
= \lim_{c\rightarrow \infty} \frac{1-\cos(cx)}{\pi c x^2}
= \lim_{c\rightarrow \infty} \frac{\sin^2(cx/2)}{2\pi c (x/2)^2}
\label{B.8a} \\ [2mm]
&=& \lim_{\epsilon \rightarrow 0} \frac{1}{\pi}
\int_0^\infty \exp(-\epsilon x) \cos(kx) \d k
\rule{10mm}{0mm}
\label{B.8b} \\ [2mm]
&=& \lim_{c\rightarrow \infty} \frac{1}{2\pi} \int_{-c}^c
\exp({\rm i}kx)\, \d k =
\lim_{c\rightarrow \infty}
\frac{1}{2\pi} \int_{-c}^c \cos(kx)\, \d k.
\label{B.8c}
\eeqa
\end{subequations}
Although the delta distribution is not an ordinary function, it is
frequently called the ``delta function''.

Owing to its singular character, the delta function is meaningful only
in integrals over its argument. The following relations between delta
functions can be proved by multiplying the two sides of the equality by
an arbitrary function $g(x)$, continuous at $x=0$, and integrating over
$x$ in the interval $(-\infty,\infty)$.
\begin{subequations}
\beq
\delta(x)= - x \frac{\d \delta(x)}{\d x},
\label{B.9a}\eeq
\beq
x \delta(x)= 0, \quad x^2 \frac{\d \delta(x)}{\d x} = 0 ,
\label{B.9b}\eeq
\beq
\frac{\d \delta(x)}{\d x} = - \frac{\d \delta(-x)}{\d x} ,
\label{B.9c}\eeq
\beq
\delta(ax) = |a|^{-1} \delta(x) ,
\label{B.9d}\eeq
\beq
\delta(x^2-a^2) = (2a)^{-1} [\delta(x-a) + \delta(x+a)],
\label{B.9e}\eeq
\beq
f(x) \delta(x-a) = f(a) \delta(x-a) ,
\label{B.9f}\eeq
\beq
\int \delta(a-x) \delta(x-b) \, \d x = \delta(a-b) ,
\label{B.9g}\eeq
\beq
\int f(x) \frac{\d^n \delta(x)}{\d x^n} \, \d x = (-1)^n f^{(n)}(0).
\label{B.9h}\eeq
\end{subequations}

The delta function of a function $g(x)$ can be expressed in a simple
form. Clearly, $\delta(g(x))$ vanishes except in the neighborhood of the
zeros $x_i$ of $g(x)$. If $g(x)$ is analytical near those zeros, we can
consider the approximation $g(x) \sim g'(x_i) (x-x_i)$ for $x$
sufficiently close to $x_i$. Then, from the definition \req{B.4} and the
property \req{B.9d} it follows that
\beq
\delta(g(x)) = \sum_i \frac{1}{\left| g'(x_i) \right|} \, \delta(x-x_i),
\label{B.10}\eeq
provided $g'(x_i)\ne 0$, that is, whenever $x_i$ is a simple root of $g(x)$.

The identity
\beq
\frac{1}{x\pm {\rm i}\epsilon} =
\frac{x}{x^2+\epsilon^2} \mp
\frac{{\rm i} \epsilon}{x^2+\epsilon^2}
\label{B.11}\eeq
allows deriving a useful formula. Let us consider the integral
\beq
\lim_{\epsilon\rightarrow 0} \int_{-\infty}^{\infty} f(x)
\frac{1}{x\pm {\rm i}\epsilon} \, \d x =
\lim_{\epsilon\rightarrow 0} \int_{-\infty}^{\infty} f(x)
\frac{x}{x^2+\epsilon^2}\, \d x \mp
{\rm i}\, \lim_{\epsilon\rightarrow 0} \int_{-\infty}^{\infty} f(x)
\frac{\epsilon}{x^2+\epsilon^2}
\, \d x.
\label{B.12}\eeq
By virtue of \req{B.5b}, the integral in the second term on the
right-hand side is
$$
\lim_{\epsilon\rightarrow 0} \int_{-\infty}^{\infty} f(x)
\frac{\epsilon}{x^2+\epsilon^2}
\, \d x = \pi \int_{-\infty}^\infty f(x) \, \delta(x) \, \d x.
$$
When $\epsilon$ is small, the integrand in the first term on the
right-hand side of the equality \req{B.12} reduces to $1/x$, except near
$x=0$. If the function $f(x)$ is well behaved, we can write
$$
\lim_{\epsilon\rightarrow 0} \int_{-\infty}^{\infty} f(x)
\frac{x}{x^2+\epsilon^2} \, \d x =
\lim_{\epsilon\rightarrow 0} \int_{-\infty}^{-\epsilon} \frac{f(x)}{x}
\,
\d x +
\lim_{\epsilon\rightarrow 0} \int_{-\epsilon}^{\epsilon} f(x)
\frac{x}{x^2+\epsilon^2} \, \d x +
\lim_{\epsilon\rightarrow 0} \int_{\epsilon}^{\infty} \frac{f(x)}{x} \,
\d x
$$
$$
=
\lim_{\epsilon\rightarrow 0} \int_{-\infty}^{-\epsilon} \frac{f(x)}{x}
\,
\d x +
\lim_{\epsilon\rightarrow 0} \int_{\epsilon}^{\infty} \frac{f(x)}{x} \,
\d x + f(0)
\lim_{\epsilon\rightarrow 0} \int_{-\epsilon}^{\epsilon}
\frac{x}{x^2+\epsilon^2} \, \d x.
$$
The last integral vanishes because the integrand is an odd function of
$x$. Hence,
$$
\lim_{\epsilon\rightarrow 0} \int_{-\infty}^{\infty} f(x)
\frac{x}{x^2+\epsilon^2} \, \d x = \lim_{\epsilon\rightarrow 0} \left(
\int_{-\infty}^{-\epsilon} \frac{f(x)}{x} \, \d x +
\int_{-\epsilon}^{\infty} \frac{f(x)}{x} \, \d x \right) \equiv
{\cal P} \int_{-\infty}^{\infty} \frac{f(x)}{x} \, \d x,
$$
where ${\cal P}$ indicates the principal value. Therefore,
\beq
\lim_{\epsilon\rightarrow 0}\, \frac{1}{x\pm {\rm i}\epsilon} =
{\cal P}\, \frac{1}{x} \mp {\rm i} \pi \delta(x).
\label{B.13}\eeq

\index{step function}

The Heaviside unit step function can be defined as\index{Heaviside unit
step function}
\beq
{\cal S}(x) \equiv \int_{-\infty}^{x} \delta(x') \, \d x' =
\left\{
\begin{array}{l} \mbox{0 \; \; if $x<0$,} \\ [3mm] \mbox{1 \; \; if $x>0$.}
\end{array} \right.
\label{B.14}\eeq
Evidently,
\beq
\delta(x) = \d {\cal S}(x) / \d x\, .
\label{B.15}\eeq
Since $\delta(x)$ is an even function of $x$, we have
\beq
\int_0^\infty \delta(x) \, \d x  = \frac{1}{2}
\int_{-\infty}^\infty \delta(x) \, \d x  = \frac{1}{2}
\qquad \mbox{and} \qquad {\cal S}(0)=\frac{1}{2}.
\label{B.16}\eeq
It is also convenient to define the {\it unit window function} \index{window
function}  \index{unit window function}
\beq
{\cal W}(x_1,x_2; x) \equiv {\cal S}(x-x_1) \, {\cal S} (x_2-x)
= \left\{
\begin{array}{l} \mbox{1 \; \; if $x_1<x< x_2$,} \\ [3mm]
\mbox{0 \; \; if $x<x_1$ or $x > x_2$.}
\end{array} \right.
\label{B.17}\eeq

\index{Dirac delta function! three-dimensional}
The three-dimensional delta function is defined by the property
\req{B.1},
\beq
\int f({\bf r}) \, \delta({\bf r}-{\bf r}') \, \d {\bf r} = f({\bf r}'),
\label{B.18}\eeq
if the point ${\bf r}'$ is in the interior of the integration volume. We have
\beq
\delta({\bf r}) = \delta(x) \, \delta(y) \, \delta(z) =
\frac{1}{(2\pi)^3} \int_{-\infty}^{\infty}
\exp({\rm i} {\bf k}\dotprod {\bf r}) \, \d {\bf k},
\label{B.19}\eeq
where $(x,y,z)$ are the Cartesian coordinates of the point ${\bf r}$.

\index{spherical polar coordinates}
\vspace{2mm}
\begin{figure}[htb] \begin{center}
\includegraphics*[width=6.0cm]{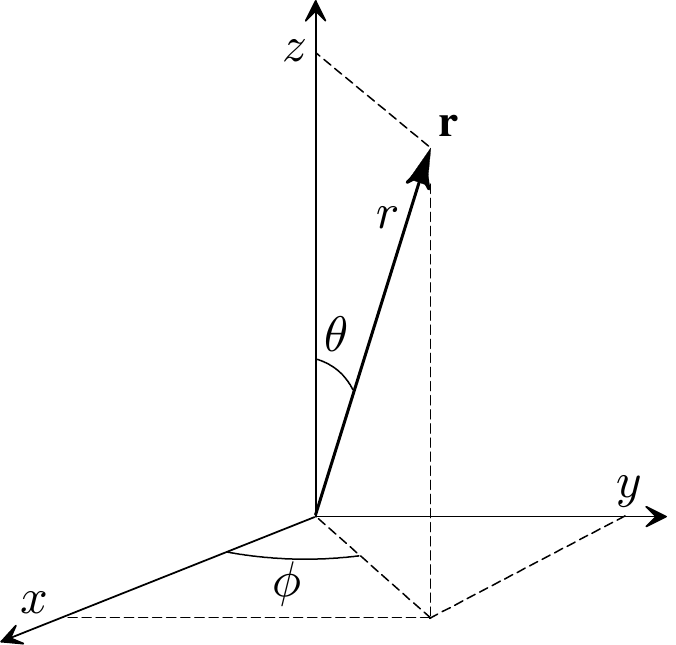}
\caption{Cartesian and spherical coordinates. The $z$ axis is the polar
axis. $\theta$ and
$\phi$ are the polar angle and the azimuthal angle, respectively.
\label{figB.1}}
\end{center} \end{figure}

In the calculations it may be convenient to represent the position
vector in spherical coordinates $(r,\theta,\phi)$ defined by
[see Fig.\ \ref{figB.1}]:
\beq
r = \left( x^{2}+y^{2}+z^{2} \right)^{1/2}, \quad
\theta = \cos^{-1}(z/r), \quad \phi = \tan^{-1}(y/x).
\label{B.20}\eeq
The polar angle $\theta$ takes values in the interval
$(0,\pi)$ and is unambiguously determined by the value of $\cos\theta$.
The azimuthal angle
$\phi$ takes values in the interval $(0,2\pi)$; the
definition $\phi = \tan^{-1}(y/x)$ involves an ambiguity of $\pm \pi$,
which is resolved by the signs of the $x$ and $y$ components of ${\bf
r}$; in programming languages, $\phi$ is given unambiguously
by the two-argument
arctangent function ${\rm atan2}(y,x)$. The inverse transformation is
\beq
x = r\sin\theta\cos\phi, \quad y = r\sin\theta\sin\phi, \quad
z = r\cos\theta.
\label{B.21}\eeq
It is worth noticing that the spherical coordinates of the point $-{\bf
r}$ are $(r,\pi-\theta,\pi+\phi)$.
Because the volume element $d {\bf r}$ about the position ${\bf r}$ can
be expressed as
\begin{subequations}
\label{B.22}
\beq
\d {\bf r} = r^2 \sin\theta \, \d r \, \d \theta \, \d \phi = r^2 \, \d
r \, \d(\cos \theta) \, \d \phi,
\label{B.22a}\eeq
the three-dimensional delta function in spherical coordinates is
\beq
\delta({\bf r}-{\bf r}') = \frac{1}{r^2} \, \delta(r-r') \,
\delta(\cos\theta-\cos\theta') \, \delta(\phi-\phi').
\label{B.22b}\eeq
\end{subequations}
We also have\footnote{If the delta function is at an endpoint of the
integration interval, it contributes half the function value. Thus,
$$
\int_0^\infty f(r) \, \delta(r) \, \d r = \frac{1}{2} \, f(0).
$$
}
\beq
\delta({\bf r}) = \frac{1}{2\pi}
\, \frac{\delta(r)}{r^2}.
\label{B.23}\eeq

A straightforward calculation shows that $\nabla^2 (1/r)$ vanishes for
all ${\bf r}\ne 0$ but it is singular at the origin of coordinates. The
divergence theorem applied to a sphere ${\cal S}$ of radius $R$ centered at the
origin, gives
\beqa
\int_{\cal V} \nabla^2 \left( \frac{1}{r} \right)  \, \d {\bf r} &=&
\int_{\cal V} \nablab \dotprod \left[ \nablab \left( \frac{1}{r}
\right)\right]  \, \d {\bf r} =
\int_{\cal S} \nablab \left( \frac{1}{r} \right) \dotprod \hat{\bf n} \, \d s
=- \int_{\cal S} \frac{1}{r^2}\, \hat{\bf r} \dotprod \hat{\bf n} \, \d s =
-4\pi. \rule{10mm}{0mm}
\nonumber\eeqa
That is, $\nabla^2 (1/r)$ vanishes everywhere except at the origin and
its volume integral equal $4 \pi$. Hence,
\beq
\nabla^2 \left( \frac{1}{r} \right) = - 4\pi \delta({\bf r}) \, .
\label{B.24}\eeq
This result implies that the Laplacian of a function of only the
radial distance $r$ is
\beq
\nabla^2 f(r) = \frac{1}{r} \frac{\d^2}{\d r^2} \, r f(r) - 4 \pi \,
\delta({\bf r}) \, r f(r).
\label{B.25}\eeq

\index{Bethe integral}
The equality \req{B.24} can be used to evaluate the following integral
\beqa
\int_{{\mathbb R}^3} \frac{\exp({\rm i}{\bf q}\dotprod{\bf r})}{|{\bf r}-{\bf
r}'|} \, \d{\bf r} &=&
-\frac{1}{q^2} \int \frac{1}{|{\bf r}-{\bf r}'|}
 \nabla^{2} \exp({\rm i}{\bf q}\dotprod{\bf r}) \, \d{\bf r}
\nonumber \\[2mm]
&=& - \frac{1}{q^2}
\int \exp({\rm i}{\bf q}\dotprod{\bf r}) \nabla^{2}
\left( \frac{1}{|{\bf r}-{\bf r}'|} \right)
\d{\bf r}
\nonumber \\[2mm]
&=& -\frac{1}{q^2}
\int \exp({\rm i}{\bf q}\dotprod{\bf r}/\hbar)
\left[ -4\pi\delta({\bf r}-{\bf r}') \right] \d{\bf r},
\nonumber \eeqa
where we have integrated twice by parts. Hence,
\beq
\int_{{\mathbb R}^3}
\frac{\exp({\rm i}{\bf q}\dotprod{\bf r})}{|{\bf r}-{\bf
r}'|} \, \d{\bf r}
= \frac{4\pi}{q^2} \exp({\rm i}{\bf q}\dotprod{\bf r}').
\label{B.26}\eeq
This important identity is known as the {\it Bethe integral
formula}.
\index{Dirac delta function|)}


\section{Spherical harmonics \label{appB.2}}

\index{spherical harmonics|(}
In coordinate representation, the orbital angular momentum operator of
a particle is defined as
\beq
{\bf L} \equiv \frac{1}{\hbar} {\bf r} \times {\bf p} =
- {\rm i} {\bf r} \times \nablab,
\label{B.27}\eeq
where ${\bf r}$ is the position vector and ${\bf p}=-{\rm i}\hbar\nablab$
is the linear momentum operator. Strictly speaking, ${\bf L}$ is the
angular momentum in units of $\hbar$, the reduced Planck constant. The
Cartesian components of ${\bf L}$ are
\beq
L_{x} = (yp_{z}-zp_{y})/\hbar, \quad
L_{y} = (zp_{x}-xp_{z})/\hbar, \quad
L_{z} = (xp_{y}-yp_{x})/\hbar.
\label{B.28}\eeq

For calculation purposes it is convenient to express ${\bf L}$ in
spherical coordinates $(r,\theta,\phi)$. We have
\begin{subequations}
\label{B.29}
\beq
L_{x} = {\rm i} \left( \sin\phi \frac{\partial}{\partial\theta} +
\cot\theta \cos\phi \frac{\partial}{\partial\phi} \right),
\label{B.29a}\eeq
\beq
L_{y} = {\rm i} \left( -\cos\phi \frac{\partial}{\partial\theta} +
\cot\theta \sin\phi \frac{\partial}{\partial\phi} \right),
\label{B.29b}\eeq
\beq
L_{z} = -{\rm i} \frac{\partial}{\partial\phi}.
\label{B.29c}\eeq
In addition,
\beq
{\bf L}^{2} = L_{1}^{2} + L_{2}^{2} + L_{3}^{2} =
- \left[ \frac{1}{\sin\theta} \frac{\partial}{\partial\theta}
\left( \sin\theta \frac{\partial}{\partial\theta} \right) +
\frac{1}{\sin^{2}\theta} \frac{\partial^{2}}{\partial\phi^{2}} \right].
\label{B.29d}\eeq
It is also worth noticing that \citep[see][p. 109]{Arfken1985}
\beq
\nabla^2 = \frac{1}{r} \, \frac{\partial^2}{\partial r^2} \, r -
\frac{L^2}{r^2} \, .
\label{B.29e}\eeq
\end{subequations}

The four operators $L_{x}$, $L_{y}$, $L_{z}$, and ${\bf L}^{2}$ are
Hermitian. Although the Cartesian components of ${\bf L}$ do not commute
\citep[see, \eg,][]{Schiff1968}, each of them commutes with $L^2$. That
is, $[L_z,L^2]=0$ and, consequently, there is a basis of common
eigenfunctions of the operators $L_z$ and $L^2$,
\begin{subequations}
\label{B.30}
\beq
{\bf L}^{2}\, Y_{\ell m} = \ell(\ell+1)\, Y_{\ell m},
\label{B.30a}\eeq
\beq
L_{3}\, Y_{\ell m} = m \, Y_{\ell m}.
\label{B.30b}\eeq
\end{subequations}
Expressing the operators $L_z$ and $L^2$ in spherical coordinates, Eqs.\
\req{B.29c} and \req{B.29d}, the eigenvalue equations \req{B.30} take
the form of partial differential equations. Their solutions
$Y_{\ell m}(\theta,\phi)$, considered as functions of the polar and
azimuthal angles, are the {\it spherical harmonics}. Writing
\beq
Y_{\ell m}(\theta,\phi) = \Theta_{\ell m}(\theta) \, \Phi_{m}(\phi),
\label{B.31}\eeq
the Eq.\ \req{B.30b} reduces to
\beq
-{\rm i}\, \frac{\d\Phi_{m}}{\d\phi} = m\, \Phi_{m},
\label{B.32}\eeq
whose general solution, normalized to unity, is
\beq
\Phi_{m}(\phi) = (2\pi)^{-1/2} \, \exp({\rm i}m\phi).
\label{B.33}\eeq
Because the functions $Y_{\ell m}(\theta,\phi)$ must be continuous and
single-valued, it is required that $\Phi_{m}(\phi) =
\Phi_{m}(\phi+2\pi)$ and, therefore, $m$ can only take integer values.

Introducing the factorization \req{B.31} and the expression \req{B.33},
Eq.\ \req{B.30a} becomes
\beq
\left[ - \frac{1}{\sin\theta} \frac{\d}{\d\theta}
\left( \sin\theta \frac{\d}{\d\theta} \right) +
\frac{m^{2}}{\sin^{2}\theta} \right] \Theta_{\ell m} (\theta) =
\ell(\ell+1)\, \Theta_{\ell m}(\theta).
\label{B.34}\eeq
This equation admits physically acceptable solutions (that is, solutions
that are finite in the interval $0\leq\theta\leq\pi$) only for
non-negative integer values of $\ell$ and for
$m=-\ell,-\ell+1,\ldots,\ell-1,\ell$. These solutions can be expressed
in terms of the associated Legendre functions $P_\ell^m(\cos\theta)$
\citep[see][]{Olver2010, Arfken1985}.

\index{Legendre polynomials}
Let us introduce the variable $\omega=\cos\theta$. The Legendre
polynomial of degree $\ell$ ($\geq 0$) is defined by
\beq
P_{\ell}(\omega) = \frac{1}{2^{\ell}\ell!} \;
\frac{\d^{\ell}}{\d\omega^{\ell}} \; (\omega^{2}-1)^{\ell},
\label{B.35}\eeq
These polynomials satisfy the ordinary differential equation
\beq
\left[ (1-\omega^{2})\frac{\d^{2}}{\d\omega^{2}} -
2\omega\frac{\d}{\d\omega} + \ell(\ell+1) \right] P_{\ell}(\omega) = 0,
\label{B.36}\eeq
which is equivalent to Eq.\ \req{B.34} with $m=0$. The Legendre
polynomials are orthogonal in the interval $(-1,1)$,
\beq
\int_{-1}^{1} P_{\ell'}(\omega)P_{\ell}(\omega) \, \d\omega =
\frac{2}{2\ell+1} \delta_{\ell'\ell}
\label{B.37}\eeq
and constitute a complete set for the functions defined in this interval.
Any function $f(\omega)$ admits the following expansion in terms of
Legendre polynomials,
\beq
f(\omega) = \sum_{\ell=0}^{\infty} \left[ \frac{2\ell+1}{2}
\int_{-1}^{1} f(\omega')P_{\ell}(\omega') \, \d\omega' \right]
P_{\ell}(\omega).
\label{B.38}\eeq
In particular,
\beq
\delta(\omega-\omega_{0}) = \sum_{\ell=0}^{\infty} \frac{2\ell+1}{2}
P_{\ell}(\omega_{0}) P_{\ell}(\omega).
\label{B.39}\eeq
The recurrence relation
\beq
(2\ell+1)\omega P_{\ell}(\omega) - (\ell+1)P_{\ell+1}(\omega) -
\ell P_{\ell-1}(\omega) = 0,
\label{B.40}\eeq
allows computing $P_{\ell+1}$ from $P_{\ell}$ and $P_{\ell-1}$.
The Legendre polynomials of lower degrees are
\beq
P_{0}(\omega) = 1, \quad
P_{1}(\omega) = \omega, \quad
P_{2}(\omega) = \frac{1}{2}(3\omega^{2}-1).
\label{B.41}\eeq
The polynomials $P_{\ell}(\omega)$ have definite parity,
\beq
P_{\ell}(-\omega) = (-1)^{\ell} P_{\ell}(\omega).
\label{B.42}\eeq
From \req{B.40} and \req{B.41} it follows that
\begin{subequations}
\label{B.43}
\beq
P_{\ell}(1) = 1
\label{B.43a} \eeq
and, by virtue of \req{B.42},
\beq
P_{\ell}(-1) = (-1)^{\ell}.
\label{B.43b}
\eeq
\end{subequations}

\index{associated Legendre functions}
The associated Legendre functions $P_{\ell}^{|m|}(\omega)$ are defined
by
\beq
P_{\ell}^{|m|}(\omega) = (-1)^{|m|} (1-\omega^{2})^{|m|/2} \;
\frac{\d^{|m|}}{\d\omega^{|m|}} P_{\ell}(\omega).
\label{B.44}\eeq
These functions satisfy the differential equation \req{B.34}, they are
orthogonal in the interval $(-1,1)$,
\beq
\int_{-1}^{1} P_{\ell'}^{|m|}(\omega) \; P_{\ell}^{|m|}(\omega) \,
\d\omega = \frac{2}{2\ell+1} \, \frac{(\ell+m)!}{(\ell-m)!} \;
\delta_{\ell'\ell},
\label{B.45}\eeq
and obey the recurrence relations
\begin{subequations}
\label{B.46}
\beq
(2\ell-1)\omega P_{\ell-1}^{|m|}(\omega) -
(\ell-|m|)P_{\ell}^{|m|}(\omega) -
(\ell-1+|m|)P_{\ell-1}^{|m|}(\omega) = 0 \, ,
\label{B.46a}\eeq
\beq
(2\ell+1)(1-\omega^{2})^{1/2} P_{\ell}^{|m|-1}(\omega) +
P_{\ell+1}^{|m|}(\omega) - P_{\ell-1}^{|m|}(\omega) = 0\, .
\label{B.46b}\eeq
\end{subequations}
Those of lower orders are
\beq
P_{1}^{1}(\omega) = - (1-\omega^{2})^{1/2}, \quad
P_{2}^{1}(\omega) = - 3\omega(1-\omega^{2})^{1/2}, \quad
P_{2}^{2}(\omega) = 3(1-\omega^{2}).
\label{B.47}\eeq

The solutions $\Theta_{\ell m}(\theta)$ of Eq.\ \req{B.34}, normalized
so that
\beq
\int_{0}^{\pi} \Theta_{\ell' m}^{\ast}(\theta) \,
\Theta_{\ell m}(\theta) \sin\theta \, \d\theta = \delta_{\ell'\ell},
\label{B.48}\eeq
can be written in terms of the associated Legendre functions as
\beqa
\Theta_{\ell m}(\theta) &=&
\left[ \frac{(2\ell+1)(\ell-m)!}{2(\ell+m)!} \right]^{1/2}
P_{\ell}^{m}(\cos\theta) \qquad {\rm if}\ m \geq 0 \nonumber \\ [2mm]
&=&\Theta_{\ell,-m}(\theta) \quad {\rm if}\ m < 0.
\label{B.49}\eeqa

Introducing \req{B.33} and \req{B.49}, from Eq.\ \req{B.31} we have
\beqa
Y_{\ell m}(\theta,\phi) &=&
\left[ \frac{2\ell+1}{4\pi} \, \frac{(\ell-m)!}{(\ell+m)!} \right]^{1/2}
P_{\ell}^{m}(\cos\theta) \exp({\rm i}m\phi) \quad {\rm if}\ m \geq 0,
\nonumber \\ [2mm]
&=&
(-1)^{-m} \, Y^{\ast}_{\ell,-m}(\theta,\phi) \quad {\rm if}\ m < 0.
\label{B.50}\eeqa
Table \ref{tabB.1} gives the expressions of the spherical harmonics
with $\ell \le 3$. Notice that the functions $|Y_{\ell
m}(\theta,\phi)|^2$ are independent of $\phi$.

\begin{table}[htb]
\caption{\rm Spherical harmonics.\label{tabB.1}}
\vskip 4mm
\begin{center}
\begin{tabular}{lrcc} \hline\hline
$\ell$ & $m$ & $Y_{\ell m}(\theta,\phi)$ & $r^\ell Y_{\ell m}
(\theta,\phi)$ \rule[-3mm]{0mm}{9mm}\\ \hline
0 & 0 & $\sqrt{\frac{1}{4\pi}}$ \rule[5mm]{0mm}{2mm}&
$\sqrt{\frac{1}{4\pi}}$ \\ [3mm]
1 & 0 & $\sqrt{\frac{3}{4\pi}} \, \cos\theta$ &
$\sqrt{\frac{3}{4\pi}} \, z$ \\ [3mm]
1 & $\pm$1 & $\mp \sqrt{\frac{3}{8\pi}} \, \sin\theta \exp (\pm {\rm
i}\phi)$ & $\mp \sqrt{\frac{3}{8\pi}} \, (x \pm {\rm i}y)$ \\ [3mm]
2 & 0 & $\sqrt{\frac{5}{4\pi}} \sqrt{\frac{1}{4}} \, \left( 3
\cos^2\theta - 1\right)$ &
$\sqrt{\frac{5}{4\pi}} \sqrt{\frac{1}{4}} \, ( 3z^2 - r^2)$ \\ [3mm]
2 & $\pm$1 & $\mp \sqrt{\frac{5}{4\pi}} \sqrt{\frac{3}{2}} \,
\cos\theta \sin\theta
\exp (\pm {\rm i}\phi)$ & $\mp \sqrt{\frac{5}{4\pi}} \sqrt{\frac{3}{2}}
\, z(x \pm {\rm i}y)$ \\ [3mm]
2 & $\pm$2 & $ \sqrt{\frac{5}{4\pi}} \sqrt{\frac{3}{8}} \, \sin^2 \theta
\exp (\pm 2{\rm i}\phi)$ & $\sqrt{\frac{5}{4\pi}} \sqrt{\frac{3}{8}} \,
(x \pm {\rm i}y)^2$ \\ [3mm]
3 & 0 & $ \sqrt{\frac{7}{4\pi}} \sqrt{\frac{1}{4}} \, \cos\theta \,
\left( 5 \cos^2\theta - 3 \right) $ & $
\sqrt{\frac{7}{4\pi}} \sqrt{\frac{1}{4}} \, z\left( 5z^2 - 3 r^2\right)$
\\ [3mm]
3 & $\pm$1 & $\mp \sqrt{\frac{7}{4\pi}} \sqrt{\frac{3}{16}} \,
\left( 5 \cos^2\theta -1 \right)\, \sin\theta  \exp(\pm
{\rm i} \phi)$ & $\mp \sqrt{\frac{7}{4\pi}} \sqrt{\frac{3}{16}}
\, (5z^2-r^2)\, (x \pm {\rm i} y) $ \\ [3mm]
3 & $\pm$2 & $\sqrt{\frac{7}{4\pi}} \sqrt{\frac{15}{8}} \,
\cos\theta \sin^2 \theta \exp(\pm 2 {\rm i} \phi) $ &
$\sqrt{\frac{7}{4\pi}} \sqrt{\frac{15}{8}} \, z(x \pm {\rm i}
y)^2$ \\ [3mm]
3 & $\pm$3 & $\; \; \mp \sqrt{\frac{7}{4\pi}} \sqrt{\frac{5}{16}} \,
\sin^3 \theta \exp(\pm 3 {\rm i} \phi) \; \; $ &
$\; \; \mp \sqrt{\frac{7}{4\pi}} \sqrt{\frac{5}{16}} \, (x \pm {\rm i}
y)^3 \; \; $
\rule[-4mm]{0mm}{4mm} \\ \hline\hline
\end{tabular} \end{center} \end{table}

The angles $\theta$ and $\phi$ give the direction of the unit vector
$\hat{\bf r} = {\bf r}/r \equiv (\theta,\phi)$, which determines a point
on the surface of the sphere of unit radius centered at the origin (the
unit sphere). We can consider the spherical harmonics as functions
defined on the surface of the unit sphere, and write $Y_{\ell
m}(\theta,\phi) \equiv Y_{\ell m}(\hat{\bf r})$. We recall that the
solid angle element (= area on the unit sphere) about the direction
$\hat{\bf r}$ is $\d \hat{\bf r} = \sin\theta\, \d \theta \, \d \phi =
\d(\cos\theta) \, \d \phi$. The $\delta$ function on the unit sphere is
defined by
\beq
\delta (\hat{\bf r}-\hat{\bf r}') =
\delta(\cos\theta-\cos\theta') \delta(\phi-\phi').
\label{B.51}\eeq
Evidently [see Eq.\ \req{B.22b}]
\beq
\delta({\bf r}-{\bf r}') = \frac{1}{r^2} \, \delta(r-r') \,
\delta(\hat{\bf r}-\hat{\bf r}').
\label{B.52}\eeq

The spherical harmonics \req{B.50} constitute a complete orthonormal
basis, that is, they are orthogonal,
\beq
\int Y^{\ast}_{\ell' m'}(\hat{\bf r}) \, Y_{\ell m}(\hat{\bf r}') \,
\d\hat{\bf r} = \int_0^\pi \sin \theta \, \d \theta \int_0^{2\pi} \,
\d \phi \, Y^{\ast}_{\ell' m'}(\theta,\phi) \, Y_{\ell m}(\theta,\phi) =
\delta_{\ell' \ell} \delta_{m'm},
\label{B.53}\eeq
and satisfy the closure relation
\beq
\sum_{\ell=0}^{\infty} \sum_{m=-\ell}^{\ell}
Y^{\ast}_{\ell m}(\hat{\bf r}') \, Y_{\ell m}(\hat{\bf r}) =
\delta(\hat{\bf r}' - \hat{\bf r}).
\label{B.54}\eeq
The last property implies that any function
$f(\hat{\bf r})$ defined on the unit sphere can be expanded as a series
of spherical harmonics,
\beqa
f(\hat{\bf r}) &=& \int \delta(\hat{\bf r}' - \hat{\bf r})
f(\hat{\bf r}')
\d \hat{\bf r}'
\nonumber \\ [2mm]
&=& \int
\left(
\sum_{\ell=0}^\infty \sum_{m=-\ell}^\ell
Y^\ast_{\ell m} (\hat{\bf r}')
Y_{\ell m} (\hat{\bf r}) \right)
f(\hat{\bf r}')
\d \hat{\bf r}'
\nonumber \\ [2mm]
&=& \sum_{\ell=0}^\infty \sum_{m=-\ell}^\ell
\left( \int Y_{\ell m}^\ast (\hat{\bf r}') \,
f(\hat{\bf r}') \d \hat{\bf r}' \right)
Y_{\ell m} (\hat{\bf r}) .
\label{B.55}\eeqa

It can be readily verified that
\begin{subequations}
\label{B.56}
\beq
Y^{\ast}_{\ell m}(\hat{\bf r}) = (-1)^m Y_{\ell, -m}(\hat{\bf r}),
\label{B.56a}\eeq
\beq
Y_{\ell 0}(\theta,\phi) =
\left[ \frac{2\ell+1}{4\pi} \right]^{1/2} P_{\ell}(\cos\theta),
\label{B.56b}\eeq
\beq
Y_{\ell m}(0,\phi) =
\left[ \frac{2\ell+1}{4\pi} \right]^{1/2} \delta_{m0}.
\label{B.56c}\eeq
The spherical harmonics have definite
parity under space inversion, that is,
\beq
Y_{\ell m}(\hat{\bf r}) =
(-1)^{\ell} Y_{\ell m}(- \hat{\bf r})
\qquad \mbox{or} \qquad
Y_{\ell m}(\theta,\phi) =
(-1)^{\ell} Y_{\ell m}(\pi-\theta,\pi+\phi).
\label{B.56d}\eeq
\end{subequations}

\index{spherical harmonics!addition theorem}
\index{spherical harmonics! Uns\"{o}ld's theorem}
The spherical harmonics satisfy the following addition theorem
\citep[see][]{Olver2010, Arfken1985}
\beq
\sum_{m=-\ell}^{\ell}  Y_{\ell m}^\ast (\hat{\bf r}_1)
Y_{\ell m} (\hat{\bf r}_2) =
\frac{2\ell+1}{4\pi} \, P_\ell (\cos \alpha),
\label{B.57}\eeq
where $\alpha$ is the angle between the vectors $\hat{\bf r}_1$ and
$\hat{\bf r}_2$, that is, $\cos\alpha = \hat{\bf r}_1 \cdot
\hat{\bf r}_2$. In particular, when $\hat{\bf r}_1 = \hat{\bf r}_2$, we
have the equality
\beq
\sum_{m=-\ell}^{\ell} \left| Y_{\ell m}(\theta,\phi) \right|^{2} =
\frac{2\ell+1}{4\pi},
\label{B.58}\eeq
which is known as the {\it Uns\"old theorem}. This theorem plays an
important role in calculations of atomic and nuclear structure; it
implies that the density of electrons or nucleons in a closed shell
has spherical symmetry (See Chapter \ref{chapt3}).
\index{spherical harmonics|)}


\section{Modified Bessel functions \label{appB.3}}

\index{modified Bessel functions}
The modified Bessel functions $K_\nu (x)$ are solutions of the
differential equation
\beq
z^2 \, \frac{\d^2 w}{\d z^2} + z \, \frac{\d w}{\d z} - (z^2+\nu^2) w = 0
\label{B.59}\eeq
with the asymptotic behavior
\beq
K_n(z) = \sqrt{\frac{\pi}{2z}} \; \exp(-z) \left[1 + \frac{4n^2-1}{8z} +
{\cal O} \left(
z^{-2} \right) \right],  \qquad |z| \gg 1, \; \; |{\rm arg} z| < \pi.
\label{B.60}\eeq
These functions are irregular at $z=0$. They satisfy the relations,
\citep{AbramowitzStegun1974}
\begin{subequations}
\label{B.61}
\beq
K_{-\nu}(z)=K_\nu(z), \qquad K_\nu(z^\ast) = K_\nu^\ast(z),
\label{B.61a}\eeq
\beq
K_{\nu+1}(z) - K_{\nu-1}(z) = \frac{2\nu}{z} \, K_{\nu}(z),
\label{B.61b}\eeq
\beq
\frac{\d}{\d z} \left[ z^{-\nu} K_\nu (z) \right] = - z^{-\nu} K_{\nu+1}(x),
\qquad
\frac{\d}{\d z} \left[ z^{\nu} K_\nu (z) \right] = - z^{\nu}
K_{\nu-1}(x).
\label{B.61c}\eeq
\end{subequations}

The function $K_0(z)$ admits the following series expansion
\citep{AbramowitzStegun1974}
\beqa
K_0(z) &=& - \left[ \ln(z/2) + g \right] \left\{ 1 +
\frac{z^2/4}{(1!)^2}
+\frac{(z^2/4)^2}{(2!)^2}
+\frac{(z^2/4)^3}{(3!)^2} + \cdots \right\}
\nonumber \\ [2mm]
&& \mbox{} +
\frac{z^2/4}{(1!)^2}
+\left( 1 + \frac{1}{2} \right) \frac{(z^2/4)^2}{(2!)^2}
+ \left( 1 + \frac{1}{2} + \frac{1}{3} \right)\frac{(z^2/4)^3}{(3!)^2}
+ \cdots\, ,
\label{B.62}\eeqa
where $g= 0.577\, 215\, 665$ is Euler's constant. \index{Euler constant}
The corresponding
expansion of the function $K_1(z)$ may be obtained by using the
relation
\beq
K_1(z) = - \frac{\d}{\d z} K_0(z).
\label{B.63}\eeq
For small values of the argument ($|z| \ll 1$),
\beq
K_0(z) = - \ln(z/2) - g + {\cal O}(z), \qquad
K_1(z) = z^{-1} + {\cal O}(z).
\label{B.64}\eeq

\index{modified Bessel functions! integrals involving}
The calculation of the Fourier components of the electromagnetic field
of a moving charged particle involves the integrals
\begin{subequations}
\label{B.65}
\beq
\int_{0}^\infty \frac{x^{2m} \cos(x y)}
{(\alpha^2 + x^2)^{\nu+1/2}} \, \d x
= \frac{(-1)^m \pi^{1/2}}{(2\alpha)^\nu \Gamma(\nu+1/2)}
\, \frac{\d^{2m}}{\d y^{2m}} \left[ y^\nu K_\nu (\alpha y) \right]
\label{B.65a}\eeq
valid for $0 \le m < \nu+1/2$ and ${\rm Re}(\alpha)>0$ [\citet{Erdelyi1954},
Vol. I, Table 1.3(28)], and
\beq
\int_{0}^\infty \frac{x^{2m+1} \sin(x y)}
{(a^2 + x^2)^{\nu+1/2}} \, \d x
= \frac{(-1)^{m+1} \pi^{1/2}}{(2\alpha)^\nu \Gamma(\nu+1/2)}
\, \frac{\d^{2m+1}}{\d y^{2m+1}} \left[ y^\nu K_\nu (\alpha y) \right]
\label{B.65b}\eeq
\end{subequations}
valid for $-2 \le 2m < 2\nu$  and ${\rm Re}(\alpha)>0$
[\citet{Erdelyi1954}, Vol. I, Table 2.2(37)]. $\Gamma(x)$ is the gamma
function of the real argument $x$, which satisfies
\begin{subequations}
\label{B.66}
\beq
\Gamma(1/2) = \sqrt{\pi}, \qquad  \Gamma(1)=1,
\qquad \Gamma(x+1) = x \, \Gamma(x),
\label{B.66a}\eeq
and
\beq
\Gamma(n+1)=n! \qquad \mbox{if $n=0,1,2, \ldots$, with 0!=1.}
\label{B.66b} \eeq
\end{subequations}

In the text we make explicit use of the definite integrals,
\beq
\int_0^\infty \frac{\cos(xy)}{(\alpha^2 + x^2)^{1/2}} \, \d x =
K_0(\alpha y), \qquad
\int_0^\infty \frac{x\, \sin(xy)}{(\alpha^2 + x^2)^{1/2}} \, \d x =
\alpha \, K_1(\alpha y)
\label{B.67}\eeq
and
\beq
\int_0^\infty \frac{\cos(xy)}{(\alpha^2 + x^2)^{3/2}} \, \d x =
\frac{y}{\alpha} \, K_1(\alpha y), \qquad
\int_0^\infty \frac{x\, \sin(xy)}{(\alpha^2 + x^2)^{3/2}} \, \d x =
y \, K_0(\alpha y),
\label{B.68}\eeq
and the indefinite integrals [\citet{GradshteynRyzhik2007}, Eqs.\ 5.54]
\beq
\int x \, K_0^2(x) \, \d x =
\frac{x^2}{2} \left[ K_0^2(x) - K_1^2 (x) \,
\right],
\label{B.69}\eeq
\beq
\int x \, K_1^2(x) \, \d x =
\frac{x}{2} \left[ x \, K_1^2(x) - x \, K_0^2 (x) - 2 K_0(x) \, K_1(x)
\right],
\label{B.70}\eeq
\beq
\int x \, K_0(\alpha x) \, K_0(\beta x) \, \d x =
- \, \frac{\alpha x \, K_1(\alpha x) \, K_0(\beta x) -
\beta x \, K_0(\alpha x) \, K_1(\beta x)}{\alpha^2 - \beta^2}\, ,
\label{B.71}\eeq
\beq
\int x \, K_1(\alpha x) \, K_1(\beta x) \, \d x =
\frac{\beta x \, K_1(\alpha x) \, K_0(\beta x) -
\alpha x \, K_0(\alpha x) \, K_1(\beta x)}{\alpha^2 - \beta^2}.
\label{B.72}\eeq



\chapter{Fundamental constants and conversion factors \label{appC}}

\index{fundamental constants}
The Committee on Data for Science and Technology (CODATA) recommends
self-consistent sets of values of the fundamental physical constants.
The recommended constants are obtained from a lest-squares adjustment
that takes into account all data available at the time of publication of
each CODATA report. After publication of the 2014 adjustment
\citep{Mohr2016}, the International System of Units (SI) was revised and
new definitions of the units of fundamental magnitudes were adopted by
the Bureau International des Poids et Mesures, with effect from 20 May,
2019. CODATA Recommended Values based on the revised SI and obtained
from the latest least-squares adjustment are available on the NIST
website on Fundamental Physical Constants
(\url{https://physics.nist.gov/cuu/Constants/index.html}).
Table \ref{tabC.1} displays the values of fundamental constants adopted
in the text, which were taken from the NIST website on 30 December,
2019.

\index{classical electron radius} \index{Avogadro's constant}
\index{speed of light!in vacuum}

\begin{table}
\caption{Fundamental constants. The last two digits of the tabulated
values are usually affected by statistical uncertainties
(see \url{https://physics.nist.gov/cuu/Constants/index.html} and
\citeauthor{Tiesinga2021}, \citeyear{Mohr2016}).
\label{tabC.1}}
\vskip 4mm
\begin{center}
\begin{tabular}{p{6.5cm}r@{$=\,$}l} \hline\hline
Constant & \multicolumn{2}{c}{Symbol and value}
\rule[-2mm]{0mm}{6mm} \\ \hline
Avogadro's number (exact) \rule{0mm}{5mm} & $N_{\rm A}\;$ & 6.022 140
76 $\times$ $10^{23}$ mol$^{-1}$ \\
Speed of light in vacuum (exact) \rule{0mm}{5mm} & $c\;$ & 2.997 924
58 $\times 10^{8}$ m s$^{-1}$ \\
Planck's constant (exact) \rule{0mm}{5mm} & $h\;$ & 6.626 070 15
$\times$ $10^{-34}$ J s \\
& & 4.135 667 696 $\times$ $10^{-15}$ eV s \\
Reduced Planck's constant \rule{0mm}{5mm} &
$\hbar=h/2\pi\;$ & 1.054 571 817 $\times$ $10^{-34}$ J s \\
& & 6.582 119 569 $\times$ $10^{-16}$ eV s \\
\rule{0mm}{5mm} & $\hbar c\;$ & 1.973 269 804 $\times$
$10^{-7}$ eV m \\
Gravitation constant \rule{0mm}{5mm} & $G \;$ &
6.674 30 $\times$ $10^{-11}$ m$^3$ kg$^{-1}$ s$^{-2}$ \\
Boltzmann constant (exact) \rule{0mm}{5mm} & $k_{\rm B} \;$ & 1.380
649 $\times$ $10^{-23}$ J K$^{-1}$ \\
 & & 8.617 333 262 $\times$ $10^{-5}$ eV K$^{-1}$ \\
Atomic mass unit \rule{0mm}{5mm} & ${\rm u} \;$ &
$1{\rm g}/N_{\rm A} = {\rm m}(^{12}{\rm C})/12$ \\ & & 1.660 539 066 60
$\times$ $10^{-27}$ kg \\
& & 931.494 102 42 MeV \\
Elementary charge (exact) \rule{0mm}{5mm} & $e \;$ & 1.602 176 634
$\times$ $10^{-19}$ C \\
Electron mass \rule{0mm}{5mm} & $m_{\rm e} \;$
  & 9.109 383 7015 $\times$ $10^{-31}$ kg \\
& & 5.485 799 090 65 $\times$ $10^{-4}\; {\rm u}$ \\
& & 510.998 950 00 keV \\
Proton mass \rule{0mm}{5mm} & $m_{\rm p} \;$
  & 1.672 621 923 69 $\times$ $10^{-27}$ kg \\
& & 1.007 276 466 621 ${\rm u}$ \\
& & 1836.152 673 43 m$_{\rm e}$ \\
& & 938.272 088 16 MeV \\
Neutron mass \rule{0mm}{5mm} & $m_{\rm n} \;$
  & 1.674 927 498 04 $\times$ $10^{-27}$ kg \\
& & 1.008 664 915 95 ${\rm u}$ \\
& & 1838.683 661 73 m$_{\rm e}$ \\
& & 939.565 420 52 MeV \\
Bohr magneton \rule{0mm}{5mm} & $\mu_{\rm B}\; $ & $e\hbar/2 \me c$
(Gauss-cgs) \\
& & 9.274 010 0783 $\times$ $10^{-24}$ J T$^{-1}$ \\
& & 5.788 381 8060 $\times$ $10^{-5}$ eV T$^{-1}$ \\
Magnetic moment of the electron \rule{0mm}{5mm} & ${\cal M}_{\rm e} \;$
& $-$9.284 764 7043 $\times$ $10^{-24}$ J T$^{-1}$ \\
& & $-$1.001 159 652 181 28 $\mu_{\rm B}$
\rule[-2mm]{0mm}{4mm} \\ \hline\hline
\end{tabular} \end{center} \end{table}
\addtocounter{table}{-1}

\begin{table}[t]
\caption{Continued.}
\vskip 4mm
\begin{center}
\begin{tabular}{p{6.5cm}r@{$=\,$}l} \hline\hline
Constant & \multicolumn{2}{c}{Symbol and value}
\rule[-2mm]{0mm}{6mm} \\ \hline
Nuclear magneton \rule{0mm}{5mm} & $\mu_{\rm N}\; $ & $e\hbar/2
m_{\rm p} c $ (Gauss-cgs) \\
& & 5.050 783 7461 $\times$ $10^{-27}$ J T$^{-1}$ \\
& & 3.152 451 258 44 $\times$ $10^{-8}$ eV T$^{-1}$ \\
Magnetic moment of the proton \rule{0mm}{5mm} & ${\cal M}_{\rm p} \;$ &
1.410 606 797 36 $\times$ $10^{-26}$ J T$^{-1}$ \\
& & 2.792 847 344 63 $\mu_{\rm N}$ \\
Magnetic moment of the neutron \rule{0mm}{5mm} & ${\cal M}_{\rm n} \;$ &
$-$9.662 3651 $\times$ $10^{-27}$ J T$^{-1}$ \\
& & $-$1.913 042 73 $\mu_{\rm N}$ \\
Fine-structure constant \rule{0mm}{5mm} & $\alpha \;$
& $e^2/\hbar c$ \\ & & 7.297 352 5693 $\times$ $10^{-3}$ \\
& & 1$/$137.035 999 084 \\
Bohr radius \rule{0mm}{5mm} & $a_0 \;$ & $\hbar^2/\me e^2$
\\ & & 5. 291 772 109 03 $\times$ $10^{-11}$ m \\
Hartree energy
\rule{0mm}{5mm} & $E_{\rm h} \;$ & $\me e^4 /\hbar^2 =
e^2/a_0$ \\ & & 27.211 386 245 988 eV \\
Rydberg energy \rule{0mm}{5mm} & $\mbox{Ry} \;$&
$\me e^4 /2 \hbar^2 = E_{\rm h}/2$
\\ & & 13.605 693 122 994 eV \\
Compton wavelength \rule{0mm}{5mm} & $\lambda_{\rm C} \;$
& $h/\me c=2\pi\, \alpha\, a_0$ \\
& & 2.426 310 238 67 $\times$ $10^{-12}$ m \\
Classical electron radius \rule{0mm}{5mm} &
$r_{\rm e} \;$
& $e^2/\me c^2=\alpha^2 \, a_0$ \\
& & 2.817 940 3262 $\times$ $10^{-15}$ m
\rule[-2mm]{0mm}{4mm} \\ \hline\hline
\end{tabular} \end{center} \end{table}
%

\index{Gaussian system of units}
To simplify the formulas, in atomic physics and stopping calculations
its is customary to use the Gaussian system of units
\citep[see, \eg,][]{Jackson1975}, which is based on the cgs system of
units for mechanical quantities. In this system, the force between two
charges $q_1$ and $q_2$ at a distance $r$ (Coulomb's law) is written in
the simple form
\beq
F = q_1 q_2/r^2,
\label{C.1}\eeq
which defines the unit of electrical charge, the statcoulomb (statC) as
a combination of the mechanical units (centimeter, gram, second), 1
statC = 1 cm$^{3/2}$ g$^{1/2}$ s$^{-1}$. A convenient feature of the
Gaussian system is that the electric field, the magnetic induction, the
electric displacement, the magnetic field, the dielectric polarization
and the magnetization have all the same dimensions. Quantities in the SI
can be converted to the Gaussian system of units by means of the
following conversion factors
\begin{subequations}
\label{C.2}
\beq
1\; \mbox{\rm N} = 10^{5}\; {\rm dyn}, \quad
1\; \mbox{\rm J} = 10^{7}\; {\rm erg},
\label{C.2a}\eeq
\beq
1\; \mbox{\rm C} = \mbox{2.997\ 924\ 58 $\times$ 10}^{9}\; {\rm statC},
\quad
1\; \mbox{\rm V} = 1\; \mbox{statV}/299.792\ 458, \quad
1\; \mbox{\rm T} = 10^{4}\; {\rm G}.
\label{C.2b}\eeq
Very frequently, the energies of photons, electrons and other charged
particles are given in eV (or in multiples of this unit),
\beq
1 \; \mbox{\rm eV}
= 1.602\; 176 \; 634 \times 10^{-19}\; {\rm J}
= 1.602\; 176 \; 634 \times 10^{-12}\; {\rm erg}.
\label{C.2c}\eeq
In spectroscopic studies, it is common to give the photon frequency $\nu
= E/h$, or the inverse of the photon wavelength $\lambda^{-1} = E/hc$
(frequently referred to as the photon wave number)
instead of the photon energy $E$. We have
\beq
1\; \mbox{eV}
= 2.417\; 989\; 242 \times 10^{14}\; \mbox{Hz}
= 8065.543\; 937 \; \mbox{cm}^{-1}.
\label{C.2d}\eeq
The Rydberg constant $R_\infty$ is the wave number of photons
having the energy of 1 Rydberg (see Table C.1), that is
\beq
R_\infty = \frac{\mbox{Ry}}{hc} = 1.097\; 373\; 156\; 8160 \;
\times 10^{5}\;
\mbox{cm}^{-1}.
\label{C.2e}\eeq
\end{subequations}


\section{Atomic units}

\index{atomic units}
In atomic, molecular and solid state physics it is convenient to use the
system of atomic units (a.u.) in which the reduced Planck constant, the
electron mass, and the elementary charge are set equal to
(dimensionless) unity, that is,
\beq
\hbar = \me = e = 1.
\label{C.3}\eeq
In this system of units the constants \req{C.3} disappear from the
quantum wave equations, which adopt relatively simple forms. In
addition, various characteristic magnitudes of the hydrogen atom take
values of the order of unity.
\index{Bohr radius} \index{Hartree energy}

\begin{table}[t!]
\caption{Values of the atomic units of relevant quantities.
\label{tabC.2}}
\vskip 4mm
\begin{center}
\begin{tabular}{lcl} \hline\hline
Quantity & Unit & Value of the unit \rule[-2mm]{0mm}{7mm} \\ \hline
Mass & $\me$ & 9.109 383 7015 $\times \; 10^{-28}$ g
\rule{0mm}{5mm}
\\
 & & 510.998 950 keV $c^{-2}$ \\ [2mm]
Electric charge & $e$ & 4.803 204 714 $\times \; 10^{-10}$ statC \\
 & & 1.602 176 634 $\times \; 10^{-19}$ C \\  [2mm]
Angular momentum & $\hbar$ & 1.054 571 818 $\times \; 10^{-27}$ erg s
\\
 & & 6.582 119 570 $\times \; 10^{-16}$ eV s \\  [2mm]
Length & $\hbar^2/\me e^2$ & 5.291 772 109 $\times \;
10^{-9}$ cm \\
[2mm] Time & $\hbar^3/\me e^4$ & 2.418 884 327
$\times \; 10^{-17}$ s
\\ [2mm]
Velocity & $e^2/\hbar$ & 2.187 691 264 $\times \; 10^{8}$ cm s$^{-1}$
\\ [2mm]
Energy & $\me e^4/\hbar^2$ & 4.359 744 722 $\times
\; 10^{-11}$ erg
\\  & & 27.211 386 246 eV \\  [2mm]
Force & $\me^2e^6/\hbar^4$ & 8.238 723 498 $\times \;
10^{-3}$ dyn
\\ [2mm]
Electric potential & $\me e^3/\hbar^2$ & 9.076 741 430 $\times
\; 10^{-2}$ statV \\
& & 27.211 386 246 V \\ [2mm]
Electric field & $\me^2e^5/\hbar^4$ &
1.715 255 541 $\times \; 10^{7}$ statV cm$^{-1}$ \\
& & 5.142 206 748 $\times \; 10^{11}$ V m$^{-1}$ \\ [2mm]
Electric dipole moment & $\hbar^2/\me e$ & 2.541
746 474 $\times \; 10^{-18}$ statC cm \\
& & 8.478 353 625 $\times \; 10^{-30}$ C m
\\ [2mm]
Magnetic induction & $\me^2e^5/\hbar^4$ & 1.715 255
541 $\times \; 10^{7}$ G \\ [2mm]
Magnetic dipole moment & $\hbar^2/\me e$ &
2.541 746 474 $\times \; 10^{-18}$ erg G$^{-1}$ \\
& & 1.586 433 368 $\times \; 10^{-6}$ eV G$^{-1}$
\rule[-2mm]{0mm}{4mm} \\ \hline\hline
\end{tabular} \end{center} \end{table}

\index{Bohr radius}
The atomic unit of length is the radius of the first Bohr orbit,
\begin{subequations}
\label{C.4}
\beq
a_{0} = \frac{\hbar^{2}}{\me e^{2}} = 5.291 \; 772 \; 109 \;
03 \; \times 10^{-9} \; \mbox{cm},
\label{C.4a}\eeq
the atomic unit of mass is the electron mass
\beq
\me = 9.109 \; 383 \; 7015 \; \times 10^{-28} \; \mbox{g},
\label{C.4b}\eeq
and the unit of time is
\beq
t_{0} = \frac{\hbar^{3}}{\me e^{4}} = 2.418 \; 884 \; 326 \;
5857 \; \times 10^{-17} \; \mbox{s}.
\label{C.4c}\eeq
\end{subequations}

\index{Hartree energy}
It is worth noticing that the system of atomic units derives from the
Gaussian system through the replacement of the fundamental quantities
\{length, mass time\} with \{angular momentum, mass, charge\}. With the
aid of the relations \req{C.4}, we can obtain the atomic unit of any
quantity from its dimensionality in the Gaussian system. Thus, the
atomic unit of energy ($[E] = \mathsf{M} \mathsf{L}^{2} \mathsf{T}^{-2}$)
is
\beq
E_{\rm h} = \me \, a_{0}^{2} \, t_{0}^{-2} =
\frac{\me e^{4}}{\hbar^{2}} = 27.211 \; 386 \; 245  \; 988
\; \mbox{eV},
\label{C.5}\eeq
which is called the {\it hartree}. The values of the atomic units of
relevant quantities are given in Table \ref{tabC.2}. A modified system
of atomic units in which the unit of energy is the Rydberg, ${\rm
Ry}=E_{\rm h}/2$, is occasionally used in the literature.

\index{fine-structure constant}
In atomic units, the speed of light in vacuum equals the reciprocal of
the fine-structure constant,
\beq
c = \frac{c}{a_{0}/t_{0}} = \frac{c\hbar}{e^{2}} = \alpha^{-1} \approx
137,
\label{C.6}\eeq
and the Bohr magneton $\mu_{\rm B}$ equals $\alpha/2$.


   \newpage \clearpage

 \addcontentsline{toc}{chapter}{References}
 \renewcommand{\rightmark}{\sl References}
 \renewcommand{\leftmark}


{\sl References}
 \begin{thebibliography}{228}
\newcommand{\enquote}[1]{``#1''}
\providecommand{\natexlab}[1]{#1}

\bibitem[{Abramowitz and Stegun(1972)}]{AbramowitzStegun1974}
Abramowitz, M. and I.~A. Stegun (1972), \emph{{Handbook of Mathematical
  Functions}} (Dover, New York).

\bibitem[{Abril \emph{et~al.}(1998)Abril, Garc\'{\i}a-Molina, Denton,
  P\'{e}rez-P\'{e}rez, and Arista}]{Abril1998}
Abril, I., R.~Garc\'{\i}a-Molina, C.~D. Denton, F.~J. P\'{e}rez-P\'{e}rez, and
  N.~R. Arista (1998), \enquote{{Dielectric description of wakes and stopping
  powers in solids},} \emph{Phys.\ Rev.\ A} \textbf{58}, 357--366.

\bibitem[{Ahlen(1980)}]{Ahlen1980}
Ahlen, S.~P. (1980), \enquote{Theoretical and experimental aspects of the
  energy loss of relativistic heavily ionizing particles,} \emph{Rev.\ Mod.\
  Phys.} \textbf{52}, 121--173.

\bibitem[{Al-Ahmad and Watt(1983)}]{AlAhmad1983}
Al-Ahmad, K.~O. and D.~E. Watt (1983), \enquote{{Stopping powers and
  extrapolated ranges for electrons (1-10 keV) in metals},} \emph{J. Phys.\ D:
  Appl.\ Phys.} \textbf{16}, 2257--2267.

\bibitem[{Arfken(1985)}]{Arfken1985}
Arfken, G. (1985), \emph{Mathematical Methods for Physicists} (Academic Press,
  Inc., San Diego, California), 3rd edition.

\bibitem[{Ashcroft and Mermin(1976)}]{AshcroftMermin1976}
Ashcroft, N.~W. and N.~D. Mermin (1976), \emph{Solid State Physics} (Holt,
  Rinehart and Winston, New York).

\bibitem[{Ashley \emph{et~al.}(1972)Ashley, Ritchie, and Brandt}]{Ashley1972}
Ashley, J., R.~H. Ritchie, and W.~Brandt (1972), \enquote{{$Z_1^3$} effect in
  the stopping power of matter for charged particles,} \emph{Phys.\ Rev.\ B}
  \textbf{5}, 2393--2397.

\bibitem[{Ashley \emph{et~al.}(1973)Ashley, Ritchie, and Brandt}]{Ashley1973}
Ashley, J., R.~H. Ritchie, and W.~Brandt (1973), \enquote{{$Z_1^3$}-dependent
  stopping power and range contributions,} \emph{Phys.\ Rev.\ A} \textbf{8},
  2402--2408.

\bibitem[{Ashley(1991)}]{Ashley1991b}
Ashley, J.~C. (1991), \enquote{Optical-data model for the stopping power of
  condensed matter for protons and antiprotons,} \emph{J.\ Phys.:\ Condens.\
  Matter} \textbf{3}, 2741--2753.

\bibitem[{Baym(1974)}]{Baym1974}
Baym, G. (1974), \emph{Lectures in Quantum Mechanics} (Westview Press, Boulder,
  Colorado).

\bibitem[{Becchetti and Greenlees(1969)}]{BecchettiGreenlees1969}
Becchetti, F.~D. and G.~W. Greenlees (1969), \enquote{{Nucleon-nucleus
  optical-model parameters, $A>40$, $E<50$ MeV},} \emph{Phys.\ Rev.}
  \textbf{182}, 1190--1209.

\bibitem[{Berger(1963)}]{Berger1963}
Berger, M.~J. (1963), \enquote{{Monte Carlo calculation of the penetration and
  diffusion of fast charged particles},} in B.~Alder, S.~Fernbach, and
  M.~Rotenberg (editors), \enquote{{\it Methods in Computational Physics},}
  volume~1, pages 135--215 (Academic Press, New York).

\bibitem[{Berger(1992)}]{Berger1992}
Berger, M.~J. (1992), \emph{{ESTAR, PSTAR and ASTAR: computer programs for
  calculating stopping-power and range tables for electrons, protons and helium
  ions}}, Technical Report NISTIR 4999, National Institute of Standards and
  Technology, Gaithersburg, MD.

\bibitem[{Berger and Bichsel(1994)}]{BergerBichsel1994}
Berger, M.~J. and H.~Bichsel (1994), \enquote{{BEST, BEthe STopping power
  program},} unpublished.

\bibitem[{Berger and Seltzer(1982)}]{BergerSeltzer1982}
Berger, M.~J. and S.~M. Seltzer (1982), \emph{Stopping Power of Electrons and
  Positrons}, Technical Report NBSIR 82-2550, National Bureau of Standards,
  Gaithersburg, MD.

\bibitem[{Bethe(1930)}]{Bethe1930}
Bethe, H.~A. (1930), \enquote{{Zur Theorie des Durchgangs schneller
  Korpuskularstrahlen durch Materie},} \emph{Ann.\ Physik} \textbf{397},
  325--400.

\bibitem[{Bethe(1932)}]{Bethe1932}
Bethe, H.~A. (1932), \enquote{{Bremsformel f\"ur Elektronen relativistischer
  Geschwindigkeit},} \emph{Z. Physik} \textbf{76}, 293--299.

\bibitem[{Bethe(1953)}]{Bethe1953}
Bethe, H.~A. (1953), \enquote{{Moli\`{e}re's} theory of multiple scattering,}
  \emph{Phys.\ Rev.} \textbf{89}, 1256--1266.

\bibitem[{Bethe and Heitler(1934)}]{BetheHeitler1934}
Bethe, H.~A. and W.~Heitler (1934), \enquote{On the stopping of fast particles
  and on the creation of positive electrons,} \emph{Proc.\ R.\ Soc.\ A}
  \textbf{146}, 83--112.

\bibitem[{Bethe and Jackiw(1997)}]{BetheJackiw1997}
Bethe, H.~A. and R.~Jackiw (1997), \emph{Intermediate Quantum Mechanics}
  (Westview Press, Boulder, CO).

\bibitem[{Bethe and Salpeter(1957)}]{BetheSalpeter1957}
Bethe, H.~A. and E.~E. Salpeter (1957), \emph{Quantum Mechanics of One- and
  Two-Electron Atoms} (Springer-Verlag, Berlin).

\bibitem[{Bhabha(1936)}]{Bhabha1936}
Bhabha, H.~J. (1936), \enquote{{The scattering of positrons by electrons with
  exchange on Dirac's theory of electrons},} \emph{Proc.\ Phys.\ Soc.\ A}
  \textbf{154}, 195--196.

\bibitem[{Bichsel(1983)}]{Bichsel1983}
Bichsel, H. (1983), \enquote{{Stopping power of M-shell electrons for heavy
  charged particles},} \emph{Phys.\ Rev.\ A} \textbf{28}, 1147--1150.

\bibitem[{Bichsel(1988)}]{Bichsel1988}
Bichsel, H. (1988), \enquote{Straggling in thin silicon detectors,} \emph{Rev.\
  Mod.\ Phys.} \textbf{60}, 663--699.

\bibitem[{Bichsel(2002)}]{Bichsel2002}
Bichsel, H. (2002), \enquote{{Shell corrections in stopping powers},}
  \emph{Phys.\ Rev.\ A} \textbf{65}, 052709.

\bibitem[{Bichsel and Saxon(1975)}]{BichselSaxon1975}
Bichsel, H. and R.~P. Saxon (1975), \enquote{Comparison of calculational
  methods for straggling in thin absorbers,} \emph{Phys.\ Rev.\ A} \textbf{11},
  1286--1296.

\bibitem[{Bister and Hautala(1978)}]{BisterHautala1978}
Bister, M. and M.~Hautala (1978), \enquote{{Calculation of the Lenz-Jensen
  potential using the third order approximation},} \emph{Phys.\ Lett.}
  \textbf{68A}, 98--100.

\bibitem[{Bloch(1933)}]{Bloch1933}
Bloch, F. (1933), \enquote{{Zur Bremsung rasch bewegter Teilchen beim Durchgang
  durch Materie},} \emph{Ann.\ Phys.\ (Leipzig)} \textbf{16}, 285--320.

\bibitem[{Blunck and Leisegang(1950)}]{BlunckLeisegang1950}
Blunck, O. and S.~Leisegang (1950), \enquote{{Zum Energieverlust schneller
  Elektronen in d{\"{u}}nnen Schichten},} \emph{Z. Physik} \textbf{128},
  500--505.

\bibitem[{Bohr(1913)}]{Bohr1913}
Bohr, N. (1913), \enquote{On the theory of the decrease of velocity of moving
  electrified particles on passing through matter,} \emph{Phil.\ Mag.}
  \textbf{25}, 10--31.

\bibitem[{Bohr(1915)}]{Bohr1915}
Bohr, N. (1915), \enquote{On the decrease of velocity of swiftly moving
  electrified particles in passing through matter,} \emph{Phil.\ Mag.}
  \textbf{30}, 581--612.

\bibitem[{Bohr(1948)}]{Bohr1948}
Bohr, N. (1948), \enquote{The penetration of atomic particles through matter,}
  \emph{K.\ Dan.\ Vidensk.\ Selsk.\ Mat.\ Fys.\ Medd.} \textbf{18}, 1--144.

\bibitem[{Bonderup(1967)}]{Bonderup1967}
Bonderup, E. (1967), \enquote{Stopping of swift protons evaluated from
  statistical atomic model,} \emph{Mat.\ Fys.\ Medd.\ Dan.\ Vid.\ Selsk.}
  \textbf{35}, 1--19.

\bibitem[{Born and Wolf(2002)}]{BornWolf2002}
Born, M. and E.~Wolf (2002), \emph{Principles of Optics} (Cambridge University
  Press, Cambridge), 7th (expanded) edition.

\bibitem[{Bote and Salvat(2008)}]{BoteSalvat2008}
Bote, D. and F.~Salvat (2008), \enquote{{Calculations of inner-shell ionization
  by electron impact with the distorted-wave and plane-wave Born
  approximations},} \emph{Phys.\ Rev.\ A} \textbf{77}, 042701.

\bibitem[{Bransden(1970)}]{Bransden1970}
Bransden, B.~H. (1970), \emph{Atomic Collision Theory} (W.A. Benjamin, Inc.,
  New York).

\bibitem[{Bransden and Joachain(1983)}]{BransdenJoachain1983}
Bransden, B.~H. and C.~J. Joachain (1983), \emph{Physics of Atoms and
  Molecules} (Longman, Essex, England).

\bibitem[{Breit and Bethe(1954)}]{BreitBethe1954}
Breit, G. and H.~A. Bethe (1954), \enquote{Ingoing waves in final state of
  scattering problems,} \emph{Phys.\ Rev.} \textbf{93}, 888--890.

\bibitem[{Bush and Caldwell(1931)}]{BushCaldwell1931}
Bush, V. and S.~H. Caldwell (1931), \enquote{{Thomas-Fermi equation solution by
  the differential analyzer},} \emph{Phys.\ Rev.} \textbf{38}, 1898--1902.

\bibitem[{Byron and Joachain(1977)}]{ByronJoachain1977}
Byron, F.~W. and C.~J. Joachain (1977), \enquote{Elastic scattering of
  electrons and positrons by complex atoms at intermediate energies,}
  \emph{Phys.\ Rev.\ A} \textbf{15}, 128--146.

\bibitem[{Carlson(1975)}]{Carlson1975}
Carlson, T.~A. (1975), \emph{Photoelectron and Auger Spectroscopy} (Plenum
  Press, New York).

\bibitem[{Case and Zweifel(1967)}]{CaseZweifel1967}
Case, K.~M. and L.~F. Zweifel (1967), \emph{Linear Transport Theory}
  (Addison-Wesley, London).

\bibitem[{Cohen(2003)}]{Cohen2003}
Cohen, S.~M. (2003), \enquote{Bethe stopping power theory for heavy-element
  targets and relativistic projectiles,} \emph{Phys.\ Rev.\ A} \textbf{68},
  012720.

\bibitem[{Cohen and Leung(1998)}]{CohenLeung1998}
Cohen, S.~M. and P.~T. Leung (1998), \enquote{General formulation of the
  semirelativistic approach to atomic sum rules,} \emph{Phys.\ Rev.\ A}
  \textbf{57}, 4994--4997.

\bibitem[{Colgan \emph{et~al.}(2006)Colgan, Fontes, and Zhang}]{Colgan2006}
Colgan, J., C.~J. Fontes, and H.~L. Zhang (2006), \enquote{Inner-shell
  electron-impact ionization of neutral atoms,} \emph{Phys.\ Rev.\ A}
  \textbf{73}, 062711.

\bibitem[{Condon(1930)}]{Condon1930}
Condon, E.~U. (1930), \enquote{The theory of complex spectra,} \emph{Phys.\
  Rev.} \textbf{36}, 1121--1133.

\bibitem[{Condon and Odaba\c{s}i(1980)}]{CondonOdabasi1980}
Condon, E.~U. and H.~Odaba\c{s}i (1980), \emph{Atomic Structure} (Cambridge
  University Press, Cambridge, UK).

\bibitem[{Coursey \emph{et~al.}(2015)Coursey, Schwab, Tsai, and
  Dra}]{Coursey2015}
Coursey, J.~S., D.~J. Schwab, J.~J. Tsai, and R.~A. Dra (2015),
  \enquote{{Atomic and isotopic compositions for all elements, {\rm NIST
  Standard Reference Database 144}},} National Institute of Standards and
  Technology, Gaithersburg, MD, available from
  \url{www.nist.gov/srd/chemistry}.

\bibitem[{{Cram\'{e}r}(1962)}]{Cramer1962}
{Cram\'{e}r}, H. (1962), \emph{Mathematical Methods of Statistics} (Asia
  Publishing House, Bombay).

\bibitem[{Da \emph{et~al.}(2014)Da, Shinotsuka, Yoshikawa, Ding, and
  Tanuma}]{Da2014}
Da, B., H.~Shinotsuka, H.~Yoshikawa, Z.~Ding, and S.~Tanuma (2014),
  \enquote{{Extended Mermin method for calculating the electron inelastic mean
  free path},} \emph{Phys.\ Rev.\ Lett.} \textbf{113}, 063201.

\bibitem[{Dennery and Krzywicki(1996)}]{DenneryKrzywicki1996}
Dennery, P. and A.~Krzywicki (1996), \emph{Mathematics for Physicists} (Dover
  Publications, Mineola, NY).

\bibitem[{Desclaux(1975)}]{Desclaux1975}
Desclaux, J.~P. (1975), \enquote{{A multiconfiguration relativistic Dirac-Fock
  program},} \emph{Comput.\ Phys.\ Commun.} \textbf{9}, 31--45, {Erratum
  (1977), {\it Ibidem}, {\bf 13}, 71}.

\bibitem[{DeVries(1994)}]{DeVries1994}
DeVries, P.~L. (1994), \emph{A First Course in Computational Physics} (John
  Wiley and Sons, Inc., New York).

\bibitem[{Echenique \emph{et~al.}(1986)Echenique, Nieminen, Ashley, and
  Ritchie}]{Echenique1986}
Echenique, P.~M., R.~M. Nieminen, J.~C. Ashley, and R.~H. Ritchie (1986),
  \enquote{{Nonlinear stopping power of an electron gas for slow ions},}
  \emph{Phys.\ Rev.\ A} \textbf{33}, 897--904.

\bibitem[{Echenique \emph{et~al.}(1979)Echenique, Ritchie, and
  Brandt}]{Echenique1979}
Echenique, P.~M., R.~H. Ritchie, and W.~Brandt (1979), \enquote{{Spatial
  excitation patterns induced by swift ions in condensed matter},} \emph{Phys.\
  Rev.\ B} \textbf{20}, 2567--2580.

\bibitem[{Edmonds(1960)}]{Edmonds1960}
Edmonds, A.~R. (1960), \emph{Angular Momentum in Quantum Mechanics} (Princeton
  University Press, Princeton, NJ).

\bibitem[{Egerton(2009)}]{Egerton2009}
Egerton, R.~F. (2009), \enquote{{Electron energy-loss spectroscopy in the
  TEM},} \emph{Rep.\ Prog.\ Phys.} \textbf{72}, 016502.

\bibitem[{Egerton(2011)}]{Egerton2011}
Egerton, R.~F. (2011), \emph{{Electron Energy-loss Spectroscopy in the Electron
  Microscope}} (Springer, New York), 3 edition.

\bibitem[{{Erd\'{e}lyi}(1954)}]{Erdelyi1954}
{Erd\'{e}lyi}, A. (1954), \emph{Tables of Integral Transforms} (McGraw-Hill
  Book Co., New York).

\bibitem[{Everhart \emph{et~al.}(1955)Everhart, Stone, and
  Carbone}]{Everhart1955}
Everhart, E., G.~Stone, and R.~J. Carbone (1955), \enquote{{Classical
  calculation of differential cross section for scattering from a Coulomb
  potential with exponential screening},} \emph{Phys.\ Rev.} \textbf{99},
  1287--1290.

\bibitem[{Eyges(1948)}]{Eyges1948}
Eyges, L. (1948), \enquote{Multiple scattering with energy loss,} \emph{Phys.\
  Rev.} \textbf{74}, 1534--1535.

\bibitem[{Fano(1956{\natexlab{a}})}]{Fano1956}
Fano, U. (1956{\natexlab{a}}), \enquote{Atomic theory of electromagnetic
  interactions in dense materials,} \emph{Phys.\ Rev.} \textbf{103},
  1202--1218.

\bibitem[{Fano(1956{\natexlab{b}})}]{Fano1956b}
Fano, U. (1956{\natexlab{b}}), \enquote{Differential inelastic scattering of
  relativistic charged particles,} \emph{Phys.\ Rev.} \textbf{103}, 385--387.

\bibitem[{Fano(1963)}]{Fano1963}
Fano, U. (1963), \enquote{Penetration of protons, alpha particles and mesons,}
  \emph{Ann.\ Rev.\ Nucl.\ Sci.} \textbf{13}, 1--66.

\bibitem[{Fathers and Rez(1984)}]{FathersRez1984}
Fathers, D. and P.~Rez (1984), \enquote{{A transport theory of electron
  scattering in solids},} in D.~Kyser, H.~Niedrig, D.~Newbury, and R.~Shimizu
  (editors), \enquote{{Electron Beam Interactions with Solid},} pages 193--208
  ({Scanning Electron Microscopy, Inc.}).

\bibitem[{Fermi(1940)}]{Fermi1940}
Fermi, E. (1940), \enquote{The ionization loss of energy in gases and in
  condensed materials,} \emph{Phys.\ Rev.} \textbf{57}, 485--493.

\bibitem[{Fern{\'a}ndez-Varea
  \emph{et~al.}(1993{\natexlab{a}})Fern{\'a}ndez-Varea, Mayol, Bar\'{o}, and
  Salvat}]{FernandezVarea1993}
Fern{\'a}ndez-Varea, J.~M., R.~Mayol, J.~Bar\'{o}, and F.~Salvat
  (1993{\natexlab{a}}), \enquote{On the theory and simulation of multiple
  elastic scattering of electrons,} \emph{Nucl.\ Instrum.\ Meth.\ B}
  \textbf{73}, 447--473.

\bibitem[{Fern{\'a}ndez-Varea
  \emph{et~al.}(1993{\natexlab{b}})Fern{\'a}ndez-Varea, Mayol, Liljequist, and
  Salvat}]{FernandezVarea1993b}
Fern{\'a}ndez-Varea, J.~M., R.~Mayol, D.~Liljequist, and F.~Salvat
  (1993{\natexlab{b}}), \enquote{Inelastic scattering of electrons in solids
  from a generalized oscillator strength model using optical and photoelectric
  data,} \emph{J.\ Phys.: Condens.\ Matter} \textbf{5}, 3593--3610.

\bibitem[{Fern{\'a}ndez-Varea \emph{et~al.}(2005)Fern{\'a}ndez-Varea, Salvat,
  Dingfelder, and Liljequist}]{FernandezVarea2005}
Fern{\'a}ndez-Varea, J.~M., F.~Salvat, M.~Dingfelder, and D.~Liljequist (2005),
  \enquote{A relativistic optical-data model for inelastic scattering of
  electrons and positrons in condensed matter,} \emph{Nucl.\ Instrum.\ Meth.\
  B} \textbf{229}, 187--218.

\bibitem[{Feynman \emph{et~al.}(1949)Feynman, Metropolis, and
  Teller}]{Feynman1949}
Feynman, R., N.~Metropolis, and E.~Teller (1949), \enquote{{Equations of state
  based on the generalized Fermi--Thomas theory},} \emph{Phys.\ Rev.}
  \textbf{75}, 1561--1573.

\bibitem[{Furry(1951)}]{Furry1951}
Furry, W.~H. (1951), \enquote{On bound states and scattering in positron
  theory,} \emph{Phys.\ Rev.} \textbf{81}, 115--124.

\bibitem[{Garber \emph{et~al.}(1971)Garber, Nakai, Harter, and
  Birkhoff}]{Garber1971}
Garber, F.~W., M.~Y. Nakai, J.~A. Harter, and R.~D. Birkhoff (1971),
  \enquote{{Low-energy electron beam studies in thin aluminum foils},}
  \emph{J.\ Appl.\ Phys.} \textbf{42}, 1149.

\bibitem[{Gervais(1991)}]{Gervais1991}
Gervais, F. (1991), \enquote{{Aluminium oxide (Al$_2$O$_3$)},} in E.~D. Palik
  (editor), \enquote{Handbook of Optical Constants of Solids,} volume~2, pages
  761--776 (Academic Press, San Diego, CA).

\bibitem[{Goldberger and Watson(1964)}]{GoldbergerWatson1964}
Goldberger, M.~L. and K.~M. Watson (1964), \emph{Collision Theory} (John Wiley
  and Sons, New York).

\bibitem[{Goldstein(1980)}]{Goldstein1980}
Goldstein, H. (1980), \emph{Classical Mechanics} (Addison-Wesley, Reading, MA).

\bibitem[{Goudsmit and Saunderson(1940{\natexlab{a}})}]{GoudsmitSaunderson1940}
Goudsmit, S. and J.~L. Saunderson (1940{\natexlab{a}}), \enquote{Multiple
  scattering of electrons,} \emph{Phys.\ Rev.} \textbf{57}, 24--29.

\bibitem[{Goudsmit and
  Saunderson(1940{\natexlab{b}})}]{GoudsmitSaunderson1940b}
Goudsmit, S. and J.~L. Saunderson (1940{\natexlab{b}}), \enquote{{Multiple
  scattering of electrons. II},} \emph{Phys.\ Rev.} \textbf{58}, 36--42.

\bibitem[{Gradshteyn and Ryzhik(2007)}]{GradshteynRyzhik2007}
Gradshteyn, I.~S. and I.~M. Ryzhik (2007), \emph{Table of Integrals, Series,
  and Products} (Elsevier - Academic Press, London), 3rd edition.

\bibitem[{Grant(1970)}]{Grant1970}
Grant, I.~P. (1970), \enquote{{Relativistic calculations of atomic
  structures},} \emph{Advances in Physics} \textbf{19}, 747--811.

\bibitem[{Griffiths(1995)}]{Griffiths1995}
Griffiths, D. (1995), \emph{Introduction to Quantum Mechanics} (Prentice Hall,
  Inc., Upper Saddle River, NJ).

\bibitem[{Groom \emph{et~al.}(2001)Groom, Mokhov, and Striganov}]{Groom2001}
Groom, D.~E., N.~V. Mokhov, and S.~I. Striganov (2001), \enquote{{Muon Stopping
  power and range tables 10 MeV-100 TeV},} \emph{At.\ Data Nucl.\ Data Tables}
  \textbf{78}, 183--356.

\bibitem[{Gryzinski(1957)}]{Gryzinski1957}
Gryzinski, M. (1957), \enquote{Stopping power of a medium for heavy, charged
  particles,} \emph{Phys.\ Rev.} \textbf{107}, 1471--1475.

\bibitem[{Gryzinski(1959)}]{Gryzinski1959}
Gryzinski, M. (1959), \enquote{Classical theory of electronic and ionic
  inelastic collisions,} \emph{Phys.\ Rev.} \textbf{115}, 374--383.

\bibitem[{Hahn \emph{et~al.}(1956)Hahn, Ravenhall, and Hofstadter}]{Hahn1956}
Hahn, B., D.~G. Ravenhall, and R.~Hofstadter (1956), \enquote{High-energy
  electron scattering,} \emph{Phys.\ Rev.} \textbf{101}, 1131--1142.

\bibitem[{Hanson \emph{et~al.}(1951)Hanson, Lanzl, Lyman, and
  Scott}]{Hanson1951}
Hanson, A.~O., L.~H. Lanzl, E.~M. Lyman, and M.~B. Scott (1951),
  \enquote{Measurement of multiple scattering of 15.7-{MeV} electrons,}
  \emph{Phys.\ Rev.} \textbf{84}, 634--637.

\bibitem[{Hartree(1957)}]{Hartree1957}
Hartree, D.~R. (1957), \emph{The Calculation of Atomic Structures} (John Wiley
  and Sons, New York).

\bibitem[{Haug(1975)}]{Haug1975}
Haug, E. (1975), \enquote{Bremsstrahlung and pair production in the field of
  free electrons,} \emph{Z. Naturforsch.} \textbf{30a}, 1099--1113.

\bibitem[{Haug and Nakel(2004)}]{HaugNakel2004}
Haug, E. and W.~Nakel (2004), \emph{{The Elementary Process of Bremsstrahlung}}
  (World Scientific, Singapore).

\bibitem[{Henke \emph{et~al.}(1993)Henke, Gullikson, and Davis}]{Henke1993}
Henke, B.~L., E.~M. Gullikson, and J.~C. Davis (1993), \enquote{{X-ray
  interactions: photoabsorption, scattering, transmission, and reflection at
  {$E=$ 50--30,000 eV, $Z=$ 1--92}},} \emph{At.\ Data Nucl.\ Data Tables}
  \textbf{54}, 181--342.

\bibitem[{Henke \emph{et~al.}(2010)Henke, Gullikson, and Davis}]{Henke2010}
Henke, B.~L., E.~M. Gullikson, and J.~C. Davis (2010), \enquote{{Low-energy
  X-ray Interaction Coefficients: Photoabsorption, Scattering, and Reflection,
  $E=$ 30--30,000 eV, $Z=$ 1--92},} Lawrence Berkeley Laboratory, Berkeley, CA,
  available from \url{https://github.com/praxes/henke_physical_reference}.

\bibitem[{Heras(2011)}]{Heras2011}
Heras, J.~A. (2011), \enquote{{A short proof that the Coulomb-gauge potentials
  yield the retarded fields},} \emph{Eur.\ J.\ Phys.} \textbf{32}, 213--216.

\bibitem[{Heredia-Avalos \emph{et~al.}(2005)Heredia-Avalos, Garcia-Molina,
  Fern{\'a}ndez-Varea, and Abril}]{Heredia-Avalos2005}
Heredia-Avalos, S., R.~Garcia-Molina, J.~M. Fern{\'a}ndez-Varea, and I.~Abril
  (2005), \enquote{{Calculated energy loss of swift He, Li, B, and N ions in
  SiO$_2$, Al$_2$O$_3$, and ZrO$_2$},} \emph{Phys.\ Rev.\ A} \textbf{72},
  052902.

\bibitem[{Hodgson(1971)}]{Hodgson1971}
Hodgson, P.~E. (1971), \enquote{The nuclear optical model,} \emph{Rep.\ Prog.\
  Phys.} \textbf{34}, 765--819.

\bibitem[{Hoerni and Ibers(1953)}]{HoerniIbers1953}
Hoerni, J.~A. and J.~A. Ibers (1953), \enquote{{Complex amplitudes for electron
  scattering by atoms},} \emph{Phys.\ Rev.} \textbf{91}, 1182--1185.

\bibitem[{{ICRU Report 37}(1984)}]{ICRU37}
{ICRU Report 37} (1984), \emph{Stopping Powers for Electrons and Positrons}
  (ICRU, Bethesda, MD).

\bibitem[{{ICRU Report 49}(1993)}]{ICRU49}
{ICRU Report 49} (1993), \emph{Stopping Powers and Ranges for Protons and Alpha
  Particles} (ICRU, Bethesda, MD).

\bibitem[{{ICRU Report 77}(2007)}]{ICRU77}
{ICRU Report 77} (2007), \emph{Elastic Scattering of Electrons and Positrons}
  (ICRU, Bethesda, MD).

\bibitem[{{ICRU Report 85}(2011)}]{ICRU85}
{ICRU Report 85} (2011), \emph{Fundamental Quantities and Units for Ionizing
  Radiation} (ICRU, Bethesda, MD).

\bibitem[{{ICRU Report 90}(2016)}]{ICRU90}
{ICRU Report 90} (2016), \emph{Key Data for Ionizing-Radiation Dosimetry:
  Measurement Standards and Applications} (ICRU, Bethesda, MD).

\bibitem[{Inokuti(1971)}]{Inokuti1971}
Inokuti, M. (1971), \enquote{{Inelastic collisions of fast charged particles
  with atoms and molecules --- {T}he {B}ethe theory revisited},} \emph{Rev.\
  Mod.\ Phys.} \textbf{43}, 297--347.

\bibitem[{Inokuti and Smith(1982)}]{InokutiSmith1982}
Inokuti, M. and D.~Y. Smith (1982), \enquote{Fermi density effect on the
  stopping power of metallic aluminum,} \emph{Phys.\ Rev.} \textbf{25}, 61--66.

\bibitem[{Jackson(1975)}]{Jackson1975}
Jackson, J.~D. (1975), \emph{Classical Electrodynamics} (John Wiley and Sons,
  New York), 2nd edition.

\bibitem[{Jackson and McCarthy(1972)}]{JacksonMcCarthy1972}
Jackson, J.~D. and R.~L. McCarthy (1972), \enquote{{$Z_1^3$} corrections to
  energy loss and range,} \emph{Phys.\ Rev.\ B} \textbf{6}, 4131--4141.

\bibitem[{Jancovici(1962)}]{Jancovici1962}
Jancovici, B. (1962), \enquote{On the relativistic degenerate eelctron gas,}
  \emph{Il Nuovo Cimento} \textbf{25}, 428--455.

\bibitem[{Jenkins \emph{et~al.}(1988)Jenkins, Nelson, and Rindi}]{Jenkins1988}
Jenkins, T.~M., W.~R. Nelson, and A.~Rindi (1988), \emph{Monte Carlo Transport
  of Electrons and Photons} (Plenum, New York).

\bibitem[{Jensen(1932)}]{Jensen1932}
Jensen, H. (1932), \enquote{{Die Ladungsverteilung in Ionen und die
  Gitterkonstante des Rubidiumbromids nach der statistischen Methode},}
  \emph{{Zeitschrift f\"{u}r Physik}} \textbf{77}, 722--745.

\bibitem[{Joachain(1975)}]{Joachain1975}
Joachain, C.~J. (1975), \emph{Quantum Collision Theory} (North Holland,
  Amsterdam).

\bibitem[{Johnson and Inokuti(1983)}]{JohnsonInokuti1983}
Johnson, R.~E. and M.~Inokuti (1983), \enquote{The local-plasma approximation
  to the oscillator-strength spectrum: How good is it and why?} \emph{Comments
  At.\ Mol.\ Phys.} \textbf{14}, 19--31.

\bibitem[{Joy(2008)}]{Joy2008}
Joy, D.~C. (2008), \emph{{A database of electron-solid interactions}},
  Technical report, University of Tennessee, (unpublished).

\bibitem[{Kawrakow and Bielajew(1998)}]{KawrakowBielajew1998}
Kawrakow, I. and A.~F. Bielajew (1998), \enquote{On the condensed history
  technique for electron transport,} \emph{Nucl.\ Instrum.\ Meth.\ B}
  \textbf{142}, 253--280.

\bibitem[{Khandelwal and Merzbacher(1966)}]{KhandelwalMerzbacher1966}
Khandelwal, G.~S. and E.~Merzbacher (1966), \enquote{Stopping power of {M}
  electrons,} \emph{Phys.\ Rev.} \textbf{144}, 349--352.

\bibitem[{Kim \emph{et~al.}(1986)Kim, Pratt, Seltzer, and Berger}]{Kim1986}
Kim, L., R.~H. Pratt, S.~M. Seltzer, and M.~J. Berger (1986), \enquote{Ratio of
  positron to electron bremsstrahlung energy loss: an approximate scaling law,}
  \emph{Phys.\ Rev.\ A} \textbf{33}, 3002--3009.

\bibitem[{Kittel(1976)}]{Kittel1976}
Kittel, C. (1976), \emph{Introduction to Solid State Physics} (John Wiley and
  Sons, New York).

\bibitem[{Kliewer and Fuchs(1969)}]{KliewerFuchs1969}
Kliewer, K.~L. and R.~D. Fuchs (1969), \enquote{Lindhard dielectric functions
  with a finite electron lifetime,} \emph{Phys.\ Rev.} \textbf{181}, 552--558.

\bibitem[{Kohn and Sham(1965)}]{KohnSham1965}
Kohn, W. and L.~J. Sham (1965), \enquote{Self-consistent equations including
  exchange and correlation effects,} \emph{Phys.\ Rev.} \textbf{140},
  A1133--A1138.

\bibitem[{Koning and Delaroche(2003)}]{KoningDelaroche2003}
Koning, A. and J.~Delaroche (2003), \enquote{{Local and global nucleon optical
  models from 1 keV to 200 MeV},} \emph{Nucl.\ Phys.\ A} \textbf{713},
  231--310.

\bibitem[{Landau(1944)}]{Landau1944}
Landau, L.~D. (1944), \enquote{On the energy loss of fast particles by
  ionization,} \emph{Journal of Physics-USSR} \textbf{8}, 201--205.

\bibitem[{Landau and Lifshitz(1971)}]{LandauLifshitz1971}
Landau, L.~D. and E.~M. Lifshitz (1971), \emph{{The Classical Theory of
  Fields}} (Pergamon Press, Oxford), {3rd revised English} edition.

\bibitem[{Langer(1937)}]{Langer1937}
Langer, R.~E. (1937), \enquote{On the connection formulas and the solutions of
  the wave equation,} \emph{Phys.\ Rev.} \textbf{51}, 669--676.

\bibitem[{Latter(1955)}]{Latter1955}
Latter, R. (1955), \enquote{{Atomic energy levels for the Thomas--Fermi and
  Thomas--Fermi--Dirac potential},} \emph{Phys.\ Rev.} \textbf{99}, 510--519.

\bibitem[{Lenz(1932)}]{Lenz1932}
Lenz, W. (1932), \enquote{{\"{U}ber die Anwendbarkeit der Statistischen Methode
  auf Ionengitter},} \emph{{Zeitschrift f\"{u}r Physik}} \textbf{77}, 713--721.

\bibitem[{Levine and Louie(1982)}]{LevineLouie1982}
Levine, Z.~H. and S.~G. Louie (1982), \enquote{{New model dielectric function
  and exchange-correlation potential for semiconductors and insulators},}
  \emph{Phys.\ Rev.\ B} \textbf{25}, 6310--6316.

\bibitem[{Levinger \emph{et~al.}(1957)Levinger, Rustgi, and
  Okamoto}]{Levinger1957}
Levinger, J.~S., M.~L. Rustgi, and K.~Okamoto (1957), \enquote{{Relativistic
  corrections to the dipole sum rule},} \emph{Phys.\ Rev.} \textbf{106},
  1191--1194.

\bibitem[{Lewis(1950)}]{Lewis1950}
Lewis, H.~W. (1950), \enquote{Multiple scattering in an infinite medium,}
  \emph{Phys.\ Rev.} \textbf{78}, 526--529.

\bibitem[{Liberman \emph{et~al.}(1971)Liberman, Cromer, and
  Waber}]{Liberman1971}
Liberman, D., D.~T. Cromer, and J.~T. Waber (1971), \enquote{Relativistic
  self-consistent field program for atoms and ions,} \emph{Comput.\ Phys.\
  Commun.} \textbf{2}, 107--113.

\bibitem[{Liberman \emph{et~al.}(1965)Liberman, Cromer, and
  Waber}]{Liberman1965}
Liberman, D.~A., D.~T. Cromer, and J.~T. Waber (1965),
  \enquote{{Self--Consistent--field Dirac wave functions for atoms and ions. I.
  Comparison with previous calculations},} \emph{Phys.\ Rev.} \textbf{137},
  A27--A34.

\bibitem[{Liljequist(1983)}]{Liljequist1983}
Liljequist, D. (1983), \enquote{{A simple calculation of inelastic mean free
  path and stopping power for 50 eV -- 50 keV electrons in solids},} \emph{J.\
  Phys.\ D:\ Appl.\ Phys.} \textbf{16}, 1567--1582.

\bibitem[{Lindhard(1954)}]{Lindhard1954}
Lindhard, J. (1954), \enquote{On the properties of a gas of charged particles,}
  \emph{Dan.\ Mat.\ Fys.\ Medd.} \textbf{28}, 1--57.

\bibitem[{Lindhard(1976)}]{Lindhard1976}
Lindhard, J. (1976), \enquote{{The Barkas effect -- or $Z_1^3$,
  $Z_1^4$-corrections to stopping of swift vharged particles},} \emph{Nucl.\
  Instrum.\ Meth.} \textbf{132}, 1--5.

\bibitem[{Lindhard and Scharff(1953)}]{LindhardScharff1953}
Lindhard, J. and M.~Scharff (1953), \enquote{Energy loss in matter by charged
  particles of low charge,} \emph{Dan.\ Mat.\ Fys.\ Medd.} \textbf{27}, 1--31.

\bibitem[{Lindhard and {S{\o}rensen}(1996)}]{LindhardSorensen1996}
Lindhard, J. and A.~H. {S{\o}rensen} (1996), \enquote{Relativistic theory of
  stopping for heavy ions,} \emph{Phys.\ Rev.\ A} \textbf{53}, 2443--2456.

\bibitem[{Llovet \emph{et~al.}(2014)Llovet, Powell, Jablonski, and
  Salvat}]{Llovet2014}
Llovet, X., C.~J. Powell, A.~Jablonski, and F.~Salvat (2014), \enquote{Cross
  sections for inner-shell ionization by electron impact,} \emph{J.\ Phys.\
  Chem.\ Ref.\ Data} \textbf{43}, 013102.

\bibitem[{Lorentz(1909)}]{Lorentz1909}
Lorentz, H.~A. (1909), \emph{The theory of electrons} (Columbia University
  Press, New York).

\bibitem[{Luo \emph{et~al.}(1991)Luo, Zhang, and Joy}]{Luo1991}
Luo, S., X.~Zhang, and D.~C. Joy (1991), \enquote{{Experimental determinations
  of electron stopping power at low energies},} \emph{Radiation Effects and
  Defects in Solids} \textbf{117}, 235--242.

\bibitem[{MacPherson(1998)}]{MacPherson1998}
MacPherson, M.~S. (1998), \emph{Accurate measurements of the collision stopping
  powers for 5 to 30 MeV electrons}, Ph.D. thesis, Carleton University, Ottawa,
  Ontario, ({Also available as document PIRS-0626, National Research Council,
  Canada}).

\bibitem[{Manson(1972)}]{Manson1972}
Manson, S.~T. (1972), \enquote{{Inelastic collisions of fast charged particles
  with atoms: ionization of the aluminum L shell},} \emph{Phys.\ Rev.\ A}
  \textbf{6}, 1013--1024.

\bibitem[{Maron(1982)}]{Maron1982}
Maron, M.~J. (1982), \emph{Numerical Analysis: A Practical Approach}
  (Macmillan, New York).

\bibitem[{Martin(2004)}]{Martin2004}
Martin, R.~M. (2004), \emph{{Electronic Structure: Basic Theory and Practical
  Methods}} (Cambridge Univ. Press, Cambridge, UK).

\bibitem[{McKinley and Feshbach(1948)}]{McKinleyFeshbach1948}
McKinley, W.~A. and H.~Feshbach (1948), \enquote{{The Coulomb scattering of
  relativistic electrons by nuclei},} \emph{Phys.\ Rev.} \textbf{74},
  1759--1763.

\bibitem[{Mermin(1970)}]{Mermin1970}
Mermin, N.~D. (1970), \enquote{Lindhard dielectric function in the
  relaxation-time approximation,} \emph{Phys.\ Rev.\ B} \textbf{1}, 2362--2363.

\bibitem[{Merzbacher(1970)}]{Merzbacher1970}
Merzbacher, E. (1970), \emph{Quantum Mechanics} (John Wiley and Sons, New
  York), 3rd edition.

\bibitem[{Messiah(1999)}]{Messiah1999}
Messiah, A. (1999), \emph{Quantum Mechanics} (Dover Publications Inc., New
  York).

\bibitem[{{M\o ller}(1932)}]{Moller1932}
{M\o ller}, C. (1932), \enquote{{Zur Theorie des Durchgangs schneller
  Elektronen durch Materie},} \emph{Ann. Physik} \textbf{14}, 531--585.

\bibitem[{Mohr \emph{et~al.}(2016)Mohr, Newell, and Taylor}]{Mohr2016}
Mohr, P.~J., D.~B. Newell, and B.~N. Taylor (2016), \enquote{{CODATA}
  recommended values of the fundamental physical constants: 2014,} \emph{Rev.\
  Mod.\ Phys.} \textbf{88}, 035009.

\bibitem[{Moli\`{e}re(1947)}]{Moliere1947}
Moli\`{e}re, G. (1947), \enquote{{Theorie der Streuung schneller geladener
  Teilchen I: Einzelstreuung am abgeschirmten Coulomb-Feld},} \emph{Z.
  Naturforsch.} \textbf{2a}, 133--145.

\bibitem[{Moli\`{e}re(1948)}]{Moliere1948}
Moli\`{e}re, G. (1948), \enquote{{Theorie der Streuung schneller geladener
  Teilchen II: Mehrfach- und Vielfachstreuung},} \emph{Z. Naturforsch.}
  \textbf{3a}, 78--97.

\bibitem[{Montanari and Dimitriou(2017)}]{Montanari2017}
Montanari, C.~C. and P.~Dimitriou (2017), \enquote{{The IAEA stopping power
  database, following the trends in stopping power of ions in matter},}
  \emph{Nucl.\ Instrum.\ Meth.\ B} \textbf{408}, 50--55.

\bibitem[{Morse(1932)}]{Morse1932}
Morse, P.~M. (1932), \enquote{{Unelastische Streuung von Kathodenstrahlen},}
  \emph{Physik.\ Zeitschr.} \textbf{33}, 443--445.

\bibitem[{Mott(1929)}]{Mott1929}
Mott, N.~F. (1929), \enquote{The scattering of fast electrons by atomic
  nuclei,} \emph{Proc.\ Roy.\ Soc.\ London A} \textbf{124}, 425--442.

\bibitem[{Mott and Massey(1965)}]{MottMassey1965}
Mott, N.~F. and H.~S.~W. Massey (1965), \emph{The Theory of Atomic Collisions}
  (Oxford University Press, London).

\bibitem[{Negreanu \emph{et~al.}(2005)Negreanu, Llovet, Chawla, and
  Salvat}]{Negreanu2005}
Negreanu, C., X.~Llovet, R.~Chawla, and F.~Salvat (2005), \enquote{Calculation
  of multiple-scattering angular distributions of electrons and positrons,}
  \emph{Radiat.\ Phys.\ Chem.} \textbf{74}, 264--281.

\bibitem[{Nelder and Mead(1965)}]{NelderMead1965}
Nelder, J. and R.~Mead (1965), \enquote{A simplex method for function
  minimization,} \emph{The Computer Journal} \textbf{7}, 308--313.

\bibitem[{Ochkur(1964)}]{Ochkur1964}
Ochkur, V.~I. (1964), \enquote{The {B}orn-{O}ppenheimer method in the theory of
  atomic collisions,} \emph{Sov.\ Phys.\ JETP} \textbf{18}, 503--508.

\bibitem[{Ochkur(1965)}]{Ochkur1965}
Ochkur, V.~I. (1965), \enquote{Ionization of the hydrogen atom by electron
  impact with allowance for the exchange,} \emph{Sov.\ Phys.\ JETP}
  \textbf{20}, 1175--1178.

\bibitem[{Olver \emph{et~al.}(2010)Olver, Lozier, Boisvert, and
  Clark}]{Olver2010}
Olver, F., D.~Lozier, R.~Boisvert, and C.~Clark (2010), \emph{{NIST Handbook of
  Mathematical Functions}} (Cambridge University Press, New York), print
  companion to the NIST Digital Library of Mathematical Functions (DLMF),
  \url{http://dlmf.nist.gov/}.

\bibitem[{Palik(1985)}]{Palik1985}
Palik, E.~D. (editor) (1985), \emph{Handbook of Optical Constants of Solids}
  (Academic Press, San Diego, CA).

\bibitem[{Palik(1991)}]{Palik1991}
Palik, E.~D. (editor) (1991), \emph{Handbook of Optical Constants of Solids
  {II}} (Academic Press, San Diego, CA).

\bibitem[{Palik(1998)}]{Palik1998}
Palik, E.~D. (editor) (1998), \emph{Handbook of Optical Constants of Solids
  {III}} (Academic Press, San Diego, CA).

\bibitem[{Parzen(1950)}]{Parzen1950}
Parzen, G. (1950), \enquote{{On the scattering theory of the Dirac equation},}
  \emph{Phys.\ Rev.} \textbf{80}, 261--268.

\bibitem[{Penn(1987)}]{Penn1987}
Penn, D.~R. (1987), \enquote{Electron mean-free-path calculations using a model
  dielectric function,} \emph{Phys.\ Rev.\ B} \textbf{35}, 482--486.

\bibitem[{Pines and Nozi{\`e}res(1989)}]{PinesNozieres1989}
Pines, D. and P.~Nozi{\`e}res (1989), \emph{Theory of Quantum Liquids}
  (Addison-Wesley Pub. Co., Advanced Book Program, Reading, Mass.).

\bibitem[{Pratt \emph{et~al.}(1973)Pratt, Ron, and Tseng}]{Pratt1973}
Pratt, R.~H., A.~Ron, and H.~K. Tseng (1973), \enquote{{Atomic photoelectric
  effect above 10 keV},} \emph{Rev.\ Mod.\ Phys.} \textbf{45}, 273--325,
  {Erratum (1973), {\it Ibidem}, {\bf 45}, 663--664}.

\bibitem[{Pratt \emph{et~al.}(1977)Pratt, Tseng, Lee, and Kissel}]{Pratt1977}
Pratt, R.~H., H.~K. Tseng, C.~M. Lee, and L.~Kissel (1977),
  \enquote{{Bremsstrahlung energy spectra from electrons of kinetic energy 1
  keV {$\le T_{1}\le$} 2000 keV incident on neutral atoms 2 {$\le Z\le$} 92},}
  \emph{At.\ Data Nucl.\ Data Tables} \textbf{20}, 175--209, {Erratum (1981),
  {\it Ibidem}, {\bf 26}, 477--481}.

\bibitem[{Press \emph{et~al.}(1992)Press, Teukolski, Vetterling, and
  Flannery}]{Press1992}
Press, W.~H., S.~A. Teukolski, W.~Vetterling, and B.~Flannery (1992),
  \emph{Numerical Recipes in Fortran 77. The art of Scientific Computing}
  (Cambridge University Press, New York), 2nd edition.

\bibitem[{Rao-Sahib and Wittry(1974)}]{Rao-SahibWittry1974}
Rao-Sahib, T.~S. and D.~B. Wittry (1974), \enquote{X-ray continuum from thick
  elemental targets for 10-50 kev electrons,} \emph{J. Appl.\ Phys.}
  \textbf{45}, 5060--5068.

\bibitem[{Rehr and Albers(2000)}]{RehrAlbers2000}
Rehr, J.~J. and R.~C. Albers (2000), \enquote{Theoretical approaches to x-ray
  absorption fine structure,} \emph{Rev.\ Mod.\ Phys.} \textbf{72}, 621--654.

\bibitem[{Ritchie(1959)}]{Ritchie1959}
Ritchie, R.~H. (1959), \enquote{{Interaction of charged particles with a
  degenerate Fermi-Dirac electron gas},} \emph{Phys.\ Rev.} \textbf{114},
  644--654.

\bibitem[{Ritchie \emph{et~al.}(1976)Ritchie, Brandt, and
  Echenique}]{Ritchie1976}
Ritchie, R.~H., W.~Brandt, and P.~M. Echenique (1976), \enquote{{Wake potential
  of swift ions in solids},} \emph{Phys.\ Rev.\ B} \textbf{14}, 4808--4812.

\bibitem[{Rohrlich and Carlson(1954)}]{RohrlichCarlson1954}
Rohrlich, F. and B.~C. Carlson (1954), \enquote{Positron-electron differences
  in energy loss and multiple scattering,} \emph{Phys.\ Rev.} \textbf{93},
  38--44.

\bibitem[{Rose(1961)}]{Rose1961}
Rose, M.~E. (1961), \emph{Relativistic Electron Theory} (John Wiley and Sons,
  New York).

\bibitem[{Rose(1995)}]{Rose1995}
Rose, M.~E. (1995), \emph{Elementary Theory of Angular Momentum} (Dover, New
  York).

\bibitem[{Rossi and Greisen(1941)}]{RossiGreisen1941}
Rossi, B. and K.~Greisen (1941), \enquote{Cosmic-ray theory,} \emph{Rev.\ Mod.\
  Phys.} \textbf{13}, 240--309.

\bibitem[{Royer and Gautier(2006)}]{RoyerGautier2006}
Royer, G. and C.~Gautier (2006), \enquote{{Coefficients and terms of the liquid
  drop model and mass formula},} \emph{Phys.\ Rev.\ A} \textbf{73}, 067302.

\bibitem[{Rudge(1968)}]{Rudge1968}
Rudge, M. R.~H. (1968), \enquote{Theory of the ionization of atoms by electron
  impact,} \emph{Rev.\ Mod.\ Phys.} \textbf{40}, 564--590.

\bibitem[{Rutherford(1911)}]{Rutherford1911}
Rutherford, E. (1911), \enquote{{LXXIX. The scattering of $\alpha$ and $\beta$
  particles by matter and the structure of the atom},} \emph{Phil.\ Mag.\ S. 6}
  \textbf{21:125}, 669--688.

\bibitem[{Sabbatucci and Salvat(2016)}]{SabbatucciSalvat2016}
Sabbatucci, L. and F.~Salvat (2016), \enquote{{Theory and calculation of the
  atomic photoeffect},} \emph{Radiat.\ Phys.\ Chem.} \textbf{121}, 122--140.

\bibitem[{Sakurai(1967)}]{Sakurai1967}
Sakurai, J.~J. (1967), \emph{Advanced Quantum Mechanics} (Addison and Wesley,
  New York).

\bibitem[{Salvat(2003)}]{Salvat2003}
Salvat, F. (2003), \enquote{Optical-model potential for electron and positron
  elastic scattering by atoms,} \emph{Phys.\ Rev.\ A} \textbf{68}, 012708.

\bibitem[{Salvat(2013)}]{Salvat2013}
Salvat, F. (2013), \enquote{{A generic algorithm for Monte Carlo simulation of
  proton transport},} \emph{Nucl.\ Instrum.\ Meth.\ B} \textbf{316}, 144--159.

\bibitem[{Salvat(2022)}]{Salvat2022c}
Salvat, F. (2022), \enquote{{Bethe stopping-power formula and its
  corrections},} \emph{Phys.\ Rev.\ A} \textbf{106}, 032809.

\bibitem[{Salvat(2025)}]{Salvat2025}
Salvat, F. (2025), \emph{{\sc penelope}-2024: A Code System for Monte Carlo
  Simulation of Electron and Photon Transport} (OECD Nuclear Energy Agency,
  document NEA/MBDAV/R(2024)1, OECD Publishing, Paris),
  \url{https://doi.org/10.82155/1vk5-0513}.

\bibitem[{Salvat and Andreo(2023)}]{SalvatAndreo2023}
Salvat, F. and P.~Andreo (2023), \enquote{{SBETHE: Stopping powers of materials
  for swift charged particles from the corrected Bethe formula},}
  \emph{Comput.\ Phys.\ Commun.} \textbf{287}, 108697, {New version
  announcement: (2025), {\it Ibidem}, {\bf 316}, 109796}.

\bibitem[{Salvat \emph{et~al.}(2022{\natexlab{a}})Salvat, Barjuan, and
  Andreo}]{Salvat2022a}
Salvat, F., L.~Barjuan, and P.~Andreo (2022{\natexlab{a}}), \enquote{{Inelastic
  collisions of fast charged particles with atoms. Bethe asymptotic formulas
  and shell corrections},} \emph{Phys.\ Rev.\ A} \textbf{105}, 042813.

\bibitem[{Salvat and Fern\'{a}ndez-Varea(2019)}]{SalvatFernandezVarea2019}
Salvat, F. and J.~M. Fern\'{a}ndez-Varea (2019), \enquote{{RADIAL: a Fortran
  subroutine package for the solution of the radial Schr\"{o}dinger and Dirac
  wave equations},} \emph{Comput.\ Phys.\ Commun.} \textbf{240}, 165--177, see
  also the manual of the computer code.

\bibitem[{Salvat \emph{et~al.}(1995)Salvat, Fern\'{a}ndez-Varea, and
  {Williamson Jr}}]{Salvat1995}
Salvat, F., J.~M. Fern\'{a}ndez-Varea, and W.~{Williamson Jr} (1995),
  \enquote{{Accurate numerical solution of the radial Schr{\"o}dinger and Dirac
  wave equations},} \emph{Comput.\ Phys.\ Commun.} \textbf{90}, 151--168.

\bibitem[{Salvat and Heredia(2023)}]{SalvatHeredia2023}
Salvat, F. and C.~Heredia (2023), \enquote{{Electromagnetic interaction models
  for Monte Carlo simulation of protons and alpha particles},} \emph{Nucl.\
  Instrum.\ Meth.\ B} \textbf{546}, 165157.

\bibitem[{Salvat \emph{et~al.}(2005)Salvat, Jablonski, and Powell}]{Salvat2005}
Salvat, F., A.~Jablonski, and C.~J. Powell (2005), \enquote{{ELSEPA}--{D}irac
  partial-wave calculation of elastic scattering of electrons and positrons by
  atoms, positive ions and molecules,} \emph{Comput.\ Phys.\ Commun.}
  \textbf{165}, 157--190.

\bibitem[{Salvat \emph{et~al.}(2022{\natexlab{b}})Salvat, Llosa, Lallena, and
  Almansa}]{Salvat2022b}
Salvat, F., J.~Llosa, A.~M. Lallena, and J.~Almansa (2022{\natexlab{b}}),
  \enquote{{ECCPA: calculation of classical and quantum cross sections for
  elastic collisions of charged particles with atoms},} \emph{Comput.\ Phys.\
  Commun.} \textbf{277}, 108368.

\bibitem[{Salvat \emph{et~al.}(1987{\natexlab{a}})Salvat, Mart{\'\i}nez, Mayol,
  and Parellada}]{Salvat1987}
Salvat, F., J.~D. Mart{\'\i}nez, R.~Mayol, and J.~Parellada
  (1987{\natexlab{a}}), \enquote{{Analytical Dirac-Hartree-Fock-Slater
  screening function for atoms ($Z$ = 1--92)},} \emph{Phys.\ Rev.\ A}
  \textbf{36}, 467--474.

\bibitem[{Salvat \emph{et~al.}(1987{\natexlab{b}})Salvat, Mayol, and
  Mart\'{\i}nez}]{Salvat1987c}
Salvat, F., R.~Mayol, and J.~D. Mart\'{\i}nez (1987{\natexlab{b}}),
  \enquote{{Elastic scattering of electrons by atoms: a phenomenological
  approach},} \emph{{J.\ Phys.\ B: Atom.\ Mol.\ Opt.\ Phys.}} \textbf{20},
  6597--6612.

\bibitem[{Salvat and Quesada(2020)}]{SalvatQuesada2020}
Salvat, F. and J.~M. Quesada (2020), \enquote{{Nuclear effects in proton
  transport and dose calculations},} \emph{Nucl.\ Instrum.\ Meth.\ B}
  \textbf{475}, 49--62.

\bibitem[{Schiff(1968)}]{Schiff1968}
Schiff, L.~I. (1968), \emph{Quantum Mechanics} (McGraw-Hill, Tokyo).

\bibitem[{Scofield(1973)}]{Scofield1973}
Scofield, J.~H. (1973), \emph{{Theoretical photoionization cross sections from
  1 to 1500 keV}}, Technical Report UCRL-51326, Lawrence Livermore Laboratory,
  Livermore, California.

\bibitem[{Segui \emph{et~al.}(2002)Segui, Dingfelder, Fern{\'{a}}ndez-Varea,
  and Salvat}]{Segui2002}
Segui, S., M.~Dingfelder, J.~M. Fern{\'{a}}ndez-Varea, and F.~Salvat (2002),
  \enquote{{The structure of the Bethe ridge. Relativistic Born and impulse
  approximations},} \emph{J. Phys.\ B: At.\ Mol.\ Opt.\ Phys.} \textbf{35},
  33--53.

\bibitem[{Segui \emph{et~al.}(2003)Segui, Dingfelder, and Salvat}]{Segui2003}
Segui, S., M.~Dingfelder, and F.~Salvat (2003), \enquote{Distorted-wave
  calculation of cross sections for inner-shell ionization by electron and
  positron impact,} \emph{Phys.\ Rev.\ A} \textbf{67}, 062710.

\bibitem[{Seltzer and Berger(1985)}]{SeltzerBerger1985}
Seltzer, S.~M. and M.~J. Berger (1985), \enquote{Bremsstrahlung spectra from
  electron interactions with screened atomic nuclei and orbital electrons,}
  \emph{Nucl.\ Instrum.\ Meth.\ B} \textbf{12}, 95--134.

\bibitem[{Seltzer and Berger(1986)}]{SeltzerBerger1986}
Seltzer, S.~M. and M.~J. Berger (1986), \enquote{{Bremsstrahlung energy spectra
  from electrons with kinetic energy 1 keV--10 GeV incident on screened nuclei
  and orbital electrons of neutral atoms with {$Z=$} 1--100},} \emph{At.\ Data
  Nucl.\ Data Tables} \textbf{35}, 345--418.

\bibitem[{Shiles \emph{et~al.}(1980)Shiles, Sasaki, Inokuti, and
  Smith}]{Shiles1980}
Shiles, E., T.~Sasaki, M.~Inokuti, and D.~Y. Smith (1980),
  \enquote{{Self-consistency and sum-rule tests in the Kramers-Kronig analysis
  of optical data: applications to aluminum},} \emph{Phys.\ Rev.\ B}
  \textbf{22}, 1612--1628.

\bibitem[{Slater(1951)}]{Slater1951}
Slater, J.~C. (1951), \enquote{{A simplification of the Hartree-Fock method},}
  \emph{Phys.\ Rev.} \textbf{81}, 385--390.

\bibitem[{Spencer and Fano(1954)}]{SpencerFano1954}
Spencer, L.~V. and U.~Fano (1954), \enquote{Energy spectrum resulting from
  electron slowing down,} \emph{Phys.\ Rev.} \textbf{93}, 1172--1181.

\bibitem[{Stabler(1964)}]{Stabler1964}
Stabler, R.~C. (1964), \enquote{Classical impulse approximation for inelastic
  electron-atom collisions,} \emph{Phys.\ Rev.} \textbf{133}, A1268--A1273.

\bibitem[{Sternheimer(1952)}]{Sternheimer1952}
Sternheimer, R.~M. (1952), \enquote{The density effect for the ionization loss
  in various materials,} \emph{Phys.\ Rev.} \textbf{88}, 851--859, {Erratum
  (1953), {\it Ibidem}, {\bf 89}, 1309}.

\bibitem[{Sternheimer \emph{et~al.}(1982)Sternheimer, Seltzer, and
  Berger}]{Sternheimer1982}
Sternheimer, R.~M., S.~M. Seltzer, and M.~J. Berger (1982), \enquote{Density
  effect for the ionization loss of charged particles in various substances,}
  \emph{Phys.\ Rev.\ B} \textbf{26}, 6067--6076, {Erratum (1983), {\it Ibidem},
  {\bf 27}, 6971}.

\bibitem[{Sternheimer \emph{et~al.}(1984)Sternheimer, Seltzer, and
  Berger}]{Sternheimer1984}
Sternheimer, R.~M., S.~M. Seltzer, and M.~J. Berger (1984), \enquote{Density
  effect for the ionization loss of charge particles in various substances,}
  \emph{At.\ Data and Nucl.\ Data Tables} \textbf{26}, 261--271.

\bibitem[{Strange(1998)}]{Strange1998}
Strange, P. (1998), \emph{Relativistic Quantum Mechanics} (Cambridge University
  Press, Cambridge, UK).

\bibitem[{Thomson(1912)}]{Thomson1912}
Thomson, J.~J. (1912), \enquote{{XLII}. {I}onization by moving electrified
  particles,} \emph{Phil.\ Mag.\ Series 6} \textbf{23}, 449--457.

\bibitem[{Tiesinga \emph{et~al.}(2018)Tiesinga, Mohr, Newell, and
  Taylor}]{Tiesinga2021}
Tiesinga, E., P.~J. Mohr, D.~B. Newell, and B.~N. Taylor (2018),
  \enquote{{CODATA} recommended values of the fundamental physical constants:
  2018,} \emph{Rev.\ Mod.\ Phys.} \textbf{93}, 025010.

\bibitem[{Tomiki \emph{et~al.}(1993)Tomiki, Ganaha, Shikenbaru, Futemma, Yuri,
  Aiura, Sato, Fukutani, Kato, Miyahara, Yonesu, and Tamashiro}]{Tomiki1993}
Tomiki, T., Y.~Ganaha, T.~Shikenbaru, T.~Futemma, M.~Yuri, Y.~Aiura, S.~Sato,
  H.~Fukutani, H.~Kato, T.~Miyahara, A.~Yonesu, and J.~Tamashiro (1993),
  \enquote{{Anisotropic optical spectra of $\alpha$-Al$_2$O$_3$ single crystals
  in the vacuum ultraviolet region. I. Spectra of absorption tail and
  reflectivity},} \emph{Journal of the Physical Society of Japan} \textbf{62},
  573--584.

\bibitem[{Tropf and Thomas(1998)}]{TropfThomas1998}
Tropf, W.~J. and M.~E. Thomas (1998), \enquote{Aluminium oxide (al$_2$o$_3$)
  revisited,} in E.~D. Palik (editor), \enquote{Handbook of Optical Constants
  of Solids,} volume~3, pages 653--682 (Academic Press, San Diego, CA).

\bibitem[{Tsai(1974)}]{Tsai1974}
Tsai, Y.~S. (1974), \enquote{Pair production and bremsstrahlung of charged
  leptons,} \emph{Rev.\ Mod.\ Phys.} \textbf{46}, 815--851, {Erratum (1977),
  {\it Ibidem}, {\bf 49}, 421--423}.

\bibitem[{Tucker \emph{et~al.}(1969)Tucker, Roberts, Nestor, Carlson, and
  Malik}]{Tucker1969}
Tucker, T.~C., L.~D. Roberts, C.~W. Nestor, T.~A. Carlson, and F.~B. Malik
  (1969), \enquote{{Relativistic self-consistent-field calculation of the wave
  functions, eigenvalues, isotope shifts, and the $6s$ hyperfine-structure
  coupling constant as a function of pressure for metallic gold in the
  Wigner-Seitz model},} \emph{Phys.\ Rev.} \textbf{178}, 998--1008.

\bibitem[{Tung \emph{et~al.}(1988)Tung, Shyu, and Kwei}]{Tung1988}
Tung, C.~J., R.~L. Shyu, and C.~M. Kwei (1988), \enquote{Mean excitation
  energies of atoms using the local plasma approximation,} \emph{J.\ Phys.\ D:
  Appl.\ Phys.} \textbf{21}, 1125--1129.

\bibitem[{Vavilov(1957)}]{Vavilov1957}
Vavilov, P.~V. (1957), \enquote{Ionization losses of high-energy heavy
  particles,} \emph{Sov.\ Phys.\ JETP} \textbf{5}, 749--751.

\bibitem[{Vriens(1966)}]{Vriens1966}
Vriens, L. (1966), \enquote{Binary-encounter electron-atom collision theory,}
  \emph{Phys.\ Rev.} \textbf{141}, 88--92.

\bibitem[{Vriens(1968)}]{Vriens1968}
Vriens, L. (1968), \enquote{{Differential cross sections for ionization of the
  hydrogen atom by fast charged particles in the binary-encounter theory and
  Bethe theory},} \emph{J.\ Phys.\ B (Proc.\ Phys.\ Soc.)} \textbf{1},
  1123--1130.

\bibitem[{Walker(1971)}]{Walker1971}
Walker, D.~W. (1971), \enquote{Relativistic effects in low energy electron
  scattering from atoms,} \emph{Adv.\ Phys.} \textbf{20}, 257--323.

\bibitem[{Wallace(1971)}]{Wallace1971}
Wallace, S.~J. (1971), \enquote{Eikonal expansion,} \emph{Phys.\ Rev.\ Lett.}
  \textbf{27}, 622--625.

\bibitem[{Waller and Hartree(1929)}]{WallerHartree1929}
Waller, I. and D.~R. Hartree (1929), \enquote{{On the intensity of total
  scattering of X-rays},} \emph{Proc.\ R.\ Soc.\ London A} \textbf{124},
  119--142.

\bibitem[{Walske(1952)}]{Walske1952}
Walske, M.~C. (1952), \enquote{{The stopping power of K-electrons},}
  \emph{Phys.\ Rev.} \textbf{88}, 1283--1289.

\bibitem[{Walske(1956)}]{Walske1956}
Walske, M.~C. (1956), \enquote{{Stopping power of L-electrons},} \emph{Phys.\
  Rev.} \textbf{101}, 940--944.

\bibitem[{Wentzel(1927)}]{Wentzel1927}
Wentzel, G. (1927), \enquote{{Zwei Bemerkungen {\"{u}}ber die Zerstreuung
  korpuskularer Strahlen als Beugungserscheinung},} \emph{Z. Phys.}
  \textbf{42}, 590--593.

\bibitem[{Werner \emph{et~al.}(2009)Werner, Glantschnig, and
  Ambrosch-Draxl}]{Werner2009}
Werner, W. S.~M., K.~Glantschnig, and C.~Ambrosch-Draxl (2009),
  \enquote{Optical constants and inelastic electron-scattering data for 17
  elemental metals,} \emph{J. Phys.\ Chem.\ Ref.\ Data} \textbf{38},
  1013--1092.

\bibitem[{Wheeler and Lamb(1939)}]{WheelerLamb1939}
Wheeler, J.~A. and W.~E. Lamb (1939), \enquote{Influence of atomic electrons on
  radiation and pair production,} \emph{Phys.\ Rev.} \textbf{55}, 858--862,
  {Erratum (1956), {\it Ibidem}, {\bf 101}, 1836}.

\bibitem[{Yennie \emph{et~al.}(1954)Yennie, Ravenhall, and Wilson}]{Yennie1954}
Yennie, D.~R., D.~G. Ravenhall, and R.~N. Wilson (1954), \enquote{Phase-shift
  calculation of high-energy electron scattering,} \emph{Phys.\ Rev.}
  \textbf{95}, 500--512.

\bibitem[{Zeitler and Olsen(1967)}]{ZeitlerOlsen1967}
Zeitler, E. and H.~Olsen (1967), \enquote{Complex scattering amplitudes in
  elastic electron scattering,} \emph{Phys.\ Rev.} \textbf{162}, 1439--1447.

\bibitem[{Zheng-Ming and Brahme(1993)}]{Zheng-MingBrahme1993}
Zheng-Ming, L. and A.~Brahme (1993), \enquote{{An overview of the transport
  theory of charged particles},} \emph{Radiat.\ Phys.\ Chem.} \textbf{41},
  673--703.

\bibitem[{Ziegler(1999)}]{Ziegler1999}
Ziegler, J.~F. (1999), \enquote{The stopping of energetic light ions in
  elemental matter,} \emph{J.\ Appl.\ Phys. / Rev.\ Appl.\ Phys.} \textbf{85},
  1249--1272.

\bibitem[{Ziegler \emph{et~al.}(2012)Ziegler, Biersack, and
  Ziegler}]{Ziegler2012}
Ziegler, J.~F., J.~P. Biersack, and M.~D. Ziegler (2012), \emph{SRIM The
  Stopping and Range of Ions in Matter} (\url{www.srim.org},
  \url{www.lulu.com}).

\end{thebibliography}
 \clearpage


 \addcontentsline{toc}{chapter}{Index}
 \renewcommand{\rightmark}{\sl Index}
 \renewcommand{\leftmark}{\sl Index}
 \begin{theindex}

  \item addition of velocities, \hyperpage{464}
  \item additivity approximation, \hyperpage{254}, \hyperpage{301}
  \item associated Legendre functions, \hyperpage{484}
  \item atomic configuration, \hyperpage{80}, \hyperpage{82}, 
		\hyperpage{190}, \hyperpage{192}, \hyperpage{200}, 
		\hyperpage{233}, \hyperpage{293}
    \subitem ground state, \hyperpage{95}
  \item atomic number, \hyperpage{72}
  \item atomic potential, \hyperpage{101--105}
    \subitem analytical, \hyperpage{103}
  \item atomic screening function, \hyperpage{101}
    \subitem DHFS, \hyperpage{104}
    \subitem Thomas--Fermi, \hyperpage{104}
    \subitem Wentzel, \hyperpage{105}
  \item atomic units, \hyperpage{491}
  \item Avogadro's constant, \hyperpage{489}

  \indexspace

  \item Barkas $Z_1^3$ correction, \hyperpage{347--355}
  \item barycentric frame of reference, \hyperpage{465}
  \item Bethe formula, \hyperpage{357}
  \item Bethe integral, \hyperpage{481}
  \item Bethe ridge, \hyperpage{226}
  \item Bethe stopping power formula, \hyperpage{249}, \hyperpage{252}
    \subitem derivation, \hyperpage{249--256}
    \subitem electrons and positrons, \hyperpage{253}
    \subitem low-energy extrapolation, \hyperpage{256}
    \subitem restricted stopping power, \hyperpage{253}
  \item Bethe sum rule, \hyperpage{41}, \hyperpage{204}
    \subitem for atoms, \hyperpage{210}
  \item Bethe--Bloch formula, \hyperpage{345}
  \item Bhabha cross section, \hyperpage{240}
  \item binary encounter approximation, \hyperpage{218--227}
    \subitem electron gas, \hyperpage{222}
    \subitem hydrogenic ions, \hyperpage{222}
  \item Bloch $Z_1^4$ correction, \hyperpage{342}
  \item Blunk--Leisegang distribution, \hyperpage{397}
  \item Bohr radius, \hyperpage{43}, \hyperpage{492}
  \item Bohr's relativistic stopping power formula, \hyperpage{330}
  \item Bohr's stopping power formula, \hyperpage{328}
  \item Boltzmann transport equation, \hyperpage{386--389}
  \item Born phase shifts, \hyperpage{159}, \hyperpage{177}
  \item Born scattering amplitude, \hyperpage{158}

  \indexspace

  \item causality principle, \hyperpage{36}
  \item center-of-mass frame of reference, \hyperpage{124, 125}, 
		\hyperpage{465}
  \item central-field orbitals, \hyperpage{49}, \hyperpage{58}, 
		\hyperpage{67}
  \item charge density, \hyperpage{8}
  \item Cherenkov radiation, \hyperpage{337}
  \item classical collisions, \hyperpage{123}, \hyperpage{131}
    \subitem deflection angle, \hyperpage{113}
    \subitem relative motion, \hyperpage{124}, \hyperpage{133}
  \item classical electron radius, \hyperpage{489}
  \item classical scattering
    \subitem Coulomb potential, \hyperpage{115}
    \subitem differential cross section, \hyperpage{113}
    \subitem Rutherford DCS, \hyperpage{116}
    \subitem screened Coulomb potential, \hyperpage{116}
    \subitem validity, \hyperpage{119}
  \item classical stopping power, \hyperpage{319}
  \item classical stopping theory, \hyperpage{32}
  \item classical trajectories, \hyperpage{110}
  \item Clausius--Mossotti equation, \hyperpage{28}
  \item collisions in the center-of-mass frame, \hyperpage{135}
  \item collisions in the laboratory frame, \hyperpage{140}, 
		\hyperpage{144}
  \item collisions of electrons, \hyperpage{237}, \hyperpage{312}
  \item collisions of identical particles, \hyperpage{184}
    \subitem Coulomb collisions, \hyperpage{186}
  \item collisions of positrons, \hyperpage{240}, \hyperpage{312}
  \item complex refractive index, \hyperpage{25}
  \item Compton profile, \hyperpage{220}
  \item computer code
    \subitem {\sc elastic}, \hyperpage{431--438}, \hyperpage{444}
    \subitem {\sc gosan}, \hyperpage{409}
    \subitem {\sc sbethe}, \hyperpage{431}, \hyperpage{438--444}
  \item continuity equation, \hyperpage{1}, \hyperpage{11}
  \item continuous slowing down approximation, \hyperpage{392}
  \item corrected Bethe formula, \hyperpage{357}
    \subitem low-energy extrapolation, \hyperpage{361}
  \item Coulomb distorted plane waves, \hyperpage{53}
  \item Coulomb gauge, \hyperpage{2}, \hyperpage{12}
  \item Coulomb scattering amplitude, \hyperpage{54}
  \item cross sections in the center-of-mass frame, \hyperpage{124}, 
		\hyperpage{135}
  \item cross sections in the laboratory frame, \hyperpage{127}
  \item CSDA range, \hyperpage{392}
  \item current density, \hyperpage{8}
    \subitem longitudinal, \hyperpage{8}
    \subitem transverse, \hyperpage{10}

  \indexspace

  \item delta rays, \hyperpage{385}
  \item density-effect correction, \hyperpage{339}
  \item DHFS self-consistent model, \hyperpage{95}, \hyperpage{98}
    \subitem energy eigenvalues, \hyperpage{101}
    \subitem Latter-tail correction, \hyperpage{97}
    \subitem Wigner--Seitz boundary conditions, \hyperpage{97}
  \item DHFS-model OOS, \hyperpage{300}, \hyperpage{302, 303}
  \item dielectric functions
    \subitem and generalized oscillator strengths, \hyperpage{243}
    \subitem classical Lorentz--Drude model, \hyperpage{27--32}
    \subitem definitions, \hyperpage{14--20}
    \subitem of semiconductors and insulators, \hyperpage{282}
    \subitem of the electron gas, \hyperpage{261}
  \item dielectric polarization, \hyperpage{331}
  \item differential inverse mean free path, \hyperpage{384}
  \item Dirac delta function, \hyperpage{477--481}
    \subitem  three-dimensional, \hyperpage{480}
  \item Dirac matrices, \hyperpage{56}
  \item Dirac wave equation, \hyperpage{55}
    \subitem central potentials, \hyperpage{58}
    \subitem distorted plane waves, \hyperpage{63}
    \subitem probability current density, \hyperpage{57}
    \subitem probability density, \hyperpage{57}
    \subitem stationary states, \hyperpage{57}
  \item direct scattering amplitude, \hyperpage{65}
  \item dispersion relation of electromagnetic radiation, 
		\hyperpage{25}
  \item dissipated electromagnetic energy, \hyperpage{21}
  \item distorted plane waves
    \subitem Dirac wave equation, \hyperpage{63}
      \subsubitem partial-wave series, \hyperpage{64}
    \subitem Schr\"{o}dinger wave equation, \hyperpage{51}
      \subsubitem partial-wave series, \hyperpage{52}, \hyperpage{55}

  \indexspace

  \item eikonal approximation, \hyperpage{164--171}
    \subitem optical theorem, \hyperpage{169}
    \subitem phase function, \hyperpage{167}
    \subitem phase shifts, \hyperpage{168}
    \subitem scattering amplitude, \hyperpage{166}, \hyperpage{168}
  \item electric conductivity, \hyperpage{19}, \hyperpage{30}, 
		\hyperpage{39}
    \subitem longitudinal, \hyperpage{19}
    \subitem transverse, \hyperpage{19}
  \item electromagnetic field
    \subitem in material media, \hyperpage{17}
    \subitem of a moving charge, \hyperpage{2--5}, \hyperpage{323}
  \item electromagnetic potentials, \hyperpage{2}, \hyperpage{13}
    \subitem in material media, \hyperpage{17}
  \item electromagnetic radiation, \hyperpage{24}
  \item electron capture by positive ions, \hyperpage{373}
  \item energy of electromagnetic fields, \hyperpage{21}
  \item energy straggling, \hyperpage{389--402}
    \subitem Blunk--Leisegang distribution, \hyperpage{397}
    \subitem continuous slowing down approximation, \hyperpage{392}
    \subitem Landau distribution, \hyperpage{393--397}
    \subitem numerical distributions, \hyperpage{399--402}
  \item energy straggling cross section, \hyperpage{214}, 
		\hyperpage{237}
  \item energy transfer in Coulomb collisions, \hyperpage{320}
  \item energy transfer to classical oscillators, \hyperpage{322}
  \item energy-loss DCS, \hyperpage{130}, \hyperpage{146}
    \subitem Barkas correction to the, \hyperpage{354}
  \item energy-momentum four-vector, \hyperpage{463}
  \item energy-range curve, \hyperpage{392}
  \item Euler constant, \hyperpage{327}, \hyperpage{342}, 
		\hyperpage{355}, \hyperpage{487}
  \item external electromagnetic fields, \hyperpage{13}
  \item extinction coefficient, \hyperpage{25}

  \indexspace

  \item Fermi density effect, \hyperpage{331}
  \item Fermi's distant stopping power, \hyperpage{336}
  \item Fermi--Eyges theory, \hyperpage{418}
  \item fine-structure constant, \hyperpage{493}
  \item finite mass of nucleus, \hyperpage{73}
  \item finite size of nucleus, \hyperpage{73}
  \item fluence, \hyperpage{380}
    \subitem distribution with respect to energy, \hyperpage{382}
  \item flux density, \hyperpage{378, 379}
  \item folding theorem, \hyperpage{405}
  \item Fortran programs, \hyperpage{431--448}
  \item four vectors, \hyperpage{461}
    \subitem inner product of, \hyperpage{461}
  \item four-momentum, \hyperpage{463}
  \item Fourier transform, \hyperpage{5--10}, \hyperpage{13, 14}, 
		\hyperpage{17--20}, \hyperpage{32}, \hyperpage{222}, 
		\hyperpage{229}, \hyperpage{241}, \hyperpage{263}, 
		\hyperpage{265}, \hyperpage{267, 268}, \hyperpage{325}, 
		\hyperpage{331}, \hyperpage{335}, \hyperpage{350}, 
		\hyperpage{420}
    \subitem convolution theorem, \hyperpage{398}
    \subitem of the screened Coulomb potential, \hyperpage{6}
  \item Free-atom boundary conditions, \hyperpage{97}
  \item fundamental constants, \hyperpage{489}

  \indexspace

  \item gapped dielectric function, \hyperpage{282}
  \item Gauss--Legendre quadrature, \hyperpage{458}
    \subitem adaptive, \hyperpage{118}, \hyperpage{137}, 
		\hyperpage{286}, \hyperpage{352}, \hyperpage{432}, 
		\hyperpage{459}
  \item Gaussian system of units, \hyperpage{1}, \hyperpage{489}
  \item generalized oscillator strength, \hyperpage{41}, 
		\hyperpage{201}
    \subitem and dielectric functions, \hyperpage{243}
    \subitem calculation for atoms, \hyperpage{207}
  \item gnuplot, \hyperpage{432}, \hyperpage{438}, \hyperpage{448}
  \item Goudsmit--Saunderson distribution, \hyperpage{406}, 
		\hyperpage{438}

  \indexspace

  \item Hamiltonian of atomic systems, \hyperpage{72}
  \item Hamiltonian of the relative motion, \hyperpage{76}
  \item Hartree energy, \hyperpage{43}, \hyperpage{492, 493}
  \item Heaviside unit step function, \hyperpage{479}
  \item Helmholtz vector equation, \hyperpage{10}
  \item hydrogenic ions, \hyperpage{77}
    \subitem energy levels, \hyperpage{77}
    \subitem generalized oscillator strength, \hyperpage{209}
      \subsubitem binary encounter approximation, \hyperpage{222}
    \subitem radial wave functions, \hyperpage{77}

  \indexspace

  \item impulse approximation, \hyperpage{218--227}
  \item incoherent scattering function, \hyperpage{217}
  \item independent-electron approximation, \hyperpage{78}
  \item independent-electron models
    \subitem non-relativistic, \hyperpage{79}
    \subitem relativistic, \hyperpage{81}
  \item index of refraction, \hyperpage{25}
  \item induced electromagnetic fields, \hyperpage{13}
  \item inelastic collisions
    \subitem  kinematics of , \hyperpage{469--475}
    \subitem angular DCS, \hyperpage{216}
    \subitem energy-loss DCS, \hyperpage{214}
    \subitem integrated cross sections, \hyperpage{212}
    \subitem with large recoil energies, \hyperpage{246}
  \item inelastic collisions in dense media, \hyperpage{309}
    \subitem integrated cross sections, \hyperpage{310}
    \subitem polarization correction, \hyperpage{311}
  \item interpolation, \hyperpage{448}
    \subitem cubic splines, \hyperpage{453--458}
  \item inverse dielectric function, \hyperpage{20}
    \subitem analyticity, \hyperpage{38}
  \item irrotational vector field, \hyperpage{7}

  \indexspace

  \item Kramers--Kronig relations, \hyperpage{36--42}
    \subitem sum rules derived from, \hyperpage{40}
  \item Kronecker delta, \hyperpage{9}, \hyperpage{47}
  \item Kummer function, \hyperpage{53}

  \indexspace

  \item laboratory frame of reference, \hyperpage{124}, \hyperpage{465}
  \item Landau distribution, \hyperpage{393}
  \item Latter-tail correction, \hyperpage{97}
  \item Legendre functions of the second kind, \hyperpage{159}, 
		\hyperpage{412}
  \item Legendre polynomials, \hyperpage{483}
    \subitem recurrence relation, \hyperpage{410}
  \item Lewis theory, \hyperpage{423}
  \item Li\'{e}nard--Wiechert field, \hyperpage{3}
  \item Lindhard dielectric function, \hyperpage{262--281}
  \item Lindhard--S\o rensen relativistic correction, \hyperpage{346}
  \item local-plasma approximation, \hyperpage{293}
  \item localized charge distribution, \hyperpage{9}
    \subitem scalar potential of a, \hyperpage{10}
      \subsubitem multipole expansion, \hyperpage{10}
  \item longitudinal dielectric function, \hyperpage{15}
  \item longitudinal vector field, \hyperpage{7}
    \subitem integral formula, \hyperpage{7}
  \item Lorentz force, \hyperpage{2}, \hyperpage{12}
  \item Lorentz transformation, \hyperpage{461}
  \item Lorenz--Lorentz correction, \hyperpage{30}
  \item low-energy extrapolation, \hyperpage{361}

  \indexspace

  \item M\o ller cross section, \hyperpage{231}, \hyperpage{239}
    \subitem non-relativistic, \hyperpage{187}
  \item macroscopic cross sections, \hyperpage{214}
  \item magnetic permeability, \hyperpage{16}
  \item mass number of the nucleus, \hyperpage{72}
  \item mass stopping power, \hyperpage{215}
  \item Maxwell equations
    \subitem in material media, \hyperpage{11--20}
    \subitem in vacuum, \hyperpage{1--2}
  \item mean excitation energy, \hyperpage{250}, \hyperpage{254}, 
		\hyperpage{327}
  \item mean excitation potential, \hyperpage{250}, \hyperpage{254}, 
		\hyperpage{327}
  \item Mermin dielectric function, \hyperpage{281}
  \item modified Bessel functions, \hyperpage{486}
    \subitem  integrals involving, \hyperpage{487}
  \item modified Coulomb potentials, \hyperpage{50}
  \item Moli\`{e}re distribution, \hyperpage{412}
  \item monochromatic plane waves, \hyperpage{25}
  \item Mott DCS, \hyperpage{176}
    \subitem McKinley-Fesbach formula, \hyperpage{176}
  \item multiple scattering, \hyperpage{402--417}
    \subitem Goudsmit--Saunderson distribution, \hyperpage{406}
      \subsubitem Wentzel model, \hyperpage{411}
    \subitem Moli\`{e}re distribution, \hyperpage{412--417}
  \item multiple scattering with energy loss, \hyperpage{417--429}
    \subitem Fermi--Eyges distribution, \hyperpage{418--423}
    \subitem Lewis theory, \hyperpage{423--429}

  \indexspace

  \item nuclear mass formula, \hyperpage{72}
  \item nuclear stopping, \hyperpage{371}
  \item numerical integration, \hyperpage{448}
    \subitem cubic splines, \hyperpage{457}
  \item numerical methods, \hyperpage{448--460}
    \subitem cubic spline interpolation, \hyperpage{453--458}
    \subitem Gauss--Legendre quadrature, \hyperpage{458}
      \subsubitem adaptive bisection, \hyperpage{459}
    \subitem integration, \hyperpage{449}
    \subitem interpolation, \hyperpage{448}

  \indexspace

  \item Ochkur approximation, \hyperpage{228--231}
  \item one-active-electron approximation, \hyperpage{193}
  \item optical dielectric function, \hyperpage{20}
  \item optical oscillator strength, \hyperpage{41}, \hyperpage{203}
  \item optical oscillator strength of materials, \hyperpage{292}
    \subitem DHFS shell contributions, \hyperpage{297}
    \subitem empirical, \hyperpage{300}
    \subitem local-plasma approximation, \hyperpage{293}
  \item optical oscillator strength of Thomas--Fermi atoms, 
		\hyperpage{294}
  \item optical theorem, \hyperpage{155}, \hyperpage{169}
  \item optical-data models, \hyperpage{305}
    \subitem energy-straggling parameter, \hyperpage{313}
    \subitem inelastic mean free path, \hyperpage{313}
    \subitem Lindhard--Liljequist extension scheme, \hyperpage{309}
    \subitem Penn model, \hyperpage{305}
    \subitem stopping power, \hyperpage{313}

  \indexspace

  \item Pauli spin matrices, \hyperpage{56}
  \item Pauli wave equation, \hyperpage{65}
    \subitem central potentials, \hyperpage{67}
  \item phase-shifts, \hyperpage{51}
  \item photoabsorption cross section, \hyperpage{26}
  \item photon line, \hyperpage{474}
  \item plane waves, \hyperpage{6}
    \subitem continuum normalization, \hyperpage{48}
    \subitem density of states, \hyperpage{47}, \hyperpage{49}
    \subitem periodic boundary conditions, \hyperpage{46}
  \item plasma resonance frequency, \hyperpage{29}, \hyperpage{242}, 
		\hyperpage{272}, \hyperpage{292}, \hyperpage{320}
  \item plasmon excitation, \hyperpage{277}
  \item plasmon-pole approximation, \hyperpage{280}
  \item Poisson equation, \hyperpage{2}, \hyperpage{13}
  \item polar angles, \hyperpage{376}
  \item polar coordinates in a plane, \hyperpage{108}
  \item Poynting vector, \hyperpage{21}

  \indexspace

  \item quantum inelastic collisions, \hyperpage{191}
    \subitem collisions of electrons, \hyperpage{197}
    \subitem collisions with electrons at rest, \hyperpage{202}
    \subitem distorted-wave Born approximation, \hyperpage{194}
    \subitem one-electron approximation, \hyperpage{192}
    \subitem plane-wave Born approximation, \hyperpage{198}
  \item quantum scattering
    \subitem Born approximation, \hyperpage{177}
    \subitem Coulomb scattering, \hyperpage{155}
    \subitem DCS, \hyperpage{153}, \hyperpage{172}, \hyperpage{181}
    \subitem electrons and positrons, \hyperpage{175}
    \subitem partial-wave expansion method, \hyperpage{153}
    \subitem plane-wave Born approximation, \hyperpage{156}
    \subitem WKB phase shifts, \hyperpage{178}
  \item quantum scattering states, \hyperpage{150}

  \indexspace

  \item radial Dirac wave equation, \hyperpage{60}
    \subitem bound states, \hyperpage{61}
    \subitem free states, \hyperpage{62}
      \subsubitem phase shifts, \hyperpage{62}
    \subitem non-relativistic limit, \hyperpage{63}
  \item radial Schr\"{o}dinger wave equation, \hyperpage{49}, 
		\hyperpage{67}, \hyperpage{69}
    \subitem bound states, \hyperpage{50}
    \subitem free states, \hyperpage{50}
      \subsubitem phase shifts, \hyperpage{51}
  \item radiation fields, \hyperpage{376--384}
    \subitem angular current density, \hyperpage{378}
    \subitem angular flux density, \hyperpage{379}
    \subitem average distribution of fluence with respect to energy, 
		\hyperpage{383}
    \subitem average energy fluence in a volume, \hyperpage{381}
    \subitem average fluence in a volume, \hyperpage{380}
    \subitem distribution of fluence with respect to energy, 
		\hyperpage{382}
    \subitem electromagnetic, \hyperpage{10}
    \subitem energy fluence, \hyperpage{381}
    \subitem fluence, \hyperpage{380}
    \subitem flux of particles across a surface, \hyperpage{383}
    \subitem particle density, \hyperpage{376}
    \subitem trajectory picture, \hyperpage{377}
  \item radiative stopping power for electrons and positrons, 
		\hyperpage{366}
  \item range of charged particles, \hyperpage{392}
  \item reduced mass, \hyperpage{76}, \hyperpage{124}
    \subitem relativistic, \hyperpage{134}
  \item reduced series method, \hyperpage{180}
  \item relativistic Born approximation, \hyperpage{231}
  \item relativistic kinetic energy, \hyperpage{463}
  \item relativistic total energy, \hyperpage{463}
  \item relativistic transformation of angles , \hyperpage{467--469}
  \item relativistic velocity, \hyperpage{464}
  \item rest energy of a particle, \hyperpage{463}
  \item restricted stopping power, \hyperpage{252}, \hyperpage{394}
  \item Rutherford cross section, \hyperpage{321}
  \item Rutherford DCS, \hyperpage{156}, \hyperpage{158}

  \indexspace

  \item scattering amplitude, \hyperpage{52}
    \subitem partial-wave series, \hyperpage{53}
  \item scattering plane, \hyperpage{108}
  \item Schr\"{o}dinger wave equation, \hyperpage{44}
    \subitem central potentials, \hyperpage{49}
    \subitem distorted plane waves, \hyperpage{51}
    \subitem plane waves, \hyperpage{46}
    \subitem probability current density, \hyperpage{45}
    \subitem probability density, \hyperpage{44}
    \subitem stationary states, \hyperpage{45}
  \item semi-classical dielectric theory, \hyperpage{241--246}
  \item shell correction, \hyperpage{256--260}
  \item solenoidal vector field, \hyperpage{7}
  \item Sommerfeld parameter, \hyperpage{51}, \hyperpage{62}, 
		\hyperpage{121}, \hyperpage{154, 155}, \hyperpage{176}, 
		\hyperpage{186}, \hyperpage{321}, \hyperpage{342}
  \item speed of light
    \subitem in a dielectric material, \hyperpage{26}
    \subitem in vacuum, \hyperpage{489}
  \item spherical harmonics, \hyperpage{482--486}
    \subitem  Uns\"{o}ld's theorem, \hyperpage{486}
    \subitem addition theorem, \hyperpage{486}
  \item spherical polar coordinates, \hyperpage{34}, \hyperpage{480}
  \item spin-flip scattering amplitude, \hyperpage{65}
  \item step function, \hyperpage{479}
  \item stopping cross section, \hyperpage{214}, \hyperpage{237}
  \item stopping force, \hyperpage{33}
  \item stopping logarithm, \hyperpage{342}, \hyperpage{357}
  \item stopping power
    \subitem Barkas $Z_1^3$ correction, \hyperpage{347--355}
    \subitem Bloch $Z_1^4$ correction, \hyperpage{342}
    \subitem Bohr formula, \hyperpage{319--330}
      \subsubitem relativistic close collisions, \hyperpage{328}
    \subitem classical theory, \hyperpage{33}
    \subitem corrected Bethe formula, \hyperpage{357--361}
    \subitem density effect correction, \hyperpage{339}
    \subitem Lindhard--S\o rensen relativistic correction, 
		\hyperpage{346}
  \item stopping power of the electron gas, \hyperpage{285}
  \item straggling cross section, \hyperpage{214}, \hyperpage{237}
  \item subshell ionization energies, \hyperpage{101}

  \indexspace

  \item Thomas--Fermi equation, \hyperpage{89}
  \item Thomas--Fermi model of the atom, \hyperpage{86}
    \subitem binding energy, \hyperpage{93}
  \item Thomas--Fermi radius, \hyperpage{89}
  \item Thomas--Fermi screening function
    \subitem Lenz--Jensen, \hyperpage{92}
    \subitem Moli\`{e}re, \hyperpage{92}
  \item Thomson cross section, \hyperpage{130}, \hyperpage{214}, 
		\hyperpage{226}
  \item threshold energy of a reaction, \hyperpage{467}
  \item time-independent Schr\"{o}dinger equation, \hyperpage{45}
  \item total inelastic cross section, \hyperpage{237}
  \item trajectory equation, \hyperpage{112}
  \item transport equation, \hyperpage{386--389}
  \item transverse dielectric function, \hyperpage{15}
  \item transverse vector field, \hyperpage{7}
    \subitem integral formula, \hyperpage{7}

  \indexspace

  \item unit window function, \hyperpage{480}

  \indexspace

  \item vacuum photon line, \hyperpage{474}
  \item virial theorem, \hyperpage{78}, \hyperpage{95}, \hyperpage{229}

  \indexspace

  \item wave equation
    \subitem Dirac, \hyperpage{55}
    \subitem of a radiation field, \hyperpage{24}
    \subitem of the electromagnetic field, \hyperpage{8}
    \subitem Pauli, \hyperpage{65}
    \subitem Schr\"{o}dinger, \hyperpage{44}
  \item Wentzel DCS, \hyperpage{158}, \hyperpage{411}
  \item Wentzel potential, \hyperpage{105}, \hyperpage{121}, 
		\hyperpage{157}
  \item Wigner--Seitz boundary conditions, \hyperpage{97}
  \item window function, \hyperpage{480}
  \item WKB phase shifts, \hyperpage{160}, \hyperpage{178}

\end{theindex}


 \end{document}